\documentclass[
		twoside,openright,titlepage,numbers=noenddot,headinclude,
	 	footinclude=true,cleardoublepage=empty,
		dottedtoc, 
		BCOR=5mm,paper=a4,fontsize=11pt, 
		french,american 
		]{scrreprt} 
                
\PassOptionsToPackage{utf8}{inputenc} 
\usepackage{inputenc}

\PassOptionsToPackage{eulerchapternumbers,listings, pdfspacing, subfig,beramono,eulermath,parts}{classicthesis}

\newcommand{\myTitle}{Caractérisation et modélisation de la co-évolution des réseaux de transport et des territoires\xspace}
\newcommand{\myTitleEn}{Characterising and modeling the co-evolution of transportation networks and territories\xspace}
\newcommand{\mySubtitle}{Mémoire de Thèse de Doctorat\xspace}
\newcommand{\mySubtitleEn}{PhD Thesis\xspace}

\newcommand{\myName}{\textsc{Juste Raimbault}\xspace}
\newcommand{\myProf}{\noun{Arnaud Banos}\xspace}
\newcommand{\myOtherProf}{\noun{Florent Le Néchet}\xspace}

\newcommand{\myFaculty}{Universit{\'e} Paris Diderot - Paris 7\xspace}

\newcommand{\myUni}{\xspace}

\newcommand{\myTime}{16 Février 2018\xspace}
\newcommand{\myVersion}{version 3.5.1\xspace}

\newcounter{dummy} 
\providecommand{\mLyX}{L\kern-.1667em\lower.25em\hbox{Y}\kern-.125emX\@}

\usepackage{lipsum} 

\PassOptionsToPackage{french,american}{babel}  
\usepackage{babel}

\usepackage{csquotes}
\PassOptionsToPackage{%
backend=bibtex8,bibencoding=ascii,%
language=auto,%
style=authoryear-comp, 
bibstyle=authoryear,dashed=true, 
sorting=nyt, 
maxbibnames=2, 
natbib=true, 
url=false,doi=false,isbn=false,eprint=false
}{biblatex}
\usepackage{biblatex}
\renewbibmacro{in:}{}
\DeclareFieldFormat[article,inbook,incollection,inproceedings,patent,thesis,unpublished]{title}{#1}
 
\PassOptionsToPackage{fleqn}{amsmath} 
 \usepackage{amsmath}
 
\PassOptionsToPackage{T1}{fontenc} 
\usepackage{fontenc}

\usepackage{textcomp} 

\usepackage{scrhack} 

\usepackage{xspace} 

\usepackage{mparhack} 

\usepackage{fixltx2e} 

\PassOptionsToPackage{smaller}{acronym} 
\usepackage{acronym} 

\PassOptionsToPackage{pdftex}{graphicx}
\usepackage{graphicx} 

\usepackage{tabularx} 
\usepackage{caption}
\usepackage{subfig}  

\usepackage{listings} 
\PassOptionsToPackage{pdftex,
  hyperfootnotes=false,
  pdfpagelabels
}{hyperref}
\usepackage{hyperref}  
\hypersetup{
colorlinks=true, linktocpage=true, pdfstartpage=3, pdfstartview=FitV,
breaklinks=true, pdfpagemode=UseNone, pageanchor=true, pdfpagemode=UseOutlines,%
plainpages=false, bookmarksnumbered, bookmarksopen=true, bookmarksopenlevel=1,%
hypertexnames=true, pdfhighlight=/O,
urlcolor=webbrown, linkcolor=RoyalBlue, citecolor=webgreen, 
pdftitle={\myTitle},
pdfauthor={\textcopyright\ \myName, \myUni, \myFaculty},
pdfsubject={},
pdfkeywords={},
pdfcreator={},
pdfproducer={}
}

\makeatletter
\@ifpackageloaded{babel}
{
\addto\extrasamerican{

}
\addto\extrasngerman{

}
}{\relax}
\makeatother

\usepackage{classicthesis} 

\def \thelanguage {0}

\def \draft {1}

\newcommand{\noun}[1]{\textsc{#1}}

\newcommand{\headercit}[3]{
\begin{multicols}{2}
\phantom{}
\columnbreak
\textit{#1}

 - \noun{#2}~#3
\end{multicols}
}

\DeclareMathOperator{\Cov}{Cov}
\DeclareMathOperator{\Var}{Var}
\DeclareMathOperator{\E}{\mathbb{E}}
\DeclareMathOperator{\Proba}{\mathbb{P}}

\newcommand{\Covb}[2]{\ensuremath{\Cov\!\left[#1,#2\right]}}
\newcommand{\Eb}[1]{\ensuremath{\E\!\left[#1\right]}}
\newcommand{\Pb}[1]{\ensuremath{\Proba\!\left[#1\right]}}
\newcommand{\Varb}[1]{\ensuremath{\Var\!\left[#1\right]}}
\newcommand{\rhob}[2]{\ensuremath{\rho\!\left[#1,#2\right]}}

\newcommand{\norm}[1]{\| #1 \|}

\newcommand{\indep}{\rotatebox[origin=c]{90}{$\models$}}

\DeclareMathOperator*{\argmin}{\arg\!\min}

\newtheorem{definition}{Définition}
\newtheorem{proposition}{Proposition}
\newtheorem{assumption}{Hypothèse}
\newtheorem{lemma}{Lemme}

\newenvironment{proof}[1][Preuve]{\begin{trivlist}
\item[\hskip \labelsep {\bfseries #1}]}{\end{trivlist}}

\newcommand{\qed}{\nobreak \ifvmode \relax \else
      \ifdim\lastskip<1.5em \hskip-\lastskip
      \hskip1.5em plus0em minus0.5em \fi \nobreak
      \vrule height0.75em width0.5em depth0.25em\fi}

\usepackage{mdframed}

\usepackage[angle=0,scale=1,opacity=1,color=black]{background}

\usepackage{tikz}
\usetikzlibrary{calc}

\newlength{\arrowht}
\newcommand\tikzmark[1]{\tikz[remember picture, baseline=(#1.base)] \node[anchor=base,inner sep=0pt, outer sep=0pt] (#1) {#1};}
\tikzset{
    ncbar angle/.initial=90,
    ncbar/.style={
        to path=(\tikztostart)
        -- ($(\tikztostart)!#1!\pgfkeysvalueof{/tikz/ncbar angle}:(\tikztotarget)$)
        -- ($(\tikztotarget)!($(\tikztostart)!#1!\pgfkeysvalueof{/tikz/ncbar angle}:(\tikztotarget)$)!\pgfkeysvalueof{/tikz/ncbar angle}:(\tikztostart)$)
        -- (\tikztotarget)
    },
    ncbar/.default=0.5cm,
}
\newcommand{\arrow}[2]{
\shorthandoff{:!}
\begin{tikzpicture}[remember picture,overlay]
\draw[->,shorten >=3pt,shorten <=3pt] (#1.base) to [ncbar=\arrowht] (#2.base);
\end{tikzpicture}
\shorthandon{:!}
}

\usepackage{amssymb}
\usepackage{bbm}

\usepackage{CJKutf8}
\newcommand{\cn}[1]{
  \begin{CJK*}{UTF8}{gbsn}
  #1
  \end{CJK*}
}

\usepackage[flushleft]{threeparttable}
\usepackage{multirow}
\usepackage{multicol}
\usepackage{makecell}

\usepackage{xparse}
\usepackage{ifthen}

\newcommand{\bpar}[2]{
    \ifthenelse{\thelanguage=0}{#1}{}
    \ifthenelse{\thelanguage=1}{#2}{}
}

\let\oldsection\section

\RenewDocumentCommand\section {s o o m m}
   {
    \IfBooleanTF{#1}                   
        {
        \ifthenelse{\thelanguage=0}{\oldsection*{#4}}{}
        \ifthenelse{\thelanguage=1}{\oldsection*{#5}}{}
          \IfNoValueF{#2}            
            {          
              \ifthenelse{\thelanguage=0}{\addcontentsline{toc}{section}{#2}}{}
              \ifthenelse{\thelanguage=1}{\addcontentsline{toc}{section}{#3}}{}
            }
        }  
      {                              
        \IfNoValueTF{#2}
          {
            \ifthenelse{\thelanguage=0}{\oldsection{#4}}{}
            \ifthenelse{\thelanguage=1}{\oldsection{#5}}{}
           }       
          { 
            \ifthenelse{\thelanguage=0}{\oldsection[#2]{#4}}{}
            \ifthenelse{\thelanguage=1}{\oldsection[#3]{#5}}{}
           }
      }   
  }

\let\oldsubsection\subsection

\RenewDocumentCommand\subsection {s o o m m}
   {
    \IfBooleanTF{#1}                   
        {
        \ifthenelse{\thelanguage=0}{\oldsubsection*{#4}}{}
        \ifthenelse{\thelanguage=1}{\oldsubsection*{#5}}{}
          \IfNoValueF{#2}            
             {          
              \ifthenelse{\thelanguage=0}{\addcontentsline{toc}{subsection}{#2}}{}
              \ifthenelse{\thelanguage=1}{\addcontentsline{toc}{subsection}{#3}}{}
            }
        }  
      {                              
        \IfNoValueTF{#2}
          {
           \ifthenelse{\thelanguage=0}{\oldsubsection{#4}}{}
           \ifthenelse{\thelanguage=1}{\oldsubsection{#5}}{}
           }       
          { 
           \ifthenelse{\thelanguage=0}{\oldsubsection[#2]{#4}}{}
           \ifthenelse{\thelanguage=1}{\oldsubsection[#3]{#5}}{}
           }
      }   
  }

\let\oldsubsubsection\subsubsection

\RenewDocumentCommand\subsubsection {s o o m m}
   {
    \IfBooleanTF{#1}   
        {
        \ifthenelse{\thelanguage=0}{\oldsubsubsection*{#4}}{}
        \ifthenelse{\thelanguage=1}{\oldsubsubsection*{#5}}{}
          \IfNoValueF{#2}            
             {          
              \ifthenelse{\thelanguage=0}{\addcontentsline{toc}{subsubsection}{#2}}{}
              \ifthenelse{\thelanguage=1}{\addcontentsline{toc}{subsubsection}{#3}}{}
            }
        }  
      {                              
        \IfNoValueTF{#2}
          {
           \ifthenelse{\thelanguage=0}{\oldsubsubsection{#4}}{}
           \ifthenelse{\thelanguage=1}{\oldsubsubsection{#5}}{}
           }       
          { 
           \ifthenelse{\thelanguage=0}{\oldsubsubsection[#2]{#4}}{}
           \ifthenelse{\thelanguage=1}{\oldsubsubsection[#3]{#5}}{}
           }
      }   
  }

\let\oldparagraph\paragraph

\RenewDocumentCommand\paragraph {s m m}
   {
    \IfBooleanTF{#1}
        {
        \ifthenelse{\thelanguage=0}{\oldparagraph*{#2}}{}
        \ifthenelse{\thelanguage=1}{\oldparagraph*{#3}}{}
        }  
      { 
           \ifthenelse{\thelanguage=0}{\oldparagraph{#2}}{}
           \ifthenelse{\thelanguage=1}{\oldparagraph{#3}}{}   
      }   
  }

\usepackage[strict]{changepage}

\let\oldfigure\figure
\let\oldendfigure\endfigure          
               
\RenewDocumentEnvironment{figure}{o}
 {%
  \IfNoValueTF{#1}
   {
     \oldfigure
   }%
    {
     \oldfigure[#1]  
    }
    \begin{adjustwidth*}{-2cm}{-4cm}
    \centering
 }
 {\end{adjustwidth*}
 \oldendfigure}

\let\oldtable\table
\let\oldendtable\endtable         
               
\RenewDocumentEnvironment{table}{o}
 {%
  \IfNoValueTF{#1}
   {
     \oldtable
   }%
    {
     \oldtable[#1]  
    }
    \begin{adjustwidth*}{-2cm}{-4cm}
    \centering
 }
 {\end{adjustwidth*}
 \oldendtable}

\newcommand{\appcaption}[2]{
    \begin{minipage}{1.9cm}\end{minipage}
    \begin{minipage}{\linewidth}
	\refstepcounter{figure}
	\paragraph{Figure \thefigure :}{Figure \thefigure :}
	\bpar{#1}{#2}
	\end{minipage}
}

\newcommand{\apptabcaption}[2]{
    \begin{minipage}{1.9cm}\end{minipage}
    \begin{minipage}{\linewidth}
	\refstepcounter{table}
	\paragraph{Table \thetable :}{Table \thetable :}
	\bpar{#1}{#2}
	\end{minipage}
}

\newcommand{\framecaption}[2]{
    \begin{minipage}{1.9cm}\end{minipage}
    \begin{minipage}{\linewidth}
	\refstepcounter{frame}
	\paragraph{Frame \theframe : }{Encadré \theframe : }
	\bpar{#1}{#2}
	\end{minipage}
}

\newcounter{frame}

\let\oldcite\cite
\renewcommand{\cite}[1]{[\oldcite{#1}]}

\newcommand{\stars}{\vspace{1cm}\noindent\hspace{0.45\textwidth}$\star$\hspace{0.1\textwidth} $\star$\\

\noindent\hspace{0.505\textwidth} $\star$\vspace{1cm}
}

\usepackage{todonotes}

\let\marginpar\oldmarginpar

\DeclareDocumentCommand{\comment}{o m o o o o}
{\ifthenelse{\draft=1}{
  \IfValueT{#1}{
      \todo[size=\tiny]{
        \textbf{C (#1) : }#2
        \IfValueT{#3}{\textcolor{blue}{ - \textbf{A1 : }#3}}
        \IfValueT{#4}{\textcolor{ForestGreen}{ - \textbf{A2 : }#4}}
        \IfValueT{#5}{\textcolor{red!50!blue}{ - \textbf{A3 : }#5}}
        \IfValueT{#6}{\textcolor{Aquamarine}{ - \textbf{A4 : }#6}}
      }
    }
    \IfNoValueT{#1}{
      \todo[size=\tiny]{
        \textbf{C : }#2
        \IfValueT{#3}{\textcolor{blue}{ - \textbf{A1 : }#3}}
        \IfValueT{#4}{\textcolor{ForestGreen}{ - \textbf{A2 : }#4}}
        \IfValueT{#5}{\textcolor{red!50!blue}{ - \textbf{A3 : }#5}}
        \IfValueT{#6}{\textcolor{Aquamarine}{ - \textbf{A4 : }#6}}
      }
    }
 }{}
}

\makeatletter
\def\blx@maxline{77}
\makeatother

\begin{document}

\frenchspacing 

\raggedbottom 

\ifthenelse{\thelanguage=0}{\selectlanguage{american}}{\selectlanguage{french}}


\pagenumbering{roman} 

\pagestyle{plain} 




\backgroundsetup{%
    contents={%
        \begin{tikzpicture}
            \node[opacity=0.35] {\includegraphics[width=\paperwidth]{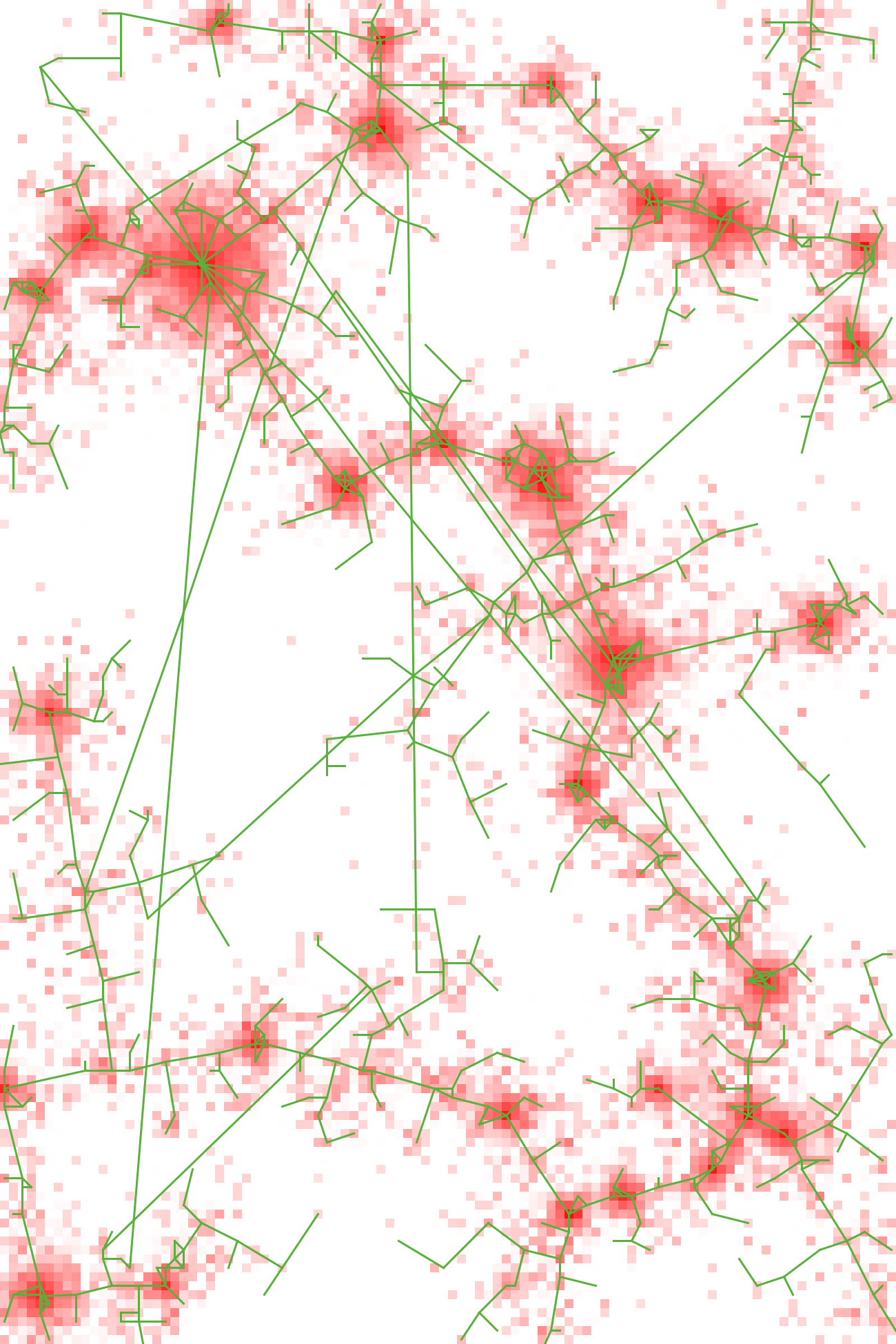}};
        \end{tikzpicture}
    }
}

\begin{titlepage}


\begin{addmargin}[-1cm]{-3cm}
\begin{center}
\large

\hfill

\vspace{-1.5cm}
\includegraphics[width=5cm]{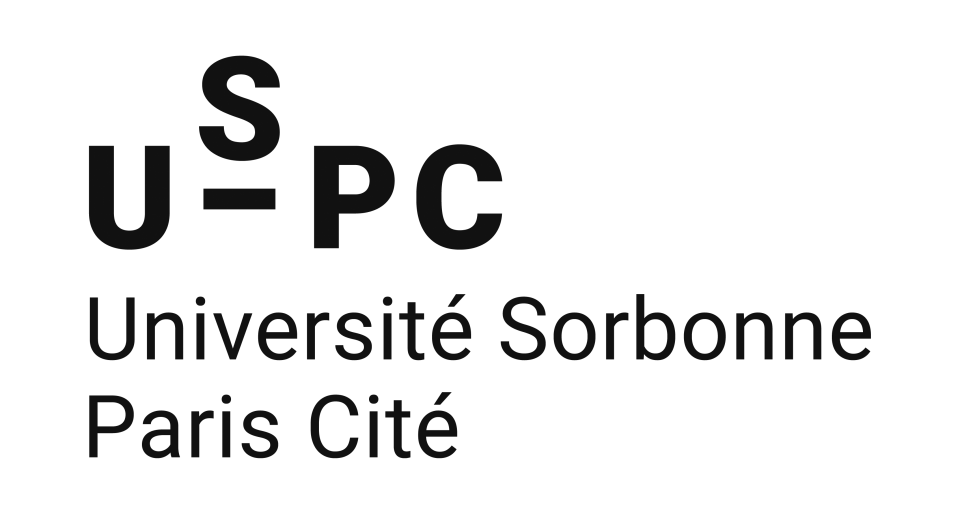}
\hfill
\includegraphics[height=4cm]{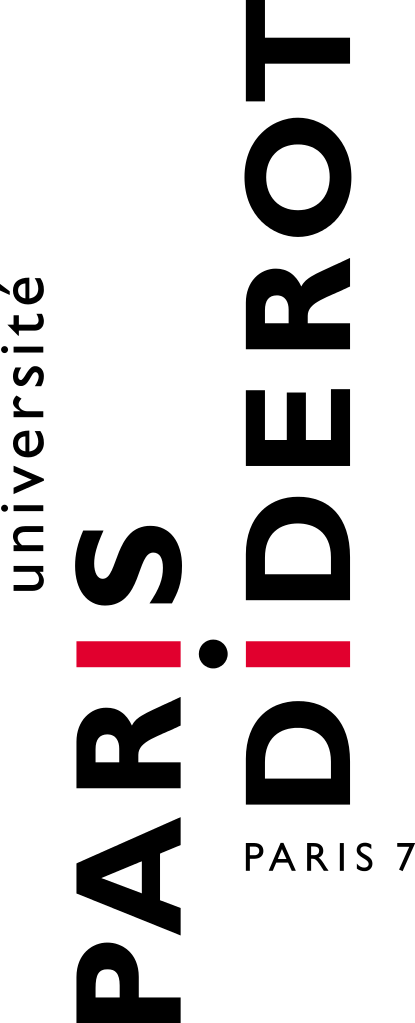}

\vspace{0.5cm}

\begingroup
\bpar{
\spacedallcaps{PhD Thesis}\\\medskip
\spacedallcaps{from Université Sorbonne Paris Cité}\\\medskip
\spacedallcaps{Prepared at Université Paris Diderot}\\\medskip
\spacedallcaps{École Doctorale de Géographie de Paris (ED 434)}\\\vspace{0.5cm}
\textit{UMR CNRS 8504 Géographie-cités / Équipe P.A.R.I.S.}\\
\textit{UMR-T IFSTTAR 9403 LVMT}
}{
\spacedallcaps{Thèse de Doctorat}\\\medskip
\spacedallcaps{de l'Université Sorbonne Paris Cité}\\\medskip
\spacedallcaps{Préparée à l'Université Paris Diderot}\\\medskip
\spacedallcaps{École Doctorale de Géographie de Paris (ED 434)}\\\vspace{0.5cm}
\textit{UMR CNRS 8504 Géographie-cités / Équipe P.A.R.I.S.}\\
\textit{UMR-T IFSTTAR 9403 LVMT}
}
\endgroup

\vspace{1.5cm}

\begingroup
\bpar{
\textbf{\spacedallcaps{\myTitleEn}} \\ \bigskip 
}{
\textbf{\spacedallcaps{\myTitle}} \\ \bigskip 
}
\endgroup

\vspace{1.5cm}



\bpar{
\textit{Presented by} \spacedlowsmallcaps{\myName}\\
}{
\textit{Présentée par} \spacedlowsmallcaps{\myName}\\ 
}
\bigskip

\bpar{
PhD Thesis in Geography
}{
Thèse de doctorat de Géographie
}

\bpar{
Under the supervision of \myProf and \myOtherProf \\ \medskip
}{
Dirigée par \myProf et \myOtherProf \\ \medskip
}




\vspace{0.5cm}

\bpar{
\textit{Presented and defended publicly at the Institut des Systèmes Complexes (Paris) on June 11th 2018, in front of a jury composed by:}
}{
\textit{Présentée et soutenue publiquement à l'Institut des Systèmes Complexes (Paris) le 11 juin 2018, devant le jury composé de :}
}

\bigskip

\begin{adjustwidth*}{-0.5cm}{-2cm}
\begin{minipage}{0.28\linewidth}
\raggedright
\textbf{\noun{Denise Pumain}}\\
\textbf{\noun{Didier Josselin}}\\
\textbf{\noun{Catherine Morency}}\\
\textbf{\noun{Olivier Bonin}}\\
\textbf{\noun{Anne Ruas}}\\
\textbf{\noun{Arnaud Banos}}\\
\textbf{\noun{Florent Le Néchet}}
\end{minipage}
\begin{minipage}{0.7\linewidth}
\raggedright
Professeure, Université Paris 1 (Présidente du Jury)\\
Directeur de Recherche, CNRS (Rapporteur)\\
Professeure, Ecole Polytechnique de Montréal (Rapporteuse)\\
Chargé de Recherche, IFSTTAR (Examinateur)\\
Directrice de Recherche, IFSTTAR (Examinatrice)\\
Directeur de Recherche, CNRS (Directeur)\\
Maître de Conférences, Université Paris-Est (Directeur)\\
\end{minipage}
\end{adjustwidth*}

\vspace{2cm}

\begin{minipage}{0.28\linewidth}
	\includegraphics[width=\textwidth]{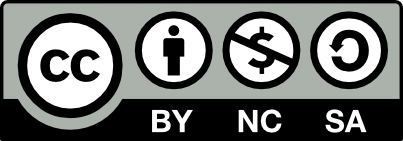}
\end{minipage}
\hfill
\begin{minipage}{0.7\linewidth}
This work is licensed under a \href{http://creativecommons.org/licenses/by-nc-sa/4.0/}{Creative Commons Attribution-NonCommercial-ShareAlike 4.0 International License}.
\end{minipage}

\end{center}
\end{addmargin}

\end{titlepage}


\backgroundsetup{%
    contents={}
}


\thispagestyle{empty}

\hfill

\vfill

\bpar{
\noindent\myName: \textit{\myTitleEn,} \mySubtitleEn, 
\textcopyright\ \myTime
}{
\noindent\myName: \textit{\myTitle,} \mySubtitle, 
\textcopyright\ \myTime
}








\cleardoublepage


\pdfbookmark[1]{Résumé}{Résumé} 

\begingroup
\let\clearpage\relax
\let\cleardoublepage\relax
\let\cleardoublepage\relax

\begin{adjustwidth*}{-1cm}{-1cm}

\section*{Title}{Title}

Characterising and modeling the co-evolution of transportation networks and territories

\section*{Abstract}{Abstract}

The identification of structuring effects of transportation infrastructure on territorial dynamics remains an open research problem. This issue is one of the aspects of approaches on complexity of territorial dynamics, within which territories and networks would be co-evolving. The aim of this thesis is to challenge this view on interactions between networks and territories, both at the conceptual and empirical level, by integrating them in simulation models of territorial systems. The intrinsically multidisciplinary nature of the question requires first to proceed to a quantitative epistemology analysis, that allow us to draw a map of the scientific landscape and to give a description of common features and specificities of models studying the co-evolution between network and territories within each discipline. We propose consequently a definition of co-evolution and an empirical method for its characterization, based on spatio-temporal correlation analysis. Two complementary modeling approaches, that correspond to different scales and ontologies, are then explored. At the macroscopic scale, we build a family of models inheriting from interaction models within system of cities, developed by the Evolutive Urban Theory (Pumain, 1997). Their exploration shows that they effectively capture co-evolutionary dynamics, and their calibration on demographic data for the French system of cities (1830-1999) quantifies the evolution of interaction processes such as the tunnel effect or the role of centrality. At the mesoscopic scale, a morphogenesis model captures the co-evolution of the urban form and of network topology. It is calibrated on corresponding indicators for local form and topology, computed for all Europe. Multiple network evolution processes are shown complementary to reproduce the large variety of observed configurations, at the level of indicators but also interactions between indicators. These results suggest new research directions for urban models integrating co-evolutive dynamics in a multi-scale perspective.

\section*{Keywords}{Keywords}

Territories; Transportation Networks; Co-evolution; Morphogenesis; Evolutive Urban Theory; Quantitative Epistemology; Systems of Cities; Urban Morphology; Greater Paris; Pearl River Delta

\end{adjustwidth*}

\newpage


\begin{adjustwidth*}{-1cm}{-1cm}


\section*{Titre}{Titre}

Caractérisation et modélisation de la co-évolution des réseaux de transport et des territoires

\section*{Résumé}{Résumé}

L'identification d'effets structurants des infrastructures de transports sur la dynamique des territoires reste un défi scientifique ouvert. Cette question est une des facettes de recherches sur la complexité des dynamiques territoriales, au sein desquelles territoires et réseaux de transport seraient en co-évolution. L'objectif de cette thèse est de mettre à l'épreuve cette vision des interactions entre réseaux et territoires, autant sur le plan conceptuel que sur le plan empirique, en les intégrant au sein de modèles de simulation des systèmes territoriaux. La nature intrinsèquement pluri-disciplinaire de la question nous conduit à mener un travail d'épistémologie quantitative, qui permet de dresser une carte du paysage scientifique et une description des éléments communs et des spécificités des modèles traitant la co-évolution entre réseaux et territoires dans chaque discipline. Nous proposons ensuite une définition de la co-évolution, ainsi qu'une méthode de caractérisation empirique, basée sur une analyse de corrélations spatio-temporelles. Deux pistes complémentaires de modélisation, correspondant à des ontologies et des échelles différentes sont alors explorées. A l'échelle macroscopique, nous construisons une famille de modèles dans la lignée des modèles d'interaction au sein des systèmes de villes développés par la Théorie Evolutive des Villes (Pumain, 1997). Leur exploration montre qu'ils capturent effectivement des dynamiques de co-évolution, et leur calibration sur des données démographiques pour le système de villes français (1830-1999) quantifie l'évolution des processus d'interaction comme l'effet tunnel ou le rôle de la centralité. A l'échelle mésoscopique, un modèle de morphogenèse capture la co-évolution de la forme urbaine et de la topologie du réseau. Il est calibré sur les indicateurs correspondants pour la forme et la topologie locales calculés pour l'ensemble de l'Europe. De multiples processus d'évolution du réseau s'avèrent être complémentaires pour reproduire la grande variété des configurations observées, au niveau des indicateurs ainsi que des interactions entre indicateurs. Ces résultats suggèrent de nouvelles pistes d'exploration des modèles urbains intégrant les dynamiques co-évolutives dans une perspective multi-échelles.

\section*{Mots-clefs}{Mots-clés}

Territoires ; Réseaux de Transport ; Co-évolution ; Morphogenèse ; Théorie Évolutive des Villes ; Épistémologie Quantitative ; Systèmes de Villes ; Morphologie Urbaine ; Grand Paris ; Delta de la Rivière des Perles

\end{adjustwidth*}

\newpage

\begin{adjustwidth*}{-1cm}{-1cm}

\cn{\textbf{标题}}

\cn{建模交通网络和地域的共同演变}\\

\cn{\textbf{摘要}}

\cn{运输基础设施对领土体系结构效应存在的问题远未得到解决。
这是复杂的地域动态的一个方面，其中领土和交通网络正在共同演变。
这篇论文的目的是测试网络和地域之间的相互作用。
它将在概念和经验上做到这一点，目的是将其整合到地域系统的模拟模型中。
我们正在处理的问题本质上是多学科的。
出于这个原因，我们首先进行量化的认识论分析。
它可以绘制科学的景观图，并精确地描述每个学科不同模型的结构。
我们制定了一个共同进化的定义，并开发了一个基于时空相关分析的经验表征方法。
探索两个互补的建模轨道。 它们对应于不同的本体和尺度。
在宏观层面上，我们根据城市演变理论发展起来的城市体系内的相互作用模型发展了一个模型家族。
他们的探索表明，他们实际上捕捉到共同演化的动力。 他们对法国城市系统（1830-1999）的人口统计数据的校准量化了互动过程的演变。 这些例如是隧道效应或网络中心性的影响。
在介观尺度上，形态演化模型捕捉城市形态和网络拓扑的共同演化。
根据整个欧洲计算的局部形态和拓扑结构的相应指标进行校准。
网络演进的多个过程被考虑到：成本效益计划，潜在的突破，自组织。 它们似乎是互补的，可以产生所有的真实配置。
校准也是按照第二顺序进行的，也就是指标之间的相互作用，模型重现了现有情况的多样性。
这些结果一方面表明了把城市演变理论与形式演变相结合的理论建构。 另一方面，他们开辟了新一代城市模式的探索，这些模型将不得不整合多尺度协同进化动力学。
}\\

\cn{\textbf{关键字}}

\cn{地域; 交通网络; 共同演变; 形态; 演变城市理论; 量化认识论; 城市系统; 城市形态; 大巴黎; 珠江三角洲}


\end{adjustwidth*}

\endgroup			

\vfill

\cleardoublepage



\pdfbookmark[1]{Notes de Lecture}{Notes de Lecture}

\begingroup
\let\clearpage\relax
\let\cleardoublepage\relax
\let\cleardoublepage\relax


\bpar{
\chapter*{Reading Notes}
}{
\chapter*{Notes de Lecture}
}

\bpar{
This thesis was initially intended to be written in English. A first third and most of papers were, to be then adapted and translated into French, in order to fulfil an administrative constraint from an other age. The rest was written in French and translated for this English version. It has also been thought as a ``Paper Thesis'', but the strong recommendations of CNU have rapidly swept this ambition. Therefore, the current version has gone through several transformations and ``smoothing'', in order to give it a ``classical'' form, background and identity. We apologize in advance to the reader if translation or articulation issues remain and disturb the fluidity of the reading, since this English version was moreover fully translated again back from French. 
}{
Cette thèse devait initialement être rédigée en anglais. Un premier tiers et la majorité des articles l'ont été, pour être repris et traduits par la suite, afin de répondre à une contrainte administrative d'un autre âge. Elle avait également été conçue comme une ``thèse à articles'', mais les fortes recommandations du CNU ont vite eu vent de cette ambition. Ainsi, la version courante est passée par maintes transformations et ``lissages'', afin de lui donner une forme, un fond et une identité ``classiques''. Nous nous excusons préalablement auprès du lecteur si des écueils de traduction ou d'articulation subsistent et perturbent la fluidité de la lecture.
}

\bpar{
The layout is designed to be narrow in order to allow the reader to write notes on the manuscript where he wants, on the digital or paper version: maybe the dream of all manuscript is to become interactive.
}{
La mise en page est voulue étroite pour permettre au lecteur d'annoter à loisir ce manuscrit, de manière informatique ou papier : peut être que le rêve de tout manuscrit est de devenir interactif.
}

\bpar{
All the figures in main text are produced by the author, at the exception of Fig.~\ref{fig:computation:xkcd} (source xkcd \url{https://xkcd.com/}) and two illustrations in the Frame~\ref{frame:interdiscmorphogenesis:examples}. A large majority of figures are \emph{directly} reproducible, i.e. can be obtained by executing the scripts. All source code, from models to the interpretation of results and to this proper writing, is available openly with all its atomic history (\emph{commits}) on the repository of the project\footnote{at \url{https://github.com/JusteRaimbault/CityNetwork}}. All the datasets produced in that frame are open, and all data used are open or made open (in an aggregated way corresponding to the level of use by models in the case of a third-party closed database).  
}{
L'ensemble des figures du texte principal est produit par l'auteur, sauf la Fig.~\ref{fig:computation:xkcd} (source xkcd \url{https://xkcd.com/}) et deux illustrations dans l'Encadré~\ref{frame:interdiscmorphogenesis:examples}. La grande majorité des figures est \emph{directement} reproductible, c'est-à-dire pouvant être obtenue par exécution des scripts. L'ensemble des codes sources, des modèles à l'interprétation des résultats et à cette propre rédaction, est disponible de manière ouverte avec l'ensemble de son historique atomique (\emph{commits}) sur le dépôt du projet\footnote{à \url{https://github.com/JusteRaimbault/CityNetwork}}. L'ensemble des jeux de données produits dans ce cadre est ouvert, et l'ensemble des données utilisées sont ouvertes ou rendues ouvertes (de manière agrégée correspondant au niveau d'utilisation par les modèles dans le cas d'une base tierce fermée).
}

\bpar{
This memoir in itself has been proofread by the following readers (in alphabetical order): Arnaud Banos (AB), Clémentine Cottineau (CC), Florent Le Néchet (FL), Cinzia Losavio (CL), Sébastien Rey (SR), Hélène Serra (HS) in the spirit of an open review. By following the successive commits at \url{https://github.com/JusteRaimbault/ThesisMemoire}, the use of specific commands for the review remarks allows to track the full review process.
}{
Ce mémoire en lui-même a été relu par les lecteurs suivants (ordre alphabétique) : Arnaud Banos (AB), Clémentine Cottineau (CC), Florent Le Néchet (FL), Cinzia Losavio (CL), Sébastien Rey (SR), Hélène Serra (HS) dans l'esprit d'une revue ouverte. En suivant les commits successifs à \url{https://github.com/JusteRaimbault/ThesisMemoire}, l'utilisation de commandes spécifiques pour les retours de relecture permet de retracer l'ensemble du processus de revue.
}

\bpar{
Names in Mandarin (cities, places, people, etc.) are transcribed using the \emph{pinyin} system. 
}{
Les noms en Mandarin (villes, lieux, personnes, etc.) sont transcrits en système \emph{pinyin}.
}

\endgroup			

\vfill



\cleardoublepage 


\pdfbookmark[1]{Publications}{Publications} 

\bpar{
\chapter*{Publications}
}{
\chapter*{Publications} 
}





\bpar{
The following publications and communications contain most content of this thesis. Sources are precisely mentioned at the beginning of each chapter. Translations were ensured by the author when needed.
}{
Les publications et communications suivantes contiennent la majorité du contenu de cette thèse. Les sources sont précisément mentionnées en introduction de chaque chapitre. Les traductions sont assurées par l'auteur le cas échéant.
}

\bpar{
The English translation of this dissertation, deposited on arXiv, has consequently substantial overlap with the following arXiv preprints: 1605.08888, 1608.00840, 1608.05266, 1612.08504, 1706.07467, 1706.09244, 1708.06743, 1709.08684, 1712.00805, 1803.11457, 1804.09416, 1804.09430, 1805.05195, 1808.07282, 1809.00861, 1811.04270, 1812.01473, 1812.06008, 1908.02034, 2012.13367, 2102.13501, 2106.11996. 
}{

}

\section*{Journal papers}{Articles}

\noindent Raimbault, J., \& Le Néchet, F. (2021). Introducing endogenous transport provision in a LUTI model to explore polycentric governance systems. \textit{Journal of Transport Geography}, 94, 103115.

\bigskip

\noindent Raimbault, J., Chasset, P. O., Cottineau, C., Commenges, H., Pumain, D., Kosmopoulos, C., \& Banos, A. (2021). Empowering open science with reflexive and spatialised indicators. \textit{Environment and Planning B: Urban Analytics and City Science}, 48(2), 298-313.

\bigskip

\noindent Bergeaud, A., \& Raimbault, J. (2020). An empirical analysis of the spatial variability of fuel prices in the United States. \textit{Transportation Research Part A: Policy and Practice}, 132, 131-143.

\bigskip

\noindent Raimbault, J. (2020). Indirect evidence of network effects in a system of cities. \textit{Environment and Planning B: Urban Analytics and City Science}, 47(1), 138-155.

\bigskip

\noindent Raimbault, J., Cottineau, C., Le Texier, M., Le Néchet, F., \& Reuillon, R. (2019). Space Matters: Extending Sensitivity Analysis to Initial Spatial Conditions in Geosimulation Models. \textit{Journal of Artificial Societies and Social Simulation}, 22(4).

\bigskip

\noindent Raimbault, J. (2019). Exploration of an interdisciplinary scientific landscape. \textit{Scientometrics}, 119(2), 617-641.

\bigskip

\noindent Raimbault, J. (2019). Second-order control of complex systems with correlated synthetic data. \textit{Complex Adaptive Systems Modeling}, 7(1), 1-19.

\bigskip

\noindent Raimbault, J. (2018). Calibration of a density-based model of urban morphogenesis. \textit{PloS one}, 13(9), e0203516.

\bigskip

\noindent Bergeaud, A., Potiron, Y., \& Raimbault, J. (2017). Classifying patents based on their semantic content. \textit{PloS ONE}, 12(4), e0176310.

\section*{Book chapters}{Chapitres d'ouvrage}

\noindent Raimbault, J. (2021). Modeling the co-evolution of cities and networks. In \textit{Handbook of Cities and Networks}. Edward Elgar Publishing.

\bigskip

\noindent Raimbault, J. (2020). Relating complexities for the reflexive study of complex systems. In \textit{Theories and models of urbanization} (pp. 27-41). Springer, Cham.

\bigskip

\noindent Raimbault, J. (2020). Unveiling co-evolutionary patterns in systems of cities: a systematic exploration of the simpopnet model. In \textit{Theories and Models of Urbanization} (pp. 261-278). Springer, Cham.

\bigskip

\noindent Raimbault, J. (2019). An urban morphogenesis model capturing interactions between networks and territories. In \textit{The mathematics of urban morphology} (pp. 383-409). Birkhäuser, Cham.

\bigskip

\noindent Raimbault, J. (2019). Evolving accessibility landscapes: mutations of transportation networks in China. In \textit{Pathways of sustainable urban development across China: the cases of Hangzhou, Datong and Zhuhai} (pp.89-108). Imago Editor.

\section*{Conference proceedings}{Actes de conferences}

\noindent Raimbault, J. (2017). An applied knowledge framework to study complex systems, \textit{Complex Systems Design \& Management} (pp.31-45).

\bigskip

\noindent Raimbault, J. (2017). Identification de causalités dans des données spatio-temporelles, \textit{Spatial Analysis and GEOmatics 2017.}

\bigskip

\noindent Raimbault, J. (2017). A discrepancy-based framework to compare robustness between multi-attribute evaluations, \textit{Complex Systems Design \& Management} (pp. 141-154). Springer International Publishing. 

\bigskip

\noindent Raimbault, J. (2017). Investigating the empirical existence of static user equilibrium. \textit{Transportation Research Procedia}, 22, 450-458. 

\bigskip

\noindent Raimbault, J. (2017). Models coupling urban growth and transportation network growth: An algorithmic systematic review approach. \textit{Plurimondi}, (17), \textit{ECTQG 2015 Proceedings}.

\section*{Working papers}{Documents de travail}

\noindent Raimbault, J. (2018). Co-evolution and morphogenetic systems. \textit{Rejected for Artificial Life 2018.} arXiv preprint arxiv:1803.11457.







\bigskip

\noindent Antelope, C., Hubatsch, L., Raimbault, J., and Serna, J. M. (2016). An interdisciplinary approach to morphogenesis. \textit{Working Paper, Santa Fe Institute CSSS 2016.}

\section*{Communications}{Communications}


\noindent Multi-modeling the morphogenesis of transportation networks, \textit{extended abstract forthcoming in Proceedings of ALife 2018, Tokyo, July 2018.}

\bigskip

\noindent Modeling Urban Morphogenesis: towards an integration of territories and networks, \textit{GoPro 2017, Lyon, Dec. 2017.}

\bigskip

\noindent Modeling the Co-evolution of Urban Form and Transportation Networks, \textit{Conference on Complex Systems 2017, Cancun, Sept. 2017.}

\bigskip

\noindent Raimbault J. \& Baffi S. (2017). Structural Segregation: Assessing the impact of South African Apartheid on Underlying Dynamics of Interactions between Networks and Territories, \textit{ECTQG 2017, York, Sept. 2017.}

\bigskip

\noindent Invisible Bridges ? Scientific landscapes around similar objects studied from Economics and Geography perspectives, \textit{ECTQG 2017, York, Sept. 2017.}

\bigskip

\noindent Raimbault, J. \& Bergeaud, A. (2017). The Cost of Transportation: Spatial Analysis of Fuel Prices in the US, \textit{EWGT 2017, Budapest, Sept. 2017.}

\bigskip

\noindent Cottineau C., Raimbault J., Le Texier M., Le N{\'e}chet F. \& Reuillon R. (2017). Initial spatial conditions in simulation models: the missing leg of sensitivity analyses?, \textit{Geocomputation 2017, Leeds, Sept. 2017}

\bigskip

\noindent A macro-scale model of co-evolution for cities and transportation networks, \textit{Medium International Conference, Guangzhou, June 2017.}

\bigskip

\noindent Losavio C. \& Raimbault J. (2017). Agent-based Modeling of Migrant Workers Residential Dynamics within a Mega-city Region: the Case of Pearl River Delta, China, \textit{Urban China Development International Conference, London, May 2017.}

\bigskip

\noindent Co-construire Modèles, Etudes Empiriques et Théories en Géographie Théorique et Quantitative: le cas des Interactions entre Réseaux et Territoires, \textit{Treizièmes Rencontres de ThéoQuant, Besançon, May 2017.}

\bigskip

\noindent Un Cadre de Connaissances pour une Géographie Intégrée,  \textit{Journée des jeunes chercheurs de l'Institut de Géographie de Paris, Paris, April 2017.}

\bigskip

\noindent Towards a Theory of Co-evolutive Networked Territorial Systems: Insights from Transportation Governance Modeling in Pearl River Delta, China, \textit{MEDIUM Seminar : Sustainable Development in Zhuhai, Guangzhou, Dec 2016.}

\bigskip

\noindent Models of growth for system of cities : Back to the simple, \textit{Conference on Complex Systems 2016, Amsterdam, Sept. 2016.}

\bigskip

\noindent Raimbault J., Bergeaud A. and Potiron Y. (2016). Investigating Patterns of Technological Innovation. \textit{Conference on Complex Systems 2016, Amsterdam, Sep 2016.}

\bigskip

\noindent For a Cautious Use of Big Data and Computation. \textit{Royal Geographical Society - Annual Conference 2016 - Session : Geocomputation, the Next 20 Years (1), London, Aug 2016.}

\bigskip

\noindent Indirect Bibliometrics by Complex Network Analysis. \textit{20e Anniversaire de Cybergeo, Paris, May 2016.}

\bigskip

\noindent Raimbault, J. \& Serra, H. (2016). Game-based Tools as Media to Transmit Freshwater Ecology Concepts, \textit{poster corner at SETAC 2016 (Nantes, May 2016).}

\bigskip

\noindent Le Néchet, F. \& Raimbault, J. (2015). Modeling the emergence of metropolitan transport authority in a polycentric urban region, \textit{ECTQG 2015, Bari, Sept. 2015).}

\bigskip

\noindent Hybrid Modeling of a Bike-Sharing Transportation System, \textit{poster presented at ICCSS 2015, Helsinki, June 2015.}

\bigskip

\noindent Raimbault, J. \& Gonzales, J. (2015). Application de la Morphog{\'e}n{\`e}se de R{\'e}seaux Biologiques {\`a} la Conception Optimale d'Infrastructures de Transport, \textit{poster presented at Rencontres du Labex Dynamite, Paris, May 2015.}

\cleardoublepage 


\pdfbookmark[1]{Acknowledgements}{Acknowledgements} 




\begingroup

\let\clearpage\relax
\let\cleardoublepage\relax
\let\cleardoublepage\relax

\chapter*{Acknowledgements}



\bpar{
A significant part of results obtained in this thesis have been computed on the virtual organisation vo.complex-system.eu of the European Grid Infrastructure (http://www.egi.eu). I thank the European Grid Infrastructure and its National Grid Initiatives (France-Grilles in particular) to provide technical support and infrastructure.
}{
Une grande partie des résultats obtenus dans cette thèse ont été calculés sur l'organisation virtuelle vo.complex-system.eu de l'European Grid Infrastructure (http://www.egi.eu). Je remercie l'European Grid Infrastructure et ses National Grid Initiatives (France-Grilles en particulier) pour fournir le support technique et l'infrastructure.
}

\bigskip


\bpar{
This research work was conducted in the context of the MEDIUM project (New pathways for sustainable urban development in China’s MEDIUM sized-cities). I thus thank CNRS and UMR 8504 Géographie-cités for their support and MEDIUM partners, in particular Sun-Yat-Sen University. The MEDIUM project was funded by the EU, grant contract ICI+/2014/348-005.
}{
Ce travail de recherche a été mené dans le cadre du project MEDIUM (New pathways for sustainable urban development in China’s MEDIUM sized-cities). Je remercie donc le Centre National de la Recherche Scientifique (CNRS) et l’UMR 8504 Géographie-cités pour leurs soutien ainsi que les partenaires de MEDIUM, en particulier la Sun-Yat-Sen University. Le projet MEDIUM a été cofinancé par l’Union européenne au titre de l’Action Extérieure de l’UE – Contrat de subvention ICI+/2014/348-005.
}

\bigskip


\bpar{
I would like to thank Denise Pumain for the honour she gave to accept presiding the Jury, the rest of the jury Olivier Bonin and Anne Ruas to accept evaluating this work, and reviewers Didier Josselin and Catherine Morency to ensure the consequent work to digest each word of this manuscript.
}{
Je tiens à remercier Denise Pumain pour l'honneur qu'elle me fait d'accepter la présidence du Jury, les examinateurs Olivier Bonin et Anne Ruas pour accepter d'évaluer ce travail, et les rapporteurs Didier Josselin et Catherine Morency pour avoir assuré le conséquent travail de digérer chaque mot de ce manuscrit.
}

\bigskip

\bpar{
A PhD thesis is both unexpected and obvious. Unexpected as we cannot imagine a few years back what could be its detailed subject. Unexpected as we consider the discrepancy between the initial project and what was finally explored. But also obvious within this same project containing seeds of main developments, suggesting a morphogenesis of knowledge. Also strangely obvious through introspection, as I once had a vocation of becoming train driver and later mapper. Obvious and unexpected when considering the paths open. As much a beginning as an end, a mean as a target, a trajectory as a position. I will try to acknowledge here all that contributed to realising this complexity.
}{
Un travail de thèse est à la fois improbable et évident. Improbable car on ne s'imaginait pas quelques années en arrière quel pourrait bien être le sujet avec lequel on devrait s'obséder pendant trois longues années. Improbable quand on regarde l'écart entre le projet initial et les cavités et les arêtes effilées qui ont finalement été explorées. Mais aussi évident quand on regarde ce même projet, et qu'on retrouve les graines de chacun des développements fondamentaux, suggérant une morphogenèse de la connaissance. Étrangement évident par un travail d'introspection : les métiers des rêves de mon enfance ont été conducteur de métro puis cartographe, peut être n'est-ce pas une coïncidence si le coeur du sujet ici rassemble les transports et les territoires. Évident et improbable quand on contemple les futurs possibles, la méta-structure qui s'en dégage finalement. Autant un commencement qu'une fin, un moyen qu'une finalité, une trajectoire qu'une position, une fête qu'une tristesse, une poésie qu'un rapport technocratique. Je vais tenter de remercier ici tout ceux qui ont permis la concrétisation de cette complexité.
}

\bigskip


\bpar{
My deep recognition goes naturally to my supervisors, which made this possible all along the journey. I met Arnaud Banos for the first time in October 2012 at the Complex Systems Institute. It was for a supervision of a master project, and we had at the time dived with my colleague Jorge into the world of multi-scale, multi-objective problems, self-oragnised biological networks (project which implementation was in fact reused here). Thanks to Arnaud to have guided us there. I kept until now most fundamental research paradigms discovered during this period.
}{
Ma profonde reconnaissance va naturellement à mes directeurs, qui ont rendu cette aventure possible et ont permis sa forme finale, par un pilotage subtil du système complexe que formaient objets, modèles, idées. J'ai rencontré pour la première fois Arnaud Banos en octobre 2012 à l'ISC qu'il dirigeait, alors toujours rue Lhomond. C'était dans le cadre d'une supervision des \emph{Open Problems} du PA Systèmes Complexes, et nous nous étions immergés avec mon collègue Jorge dans le monde du multi-échelle, de l'optimisation multi-objectif, des réseaux biologiques auto-organisés (projet dont l'implémentation originale a d'ailleurs été reprise ici). Ou plutôt jetés inconsciemment à l'eau au risque de se noyer, merci à Arnaud de nous avoir repêchés. Je garde un certain nombre de paradigmes fondamentaux qu'il nous avait transmis dès le premier contact avec la recherche. Cette bifurcation coïncide étrangement avec une autre plus personnelle, peut être ironiquement pour rappeler la place du sujet dont l'objectivité de la recherche ne fait aucun sens.
}

\bpar{
My first meeting with Florent Le Néchet was in March 2014 at Ecole des Ponts, to discuss of this project. I naively presented my work with the RBD model, and he immediately gave depth to the project. His various ideas and suggestions has been crucial for this work, and without him the social science component of the thesis would be significantly reduced.
}{
Ma première rencontre avec Florent Le Néchet a eu lieu en mars 2014, à la cafétéria des Ponts, pour discuter de ce projet de thèse. Naïvement, je lui présentais mon modèle RBD ainsi que des idées floues sur les ruptures de potentiel. Il a alors immédiatement donné de la profondeur au projet, en évoquant les Mega-city Regions, les nouveaux régimes urbains, la Chine : vision finalement prémonitoire (ou prophétie auto-réalisatrice ?). La richesse de ses idées n'a cessé d'irriguer ce travail mais aussi mes reflexions de manière plus générale. Sans lui, cette thèse n'aurait de géographie que le nom, et je lui suis fortement reconnaissant d'avoir été patient devant mes difficultés à appréhender les Sciences Humaines et Sociales.
}

\bpar{
Furthermore, although Denise Pumain did not officially supervised the thesis, her advice was of great value, as much on thematic as on epistemological issues. Her implications in the different projects was a considerable source of motivation, as for the current projects. Her academic support was also precious.
}{
Par ailleurs, même si Denise Pumain n'a pas officiellement dirigé cette thèse, son conseil a été d'une valeur inestimable, autant sur le plan thématique qu'épistémologique. Son intérêt pour les différents projets a été une source de motivation considérable, comme pour les nombreux projets futurs en perspective. Enfin, son soutien académique a été précieux.
}


\bigskip


\bpar{
I thank the academic actors who accepted to be interviewed for research material: Denise Pumain, Romain Reuillon, Clémentine Cottineau et Alain Bonnafous.
}{
Je remercie les acteurs académiques ayant accepté de mener des entretiens qui ont servi de matériau de recherche : Denise Pumain, Romain Reuillon, Clémentine Cottineau et Alain Bonnafous.
}

\bigskip


\bpar{
Technical support was crucial, I thank the OpenMOLE team and in particular Romain Reuillon for his efficiency in technical support. I thank Maziyar Panahi for the technical support with Zebulon.
}{
Le soutien technique a été crucial, je remercie l'équipe d'OpenMole et en particulier Romain Reuillon pour sa rapidité de réponse et de résolution des problèmes. Je remercie Maziyar Panahi pour le soutien technique sur Zebulon.
}

\bigskip


\bpar{
I thank the different reviewers who contributed to make the manuscript readable: Arnaud Banos, Clémentine Cottineau, Florent Le Néchet, Cinzia Losavio, Sébastien Rey, Hélène Serra. I also thank the persons who were determining at the end of the writing period: Nicolas Coulombel, Hadrien Commenges, Caroline Gallez.
}{
Je remercie les différents relecteurs de ce mémoire qui ont grandement contribué à le rendre lisible : Arnaud Banos, Clémentine Cottineau, Florent Le Néchet, Cinzia Losavio, Sébastien Rey, Hélène Serra. Je remercie également ceux avec qui les discussions ont été déterminantes dans la fin de la rédaction : Nicolas Coulombel, Hadrien Commenges, Caroline Gallez.
}

\bigskip


\bpar{
This research trajectory would not have been possible without the key people in my education. I thus thank Paul Bourgine and Kashayar Pakdaman for having introduced me to complex systems, and the rest of the team of the Complex Systems Master, in particular René Doursat. I thank Eric Marandon for the quality of the research internship at L2 technologies. I also thank the VET department at Ecole des ponts, in particular Nicolas Coulombel, Fabien Leurent, Zoi Cristoforou, Antoine Picon.
}{
Cette trajectoire de recherche n'aurait pas été possible sans les personnes qui ont joué un rôle clé dans ma formation. Je tiens ainsi à remercier Paul Bourgine et Kashayar Pakdaman pour m'avoir introduit aux systèmes complexes, et l'ensemble de l'équipe pédagogique du Master pour la qualité de l'enseignement, en particulier René Doursat pour son efficacité de formation à la recherche. Je remercie Eric Marandon pour la qualité scientifique et la stimulation intellectuelle pendant le stage chez L2. Je remercie également l'équipe du Département Ville, Environnement, Transport des Ponts, en particulier Nicolas Coulombel, Fabien Leurent, Zoi Cristoforou, Antoine Picon. 
}

\bpar{
The Medium project was already ``officially'' mentioned, but I must personaly thank Natacha Aveline for the opportunity, Chenyi Shi and Ming for their precious help in Zhuhai, Florent Resche-Rigon for the management, Céline Rozenblat for the Medium conference, and other participants for the moments in China: Cinzia Losavio, Valentina Ansoize, Judith Audin, Yinghao Li.
}{
Le projet Medium a déjà été mentionné ``officiellement'', mais je me dois de remercier personnellement Natacha Aveline pour m'avoir donné l'opportunité d'y participer, Chenyi Shi et Ming pour leur aide précieuse à Zhuhai, Florent Resche-Rigon pour la supervision, Céline Rozenblat pour l'organisation de la session modélisation à la conférence Medium, et les participants Cinzia Losavio, Valentina Ansoize, Judith Audin, Yinghao Li pour les moments passés à Zhuhai.
}

\bigskip


\bpar{
Summer school also had an important role in my education. I thank the team and partcipants of the 2016 SFI summer school, and of the 2014 Labex Dynamite school.
}{
Les écoles d'été ont également pris une place importante dans ma formation. Je remercie l'ensemble de l'équipe pédagogique et des participants de la SFI Complex Systems Summer School 2016 à Santa Fe et ceux de l'École d'été du Labex Dynamite 2014 à Florence.
}

\bigskip


\bpar{
I also thank my co-author and collaborators on the different projects implied:
\begin{itemize}
	\item the SpaceMatters team Clémentine Cottineau, Florent Le Néchet, Marion Le Texier, Romain Reuillon;
	\item the CybergeoNetworks team Arnaud Banos, Pierre-Olivier Chasset, Clémentine Cottineau, Hadrien Commenges, Denise Pumain;
	\item the PatentsMining team Antonin Bergeaud and Yoann Potiron;
	\item Antonin Bergeaud for EnergyPrice;
	\item Hélène Serra for the project on games for scientific mediation;
	\item Cinzia Losavio for migration dynamics in China;
	\item Solène Baffi for structural dynamics in South Africa;
	\item the Morphogenesis team Chenling Antelope, Lars Hubatsch, Jesus Mario Serna;
	\item Florent Le Néchet for Lutecia.
\end{itemize}
}{
Je remercie également mes co-auteurs et collaborateurs sur les différents projets reliés de près ou de loin à cette thèse :
\begin{itemize}
	\item l'équipe SpaceMatters Clémentine Cottineau, Florent Le Néchet, Marion Le Texier, Romain Reuillon ;
	\item l'équipe CybergeoNetworks Arnaud Banos, Pierre-Olivier Chasset, Clémentine Cottineau, Hadrien Commenges, Denise Pumain ;
	\item l'équipe de PatentsMining Antonin Bergeaud et Yoann Potiron ;
	\item Antonin Bergeaud pour EnergyPrice ;
	\item Hélène Serra pour le projet de communication scientifique ;
	\item Cinzia Losavio pour les dynamiques migratoires en Chine ;
	\item Solène Baffi pour les dynamiques structurelles en Afrique du Sud ;
	\item l'équipe Morphogenesis Chenling Antelope, Lars Hubatsch, Jesus Mario Serna ;
	\item Florent Le Néchet pour Lutecia.
\end{itemize}
}

\bpar{
I thank Benjamin Carantino for the joint organisation of the special session at ECTQG2017, and the invited participants Antonin Bergeaud, Clémentine Cottineau, Olivier Finance, Céline Rozenblat, Medhi Bida, Elfie Swerts, Denise Pumain.
}{
Je remercie Benjamin Carantino pour l'organisation conjointe de la session Eco-geo à ECTQG2017, et les participants invités Antonin Bergeaud, Clémentine Cottineau, Olivier Finance, Céline Rozenblat, Medhi Bida, Elfie Swerts, Denise Pumain, d'avoir accepté d'y participer.
}

\bpar{
I thank Céline Rozenblat, Luca D'Acci and Denise Pumain to have invited me to write several book chapters included in this thesis.
}{
Je remercie Céline Rozenblat, Luca D'Acci et Denise Pumain de m'avoir invité à rédiger divers chapitres d'ouvrage rendant compte de ce travail de thèse.
}

\bigskip


\bpar{
Learning is also learning to teach - I thank the pedagogical team of Paris 7 who made the teaching experience smooth. I also thank the motivated students.
}{
Apprendre c'est aussi apprendre à apprendre, et donc à enseigner. Je remercie les membres de l'équipe pédagogique de Paris 7 qui ont rendu cette expérience agréable, et pardonne ceux qui m'ont fait souffrir par psycho-rigidité. Dans les moments difficiles, la curiosité des élèves a été vraiment porteuse de sens, et je remercie tout ceux qui étaient motivés et qui ont aimé appréhender la multi-modélisation.
}

\bigskip

\bpar{
Research laboratories were also crucial in the success of this thesis. I thank the members of Géographie-cités and LVMT, and more particularly Thibault Le Corre, Julien Migozzi, Paul Gourdon, Pierre-Olivier Chasset, Daphnée Caillol, Mathieu Pichon, Anne-Cécile Ott, Flora Hayat, Anaïs Dubreuil, Laetita Verhaeghe, Ryma Hachi, Natalia Zdanowska, Cinzia Losavio, Eugenia Viana, Solène Baffi, Brenda Le Bigot, Olivier Finance, Julie Gravier, Lucie Nahassia, Robin Cura, Etienne Toureille, Thomas Louail, Clémentine Cottineau, Paul Chapron, Hadrien Commenges, François Queroy.
}{
Les laboratoires qui m'ont accueilli ont joué un rôle déterminant dans la réussite de cette thèse (malgré les difficultés récurrentes de financement qui laissent pessimiste sur l'avenir de la recherche publique). Je remercie les différents membres de Géographie-cités (Oven Street et Olympe) et du LVMT qui ont rendu l'accueil toujours chaleureux. Je remercie en particulier parmi les doctorants Thibault Le Corre pour le soutien intellectuel et logistique, Julien Migozzi pour le soutien théâtral, Paul Gourdon pour le soutien poétique, Pierre-Olivier Chasset pour le soutien informatique, Daphnée Caillol pour le soutien qualitatif, Mathieu Pichon pour le soutien épistémologique, Anne-Cécile Ott pour le soutien pédagogique, Flora Hayat pour le soutien cartographique, Anaïs Dubreuil pour le soutien alpin, Laetita Verhaeghe pour le soutien territorial, Ryma Hachi pour le soutien réticulaire, Natalia Zdanowska pour le soutien sportif, Cinzia Losavio pour le soutien ethnographique, Eugenia Viana pour le soutien moral ; les anciens Solène Baffi, Brenda Le Bigot, Olivier Finance, Julie Gravier, Lucie Nahassia, Robin Cura, Etienne Toureille ; les titulaires Thomas Louail, Clémentine Cottineau, Paul Chapron, Hadrien Commenges, François Queroy pour les discussions stimulantes ; et tous ceux que j'oublie.
}

\bigskip

\bpar{
I thank Joris, Mario, Marius for the scientific and friendship experiences.
}{
Je remercie Joris, Mario, Marius pour les expériences autant scientifiques que d'amitié, dédicace circulaire.
}

\bigskip

\bpar{
I thank Cinzia, Chenyi, Ming, Jing Jing, Xing et Meng for the Chinese dream.
}{
Je remercie Cinzia, Chenyi, Ming, Jing Jing, Xing et Meng pour l'expérience chinoise et la patience devant mes difficultés linguistiques.
}

\bpar{
Research is a life without forgetting one's life - I thank my friends who helped me keep one: Alexis, Emmanuel, les SFR, Antonin, Yoann, Maximilien, Simon, Arnaud, Hélène, Axel, Jonas, Nihal, Fabrice. I finally thank my family whose presence was essential.
}{
La recherche c'est une vie et malheureusement souvent oublier sa vie, je suis extrêmement reconnaissant à mes amis qui m'ont permis d'en garder un semblant : Alexis, Emmanuel, les SFR, Antonin, Yoann, Maximilien, Simon, Arnaud, Hélène, Axel, Jonas, Nihal, Fabrice. Je remercie (partiellement seulement, pour la quantité de vie injectée dans ce travail en conséquence) celle qui m'a laissé rapidement seul avec ce démon de thèse, et celles et ceux qui m'ont permis de me sentir moins seul par moments. Enfin je remercie ma famille dont la présence a été indispensable.
}


















\endgroup

\pagestyle{scrheadings} 



\refstepcounter{dummy}

\pdfbookmark[1]{\contentsname}{tableofcontents} 

\setcounter{tocdepth}{1} 

\setcounter{secnumdepth}{2} 


\begin{adjustwidth*}{-1cm}{-1cm}
\tableofcontents 
\end{adjustwidth*}

\automark[section]{chapter}
\renewcommand{\chaptermark}[1]{\markboth{\spacedlowsmallcaps{#1}}{\spacedlowsmallcaps{#1}}}
\renewcommand{\sectionmark}[1]{\markright{\thesection\enspace\spacedlowsmallcaps{#1}}}

\clearpage

\begingroup 
\let\clearpage\relax
\let\cleardoublepage\relax
\let\cleardoublepage\relax


\refstepcounter{dummy}
\pdfbookmark[1]{\listfigurename}{lof} 

\begin{adjustwidth*}{-1cm}{-1cm}

\listoffigures

\end{adjustwidth*}

\vspace{8ex}


\refstepcounter{dummy}
\pdfbookmark[1]{\listtablename}{lot} 


\begin{adjustwidth*}{-1cm}{-1cm}

\phantomsection

\listoftables

\end{adjustwidth*}
        
\vspace{8ex}
\newpage
    



       






\endgroup

\cleardoublepage

\pagenumbering{arabic} 

\cleardoublepage 




%
\ctparttext{}
\bpar{
\part*{Introduction}
}{
\part*{Introduction}
}
%


%
%
%
%

\chapter*{Introduction}

\markboth{Introduction}{Introduction}


\bigskip

\bpar{
\textit{Would the fog machine on the Saclay plateau be the only atemporal artefact in this metropolitan environment still searching for its own identity? Let's project us in 2100, in this southern suburb of what will still be Paris. Local transformations have indeed happened, but not in the expected way, the local climate being still fond of this well-known fog. However, the urban environment and the relation to the city are entirely conditioned by a proximity to structural transportation lines: the disappearance of fossil fuel transportation modes, then of all light vehicles through the technological failure of electric alternatives, have exacerbated the role of existing train or metro lines. Densities have progressively increased around stations to produce impressing tower compounds, whereas the peri-urban space became progressively empty. Concerning transportation infrastructures, they stayed quite at the identical after 2030, the few available resources being dedicated to their maintenance, and their extension became conjointly rapidly out of the political agendas. This plateau is therefore filled with abandoned buildings, since it still expects this line of the Grand Paris Express which finally would never have been realized. Nature progressively finds its way again.}
}{
\textit{La machine à brouillard du Plateau de Saclay serait-elle le seul artefact intemporel dans cet environnement métropolitain qui se cherche toujours ? Projetons nous en 2100, dans cette banlieue sud de ce qui sera toujours Paris. Les bouleversements locaux ont bien eu lieu, mais pas de la façon attendue, le climat local étant toujours féru de ce fameux brouillard. Par contre, l'environnement urbain et la relation à la ville sont entièrement conditionnés par une grande proximité aux lignes de transport lourd : la disparition des moyens de transport thermiques, puis de l'ensemble des véhicules légers par échec technologiques des alternatives électriques, ont exacerbé le rôle des lignes de train ou de metro existantes. Les densités ont progressivement augmenté autour des gares pour produire d'impressionnants complexes de tours, tandis que les espaces péri-urbains se vidaient progressivement. Les infrastructures de transport sont quant à elles restées quasiment à l'identique après 2030, le peu de ressources disponibles étant dédié à leur entretien, et leur extension étant conjointement sortie rapidement des agendas politiques. Ce plateau est alors rempli de bâtiments à l'abandon, puisqu'il attend toujours ce tronçon du Grand Paris Express qui n'aura finalement jamais été réalisé. La nature reprend peu à peu ses droits.}
}

\bpar{
This scenario for a low budget anticipation film has the advantage of revealing the existence of complex processes entangled at different space and time scales in the production of cities: the historical development of the railway network in the Parisian region conditioned the future evolutions, the RER B followed the old Ligne de Sceaux; the masterplan by \noun{Delouvrier} for regional development and its incomplete realization are elements explaining the structure of the Parisian public transportation network which strongly condition urban development in our scenario; relocation processes within the metropolitan space, related to a more or less strong need for proximity or accessibility depending on transportation modes used, play they role in the urban evolution; in the case of the plateau of Saclay specific planning processes at different levels play a crucial role in the differentiation of the territory.
}{
Ce pitch pour film d'anticipation à petit budget a pour avantage de nous révéler l'existence de processus complexes intriqués à différentes échelles de temps et d'espace dans la fabrique des villes : le développement historique du réseau ferroviaire en région parisienne a conditionné les évolutions futures, le RER B a suivi l'ancienne Ligne de Sceaux ; le plan de \noun{Delouvrier} pour le développement régional et son execution partielle sont des éléments d'explication de la structure du réseau parisien de transports en commun qui conditionnent fortement le développement urbain dans notre scenario ; les processus de relocalisation au sein de l'espace de la métropole, liés à une plus ou moins grande nécessité de proximité ou d'accessibilité selon les modes de transports utilisés, participent à l'évolution urbaine ; dans le cas du plateau de Saclay des processus de planification spécifiques à différents niveaux jouent un rôle crucial dans la différentiation du territoire.
}

\bpar{
The list could be much further developed, since each approach brings its mature vision related to a scientific body of knowledge in different disciplines such as geography, urban economics, transportation. This anticipation scenario is enough to give a glance on the complexity of territorial systems we will study. Our aim here is to dive within this complexity, and more particularly to give an original viewpoint on the study of relations between transportation networks and territories. The choice of this positioning will be largely discussed in a thematic part, and we now concentrate on the originality of the point of view we will take.
}{
La liste pourrait être ainsi continuée indéfiniment, chaque approche apportant sa vision mature correspondant à un corpus de connaissances scientifiques dans des disciplines diverses comme la géographie, l'économie urbaine, les transports. Ce scénario d'anticipation est suffisant pour faire ressentir la complexité des systèmes territoriaux que nous étudierons. Notre but ici est de se plonger dans cette complexité, et en particulier de donner un point de vue original sur l'étude des relations entre réseaux de transport et territoires. Le choix de cette position sera largement discuté dans une partie thématique, et nous nous concentrons à présent sur l'originalité du point de vue que nous allons prendre.
}

\section*{On General Positioning}{De la position générale}

\bpar{
\emph{The ambition of this thesis is to have no a priori ambition.} Such an introduction, although seeming rash, contains at all levels the implicit logics behind our research process. At the first degree, we try as much as possible to take a exploratory and constructive approach, as much on theoretical and methodological domains than thematic domain, but also proto-methodological (tools applying the method) : if unidimensional or integrated ambitions should emerge, they would be conditioned by the arbitrary choice of a time sampling among the continuity of the dynamic that structures any research project. In the structural sense, the self-reference that underlines an apparent contradiction points out the central aspect of reflexivity in our constructive approach, as much in the sense of the recursion of theoretical apparels, than for application of tools and methods developed to the work itself, or in the sense of the co-construction of the different approaches and of the different thematic axis. The processus of knowledge production can this way be understood as a metaphor of studied processes. Finally, from a point of view closer to the interpretation, it suggests the intention of a delicate positioning linking a political positioning which necessity is intrinsic to humanities (for example here against the technocratic application of models, or for the development of tools for an Open Science) with a rigor of objectivity coming more from other fields used, position that impose an increased prudence.
}{
\emph{L'ambition de cette thèse est de ne pas avoir d'ambition a priori.}
 Cette entrée en matière, rude en apparence, contient à différents niveaux les logiques sous-jacentes à notre processus de recherche. Au sens propre, nous nous plaçons tant que possible dans une démarche constructive et exploratoire, autant sur les plans théorique et méthodologique que thématique, mais encore proto-méthodologique (outils appliquant la méthode) : si des ambitions unidimensionnelles ou intégrées devaient émerger, elles seraient conditionnées par l'arbitraire choix d'un échantillon temporel parmi la continuité de la dynamique qui structure tout projet de recherche. Au sens structurel, l'auto-référence qui soulève une contradiction apparente met en exergue l'aspect central de la réflexivité dans notre démarche constructive, autant au sens de la récursivité des appareils théoriques, de celui de l'application des outils et méthodes développés au travail lui-même ou que de celui de la co-construction des différentes approches et des différents axes thématiques. Le processus de production de connaissance pourra ainsi être lu comme une métaphore des processus étudiés. Enfin, sur un plan plus enclin à l'interprétation, cela suggérera la volonté d'une position délicate liant une conscience politique dont la nécessité est intrinsèque aux sciences humaines (par exemple ici contre l'application technocratique des modèles, ou pour le développement d'outils luttant pour une science ouverte) à une rigueur d'objectivité plus propre aux autres champs abordés, position forçant à une prudence accrue.
}


\section*{Scientific context : paradigms of complexity}{Contexte scientifique : paradigmes de la complexité}


\bpar{
To better introduce our subject, it is necessary to develop the scientific context we will be integrated into. This context is crucial both to understand the general epistemology underlying research questions, and to be aware of the variety of methods and tools used.
}{
Pour une meilleure introduction du sujet, il est nécessaire d'insister sur le cadre scientifique dans lequel nous nous positionnons. Ce contexte est crucial à la fois pour comprendre les concepts épistémologiques implicites dans nos questions de recherche, et aussi pour être conscient de la variété de méthodes et outils utilisés.
}

\bpar{
Contemporaneous science is progressively taking the shift of complexity in many fields that we will illustrate in the following, what implies an epistemological mutation to abandon strict reductionism\footnote{In a schematic way, reductionism consists in the epistemological positioning that systems are entirely understandable from the fundamental elements they are constituted of and from the laws driving their evolution. Superior levels have neither an autonomy nor irreducible causal powers.} that failed in most of its synthesis attempts~\cite{anderson1972more}. \cite{arthur2015complexity} recently recalled that a mutation of methods and paradigms was also at stake, through the increasing role of computational approaches replacing purely analytical techniques generally limited in their modeling and resolution scope. Capturing \emph{emergent properties} in models of complex systems is one of the ways to understand the essence of these approaches.
}{
La science contemporaine prend progressivement le tournant de la complexité dans de nombreux champs que nous illustrerons par la suite, ce qui implique une mutation épistémologique pour abandonner le réductionnisme\footnote{De manière schématique, le réductionnisme consiste en la position épistémologique que les systèmes sont entièrement compréhensibles à partir des éléments fondamentaux les constituant et des lois régissant leur évolution. Les niveaux supérieurs n'ont ni autonomie ni pouvoirs causaux irréductibles.} strict qui a échoué dans la majorité de ses tentatives de synthèse~\cite{anderson1972more}. \cite{arthur2015complexity} a rappelé récemment qu'une mutation des méthodes et paradigmes en était également un enjeu, par la place grandissante prise par les approches computationnelles qui remplacent les résolutions purement analytiques généralement limitées en possibilités de modélisation et de résolution. La capture des \emph{propriétés émergentes} par des modèles de systèmes complexes est une des façons d'interpréter la philosophie de ces approaches.
}

\bpar{
These considerations are well known in Social Sciences and Humanities (both quantitative and qualitative), for which the complexity of studied agents and systems is one of the justifications of their existence: if humans were indeed particles, we could expect that most fields studying them would have never emerged, as thermodynamics would have solved most of social issues. \footnote{Even if this affirmation can also be discussed, since classical physics also failed in their attempts to include irreversibility and evolutions of Complex Adaptive Systems as \cite{prigogine1997end} points out.}. They are however less known nor accepted in more ``hard'' sciences such as physics: \cite{laughlin2006different} develops a view of physics at a similar position of a ``frontier of knowledge'' compared to other more recent fields that could appear as being still in their genesis. Most of knowledge concerns classical simple structures, whereas a large number of systems appear as \emph{self-organized}, in the sense that the single microscopic laws are not enough to determine macroscopic properties unless system evolution is entirely simulated (more precisely this view can be taken as a definition of emergence on which we will come back later, and self-organized properties are indeed emergent). This corresponds to the first nightmare of Laplace's Deamon developed in~\cite{deffuant2015visions}.
}{
Ces considérations sont bien connues des Sciences Humaines et Sociales (qualitatives et quantitatives) pour lesquelles la complexité des agents et systèmes étudiés est une des justifications de leur existence : si les humains étaient effectivement des particules, on pourrait s'attendre à ce que la majorité des disciplines les prenant comme objet d'étude n'aient jamais émergé puisque la thermodynamique aurait alors résolu la majorité des problèmes sociaux\footnote{Bien que cette affirmation soit elle-même discutable, les sciences physiques classiques ayant également échoué à prendre en compte l'irréversibilité et l'évolution de systèmes complexes adaptatifs comme le soulignent~\cite{prigogine1997end}.}. Elles sont au contraire moins connues et acceptées en sciences ``dures'' comme la physique : \cite{laughlin2006different} développe une vision de la physique à la même position de ``frontière des connaissances'' que d'autres champs plus récents qui pourraient sembler en être encore à leur genèse. La plupart des connaissances actuelles concernent des structures classiques simples, alors qu'un grand nombre de systèmes présentent des propriétés \emph{d'auto-organisation}, au sens où les lois microscopiques ne sont pas suffisantes pour inférer les propriétés macroscopiques du système à moins que son évolution ne soit entièrement simulée (plus précisément cette vision peut être prise comme une définition de l'émergence sur laquelle nous reviendrons par la suite, or des propriétés auto-organisées sont par nature émergentes). Cela correspond au premier cauchemar du Démon de Laplace développé dans~\cite{deffuant2015visions}. 
}


\bpar{
At the crossroads of epistemological positions, methods, and fields of applications, sciences of \emph{Complexity} focus on the importance of emergence and self-organization in most of phenomena of the real world, which make it lie closer to a frontier of knowledge closer than we can imagine for classical disciplines \cite{laughlin2006different}. These concepts are indeed not recent and had already been shown by~\cite{anderson1972more}. We can also interpret Cybernetics as a precursor of Complexity Sciences, by reading it as a bridge between technology and cognitive sciences~\cite{wiener1948cybernetics}, and moreover by developing the notions of feedback and control.
}{
A la croisée de positionnements épistémologiques, de méthodes et de champs d'application, les \emph{Sciences de la complexité} se concentrent sur l'importance de l'émergence et de l'auto-organisation dans la plupart des phénomènes réel, ce qui les place plus proche de la frontière des connaissances que ce que l'on peut penser pour des disciplines classiques \cite{laughlin2006different}. Ces concepts ne sont pas récents et avaient déjà été mis en valeur par~\cite{anderson1972more}. On peut aussi interpréter la Cybernétique comme un précurseur des Sciences de la Complexité en la lisant comme un pont entre technologie et sciences cognitives~\cite{wiener1948cybernetics}, et surtout en développant les notions de rétroaction et de contrôle.
}

\bpar{
Later, Synergetics~\cite{haken1980synergetics} paved the way for a theoretical approach of collective phenomena in physics. Possible reasons for the recent growth of works claiming a complexity approach are numerous. The explosion of computing power is surely one of these because of the central role of numerical simulations~\cite{varenne2010simulations}. They could also be related to epistemological progresses: introduction of the notion of perspectivism~\cite{giere2010scientific}, finer reflexions around the nature of models~\cite{varenne2013modeliser}\footnote{In that frame scientific and epistemological progresses can not be dissociated and can be seen as co-evolving, in the sense of a strong interdependency and a mutual adaptation}.
}{
Plus tard, la Synergétique~\cite{haken1980synergetics} a posé les bases d'approches théoriques des phénomènes collectifs en physique. Les causes possibles de la croissance récente du nombre de travaux se réclamant d'approches complexes sont nombreuses. L'explosion de la puissance de calcul en est certainement une vu le rôle central que jouent les simulations numériques~\cite{varenne2010simulations}. Elles peuvent aussi être à chercher auprès de progrès en épistémologie : introduction de la notion de perspectivisme~\cite{giere2010scientific}, reflexions plus fine autour de la nature des modèles~\cite{varenne2013modeliser}\footnote{Dans ce cadre, les progrès scientifiques et épistémologiques ne peuvent pas être dissociés et peuvent être vus comme étant en co-évolution, au sens d'une forte interdépendance et d'une adaptation mutuelle.}.
}

\bpar{
The theoretical and empirical potentialities of such approaches play surely a role in their success\footnote{Although the adoption of new scientific practices may be strongly biased by imitation and lack of originality~\cite{dirk1999measure}, or in a more ambivalent way, by marketing strategies independent of knowledge strategies, as the fight for funds is becoming a huge obstacle for research~\cite{bollen2014funding}.}, as confirmed by the various domains of application (see~\cite{newman2011complex} for a general survey), as for example Network Science~\cite{barabasi2002linked}; Neuroscience~\cite{koch1999complexity}; Social Sciences including Geography~\cite{manson2001simplifying,pumain1997pour}; Finance with econophysics approaches~\cite{stanley1999econophysics}; Ecology~\cite{grimm2005pattern}. The Complex Systems Roadmap~\cite{2009arXiv0907.2221B} proposes a double entry to studies on Complex Systems: an horizontal approach connecting fields of study with transversal questions on theoretical foundations of complexity and empirical common stylized facts, and a vertical approach to disciplines, with the aim at constructing integrated disciplines and corresponding multi-scale heterogeneous models. Interdisciplinarity is thus central in our scientific background.
}{
Les potentialités théoriques et empiriques de telles approchent jouent nécessairement un rôle dans leur succès\footnote{Même si l'adoption de nouvelles pratiques scientifiques peut par ailleurs être biaisé par l'imitation et le manque d'originalité~\cite{dirk1999measure}, ou de façon plus ambivalente, par des stratégies de positionnement indépendante des stratégies de connaissance, puisque le combat pour les fonds est un obstacle croissant à une recherche saine~\cite{bollen2014funding}.}, comme le confirme les domaines très variés d'application (voir~\cite{newman2011complex} pour une revue très générale), comme par exemple la Science des Réseaux~\cite{barabasi2002linked}; les Neurosciences~\cite{koch1999complexity}; les Sciences Humaines et Sociales,  dont la Géographie~\cite{manson2001simplifying,pumain1997pour}; la Finance avec les approches éconophysiques~\cite{stanley1999econophysics}; l'Ecologie~\cite{grimm2005pattern}. La Feuille de Route des Systèmes Complexes~\cite{2009arXiv0907.2221B} propose une double lecture des travaux en Complexité: une approche horizontale faisant la connexion entre champs d'étude par des questions transversales sur les fondations théoriques de la complexité et des faits stylisés empiriques communs, et une approche verticale, dans le but de construire des disciplines intégrées et les modèles multi-scalaires hétérogènes correspondants. L'interdisciplinarité est ainsi cruciale pour notre contexte scientifique.
}

\section*{Interdisciplinarity}{Interdisciplinarité}

\bpar{
We must further insist on the role of interdisciplinarity in the research positioning taken here. This is as much a work in Theoretical and Quantitative Geography than in Complex Systems Modeling, being finally both depending on the point of view taken by the reader. In that sense, we claim it to belong to \emph{Complex Systems Science} that we aim at positioning as a proper discipline through this precise implementation\footnote{An abstract level of reading of the work in its entirety will bring informations on knowledge production itself, as we will develop in~\ref{sec:knowledgeframework}.}. There are risks of being read with mistrust or even defiance by scholars of various concerned disciplines, as recent examples of misunderstandings and conflicts have illustrated~\cite{dupuy2015sciences}. We need to recall the importance of \noun{Banos}' virtuous circle between disciplinarity and interdisciplinarity~\cite{banos2013pour}. It must necessarily imply different scientific agents, and it is complicated for an agent to be positioned in the two branches; our scientific background will have to allow us to not be positioned only within \emph{geographical disciplinarity} (even if it will simultaneously be a crucial component) but as much within Complex Systems (which is interdisciplinary, see~\ref{sec:epistemology} to go beyond the apparent contradiction), and our scientific and epistemological sensitivity leads us to do the same.
}{
Il est important d'insister sur le rôle de l'interdisciplinarité dans la position de recherche prise ici. Il s'agit autant d'un travail en Géographie Théorique et Quantitative qu'en Modélisation de Systèmes Complexes, étant finalement les deux à la fois selon le point de vue que prendra le lecteur. En ce sens, nous le réclamons de la \emph{Science des Systèmes Complexes} que nous tenterons de positionner comme discipline propre à travers cette implémentation précise\footnote{Un niveau de lecture abstrait du travail dans son ensemble apportera des informations sur la production de connaissance elle-même, comme nous le développerons en~\ref{sec:knowledgeframework}.}. Ce n'est pas sans risques d'être lu avec méfiance voir défiance par les tenants des disciplines classiques, comme des exemples récents de malentendus ou conflits ont récemment illustré~\cite{dupuy2015sciences}. Il faut se rappeler l'importance de la spirale vertueuse de \noun{Banos} entre disciplinarité et interdisciplinarité~\cite{banos2013pour}. Celle-ci doit nécessairement impliquer différents agents scientifiques, et il est compliqué pour un agent de se positionner dans les deux branches ; notre fond scientifique devra nous permettre de ne pas de nous positionner uniquement dans la \emph{disciplinarité géographique} (même si celle-ci sera simultanément une composante cruciale) mais bien aussi dans celle des Systèmes Complexes (qui est interdisciplinaire, voir~\ref{sec:epistemology} pour contourner la contradiction apparente), et notre sensibilité scientifique et épistémologique nous pousse à faire de même.
}

\bpar{
The scientific evolution of complexity sciences, that some see as a revolution~\cite{colander2003complexity}, or even as \emph{a new kind of science}~\cite{wolfram2002new}, could indeed face intrinsic difficulties due to behaviors and a-priori of researchers as human beings. More precisely, the need for interdisciplinarity which makes the strength of Complexity Science may be one of its greatest weaknesses, since the highly partitioned structure of the organization of science may have negative impacts on works involving different disciplines. We do not tackle the issue of over-publication, quantification, competition, which is more linked to a question of Open Science and its ethics, also of high importance but of an other nature. That barrier haunting us and that we might struggle to triumph of, has as the most obvious symptom \emph{cultural disciplinary differences}, and resulting opinion conflicts. The drama of scientific misunderstandings is that they can indeed totally annihilate progresses by interpreting as a falsification some works that answer a totally different question.
}{
L'évolution scientifique des sciences de la complexité, qui est vue par certains comme une révolution~\cite{colander2003complexity}, ou même comme \emph{un nouveau type de science}~\cite{wolfram2002new}, pourrait affronter des difficultés intrinsèques dues aux comportements et a-priori des chercheurs en tant qu'êtres humains. Plus précisément, le besoin d'interdisciplinarité qui fait la force des Sciences de la Complexité pourrait devenir une de ses grandes faiblesses, puisque la structure fortement en silo de la science peut avoir des impacts négatifs sur les initiatives impliquant des disciplines variées. Nous n'évoquons pas les problèmes de sur-publication, quantification, compétition, qui sont plus liés à des questions de Science Ouverte et de son éthique, de toute aussi grande importance mais d'une autre nature. Cette barrière qui nous hante et que nous pourrions ne pas surmonter, a pour plus évident symptôme des \emph{divergences culturelles disciplinaires}, et les conflits d'opinion en résultant. Ce drame du malentendu scientifique est d'autant plus grave qu'il peut en effet détruire totalement certains progrès en interprétant comme une falsification des travaux qui traitent une question toute différente.
}

\bpar{
The recent example in economics of a work on top-income inequalities presented in~\cite{aghion2015innovation}, which conclusions are presented as opposed to the ones obtained by~\cite{piketty2013capital}, is typical of this scheme. The latest focuses on the construction of long-time clean databases for income data and shows empirically a recent acceleration of income inequalities, his simple model aiming to link this stylized fact with the accumulation of capital has been criticized as oversimplified. On the other hand, \cite{aghion2015innovation} show with econometric analyses that there indeed exist a causality link from innovation to top-income inequalities, the innovation however increasing social mobility, being thus also a driver of inequalities reductions. Therefore do they obtain divergent conclusion on the role of capitals in an economy, in particular on their ambiguous relation to innovation. But diverging \emph{points of view} or \emph{interpretations} do not imply a scientific incompatibility, and one could even imagine to try gathering both approaches in an unified framework and model, yielding possibly similar or different interpretations. Such an integrated approach will have chances to contain more information (depending on how coupling is done) and to be a scientific progress.
}{
L'exemple récent en économie d'un travail sur les inégalités liées aux hauts revenus présenté dans~\cite{aghion2015innovation}, et dont les conclusions ont été commentées comme s'opposant aux thèses de~\cite{piketty2013capital}, est typique de ce schéma. Ce second se concentre sur la construction de bases de données propres sur le temps long pour les revenus et montre empiriquement une récente accélération des inégalités de revenus, son modèle visant à lier ce fait stylisé avec l'accumulation de capital a été critiqué comme trop simpliste. D'autre part, \cite{aghion2015innovation} montrent par des analyses économétriques que s'il existe bien un lien de causalité de l'innovation vers les inégalités de haut salaires, l'innovation accroit cependant la mobilité sociale, étant donc également moteur de réduction des inégalités. D'où des conclusions divergentes sur le rôles des capitaux personnels dans une économie, notamment sur leur relation ambigüe à l'innovation. Mais des \emph{point de vue} ou \emph{interprétations} différentes ne signifient pas une incompatibilité scientifique, et on pourrait même imaginer rassembler ces deux approches dans un cadre et modèle unifié, produisant des interprétations possiblement similaires et potentiellement encore nouvelles. Une telle approche intégrée aura de grandes chances de contenir plus d'information (selon la façon dont le couplage est opéré) et d'être une avancée scientifique.
}

\bpar{
This thought experiment illustrates the potentialities and the necessity of interdisciplinarity. In an other but similar vein, \cite{2017arXiv170105627H} reanalyses biological data from a 1943 experiment that claimed to rule out Lamarckian over Darwinian evolution processes, and show that the conclusions do not hold in the current context of data analysis (enormous advances in theoretical and processing techniques) and scientific context (with numerous other proofs today of Darwinian processes): this is a good example of a misunderstanding on the context and how conclusions strongly depend on both technical and thematic frameworks. We shall now briefly develop other examples to give an overview how conflicts between disciplines can be damaging.
}{
Cette expérience de pensée illustre les potentialités et la nécessité de l'interdisciplinarité. Dans une autre veine assez similaire, \cite{2017arXiv170105627H} ré-analyse des données biologiques d'une expérience de 1943 qui prétendait confirmer l'hypothèse des processus d'évolution Darwiniens par rapport aux processus Lamarckiens, et montrent que les conclusions ne tiennent plus dans le contexte actuel d'analyse de données (avances énormes sur la théorie et les possibilités de traitement) et scientifique (avec d'autres nombreuses preuves de nos jours des processus Darwiniens) : c'est un bon exemple de malentendu sur le contexte, et la manière selon laquelle le cadre de travail à la fois technique et thématique influence fortement les conclusions scientifiques. Nous développons à présent divers exemples révélateurs de la manière dont des conflits entre disciplines peuvent être dommageables.
}

\bpar{
As already mentioned, \noun{Dupuy} and \noun{Benguigui} point out in \cite{dupuy2015sciences} the fact that in the field of urban studies, have recently appeared open conflicts between classical heres of dicsciplines and new incomers, in particular physicists, even if their entry in this domain is not new. The availability of large datasets for new types of data (social networks, data from new information and communication technologies) have drawn an increased attention towards the study of objects traditionally studied by human sciences, as analytical and computational methods of statistical physics became applicable. Although these studies are generally presented as the construction of a scientific approach to cities, discussing the scientific character of existing approach, the effective novelty of the results obtained and the discredit of ``classical'' approaches are discussable. To give a few examples, \cite{barthelemy2013self} conclude that Paris has followed a transition during the Haussman period and it global planning operations, which are well-known facts for a long time in urban history and urban geography. \cite{chen2009urban} rediscovers that the gravity model can be improved by adding lags in interactions and theoretically derives the expression of the force of interaction between cities, without any thematic theoretical or themartical background. Similar examples could be multiplied, confirming the current discomfort between physicists and urban geographers. Significant benefices could results from a wise integration of disciplines~\cite{o2015physicists} but the road seems to be still long. 
}{
Comme déjà mentionné, \noun{Dupuy} et \noun{Benguigui} soulignent dans~\cite{dupuy2015sciences} le fait que dans le domaine de l'urbanisme, ont récemment éclaté des conflits ouverts entre les tenants classiques des disciplines et des nouveaux arrivants, en particulier les physiciens, même si leur entrée dans le domaine n'est pas nouvelle. La disponibilité de grands jeux de données d'un nouveau type (réseaux sociaux, données des nouvelles technologies de la communication) ont attiré l'attention d'un plus grand nombre sur des objets plus traditionnellement étudiés par les sciences humaines, puisque les méthodes analytiques et computationnelles de la physique statistique sont devenues applicables. Bien que ces travaux soient généralement présentés comme la construction d'une approche scientifique des villes, tout en discutant la nature scientifique des approches existantes, la nouveauté réelle des résultats obtenus et la non-légitimation des approches ``classiques'' sont discutables. Pour citer quelques exemples, \cite{barthelemy2013self} concluent que Paris a subit une transition pendant la période d'Haussman et ses opérations de planification globale, qui sont des faits naturellement connus depuis longtemps en Histoire Urbaine et Géographie Urbaine. \cite{chen2009urban} redécouvre que le modèle gravitaire est amélioré par l'introduction de décalages dans les interactions et dérive analytiquement l'expression d'une force d'interaction entre les villes, sans se placer dans un cadre théorique ou thématique. De tels exemples peuvent être multipliés, confirmant l'inconfort courant entre physiciens et géographes. Des bénéfices significatifs pourraient résulter d'une intégration raisonnée des disciplines~\cite{o2015physicists} mais la route semble être bien longue encore.
}


\bpar{
Similar conflict can be found at the interface of relations between economics and geography: as \cite{marchionni2004geographical} describes, the discipline of geographical economics, traditionally close to geography, has heavily criticized at its emergence the relatively recent approach of the \emph{New Economic Geography}. This approach comes from economics and its purpose is to take space into account in classical economic methods. They have indeed not the same purposes and intentions, and the conflict appears as a complete misunderstanding when seen from an external eye. For exemple, the New Economic Geography will privilege explications that imply universal economic processes and independent of scales, whereas Geographical Economics will base its arguments on local particularities and the contingency of processes. Underlying epistemological assumptions are also very different, such as for exemple the relation to realism, the first being founded on an abstract realism which is not necessary concretely realistic (use of abstract processes), whereas the second will be more pragmatic. The extent in which these approaches are complementary or incompatible remains however an open question according to \cite{marchionni2004geographical}. Similar disciplinary relations will be encountered in our work, such as between physics and geography. We furthermore illustrate this question in~\ref{app:sec:ecogeo} by an exploration of links between economics and geography from the point of view of modeling.
}{
Des conflits similaires se rencontrent à l'interface des relations entre économie et géographie : comme le décrit \cite{marchionni2004geographical}, la discipline de la géographique économique, traditionnellement proche de la géographie, a fortement critiqué à son émergence l'approche relativement récente de la \emph{Nouvelle Economie Géographique}. Celle-ci provient de l'économie et son but est la prise en compte de l'espace par les méthodes économiques classiques. Elles n'ont en fait pas les mêmes desseins et buts, et le conflit apparaît comme un malentendu complet vu d'un oeil extérieur. Par exemple, la Nouvelle Economie Géographique privilégiera des explications impliquant des processus économiques universels et indépendant des échelles, tandis que la Géographie Economique basera son argumentation sur les particularité locales et la contingence des processus. Les hypothèses épistémologiques sous-jacentes sont également très différentes, comme par exemple la relation au réalisme, la première étant fondée sur un réalisme abstrait pas forcément concrètement réaliste (utilisation de processus abstraits), tandis que la deuxième sera plus pragmatique. La mesure dans laquelle ces deux approches sont complémentaires ou incompatibles reste toutefois une question ouverte d'après \cite{marchionni2004geographical}. Des relations disciplinaires similaires seront rencontrées dans notre travail, comme entre la physique et la géographie. Nous illustrons par ailleurs cette question en~\ref{app:sec:ecogeo} par une exploration des liens entre économie et géographie du point de vue de la modélisation.
}


\bpar{
Disciplinary conflicts may also emerge under the form of a reject of novel methods by dominating currents. According to \cite{farmer2009economy}, the operational failure of most classic economic approaches could be compensated by a broader use of agent-based modeling and simulation practices. The lack of analytical resolution which is inevitable for the study of most complex adaptive systems, seen to repel most of economists. However, \cite{barthelemy2016structure} insists on the exacerbated non-connection between numerous economic models and theories and empirical observation, at least in the field of urban economics. This could be a symptom of the disciplinary non-connection evoked above. Still in economics, \cite{storper2009rethinking} also propose paradigms shifts for a return to the agent and an associated construction of \emph{evidence-based theories}.
}{
Des conflits disciplinaires peuvent aussi se manifester sous la forme d'un rejet de méthodes nouvelles par les courants dominants. Suivant \cite{farmer2009economy}, l'échec opérationnel de la plupart des approches économiques classiques pourrait être compensé par un usage plus systématique de la modélisation et simulation basées sur les agents. L'absence de résolution analytique qui est inévitable pour l'étude de la plupart des systèmes complexes adaptatifs semble rebuter une grande partie des économistes. Or, \cite{barthelemy2016structure} insiste sur la déconnexion exacerbée entre de nombreux modèles et théories économiques et les observations empiriques, du moins dans le domaine de l'économie urbaine. Celle-ci pourrait être un symptôme de la déconnexion disciplinaire évoquée ci-dessus. Toujours en économie, \cite{storper2009rethinking} proposent aussi des changements de paradigmes par un retour à l'agent et une construction associée de théories \emph{evidence-based}.
}


\bpar{
Quantitative finance can be instructive for our purpose and subject, through the similarities of its interdisciplinary kitchen with our domain (relations with physics and economics, fields more or less ``rigorous'', etc.). In this domain coexist various fields of research having very few interactions between them. We can consider two example. On the one hand, statistics and econometrics are highly advanced in theoretical mathematics, using for example stochastic calculus and probability theory to obtain very refined estimators of parameters for a given model (see e.g. \cite{barndorff2011multivariate}). On the other hand, Econophysics aims at studying empirical stylized facts and infer empirical laws to explain economic phenomena, for example the ones linked to complexity of financial markets~\cite{stanley1999econophysics}. They include cascades leading to market crashes, fractal properties of asset signals, complex structure of correlation networks. Both have their advantages in a particular context and each would benefit from increased interactions between the fields.
}{
La finance quantitative peut être instructive pour notre propos et sujet, de par les similarités de la cuisine interdisciplinaire avec notre domaine (rapport avec la physique et l'économie, champs plus ou moins ``rigoureux'', etc.). Dans ce domaine coexistent divers champs de recherche ayant très peu d'interactions entre eux. On peut considérer deux exemples. D'une part, les statistiques et l'économétrie sont extrêmement avancées en mathématiques théoriques, utilisant par exemple des méthodes de calcul stochastique et de théorie des probabilités pour obtenir des estimateurs très raffinés de paramètres pour un modèle donné (voir par exemple \cite{barndorff2011multivariate}). D'autre part, l'éconophysique a pour but d'étudier des faits stylisés empiriques et inférer les lois correspondantes pour tenter d'expliquer des phénomènes économiques, par exemple ceux liés à la complexité des marchés financiers~\cite{stanley1999econophysics}. Ceux-ci incluent les cascades menant aux ruptures de marché, les propriétés fractales des signaux des actifs, la structure complexe des réseaux de corrélation. Chacun a ses avantages dans un contexte particulier et gagnerait à des interactions accrues entre les deux domaines.
}

\bpar{
These diverse examples caught in the wind give short illustrations of how crucial interdisciplinarity is and how it is difficult to achieve. Without being close to exaggerating, we could imagine all researchers complaining about bad or difficult experiences in interdisciplinarity, with a largely positive return in the rare cases of a success. We will in the following try to follow that narrow path, borrowing ideas, theories and methods from diverse disciplines, in the spirit of the construction of an integrated knowledge.
}{
Ces divers exemples pris au fil du vent sont de brèves illustrations du caractère crucial de l'interdisciplinarité et de la difficulté à la pratiquer. Sans presque exagérer, on pourrait imaginer l'ensemble des chercheurs se plaindre de mauvaises ou difficiles expériences d'interdisciplinarité, avec un retour largement positif lors des rares succès. Nous allons tenter par la suite d'emprunter ce chemin étroit, empruntant des idées, théories et méthodes de diverses disciplines, dans l'idéal de la construction d'une connaissance intégrée.
}


\section*{Complexity paradigms in Geography}{Paradigmes de la Complexité en Géographie}

\bpar{
Coming back to our introducing anecdote, we will focus on the study of a thematic object that will be territorial systems: at the microscopic scale, agents can indeed be seen as fundamental elements constituting the territory, which will emerge as a complex process at different scales. More generally, we propose to begin with sketching an overview of the role of complexity in geography. Geographers are naturally familiar with complexity, since the study of spatial interactions is one of their preferred object. The variety of fields in geography (geomorphology, physical geography, environmental geography, human geography, health geography, etc. to give a few) has certainly played a key role in the constitution of a subtle geographical thinking, which considers heterogeneous and multi-scalar processes.
}{
Pour revenir à notre anecdote introductive, nous nous concentrons sur l'étude d'un objet thématique qui sera les systèmes territoriaux : à l'échelle microscopique, les agents peuvent bien être vus comme éléments constitutifs fondamentaux du territoire, qui émergera comme processus complexe à différentes échelles. Plus généralement, il s'agit par commencer de brosser une revue du rôle de la complexité en géographie. Les géographes sont naturellement familiers avec la complexité, puisque l'étude des interactions spatiales est l'un de leurs objets de prédilection. La variété de champs en géographie (géomorphologie, géographie physique, géographie environnementale, géographie humaine, géographie de la santé, etc. pour en nommer certains) a sûrement joué un rôle clé dans la constitution d'une pensée géographique subtile, qui considère des processus hétérogènes et multi-scalaires.
}

\bpar{
\cite{pumain2003approche} gives a subjective history of the emergence of complexity paradigms in geography, that we synthesize here. Cybernetics yielded system theories such as the one used for first system dynamics models aiming at simulating the evolution of variables characterizing a territory, under the form of coupled differential equations, as \cite{chamussy1984dynamique} illustrate for a model coupling population, employments and housing stock. Later, the shift towards concepts of self-organized criticality and self-organisation in physics lead to corresponding developments in geography, as \cite{sanders1992systeme} which witnesses the application of concepts from synergetics to the dynamics of urban systems.
}{
\cite{pumain2003approche} donne une histoire subjective de l'émergence des paradigmes de la complexité en géographie, que nous restituons ici. La cybernétique a produit des théories des systèmes comme celle utilisée pour les premiers modèles de dynamique des systèmes visant à simuler l'évolution de variables caractérisant un territoire, sous la forme d'équations différentielles couplées, comme \cite{chamussy1984dynamique} l'illustrent pour un modèle couplant population, emplois et stock de logements. Plus tard, le glissement vers les concepts de criticalité auto-organisée et d'auto-organisation en physique ont conduit aux développements correspondants en géographie, comme~\cite{sanders1992systeme} qui témoigne de l'application des concepts de la synergétique aux dynamiques des systèmes urbains.
}

\bpar{
Finally, current paradigms of complex systems have been introduced through several relatively independent entries. We can exhibit among them concepts from fractals, cellular automatons, \emph{Scaling} concepts, and the evolutive urban theory. We briefly review these approaches below.
}{
Enfin, les paradigmes actuels des systèmes complexes ont été introduits par plusieurs entrées relativement indépendantes. On peut nommer parmi celles-ci les concepts issus des fractales, les automates cellulaires, le \emph{Scaling}, et la théorie évolutive des villes. Nous revoyons brièvement ces approches ci-dessous.
}

\bpar{
The study of the fractal nature of urban form was introduced by~\cite{batty1994fractal}, has been later syntesized by~\cite{batty1994fractal} and had numerous application including more recent developments such as~\cite{keersmaecker2003using} for analyzing the urban form or \cite{tannier:hal-00860260} for the conception of sustainable urban planning.
}{
L'étude de la nature fractale de la forme urbaine a été introduite par~\cite{batty1986fractal}, plus tard synthétisée par~\cite{batty1994fractal} et a eu de nombreuses applications jusqu'à des développements plus récents comme~\cite{keersmaecker2003using} pour l'analyse de la forme urbaine ou~\cite{tannier:hal-00860260} pour l'élaboration de planifications urbaines durables.
}


\bpar{
The theory of \emph{Scaling} has furthermore been imported from physics and biology (allometric relations) to explain urban scaling laws as universal properties linked to the type of activity: infrastructure and economies of scale (infralinear scaling) or resulting from a process of social interactions (supralinear scaling), and assumes cities as scaled versions of each other~\cite{bettencourt2007growth}. We will not explicitly use these two approaches but they remain underlying in the paradigms we will use\footnote{For example, scaling laws have a privileged role in the application of the evolutive urban theory~\cite{pumain2006evolutionary}.}.
}{
La théorie du \emph{Scaling} a par ailleurs été importée de la physique et de la biologie (relations allométriques) pour expliquer les lois d'échelle urbaines comme propriétés universelles liées au type d'activité : infrastructure et économies d'agglomération (\emph{scaling} infralinéaire) ou résultante d'un processus d'interactions sociales (\emph{scaling} supralinéaire), et suppose les villes comme versions à l'échelle l'une de l'autre~\cite{bettencourt2007growth}. Nous n'utiliserons pas explicitement ces deux approches mais celles-ci restent sous-jacentes dans les paradigmes utilisés\footnote{Par exemple, les lois d'échelles ont une place privilégiée dans l'application de la théorie évolutive des villes~\cite{pumain2006evolutionary}.}.
}

\bpar{
Cellular automatons, introduced in geography by \noun{Tobler}~\cite{couclelis1985cellular}, are an other entry of complex approaches for urban modeling. \noun{Batty} proposes a joint synthesis of it with agent-based models and fractals in~\cite{batty2007cities}. This type of model will take a modest but not negligible place in our work.
}{
Les automates cellulaires, introduits en géographie par \noun{Tobler}~\cite{couclelis1985cellular}, sont une autre entrée des approches complexes pour la modélisation urbaine. \noun{Batty} en propose une synthèse jointe avec les modèles basés agents et les fractales dans~\cite{batty2007cities}. Ce type de modèle prendra une place modeste mais non négligeable dans notre travail.
}

\bpar{
An other incursion of complexity in geography was for the case of urban systems through the evolutive urban theory of \noun{Pumain}. We will position more particularly within its heritage and will develop it with more details. In close relation with modeling from the beginning (the first Simpop model described in~\cite{sanders1997simpop} enters the theoretical framework of \cite{pumain1997pour}), this theory aims at understanding systems of cities as systems of co-evolving adaptive agents, interacting in many ways, with particular features emphasized such as the importance of the diffusion of innovations.
}{
Une autre introduction de la complexité en géographie fut pour le cas des systèmes urbains à travers la théorie évolutive des villes de \noun{Pumain}. Nous nous placerons plus particulièrement dans la lignée de celle-ci et la développons ainsi avec plus de détails. En interaction intime avec la modélisation dès ses débuts (le premier modèle Simpop décrit par~\cite{sanders1997simpop} rentre dans le cadre théorique de~\cite{pumain1997pour}), cette théorie vise à comprendre les systèmes de villes comme des systèmes d'agents adaptatifs en co-évolution, aux interactions multiples, avec différents aspects mis en valeur comme l'importance de la diffusion des innovations.
}

\bpar{
The series of Simpop models~\cite{pumain2012multi} was conceived to test various assumptions of the theory, such as the role of innovation diffusion processes in the organisation of the urban system. Thus, different underlying regimes were revealed for systems of cities in Europe and in the United States~\cite{bretagnolle2010comparer}.
}{
La série des modèles Simpop~\cite{pumain2012multi} a été conçue pour tester différentes hypothèses de la théorie, comme par exemple le rôle des processus de diffusion de l'innovation dans l'organisation du système urbain. Ainsi, des régimes sous-jacent différents ont été mis en évidence pour les systèmes de ville en Europe et aux Etats-unis~\cite{bretagnolle2010comparer}.
}

\bpar{
At other time scales and in other contexts, the SimpopLocal model~\cite{schmitt2014modelisation} aims at investigating the conditions for the emergence of hierarchical urban systems from disparate settlements. A minimal model (in the sense of sufficient and necessary parameters) has been isolated through to the use of intensive computation with the model exploration software OpenMole~\cite{schmitt2014half}, what was a result impossible to obtain analytically for such a kind of complex model. The technical progresses of OpenMole~\cite{reuillon2013openmole} were done simultaneously with theoretical and empirical advancements.
}{
A d'autres échelles de temps et dans d'autres contextes, le modèle SimpopLocal~\cite{schmitt2014modelisation} a pour but d'étudier les conditions pour l'émergence de systèmes urbains hiérarchiques à partir d'établissements disparates. Un modèle minimal (au sens de paramètres nécessaires et suffisants) a été isolé grace à l'utilisation de calcul intensif via le logiciel d'exploration de modèles OpenMole~\cite{schmitt2014half}, ce qui était un résultat impossible à atteindre de manière analytique pour un tel type de modèle complexe. Les progrès techniques d'OpenMole~\cite{reuillon2013openmole} ont été menés simultanément avec les avances théoriques et empiriques.
}

\bpar{
Epistemological advances were also crucial to this framework, as \cite{rey2015plateforme} develops, and new concepts such as incremental modeling~\cite{cottineau2015incremental} were discovered, with powerful concrete applications: \cite{cottineau2014evolution} applies it on the soviet system of cities and isolates dominating socio-economic processes, by systematic testing of thematic assumptions and implementation functions. Directions for the development of such modeling and simulation practices in quantitative geography were recently introduced by \cite{banos2013pour}. He concludes with nine principles\footnote{Must it become the ten commandments ? \noun{Ren{\'e} Doursat} underlined the absence of the last Banos' commandment, the intrinsic essence of our enterprise may be linked to its pursuit.}, among which we can cite the importance of intensive exploration of computational models and the importance of heterogeneous models coupling, that are among other principles such as reproducibility at the center of the study of complex geographical systems from the point of view described before. We will be positioned mainly within the legacy of this line of research, working conjointly in the theoretical, empirical, epistemological and modeling aspects.
}{
Les avancées épistémologiques ont également été cruciales dans ce cadre, comme \cite{rey2015plateforme} le développe, et de nouveaux concepts comme la modélisation incrémentale~\cite{cottineau2015incremental} ont été découverts, avec de puissantes applications concrètes : \cite{cottineau2014evolution} l'applique sur le système de villes soviétiques et isole les processus socio-économiques dominants, par un test systématique des hypothèses thématiques et des fonctions d'implémentation. Des directions pour le développement de telles pratiques de modélisation et simulation en géographie quantitative ont récemment été introduits par \cite{banos2013pour}. Il conclut par neuf principes\footnote{Cela doit-il devenir les dix commandements ? \noun{Ren{\'e} Doursat} soulignait l'absence du dernier commandement de \noun{Banos}, l'essence intrinsèque de notre entreprise est peut être en partie liée à sa recherche.}, parmi lesquels nous pouvons citer l'importance de l'exploration intensive des modèles computationnels et l'importance du couplage de modèles hétérogènes, qui sont avec d'autre principes tel la reproductibilité au centre de l'étude des systèmes complexes géographiques selon le point de vue décrit précédemment. Nous nous positionnerons en grande partie dans l'héritage de cette ligne de recherche, travaillant de manière conjointe sur les aspects théoriques, empiriques, épistémologiques et de modélisation.
}


\section*{Cities, Systems of Cities, Territories}{Villes, Systèmes de Villes, Territoires}

\bpar{
We can enter now the heart of the matter to progressively construct the precise problematic which will enter the global context developed up to here. Our elementary geographical objects (in the sense of precursors in our theoretical genesis) will be the \emph{City}, the \emph{System of cities}, and the \emph{Territory}, that we will now define.
}{
Entrons à présent dans le vif du sujet pour construire progressivement la problématique précise qui s'inscrira dans le contexte global développé jusqu'ici. Nos objets géographiques élémentaires (au sens de précurseurs dans notre genèse théorique) sont la \emph{Ville}, le \emph{Système de Villes}, et le \emph{Territoire}, que nous allons à présent définir.
}

\bpar{
A central element of socio-geographical systems is the \emph{City} object, on which we position for a proper epistemological consistence. The question of the definition of the city has fostered numerous contributions. \cite{robic1982cent} shows for example that \noun{Reynaud} had already conceptualized the city as a central place of a geographical space, allowing aggregation and exchanges, theory that will be reformulated by \noun{Christaller} as the \emph{Central Place Theory}. This theoretical definition is rejoined by the conception of \noun{Pumain} which considers the city as a clearly identifiable spatial entity, constituted by social agents (that may be elementary or not) and of technical artifacts, and which is the incubator of social change and innovation~\cite{pumain2010theorie}. We will use this definition in our work. We must however keep in mind that the concrete definition of a city in terms of geographical entities and spatial extent is problematic: morphological definitions (i.e. based on the shape and the distribution of the built environment), functional definitions (based on the use of urban functions by agents, for example through area of dominating daily commuting), administrative definitions, etc., are partly orthogonal and more or less adapted to the problem studied~\cite{guerois2002commune}. Recently, several studies have shown the strong sensitivity of urban scaling laws\footnote{Scaling laws consist in a statistical regularity which can be observed within a system of cities, linking for example a characteristic variable $Y_i$ to the population $P_i$ under the form of a power law $Y_i = Y_0\cdot \left(P_i / P_0\right)^{\alpha}$.} to the delineation chosen for the estimation, leading sometimes to an inversion of expected qualitative properties (see for example~\cite{arcaute2015constructing}). Variations of estimated exponents as a function of parameters of the definition, as done by~\cite{2015arXiv150707878C}, can be interpreted as a more global property and a signature of the urban system.
}{
Un élément central des systèmes socio-géographiques est l'objet \emph{Ville}, sur lequel nous nous positionnons pour une cohérence épistémologique propre. La question de la définition de la ville a fait couler beaucoup d'encre. \cite{robic1982cent} montre par exemple que \noun{Reynaud} avait déjà conceptualisé la ville comme lieu central d'un espace géographique, permettant agrégation et échanges, théorie qui sera reformulée par \noun{Christaller} comme \textit{Théorie des Lieux Centraux}. Cette définition théorique est rejointe par la conception de \noun{Pumain} qui considère la ville comme une entité spatiale clairement identifiable, constituée d'agents sociaux (élémentaires ou non) et d'artefacts techniques, et qui est l'incubateur du changement social et de l'innovation~\cite{pumain2010theorie}. Nous prendrons cette définition dans notre travail. Il faut toutefois garder à l'esprit que la définition concrète d'une ville en terme d'entités géographiques et d'étendue spatiale est problématique : des définitions morphologiques (c'est-à-dire se basant sur la forme et la distribution du bâti), fonctionnelles (se basant sur l'utilisation des fonctions urbaines par les agents, par exemple par aire de déplacement domicile-travail dominant), administratives, etc., sont partiellement orthogonales et plus ou moins adaptées au problème étudié~\cite{guerois2002commune}. Récemment, un certain nombre d'études ont montré la forte sensibilité des lois d'échelles urbaines\footnote{Les lois d'échelle consistent en une régularité statistique observable au sein d'un ensemble de ville, reliant par exemple une variable caractéristique $Y_i$ à la population $P_i$ sous la forme d'une loi puissance $Y_i = Y_0\cdot \left(P_i / P_0\right)^{\alpha}$.} aux délimitations choisies pour l'estimation, pouvant entrainer une inversion des propriétés qualitatives attendues (voir par exemple~\cite{arcaute2015constructing}). Les variations des exposants estimés en fonction de paramètres de définition, comme effectué par~\cite{2015arXiv150707878C}, peut être interprété comme une propriété plus globale et une signature du système urbain.
}


\bpar{
This confirms the necessity to consider cities within their system, and the importance of the notion of \emph{Urban System}\footnote{Concerning the definition of a system, we can take it in all generality as a set of elements in interaction, presenting a certain structure determined by it, and which posses a certain level of autonomy in its environment. It can be a mainly ontological autonomy in the case of an open system, or a real autonomy in the case of a closed system.}. An urban system can be considered as a set of cities in interaction, which dynamics will be more or less strongly coupled. \cite{berry1964cities} considers cities as ``\textit{systems within systems of cities}'', insisting on the multi-scalar nature (in the sense of intricate scales with a certain level of autonomy)\footnote{The definition of scale is ambiguous in geography, since according to \cite{hypergeo}, the scale designates simultaneously a spatial and/or temporal extent (scale of the map) and an abstract representation of ``levels which make sense regarding a particular problem''. As \cite{manson2008does} indicates, scale is indeed placed within an epistemological continuum, from realistic conceptions to constructivist conceptions, and the ones making it correspond to intrinsic levels of self-organization of the system considered. We will position in a privileged way in this latest logic of complexity.} and necessarily complex, conception which is adopted and extended by the evolutive urban theory previously detailed. The term of \emph{System of Cities} will be used when we will be able to clearly identify cities as sub-systems, and we will use the term of urban system more generally (a city being itself an urban system).
}{
Cela confirme la nécessité de considérer les villes dans leur système, et l'importance de la notion de \emph{Système Urbain}\footnote{Concernant la définition d'un système, nous pourrons la prendre en toute généralité comme un ensemble d'éléments en interaction, présentant une certaine structure déterminée par celle-ci, et possédant un certain niveau d'autonomie avec son environnement. Il peut s'agir d'une autonomie majoritairement ontologique dans le cas d'un système ouvert, ou d'une autonomie réelle dans le cas d'un système fermé.}. Un système urbain peut être considéré comme un ensemble de villes en interaction, dont les dynamiques seront plus ou moins fortement couplées. \cite{berry1964cities} considère les villes comme ``\textit{systèmes dans des systèmes de villes}'', appuyant sur le caractère multi-scalaire (au sens d'échelles emboîtées ayant un certain niveau d'autonomie)\footnote{La définition de l'échelle est ambigüe en géographie, puisque selon \cite{hypergeo}, l'échelle désigne à la fois une étendue spatiale et/ou temporelle (échelle de la carte) et une représentation abstraite de ``niveaux qui ont sens par rapport à une problématique particulière''. Comme l'indique \cite{manson2008does}, l'échelle se place en fait dans un continuum épistémologique, des conceptions réalistes à celles constructivistes, et celles la faisant correspondre aux niveaux d'auto-organisation intrinsèque du système considéré. Nous nous placerons de manière privilégiée dans cette dernière logique de complexité.} et nécessairement complexe, conception reprise et étendue par la théorie évolutive des villes détaillée précédemment. Le terme de \emph{Système de Villes} sera utilisé lorsque nous pourrons clairement identifier des villes comme sous-systèmes, et nous parlerons de système urbain de manière plus générale (une ville elle-même étant un système urbain).
}

\bpar{
Finally, underlying to the understanding of urban systems dynamics intervenes the notion of \emph{Territory}. Polymorphic and corresponding to multiple visions, as we will develop deeply in~\ref{sec:networkterritories}, it can be simply defined in a preliminary way. The territory thus designates the spatial distribution of urban activities, of agents practicing or developing them, and of technical artifacts, including infrastructure, supporting them, and also the superstructure\footnote{We understand the superstructure in its marxist sense, i.e. the organizational structure and the ideas of a society, including political structures.} which is associated to it\footnote{The link between the Territory and the City, or the System of Cities, will be also developed more deeply further when the concept will be constructed.}.
}{
Enfin, sous-jacente à la compréhension des dynamiques des systèmes urbains intervient la notion de \emph{Territoire}. Polymorphe et correspondant à des visions multiples, celle-ci, que nous développerons en profondeur en~\ref{sec:networkterritories}, peut être définie de manière préliminaire simplement. Le territoire désigne alors la distribution spatiale des activités urbaines, des agents les exerçant ou les développant, et des artefacts techniques, dont l'infrastructure, les supportant, ainsi que la superstructure\footnote{Nous comprenons la superstructure au sens marxiste, c'est-à-dire la structure organisationnelle et l'ensemble des idées d'une société, incluant les structures politiques.} qui leur est associée\footnote{Le lien entre le Territoire et la Ville, ou le Système de Villes, sera également creusé plus loin lors de la construction approfondie du concept.}.
}

\section*{Networks, Interactions and Co-evolution}{Réseaux, Interactions et Co-évolution}

\bpar{
A fundamental characteristic of urban systems and territories is their simulatneous inscription in space and time, that is contained in spatio-temporal dynamics, at multiple scales. The notion of \emph{process} in the sense of \cite{hypergeo}, i.e. a dynamical chain of facts with causal properties\footnote{We will understand causality in the sense of circular causality in complex systems, which considers fostering cycles between phenomenons, or more complex structures. Linear causality, i.e. a phenomenon driving an other, is an idealized particular case of this. We will come back with more details on the notion of causality and on its different approaches by geographers in section~\ref{sec:causalityregimes}.}, allows to capture relationships between components of these dynamics, and is thus an interesting approach for a partial understanding of such systems. Any partial understanding will be associated to the choice of \emph{scales} and an \emph{ontology} which corresponds to the specification of real objects studied\footnote{More precisely, we use the definition of~\cite{livet2010} which couples the ontological approach from the point of view of philosophy, i.e. ``\textit{the study of what can exist}'', and the one from computer science which consists in defining classes, objects and their relations which constitute the knowledge of a domain. This use of the notion of ontology naturally biases our research towards modeling paradigms, but we take the position (developed in more details later) to understand any scientific construction as a \emph{model}, making the boundary between theory and models less relevant than for more classical visions. Any theory has to make choices on described objects, their relations, and the implicated processes, and contain thus an ontology in that sense.}. We will now specify these abstract concepts, by introducing \emph{networks}, their \emph{interactions} with territories and their approach through \emph{co-evolution}.
}{
Une caractéristique fondamentale des systèmes urbains et des territoires est leur inscription simultanée dans l'espace et le temps, qui transparaît dans leurs dynamiques spatio-temporelles, à de multiples échelles. La notion de \emph{processus} au sens de~\cite{hypergeo}, c'est-à-dire l'enchainement dynamique de faits aux propriétés causales\footnote{Nous prendrons la causalité au sens de causalité circulaire dans les systèmes complexes, qui considère des cycles d'entrainement entre phénomènes, ou des structures plus complexes. La causalité linéaire, c'est-à-dire un phénomène entrainant un autre, est un cas particulier idéalisé de celle-ci. Nous reviendrons en détail sur la notion de causalité et sur ses différentes approches par les géographes en section~\ref{sec:causalityregimes}.}, permet de capturer les relations entre composantes de ces dynamiques, et est ainsi une notion clé pour une compréhension partielle de ces systèmes. Toute compréhension partielle sera associée au choix d'\emph{échelles} et d'une \emph{ontologie} qui correspond à la spécification des objets réels étudiés\footnote{Plus précisément, nous utilisons la définition de~\cite{livet2010} qui couple l'approche ontologique du point de vue de la philosophie, c'est-à-dire ``\textit{l'étude de ce qui peut exister}'', et celui de l'informatique qui consiste à définir les classes, les objets et leurs relations qui constituent la connaissance d'un domaine. Cet usage de la notion d'ontologie biaise naturellement notre recherche vers des paradigmes de modélisation, mais nous prenons la position (développée en détails plus loin) de comprendre toute construction scientifique comme un \emph{modèle}, rendant la frontière entre théories et modèles moins pertinentes que pour des visions plus classiques. Toute théorie doit faire des choix sur les objets décrits, leur relations et les processus impliqués, et contient donc une ontologie dans ce sens.}. Nous allons à présent spécifier ces concepts abstraits, en introduisant les \emph{réseaux}, leurs \emph{interactions} avec les territoires et leur approche par la \emph{co-évolution}.
}

\bpar{
A particular ontology will hold our attention: within territories emerge \emph{Physical Networks}, which can be understood according to \cite{dupuy1987vers} as the materialization of a set of potential connections between agents of the territory. The question of the implication of these networks and their dynamics in territorial dynamics, which we can synthesize as \emph{interactions between networks and territories}, has been the subject of numerous technical and scientific debates, in particular in the case of transportation networks. We will come back on their nature and positioning in Chapter~\ref{ch:thematic}, but we can already take some of the underlying difficulties as a starting point for our questioning. One recurring aspect is the \emph{myth of structuring effects}, suggested by \cite{offner1993effets} when criticizing an exaggerated use by planners and politics of a scientific concept which empirical basis are still discussed. The fundamental underlying question that we reformulate is the following: \textit{to what extent is it possible to associate territorial dynamics to an evolution of the transportation infrastructure ?} We can ask the question reciprocally, and even generalize it: what are the processes capturing the interactions between these two objects ?
}{
Une ontologie particulière retiendra notre attention : au sein des territoires émergent des \emph{Réseaux Physiques}, qui peuvent être compris selon \cite{dupuy1987vers} comme la matérialisation d'un ensemble de connexions potentielles entre agents du territoire. La question de l'implication de ces réseaux et de leur dynamique dans les dynamiques territoriales, qu'on peut synthétiser comme \emph{interactions entre réseaux et territoires}, a fait l'objet d'abondants débats scientifiques et techniques, notamment dans le cas des réseaux de transport. Nous reviendrons sur la nature et le positionnement de ceux-ci au Chapitre~\ref{ch:thematic}, mais nous pouvons d'ores et déjà prendre certaines des difficultés soulevées comme point de départ de notre questionnement. L'un des aspects récurrents est celui du \emph{mythe des effets structurants}, consacré par \cite{offner1993effets} en critique d'une utilisation exagérée par les planificateurs et les politiques d'un concept scientifique dont les fondements empiriques sont encore discutés. La question fondamentale sous-jacente que nous reformulons est la suivante : \textit{dans quelle mesure est-il possible d'associer des dynamiques territoriales à une évolution de l'infrastructure de transport ?} Nous pouvons poser la question de manière réciproque, et même la généraliser : quels sont les processus capturant les interactions entre ces deux objets ?
}

\bpar{
An approach allowing to consider the problem from an other angle is the notion of \emph{co-evolution}, used in the evolutive urban theory to designate strongly coupled processes\footnote{We will use the term of \emph{coupling} systems or processes to designate the constitution of a system including the coupled elements, through the emergence of new interactions or new elements. The definition of the nature and the strength of a coupling is an open question, and we will use the notion in an intuitive way, to designate a more or less high level of interdependency between coupled sub-systems.} of evolution of cities as used by~\cite{paulus2004coevolution}, and applied to the relations between networks and cities by~\cite{bretagnolle:tel-00459720}\footnote{\cite{paulus2004coevolution} directly transfers the biological concept of co-evolution (which consists in a strong interdependency between two species in their evolutionary trajectories, and which in fact corresponds to the existence of an \emph{ecological niche} constituted by species as we will further develop in~\ref{sec:theory}), and studies cities which ``are in concurrence, imitate themselves, and cooperate''. This transfer remains fuzzy (on temporal scales implied, the status of objects which co-evolve) and finally not explored. Similar trajectories can not be enough to exhibit strong interdependencies as he states in conclusion, since these can be spurious. Furthermore, the transfer of concepts between disciplines is an operation on which one must remain cautious (we will illustrate this through the interdisciplinary study of morphogenesis, concept which is initially from biology, in Chapter~\ref{ch:morphogenesis}).}. This last work distinguishes a phase of ``mutual adaptation'' between networks and cities, corresponding to a dynamic in which causal effects can clearly be attributed to one on the development of the other (for example, new transportation lines answer to a growing demand inducted by urban growth, or inversely urban growth is favored by a new connectivity to the network), from the phase of co-evolution, which is defined as a ``strong interdependency'' (p.~150) in which retroactions play a privileged role and ``the dynamic of the system of cities is not anymore constrained by the development of transportation networks'' (p.~170). These feedback loops and this mutual interdependency, seen in their dynamical perspective, correspond to circular causal relationships (in the sense given above) that are difficult to disentangle. We will take as preliminary definition of co-evolution between two components of a system \emph{the existence of a strong coupling, corresponding generally to circular causal relationships}.
}{
Une approche permettant de poser différemment le problème est la notion de \emph{co-evolution}, utilisée en théorie évolutive des villes pour qualifier les processus fortement couplés\footnote{On parlera de \emph{couplage} de systèmes ou de processus pour désigner la constitution d'un système englobant les éléments couplés, par l'émergence de nouvelles interactions ou de nouveaux éléments. La définition de la nature et de la force d'un couplage est une question ouverte, et nous utiliserons la notion de manière intuitive, pour désigner un plus ou moins grand niveau d'interdépendance entre les sous-systèmes couplés.} d'évolution des villes comme utilisé par~\cite{paulus2004coevolution}, et appliqué aux relations entre réseaux et villes par~\cite{bretagnolle:tel-00459720}\footnote{\cite{paulus2004coevolution} transfère directement le concept biologique de co-évolution (qui consiste en une interdépendance forte entre deux espèces dans leurs trajectoires évolutives, et qui en fait correspond à l'existence d'une \emph{niche écologique} constituée par les espèces comme nous le développerons plus loin en~\ref{sec:theory}), et parle de villes qui ``se concurrencent, s'imitent, coopèrent''. Ce transfert reste flou (sur les échelles temporelles impliquées, le statut des objets qui co-évoluent) et finalement non exploré. Des trajectoires similaires ne peuvent suffire à exhiber des interdépendances fortes comme il affirme en conclusion, celles-ci pouvant être fortuites. De plus, le transfert de concepts entre disciplines est une opération pour laquelle prudence doit être de mise (nous illustrerons cela par l'étude interdisciplinaire de la morphogenèse, concept initialement biologique, en Chapitre~\ref{ch:morphogenesis}).}. Cette dernière distingue une phase ``d'adaptation mutuelle'' entre réseaux et villes, correspondant à une dynamique dans laquelle des effets causaux sont clairement attribuables à l'un sur le développement de l'autre (par exemple, les nouvelles lignes de transport répondent à une demande croissante induite par la croissance urbaine, ou inversement la croissance urbaine est favorisée par une nouvelle connectivité au réseau), de la phase de co-évolution, qu'elle définit comme une ``interdépendance forte'' (p.~150) dans laquelle les rétroactions jouent un rôle privilégié  et ``la dynamique du système de villes n'est plus contrainte par le développement des réseaux de transport'' (p.~170). Ces boucles de rétroaction et cette interdépendance mutuelle, vus dans leur perspective dynamique, correspondent à des relations causales circulaires (au sens donné plus haut) difficiles à séparer. Nous prendrons comme définition préliminaire de la co-évolution entre deux composantes d'un système \emph{l'existence d'un couplage fort, correspondant généralement à des relations causales circulaires}.
}


\section*{Problematic}{Problématique}

\bpar{
This framework allows to capture a certain degree of complexity, but however remains fuzzy or too general in its characterization, both theoretically and empirically. We will try here to challenge and to deepen this approach, to shed a light on its potential contributions for the understanding of interactions between networks and territories. The clarification on the one hand of what it means and on the other hand of its empirical existence will be a Gordian knot of our approach. Our general problematic is thus decomposed into two complementary axis:
}{
Ce cadre permet de capturer un certain degré de complexité, mais reste cependant flou ou trop général dans sa caractérisation, à la fois théorique et empirique. Nous ferons ici le pari de mettre à l'épreuve et d'approfondir cette approche, pour éclaircir ses apports potentiels pour la compréhension des interactions entre réseaux et territoires. La clarification d'une part de ce qu'elle signifie et d'autre part de son existence empirique sera un noeud gordien de notre démarche. Notre problématique générale se décompose alors en deux axes complémentaires :
}

\bpar{
\begin{enumerate}
	\item How to define and/or characterize co-evolution processes between transportation networks and territories ?
	\item How to model these processes, at which scales and through which ontologies ?
\end{enumerate}
}{
\begin{enumerate}
	\item Comment définir et/ou caractériser les processus de co-évolution entre réseaux de transports et territoires ?
	\item Comment modéliser ces processus, à quelles échelles et par quelles ontologies ?
\end{enumerate}
}

\bpar{
The second aspect is a consequence of our scientific positioning, which postulates the use of modeling, and more particularly of simulation of models, as a fundamental tool for the knowledge of processes within complex systems.
}{
Le deuxième aspect découle de notre positionnement scientifique, qui postule l'utilisation de la modélisation, et plus particulièrement de la simulation de modèle, comme un instrument fondamental de connaissance des processus au sein des systèmes complexes.
}


\section*{General Organization}{Organisation Générale}

\bpar{
We propose to answer to the above problematic through the following strategy. A first part will build the necessary foundations, by detailing definitions, studied concepts and objects, by sketching the scientific landscape gravitating around our question, and by refining the epistemological positioning. This part is composed by three chapters:
\begin{enumerate}
	\item A first chapter develops the question of interactions between networks and territories, from a theoretical point of view but also by illustrating them by case studies and fieldwork elements. It allows to situate the notion of co-evolution both from a concrete and abstract point of view.
	\item A second chapter aims in a similar way at clarifying the positioning regarding the modeling of co-evolution. The state of the art is completed by a mapping of concerned scientific disciplines and by a modelography, i.e. a classification and systematic decomposition of a corpus of models in order to understand the ontologies used and possible determinants of these.
	\item A third chapter develops our epistemological positioning, which appears to have a considerable influence on modeling choices that will be taken in the following. We develop therein issues linked to modeling practices, to datamining and intensive computation, to reproducibility and open science, and more general epistemological considerations that are intrinsic to the systems studied.
\end{enumerate}
}{
Nous proposons de répondre à la problématique ci-dessus par la stratégie suivante. Une première partie posera les fondations nécessaires, en précisant les définitions, concepts et objets étudiés, en dessinant le paysage scientifique gravitant autour de la question, et en raffinant le positionnement épistémologique. Cette partie est composée de trois chapitres :
\begin{enumerate}
	\item Un premier chapitre développe la question des interactions entre réseaux et territoires, d'un point de vue théorique mais aussi en les illustrant par des études de cas et des éléments de terrain. Il permet de situer la notion de co-évolution à la fois de manière concrète et abstraite.
	\item Un deuxième chapitre se charge d'une manière similaire de clarifier le positionnement au regard de la modélisation de la co-évolution. L'état de l'art est complété par une cartographie des disciplines scientifiques concernées et par une modélographie, c'est-à-dire une classification et décomposition systématique d'un corpus de modèles afin de comprendre les ontologies utilisées et de possibles déterminants de celles-ci.
	\item Un troisième chapitre développe notre positionnement épistémologique, qui s'avère avoir une influence considérable sur les choix de modélisation qui seront opérés par la suite. Nous y développons les questions liées au pratiques de modélisation, de \emph{datamining} et de calcul intensif, des questions de reproductibilité et d'ouverture, et des considérations épistémologiques plus générales intrinsèques aux systèmes étudiés. 
\end{enumerate}
}

\bpar{
From these complementary analyses emerge two thematic positioning that correspond to two modeling scales, that remain poorly explored for our particular question: the evolutive urban theory which induces a macroscopic modeling at the level of the system of cities, and urban morphogenesis which allows to consider the links between form and function at the mesoscopic scale. The second part will aim thus at constructing elementary bricks from these approaches, which will be used in the following to construct models:
\begin{enumerate}\setcounter{enumi}{3}
	\item The fourth chapter deals with different aspects implied by the evolutive urban theory. The non-stationary character of processes in space is a crucial element, that we empirically demonstrate in a first section through the study of spatial correlations between urban form and road network topology for Europe and China. Then, the notion of circular causality is explored, and we develop a method allowing to isolate what we call \emph{causality regimes}, i.e. typical configurations of interaction captured by lagged correlation patterns. It is tested on synthetic data and observed data in the case of South Africa, for which we demonstrate an effect of segregation policies on the interactions between networks and territories themselves. This first part of the chapter complements in an empirical way the characterization of co-evolution sketched in the first part. Finally, we construct a model of an urban system based on interactions between cities, which allows to indirectly demonstrate the existence of network effects.
	\item The fifth chapter will deepen the notion of \emph{morphogenesis}, by beginning with proposing a point of view consistent across disciplines using it, in order to exhibit a characterization based on the emergence of an architecture through causal circular relations between form and function. This precision will be crucial for the nature of models we will elaborate. A second section develops a simple model of urban growth taking into account the distribution of population alone, and capturing the contradictory forces of concentration and dispersion. We demonstrate its ability to reproduce existing urban forms using urban form data previously computed. It is then coupled in a sequential manner to a network generation model, what allows to exhibit a large spectrum of potentially generated correlations.
\end{enumerate}
}{
De ces analyses complémentaires se dégagent deux positionnements thématiques correspondant à deux échelles de modélisation, peu explorés pour notre question particulière : la théorie évolutive des villes qui induit une modélisation macroscopique au niveau du système de ville, et la morphogenèse urbaine qui permet de considérer les liens entre forme et fonction à l'échelle mesoscopique. La deuxième partie s'attèlera donc à construire les briques élémentaires à partir de ces approches, qui serviront par la suite à la construction des modèles :
\begin{enumerate}\setcounter{enumi}{3}
	\item Le quatrième chapitre traite de différents aspects impliqués par la théorie évolutive des villes. Le caractère non-stationnaire des processus dans l'espace est un élément crucial, que nous démontrons empiriquement dans une première section par l'étude des corrélations spatiales entre forme urbaine et topologie du réseau routier pour l'Europe et la Chine. Ensuite, la notion de causalité circulaire est explorée, et nous développons une méthode permettant d'isoler ce qu'on appelle des \emph{régimes de causalité}, c'est-à-dire des configurations typiques d'interaction capturées par les motifs de corrélation retardée. Celle-ci est testée sur données synthétiques et données observées dans le cas de l'Afrique du Sud, où l'on démontre un effet des politiques de ségrégation sur les interactions réseaux-territoires elles-mêmes. Cette première partie du chapitre complète de manière empirique la caractérisation de la co-évolution ébauchée en première partie. Enfin, nous construisons un modèle de système urbain basé sur les interactions entre villes, qui permet de démontrer indirectement l'existence d'effets de réseau.
	\item Le cinquième chapitre creusera la notion de \emph{morphogenèse}, en commençant par en proposer un point de vue cohérent au travers de différentes disciplines la mobilisant, afin d'en dégager une caractérisation se reposant sur l'émergence d'une architecture par relations causales circulaires entre forme et fonction. Cette précision sera cruciale dans la nature des modèles mis en place. Une deuxième section développe un modèle simple de croissance urbaine prenant en compte la distribution de la population seule, et capturant les forces contradictoires de concentration et de dispersion. Nous démontrons sa capacité à reproduire des formes urbaines existantes à partir des données de forme urbaine calculées précédemment. Il est ensuite couplé séquentiellement à un modèle de génération de réseau, ce qui permet d'exhiber un large spectre de corrélations potentiellement générées.
\end{enumerate}
}

\bpar{
At this stage, we build in the third part from the foundations and with elementary bricks our fundamental construction, which consists in different models of co-evolution, that we differentiate according to the two approaches considered. Still within a logic of parallel and complementary approaches, we elaborate developments of the two previous chapters, in two chapters modeling co-evolution:
\begin{enumerate}\setcounter{enumi}{5}
	\item The sixth chapter develops a co-evolution model at the macroscopic scale. Firstly, we explore systematically the unique existing analog model. We then develop the model by extending the interaction model already introduced. Its systematic exploration reveals its ability to produce different regimes of co-evolution, some witnessing circular causalities. It is also calibrated on the French system of cities on a long time period, on population and railway network data, which allows to infer indirect informations on implied processes.
	\item The seventh chapter deals with urban morphogenesis models which capture co-evolution processes. The question of network generation heuristics is first tackled, by comparing the potentialities of diverse methods. In an approach of multi-modeling, these are then integrated in a family of morphogenesis models, which are calibrated on urban form and network topology indicators, at the first order (values of indicators) and at the second order (correlations matrices). We then sketch a more complex model, aiming at integrating governance processes in the growth of the transportation network. It is explored in a preliminary way.
\end{enumerate}
}{
A ce stade, nous bâtissons dans la troisième partie sur les fondations et avec les briques élémentaires notre construction fondamentale, qui consiste en différents modèles de co-évolution, que nous différencions selon les deux approches considérées. Toujours dans une logique d'approches parallèles et complémentaires, nous élaborons les développements des deux chapitres précédents, dans deux chapitres modélisant la co-évolution :
\begin{enumerate}\setcounter{enumi}{5}
	\item Le sixième chapitre développe un modèle de co-évolution à l'échelle macroscopique. Dans un premier temps, nous explorons de manière systématique l'unique modèle analogue existant. Nous développons ensuite le modèle par extension du modèle d'interaction déjà introduit. Son exploration systématique révèle sa capacité à produire différents régimes de co-évolution, certains témoignant de causalités circulaires. Il est également calibré sur le système de villes français sur le temps long, sur données de population et de réseau ferroviaire, ce qui permet d'inférer des informations indirectes sur les processus impliqués.
	\item Le septième chapitre s'intéresse aux modèles de morphogenèse urbaine capturant les processus de co-évolution. La question des heuristiques de génération de réseau est d'abord traitée, en comparant les potentialités de diverses méthodes. Dans une démarche de multi-modélisation, celles-ci sont ensuite intégrées dans une famille de modèles de morphogenèse, qui sont calibrés sur les indicateurs de forme urbaine et de topologie de réseau, au premier ordre (valeurs des indicateurs) et au second ordre (matrices des corrélations). Nous ébauchons ensuite un modèle plus complexe, visant à intégrer les processus de gouvernance dans la croissance du réseau de transport. Celui-ci est exploré de manière préliminaire.
\end{enumerate}
}

\bpar{
After having demonstrated the potentialities of our two approaches to capture some aspects of co-evolution and to inform corresponding processes, we finally proceed with an opening:
\begin{enumerate}\setcounter{enumi}{7}
	\item The eighth and last chapter consists in an theoretical and epistemological opening. We first draw a bilan of our contributions and put them into perspective. We then sketch a theoretical reconciliation of morphogenesis and the evolutive urban theory, in which co-evolution is central. This development could be the basis of a theory and multi-scalar models for co-evolution. We finally develop in a reflexive manner a knowledge framework for the study of complex systems, both product and precursor of all our work.
\end{enumerate}
}{
Après avoir démontré les capacités de nos deux approches à capturer certains aspects de la co-évolution et d'informer les processus correspondants, nous procédons finalement à une ouverture :
\begin{enumerate}\setcounter{enumi}{7}
	\item Le huitième et dernier chapitre consiste en une ouverture théorique et épistémologique. Nous faisons dans un premier temps un bilan de nos contributions et les mettons en perspective. Nous esquissons ensuite une réconciliation théorique de la morphogenèse et de la théorie évolutive, dans laquelle la co-évolution est centrale. Ce développement pourrait poser les bases d'une théorie et de modèles multi-échelle pour la co-évolution. Nous développons enfin dans une démarche réflexive un cadre de connaissance pour l'étude des systèmes complexes, à la fois produit et précurseur de l'ensemble de notre démarche.
\end{enumerate}
}


\bpar{
We summarize this organisation, and also direct or indirect dependencies between the different chapters, in the Frame~\ref{frame:intro:organisation} on the following page.
}{
Nous résumons cette organisation, ainsi que les dépendances directes ou indirectes entre les différents chapitres, dans l'Encadré~\ref{frame:intro:organisation} page suivante.
}

\begin{figure}[h!]
	\begin{mdframed}
		
		  \includegraphics[width=\linewidth]{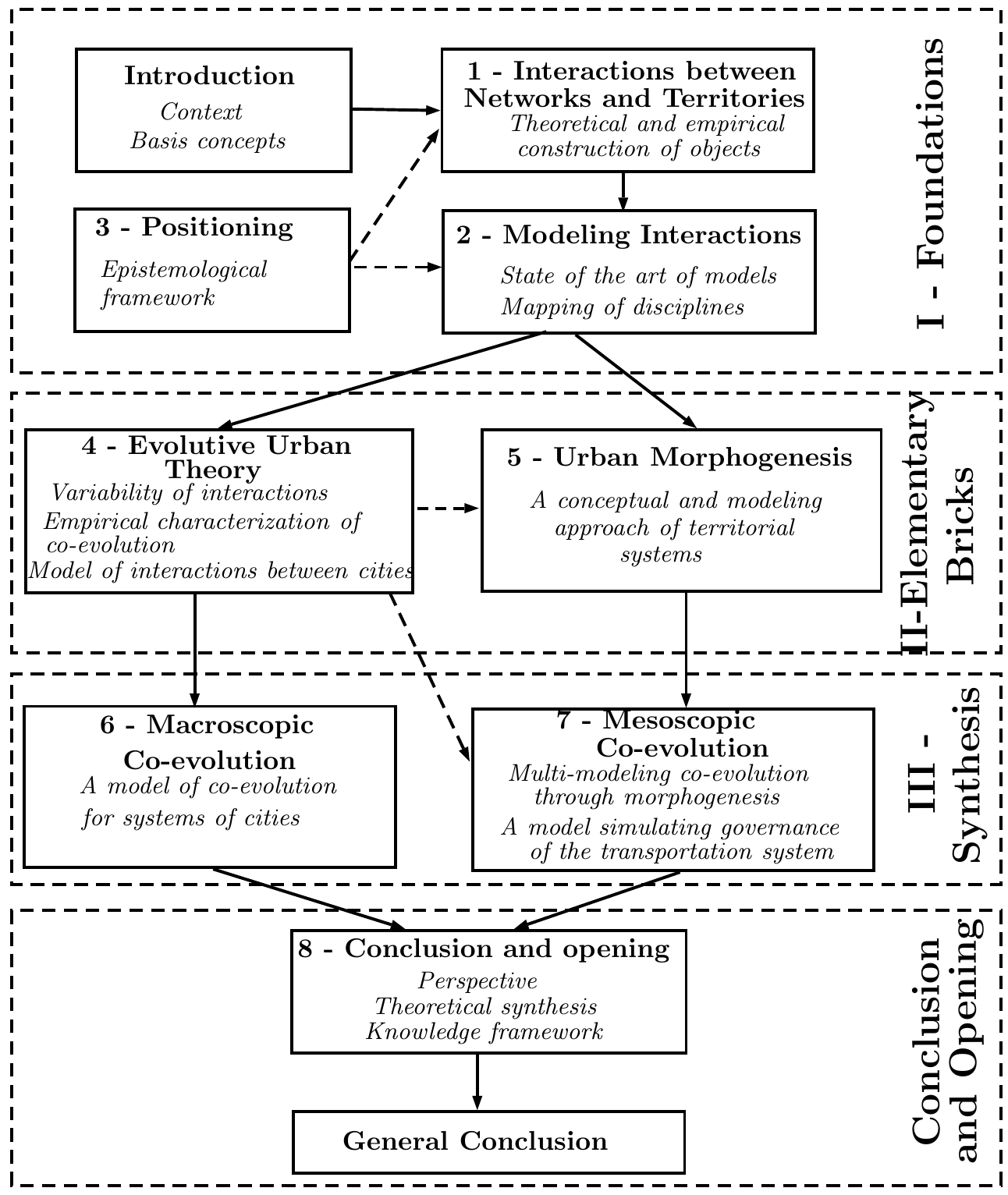}
		
		\medskip
		
		\framecaption{\textbf{General organisation of the memoire.} Full arrows give a direct dependency (logical chaining or extensions), dotted arrows an indirect dependency (reuse of data or methods).\label{frame:intro:organisation}}{\textbf{Organisation générale du mémoire.} Les flèches pleines donnent une dépendance directe (enchainement logique ou extensions), les flèches pointillées une dépendance indirecte (réutilisation de données ou de méthodes).\label{frame:intro:organisation}}
	\end{mdframed}
\end{figure}





\stars



%
\ctparttext{This part builds the foundations of our work, by reconstructing the question in a theoretical way and through the illustration of case studies, and then by describing the scientific landscape of its existing approaches in modeling. We also develop our epistemological position	ing with important practical implications.}

\part{Foundations}
%

\bpar{
\chapter*{Introduction of Part I}
}{
\chapter*{Introduction de la Partie I}
}

\markboth{Introduction of Part I}{Introduction of Part I}



\bigskip

\bpar{
\textit{A journey, discovering a city, new encounters, sharing ideas: as much processes which imply a cognitive generativity and a complex interaction between our representations, our actions, and the environment. The construction of a scientific knowledge does not escape these rules. We could then see in the studied object itself, let take the city and its agents, an allegory of the knowledge production process on the object. As Romain Duris which lands in \emph{l'Auberge Espagnole}, and discovers these unknown streets that later we will have walked a hundred times, where we will have lived a thousand things: we land in a world of complementary concepts, approaches, points of view, on things that are not the same thing. This ontological discrepancy is indeed as much present in our representations of the urban space: \emph{Oven Street} is one center of knowledge for the member of \emph{Géocités}; it is the center of Paris, thus of France, thus of the World for the proud native of the 6th \emph{arrondissement} ; it is the \emph{Saint-Germain} market and globalized luxury shopping for the international tourist ; it is a piece of history for the student of \emph{Ecole des Ponts} to which it reminds the era of \emph{Saint-Pères}. Objects, concepts, understood and defined by multiple disciplines and agents that produce knowledge: do we finally designate the same thing ? How to benefit from this wealth of viewpoints, how to integrate the complexity allowed by this diversity ? To bring elements of answer requires a constructive, generative, and as much inclusive as possible approach. Choices are always more enlighten if we have a grasp on a maximum of alternatives. The trader living in his loft at the top of \emph{mid-levels} and works in his close building between two rails, knows well Hong-Kong, but only one among its multiple faces, and it will be difficult to conceive the existence of a misery in Kwoloon, which inhabitants do not conceive the ephemeral but sometimes cyclic Hong-Kong of temporary workers from mainland, which them do not conceive the administrative and financial difficulties of migrants from Thailand or India, the whole picture being even less conceivable for a lost Parisian student. But it is indeed the loss, which in appropriate doses, will be source of a broader knowledge: ants establish their very precise optimizations from a walk that can be considered as random. Genetic algorithms, and even more biological evolution processes anchored in the physical, rely on a subtle compromise between order and disorder, between signal and noise, between stability and perturbations. To loose oneself to better find oneself makes the essence and the charm of the journey, let it be physical, conceptual, social. Finally, no possible comparison between orienteering in \emph{Le Caylar} or \emph{Montagne de Bange} to a rectilinear boredom in the \emph{Orléans} forest.}
}{
\textit{Un voyage, la découverte d'une ville, de nouvelles rencontres, un partage d'idées : autant de processus qui impliquent une générativité cognitive et une interaction complexe entre nos representations, nos actions et l'environnement. La construction d'une connaissance scientifique n'échappe pas à ces règles. On pourrait alors voir dans l'objet étudié lui-même, prenons la ville et ses agents, une allégorie du processus de production de connaissance sur l'objet. Comme Romain Duris qui débarque dans l'Auberge Espagnole, et découvre ces rues inconnues que plus tard on aura parcouru cent fois, où on aura vécu mille choses : on débarque dans un monde de concepts, d'approches, de point de vues complémentaires sur des choses qui ne sont pas la même chose. Cette discrépance ontologique est finalement tout aussi présente dans nos représentations de l'espace urbain : \emph{Oven Street} c'est un des centres de la connaissance pour le membre de Géocités ; c'est le centre de Paris, donc de la France, donc du Monde pour le fier autochtone du 6ème ; c'est le marché Saint-Germain et le shopping de luxe globalisé pour le touriste international; c'est un morceau d'histoire pour l'élève des Ponts pour qui cela évoque le temps des Saint-pères. Des objets, des concepts, compris et définis par de multiples disciplines et agents producteurs de connaissance : parle-t-on finalement vraiment de la même chose ? Comment tirer parti de cette richesse de points de vue, comment intégrer la complexité permise par cette diversité ? Apporter des éléments de réponse suppose une démarche constructive, générative et autant inclusive que possible. Les choix sont toujours plus éclairés si on a un aperçu d'un maximum d'alternatives. Le trader qui habite son loft en haut des \emph{mid-levels} et travaille dans son building à deux pas entre deux rails, connait bien Hong-Kong, mais un seul parmi ses multiples visages, et il lui sera difficilement concevable qu'existe une misère à Kwoloon, dont les habitants ne conçoivent pas le Hong-Kong éphémère mais parfois cyclique des travailleurs temporaires du mainland, qui eux ne conçoivent pas les difficultés administratives et financières de migrants de Thaïlande ou d'Inde, l'ensemble étant encore moins concevable pour un étudiant parisien égaré. Mais c'est justement l'égarement qui à dose appropriée sera source d'une connaissance plus large : les fourmis établissent leurs optimisations extrêmement précises à partir d'une marche qu'on peut considérer comme aléatoire. Les algorithmes génétiques, mais encore plus les processus d'évolution biologiques ancrés dans le physique, reposent sur un subtil compromis entre ordre et désordre, entre signal et bruit, entre stabilités et perturbations. Se perdre pour mieux se retrouver fait l'essence et le charme du voyage, qu'il soit physique, conceptuel, social. Finalement, pas de comparaison possible entre une orientation au Caylar ou sur la montagne de Bange à un ennui rectiligne en forêt d'Orléans.}
}

\bigskip


\bpar{
This literary interlude raises fundamental issues induced by a demand of interdisciplinarity and the will to construct a complex integrative knowledge. First, reflexivity and making a relation between a perspective taken with a certain number of other existing perspectives is necessary for its relevance. It is thus about constructing concepts in a solid way and to specify empirical references, in order to precise the problematic and its objectives \emph{endogenously}. Secondly, the epistemological frame of the approach must be given. Above is indeed pictures a \emph{perspectivist} approach, which is a particular epistemological positioning that we will detail here. Furthermore, the status of proofs is conditioned by the conception of methods and tools, which is particular in the case of simulation models.
}{
Cet intermède littéraire soulève des problèmes fondamentaux induits par une exigence d'interdisciplinarité et la volonté de construction d'une connaissance complexe intégrative. Dans un premier temps, la réflexivité et la mise en relation d'une perspective prise avec un certain nombre d'autres perspectives existantes est nécessaire pour la pertinence de celle-ci. Il s'agit donc de construire solidement les concepts et spécifier les références empiriques, afin de préciser la problématique et ses objectifs \emph{de manière endogène}. D'autre part, le cadre épistémologique de la démarche se doit d'être précisé. Ci-dessus est finalement imagée une approche \emph{perspectiviste}, qui est une position épistémologique particulière que nous détaillerons ici. De plus, le statut des démonstrations est conditionné par la conception des méthodes et des outils, qui est particulière dans le cas des modèles de simulation.
}

\bpar{
This part respond to these constraints, by building the \emph{foundations} necessary to the following of our work. In a relatively shifting terrain, these will have in some cases to be particularly deep for the global stability of the construction: this will for example be the case of the state of the art which will use techniques in quantitative epistemology. We recall that it is organized the following way:
\begin{enumerate}
	\item The first chapter constructs concepts and objects from a theoretical point of view, and unveils a broad spectrum of possible approaches to interactions between transportation networks and territories.
	\item The second chapter develops the different approaches in modeling interactions between networks and territories. It establishes the state of the art, structured by a typology previously obtained. It then describes the scientific landscape of concerned disciplines, and suggests the characteristics of models proper to each discipline and also possible determinants for it in a modelography.
	\item The third chapter is relatively independent and precises our epistemological positioning. It allows in particular to situate the complexity which we aim at reaching, to specify what can be expected from a modeling approach, and to give a broader definition of the concept of co-evolution.
\end{enumerate}
}{
Cette partie répond à ces contraintes, en posant les \emph{fondations} nécessaires à la suite de notre démarche. En terrain relativement mouvant, celles-ci devront dans certains cas être particulièrement profondes pour une stabilité de l'édifice global : ce sera par exemple le cas de l'état de l'art qui mobilisera des techniques d'épistémologie quantitative. Nous rappelons qu'elle s'organise de la manière suivante :
\begin{enumerate}
	\item Le premier chapitre construit les concepts et objets de manière théorique, et dégage un large éventail d'approches possibles aux interactions entre réseaux de transport et territoires.
	\item Le second chapitre développe les différentes approches de modélisation des interactions entre réseaux et territoires. Il établit un état de l'art, structuré par une typologie établie précédemment. Il dresse ensuite le paysage scientifique des disciplines concernées, et cherche les caractéristiques des modèles propres à chaque discipline ainsi que des possibles déterminants de celles-ci dans une modélographie.
	\item Le troisième chapitre est relativement indépendant et précise nos positions épistémologiques. Il permet notamment de situer la complexité dans laquelle nous cherchons à nous placer, de spécifier ce qui peut être attendu d'une démarche de modélisation, et de donner une définition plus large du concept de co-évolution.
\end{enumerate}
}

\stars

\bpar{
\chapter{Interactions between networks and territories}
}{
\chapter{Interactions entre réseaux et territoires}
}

\label{ch:thematic} 





\bpar{
Networks and territories seem to be interlaced in complex causal relationships. In order to better understand notions of circular causalities within complex systems, and why these can lead to apparent paradoxes, the image given by \noun{Diderot} in~\cite{diderot1965entretien} is enlightening: ``\textit{If you are embarrassed by the precedence of the chicken by the egg or of the egg by the chicken, it is because you are assuming that animals have always be the way they are now}''. By trying to naively tackle similar questions induced by our problematic previously introduced, causalities within geographical complex systems can be presented as a ``chicken-and-egg'' problem: if one effect seem to cause the other and reciprocally, is it possible and even relevant to try to isolate corresponding processes, if they are indeed part of a larger system which evolve at other scales ?
}{
Les réseaux et les territoires semblent s'entrelacer dans des relations causales complexes. Pour mieux appréhender les notions de causalités circulaires dans les systèmes complexes, et pourquoi celles-ci peuvent conduire à des paradoxes en apparence, l'image fournie par \noun{Diderot} dans~\cite{diderot1965entretien} est éclairante : ``\textit{Si la question de la priorit{\'e} de l'\oe{}uf sur la poule ou de la poule sur l'\oe{}uf vous embarrasse, c'est que vous supposez que les animaux ont {\'e}t{\'e} originairement ce qu'ils sont {\`a} pr{\'e}sent}''. En voulant traiter naïvement des questions similaires induites par notre problématique introduite précédemment, les causalités au sein de systèmes complexes géographiques peuvent être présentées comme un problème ``de poule et {\oe}uf'' : si un effet semble causer l'autre et réciproquement, est-il possible et même pertinent de vouloir isoler les processus correspondants, s'ils font en fait partie d'un système plus large qui évolue à d'autres échelles ?
}

\bpar{
A reducing approach, which would consist in attributing systematic roles to one component or the other, is opposed to the idea suggested by \noun{Diderot} which rejoins the one of \emph{co-evolution}. One of the issues is thus to give an overview of interaction processes between networks and territories, in order to precise the definition of co-evolution, what will be after a similar work for modeling approaches, at the end of the first part.
}{
Une vision réductrice, qui consisterait à attribuer des rôles systématiques à l'une composante ou l'autre, s'oppose à l'idée suggérée par \noun{Diderot} qui rejoint celle de \emph{co-évolution}. L'un des enjeux est donc de dresser un aperçu des processus d'interactions entre réseaux et territoires, afin de préciser la définition de la co-évolution, ce qui sera fait à l'issue d'un travail similaire pour les approches par la modélisation, à la fin de la première partie.
}

\bpar{
This chapter must be read as the construction introducing our objects and positions of study, and will be completed by an exhaustive literature review on the precise subject of modeling interactions, which will be the object of chapter~\ref{ch:modelinginteractions}.
}{
Ce chapitre doit être lu comme la construction introduisant nos objets et positions d'étude, et sera complété par une revue de littérature exhaustive sur le sujet précis de la modélisation des interactions, qui fera l'objet du chapitre~\ref{ch:modelinginteractions}.
}

\bpar{
In a first section~\ref{sec:networkterritories}, we will precise the approach we take of the territory object, and to what extent it implies to consider transportation networks for the understanding of coupled dynamics. This allows to construct a framework which gives a definition of territorial systems, and which is particularly suited to our approach through co-evolution.
}{
Dans une première section~\ref{sec:networkterritories}, nous préciserons l'approche prise de l'objet territoire, et dans quelle mesure celui-ci implique la considération des réseaux de transport pour la compréhension des dynamiques couplées. Cela permet de construire un cadre de lecture définissant les systèmes territoriaux, particulièrement adapté à notre approche par la co-évolution.
}

\bpar{
These abstract considerations will be illustrated by empirical case studies in the second section~\ref{sec:casestudies}, chosen as very different to understand the underlying universality issues: the Greater Paris metropolitan area and Pearl River Delta in China.
}{
Ces considérations abstraites seront illustrées par des cas d'étude empiriques dans la deuxième section~\ref{sec:casestudies}, choisis très différents pour comprendre les enjeux d'universalité sous-jacents : la métropole du Grand Paris et le Delta de la rivière des Perles en Chine.
}

\bpar{
Finally, in the last section~\ref{sec:qualitative}, fieldwork observation elements obtained in China will precise and make more complex the construction of this theoretical and empirical framework.
}{
Enfin, dans la troisième section~\ref{sec:qualitative}, des éléments d'observation de terrain effectués en Chine préciseront et complexifieront la construction de ce cadre théorique et empirique.
}

\stars


\bpar{
\textit{This chapter is fully unpublished.}
}{
\textit{Ce chapitre est entièrement inédit.}
}


%

\newpage


\section{Territories and networks}{Territoires et réseaux}

\label{sec:networkterritories}


\bpar{
We begin by constructing more precisely the concepts we will use. This construction helps to understand how the concepts of territory and network are rapidly in strong interaction, implying an ontological importance of interactions between corresponding objects. We will see that territories imply the existence of networks, but that reciprocally they are also influenced by them. A refined focus on properties of transportation networks allow to progressively a precise vision of \emph{co-evolution}, that we will take up to there in its preliminary sense given before, i.e. the existence of circular causal relationships between transportation networks and territories.
}{
Nous commençons par une construction plus précise des concepts mobilisés, qui permet de comprendre comment les concepts de territoire et de réseau sont rapidement en interdépendance forte, impliquant une importance ontologique des interactions entre les objets correspondants. Nous verrons que les territoires impliquent l'existence de réseaux, mais que réciproquement ceux-ci les influencent également. Un développement plus particulier sur les propriétés des réseaux de transport permet d'amener progressivement une vision précise de la \emph{co-évolution}, que nous prendrons jusque là dans son sens préliminaire donné précédemment, c'est-à-dire l'existence de relations causales circulaires entre réseaux de transports et territoires.
}

\subsection{Territories and networks, closely linked since their definition}{Territoires et Réseaux, intimement liés dès leur définition}

\subsubsection{Territories: an approach by systems of cities}{Territoires : une approche par les systèmes de villes}

\bpar{
The concept\footnote{We will use the term \emph{concept} for constructed knowledge, more than \emph{notion}, which following \cite{raffestin1978construits} is closer to an empirical information.} of \emph{territory}, that we introduced before through cities and systems of cities, will be central to our reasoning and must be depthen and enriched. In ecology, a territory corresponds to a spatial extent occupied by a group of agents or more generally an ecosystem \cite{tilman1997spatial}. Territories of human societies imply supplementary dimensions, for example through the importance of their semiotic representations\footnote{In the sense signs marking the territory and their meaning, but also their representations, as maps for example.}. These play a significant role in the emergence of social constructions, which genesis is profoundly linked to the one of urban systems. According to~\cite{raffestin1988reperes}, the \emph{Human Territoriality} is the ``conjonction of a territorial process with an informational process'', what means that the physical occupation and exploitation of space by human societies can not be dissociated from the representations (cognitive and material) of these territorial processes, driving in return its further evolutions.
}{
Le concept\footnote{Nous utiliserons le terme \emph{concept} pour des connaissances construites, plutôt que celui de \emph{notion}, qui suivant~\cite{raffestin1978construits} est plus proche d'une information empirique.} de \emph{territoire}, que nous avons introduit précédemment par ceux de ville et de système de ville, sera central à nos raisonnements et nécessite d'être approfondi et enrichi. En écologie, un groupe d'agents ou plus généralement un écosystème occupe une certaine étendue spatiale~\cite{tilman1997spatial}, qu'on peut identifier comme notion de territoire. Les territoires des sociétés humaines impliquent des dimensions supplémentaires, par exemple par l'importance de leur représentations sémiotiques\footnote{C'est-à-dire des signes marquants les territoires et leur sens, mais aussi leur représentations, cartographiques par exemple.}. Celles-ci jouent un rôle significatif dans l'émergence des constructions sociétales, dont la genèse est profondément liée à celle des systèmes urbains. Selon~\cite{raffestin1988reperes}, la \emph{Territorialité Humaine} est ``la conjonction d'un processus territorial avec un processus informationnel'', ce qui implique que l'occupation physique et l'exploitation de l'espace par les sociétés humaines sont complémentaires des représentations (cognitives et matérielles) de ces processus territoriaux, qui influent en retour sur leur évolution.
}

\bpar{
In other words, as soon as social constructions are implied in the constitution of human settlements, concrete and abstract social structures will play a role in the evolution of territories, and these two objects will be intimately binded. Examples of such links are for example the propagation of information and representations, political processes, or the conjunction or disjunction between lived and perceived territory. A territory is thus understood as a social structure organized in space, which includes its concrete abstract artifacts.
}{
En d'autres termes, à partir de l'instant où les constructions sociales déterminent la constitution des établissements humains, les structures sociales abstraites et concrètes joueront un rôle dans l'évolution des territoires, et ces deux objets seront intimement liés. Des exemples de tels liens se retrouvent à travers la propagation d'informations et de représentations, par des processus politiques, ou encore par la correspondance plus ou moins effective entre territoire vécu et territoire perçu. Un territoire est ainsi compris comme une structure sociale organisée dans l'espace, qui comprend ses artefacts concrets et abstraits.
}

\bpar{
This approach of the territory rejoin the preliminary definition we took and reinforces it. The approach of \noun{Raffestin} insists on the role of cities as places of power (in the sense of a place gathering decision processes and of socio-economic control) and of wealth creation through social and economical exchanges and interactions\footnote{An interaction will be taken in its broader meaning, as a reciprocal action of several entities one on the other. It can be physical, informational, transform the entities, etc. See \cite{morin1976methode} for a complete and complex construction of the concept, closely linked with the concept of organisation.}. The city has however no existence without its hinterland, that can be interpreted as the \emph{territory of a city}\footnote{Although an exact correspondance between territories and cities is probably only a simplification of reality, since territories can be entangled at different scales, along different dimensions. A reading through central places typical of \noun{Christaller}~\cite{banos2011christaller} gives a conceptual idea of this correspondance. Functional definitions such as \emph{Insee}'s urban areas, that defines the area around a center above a critical size (10000 jobs) by the cities for which a minimal threshold of actives work in that center (40\%) - see \url{https://www.insee.fr/fr/metadonnees/definition/c2070}, is a possible approach. The sensitivity of the properties of the urban system to these parameters is tested by~\cite{2015arXiv150707878C}. The definition of the city is therefore intimely linked to the one of territories, and the definition of the urban system to the set of territories.}. This correspondence sheds a light on all territories from the point of view of systems of cities, as developed by the evolutive urban theory~\cite{pumain2010theorie}. This theory interprets cities as complex self-organized systems, which act as mediators of social change: for example, innovation cycles initialize within cities and propagate between them (see~\ref{app:sec:patentsmining} for an empirical entry on the notion of innovation). It yield a vision of the territory as a space of flows, what will introduce the notion of network as we will see further. Cities are furthermore seen as competitive agents that co-evolve \cite{paulus2004coevolution}, what already suggests the importance of co-evolution for territorial dynamics.
}{
Cette approche du territoire rejoint la définition préliminaire que nous en avions prise, et vient alors la renforcer. L'approche de \noun{Raffestin} insiste sur le rôle des villes comme lieu de pouvoir (au sens d'un lieu rassemblant des processus décisionnels et de contrôle socio-économique) et de création de richesse au travers des échanges et interactions\footnote{Une interaction sera comprise dans son sens le plus général, comme une action réciproque de plusieurs entités l'une sur l'autre. Celle-ci peut être physique, informationnelle, transformer les entités, etc. Voir~\cite{morin1976methode} pour une construction complète et complexe du concept, en lien intime avec celui d'organisation.} (sociaux, économiques). La ville n'a cependant pas d'existence sans son hinterland, ce que nous pouvons interpréter comme le \emph{territoire d'une ville}\footnote{Même si une correspondance exacte entre territoires et villes n'est probablement qu'une simplification de la réalité, puisque les territoires peuvent s'entremêler à différentes échelles, selon différentes dimensions. Une lecture par lieux centraux de type \noun{Christaller}~\cite{banos2011christaller} permet de se faire une image conceptuelle de cette correspondance. Des définitions fonctionnelles comme celles des aires urbaines de l'Insee, qui définit l'aire autour d'un pôle dépassant une taille critique (10000 emplois) par les communes dont un seuil minimal d'actifs travaillent dans le pôle (40\%) - voir \url{https://www.insee.fr/fr/metadonnees/definition/c2070}, est une approche possible. La sensibilité des propriétés du système urbain à ces paramètres est testée par~\cite{2015arXiv150707878C}. La définition de la ville est alors intimement liée à celle de ses territoires, et celle du système urbain à l'ensemble des territoires.}. Cette correspondance permet de lire l'ensemble des territoires au prisme du système de villes, comme développé par la théorie évolutive des villes~\cite{pumain2010theorie}. Celle-ci interprète les villes comme des systèmes complexes auto-organisés, qui agissent comme des médiateurs du changement social : par exemple, les cycles d'innovation s'initialisent au sein des villes et se propagent entre elles (voir~\ref{app:sec:patentsmining} pour une entrée empirique sur la notion d'innovation). Cela permet de comprendre le territoire comme un espace des flux, ce qui permettra d'introduire la notion de réseau comme nous le verrons plus loin. Les villes sont par ailleurs vues comme des agents compétitifs qui co-évoluent~\cite{paulus2004coevolution}, ce qui permet de préfigurer également l'importance de la co-évolution pour les dynamiques territoriales.
}


\bpar{
We have thus two complementary approaches of the territory that allow us to consider human territories structured by systems of cities\footnote{These complementary views on the territory can also be enriched with an historical perspective. \cite{di1998espace} gives an historical analysis of the different conceptions of space (that lead in particular to the lived space, the social space and the classical space of geography) and shows how their combination yields what \noun{Raffestin} describes as territories. \cite{giraut2008conceptualiser} recalls the different recent uses that have been done of the concept of territory, from cultural geography where it was used more as a scientific fashion, to geopolitics where it is a very specific term linked to governance structures, to uses where it is more an abstract concept, and highlights therein the interdisciplinary aspect of an object capturing a certain level of complexity of the systems studied.}.
}{
On a ainsi deux approches complémentaires du territoire qui nous permettent de considérer des territoires humains structurés par les systèmes de villes\footnote{Ces visions complémentaires du territoire peuvent également être enrichies par une perspective historique. \cite{di1998espace} procède à une analyse historique des différentes conceptions de l'espace (qui aboutissent entre autres à l'espace vécu, l'espace social et l'espace classique de la géographie) et montre comment leur combinaison forme ce que \noun{Raffestin} décrit comme territoires. \cite{giraut2008conceptualiser} rappelle les différents usages récents qui ont été faits du concept de territoire, de la géographie culturelle où il a plus été utilisé par effet de mode, à la géopolitique où c'est un terme bien spécifique lié aux structures de gouvernance, en passant par des utilisations où il sert plus de concept abstrait, et dégage l'aspect interdisciplinaire d'un objet capturant une certaine complexité des systèmes étudiés.}.
}

\bpar{
Moreover, a central aspect of human settlements that were studied in geography for a long time, and that relates directly to the concept of territory, is the one of \emph{networks}. We will detail their definition and show how switching from one to the other is intrinsic to the approaches we take on these.
}{
Par ailleurs, un aspect central des établissements humains qui a une longue tradition d'étude en géographie, et qui est directement relié au concept de territoire, est celui des \emph{réseaux}. Nous allons préciser leur définition et voir comment le passage de l'un à l'autre est intrinsèque aux approches que nous en prenons.
}

\subsubsection{Definition of networks}{Définition des réseaux}


\bpar{
A \emph{network} must be understood in the broad sense of the establishment of relations between entities of a system, that can be seen as abstract relations, links, interactions. \cite{haggett1970network} postulates that the existence of a network is necessarily linked to the existence of flows\footnote{Flows are defined as a material exchange (people, goods, raw materials) or immaterial (information) between two entities.}, and recalls the topological representation as a graph of any geographical system in which flows circulate between entities or places that are abstracted as nodes, linked by edges. Edges of the graph have then a \emph{capacity}, which translate their ability to transport flows (that can be defined in a similar way as an \emph{impedance}). The topological analysis already unveils a certain number of system properties, but \cite{haggett1970network} precises the importance of the network spatialization, included in the properties of its nodes (localization) and of its links (localization, impedance), for the understanding of dynamics within the network (flows) or of the network itself (network growth). This specificity is recalled by~\cite{barthelemy2011spatial} which puts into perspective empirical domains that relate to spatial networks, some network growth models, and some models of processes within networks: for example, topological structures, or diffusion processes will be strongly constrained by the spatial dimension.
}{
Un \emph{réseau} doit être compris au sens large d'une mise en relation entre entités d'un système, qui peuvent être vus comment relations abstraites, liens, interactions. \cite{haggett1970network} postule que l'existence d'un réseau est nécessairement liée à celle de flux\footnote{On définit le flux comme un échange matériel (personnes, marchandises, matières premières) ou immatériel (information) entre deux entités.}, et rappelle la représentation topologique sous forme de graphe de tout système géographique dans lequel circulent des flux entre des entités ou des lieux qui sont abstraits sous la forme de noeuds, reliés par des liens. Les liens du réseau disposent alors d'une \emph{capacité}, qui traduit leur capacité à transporter les flux (qui peut également être définie de manière équivalence comme \emph{impédance}). L'analyse topologique révèle déjà un certain nombre de propriétés du système, mais \cite{haggett1970network} précise l'importance de la spatialisation du réseau, incluse dans les propriétés de ses noeuds (localisation) et de ses liens (localisation, impédance), pour la compréhension des dynamiques dans le réseau (flux) ou du réseau lui-même (croissance du réseau). Cette spécificité est rappelée par~\cite{barthelemy2011spatial} qui met en perspective les domaines empiriques concernés par les réseaux spatiaux, certains modèles de croissance de réseau, et certains modèles de processus dans les réseaux : par exemple, les structures topologiques, ou les processus de diffusion seront très contraints par le caractère spatial.
}


\bpar{
To study more thoroughly the concept of network by focusing on its strong interdependency with the concept of territory, we follow~\cite{dupuy1987vers} which proposes elements for ``a territorial theory of networks'' inspired by the concrete case of an urban transportation network. This theory distinguishes \emph{real networks}\footnote{Real networks include a category that can be described as concrete, material or physical networks - we will use these terms in an interchangeable manner in the following, to which transportation networks belong; other categories such as social networks are also real networks that we will not study.} and \emph{virtual networks}, that are themselves induced partly by the territorial configuration. Real networks are the materialization of virtual networks. More precisely, a territory is characterized by strong spatio-temporal discontinuities induced by the non-uniform distribution of agents and ressources. These discontinuities naturally induce a network of of potential interactions between the elements of the territorial system, namely agents and ressources. \cite{dupuy1987vers} designates these potential interactions as \emph{transactional projects}. These induce the notion of \emph{potential of interaction}, i.e. a property of space from which the interactions derive\footnote{Given any vectorial field of class $\mathcal{C}^1$ on $\mathbb{R}^3$, the \noun{Helmoltz} theorem yields a vector potential and a scalar potential from which this field derives as a rotational and a gradient. It justifies in the particular case of such a viewpoint the correspondence between an interaction field between agents and a potential field.}. For example nowadays people need to access the ressource of employments, economic exchanges operate between different territories that can be more or less specialized in different types of production.
}{
Pour approfondir le concept de réseau en appuyant sur sa forte interdépendance avec celui de territoire, nous reprenons~\cite{dupuy1987vers} qui propose des éléments pour une ``théorie territoriale des réseaux'' s'inspirant du cas concret d'un réseau de transport urbain. Cette théorie distingue les \emph{réseaux réels}\footnote{Les réseaux réels contiennent une catégorie qu'on peut désigner comme réseaux concrets, matériels ou physiques - nous utiliserons ces termes de manière interchangeable par la suite, à laquelle les réseaux de transport appartiennent ; d'autres catégories comme les réseaux sociaux sont également des réseaux réels sur lesquels nous ne nous attarderons pas.} et les \emph{réseaux virtuels}, eux-mêmes induits entre autres par la configuration territoriale. Les réseaux réels sont la matérialisation de réseaux virtuels. Plus précisément, un territoire est caractérisé par de fortes discontinuités spatio-temporelles induites par la distribution non-uniforme des agents et des ressources. Ces discontinuités induisent naturellement un réseau d'interactions potentielles entre les éléments du système territorial, notamment des agents et des ressources. \cite{dupuy1987vers} désigne ces interactions potentielles comme \emph{projets transactionnels}. Celles-ci induisent la notion de \emph{potentiel d'interaction}, c'est-à-dire une propriété de l'espace dont les interactions dérivent\footnote{Étant donné tout champ vectoriel de classe $\mathcal{C}^1$ sur $\mathbb{R}^3$, le théorème d'\noun{Helmoltz} fournit un potentiel vecteur et un potentiel scalaire dont ce champ dérive par rotationnel et gradient. Cela justifie dans le cas particulier d'un tel point de vue formel le passage d'un champ d'interactions entre agents à un champ de potentiel.}. Par exemple, de nos jours les actifs ont besoin d'accéder à la ressource qu'est l'emploi, et des échanges économiques s'effectuent entre les différents territoires qui peuvent être plus ou moins spécialisés dans les productions de différents types.
}

\subsubsection{From networks to real networks}{Des réseaux aux réseaux réels}


\bpar{
In some cases, a potential network is materialized into a real network. The underlying question is then to determine if the potential field of territories is partly at the origin of this materialization, if it is totally independent, or if the dynamic of the two is strongly coupled, in other terms in co-evolution. The materialization will generally result of the combination of economic and geographical constraints with demand patterns, in a non-linear way. Such a process is not immediate, leading to strong non-stationarity and path-dependancy effects\footnote{Spatial non-stationarity consists in the dependancy of the covariance structure of processes to space, whereas path-dependency corresponds to the fact that trajectories taken in the past strongly influence the current trajectories of the system.}: the extension of an existing network will depend on previous configurations, and depending on involved time scales, the logic and even the nature of operators, i.e. agents participating to its production, may have evolved.
}{
Dans certains cas, un réseau potentiel peut se matérialiser en réseau réel. La question sous-jacente est alors de savoir si le champ de potentiel des territoires est en partie à l'origine de cette matérialisation, si celle-ci est totalement indépendante, ou si la dynamique des deux est fortement couplée, en d'autres termes en co-évolution. La matérialisation résultera généralement de la combinaison de contraintes économiques et géographiques avec des motifs de demande, de manière non-linéaire. Un tel processus est loin d'être immédiat, et conduit à de forts effets de non-stationnarité et de dépendance au chemin\footnote{La non-stationnarité spatiale consiste en la dépendance de la structure de covariance des processus à l'espace, tandis que la dépendance au chemin traduit le fait que les trajectoires prises par le passé influencent fortement les trajectoires actuelles du système.} : l'extension d'un réseau existant dépendra de la configuration précédente, et selon les échelles de temps impliquées, la logique et même la nature des opérateurs, c'est-à-dire des agents participant à sa production, peut avoir évolué.
}

\bpar{
Examples of concrete trajectories can be quite varied: \cite{kasraian2015development} show for example, in the case of Randstad on long time, a first period during which the railway network has developed to follow urban development, whereas opposite effects has been more recently observed. At a urban scale on long time, the path-dependency is shown for Boston by~\cite{block2012hysteresis} since the built environment and the distribution of population appear as highly dependant of past tramway lines even when they do not exist anymore: the way the transportation line changes the urban space acts on immediate dynamics but also on a longer time through reinforcement effects or because of the inertia of the built environment for example. 
}{
Les exemples de trajectoires concrètes peuvent être très variées : \cite{kasraian2015development} montrent par exemple dans le cas de la Randstad sur le temps long, une première période pendant laquelle le réseau ferré s'est développé pour suivre le développement urbain, tandis que des effets inverses ont été constatés plus récemment. À une échelle urbaine sur le temps long, la dépendance au chemin est montrée pour Boston par~\cite{block2012hysteresis} puisque l'environnement bâti et la distribution de la population apparaissent comme fortement dépendants des lignes de tramway antérieures même lorsqu'elles n'existent plus : la façon dont la ligne de transport change l'espace urbain s'opère dans les dynamiques immédiates mais aussi sur le temps long par des effets de renforcement ou à cause de l'inertie du bâti par exemple.
}

\bpar{
Therefore, the existence of a human territory necessarily imply the presence of abstract interaction networks, and concrete networks are crucial for the transport of people and ressources (including communication networks as information is a crucial ressource~\cite{morin1976methode}), but the processes through which they are established are difficult to identify generally. Our ontological choice of positioning within \noun{Dupuy}'s theory, gives a privileged place to the relations between networks and territories, since it induces in the construction of the objects themselves a complex entanglement between these.
}{
Ainsi, l'existence d'un territoire humain implique nécessairement la présence de réseaux d'interactions abstraites, et les réseaux concrets sont cruciaux pour transporter les individus et les ressources (incluant les réseaux de communication puisque l'information est une ressource essentielle~\cite{morin1976methode}), mais les processus d'établissement de ceux-ci sont difficiles à identifier de manière générale. Notre choix ontologique de positionnement dans la théorie de \noun{Dupuy}, donne une place privilégiée aux relations entre réseaux et territoires, puisqu'il induit dans la construction des objets même une imbrication complexe entre ceux-ci.
}


\bpar{
The status of the network in relation with the territory is moreover highly conditioned by the socio-economical and technological context. Following~ \cite{duranton1999distance}, a factor influencing the form of pre-industrial cities was the performance of transportation networks. Technological progresses, leading to a decrease in transportation costs, have inducted a regime change, what conducted to a preponderance of land markets in shaping cities (and thus a role of transportation network since they influence prices through accessibility), and more recently to the rising importance of telecommunication networks what induced a ``tyranny of proximity'', since a physical presence can not be replaced by virtual communications \cite{duranton1999distance}.
}{
Le statut du réseau par rapport au territoire est d'autre part fortement conditionné par le contexte socio-économique et technologique. Selon \cite{duranton1999distance}, un facteur influençant la forme des villes pré-industrielles était la performance des réseaux de transport. Les progrès technologiques, conduisant à une baisse des coûts de transport, ont induit un changement de régime, ce qui a mené à une prépondérance du marché foncier dans la formation des villes (et par conséquent un rôle des réseaux de transport qui déterminent les prix par l'accessibilité), et plus récemment à une importance croissante des réseaux de télécommunication ce qui a induit une ``tyrannie de la proximité'' puisque la présence physique n'est pas remplaçable par une communication virtuelle \cite{duranton1999distance}.
}


\bpar{
This territorial approach to networks seems natural in geography, since networks are studied conjointly with geographical objects they connect, in opposition to theoretical works on complex networks which study them in a relatively disconnected way from their thematic background~\cite{ducruet2014spatial}.
}{
Cette approche territoriale des réseaux semble naturelle en géographie, puisque les réseaux sont étudiés conjointement avec des objets géographiques qu'ils connectent, en opposition aux travaux théoriques sur les réseaux complexes qui les étudient de manière relativement déconnectée de leur fond thématique~\cite{ducruet2014spatial}.
}

\subsubsection{Networks shaping territories ?}{Des réseaux qui façonnent les territoires ?}


\bpar{
However networks are not only a material manifestation of territorial processes, but play their role in these processes since their evolution may influence the evolution of territories in return. Here comes an intrinsic difficulty: it is far from evident to attribute territorial mutations to an evolution of the network, and reciprocally the materialization of a network to precise territorial dynamics. Different exogenous factors are furthermore important, such as the price of energy or existing technologies in the case of the effect of the network on territories for example. In the case of \emph{technical networks}, an other designation of concrete networks given in~\cite{offner1996reseaux}, many examples of such feedbacks can be found: an increased accessibility may shape urban growth, or the interconnectivity of different transportation networks allows a significant extension of mobility ranges. At a smaller scale, changes in accessibility may induce relocalizations of different urban components. These retroactions of networks on territories does not necessarily act on concrete components: \cite{claval1987reseaux} shows that transportation and communication networks contribute to the collective representation of a territory by acting on the sentiment to belong to the territory, that can then play a crucial role in the emergence of a strongly coherent regional dynamic. We first develop with more details the possible influences of networks on territories.
}{
Cependant les réseaux ne sont pas seulement une manifestation matérielle de processus territoriaux, mais jouent également leur rôle dans ces processus puisque leur évolution peut influencer l'évolution des territoires en retour. Il emerge alors une difficulté intrinsèque : il n'est pas évident d'attribuer des mutations territoriales à une évolution du réseau et réciproquement la matérialisation d'un réseau à des dynamiques territoriales précises. Différents facteurs exogènes rentrent par ailleurs en compte, comme le prix de l'énergie ou les technologies existantes dans le cas de l'effet du réseau sur les territoires par exemple. Dans le cas des \emph{réseaux techniques}, une autre désignation des réseaux concrets donnée dans~\cite{offner1996reseaux}, de nombreux exemples de tels retroactions peuvent être mis en évidence : une accessibilité accrue peut être un facteur favorisant la croissance urbaine, ou bien l'interconnexion de différents réseaux de transport permet une extension significative de la portée des déplacements. À une plus petite échelle, des changements de l'accessibilité peuvent induire des relocalisations de différentes composantes urbaines. Ces rétroactions des réseaux sur les territoires n'agissent pas nécessairement sur des composantes concretes : \cite{claval1987reseaux} montre que les réseaux de transport et de communication contribuent à la représentation collective d'un territoire en agissant sur un sentiment d'appartenance, qui peut alors jouer un rôle crucial dans l'émergence d'une dynamique régionale fortement cohérente. Développons d'abord plus en détail les possibles influences des réseaux sur les territoires.
}

\bpar{
The confusion on possible simple causal relationships has fed a scientific debate that is still active nowadays. The underlying question relies on more or less deterministic attributions of impacts to transportation infrastructures or to a new transportation mode on territorial transformations. Precursors of such a reasoning can be tracked back in the twenties: \noun{McKenzie}, from the Chicago school, mentions in~\cite{burgess1925city} some ``modifications of forms of transportation and communication as determining factors of growth and decline cycles [of territories]'' (p.~69). Methodologies to identify what is then called \emph{structuring effects} of transportation networks has been developed for planning in the seventies: \cite{bonnafous1974methodologies} situates the concept of structuring effect in the perspective of using the transportation offer as a planning tool (the alternatives are the development of an offer to answer to a congestion of the network, and the simultaneous development of associated offer and planning). These authors identify from an empirical viewpoint direct effects of a novel offer on the behavior of agents, on transportation flows and possible inflexions on socio-economic trajectories of concerned territories. \cite{bonnafous1974detection} develop a method to identify such effects through the modification of the class of cities in a typology established a posteriori. More recently, \cite{bonnafous2014observatoires} recalls that the institution of \emph{permanent observatories} for territories makes such analyses more robust, allowing a continuous monitoring of the territories that are the most concerned by the extent of a new infrastructure.
}{
La confusion autour de possibles relations causales simples a nourri un débat scientifique encore actif aujourd'hui. La question sous-jacente repose sur des attributions plus ou moins déterministes d'impact d'infrastructures ou d'un nouveau mode de transport sur des transformations territoriales. Nous pouvons trouver des précurseurs de ce raisonnement dès les années 1920 : \noun{McKenzie}, de l'école de Chicago, parle dans~\cite{burgess1925city} des ``modifications des formes du transport et de la communication comme facteurs déterminants des cycles de croissance et de déclin [des territoires]'' (p.~69). Des méthodologies pour identifier ce qui est alors nommé \emph{effets structurants} des réseaux de transport ont été développées pour la planification dans les années 1970 : \cite{bonnafous1974methodologies} situe le concept d'effet structurant dans le cadre d'une logique d'utilisation de l'offre de transport comme outil d'aménagement (les alternatives étant le développement d'une offre pour répondre à une congestion du réseau, et le développement simultané d'une offre et d'un aménagement associé). Ces auteurs identifient du point de vue empirique des effets directs d'une nouvelle offre sur le comportement des agents, sur les flux de transport et des possibles inflexions sur les trajectoires socio-économiques des territoires concernés. \cite{bonnafous1974detection} développent une méthode pour identifier de tels effets par modifications de la classe des communes dans une typologie établie a posteriori. Plus récemment, \cite{bonnafous2014observatoires} rappelle que la mise en place \emph{d'observatoires permanents} des territoires permet de rendre plus robustes ce type d'analyse, en permettant un suivi continu de l'évolution des territoires les plus concernés par l'emprise d'une nouvelle infrastructure.
}

\bpar{
According to \cite{offner1993effets} which follows ideas already given by~\cite{franccois1977autoroutes} for example, a not reasoned and out-of-context use of these methods has then been developed by planners and politicians which generally used them to justify transportation projects in a technocratic manner: through the argument of a direct effect of a new infrastructure on local development (for example economic), politics are able to ask for subsidies and to legitimate their action in front of the people. \cite{offner1993effets} insists on the necessity of a critical positioning on these issues, recalling that there exists no scientific demonstration of an effect that would be systematic. A special issue of the journal \emph{L'Espace Géographique}~\cite{espacegeo2014effets} on that debate recalled that on the one hand misconceptions and misuses were still greatly present in operational and planning communities, which can be explained for example by the need to justify public actions, and on the other hand that a scientific understanding of relations between networks and territories is still in construction. \noun{A. Bonnafous} (interview on the 09/01/2018, see Appendix~\ref{app:sec:interviews}) gives the current example of the project of the Seine-Nord-Europe canal\footnote{The canal project links the Oise at Compiègne to the Dunkerque-Escault canal in the north, see \url{https://www.canal-seine-nord-europe.fr/Projet}.} as a transportation project for which traffic previsions were largely overestimated and that politics of concerned territories have largely instrumentalized.
}{
Selon \cite{offner1993effets} qui reprend des idées déjà évoquées par~\cite{franccois1977autoroutes} par exemple, il s'est par la suite développé un usage non raisonné et hors contexte de ces méthodes par les planificateurs et les politiques qui les mobilisaient généralement pour justifier des projets de transports de manière technocratique : par l'argument d'un effet direct d'une nouvelle infrastructure sur le développement local (par exemple économique), les élus sont en mesure de demander des financements et de légitimer leur action auprès des contribuables. \cite{offner1993effets} insiste sur la nécessité d'un positionnement critique sur ces enjeux, rappelant qu'il n'existe pas de démonstration scientifique d'un effet qui serait systématique. Une édition spéciale de l'Espace Géographique sur ce débat~\cite{espacegeo2014effets} a rappelé d'une part que de telles croyances était encore largement présentes aujourd'hui dans les milieux opérationnels de la planification, ce qui peut s'expliquer par exemple par le besoin de justifier l'action publique, et d'autre part qu'une compréhension scientifique des relations entre réseaux et territoires est encore en pleine construction. \noun{A. Bonnafous} (entretien du 09/01/2018, voir Annexe~\ref{app:sec:interviews}) donne l'exemple actuel du projet du canal Seine-Nord-Europe\footnote{Le projet de canal relie l'Oise à Compiègne au canal Dunkerque-Escault au nord, voir \url{https://www.canal-seine-nord-europe.fr/Projet}.} comme projet de transport pour lequel les prévisions de trafic ont été largement surestimées et que les élus des territoires concernés ont largement instrumentalisé.
}

\bpar{
An other concrete illustration in the actuality gives an idea of this instrumentalization: debates in July of 2017 concerning the opening of the \emph{LGV Bretagne} and the \emph{LGV Sud-Ouest} have shown the full ambiguity of positions, conceptions, imaginaries both of politics but also of the public: worries on the speculation on real estate in stations neighborhoods, questionings on daily mobility but also social mobility\footnote{See for example \url{http://www.liberation.fr/futurs/2017/07/02/immobilier-plus-de-parisiens-comment-les-bordelais-voient-l-arrivee-de-la-lgv_1580776}, or \url{http://www.lemonde.fr/big-browser/article/2017/10/24/a-bordeaux-une-fronde-anti-parisiens-depuis-l-ouverture-de-la-ligne-a-grande-vitesse_5205282_4832693.html} for an immediate reaction of diverse local actors, witnessing at least an impact on representations. For example, people in Bordeaux seem to fear the arrival of Parisians searching for cheaper housing and better living conditions, what could increase prices in the surroundings of the station.}. The complexity and the reach of these subjects show well the difficulty of a systematic understanding of effects of transportation on territories.
}{
Une autre illustration concrète d'actualité permet de se faire une image de cette instrumentalisation : les débats en juillet 2017 relatifs à l'ouverture des LGV Bretagne et Sud-Ouest ont montré toute l'ambiguïté des positions, des conceptions, des imaginaires à la fois des politiques mais aussi du public : inquiétude quant à la spéculation sur l'immobilier dans les quartiers de gare, questionnements des pratiques de mobilité quotidienne mais aussi sociale\footnote{Voir par exemple \url{http://www.liberation.fr/futurs/2017/07/02/immobilier-plus-de-parisiens-comment-les-bordelais-voient-l-arrivee-de-la-lgv_1580776}, ou \url{http://www.lemonde.fr/big-browser/article/2017/10/24/a-bordeaux-une-fronde-anti-parisiens-depuis-l-ouverture-de-la-ligne-a-grande-vitesse_5205282_4832693.html} pour une réaction ``à chaud'' de divers acteurs locaux, témoignant d'un impact au minimum sur les représentations. Par exemple, les Bordelais semblent craindre l'arrivée de Parisiens en recherche d'un logement moins cher et de meilleures conditions de vie, ce qui pourrait augmenter les prix aux environs de la gare.}. La complexité et la portée des sujets montrent bien la difficulté d'une compréhension systématique d'effets du transport sur les territoires.
}

\subsubsection{An integrative approach: Territorial Systems}{Une vision intégrative : les Systèmes Territoriaux}

\bpar{
This overview as an introduction, from territories to networks, allows us thus to clarify our approach of territorial systems that will be underlying all the following. Taking into account diverse potential feedbacks of networks for the understanding of territories is suggested when coming back to the citation by Diderot that introduced the subject, in the sense that we must consider neither the network nor territories as independent systems that would influence themselves through one directional causal relations, but as strongly coupled components of a broader system, and thus being in a circular causal relationship. Depending on components and the scale that are considered, different manifestations of these will be observable, and there will exist some cases where there is apparently the influence of one on the other, other where influences are simultaneous, or moreover others where no relationship can be observed in a significant way. 
}{
Cet aperçu introductif, des territoires aux réseaux, nous permet ainsi de clarifier notre approche des systèmes territoriaux qui sera sous-jacente dans l'ensemble de la suite. Une prise en compte des diverses rétroactions potentielles des réseaux pour la compréhension des territoires est suggérée par un retour à la citation de Diderot ayant introduit le sujet, au sens où il ne faut pas considérer le réseau ni les territoires comme des systèmes indépendants qui s'influenceraient soit l'un soit l'autre par des relations causales en sens unique, mais comme des composantes fortement couplées d'un système plus large, et donc étant en relations causales circulaires. Selon les composantes ainsi que l'échelle considérées, différentes manifestations de celles-ci pourront être observables, et il existera des cas où il y a apparemment influence de l'une sur l'autre, d'autres où les influences sont simultanées, ou encore d'autres où aucune relation n'est observable de manière significative.
}

\bpar{
Since we have highlighted the role of networks in several aspects of territorial dynamics, we propose a definition of territorial systems that explicitly includes them. We consider a \emph{Territorial System} as a \emph{human territory that contains both interactions networks and real networks}. Real networks, and more particularly concrete networks\footnote{Which are as we previously saw materialized real networks.}, are an entire component of the system, influencing evolution processes, through multiple feedbacks with other components at many spatial and temporal scales. 
}{
Comme nous avons mis en exergue le rôle des réseaux dans de nombreux aspects des dynamiques territoriales, nous proposons une définition des systèmes territoriaux les incluant explicitement. Nous considérons un \emph{Système Territorial} comme un \emph{territoire humain qui contient à la fois des réseaux d'interactions et des réseaux réels}. Les réseaux réels, et plus particulièrement les réseaux concrets\footnote{Qui comme nous l'avons vu précédemment sont des réseaux réels matérialisés.}, sont une composante à part entière du système, jouant dans les processus d'évolution, au travers de multiples rétroactions avec les autres composantes à plusieurs échelles spatiales et temporelles.
}

\bpar{
The network is not necessarily a component in itself of the territory, but indeed of the \emph{Territorial System} in our sense\footnote{This ontological choice is not innocent and reinforces the dialectic between networks and territories. Starting from the distant past where physical networks did not exist, the emergence of a human territory, that we assume equivalent to a network of interactions, induces the establishment of the complex diachronic dialectic between physical networks and human territories. We can thus read the genesis of a territorial system as a morinian loop~\cite{morin1976methode}, in which we enter by the initial territory and which then loops from the physical network to territorial components to produce the territorial system (thus the territory in most cases) in the following recursive way:\\Initial territory $\rightarrow$ Territory $=$ \tikzmark{Territorial} configuration $\rightarrow$ Physical \tikzmark{network}\arrow{network}{Territorial}\\}. This view rejoins the positioning of \cite{dupuy1985systemes} which introduces the territory as the ``product of a dialectic'' between territorial components and networks. We remark the semantic shortcut to designate components of the territorial system that are not the network and which interact with it, through the term of territory. These depend on ontologies and scales considered, as we will see in the following, and can span from microscopic agents to cities themselves. As we will also see in the following (see~\ref{sec:modelingsa}), there exists some paradigms in which this simplification is not done, such as in the particular case of interactions between transportation and land-use where entities are specific. But it is done if we stay in a more general framework, as witnesses one of the reference works on the subject~\cite{offner1996reseaux}\footnote{When \cite{amar1985essai} proposes a conceptual model of network morphogenesis, he designates the territorial components as ``The World'', what does not solve the semantic issue. The choice to keep the term of territory, within the territory, suggests a recursivity, and thus a complexity in the generativity of the system~\cite{morin1976methode}. The use of the concept of morphogenesis starting from chapter~\ref{ch:morphogenesis} suggests that this recursivity would not be spurious, but indeed intrinsic to the problem.}. We will similarly postulate this semantic simplification, when designating by \emph{interactions between networks and territories} or \emph{co-evolution between networks and territories}, the interactions or the co-evolution between physical networks and components they connect, within a territorial system and thus a territory.
}{
Le réseau n'est pas nécessairement une composante en tant que telle du territoire, mais bien du \emph{Système Territorial} en notre sens\footnote{Ce choix ontologique n'est pas anodin et appuie la dialectique entre réseaux et territoires. Partant de l'époque lointaine où les réseaux physiques n'existaient pas, l'émergence d'un territoire humain, que nous supposons équivalent à un réseau d'interactions, induit la mise en place de la dialectique diachronique complexe entre réseaux physiques et territoires humains. On peut ainsi lire la genèse du système territorial comme une boucle morinienne~\cite{morin1976methode}, dans laquelle on entre par le territoire initial puis qui se boucle du réseau physique aux composantes territoriales pour former le système territorial (donc le territoire dans la majorité des cas) de la manière récursive suivante :\\Territoire initial $\rightarrow$ Territoire $=$ \tikzmark{Configuration} territoriale$\rightarrow$ Réseau \tikzmark{physique}\arrow{physique}{Configuration}\\}. Cette vision rejoint le positionnement de \cite{dupuy1985systemes} qui introduit le territoire comme ``produit d'une dialectique'' entre composantes territoriales et réseaux. Notons le raccourci sémantique pour désigner les composantes du système territorial qui ne sont pas les réseaux et qui interagissent avec celui-ci, par le terme de territoire. Celles-ci dépendent des ontologies et des échelles considérées, comme nous le verrons par la suite, et peuvent aller des agents microscopiques aux villes elle-mêmes. Comme nous le verrons aussi par la suite (voir~\ref{sec:modelingsa}), il existe des paradigmes où ce raccourci n'est pas fait, comme dans le cas particulier des interactions entre transport et usage du sol où les entités sont spécifiques. Mais il est fait si nous restons dans un cadre plus général, comme en témoigne l'un des ouvrages de référence sur le sujet~\cite{offner1996reseaux}\footnote{Lorsque \cite{amar1985essai} propose un modèle conceptuel de morphogenèse des réseaux, il désigne les composantes territoriales par ``Le Monde'', ce qui n'apporte pas de solution au problème sémantique. Le parti pris de garder le territoire, au sein du territoire, suggère une récursivité, et donc une complexité dans la générativité du système~\cite{morin1976methode}. La mobilisation du concept de morphogenèse à partir du chapitre~\ref{ch:morphogenesis} suggère que cette récursivité serait plus que fortuite, mais bien intrinsèque au problème.}. Nous assumerons également ce raccourci de langage, en désignant par \emph{interactions entre réseaux et territoires} ou \emph{co-évolution entre réseaux et territoires}, les interactions ou la co-évolution entre les réseaux physiques et les composantes qu'ils relient, au sein d'un système territorial et donc d'un territoire.
}

\subsection{Transportation networks, specific carriers of interactions}{Les réseaux de transport, catalyseurs privilégiés des interactions}

\bpar{
We now precise the particular case of transportation networks and develop associated specific concepts that will play an important role in the precision of our problematic.
}{
Nous précisons à présent le cas particulier des réseaux de transport et développons des concepts spécifiques associés qui joueront un rôle prépondérant dans la précision de notre problématique.
}

\subsubsection{Characteristics and specificities of transportation networks}{Caractéristiques et spécificités des réseaux de transport}

\bpar{
Central to the already evoked debates on structuring effects of networks, transportation networks play a significant role in the evolution of territories, but it is of course out of question to give them deterministic causal effects. We will generally use the term of transportation network to designate the functional entity allowing a movement of agents and resources within and between territories\footnote{We designate thus simultaneously the infrastructure, but also its exploitation conditions, the rolling stock, the exploitation agents.}. Even if other types of networks are also strongly implicated in the evolution of territorial systems (see for example the debates on the impact of communication networks on the localization of economic activities), transportation networks condition other types of networks (logistic, commercial exchanges, concrete social interactions to give a few examples) and are a privileged entry regarding patterns of territorial evolution, in particular in our contemporary societies for which transportation networks play a crucial role~\cite{bavoux2005geographie}. We will therefore focus in the following only on transportation networks.
}{
Centraux aux discussions déjà évoquées sur les effets structurants des réseaux, les réseaux de transports jouent un rôle significatif dans l'évolution des territoires, mais il n'est évidemment pas question de leur attribuer des effets causaux déterministes. Nous parlerons de manière générale de réseau de transport pour désigner l'entité fonctionnelle permettant un déplacement des agents et des ressources au sein et entre les territoires\footnote{On désigne ainsi à la fois l'infrastructure, mais aussi ses conditions d'exploitation, le matériel roulant, les agents exploitants.}. Même si d'autres types de réseaux sont également fortement impliqués dans l'évolution des systèmes territoriaux (voir par exemple les débats sur l'impact des réseaux de communication sur la localisation des activités économiques), les réseaux de transport conditionnent d'autres types de réseaux (logistique, échanges commerciaux, interactions sociales concrètes pour donner quelques exemples) et sont une entrée privilégiée en rapport aux motifs d'évolution territoriale, en particulier dans nos sociétés contemporaines pour lesquelles les réseaux de transport jouent un rôle crucial~\cite{bavoux2005geographie}. Nous nous concentrerons ainsi par la suite uniquement sur les réseaux de transport.
}

\bpar{
The development of the French high speed rail network is an illustration of the role of transportation networks on policies of territorial development. Presented as a new era of railway transportation, it consisted in a top-down planning of totally novel lines, relatively independent through they two times higher speed, as~\cite{zembri1997fondements} puts it. High speed has been defended by political actors among other things as central for the development.
The weak integration of these new networks with the existing network and with local territories is now understood as a structural weakness~\cite{zembri1997fondements} (i.e. that is a consequence of network structure such as it was planned in the \emph{Scéma Directeur} of 1990), and negative impacts on some territories, such as the suppression of intermediate stops on classical lines used by the TGV, what contributes to an increase of the tunnel effect\footnote{The tunnel effect designates the process of telescoping the territory traversed by the infrastructure, when it is not accessible from this territory.} have been shown~\cite{zembri2008contribution}. A review done in~\cite{bazin2011grande} confirms that no general conclusions on local effects of a connection to a high speed line could be drawn, although it keeps a strong place in imaginaries of politics\footnote{But particular conclusions exist in some cases: for example a positive effect of the LGV Sud-Est on the touristic intensity in intermediate medium-sized cities such as Montbard or Beaune~\cite{bonnafous1987regional}; or the positioning of Lille as an European metropolis in which the connexions to the LGV have played a role~\cite{giblin2004lille}.}. The development of different high speed lines takes place in very different territorial contexts, and it is in any case difficult to interpret processes out of context: for example, the LGV Nord and LGV Est lines are situated within European scales that are broader than for the LGV Bretagne opened in July 2017\footnote{The LGV Nord line links Paris to Lille then Calais (entirely opened in 1997), and is used for the link with London, Brussels, Amsterdam. The LGV Est line links Paris to Strasbourg (partially opened in 2007, fully in 2016) and allows to serve Luxembourg and Germany. The LGV Bretagne line, opened in 2017, is the branch of the LGV Ouest towards Rennes and its service is uniquely to Britanny~\cite{zembri2010new}.}. The effects of the opening of a line can extend beyond the directly concerned territories: \cite{l2014contribution} show through the use of indicators from \emph{Time Geography}\footnote{The \emph{Time Geography}, introduced by the Swedish geographer \noun{Hägerstrand}, focuses mainly on trajectories of individuals in time and space, and of their implications in interactions with the environment~\cite{chardonnel2007time}.} (measuring an available working time in the context of a return journey within the day) that the Tours-Bordeaux line has potential impacts in the North and East of France. These examples illustrate well the way transportation networks can have effects both directly and indirectly, positive or negative, at different scales, or no effect at all on territorial dynamics.
}{
Le développement du réseau français à grande vitesse est une illustration du rôle des réseaux de transport sur les politiques de développement territorial. Présenté comme une nouvelle ère de transport sur rail, il s'agit d'une planification au niveau de l'État de lignes totalement nouvelles et relativement indépendantes de par leur vitesse deux fois plus élevée, selon la lecture de~\cite{zembri1997fondements}. La grande vitesse a été défendue par les acteurs politiques entre autres comme central pour le développement. L'articulation faible de ces nouveaux réseaux avec le réseau classique et avec les territoires locaux est à présent observé comme une faiblesse structurelle~\cite{zembri1997fondements} (c'est-à-dire conséquence de la structure du réseau tel qu'il a été planifié dans le Schéma Directeur de 1990), et des impacts négatifs sur certains territoires, comme par la suppression de dessertes intermédiaires sur les lignes classiques empruntées par le TGV, qui contribue à un accroissement de l'effet tunnel\footnote{L'effet tunnel désigne le processus de télescopage du territoire traversé par une infrastructure, celle-ci n'étant utilisable à partir de celui-ci.} ont été montrés~\cite{zembri2008contribution}. Une revue faite dans~\cite{bazin2011grande} confirme qu'aucune conclusion générale sur des effets locaux d'une connection à une ligne à grande vitesse ne peut être tirée, bien que ce sésame garde une place conséquente dans les imaginaires des élus\footnote{Mais des conclusions particulières existent dans certains cas : par exemple un effet positif de la LGV Sud-Est sur la fréquentation touristique de villes moyennes intermédiaires comme Montbard ou Beaune~\cite{bonnafous1987regional} ; ou le positionnement de Lille comme métropole européenne dans lequel les connexions LGV ont joué~\cite{giblin2004lille}.}. Le développement des différentes Lignes à Grande Vitesse s'inscrit dans des contextes territoriaux très différents, et il est dans tous les cas délicat d'interpréter des processus en les sortant de leur contexte : par exemple, les lignes LGV Nord et LGV Est s'inscrivent dans des échelles européennes plus vastes que la LGV Bretagne ouverte en juillet 2017\footnote{La ligne LGV Nord relie Paris à Lille puis Calais (ouverte entièrement en 1997), et s'inscrit dans la liaison avec Londres, Bruxelles et Amsterdam. La LGV Est relie Paris à Strasbourg (ouverte partiellement en 2007, puis entièrement en 2016) et permet de desservir le Luxembourg et l'Allemagne. La LGV Bretagne, ouverte en 2017, est le tronçon de la LGV Ouest vers Rennes et sa desserte est uniquement bretonne~\cite{zembri2010new}.}. Les effets de l'ouverture d'une ligne peuvent s'étendre au delà des seuls territoires directement concernés : \cite{l2014contribution} montrent par l'utilisation d'indicateurs issus de la \emph{Time Geography}\footnote{La \emph{Time Geography}, introduite par le géographe suédois \noun{Hägerstrand}, s'intéresse majoritairement aux trajectoires des individus dans le temps et l'espace, et de leurs implications dans les interactions avec l'environnement~\cite{chardonnel2007time}.} (mesurant une quantité de temps de travail disponible dans le cadre d'un aller-retour journalier) que la ligne Tours-Bordeaux a des répercussions potentielles dans le Nord et l'Est de la France. Ces exemples illustrent la manière dont les réseaux de transport peuvent avoir des effets à la fois directs et indirects, positifs ou négatifs, et à différentes échelles, ou bien aucun effet sur les dynamiques territoriales.
}

\subsubsection{Processes depending on scales}{Des processus dépendant des échelles}

\bpar{
The question of concerned temporal and spatial scales has until now been tackled only on a secondary plan compared to the concepts introduced. We propose now to integrate them to our reasoning in a structural way, i.e. guiding the developments of new concepts. Therefore, the concepts of \emph{Mobility}, \emph{Accessibility}\footnote{The accessibility, as we will see, can be defined at different scales, but we will use this term in a privileged way for accessibility landscapes at the metropolitan scale.}, and \emph{Structural Dynamics on long time}, correspond each to decreasing scales in time and space: intra-urban and daily, metropolitan and decennial, regional (in the broad and flexible sense of the range of a system of cities) and centennial. The correspondence we postulate here between time scales and spatial scales, far from being an evidence, will be shown during the development of each of these concepts. However, to take into account multiple scales is important, as shows \cite{RIETVELD1994329} with a review of economic approaches to interactions, which insists on the difference between intra-urban and intra-regional: at a large scale, different methods (models or qualitative approaches) give very different results concerning the impact of the infrastructure stock, whereas at a small scale, the positive impact of the global stock on productivity can not a priori been discussed.
}{
La question des échelles temporelles et spatiale concernées a été jusqu'ici abordée de manière auxiliaire aux concepts introduits. Nous proposons à présent de les intégrer de manière structurelle à notre raisonnement, c'est-à-dire guidant le développements de nouveaux concepts. Ainsi, les concepts de \emph{Mobilité}, d'\emph{Accessibilité}\footnote{L'accessibilité, comme nous le verrons, se définit à plusieurs échelles, mais nous privilégierons ce terme pour les paysages d'accessibilité à l'échelle métropolitaine.}, puis de \emph{Dynamique structurelle sur le temps long}, correspondent chacun à des échelles de temps et d'espace décroissantes : intra-urbain et journalier, métropolitain et décennal, régional (au sens large et flexible de la portée d'un système de villes) et centennal. La correspondance que nous postulons ici entre échelles de temps et échelles d'espace, loin d'être évidente, sera montrée lors du développement de chacun de ces concepts. Par contre, la prise en compte d'échelles multiples est importante, comme le montre \cite{RIETVELD1994329} par une revue des approches économiques des interactions, qui appuie la différence entre l'intra-urbain et l'intra-régional : à grande échelle, différentes méthodes (modèles ou approches qualitatives) donnent des résultats très différents quant à l'impact du stock d'infrastructure, tandis qu'à petite échelle, l'impact positif du stock global sur la productivité est a priori non discutable.
}

\subsubsection{Transportation and mobility}{Transports et mobilité}

\bpar{
The notion of mobility and all the associated approaches capture partly our questionings at a large scale. We will define mobility in a broad manner as a movement of territorial agents in space and time. It is related to use patterns of transportation networks. \cite{hall2005reconsidering} introduces a theoretical framework that yields a typology of mobility practices. In particular, he shows a rapid decrease of the frequency of journeys with spatial range and duration, and thus that ``micro-micro'' patterns (for the daily temporal scale and the intra-urban spatial scale), that we designate as \emph{daily mobility}, correspond to the most of journeys. It does not however mean an absence of link with other scales: on the one hand mobility patterns are very strongly conditioned by the distribution of activities as illustrate~\cite{lee2015relating}, but are on the other hand correlated to the social structure~\cite{camarero2008exploring}, that evolve both at time scales of a different magnitude (larger than a decade, thus at least one order in magnitude). Therefore, infrastructure and superstructure determine mobility practices, giving an important role to transportation networks in these.
}{
La notion de mobilité et l'ensemble des approches associées capturent en partie nos questionnements à grande échelle. Nous définirons la mobilité de manière générale comme un déplacement d'agents territoriaux dans l'espace et le temps. Elle relève des motifs d'utilisation des réseaux de transport.  \cite{hall2005reconsidering} introduit un cadre théorique permettant une typologie des pratiques de mobilité. En particulier, il montre une décroissance rapide de la fréquence des déplacements avec la portée spatiale et la durée, et donc que les motifs ``micro-micro'' (pour échelle temporelle journalière et échelle spatiale intra-urbaine), qu'on désigne par \emph{mobilité quotidienne}, sont majoritaires. Cela ne signifie pas pour autant une absence de lien avec d'autres échelles : d'une part les motifs de mobilité sont très fortement conditionnés par la distribution des activités comme l'illustrent~\cite{lee2015relating}, mais également corrélés à la structure sociale~\cite{camarero2008exploring}, qui évoluent tous deux à des échelles de temps d'un ordre différent (supérieur à la dizaine d'année, donc au moins un ordre de grandeur de différence). Ainsi, infrastructure et superstructure déterminent pratiques de mobilité, donnant un rôle important aux réseaux de transports dans celles-ci.
}

\bpar{
Reciprocally, use patterns of transportation networks are the product of daily mobility dynamics, and they adapt to it, while inducing relocations of actives and employments: there exists a co-evolution between transportation and territorial components at the microscopic and mesoscopic scales, which are objects of study in themselves. For example, \cite{fusco2004mobilite} unveils an influence\footnote{Which is interpreted as causal in the sense of bayesian networks.} of mobility on the urban structure, whereas the offer in infrastructure and its properties have however simultaneous effects on mobility and on the urban structure. In the case of freeway networks, \cite{faivr2003} recalls the necessity to construct a framework going beyond the logic of structuring effects on long times, and exhibits also interactions at a large scale that are typical of mobility on which more systematic conclusions can be established, such as an evolution of mobility practices implying a different use of the transportation network. We have thus at a large scale a first strong interdependency between transportation networks and territories, a first scale of co-evolution.
}{
Réciproquement, les motifs d'utilisation des réseaux de transport sont le produit des dynamiques de mobilité quotidiennes, et ceux-ci s'y adaptent, tout en induisant des relocalisations des actifs et emplois : il existe une co-évolution entre transports et composantes territoriales aux échelles microscopiques et mesoscopiques, qui sont un objet d'étude à part entière. Par exemple, \cite{fusco2004mobilite} révèle une influence\footnote{Qui est interprétée comme causale au sens des réseaux Bayesiens.} de la mobilité sur la structure urbaine, l'offre d'infrastructure et ses propriétés ayant cependant des effets simultanément sur la mobilité et sur la structure urbaine. Dans le cas des réseaux autoroutiers, \cite{faivr2003} rappelle la nécessité de construire un cadre d'analyse dépassant la logique des effets structurants sur le temps long, et montre également des interactions à grande échelle propres à la mobilité sur lesquelles des conclusions plus systématiques peuvent être établies, comme une évolution des pratiques de mobilité impliquant une utilisation différente du réseau de transport. Nous avons donc à grande échelle une première interdépendance forte entre réseaux de transports et territoires, une première échelle de co-évolution.
}

\bpar{
It is important to keep in mind the strong contingency of concepts we use here. The co-construction of the concept of mobility with technical solutions that model it with an operational purpose, has been illustrated by~\cite{commenges:tel-00923682} for the French context, which reveals among other things an application of frameworks and methods imported from the United States which were not well adapted to the French context. This contingency means that even the choice of concepts depends of broader conditions than their direct utility, and suggests a global systemic insertion within the \emph{Territorial System}.
}{
Il est important de garder à l'esprit la forte contingence des concepts mobilisés ici. La co-construction du concept de mobilité et des solutions techniques modélisant celle-ci dans un but opérationnel, a été illustrée par~\cite{commenges:tel-00923682} pour le contexte français, qui révèle entre autres une application peu adaptée au contexte français de cadres et méthodes importés des États-Unis. Cette contingence signifie que le choix des concepts même dépend de déterminants plus larges que leur utilité directe, et suggère une inscription systémique globale dans le \emph{Système Territorial}.
}


\bpar{
Finally, we have to remark that our approach of mobility is necessarily in a way reductionist, and overshadows socio-economic problematics for example: following \cite{remy2000metropolisation} mobility is indeed a ``virtual field'', i.e. it increases the potentialities offered to individuals, but in a way strongly dependent to the social class and to the socio-economic status. Indeed, mobility practices and political measures acting on transportation are closely linked and can lead to high socio-spatial inequalities in access to urban amenities \cite{gallez2015mobilite} (p.~236). Mobility practices will be indeed indirectly studied in an empirical preliminary study of traffic flows in~\ref{sec:reproducibility}, but we will not be able to treat of their socio-economic aspect: we must stay conscious that this aspect is not taken into account in our work. 
}{
Enfin, nous devons noter que notre approche de la mobilité est nécessairement réductrice, et occulte par exemple des problématiques socio-économiques : selon \cite{remy2000metropolisation} la mobilité est bien un ``champ de virtualité'', c'est-à-dire qu'elle accroit les potentialités offertes aux individus, mais de manière fortement dépendante à la classe sociale et au statut socio-économique. En effet, les pratiques de mobilité et les mesures politiques agissant sur le transport sont en lien étroit et peuvent conduire à de fortes inégalités socio-spatiales d'accès au aménités urbaines \cite{gallez2015mobilite} (p.~236). Les pratiques de mobilité seront bien abordées indirectement dans une étude empirique préliminaire des flux de trafic en~\ref{sec:reproducibility}, mais nous ne serons pas en mesure d'aborder leur aspect socio-économique : il faut ainsi rester conscient que cet aspect n'est pas pris en compte dans notre travail.
}


\subsubsection{Transportation and accessibility}{Transports et accessibilité}


\bpar{
The concept of \emph{accessibility} is fundamental to our question, since it is positioned at the exact crossroad of networks and territories. Based on the ability to access a place through a transportation network (that can take into account the speed, the difficulty to travel), it is generally defined as a spatial interaction potential\footnote{And often generalized as a \emph{functional accessibility}, for example employments accessible to the actives in one place. Spatial interaction potentials that are expressed in gravity laws can also be understood in the same way.}~\cite{bavoux2005geographie}. It was initially introduced in this form by~\cite{hansen1959accessibility}, with the aim to be applied to planning. Various formulations and formalizations of corresponding indicators have been proposed. It was shown that these enter the same theoretical frame. Indeed, \cite{weibull1976axiomatic} develops an axiomatic approach to accessibility, i.e. proposing to characterize it starting from a minimal number of fundamental hypothesis (axioms). \cite{miller1999measuring} takes the same frame and shows that it includes three classical ways to view accessibility. These are respectively the one based on \emph{Time Geography} and constraints, the one on utility measures for the user, and the one on an average travel time. Corresponding measures are derived within an unified mathematical framework, what allows both a theoretical and operational link between approaches of the concept that are a priori different.
}{
Le concept d'\emph{accessibilité} est fondamental pour notre question, puisqu'il se positionne à la croisée même des réseaux et des territoires. Basée sur la possibilité d'accéder un lieu par un réseau de transport (pouvant prendre en compte la vitesse, la difficulté de se déplacer), elle est généralement définie comme un potentiel d'interaction spatiale\footnote{Et souvent généralisée comme une \emph{accessibilité fonctionnelle}, par exemple les emplois accessibles aux actifs d'un lieu. Les potentiels d'interaction spatiale s'exprimant dans les lois gravitaires peuvent aussi être compris de cette façon.}~\cite{bavoux2005geographie}. Elle a été introduite sous cette forme initialement par~\cite{hansen1959accessibility}, dans un but d'application à la planification. Diverses formulations et formalisations d'indicateurs correspondants ont été proposées. Il a été montré que celles-ci rentrent dans le même cadre théorique. En effet, \cite{weibull1976axiomatic} développe une approche axiomatique de l'accessibilité, c'est-à-dire proposant de la caractériser à partir d'un nombre minimal d'hypothèses fondamentales (les axiomes). \cite{miller1999measuring} reprend ce cadre et montre qu'il englobe trois façons classiques de comprendre l'accessibilité. Celles-ci sont respectivement celle basée sur la \emph{Time Geography} et les contraintes, celle sur les mesures d'utilité pour l'utilisateur, et celle sur un temps de trajet moyen. Les mesures correspondantes sont dérivées dans un cadre mathématique unifié, ce qui permet un lien à la fois théorique et opérationnel entre des approches du concept a priori différentes.
}

\bpar{
We can first see to what extent accessibility patterns induce a evolution of the network. This concept is often used as a planning tool or as an explicative variable for the localization of agents, since it is for example a good indicator of the quantity of people concerned by a transportation project. 
}{
Nous pouvons voir dans un premier temps dans quelle mesure des motifs d'accessibilité induisent une évolution du réseau. Ce concept est souvent utilisé comme un outil de planification ou comme une variable explicative de localisation des agents, puisqu'il s'agit par exemple d'un bon indicateur pour la quantité de personnes affectées par un projet de transport.
}

\bpar{
Recent debates on the planning of \emph{Grand Paris Express}~\cite{mangin2013paris}, this new metropolitan transportation infrastructure planned for the next twenty years, has revealed the opposition between a vision of accessibility as necessary to open up disadvantaged territories, and a vision of accessibility as a driver of economic development for already dynamic areas, both being not necessarily compatible since they correspond to different transportation corridors. One was initially defended by the state in the perspective of competitive clusters, the other by the region in a perspective of territorial equity. These two logics answer naturally to different objective at various levels, and the chosen solution must be a compromise. We will come back on this precise example of the greater Paris in details in the following.
}{
Les débats récents sur la planification du \emph{Grand Paris Express}~\cite{mangin2013paris}, cette nouvelle infrastructure de transport métropolitaine planifiée pour les vingts prochaines années, a révélé l'opposition entre une vision de l'accessibilité comme nécessaire pour désenclaver des territoires désavantagés, et une vision de l'accessibilité comme moteur du développement économique pour des zones déjà dynamiques, les deux n'étant pas forcément compatibles car correspondent à des corridors de transport différents. L'un était initialement porté par l'Etat dans la perspective des pôles de compétitivité, l'autre par la région dans une perspective d'équité territoriale. Ces deux logiques répondent bien sûr à des objectifs différents à plusieurs niveaux, et la solution choisie doit former un compromis. Nous reviendrons sur cet exemple précis du Grand Paris en détails par la suite.
}

\bpar{
This example allows us to suggest an effect of patterns of potential on network evolution: even if this goes through complex social structures (we will also come back on this point in details further), there exists numerous situations where a growth of the transportation network (that can correspond to a topological evolution, i.e. the addition of a link, but also an evolution of link capacities) is directly or indirectly induced by a distribution of the accessibility~\cite{zhang2007economics}. This phenomenon can concern fundamental modifications of the networks or minor modifications: \cite{rouleau1985villages} studies the evolution on long times (from 1800 to 1980) of satellite villages around Paris that have progressively been integrated to its urban fabric and shows both a persistence of the roads and parcels frame, but also local evolutions answering to a logic of connectivity for example, while being part of a more complex evolution context (as in the case of Haussmann). We will designate this abstract process of an answer of the network to a connectivity demand as \emph{potential breakdown}\footnote{In analogy with the phenomenon of \emph{dieletric breakdown} which corresponds to the breakthrough of electrical current in a insulator when the difference of electrical potential is too high.}.
}{
Cet exemple permet de suggérer un effet des motifs de potentiels sur l'évolution du réseau : même si celui-ci passe par des structures sociales complexes (nous y reviendrons aussi en détail plus loin), il existe de nombreuses situations où une croissance du réseau de transport (qui peut se manifester par une évolution topologique, c'est-à-dire l'ajout d'un lien, mais aussi une évolution des capacités des liens) est directement ou indirectement induite par une distribution d'accessibilité~\cite{zhang2007economics}. Ce phénomène peut concerner des modifications fondamentales du réseau comme des modifications mineures : \cite{rouleau1985villages} étudie l'évolution sur le temps long (de 1800 à 1980) des villages satellites à Paris qui ont été progressivement intégrés à son tissu urbain et montre à la fois une persistance de la trame viaire et parcellaire, mais aussi des évolutions locales répondant à des logiques de connectivité par exemple, tout en s'inscrivant dans un cadre d'évolution globale plus complexe (comme dans le cas d'Haussmann). Nous désignerons ce processus abstrait de réponse du réseau à une demande de connectivité par \emph{rupture de potentiel}\footnote{En analogie avec le phénomène de \emph{dieletric breakdown}, ou décharge partielle, qui correspond au passage du courant dans un isolant quand la différence de potentiel électrique est trop grande.}.
}

\bpar{
An other significant process is the impact of an evolution of accessibility through relocations on network use patterns, and more particularly congestion, inducing a modification of capacity (flow that can be carried by network links): this phenomenon is shown in the case of Beijing by~\cite{yang2006transportation}, which unveils modification of the impedance (effective speed in the transportation network) up to 30\%. This can be put in correspondence with processes linked to mobility, even if we are more within meso-meso scales here, i.e. an evolution of the network and relocations on time scales of the order of the decade (the network being slower, of the order of two decades), and on spatial metropolitan scales\footnote{Which correspond to spatial extents from 100 to 200km, but to various urban realities. A metropolis will be a city of importance in a system of cities at a small scale, and will be seen with its functional territory (for example Paris and a consequent part of Ile-de-France). The emergence of new metropolitan forms, such as \emph{Mega-city-regions} which are composed by metropolis of comparable sizes, on a small spatial extent, and with strong interactions, makes this question of the scale more complicated. We will come back on these objects in~\ref{sec:casestudies}.}.
}{
Un autre processus significatif est l'impact d'une évolution de l'accessibilité par relocalisations sur les motifs d'utilisation du réseau, et particulièrement la congestion, induisant une modification de la capacité (flux pouvant être porté par les liens du réseau) : ce phénomène est montré dans le cas de Beijing par~\cite{yang2006transportation}, qui révèle des modifications d'impédance (vitesse effective dans le réseau routier) allant jusqu'à 30\%. Il peut être mis en correspondance avec les processus liés à la mobilité, même si on se situe ici plutôt dans des échelles meso-meso, c'est-à-dire une évolution du réseau et des relocalisations sur des temporalités de l'ordre de la dizaine d'année (le réseau étant plus lent, de l'ordre de la vingtaine d'années), et sur des échelles spatiales métropolitaines\footnote{Qui correspondent à des étendues spatiales de 100 à 200km, mais à diverses réalités urbaines. Une métropole sera une ville d'importance dans un système de villes à petite échelle, et sera vue avec son territoire fonctionnel (par exemple Paris et une grande partie de l'Ile-de-France). L'émergence de nouvelles formes métropolitaines, comme les \emph{Mega-city-regions} qui sont composées de métropoles de taille comparable, sur une faible étendue spatiale, et en très forte interaction, complique cette question de l'échelle. Nous reviendrons sur ces objets en~\ref{sec:casestudies}.}.
}

\bpar{
Reciprocally, an evolution of the network implies an immediate reconfiguration of the spatial distribution of accessibilities (in the sense of all existing approaches, since all take the network into account), and also potentially of territorial transformations on a longer time: we finally come back to the debate of structuring effects we already commented on. We have seen that accessibility co-evolves\footnote{The concept applies a priori at different scales, what will be confirmed by the more precise definition we will take at the end of this first part.} with mobility practices, what suggests an effect at this scale. Concerning relocations and distribution of populations, there exists some cases where it is indeed possible to attribute some territorial dynamics to network growth, that we will develop in the following.
}{
Réciproquement, une évolution du réseau implique une reconfiguration immédiate de la distribution spatiale des accessibilités (au sens de l'ensemble des approches existantes, puisque toutes mobilisent le réseau), et aussi potentiellement des transformations territoriales sur une plus longue durée : on rejoint finalement le débat des effets structurants que nous avons déjà commenté. Nous avons vu que l'accessibilité co-évolue\footnote{Le concept s'applique a priori à diverses échelles, ce qui sera confirmé par la définition plus précise que nous prendrons à la fin de cette première partie.} avec les pratiques de mobilité, ce qui suppose un effet à cette échelle. Concernant les relocalisations et la distribution des populations, il existe des cas où il est en effet possible d'attribuer à la croissance du réseau des dynamiques des territoires, que nous allons développer par la suite.
}

\bpar{
\cite{duranton2012urban} show thus at a medium time scale of 20 years for the United States, through the use of instrumental variables\footnote{The method of instrumental variables aims at unveiling causal relations between an explicative and an explicated variable. The choice of a third variable, called the instrumental variable, must be done such that it influences only the explicative variable but not the explicated variable, in a sense an exogenous shock.}, that accessibility growth in a city causes the growth of employments. On a similar time scale, but at the spatial scale of the country for Sweden, \cite{johansson1993infrastructure} show the local accessibility (``intra-regional'') and global accessibility (``inter-regional'') explains the growth of production and of the productivity of companies. \cite{doi:10.1080/01441647.2016.1168887} proceed to a systematic review of empirical studies of impacts at a medium term of transportation infrastructures, and show that an urban densification at the proximity of new infrastructures is highly probable, being residential in the case of a railway infrastructure and for employments and industrial activity in the case of a road infrastructure\footnote{The studies reviewed cover mainly the second half of the 20th century and Europe, the United States and East Asia. It is important to keep in mind that even if they are relatively general, conclusions must always be contextualized.}. Similarly, it is possible to show strong effects of the presence of infrastructures for particular types of land-use: \cite{nilsson2016measuring} show it for example for fast foods in two cities in the United States, by showing statistically that the access to an important infrastructure induces a spatial aggregation of commerces. 
}{
\cite{duranton2012urban} montrent ainsi à une échelle de temps moyenne de 20 ans pour les États-unis, par l'utilisation de variables instrumentales\footnote{La méthode des variables instrumentales permet de dégager des relations causales entre une variable explicative et une variable expliquée. Le choix d'une troisième variable, appelée variable instrumentale, doit être fait tel que celle-ci n'influence que la variable explicative mais pas la variable expliquée, en quelque sorte un choc exogène.}, que la croissance de l'accessibilité dans une ville cause une croissance de l'emploi. Sur une échelle temporelle similaire, mais à l'échelle spatiale du pays pour la Suède, \cite{johansson1993infrastructure} montre que l'accessibilité locale (``intra-régionale'') et globale (``inter-régionale'') explique la croissance de la production et la productivité des entreprises. \cite{doi:10.1080/01441647.2016.1168887} procèdent à une revue systématique des études empiriques des impacts à moyen terme des infrastructures de transport, et montrent qu'une densification urbaine à proximité des nouvelles infrastructures est très probable, celle-ci étant résidentielle dans le cas d'une infrastructure ferroviaire et pour les emplois et l'activité industrielle et commerciale dans le cas d'une infrastructure routière\footnote{Les études revues couvrent majoritairement la seconde moitié du 20ème siècle et l'Europe, les Etats-Unis et l'Asie de l'Est. Il est donc important de garder à l'esprit que même relativement générales, les conclusions doivent toujours être contextualisées.}. De même, il est possible de montrer des effets forts de la présence d'infrastructures pour des types particuliers d'usage du sol : \cite{nilsson2016measuring} l'illustrent par exemple pour les fast food dans deux villes aux Etats-Unis, en montrant statistiquement que l'accès à une infrastructure importante induit une agrégation spatiale des commerces.
}

\bpar{
The latest examples suggest the potential existence of effects of accessibility, and thus of the network, on territorial dynamics. In some cases, structuring effects are thus present. But these are always links to the precise context and also to scales. This allows us to make the transition to concepts linked by dynamics of urban systems on long times.
}{
Ces derniers exemples suggèrent l'existence potentielle d'effets de l'accessibilité, et donc du réseau, sur les dynamiques territoriales. Dans certain cas, les effets structurants sont ainsi présents. Mais ceux-ci sont toujours liés au contexte précis ainsi qu'aux échelles. Cela nous permet de faire la transition vers les concepts liés aux dynamiques des systèmes urbains sur le temps long.
}

\subsubsection{Transportation and urban systems}{Transports et systèmes urbains}

\bpar{
The third conceptual entry on interactions between networks and territories, and which will be particularly linked to the idea of co-evolution, is the one of urban systems, at a small spatial scale and on long times. We will designate the concept by \emph{structural dynamics of the urban system}.
}{
La troisième entrée conceptuelle sur les interactions entre réseaux et territoires, et qui sera particulièrement liée à l'idée de co-évolution, est celle par les systèmes urbains, à petite échelle spatiale et sur le temps long. Nous désignerons le concept par \emph{dynamique structurelle du système urbain}.
}

\bpar{
The evolutive urban theory considers systems of cities as systems of systems at multiple scale, from the intra-urban microscopic level, to the macroscopic level of the whole system, through the mesoscopic level of the city~\cite{pumain2008socio}. These systems are complex, dynamical, and adaptive: their components \emph{co-evolve} and the system answers to internal or external perturbations by modifying its structure and its dynamics. We will largely develop the multiple implications of this approach all along our work, and retain here processes of interactions between cities. These interactions consist in material or informational exchanges, and the diffusion of innovation is therein a crucial component~\cite{pumain2010theorie}. These are necessarily carried by physical networks, and more particularly transportation networks. We expect thus from a theoretical point of view strong interdependencies between cities and transportation networks at these scales, i.e. a co-evolution.
}{
La théorie évolutive des villes considère les systèmes de villes comme des systèmes de systèmes à de multiples échelles, du niveau microscopique intra-urbain, au niveau macroscopique du système entier, par le niveau mesoscopique de la ville~\cite{pumain2008socio}. Ces systèmes sont complexes, dynamiques, et adaptatifs : leur composants \emph{co-évoluent} et le système répond à des perturbations intérieures ou extérieures par des modifications de sa structure et de sa dynamique. Nous développerons longuement les multiples implications de cette approche tout au long de notre travail, et retenons ici les processus d'interactions entre villes. Ces interactions consistent en des échanges informationnels ou matériels, et la diffusion de l'innovation en est une composante cruciale~\cite{pumain2010theorie}. Elles sont nécessairement portées par les réseaux physiques, et plus particulièrement les réseaux de transport. On s'attend ainsi du point de vue théorique à une interdépendance forte entre villes et réseaux de transport à ces échelles, c'est-à-dire à une co-évolution.
}

\bpar{
From the empirical point of view, it has already been shown: \cite{bretagnolle:tel-00459720} reveals an increasing correlation in time between urban hierarchy and the hierarchy of temporal accessibility for the French railway network (which is a priori clearer for this measure than for integrated measures of accessibility that are prone to auto-correlation as we will see in~\ref{sec:causalityregimes}). This correlation is a witness of positive feedbacks between urban ranks and network centralities. Different regimes in space and times has been identified: for the evolution of the French railway network, a first phase of adaptation of the network to the existing urban configuration was followed by a phase of co-evolution, in the sense that causal relations became difficult to identify. The impact of the contraction of space-time by networks on patterns of growth potential had already been shown for Europe with an exploratory analysis in~\cite{bretagnolle1998space}.
}{
Du point de vue empirique, celle-ci a déjà été mise en valeur : \cite{bretagnolle:tel-00459720} souligne une corrélation croissante dans le temps entre la hiérarchie urbaine et la hiérarchie de l'accessibilité temporelle pour le réseau ferroviaire français (a priori plus claire pour cette mesure que pour les mesures intégrées d'accessibilité soumises à l'auto-corrélation comme nous le verrons en~\ref{sec:causalityregimes}). Celle-ci est un marqueur de rétroactions positives entre le rang urbain et la centralité de réseau. Différents régimes dans le temps et l'espace ont été identifiés : pour l'évolution du réseau ferroviaire français, une première phase d'adaptation du réseau à la configuration urbaine existante a été suivie par une phase de co-évolution, au sens où les relations causales sont devenues difficiles à identifier. L'impact de la contraction de l'espace-temps par les réseaux sur le potentiel de croissance des villes avait déjà été montré pour l'Europe par des analyses exploratoires dans~\cite{bretagnolle1998space}.
}

\bpar{
Modeling results by~\cite{bretagnolle2010comparer}, and more particularly the different parametrizations of the Simpop2 model\footnote{The generic structure of the Simpop2 model is the following~\cite{pumain2008socio}: cities are characterized by their population ad their wealth; they product goods according to their economic profile; interactions between cities produce exchanges, determined by the offer and demand functions; populations evolve according to wealth after exchanges.}, show that the evolution of the railway network in the United States has followed a rather different dynamic, without hierarchical diffusion, shaping locally urban growth in some cases. This particular context of conquest of a space empty of infrastructures implies a specific regime for the territorial system. Other contexts reveal different impacts of the network at short and long term: \cite{berger2017locomotives} study the impact of the construction of the Swedish railway network on the growth of urban populations, from 1800 to 2010, and find an immediate causal effect of the accessibility increase on population growth, followed on long times of a strong inertia for population hierarchy. In each case, we indeed observe the existence of \emph{structural dynamics} on long times, which correspond to the slow dynamics of the urban system structure, and witness in that sense of \emph{structuring effects on long times} as~\cite{pumain2014effets} puts it.
}{
Les résultats de modélisation par~\cite{bretagnolle2010comparer}, et plus particulièrement les paramétrisations différentes du modèle Simpop2\footnote{La structure générique du modèle Simpop2 est la suivante~\cite{pumain2008socio} : les villes sont caractérisées par leur population et leur richesse ; produisent des biens selon leur profil économique ; les interactions entre villes produisent des échanges, déterminés par les fonctions d'offre et demande ; les populations évoluent selon la richesse après échanges.}, montrent que l'evolution du réseau ferroviaire aux États-unis a suivi une dynamique bien différente, sans diffusion hiérarchique, donnant forme localement à la croissance urbaine dans certains cas. Ce contexte particulier de conquête d'un espace vierge d'infrastructures implique un régime spécifique pour le système territorial. D'autres contextes révèlent des impacts différents du réseau à court et long terme : \cite{berger2017locomotives} étudient l'impact de l'établissement du réseau ferroviaire suédois sur la croissance des populations urbaines, de 1800 à 2010, et trouvent un effet causal immédiat de la croissance de l'accessibilité sur la croissance de la population, suivi sur le temps long d'une forte inertie de la hiérarchie des populations. Dans chaque cas, on a bien existence de \emph{dynamiques structurelles} sur le temps long, qui correspondent aux dynamiques lentes de la structure du système urbain, et témoignent en ce sens d'\emph{effets structurants sur le temps long} comme le souligne~\cite{pumain2014effets}.
}

\bpar{
We must be careful to differentiate the latest from the structuring effects previously mentioned which are subject to debates. At the level of the urban system, it is relevant to globally follow trajectories that were possible, and locally the effect has necessarily a probabilistic aspect. Moreover, we insist on the role of path-dependency for trajectories of urban systems: for example the existence in France of a previous system of cities and network (postal roads) has strongly influenced the development of the railway network, or as \cite{berger2017locomotives} showed for Sweden. The same way, \cite{doi:10.1068/b39089} highlight the importance of historical events in coupled dynamics of the road network and territories, historical shocks that can be seen as exogenous and inducing bifurcations of the system that accentuate the effect of path-dependency. Therefore, for these structural dynamics on long times, forecasting can difficultly be considered.
}{
Il s'agit bien de différencier ces derniers des effets structurants sujets des débats mentionnés précédemment. Au niveau du système urbain, il est pertinent de suivre globalement des trajectoires qui étaient possibles, et localement l'effet a nécessairement un aspect probabiliste. D'autre part, il faut mettre l'accent sur le rôle de la dépendance au chemin pour les trajectoires des systèmes urbains : par exemple la présence en France d'un système préalable de villes et de réseau (routes postales) a fortement influencé le développement du réseau ferré, ou comme \cite{berger2017locomotives} l'ont montré pour la Suède. De même, \cite{doi:10.1068/b39089} soulignent l'importance des évènements historiques dans les dynamiques couplées du réseau routier et des territoires, choc historiques pouvant être vus comme exogènes et induisant des bifurcations du système qui accentuent l'effet de la dépendance au chemin. Ainsi, pour ces dynamiques de structure sur le temps long, des prévisions ne sont guère envisageables.
}

\bpar{
This third approach allowed us to unveil a complementary point of view on co-evolution, at an other scale.
}{
Cette troisième approche nous a permis de dégager un point de vue complémentaire de la co-évolution, à une autre échelle.
}


\subsubsection{Links between scales suggested by Scaling Laws}{Des liens entre échelles suggérés par les Lois d'Échelle}


\bpar{
Our framework with successive scales, that yield a reasonable correspondence between spatial and temporal scales, and also to associate the corresponding concepts, does naturally not capture the full range of possible processes: these that would fundamentally be multi-scalar, for example by implying the emergence of their own intermediate level, are not evoked. These are important and we will come back to them below. First we propose to establish a conceptual link between scales by the intermediary of \emph{scaling laws} (that we understand in the general sense given in introduction). This link aims in particular at going beyond a reductionist reading through the compartmentalization of scales.
}{
Notre grille de lecture par échelles progressives, qui permet de dégager une assez bonne correspondance entre échelle spatiale et temporelle, ainsi que d'y associer les concepts adaptés, ne capture bien sûr pas l'ensemble des processus possibles : ceux qui seraient fondamentalement multi-échelles, par exemple en impliquant l'émergence de leur propre niveau intermédiaire, ne sont pas évoqués. Ceux-ci sont importants et nous y reviendrons ci-dessous. Dans un premier temps, nous proposons d'effectuer un lien conceptuel entre les échelles par l'intermédiaire des \emph{lois d'échelles} (que nous comprenons au sens général donné en introduction). Ce lien permet en particulier de dépasser une lecture réductrice par cloisonnement d'échelle.
}

\bpar{
Transportation networks are by essence hierarchical, this property depending on scales they are embedded in, and leading to the emergence of scaling laws for their properties. For example, \cite{10.1371/journal.pone.0102007} show empirical scaling properties for a consequent number of metropolitan areas across the world. Indeed, scaling laws reveal the presence of hierarchy within a system, as for size hierarchy for systems of cities expressed by Zipf's law~\cite{nitsch2005zipf} or other urban scaling laws~\cite{arcaute2015constructing,bettencourt2016urban}, what suggests a particular structure for these systems. We can expect to find it again in interaction processes themselves. Transportation network topology follows such laws for the distribution of its local measures such as centrality~\cite{samaniego2008cities}, these being directly linked to accessibility patterns at different scales. Furthermore, network topology is among the factors inducing the hierarchy of use, since it influences congestion negative externalities, in relation with the spatial distribution of land-use~\cite{Tsekeris20131}. Thus, considering scaling laws for transportation networks, and more generally for territorial systems, is first a signature of the complexity of these systems, and secondly yields an implicit link between scales.
}{
Les réseaux de transport sont par essence hiérarchiques, cette propriété dépendant des échelles dans lesquelles ils sont intégrés, et se manifestant par l'émergence de lois d'échelles pour leurs propriétés. Par exemple, \cite{10.1371/journal.pone.0102007} montrent empiriquement des propriétés de loi d'échelle pour un nombre conséquent d'aires métropolitaines à travers la planète. Or les lois d'échelle révèlent la présence de hiérarchies dans un système, comme pour la hiérarchie de tailles dans les systèmes de villes exprimée par la loi de Zipf~\cite{nitsch2005zipf} ou d'autres lois d'échelles urbaines~\cite{arcaute2015constructing,bettencourt2016urban}, ce qui suggère une structure particulière pour ces systèmes. Nous pouvons nous attendre à la retrouver dans les processus d'interaction eux-mêmes. La topologie du réseau de transport suit de telles lois pour la distribution de ses mesures locales comme la centralité~\cite{samaniego2008cities}, celles-ci étant directement liées au motifs d'accessibilité à différentes échelles. De plus, la topologie du réseau fait partie des facteurs induisant la hiérarchie d'usage, se retrouvant dans les externalités négatives de congestion, en relation avec la distribution spatiale de l'usage du sol~\cite{Tsekeris20131}. Ainsi, la considération des lois d'échelles pour les réseaux de transport, et plus généralement pour les systèmes territoriaux, est dans un premier temps une signature de la complexité de ces systèmes, et permet dans un second temps un lien implicite entre les échelles.
}


\subsubsection{Scales: a synthesis}{Echelles : synthèse}

\bpar{
To recall our framework by scales, we propose the Table~\ref{tab:networkterritories:scales}. Designations and orders of magnitude of temporal and spatial scales are of course indicative, such as key concept that are indeed the ones that allowed us to enter these scales. We also give references that illustrate corresponding conceptual frameworks. This table will however be useful to keep in mind  the typical scales to which we refer.
}{
Pour rappeler notre cadre de lecture par échelles, nous proposons la Table~\ref{tab:networkterritories:scales}. Les appellations ainsi que les ordres de grandeur des échelles temporelles et spatiales sont évidemment indicatifs, de même que les concepts clés qui sont en fait ceux qui nous ont permis d'entrer dans ces échelles. Nous donnons également des références illustrant des cadres conceptuels correspondant. Ce tableau nous sera toutefois utile pour garder à l'esprit les échelles typiques auxquelles nous ferons référence.
}

\begin{table}
\caption[Synthesis of the approach by scales]{\textbf{Synthesis of the approach by scales of interactions between transportation networks and territories.} References give a possible theoretical frame for each scale.\label{tab:networkterritories:scales}}
\bpar{
\begin{tabular}{|p{1.7cm}|p{3.7cm}|p{3.7cm}|p{3.7cm}|p{2.7cm}|}\hline
	Scale & Spatial scale & Temporal scale & Concept & Reference \\ \hline
	Micro & Intra-urban (10km) & daily (1d) & Mobility practices & \cite{hall2005reconsidering} \\ \hline
	Meso & Metropolitan (100km) & Decade (10y) & Metropolitan reconfiguration & \cite{wegener2004land} \\\hline
	Macro & Regional (500km) & Century (100y) & Structural dynamic on long times & \cite{pumain1997pour} \\\hline
\end{tabular}
}{
\begin{tabular}{|p{1.7cm}|p{3.7cm}|p{3.7cm}|p{3.7cm}|p{2.7cm}|}\hline
	Echelle & Echelle spatiale & Echelle temporelle & Concept & Référence \\ \hline
	Micro & Intra-urbaine (10km) & Journalière (1j) & Pratiques de mobilité & \cite{hall2005reconsidering} \\ \hline
	Meso & Métropolitaine (100km) & Décade (10ans) & Reconfiguration métropolitaine & \cite{wegener2004land} \\\hline
	Macro & Régionale (500km) & Siècle (100ans) & Dynamique structurelle sur le temps long & \cite{pumain1997pour} \\\hline
\end{tabular}
}
\end{table}

\subsubsection{Processus: a synthesis}{Processus : synthèse}

\bpar{
At this stage, we can already propose a preliminary of the interaction processes we introduced. A more exhaustive typology will be possible at the end of this chapter.
}{
A ce stade, nous pouvons d'ores et déjà proposer une synthèse préliminaire des processus d'interaction que nous avons introduit. Une typologie plus exhaustive sera possible à l'issue du chapitre.
}

\bpar{
Thus, territorial components can act on networks by:
}{
Ainsi, des composantes territoriales peuvent agir sur les réseaux de transport par :
}

\bpar{
\begin{itemize}
	\item Impact of mobility patterns on impedances and capacities
	\item Potential breakdown, emergence of centralities
	\item Hierarchical selection of accessibility
	\item Systemic structural effects and bifurcations 
\end{itemize}
}{
\begin{itemize}
	\item Impact des motifs de mobilité sur les impédances et les capacités
	\item Rupture de potentiel, émergence de centralités
	\item Sélection hiérarchique de l'accessibilité
	\item Effets systémiques structurels et bifurcations
\end{itemize}
}

\bpar{
Reciprocally, processes where network properties act on territories include:
}{
Réciproquement, des processus où les propriétés des réseaux agissent sur les territoires incluent :
}

\bpar{
\begin{itemize}
	\item Relocations induced by mobility constraints
	\item Land-use changes due to a transportation infrastructure
	\item Accessibility patterns induced by networks, that can induce relocations
	\item Interactions between territories carried by network, including the tunnel effect when these are telescoped
\end{itemize}
}{
\begin{itemize}
	\item Relocalisations induite par des contraintes de mobilité
	\item Changement d'usage du sol dû à une infrastructure de transport
    \item Motifs d'accessibilité induits par les réseaux, pouvant induire des relocalisations
	\item Interactions entre territoires portées par les réseaux, incluant l'effet tunnel lorsque celles-ci sont télescopées
\end{itemize}
}

\bpar{
These different processes do not all have the same level of abstraction neither the same scales. We have furthermore hidden some processes already evoked, within which the coupling is stronger and for which the circularity is already present in the ontology, such as processes linked to planning. We will now detail these, what will allow us then to refine the list above and to present it as a typology after having enriched it with empirical studies.
}{
Ces différents processus n'ont pas tous le même statut d'abstraction ni les mêmes échelles. Nous avons de plus volontairement occulté des processus déjà évoqués, au sein desquels le couplage est plus fort et pour lesquels la circularité est déjà présente dans l'ontologie, comme les processus liés à la planification. Nous allons détailler à présent ceux-ci, ce qui nous permettra par la suite de raffiner la liste ci-dessus et de la présenter sous forme de typologie après l'avoir enrichie par des études empiriques.
}

\subsection{From interactions to co-evolution}{Des interactions à la co-évolution}

\bpar{
At this stage, we have identified processes of interaction between transportation networks and territories that play a significant role in the complexity of territorial systems. In the frame of our preliminary definition of a territorial system, this question can be reformulated as the study of networked territorial systems with an emphasis on the role of transportation networks. We have seen that the extent of spatial and temporal scales spans from daily mobility (micro-micro) to processes on long time in systems of cities (macro-macro), with the possibility of intermediate combinations. The precision of scales that are particularly relevant will be the subject of most of preliminaries (Part 1) and of foundations (Part 2), until chapter~\ref{ch:morphogenesis} that concludes foundations. We now extend this list and give concrete examples in terms of the complexity of interactions.
}{
A ce stade, nous avons identifié des processus d'interaction entre réseaux de transport et territoires jouant un rôle significatif dans la complexité des systèmes territoriaux. Dans le cadre de l'approche d'un système territorial par la définition donnée lors de la construction première des concepts, cette question peut être reformulée comme l'étude de systèmes territoriaux réticulaires, avec une emphase sur le rôle des systèmes de transports. Nous avons vu que l'étendue des échelles spatiales et temporelles va de celle de la mobilité quotidienne (micro-micro) à des processus sur le temps long dans les systèmes de villes (macro-macro), avec la possibilité de combinaisons intermédiaires. La précision des échelles particulièrement pertinentes fera l'objet de la majorité des préliminaires (Partie 1) et des fondations (Partie 2), jusqu'au chapitre~\ref{ch:morphogenesis} qui conclura les fondations. Étendons à présent cette liste et donnons des exemples concrets précisant la complexité des interactions.
}

\subsubsection{Importance of the geographical context}{Importance du contexte géographique}

\bpar{
The contextualization of our question in a particular frame reveals the importance of taking into account the geographical context. The exemple of mountain territories, where constraints on ressources and travel are stronger, shows the richness of possible situations when a generic frame is put in context of a particular case.
}{
La mise en contexte de notre question dans un cadre bien particulier révèle l'importance de la prise en compte du contexte géographique. L'exemple des territoires de montagne, où les contraintes de ressources et de déplacement sont fortes, montre la richesse des situations possibles lorsqu'un schéma générique est mis en contexte dans un cas particulier.
}

\bpar{
For example, on comparable French mountain territories, \cite{berne2008ouverture} shows that reactions to a same context of evolution of the transportation network can lead to very different territorial dynamics, some territories highly benefiting of the increased accessibility, others in the contrary becoming more closed. In the same frame, these possible opposed processes are scrutinized with more details by~\cite{bernier2007dynamiques}, for which he proposes a typology based on the opening potential both of territorial dynamics and network dynamics: for example, a territory can exhibit rich opportunities to be attractive, such as touristic opportunities, but keep a low accessibility. Reciprocally, he gives the illustration of custom constraints that can impede the opening potential of a performant infrastructure.
}{
Par exemple, sur des territoires de montagne français comparables, \cite{berne2008ouverture} montre que les réactions à un même contexte d'évolution du réseau de transport peuvent mener à des dynamiques territoriales très diverses, certains trouvant de forts bénéfices de l'accessibilité accrue, d'autres au contraire devenant plus fermés. Dans le même cadre, ces potentiels processus antagonistes sont examinés plus en détail par~\cite{bernier2007dynamiques}, pour lesquels il propose une typologie basée sur le potentiel d'ouverture à la fois des dynamiques territoriales et des dynamiques des réseaux : par exemple, un territoire peut présenter de riches opportunités d'attractivité, comme des opportunités touristiques, tout en gardant une faible accessibilité. Réciproquement, il donne l'illustration des contraintes douanières pouvant freiner le potentiel d'ouverture d'une infrastructure performante.
}

\bpar{
Similarly to approaches considering systems of cities, \cite{torricelli2002traversees} shows how it is possible in that context to establish a link between the nature of transportation flows and the local development of the urban system: cities in the mountains have first emerged as waypoints on paths to mountain passes, then have lost their importance when roads came into existence. The construction of railways gave them a new dynamic, through tourism and industry, and finally freeways has more recently inducted a loss of urban structure through peri-urbanization for example. Thus, structural dynamics on long time are particular, as a consequence of the geographical context.
}{
En écho aux approches par systèmes de villes, \cite{torricelli2002traversees} montre comment dans ce contexte il est possible de faire un lien entre nature des flux de transport et développement local du système urbain : les villes de montagne ont d'abord émergé comme point de passage sur les chemins de col, puis ont perdu de leur importance avec l'avènement des routes. L'arrivée du chemin de fer a pu les re-dynamiser, par le tourisme et l'industrie, et enfin l'autoroute a encore plus récemment induit une déstructuration par des effets de périurbanisation par exemple. Ainsi, les dynamiques structurelles sur le temps long sont particulières, en conséquence du contexte géographique.
}

\subsubsection{Planification processes}{Processus de planification}

\bpar{
As we already suggested, potential impacts of territorial dynamics on networks imply processes at different levels. This way, infrastructure projects are generally planned\footnote{We will use the term \emph{planning} in general, territorial or urban, of an infrastructure project, when a project and its plan is willingly elaborated by a planning stakeholder, with an aim at transforming space according to some motivations depending on the stakeholder and on its interactions with other stakeholders.}, in order to fulfil some objectives fixed generally by institutional actors. These objects bring progressively the concept of governance, but let first give some illustrations of planned projects. 
}{
Comme nous l'avons déjà suggéré, les potentiels impacts des dynamiques territoriales sur les réseaux impliquent des processus à plusieurs niveaux. Ainsi, les projets d'infrastructure sont généralement planifiés\footnote{Nous parlerons de \emph{planification} en général, urbaine, territoriale, d'un projet d'infrastructure, pour désigner la conception volontaire d'un projet et d'un plan par un acteur d'aménagement, dans le but de transformer l'espace selon certaines motivations propres à l'acteur et à ses interactions avec les autres acteurs.}, afin de répondre à certains objectifs fixés par des acteurs souvent institutionnels. Ces objets nous amènent progressivement vers le concept de gouvernance, mais prenons d'abord un instant pour illustrer des projets planifiés.
}

\bpar{
The example of the failure in the planning of the Ciudad Real airport in Spain shows that the answer to a planned infrastructure is far from systematic. The explanations to it are probably a complex combination of diverse factors, difficult to disentangle. \cite{otamendi2008selection} predicted before the opening of the airport a complex management due to the dimension of expected flows and proposed a suited model, but the order of magnitude of effective flows where closer to thousands than millions that were planned and the airport rapidly closed. It is complicated to know the reason of the failure, if it is an optimism of the regional level of polycentricity (the airport is halfway between Madrid and Seville), the lack of construction of a train station on the high speed line, or just purely economical factors.
}{
L'exemple de l'échec de planification de l'aéroport de Ciudad Real en Espagne montre que la réponse à une infrastructure planifiée n'est pas systématique. Les explications à celui-ci découlent très probablement d'une combinaison complexe de multiples facteurs, difficiles à séparer. \cite{otamendi2008selection} prédisaient avant l'ouverture de l'aéroport une gestion complexe due à la dimension des flux attendus et proposaient un modèle approprié, or les ordres de grandeurs de flux effectifs étaient plus proches des milliers que des millions planifiés et l'aéroport a rapidement fermé. Il est compliqué de savoir la raison de l'échec, s'il s'agit de l'optimisme quand au polycentrisme régional (l'aéroport est à mi-chemin de Madrid et Séville), la non-réalisation de la gare sur la ligne à grande vitesse, ou des facteurs purement économiques.
}

\bpar{
\cite{heddebaut:hal-01355621}\footnote{The possible pun with the ambiguous title on the existence of the ``Tunnel effect'' recalls the effect through which an infrastructure traversing a territory has no interaction with it.} show for the impact of infrastructures on the long term, in the case of the Channel tunnel\footnote{Put into service in 1994 between Calais in France and Folkestone in the United Kingdom, this railway underwater tunnel with a length of 50km establishes a physical link between the continent and the UK.}, through an analysis of investments and political actions in time, that the effect effectively observed for the Nord-Pas-de-Calais region such as a gain in centrality and in European visibility, are in strong distorsion with the initial justifications of the project, and that the renewing of stakeholders implies that the project is not accompanied on the long time what makes its impact more uncertain. We rejoin the idea advocated by \cite{espacegeo2014effets} according to which some ``structure effects'' effectively exist but that these can be observed on the long time in terms of the dynamic of the system for which a short time local vision does not make much sense. At the intra-urban scale, \cite{fritsch2007infrastructures} takes the example of the tramway in Nantes to show, through a localized study of urban transformations in the nieghborhood of a new line, that urban densification dynamics are far from what was expected from deciders and planners, i.e. a strong correspondence between the proximity to the line and a densification.
}{
\cite{heddebaut:hal-01355621}\footnote{Le possible jeu de mot par le titre ambigu sur l'existence du ``Tunnel effect'' rappelle l'effet tunnel, qui réside en la non-interaction d'une infrastructure sur un territoire le traversant sans s'y arrêter.} montrent pour l'impact des infrastructures sur le long terme, dans le cas du tunnel sous la Manche\footnote{Mis en service en 1994 entre Calais en France et Folkestone au Royaume-uni, ce tunnel ferroviaire sous-marin de 50km permet une liaison physique entre l'Europe continentale et le Royaume-uni.}, par une analyse des investissements et des politiques dans le temps, que les effets effectivement constatés pour la région Nord-Pas-de-Calais comme un gain de centralité et de visibilité au niveau Européen, sont en fort décalage avec les discours justifiant le projet, et que le renouvellement des acteurs implique un non-accompagnement du projet sur le long terme, rendant son impact plus hasardeux. Nous rejoignons l'idée défendue par \cite{espacegeo2014effets} selon laquelle des ``effets de structure'' effectivement existent mais que ceux-ci se manifestent sur le temps long en termes de dynamiques systémiques pour lesquelles une vision locale courte n'a que peu de sens. A l'échelle intra-urbaine, \cite{fritsch2007infrastructures} prend l'exemple du tramway de Nantes pour montrer, par une étude localisée des transformations urbaines à proximité d'une nouvelle ligne, que les dynamiques de densification urbaine sont en décalage avec ce qu'en attendaient les élus et planificateurs, c'est-à-dire une association forte entre proximité à la ligne et densification.
}

\bpar{
These exemples confirm that the understanding of effects of territories on infrastructures imply to take into the concept of \emph{governance}.
}{
Ces exemples confirment que la compréhension des effets des territoires sur les infrastructures impliquent la prise en compte du concept de \emph{gouvernance}.
}

\subsubsection{Governance}{Gouvernance}

\bpar{
The development of a transportation network necessitate actors disposing of both concrete and economic capabilities to proceed to the construction, and furthermore having the legitimacy to lead this development. This must thus necessarily be actors of the social superstructure, that can be different levels of public governance, sometimes associated with private actors. The concept of \emph{governance}, that we understand as the management of an organisation with common ressources with targets linked to the interest of the concerned community (these can be defined in different ways, for example in a \emph{top-down} manner by governance actors or in a \emph{bottom-up} manner by consulting the agents concerned with the decision), is then crucial to understand the evolution of transportation projects and thus of transportation networks. We will use the term of \emph{territorial governance} when decision imply directly or indirectly components of territorial systems. 
}{
Le développement d'un réseau de transport nécessite des acteurs disposant à la fois des moyens concrets et économiques de mener à bien la construction, et d'autre part ayant la légitimité de mener ce développement. Il s'agit donc nécessairement d'acteurs de la superstructure sociale, pouvant être différents niveaux de pouvoirs publics, parfois associés à des acteurs privés. Le concept de \emph{gouvernance}, que nous comprenons comme la gestion d'une organisation disposant de ressources communes dans des buts liés à l'intérêt de la communauté concernée (pouvant être définis de différentes façons, par exemple de manière \emph{top-down} par les acteurs de gouvernance ou de manière \emph{bottom-up} par consultation des agents concernés par la décision), est alors essentiel pour comprendre l'évolution des projets de transports et donc des réseaux de transport. Nous parlerons de \emph{gouvernance territoriale} lorsque les décisions concernent directement ou indirectement des composantes de systèmes territoriaux. 
}

\bpar{
For example, \cite{offner2000territorial} illustrates the difficulties posed by the deregulation of some networked public services concerning the territorial competences of authorities, and proposes the emergence of a new local regulation for a new compromise between networks and territories.
}{
Par exemple, \cite{offner2000territorial} illustre les difficultés posées par la dérégulation de certains services publics en réseau quant aux compétences territoriales des autorités, et propose l'émergence d'une régulation locale pour un nouveau compromis entre réseaux et territoires.
}

\bpar{
Some aspects of territorial governance can have a significant impact on the development of transportation infrastructures. We can illustrate some for particular cases of the application of \emph{urban models}\footnote{In the sense of planning, i.e. conceptual generic schemas acting as a guide to the planification.}. \cite{deng2007potential} show in the case of Chinese cities that new directives in terms of housing can significantly deteriorate the performance of infrastructures, and that specific actions must be taken to anticipate these negative externalities. These concern in particular the dispositions in terms of \emph{Transit Oriented Development} (TOD). TOD is a particular approach to urban planning that aims at articulating the development of public transportation and urban development. It can be understood as a voluntary co-evolution by developers (administrative authorities and/or planning authorities), in which the articulation is thought and planned. We will come back on TOD during empirical studies in the following.
}{
Certains aspects de la gouvernance territoriale peuvent avoir un impact déterminant sur le développement des infrastructures de transport. Illustrons ceux-ci pour des cas particuliers d'application de \emph{modèles urbains}\footnote{Au sens de la planification, c'est-à-dire de schémas conceptuels génériques permettant de guider une démarche de planification.}. \cite{deng2007potential} montrent dans le cas des villes Chinoises que les nouvelles directives en terme de logement peuvent fortement détériorer la performance des infrastructures, et que des dispositions spécifiques doivent être prises pour anticiper ces externalités négatives. Celles-ci concernent notamment les dispositions en termes de \emph{Transit Oriented Development} (TOD). Le TOD est une approche particulière de l'aménagement urbain visant à articuler développement de l'offre de transport en commun et développement urbain. Il s'agit en quelque sorte d'une co-évolution volontaire de la part des développeurs (autorités administratives et/ou de planification), dans laquelle l'articulation est pensée et planifiée. Nous reviendrons sur le TOD lors d'études empiriques par la suite.
}

\bpar{
These concepts are not new, since they were for example implicit in the planning of new towns in \emph{Ile-de-France}, under a different form since these were strongly zoned (i.e. planned into relatively isolated mono-functional areas) and dependant on the automotive for some districts~\cite{es119}. \cite{l2012ville} give an example of an European project that has explored some implementations of TOD paradigms: planning details such as a quality of the network for active mobility modes at a short range are crucial for the concretization of principles. For example, \cite{lhostis:hal-01179934} use a multi-criteria analysis\footnote{In the frame of decision making for the planning of transportation infrastructures, multi-criteria analysis is an alternative to cost-benefit analysis (that compare projects by aggregating a generalized cost) which allows to take into account multiple dimensions, that are often contradictory (for example construction cost and robustness for a network), and obtain optimal solutions in the Pareto sense.} to understand determining factors in the selection	of stations for the planned city, including urban density and access time to stations. \cite{LIU2014120} show that even if some planning policies do not directly take a positioning as such, particularly in France, they exhibit very similar characteristics as shows the case of Lille.
}{
Ces concepts ne sont pas nouveaux, puisqu'ils étaient implicites par exemple dans l'aménagement des villes nouvelles en Ile-de-France, sous une forme différente car celles-ci étaient également fortement zonées (c'est-à-dire planifiées en zones relativement cloisonnées et mono-fonctionnelles) et dépendantes de l'automobile pour certains quartiers~\cite{es119}. \cite{l2012ville} donnent un exemple de projet européen ayant exploré des mises en pratiques de paradigmes du TOD : des détails d'aménagement comme un réseau de qualité pour les modes actifs à courte portée sont cruciaux pour une concrétisation des principes. Par exemple, \cite{lhostis:hal-01179934} utilisent une analyse multi-critères\footnote{Dans le cadre de l'aide à la décision pour la planification des infrastructures de transport, l'analyse multi-critère est une alternative aux analyses coût-bénéfices (qui comparent des projets en agrégeant un coût généralisé) qui permet de prendre en compte de multiples dimensions, souvent contradictoires (par exemple coût de construction et robustesse pour un réseau), et obtenir des solutions optimales au sens de Pareto.} pour comprendre les facteurs déterminants dans la sélection des stations de la ville planifiée, incluant densité urbaine et temps d'accès aux stations. \cite{LIU2014120} montrent que si certaines politiques de planification, en particulier en France, ne se réclament pas directement de cette approche, leurs caractéristiques sont très similaires comme le révèle le cas de Lille.
}

\bpar{
The articulation between transportation and urban planning must often be operated in a strongly coupled manner to attain the expected objectives, even more when the project is specialized: \cite{larroque2002paris} recall the case of the SK metro in Noisy-le-Grand which unveils a case of a complete dependency of the functionality of the transport to local development. In order to serve a project of a office complex, a specific line with a lightweight equipment is constructed to make a link with the RER station of Mont-d'Est. The real estate project will fail whereas the line is inaugurated in 1993, it will be first regularly maintained and then abandoned without having never been opened to the public. 
}{
L'articulation entre transport et aménagement doit souvent être opérée de façon fortement couplée pour parvenir aux objectifs recherchés, d'autant plus que le projet est spécialisé : \cite{larroque2002paris} rappellent l'anecdote du metro SK de Noisy-le-Grand qui montre un cas de dépendance complète de la fonctionnalité du transport à l'aménagement local. Pour desservir un projet de complexe de bureaux, une ligne spécifique avec une matériel roulant léger est construite pour faire le lien avec la gare RER de Mont-d'Est. Le projet immobilier avortera alors que la ligne est inaugurée en 1993, celle-ci sera d'abord entretenue régulièrement puis laissée a l'abandon sans jamais avoir été ouverte au public.
}

\bpar{
Therefore, governance processes, that manifest themselves in different ways, such as planning, or more particularly as TOD, play an important role in interactions between transportation networks and territories. These add up to our panorama, being of a particular type since they imply their own level of emergence and a strong autonomy.
}{
Ainsi, les processus de gouvernance, qui peuvent se décliner de plusieurs manières, comme ceux de planification, ou plus spécifiques de TOD, jouent un rôle important dans les interactions entre réseaux de transports et territoires. Ceux-ci s'ajoutent à notre panorama, étant d'un type particulier car impliquant leur propre niveau d'émergence et une forte autonomie.
}

\subsubsection{Co-evolution of networks and territories}{Co-évolution des réseaux de transport et des territoires}

\bpar{
This progressive construction allowed us to highlight the complexity of interactions between networks and territories, what suggests the relevance of the particular ontology of \emph{co-evolution} as we defined in introduction. \cite{levinson2011coevolution} insists on the difficulty of understanding the co-evolution between transport and land-use in terms of circular causalities, partly because of the different time scales implied, but also because of the heterogeneity of components. \cite{offner1993effets} uses the term of congruence, that can be understood as systemic dynamics implying correlations that can be spurious or not, that would be a preliminary vision of co-evolution.
}{
Cette construction progressive nous a permis de souligner la complexité des interactions entre réseaux et territoires, ce qui suggère la pertinence de l'ontologie particulière de la \emph{co-évolution} comme nous l'avons définie en introduction. \cite{levinson2011coevolution} insiste sur la difficulté de la compréhension de la co-évolution entre transport et usage du sol en termes de causalités circulaires, en partie à cause des différentes échelles de temps impliquées, mais aussi par l'hétérogénéité des composantes. \cite{offner1993effets} parle de congruence, qu'on peut comprendre comme une dynamique systémique impliquant des corrélations fortuites ou non, ce qui serait une vision préliminaire de la co-évolution.
}

\bpar{
The necessity to go past reducing approaches of structuring effects, together with the capture of the complexity of interactions between networks and territories through their co-evolution, is confirmed by the case of economic effects of high speed lines: \cite{Blanquart2017} proceeds to a both empirical and theoretical review, including grey literature, of studies of this specific case, and concludes, beyond the direct effects linked to the construction on which there is a consensus, that proper effects on a long time seem to be random. This witnesses in fact complex local situations, a large number of conjunctural aspects playing a role in the production of effects, that can then not be attributed to transport only. This review confirms moreover the gap between political and technical narratives preceding transportation projects and the effective posterior analysis, revealed by~\cite{bazin:hal-00615196}. \cite{bazin2007evolution} conduct also a targeted study of the real estate market in Reims in anticipation to the arrival of the \emph{TGV Est}. Through a diachronic analysis for each year between 1999 and 2005, for each district, of the real estate prices and the origin of buyers (locals or from the region of Paris), they conclude that only very localized operations can be directly linked to the TGV, the whole market following a global independent dynamic.
}{
La nécessité de dépasser les approches réductrices des effets structurants, tout en capturant la complexité des interactions entre réseaux et territoires par leur co-évolution, est confirmée par le cas des effets économiques des lignes à grande vitesse : \cite{Blanquart2017} procèdent à une revue à la fois théorique et empirique, incluant la littérature grise, des études de ce cas spécifique, et conclut, au delà des retombées directes liées à la construction sur lesquelles il y a consensus, que les effets propres sur un temps plus longs paraissent aléatoires. Cela témoigne en fait de situations locales complexes, un grand nombre d'aspects conjoncturels entrant en jeu dans la production d'effets, qu'on ne peut alors pas attribuer seulement au transport. Cette revue confirme par ailleurs le décalage entre les discours politiques et techniques prévalant aux projets de transports et les analyses effectives a posteriori, révélé par~\cite{bazin:hal-00615196}. \cite{bazin2007evolution} mènent d'autre part à une étude ciblée du marché immobilier à Reims en anticipation de l'arrivée du TGV Est. En procédant à une analyse diachronique pour chaque année entre 1999 et 2005, par quartier, des prix immobiliers et de la provenance des acheteurs (franciliens ou locaux), ils concluent que seulement des opérations très localisées pouvaient être directement reliées au TGV, l'ensemble du marché répondant à une dynamique globale indépendante.
}

\stars

\bpar{
Thus, our constructive overview, broad and conceived as circular, of interactions between transportation networks and territories, confirms the relevance of the concept of \emph{co-evolution} on the one hand, but suggests on the other hand a more thorough investigation and clarification for it.
}{
Ainsi, notre aperçu constructif, large et voulu circulaire, des interactions entre réseaux de transports et territoires, confirme la pertinence de ce concept de \emph{co-évolution} d'une part, mais suggère d'autre part un approfondissement et une clarification de celui-ci.
}

\bpar{
We have therefore seen in this section that (i) the concept of territory naturally yields the concept of network; (ii) reciprocally, networks can transform territories, following different processes more or less established depending on scales; (iii) there exists a large number of cases and of particular processes for which the relation between networks and territories is imbricated, and for which we can use the term of \emph{co-evolution}.
}{
Nous avons ainsi vu dans cette section que (i) le concept de territoire amène naturellement celui de réseau ; (ii) réciproquement, le réseau peut transformer les territoires, selon différents processus plus ou moins établis selon les échelles ; (iii) qu'il existe un grand nombre de cas et de processus particuliers pour lesquels la relation entre réseau et territoire est imbriquée, et où on peut parler de \emph{co-évolution}.
}

\bpar{
We will aim in the following section at studying more thoroughly in an empirical way various aspects evoked here, to put into perspective and refine the questions we aim at answering here.
}{
Nous nous appliquerons dans la section suivante à approfondir de manière empirique différents aspects abordés ici, pour une mise en situation et un raffinement des questions que nous nous posons.
}

\stars

%


\newpage

\section{Transportation projects from Paris to Zhuhai}{Projets de transport de Paris à Zhuhai}

\label{sec:casestudies}


\bpar{
We develop in this section some geographical case studies at the metropolitan scale as we previously defined. We choose them to be very different to maximize the diversity of processes that can potentially be identified (since as we showed the geographical context is crucial). These are the Greater Paris metropolitan area, and the mega-city-region of Pearl River Delta in the South of China.
}{
Nous développons dans cette section des cas d'étude géographique à l'échelle métropolitaine comme nous l'avons définie précédemment. Nous les choisissons très différents pour maximiser la diversité des processus potentiellement identifiables (puisque comme nous l'avons montré le contexte géographique est crucial). Il s'agit de la métropole du Grand Paris, et de la mega-région urbaine du Delta de la Rivière des Perles dans le sud de la Chine.
}


\bpar{
The objective of this section is to specify, precise, illustrate, enrich, the overview of co-evolution processes that we established in a general manner. Geography can not draw general conclusions, in the cases where these are relevant, without very precise and particular case studies. When applying a generic model to a set of territories, we will investigate the deviation to the model, that must then be explained through geographical reasoning, meaning a strong implication with the place in particular. Our approach is similar: if we can link several developed concepts to a case study, these will be necessarily enriched\footnote{And possibly connected through the transfer of the structure of the particular system to the structure of knowledge.}.
}{
L'objectif de cette section est de spécifier, préciser, illustrer, enrichir, l'aperçu des processus de co-évolution que nous avons établi de manière générale. La géographie ne peut tirer de conclusion générales, dans les cas où celles-ci sont pertinentes, sans études de cas particuliers bien précis. Dans l'application d'un modèle générique à un ensemble de territoires, on cherchera les déviations au modèle, qu'il s'agira alors d'expliquer par des raisonnements géographiques, signifiant une forte implication avec le lieu en particulier. Notre démarche est similaire : si nous pouvons raccrocher nombre de concepts développés à un cas d'étude, ceux-ci seront nécessairement enrichis\footnote{Et possiblement connectés par le transfert de la structure du système particulier à la structure de la connaissance.}.
}


\subsection{Greater Paris: history and issues}{Le Grand Paris : histoire et enjeux}

\bpar{
The Parisian region is a good illustration of the complexity of interactions between transportation networks and territories. The relevant time period for our question ranges from the end of the 19th century to nowadays. We propose, after a bief presentation of the context, to recall the history of the development of public transportation in \emph{Ile-de-France}, which allows to reveals its articulations with urbanism, in particular the issues linked to transportation network planning. We will then study the present and future of \emph{Grand Paris}, first concerning the emergence of a new governance structure at the level of the metropolitan area, and then the implied recent transportation projects, putting the example at the core of our problematic. We will finally make a more detailed incursion within an empirical analysis of relations between territorial variables and accessibility differentials for transportation projects, sketching some of the methodological developments we will develop in the following.
}{
La région parisienne est une bonne illustration de la complexité des interactions entre réseaux de transports et territoires. La période temporelle pertinente pour notre question court de la fin du 19ème siècle à nos jours. Nous proposons, après une présentation brève du contexte, de rappeler l'histoire du développement des transports en Ile-de-France, qui permet de révéler ses articulations avec l'urbanisme, en particulier les enjeux liés à la planification du réseau de transport. Nous traiterons ensuite le présent et le futur du Grand Paris, d'abord concernant l'émergence d'une nouvelle structure de gouvernance au niveau de la métropole, puis les projets de transport récents impliqués, mettant l'exemple au coeur de notre problématique. Nous ferons finalement une incursion plus détaillée dans une analyse empirique des relations entre variables territoriales et différentiels d'accessibilité pour les projets de transport, préfigurant certains des développements méthodologiques que nous mènerons par la suite.
}


\subsubsection{Context}{Contexte}

\bpar{
The spatial context is the intermediate scale of a globally monocentric metropolitan area. Let precise this spatial structure. If the metropolis taken up to the \emph{moyenne couronne} (i.e. the extent corresponding roughly to the central urban core with continuous built environment) exhibits a certain level of polycentrism\footnote{Polycentrism, by opposition to monocentrism, means that it is possible to identify different centers in an urban system. The way to define a center will depend on the scale and on the phenomenons considered: it can for example be the existence of different employment poles of comparable size at the infra-metropolitan scale. The same way that the concept is polymorphic, the way to measure it quantitatively are multiple and complementary~\cite{servais2004polycentrisme}.}, in particular through the effect of new towns, which became important local employment centers~\cite{berroir2005contribution}.
}{
Le contexte spatial est l'échelle intermédiaire d'une région métropolitaine globalement monocentrique. Précisons cette structure spatiale. Si la métropole prise jusqu'à la moyenne couronne (c'est-à-dire l'étendue correspondant environ au noyau urbain central bâti de manière continue) possède un certain niveau de polycentrisme\footnote{Le polycentrisme, en opposition au monocentrisme, signifie qu'il est possible d'identifier différents centres dans un système urbain. La façon de définir un centre dépendra de l'échelle et des phénomènes considérés : il peut s'agir par exemple de l'existence de différents pôles d'emplois de taille comparable à l'échelle intra-métropolitaine. De la même façon que le concept est polymorphe, les façons de le mesurer quantitativement sont multiples et complémentaires~\cite{servais2004polycentrisme}.}, notamment grâce à l'effet des villes nouvelles, devenues d'importants pôles d'emplois locaux~\cite{berroir2005contribution}.
}

\bpar{
The role of different transportation infrastructures in the different economical dynamics in \emph{Ile-de-France} is not trivial, as shows~\cite{PADEIRO201344} which aims at statistically explicating employment growth between 1993 and 2008 in medium-sized and small communes in the Parisian region as a function of the proximity to an infrastructure: effects depends both on transportation mode (highway or airport) but also on the economic sector considered. Reciprocally, successive developments of transportation projects, generally operate in a discontinuous way in time. As we will detail in the following, they are linked to planning dynamics and governance processes that must be understood conjointly to territorial dynamics. The Parisian metropolis thus witnesses of complex relations between territories and networks.
}{
Le rôle des différentes infrastructures de transport dans les différentes dynamiques économiques en Ile-de-France n'est pas trivial, comme le montre~\cite{PADEIRO201344} qui cherche à expliquer statistiquement la croissance de l'emploi entre 1993 et 2008 dans les moyennes et petites communes franciliennes en fonction de la proximité à une infrastructure : les effets dépendent à la fois du mode (autoroute ou aéroport) mais aussi du secteur économique considéré. Réciproquement, les développements successifs des projets de transport, s'opèrent de manière généralement discontinue dans le temps. Comme nous le détaillerons par la suite, ils sont liés à des dynamiques de planification et des processus de gouvernance qu'il convient de comprendre de manière conjointe aux dynamiques territoriales. La métropole parisienne témoigne ainsi de relations complexes entre territoires et réseaux.
}

\subsubsection{Greater Paris transportation network}{Réseau de transport du Grand Paris}

\bpar{
The history of the development of the transportation network of Parisian metropolitan area is recalled in~\cite{larroque2002paris}. The French particularity with centralization lead to a particular structure for the railway network at the national scale, but also at the regional scale. The domination of Paris has indeed strongly shaped the structuration of the transportation network during the different historical periods during which it underwent significant evolutions. \cite{larroque2002paris} decompose the second half of the twentieth century in three periods.
}{
L'histoire du développement du réseau de transport de la métropole francilienne est rappelée dans~\cite{larroque2002paris}. La particularité centralisatrice française a conduit à une structure particulière du réseau ferré à l'échelle nationale, mais aussi à l'échelle régionale. La domination de Paris a en effet fortement marqué la structuration du réseau de transport au cours des différentes périodes historiques où il a subi des évolutions conséquentes. \cite{larroque2002paris} décomposent la seconde moitié du vingtième siècle en trois périodes. 
}

\bpar{
Before 1975, the distribution of accessibility of actives to employments is clearly centralized and the center of Paris exhibits a strong congestion. The establishment of the RER network between 1975 and 1988 allows, thanks to the conjoint construction of \emph{Villes Nouvelles}, an articulation between transportation and urbanism and a certain degree of polycentrism. \cite{larroque2002paris} however recall that realizations during this period show an increasing gap with the real demand for transportation. The period following 1988 until 2000, year of a political alternance, will mostly consist in the renewing of actors and the elaboration of new strategies, as witnesses the \emph{Schéma Directeur} in 1994. Network developments during this period do not induce any major change in the spatial distribution of accessibility, despite the realization of the central interconnexion for RER D, of the line 14 and of RER E. 
}{
Avant 1975, la distribution de l'accessibilité des actifs aux emplois est clairement centralisée et le centre de Paris fortement congestionné. La mise en place du réseau RER entre 1975 et 1988 permet grâce à la construction conjointe des Villes Nouvelles une articulation entre transport et urbanisme et un certain niveau de polycentrisme. \cite{larroque2002paris} rappellent toutefois que les réalisations dans cette période sont en décalage croissant avec la demande réelle de transport. La période qui suivra 1988 jusqu'à 2000, année marquée par l'alternance politique, consistera surtout en le renouvellement des acteurs et l'élaboration de nouvelles stratégies, comme en témoigne le Schéma Directeur de 1994. Les développements du réseau sur cette période n'induisent aucun changement majeur de la distribution spatiale de l'accessibilité, malgré la réalisation de l'interconnexion centrale du RER D, de la ligne 14 et du RER E.
}

\bpar{
The successive planning schemes lead to the SDRIF of 2013 \cite{sdrif2013}. They present early signs of the future network of the \emph{Grand Paris Express}, of which a strong impact is expected in terms of territorial cohesion by favouring links between suburbs which are the most problematic in the current network. Furthermore, the plan is voluntary integrated, by densification around stations and an articulation between urban operations and new infrastructures. This aspect of network integration within territories and of territories by networks can be indeed observed in the public communication of the transportation organisation authority (former STIF, which became \emph{Ile-de-France Mobilités})\footnote{See for example the actuality of the 4th October 2017 at \url{https://www.iledefrance-mobilites.fr/actualites/un-reseau-de-transports-qui-grandit/} which underlines that ``\textit{With 29km of additional network length and the opening of 28 desserve points, territories are getting closer}'', witnessing the importance of accessibility for territories, notion which is furthermore fuzzy. Similar orientations in discourse can be found for the different projects of extension or construction of new lines.}. We therefore find again the importance of governance processes in the articulation between transportation networks and territories for the example of Ile-de-France in time.
}{
Les schémas directeurs successifs conduisent au SDRIF de 2013 \cite{sdrif2013}. Ceux-ci préfigurent le futur réseau du \emph{Grand Paris Express}, dont un fort impact est attendu en termes de cohésion territoriale en favorisant les liaisons de banlieue à banlieue qui sont les plus problématiques dans le réseau actuel. De plus, le schéma est volontairement intégré, par densification autour des gares et articulation des opérations d'aménagement et des nouvelles infrastructures. Cet aspect d'intégration des réseaux dans les territoires et des territoires par les réseaux se retrouve bien dans la communication publique de l'Autorité Organisatrice des Transports (ancien STIF, devenu Ile-de-France Mobilités)\footnote{Voir par exemple l'actualité du 4 octobre 2017 sur \url{https://www.iledefrance-mobilites.fr/actualites/un-reseau-de-transports-qui-grandit/} qui souligne que ``\textit{Avec 29 km de réseau supplémentaires et l’ouverture de 28 points de desserte, les territoires se rapprochent}'', témoignant de l'importance de l'accessibilité pour les territoires, notion par ailleurs floue. Les mêmes orientations de discours se retrouvent pour les différents projets d'extension ou de construction de nouvelles lignes.}. Nous retrouvons donc l'importance des processus de gouvernance dans l'articulation des réseaux de transport et des territoires dans l'exemple de l'Ile-de-France au cours du temps.
}

\bpar{
Other processes already mentioned also manifest themselves, under different forms. For example, the role of path-dependency in trajectories of the territorial system is illustrated by \cite{larroque2002paris} which shows the inertia due to successive technical choices when they are successful: the initial choice of a metropolitan network within Paris' walls, the realization of the RER network, the tarification politic by areas for the \emph{carte orange} at the end of the nineties, are different decisions in diverse domains but having each their significant part in the possible posterior developments. These authors also show how decisions concerning the public transportation network can induce, through a bad covering or performance of the public transportation network, the emergence of interaction processes where the couple use of the car and periurbanization\footnote{The periurban belongs to the new forms of urbanization, and consists in intermediate territories between the rural and the urban, benefiting from a good accessibility but exhibiting low densities and mostly individual dwellings.} is favored, in a way similar to the \emph{automobile city} described by~\cite{newman1996land}.
}{
D'autres processus déjà mentionnés se manifestent également, sous différentes formes. Par exemple, le rôle de la dépendance au chemin dans les trajectoires du système territorial est illustré par \cite{larroque2002paris} qui montre l'inertie due aux choix techniques successifs lorsque ceux-ci rencontrent un succès : le choix initial d'un réseau métropolitain intra-muros, la mise en place du réseau RER, la politique de tarification par zones de la carte orange à la fin des années 90, sont autant de décisions sur des domaines divers mais ayant chacune leur part significative dans les développements postérieurs possibles. Ces auteurs montrent également comment les décisions concernant le réseau de transport en commun peuvent induire, par mauvaise couverture ou performance du réseau de transport en commun, l'émergence de processus d'interactions où le couple usage de la voiture et périurbanisation\footnote{Le périurbain fait partie des nouvelles formes d'urbanisation, et consiste en des territoires intermédiaires entre urbain et rural, bénéficiant d'une bonne accessibilité mais présentant un faible densité et des habitats individuels majoritairement.} est favorisé, à l'image de l'\emph{automobile city} décrite par~\cite{newman1996land}. 
}

\bpar{
\cite{padeiro:tel-00438092} recalls that the extension of metro lines to the close suburbs has always been restricted, reinforcing the role of Paris' city in the relation between the metropolitan territory and networks. Furthermore, he shows that urban polarizations (adaptation of the built environment and of the socio-economical composition) around stations beyond the limits of Paris are for their socio-economical part anterior dynamics that the arrival of the metro then accompanies: in that case, there is no structuring effect in the proper sense.
}{
\cite{padeiro:tel-00438092} rappelle que le prolongement des lignes de métro en proche banlieue a toujours été restreint, renforçant le rôle de la ville Paris dans les relations entre le territoire métropolitain et les réseaux. Par ailleurs, il montre que les polarisations urbaines (adaptation du bâti et de la composition socio-économique) autour des stations au delà du périphérique sont pour leur partie socio-économique des dynamiques antérieures qu'accompagne alors l'arrivée du métro : dans ce cas, il n'y a pas d'effet structurant à proprement parler.
}


\subsubsection{Towards a metropolitan governance}{Vers une gouvernance métropolitaine}

\bpar{
To the metropolitan context previously described corresponds a complexity of the governance structure. In particular, current developments, both of the transportation network and of urban projects, coincide with the emergence of a new level of governance, an intermediary between \emph{communes} and \emph{départements} on one side, and the Region and the State on the other side. We can ask to what extent this emergence is linked to dynamics of interactions between territories, and how it will influence the interactions between territories and networks. \cite{gilli2009paris} propose in 2009 a diagnostic of the institutional situation of the Parisian region, and directions for a coupled approach between governance and planning. They highlight the early signs of the ``establishment of a collective metropolitan actor'', which corresponds to the \emph{métropole du Grand Paris} which will be inaugurated 7 years later, since the metropolitan council in put into place in the end of 2016.
}{
Au contexte métropolitain décrit précédemment correspond une complexité de la structure de gouvernance. En particulier, les développements actuels, à la fois du réseau de transport et des projets d'aménagement, coincident avec l'émergence d'un nouveau niveau de gouvernance, intermédiaire entre communes et départements d'une part et Région et État d'autre part. On peut se demander dans quelle mesure cette émergence est reliée aux dynamiques d'interactions entre territoires, et comment celle-ci influera sur les interactions entre territoires et réseaux. \cite{gilli2009paris} proposent en 2009 un diagnostic de la situation institutionnelle de la région parisienne, et des pistes pour une approche couplée entre gouvernance et aménagement. Ils mettent en valeur la préfiguration de ``l'instauration d'un acteur collectif métropolitain'', qui correspond à la métropole du Grand Paris qui sera inaugurée 7 ans plus tard, puisque le conseil métropolitain est mis en place fin 2016.
}

\bpar{
The establishment of this new level of governance has been studied more recently still by~\cite{gilli2014gouverner}, which situates it within a broader socio-economical context and of other levels of governance (State, Region, \emph{intercommunalités}). It allows him to sketch a territorial diagnosis which gives elements explaining its emergence: gaining retard in the domain of planification compared to its past dynamics, but also in the social domain given very high local socio-economical inequalities, the metropolis needs to reinvent itself, and this new dynamics naturally crystallize in the \emph{Grand Paris}, what means that, as he concludes, ``the future of Paris are its suburbs''. This initiative is made concrete by the convergence on the one hand of initiatives and the voluntarism of local politics, and on the other hand of a redefinition of the role of the State, wanted with a centralization until 2012 and freeing the stage to metropolitan governance with the political alternance in 2012. The projects launched and financing remain roughly the same: the project of the \emph{Grand Paris Express} is a compromise between the solution wanted by the State and the one defended by the Region. Following~\cite{desjardins2016grand}, although the metropolitan governance structure has still today relatively no power, and although the negligence of the social aspect of metropolitan development is always highly present, these mutations however witness a deep structural change in the organisation of the region. We now detail the transportation project of the \emph{Grand Paris Express}.
}{
La mise en place de ce nouveau niveau de gouvernance a été disséquée plus récemment toujours par~\cite{gilli2014gouverner}, qui la situe dans un contexte socio-économique et des autres niveaux de gouvernance (État, Région, intercommunalités) plus large. Cela lui permet de dresser un diagnostic territorial qui fournit des éléments explicatifs à son émergence : en perte de vitesse sur le plan de l'aménagement par rapport à ses dynamiques passées, mais aussi sur le plan social au vu d'inégalités socio-économiques locales très fortes, la métropole a besoin de se réinventer, et ce nouveau souffle se cristallise naturellement dans le Grand Paris, c'est-à-dire que, comme il conclut, ``l'avenir de Paris est sa banlieue''. Cette initiative se concrétise par la convergence d'une part des initiatives et du volontarisme des élus locaux, et d'autre part d'une redéfinition du rôle de l'Etat, voulue centralisatrice jusqu'en 2012 puis laissant la place libre à la gouvernance métropolitaine avec l'alternance politique en 2012. Les projets lancés et les financements restent les mêmes dans les grandes lignes : le projet du Grand Paris Express est un compromis entre la solution voulue par l'Etat et celle poussée par la région. Suivant~\cite{desjardins2016grand}, si la structure de gouvernance métropolitaine est aujourd'hui toujours relativement impuissante, et si l'oubli de l'aspect social du développement métropolitain est toujours très présent, ces mutations témoignent toutefois d'un changement structurel profond dans l'organisation de la région. Nous détaillons à présent le projet de transport du Grand Paris Express.
}


\subsubsection{Project of the Grand Paris Express: towards a rebalancing of accessibilities?}{Projet du Grand Paris Express : vers un rééquilibrage des accessibilités ?}

\begin{figure}
\includegraphics[width=\linewidth]{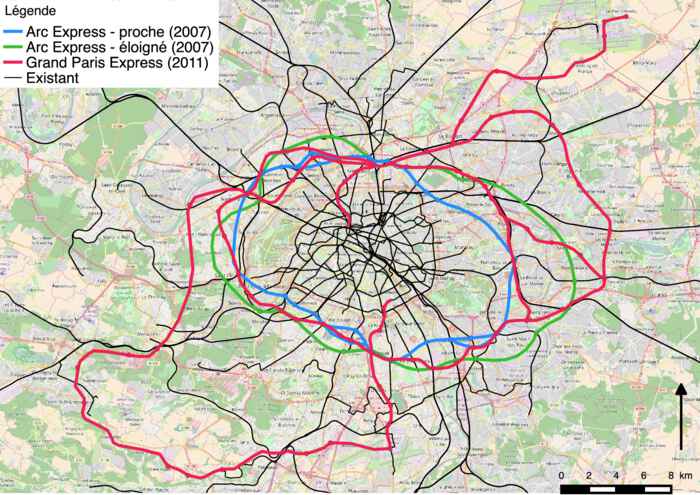}
\caption[Successive transportation projects for Greater Paris]{\textbf{Successive transportation network projects for the Grand Paris metropolitan area.} We show the two alternatives for the \emph{Arc Express} project elaborated by the Region, and the \emph{Grand Paris Express} (GPE) advocated by the State. The \emph{Réseau du Grand Paris}, a precursor for GPE, is not shown here for visibility reasons because of its proximity with it. The source of the map background, given to situate the lines, is OpenStreetMap.\label{fig:casestudies:projects}}
\end{figure}

\begin{figure}
	\includegraphics[width=\linewidth]{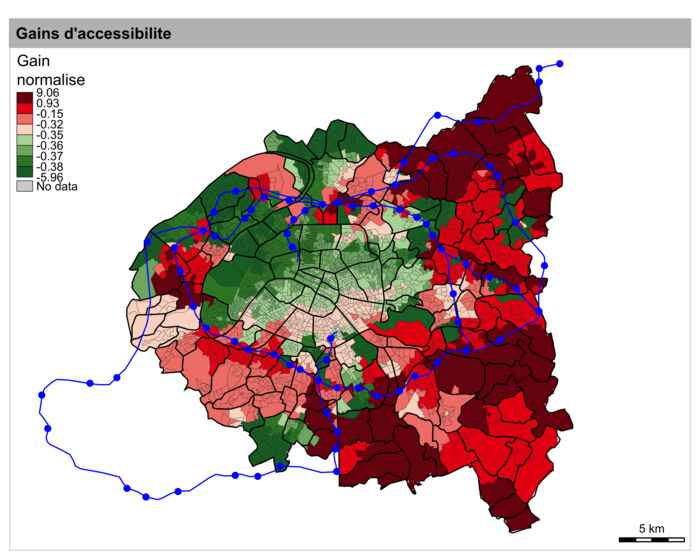}
	\caption[Impact of \emph{Grand Paris Express} on accessibility]{\textbf{Impact of GPE lines on temporal accessibility.} The map gives, for the \emph{départements de petite couronne} and Paris (75, 92, 93, 94) the temporal accessibility gains, defined for each Iris (elementary infra-communal statistical unit) as the average travel time with public transport to all centroids of other communes, weighted by destination population. The gain is computed as the accessibility difference with and without \emph{Grand Paris Express}. We show a normalized gain, i.e. centered (with a null average) and reduced (unit standard deviation). In blue, the lines and new stations of GPE. We observe the strongest gains mostly in the East, in consistence with the existing literature such as~\cite{beaucire2013grand}. The territorial imprints of RER lines (A in the West, D and B in North, B in the South) exhibit relatively low gains since they are already very accessible.\label{fig:casestudies:gpe}}
\end{figure}

\bpar{
The metropolitan region of Paris is currently undergoing significant transformations, with the constitution of a metropolitan governance and new transportation infrastructures. The construction of a ring metro network allowing suburbs to suburbs links answers to an ancient need, and lead to several proposals on which the State and the Region have been in conflict around 2010~\cite{desjardins2010bataille}. The \emph{Arc Express} project~\cite{stif2007arc}, advocated by the Region and more focused on territorial equity, can be contrasted with initial proposals for a \emph{Réseau du Grand Paris} aimed at linking ``excellence clusters'' despite a potential tunnel effect. The solution finally adopted (see the last \emph{Schéma Directeur}~\cite{sdrif2013}) is a compromise and allows a rebalancing of accessibility between the west and the east~\cite{beaucire2013grand}. The Fig.~\ref{fig:casestudies:projects} maps the different projects.
}{
La région métropolitaine de Paris est en train de connaître de grandes mutations, avec la mise en place d'une gouvernance métropolitaine et de nouvelles infrastructures de transport. La construction d'un réseau de métro en rocade permettant des liaisons de banlieue à banlieue répond à un besoin ancien, et a mené à plusieurs propositions sur lesquelles se sont opposés l'Etat et la Région au tournant des années 2010~\cite{desjardins2010bataille}. Le projet Arc Express~\cite{stif2007arc}, porté par la Région et plus axé sur une égalité des territoires, contrastait avec les propositions initiales de Réseau du Grand Paris visant à relier des ``clusters d'excellence'' en dépit d'un possible effet tunnel. La solution finalement adoptée (voir le dernier schéma directeur \cite{sdrif2013}) est un compromis et permet un rééquilibrage est-ouest de l'accessibilité~\cite{beaucire2013grand}. La Fig.~\ref{fig:casestudies:projects} cartographie les différents projets.
}

\bpar{
The immediate impacts of a new transportation infrastructure in terms of accessibility, i.e. of the transformation of the spatial distribution of different accessibilities, generally occur for much larger territories than the areas in which the line and its stations are constructed: accessibility patterns are a consequence of topological properties of the network and these are strongly discontinuous as a function of graph structure. We can illustrate the case of \emph{Grand Paris Express} lines and of their direct impact on regional accessibility. We map in Fig.~\ref{fig:casestudies:gpe} the temporal accessibility gains allowed by the \emph{Grand Paris Express} for metropolitan \emph{départements} (75, 92, 93 and 94). The temporal accessibility is computed for each Iris $i$ the following way: with $P_j$ the populations of \emph{communes}, $t_0$ a parameter giving the typical commuting duration (that we fix at one hour~\cite{zahavi1980regularities}), $t_{ij}$ the travel time with public transport between the centroid of $i$ and the one of commune $j$, we take a weighted average defined by
}{
Les impacts immédiats d'une nouvelle infrastructure de transport en termes d'accessibilité, c'est-à-dire la transformation de la distribution spatiale des différentes accessibilités, concernent généralement des territoires bien plus larges que les zones où la ligne et ses stations sont implantées : les motifs d'accessibilité sont dus aux propriétés topologiques du réseau et celles-ci sont fortement discontinues en fonction de la structure du graphe. Illustrons le cas des lignes du Grand Paris Express et de leur impact direct sur l'accessibilité régionale. Nous cartographions en Fig.~\ref{fig:casestudies:gpe} les gains d'accessibilité temporelle permis par le projet du Grand Paris Express sur les départements métropolitains (75, 92, 93 et 94). L'accessibilité temporelle est calculée pour chaque Iris $i$ de la manière suivante : avec les populations des communes $P_j$, $t_0$ un paramètre de durée typique d'un déplacement (que nous fixons à une heure~\cite{zahavi1980regularities}), $t_{ij}$ le temps de trajet en transports en commun entre le centroïde de $i$ et celui de la commune $j$, nous prenons une moyenne pondérée définie comme
}

\[
Z_i = \sum_j \left(\frac{P_j}{\sum_k P_k}\right)\cdot \exp\left(- t_{ij}/t_0\right)
\]

\bpar{
This expression indeed allows to have an accessibility potential, and the weighting be population should remove some bias due to potentially negligible trajectories as a proportion of total travels. We recall that this is a normative accessibility in the sense of \cite{paez2012measuring} since the gravity parameter is fixed in a stylized way.
}{
Cette expression permet de bien avoir un potentiel d'accessibilité, et la pondération par la population permet de ne pas biaiser l'indicateur par des trajets potentiellement négligeables en proportion des trajets totaux. Notons qu'il s'agit d'une accessibilité normative au sens de \cite{paez2012measuring} puisque le paramètre gravitaire est fixé de manière stylisée.
}

\bpar{
We observe, in accordance with the analysis by~\cite{beaucire2013grand}, a rebalancing of accessibility differentials between the East and the West. At an equal distance of the center, accessibility is lower for Seine-Saint-Denis and Val-de-Marne that for Hauts-de-Seine, i.e. that these \emph{départements} have potentially more difficulties to access the rest of the metropolis. The map of average time gains also exhibits the highest gains for this two \emph{départements}. Some \emph{communes} that are socially and economically disadvantaged as Aulnay benefit from the highest time gains. The line 16 indeed allows a significant opening up of the North-east of Seine-Saint-Denis~\cite{desjardins2016grand}. The creation of links from suburbs to suburbs is a crucial aspect of this opening up and is conceived as a motor of the emergence of new centralities, towards an always more polycentric metropolis, in the inheritance of the planning policy of \emph{Villes Nouvelles}, in order to obtain not neighboring suburbs anymore but districts that are a full part of Greater Paris. The effects can remain however mitigated depending on the areas: \cite{l2013grand} show that the \emph{Grand Paris Express} will induce a direct access to a larger number of employments for a significant number of unemployed within the \emph{Petite Couronne}, but that inequalities with \emph{Grande Couronne} will increase and that there exists some risks of dropping out for far away \emph{communes} with a low accessibility.
}{
Nous observons, conformément à l'analyse de~\cite{beaucire2013grand}, un rééquilibrage des différentiels d'accessibilité entre Est et Ouest. A distance égale du centre, l'accessibilité est plus basse pour la Seine-Saint-Denis et le Val-de-Marne que pour les Hauts-de-Seine, c'est-à-dire que ces départements ont potentiellement plus de difficultés pour accéder au reste de la métropole. La carte des gains temporels moyens montre les gains plus grands également pour ces deux départements. Des communes socio-économiquement défavorisées comme Aulnay sont bénéficiaires des plus grands gains de temps. La ligne 16 permet en effet un désenclavement significatif du nord-est de la Seine-Saint-Denis~\cite{desjardins2016grand}. La création de liaisons de banlieue à banlieue est un aspect majeur de ce désenclavement et est voulue comme un moteur de l'émergence de nouvelles centralités, vers une métropole toujours plus polycentrique, dans la lignée de la politique d'aménagement des villes nouvelles, pour ne plus parler de proche banlieue mais de quartiers faisant partie intégrante du Grand Paris. Les effets peuvent cependant être mitigés selon les zones : \cite{l2013grand} montrent que le Grand Paris Express induira un accès direct à un plus grand nombre d'emplois pour un nombre significatif de chômeurs en petite couronne, mais que les écarts avec la grande couronne seront accentués et qu'il existe des risques de décrochage de certaines communes lointaines mal desservies.
}

\bpar{
One of the crucial issues for the construction of Greater Paris is to stay careful on not obtaining a metropolis with multiple separated levels, and to exploit the increased connectivity at different scales (international, national, regional, metropolitan) in order to reduce territorial inequalities instead of increasing them\footnote{We recall that an unequal distribution of agents and resources will generate differences in potential larger than a uniform distribution, these can then be linked to the evolution of the network.}. The novel network seems to contribute to this dynamic, under the condition of a coordinated territorial development, allowing the realization of immediate accessibility gains in terms of territorial transformations. There exists no method that can forecast it in a deterministic way as we already developed. It is however possible to retrospectively analyze from an empirical point of view the couplings between territorial variables and network variables, in order to quantitatively unveil co-evolution phenomena. We propose now to illustrate this approach.  
}{
L'un des enjeux cruciaux pour la construction du Grand Paris est de veiller à ne pas obtenir une métropole à plusieurs vitesses, et de tirer parti de la connectivité accrue à plusieurs échelles (internationale, nationale, régionale, métropolitaine) pour réduire les inégalités territoriales plutôt que les accroitre\footnote{Rappelons qu'une inégale répartition des agents et des ressources générera des différences de potentiel plus grandes qu'une distribution uniforme, celles-ci pouvant alors être liées à l'évolution du réseau.}. Le nouveau réseau semble contribuer à cette dynamique, sous condition d'un développement territorial coordonné, permettant la concrétisation des gains immédiats d'accessibilité en terme de transformation territoriale. Il n'existe pas de méthode pouvant prévoir celle-ci de manière déterministe comme nous l'avons déjà développé. Il est cependant possible d'analyser rétrospectivement de manière empirique les couplages entre variables territoriales et variables de réseau, pour essayer de mettre en valeur quantitativement les phénomènes de co-évolution. Nous proposons à présent d'illustrer cette démarche.
}

\subsubsection{Linking territorial dynamics and construction of the Grand Paris Express}{Lier dynamiques territoriales et construction du Grand Paris Express}

\bpar{
One of the aims of our work in the following will be to empirically clarify situations in which strongly coupled dynamics linked to our problematic can be exhibited, and then through models to isolate processes and conditions allowing one or the other situation. We propose to deepen the illustration of GPE, while introducing a potential approach to link a territorial dynamic with the one of the anticipated network.
}{
L'un des enjeux de notre travail par la suite sera de clarifier empiriquement des situations dans lesquelles des dynamiques fortement couplées relevant de cette problématique pourront être mises en évidence, puis à travers des modèles d'isoler des processus et des conditions permettant telle ou telle situation. Nous proposons d'approfondir l'illustration du GPE, tout en introduisant une approche possible pour lier dynamique territoriale et celle du nouveau réseau anticipé.
}

\bpar{
Various aspects of territories are concerned by interactions with networks. In previous empirical studies, no socio-economic attributes of populations inhabiting the territory nor economic values for land and real estate was considered. Both are however crucial elements of territorial dynamics and are extensively studied in fields such as territorial analysis or urban economics : for example, \cite{homocianu:tel-00359302} studies households residential choices to understand land-use transportation interactions. We propose here to use a database of Real Estate transactions for Parisian region on the last 20 years, with 2 years temporal granularity and exact spatial coordinates. \cite{guerois2009dynamique} used it for example to obtain typologies of spatial dynamics of the Parisian real estate market.
}{
Des aspects très variés des territoires sont concernés par l'interaction avec les réseaux. Dans nos études précédentes, les aspects économiques et financiers du foncier et l'immobilier n'ont pas été considérés. Il s'agit cependant d'éléments cruciaux des dynamiques territoriales et sont étudiés de manière intensive dans des champs comme l'analyse territoriale ou l'économie urbaine : par exemple, \cite{homocianu:tel-00359302} étudie les choix résidentiels des ménages pour comprendre les interactions entre usage du sol et transport. Nous proposons ici d'utiliser entre autres une base de données de transactions immobilières pour la région parisienne sur les 20 dernières années, avec une granularité temporelle de 2 ans et coordonnées spatiales exactes. \cite{guerois2009dynamique} l'utilise par exemple pour établir une typologie des dynamiques spatiales du marché immobilier parisien.
}

\bpar{
This more precise study can be understood as a research of early warnings of network potential breakdowns: indeed, if intrinsic territorial dynamics anticipate the arrival of a new public transportation station, the implications will be much different to the case where it will then drive these variables after its construction. The interpretation in terms of ``structuring effects'' will indeed be significantly different. We apply here the method of spatio-temporal causalities developed in~\ref{sec:causalityregimes}. We propose to study the relations between the accessibility differential for each project, and variables linked to land (real estate transactions) and socio-economical, in order to see if it is possible to capture a link between accessibility differentials and differentials in territorial variables. Indeed, the links between new lines and real estate value evolution are sometimes dramatic~\cite{damm1980response}. 
}{
Cette étude plus précise peut être comprise comme une recherche de signes précurseurs de rupture de potentiels du réseau : en effet, si des dynamiques territoriales intrinsèques anticipent l'arrivée d'une nouvelle station de transports en commun, les implications seront bien différentes du cas où celle-ci conduit ces variables après sa construction. L'interprétation en termes ``d'effets structurants'' sera notamment très différente. Nous appliquons ici la méthode de causalités spatio-temporelles développée en~\ref{sec:causalityregimes}. Nous proposons d'étudier les relations entre différentiel d'accessibilité pour chaque projet, et variables liées au foncier (transactions immobilières) et socio-économiques, afin de voir s'il est possible de capturer un lien entre les différentiels d'accessibilité et les différentiels des variables territoriales. En effet, les liens entre nouvelles lignes et évolution du foncier sont parfois remarquables~\cite{damm1980response}.
}



\bpar{
Data for real estate transactions are provided by the BIENS database (\emph{Chambre des Notaires d'Ile de France}, proprietary database). The number of transactions that can be used after cleaning is 862360, distributed across all IRIS areas (basic census units in France), for a temporal span covering the years 2003 to 2012 included. The data at the IRIS level for population and income (median income and Gini index) come from INSEE. Network data have been vectorialized from projects maps (see figure~\ref{fig:casestudies:projects} for the different projects). Travel times are computed by public transportation only, with standard values for average speeds of different modes\footnote{That we take as the following: RER 60km.h\textsuperscript{-1}, Transilien 100km.h\textsuperscript{-1}, Metro 30km.h\textsuperscript{-1}, Tramway 20km.h\textsuperscript{-1}.}.
}{
Les données des transactions immobilières sont fournies par la base BIENS (Chambre des Notaires d'Ile de France, base propriétaire). Le nombre de transactions utilisables après nettoyage est de 862360, se répartissant sur l'ensemble des IRIS, pour une plage temporelle couvrant de 2003 à 2012 incluses. Les données par IRIS pour population et revenu (revenu médian et indice de Gini) proviennent de l'INSEE. Les données de réseau ont été vectorialisées à partir des cartes des projets (voir Fig.~\ref{fig:casestudies:projects} pour les projets). Les temps de trajets sont calculés par transport en commun uniquement, avec des valeurs standard pour les vitesses moyennes des différents modes~\cite{larroque2002paris}\footnote{Que nous prenons les suivantes : RER 60km.h\textsuperscript{-1}, Transilien 100km.h\textsuperscript{-1}, Metro 30km.h\textsuperscript{-1}, et Tramway 20km.h\textsuperscript{-1}.}.
}


\bpar{
The travel time matrix is computed from all the centroids of IRIS to all the centroids of \emph{Communes} (above aggregation level). These are linked to the network with abstract connectors to the closest station, with a speed of 50km.h\textsuperscript{-1} (travel by car). Analysis are implemented in R~\cite{R-Core-Team:2015fk} and all data, source code and results are available on an open git repository\footnote{At\\\texttt{https://github.com/JusteRaimbault/CityNetwork/tree/master/Models/SpatioTempCausality/GrandParis}. Data for the BIENS database are given only at the aggregated level of IRIS and for price and mortgage variables, for contractual reasons closing the database.}.
}{
La matrice des temps est calculée depuis l'ensemble des centroïdes des IRIS vers l'ensemble des centroïdes des communes. Ceux-ci sont reliés au réseau par des connecteurs à la gare la plus proche, de vitesse 50km.h\textsuperscript{-1} (trajet en voiture). Les analyses sont implémentées intégralement en langage R~\cite{R-Core-Team:2015fk} et l'ensemble des données, du code source et des résultats sont disponibles sur un dépôt git ouvert\footnote{A l'adresse \url{https://github.com/JusteRaimbault/CityNetwork/tree/master/Models/SpatioTempCausality/GrandParis}. Les données de la base BIENS ne peuvent être fournies pour raison de fermeture contractuelle de la base.}.
}

\begin{figure}
\includegraphics[width=\linewidth]{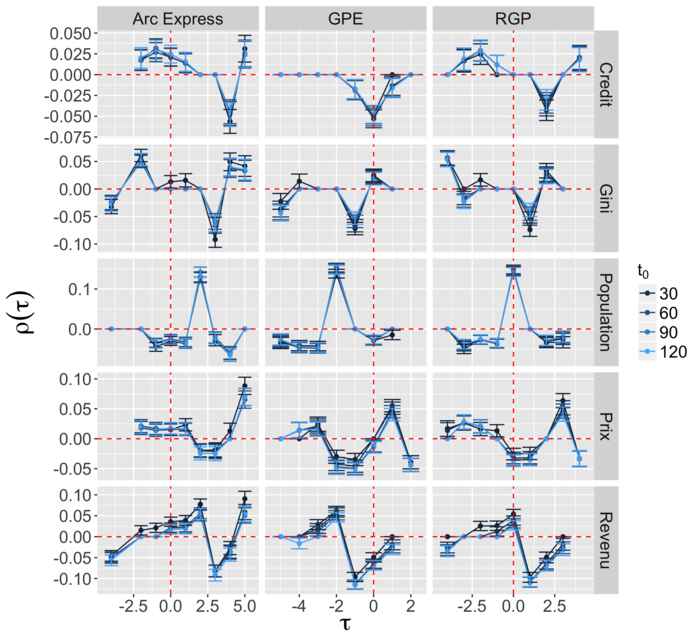}
\caption[Empirical lagged correlations between accessibility gain and territorial variables]{\textbf{Empirical lagged correlations between accessibility differential and territorial variables.} Plots show the value of the lagged correlation between differentials of accessibility $\rho(\tau)$ as a function of the lag $\tau$, in terms of average travel time $\Delta T_i$, for each project (in colunms: Arc Express, Grand Paris Express (GPE), Réseau du Grand Paris (RGP)) and the differential of the different socio-economic and real estate variables $\Delta Y_i$ (in rows: values of real estate mortgages (Credit), Average price of real estate transactions (Price), Median income (Income), Gini index for incomes (Gini), Population), for different values of the decay parameter $t_0$. Error bars give the 95\% confidence interval. Dotted red lines are a reading guide: they allow horizontally to check if correlations are significant, and vertically to check the value of the optimal lag. For example, an interpretation of the first row suggests that the older projects have caused a decrease in granted real estate mortgages in Iris which accessibility had a positive growth, and that these variables are synchronized for GPE.\label{fig:casestudies:empiricalres}}
\end{figure}


\bpar{
We compute for each project, the accessibility differentials $\Delta T_i$ in average travel time from each IRIS, in comparison with the network without the project. Average travel time accessibility is defined as $T_i = \sum_k \exp{-t_{ik}/t_0}$ with $k$ \emph{Communes}, $t_{ik}$ travel time, and $t_0$ a decay parameter. We do not weight here by the population of destination communes, on the contrary to the accessibility $Z_i$ we used previously, to ensure we do not capture any auto-correlation for population or correlations between population and the territorial variables we study. To each project is associated a date\footnote{2006 for \emph{Arc Express}, 2008 for \emph{Réseau du Grand Paris} and 2010 for \emph{Grand Paris Express}}, corresponding roughly to the mature announcement of the project, what remains a bit arbitrary as it is difficult on the one hand to determine precisely as a planning project does not emerge from nothing in one day, and on the other hand it may correspond to different realities of learning about the project by economic agents (we do therefore the limiting but necessary assumption of a diffusion of information for the majority of agents in a time smaller than a year).
}{
Nous calculons pour chaque projet, le différentiel $\Delta T_i$ d'accessibilité temporelle de trajet à partir de chaque IRIS en comparaison à celui dans le réseau sans le projet, où accessibilité temporelle est définie par $T_i = \sum_j \exp{-t_{ij}/t_0}$ avec $j$ communes, $t_{ij}$ temps de trajet, et $t_0$ paramètre d'atténuation. Nous ne pondérons pas ici par la population des communes de destination contrairement à l'accessibilité $Z_i$ utilisée précédemment, pour être certain de ne pas capturer d'auto-corrélation pour la population ou de corrélations entre population et variables territoriales que nous étudions. À chaque projet est associée une date\footnote{2006 pour Arc Express, 2008 pour le Réseau du Grand Paris, 2010 pour le Grand Paris Express.}, correspondant environ à l'année d'annonce mature du projet, restant toutefois arbitraire car difficile d'une part à déterminer précisément, un projet n'émergeant pas d'un coup du jour au lendemain, et d'autre part pouvant correspondre à des réalités différentes d'apprentissage du projet par les différents agents économiques (nous faisons donc l'hypothèse réductrice mais nécessaire d'une diffusion sur la majorité des agents dans un temps inférieur à l'année).
}

\bpar{
The link between accessibility differentials and variations of territorial variables is done through the study of lagged correlations. This method will be developed in details in~\ref{sec:causalityregimes}, but we do not need to enter into technical details here. The idea is the following: if two variables exhibit a strong correlation at a given temporal lag, there is a weak notion of causality, and the variation of the upstream variable may be at the origin of the ones of the variable which is not lagged in time (we use the term weak, since it is of course always possible that correlations are spurious).
}{
Le lien entre différentiels d'accessibilité et variations des variables territoriales est effectué par l'étude des corrélations retardées. Cette méthode sera développée en détails en~\ref{sec:causalityregimes}, mais nous n'avons pas besoin d'entrer dans les détails techniques ici. L'idée est la suivante : si deux variables présentent une forte corrélation avec un certain retard temporel, il y a une notion faible de causalité, les variations de la variable précurseur pouvant être à l'origine de celles de la variable non-décalée dans le temps (on dit faible, car il est toujours possible que les corrélations soient fortuites bien sûr).
}

\bpar{
We study the lagged correlations of $\Delta T_i$ with the variations $\Delta Y_{ij}$ of the following socio-economic variables: population, median income, Gini index for income, average price of real estate transactions and average value of real estate mortgages. Correlation is estimated by lagging accessibility, i.e. by estimating $\rho\left[\Delta T_i(t-\tau),\Delta Y_{i}(t)\right]$. A Fisher test is done for each estimation and the value is set to 0 if it is not significant ($p<0.05$ in a classical manner). The study with generalized accessibility in the sense of Hansen~\cite{hansen1959accessibility} (weighted by populations at destination, or with populations at the origin and employments at destination) has also been conducted but is less interesting as it has a very low sensitivity to the mobility component (network and decay) compared to the variables themselves. It informs therefore only on relations between these and is not presented here.
}{
Nous étudions les corrélations retardées de $\Delta T_i$ avec les variations $\Delta Y_{i}$ des variables socio-économiques suivantes : population, revenu médian, indice de Gini des revenus, prix moyen des transactions immobilières et montant moyen des crédits immobiliers. La corrélation est estimée en retardant l'accessibilité, c'est-à-dire en estimant $\rho\left[\Delta T_i(t-\tau),\Delta Y_{i}(t)\right]$. Un test de Fisher est effectué pour chaque estimation, et la valeur est fixée nulle si celui-ci n'est pas significatif ($p<0.05$ de manière classique). L'étude avec accessibilité généralisée au sens de Hansen~\cite{hansen1959accessibility} (pondérée par les populations à la destination, ou les populations à l'origine et les emplois à la destination) a également été menée mais moins intéressante car très peu sensible à la composante mobilité (réseau et atténuation) par rapport aux variables elle-mêmes, informe uniquement sur des relations entre celles-ci et n'est donc pas présentée ici.
}

\bpar{
We show in figure~\ref{fig:casestudies:empiricalres} the results for all networks and variables. The interpretation can be done the following way: for a variable and a given project, the curve $\rho(\tau)$ can exhibit maxima for a value $\tau_m > 0$ or $\tau_m <0$. This maximal correlation corresponds to a lag giving a ``maximal synchronization'' between the two variables, and the sign of the lag gives the sense of causality between the two variables.
}{
Nous présentons en Fig.~\ref{fig:casestudies:empiricalres} les résultats pour l'ensemble des réseaux et variables. La lecture s'effectue de la façon suivante : pour une variable et un projet donnés, la courbe $\rho(\tau)$ peut présenter des maxima pour une valeur $\tau_m >0$ ou $\tau_m<0$. Cette corrélation maximale correspond à un retard donnant une ``synchronisation maximale'' entre les deux variables, et le signe du retard donne le sens de la causalité entre les deux variables.
}

\bpar{
It is first remarkable to note the presence of significant effects (in the sens of significant correlations and a 95\% confidence interval which does not contains 0) for all variables. Lower values for the parameter $t_0$ give correlations higher in absolute value, unveiling a possible higher importance of local accessibility on territorial dynamics. The behavior of population shows a clearly detached peak corresponding to 2008, what suggests an impact of the older project \emph{Arc Express} on population growth. Under this assumption, the effect of other projects would then be spurious from their proximity in the most important branches. It would imply that areas where they are fundamentally different such as \emph{Plateau de Saclay} are less sensitive to transportation projects, what would confirm the artificial planned aspect of the development of this territory.
}{
Il est remarquable tout d'abord de noter l'existence d'effets significatifs (au sens de corrélations significatives et d'un intervalle de confiance à 95\% ne contenant pas 0) pour l'ensemble des variables. Des valeurs plus basses du paramètre $t_0$ donnent des corrélations plus fortes en valeur absolue, révélant une possible plus grande importance de l'accessibilité locale sur les dynamiques territoriales. Le comportement de la population montre un pic très détaché correspondant à 2008, laissant supposer un impact du plus vieux projet d'Arc Express sur la croissance de la population. Sous cette hypothèse, l'effet des autres projets serait alors fallacieux de par leur proximité dans les grands tronçons. Cela impliquerait d'ailleurs que les zones où ils diffèrent fondamentalement comme le Plateau de Saclay ne soient que très peu sensibles au projet de transport, confirmant l'aspect artificiel planifié du développement de ce territoire.
}

\bpar{
Concerning income, we observe a similar behavior but in a negative way, what would imply a decrease of wealth linked to the increase of accessibility, however accompanied by a decrease of inequalities since the Gini coefficient also presents a negative correlation in positive lags. Finally, real estate prices are as expected driven by the potential arrival of new networks, suggesting a temporal speculation bubble. We demonstrate thus the existence of complex lagged correlation links, that we call causalities in this sense, between territorial dynamics and anticipated dynamics of networks. A finer understanding of implied processes is beyond the scope of this preliminary study and would imply for example qualitative fieldwork or targeted case studies.
}{
Concernant les revenus, on observe un comportement similaire mais négatif, ce qui impliquerait un appauvrissement lié à l'augmentation de l'accessibilité, mais qui semble toutefois s'accompagner d'une baisse des inégalités puisque le coefficient de Gini présente également une corrélation négative dans les retards positifs. Enfin, comme attendu les prix immobiliers sont tirés par l'arrivée potentielle des nouveaux réseaux, effet qui disparait à deux ans pour le Grand Paris Express, suggérant une bulle immobilière passagère dans les quartiers autour des gares. Nous démontrons ainsi l'existence de liens de correlations retardées complexes qu'on nomme causalités en ce sens, entre dynamiques territoriales et dynamiques anticipées des réseaux. Une compréhension plus fine des processus à l'oeuvre est au delà de la portée de cet étude préliminaire, car supposerait par exemple des études de terrain qualitatives ou des études de cas ciblées.
}

\bpar{
This study suggests potential effects of the modification of accessibility due to Greater Paris projects, since some effects that were revealed can be linked to planning policing that also anticipate the new network. We thus suggest an effective existence of processes implying an effect of the network on territories, since most optimal lags are positive.
}{
Cette étude suggère des effets potentiels de la modification d'accessibilité due au projets du Grand Paris, puisque certains effets révélés peuvent être liés à des politiques d'aménagement anticipant également le nouveau réseau. On suggère ainsi une existence réelle des processus d'effet du réseau sur les territoires, puisque la majorité des retards optimaux sont positifs.
}


\subsection{Pearl River Delta}{Le Delta de la Rivière des Perles}

\bpar{
We now switch the geographical region, the urban structure, and the time period, in order to describe an other relevant case study in China. The extended Parisian region can be read as a consistent entity\footnote{\cite{gilli2005bassin} recalls the importance of the hinterland of Bassin Parisien and the importance of not considerating the hypercenter in an isolated way, and thus considerate the MCR which includes a certain number of important urban centers at one hour of Paris: Chartres, Orléans, Rouen, Reims and Lille thanks to High Speed Lines.}: it would be a \emph{mega-city region}, concept that we will now define and develop for the particular instance of Pearl River Delta.
}{
Nous changeons à présent de région géographique, de structure urbaine, de période temporelle, pour évoquer un autre cas d'étude pertinent en Chine. La région parisienne étendue peut être lue comme un ensemble cohérent\footnote{\cite{gilli2005bassin} rappelle l'importance de l'hinterland du Bassin Parisien et l'importance de ne pas considérer l'hypercentre de manière isolée, et considérer ainsi la MCR qui inclut un certain nombre de centres urbains importants à une heure de Paris : Chartres, Orléans, Rouen, Reims et Lille grâce à la grande vitesse.} : il s'agirait d'une \emph{méga-région urbaine}, concept que nous allons à présent définir et développer pour l'instance particulière du Delta de la Rivière des Perles.
}


\subsubsection{New urban regimes and mega-city regions}{Nouveaux régimes urbains et méga-régions urbaines}

\bpar{
The notion of megalopolis has been introduced by~\cite{gottmann1961megalopolis} to designate the emergence of urban agglomerates at a scale that did not exist before. It is at the origin of the concept of \emph{Mega-city Region} (MCR) which was consecrated by~\cite{hall2006polycentric}. For the European case, they unveil assemblies of metropolis that are strongly connected regarding mobility flows, connections between companies, which form what they call polycentric \emph{Mega-city Regions} (for example Randstad in Netherlands, the Rhine-Rhur region in Germany). Their characteristics are a certain geographical proximity of centers, a strong integration through flows, and a certain level of polycentrism. It consists in an urban form that did not exist before, which emergence seems linked to globalization processes.
}{
La notion de megalopolis a été introduite par~\cite{gottmann1961megalopolis} pour désigner l'émergence d'agglomérats urbains à une échelle non-existante auparavant. Elle est à l'origine du concept de \emph{Mega-city Region} (MCR) consacré par~\cite{hall2006polycentric}. Sur le cas Européen, ils dégagent des ensembles de métropoles fortement connectées par rapport aux flux de mobilité, aux connections entre entreprises, qui forment ce qu'il appellent des \emph{Mega-city Regions} polycentriques (par exemple la Randstad aux Pays-bas, la région Rhin-Ruhr en Allemagne). Les caractéristiques sont une certaine proximité géographique des centres, une forte intégration par les flux, et un certain niveau de polycentrisme. Il s'agit d'une forme urbaine inédite par le passé, dont l'émergence semble liée aux processus de globalisation.
}


\bpar{
This concept is even more relevant with the recent emergence of new types of urbanization, in particular through the accelerated urbanization in countries with a strong economic growth and undergoing a very rapid mutation such as China~\cite{swerts2015megacities}.
}{
Ce concept est toujours plus d'actualité avec l'apparition récente de nouveaux types d'urbanisation, notamment par l'urbanisation accélérée dans des pays à forte croissance économique et en mutation très rapide comme la Chine~\cite{swerts2015megacities}.
}

\bpar{
The second case that we develop here enters this category: Pearl River Delta (PRD) is one of the classical illustrations of the structure of a strongly polycentric MCR. Historically initially only composed by Guangzhou, the development of Hong-Kong and the establishment of Special Economic Zones (SEZ) in the context of opening policies by \noun{Deng Xiaoping}, lead to an extremly rapid development of Shenzhen, and in a less proportion of Zhuhai\footnote{Shenzhen and Zhuhai were among the first Special Economic Zones, created in 1979 to attract foreign investments in these areas with flexible economic rules. The development model of Zhuhai was different of Shenzhen, since heavy industry was forbidden.}. Guangdong province in which PRD is fully located has currently the highest regional GDP within China, and the MCR contains a population of around 60 millions (estimations strongly fluctuate depending on the definition of the MCR which is taken, and the inclusion of the floating population). The phenomenon of migrations from rural areas is highly present in the region and a city such as Dongguan has for example based its economy on factories employing these migrant workers.
}{
Le second cas que nous développons ici rentre dans cette catégorie : le Delta de la Rivière des Perles (PRD) est une des illustrations classiques de la structure d'une MCR fortement polycentrique. Historiquement initialement composé de Guangzhou uniquement, le développement de Hong-Kong puis la mise en place des Zones Économiques Spéciales (ZES) dans le cadre des politiques d'ouverture de \noun{Deng Xioaping}, a conduit à un développement extrêmement rapide de Shenzhen, et dans une moindre mesure de Zhuhai\footnote{Shenzhen et Zhuhai ont été parmi les premières Zones Économiques Spéciales, instaurées en 1979 pour attirer les investissements étrangers dans ces zones aux règles économiques flexibles. Le modèle de développement de Zhuhai a été différent de celui de Shenzhen, puisque l'industrie lourde y était interdite.}. La province du Guangdong dans lequel le PRD se situe intégralement a actuellement le plus fort PIB régional de Chine, et la MCR regroupe une population d'environ 60 millions (les estimations fluctuant fortement selon la définition prise de la MCR et la prise en compte de la population flottante). Le phénomène de migration des campagnes est très présent dans la région et une ville comme Dongguan a par exemple basé son économie sur des manufactures employant ces travailleurs migrants.
}

\subsubsection{Governance of the mega-city region}{Gouvernance de la méga-région urbaine}

\bpar{
\cite{Ye2014200} analyzes the actions of metropolitan governance at the scale of centers of the MCR, and more particularly how municipalities of Guangzhou and Foshan have progressively increased their cooperation to form an integrated metropolitan area, what can thus strongly influence the development of transportation for example and allowing the construction of a connected network. A strong tension between bottom-up processes, and a state control which is relatively strong in China, which originates from the Central State, to the province government and local government, has allowed the emergence of such a structure. The competition with other cities in the MCR remains strong, and the logic of integration (in the sense of articulation between the different centers, of interactions and of flows between these) of the MCR is only partly guided by the region. The particular nature of SEZ of Shenzhen and Zhuhai, linked to the privileged relations with the Special Administrative Zones of Hong-Kong and Macao, which have returned to the Popular Republic only at the end of the last millenium and keep a certain level of independence in terms of governance, complicates even more the relations between actors within the region. The issue of a correspondence between some levels of governance and urban processes is a tricky one: \cite{liao2017ouverture} interprets the progressive transfers of economic initiatives from the central power to local authorities as a form of a multi-level governance.
}{
\cite{Ye2014200} analyse les actions de gouvernance métropolitaine à l'échelle des centres de la MCR, et plus particulièrement comment les communes de Guangzhou et Foshan ont progressivement accru leur coopération pour former une zone métropolitaine intégrée, pouvant ainsi fortement influencer le développement des transports par exemple et permettant la mise en place d'un réseau connecté. Une forte tension entre des processus émergents par le bas, et un dirigisme d'état relativement fort en Chine, se répercutant de l'État central, au gouvernement provincial jusqu'aux gouvernements locaux, a permis la mise en place d'une telle structure. La compétition avec les autres villes de la MCR reste très forte, et la logique d'intégration (au sens d'articulation entre les différents centres, d'interactions et de flux entre ceux-ci) de la MCR est seulement partiellement guidée par la région. La nature particulière des ZES de Shenzhen et Zhuhai, liée aux relations privilégiées avec les Zones Administratives Spéciales de Hong-Kong et Macao, qui n'ont été réintégrées à la République Populaire qu'à la fin du millénaire et conservent un certain niveau d'indépendance en termes de gouvernance, complique encore les jeux d'acteurs au sein de la région. La question de la correspondance entre certains niveaux de gouvernance et des processus urbains est épineuse : \cite{liao2017ouverture} interprète les transferts progressifs des initiatives économiques du pouvoir central vers les autorités locales comme une forme de gouvernance multi-niveaux.
}



\subsubsection{Transportation Governance}{Gouvernance des transports}

\bpar{
In the frame of transportation within the MCR, there is no specific authority at this scale for the organization of transportation (but indeed entities at the level of the State, of the province and of municipalities), and each municipality manages independently the local network, whereas the connections between cities are ensured by the national train network. This leads to particular situations in which some areas have a very low accessibility, with a very strong heterogeneity locally. Therefore, the southern part of the city of Guangzhou which constitutes a direct access to the sea, is geographically closer to the center of Zhongshan, but a direct link by public transport is difficult to imagine, whereas the area is well linked to the center of Guangzhou by the metro line. A similar situation can be observed at the terminus of line 11 in Shenzhen, for the neighbor district of Dongguan, the latest having a very low accessibility by public transport\footnote{See the map~\ref{fig:casestudies:prd} for locations, the map giving also the accessibility with the road network.}. This situation could however be transitory, given the infrastructures already being built and the ones planned on a longer term: the Shenzhen metro, which covers today 285km, is planned to reach 30 lines and a length of around 1100km\footnote{For comparison, the Transilien network have a length around 1300km with RER lines included, what could make them comparable, but one must keep in mind that Ile-de-France has a surface of 12000km$^2$ against 2000km$^2$ for Shenzhen. This implies for Shenzhen a much higher transport density, corresponding to high urban density areas, such that the plan anticipates 70\% of commuting by metro at the horizon 2030.} in 2030 as declared by the official plan of the city~\cite{shenzhen2016plan}. It is clear that these developments mostly follow an existing urban development, a crucial issue is the voluntarism and the capacity to contain urban sprawl and to structure future developments around this new network, in the spirit of a voluntary integration between urbanism and transportation of the type \emph{Transit Oriented Development} that we introduced before. Different final stations will be connected to the Dongguan metro, and new intercity lines will structure the longer range mobility, what will make the Delta a relatively well integrated in terms of public transport in a close temporal horizon. To have an idea of the development of the network in the coming years, the Table~\ref{tab:casestudies:stats} gives the size of the planed networks in the different cities for 2030.
}{
Dans le cadre des transports pour la MCR, il n'existe pas d'autorité spécifique à cette échelle pour l'organisation des transports (mais bien des entités au niveau de l'État, de la province et des communes), et chaque commune gère indépendamment le réseau local, tandis que les connections entre villes sont assurées par le réseau de train national. Cela conduit à des situations particulières dans lesquelles des zones se retrouveront très enclavées, avec une hétérogénéité très forte localement. Ainsi, la pointe sud de la ville de Guangzhou qui sert d'accès direct à la mer, est plus proche géographiquement du centre de Zhongshan, mais un lien direct par transports en commun est difficile à envisager, alors que la zone est bien reliée au centre de Guangzhou par la ligne de métro. Une situation similaire s'observe au terminus de la ligne 11 de Shenzhen, pour le quartier limitrophe de Dongguan, ce dernier étant très peu accessible en transports en commun\footnote{Voir la carte~\ref{fig:casestudies:prd} pour les localisations, celle-ci donnant par ailleurs les accessibilités par réseau routier.}. Cette situation serait cependant transitoire, étant donné les infrastructures déjà en construction et celles planifiées sur un plus long terme : le métro de Shenzhen, qui couvre aujourd'hui 285km, est planifié pour atteindre jusqu'à 30 lignes et une longueur d'environ 1100km\footnote{A titre de comparaison, le réseau Transilien a une longueur avoisinant les 1300km en incluant les lignes RER, ce qui pourrait les rendre comparable, mais il faut garder à l'esprit que l'Ile-de-France a une surface de 12000km$^2$ contre 2000km$^2$ pour Shenzhen. Cela implique pour Shenzhen une densité de desserte bien plus haute, correspondant aux zones de fortes densité urbaine, si bien que le plan prévoit 70\% de transit par métro à l'horizon 2030.} en 2030 comme déclaré par le plan officiel de la ville~\cite{shenzhen2016plan}. Il est clair que ces développements suivent pour la majorité un développement urbain existant, une question cruciale est la volonté et la capacité à contenir l'étalement urbain et structurer les futurs développements autour de ce nouveau réseau, dans l'esprit d'une intégration volontaire entre urbanisme et transport de type \emph{Transit Oriented Development} que nous avons introduit précédemment. Différents terminus seront connectés au metro de Dongguan, et de nouvelles lignes intercités structureront les déplacements de plus longue portée, ce qui fera du Delta dans un horizon temporel proche une MCR relativement bien intégrée en termes de transports en communs. Pour se donner une idée du développement du réseau dans les années à venir, la Table~\ref{tab:casestudies:stats} donne la taille des réseaux planifiés dans les différentes villes d'ici 2030.
}

\begin{table}
\caption[Public transportation in Pearl River Delta]{\textbf{Public transportation in Pearl River Delta.} We give populations in 2010 taken from \cite{yearbook2013guangdong}. Network lengths are taken from the different planning documents for the Guangzhou metro~\cite{guangzhou2016metro}, the Shenzhen metro~\cite{shenzhen2016plan} and the Dongguan metro~\cite{dongguan2017ditie}, and for the Zhuhai tramway~\cite{zhuhai2016tram}. Zhongshan is not included since it exploits a BRT system but no heavy infrastructure.\label{tab:casestudies:stats}}
\begin{tabular}{|c|c|c|c|}\hline
\bpar{Ville & Population & Réseau 2016 & Réseau 2030}{City & Population & Network 2016 & Network 2030} \\\hline
	Guangzhou - Foshan & 18.9 Mio & 390km & 800km \\\hline
	Shenzhen & 10.4Mio & 286km & 1124km \\\hline
	Dongguan & 8.2Mio & 38km & 195km \\\hline
	Zhuhai (Tramway) & 1.6Mio & 10km & 173km \\\hline
\end{tabular}	
\end{table}



\subsubsection{Impact of the Zhuhai-Hong-Kong-Macao bridge}{Impact du pont Zhuhai-Hong-Kong-Macao}

\begin{figure}
	\includegraphics[width=\linewidth]{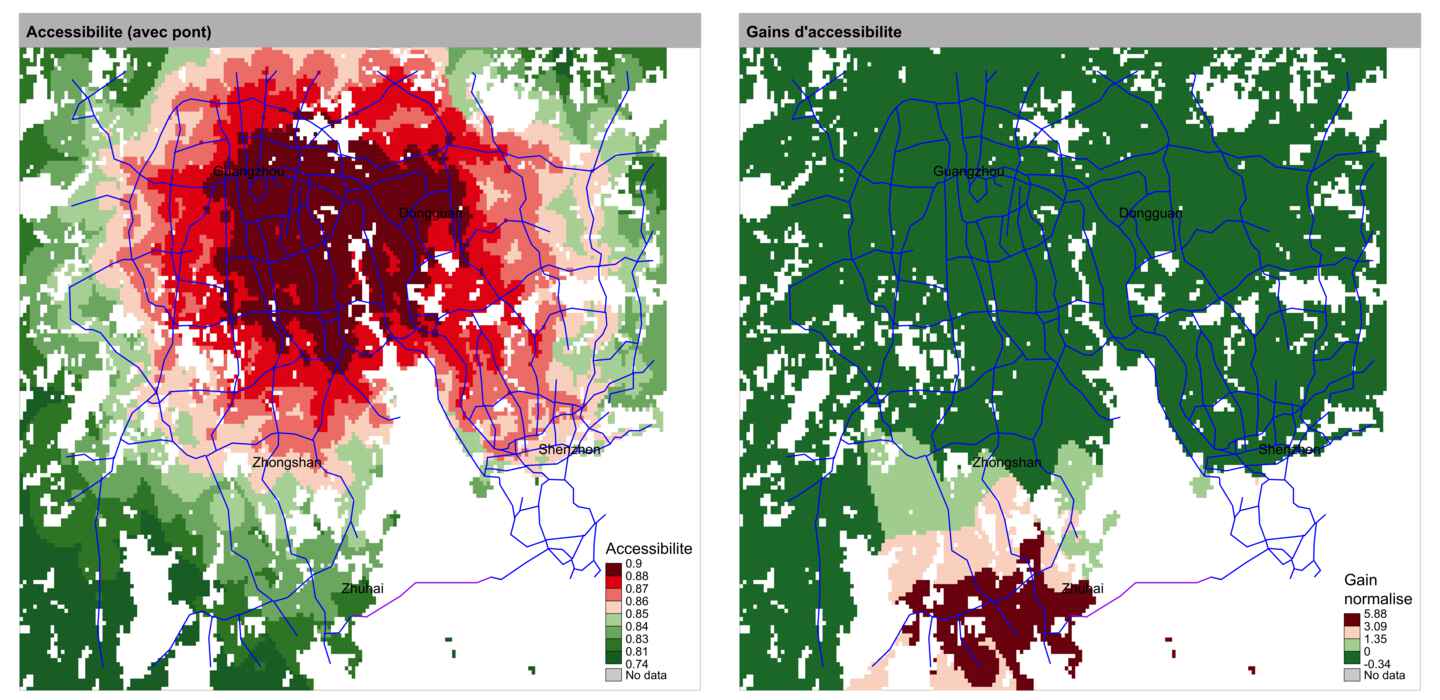}
	\caption[Accessibility gain induced by the Hong-Kong-Zhuhai-Macao gain]{\textbf{Accessibility gain induced by the HZMB in Pearl River Delta, for the territory of mainland China.} (\textit{Left}) Accessibility to population $Z_i$; (\textit{Right}) Normalized accessibility gains. The population of Hong-Kong is taken into account in destination points. The highway network (2017) is mapped in blue and and the new link of the bridge in purple.\label{fig:casestudies:prd}}
\end{figure}

\bpar{
A major transportation infrastructure project in the region is the bridge-tunnel closing the mouth of the Delta, linking Zhuhai and Macao to Hong-Kong (HZMB). The length of the crossing is 36.5km, what makes it an exceptional infrastructure~\cite{hussain2011hong}. The opening to traffic was delayed of several years and is finally planned for 2018\footnote{See the official website at \url{http://www.hzmb.org/cn/default.asp}.}. \cite{zhou2016medium} shows that the expected changes in accessibility patterns for the West of the Delta are relatively strong, and these can potentially induce strong bifurcations in the trajectories of cities. The necessity of the project is advocated by the different stakeholders of the project (Guangdong province, Hong-Kong Special Administrative Region, Macao Special Administrative Region) using an argumentation of a strong economic benefit in the frame of opening policies, and also through a social benefit for the West in particular. For example, Zhuhai is positioned as a new pivot between Hong-Kong and the West. The balancing of accessibility, in the sense of a diminution of spatial accessibility inequalities, operates however only for the private car transportation mode, what conducts to question its potential impacts: on the one hand the access to automotive remains reserved to a part of the population only, on the other hand the negative impacts of congestion can rapidly moderate the accessibility gains. These accessibility gains are mapped following the same method as previously, and shown with accessibility $Z_i$ itself in Fig.~\ref{fig:casestudies:prd}.
}{
Un projet majeur d'infrastructure de transport dans la région est le pont-tunnel fermant l'embouchure du Delta, reliant Zhuhai et Macao à Hong-Kong (HZMB). La longueur de la traversée est de 36.5km, ce qui en fait un ouvrage d'art exceptionnel~\cite{hussain2011hong}. L'ouverture au traffic a été retardée de plusieurs années et est prévue finalement en 2018\footnote{Voir le site officiel à \url{http://www.hzmb.org/cn/default.asp}.}. \cite{zhou2016medium} montre que les changements de motifs d'accessibilité attendus pour l'Ouest du Delta sont relativement forts, et ceux-ci peuvent potentiellement induire de fortes bifurcations dans les trajectoires des villes. La nécessité du projet est défendue par les différents porteurs du projet (Province du Guangdong, Région Administrative Spéciale de Hong-Kong, Région Administrative Spéciale de Macao) par des arguments de fort bénéfice économique dans le cadre des politiques d'ouverture, ainsi que par un bénéfice social pour l'Ouest notamment. Par exemple, Zhuhai se positionne comme un nouveau pivot entre Hong-Kong et l'ouest. L'équilibrage d'accessibilité, au sens de la diminution des inégalités spatiales d'accessibilité, s'opère cependant pour le mode routier uniquement, ce qui conduit à questionner ses impacts potentiels : d'une part l'accès à l'automobile reste réservé à une partie de la population seulement, d'autre part les effets négatifs de la congestion peuvent rapidement modérer les gains d'accessibilité. Ces gains d'accessibilité sont cartographiés suivant la même méthode que précédemment, et montrés avec l'accessibilité $Z_i$ elle-même en Fig.~\ref{fig:casestudies:prd}.
}

\bpar{
The medium and long term impacts of the bridge are difficult to estimate. \cite{wu2012impact} finds patterns similar to the ones we estimate, i.e. a significant benefit for Zhuhai (and Hong-Kong that we did not take into account), and also immediate effects of traffic modification and economic impacts due to the toll or the increase of tourism. They mostly postulate the position of Zhuhai-Macao as a new pivot in the region. Even if it can be directly verified in terms of centrality and accessibility, it is not evident that this new position will influence particularly the socio-economic trajectory of Zhuhai. An increased particular political accompaniment implying an increased collaboration between Hong-Kong, Zhuhai and Macao will be important~\cite{zhou2016medium}. Immediate economic effects are expected, as an increase of Zhuhai residents working in Hong-Kong (Zhuhai inhabitants are the only ones in the region to benefit of a special card allowing them to regularly visit the Special Administrative Areas\footnote{Source: fieldwork on 06/11/2016 with C. Losavio (see~\ref{app:sec:qualitative}).}), but cases showing the contrary, such as investments from Hong-Kong towards the West of the Delta, have no reason to be systematic: the first case extends the already existing dynamic with Macao, the second is mostly to be constructed. Thus, this example is a typical case of our general problematic.
}{
Les impacts à moyen et long terme du pont sont ainsi difficiles à estimer. \cite{wu2012impact} trouve des motifs similaires à ceux que nous estimons, c'est-à-dire un bénéfice significatif pour Zhuhai (et Hong-Kong que nous n'avons pas pris en compte), ainsi que des effets immédiats de modification de traffic et des impacts économiques liés au péage ou à l'accroissement du tourisme. Ils postulent surtout la position de Zhuhai-Macao comme un nouveau pivot dans la région. Si cela est vérifiable immédiatement en termes de centralité et d'accessibilité, il n'est pas dit que cette nouvelle position influence particulièrement la trajectoire socio-économique de Zhuhai. Un accompagnement politique particulier passant par une collaboration accrue entre Hong-Kong, Zhuhai et Macao sera importante~\cite{zhou2016medium}. Des effets économiques immédiats sont attendus, comme une augmentation des résidents de Zhuhai travaillant à Hong-Kong (les habitants de Zhuhai sont les seuls de la région à bénéficier d'une carte spéciale leur permettant de se rendre régulièrement dans les Zones Administratives Spéciales\footnote{Source : sortie de terrain du 06/11/2016 avec C. Losavio (voir~\ref{app:sec:qualitative}).}), mais des cas contraires, comme des investissements de Hong-Kong vers l'ouest du Delta, n'ont pas de raison d'être systématiques : le premier cas prolonge la dynamique déjà existante avec Macao, le second est à construire en grande partie. Ainsi, cet exemple est un cas typique de notre problématique générale.
}

\subsubsection{Perspectives}{Perspectives}

\bpar{
A direction of exploration through modeling consists in considering the problem differently and to try to understand the dynamics of the metropolitan system in an integrated way, i.e. as a territorial system in our sense, in which the strong coupling between territory and network is operated through a proper ontology for governance entities. It will be the object of section~\ref{sec:lutecia}.
}{
Une piste d'exploration passant par la modélisation consiste à poser le problème différemment et de chercher comprendre la dynamique du système métropolitain de manière intégrée, c'est-à-dire comme un système territorial en notre sens, dans lequel le couplage fort entre territoire et réseau est opéré par une ontologie propre des entités de gouvernance. Celle-ci sera l'objet de la section~\ref{sec:lutecia}.
}

\bpar{
This second shorter study has allowed us to emphasis a fundamentally different governance structure, but the same idea of a considerable transportation project which deeply modifies accessibility patterns. The expectations of actors regarding territorial mutations potentially induced are comparable in the sense that a high expectation is put in the project.
}{
Cette deuxième étude plus brève nous a permis de mettre en valeur une structure de gouvernance fondamentalement différente, mais la même idée d'un projet de transport considérable modifiant profondément les motifs d'accessibilité. Les attentes des acteurs quant aux mutations territoriales potentiellement induites sont comparables au sens qu'une forte attente est mise dans le projet.
}


\subsection{Comparability of case studies}{Comparabilité des études de cas}

\bpar{
We have studied here two cases of metropolitan development and infrastructure projects in their frame. The possibility of transfer of urban models (such as TOD), in the sense of the applicability of generic frameworks to different geographical contexts, is generally difficult. The synthesis of empirical conclusions obtained from very diverse case studies is also difficult.
}{
Nous avons étudié ici deux cas de développement métropolitain et de projets d'infrastructures dans leurs cadres. La possibilité de transfert des modèles urbains (comme le TOD), au sens de l'applicabilité de cadres génériques à des contextes géographiques différents, est généralement délicate. La synthèse de conclusions empiriques issus de cas d'étude très éloignés l'est également.
}

\bpar{
The East-Asian particularity has already been shown for the economic structure, and how it can not be interpreted in a simple way by a separation of microscopic and macroscopic processes as some quick and ideologically oriented readings may have done, such as the approach of the World Bank~\cite{amsden1994isn}. The comparability of urban systems is an open question at the core of issues for the Evolutive Urban Theory. It is linked to the ergodic character of these systems: the ergodicity assumption postulates that the trajectory of a city in time captures the set of possible urban states, and also that different cities are different manifestations of the same stochastic process at different periods. In that case, an ensemble of cities would allow to understand their temporal trajectories. It is intuitively not the case, and urban systems would rather be non-ergodic~\cite{pumain2012urban}. Empirically, this non-correspondence between global statistics and individual dynamics of cities is shown for traffic data by~\cite{2017arXiv171009559D}. Thus we will have to remain cautious for the generalization of conclusions, as much as empirical as theoretical, or obtained through modeling.
}{
La particularité Est-asiatique a déjà été montrée pour la structure économique, et comment celle-ci ne peut être interprétée de manière simple par une séparation des processus microscopiques et macroscopiques comme certaines lectures rapides et idéologiquement orientée ont pu le faire, comme la vision de la Banque Mondiale~\cite{amsden1994isn}. La comparabilité de systèmes urbains est une question ouverte au centre des enjeux de la Théorie Evolutive Urbaine. Celle-ci est liée au caractère ergodique de ces systèmes : l'hypothèse d'ergodicité postule que la trajectoire d'une ville dans le temps capture l'ensemble des états urbains possibles, et ainsi que les différentes villes sont différentes manifestations du même processus stochastique à différentes périodes. Dans ce cas, un ensemble de villes permettrait de se faire une idée des trajectoires temporelles. Intuitivement ce n'est pas le cas, et les systèmes urbains seraient plutôt non-ergodiques~\cite{pumain2012urban}. Empiriquement, cette non-correspondance entre statistiques globales et dynamiques individuelles des villes est montrée pour des données de traffic par~\cite{2017arXiv171009559D}. Ainsi il s'agira de rester prudent pour la généralisation des conclusions, autant empiriques que théoriques, ou issues de la modélisation.
}




\stars

\bpar{
We have thus seen in this section, from two very different case studies, but having the common feature to exhibit significant transportation infrastructure projects, that the immediate impacts of these in terms of accessibility can be consequent, but that it is complicated to associate these gains to possible future mutations. We begin to foresee the difficulty to characterize co-evolution.
}{
Nous avons donc vu dans cette section, à partir de deux études de cas très différentes, mais ayant le point commun de présenter des projets significatifs d'infrastructures de transport, que les impacts immédiats de celles-ci en termes d'accessibilité peuvent être conséquents, mais qu'il est compliqué d'associer ces gains à de possibles mutations futures. Nous commençons à entrevoir la difficulté de caractériser la co-évolution.
}

\bpar{
We will in the next section even more diversify our examples, from fieldwork observations, and thus from a more subjective and complementary point of view.
}{
Nous allons dans la section suivante encore diversifier nos exemples, à partir d'observations de terrain, et donc selon un point de vue plus subjectif et complémentaire.
}

\stars

%


\newpage

\section{Fieldwork observations of interactions}{Observations de terrain des interactions}

\label{sec:qualitative}


\bpar{
This section proposes to illustrate the issue of interactions between transportation networks and territories, and more particularly their complexity and the diversity of possible situations already perceptible in a qualitative way (and also subjective in a second time) at the microscopic scale, through concrete fieldwork examples. The geographical subject is Pearl River Delta, in Guangdong province, that we already described before, and more particularly mostly the city of Zhuhai. The objective is to enrich our repertory with concrete situations, to understand if these can be associated to the generic processes we have already exhibited, or if others can be observed at the scales of observation.
}{
Cette section propose d'illustrer la problématique des interactions entre réseaux de transports et territoires, et plus particulièrement leur complexité et la diversité des situations possibles déjà perceptibles de manière qualitative (et également subjective dans un second temps) à l'échelle microscopique, par des exemples concrets de terrain. Le terrain géographique est le Delta de la Rivière des Perles en Chine, dans la province du Guangdong, que nous avons décrit ci-dessus, et plus particulièrement en grande partie la ville de Zhuhai. L'objectif est d'enrichir notre répertoire par des situations concrètes, de voir si celles-ci peuvent être associées aux processus génériques que nous avons déjà dégagé, ou si d'autres se manifestent aux échelles d'observation.
}

\bpar{
We assume the term of \emph{Geographical Fieldwork}, with all knowledge of epistemological debates its use can raise. Indeed, we extract observations from places that were experimented, in the context of a given problematic~\cite{retaille2010terrain}. Our approach will also highlight the role of representations, underlined as a type of fieldwork in itself by~\cite{lefort2012terrain}, when we will give a subjective view.
}{
Nous assumons le terme de \emph{Terrain géographique}, en toute conscience des débats épistémologiques que peuvent poser l'utilisation de celui-ci. En effet, nous extrayons des observations de lieux expérimentés, dans le cadre d'une problématique particulière~\cite{retaille2010terrain}. Notre démarche appuiera aussi sur le rôle des représentations, souligné comme forme à part entière de terrain par~\cite{lefort2012terrain}, lorsque nous prendrons une position subjective.
}


\bpar{
In the frame of the European project Medium\footnote{The Medium project, which establishes a partnership between European and Chinese universities, is entitled ``\textit{New pathways for sustainable urban development in China’s medium-sized cities}''. It aims at studying sustainability through an interdisciplinary and multidimensional viewpoint, in the case of rapidly growing urban areas. Three medium-sized Chinese cities were chosen as a case study. See \url{http://mediumcities-china.org/} for more information.}, aiming at an interdisciplinary approach of sustainability for Chinese cities by concentrating on medium-sized cities\footnote{The definition of medium-sized cities considered for the project is broader than the official statistical definition of the Chinese government, and covers cities from 1 to 10 millions of inhabitants.}, this city was chosen as a case study. When the source is not explicitly precised, observations come from fieldwork, for which narrative reports are available in Appendix\ref{app:sec:qualitative}. The format of narrative reports is ``on-the-fly'' following the recommendations of \cite{goffman1989fieldwork} for taking notes in an immersive fieldwork in particular, whereas the voluntary subjective position rejoins \cite{ball1990self} which recalls the importance of reflexivity in order to draw rigorous conclusions from qualitative fieldwork observations of which the researcher is a part in itself\footnote{The consideration of the researcher as a \emph{subject} in relation with its object of study does not imply in our case a feedback of the researcher on the system because of its size in the case of a transportation network at the scale of the city, and indeed a conditioning of observations by a subjectivity of which we must detach in the posterior exploitation of the observation material, but which ignoring can only increase the biases.}.
}{
Dans le cadre du projet européen Medium\footnote{Le projet Medium, qui met en partenariat des université européennes et chinoises, s'intitule ``\textit{New pathways for sustainable urban development in China’s medium-sized cities}''. Il vise à étudier la soutenabilité selon un prisme interdisciplinaire et multidimensionnel, dans le cas de zones urbaines en forte croissance. Trois villes moyennes chinoise ont été choisies comme cas d'étude. Voir \url{http://mediumcities-china.org/} pour plus d'informations.}, visant à une approche interdisciplinaire de la soutenabilité pour les villes Chinoises en se concentrant sur les villes moyennes\footnote{La définition des villes moyennes considérées par le projet est plus large que la définition statistique officielle du gouvernement Chinois, et couvre des villes de 1 à 10 millions d'habitants.}, cette ville a été choisie comme cas d'étude. Lorsque la source n'est pas explicitement précisée, les observations proviennent du travail de terrain, pour lequel des compte-rendus narratifs sont disponibles en Annexe~\ref{app:sec:qualitative}. Le format des compte-rendus narratifs est ``à-la-volée'' suivant les recommandations de \cite{goffman1989fieldwork} pour la prise de notes en terrain d'immersion notamment, tandis que la position volontairement subjective rejoint \cite{ball1990self} qui souligne l'importance de la réflexivité pour tirer des conclusions rigoureuses à partir d'observations qualitatives de terrain duquel le chercheur est partie intégrante\footnote{La considération du chercheur comme \emph{sujet} en relation avec son objet d'étude n'implique pas dans notre cas de rétroaction du chercheur sur le système vu l'ampleur de celui-ci dans le cas d'un réseau de transport à l'échelle d'une ville, mais bien un conditionnement des observations par une subjectivité dont il s'agit de se détacher dans l'exploitation postérieure du matériau d'observation, mais qu'ignorer ne peut qu'augmenter les biais.}.
}



\subsection{Development of a transportation network}{Développement d'un réseau de transport}

\bpar{
The objective of fieldwork is thus to observe the multiple facets and layers of a complex public transport system which is always transforming, its links with observable urban operations, and to what extent these witness of interaction processes between networks and territories. The spatial extent of observations spans on Zhuhai as an illustration of local transportation but also punctually on other regions in China. These observations have their proper logic in comparison to the modeling of transportation networks or data analysis, such as accessibility studies or interaction models between land-use and transportation, that will be done in the following. Indeed, these fail generally in capturing aspects at a large scale, which are often directly linked to the user, and which can become crucial regarding the effective use of the network. For example, multi-modality\footnote{Multi-modality consists in the combination of different transportation modes: road, train, metropolitan, tramway, bus, peaceful modes, etc., in a mobility pattern. A multimodal transportation system consists in the superposition of modal layers, and these can be more or less well articulated for the production of optimal routes following multiple objectives (cost, time, generalized cost, comfort, etc.) which themselves depend on the user, and of the mobility pattern.} can be in practice made efficient through the emergence of self-organized informal transportation modes, or the establishment of new modes such as bike-sharing, what solves the ``last-mile problem''~\cite{liu2012solving}, which seems to be often neglected in the planning of newly developed areas in China. On the contrary, practical details such as tickets reservation or check-in delays at boarding can considerably influence use patterns.
}{
L'objectif du travail de terrain est donc d'observer les multiples facettes et couches d'un système de transport public complexe et en mutation permanente, ses liens avec les opérations urbaines visibles, et dans quelle mesure ceux-ci témoignent de processus d'interaction entre réseaux et territoires. La portée des observations s'étend sur Zhuhai comme illustration des transports locaux mais aussi ponctuellement sur d'autres régions en Chine. Ces observations ont une logique propre en comparaison à la modélisation des réseaux de transport ou l'analyse de données, comme des études d'accessibilité ou des modèles d'interaction entre usage du sol et transport, qui seront menés par la suite. En effet, celles-ci échouent généralement à capturer des aspects à une grande échelle, souvent directement liés à l'utilisateur, qui peuvent devenir cruciaux au regard de l'utilisation effective du réseau. Par exemple, la multi-modalité\footnote{La multi-modalité consiste en la combinaison de différents modes de transports : routier, train, métropolitain, tramway, bus, modes doux, etc., dans un motif de mobilité. Un système de transport multimodal consiste en la superposition des couches modales, et celles-ci peuvent plus ou moins bien s'articuler pour la production de trajets optimaux selon de multiples objectifs (coût, temps, coût généralisé, confort, etc.) qui eux-même dépendent de l'individu, du motif de déplacement.} peut être rendue efficace en pratique par l'emergence de modes de transports auto-organisés informels, ou la mise en place de nouveaux modes comme le vélo en partage, ce qui résout le ``problème du dernier kilomètre''~\cite{liu2012solving}, qui semble être souvent négligé dans la planification de zones nouvellement développées en Chine. Au contraire, des détails pratiques comme la réservation des tickets ou les délais d'enregistrement à l'embarquement peuvent influencer considérablement les motifs d'usage.
}



\bpar{
Several trips on the Chinese territory were made to observe the concrete manifestations of the high speed network development. Since 2008, China has established the larger HSR network in the world from scratch, which has a great success and which lines are currently saturated. It answers primary demand patterns in terms of city size, showing that it was planned such that the network answers to territorial dynamics. Its high usage shows the impact of network on mobility, what is a possible precursor of territorial mutations.
}{
Différents voyages sur le territoire Chinois ont été effectués pour observer les manifestations concrètes du développement du réseau à grande vitesse. Depuis 2008, la Chine a établi le plus grand réseau de HSR du monde à partir de zéro, qui a connu un grand succès et dont les lignes sont actuellement saturées. Celui-ci répond à des motifs de demande primaires en termes de taille de ville, montrant qu'il a été planifié de telle façon que le réseau réponde à des dynamiques territoriales. Son fort usage montre l'impact du réseau sur la mobilité, possible précurseur de mutations territoriales.
}

\bpar{
To show to what extent territories can influence the development of network in diverse ways, we can take a particular example, linked to the development of tourism, which corresponds to a particular dimension taken into account in planning. Thus, the line between Guangzhou and Guiyang (North-West axis which is precursor of the future direct link Guangzhou-Chengdu) have witnessed the opening of stations specifically for the development of tourism, such as Yangshuo in Guangxi, which number of visits has then strongly increased (see maps in Appendix~\ref{app:sec:qualitative}). One year after the opening of the station, the main road link with the city is still under construction, showing that the different networks react differently to constraints at different levels. A higher number of trains stops on week-ends - more than one each hour, are are full more than two weeks in advance. New mobility patterns can be induced by this new offer, as illustrate the interview of an inhabitant of Guangzhou done in Yangshuo, which came for a short week-end with her colleagues, in the context of a ``team-building'' trip financed by her startup in information technology. These new mobility practices are shown in a second interview of an inhabitant of Beijing met at Emeishan, sent by her company in Industrial Design for a short stay in Chengdu for a training in a local subsidiary. The company prefers the high speed train, and it recently increased the mobility practices for its employees.
}{
Pour montrer dans quelle mesure les territoires peuvent affecter le développement des réseaux de manière diverse, prenons un exemple particulier, lié au développement du tourisme, qui correspond à une dimension particulière qui a été prise en compte dans la planification. Ainsi, la ligne entre Guangzhou et Guiyang (axe nord-ouest précurseur de la future liaison directe Guangzhou-Chengdu) a vu la construction de stations spécifiques au développement du tourisme, comme Yangshuo dans le Guangxi, dont la fréquentation a alors fortement augmenté (voir cartes en Annexe~\ref{app:sec:qualitative}). Un an après l'ouverture de la gare, le lien routier majeur avec la ville est toujours en construction, montrant que les différents réseaux réagissent différemment aux contraintes à différents niveaux. Un grand nombre de trains s'y arrêtent toutefois le week-end - plus d'un par heure, et sont remplis plus de deux semaines en avance. De nouveaux motifs de mobilité peuvent être induits par cette nouvelle offre, comme l'illustre l'interview d'une habitante de Guangzhou faite a Yangshuo, qui était venue pour un court week-end avec ses collègues, dans le cadre d'un voyage de ``team-building'' financé par sa startup en technologie de l'information. Ces nouvelles pratiques de mobilité sont montrées par une deuxième interview d'une habitante de Beijing rencontrée à Emeishan, envoyée par son entreprise de Design Industriel pour un court passage à Chengdu pour une formation dans une filiale locale. L'entreprise privilégie le train à grande vitesse, et celle-ci a récemment accru ses pratiques de mobilité pour ses salariés.
}



\bpar{
A similar strategy can be observed concerning the connection of touristic destinations for the line Chengdu-Emeishan. The principal objective of this line is for now to serve the highly frequented touristic destinations of Emeishan and Leshan. However, the missing link between Leshan and Guiyang is already well advanced in its construction and will complete the direct link between Guangzhou and Chengdu. This reveals diachronic and complementary dynamics of network development following properties of territories. This line is a part of the structuring skeleton of the ``8+8'' recently reformulated by the central government\footnote{It corresponds to the general plan for future high speed lines, recently actualized to include 8 North-South parallels and 8 East-West others, completing the 4+4 already realized.}, and the traversed territories expect a lot from it as shows \cite{lu2012chengdu} for the city of Yibin halfway between Chengdu and Guiyang.
}{
Une stratégie similaire peut s'observer concernant la desserte de destinations touristiques pour la ligne Chengdu-Emeishan. L'objectif principal de cette ligne est pour l'instant de desservir les destinations touristiques très fréquentées d'Emeishan et de Leshan. Cependant, le lien manquant entre Leshan et Guiyang est deja bien avancé dans sa construction et complétera le lien direct entre Guangzhou et Chengdu. Cela révèle des dynamiques diachroniques et complémentaires de développement du réseau en fonction des propriétés de territoires. Cette ligne fait partie du squelette structurant des ``8+8'' reformulées récemment par le gouvernement central\footnote{Il s'agit du plan général pour les futures lignes à grande vitesse, réactualisé récemment pour comprendre 8 parallèles nord-sud et 8 autres est-ouest, complétant les 4+4 déjà réalisées.}, et les territoires traversés en attendent beaucoup comme le montre \cite{lu2012chengdu} pour la ville de Yibin à mi-chemin entre Chengdu et Guiyang.
}

\begin{figure}
	\includegraphics[width=\linewidth]{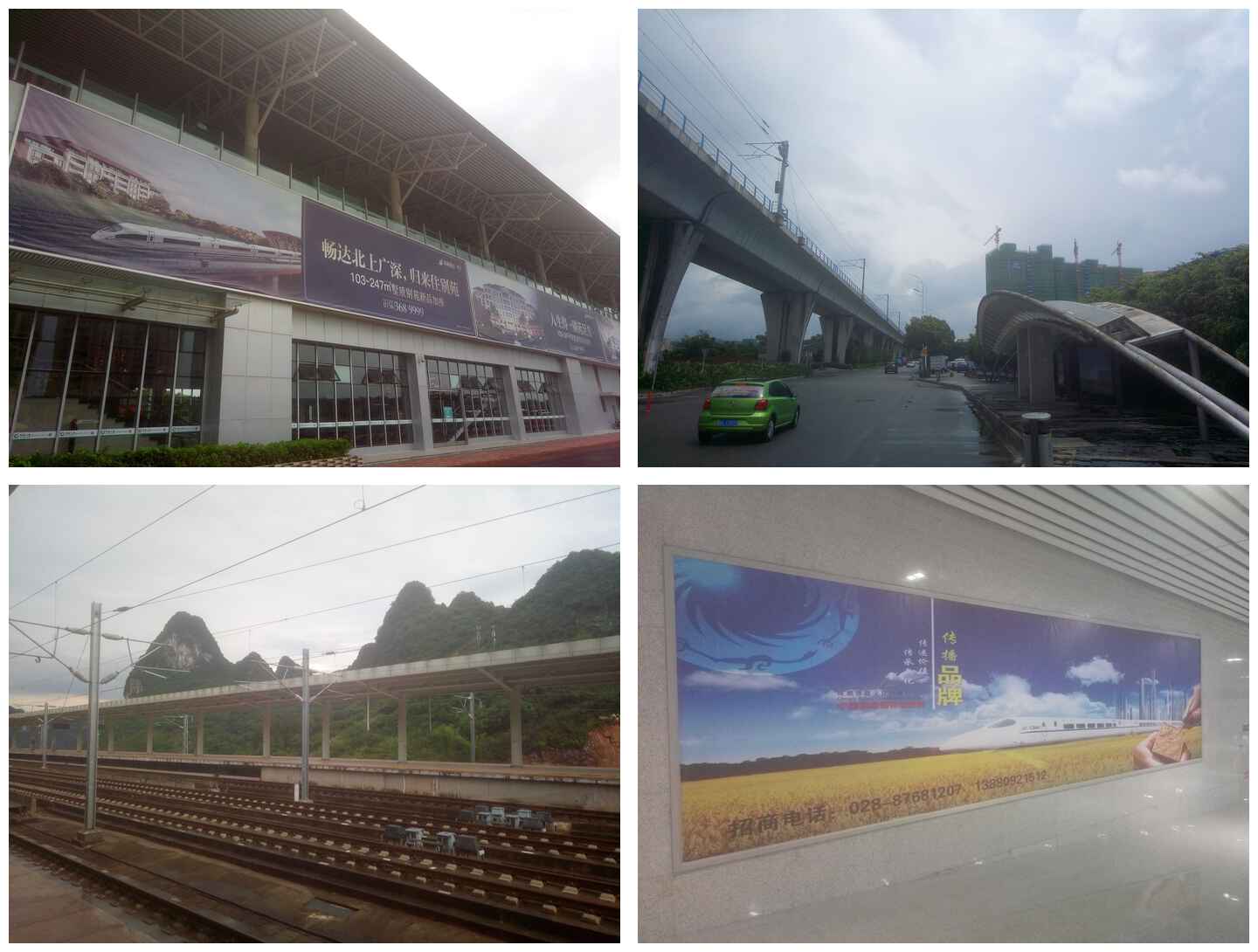}
	\caption[High speed network in China]{\textbf{Local manifestations of the mutations induced by the new high speed network.} \textit{(Top Left)} High speed station of Tangjia, in Zhuhai city. The monumental advertisement for a real estate operation praises the merits of the proximity to the network, which is also used as an argument for higher prices; \textit{(Top Right)} High speed line in Zhuhai, deserted bus stop and real estate project being realized in a difficultly accessible area: this urban fringe is in direct contact with the rural environment on the other side of the line, and eccentric from the city; \textit{(Bottom Left)} Yangshuo station on the Guangzhou-Guiyang line, which principal function is the development of this touristic destination which bases most of its economy on that field; \textit{(Bottom Right)} Advertisement for high speed in Sichuan, at the station of the international Chengdu airport on the line to Leshan and Emeishan. The train departs from the futurist city to fly over the countryside, recalling the tunnel effect of territories telescoped by high speed.\label{fig:qualitative:hsr}}
\end{figure}

\bpar{
We also observe join mutations of the railway network and of the city. We illustrate thus in Fig.~\ref{fig:qualitative:hsr} the insertion of the HSR in its territories. Direct effects of the network are linked to the development of totally new districts in the neighborhood of new stations, sometimes in an approach of type ``\emph{Transit Oriented Development}'' (TOD)\footnote{As we defined in~\ref{sec:networkterritories}, this planning paradigm aims at articulating the development of an heavy transportation infrastructure with urbanization, typically through a densification around stations.} - we will come back to it with more details. Furthermore, more subtle indirect effects are suggested by clues such as the promotion of operations through advertisement. It shows the socio-economic expectations regarding the network and the local agents which have to contribute to its success: advertisements claiming the merits of high speed, and the selling of appartements in the associated real estate operations. This dynamic seems to contribute to the construction of a ``middle class'' and of the role it has to play in the dynamism of territories~\cite{rocca2008power}\footnote{Construction which is, as \noun{Jean-Louis Rocca} emphasizes, as much concrete since it depends on objective realities, as imaginary in the academic and political discourse, which construct the object simultaneously to its study or use.}. The insertion of lines in territories seems in some case to be forced, as shows the Yangshuo station which exploits the tourism opportunity offered by the passage of the line in a low populated area but which is very attractive by its landscapes, or the new real estate operations in Zhuhai which are not very accessible because of their price.
}{
Nous observons également des mutations conjointes du réseau ferroviaire et de la ville. Nous illustrons ainsi en Fig.~\ref{fig:qualitative:hsr} l'insertion du HSR dans ses territoires. Des effets directs du réseau sont liés au développement de quartiers totalement neufs aux alentours des nouvelles gares, parfois dans une démarche de type ``\emph{Transit Oriented Development}'' (TOD)\footnote{Comme nous l'avons défini en~\ref{sec:networkterritories}, ce paradigme de planification vise à articuler développement d'une infrastructure de transport lourde et urbanisation, typiquement par une densification autour des gares.} - nous y reviendrons plus en détail. De plus, des effets indirects plus subtils sont suggérés par des indices comme la promotion des opérations par la publicité. Celle-ci montre les attentes socio-économiques envers le réseau et les agents locaux qui se doivent contribuer à son succès : les publicités vantant les mérites de la grande vitesse, et la vente d'appartements dans des opérations immobilières associées. Cette dynamique semble contribuer à la construction d'une ``classe moyenne'' et au rôle qu'elle doit jouer dans le dynamisme des territoires~\cite{rocca2008power}\footnote{Construction, comme le souligne \noun{Jean-Louis Rocca}, autant concrète car relevant de certaines réalités objectives, qu'imaginaire dans les discours universitaires et politiques, qui construisent l'objet simultanément à son étude ou son utilisation.}. L'insertion des lignes dans les territoires semble dans certains cas forcée, comme le montre la gare de Yangshuo qui exploite l'opportunité du tourisme offerte par le passage de la ligne dans une zone très peu peuplée mais très attractive par ses paysages, ou les nouvelles opérations immobilières peu accessibles par leur prix à Zhuhai.
}

\bpar{
Finally, it is important to remark the network development answers simultaneously to different types of territorial contexts. Branches of the new high speed network with a short range, such as the line Guangzhou-Zhuhai, can be seen as being at the intermediary between a long range service and a proximity regional transport, depending on the modularity of serving patterns. This line is thus placed within long range urban interactions (the service Zhuhai-Guiyang being for example ensured) and within interactions in the mega-city region, most of the service being trains to Guangzhou. To this can be added the classical train network which keeps a certain role in territorial interactions: some connections require the use of both networks and of urban transportation, such as the link between Zhuhai and Hong-Kong, experimented through terrestrial transportation modes only\footnote{Following the Hato Typhoon on 23/08/2017, maritime links with the center of Hong-Kong and the international airport has been interrupted for a significant part of the delta, and has been reopened for Zhuhai in the beginning of November 2017 only.}.
}{
Enfin, il est important de noter que le développement du réseau répond simultanément à différents types de contextes territoriaux. Des branches à courte portée du nouveau réseau à grande vitesse, comme la ligne Guangzhou-Zhuhai, peuvent être vues comme à l'intermédiaire entre un service à longue distance et un transport régional de proximité, en fonction de la modularité des motifs de desserte. Cette ligne s'inscrit ainsi dans des interactions urbaines à longue portée (le service Zhuhai-Guiyang étant par exemple assuré) et dans des interactions au sein de la méga-région urbaine, l'essentiel de la desserte étant des trains pour Guangzhou. À cela s'ajoute le réseau de train classique qui conserve un certain rôle dans les interactions territoriales : certaines connexions requièrent l'utilisation des deux réseaux et des transports urbains, comme la liaison entre Zhuhai et Hong-Kong expérimentée par voie terrestre seulement\footnote{À la suite du Typhoon Hato le 23/08/2017, les liaisons maritimes avec le centre de Hong-Kong et l'aéroport international ont été interrompues pour une grande partie du delta, et n'ont été rétablies pour Zhuhai que début novembre 2017.}.
}

\subsection{Implementing TOD: contrasted illustrations}{Implémentation du TOD : des illustrations contrastées}

\bpar{
The simultaneous development of the transportation network and the urban environment can be directly observed on the field. The local urban network and real estate development operations are planned closely with the new train network: the Zhuhai tramway, for which a single line is open at the current time and still being tested, is thought to participate in a TOD approach\footnote{See preliminary works of planning consulting, such as for example \url{https://wenku.baidu.com/view/b1526461ff00bed5b8f31d01.html} for the context of the new Xiaozhen district, in the West of Xiangzhou.} to urban development which aims at favoring the use of public transportation and a city with less cars, such as wanted for example by the Planning Committee of the \emph{High-Tech Zone} in charge of the development around Zhuhai North station. The observation of the surroundings of Tangjia station, also built in the same spirit, reveals a certain atmosphere of desertion and an unpractical organisation can lead to questioning the efficacy of the approach. This also suggests a certain self-fulfilling nature of the project, as suggested by advertisements for new real estate for sale, insisting on the importance of the presence of the railway line. A full narrative encouraging local actors and individuals to be involved around TOD seems to be used by different actors of development.
}{
Le développement simultané du réseau de transport et de l'environnement urbain est directement observable sur le terrain. Le réseau urbain local et les opérations de développement immobilières sont planifiés en étroite conjonction avec le nouveau réseau de train : le tramway de Zhuhai, pour lequel une unique ligne est aujourd'hui ouverte et en phase de test, vise à participer à une approche par TOD\footnote{Voir les travaux préliminaires de consultation pour la planification, comme par exemple \url{https://wenku.baidu.com/view/b1526461ff00bed5b8f31d01.html} pour le contexte du nouveau quartier de Xiaozhen, à l'ouest de Xiangzhou.} du développement urbain qui vise à favoriser l'utilisation des transports publics et une ville avec moins d'automobiles, comme voulu par exemple par le Comité de Planification de la \emph{High-Tech Zone} en charge du développement autour de la gare nord de Zhuhai. L'observation des alentours de la gare de Tangjia, également construite dans le même esprit, révèle une certaine atmosphère de désertion et une organisation peu pratique peut mener au questionnement de l'efficacité de l'approche. Cela suggère également une certaine nature auto-prophétique du projet, comme suggéré par les publicités pour un nouvel immobilier à vendre, appuyant sur l'importance de la présence de la ligne ferroviaire. Toute une narration incitant les acteurs locaux et les individus à s'impliquer autour du TOD semble être utilisée par différents acteurs du développement.
}

\begin{figure}
	\includegraphics[width=\linewidth]{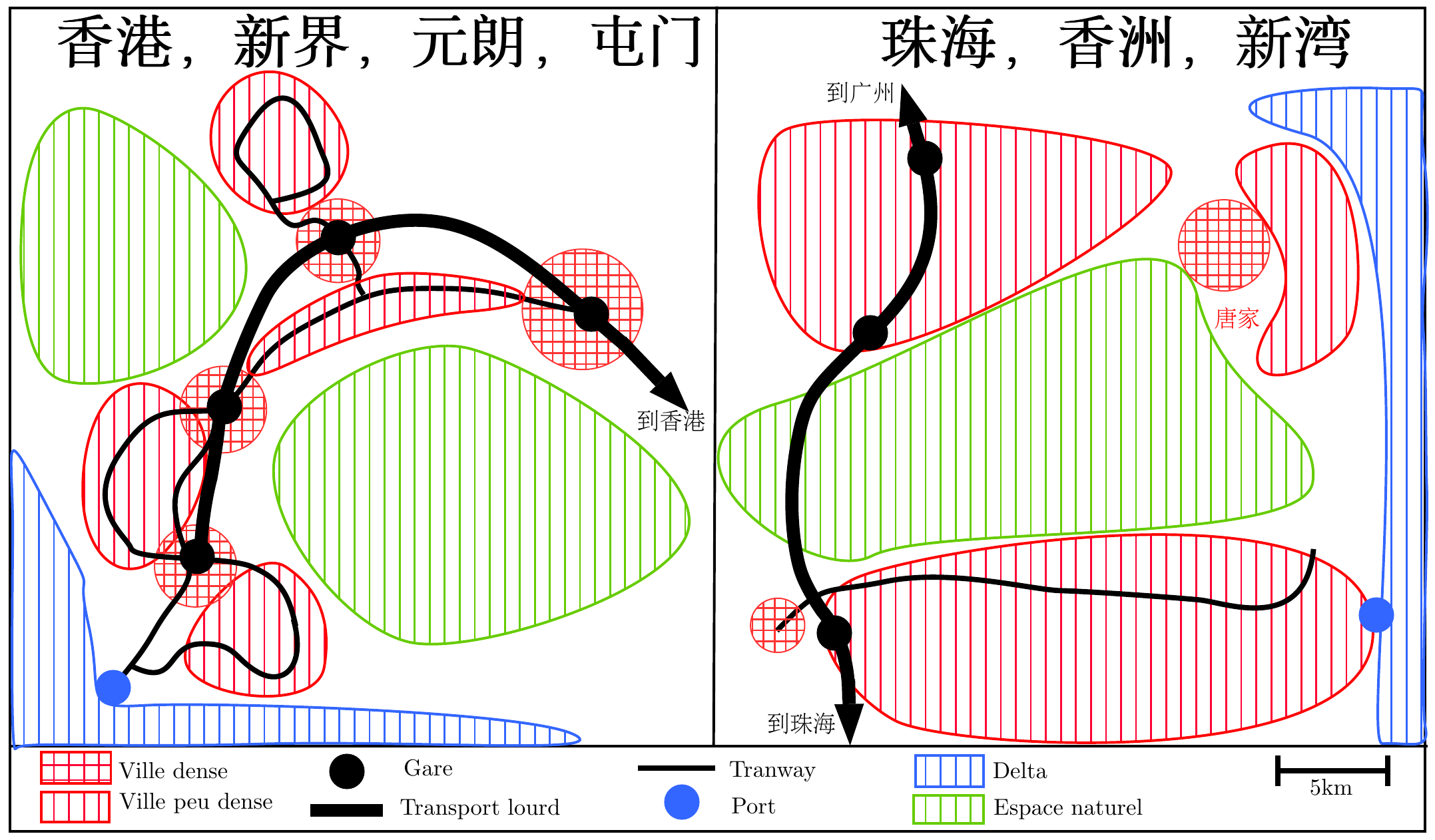}
	\caption[Transit-oriented development in Hong-Kong and Zhuhai]{\textbf{Comparative analysis of two implementations of TOD in PRD.} At a comparable scale, we synthesize the urban configuration of Yuenlong (\cn{元朗}) and Tuenmun (\cn{屯门}), Hong-Kong New Territories (\cn{香港，新界}), on the left, and of Xinwan, Xiangzhou, Zhuhai (\cn{珠海，香洲，新湾}), on the right, which contains the Zhuhai High-tech zone in its nothern part in particular. The configurations illustrate different dynamics of articulation, and shifted construction temporalities, unveiling thus different realities under the notion of TOD. A first interpretation would be that it is effective if the trajectory of the full territorial system (urban development and transportation network) is modified early in its genesis, whereas a system with a higher level of maturity will have more inertia. \textit{Trans. : } \cn{到香港} - towards Hong-Kong ; \cn{到广州} - towards Guangzhou ; \cn{到珠海} - towards Zhuhai.\label{fig:qualitative:schema}}
\end{figure}

\bpar{
Other fieldwork observations, such as in the \emph{New Territories} in Hong-Kong, witness of an efficient TOD which fulfils its objective, with a complementarity between heavy rail and local light tramway, and also a high urban density around stations. These observations recall the complexity of urban trajectories coupled to the development of the network, and that we must remain cautious before drawing any general conclusion from particular cases. We summarize in Fig.~\ref{fig:qualitative:schema} the comparison between the two TOD cases detailed above, as synthetic schemes of urban structures of each area. In Hong-Kong, urban areas have been conjointly planned with the MTR line (heavy transport) and the multiple light tramway lines~\cite{hui2005study}. The infrastructure of light rail and the organisation of missions allow to rapidly connect with the closest station, distributing a highly uniform accessibility for all districts of the territory. On the contrary in Zhuhai, the village of Tangjia is old, even anterior to the rest of Zhuhai, and has developed without any particular articulation with transportation infrastructures. The location of the tramway, which just opened, completes the trajectory of the new railway line, with an objective of reorganizing the North of Zhuhai, and in particular the High-tech Zone which extends from the North railway station (Zhuhai Bei) to Tangjia. Currently, the urban organisation is strongly imprinted with this unsynchronized development, since public transportation accessibility is still relatively low, bus lines being subject to an increasing congestion due to the strong increase in the number of cars. Furthermore, the exploitation of the tramway has been difficult, since the technology used with a third rail in the ground has been imported from Europe and had never been tested in such humidity conditions\footnote{Source: personal communication with \noun{Yinghao Li}, July 2017.}, what lead to a questioning of the network plan in its entirety. 
}{
D'autres observations de terrain, comme dans les Nouveaux Territoires (\emph{New Territories}) à Hong-Kong, témoignent d'un TOD efficace et réalisant son objectif, avec une complémentarité entre transport lourd et tramway local léger, ainsi qu'une grande densité urbaine autour des gares. Ces observations rappellent la complexité des trajectoires urbaines couplées au développement du réseau, et qu'il s'agit d'être prudent avant de tirer toute conclusion générale à partir de cas particuliers. Nous résumons en Fig.~\ref{fig:qualitative:schema} la comparaison des deux cas de TOD évoqués ci-dessus, sous forme de schéma synthétique des grandes lignes urbanistiques de chacune des zones. À Hong-Kong, les zones urbaines ont été planifiées conjointement avec la ligne du MTR (transport lourd) et les multiples lignes de tramway léger~\cite{hui2005study}. L'infrastructure du transport léger et l'organisation des missions permettent de rejoindre rapidement la gare la plus proche, distribuant une accessibilité très uniforme pour l'ensemble des quartiers du territoire. Au contraire à Zhuhai, le village de Tangjia est ancien, antérieur même à l'ensemble du reste de Zhuhai, et s'est développé sans articulation particulière avec les infrastructures de transport. Le tracé du tramway, qui vient d'ouvrir, complète le tracé de la nouvelle ligne ferroviaire, dans un but de reorganisation du nord de Zhuhai, et en particulier la High-tech Zone qui s'étend de la gare du Nord (Zhuhai Bei) à Tangjia. Actuellement, l'organisation urbaine est fortement marquée par cette mise en place déphasée, puisque l'accessibilité en transport en commun est toujours relativement faible, les lignes de bus étant sujettes à une congestion croissante due à la forte augmentation du nombre d'automobiles. Par ailleurs, la mise en place du Tramway a été laborieuse, de par l'utilisation d'une technologie par troisième rail au sol importée d'Europe, et qui n'avait jamais été testée dans de telles conditions d'humidité\footnote{Source : communication personnelle avec \noun{Yinghao Li}, juillet 2017.}, ce qui a poussé à une remise en question du plan du réseau dans son ensemble.
}


\bpar{
This fieldwork example thus shows us that (i) under the same designation very different processes exist, and are extremely dependant to geographical, political and economical particularities; and that (ii) the development of a territory which is functional in terms of accessibility necessitates a fine articulation which seems to be the outcome of an integrated planning approach on the long time.
}{
Cet exemple de terrain nous démontre ainsi que (i) sous la même qualification existent des processus très différents, extrêmement dependants aux particularités géographiques, politiques, économiques ; et que (ii) la mise en place d'un territoire fonctionnel en termes d'accessibilité nécessite une articulation fine qui semble résulter d'une approche de planification intégrée réalisée sur le temps long.
}

\subsection{An experiment in floating observation}{Une expérience en observation flottante}

\bpar{
We finally propose to sketch a qualitative and subjective entry of a certain type, to suggest a way to complete our knowledge and better define the shape processes in a concrete way.
}{
Nous proposons finalement d'ébaucher une entrée qualitative et subjective d'un certain type, pour suggérer une façon de compléter nos connaissances et mieux cerner les processus de manière concrète.
}

\bpar{
The entry we take follows the method of \emph{floating observation}, introduced at the interface of anthropology and sociology by~\cite{petonnet1982observation}, with the ambition to lay the basis of an urban anthropology, in the sense of a study of human behaviors with an urban environment. This does not correspond exactly to the same idea than \noun{Choay}'s anthropology of space~\cite{choay2009pour} which explores the opposite direction, i.e. the particular feature of human societies to shape space, and the ability to construct a built environment at different scales through architecture and urbanism. Our methodological context is the following. Answering to a need of mobility that a sedentary person easily feels, the researcher is placed at the center of the knowledge production process, we quote, by ``\textit{remaining in any circonstance vacant and available, to not focus the attention on a precise object, but to let it float in order that information penetrate it without any filter, any a priori, until landmarks, convergences, appear and we then manage to discover underlying rules}''. This method can be used as a preliminary study to build precise protocols and interview questionnaires: it is for example used furthermore in the case of transportation by~\cite{de2012deplacements}. We use it in our case as a method to extract stylized facts, in order to inform examples of interaction processes that are directly observable.
}{
L'entrée prise suit la méthode \emph{d'observation flottante}, introduite à l'interface de l'anthropologie et la sociologie par~\cite{petonnet1982observation}, avec l'ambition de fonder une anthropologie urbaine, au sens de l'étude des comportements humains au sein d'un environnement urbain. Il ne s'agit pas exactement de la même idée que l'anthropologie de l'espace de \noun{Choay}~\cite{choay2009pour} qui explore la direction inverse, c'est-à-dire le propre des sociétés humaines de façonner l'espace, et la capacité de construire un environnement bâti à différentes échelles par l'architecture et l'urbanisme. Notre contexte méthodologique est le suivant. Répondant à un besoin de mouvement que le sédentaire éprouve facilement, le chercheur se place au centre du processus de production de connaissances, nous citons, en ``\textit{rest[ant] en toute circonstance vacant et disponible, à ne pas mobiliser l'attention sur un objet précis, mais à la laisser flotter afin que les informations la pénètrent sans filtre, sans a priori, jusqu'à ce que des points de repère, des convergences, apparaissent et que l'on parvienne alors à découvrir des règles sous-jacentes}''. Cette méthode peut servir d'étude préliminaire pour fixer des protocoles et grilles précises d'entretien : elle est par exemple utilisée justement au sujet du transport par~\cite{de2012deplacements}. Nous nous en servons dans notre cas comme méthode d'extraction de faits stylisés, afin d'informer des exemples de processus d'interactions directement visibles.
}

\paragraph{Method}{Méthode}

\bpar{
The commuting movements at an intra-metropolitan scale are necessarily lived in a particular way in comparison to other geographical spaces or other scales on the same place. And is one way to capture particular stylized facts would then be to proceed to the analog of a perturbation study on the system, but taking as a referential the observer himself ? It would consist in generating a shock on an ``equilibrium'' situation, and then to let himself float following the current to apprehend the reaction and some mechanisms that would have been difficult to consider when following a routine. A natural experiment caused by a perturbation of transportation (which in Paris region is rather frequent, at least much more than in China) is an event provoking a natural experiment, in the sense that the researcher can capture particular situations and individual reactions. Our methodology is relatively simple: strolling within public transportation, with or without an aim and in a random way or not, but trying on each journey to maximize opportunities of being in a situation or capturing an event, typically by avoiding a routine journey\footnote{This constraint will be in our case respected for Guangdong, but not for Ile-de-France.}. The repetition of the experiment will also aim a maximizing the spatial, temporal and situational extent. A traceable production is in theory necessary at each iteration, let it be a factual description, a perceived description, a semi-synthesis. This allows a posteriori to see the successive stratifications of what has been lived and observation experiments progressively refined in their context, and thus to trace the genesis of induced ideas. We make the choice to retranscribe the subjective aspect, even to maximize it in the general synthesis of observations, in order to underline this aspect in contrast to the following of our work which will be relatively disconnected from the subject leading the research, and in echo with recommendations by \cite{ball1990self} for the role of subjectivity in fieldwork ethnographical research.
}{
Les mouvements pendulaires à échelle intra-métropolitaine sont nécessairement vécus d'une façon particulière en comparaison à d'autres lieux géographiques et à d'autres échelles sur le même lieu. Et si une façon d'appréhender des faits stylisés particuliers était alors d'effectuer l'analogue d'une étude de perturbation sur le système, mais en prenant comme référentiel l'observateur lui-même ? Il s'agirait de faire porter un choc sur une situation ``d'équilibre'', puis de se laisser flotter au gré du courant pour appréhender la réaction et certains mécanismes qu'il aurait été difficile de considérer en suivant sa routine. Une expérience naturelle causée par une perturbation des transports (qui en région francilienne est bien courante, dans tous les cas plus qu'en Chine) est un événement provoquant une expérience naturelle, au sens où le chercheur peut capturer des situations et réactions individuelles particulières. Notre méthodologie est relativement simple : déambuler dans les transports en commun, avec ou sans but et de manière ou non aléatoire, mais en essayant sur chaque trajet de maximiser les opportunités de mise en situation ou de capture d'évènement, typiquement en évitant un trajet de routine\footnote{Cette contrainte sera respectée dans notre cas pour le Guangdong, mais pas pour l'Ile-de-France.}. La répétition de l'expérience visera également à maximiser la couverture spatiale, temporelle, de situation. Une production traçable est en théorie nécessaire à chaque itération, qu'il s'agisse de description factuelle, de description perçue, de semi-synthèse. Celle-ci permet a posteriori de voir les stratifications successives du vécu et des expériences d'observation progressivement raffinées dans leur contexte, et de tracer ainsi la genèse des idées induites. Nous faisons le choix de retranscrire l'aspect subjectif, voir maximiser celui-ci dans les synthèses générales des observations, afin d'appuyer cet aspect en contraste avec la suite de notre travail qui sera relativement déconnecté du sujet menant la recherche, et en écho avec les recommandations de \cite{ball1990self} pour la place de la subjectivité dans la recherche ethnographique de terrain.
}


\bpar{
By the choice of the method, results in this subsection deal mainly with transportation. Interactions with territories will be mostly perceived within observed mobility practices.
}{
De par le choix de la méthode, les résultats de cette sous-section portent majoritairement sur les transports. Les interactions avec les territoires seront perçues majoritairement dans les pratiques de mobilité observées.
}

\begin{figure}[h!]
\begin{mdframed}
\bpar{
The sky is grey and not a single smile on faces, this northern Sun has indeed as a light only its name. The initiated will not be fooled and will feel deeply within himself this banal routine of a daily return trip with the RER. He will either curse the successive plans with temporal stratifications have let this incongruous territorial organization decant, neither surprise himself to dream of an alternative life trajectory since choosing is a bit dying and he does not have a Phoenix soul today. Perhaps the beauty of the city is finally in these tensions which shape it at all levels and in all domains, these paradoxes which become a living environment to the point of daily imposing a truth. This subway corridor philosophy, the Parisian makes it his workhorse, since after all he lives in the city he must know it. Again a broken rail on line A, ``everything is badly managed, and this network is badly designed'' vociferates a daily user, improvizing himself as an expert in planning; others that are more patient take their troubles patiently but present themselves with as much knowledge of an illusory global vision of a territory with multiple faced. These users \emph{are} however the system, in a concrete way at their space and time scale, by induction and emergence at higher scales. The ant is assumed not to be conscient of the collective intelligence of which it is a fundamental component. They similarly have only limited perception of the self-desorganization of which they are the source, maybe the cause, and which highly probably undergo the unpleasantness of its dynamics. Let himself float in Parisian transportation is an timeless experience. Even therapeutic sometimes, when one begins to loose his optimism regarding the advantages of an urban life, a random excursion in the metro rapidly recalls the richness and diversity which are one of the greatest success of cities. This apparent variety of profiles will mainly be retained by the researchers in these perambulations, and he will keep in mind that there is no scale at which a specific treatment of each geographical object is not necessary: in a few stations on line 4 the socio-economic profile of districts deeply changes and often without any transition at least three times, as on the north of line 13 where temporal patterns reveal even more difficult socio-economic realities which are indeed geographical in this \emph{produced space} of the metropolis. When it comes to modeling, taking into account the limits of any attempt to generalize is even more crucial since each model is a fragile equilibrium between specificity and generality.
}{
Le ciel est gris et les visages fermés, ce Soleil du Nord n'a bien de lumière que le nom. L'initié ne saura s'y tromper et ressentira au fond de lui-même cette banale routine d'un aller-retour quotidien en RER. Il ne cherchera ni à maudire les planifications successives dont les stratifications temporelles ont laissé décanter cette organisation territoriale incongrue, ni à se prendre à rêver d'une trajectoire de vie alternative puisque choisir c'est un peu mourir et qu'il ne se sent pas une âme de Phoenix aujourd'hui. Peut être que la beauté de la ville est finalement dans ces tensions qui la façonnent à tous les niveaux et dans tous les domaines, ces paradoxes qui deviennent cadre de vie au point d'asséner quotidiennement une vérité. Cette philosophie de couloir de métro, le francilien en fait son cheval de bataille car après tout s'il vit en ville il doit bien la connaître. Encore un rail cassé sur le A, ``tout cela est mal géré, et ce réseau est mal conçu'' vocifère un utilisateur journalier, s'improvisant expert en planification ; d'autres plus patients prennent leur mal en patience mais se présentent tout aussi connaisseurs d'une illusoire vision d'ensemble d'un territoire aux multiples visages. Ces usagers \emph{sont} pourtant le système, de manière concrète à leur échelle d'espace et de temps, par induction et émergence aux échelles supérieures. La fourmi est supposée ne pas avoir conscience de l'intelligence collective dont elle est une des composantes fondamentales. Ils n'ont de la même manière que peu de perception de l'auto-désorganisation dont ils sont la source, peut-être la cause, et qui très sûrement subissent les désagréments de ses dynamiques. Se laisser flotter dans les transports franciliens est une expérience intemporelle. Presque thérapeutique parfois, quand l'un commence à perdre son optimisme quant à l'intérêt d'une vie urbaine, une excursion aléatoire en métro rappelle rapidement la richesse et la diversité qui sont un des plus grand succès des villes. C'est cette variété apparente de profils que le chercheur retiendra principalement de ces errements, et il gardera à l'esprit qu'il n'existe pas d'échelle où un traitement spécifique de chaque objet géographique n'est pas nécessaire : en quelques stations sur la ligne 4 le profil socio-économique des quartiers change profondément et souvent sans transition au moins trois fois, comme sur la ligne 13 nord où les motifs horaires soulignent d'autant plus de dures réalités socio-économiques qui sont en fait géographiques dans cet \emph{espace produit} de la métropole. Lorsqu'il s'agit de modéliser, prendre en compte les limites de toute tentative de généralisation est d'autant plus cruciale comme chaque modèle est un équilibre fragile entre spécificité et généralité.
}

\medskip


\framecaption{\textbf{A floating observation experiment in the Paris region.}}{\textbf{Une expérience en observation flottante en région parisienne.}}
\end{mdframed}
\end{figure}

\begin{figure}[h!]
\begin{mdframed}
\bpar{
The journey will be long. The chosen perturbation is the simulation of the luckless event, ``\cn{我的护照丢了，我得去法国的领事馆在广州}'', i.e. the loss of his passport, which obliges to take the transportation to go to the consulate. Such an event in China is indeed unfortunate, since the totality of inter-urban transportation is conditioned to it. Crossing the urban mega-region from the South to the North to rejoin Guangzhou in this situation is indeed a challenge. From urban bus to urban bus, terminuses that are more or less well articulated. A fake traditional village has been build for the happiness of tourists, not far from Zhongshan birthplace, not credible given the accessibility. Striking contrasts and a highly heterogenous landscape, poverty enclaves within areas in view since recently. More or less voluntary relocations towards fringes shape a new landscape of geographical inequality  that is already well known in Europe. Similarly to this continuous traffic jam, the reinvention of the city which is already well advanced here has to make crucial choices to be the example of a sustainable trajectory. An impressive resilience of users to a major perturbation, a local ability to self-organize giving functionality to projects which could have been not functional at all: from Shenzhen, Baoan to Zhuhai, Tangjia or Zhongshan, Xiaolan, the fleet of informal moto-taxis saves the local accessibility, as confirms to me Jingzi living in the South of Zhongshan and studying in the North of Zhuhai and for which the train is a mobility solution even not considered. From tramway to BRT, equivalent choices and compromises ? The first surprises more the new users. Maybe also a percussive argument to valorize the complex specifically constructed around the terminus. Local choices make even more differences when it is more difficult to go from one area to the other. Blocked not far from Guangzhou, the bridge is closed, the metro is on the other side but impossible to rejoin it. Just the time to go to Xiaolan station and back to the starting point, challenge far from being realized. Observing adaptability is not enough to develop it ? Mobility practices adapted very fast by users: high speed trains that are full at any time of the week, it seems for very diverse purposes. An apparent territorial development, middle term impacts that we can bet to be non discussable. If the structure is integrated and flexible, discussing of structuring effects become a tautology since the trajectory of the urban system then becomes the more or less controllable aspect, depending on time and spatial scales.
}{
Le trajet sera long. La perturbation choisie est la simulation de l'événement malencontreux, ``\cn{我的护照丢了，我得去法国的领事馆在广州}'', c'est-à-dire la perte de son passeport, qui oblige à prendre les transports pour se rendre au consulat. Celle-ci en Chine est assurément malencontreuse, puisque l'intégralité des trajets interurbains y est conditionnée. Traverser la mega-région urbaine du sud vers le nord pour rejoindre Guangzhou dans cette situation relève du défi. De bus urbain en bus urbain, des terminus plus ou moins bien articulés. Un village traditionnel factice est sorti de terre pour faire le bonheur des touristes, non loin de la maison natale de Zhongshan, peu crédible vu l'accessibilité. Des contrastes saisissants et un paysage très hétérogène, des enclaves de pauvreté dans des zones nouvellement prisées. Les relocalisations plus ou moins volontaires vers les franges façonnent un nouveau paysage d'inégalité géographique que l'on connait déjà bien en Europe. À l'image de cet embouteillage continu, la réinvention de la ville déjà bien avancée ici se doit de faire des choix cruciaux pour être l'exemple d'une trajectoire durable. Une résilience impressionnante des usagers à une perturbation majeure, une capacité d'auto-organisation locale rendant fonctionnels des aménagements qui auraient pu ne pas l'être : de Shenzhen, Baoan à Zhuhai, Tangjia ou à Zhongshan, Xiaolan, la flotte de moto-taxis informels sauve l'accessibilité locale, comme me le confirme Jingzi habitant le sud de Zhongshan et étudiant au nord de Zhuhai et pour qui le train est une solution de mobilité même pas envisagée. Du tramway au BRT, choix et compromis équivalents ? Le premier étonne plus les nouveaux usagers. Peut être aussi un argument percutant pour valoriser le complexe spécialement conçu autour du terminus. Les choix locaux sont d'autant plus différentiables qu'il est difficile de passer d'une zone à l'autre. Bloqué non loin de Guangzhou, le pont est fermé, le métro est en face mais impossible de le rejoindre. Juste le temps pour se rabattre sur la gare de Xiaolan et retour à la case départ, défi bien loin d'être réalisé. Observer l'adaptabilité ne suffit pas à la développer ? Des pratiques de mobilité très vite adaptées par les usagers : des trains à grande vitesse bondés en toute heure de la semaine, semble-t-il pour des motifs très divers. Un développement territorial apparent, des impacts à moyen terme qu'on peut parier non discutables. Si la structure est intégrée et flexible, discuter d'effets structurants devient une tautologie puisque la trajectoire du système urbain devient alors l'aspect plus ou moins contrôlable, selon les échelles de temps et d'espace.
}

\medskip

\framecaption{\textbf{A floating observation experiment in Guangdong, Zhuhai.}}{\textbf{Une expérience en observation flottante, Guangdong, Zhuhai.}}
\end{mdframed}
\end{figure}

\paragraph{Results}{Résultats}

\bpar{
Our fieldwork observation sequences have taken place on the one hand in China, mostly in Guangdong, Zhuhai, during dedicated sessions. Observations span between the 10/10/2016 and the 23/01/2017 and also between the 08/06/2017 and the 01/09/2017. The main transportation mode is urban bus, followed by regional train, and high-speed train and ferry; the range of journeys corresponds to the range of modes. Detailed reports, written on the fly in a subjective way and edited a posteriori the less possible, as previously explained, are available in Appendix~\ref{app:sec:qualitative}. Observations for the Parisian region are at a quasi-daily frequency and are not recorded; these have mostly taken place on metro line 4 and on RER line A between February 2016 and October 2016, on Transilien line R and RER line A between November 2016 and September 2017 and between February 2017 and May 2017, and then on metro line 9 and line 4 between September 2017 and October 2017.
}{
Nos séquences d'observation de terrain ont eu lieu d'une part en Chine, majoritairement dans le Guangdong à Zhuhai, lors de sessions dédiées. Les observations s'étendent entre le 10/10/2016 et le 23/01/2017 ainsi qu'entre le 08/06/2017 et le 01/09/2017. Le mode de transport majoritaire est le bus de ville, suivi par le train régional, puis le train à grande vitesse et le ferry ; la portée des déplacements correspondent à celle des modes. Les compte-rendus détaillés, écrits à la volée de manière subjective et édités a posteriori le moins possible, comme expliqué précédemment, sont disponibles en Annexe~\ref{app:sec:qualitative}. Les observations pour la région parisienne sont quasi-quotidiennes et non consignées ; celles-ci ont eu lieu en plus grande partie sur la ligne 4 du métro et sur la ligne A du RER entre février 2016 et octobre 2016, sur la ligne R du Transilien et la ligne A du RER entre novembre 2016 et septembre 2017 puis entre février 2017 et mai 2017, puis sur la ligne 9 et la ligne 4 entre septembre 2017 et octobre 2017.
}



\bpar{
The two floating observation synthesis for each region, which are materials produced from raw reports, are presented in the frames above. These illustrate in particular through subjective examples some instances of interactions between networks and territories, mostly at microscopic and mesoscopic scales, for processes related to mobility. Subjectivity and interpretation allows also to extrapolate on processes at smaller scales, in termes of accessibility for example. These can however not be taken as more than a thematic illustration and introduction. By taking a step back, we propose to list some learnings which can be drawn from this experiment at a high level of synthesis, in contrast with the subjective and specific aspect of the product of the experiment. They are the following:
}{
Les deux synthèses d'observation flottante pour chacune des régions, matériaux produit à partir des notes brutes, sont présentées dans les encadrés ci-dessus. Celles-ci illustrent entre autres par des exemples subjectifs certaines instances d'interactions entre réseaux et territoires, majoritairement aux échelles microscopique et mesoscopique, pour des processus touchant à la mobilité. La subjectivité et l'interprétation permet aussi d'extrapoler sur des processus à plus petite échelle, en terme d'accessibilité par exemple. Ceux-ci ne peuvent toutefois être pris plus que comme une illustration et introduction thématique. Par une prise de recul, nous proposons de lister certains enseignements qui peuvent être tirés de cette expérience à un niveau de synthèse élevé, en contraste avec l'aspect subjectif et spécifique du produit de l'experience. Ils sont les suivants :
}

\bpar{
\begin{enumerate}
	\item The complexity of the transportation system and as a consequence its integration with urbanism within the territorial system, can have divergent consequences in terms of final performance, and for example sustainability. In the Chinese case, self-organization and local adaptability are assets of the local performance of new stations, whereas in France the complexity seems to be a source of constraints and finally of negative externalities\footnote{This effect being furthermore necessarily in strong interdependency with cultural properties, what makes of it a fundamental component of territories.}.
	\item The adaptability of territories, of which one component is for example the speed of mutation of mobility practices and linked to adaptability, seems also highly sensitive to geographical particularities.
	\item The question of observable time and space scales, what will partly condition the ones that can be modeled, is ambiguous within the observation, as witnesses the joint observation of mobility and manifestations of accessibility patterns.
	\item The comparability of cases and geographical situations is, in our case, but a priori more generally, a difficult point to which there does not exist an ideal solution. The compromise between generality and particularity is then determining in the construction of a geographical theory and models. This conclusion drawn from empirical studies should also apply on models, but to what extent remains an open question.
\end{enumerate}
}{
\begin{enumerate}
	\item La complexité du système de transport et en conséquence de son intégration avec l'urbanisme dans le système territorial, peut avoir des conséquences divergentes en termes de performance finale, et par exemple de soutenabilité. Dans le cas Chinois, l'auto-organisation et l'adaptabilité locale sont des atouts de la performance locale des nouvelles gares, tandis qu'en France la complexité semble être source de freins et finalement d'externalités négatives\footnote{Cet effet étant par ailleurs nécessairement en interdépendance forte avec les propriétés culturelles, qui est en fait une composante fondamentale des territoires.}.
	\item L'adaptabilité des territoires, dont l'une des composantes est par exemple la vitesse de mutation des pratiques de mobilité et reliée à l'adaptabilité, semble également très sensible aux particularités géographiques.
	\item La question des échelles de temps et d'espace observables, ce qui conditionnera partiellement celles qu'on peut modéliser, est ambiguë dans l'observation, comme le témoigne l'observation conjointe de la mobilité et de manifestation de motifs d'accessibilité.
	\item La comparabilité des cas et des situations géographiques est, dans notre cas, mais a priori plus généralement, un point épineux auquel il n'existe pas de solution idéale. Le compromis entre généralité et particularité est alors déterminant dans la construction d'une théorie et de modèles géographiques. Cette conclusion tirée sur des études empiriques devrait s'appliquer aussi aux modèles, mais dans quelle mesure il s'agit d'une question ouverte.
\end{enumerate}
}

\bpar{
These considerations will participate to the orientation of ontological and epistemological positioning we will take in the following.
}{
Ces considérations participeront à l'orientation des postures ontologiques et épistémologiques que nous prendrons par la suite.
}

\stars

%


\newpage

\section*{Synthesis of studied processes}{Synthèse des processus étudiés}

\bpar{
We conclude this introducing chapter by a synthesis and a perspective on interaction processes that have been identified in the theoretical and empirical analysis and in the literature. This will allow to situate the reviews of modeling entreprises to which we will proceed in chapter~\ref{ch:modelinginteractions}, and then will be compared to the one we will establish in the case of models.
}{
Nous concluons ce chapitre introductif par une synthèse et une mise en perspective des processus d'interaction identifiés par l'analyse théorique, empirique et la littérature. Celle-ci permettra de situer les revues des entreprises de modélisation auxquelles nous procéderons dans le chapitre~\ref{ch:modelinginteractions}, puis pourra être comparée à celle que nous établirons dans le cas des modèles.
}

\subsubsection*{A entry by scales}{Une entrée par les échelles}

\bpar{
A first entry to synthesize the processes considered consists in considering then by scale. We have seen that a multi-scale reading was relevant, and that it allowed globally to isolate characteristic spatial and temporal scales: microscopic, mesoscopic and macroscopic, with a rather good correspondence between temporal and spatial scales. This typology remains of course reduced, since it simplifies the class of processes that could result of these correspondences, for exemple a mobility at a large scale, or a bifurcation of the urban system which happens quickly, Similarly, processes that are themselves multi-scalar (the governance of Greater Paris is a good illustration, since it relates to governance levels and territorial issues at different scales) are taken into account in a simplified way. The axis complementary to the one of scales is based on ``effects and causes'': although we still remain within the frame of a complex causality as presented in introduction, we have unveiled processes for which it is possible to identify a precursor among the network or the territory (we will then denote them by $A \rightarrow B$), others are intrinsically complex and already contain circular causalities (for example in the case of governance processes), we will denote them by Networks $\leftrightarrow$ Territories. The synthesis table is then given in Table~\ref{tab:thematic:processes}.
}{
Une première entrée pour synthétiser les processus abordés consiste à les considérer par échelle. Nous avons vu qu'une lecture multi-échelle était pertinente, et que celle-ci permettait globalement de dégager des échelles spatiales et temporelles caractéristiques : microscopique, mesoscopique et macroscopique, avec une assez bonne correspondance des échelles spatiales et temporelles. Cette typologie est bien sûr réduite, puisqu'elle simplifie la classe des processus qui pourraient sortir de ces correspondances, par exemple une mobilité à grande échelle, ou une bifurcation du système urbain qui se manifeste rapidement. De même, les processus eux-mêmes multi-échelles (la gouvernance du Grand Paris en est une bonne illustration, mobilisant des niveaux de gouvernance et des enjeux territoriaux à différentes échelles) sont pris en compte de manière simplifiée. L'axe complémentaire à celui des échelles se base sur les ``effets et causes'' : bien que nous restions toujours dans le cadre d'une causalité complexe comme présenté en introduction, nous avons mis en évidence des processus pour lesquels il est possible d'identifier un précurseur parmi le réseau ou le territoire (nous les noterons alors $A \rightarrow B$), d'autres sont intrinsèquement complexes et contiennent déjà des causalités circulaires (par exemple dans le cas des processus de gouvernance), nous les noterons Réseaux $\leftrightarrow$ Territoires. Le tableau de synthèse est alors donné en Table~\ref{tab:thematic:processes}.
}

\begin{table}
\caption[Interaction processes between networks and territories]{\textbf{Interaction processes between networks and territories.} We synthesize the processes according to scales and the precursor typology.\label{tab:thematic:processes}}
\bpar{
\begin{tabular}{|l|p{5cm}|p{5cm}|p{5cm}|}
\hline
 & Networks $\rightarrow$ Territories & Territories $\rightarrow$ Networks & Networks $\leftrightarrow$ Territories\\ \hline
Micro & Mobility patterns & Network congestion ; Negative externalities & Mobility and social structure \\ \hline
Meso & Relocations ; Local effects of infrastructures & Potential breakdown & Metropolitan planning ; TOD \\ \hline
Macro & Interactions between cities ; Tunnel effect & Hierarchical differentiation of accessibility & Large scale planning ; Structural dynamics ; Bifurcations\\ \hline
\end{tabular}
}{
\begin{tabular}{|l|p{5cm}|p{5cm}|p{5cm}|}
\hline
 & Réseaux $\rightarrow$ Territoires & Territoires $\rightarrow$ Réseaux & Réseaux $\leftrightarrow$ Territoires\\ \hline
Micro & Motifs de mobilité & Congestion du réseau ; Externalités négatives & Mobilité et structure sociale \\ \hline
Meso & Relocalisations ; Effets locaux des infrastructures & Rupture de potentiel & Planification métropolitaine ; TOD \\ \hline
Macro & Interactions entre villes ; Effet tunnel & Différenciation hiérarchique de l'accessibilité & Planification à grande échelle ; Dynamique structurelle ; Bifurcations\\ \hline
\end{tabular}
}
\end{table}

\subsubsection*{A entry by actors}{Une entrée par les acteurs}


\bpar{
A second entry privileges the role of \emph{actors}, i.e. of agents which make the territory. Indeed, the problematics linked to mobility concern the microscopic agents, the ones linked to accessibility urban and economic actors, the ones linked to planning governance actors. This aspect can be summarized by the scheme in Frame~\ref{frame:networkterritories:acteurs}.
}{
Une deuxième entrée privilégie le rôle des \emph{acteurs}, c'est-à-dire des agents qui font le territoire. En effet, les problématiques liées à la mobilité concernent les agents microscopiques, celles liées à l'accessibilité des acteurs urbains et économiques, celles liées à la planification des acteurs de gouvernance. Cet aspect peut être résumé par le schéma en Encadré~\ref{frame:networkterritories:acteurs}.
}

\begin{figure}[h!]
\begin{mdframed}
\includegraphics[width=\textwidth]{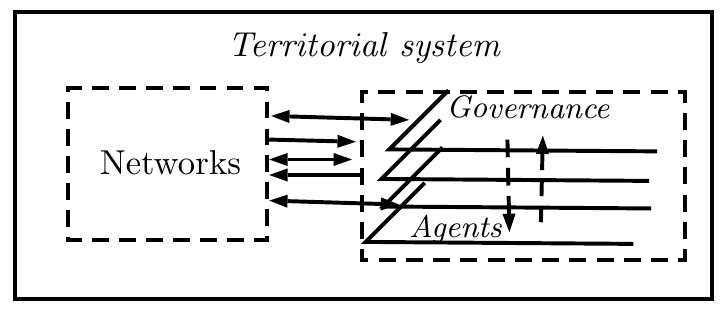}
\medskip
\framecaption{\textbf{An entry by actors on territorial systems.}\label{frame:networkterritories:acteurs}}{\textbf{Une entrée sur les systèmes territoriaux par les acteurs.}\label{frame:networkterritories:acteurs}}
\end{mdframed}
\end{figure}

\bpar{
In this scheme, we identify the territorial actors within the territorial system, which can be schematically formulated on two scales: agents at the microscopic scale which will be central for mobility processes, and governance actors at larger scales, which lead the governance processes. They interact between them in a complex way, and are here conceptually separated by dashed lines from others aspects of the territory to which thay are also strongly coupled.
}{
Dans ce schéma, on identifie les acteurs territoriaux au sein du système territorial, qui se déclinent schématiquement sur deux échelles : les agents à l'échelle microscopique qui seront centraux pour les processus de mobilité, et les acteurs de gouvernance à des échelles supérieures, qui mènent les processus de gouvernance. Ils interagissent entre eux de manière complexe, et sont séparés ici conceptuellement par les pointillés d'autres aspects du territoire avec lesquels ils sont aussi couplés fortement.
}

\bpar{
This entry can be put in perspective with the conceptual frame of~\cite{le2010approche}, which studies the links between the urban form and mobility practices within metropolitan contexts. This framework understands the urban system as a strong coupling between the location system, the activity system and the transportation system, by recalling the influence of demand agents (micro-economic agents) and planning agents (governance agents) on each system. The transportation system corresponds to our networks and the two other systems to an aspect of territorial agents, which also contain agents formulated within this frame. This parallel must be nuanced when changing scales: at the scale of the system of cities, when agents are cities, the location system has no meaning anymore, since it is adapted to a scale at most metropolitan, and specifically to corresponding ontologies.
}{
Cette entrée peut être mise en perspective avec le cadre conceptuel de~\cite{le2010approche}, qui étudie les liens entre forme urbaine et pratiques de mobilité dans des contextes métropolitains. Celui-ci comprend le système urbain comme un couplage fort entre système de localisation, système d'activités et système de transport, en précisant l'influence des agents demandeurs (agents micro-économiques) et des agents aménageurs (agents de gouvernance) sur chaque système. Le système de transport correspond à nos réseaux et les deux autres systèmes à un aspect des agents territoriaux, qui contiennent aussi les agents précisés dans ce cadre. Ce parallèle reste à nuancer lorsqu'on change d'échelle : à celle du système de villes, lorsque les agents sont les villes, le système de localisation n'a plus de sens, puisque celui-ci est adapté à une échelle au plus métropolitaine, et surtout aux ontologies correspondantes.
}

\bigskip

\bpar{
This double entry to read interaction processes between networks and territories will on the one hand condition the literature review of models done in chapter~\ref{ch:modelinginteractions}, and will on the other hand be completed and specified after it. 
}{
Cette double entrée de lecture des processus d'interaction entre réseaux et territoires conditionnera d'une part la revue de littérature des modèles faite en chapitre~\ref{ch:modelinginteractions}, et sera d'autre part complétée et précisée à l'issue de celui-ci.
}

\stars


\newpage

\section*{Chapter Conclusion}{Conclusion du Chapitre}

\bpar{
Territories interact in a complex way with networks, in particular transportation networks, as shown by the numerous empirical examples or the theoretical constructions we reviewed. At different typical temporal scales (the day, the decade, and the century), correspond spatial scales (urban, metropolitan and system of cities), and also processes (mobility, accessibility and relocations, systemic structural effects and bifurcations). Concrete situations witness of local realities expressed with different nuances, and processes carrying these abstract processes with different roles and interactions between them.
}{
Les territoires interagissent de manière complexe avec les réseaux, en particulier ceux de transport, comme montré par les nombreux exemples empiriques ou les constructions théoriques passés en revue. À différentes échelles temporelles typiques (le jour, la décennie et le siècle), correspondent des échelles spatiales (urbaine, métropolitaine et système de villes), ainsi que des processus (mobilité, accessibilité et relocalisations, effets systémiques structurels et bifurcations). Les situations concrètes témoignent de réalités locales déclinés avec différentes nuances, et des processus portant ces processus abstraits avec différents rôles et interactions entre eux.
}

\bpar{
We have in a first section clarified this notion of interaction between transportation networks and territories by constructing a theoretical frame which allows to consider them as components of the territorial system in it entirety. We have then suggested an approach by co-evolution to take into account this complexity. In order to better identify these notions on concrete geographical examples, we have developed in~\ref{sec:casestudies} two metropolitan case study which are current issues, and underlined the certainties in terms of accessibility impact for major infrastructure projects which are systematically accompanied by an uncertainty in terms of the trajectory of the system on a longer term. Finally, we propose in~\ref{sec:qualitative} an excursion through fieldwork elements in China, Guangdong.
}{
Nous avons dans une première section clarifié cette notion d'interaction entre réseaux de transports et territoires en construisant un cadre théorique qui permet de les considérer comme des composantes du système territorial dans son ensemble. Nous avons alors suggéré une approche par la co-évolution pour tenir compte de cette complexité. Afin de mieux cerner ces notions sur des exemples géographiques concrets, nous avons développé en~\ref{sec:casestudies} deux cas d'étude métropolitain d'actualité, et souligné les certitudes en termes d'impact d'accessibilité pour des projets majeurs d'infrastructures qui s'accompagnent systématiquement d'incertitude en terme de trajectoire du système à plus long terme. Enfin, nous proposons en~\ref{sec:qualitative} une excursion par des éléments de terrain dans le Guangdong, Chine.
}

\bpar{
At this stage, having introduced the thematic object of study, we propose more particularly to focus on approaches implying modeling, making the choice of a fundamental role of the \emph{model} (on which we will come back with more details) in the production of knowledge.
}{
À ce stade, ayant introduit l'objet d'étude thématique, nous proposons de nous intéresser plus particulièrement aux approches impliquant une modélisation, faisant le choix d'un rôle fondamental du \emph{modèle} (sur lequel nous reviendrons plus en détails par la suite) dans la production de connaissance.
}

\stars




\bpar{
\chapter{Modeling interactions between networks and territories}
}{
\chapter{Modéliser les interactions entre réseaux et territoires}
}

\label{ch:modelinginteractions}


\bpar{
The empirical and thematic literature, together with the case studies previously developed, seem to converge towards a consensus on the complexity of relations between transportation networks and territories. In some configurations and at some scales, it is possible to exhibit circular causal relationships between territorial dynamics and transportation networks dynamics. We designate their existence through the concept of \emph{co-evolution}. It seems to be difficult to introduce simple or systematic explanations for these dynamics, as recall for example the debates around structuring effects of infrastructures~\cite{offner1993effets}.
}{
La littérature empirique et thématique, ainsi que les cas d'études développés précédemment, semblent converger vers un consensus sur la complexité des relations entre réseaux de transport et territoires. Dans certaines configurations et à certaines échelles, il est possible de mettre en valeur des relations circulaires causales entre dynamiques territoriales et dynamiques des réseaux de transports. Nous désignons leur existence par le concept de \emph{co-évolution}. Il semble difficile d'introduire des explications simples ou systématiques de ces dynamiques, comme le rappelle par exemple les débats autour des effets structurants des infrastructures~\cite{offner1993effets}.
}

\bpar{
Furthermore, the multiple geographical situations suggest a strong dependency to the context, giving a relevance to fieldwork and to targeted studies. But geographical explanation and the understanding of processes remains quickly limited in this approach, and intervenes a need for a certain level of generality. Its on such a point that the evolutive urban theory is focused in particular, since it allows to combine schemes and general models to the geographical particularities. On the contrary, some theories coming from physics applying to the study of urban systems~\cite{west2017scale} can be more difficult to accept for geographers because of their universality positioning which is on the opposite of their ordinary epistemologies.
}{
Par ailleurs, les multiples situations géographiques suggèrent une forte dépendance au contexte, donnant une pertinence au travail de terrain et aux études ciblées. Or l'explication géographique et la compréhension des processus est très vite limitée dans cette approche, et intervient un besoin d'un certain niveau de généralisation. C'est sur un tel point que la théorie évolutive des villes se concentre en particulier, puisqu'elle permet de combiner des schémas et modèles généraux aux particularités géographiques. Au contraire, certaines théories issues de la physique s'appliquant à l'étude des systèmes urbains~\cite{west2017scale} peuvent être plus difficiles à accepter pour les géographes de par leur positionnement d'universalité qui est à l'opposé de leurs épistémologies habituelles.
}

\bpar{
In any case, the \emph{medium} which allows to gain in generality on processes and structures of systems is always the model. As \noun{J.P. Marchand} puts it\footnote{Personal communication, May 2017.}, ``\textit{our generation has understood that there was a co-evolution, yours aims at understanding it}'', what insists on the power of understanding brought by modeling and simulation that we judge to be today still with a very high potential for development.
}{
Dans tous les cas, le \emph{medium} qui permet de gagner en généralité sur les processus et structures des systèmes est toujours le modèle. Comme le rappelle \noun{J.P. Marchand}\footnote{Communication personnelle, Mai 2017.}, ``\textit{notre génération a compris qu'il y avait une co-évolution, la votre cherche à la comprendre}'', ce qui appuie le pouvoir de compréhension apporté par la modélisation et la simulation que nous jugeons être encore aujourd'hui à très fort potentiel de développement.
}

\bpar{
Without developing for now the numerous functions that a model can have, we will rely on the positioning of \noun{Banos} which states that ``modeling is learning'', and following our positioning within a complex systems science suggested in introduction, we will thus make \emph{modeling interactions between networks and territories} our principal subject of study, tool, object\footnote{Even if after a rereading of this positioning at the light of~\ref{sec:knowledgeframework}, it has no meaning since our appproach already contained models as soon as it was scientific.}. This chapter must be taken as a ``state-of-the-art'' of approaches modeling interactions between networks and territories. It aims in particular at capturing different dimensions of knowledge: therefore, we will uses quantitative epistemology analyses.
}{
Sans développer pour le moment les nombreuses fonctions que peut avoir un modèle, nous nous baserons sur la position de \noun{Banos} qui soutient que ``modéliser c'est apprendre'', et suivant notre positionnement dans une science des systèmes complexes suggéré en introduction, nous ferons ainsi de la \emph{modélisation des interactions entre réseaux et territoires} notre principal sujet d'étude, outil, objet\footnote{Même si dans une relecture à la lumière de~\ref{sec:knowledgeframework} de ce positionnement n'a pas de sens puisque notre démarche contenait déjà des modèles à partir du moment où elle était scientifique.}. Ce chapitre doit être pris comme un ``état de l'art'' des démarches de modélisation des interactions entre réseaux et territoires. Il vise en particulier à capturer différentes dimensions des connaissances : pour cela, nous mobiliserons des analyses en épistémologie quantitative.
}

\bpar{
In a first section~\ref{sec:modelingsa}, we review in an interdispclinary perspective the models that can be concerned, even remotely, without a priori of temporal or spatial scale, of ontologies, of structure, or of application context. This overview is possible thanks to the diverse disciplinary entries revealed in the previous chapter: for example geography, transportation geography, planning. This overview suggests relatively independent knowledge structures and disciplines that rarely communicate.
}{
Dans une première section~\ref{sec:modelingsa}, nous passons en revue de manière interdisciplinaire les modèles pouvant être concernés, même de loin, sans a priori d'échelle temporelle ou spatiale, d'ontologies, de structure, ou de contexte d'application. Cet aperçu est possible par les entrées disciplinaires diverses révélées au chapitre précédent : par exemple géographie, géographie des transports, planification. Cet aperçu suggère des structures de connaissances assez indépendantes et des disciplines ne communiquant que rarement.
}

\bpar{
We proceed in~\ref{sec:quantepistemo} to an algorithmic systematic review, which corresponds to a reconstruction by iterative exploration of a scientific landscape. Its results tend to confirm this compartmentalization. The study is completed by a multilayer network analysis, combining citation network and semantic network obtained through text-mining, which allows to better grasp the relations between disciplines, their lexical field and their interdisciplinarity patterns.
}{
Nous procédons dans~\ref{sec:quantepistemo} à une revue systématique algorithmique, qui correspond à une reconstruction par exploration itérative d'un paysage scientifique. Ses résultats tendent à confirmer ce cloisonnement. L'étude est complétée par une analyse de réseau multi-couches, combinant réseau de citations et réseau sémantique issu d'analyse textuelle, qui permet de mieux cerner les relations entre disciplines, leur champs lexicaux et leur motifs d'interdisciplinarité.
}

\bpar{
This study allows the construction of a corpus used for the modelography (typology of models) and the meta-analysis (characterization of this typology) done in the last section~\ref{sec:modelography}. It dissects the nature of several models and link it to the disciplinary context, what sets up the foundations and the precise frame of the modeling efforts that will be developed in the following.
}{
Cette étude permet la constitution d'un corpus utilisé pour la modélographie (typologie de modèles) et la méta-analyse (caractérisation de cette typologie) effectuée en dernière section~\ref{sec:modelography}. Celle-ci dissèque la nature d'un certain nombre de modèles et la relie au contexte disciplinaire, ce qui pose les bases et le cadre précis des efforts de modélisation qui seront développés par la suite.
}


\stars

\bpar{
\textit{This chapter is unpublished for its first section; uses in its second section the text of~\cite{raimbault2015models}, and then in its second subsection the methodology introduced by \cite{raimbault2016indirect} and developed in~\cite{raimbault2017exploration} and also the tools of \cite{bergeaud2017classifying}; it is finally unpublished for its last part.}
}{
\textit{Ce chapitre est inédit pour sa première section ; reprend dans sa deuxième section le texte traduit de~\cite{raimbault2015models}, puis pour sa deuxième partie la méthodologie introduite par \cite{raimbault2016indirect} et développée dans~\cite{raimbault2017exploration} ainsi que les outils de \cite{bergeaud2017classifying} ; et enfin est inédit pour sa dernière partie.}
}

%


\newpage

\section{Modeling Interactions}{Modéliser les interactions}
\label{sec:modelingsa}


\subsection{Modeling in Quantitative Geography}{Modélisation en Géographie Quantitative}

\subsubsection{History}{Histoire}

\bpar{
Modeling has in Theoretical and Quantitative Geography (TQG) a privileged role. \cite{cuyala2014analyse} proposes an analysis of the spatio-temporal development of French speaking TQG scientific movement and underlines the emergence of the discipline as the combination between quantitative analysis (e.g. spatial analysis or modeling and simulation practices) and theoretical constructions. This dynamic can be tracked back to the end of the seventies, and is closely linked to the growing use and appropriation of mathematical tools~\cite{pumain2002role}. The integration of these two components allows to construct theories from empirical stylized facts, which then produce theoretical hypothesis that can be tested on empirical data. This approach is born under the influence of the \emph{New Geography} in Anglo-Saxon countries and in Sweden.
}{
La modélisation joue en géographie théorique et quantitative un rôle fondamental. \cite{cuyala2014analyse} procède à une analyse spatio-temporelle du mouvement de la géographie théorique et quantitative en langue française et souligne l'émergence de la discipline comme une combinaison d'analyses quantitatives (e.g. analyse spatiale et pratiques de modélisation et de simulation) et de construction théoriques. Cette dynamique est datée à la fin des années 1970, et est intimement liée à l'utilisation et l'appropriation des outils mathématiques~\cite{pumain2002role}. L'intégration de ces deux composantes permet la construction de théories à partir de faits stylisés empiriques, qui produisent à leur tour des hypothèses théoriques pouvant être testées sur les données empiriques. Cette approche est née sous l'influence de la \emph{New Geography} dans les pays Anglo-saxons et en Suède.
}

\bpar{
Concerning urban modeling in itself, other fields than geography have proposed simulation models approximatively at the same period. For example, the \noun{Lowry} model, developed by~\cite{lowry1964model} with the objective to be applied directly to the Pittsburg metropolitan region, assumes a system of equations for the localization of actives and employments in different areas. This model has been a cornerstone of urban modeling, since as shows \cite{goldner1971lowry} it already had less than ten years after a broad heritage of conceptual and operational developments\footnote{\cite{goldner1971lowry} makes the hypothesis that this success is due to the combination of three factors: a possibility of an immediate operational application, a causal structure of the model easy to grasp (actives relocate depending on employments), and a flexible frame that can be extended or adapted.}. Relatively similar models are still largely used nowadays.
}{
Concernant la modélisation urbaine en elle-même, d'autre champs que la géographie ont proposé des modèles de simulation à peu près à la même période. Par exemple, le modèle de \noun{Lowry}, développé par~\cite{lowry1964model} dans un but appliqué immédiat à la région métropolitaine de Pittsburg, suppose un système d'équations pour la localisation des actifs et des emplois dans différentes zones. Ce modèle a été une pierre angulaire de la modélisation urbaine, puisque comme le montre~\cite{goldner1971lowry} il avait déjà moins d'une dizaine d'années plus tard un conséquent héritage de développements conceptuels et opérationnels\footnote{\cite{goldner1971lowry} fait l'hypothèse que ce succès est du à la combinaison de trois facteurs : une possibilité d'application opérationnelle directe, une structure causale du modèle simple à appréhender (les actifs se localisent en fonction des emplois), et un cadre flexible pouvant être étendu ou adapté.}. Des modèles relativement similaires sont toujours largement utilisés aujourd'hui.
}


\subsubsection{Simulation of models and intensive computation}{Simulation de modèle et calcul intensif}

\bpar{
A broad history of the genesis of models of simulation in geography is done by~\cite{rey2015plateforme} with a particular emphasis on the notion of validation of models (we will come back on the role of these aspects in our work in~\ref{ch:positioning}). The use of computation ressources for the simulation of models is anterior to the introduction of current paradigms of complexity, coming back for exemple to \noun{Forrester}, a computer scientist pioneer in spatial economics models inspired by cybernetics\footnote{Which was, together with the systemic trend, precursors of current paradigms of complexity as we already developed.}. With the increase of computational capabilities, epistemological transformations have also occurred, with the emergence of explicative models as experimental tools. \noun{Rey} compares the dynamism of seventies when computation centers were opened to geographers to the current democratization of High Performance Computing\footnote{The development of the first urban simulation models coincides with the opening of the first computation centers to social sciences and humanities, as recalls also \noun{Pumain} (interview on 31/03/2017, see Appendix~\ref{app:sec:interviews}) for example for the implementation of the \noun{Allen} entropy model.}. Today, this ease of use is in particular exemplified by grid computing with a transparent use, i.e. without the need for advanced technical skills related to mechanisms of computation distribution. This way, \cite{schmitt2014half} givef an exemple of the possibilities offered in terms of model validation and calibration, reducing the computational time from 30 years to one week - these techniques will play a crucial role in the results we will obtain in the following. This evolution is also accompanied by an evolution of modeling practices~\cite{banos2013pour} and techniques~\cite{10.1371/journal.pone.0138212}.
}{
Une histoire étendue de la genèse des modèles de simulation en géographie est faite par \cite{rey2015plateforme} avec une attention particulière pour la notion de validation de modèles (nous reviendrons sur la place de ces aspects dans notre travail en~\ref{ch:positioning}). L'utilisation de ressources de calcul pour la simulation de modèles est antérieure à l'introduction des paradigmes de la complexité actuels, remontant par exemple à \noun{Forrester}, informaticien qui a été pionnier des modèles d'économie spatiale inspirés par la cybernétique\footnote{Celle-ci, ainsi que le courant systémique, sont comme nous l'avons déjà développé précurseurs des paradigmes actuels de la complexité.}. Avec l'augmentation des potentialités de calcul, des transformations épistémologiques ont également suivi, avec l'apparition de models explicatifs comme outils expérimentaux. \noun{Rey} compare le dynamisme des années soixante-dix quand les centres de calcul furent ouverts aux géographes à la démocratisation actuelle du Calcul Haute Performance\footnote{Le développement des premiers modèles de simulation urbaine coincide avec l'ouverture des premiers centres de calcul aux sciences humaines, comme le rappelle par ailleurs \noun{Pumain} (entretien du 31/03/2017, voir Annexe~\ref{app:sec:interviews}) par exemple pour l'implémentation du modèle d'entropie d'\noun{Allen}.}. Aujourd'hui, cette facilité d'accès consiste entre autres à du calcul sur grille dont l'utilisation est rendue transparente, c'est-à-dire sans besoin de compétences techniques pointues liées au mécanismes de la distribution des calculs. Ainsi~\cite{schmitt2014half} donnent un exemple des possibilités offertes en termes de calibration et de validation de modèle, réduisant le temps de calcul nécessaire de 30 ans à une semaine - ces techniques jouent un rôle clé pour les résultats que nous obtiendrons par la suite. Cette évolution est également accompagnée par une évolution des pratiques~\cite{banos2013pour} et techniques~\cite{10.1371/journal.pone.0138212} de modélisation.
}

\bpar{
Modeling, and in particular computational models of simulation, is seen by many as a fundamental building brick of knowledge: \cite{livet2010} recalls the combination of empirical, conceptual (theoretical) and modeling domains, with constructive feedbacks between each domain. A model can be an exploration tool to test assumptions, an empirical tool to validate a theory against datasets, an explicative tool to reveal causalities and internal processes of a system, a constructive tool to iteratively build a theory jointly with associated models. These are examples among others: \cite{varenne2010simulations} proposes a classification of diverse functions of a model. We will consider modeling as a fundamental instrument of knowledge on processes within systems, and more particularly in our case within complex adaptive systems. We recall thus that our research question will focus on \emph{models which ontology is mainly composed by interactions between transportation networks and territories}.
}{
La modélisation, et en particulier les modèles de simulation, est vue par beaucoup comme une brique fondamentale de la connaissance : \cite{livet2010} rappelle la combinaison des domaines empirique, conceptuel (théorique) et de la modélisation, avec des retroactions constructives entre chaque. Un modèle peut être un outil d'exploration pour tester des hypothèses, un outil empirique pour valider une théorie sur des jeux de données, un outil explicatif pour révéler des causalités et ainsi des processus internes au système, un outil constructif pour construire itérativement une théorie conjointement avec celle des modèles associés. Ce sont des exemples de fonctions parmi d'autres : \cite{varenne2010simulations} propose une classification des diverses fonctions d'un modèle. Nous considérons la modélisation comme un instrument fondamental de connaissance des processus au sein d'un système, plus particulièrement dans notre cas au sein d'un système complexe adaptatif. Nous rappelons ainsi que notre question de recherche s'intéressera aux \emph{modèles dont l'ontologie contient une part non négligeable d'interactions entre réseaux et territoires}.
}

\subsection{Modeling networks and territories}{Modéliser les territoires et réseaux}

\bpar{
We develop now an overview of different approaches modeling interactions between networks and territories. First of all, we need to notice a high contingency of scientific constructions underlying these. Indeed, according to~\cite{bretagnolle2002time}, the ``\textit{ideas of specialists in planning aimed to give definitions of city systems, since 1830, are closely linked to the historical transformations of communication networks}''. The historical context (and consequently the socio-economical and technological contexts) conditions strongly the formulated theories. It implies that ontologies and corresponding models addressed by geographers and planners are closely linked to their current historical preoccupations, thus necessarily limited in scope and/or operationnal purpose. In a perspectivist vision of science~\cite{giere2010scientific}, such boundaries are the essence of the scientific entreprise, and as we will argue in chapter~\ref{ch:theory} their combination and coupling in the case of models is generally a source of knowledge.
}{
Développons à présent un aperçu des différentes approches modélisant des interactions entre réseaux de transport et territoires. Remarquons de manière préliminaire une forte contingence des constructions scientifiques sous-jacentes à celles-ci. En effet, selon~\cite{bretagnolle2002time}, ``\textit{les idées des spécialistes de la planification cherchant à donner des définitions des systèmes de ville, depuis 1830, sont étroitement liées aux transformations des réseaux de communication}''. Le contexte historique (et donc socio-économique et technologique) conditionne fortement les théories formulées. Cela implique que les ontologies et les modèles correspondants proposés par les géographes et les planificateurs sont fortement liés aux préoccupations historiques courantes, ce qui limite nécessairement leur portée théorique et/ou opérationnelle. Au delà de la question de la définition du système qui joue également un rôle central, on comprend bien l'impact que peut avoir cette influence sur la portée des modèles développés. Dans une vision perspectiviste de la science~\cite{giere2010scientific} de telles limites sont l'essence de l'entreprise scientifique, et comme nous suggèrerons dans le chapitre~\ref{ch:theory} leur combinaison et couplage dans le cas de modèles est généralement une source de connaissance.
}



\bpar{
The entry we take here to sketch an overview of models is complementary to the one taken in chapter~\ref{ch:thematic}, by declining them through their main object (i.e. the relations Network $\rightarrow$ Territory, Territory $\rightarrow$ Network and Territory $\leftrightarrow$ Network)\footnote{We recall the meaning of this notation introduced in chapter~\ref{ch:thematic}: a cirect arrow correspond to processes that we can relatively univocally attribute to the origin, whereas a reciprocal arrow assumes the intrinsic existence of reciprocal interaction, generally in coincidence with the emergence of entities playing a role in these.}.
}{
L'entrée que nous proposons ici pour dresser un aperçu des modèles est complémentaire à celle prise au chapitre~\ref{ch:thematic}, en regardant par objet principal (c'est-à-dire les relations Réseau $\rightarrow$ Territoire, Territoire $\rightarrow$ Réseau et Territoire $\leftrightarrow$ Réseau)\footnote{Nous rappelons la signification de cette notation introduite au chapitre~\ref{ch:thematic} : une flèche directe signifie des processus qu'on peut attribuer relativement de manière univoque à l'origine, tandis qu'une flèche réciproque suppose intrinsèquement l'existence d'interactions réciproques, généralement en coincidence avec l'émergence d'entités jouant un rôle dans celles-ci.}.
}

\bpar{
The reference frame for scales is also the one introduced in chapter~\ref{ch:thematic}, knowling that we do not consider the microscopic scales by choice to discard daily mobility. We have therefore roughly mesoscopic and macroscopic temporal and spatial scales.
}{
Le cadre de lecture des échelles est également celui proposé au chapitre~\ref{ch:thematic}, sachant que nous ne nous intéressons pas aux échelles microscopiques par choix de ne pas considérer la mobilité quotidienne. On a ainsi schématiquement des échelles temporelles et spatiales mesoscopiques et des échelles macroscopiques.
}

\bpar{
We have seen that the correspondence to temporal and spatial scales is not systematic (see the provisional double entry typology for processes). On the contrary, the correspondence to fields of study and types of stakeholders is more systematic. This literature review is thus done following the latest logic.
}{
Nous avons vu que la correspondance à des échelles temporelles et spatiales n'est pas systématique (voir la typologie provisoire à double entrée des processus). Par contre, celle à des domaines particuliers et à des acteurs l'est plus. Cette revue de littérature est donc orientée dans cette seconde direction.
}

\subsubsection{Territories}{Territoires}

\bpar{
The main current dealing with the modeling of the influence of transportation networks on territories lies in the field of planning, at medium temporal and spatial scales (the scales of metropolitan accessibility we developed before). Models in geography at other scales, such as the Simpop models already described~\cite{pumain2012multi}, do not include a particular ontology for transportation networks, and even if they include networks between cities as carriers of exchanges, they do not allow to study in particular the relations between networks and territories. We will come back later on extensions that are relevant for our question. First, let recall the context of models closer to planning studies.
}{
Le courant principal s'intéressant à la modélisation de l'influence du réseau de transport sur les territoires se trouve dans le champ de la planification, à des échelles spatiales et temporelles moyennes (les échelles de l'accessibilité métropolitaine que nous avons développées ci-dessus). Des modèles en géographie à d'autres échelles, comme les modèles Simpop déjà évoqués~\cite{pumain2012multi}, ne supposent pas une ontologie particulière pour le réseau de transport, et s'ils incluent des réseaux entre les villes comme porteur des échanges, ne permettent cependant pas d'étudier en particulier les relations entre réseaux et territoires. Nous reviendrons plus loin sur des extensions pertinentes pour notre question. Revoyons pour commencer un contexte de modèles plus proches des études de planification.
}

\paragraph{LUTI models}{Modèles LUTI}

\bpar{
These approaches are generally named as \emph{models of the interaction between land-use and transportation} (\emph{LUTI}, for \textit{Land-Use Transport Interaction}). Land-use generally means the spatial distribution of territorial activities, generally classified into more or less precise typologies (for example housing, industry, tertiary, natural space). These works can be difficult to apprehend as they relate to different scientific disciplines\footnote{We make here the choice to gather numerous approaches having the common characteristic to principally model the evolution of land-use, on medium temporal and spatial scales. The unity and the relative positioning of these approaches covering from economics to planning, remain an open question that to the best of our knowledge has never been frontally tackled. The work done in~\ref{sec:quantepistemo} introduces elements of answer through an approach in quantitative epistemology.}. Their general principle is to model and simulate the evolution of the spatial distribution of activities, taking transportation networks as a context and significant drivers of localizations. To understand the underlying conceptual frame to most approaches, the Frame~\ref{frame:modelingsa:wegener} sums up the one given by \cite{wegener2004land}\footnote{A more general frame that we already developed, that allows to bridge it with our frame, is the one given by~\cite{le2010approche}, which situates the triad Transportation system/Localization system/Activities system within the relation with agents: agents creating demand, agents building the city, external factors.}.
}{
Ces approches sont désignées de manière générale comme \emph{modèles d'interaction entre usage du sol et transport} (\emph{LUTI}, pour \textit{Land-Use Transport Interaction}). Il est entendu par usage du sol la répartition des activités territoriales, généralement réparties en typologies plus ou moins précises (par exemple logements, industrie, tertiaire, espace naturel). Ces travaux peuvent être difficiles à cerner car liés à différentes disciplines scientifiques\footnote{Nous prenons le parti ici de rassembler de nombreuses approches ayant la caractéristique commune de modéliser principalement l'évolution de l'usage du sol, sur des échelles temporelles et spatiales moyennes. L'unité et le positionnement relatif de ces approches couvrant de l'économie à la planification, sont une question ouverte qui n'a à notre connaissance jamais été traitée de front. Le travail mené en~\ref{sec:quantepistemo} donne des pistes de réponse par une approche d'épistémologie quantitative.}. Leur principe général est de modéliser et simuler l'évolution de la distribution spatiale des activités, en prenant le réseau de transport comme contexte et déterminant significatif des localisations. Pour comprendre le cadre conceptuel sous-jacent à la majorité des travaux, l'Encadré~\ref{frame:modelingsa:wegener} résume celui issu de \cite{wegener2004land}\footnote{Un cadre plus général que nous avons déjà développé, qui permet de faire le pont avec notre cadre, est celui de~\cite{le2010approche}, qui replace le triptyque Système de transport/Système de localisation/Système d'activités en relation avec les agents : agents demandeurs, agents aménageurs, facteurs externes.}.
}

\bpar{
For example, from the point of view of urban economics, propositions for such models have existed for a relatively long time: \cite{putman1975urban} recalls the frame of urban economics in which main components are employments, demography and transportation, and reviews economic models of localization that relate to the \noun{Lowry} model already mentioned.
}{
Par exemple, du point de vue de l'économie urbaine, les propositions de tels modèles existent depuis un certain temps : \cite{putman1975urban} rappelle le cadre d'économie urbaine où les principales composantes sont les emplois, la démographie et le transport, et passe en revue des modèles économiques de localisation qui s'apparentent au modèle de \noun{Lowry} déjà mentionné.
}

\begin{figure}[h!]
	\begin{mdframed}
		
		\bpar{
		\cite{wegener2004land} introduces a general theoretical and empirical frame for land-use transport interaction models. The four concepts included are land-use, localization of activities, the transportation system and the distribution of accessibility. A cycle of circular effects are summed up in the following loop:
		}{
		\cite{wegener2004land} pose un cadre général théorique et empirique pour les modèles d'interaction entre transport et usage du sol. Les quatre concepts mobilisés sont l'usage du sol, la localisation des activités, le système de transport et la distribution de l'accessibilité. Un cycle d'effets circulaires sont résumés dans la boucle suivante :
		}		
		
		\begin{center}
		\bigskip
		\bpar{
		\tikzmark{Activities} $\longrightarrow$ Transportation system $\longrightarrow$ Accessibility $\longrightarrow$ Land-\tikzmark{use}\arrow{use}{Activities}
		}{
		 \tikzmark{Activités} $\longrightarrow$ Système de Transport $\longrightarrow$ Accessibilité $\longrightarrow$ Usage \tikzmark{du} sol\arrow{du}{Activités}
		}		
		\bigskip
		\end{center}
		
		\bpar{
		The transportation system is assumed with a \emph{fixed infrastructure}, i.e. effects of the distribution of activities are effects on the \emph{use} of the transportation system (and thus link to \emph{mobility} in our more general frame): modal choice, frequency of trips, length of travels.
		}{
		Le système de transport est supposé \emph{à infrastructure fixe}, c'est-à-dire que les effets de la distribution des activités sont ceux sur \emph{l'utilisation} du système de transport (donc liés à la \emph{mobilité} dans notre cadre plus général) : choix modal, fréquence des voyages, longueur des voyages.
		}
		
		\bpar{
		The theoretically expected effects are classified according to the direction of the relation (\textit{Land-use}$\rightarrow$\textit{Transport} or \textit{Transport}$\rightarrow$\textit{Land-use}, and a loop \textit{Transport}$\rightarrow$\textit{Transport} that is not taken into account in our case), and according to the acting factor (residential density, of employments, localization, accessibility, transportation costs) and also by the aspect that is modified (length and frequency of trips, modal choice, densities, localizations). We can for example take:
		\begin{itemize}
		\item \textit{Land-use}$\rightarrow$\textit{Transport}: a minimal residential density is necessary for the efficiency of public transportation, a concentration of employments implies longer trips, larger cities have a greater proportion of the modal part of public transportation.
		\item \textit{Transport}$\rightarrow$\textit{Land-use}: a high accessibility implies higher prices and an increased development of residential housing, companies locate for a better accessibility to transportation at a larger scale.
		\item \textit{Transport}$\rightarrow$ \textit{Transport}: places with a good accessibility will produce more and longer trips, modal choice and transportation cost are highly correlated. 
		\end{itemize}
		}{
		Les effets théoriquement attendus sont classés selon les directions de la relation (\textit{Usage du sol}$\rightarrow$\textit{Transport} ou \textit{Transport}$\rightarrow$\textit{Usage du sol}, ainsi qu'une boucle \textit{Transport}$\rightarrow$\textit{Transport}, qui fait partie des éléments ignorés dans notre cas), et par ailleurs par facteur agissant (densité résidentielle, d'emplois, localisation, accessibilité, coûts de transport) ainsi que par aspect affecté (longueur et fréquence des voyages, choix de mode, densités, localisations).  On peut par exemple citer :
		\begin{itemize}
			\item \textit{Usage du sol}$\rightarrow$\textit{Transport} : une densité résidentielle minimale est nécessaire pour l'efficience du transport public, une concentration des emplois implique des voyages plus long, les villes plus grandes ont une part modale plus importante pour les transports en commun.
			\item \textit{Transport}$\rightarrow$\textit{Usage du sol} : une forte accessibilité implique des prix plus élevés et un développement accru pour le résidentiel, les entreprises se localisent pour une meilleure accessibilité aux moyens de transport à grande échelle. 
			\item \textit{Transport}$\rightarrow$\textit{Transport} : les lieux avec une bonne accessibilité produiront plus et de plus longs voyages, le choix modal et le coût de transport sont fortement corrélés.
		\end{itemize}
		}
		
		\bpar{
		These theoretical effects are then compared to empirical observations, which for most of them give the way processes are implemented. Some are not observed in practice, whereas most converge with theoretical expectations.		
		}{
		Ces effets théoriques sont par ailleurs comparés aux observations empiriques, qui pour la plupart donnent la manière d'implémentation du processus. Certains ne sont pas observés en pratique, tandis que la plupart sont en accord avec les attentes théoriques.
		}		
		
		\bigskip
		
		\bpar{
		\textit{Comment 1: An uniscalar framework ?} This framework takes schematically into account two main scales, the scale of daily mobility and the scale of the localization of activities. Knowing that in practice mobility behaviors are generally taken into account as average flows, it often reduces to a unique mesoscopic scale. All in all, it does not allow to take into account dynamics on longer time scales, that would include the evolution of the transportation network infrastructure or structural dynamics of systems of cities on long time periods.
		}{
		\textit{Commentaire 1 : Un cadre uniscalaire ?} Ce cadre prend en compte schématiquement deux échelles principales, celle de la mobilité quotidienne et celle de la localisation des activités. Sachant qu'en pratique les comportement de mobilité sont généralement pris en compte sous forme de flux moyens, il se réduit souvent à une unique échelle mesoscopique. Dans tous les cas, il ne permet pas de tenir compte de dynamiques sur le temps plus long comprenant l'évolution de l'infrastructure du système de transport ou des dynamiques structurelles des systèmes de villes sur le temps long.
		}		
		
		\bigskip
		
		\bpar{
		\textit{Comment 2: A systematic view of structuring effects ?} Furthermore, critics of the rhetoric of structuring effects may find in this framework its strong presence, since direct effects of accessibility on land-use and then the localization of activities are assumed here. These critics can be undermined by observing that these are theoretical expected effects, and that the framework is put into perspective of empirical effects indeed observed. We will however always take it with caution, by situating it in terms of context and scales.
		}{
		\textit{Commentaire 2 : Une vision systématique des effets structurants ?} Par ailleurs, les pourfendeurs de la rhétorique des effets structurants trouveront en ce cadre leur bête noire, puisque les effets directs de l'accessibilité sur l'usage du sol puis sur la localisation des activités sont postulés ici. Ces critiques pourront être refoulées par l'observation qu'il s'agit des effets attendus théoriques, et que le cadre est mis en perspective des effets empiriques effectivement observés. Il sera à cependant à prendre avec précaution, en le situant toujours en terme de contexte et d'échelles.
		}

		\bigskip
		
		\framecaption{\textbf{Conceptual framework of land-use transport interactions according to \cite{wegener2004land}.}\label{frame:modelingsa:wegener}}{\textbf{Cadre conceptuel des interactions entre transport et usage du sol selon \cite{wegener2004land}.}\label{frame:modelingsa:wegener}}		
	\end{mdframed}
\end{figure}

\bpar{
\cite{wegener2004land} give more recently a state of the art of empirical studies and in modeling on this type of approach of interactions between land-use and transport. The theoretical positioning is closer of disciplines such as transportation socio-economics and planning (see the disciplinary landscapes described in~\ref{sec:quantepistemo}). \cite{wegener2004land} compare and classify seventeen models, among which no one includes an endogenous evolution of the transportation network on relatively short time scales for simulations (of the order of the decade). We find again indeed the correspondance with typically mesoscopic scales previously established. A complementary review is done by~\cite{chang2006models}, broadening the context with the inclusion of more general classes of models, such as spatial interactions models (which contain trafic assignment and four steps models), planing models based on operational research (optimization of locations of different activities, generally homes and employments), the microscopic models of random utility, and models of the real estate market.
}{
\cite{wegener2004land} donnent plus récemment un état de l'art des études empiriques et de modélisation sur ce type d'approche des interactions entre usage du sol et transport. Le positionnement théorique est plutôt proche des disciplines de la socio-économie des transports et de la planification (voir les paysages disciplinaires dressés en~\ref{sec:quantepistemo}). \cite{wegener2004land} comparent et classifient dix-sept modèles, parmi lesquels aucun n'inclut une évolution endogène du réseau de transport sur les échelles de temps relativement courtes (de l'ordre de la décade) des simulations. On retrouve bien la correspondance avec les échelles typiquement mesoscopiques établies précédemment. Une revue complémentaire est faite par~\cite{chang2006models}, élargissant le contexte avec l'inclusion de classes plus générales de modèles, comme des modèles d'interactions spatiales (parmi lesquels l'attribution du traffic et les modèles à quatre temps), les modèles de planification basés sur la recherche opérationnelle (optimisation des localisations des différentes activités, généralement résidences et emplois), les modèles microscopiques d'utilité aléatoire, et les modèles de marché foncier.
}

\begin{figure}
	\begin{mdframed}
	
	\bpar{
	The Pirandello\textregistered model\footnote{The origin of the name is not given, but strongly suggests the influence of its original creators \noun{V. Piron} and \noun{J. Delons}.} is presented in \cite{delons:hal-00319087} as one of the first attempts to develop an operational Luti model in France. The model is based on four fundamental economic processes: the real estate market and the dwellings offer, the residential mobility of households, the attribution of travel destinations, the model choice. The model is static, i.e. computes n equilibrium for spatial distributions of actives and employments, and also for transportation flows.
	}{
	Le modèle Pirandello\textregistered\footnote{L'origine du nom n'est pas donnée, mais suggère fortement l'influence de ses créateurs originaux \noun{V. Piron} et \noun{J. Delons}.} est présenté dans \cite{delons:hal-00319087} comme l'une des premières tentatives de développement de modèle Luti opérationnel en France. Le modèle se base sur quatre processus économiques fondamentaux : le marché du foncier et l'offre de logement, la mobilité résidentielle des ménages, l'attribution des destinations de déplacement, le choix modal. Le modèle est statique, c'est-à-dire calcule un équilibre pour les distributions spatiales des actifs et des emplois, ainsi que pour les flux de transport.
	}

	\bpar{
	The fundamental processes taken into account and their implementation are the following:
	\begin{itemize}
		\item Residential choices of households are based on a utility function taking into account (i) a confort term as a Cobb-Douglas of housing surface and income, corrected by a linear preference for individual dwellings; (ii) an accessibility term based on generalized cost (aggregation of transportation cost and time, with a value of time); (iii) the dwelling price and the local tax as a function of the housing surface; (iv) a fixed effect by income and by area; and (v) a random term assumed to follow a Gumbel law. Location probabilities for an income group are then given by a discrete choice model given this utility.
		\item The housing prices are formed following a scaling law of population.
		\item A local bidding mechanism answers to the demand previously obtained, as a function of an exogenous dwelling offer.
		\item Companies locate by maximizing their profit, function of the productivity (Cobb-Douglas in the salary and the accessibility) and the real estate price, under the constraint of a fixed spatial distribution of the number of employments, of the office surface, and of the total production of the region.
		\item Transportation is taken into account through a four steps model, which distributes model choices and destination choices with a discrete choice model, and flows are assigned according to a Wardrop equilibrium (see~\ref{sec:reproducibility}), what allows to adjust the values of accessibility given a spatial distribution of activities.
	\end{itemize}
	}{
	Les processus fondamentaux pris en compte et leur implémentation sont les suivants :
	\begin{itemize}
		\item Les choix résidentiels des ménages se basent sur une fonction d'utilité prenant en compte (i) un terme de confort en Cobb-Douglas de la surface et du revenu, corrigée par une préférence linéaire pour les logements individuels ; (ii) un terme d'accessibilité basé sur le coût généralisé (agrégation du coût de transport et du temps, avec un prix du temps) ; (iii) le prix du logement et de la taxe locale en fonction de la surface de logement ; (iv) un effet fixe par revenu et par zone ; et (v) un terme aléatoire supposé suivre une loi de Gumbel. Les probabilités de localisation pour une tranche de revenus sont alors données par un modèle de choix discret étant donné cette utilité.
		\item Les prix du logement sont formés selon une loi d'échelle de la population.
		\item Un système d'enchère local répond à la demande obtenue précédemment, en fonction d'une offre de logement exogène.
		\item Les entreprises se localisent en maximisant leur profit, fonction de la productivité (Cobb-Douglas du salaire et de l'accessibilité) et du prix du foncier, sous contrainte d'une distribution spatiale fixée du nombre d'emplois, de la surface de bureaux, et de la production totale de la région.
		\item Le transport est pris en compte par un modèle à quatre étapes, qui distribue les choix modaux et les choix de destination par un modèle de choix discrets, et les flux assignés selon un équilibre de Wardrop (voir~\ref{sec:reproducibility}), ce qui permet d'ajuster les valeurs de l'accessibilité étant donné une distribution spatiale des activités.
	\end{itemize}
	}

	\bpar{
	The mechanism to combine these different processes to obtain a global equilibrium is detailed by~\cite{kryvobokov2013comparison}, and consists in the establishment of three sub-equilibriums at different scales: transportation flows (giving costs) on a short term, location and real estate prices on the middle term, land prices and available terrains (fixed in an exogenous way for all the modeled period).
	}{
	Le mécanisme de combinaison de ces différents processus pour obtenir un équilibre global est détaillé par~\cite{kryvobokov2013comparison}, et consiste à l'établissement de trois sous-équilibres à différentes échelles : flux de transport (donnant les coûts) sur le court terme, localisation et prix de l'immobilier sur le moyen terme, prix du foncier et terrain disponibles (fixés de manière exogène pour l'ensemble de la période modélisée).
	}
	
	\bigskip

	\bpar{
	\textit{Commentary: Equilibrium, operational model and calibration.} A certain number of remarks can be done concerning this model, the most important for our approach are: (i) the equilibrium assumption can be a powerful tool to understand the structure of the attractors of the system, but has no empirical foundation, and even less for the coupling of equilibriums at different scales; (ii) thus, the operational nature of the model can be discussed, since the study of the impact of scenarios on the movements of attractors can difficultly allow to infer on local dynamics of the system; and (iii) sub-models are calibrated more or less rigorously and relatively separately, but the conditions of a calibration by decomposition are an open question still not well explored and linked to the nature of model coupling. In our sense, such a micro-based model would in any case be in better consistence with a philosophy of dynamical generative modeling and parsimony (see\ref{sec:computation}).
	}{
	\textit{Commentaire : Équilibre, modèle opérationnel et calibration.} Un certain nombre de remarques peuvent être faites à ce modèle, les plus importantes pour notre approche étant : (i) l'hypothèse d'équilibre peut être un outil puissant pour comprendre la structure des attracteurs du système, mais n'a pas de fondement empirique, et encore moins pour le couplage d'équilibres à différentes échelles ; (ii) ainsi, la nature opérationnelle du modèle est discutable, puisque l'étude de l'impact de scenarios sur les déplacements des attracteurs permet difficilement d'inférer sur les dynamiques locales du système ; et (iii) les sous-modèles sont calibrés plus ou moins rigoureusement et relativement séparément, or les conditions d'un calibrage par décomposition sont une question ouverte encore peu explorée et liée à la nature du couplage de modèles. En notre sens, un tel modèle micro-fondé serait dans tous les cas en meilleure cohérence avec une philosophie de modélisation générative dynamique et de parcimonie (voir~\ref{sec:computation}).
	}
	
	\medskip
	
	\framecaption{\textbf{The Pirandello model.}\label{frame:modelingsa:pirandello}}{\textbf{Le modèle Pirandello.}\label{frame:modelingsa:pirandello}}
	
	\end{mdframed}
\end{figure}

\begin{figure}
	\begin{mdframed}
	
	\bpar{
	The Nedum2D model, described in details by~\cite{viguie2014downscaling}, is focused on the localization of actives and their interaction with land rent and real estate promoters: it is a model inspired by the Fujita-Ogawa model~\cite{fujita1982multiple}, inheriting from the literature in Urban Economics.
	}{
	Le modèle Nedum2D, décrit en détail dans~\cite{viguie2014downscaling}, se concentre sur la localisation des actifs et leur interaction avec la rente foncière et les promoteurs immobiliers : il s'agit d'un modèle inspiré du modèle de Fujita-Ogawa~\cite{fujita1982multiple}, héritant de la littérature en Economie Urbaine.
	}
	
	\bpar{
	The processes included in the model are, with each its own time scale fixed by a parameter:
	\begin{itemize}
		\item Households make a compromise between housing surface and available budget without transportation costs and rent, following a Cobb-Douglas function for the corresponding utility. This process induces a dynamic for housing surface as a function of the distance to the center.
		\item They relocate in order to have an expected utility larger than the average.
		\item Rents evolve to maximize the occupation or in response to an external demand.
		\item New buildings are built by promoters that aim at maximizing their profits.
	\end{itemize}
	}{
	Les processus inclus dans le modèle sont, chacun ayant une échelle de temps particulière fixée par un paramètre :
	\begin{itemize}
		\item Les ménages font un compromis entre surface de logement et budget disponible hors coûts de transports et loyer, suivant une fonction de Cobb-Douglas pour l'utilité correspondante. Ce processus induit une dynamique pour la surface des logements en fonction de la distance au centre.
		\item Ils se relocalisent pour avoir une utilité moyenne plus grande que la moyenne.
		\item Les loyers évoluent pour maximiser l'occupation ou en réponse à une demande extérieure.
		\item De nouveaux bâtiments sont construits par des promoteurs cherchant à maximiser leur profits.
	\end{itemize}
	}
	
	\bpar{
	This model is dynamical and simulates the evolution of these different variables in space (the formulation above is monocentric, a polycentric extension and one taking into account an exogenous distribution of employments exist) and time. Its spatial scale is metropolitan, and the time scale can range from a medium scale (decade) to longer time-periods (century), knowing that the latest has a low credibility since it keeps static numerous other components of the urban system.
	}{
	Ce modèle est dynamique et simule l'évolution de ces différentes variables dans l'espace (la formulation ci-dessus est monocentrique, une variante polycentrique et prenant en compte une distribution exogène d'emplois existent) et dans le temps. Son échelle spatiale est métropolitaine, et l'échelle de temps peut s'étendre d'une échelle moyenne (décade) à du temps plus long (siècle), sachant que cette dernière est peu crédible puisque qu'elle garde statique de nombreuses autres composantes du système urbain.
	}
	
	\medskip
	
	\bpar{
	\textit{Comment: extension of ontologies.} The coupling of Nedum with a model for traffic assignment, the Modus model\footnote{In the frame of the current research project ANR VITE! (see \url{http://www.agence-nationale-recherche.fr/Projet-ANR-14-CE22-0013}).}, aims at including the feedback of congestion in the transportation system on costs, and thus on the localization and on the urban structure. Fundamental questions arise from the first coupling experiments:
	\begin{itemize}
		\item Is the masterplan \emph{Schéma Directeur} really useful, since is seems to only accompany already existing dynamics ? In other words, \textit{is the governance process endogenous} ? Does the Sdrif in fact capture an intrinsic dynamic on a longer time ?
		\item The coupling of models raises in itself technical difficulties, for communication between modules already implemented in different languages and for convergence of the coupled model in a reasonable number of iterations.
		\item It furthermore raises ontological difficulties: each model includes opposite mechanisms for the same ontology (aggregation effect against congestion effect for the distribution of population). The question is then if a specific coupling ontology is necessary (for example with specific equations integrating these contradictory effects), to allow on the one hand a better convergence, on the other hand a better ontological consistency.
	\end{itemize}
	}{
	\textit{Commentaire : extension des ontologies.} Le couplage de Nedum avec un modèle d'attribution de transport, le modèle Modus\footnote{Dans le cadre du projet en cours de réalisation ANR VITE! (voir \url{http://www.agence-nationale-recherche.fr/Projet-ANR-14-CE22-0013}).}, vise à inclure la retroaction de la congestion dans le système de transport sur les coûts, et donc sur la localisation et la structure urbaine. Des questions fondamentales se dégagent des premières expériences de couplage :
	\begin{itemize}
		\item Le schéma directeur est-il vraiment utile, puisqu'il ne semble qu'accompagner des dynamiques déjà présentes ? En d'autres termes, \textit{le processus de gouvernance est-il endogène} ? Le Sdrif capture-t-il en fait une dynamique intrinsèque sur le temps plus long ?
		\item Le couplage des modèles pose en lui-même des difficultés techniques, de communication entre des modules déjà implémentés dans différents langages et de convergence du modèle couplé en un nombre raisonnable d'itérations.
		\item Il pose d'autre part des difficultés ontologiques : chaque modèle inclut des mécanismes opposés pour la même ontologie (effet d'agrégation contre congestion pour la distribution de la population). La question se pose s'il faut rajouter spécifiquement une ontologie de couplage (par exemple des equations spécifiques intégrant ces effets contradictoires), pour permettre d'une part une meilleure convergence, d'autre part une meilleure cohérence ontologique.
	\end{itemize}
	}

	
	\medskip
	
	\framecaption{\textbf{The Nedum model.}\label{frame:modelingsa:nedum}}{\textbf{Le modèle Nedum.}\label{frame:modelingsa:nedum}}
	
	\end{mdframed}
\end{figure}

\bpar{
In order to give a better intuition of the logic underlying some Luti models, we detail in Frame~\ref{frame:modelingsa:pirandello} and in Frame~\ref{frame:modelingsa:nedum} the structures, the ontologies, and assumptions of two models developed in the specific case of Ile-de-France (allowing on the one hand a comparison between both and on the other hand echoing the thematic development of~\ref{sec:casestudies}). Even for very close ontologies (real estate prices, households localizations), we see the variety of possible assumptions and of issues raised by the models.
}{
Afin de donner une meilleure intuition de la logique sous-jacente à certains modèles Luti, nous détaillons en Encadré~\ref{frame:modelingsa:pirandello} et en Encadré~\ref{frame:modelingsa:nedum} les structures, les ontologies et les hypothèses de deux modèles développés dans le cas spécifique de l'Ile-de-France (permettant d'une part la comparaison entre les deux et d'autre part donnant un écho au développement thématique de~\ref{sec:casestudies}). Même pour des ontologies très proches (prix immobiliers, localisation des ménages), on voit la variété d'hypothèses possibles et de problématiques soulevées par les modèles.
}

\paragraph{Very different operational models}{Des modèles opérationnels très variés}

\bpar{
The variety of existing models lead to operational comparisons: \cite{paulley1991overview} synthesize a project comparing different model applied to different cities. Their result allow on the one hand to classify interventions depending on their impact on the level of interaction between transportation and land-use, and on the other hand to show that the effects of interventions strongly depend on the size of the city and on its socio-economic characteristics.
}{
La variété des modèles existants a conduit à des comparaisons opérationnelles : \cite{paulley1991overview} rendent compte d'un projet comparant différents modèles appliqués à différentes villes. Leurs résultats permettent d'une part de classifier des interventions en fonction de leur impact sur le niveau d'interaction entre transport et usage du sol, et d'autre part de montrer que l'effet des interventions dépend fortement de la taille de la ville et de ses caractéristiques socio-économiques.
}

\bpar{
Ontologies of processes, and more particularly on the question of equilibrium, are also varied. The respective advantages of a static approach (computation of a static equilibrium of households localisation for a given specification of their utility functions) and of a dynamical approach (out-of-equilibrium simulation of residential dynamics) has been studied by~\cite{kryvobokov2013comparison}, within a metropolitan frame on time scales of the order of the decade. The authors show that results are roughly comparable and that each model has its utility depending on the question asked.
}{
Les ontologies des processus, et notamment sur la question de l'équilibre, sont aussi variées. Les avantages respectifs d'une approche statique (calcul d'un équilibre statique de la localisation des ménages pour une certaine spécification de leur fonctions d'utilité) et d'une approche dynamique (simulation hors équilibre des dynamiques résidentielles) a été étudié par~\cite{kryvobokov2013comparison}, dans un cadre métropolitain sur des échelles de temps de l'ordre de la décennie. Les auteurs montrent que les résultats sont globalement comparables et que chaque modèle a son utilité selon la question posée.
}

\bpar{
Different aspects of the same system can be included within diverse models, as show for example~\cite{wegener1991one}, and traffic, residential and employments dynamics, the evolution of land-use as a consequence, also influenced by a static transportation network, are generally taken into account. \cite{iacono2008models} covers a similar horizon with an additional development on cellular automata models for the evolution of land-use and agent-based models. The temporal range of application of these models, around the decade, and their operational nature, make them useful for planning, what is rather far of our focus to obtain explicative models of geographical processes. Indeed, it is often more relevant for a model used in planning to be understandable as an anticipation tool, or even a communication tool, than to be faithful to territorial processes, at the cost of an abstraction.
}{
Différents aspects du même système peuvent être traduits par divers modèles, comme le montre par exemple~\cite{wegener1991one}, et le trafic, les dynamiques résidentielles et d'emploi, l'évolution de l'usage du sol en découlant, influencée aussi par un réseau de transport statique, sont généralement pris en compte. \cite{iacono2008models} couvre un horizon similaire avec un développement supplémentaire sur les modèles à automates cellulaires d'évolution d'usage du sol et les modèles à base d'agents. La portée temporelle d'application de ces modèles, de l'ordre de la décennie, et leur nature opérationnelle les rend utiles pour la planification, ce qui est assez loin de notre souci d'obtenir des modèles explicatifs de processus géographiques. En effet, il est souvent plus pertinent pour un modèle utilisé en planification d'être lisible comme outil d'anticipation, voire de communication, que d'être fidèle aux processus territoriaux au prix d'une abstraction.
}

\paragraph{Perspectives for LUTI models}{Perspectives pour les modèles LUTI}

\bpar{
\cite{timmermans2003saga} formulates doubts regarding the possibility of interaction models that would be really integrated, i.e. producing endogenous transportation patterns and being detached from artefacts such as accessibility for which the influence of its artificial nature remains to be established, in particular because of the lack of data and a difficulty to model governance and planning processes. It is interesting to note that current priorities for the development of LUTI models seem to be centered on a better integration of new technologies and a better integration with planning and decision-making processes, for example through visualization interfaces as proposed by~\cite{JTLU611}. They do not aim at being extended on problematics of territorial dynamics including the network on longer time scales for example, what confirms the range and the logic of use and development of this type of models.
}{
\cite{timmermans2003saga} émet des doutes quant à la possibilité de modèles d'interaction réellement intégrés, c'est-à-dire produisant des motifs de transports endogènes et se détachant d'artefacts comme l'accessibilité dont l'influence du caractère artificiel reste à établir, notamment à cause du manque de données et une difficulté à modéliser les processus de gouvernance et de planification. Il est intéressant de noter que les priorités actuelles de développement des modèles LUTI semblent centrées sur une meilleure intégration des nouvelles technologies et une meilleure intégration avec la planification et les processus de prise de décision, par exemple via des interfaces de visualisation comme le propose~\cite{JTLU611}. Ils ne cherchent pas à s'étendre à des problématiques de dynamiques territoriales incluant le réseau sur de plus longues échelles par exemple, ce qui confirme la portée et la logique d'utilisation et de développement de ce type de modèles.
}

\bpar{
A generalization of this type of approach at a smaller scale, such as the one proposed by \cite{russo2012unifying}, consists in the coupling between a LUTI at the mesoscopic scale to macroeconomic models at the macroscopic scale\footnote{\cite{russo2012unifying} indeed generalizes the framework of LUTI models to propose a framework of interaction between spatial economy and transportation (\emph{Spatial Economics and Transport Interactions}). This framework includes LUTI models at the urban scale, and at the national level macroeconomic models simulating production and consumption, competition between activities, production of the stock of the offer of transportation. Transportation models still assume a fixed network and establish equilibria within it, what implies a small spatial scale and a short time scale.}. These do not consider the evolution of the transportation network in an explicit manner but are interested only to abstract patterns of demand and offer. Urban economics have developed specific approaches that are similar in their context: \cite{masso2000} for example describes an integrated model coupling urban development, relocations and equilibrium of transportation flows.
}{
Une généralisation de ce type d'approche à une plus petite échelle, comme celle proposée par \cite{russo2012unifying}, consiste au couplage du LUTI à l'échelle mesoscopique à des modèles macroéconomiques à l'échelle macroscopique\footnote{\cite{russo2012unifying} généralise en fait le cadre des LUTI pour proposer un cadre d'interaction entre Economie Spatiale et Transports (\emph{Spatial Economics and Transport Interactions}). Celui-ci inclut les LUTI à l'échelle urbaine, et au niveau national les modèles macroéconomiques simulant production et consommation, compétition des activités, production du stock d'offre de transport. Les modèles de transport supposent toujours réseau fixe et établissent des équilibres au sein de celui-ci, ce qui implique une petite échelle spatiale et une courte échelle temporelle.}. Ceux-ci ne considèrent pas l'évolution du réseau de transport de manière explicite mais s'intéressent seulement aux motifs abstraits d'offre et demande. L'économie urbaine a développé des approches spécifiques similaires dans leur démarche : \cite{masso2000} décrit par exemple un modèle intégré couplant développement urbain, relocalisation et équilibre des flux de transports.
}

\bpar{
Thus, we can synthesize this type of approach, that we can designate through a semantic shortcut as \emph{LUTI approaches}, by the fundamental following characteristics: (i) models aiming at understanding an evolution of the territory, within the context of a given transportation network; (ii) models in a logic of planning and applicability, being themselves often implied in decision-making; and (iii) models at medium scales, in space (metropolitan scale) and in time (decade).
}{
Ainsi, nous pouvons synthétiser ce type d'approche, qu'on pourra désigner par abus de langage \emph{approche LUTI}, par les caractéristiques fondamentales suivantes : (i) modèles visant à comprendre une évolution du territoire, dans le contexte d'un réseau de transport donné ; (ii) modèles dans une logique de planification et d'applicabilité, étant souvent impliqués eux-même dans les prises de décision ; et (iii) modèles à des échelles moyennes, dans l'espace (métropole) et dans le temps (décennie).
}

\subsubsection{Network Growth}{Croissance du Réseau}

\bpar{
We can now switch to the ``opposite'' paradigm, focused on the evolution of the network. It may seem strange to consider a variable network while neglecting the evolution of the territory, when considering the overview of some potential evolution mechanisms we previously reviewed (potential breakdown, self-reinforcements, network planning) which occur at mainly longer time scales than territorial evolutions. We will see here that there is no paradox, since (i) either the modeling focuses on the evolution of \emph{network properties}, at a short scale (micro) for congestion, capacity, tarification processes, mainly from an economic point of view; (ii) or territorial components playing indeed a role on the network are stable on the long scales considered.
}{
Passons à présent au paradigme ``opposé'', centré sur l'évolution du réseau. Il peut sembler incongru de considérer un réseau variable en négligeant les variations du territoire, au regard de l'aperçu de certains des mécanismes potentiels d'évolution revus précédemment (rupture de potentiel, auto-renforcements, planification du réseau) qui se produisent à des échelles de temps majoritairement plus longues que les évolutions territoriales. On verra ici qu'il n'y a pas de paradoxe, vu que (i) soit la modélisation s'intéresse à l'évolution des \emph{propriétés du réseau}, à une courte échelle (micro) pour des processus de congestion, de capacité, de tarification, principalement d'un point de vue économique ; (ii) soit les composantes territoriales jouant en effet sur le réseau sont stables au échelles longues considérés.
}

\bpar{
Network growth is the subject of modeling approaches which aim at explaining the growth of transportation networks. They generally take a \emph{bottom-up} and endogenous point of view, i.e. aiming at unveiling local rules that would allow to reproduce the growth of the network on long time scales (often the road network). As we will see, it can be a topological growth (creation of new links) or the growth of link capacities in relation with their use, depending on scales and ontologies considered. To simplify, we distinguish broad disciplinary streams having studied the modeling of the growth of transportation networks: these are respectively linked to transportation economics, physics, transportation geography, and biology.
}{
La croissance de réseaux est l'objet de démarches de modélisation qui cherchent à expliquer la croissance des réseaux de transport. Ils prennent généralement un point de vue \emph{bottom-up} et endogène, c'est-à-dire cherchant à mettre en évidence des règles locales qui permettraient de reproduire la croissance du réseau sur de longues échelles de temps (souvent le réseau routier). Comme nous allons le voir, il peut s'agir de la croissance topologique (création de nouveaux liens) ou la croissance des capacités des liens en relation avec leur utilisation, selon les échelles et les ontologies considérées. Nous distinguons pour simplifier des grands courants disciplinaires s'étant intéressé à la modélisation de la croissance des réseaux de transport : ceux-ci sont liés respectivement à l'économie des transports, la physique, la géographie des transport et la biologie.
}

\bpar{
We thus partly rejoin the classification by~\cite{xie2009modeling}, which proposes an extended review of modeling the growth of transportation networks, in a perspective of transportation economics but broadened to other fields. \cite{xie2009modeling} distinguishes broad disciplinary streams having studied the growth of transportation networks: transportation geography has developed very early models based on empirical facts but which have focused on reproducing topology rather than mechanisms\footnote{According to \cite{xie2009modeling}, the contribution of geography consists in limited efforts at the period of \cite{haggett1970network}, we will therefore build on this review and not give a more thorough development.}; statistical models on case studies produce very limited conclusions on causal relations between network growth and demand (growth being in that case conditioned to demand data); economists have studied the production of infrastructure both from a microscopic and macroscopic point of view, generally not spatialized; network science has produced stylized models of network growth which are based on topological and structural rules rather than rules built on processes corresponding to empirical facts.
}{
On rejoint ainsi partiellement la classification de~\cite{xie2009modeling}, qui propose une revue étendue de la modélisation de croissance des réseaux, dans une perspective d'économie des transports mais en élargissant à d'autres champs. \cite{xie2009modeling} distingue des grands courants disciplinaires ayant étudié la croissance des réseaux de transport : la géographie des transports a développé très tôt des modèles basés sur des faits empiriques mais qui ont visé à reproduire la topologie plutôt que sur les mécanismes\footnote{Selon \cite{xie2009modeling}, la contribution de la géographie consiste en des efforts limités à l'époque de \cite{haggett1970network}, nous nous baserons donc sur cette revue et ne donnerons pas de développement approfondi.} ; les modèles statistiques sur des cas d'étude fournissent des conclusions très mitigées sur les relations causales entre croissance du réseau et demande (la croissance étant dans ce cas conditionnée aux données de demande) ; les économistes ont étudié la production d'infrastructure à la fois d'un point de vue microscopique et macroscopique, généralement non spatialisés ; la science des réseaux a produit des modèles stylisés de croissance de réseau qui se basent sur des règles topologiques et structurelles plutôt que des règles se reposant sur des processus correspondant à des réalités empiriques.
}

\paragraph{Economics}{Economie}

\bpar{
Economists have proposed models of this type: \cite{zhang2007economics} reviews transportation economics literature on network growth, recalling the three main features studied by economists on that subject, that are road pricing, infrastructure investment and ownership regime, and finally describes an analytical model combining the three. These three classes of processes are related to an interaction between microscopic economic agents (users of the network) and governance agents. Models can include a detailed description of planning processes, such as~\cite{levinson2012forecasting} which combines qualitative surveys with statistics to parametrize a network growth model.  \cite{xie2009jurisdictional} compares the relative influence of centralized (planning by a governance structure) and decentralized growth processes (local growth which does not enters the frame of a global planning). 
 \cite{yerra2005emergence} shows through a reinforcement economic model including investment rule based on traffic assignment that local rules are enough to make hierarchy of roads emerge for a fixed land-use. \cite{levinson2003induced} proceed to an empirical study of drivers of road network growth for \emph{Twin Cities} in the United States (Minneapolis-Saint-Paul), establishing that basic variables (length, accessibility change) have the expected behavior, and that there exists a difference between the levels of investment, implying that local growth is not affected by costs, what could correspond to an equity of territories in terms of accessibility. The same data are used by~\cite{zhang2016model} to calibrate a network growth model which superimposes investment decisions with network use patterns. \cite{yerra2005emergence} shows with an economic model based on self-reinforcement processes (i.e. that include a positive feedback of flows on capacity) and which includes an investment rule based on traffic assignment, that local rules are sufficient to make a hierarchy of the road network emerge with a fixed land-use. A synthesis of these works gravitating around \noun{Levinson} is done in~\cite{xie2011evolving}.
}{
Les économistes ont proposé des modèles de ce type : \cite{zhang2007economics} passe en revue la littérature en économie des transports sur la croissance des réseaux, rappelant les trois aspects principalement traités par les économistes sur le sujet, qui sont la tarification routière, l'investissement en infrastructures et le régime de propriété, et propose finalement un modèle analytique combinant les trois. Ces trois classes de processus relèvent d'une interaction entre les agents économiques microscopiques (utilisateurs du réseau) et les agents de gouvernance. Les modèles peuvent inclure une description détaillée des processus de planification, comme~\cite{levinson2012forecasting} qui combinent des enquêtes qualitatives et des statistiques pour paramétrer un modèle de croissance de réseau. \cite{xie2009jurisdictional} comparent l'influence relative des processus de croissance centralisés (planification par une structure de gouvernance) et décentralisés (croissance locale ne rentrant pas dans le cadre d'une planification globale). \cite{levinson2003induced} procèdent à une étude empirique des déterminants de la croissance du réseau routier pour les \emph{Twin Cities} aux Etats-Unis (Minneapolis-Saint-Paul), établissant que les variables basiques (longueur, changement dans l'accessibilité) ont le comportement attendu, et qu'il existe une différence entre les niveaux d'investissement, impliquant que la croissance locale n'est pas affectée par les coûts, ce qui peut correspondre à une équité des territoires en termes d'accessibilité. Ces données sont utilisées par~\cite{zhang2016model} pour calibrer un modèle de croissance de réseau qui superpose les décisions d'investissement aux motifs d'utilisation du réseau. \cite{yerra2005emergence} montre avec un modèle économique basé sur des processus d'auto-renforcement (c'est-à-dire incluant une rétroaction positive des flux sur la capacité) et incluant une règle d'investissement basée sur l'attribution du trafic, que des règles locales sont suffisantes pour faire émerger une hiérarchie du réseau routier à usage du sol fixé. Une synthèse de ces travaux gravitant autour de \noun{Levinson} est faite dans~\cite{xie2011evolving}.
}


\paragraph{Physics}{Physique}

\bpar{
Physics has recently introduced infrastructure network growth models, largely inspired by this economic literature: a model which is very similar to the last we described is given by~\cite{louf2013emergence} with simpler cost-benefit functions by obtaining a similar conclusion. Given a distribution of nodes (cities)\footnote{We are here in a case in which the assumption of non-evolving city populations whereas the networks is iteratively established finds little empirical or thematic support, since we showed that network and cities had comparable evolution time scales. This models is thus closer to produce in the proper sense a \emph{potential network} given a distribution of cities, and must be interpreted with caution.} which population follows a power law, two cities will be connected by a road link if a cost-benefit utility function, which linearly combines potential gravity flow and construction cost\footnote{What gives a cost function of the form $C = \beta / d_{ij}^{\alpha} - d_{ij}$, where $\alpha$ and $\beta$ are parameters.}, has a positive value. These simple local assumptions are sufficient to make a complex network emerge with phase transitions as a function of the relative weight parameter in the cost function, leading to the emergence of hierarchy. \cite{zhao2016population} apply this model in an iterative way to connect intra-urban areas, and shows that taking into account populations in the cost function significantly changes the topologies obtained.
}{
La physique a introduit récemment des modèles de croissance des réseaux d'infrastructure, en s'inspirant largement de cette littérature économique : un modèle très similaire au dernier cité est donné par~\cite{louf2013emergence} avec des fonctions coûts-bénéfices plus simples mais obtenant une conclusion similaire. Étant donné une distribution de noeuds (villes)\footnote{Nous nous trouvons ici dans un cas où l'hypothèse de non-évolution des population des villes tandis que le réseau s'établit itérativement trouve peu de support empirique ou thématique, puisqu'on a montré que réseau et villes avaient des échelles de temps d'évolution comparables. Ce modèle produit donc plus à proprement parler un \emph{réseau potentiel} étant donné une distribution de villes, et il est à interpréter avec précaution.} dont la population suit une loi puissance, deux villes seront connectées par un lien routier si une fonction d'utilité coût-bénéfice, combinant linéairement flux gravitaire potentiel et coût de construction\footnote{Ce qui donne une fonction de coût de la forme $C = \beta / d_{ij}^{\alpha} - d_{ij}$, où $\alpha$ et $\beta$ sont des paramètres.}, a une valeur positive. Ces hypothèses locales simples suffisent à faire émerger un réseau complexe et des transitions de phase en fonction du paramètre de poids relatif dans le coût, conduisant à l'apparition de la hiérarchie. \cite{zhao2016population} applique ce modèle de manière itérative pour connecter des zones intra-urbaines, et montre que la prise en compte des populations dans la fonction de coût change significativement les topologies obtenues.
}

\bpar{
An other class of models, close to procedural models in their ideas, are based on local geometric optimization processes, and aim at resembling real networks in their topology. \cite{bottinelli2017balancing} thus study a tree growth model applied to ant tracks, in which maintenance cost and construction cost both influence the choice of new links. The morphogenesis model by~\cite{courtat2011mathematics} which uses a compromise between realization of interaction potentials and construction cost, and also connectivity rules, reproduces in a stylized way real patterns of street networks. A very close model is described in~\cite{rui2013exploring}, but including supplementary rules for local optimization (taking into account degree for the connection of new links). Optimal network design, belonging more to the field of engineering, uses similar paradigms: \cite{vitins2010patterns} explore the influence of different rules of a shape grammar (in particular connection patterns between links of different hierarchical levels) on performances of networks generated by a genetic algorithm.
}{
Une autre classe de modèles, proche dans leur idée des modèles procéduraux, se basent sur des processus d'optimisation géométrique locale, et visent à ressembler à des réseaux réels dans leur topologie. \cite{bottinelli2017balancing} étudient ainsi un modèle de croissance d'arbre appliqué aux pistes de fourmis, dans lequel coût de maintenance et coût de construction influencent tous les deux les choix de nouveaux liens. Le modèle de morphogenèse de~\cite{courtat2011mathematics} qui utilise un compromis entre réalisation des potentiels d'interaction et coût de construction, ainsi que des règles de connectivité, reproduit de manière stylisée des motifs réels des réseaux de rues. Un modèle très proche est décrit dans~\cite{rui2013exploring}, tout en incluant des règles supplémentaires pour l'optimisation locale (prise en compte du degré pour la connection de nouveaux liens). La conception optimale de réseau, plutôt pratiquée par l'ingénierie, utilise des paradigmes similaires : \cite{vitins2010patterns} explorent l'influence de différentes règles d'une grammaire de formes (notamment les motifs de connection entre les liens de différents niveaux hiérarchiques) sur les performances de réseaux générés par algorithme génétique.
}

\bpar{
We can detail the mechanisms of one of these geometrical growth models. \cite{barthelemy2008modeling} describe a model based on a local optimization of energy which generates road networks with a globally reasonable shape. The model assumes ``centers'', which correspond to nodes of a road network, and road segments in space linking these centers. The model starts with initial connected centers, and proceeds by iterations to simulate network growth the following way:
\begin{enumerate}
	\item New centers are randomly added following an exogenous probability distribution, at fixed duration time steps.
	\item The network grows following a cost minimization rule: centers are grouped by projection on the network; each group makes a fixed length segment grow in the average direction towards the group starting from the projection (except if it vanishes in length, a segment then grows in the direction of each point).
\end{enumerate}
This model is adjusted in order that areas of parcels delimited by the network follow a power law with an exponent similar to the one observed for the city of Dresde. It has the advantage to be simple, to have few parameters (probability distribution for centers, length of segments built), to rely on reasonable local rules. This last point has also a dark side, since we can then expect the model to only capture few complexity, by neglecting numerous processes unveiled in chapter~\ref{ch:thematic} such as governance.
}{
Détaillons les mécanismes de l'un de ces modèles de croissance géométrique. \cite{barthelemy2008modeling} décrivent un modèle basé sur une optimisation locale de l'énergie qui génère des réseaux routiers à l'aspect globalement crédible. Le modèle suppose des ``centres'', qui correspondent à des noeuds d'un réseau routier, et des segments de route dans l'espace reliant ces centres. Le modèle part de centres initiaux connectés, et procède par itérations pour simuler la croissance du réseau de la façon suivante :
\begin{enumerate}
	\item De nouveaux centres sont ajoutés aléatoirement suivant une distribution de probabilité exogène, aux pas de temps multiple d'une durée fixée.
	\item Le réseau croit suivant une règle de minimisation de coût : les centres sont groupés par projection sur le réseau ; chaque groupe fait croître un segment de longueur fixée dans la direction moyenne vers l'ensemble du groupe à partir de la projection (sauf si celle-ci est nulle, un segment croit alors en direction de chaque point).
\end{enumerate}
Ce modèle est ajusté pour que les aires des parcelles délimitées par le réseau suivent une loi d'échelle avec un exposant similaire à celui observé pour la ville de Dresde. Il a l'avantage d'être simple, d'avoir peu de paramètres (distribution de probabilité pour les centres, taille des tronçons construits), de reposer sur des règles locales crédibles. Cette dernière propriété est à double tranchant, puisqu'on peut alors s'attendre à ce que le modèle ne puisse capturer que peu de complexité, en négligeant de nombreux processus mis en valeur au chapitre~\ref{ch:thematic} comme la gouvernance.
}






\paragraph{Biological networks}{Réseaux biologiques}

\bpar{
Finally, an interesting and original approach to network growth are biological networks. This approach belongs to the field of morphogenetic engineering, which aims at conceiving artificial complex systems inspired from natural complex systems and on which a control of emerging properties is possible~\cite{doursat2012morphogenetic}. \emph{Physarum machines}, which are models of a self-organized mould (\emph{slime mould}) have been proved to solve in an efficient way difficult problems (in the sense of their computational complexity, see~\ref{sec:epistemology}) such as routing problems~\cite{tero2006physarum} or NP-complete navigation problems such as the Traveling Salesman Problem~\cite{zhu2013amoeba}. These properties allow these systems to produce networks with Pareto-efficient properties for cost and robustness~\cite{tero2010rules} which are typical of empirical properties of real networks, and furthermore relatively close to these in terms of shape (under certain conditions, see~\cite{adamatzky2010road}).
}{
Enfin, une approche originale et intéressante pour la croissance des réseaux est le réseau biologique. Cette approche appartient au champ de l'ingénierie morphogénétique, qui vise à concevoir des systèmes complexes artificiels inspirés de systèmes complexes naturels et sur lesquels un contrôle des propriétés émergentes est possible~\cite{doursat2012morphogenetic}. Les \emph{machines Physarum}, qui sont des modèles d'une moisissure auto-organisée (\emph{slime mould}) ont été prouvés comme résolvant de manière efficiente des problèmes difficiles (au sens de leur complexité computationnelle, voir~\ref{sec:epistemology}) comme des problèmes de routage~\cite{tero2006physarum} ou des problèmes de navigation NP-complets comme le Problème du Voyageur de Commerce~\cite{zhu2013amoeba}. Ces propriétés permettent à ces systèmes de produire des réseaux ayant des propriétés de coût-robustesse Pareto-efficientes~\cite{tero2010rules} qui sont typiques des propriétés empiriques des réseaux réels, et de plus relativement proches en forme de ceux-ci (sous certaines conditions, voir~\cite{adamatzky2010road}).
}

\bpar{
This type of models can have an interest in our case since self-reinforcement processes based on flows are analogous to link reinforcement mechanisms in transportation economics. This type of heuristic has been tested to generate the French railway network by~\cite{mimeur:tel-01451164}, making an interesting bridge with investment models by \noun{Levinson} we previously described\footnote{Knowing that for this study, validation criteria that were applied remain however limited, either at a level inappropriate to the stylized facts studied (number of intersection or of branches) or too general and that can be reproduced by any model (total length and percentage of population deserved), and belong to criteria of form that are typical to procedural modeling which can only difficultly account of internal dynamics of a system as previously developed. Furthermore, taking for an external validation the production of a hierarchical network reveals an incomplete exploration of the structure and the behavior of the model, since through its preferential attachment mechanisms it must mechanically produce a hierarchy. Thus, a particular caution will have to be given to the choice of validation criteria.}.
}{
Ce type de modèles peut être d'intérêt dans notre cas puisque les processus d'auto-renforcement basés sur les flots sont analogues aux mécanismes de renforcement de lien en économie des transports. Ce type d'heuristique a été testé pour générer le réseau ferré Français par~\cite{mimeur:tel-01451164}, faisant un pont intéressant avec les modèles d'investissement de \noun{Levinson} décrits précédemment\footnote{Sachant que pour cette étude, les critères de validation appliqués restent toutefois limités, soit à un niveau inadapté aux faits stylisés étudiés (nombre d'intersection ou de branches) soit trop généraux et pouvant être produit par n'importe quel modèle (longueur totale et pourcentage de population desservie), et relèvent de critères de forme typique de la modélisation procédurale qui ne peuvent que difficilement rendre compte des dynamiques internes d'un système comme développé précédemment. De plus, prendre pour validation externe la production d'un réseau hiérarchique découle d'une exploration incomplète de la structure et du comportement du modèle, puisque celui-ci par ses mécanismes d'attachement préférentiel doit mécaniquement produire une hiérarchie. Ainsi, une attention particulière devra être donnée au choix des critères de validation.}.
}



\paragraph{Procedural modeling}{Modélisation procédurale}

\bpar{
Finally, we can mention other tentatives such as~\cite{de2007netlogo,yamins2003growing}, which are closer to procedural modeling~\cite{lechner2004procedural,watson2008procedural} and therefore have only little interest in our case since they can difficultly be used as explicative models\footnote{Following~\cite{varenne2017theories}, an explicative model allows to produce an explanation to observed regularities or laws, for example by suggesting processes which can be at their origin. If model processes are explicitly detached from a reasonable ontology, they can not be potential explanations. We will give in~\ref{sec:computation} a development of this notion in the frame of a more global reflexion on the epistemology of modeling.}. Procedural modeling consists in generating structures in a way similar to shape grammars\footnote{A shape grammar is a formal system (i.e a set of initial symbos, axioms, and a set of transformation rules) which acts on geometrical objects. Starting from initial patterns, they allow to generate classes of objects.}, but it also concentrates generally on the faithful reproduction of local form, without considering macroscopic emerging properties. Classifying them as morphogenesis models is incorrects and corresponds to a misunderstanding of mechanisms of \emph{Pattern Oriented Modeling}~\cite{grimm2005pattern}\footnote{\emph{Pattern Oriented Modeling} consists in seeking to explain observed patterns, generally at multiple scales, in a \emph{bottom-up} way. Procedural modeling does not correspond to that, since it aims at reproducing and not at explaining.} on the one hand and of the epistemology of morphogenesis on the other hand (see~\ref{sec:interdiscmorphogenesis}). We will use this type of models (exponential mixture to produce a population density for example) to generate initial synthetic data uniquely to parametrize other complex models (see~\ref{sec:computation} and \ref{sec:correlatedsyntheticdata}).
}{
Finalement, nous pouvons mentionner d'autres tentatives comme~\cite{de2007netlogo,yamins2003growing}, qui sont plus proches de la modélisation procédurale~\cite{lechner2004procedural,watson2008procedural} et pour cette raison n'ont que peu d'intérêt pour notre cas puisqu'ils peuvent difficilement être utilisés comme modèles explicatifs\footnote{Suivant~\cite{varenne2017theories}, un modèle explicatif permet de produire une explication à des régularités ou des lois observées, par exemple en suggérant des processus pouvant en être à l'origine. Si les processus du modèle sont explicitement dissociés d'une ontologie raisonnable, ceux-ci ne peuvent être explications potentielles. Nous donnerons en~\ref{sec:computation} un développement de cette notion dans le cadre d'une réflexion plus générale sur l'épistémologie de la modélisation.}. La modélisation procédurale consiste à générer des structures à la manière des grammaires de forme\footnote{Une grammaire de forme est un système formel (c'est-à-dire un ensemble de symboles initiaux, les axiomes, et un ensemble de règles de transformation) qui agit sur des objets géométriques. Partant de motifs initiaux, elles permettent de générer des classes d'objets.}, mais celle-ci se concentre généralement sur la reproduction fidèle de forme locale, sans tenir compte des propriétés macroscopiques émergentes. Les classifier comme modèles de morphogenèse n'est pas correct et correspond à une incompréhension des mécanismes du \emph{Pattern Oriented Modeling}~\cite{grimm2005pattern}\footnote{Le \emph{Pattern Oriented Modeling} consiste à chercher à expliquer des motifs observés, généralement à plusieurs échelles, dans une démarche \emph{bottom-up}. La modélisation procédurale n'en relève pas, puisqu'elle vise à reproduire et non à expliquer.} d'une part et de l'épistémologie de la morphogenèse d'autre part (voir~\ref{sec:interdiscmorphogenesis}). Nous utiliserons ce type de modèle (mélange d'exponentielles pour produire une densité de population par exemple) pour générer des données synthétiques initiales uniquement pour paramétrer d'autres modèles complexes (voir~\ref{sec:computation} et \ref{sec:correlatedsyntheticdata}).
}

\subsection{Modeling co-evolution}{Modéliser la co-évolution}

\bpar{
We can now switch to models that integrate dynamically the paradigm Territory $\leftrightarrow$ Network, which as we recall assumes that the conditioning of one by the other can not be identified. The ontologies used, as we will see, often couple\footnote{We recall the definition of model coupling, which corresponds to the one of system or process coupling given in introduction: it is the construction of a model that is simultaneously the extension of each initial model.} network elements with territorial components, but this positioning is not necessary and some elements may be hybrid (for example a governance structure for the transportation network may simultaneously belong to both aspects). In our reading of models, these different specifications will naturally arise.
}{
Nous pouvons à présent nous intéresser aux modèles intégrant dynamiquement le paradigme Territoire $\leftrightarrow$ Réseau, qui on le rappelle suppose qu'un conditionnement de l'un par l'autre n'est pas identifiable. Les ontologies utilisées, comme nous le verrons, couplent\footnote{Nous rappelons la définition du couplage de modèle, qui correspond à celle de couplage de système ou de processus donnée en introduction : il s'agit de la constitution d'un modèle qui est simultanément une extension de chacun des modèles initiaux.} souvent des éléments de réseau avec des composantes territoriales, mais cette position n'est pas une nécessité et certains éléments peuvent être hybrides (par exemple une structure de gouvernance du système de transport peut relever simultanément des deux aspects). Dans notre lecture des modèles, ces différentes spécifications se dégageront naturellement.
}


\bpar{
We will broadly designate by model of co-evolution simulation models that include a coupling of urban growth dynamics and transportation network growth dynamics. These are relatively rare, and for most of them still at the stage of stylized models. The efforts being relatively sparse and in very different domains, there is not much unity in these approaches, beside the abstraction of the assumption of an interdependency between networks and territorial characteristics in time. We propose to review them still through the prism of scales.
}{
Nous désignerons largement par modèle de co-évolution les modèles de simulation qui incluent un couplage des dynamiques de la croissance urbaine et du réseau de transport. Ceux-ci sont relativement rares, et pour la plupart au stade de modèles stylisés. Les efforts étant assez disparates et dans des domaines très variés, il y a peu d'unité dans ces approches, si ce n'est l'abstraction de l'hypothèse d'interdépendance entre réseaux et caractéristiques du territoire dans le temps. Nous proposons de les passer en revue toujours avec la grille de lecture des échelles.
}

\subsubsection{Microscopic and mesoscopic scales}{Echelle microscopique et mesoscopique}

\paragraph{Geometrical Models}{Modèles géométriques}

\bpar{
\cite{achibet2014model} describes a co-evolution model at a very large scale (scale of the building), in which evolution of both network and buildings are ruled by a same agent, influenced differently by network topology and population density, and that can be understood as an agent of urban development. The model allows to simulate an auto-organized urban extension and to produce district configurations. Even if it strongly couples territorial components (buildings) and the road network, described results do not imply any conclusion on the processes of co-evolution themselves.
 }{
\cite{achibet2014model} décrit un modèle de co-évolution à une très grande échelle (échelle du bâtiment), dans lequel l'évolution du réseau et des bâtiments sont tous les deux régis par un agent commun, influencé différemment par la topologie du réseau et la densité de population, qui peut être compris comme un agent développeur. Le modèle permet de simuler une extension urbaine auto-organisée et de produire des configurations de quartier. Bien qu'il couple fortement composantes territoriales (bâtiments) et réseau routier, les résultats présentés ne permettent pas de tirer de conclusion sur les processus de co-évolution en eux-mêmes.
}


\bpar{
A generalization of the geometrical local optimization model described before is developed in~\cite{barthelemy2009co}. It aims at capturing the co-evolution of network topology with the density of its nodes. The localization of new nodes is simultaneously influenced by density and centrality, yielding the looping of the strong coupling. More precisely, the global behavior of the model is the same, as the network extension behavior. Centers then localize following a utility function that is a linear combination of average betweenness centrality in a neighborhood and of the opposite of density (dispersion due to higher price as a function of density). This utility is used to compute the probability of localization of new centers following a discrete choices model. The model allows to show that the influence of centrality reinforces aggregation phenomena (in particular through an analytical resolution on a one-dimensional version of the model), and furthermore reproduces exponentially decreasing density profiles (Clarcke's law) which are observed empirically.  
}{
Une généralisation du modèle d'optimisation locale géométrique décrit précédemment est développée dans~\cite{barthelemy2009co}, et cherche à capturer la co-évolution entre topologie du réseau et densité de ses noeuds. La localisation de nouveaux noeuds est influencée à la fois par la densité et la centralité, permettant de boucler le couplage fort. Plus précisément, le fonctionnement global du modèle est le même, ainsi que la règle de croissance du réseau. Les centres se localisent quant à eux selon une fonction d'utilité qui est une combinaison linéaire de la centralité de chemin moyenne dans un voisinage et de l'opposée de la densité (dispersion due aux prix plus élevés en fonction de la densité). Cette utilité permet de définir la probabilité de localisation des nouveaux centres suivant un modèle de choix discrets. Le modèle permet de montrer que l'influence de la centralité accentue les phénomènes d'agrégation (notamment par une résolution analytique sur une version en une dimension du modèle), et reproduit par ailleurs des profils exponentiels décroissants pour la densité (loi de Clarke), observés empiriquement.
}

\bpar{
\cite{ding2017heuristic} introduce a model of co-evolution between different layers of the transportation network, and show the existence of an optimal coupling parameter in terms of inequalities for the centrality in network conception: if the road network is assimilated at a fine granularity to a population distribution, this model can be compared with the precedent model of co-evolution between the transportation network and the territory.
}{
\cite{ding2017heuristic} introduisent un modèle de co-évolution entre différentes couches du réseau de transport, et montrent l'existence d'un paramètre de couplage optimal en terme d'inégalités de centralité pour la conception d'un réseau : si on assimile le réseau routier à granularité très fine à une distribution de population, ce modèle se rapproche du précédent modèle de co-évolution entre réseau de transport et territoire.
}


\paragraph{Economic models}{Modèles économiques}

\bpar{
\cite{levinson2007co} take an economic approach, which is richer from the point of view of network development processes implied, similar to a four step model (i.e. including the generation of origin-destination flows and the assignment of traffic in the network) including travel cost and congestion, coupled with a road investment module simulating toll revenues for constructing agents, and a land-use evolution module updating actives and employments through discrete choice modeling. The exploration experiments show that co-evolving network and land uses lead to positive feedbacks reinforcing hierarchies. These are however far from satisfying, since network topology does not evolve as only capacities and flows change within the network, what implies that more complex mechanisms (such as the planning of new infrastructures) on longer time scales are not taken into account. \cite{li2016integrated} have recently extended this model by adding endogenous real estate prices and an optimization heuristic with a genetic algorithm for deciding agents.
}{
\cite{levinson2007co} prennent une approche économique plus riche du point de vue des processus de développement de réseau impliqués, similaire à un modèle à quatre étapes (c'est-à-dire incluant une génération de flux origine-destination et une attribution du trafic dans le réseau) qui inclut coût de transport et congestion, couplé avec un module d'investissement routier qui simule les revenus des péages pour les agents qui construisent, et un module d'évolution d'usage du sol qui simule les relocalisations des actifs et des emplois. Les expériences d'exploration de ce modèle montrent que l'usage du sol et le réseau en co-évolution conduisent à des retroactions positives renforçant les hiérarchies. Elles sont cependant loin d'être satisfaisantes puisque la topologie du réseau n'évolue pas à proprement parler puisque seules les capacités et les flux changent dans le réseau, ce qui signifie que des mécanismes plus complexes (comme la planification de nouvelles infrastructures) sur de plus longues échelles de temps ne sont pas pris en compte. \cite{li2016integrated} ont récemment étendu ce modèle par l'ajout de prix immobiliers endogènes et d'une heuristique d'optimisation par algorithme génétique pour les agents décideurs.
}

\bpar{
From an other point of view, \cite{levinson2005paving} is also presented as a model of co-evolution, but corresponds more to a predictive model based on Markov chains, and thus closer to a statistical analysis than a simulation model based on these processes. \cite{rui2011urban} describe a model in which the coupling between land-use and network topology is done with a weak paradigm, land-use and accessibility having no feedback on network topology, the land-use model being conditioned to the growth of the autonomous network.
}{
D'un autre point de vue, \cite{levinson2005paving} est aussi présenté comme un modèle de co-évolution mais repose sur un modèle prédictif à chaîne de Markov, et donc plus proche d'une analyse statistique que d'un modèle de simulation basé sur des processus. \cite{rui2011urban} décrivent un modèle dans lequel le couplage entre usage du sol et la topologie du réseau est fait par un paradigme faible, l'usage du sol et l'accessibilité n'ayant pas de retroaction sur la topologie du réseau, le modèle d'usage du sol étant conditionné à la croissance du réseau autonome.
}


\paragraph{Cellular automatons}{Automates cellulaires}

\bpar{
A simple hybrid model explored and applied to a stylized planning example of the functionnal distribution of a new district in~\cite{raimbault2014hybrid}, relies on mechanisms of accessibility to urban activities for the growth of settlements with a network adapting to the urban shape. The rules for network growth are too simple to capture more elaborated processes than just a simple systematic connection (such as potential breakdown for example), but the model produces at a large scale a broad range of urban shapes reproducing typical patterns of human settlements. This model is inspired by~\cite{moreno2012automate} for its core mechanisms but yield a much broader generation of forms by taking into account urban functions.
}{
Un modèle hybride simple exploré et appliqué à un exemple stylisé de planification de la répartition fonctionnelle d'un nouveau quartier dans~\cite{raimbault2014hybrid}, repose sur les mécanismes d'accès aux activités urbaines pour la croissance des établissements avec un réseau s'adaptant à la forme urbaine. Les règles pour la croissance du réseau sont trop simples pour capturer des processus plus élaborés qu'une simple connection systématique (comme une rupture de potentiel par exemple), mais le modèle produit à une grande échelle une large gamme de formes urbaines qui reproduisent les motifs typiques des établissements humains. Ce modèle s'inspire de~\cite{moreno2012automate} pour ses mécanismes de base mais permet une génération de formes bien plus larges par la prise en compte des fonctions urbaines.
}

\bpar{
At these relatively large scales, spanning from the urban to the metropolitan scale, mechanisms of population localization influenced by accessibility coupled to mechanisms of network growth optimizing some particular functions seem to be the rule for this kind of models: in the same way, \cite{wu2017city} couple a cellular automaton for population diffusion to a network optimizing local cost that depends on the geometry and on population distribution.
}{
A ces échelles relativement grandes, s'étendant de l'échelle urbaine à celle métropolitaine, les mécanismes de localisation de population influencée par l'accessibilité couplés à des mécanismes de croissance de réseau optimisant certaines fonctions semblent être la règle pour ces modèles : de la même façon, \cite{wu2017city} couplent un automate cellulaire de diffusion de population à un réseau optimisant un coût local dépendant de la géométrie et de la distribution de population.
}

\bpar{
Models answering to more remote questions can furthermore be linked to our problem: for example, in a conceptual way, a certain form of strong coupling is also used in~\cite{bigotte2010integrated} which by an approach of operational research propose a network design algorithm to optimize the accessibility to amenities, taking into account both network hierarchy and the hierarchy of connected centers.
}{
Des modèles répondant à des problématiques assez lointaines peuvent par ailleurs être reliés à notre question : par exemple, de manière conceptuelle, une certaine forme de couplage fort est également utilisé dans~\cite{bigotte2010integrated} qui par une approche de recherche opérationnelle proposent un algorithme de conception de réseau pour optimiser l'accessibilité aux services, prenant en compte à la fois la hiérarchie du réseau et celle des centres connectés.
}

\bpar{
This way, co-evolution models at the microscopic and mesoscopic scales globally have the following structure: (i) processes of localization or relocalization of activities (actives, buildings) influenced by their own distribution and network characteristics; (ii) network evolution, that can be topological or not, answering to very diverse rules: local optimization, fixed rules, planning by deciding agents. This diversity suggests the necessity to take into account the superposition of multiple processes ruling network evolution.
}{
Ainsi, les modèles de co-évolution aux échelles microscopique et mesoscopiques suivent globalement la structure suivante : (i) processus de localisation ou relocalisation des activités (actifs, bâtiments) influencés par leur propre distribution et les caractéristique du réseau ; (ii) evolution du réseau, topologique ou non, répondant à des règles très diverses : optimisation locale, règles fixes, planification par des agents décideurs. Cette diversité suggère la nécessité de prendre en compte la superposition de multiples processus régissant l'évolution du réseau.
}

\subsubsection{Urban systems modeling}{Modélisation de Systèmes Urbains}

\bpar{
At a macroscopic scale, co-evolution can be taken into account in models of urban systems. \cite{baptiste1999interactions} propose to couple an urban growth model based on migrations (introduced by the application of synergetics to systems of cities by~\cite{sanders1992systeme}) with a mechanism of self-reinforcement of capacities for the road network without topological modification. More precisely, the general principles of the model are the following.
\begin{itemize}
\item Attractivity and repulsion indicators allow for each city to determine emigration and immigration rates and to make populations evolve.
\item Network topology is fixed in time, but capacities of links evolve. The rule is an increase in capacity when the flow becomes greater given a fixed parameter threshold during a given number of iterations. Flows are affected with a gravity model of interaction between cities.
\end{itemize}
}{
A une échelle macroscopique, la co-évolution est parfois prise en compte dans des modèles de systèmes urbains. \cite{baptiste1999interactions} propose de coupler un modèle de croissance urbaine basé sur les migrations (introduit par l'application de la synergétique aux systèmes de villes par~\cite{sanders1992systeme}) avec un mécanisme d'auto-renforcement des capacités pour le réseau routier sans modification topologique. Plus précisément, les principes généraux du modèle sont les suivants.
\begin{itemize}
	\item Des indicateurs d'attractivité et de répulsion permettent pour chaque ville de déterminer des taux d'émigration et d'immigration et de faire évoluer les population.
	\item La topologie du réseau est fixée dans le temps, mais les capacités des liens évoluent. La règle est une augmentation de la capacité lorsque le flux dépasse celle-ci par un seuil donné comme paramètre pendant un certain nombre d'itérations. Les flux sont affectés par modèle gravitaire d'interaction entre les villes.
\end{itemize}
}


\bpar{
The last version of this model is presented by~\cite{baptistemodeling}. General conclusions that can be obtained from this work are that this coupling yield a hierarchical configuration\footnote{But we also know that simpler models, only a preferential for example, allow to reproduce this stylized fact. The model must have as an objective to answer to broader questions, such as the fine understanding of co-evolution processes, what is not done here. However, one of its operational objectives is otherwise fulfilled, through the application to France and the study of the impact of a high speed line project, recalling the multiple possible functions of a model (see~\ref{sec:computation}).} and that the addition of the network produces a less hierarchical space, allowing medium-sized cities to benefit from the feedback of the transportation network.
}{
Sa dernière version est présentée par~\cite{baptistemodeling}. Les conclusions générales qui peuvent être tirées de ce travail sont que ce couplage permet de faire émerger une configuration hiérarchique\footnote{Mais on sait par ailleurs que des modèles plus simples, un attachement préférentiel uniquement par exemple, permettent de reproduire ce fait stylisé. Le modèle doit avoir pour objectif de répondre à des problématiques plus larges, comme la compréhension fine des processus de co-évolution, ce qui n'est pas fait ici. Cependant, l'un de ses objectifs opérationnels est par ailleurs rempli, par l'application à la France et l'étude de l'impact d'un projet de Ligne à Grande Vitesse, rappelant les multiples fonctions possibles d'un modèle (voir~\ref{sec:computation}).} et que l'ajout du réseau produit un espace moins hiérarchique, permettant à des villes moyennes de bénéficier de la rétroaction du réseau de transport.
}

\bpar{
The model proposed by~\cite{blumenfeld2010network} can be seen as a bridge between the mesoscopic scale and the approaches of urban systems, since it simulates migrations between cities and network growth induced by potential breakdown when detours are too large. In the continuity of Simpop models for systems of cities, \cite{schmitt2014modelisation} describes the SimpopNet model which aims at precisely integrating co-evolution processes in systems of cities on long time scales, typically via rules for hierarchical network development as a function of the dynamics of cities, coupled with these that depends on network topology. Unfortunately the model was not explored nor further studied, and furthermore stayed at a toy-level. \cite{cottineau2014evolution} proposes an endogenous transportation network growth as the last building brick of the Marius modeling framework, but it stays at a conceptual level since this brick has not been specified nor implemented yet. To the best of our knowledge, there exists no model which is empirical or applied to a concrete case based on an approach of co-evolution by urban systems from the point of view of the evolutive urban theory.
}{
Le modèle proposé par~\cite{blumenfeld2010network} peut être vu comme un pont entre l'échelle mesoscopique et les approches de type système urbain, puisqu'il simule les migrations entre villes et la croissance du réseau induite par une rupture de potentiel lorsque les détours sont trop grands. Dans la continuité des modèles Simpop pour modéliser les systèmes de villes, \cite{schmitt2014modelisation} décrit le modèle SimpopNet qui vise à précisément intégrer les processus de co-évolution dans les systèmes de villes à longue échelle temporelle, typiquement par des règles pour un développement hiérarchique du réseau comme fonction des dynamiques des villes, couplées à celles-ci qui dépendent de la topologie du réseau. Malheureusement le modèle n'a pas été exploré ni étudié de manière plus approfondie, et de plus est resté au niveau de modèle jouet. \cite{cottineau2014evolution} propose une croissance endogène des réseaux de transport comme la dernière brique de construction du cadre de modélisation Marius, mais cela reste à un niveau conceptuel puisque cette brique n'a pas encore été spécifiée ni implémentée. Il n'existe à notre connaissance pas de modèle empirique ou appliqué à un cas concret se basant sur une approche de la co-évolution par les systèmes urbains vus par la théorie évolutive des villes.
}

\bpar{
We can see well the opposition to epistemological principles of economic geography: \cite{fujita1999evolution} introduce for example an evolutionary model able to reproduce and urban hierarchy and an organization typical of central place theory~\cite{banos2011christaller}, but that still relies on the notion of successive equilibriums, and moreover considers a ``Krugman-like'' model, i.e. a one dimensional and isotropic space, in which agents are homogeneously distributed\footnote{The absence of a real space is not an issue in this economic approach that aims at understanding processes out of their context. In our case, the structure of the geographical space is not separable, and indeed at the core of the issues we are interested in.}. This approach can be instructive on economic processes in themselves but more difficultly on geographical processes, since these imply the embedding of economic processes in the geographical space which spatial particularities not taken into account in this approach are crucial. Our work will focus on demonstrating to what extent this structure of space can be important and also explicative, since networks, and even more physical networks induce spatio-temporal processes that are path-dependent and thus sensitive to local singularities and prone to bifurcations induced by the combination of these with processes at other scales (for example the centrality inducing a flow).
}{
On voit bien l'opposition aux principes épistémologiques de l'économie géographique : \cite{fujita1999evolution} introduisent par exemple un modèle évolutionnaire capable de reproduire une hiérarchie urbaine et une organisation typique de la théorie des places centrales~\cite{banos2011christaller}, mais qui repose toujours sur la notion d'équilibres successifs, et surtout considère un modèle ``à-la-Krugman'' c'est-à-dire un espace à une dimension, isotrope, et dans lequel les agents sont répartis de manière homogène\footnote{L'absence d'espace réel n'est pas un problème dans cette approche économique qui vise à comprendre des processus hors-sol. Dans notre cas, la structure de l'espace géographique n'est pas séparable, et même au coeur des problématiques qui nous intéressent.}. Cette approche peut être instructive sur les processus économiques en eux-mêmes mais plus difficilement sur les processus géographiques, puisque ceux-ci impliquent un déroulement des processus économiques dans l'espace géographique dont les particularités spatiales qui ne sont pas prises en compte dans cette approche sont essentielles. Notre travail s'attachera à montrer dans quelle mesure cette structure de l'espace peut être importante et également explicative, puisque les réseaux, et encore plus les réseaux physiques induisent des processus spatio-temporels dépendants au chemin et donc sensibles aux singularités locales et propices aux bifurcations induites par la combinaison de celles-ci et de processus à d'autres échelles (par exemple la centralité induisant un flux).
}



\bpar{
At the macroscopic scale, existing models are based on the evolution of agents (generally cities) as a consequence of their interactions, carried by the network, whereas the evolution of the network can follow different rules: self-reinforcement, potential breakdown. The general structure is globally the same than at larger scales, but ontologies stay fundamentally different.
}{
A l'échelle macroscopique, les modèles existants se basent sur des évolutions des agents (souvent les villes)  en conséquence de leurs interactions, portées par le réseau, tandis que l'évolution du réseau peut répondre à différentes règles : auto-renforcement, rupture de potentiel. La structure générale est globalement la même qu'à des échelles plus grandes, mais les ontologies restent fondamentalement différentes.
}

\subsubsection*{Synthesis}{Synthèse}

\bpar{
It is crucial at this stage to risk a synthesis and put into perspective all models that we reviewed, since even if it will necessarily be reducing and simplifying, it gives the foundations for the analyses that will follow.
}{
Il est essentiel à ce stade de s'oser à une synthèse et une mise en perspective de l'ensemble des modèles que nous avons passé en revue, puisque même si celle-ci sera nécessairement réductrice et simplificatrice, elle donne les fondations pour les analyses qui suivront.
}


\bpar{
We will synthesize the broad types of models that we reviewed in the following table, by classing them by type (relation between networks and territories), by class (broad classes corresponding to the stratification of the review), and by giving the temporal and spatial scales concerned, the functions, the type of result obtained, the paradigms used. It is given in Table~\ref{tab:modelingsa:synthesis}.
}{
Nous synthétisons les grands types de modèles que nous avons passé en revue dans le tableau suivant, en les classant par type (relation entre réseaux et territoires), par classe (grandes classes correspondant à la stratification de la revue), et en précisant les échelles temporelles et spatiales concernées, les fonctions, le type de résultats obtenus, les paradigmes utilisés. Celle-ci est donnée en Table~\ref{tab:modelingsa:synthesis}.
}

\begin{table}
\caption[Synthesis of modeling approaches]{\textbf{Synthesis of modeling approaches.} The type gives the sense of the relation; the class is the scientific field in which the model is inserted; scales correspond to our simplified scales; functions are given in the sense of~\ref{sec:computation}; we finally give the type of results they provide and the paradigms used.\label{tab:modelingsa:synthesis}}
\bpar{
\begin{tabular}{|p{2.5cm}|p{2cm}|p{2.5cm}|p{2.5cm}|p{2.1cm}|p{2.2cm}|p{2cm}|}
\hline
Type & Class & Temporal Scale & Spatial scale & Function & Results & Paradigms\\ \hline
Networks $\rightarrow$ Territories & LUTI & Medium & Mesoscopic & Planning, Prediction & Land-use simulation & Urban economics \\ \hline
\multirow{3}{*}{Territories $\rightarrow$}& Networks Economics & Medium & Mesoscopic & Explanation & Role of economic processes & Economics, Governance\\\cline{2-7}
Networks& Geometrical growth & Long & Meso or Macro & Explanation & Reproduction of stylized shapes & Simulation models, Local optimization \\\cline{2-7}
& Biological networks & Long & Mesoscopic & Optimization & Production of optimal networks & Self-organized network \\ \hline
\multirow{2}{*}{Territories $\leftrightarrow$}& Networks Economics & Medium & Mesoscopic & Explanation & Reinforcement effects & Economics\\\cline{2-7}
Networks & Geometrical growth & Long or NA & Micro, Meso or Macro & Explanation & Reproduction of stylized shapes & Simulation models, Local optimization \\\cline{2-7}
& Urban Systems & Medium, Long & Macroscopic & Explanation, prospection & Stylized facts & Complex geography\\\hline
\end{tabular}
}{
\begin{tabular}{|p{2.5cm}|p{2cm}|p{2.5cm}|p{2.5cm}|p{2.1cm}|p{2.2cm}|p{2cm}|}
\hline
Type & Classe & Echelle Temporelle & Echelle Spatiale & Fonction & Résultats & Paradigmes\\ \hline
Réseaux $\rightarrow$ Territoires & LUTI & Moyenne & Mesoscopique & Planification, Prédiction & Simulation de l'usage du sol & Économie urbaine \\ \hline
\multirow{3}{*}{Territoires $\rightarrow$}& Économie des Réseaux & Moyenne & Mesoscopique & Explication & Rôle de processus économiques & Économie, Gouvernance\\\cline{2-7}
Réseaux& Croissance géométrique & Longue & Meso ou Macro & Explication & Reproduction de formes stylisées & Modèles de Simulation, Optimisation locale \\\cline{2-7}
& Réseaux biologiques & Longue & Mesoscopique & Optimisation & Production de réseaux optimaux & Réseau auto-organisé \\ \hline
\multirow{2}{*}{Territoires $\leftrightarrow$}& Économie des Réseaux & Moyenne & Mesoscopique & Explication & Effets de renforcement & Économie\\\cline{2-7}
Réseaux & Croissance géométrique & Longue ou NA & Micro, Meso ou Macro & Explication & Reproduction de formes stylisées & Modèles de Simulation, Optimisation locale \\\cline{2-7}
& Systèmes Urbains & Moyenne, Longue & Macroscopique & Explication, prospection & Faits stylisés & Géographie complexe\\\hline
\end{tabular}
}
\end{table}


\subsubsection*{An neglected coevolution ?}{Une co-évolution négligée ?}

\bpar{
The unbalance between the last section accounting for models integrating effectively a strongly coupled dynamic (and possibly a co-evolution) and the preceding sections leads to an interrogation: are models integrating co-evolution marginal? Is it possible then to explain this marginality?
}{
Le déséquilibre entre la dernière section rendant compte des modèles intégrant effectivement une dynamique fortement couplée (et possiblement une co-évolution) et les précédentes interroge : les modèles intégrant la co-évolution sont-ils si marginaux ? Est-il alors possible d'expliquer cette marginalité ?
}


\bpar{
The aim of the two following sections will be to propose elements of answer to these questions through epistemological analyses by increasing the knowledge on concerned fields and of the corresponding models.
}{
L'objet des deux sections qui suivent sera de proposer des éléments de réponse à ces questions par des analyses épistémologiques en accroissant la connaissance des champs concernés et des modèles correspondants.
}

\stars

\bpar{
We have thus given in this section a broad overview of models focusing on interactions between transportation networks and territories, including co-evolution models. We begin thus to foresee a refinement of the definition of the concept of co-evolution in that frame.
}{
Nous avons ainsi donné dans cette section un aperçu large des modèles s'intéressant aux interactions entre réseaux de transport et territoires, incluant les modèles de co-évolution. Nous commençons donc à entrevoir une précision de la définition du concept de co-évolution dans ce cadre.
}

\bpar{
We propose in the next section to proceed to a more systematic mapping of this scientific landscape, in order to reinforce the epistemological viewpoint and better situate the positioning we will take and the models we will introduce in the following.
}{
Nous proposons dans la section suivante de dresser une cartographie plus systématique de ce paysage scientifique, afin de renforcer le point de vue épistémologique et mieux situer la position que nous prendrons et les modèles que nous introduirons par la suite.
}

\stars

%

\newpage


\section{An epistemological approach}{Une approche épistémologique}

\label{sec:quantepistemo}


\bpar{
We gave a broad overview of different types of models taking into account interactions between networks and territories, with disciplines and problematics that are associated. These very different aspects suggest a strong compartmentalization of disciplines. It is furthermore difficult to distinguish potential models of co-evolution within this fuzzy environment. We may legitimately ask what are the existing and potential relations between the different approaches ? Which fields may have been missed although they are complementary ?
}{
Nous avons eu un aperçu large de différents types de modèles prenant en compte les interactions entre réseaux et territoires, ainsi que les disciplines et problématiques associées. Ces aspects très différents suggèrent un cloisonnement fort des disciplines. Il reste de plus difficile de situer les modèles potentiels de co-évolution dans cette nébuleuse. Il est légitime de se demander quelles sont les relations existantes et potentielles entre les différentes approches ? Quelles domaines peuvent être passés inaperçus bien que complémentaires ?
}


\bpar{
Diverse hypotheses can be proposed in order to explain the absence of investigations on co-evolution models:
\begin{itemize}
	\item Following~\cite{commenges:tel-00923682}, scientific and operational actors that would be concerned by the practical application of such models would see themselves replaced by the same models and have thus no incentive to develop them (sociological explanation).
	\item The different disciplines which possess the diverse components that are necessary to such models are compartmentalized and have divergent motivations (epistemological explanation).
	\item The construction of such models exhibits intrinsic difficulties making their development not encouraging and not well currently tackled.
\end{itemize}
}{
Diverses hypothèses peuvent être avancées pour tenter d'expliquer l'absence d'investigation des modèles de co-évolution :
\begin{itemize}
	\item Suivant~\cite{commenges:tel-00923682}, les acteurs scientifiques et opérationnels qui seraient concernés par l'application pratique de tels modèles se verraient remplacés par ces mêmes modèles et donc n'ont aucune incitation à les développer (explication sociologique).
	\item Les différentes disciplines qui détiennent les diverses composantes nécessaires à de tels modèles sont cloisonnées et ont des motivations divergentes (explication épistémologique).
	\item La construction de tels modèles comporte des difficultés intrinsèques rendant leur développement décourageant et pas parfaitement maitrisé actuellement.
\end{itemize}
}

\bpar{
We will not be able in this work to explore the first assumption (or more precisely, it would require a subject in itself, implying in particular sociological interviews). The third is either a tautology or can not be demonstrated, in a Church style as it can be put, and our whole work will allow us to bring elements of answer. The second is on the contrary as we will see more within our reach.
}{
Nous n'aurons pas les moyens d'explorer la première hypothèse (ou plutôt elle demanderait un sujet à part entière, impliquant entre autres entretiens sociologiques). La troisième est soit une tautologie soit indémontrable, à-la-Church dirait-on, et l'ensemble de notre travail permettra d'y apporter des pistes de réponse. La deuxième par contre est comme nous allons le voir plus à notre portée.
}

\bpar{
A way to explore this hypothesis and to answer to previous questions relies in an epistemological study that we propose to lead in a quantitative and systematic way. This approach is complementary to the previous literature review, and allows both to contextualize it and to systematize it. We must also recall the idea that the study of reasons for a sparsity of models will necessarily inform on models themselves and on the questions relates to their construction: the \emph{knowledge of knowledge}~\cite{morin1986methode} increases the knowledge.
}{
Une manière d'explorer cette hypothèse et de répondre aux questions précédentes consiste en une étude épistémologique que nous proposons de mener de manière quantitative et systématique. Cette approche est complémentaire de l'analyse de littérature précédente, et permet à la fois de la contextualiser et de la systématiser. Il faut par ailleurs garder en tête l'idée que l'étude des raisons de la rareté des modèles nous informera nécessairement sur les modèles eux-mêmes et les questions reliées à leur construction : la \emph{connaissance de la connaissance}~\cite{morin1986methode} accroît la connaissance.
}

\bpar{
A preliminary study aims at confirming the relevance of a quantitative epistemology approach, by suggesting a strong isolation of disciplines. This study is done using a algorithm for an algorithmic systematic review, which reconstructs corpuses of references by exploring semantic neighborhoods, i.e. an iterative collection of neighbor references in their main semantic content. We proceed then to a network analysis, coupling citation network and semantic network, to precise the shape of implied disciplines. We finally suggest possible extensions towards unsupervised learning and full-texts mining for an automatic extraction of the structure of models for example.
}{
Une étude préliminaire a pour but de confirmer la pertinence d'une approche d'épistémologie quantitative, en suggérant une forte isolation des disciplines. Celle-ci est menée par un algorithme de revue systématique algorithmique, qui reconstruit des corpus de références par exploration de voisinage sémantiques, c'est-à-dire la récupération itérative de références voisines dans leur contenu sémantique principal. Nous procédons ensuite à une analyse de réseaux, couplant réseau de citation et réseau sémantique, pour préciser les contours des disciplines impliquées. Nous suggérons finalement des possibles extensions vers de l'apprentissage non-supervisé et la fouille de texte complets pour une extraction automatique de la structure de modèles par exemple.
}

\bpar{
We shall begin by situating the context of \emph{quantitative epistemology}\footnote{We propose to use this term for works at the crossroad of bibliometrics and scientometrics, of cognitive sciences, of epistemology, and of complex systems, similarly to the \emph{Applied Epistemology} developed until 2011 by the CREA laboratory.} analyses that we propose to achieve.
}{
Commençons par situer le contexte des analyses en \emph{épistémologie quantitative}\footnote{Nous proposons d'utiliser ce terme pour des travaux à la croisée de la bibliométrie et de la scientométrie, des sciences cognitives, de l'épistémologie et des systèmes complexes, à l'image de l'\emph{Epistémologie Appliquée} développée jusqu'en 2011 par le laboratoire CREA.} que nous proposons de mener.
}


\subsection{Quantitative epistemology}{Epistémologie quantitative}

\bpar{
The possible methods for quantitative insights into epistemology are numerous. A good illustration of the variety of approaches is given by network analysis. Using citation network features, a good predicting power for citation patterns is for example obtained by~\cite{newman2014prediction}. Co-authorship networks can also be used for predictive models~\citep{sarigol2014predicting}. A multilayer network approach is proposed by~\cite{omodei2017evaluating}, using bipartites networks of papers and authors, in order to produce measures of interdisciplinarity using generalized centrality measures. Disciplines can be stratified into layers to reveal communities between them and therein collaboration patterns~\citep{battiston2016emergence}. Keyword networks are used in other fields such as economics of innovation: for example, \cite{choi2014patent} propose a method to identify technological opportunities by detecting important keywords from the point of view of topological measures. In a similar way, \cite{shibata2008detecting} use topological analysis of the citation network to detect emerging research fronts.
}{
Les méthodes possibles pour des entrées quantitatives en épistémologie sont nombreuses. Une bonne illustration de la variété des approches est donnée par l'analyse de réseau. En utilisant des caractéristiques topologiques du réseau de citations, un bon pouvoir prédictif pour les motifs de citation est par exemple obtenu par~\cite{newman2014prediction}. Les réseaux de co-auteurs peuvent également être utilisés pour des modèles prédictifs~\cite{sarigol2014predicting}. Une approche par réseau multi-couches est proposée dans~\cite{omodei2017evaluating}, qui utilise des réseaux bipartites d'articles et de chercheurs, afin de produire des mesures d'interdisciplinarité en utilisant des mesures de centralité généralisées. Les disciplines peuvent être stratifiées en couches pour révéler des communautés entre celles-ci et ainsi des motifs de collaboration~\cite{battiston2016emergence}. Les réseaux de mots-clés sont utilisés dans d'autres champs comme en économie de l'innovation : par exemple, \cite{choi2014patent} proposent une méthode pour identifier les opportunités technologiques par la détection de mots-clés importants du point de vue des mesures topologiques. De façon similaire, \cite{shibata2008detecting} utilisent une analyse topologique du réseau de citations pour détecter des fronts de recherche émergents.
}

\subsubsection{Systematic reviews}{Revues systématique}

\bpar{
With new technical means coming of age and the emergence of new data sources, the classical literature review tends to be coupled with automatic reviews. Techniques for systematic reviews have been developed, from qualitative reviews to quantitative meta-analysis which allow to produce new results by combining existing studies~\cite{rucker2012network}. Ignoring some references can even be considered as a scientific error in the context of the emergence of information systems which through an easier access to information makes the omission of key references difficult to justify~\cite{lissacksubliminal}\footnote{While remaining conscious that even with a systematic method, it is impossible to be absolutely exhaustive. The objective is to increase as much as possible the coverage, in the spirit of an approach inclusive of multiple viewpoints, as our epistemological positioning of perspectivism given in~\ref{sec:epistemology} proposes.}.
}{
Avec l'avènement des nouveaux moyens techniques et des nouvelles sources de données, la revue de littérature classique tend à se coupler à des revues automatiques. Des techniques de revue systématique ont été développées, des revues qualitatives aux méta-analyses quantitatives qui permettent de produire des nouveaux résultats par combinaison d'études existantes~\cite{rucker2012network}. Passer sous silence certaines références peut même être considéré comme une erreur scientifique dans le contexte de l'émergence des systèmes d'information qui par l'accès plus aisé à l'information rend difficilement justifiable l'omission de références clés~\cite{lissacksubliminal}\footnote{Tout en restant conscient que même avec une méthode systématique, il est impossible d'être absolument exhaustif. L'objectif est d'augmenter autant que possible la couverture, dans l'idée d'une approche inclusive de multiples points de vue, comme le propose notre positionnement épistémologique de perspectivisme donné en~\ref{sec:epistemology}.}.
}

\subsubsection{Interdisciplinarity}{Interdisciplinarité}

\bpar{
The development of interdisciplinary approaches is increasingly necessary for most of disciplines, both for further knowledge discovery but also societal impact of discoveries, as it was recently coined by the special issue of Nature~\citep{natureInterdisc}. \cite{banos2013pour} suggests that the development of such approaches must occur within a subtle spiral between and inside disciplines. An other way to understand this phenomenon is to understand it as the emergence of vertically integrated\footnote{I.e. integrating, generally between scales, different branches of a field: for example integrative biology~\cite{liu2005systems} aims at building bridges between genomics, physiology, ecology, by exploiting the integration of methods: experiments, modeling, simulation.} fields, conjointly with horizontal questions as detailed in the complex systems roadmap (\cite{2009arXiv0907.2221B}.
}{
Le développement d'approches interdisciplinaires est de plus en plus nécessaire pour la plupart des disciplines, à la fois pour la découverte de nouvelles connaissances mais aussi pour l'impact sociétal des découvertes, comme le rappelle récemment le volume spécial de la revue Nature~\citep{natureInterdisc}. \cite{banos2013pour} suggère que leur développement doit s'insérer dans une spirale subtile entre et au sein des disciplines. Une autre façon de voir ce phénomène est de le comprendre comme l'émergence de champs verticalement intégrés\footnote{C'est-à-dire intégrant, généralement entre les échelles, différentes branches d'un champ : par exemple la biologie intégrative~\cite{liu2005systems} vise à des ponts entre approches génomiques, approches physiologiques, approches écologiques, en tirant parti de l'intégration des méthodes : expérimentation, modélisation, simulation.} de manière conjointe aux questions horizontales comme détaillé dans la feuille de route des systèmes complexes \cite{2009arXiv0907.2221B}.
}

\bpar{
There are naturally multiple views on what is exactly interdisciplinarity (many other terms such as trans-disciplinarity, cross-disciplinarity also exist) and it actually depends on involved domains: recent hybrid disciplines (see e.g. the ones underlined  by \cite{bais2010praise} such as astro-biology, or others closer to our field such as geomatics) are a good illustration of the case in which entanglement is strong, whereas more loose fields such as ``urbanism'', which have multiple definitions and where integration is by essence horizontal, show to what extent horizontal integration is necessary and how transversal knowledge can be produced. Interactions between disciplines are not always smooth, as shows the misunderstandings when urban issues were recently introduced to physicists as \cite{dupuy2015sciences} recalls, misunderstandings which effects can be negative if they lead to conflicts or a neglect of knowledge already established by an other domain.
}{
Il existe naturellement de multiples points de vue sur ce qu'est exactement l'interdisciplinarité (de nombreux d'autres termes comme la trans-disciplinarité ou la cross-disciplinarité existent aussi) et cela dépend en fait des domaines impliqués : des disciplines hybrides apparues récemment (voir par exemples celles soulignées par \cite{bais2010praise} comme l'astro-biologie, ou d'autres plus proche de notre champ comme la géomatique) sont une bonne illustration du cas où les intrications sont très fortes, tandis que des champs comme ``l'urbanisme'' dont les définitions sont multiples montrent dans quelle mesure l'intégration horizontale est nécessaire et comment de la connaissance transversale peut être produite. Les interactions entre les disciplines ne sont pas toujours faciles, comme le montrent les malentendus lorsque les sujets sur la ville ont été récemment introduits aux physiciens comme \cite{dupuy2015sciences} le rappelle, malentendus dont les effets peuvent être négatifs s'ils conduisent à des conflits ou à une ignorance de connaissances déjà établies par un autre domaine.
}

\bpar{
These concerns are part of an understanding of processes of knowledge production, i.e. the \emph{knowledge of knowledge} as \cite{morin1986methode} puts it, in which evidence-based perspectives, involving quantitative approaches, play an important role. These paradigms can be understood as a \emph{quantitative epistemology}. Quantitative measures of interdisciplinarity would therefore be part of a multidimensional approach of the study of science that is in a way ``beyond bibliometrics''~\citep{cronin2014beyond}. The focus of this section is positioned within this stream of research. We first review existing approaches to the measure of interdisciplinarity.
}{
Ces questions font partie de la compréhension des processus de production de connaissance, i.e. la \emph{connaissance de la connaissance} comme \cite{morin1986methode} la présente, dans laquelle les perspectives \emph{evidence-based}, qui impliquent des approches quantitatives, jouent un rôle important. Ces paradigmes peuvent être compris comme une \emph{épistémologie quantitative}. Des mesures quantitatives de l'interdisciplinarité feraient pour cette raison partie d'une approche multidimensionnelle de l'étude de la science, qui va en quelque sorte ``au delà de la bibliométrie''~\cite{cronin2014beyond}. La préoccupation de cette section se positionne dans ce champ de recherche. Nous passons d'abord en revue les approches existantes à la mesure de l'interdisciplinarité.
}


\bpar{
Definitions of interdisciplinarity itself and indicators to measure it have already been tackled by a large body of literature. \cite{huutoniemi2010analyzing} recall the difference between \emph{multidisciplinary} (an aggregate of works from different disciplines) and \emph{interdisciplinary} (implying a certain level of integration) approaches. They construct a qualitative framework to classify types of interdisciplinarity, and for example distinguish empirical, theoretical and methodological interdisciplinarities. The multidimensionnal aspect of interdisciplinarity is confirmed even within a specific field such as literature~\citep{austin1996defining}. A first way to quantify interdisciplinarity of a set of publications is to look at the proportion of disciplines outside a main discipline in which they are published, as~\cite{rinia2002impact} do for the evaluation of projects in physics, complementary with judgement of experts. \cite{porter2007measuring} designate this measure as \emph{specialization}, and compares it with a measure of \emph{integration}, given by the spread of citations  done by a paper within the different Subject Categories (classification of the Web of Knowledge), which is also called the \emph{Rao-Stirling} index. \cite{lariviere2010relationship} uses it on a Web of Science corpus to show the existence of an optimal intermediate level of interdisciplinarity for the citation impact within a five year window. A similar work is done in~\citep{lariviere201410}, focusing on the evolution of measures on a long time range. The influence of missing data on this index is studied by \cite{moreno2016uncertainty}, providing an extended framework taking into account uncertainty. The use of networks has also been proposed : \cite{porter2009science} combine the integration index with a mapping technique which consists in visualisation of synthetic networks constructed by co-citations between disciplines. \cite{leydesdorff2007betweenness} shows that the betweenness centrality is a relevant indicator of interdisciplinarity, when considering appropriate citation neighborhood.
}{
Les définitions de l'interdisciplinarité elle-même et les indicateurs pour la mesurer ont déjà été traités par un vaste corpus de littérature. \cite{huutoniemi2010analyzing} rappelle la différence entre les approches \emph{multi-disciplinaires} (une agrégation de travaux de différentes disciplines) et \emph{interdisciplinaires} (qui implique un certain niveau d'intégration). Ils construisent un cadre qualitatif pour classifier différents types d'interdisciplinarité, et distinguent par exemple les interdisciplinarités empiriques, théoriques et méthodologique. L'aspect multi-dimensionnel de l'interdisciplinarité est confirmé même au sein d'un champ spécifique comme la littérature~\cite{austin1996defining}. Une première façon de quantifier l'interdisciplinarité d'un ensemble de publications est de regarder la proportion de disciplines hors d'une discipline principale dans lesquelles elles sont publiées, comme~\cite{rinia2002impact} fait pour l'évaluation de projets en physique, de manière complémentaire au jugement d'experts. \cite{porter2007measuring} désigne cette mesure comme \emph{spécialisation}, et la compare avec une mesure d'\emph{intégration} donnée par l'étendue des citations faites par un article au sein des différentes \emph{Subject Categories} (classification du \emph{Web of Knowledge}), qui est également appelé indice de \emph{Rao-Stirling}. \cite{lariviere2010relationship} l'utilise sur un corpus du \emph{Web of Science} pour montrer l'existence d'un niveau d'interdisciplinarité intermédiaire optimal pour l'impact en termes de citations sur une fenêtre de 5 ans post-publication. Un travail équivalent est fait dans~\cite{lariviere201410}, qui se concentre sur l'évolution des mesures sur une longue portée temporelle. L'influence des données manquantes sur cet index est étudié par~\cite{moreno2016uncertainty}, qui fournit un cadre étendu qui prend en compte l'incertitude. L'utilisation de réseaux a également été proposée : \cite{porter2009science} combine l'indice d'intégration avec une technique de cartographie qui consiste en la visualisation de réseaux synthétiques construits par les co-citations entre disciplines. \cite{leydesdorff2007betweenness} montre que la centralité de chemin est un indicateur pertinent d'interdisciplinarité, lorsqu'un environnement de citation pertinent est considéré.
}


\subsection{Algorithmic systematic review}{Revue systématique algorithmique}


\bpar{
We propose in a preliminary way to proceed to a systematic and algorithmic literature review. A formal iterative algorithm to construct corpuses of references starting from initial keywords, based on text-mining, is developed and put into practice. We study its convergence properties and proceed to a sensitivity analysis. We then apply it to requests representing our specific question, for which results tend to confirm the hypothesis of a relative isolation between disciplines.
}{
Nous proposons de procéder de manière préliminaire à une revue de la littérature systématique et algorithmique. Un algorithme itératif formel pour construire des corpus de références à partir de mots-clés initiaux, basé sur l'analyse textuelle, est développé et mis en oeuvre. Nous étudions ses propriétés de convergence et procédons à une analyse de sensibilité. Nous l'appliquons ensuite à des requêtes représentatives de notre question spécifique, pour lesquelles les résultats tendent à confirmer l'hypothèse d'isolation relative des disciplines.
}


\bpar{
Whereas most studies in bibliometrics rely on citation networks~\cite{newman2014prediction} or co-authorship networks~\cite{sarigol2014predicting}, we propose to use a less studied paradigm, based on text-mining, introduced by~\cite{chavalarias2013phylomemetic}, which produces a dynamical mapping of scientific disciplines based on their semantic content. We follow the approach of grasping the diversity of domains, introduced in~\ref{sec:modelingsa}, by this supplementary information on the scientific landscape. Methods we introduce are particularly suited for our study since we aim at understanding the structure of the content of researches on the subject.
}{
Tandis que la majorité des études en bibliométrie se reposent sur les réseaux de citations~\cite{newman2014prediction} ou les réseaux de co-auteurs~\cite{sarigol2014predicting}, nous proposons d'utiliser un paradigme moins exploré, basé sur l'analyse textuelle, introduit par~\cite{chavalarias2013phylomemetic}, qui produit une cartographie dynamique des disciplines scientifiques en se basant sur leur contenu sémantique. Nous prenons le parti d'une appréhension de la diversité des domaines, introduite en~\ref{sec:modelingsa}, par cette information supplémentaire du paysage scientifique. Les méthodes que nous introduisons sont particulièrement adaptées pour notre étude puisque nous voulons comprendre la structure du contenu des recherches sur le sujet.
}





\begin{figure}
\centering
\includegraphics[width=0.8\linewidth]{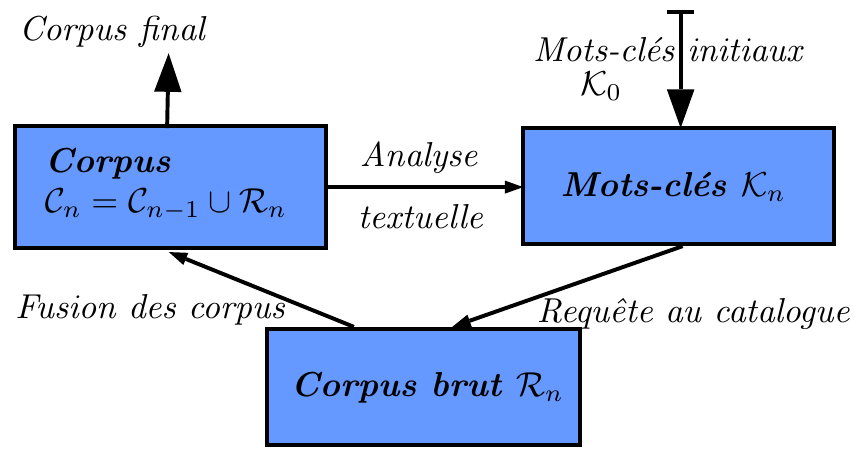}
\caption[Systematic review algorithm]{\textbf{Global architecture of the algorithm.} Starting from an initial set of keywords, we construct a corpus through a catalog request, from which new keywords are extracted by text-mining. We then iterate in loop until obtaining a fixed corpus or reaching a fixed maximal number of iterations.\label{fig:quantepistemo:algo}}
\end{figure}

\bpar{
The algorithm proceeds by iterations to obtain a stabilized corpus starting from initial keywords, reconstructing the scientific semantic horizon around a given subject. The formal description of the algorithm is detailed in Appendix~\ref{app:sec:quantepistemo}, with details of its implementation and sensitivity analyses. Its logic is given by the schema in Fig.~\ref{fig:quantepistemo:algo}: given a set of initial keywords that are gathered into a unique request, works using them are gathered, from which new keywords are extracted to iterate in a loop until eventual convergence.
}{
L'algorithme procède par itérations pour obtenir un corpus stabilisé à partir de mots-clés initiaux, reconstruisant l'horizon sémantique scientifique autour d'un sujet donné. La description formelle de l'algorithme est détaillée en Annexe~\ref{app:sec:quantepistemo}, avec les détails de son implémentation et des analyses de sensibilité. Sa logique est donnée par le schéma en Fig.~\ref{fig:quantepistemo:algo} : étant donné un ensemble de mots-clés de départ que l'on rassemble en une unique requête, on récolte des travaux qui en traitent, dont on extrait de nouveaux mots-clés pour itérer en boucle jusqu'à convergence éventuelle.
}

\bpar{
We start from five different initial requests that were manually extracted from the various domains identified in the bibliography\footnote{Which are ``cityANDsystemANDnetwork'', ``land-useANDtransportANDinteraction'', ``networkANDurbanANDmodeling'', ``populationANDdensityANDtransport'', ``transportationANDnetworkANDurbanANDgrowth''. This choice includes systems of cities approaches, LUTI approaches, network growth approaches. It can of course not be exhaustive. This study being preliminary we admit to potentially work on samples. For example, the use of ``co-evolution'' is not satisfying since too few articles use this formulation. Similarly, the question of language conditions the results: a request in French leads to linguistic niches finally relatively poor in diversity, and we thus do only requests in English. The hypernetwork approach developed later will however be multilingual.}, in order to compare corpuses obtained for each request. After having constructed the corpuses, we study their lexical consistence as an indicator to answer our initial question. Large distances should confirm the hypothesis formulated above, i.e. that self-centered disciplines may be at the origin of a lack of interest for co-evolutive models. The Table~\ref{tab:quantepistemo:lexical} shows the values of the relative lexical proximity, that we define through a weighted set similarity index given by
\[
d(I,J) = \frac{\sum_{k_i \in I, k_j \in J} \mathbbm{1}_{k_i = k_j} \cdot (s(k_i) + s(k_j))}{\sum_{k_i \in I} s(k_i) + \sum_{k_j \in J} s(k_j)}
\]
for corpuses $I,J$, and with $s$ strictly positive function giving a measure of the importance of words within corpuses, produced by the keyword extraction method (see~\ref{app:sec:quantepistemo}). Its values are significantly low in comparison to the reference value 1 for equal corpuses (the measure is interpreted as a proportion of co-occurring keywords), what tends to confirm our hypothesis\footnote{To situate these results in a relative way, we would need a null model (i.e. generating corpuses with similar semantic distributions but without a correlation structure between words) with random corpuses for example, what could be the object of future developments.}.
}{
Nous partons de cinq différentes requêtes initiales qui ont été manuellement extraites des divers domaines identifiés dans la bibliographie\footnote{Qui sont ``cityANDsystemANDnetwork'', ``land-useANDtransportANDinteraction'', ``networkANDurbanANDmodeling'', ``populationANDdensityANDtransport'', ``transportationANDnetworkANDurbanANDgrowth''. Ce choix inclut les approches par systèmes de villes, les approches LUTI, les approches de croissance de réseau. Il ne peut bien sûr être exhaustif. Cette étude étant préliminaire on admet de travailler potentiellement sur des échantillons. Par exemple, l'utilisation de ``co-evolution'' n'est pas concluante car trop peu d'articles utilisent cette formulation. De même, la question de la langue conditionne les résultats : une requête en Français conduit à des niches linguistiques finalement assez pauvres en diversité, et nous faisons ainsi uniquement des requêtes en anglais. L'approche par hyper-réseau développée plus loin sera elle multilingue.}, afin de comparer les corpus obtenus pour chaque requête. Après avoir construit les corpus, nous étudions leur cohérence lexicale comme un indicateur de réponse à notre question initiale. De grande distances devraient confirmer l'hypothèse formulée ci-dessus, i.e. que des disciplines auto-centrées pourraient être à l'origine d'un manque d'intérêt pour des modèles co-évolutifs. La Table~\ref{tab:quantepistemo:lexical} montre les valeurs de la proximité lexicale relative, que nous définissons par un indice de similarité d'ensemble pondéré donné par
\[
d(I,J) = \frac{\sum_{k_i \in I, k_j \in J} \mathbbm{1}_{k_i = k_j} \cdot (s(k_i) + s(k_j))}{\sum_{k_i \in I} s(k_i) + \sum_{k_j \in J} s(k_j)}
\]
pour les corpus $I,J$, et avec $s$ fonction strictement positive donnant une mesure d'importance des mots au sein des corpus fournie par la méthode d'extraction des mots-clés (voir~\ref{app:sec:quantepistemo}). Ses valeurs sont significativement faibles en comparaison à la valeur de référence 1 pour des corpus égaux (la mesure s'interprète comme une proportion de mots en co-occurrence), ce qui tend à confirmer notre hypothèse\footnote{Pour situer ces résultats de manière relative, il faudrait un modèle nul (c'est-à-dire générateur de corpus avec distributions sémantiques similaires mais sans structure de corrélation entre mots) avec des corpus aléatoires par exemple, ce qui pourrait faire l'objet de développements futurs.}.
}


\begin{table}
\caption[Lexical proximities between final corpuses]{\textbf{Symmetric matrix of lexical proximities between final corpuses.} These are defined as the sum of overall final keywords co-occurrences between corpuses, normalized by the total weight of final keywords. The size of final corpuses is given by $W$. The values obtained for proximities are considerably low compared to the maximal value 1, what confirms that corpus are significantly distant.\label{tab:quantepistemo:lexical}}
\includegraphics[width=0.8\linewidth]{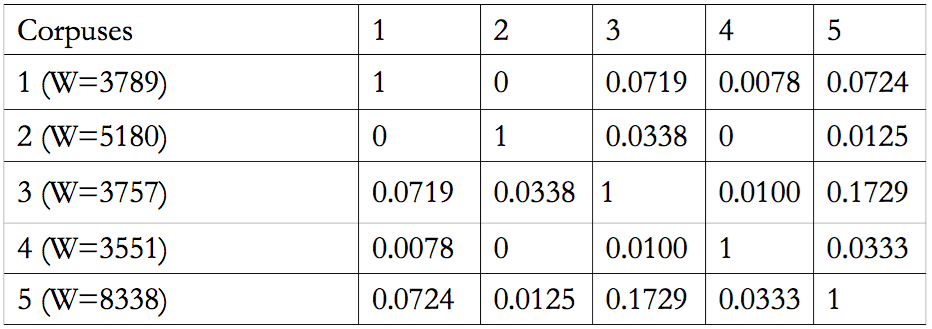}
\end{table}


\bpar{
The ascertainment of a low number of models which simulate the co-evolution between transportation networks and urban land-use could be due to the absence of communication between the scientific disciplines studying different aspects of the problem. Other possible explanations which are close can for example be the lack of concrete application cases of such models given the time scales implied and thus the absence of proper research funds - what is not so far from the absence of a discipline which would devote some of its objects to it. This question of ranges and scales of models will be the subject of the meta-analysis in the next section~\ref{sec:modelography}. To conclude, we have proposed here an algorithmic method to give elements of answer through corpus extraction based on text-mining, which numerical results seems to confirm a compartmentalization of disciplines (in the particular sense used here of a semantic distances between niche corpuses). This analyses remained relatively limited in the scope of its results, in particular because of the low number of requests and a certain amount of intrinsic uncertainties, but is \emph{sufficient} to produce a diagnosis, namely (i) a disciplinary structure strongly characterized can be extracted from corpus analysis, and (ii) the use of semantic tools allows the extraction of an endogenous information. Starting from this preliminary diagnosis, we propose to deepen the analysis by a variation and extension of the method used.
}{
La constatation d'un faible nombre de modèles qui simulent la co-évolution des réseaux de transport et de l'usage du sol urbain pourrait être due à l'absence de communication entre les disciplines scientifiques étudiant différents aspects du problème. D'autres explications possibles qui en sont proches peuvent par exemple être le manque de cas d'application concrets de tels modèles vu les échelles temporelles mises en jeu et donc l'absence de financement propre - ce qui n'est pas si loin de l'absence d'une discipline y consacrant certains de ses objets. Cette question des portées et des échelles des modèles fera l'objet de la meta-analyse à la section suivante~\ref{sec:modelography}. Ainsi, nous avons proposé ici une méthode algorithmique pour donner des éléments de réponse par l'extraction de corpus basée sur l'analyse textuelle, dont les résultats numériques semblent aller dans le sens d'une compartimentation des disciplines (au sens particulier utilisé ici d'une distance sémantique entre corpus niches). Cette analyse était relativement limitée dans la portée de ses résultats, notamment par le faible nombre de requêtes et un certain nombre d'incertitudes intrinsèques, mais est \emph{suffisante} pour produire un diagnostique, à savoir (i) une structure disciplinaire fortement marquée peut être extraite de l'analyse de corpus, et (ii) l'utilisation d'outils sémantiques permet une extraction d'information endogène. Fort de ce diagnostique préliminaire, nous proposons d'approfondir l'analyse par une variation et extension de la méthode employée.
}



\subsection{Indirect bibliometrics}{Bibliométrie indirecte}

\label{subsec:indirectbibliometrics}

\bpar{
As described before, semantic analysis of final corpus does not contain all the information on disciplinary compartmentation nor on patterns of propagation of scientific knowledge as the ones contained in citation networks for example. Furthermore, data collection in the previous algorithm is subject to convergence towards self-consistent themes because of the proper structure of the method. It could be possible to obtain more information on social patterns of ontological choices in modeling by studying communities in broader networks, that would more correspond to disciplines (or sub-disciplines depending on granularity level). We propose to reconstruct disciplines around our thematic, to obtain a more precise view of the scientific landscape on our subject and of the links between disciplines. A fundamental contribution of this section relies in the construction of ah hybrid dataset from heterogeneous sources, and the development of associated tools which can be reused and extended for similar applications. This approach can be understood as indirect bibliometrics\footnote{Bibliometrics, or scientometrics when it is applied in particular to science as in our case, consists in the measure and qualification of knowledge production patterns through the intermediary of their directly observable proxies (scientific productions, mechanisms of institutions, social relations between researchers, etc.)~\cite{cronin2014beyond}. This book recalls that this field is in complete mutation and sketches a map of new approaches.}, since we aim at reconstructing an endogenous information and at extracting relations between different dimensions.
}{
Comme décrit précédemment, l'analyse sémantique des corpus finaux ne contient pas la totalité de l'information sur les liens entre disciplines ni sur les motifs de propagation de la connaissance scientifique comme ceux contenus dans les réseaux de citations par exemple. De plus, la collection des données dans l'algorithme précédent est sujette à convergence vers des thèmes relativement auto-cohérents de par la structure propre de la méthode. Il serait possible d'obtenir plus d'information sur les motifs sociaux de choix ontologiques pour la modélisation en étudiant les communautés dans des réseaux plus larges, ce qui correspondrait plus à des disciplines (ou des sous-disciplines selon le niveau de granularité). Nous proposons de reconstruire les disciplines autour de notre thématique, pour obtenir une vue plus précise du paysage scientifique sur notre sujet et des liens entre disciplines. Une contribution fondamentale de cette section consiste en la construction de jeux de données hybrides à partir de sources hétérogènes, et les développement des outils associés qui peuvent être réutilisés et améliorés pour des applications similaires. Cette démarche peut être vue comme une bibliométrie indirecte\footnote{La bibliométrie, ou scientométrie lorsqu'elle est appliquée en particulier à la science comme dans notre cas, consiste en la mesure et la qualification des motifs de production de connaissance par l'intermédiaire de leur supports directement observables (productions scientifiques, fonctionnement des institutions, relations sociales entre chercheurs, etc.)~\cite{cronin2014beyond}. Cet ouvrage rappelle que ce domaine est en pleine mutation et dresse une carte des nouvelles approches.}, puisqu'on cherche à reconstruire une information endogène et à extraire des relations entre différentes dimensions.
}


\subsubsection{Context}{Contexte}

\bpar{
The approach developed here couples citation network exploration and analysis with text-mining, aiming at mapping the scientific landscape in the neighborhood of a particular corpus. The context is particularly interesting for the methodology developed. First of all, the subject studied is very broad and by essence interdisciplinary. Secondly, bibliographical data are difficult to obtain, raising the concern of how the perception of a scientific landscape may be shaped by actors of the dissemination and thus far from objective, making technical solutions as the ones consequently developed here crucial tools for an open and neutral science.
}{
L'approche développée ici couple exploration et analyse de réseau de citations avec analyse textuelle, dans le but de cartographier le paysage scientifique dans le voisinage d'un corpus donné. Le contexte est particulièrement intéressant pour la méthodologie développée. Premièrement, le sujet étudié est très large et par essence interdisciplinaire. Deuxièmement, les données bibliographiques sont difficiles à obtenir, soulevant la question de comment la perception d'un horizon scientifique peut être déterminée par les acteurs de la dissémination et donc loin d'être objective, rendant les solutions techniques comme celle développée ici en conséquence des outils cruciaux pour une science ouverte et neutre.
}

\bpar{
Our approach combines semantic communities analysis (as done in~\cite{palchykov2016ground} for papers in physics but without keyword extraction, or by \cite{2015arXiv151003797G} for an analysis of semantic networks of political debates) with citation network analysis, to extract e.g. interdisciplinarity measures. Our contribution differs from the previous works quantifying interdisciplinarity as it does not assume predefined domains nor classification of the considered papers, but reconstructs from the bottom-up the fields with the endogenous semantic information. \cite{nichols2014topic} already introduced a close approach, using Latent Dirichlet Allocation topic modeling\footnote{The LDA model, introduced by~\cite{blei2003latent}, assumes that documents are produced by underlying themes, with a Dirichlet distribution for their composition and also for the distribution of words by themes. Its estimation gives the composition of themes in terms of keywords.} to characterize interdisciplinarity of awards in particular sciences.
}{
Notre approche combine une analyse des communautés sémantiques (comme fait dans~\cite{palchykov2016ground} pour les articles en physique mais sans extraction des mots-clés, ou par \cite{2015arXiv151003797G} pour une analyse des réseaux sémantiques de débats politiques) avec celle du réseau de citations pour extraire par exemple des mesures d'interdisciplinarité. Cette contribution se démarque des travaux précédents quantifiant l'interdisciplinarité puisqu'elle ne suppose pas de domaines a priori ou une classification des références considérées, mais reconstruit par le bas les champs via l'information sémantique endogène. \cite{nichols2014topic} introduit une approche similaire, utilisant le modèle d'extraction de thématiques \emph{Latent Dirichlet Allocation}\footnote{Le modèle LDA, introduit par~\cite{blei2003latent}, suppose les documents comme produits par des thèmes sous-jacents, avec une distribution de Dirichlet pour leur composition ainsi que pour la distribution des mots par thèmes. Son estimation donne la composition des thèmes en termes de mots-clés.} pour caractériser l'interdisciplinarité de récompenses dans des sciences précises.
}

\subsubsection{Dataset}{Données}

\bpar{
Our approach imposes some requirements on the dataset used, namely: (i) cover a certain neighborhood of the studied corpus in the citation network in order to have a view on the scientific landscape the less biased as possible; (ii) have at least a textual description for each node. For these to be met, we need to gather and compile data from heterogeneous sources, using therefore a specific architecture and implementation, described in Appendix~\ref{app:sec:cybergeo}. For the sake of simplicity, we will denote by \emph{reference} any standard scientific production\footnote{What is of course a subject of debate, see our discussions in opening on the evolution of the modes of scientific communication.} which can be cited by another (journal paper, book, book chapter, conference paper, communication, etc.) and contains basic records (title, abstract, authors, publication year). We will work in the following on the network of references.
}{
Notre approche implique des spécifications pour le jeu de données utilisé, à savoir : (i) couvrir un voisinage conséquent du corpus étudié dans le réseau de citation afin d'avoir une vue la moins biaisée possible du paysage scientifique ; (ii) avoir au moins une description textuelle pour chaque noeud. Pour cela, nous rassemblons et compilons les données de sources hétérogènes en utilisant une architecture et implémentation spécifiques, décrites en Annexe~\ref{app:sec:cybergeo}. Pour simplifier, nous dénommons \emph{référence} toute production scientifique standard\footnote{Ce qui est bien sûr sujet à débat, voir nos discussions en ouverture sur l'évolution des modes de communication scientifique.} qui peut être citée par une autre (articles de journaux, livre, chapitre de livre, article d'actes, communication, etc.) et contient des informations de base (titre, résumé, auteurs, année de publication). Nous travaillons par la suite sur le réseau des références.
}

\paragraph{Initial Corpus}{Corpus Initial}

\bpar{
Our initial corpus is constructed starting from the state-of-the-art established in~\ref{sec:modelingsa}. Its complete composition is given in Appendix~\ref{app:sec:quantepistemo}. It consists in seven ``key'' references identified for each of the disciplines previously described. The aim here is not to be exhaustive (it will be in~\ref{sec:modelography}), but to construct a description of the neighborhood of domains we deal with. It is taken with a reasonable size (leading to a final network that can be processed without a specific method regarding the size of data), but the methods used here have been developed on massive datasets, for patents for example~\cite{bergeaud2017classifying}, and as it will be in Appendix~\ref{app:reflexivity} to our full bibliography.
}{
Notre corpus initial est construit à partir de l'état de l'art établi en~\ref{sec:modelingsa}. Sa composition complète est donnée en Annexe~\ref{app:sec:quantepistemo}. Il s'agit de 7 références ``phares'' identifiées pour chacune des disciplines abordées précédemment. Le but ici n'est pas d'être exhaustif (cela le sera en~\ref{sec:modelography}), mais de construire une description du voisinage des domaines qui nous concernent. Celui-ci est pris de taille raisonnable (conduisant à un réseau final traitable sans méthode spécifique concernant la taille des données), mais les méthodes utilisées ici ont été développées sur des données massives, pour les brevets par exemple~\cite{bergeaud2017classifying}, et comme il le sera en Annexe~\ref{app:reflexivity} à l'ensemble de notre bibliographie.
}

\paragraph{Citation data}{Données de citation}

\bpar{
Citation data is collected from \texttt{Google Scholar} which is often the only source for incoming citations~\cite{noruzi2005google} since in social sciences and humanities articles are not systematically references by database proposing (paying) services such as the citation network\footnote{For example, the Cybergeo journal is indexed by \textit{Web of Science} only since May 2016, after difficult negotiations and not without a counterpart.}. We are aware of the possible biaises using this single source (see e.g.~\cite{bohannon2014scientific})\footnote{Or \url{http://iscpif.fr/blog/2016/02/the-strange-arithmetic-of-google-scholars}.}, but these critics are more directed towards search results than citation counts. We thus retrieve \emph{citing} references at depth two, i.e. the references citing the initial corpus and the ones citing these ones. The network obtained contains $V=9462$ references corresponding to $E=12004$ citation links. Concerning languages, English covers 87\% of the corpus, French 6\%, Spanish 3\%, German 1\%, completed by other languages such as Mandarin that can be undefined (its detection has a low robustness).
}{
Le réseau de citations est reconstruit à partir de \texttt{Google Scholar} qui est souvent l'unique source des citations entrantes~\cite{noruzi2005google} puisqu'en sciences humaines les ouvrages ne sont pas systématiquement référencés par les bases fournissant des services (payants) comme le réseau de citation\footnote{Par exemple, le journal Cybergeo n'est indexé dans le \emph{Web of Science} que depuis mai 2016, suite à des négociations ardues et non sans contrepartie.}. Nous sommes conscient des biais possibles de l'utilisation de cette source unique (voir par exemple~\cite{bohannon2014scientific})\footnote{Ou \url{http://iscpif.fr/blog/2016/02/the-strange-arithmetic-of-google-scholars}.}, mais ces critiques sont dirigées vers les résultats de recherche plutôt que les comptes de citations. Nous récoltons ainsi les références \emph{citantes} à profondeur deux, c'est-à-dire les références citant le corpus initial et celles citant celles-ci. Le réseau obtenu contient $V=9462$ références correspondant à $E=12004$ liens de citation. Concernant les langues, l'anglais représente 87\% du corpus, le français 6\%, l'espagnol 3\%, l'allemand 1\%, complété par des langues comme le mandarin pouvant être indéfinies (la détection de celui-ci étant peu fiable). 
}

\paragraph{Text data}{Données textuelles}

\bpar{
To proceed to the semantic analysis, a description consequent enough is necessary. We collect therefore abstracts for the previous network. These are available for around one third of references, giving $V=3510$ nodes with a textual description.
}{
Pour mener l'analyse sémantique, une description suffisamment conséquente est nécessaire. Nous collectons pour cela les résumés pour le réseau précédent. Ceux-ci sont disponibles pour environ un tiers des références, donnant $V=3510$ noeuds avec description textuelle.
}

\subsubsection{Results}{Résultats}

\paragraph{Citation network}{Réseau de citations}

\bpar{
Basic statistics for the citation network already give interesting informations. The network has an average degree of $\bar{d}=2.53$ and a density of $\gamma=0.0013$\footnote{For reference, \cite{batagelj2003efficient} presents the characteristics of 11 scientific networks from diverse domains and with a size varying from 40 to 8851 nodes, and reports densities varying from $3.3\cdot 10^{-4}$ to $0.038$, with a median at $0.003$, close to the one of our network.}. The average in-degree (which can be interpreted as a stationary impact factor) is of $1.26$, what is relatively high for social sciences. It is important to note that it has a single weak connected component, what means that initial domains are not in total isolation: initial references are shared at a minimal degree by the different domains. We work in the following on the sub-network of nodes having at least two links, to extract the core of network structure and to remove the ``cluster'' effect (nodes with a high number of leaf neighbors). Furthermore, the network is necessarily complete between these nodes since we went up to the second level.
}{
Des statistiques basiques pour le réseau de citation donnent déjà des informations intéressantes. Le réseau a un degré moyen de $\bar{d}=2.53$ et une densité de $\gamma=0.0013$\footnote{Pour référence, \cite{batagelj2003efficient} présente les caractéristiques de 11 réseaux scientifiques de domaines divers et de taille allant de 40 à 8851 noeuds, et reporte des densités variant de $3.3\cdot 10^{-4}$ à $0.038$, avec une médiane à $0.003$, proche de celle de notre réseau.}. Le degré entrant moyen (qui peut être interprété comme un facteur d'impact stationnaire) est de $1.26$, ce qui est relativement élevé pour des sciences humaines. Il est important de noter sa connexité faible, ce qui signifie que les domaines initiaux ne sont pas en isolation totale : les références initiales sont partagées à un degré minimal par les différents domaines. Nous travaillons par la suite sur le sous-réseau des noeuds comprenant au moins deux liens, pour extraire le coeur de la structure du réseau et se débarrasser de l'effet ``grappe''. De plus, le réseau est nécessairement complet entre ces noeuds puisqu'on est remonté au deuxième niveau.
}

\bpar{
We proceed for the citation network to a community detection with the Louvain algorithm, on the corresponding non-directed network. The algorithm gives 13 communities, with a directed modularity of 0.66\footnote{Modularity is a measure of the ``level of clustering'' of a partition of a network into classes. The Louvain algorithm constructs communities by a greedy optimization of modularity.}, extremely significant in comparison to a bootstrap estimation of the same measure on the randomly rewired network with gives a modularity of $0.0005 \pm 0.0051$ on $N=100$ repetitions. Communities make sense in a thematic way, since we recover for the largest the domains presented in Table~\ref{tab:quantepistemo:citation}.
}{
Nous procédons pour le réseau de citation à une détection de communautés par l'algorithme de Louvain, sur le réseau non-dirigé correspondant. L'algorithme fournit 13 communautés, de modularité dirigée 0.66\footnote{La modularité est une mesure du ``niveau de clustering'' d'une partition d'un réseau en classes. L'algorithme de Louvain construit les communautés par optimisation gourmande de la modularité.}, extrêmement significative en comparaison à une estimation par bootstrap de la même mesure sur le graphe aléatoirement rebranché qui donne une modularité de $0.0005 \pm 0.0051$ sur $N=100$ répétitions. Les communautés font sens de manière thématique, puisqu'on retrouve pour les plus grosses les domaines présentés dans la Table~\ref{tab:quantepistemo:citation}.
}

\begin{table}
\caption[Description of citation communities]{\textbf{Description and size of citation communities.}\label{tab:quantepistemo:citation}}
\bpar{
\begin{tabular}{|l|l|}
\hline
	Domain & Size (\% of nodes)\\\hline
	LUTI & 18\% \\\hline
	Urban and Transport Geography & 16\% \\\hline
	Infrastructure planning & 12\% \\\hline
	Integrated planning - TOD & 6\% \\\hline
	Spatial Networks & 17\% \\\hline
	Accessibility stucies & 18\% \\\hline
\end{tabular}
}{
\begin{tabular}{|l|l|}
\hline
	Domaine & Taille (\% de noeuds)\\\hline
	LUTI & 18\% \\\hline
	Géographie Urbaine et des Transports & 16\% \\\hline
	Planification des infrastructures & 12\% \\\hline
	Planification intégrée - TOD & 6\% \\\hline
	Réseaux Spatiaux & 17\% \\\hline
	Études d'accessibilité & 18\% \\\hline
\end{tabular}
}
\end{table}

\bpar{
Naming of communities are done a posteriori from expert view, according to the broad fields unveiled in the literature review in~\ref{sec:modelingsa}\footnote{We note that this naming is indeed exogenous and necessarily subjective. As further developed for the semantic network, there does not exist any simple technique for an endogenous naming. We must keep this aspect in mind for the positioning of interpretations and conclusions.}.
}{
Les appellations sont à regard d'expert a posteriori, selon les grands domaines dégagés dans la revue de littérature en~\ref{sec:modelingsa}\footnote{On note que cette dénomination est bien exogène et nécessairement subjective. Comme développé plus loin pour le réseau sémantique, il n'existe pas de technique simple pour une désignation endogène. Il faut garder cet aspect en tête pour la mise en perspective des interprétations et conclusions.}.
}

\begin{figure}[!ht]
\includegraphics[width=\linewidth]{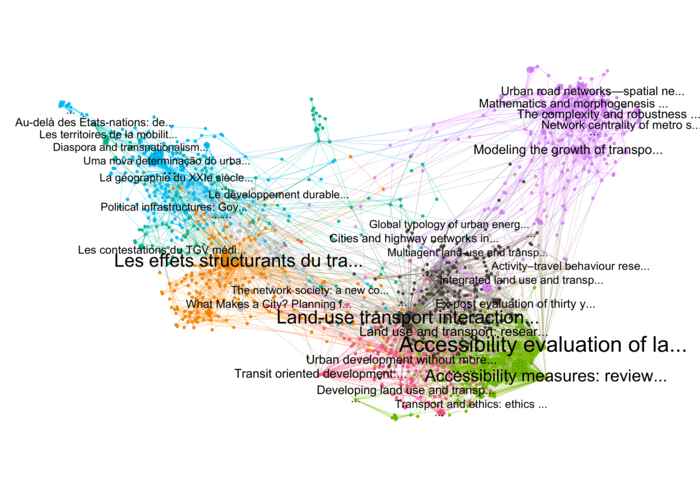}
\caption[Citation Network]{\textbf{Citation Network.} We visualize references having at least two links, using a force-atlas algorithm. Colors give communities described in text. In orange, blue, turquoise: urban geography, transport geography, political sciences; in pink, black, green: planning, accessibility, LUTI; in purple: spatial networks (physics and economics).\label{fig:quantepistemo:citnw}}
\end{figure}

\bpar{
The Fig.~~\ref{fig:quantepistemo:citnw} shows the citation network and allows to visualize the relations between these domains. It is interesting to observe that works by economists and physicists in this field fall within the same category of the study of \emph{Spatial Networks}. Indeed, the literature cited by physicists contains often a larger number of references in economics than in geography, whereas economists use network analysis techniques. Moreover, planning, accessibility, LUTI and TOD are very close but can be distinguished in their specificities: the fact that they appear as separated communities witnesses of a certain level of compartmentalization. These make the bridge between spatial network approaches and geographical approaches, which contain an important part of political science for example. Links between physics and geography remain rather low. This overview naturally depends on the initial corpus, but allows us to better understand its context in its disciplinary environment.
}{
La Fig.~\ref{fig:quantepistemo:citnw} montre le réseau de citation et permet de visualiser les relations entre ces domaines. Il est intéressant d'observer que les travaux des économistes et des physiciens dans le domaine tombent dans la même catégorie d'étude des \emph{Spatial Networks}. En effet, la littérature citée par les physiciens comporte souvent plus d'ouvrage en économie qu'en géographie, tandis que les économistes utilisent des techniques d'analyse de réseau. Ensuite, le planning, l'accessibilité, les LUTI et le TOD sont très proches mais se distinguent dans leur spécificités : le fait qu'ils apparaissent dans des communautés séparées témoigne d'un certain niveau de cloisonnement. Ceux-ci font le pont entre les approches réseaux spatiaux et les approches géographiques, qui comportent une partie importante de sciences politiques par exemple. Les liens entre physique et géographie restent très faibles. Ce panorama dépend bien sûr du corpus initial, mais nous permet de mieux comprendre le contexte de celui-ci dans son environnement disciplinaire.
}

\paragraph{Semantic communities}{Communautés sémantiques}

\bpar{
The extraction of keywords is done following an heuristic inspired by~\cite{chavalarias2013phylomemetic}. The complete description of the method and its implementation if given in Appendix~\ref{app:sec:cybergeo}. It is based on second-order relations between semantic entities, which are \emph{n-grams}, i.e. multiple keywords which can have a length up to three. These are estimated by the intermediate of the co-occurence matrix, which statistical properties yield a measure of deviation from uniform co-occurrences, which is used to evaluate the relevance of keywords. By selecting a fixed number of relevant keywords $K_W = 10000$, we can then construct a network weighted by co-occurrences.
}{
L'extraction des mots-clés est faite suivant une heuristique inspirée de~\cite{chavalarias2013phylomemetic}. La description complète de la méthode et de son implémentation est donnée en Annexe~\ref{app:sec:cybergeo}. Elle se base sur les relations au second ordre entre les entités sémantiques, qui sont des \emph{n-grams}, c'est-à-dire des mots-clés multiples pouvant avoir une longueur jusqu'à 3. Celles-ci sont estimées via la matrice de co-occurrence, dont les propriétés statistiques fournissent une mesure de déviation à des co-occurrences uniformes, qui est utilisée pour juger la pertinence des mots-clés. Sélectionnant un nombre fixe de mots-clés pertinents $K_W = 10000$, nous pouvons ensuite construire un réseau pondéré par les co-occurrences.
}

\bpar{
The topology of the raw network does not allow the extraction of clear communities, in particular because of the presence of hubs that correspond to frequent terms common to many fields (e.g. \texttt{model}, \texttt{space}). These words are used in a comparable way in all the studied fields, and do not carry information to separate them\footnote{But they will carry some if we were comparing a corpus in quantitative geography and a corpus in musicology for example.}. We make the assumption that these highest degree terms do not carry specific information on particular classes and can be thus filtered given a maximal degree threshold $k_{max}$ (we are thus interested in what makes the specificity of each domain). Similarly, edges with small weight are considered as noise and filtered according to a minimal edge weight threshold $\theta_w$. The generic method furthermore allows a preliminary filtration of keywords, according to a document frequency window $\left[ f_{min},f_{max} \right]$, to which results are not sensitive in our case. The sensitivity analysis of the characteristics of the filtered network, in particular its size, modularity and community structure, is given in Fig.~\ref{app:sec:quantepistemo}. We choose parameter values allowing a multi-objective optimization between modularity and network size, $\theta_w = 10,k_{max} = 500$, by the choice of a compromise point on a Pareto front, what gives a semantic network of size $(V=7063,E=48952)$. It is visualized in Appendix~\ref{app:sec:quantepistemo}.
}{
La topologie du réseau brut ne permet pas l'extraction claire de communautés, en particulier à cause de hubs qui correspondent à des termes fréquents commun à de nombreux champs (e.g. \texttt{model}, \texttt{space}). Ces mots sont utilisés de manière comparable dans l'ensemble des champs étudiés, et ne portent pas d'information pour les séparer\footnote{Mais en porteraient si l'on comparait un corpus de géographie quantitative et un corpus de musicologie par exemple.}. Nous faisons l'hypothèse que ces termes à fort degré ne portent pas d'information particulière sur des classes données et peuvent ainsi être filtrés étant donné un seuil de degré maximal $k_{max}$ (on s'intéresse alors à ce qui fait la spécificité de chaque domaine). De la même manière, les liens avec un poids faibles sont considérés comme du bruit et filtrés selon un seuil de poids minimal $\theta_w$. La méthode générique permet de plus une filtration préliminaire des mot-clés, complémentaire à la filtration topologique, par fréquence d'apparition dans les documents $\left[ f_{min},f_{max} \right]$, à laquelle les résultats ne sont pas sensibles dans notre cas. L'analyse de sensibilité des caractéristiques du réseau filtré, notamment de sa taille, modularité et structure des communautés, est donnée en~\ref{app:sec:quantepistemo}. Nous choisissons des valeurs de paramètres permettant une optimisation multi-objectifs entre modularité et taille du réseau, $\theta_w = 10,k_{max} = 500$, par le choix d'un point compromis sur un front de Pareto, qui donne un réseau sémantique de taille $(V=7063,E=48952)$. Celui-ci est visualisé en Annexe~\ref{app:sec:quantepistemo}.
}

\bpar{
We then retrieve communities in the network using a standard Louvain clustering on the optimal filtered network. We obtain 20 communities for a modularity of 0.58. These are examined manually to be named, the automatic naming techniques~\cite{yang2000improving} being not elaborated enough to make the implicit distinction between thematic and methodological fields for example (in fact between knowledge domains, see~\ref{sec:knowledgeframework}) which is a supplementary dimension that we do not tackle here, but necessary to have meaningful descriptions. The communities are described in Table~\ref{tab:quantepistemo:semanticdomains}. We directly see the complementarity with the citation approach, since emerge here together subjects of study (High Speed Rail, Maritime Networks), domains and methods (Networks, Remote Sensing, Mobility Data Mining), thematic domains (Policy), pure methods (Agent-based Modeling, Measuring). Thus, a reference may use several of these communities. We furthermore have a finer granularity of information. The effect of language is strong since French geography is distinguished as a separated category (advanced analyses could be considered to better understand this phenomenon and benefit from it: sub-communities, reconstruction of a specific network, studies by translation; but these are out of purpose in this exploratory study). We note the importance of networks, and of problematics in political sciences and socio-economic. We will use the first category in most models we will develop, but keeping in mind the importance of problematics linked to governance, we will proceed to a specific study in~\ref{sec:lutecia}.
}{
Nous récupérons ensuite les communautés dans le réseau par un clustering de Louvain standard sur le réseau filtré optimal. On obtient 20 communautés pour une modularité de 0.58. Celles-ci sont examinées à la main pour être nommées, les techniques de nomination automatique~\cite{yang2000improving} ne sont pas assez élaborées pour faire la distinction implicite entre champs thématiques et méthodologiques par exemple (en fait entre les domaines de connaissance, voir~\ref{sec:knowledgeframework}) qui est une dimension supplémentaire que nous ne traitons pas ici, mais nécessaire pour avoir des descriptions parlantes. Les communautés sont décrites en Table~\ref{tab:quantepistemo:semanticdomains}. On voit tout de suite la complémentarité avec l'approche par citations, puisque se dégagent ici à la fois des sujet d'étude (High Speed Rail, Maritime Networks), des domaines et méthodes (Networks, Remote Sensing, Mobility Data Mining), des domaines thématiques (Policy), des méthodes pures (Agent-based Modeling, Measuring). Ainsi, une référence peut mobiliser plusieurs de ces communautés. On a de plus une granularité plus fine de l'information. L'effet du langage est puissant puisque la géographie française se distingue en une catégorie séparée (des analyses poussées pourraient être envisagées pour mieux comprendre le phénomène et en tirer parti: sous-communautés, reconstruction d'un réseau spécifique, études par traduction ; mais celles-ci sont hors de propos dans cette étude exploratoire). On constate l'importance des réseaux, des problématiques de sciences politiques et socio-économiques. Nous mobiliserons la première catégorie dans la plupart des modèles développés, mais en gardant en tête l'importance des problématiques liées à la gouvernance, nous réaliserons un travail spécifique en~\ref{sec:lutecia}.
}

\begin{table}
\caption[Semantic communities]{\textbf{Description of semantic communities.} We give their size, their proportion in quantity of keywords (under the form of \emph{multi-stems}) cumulated on the full corpus, and representative keywords selected by maximal degree.\label{tab:quantepistemo:semanticdomains}}
\bpar{
\begin{tabular}{llll}
\hline\noalign{\smallskip}
Name & Size & Weight & Keywords  \\
\noalign{\smallskip}\hline\noalign{\smallskip}
Networks & 820 & 13.57\% & \texttt{social network, spatial network, resili} \\
Policy & 700 & 11.8\% & \texttt{actor, decision-mak, societi} \\
Socio-economic & 793 & 11.6\% & \texttt{neighborhood, incom, live} \\
High Speed Rail & 476 & 7.14\% & \texttt{high-spe, corridor, hsr} \\
French Geography & 210 & 6.08\% & \texttt{système, développement, territoire} \\
Education & 374 & 5.43\% & \texttt{school, student, collabor} \\
Climate Change & 411 & 5.42\% & \texttt{mitig, carbon, consumpt} \\
Remote Sensing & 405 & 4.65\% & \texttt{classif, detect, cover} \\
Sustainable Transport & 370 & 4.38\% & \texttt{sustain urban, travel demand, activity-bas} \\
Traffic & 368 & 4.23\% & \texttt{traffic congest, cbd, capit} \\
Maritime Networks & 402 & 4.2\% & \texttt{govern model, seaport, port author} \\
Environment & 289 & 3.79\% & \texttt{ecosystem servic, regul, settlement} \\
Accessibility & 260 & 3.23\% & \texttt{access measur, transport access, urban growth} \\
Agent-based Modeling & 192 & 3.18\% & \texttt{agent-bas, spread, heterogen} \\
Transportation planning & 192 & 3.18\% & \texttt{transport project, option, cba} \\
Mobility Data Mining & 168 & 2.49\% & \texttt{human mobil, movement, mobil phone} \\
Health Geography & 196 & 2.49\% & \texttt{healthcar, inequ, exclus} \\
Freight and Logistics & 239 & 2.06\% & \texttt{freight transport, citi logist, modal} \\
Spanish Geography & 106 & 1.26\% & \texttt{movilidad urbana, criteria, para} \\
Measuring & 166 & 1.0\% & \texttt{score, sampl, metric} \\
\noalign{\smallskip}\hline
\end{tabular}
}{
\begin{tabular}{llll}
\hline\noalign{\smallskip}
Nom & Taille & Poids & Mots-clés  \\
\noalign{\smallskip}\hline\noalign{\smallskip}
Networks & 820 & 13.57\% & \texttt{social network, spatial network, resili} \\
Policy & 700 & 11.8\% & \texttt{actor, decision-mak, societi} \\
Socio-economic & 793 & 11.6\% & \texttt{neighborhood, incom, live} \\
High Speed Rail & 476 & 7.14\% & \texttt{high-spe, corridor, hsr} \\
French Geography & 210 & 6.08\% & \texttt{système, développement, territoire} \\
Education & 374 & 5.43\% & \texttt{school, student, collabor} \\
Climate Change & 411 & 5.42\% & \texttt{mitig, carbon, consumpt} \\
Remote Sensing & 405 & 4.65\% & \texttt{classif, detect, cover} \\
Sustainable Transport & 370 & 4.38\% & \texttt{sustain urban, travel demand, activity-bas} \\
Traffic & 368 & 4.23\% & \texttt{traffic congest, cbd, capit} \\
Maritime Networks & 402 & 4.2\% & \texttt{govern model, seaport, port author} \\
Environment & 289 & 3.79\% & \texttt{ecosystem servic, regul, settlement} \\
Accessibility & 260 & 3.23\% & \texttt{access measur, transport access, urban growth} \\
Agent-based Modeling & 192 & 3.18\% & \texttt{agent-bas, spread, heterogen} \\
Transportation planning & 192 & 3.18\% & \texttt{transport project, option, cba} \\
Mobility Data Mining & 168 & 2.49\% & \texttt{human mobil, movement, mobil phone} \\
Health Geography & 196 & 2.49\% & \texttt{healthcar, inequ, exclus} \\
Freight and Logistics & 239 & 2.06\% & \texttt{freight transport, citi logist, modal} \\
Spanish Geography & 106 & 1.26\% & \texttt{movilidad urbana, criteria, para} \\
Measuring & 166 & 1.0\% & \texttt{score, sampl, metric} \\
\noalign{\smallskip}\hline
\end{tabular}
}
\end{table}

\paragraph{Measures of interdisciplinarity}{Mesures d'interdisciplinarité}

\bpar{
Distribution of keywords within communities provides an article-level interdisciplinarity. The combination of citation and semantic layers in the hyper-network provide second-order interdisciplinarity measures (semantic patterns of citing or cited), that we don't use here because of the modest size of the citation network (see \ref{app:sec:cybergeo} and \ref{app:sec:patentsmining}). More precisely, a reference $i$ can be viewed as a probability vector on semantic classes $j$, that we write in a matrix form $\mathbf{P}=(p_{ij})$. These are simply estimated by the proportions of keywords classified in each class for the reference. A classical measure of interdiscplinarity~\cite{bergeaud2017classifying} is then $I_i = 1 - \sum_j p_{ij}^2$. Let $\mathbf{A}$ be the adjacency matrix of the citation network, and let $\mathbf{I}_k$ matrices selecting rows corresponding to class $k$ of the citation classification: $Id\cdot \mathbbm{1}_{c(i)=k}$, such that $I_k \cdot A \cdot I_{k'}$ gives exactly the citations from $k$ to $k'$. The citation proximity between citation communities is then defined by $c_{kk'} = \sum \mathbf{I}_k \cdot \mathbf{A} \cdot \mathbf{I}_{k'} /  \sum \mathbf{I}_k \cdot \mathbf{A}$. We define the semantic proximity by defining a distance matrix between references by $\mathbf{D} = d_{ii'}=\sqrt{\frac{1}{2}\sum (p_{ij}-p{i'j})^2}$ and the semantic proximity by $s_{kk'} = \mathbf{I}_k \cdot \mathbf{D} \cdot \mathbf{I}_{k'} / \sum \mathbf{I}_k \sum \mathbf{I}_{k'}$.
}{
La distribution des mots clés dans les communautés permettent de définir une mesure d'interdisciplinarité au niveau de l'article. La combinaison des couches de citation et sémantique dans l'hyperréseau fournit des mesures d'interdisciplinarité au second ordre (motifs sémantiques des cités ou des citants), que nous n'utiliserons pas ici à cause de la taille modeste du réseau de citation (voir \ref{app:sec:cybergeo} et \ref{app:sec:patentsmining}). Plus précisément, une référence $i$ peut être vue comme un vecteur de probabilités sur les classes sémantiques $j$, qu'on notera sous forme matricielle $\mathbf{P}=(p_{ij})$. Celles-ci sont estimées simplement par les proportions de mots-clés classifiés dans chaque classe pour la référence. Une mesure classique d'interdisciplinarité~\cite{bergeaud2017classifying} est alors $I_i = 1 - \sum_j p_{ij}^2$. Soit $\mathbf{A}$ la matrice d'adjacence du réseau de citation, et soit $\mathbf{I}_k$ les matrices de selection des lignes correspondants à la classe $k$ de la classification de citation : $Id\cdot \mathbbm{1}_{c(i)=k}$, telle que $I_k \cdot A \cdot I_{k'}$ donne exactement les citations de $k$ vers $k'$. La proximité de citation entre les communautés de citation est alors définie par $c_{kk'} = \sum \mathbf{I}_k \cdot \mathbf{A} \cdot \mathbf{I}_{k'} /  \sum \mathbf{I}_k \cdot \mathbf{A}$. On définit la proximité sémantique en définissant une matrice de distance entre références par $\mathbf{D} = d_{ii'}=\sqrt{\frac{1}{2}\sum (p_{ij}-p{i'j})^2}$ puis la proximité sémantique par $s_{kk'} = \mathbf{I}_k \cdot \mathbf{D} \cdot \mathbf{I}_{k'} / \sum \mathbf{I}_k \sum \mathbf{I}_{k'}$.
}

\begin{figure}
\includegraphics[width=\linewidth]{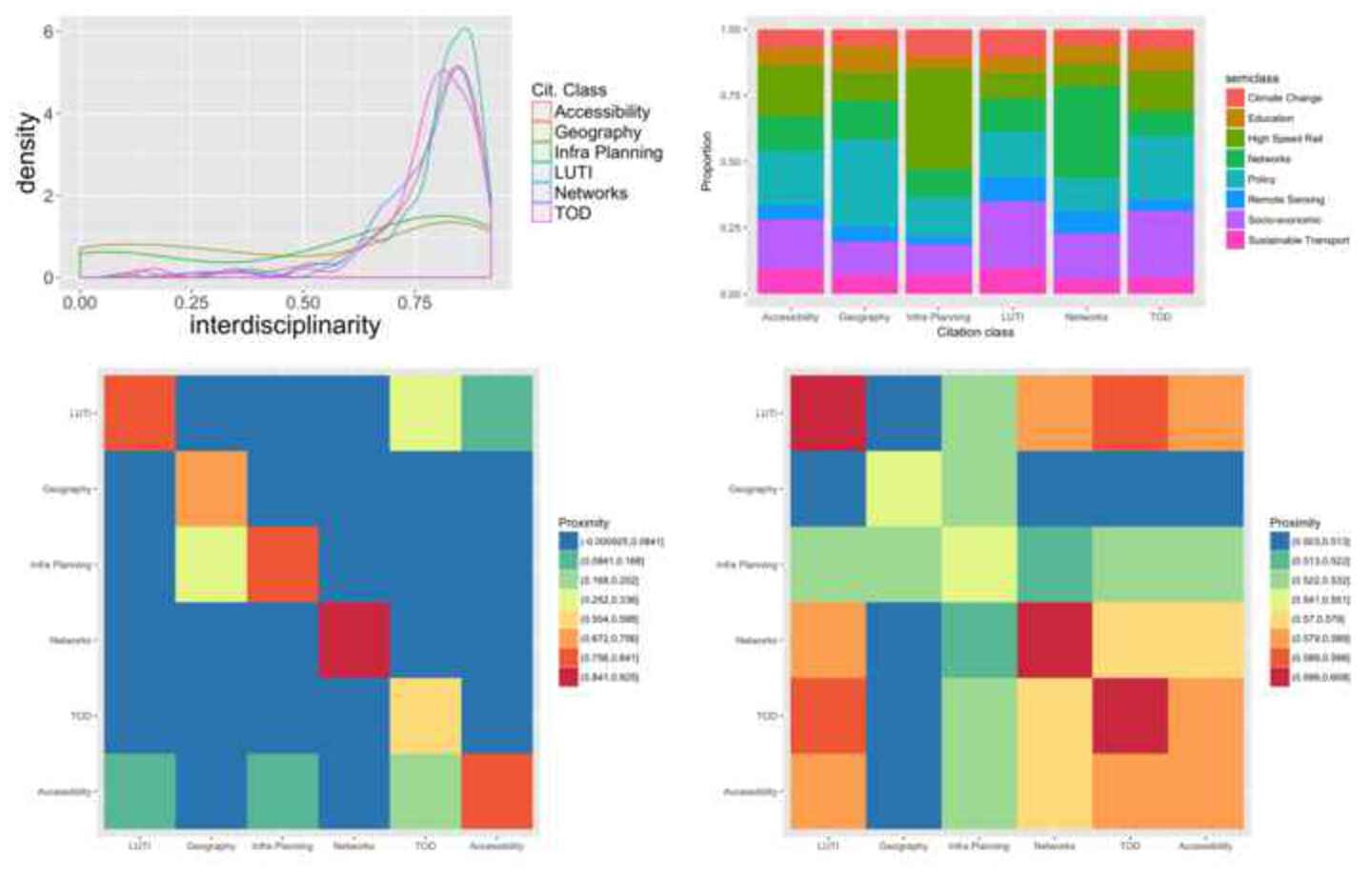}
\caption[Patterns of interdisciplinarity]{\textbf{Patterns of interdisciplinarity.} \textit{(Top Left)} Statistical distribution of $I_i$ by citation classes, in other words distribution of interdisciplinarity levels within citation classes; \textit{(Top Right)} Semantic composition of citation classes: for each citation class (in abscissa), the proportion of each semantic class (in color) is given; \textit{(Bottom Left)} Citation proximity matrix for $c_{kk'}$ between citation classes; \textit{(Bottom Right)} Semantic proximity matrix $s_{kk'}$ between citation classes. \label{fig:quantepistemo:interdisc}}
\end{figure}

\bpar{
We show in Fig.~\ref{fig:quantepistemo:interdisc} the values of these different measures, and also the semantic composition of citation communities, for the main semantic classes. The distribution of $I_i$ shows that articles orbiting in the LUTI field are the most interdisciplinary in the terms used, what could be due to their applied character. Other disciplines show similar patterns, except geography and infrastructure planning which exhibit quasi-uniform distributions, witnessing the existence of very specialized references in these classes. This is not necessarily stunning, given the targeted sub-fields exhibited (political sciences for example, and similarly prospective studies of type cost-benefit are very narrow). This first crossing of the layers confirms the specificities of each field. Regarding semantic compositions, most act as an external validation given the dominant classes. The field which is the less concerned by socio-economical issues is infrastructure planning, what could give reason to critics of technocracy. Issues on climate change and sustainability are relatively well dispatched. Finally, geographical works are mostly related to governance issues.
}{
Nous montrons en Fig.~\ref{fig:quantepistemo:interdisc} les valeurs de ces différentes mesures, ainsi que la composition sémantique des communautés de citation, pour les classes sémantiques majoritaires. La distribution de $I_i$ montre que les articles gravitant dans le domaine du LUTI sont les plus interdisciplinaires dans les termes utilisés, ce qui pourrait être lié à leur caractère appliqué. Les autres disciplines sont dans des motifs similaires, à part la géographie et la planification des infrastructures qui présentent des distributions quasi-uniformes, témoignant de l'existence de références très spécialisées dans ces classes. Ce n'est pas nécessairement étonnant vu les sous-champs pointus exhibés (sciences politiques par exemple, et de même les études prospectives type coût-bénéfices sont très étriquées). Ce premier croisement des couches nous confirme les spécificités de chaque champ. Concernant les compositions sémantiques, la plupart agissent comme validation externe vu les classes majoritaires. Le champ le moins concerné par les problèmes socio-économiques est la planification des infrastructures, ce qui donnera du grain à moudre aux détracteurs de la technocratie. Les questions de changement climatique et durabilité sont relativement bien réparties. Enfin, les ouvrages géographiques concernent en majorité des problèmes de gouvernance.
}

\bpar{
Proximity matrices confirm the conclusion obtained previously in terms of citation, the sharing being very low, the highest values being up to one fourth of planning towards geography and of LUTI towards TOD (but not the contrary, the relations can be in a unique sense). But semantic proximities show for example that LUTI, TOD, Accessibility and Networks are close in their terms, what is logical for the first three, and confirms for the last that physicists mainly rely on methods of this fields linked to planning to legitimate their works. Geography is totally isolated, its closest neighbor being infrastructure planning. This study is very useful in our context, since it shows compartmentalized domains sharing terms, and thus a priori some common problematics and subjects. Domains do not speak to each other while speaking languages that are not that far, hence the increased relevance to aim at harmonizing their music in our work: our models will have to use elements, ontologies and scales of these different fields.
}{
Les matrices de proximité confirment la conclusion obtenue précédemment en termes de citation, les partages étant très faibles, les plus hautes valeurs étant jusqu'à un quart de la planification vers la géographie et des LUTI vers le TOD (mais pas l'inverse, les relations peuvent être à sens unique). Or les proximités sémantiques montrent par exemple que LUTI, TOD, Accessibility et Networks sont proches dans leur termes, ce qui est logique pour les trois premiers, et confirme pour le dernier que les physiciens se basent majoritairement sur les méthodes des ces champs liés au planing pour légitimer leur travaux. La géographie est totalement isolée, sa plus proche voisine étant la planification des infrastructures. Cette étude est très utile pour notre propos, puisqu'elle montre des domaines cloisonnés partageant des termes et donc a priori des problématiques et sujets communs. Les domaines ne se parlent pas tout en parlant des languages pas si lointains, d'où la pertinence accrue de vouloir accorder leur partitions dans nos travaux : nos modèles devront mobiliser des éléments, ontologies et échelles de ces différents champs.
}

\bpar{
We conclude this analysis with a more robust approach to quantify proximities between the layers of the hypernetwork. It is straightforward to construct a correlation matrix between two classifications, through the correlations of their columns. We define the probabilities $\mathbf{P}_C$ all equal to 1 for the citation classification. The correlation matrix between it and $\mathbf{P}$ extends from -0.17 to 0.54 and has an average with an absolute value of 0.08, what is significant in comparison to random classifications since a bootstrap with $b=100$ repetitions with shuffled matrices gives a minimum at $-0.08 \pm 0.012$, a maximum at $0.11 \pm 0.02$ and an absolute average at $0.03 \pm 0.002$. This shows that the classifications are complementary and that this complementarity is statistically significant compared to random classifications. The adequacy of the semantic classification in relation to the citation network can also be quantified by the multi-classes modularity~\cite{nicosia2009extending} (see~\ref{app:sec:patentsmining} for a mathematical definition), which translates the likelihood that a link is due to the classification studied, taking into account the simultaneous belonging to multiple classes. Thus, the multi-class modularity of semantic probabilities for the citation network s 0.10, what on one side is a significant sign of an adequacy, a bootstrap still with $b=100$ giving a value of $0.073 \pm 0.003$, which remains limited given the maximal value fixed by citation probabilities within their own network which give a value of 0.81, what conform furthermore the complementarity of classifications.
}{
Nous concluons cette analyse par une approche plus robuste pour quantifier les proximités entre couches de l'hyperréseau. Il est aisé de construire une matrice de corrélation entre deux classifications, par les corrélations de leur colonnes. Nous définissons les probabilités $\mathbf{P}_C$ toutes égales à 1 pour la classification de citation. La matrice de correlation de celle-ci avec $\mathbf{P}$ s'étend de -0.17 à 0.54 et a une moyenne de valeur absolue de 0.08, ce qui est significatif par rapport à des classifications aléatoires puisque un bootstrap à $b=100$ répétitions avec les matrices mélangées donne un minimum à $-0.08 \pm 0.012$, un maximum à $0.11 \pm 0.02$ et une moyenne absolue à $0.03 \pm 0.002$. Cela montre que les classifications sont complémentaires et que cette complémentarité est significative statistiquement par rapport à des classifications aléatoires. L'adéquation de la classification sémantique par rapport au réseau de citation peut également être quantifiée par la modularité multi-classes~\cite{nicosia2009extending} (voir~\ref{app:sec:patentsmining} pour une définition mathématique), qui traduit la probabilité qu'un lien soit dû à la classification étudiée, en prenant en compte l'appartenance simultanée à de multiples classes. Ainsi, la modularité multi-classes des probabilités sémantiques pour le réseau de citation est de 0.10, ce qui d'une part est significativement signe d'adéquation, un bootstrap toujours à $b=100$ donnant une valeur de $0.073 \pm 0.003$, qui reste limitée vu la valeur maximale fixée par les probabilités de citations dans leur propre réseau qui donnent une valeur de 0.81, ce qui confirme d'autre part la complémentarité des classifications.
}

\bpar{
We have thus in this section sketched an overview of disciplines in relation with our subject, and also their relations. We will aim in the next section at understanding with more details their ``content'', i.e. the means used to solve the problems encountered.
}{
Nous avons ainsi dressé dans cette section un aperçu des disciplines en relation avec notre sujet, ainsi que leur relations. Il s'agira dans la section suivante de comprendre avec plus de détail leur ``contenu'', c'est-à-dire les moyens mobilisés pour résoudre les problèmes rencontrés.
}


\subsubsection{Discussion}{Discussion}

\bpar{
We briefly give directions to extend the analysis we just did and also implications for the epistemological positioning of our work.
}{
Donnons brièvement des directions d'extension de l'analyse que nous venons de mener ainsi que des implications pour le positionnement épistémologique de notre travail.
}

\subsubsection{Towards modeling themes and an automatic extraction of context}{Vers une modélisation des thèmes et une extraction automatique du contexte}

\bpar{
A possible direction to strengthen our quantitative epistemological analysis would be to work on full textes related to the modeling of interaction between networks and territories, with the aim to automatically extract thematics within articles. Methods more suited for full texts than the one used here for example include Latent Dirichlet Allocation~\cite{blei2003latent}. The idea would be to perform some kind of automatized modelography, extending the modelography methodology developed by~\cite{schmitt2013modelographie}, to extract characteristics such as ontologies, model architecture or structures, scales, or even typical parameter values. It is not clear to what extent the structure of models can be extracted from their description in papers and it surely depends on the discipline considered. For example in a framed field such as transportation planning, using a pre-defined ontology (in the sense of dictionary) and a fuzzy grammar could be efficient to extract information as the discipline has relatively strict conventions. In theoretical and quantitative geography, beyond the barrier of diversity of possible formalizations for a same ontology, the organisation of information is surely more difficult to grasp through unsupervised data-mining because of the more literary nature of the discipline: synonyms and figures of speech are generally the norm in good level human sciences writing, fuzzing a possible generic structure of knowledge description. 
}{
Une direction possible pour renforcer cette analyse en épistémologie quantitative serait de travailler sur les textes complets des références contenant des efforts de modélisation des interactions entre réseaux et territoires, avec le but d'extraire automatiquement les thématiques des articles. Des méthodes plus adaptées pour les textes complets que celle utilisée ici incluent par exemple l'Allocation Latente de Dirichlet~\cite{blei2003latent}. L'idée serait de procéder à une sorte de modélographie automatique, étendant la méthodologie de modélographie développée par~\cite{schmitt2013modelographie}, pour extraire des caractéristiques telle les ontologies, l'architecture ou la structure des modèles, les échelles ou même des valeurs typiques des paramètres. Il n'est pas clair dans quelle mesure la structure des modèles peut être extraite de leur description dans un article, et cela dépend sûrement de la discipline considérée. Par exemple dans un champ relativement cadré comme la planification des transports, l'utilisation d'une ontologie pré-définie (dans le sens d'un dictionnaire) et d'une grammaire floue pourrait être efficace vu les conventions assez strictes dans la discipline. En géographie théorique et quantitative, au-delà de la barrière de la diversité des formalisations possibles pour une même ontologie, l'organisation de l'information est sûrement plus délicate à appréhender par de l'apprentissage non-supervisé à cause de la nature plus littéraire de la discipline : les synonymes et les figures de style sont généralement la norme pour l'écriture d'un bon niveau en sciences humaines, rendant plus floue une possible structure générique de la description des connaissances.
}




\subsubsection{Reflexivity}{Réflexivité}

\bpar{
The methodology developed here is efficient to offer reflexivity instruments, i.e. it can be used to study our approach itself. One of its application, beyond the one on the scientific journal Cybergeo in a perspective of Open Science (see Appendix~\ref{app:sec:cybergeo}), will be to our own corpus of references, with the aim to reveal possible research directions or exotic issues. It is eventually possible to do it in a dynamical way, thanks to the \texttt{git} history which allows to recover any version of the bibliography at a given date on the three years elapsed. The aim will also be to understand our knowledge production patterns in order to contribute to~\ref{sec:knowledgeframework}. The detailed development is done in Appendix~\ref{app:reflexivity}.
}{
La méthodologie que nous avons développée ici est efficace pour offrir des instruments de réflexivité, c'est-à-dire qu'elle peut être utilisée pour étudier notre approche elle-même. Une de ses applications, hors de celle à la revue scientifique Cybergeo dans la perspective de Science Ouverte (voir Annexe~\ref{app:sec:cybergeo}), sera à notre propre corpus de références, dans le but de révéler des possibles directions de recherche ou problématiques exotiques. Il est éventuellement possible de le faire de manière dynamique, grâce à l'historique de \texttt{git} qui permet de récupérer n'importe quelle version de la bibliographie à une date donnée sur les trois ans écoulés. Il s'agira aussi de comprendre nos motifs de production de connaissance afin de contribuer à~\ref{sec:knowledgeframework}. Le développement détaillé est fait en Annexe~\ref{app:reflexivity}.
}

\stars

\bpar{
This section thus allowed us to sketch a landscape of disciplines in relation with our problematic, and of relations between these disciplines, in terms of citations but also of level of interdisciplinarity.
}{
Cette section nous a ainsi permis de dresser un paysage des disciplines en relation avec notre problématique, et des relations entre ces disciplines, en termes de citations mais aussi de niveau d'interdisciplinarité. 
}

\bpar{
The next section will positioned with a similar approach, but with an aim closer to exhaustivity in terms of modeling interactions: We will thus proceed to a systematic review and a modelography, in order to reinforce the typology of models obtained in section~\ref{sec:modelingsa}.
}{
La section suivante se positionnera dans une démarche voisine, mais avec un but plus marqué d'exhaustivité en termes de modélisation des interactions : nous procéderons ainsi à une revue systématique et à une modélographie, afin de renforcer la typologie des modèles obtenue en section~\ref{sec:modelingsa}.
}

\stars

%


\newpage

\section{Systematic review and modelography}{Revue systématique et modélographie}

\label{sec:modelography}


\bpar{
Whereas the studies we previously did proposed to construct a global horizon of the organization of disciplines focusing on our question, we propose now a more targeted study of characteristics of existing models. We propose therefore in a first time a systematic review, i.e. the construction of a refined corpus satisfying certain constraints, followed by a meta-analysis, i.e a tentative of explanation of some characteristics through statistical models.
}{
Tandis que les études menées précédemment proposaient de construire un horizon global de l'organisation des disciplines s'intéressant à notre question, nous proposons à présent une étude plus ciblée des caractéristiques de modèles existants. Nous proposons pour cela dans un premier temps une revue systématique, c'est-à-dire la construction d'un corpus plus précis répondant à certaines contraintes, suivie d'une méta-analyse, c'est-à-dire une tentative d'explication de certaines caractéristiques des modèles par des modèles statistiques.
}

\subsection{Systematic review}{Revue systématique}

\bpar{
Classical systematic reviews take mostly place in fields where a very targeted request, even by article title, will yield a significant number of studies studying quite the same question: typically in therapeutic evaluation, where standardized studies of a same molecule differ only by the size of samples and statistical modalities (control group, placebo, level of blinding). In this case corpus construction is easy first thanks to the existence of specialized bases allowing very precise requests, and furthermore thanks to the possibility to proceed to additional statistical analyses to confront the different studies (for example network meta-analysis, see~\cite{rucker2012network}). In our case, the exercise is much more random for the reasons exposed in the two previous sections: objects are hybrid, problematics are diverse, and disciplines are numerous. The different points we will raise in the following will often have as much thematic value as methodological value, suggesting crucial points for the realization of such an hybrid systematic review.
}{
Les revues systématiques classiques ont majoritairement lieu dans des domaines où une recherche très ciblée, même par titre d'article, fournira un certain nombre d'études étudiant quasiment la même question : typiquement en évaluation thérapeutique, où des études standardisées d'une même molécule varient uniquement par taille des effectifs et modalités statistiques (groupe de contrôle, placebo, niveau d'aveugle). Dans ce cas la construction du corpus est d'une part aisée par l'existence de bases spécialisées permettant des recherches très ciblées, et d'autre part par la possibilité de procéder à des analyses statistiques supplémentaires pour croiser les différentes études (par exemple méta-analyse par réseau, voir~\cite{rucker2012network}). Dans notre cas, l'exercice est bien plus aléatoire pour les raisons exposées dans les deux sections précédentes : les objets sont hybrides, les problématiques diverses, et les disciplines variées. Les différents points soulevés par la suite auront souvent autant de valeur thématique que de valeur méthodologique, suggérant des points cruciaux lors de la réalisation d'une telle revue systématique hybride.
}

\bpar{
We propose an hybrid methodology coupling the two methodologies previously developed with a more classical procedure of systematic review. We aim both at a representativity of all the disciplines we discovered, but also a limited noise in the references taken into account for the modelography. Therefore, we combine the corpus previously obtained and a corpus constructed through keywords requests, in a way similar to \cite{tahamtan2018core}. The protocol is thus the following:
}{
Nous proposons une méthodologie hybride couplant les deux méthodologies développées précédemment avec une procédure plus classique de revue systématique. Nous souhaitons à la fois une représentativité de l'ensemble des disciplines que l'on a découvertes, mais aussi un bruit limité dans les références prises en compte pour la modélographie. Pour cela, nous combinons le corpus obtenu précédemment et un corpus constitué par requêtes de mots-clés, de manière similaire à \cite{tahamtan2018core}. Le protocole est donc le suivant :
}

%
%
%
%
%

\bpar{
\begin{enumerate}
	\item Starting from the citation corpus isolated in~\ref{subsec:indirectbibliometrics}, we isolate a number of relevant keywords, by selecting the 5\% of links having the strongest weight (arbitrary threshold), and among the corresponding nodes the ones having a degree larger than the quantile at 0.8 of their respective semantic class. The first filtration allows to focus on the ``core'' of observed disciplines, and the second to not introduce size bias without loosing the global structure, classes being relatively balanced. A manual screening allows to remove keywords that are obviously not relevant (teledetection, tourism, social networks, \ldots), what leads to a corpus of $K=115$ keywords ($K$ is endogenous here).
	\item For each keyword, we automatically do a catalog request (scholar) while adding \texttt{model*} to it, of a fixed number $n=20$ of references. The supplementary term is necessary to obtain relevant references, after testing on samples.
	\item The potential corpus composed of obtained references, with references composing the citation network, is manually screened (review of titles) to ensure a relevance regarding the state of the art of~\ref{sec:modelingsa}, yielding the preliminary corpus of size $N_p = 297$.
	\item This corpus is then inspected for abstracts and full texts if necessary. We select articles elaborating a modeling approach, ruling out conceptual models. References are classified and characterized according to criteria described below. We finally obtain a final corpus of size $N_f = 145$, on which quantitative analyses are possible.
\end{enumerate}
}{
\begin{enumerate}
\item Partant du corpus de citation isolé en~\ref{subsec:indirectbibliometrics}, nous isolons un nombre de mots-clés pertinents, en sélectionnant les 5\% de liens ayant le plus fort poids (seuil arbitraire), puis parmi les noeuds correspondants ceux ayant un degré supérieur au quantile à 0.8 de leur classe sémantique respective. Le premier filtrage permet de se concentrer sur le ``coeur'' des disciplines observées, et le second de ne pas biaiser par la taille sans perdre la structure globale, les classes étant relativement équilibrées. Un examen manuel permet de supprimer les mots-clés clairement non-pertinents (télédétection, tourisme, réseaux sociaux, \ldots), ce qui conduit à un corpus de $K=115$ mots-clés ($K$ est endogène ici).
\item Pour chaque mot-clé, nous effectuons automatiquement une requête au catalogue (scholar) en y ajoutant \texttt{model*}, d'un nombre fixé $n=20$ de références. L'ajout du terme est nécessaire pour obtenir des références pertinentes, après test sur des échantillons.
\item Le corpus potentiel composé des références obtenues, ainsi que des références composant le réseaux de citation, est revu manuellement (passage en revue des titres) pour assurer une pertinence au regard de l'état de l'art de~\ref{sec:modelingsa}, fournissant le corpus préliminaire de taille $N_p = 297$.
\item Ce corpus est alors inspecté pour les résumés et textes complets si nécessaire. On sélectionne les articles mettant en place une démarche de modélisation, hors modèles conceptuels. Les références sont classifiées et caractérisées selon des critères décrits ci-dessous. Nous obtenons alors un corpus final de taille $N_f = 145$, sur lequel des analyses quantitatives sont possibles.
\end{enumerate}
}

\begin{figure}[h!]
\frame{\includegraphics[width=\textwidth]{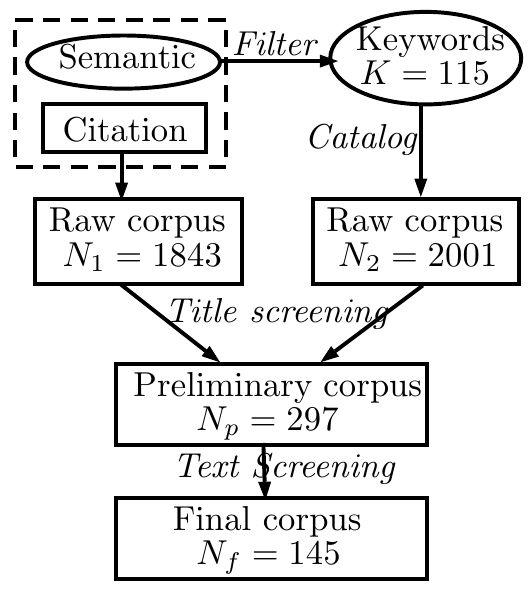}}
\caption[Methodology of the systematic review]{\textbf{Methodology of the systematic review.} Rectangles correspond to corpuses of references, ellipses to corpuses of keywords, and dashed lines to initial corpuses. At each stage the size of the corpus is given.\label{fig:modelography:systematicreview}}
\end{figure}

\bpar{
The method is summarized in Fig.~\ref{fig:modelography:systematicreview}, with parameter values and the size of the successive corpus. This exercise first of all allows to reveal several methodological points, which knowledge can be an asset to proceed to similar hybrid systematic reviewes:
}{
La méthode est résumée en Fig.~\ref{fig:modelography:systematicreview}, avec les valeurs des paramètres et la taille des corpus successifs. Cet exercice permet tout d'abord un certain nombre de points méthodologiques, dont la connaissance pourra être un atout pour mener des revues systématiques hybrides similaires :
}

\bpar{
\begin{itemize}
	\item Catalog bias seem to be inevitable. We rely on the assumption that the use of scholar allows an uniform sampling regarding catalog errors or bias. The future development of open tools for cataloging and mapping, allowing contributed efforts for a more precise knowledge of extended fields and of their interfaces, will be a crucial issue for the reliability of such methods (see~\ref{app:sec:cybergeo}).
	\item The availability of full texts is an issue, in particular for such a broad review, given the multiplicity of editors. The existence of tools to emancipate science such as Sci-hub\footnote{\url{http://sci-hub.cc/}} allow to effectively access full texts. Echoing the recent debate on the negotiations with publishers regarding the exclusivity of full texts mining, it appears to be more and more salient that a reflexive open science is totally orthogonal to the current model of publishing. We also hope for a rapid evolution of practices on this point.
	\item Journals, and indeed publishers, seem to differently influence the referencing, potentially increasing the bias during requests. Grey literature and preprints are taken into account in different ways depending on the domains.
	\item Manual screening of large corpora allows to not miss ``crucial papers'' that could have been omitted before~\cite{lissacksubliminal}. The issue of the extent to which we can expect to be informed in the most exhaustive way possible of recent discoveries linked to the subject studied is very likely to evolve given the increase of the total amount of literature produced and the separation of fields, among which some are always more refined~\cite{bastian2010seventy}. Following the previous points, we can propose that tools helping systematic analysis will allow to keep this objective as reasonable.
	\item Results of the automatized review are significantly different from the domains highlighted in the classical review: some conceptual associations, in particular the inclusion of network growth models, are not natural and do not exist much in the scientific landscape as we previously showed.
\end{itemize}
}{
\begin{itemize}
\item Les biais de catalogue semblent inévitables. Nous reposons sur l'hypothèse que l'utilisation de Scholar permet un échantillonnage uniforme au regard des erreurs ou biais de catalogage. Le développement futur d'outils ouverts de catalogage et de cartographie, permettant un effort contributif pour une connaissance plus précise de domaines étendus et de leurs interfaces, sera un enjeu crucial de la fiabilité de ce genre de méthodes (voir~\ref{app:sec:cybergeo}).
\item La disponibilité des textes complets est particulièrement un problème pour une revue si large, vu la multiplicité des éditeurs. L'existence de moyens d'émancipation de la science ouverte comme Sci-hub\footnote{\url{http://sci-hub.cc/}} permet d'effectivement accéder à l'ensemble des textes. En écho au débat sur le bras de fer récent avec les éditeurs concernant l'exclusivité de la fouille de textes complets, il parait de plus en plus évident qu'une science ouverte réflexive est totalement antagoniste au modèle actuel de l'édition. Nous espérons également une évolution rapide des pratiques sur ce point.
\item Les revues, et en fait les éditeurs, semblent influencer différemment le référencement, augmentant potentiellement le biais de requête. La littérature grise ainsi que les preprints sont pris en compte différemment selon les champs.
\item Le passage en revue manuel des grands corpora permet de pas louper des ``poids lourds'' qui auraient pu être omis en amont~\cite{lissacksubliminal}. La question de la mesure dans laquelle on peut s'attendre d'être au courant de la manière la plus exhaustive des découvertes récentes liées au sujet étudié évolue très probablement vu l'augmentation de la quantité totale de littérature produite et la fragmentation des domaines pour certains toujours plus pointus~\cite{bastian2010seventy}. Rejoignant les points précédents, on peut supposer que des outils d'aide à l'analyse systématique permettront de garder cet objectif raisonnable.
\item Les résultats de la revue automatique sont sensiblement différents des domaines dessinés dans la revue classique : certaines associations conceptuelles, notamment l'inclusion des modèles de croissance de réseaux, ne sont pas naturelles et existent peu dans le paysage scientifique comme nous l'avons montré précédemment.
\end{itemize}
}

\bpar{
Furthermore, the operation of constructing the corpus already allows to draw thematic observations that are interesting in themselves.
}{
D'autre part, l'opération de construction du corpus permet déjà de tirer des observations thématiques intéressantes en elles-mêmes.
}

\bpar{
\begin{itemize}
	\item The articles selected imply a clarification of what is meant by ``model''. We give in~\ref{sec:knowledgeframework} a very broad definition applying to all scientific perspectives. Our selection here does not retain conceptual models for example, our choice criteria being that the model must include a numerical or simulation aspect.
	\item A certain number of references consist in reviews, what is equivalent to a group of model with similar characteristics. 
	We could make the method more complicated by transcribing each review or meta-analysis, or by weighting the records of corresponding characteristics by the corresponding number of articles. We make the choice to ignore these reviews, what remains consistent in a thematic way still with the assumption of uniform sampling.
	\item A first clarification of the thematic frame is achieved, since we do not select studies uniquely linked to traffic and mobility (this choice being also linked to the results obtained in~\ref{sec:transportationequilibrium}), to pure urban design, to pedestrian flows models, to logistics, to ecology, to technical aspects of transportation, to give a few examples, even if these subjects can in an extreme view be considered as linked to interactions between networks and territories.
	\item Similarly, neighbor fields such as tourism, social aspects of the access to transportation, anthropology, were not taken into account.
	\item We observe a high frequency of studies linked to High Speed Rail (HSR), recalling the necessary association of political aspects of planning and of research directions in transportation.
\end{itemize}
}{
\begin{itemize}
\item Les articles sélectionnés supposent une clarification de ce qui est entendu par ``modèle''. Nous donnons en~\ref{sec:knowledgeframework} une définition très large s'appliquant à l'ensemble des perspectives scientifiques. Notre selection ici ne retient pas les modèles conceptuels par exemple, notre critère de choix étant que le modèle doit inclure un aspect numérique ou de simulation.
\item Un certain nombre de références consistent en des revues, ce qui revient à un groupe de modèles ayant des caractéristiques similaires. Nous pourrions compliquer la méthode en retranscrivant chaque revue ou meta-analyse, ou en pondérant par le nombre d'article correspondant les enregistrements des caractéristiques correspondants. Nous faisons le choix d'ignorer ces revues, ce qui reste cohérent de manière thématique en restant dans l'hypothèse d'échantillonnage uniforme.
\item Une première clarification du cadre thématique est opérée, puisque nous ne sélectionnons pas les études liées uniquement au traffic et à la mobilité (ce choix étant aussi lié aux résultats obtenus en~\ref{sec:transportationequilibrium}), à l'urban design pur, au modèles de flux piétons, au fret, à l'écologie, aux aspects techniques du transport, pour donner quelques exemples, même si ces sujets peuvent dans une vue extrême être considérés comme liés aux interactions entre réseaux et territoires.
\item De la même façon, des domaines annexes comme le tourisme, les aspects sociaux de l'accès aux transports, l'anthropologie, n'ont pas été pris en compte.
\item On observe une forte fréquence des études liées au Trains à Grande Vitesse (HSR), rappelant la non-dissociabilité des aspects politiques de la planification et des directions de recherche en transports.
\end{itemize}
}

\subsection{Modelography}{Modélographie}

\bpar{
We now switch to a mixed analysis based on this corpus, inspired by results of previous sections in particular for the classification. It aims at extracting and to precisely decompose ontologies, scales and processes, and then to study possible links between these characteristics of the models and the context in which they have been introduced. It is thus in a way the meta-analysis, that we will designate here as modelography. In order to not offend purists, it is indeed not a meta-analysis strictly speaking since we do not combine similar analysis to extrapolate potential results from larger samples. Our approach is close to the one of \cite{10.1371/journal.pone.0183919} which gathers references having quantitatively studied Zipf's law for cities, and then links the characteristics of studies to the methods used and the assumptions formulated.
}{
Nous passons à présent à une analyse mixte basée sur ce corpus, inspirée par les résultats des sections précédentes notamment pour la classification. Elle a pour but d'extraire et de décomposer précisément les ontologies, échelles et processus, puis d'étudier des liens possibles entre ces caractéristiques des modèles et le contexte dans lequel ils ont été introduits. Il s'agit ainsi de la méta-analyse en quelque sorte, que nous désignerons ici par modélographie. Pour ne pas froisser les puristes, il ne s'agit en effet pas d'une méta-analyse à proprement parler car nous ne combinons pas des analyses proches pour extrapoler des résultats potentiels d'échantillons plus grands. Notre démarche est proche de celle de \cite{10.1371/journal.pone.0183919} qui rassemble les références ayant étudié quantitativement la loi de Zipf pour les villes, puis lie les caractéristiques des études aux méthodes utilisées et hypothèses formulées.
}

\bpar{
The first part consists in the extraction of the characteristics of models. Automatize this work would consist in a research project in itself, as we develop in discussion below, but we are convinced of the relevance to refine such techniques (see~\ref{ch:opening}) in the frame of the development of integrated disciplines. Time being as much the enemy as the ally of research, we focus here on a manual extraction that will aim at being more precise that an approximatively convincing data mining attempt. We extract from models the following characteristics:
}{
La première partie consiste en l'extraction des caractéristiques des modèles. Automatiser ce travail constituerait un projet de recherche en lui-même, comme nous développons en discussion ci-dessous, mais nous sommes convaincu de la pertinence d'affiner de telles techniques (voir~\ref{ch:opening}) dans le cadre d'un développement de disciplines intégrées. Le temps étant autant l'ennemi que l'allié de la recherche, nous nous concentrons ici sur une extraction manuelle qui se voudra plus fine qu'une tentative peu convaincante de fouille de données. Nous extrayons des modèles les caractéristiques suivantes :
}

\begin{figure}[h!]
\includegraphics[width=\linewidth]{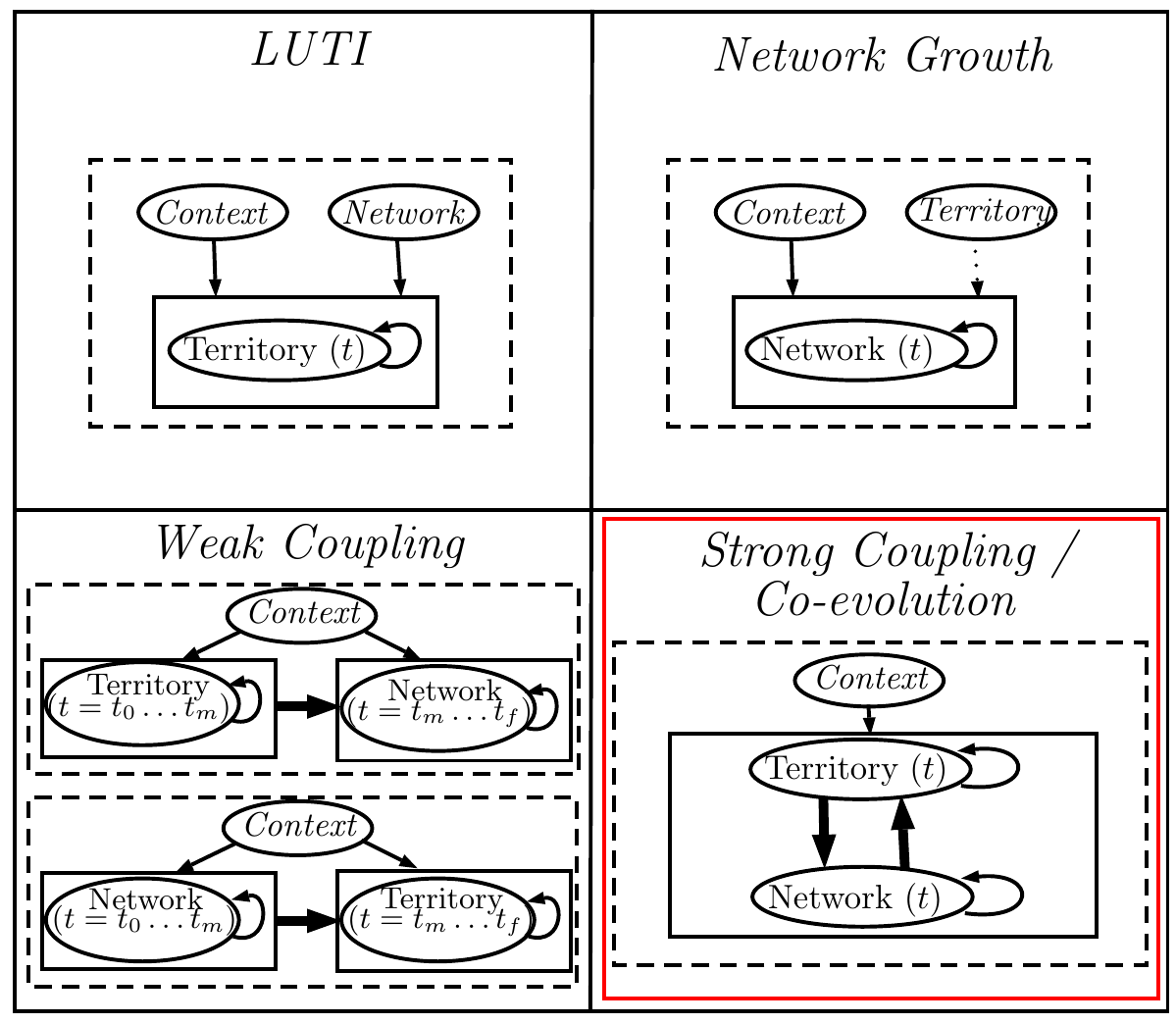}
\caption[Types of coupling in models integrating interactions between networks and territories]{\textbf{Schematic representation of the distinction between different types of models coupling networks and territories.} This typology is based on the one of Table~\ref{tab:modelingsa:synthesis}, by distinguishing approaches in which territory or network are given as a context (Luti and network growth) from a sequential coupling between a model for each. Ontologies are represented as ellippses, submodels by full boxes, models by dashed boxes, couplings by arrows. We highlight in red the approach which will be the final objective of our work.\label{fig:modelography:coevolution}}
\end{figure}

\bpar{
\begin{itemize}
	\item what is the strength of coupling\footnote{To the best of our knowledge there does not exist generic approaches to model coupling that would be not linked to a particular formalism. We will take the approach given in introduction, by distinguishing here a weak coupling as a sequential coupling (outputs of the first model become inputs of the second) from a dynamical strong coupling where the evolution is interdependent at each time step (either by a reciprocal determination of by a common ontology).} between territorial ontologies and the ones of the network, in other words is it a co-evolution model. We will therefore classify into categories following the representation of figure~\ref{fig:modelography:coevolution}: \texttt{\{territory ; network ; weak ; coevolution\}}, which results from the analysis of literature in~\ref{sec:modelingsa};
	\item maximal time scale;
	\item maximal spatial scale;
	\item domain ``a priori'', determined by the origin of authors and the domain of the journal;
	\item methodology used (statistical models, system of equations, multi-agent, cellular automaton, operational research, simulation, etc.);
	\item case study (city, metropolitan area, region or country) when relevant.
\end{itemize}
}{
\begin{itemize}
\item quelle est la force du couplage\footnote{Il n'existe à notre connaissance pas d'approche générique du couplage des modèles n'étant pas liée à un formalisme particulier. Nous prendrons l'approche donnée en introduction, en distinguant ici un couplage faible comme un couplage séquentiel (sorties du premier modèle deviennent entrées du second) d'un couplage fort dynamique où l'évolution est interdépendante à chaque instant (soit par une détermination réciproque soit par une ontologie commune).} entre les ontologies territoriales et celles du réseau, autrement dit s'agit-il d'un modèle de co-évolution. Nous classerons pour cela en catégories suivant la représentation de la figure~\ref{fig:modelography:coevolution} : \texttt{\{territory ; network ; weak ; coevolution\}}, qui résulte de l'analyse de la littérature en~\ref{sec:modelingsa} ;
\item échelle de temps maximale ; 
\item échelle d'espace maximale ; 
\item hypothèses d'équilibre ;
\item domaine ``a priori'', déterminé par l'origine des auteurs et domaine de la revue ;
\item méthodologie utilisée (modèles statistiques, système d'équations, multi-agent, automate cellulaire, recherche opérationnelle, simulation etc.) ;
\item cas d'étude (ville, métropole, région ou pays) s'il y a lieu.
\end{itemize}
}

\bpar{
We also collect in an indicative way, but without objective of objectivity or exhaustivity, the ``subject'' of the study (i.e. the main thematic question) ans also the ``processes'' included in the model. An exact extraction of processes remains hypothetical, on the one hand because it is conditioned to a rigorous definition and taking into account different levels of abstraction, of complexity, or scale, on the other hand as it depends on technical means out of reach of this modest study. We will comment these in an indicative way without including them in systematic studies.
}{
Nous collectons également de manière indicative, mais sans objectif d'objectivité ni d'exhaustivité, le ``sujet'' de l'étude (c'est-à-dire la question thématique dominante) ainsi que les ``processus'' inclus dans le modèle. Une extraction exacte des processus reste hypothétique, d'une part conditionnée à une définition rigoureuse et prenant en compte différents niveaux d'abstraction, de complexité, ou d'échelle, d'autre part dépendant de moyens techniques hors de portée de cette étude modeste. Nous commenterons ceux-ci de manière indicative sans les inclure dans les études systématiques.
}

\bpar{
We also gather scale, range and in a sense resolution to not make the extraction more complicated. Even if it would be relevant to differentiate when an element does not exist for a model (NA) to when it is badly defined by the author, this task seem to be sensitive to subjectivity and we merge the two modalities. We add to the previous characteristics the following variables:
}{
Nous confondons également échelle, portée et dans un sens résolution pour ne pas rendre plus confus l'extraction. Même s'il serait pertinent de différencier lorsque un élément n'a pas lieu d'être pour un modèle (NA) de lorsque celui-ci est mal défini par son auteur, cette tâche apparaît sujette à subjectivité et nous fusionnons les deux modalités. Nous ajoutons aux caractéristiques ci-dessus les variables suivantes :
}

\bpar{
\begin{itemize}
	\item citation domain (when available, i.e. for references initially present in the citation network, what corresponds to 55\% of references);
	\item semantic domain, defined by the domain for which the document has the highest probability;
	\item index of interdisciplinarity.
\end{itemize}
}{
\begin{itemize}
\item domaine de citation (le cas échéant, c'est-à-dire pour les références initialement présentes dans le réseau de citation, i.e. 55\% des références) ;
\item domaine sémantique, défini par le domaine pour lequel le document a la plus grande probabilité ;
\item indice d'interdisciplinarité.
\end{itemize}
}

\bpar{
Semantic domains and the interdisciplinarity measure have been recomputed for this corpus through the collection of keywords, and then extraction following the method described in~\ref{sec:quantepistemo}, with $K_W=1000$, $\theta_w=15$ and $k_{max}=500$. We obtain more targeted communities which are relatively representative of thematics and methods: Transit-oriented development (\texttt{tod}), Hedonic models (\texttt{hedonic}), Infrastructure planning (\texttt{infra planning}), High-speed rail (\texttt{hsr}), Networks (\texttt{networks}), Complex networks (\texttt{complex networks}), Bus rapid transit (\texttt{brt}).
}{
Les domaines sémantiques et la mesure d'interdisciplinarité ont été recalculés pour ce corpus par collecte des mots-clés, puis extraction selon la méthode décrite en~\ref{sec:quantepistemo}, avec $K_W=1000$, $\theta_w=15$ et $k_{max}=500$. On obtient des communautés plus ciblées et plutôt représentatives de la thématique et des méthodes : Transit-oriented development (\texttt{tod}), Hedonic models (\texttt{hedonic}), Planification des infrastructures (\texttt{infra planning}), High-speed rail (\texttt{hsr}), Réseaux (\texttt{networks}), Réseaux complexes (\texttt{complex networks}), Bus rapid transit (\texttt{brt}).
}

\bpar{
A ``good choice'' of characteristics to classify models is similar to the issue of choosing features in machine learning: in the case of supervised learning, i.e. when we aim at obtaining a good prediction of classes fixed a priori (or a good modularity of the obtained classification relatively to the fixed classification), we can select features optimizing this prediction. We will therein discriminate models that are known and judged different. If we want to extract an endogenous structure without a priori (unsupervised classification), the issue is different. We will therefore test in a second time a regression technique which allows to avoid overfitting and to select features (random forests).
}{
Un ``bon choix'' de caractéristiques pour classer les modèles est un peu le problème du choix des \emph{features} en apprentissage statistique : si on est en supervisé, c'est-à-dire qu'on veut obtenir une bonne prédiction de classe fixée a priori (ou une bonne modularité de la classification obtenue par rapport à la classification fixée), on pourra sélectionner les caractéristiques optimisant cette prédiction. On discriminera ainsi les modèles que l'on connait et que l'on juge différents. Si l'on veut extraire une structure endogène sans a priori (classification non supervisée), la question est différente. Nous testerons pour cela en second temps une technique de regression qui permet d'éviter l'overfitting et faire de la selection de caractéristiques (forêts aléatoires).
}

\subsubsection{Processes and case study}{Processus et cas d'étude}

\bpar{
Regarding the existence of a case study and its localization, 26\% of studies do not have any, corresponding to an abstract model or toy model (close to all studies in physics fall within this category). Then, they are spread across the world, with however an overrepresentation of Netherlands with 6.9\%. Processes included are too much varied (in fact as much as ontologies of concerned disciplines) to be the object of a typology, but we will observe the dominance of the notion of accessibility (65\% of studies), and then very different processes ranging from real estate market processes for hedonic studies, to relocations of actives and employments in the case of Luti, or to network infrastructure investments. We observe abstract geometric processes of network growth, corresponding to works in physics. Network maintenance appears in one study, as political history does. Abstract processes of agglomeration and dispersion are also at the core of several studies. Interactions between cities are a minority, the systems of cities approaches being drown in accessibility studies. Issues of governance and regulation also emerge, more in the case of infrastructure planning and of TOD approaches evaluation models, but remain a minority. We will stay with the fact that each domain and then each study introduces its own processes with are quasi-specific to each case.
}{
Concernant l'existence d'un cas d'étude et sa localisation, 26\% des études n'en présentent pas, correspondant à un modèle abstrait ou modèle jouet (la quasi totalité des études en physique tombant dans ce cas). Ensuite, elles sont réparties à travers le monde, avec toutefois une surreprésentation des Pays-bas avec 6.9\%. Les processus inclus sont trop variés (en fait autant que les ontologies des disciplines concernées) pour faire l'objet d'une typologie, mais nous noterons la domination de la notion d'accessibilité (65\% des études), puis des processus très variés allant de processus de marché immobilier pour les études hédoniques, aux relocalisations d'actifs et d'emplois pour les Luti, ou aux investissements d'infrastructure de réseau. Nous observons des processus abstraits géométriques de croissance de réseau, correspondant aux travaux des physiciens. La maintenance du réseau apparait dans une étude, ainsi que l'histoire politique. Les processus abstraits d'agglomération et dispersion sont aussi le coeur de quelques études. Les interactions entre villes sont minoritaires, les approches de type systèmes de villes étant noyées dans les études d'accessibilité. Les questions de gouvernance et de régulation ressortent aussi, plutôt dans le cas de planification d'infrastructures et de modèle d'évaluation de démarches TOD, mais sont aussi minoritaires. On retiendra que chaque domaine puis chaque étude introduit ses propres processus quasi-spécifiques à chaque cas.
}

\subsubsection{Corpus characteristics}{Caractéristiques du corpus}

%

\bpar{
The domains ``a priori'' (i.e. judged, or more precisely inferred from journal or institution to which authors belong), are relatively balanced for the main disciplines already identified: 17.9\% Transportation, 20.0\% Planning, 30.3\% Economics, 19.3\% Geography, 8.3\% Physics , the rest in minority being shared between environmental science, computer science, engineering and biology. Regarding the share of significant semantic domains, TOD dominates with 27.6\% of documents, followed by networks (20.7\%), hedonic models (11.0\%), infrastructure planning (5.5\%) and HSR (2.8\%). Contingence tables show that Planning does almost only TOD, physics only networks, geography is equally shared between networks and TOD (the second corresponding to articles of the type ``urban project management'', that have been classified in geography as published in geography journals) and also a smaller part in HSR, and finally economics is the most diverse between hedonic models, planning, networks and TOD. This interdisciplinarity however appears only for classes extracted for the higher probability, since average interdisciplinarity indices by discipline have comparable values (from 0.62 to 0.65), except physics which is significantly lower at 0.56 what confirms its status of ``newcomer'' with a weaker thematic depth.
}{
Les domaines ``a priori'' (i.e. jugés, ou plutôt préjugés sur la revue ou l'appartenance des auteurs), sont relativement équilibrés pour les disciplines majoritaires déjà identifiées : 17.9\% Transportation, 20.0\% Planning, 30.3\% Economics, 19.3\% Geography, 8.3\% Physics, le reste minoritaire se répartissant entre environnement, informatique, ingénierie et biologie. Concernant les poids des domaines sémantiques significatifs, le TOD domine avec 27.6\% des documents, suivi par les réseaux (20.7\%), les modèles hédoniques (11.0\%), la planification des infrastructures (5.5\%) et le HSR (2.8\%). Les tables de contingences montrent que le Planning ne fait quasiment que du TOD, la physique uniquement des réseaux, la géographie se répartit équitablement entre réseaux et TOD (le second correspondant aux articles typés ``aménagement'', qui ont été classés en géographie car dans des revues de géographie) ainsi qu'une plus faible part en HSR, enfin l'économie est la plus variée entre hédonique, planning, réseaux et TOD. Cette interdisciplinarité n'apparait cependant que pour les classes extraites pour la probabilité majoritaire, puisque les indices d'interdisciplinarité moyens par discipline ont des valeurs équivalentes (de 0.62 à 0.65), hormis la physique significativement plus basse à 0.56 ce qui confirme son statut de ``nouveau venu'' ayant une profondeur thématique plus faible.
}

\subsubsection{Models studied}{Modèles étudiés}

\bpar{
It is interesting for our problematic to answer the question ``who does what ?'', i.e. which type of models are used by the different disciplines. We give in Table~\ref{tab:modelography:what} the contingence table of the type of model as a function of disciplines a priori, of the citation class and of the semantic class. We observe that strongly coupled approaches, the closest of what is considered as co-evolution models, are mainly contained in the vocabulary of networks, what is confirmed by their positioning in terms of citations, but that the disciplines concerned are varied. The majority of studies focus on the territory only, the strongest unbalance being for studies semantically linked to TOD and hedonic models. Physics is still limited as focusing exclusively on networks.
}{
Il est intéressant pour notre question de répondre à la question ``qui fait quoi ?'', c'est-à-dire quelles types de modèles sont mobilisés par les différentes disciplines. Nous donnons en Table~\ref{tab:modelography:what} la table de contingence du type de modèle en fonction des disciplines a priori, de la classe de citation et de la classe sémantique. On constate que les approches fortement couplées, les plus proches de ce qu'on considère comme des modèles de co-évolution, sont majoritairement contenues dans le vocabulaire des réseaux, ce qui est confirmé par leur positionnement en terme de citation, mais que les disciplines concernées sont variées. La majorité des études s'intéresse au territoire uniquement, le déséquilibre le plus fort étant pour les études sémantiquement liées au TOD et à l'hédonique. La physique est encore limitée en s'intéressant exclusivement aux réseaux. 
}

\begin{table}
\caption[Type of models obtained in the modelography]{\textbf{Types of models studied according to the different classifications.} Contingence tables of the discrete variable giving the type of model (network, territory or strong coupling), for the a priori classification, the semantic classification and the citation classification.\label{tab:modelography:what}}
\begin{tabular}{|p{2.4cm}|p{2.4cm}p{2.4cm}p{2.4cm}p{2.4cm}p{2.4cm}|}
\hline
Discipline  &  economics & geography & physics & planning & transportation\\\hline
network     &     5      &      3    &   12    &    1     &         4  \\
strong      &     4      &     3     &   0     &   0      &        2  \\
territory   &    35      &    22     &   0     &    28    &         20 \\\hline  
\end{tabular} 
\medskip
\begin{tabular}{|p{2.4cm}|p{2.4cm}p{2.4cm}p{2.4cm}p{2.4cm}p{2.4cm}|}
\hline
Semantic  &  hedonic & hsr & infra planning & networks & tod\\\hline
network   &       1  & 0   &          0     &  14      & 2 \\
strong    &       0  &  0  &            0   &     5    & 1  \\
territory &      15  &  4  &            8   &    11    &  37 \\ \hline
\end{tabular}
\begin{tabular}{|p{2cm}|p{2cm}p{2cm}p{2cm}p{2cm}p{2cm}p{2cm}|}
\hline
Citation  &  accessibility & geography & infra planning & LUTI & networks & TOD \\\hline
network   &            0   &     0     &         0      &   0  &     24   &  0 \\
strong    &            0   &      0    &          0     &   2  &      5   &  0 \\
territory &           13   &      1    &          6     &  18  &      2   &  3 \\\hline
\end{tabular}
\end{table}

\subsubsection{Studied scales}{Échelles étudiées}

\bpar{
To then answer the question of the how, we can have a look at temporal and spatial typical scales of models. Planning and transportation are concentrated at small spatial scales, metropolitan or local, economics also with a strong representation of the local through hedonic studies, and a spatial range a bit larger with the existence of studies at the regional level and a few at the scale of the country (panel studies generally). Again, physics remains limited with all its contributions at a fixed scale, the metropolitan scale (which is not necessarily clear nor well specified in articles in fact since these are toy models which thematic boundaries may be very fuzzy). Geography is relatively well balanced, from the metropolitan to the continental scale. The scheme for temporal scales is globally the same. The methods used are strongly correlated to the discipline: a $\chi^2$ test gives a statistic of 169, highly significant with $p=0.04$. Similarly, spatial scale also is but in a less strong manner ($\chi^2 = 50, p = 0.08$).
}{
Pour répondre ensuite à la question du comment, on peut regarder les échelles de temps et d'espace typiques des modèles. La planification et les transports se concentrent à des petites échelles spatiales, métropolitain ou local, l'économie également avec une forte représentation du local via les études hédoniques, et une étendue un peu plus grande avec l'existence d'études au niveau régional et quelques une du pays (études de panel généralement). Encore une fois, la physique se retrouve limitée avec l'ensemble de ses contributions à une échelle fixe, métropolitaine (pas forcément claire ni bien spécifiée dans les articles d'ailleurs puisqu'il s'agit de modèles jouets dont les contours thématiques peuvent être très flous). La géographie est relativement bien équilibrée, de l'échelle métropolitaine à l'échelle continentale. Le schéma pour les échelles de temps est globalement similaire. Les méthodes utilisées sont fortement corrélées à la discipline : un test du $\chi^2$ donne une statistique de 169, très significatif avec $p=0.04$. De même, l'échelle d'espace l'est mais de manière moindre ($\chi^2 = 50, p = 0.08$).
}

\subsubsection{Classical regressions}{Régressions classiques}


\bpar{
We now study the influence of diverse factors on characteristics of models through simple linear regressions. In a multi-modeling approach, we propose to test all the possible models to explain each of the variables from the others. The number of observations for which all the variables have a value is very low, we need to take into account the number of observations used to fit each model. Furthermore, model performances can be characterized by complementary objectives. Following~\cite{igel2005multi}, we apply a multi-objective optimization, to simultaneously maximize the explained variance (adjusted R$^2$ in our case) and the information captured (corrected Akaike information criterion AICc\footnote{AIC is a measure of the information gain between two models, and allows to avoid abusive overfitting through a too large number of parameters. AICc is a version taking into account the size of the sample, the measure varying significantly for the small samples.}). It is realized conditionally to the fact of having the number of observations $N>50$ (fixed threshold regarding the distribution of $N$ on all models). The optimization procedure is detailed in Appendix~\ref{app:sec:quantepistemo} for each variable. Time scale and interdisciplinarity exhibit compromises difficult choose from, and we adjust the two candidates. Other variables exhibit dominating solutions and we adjust only a single model.
}{
Nous étudions à présent l'influence de divers facteurs sur les caractéristiques des modèles par des régressions linéaires simples. Dans une démarche de multi-modélisation, nous proposons de tester l'ensemble des modèles possibles pour expliquer chacune des variables à partir des autres. Le nombre d'observations pour lesquelles toutes les variables sont renseignées est très faible, il s'agit de prendre en compte le nombre d'observations utilisées pour ajuster chaque modèle. D'autre part, les performances du modèle peuvent être caractérisées par des objectifs complémentaires. Suivant~\cite{igel2005multi}, nous appliquons une optimisation multi-objectif, pour maximiser simultanément la variance expliquée (R$^2$ ajusté dans notre cas) et l'information capturée (Critère d'information d'Akaike corrigé AICc\footnote{L'AIC est une mesure du gain d'information entre deux modèles, et permet d'éviter l'ajustement abusif par un nombre trop grand de paramètres. L'AICc est une version prenant en compte la taille de l'échantillon, la mesure variant significativement pour les petits échantillons.}). Celle-ci est effectuée conditionnellement au fait d'avoir le nombre d'observations $N>50$ (seuil fixé au regard de la distribution de $N$ sur l'ensemble des modèles). La procédure d'optimisation est détaillée en Annexe~\ref{app:sec:quantepistemo} pour chaque variable. L'échelle de temps et l'interdisciplinarité présentent des compromis difficiles à départager, et nous ajustons les deux candidats. Les autres variables présentent des solutions dominantes et nous n'ajustons qu'un seul modèle.
}

\bpar{
Complete regression results are given in Table~\ref{tab:quantepistemo:regressions}. Temporal and spatial scales, together with year, are the variables the best explained in the sense of the variance. Time scale is very significantly influenced by the type of model: territory which decreases it, or strong coupling which increases it. The fact to be in physics also significantly influences, and broadens the time range of models. On the contrary, engineering approaches (often optimal design of a transportation network) correspond to a short time span.
}{
Les résultats complets des régressions sont donnés en Table~\ref{tab:quantepistemo:regressions}. Les échelles temporelle et d'espace, ainsi que l'année, sont les variables les mieux expliquées au sens de la variance. L'échelle de temps est influencée très significativement par le type de modèle : territoire qui diminue celle-ci, ou couplage fort qui l'augmente. Le fait d'être en physique influe également significativement, et élargit la portée temporelle des modèles. Au contraire, les approches d'ingénierie (souvent design optimal d'un réseau de transport) correspondent à une courte durée.
}

\bpar{
For the spatial scale, the fact to be in geography has a strong influence on the spatial range of models: indeed, regional studies and at the scale of the system of cities are indeed the prerogative of geography. The belonging to the field of transportation also increases slightly the spatial range (see significance in the complete regressions in Appendix~\ref{app:sec:quantepistemo}). No other variable has a significant influence.
}{
Pour l'échelle d'espace, le fait d'être en géographie a une forte influence sur la portée spatiale des modèles : en effet, les études régionales et à l'échelle du système de villes sont bien l'apanage de la géographie. L'appartenance au domaine du transport augmente aussi faiblement la portée spatiale (voir significativité dans les regressions complètes en Annexe~\ref{app:sec:quantepistemo}). Aucune autre variable n'a une influence significative.
}

\bpar{
The level of interdisciplinarity is well explained by the year, which influences it in a negative way, what confirms an increase in scientific specializations in time. Econometric studies of hedonic models appears to be very specialized. Finally, publication year is significantly and positively explained by the territory type and by the fact to be in transportation, what would correspond to a recent resurgence of a particular profile of studies. A study of the corpus suggests that this would be studies on high speed, which would appear as a recent scientific fashion.
}{
Le niveau d'interdisciplinarité est bien expliqué par l'année, qui l'influence de manière négative, ce qui confirme une augmentation des spécialisations scientifiques dans le temps. Les études économétriques des modèles hédoniques apparaissent très spécialisées. Enfin, l'année de publication est expliquée significativement et positivement par le type territoire et par le fait d'être en transports, ce qui signifierait une recrudescence récente d'un profil particulier d'études. Un examen du corpus suggère qu'il s'agit des études sur la grande vitesse, apparaissant comme une mode scientifique récente.
}

\begin{table}[!htbp]
\caption[Explanation of models characteristics]{\textbf{Explanation of models characteristics.} Results of the Ordinary Least Squares (OLS) estimation of selected linear models, for each variable to be explained: temporal scale (TEMPSCALE), spatial scale (SPATSCALE), interdisciplinarity index (INTERDISC), publication year (YEAR).\label{tab:quantepistemo:regressions}}
\begin{tabular}{lcccccc} 
\footnotesize
\\[-1.8ex]\hline 
\hline \\[-1.8ex] 
 & \multicolumn{6}{c}{\bpar{\textit{Explained variable:}}{\textit{Variable expliquée :}}} \\ 
\cline{2-7} 
 & \multicolumn{2}{c}{TEMPSCALE} & SPATSCALE & \multicolumn{2}{c}{INTERDISC} & YEAR \\ 
 & (1) & (2) & (3) & (4) & (5) & (6)\\ 
\hline \\[-1.8ex] 
 YEAR & 0.674 &  &  & $-$0.004$^{*}$ & $-$0.002$^{*}$ &  \\ 
  TYPEstrong &  & 100.271$^{***}$ &  &  & $-$0.026 &  \\ 
  TYPEterritory & $-$38.933$^{***}$ & $-$14.988 &  &  & 0.044 & 10.898$^{***}$ \\ 
  TEMPSCALE &  &  & $-$5.179 & $-$0.0003 &  & 0.035 \\ 
  FMETHODeq &  &  &  &  &  & $-$6.224 \\ 
  FMETHODmap &  &  &  &  &  & 4.747 \\ 
  FMETHODro &  &  &  &  &  & 6.128 \\ 
  FMETHODsem &  &  &  &  &  & 1.009 \\ 
  FMETHODsim &  &  &  &  &  & 5.153 \\ 
  FMETHODstat &  &  &  &  &  & $-$0.357 \\ 
  DISCIPLINEengineering & $-$52.107$^{*}$ & $-$9.609 & $-$154.461 & 0.144 &  & 13.486 \\ 
  DISCIPLINEenvironment & 17.110 & 17.886 & $-$5.878 & 0.092 &  & $-$3.668 \\ 
  DISCIPLINEgeography & 3.640 & 9.126 & 1,445.457$^{***}$ & 0.036 &  & 1.121 \\ 
  DISCIPLINEphysics & 46.879$^{*}$ & 77.897$^{***}$ & 292.559 & $-$0.103 &  & 3.392 \\ 
  DISCIPLINEplanning & 1.304 & 4.553 & $-$143.554 & $-$0.047 &  & $-$2.850 \\ 
  DISCIPLINEtransportation & $-$14.718 & 8.753 & 568.329 & 0.062 &  & 5.503$^{*}$ \\ 
  INTERDISC & 2.357 &  &  &  &  & $-$12.876 \\ 
  SEMCOMcomplex networks &  &  &  &  & $-$0.217 &  \\ 
  SEMCOMhedonic &  &  &  & $-$0.179 & $-$0.184$^{*}$ & $-$5.769 \\ 
  SEMCOMhsr &  &  &  & $-$0.100 & $-$0.122 & 6.135 \\ 
  SEMCOMinfra planning &  &  &  & $-$0.032 & $-$0.096 & $-$4.123 \\ 
  SEMCOMnetworks &  &  &  & $-$0.038 & $-$0.107 & 4.711 \\ 
  SEMCOMtod &  &  &  & $-$0.105 & $-$0.152 & $-$1.653 \\ 
  Constant & $-$1,305.126 & 22.103$^{*}$ & 235.357 & 8.962$^{**}$ & 5.531$^{**}$ & 2,004.945$^{***}$ \\ 
 \hline \\[-1.8ex] 
Observations & 64 & 94 & 94 & 64 & 98 & 64 \\ 
R$^{2}$ & 0.385 & 0.393 & 0.100 & 0.314 & 0.155 & 0.510 \\ 
R$^{2}$ ajusté & 0.282 & 0.336 & 0.027 & 0.136 & 0.068 & 0.281 \\ 
\hline 
\hline \\[-1.8ex] 
\textit{Note:}  & \multicolumn{6}{l}{$^{*}$p$<$0.1; $^{**}$p$<$0.05; $^{***}$p$<$0.01} \\ 
\end{tabular}
\end{table} 

\subsubsection{Random Forest regressions}{Régressions par Forêts Aléatoires}

\bpar{
We conclude this study by regressions and classification with random forests, which are a very flexible method allowing to unveil a structure from a dataset~\cite{liaw2002classification}. To complement the previous analysis, we propose to use it to determine the relative importances of variables for different aspects. We use each time forests of size 100000, a node size of 1 and a number of sampled variables in $\sqrt{p}$ for the classification and $p/3$ for the regression when $p$ is the total number of variables. To classify the type of models, we compare the effects of discipline, of the semantic class and of the citation class. The latest is the most important with a relative measure of 45\%, whereas the discipline accounts for 31\% and the semantic of 23\%. This way, the disciplinary compartmentalization is found again, whereas the semantic and this partly ontologies, is the most open. This encourages us in our aim at getting out of this compartmentalization. When we apply a forest regression on interdisciplinarity, still with these three variables, we obtain that they explain 7.6\% of the total variance, what is relatively low, witnessing a semantic disparity on the whole corpus independently of the different classifications. In this case, the most important variable is the discipline (39\%) followed by the semantic (31\%) and citation (29\%), what confirms that the journal targeted strongly conditions the behavior in the language used. This alerts on the risk of a decrease in semantic wealth when targeting a particular public. This way, we have unveiled certain structures and regularities of models related to our question, which implications will be useful during the construction of our models.
}{
Nous concluons cette étude par des régressions et classification par forêts aléatoires, qui sont une méthode très flexible permettant de dégager une structure d'un jeu de données~\cite{liaw2002classification}. Pour compléter les analyses précédentes, nous proposons de l'utiliser pour déterminer les importances relatives des variables pour différents aspects. Nous utilisons à chaque fois des forêts de taille 100000, une taille de noeud de 1 et un nombre de variable échantillonnée en $\sqrt{p}$ pour la classification et $p/3$ pour la régression lorsque $p$ est le nombre total de variables. Pour classifier le type de modèle, nous comparons les effets de la discipline, de la classe sémantique et de la classe de citation. Cette dernière est la plus importante avec une mesure relative de 45\%, tandis que la discipline compte pour 31\% et le sémantique pour 23\%. Ainsi, le cloisonnement disciplinaire se retrouve, tandis que le sémantique et donc en partie les ontologies, est le plus ouvert. Cela nous encourage dans notre démarche de sortir de ce cloisonnement. Lorsqu'on applique une regression de forêt sur l'interdisciplinarité, toujours avec ces trois variables, on constate qu'elles expliquent 7.6\% de la variance totale, ce qui est relativement faible, témoignant d'une disparité de sémantique sur l'ensemble du corpus indépendamment des différentes classifications. Dans ce cas, la variable la plus importante est la discipline (39\%) suivie par le sémantique (31\%) et la citation (29\%), ce qui confirme que le journal visé conditionne fortement le comportement de langage employé. Cela nous alerte sur le danger de perte de richesse sémantique lorsqu'on s'adresse à un public particulier. Ainsi, nous avons pu dégager certaines structures et régularités des modèles nous concernant, qui seront riches d'enseignements lors de la construction de nos modèles.
}

\subsection{Discussion}{Discussion}

\subsubsection{Developments}{Développements}

\bpar{
A possible development could consist in the construction of an automatized approach to this meta-analysis, from the point of view of modular modeling, combined to a classification of the aim and the scale. Modular modeling consists in the integration of heterogeneous processes and the implementation of these processes in the aim of extracting mechanisms giving the highest proximity to empirical stylized facts or to data~\cite{cottineau2015incremental}. The idea would be to be able to automatically extract the modular structure of existing models, starting from full texts as proposed in~\ref{sec:quantepistemo}, in order to classify these bricks in an endogeneous way and to identify potential couplings for new models.
}{
Un développement possible pourrait consister en la mise en place d'une approche automatique à cette méta-analyse, du point de vue de la modélisation modulaire, combiné avec une classification du but et de l'échelle. La modélisation modulaire consiste en l'intégration de processus hétérogènes et d'implémentation de ces processus dans le but d'extraire les mécanismes donnant la meilleure proximité à des faits stylisés empiriques ou à des données~\cite{cottineau2015incremental}. L'idée serait de pouvoir extraire automatiquement la structure modulaire des modèles existants, à partir des textes complets comme proposé en~\ref{sec:quantepistemo}, afin de classifier ces briques de manière endogène et identifier des couplages potentiels pour des nouveaux modèles.
}

\subsubsection{Lessons for modeling}{Leçons pour la modélisation}

\bpar{
We can summarize the main points obtained from this meta-analysis that will influence our position and our modeling choice. First of all, the interdisciplinary presence of approaches realizing a strong coupling confirms our need to build bridges and to couple approaches, and also retrospectively confirms the conclusions of~\ref{sec:quantepistemo} on the consequence of discipline compartmentalization in terms of the models formulated. Secondly, the importance of the vocabulary of networks in a large part of models will lead us to confirm this anchorage. The specificity of TOD and accessibility approaches, relatively close to the LUTI models, will be of secondary importance for us. The restricted span of works from physics, confirmed by the majority of criteria studied, suggests to remain cautious of these works and the absence of thematic meaning in the models. The wealth of temporal and spatial scales covered by geographical and economical models confirms the importance of varying these in our models, ideally to reach multi-scale models. Finally, the relative importance of classification variables on the type of model also suggest the direction of interdisciplinary bridges to cross ontologies.
}{
Nous pouvons résumer les points principaux issus de cette méta-analyse qui joueront sur notre attitude et nos choix de modélisation. Tout d'abord, la présence interdisciplinaire des approches effectuant un couplage fort confirme notre besoin de faire des ponts et de coupler les approches, et confirme également rétrospectivement les conclusions de~\ref{sec:quantepistemo} sur les conséquences du cloisonnement des disciplines en terme de modèles formulés. Ensuite, l'importance du vocabulaire des réseaux dans une grande partie des modèles nous poussera à confirmer cet ancrage. La spécificité des approches TOD et d'accessibilité, assez proches des modèles LUTI, seront secondaires pour nous. La portée restreinte des travaux issus de la physique, confirmée par la majorité des critères étudiés, nous pousse à nous méfier de ces travaux et de l'absence de sens thématique aux modèles. La richesse des échelles temporelles et spatiales couvertes par les modèles géographiques et économiques nous confirme l'importance de varier celles-ci dans nos modèles, idéalement de parvenir à des modèles multi-échelles. Enfin, les importances relatives des variables de classification sur le type de modèle vont également dans le sens de ponts interdisciplinaires pour croiser les ontologies.
}

\stars

%


\newpage

\section*{Synthesis of modeled processes}{Synthèse des processus modélisés}

\bpar{
We propose to synthesize processes taken into account by models encountered during the modelography, in order to proceed to a similar effort than the one concluding the thematic approach of chapter~\ref{ch:thematic}. We can neither have an exhaustive view (as already mentioned in the description of the modelography methodology) nor render with a high precision each model in the details, since almost each is unique in its ontology. The exercise of the synthesis allows then to take a step back from this limits and take a certain height, and have thus an overview on \emph{modeled processes}\footnote{Keeping in mind selection choices, which lead for example to not have mobility processes within this synthesis.}.
}{
Nous proposons de synthétiser les processus pris en compte par les modèles parcourus lors de la modélographie, afin de procéder à un effort similaire à celui concluant l'approche thématique du chapitre~\ref{ch:thematic}. Nous ne pouvons ni avoir une vision exhaustive (comme déjà précisé lors de la description de la méthodologie de la modélographie) ni rendre compte avec grande précision de chaque modèle en détail, puisque quasiment chacun est unique dans son ontologie. L'exercice de synthèse permet ainsi de s'extraire de ces limites et prendre un certain recul, et avoir ainsi un aperçu sur les \emph{processus modélisés}\footnote{En gardant en tête les choix de selection, qui emmènent par exemple à ne pas avoir les processus de mobilité dans cette synthèse.}.
}

\begin{table}
\caption[Synthesis of modeled processes]{\textbf{Synthesis of modeled processes.} These are classed by scale, type of model and discipline. \label{tab:modelography:processes}}
\bpar{
\begin{tabular}{|l|p{5cm}|p{5cm}|p{5cm}|}
\hline
 & Networks $\rightarrow$ Territories & Territories $\rightarrow$ Networks & Networks $\leftrightarrow$ Territories\\ \hline
\multirow{2}{*}{Micro} &
\textbf{Economics: } real estate market, relocalization, employment market & NA & \textbf{Computer Science : } spontaneous growth \\\cline{2-2}
& \textbf{Planning: } regulations, development & & \\\hline
& \textbf{Economics: } real estate market, transportation costs, amenities & \textbf{Economics: } network growth, offer and demand & \textbf{Economics: } investments, relocalizations, offer and demand, network planning\\\cline{2-4}
\multirow{2}{*}{Meso}& \textbf{Geography: } land-use, centrality, urban sprawl, network effects & \textbf{Transportation: } investments, level of governance & \textbf{Geography: } land-use, network growth, population diffusion \\\cline{2-3}
& \textbf{Planning/transportation: } accessibility, land-use, relocalization, real estate market & \textbf{Physics: } topological correlations, hierarchy, congestion, local optimization, network maintenance & \\\hline
& \textbf{Economics: } economic growth, market, land-use, agglomeration, sprawl, competition & \textbf{Economics: } interactions between cities, investments & \textbf{Economics: } offer and demand \\ \cline{2-4}
\multirow{2}{*}{Macro} & \textbf{Geography: } accessibility, interaction between cities, relocalization, political history & \textbf{Geography: } interactions between cities, potential breakdown & \textbf{Transportation: } network coverage \\\cline{2-3}
& \textbf{Transportation: } accessibility, real estate market & \textbf{Transportation: } network planing & \\\hline
\end{tabular}
}{
\begin{tabular}{|l|p{5cm}|p{5cm}|p{5cm}|}
\hline
 & Réseaux $\rightarrow$ Territoires & Territoires $\rightarrow$ Réseaux & Réseaux $\leftrightarrow$ Territoires\\ \hline
\multirow{2}{*}{Micro} &
\textbf{Economie : } marché immobilier, relocalisation, marché de l'emploi & NA & \textbf{Informatique : } croissance spontanée \\\cline{2-2}
& \textbf{Planification : } régulations, développement & & \\\hline
& \textbf{Economie : } marché immobilier, coût du transport, aménités & \textbf{Economie : } croissance du réseau, offre et demande & \textbf{Economie : } investissements, relocalisations, offre et demande, planification du réseau\\\cline{2-4}
\multirow{2}{*}{Meso}& \textbf{Géographie : } usage du sol, centralité, étalement urbain, effets de réseau & \textbf{Transports : } investissements, niveau de gouvernance & \textbf{Géographie : } usage du sol, croissance du réseau, diffusion de population \\\cline{2-3}
& \textbf{Planification/transports : } accessibilité, usage du sol, relocalisation, marché immobilier & \textbf{Physique : } corrélations topologiques, hiérarchie, congestion, optimisation locale, maintenance du réseau & \\\hline
& \textbf{Economie : } croissance économique, marché, usage du sol, agglomération, dispersion, compétition & \textbf{Economie : } interactions entre villes, investissements & \textbf{Economie : } offre et demande \\ \cline{2-4}
\multirow{2}{*}{Macro} & \textbf{Géographie : } accessibilité, interaction entre villes, relocalisation, histoire politique & \textbf{Géographie : } interactions entre villes, rupture de potentiel & \textbf{Transports : } couverture du réseau \\\cline{2-3}
& \textbf{Transports : } accessibilité, marché immobilier & \textbf{Transports : } planification de réseau & \\\hline
\end{tabular}
}
\end{table}




\bpar{
The table~\ref{tab:modelography:processes} proposes this synthesis from the 145 articles obtained from the modelography and for which a classification of the type was possible, i.e. that there existed a model entering the typology developed in~\ref{sec:modelingsa}. Being fully exhaustive would be similar to an interdisciplinary meta-modeling approach which is far out of the reach of our work\footnote{It would require to have correspondances between ontologies, without which we would obtain at least as much processes as models, even within one discipline. To the best of our knowledge such correspondence between two disciplines only does not exist. A direction for a formal approach is given in~\ref{app:sec:csframework}.}, and the list given here remain indicative.
}{
La table~\ref{tab:modelography:processes} propose cette synthèse à partir des 145 articles issus de la modélographie et pour lesquels une classification de type était possible, c'est-à-dire qu'il existait un modèle rentrant dans la typologie développée en~\ref{sec:modelingsa}. Être complètement exhaustif relèverait d'une opération de méta-modélisation interdisciplinaire qui est bien hors de la portée de notre travail\footnote{Il s'agirait pour cela d'avoir des correspondances entre les ontologies, sans lesquelles on se retrouverait avec au moins autant de processus que de modèles, même au sein d'une discipline. Il n'en existe à notre connaissance pas entre deux disciplines seulement. Une piste pour une approche formelle est donnée en~\ref{app:sec:csframework}.}, et la liste donnée ici est indicative.
}

\bpar{
We find again the correspondences between disciplines, scales and types of models obtained in the modelography in~\ref{sec:modelography}. We keep the fundamental following teachings, echoing the synthesis table obtained at the end of Chapter~\ref{ch:thematic} (Table~\ref{tab:thematic:processes}):
\begin{enumerate}
	\item The dichotomy of ontologies and processes taken into account between scales and between types is even more explicit here in models than in processes in themselves\footnote{Since this study was more detailed, it also appears stronger, since a greater precision allows then to exhibit abstract categories.}. We postulate that there indeed exists different processes at the different scales, and will make the choice to study different scales.
	\item The compartmentalization of disciplines shown in~\ref{sec:quantepistemo} can be found in a qualitative way in this synthesis: it is clear that they originally diverge in their different founding epistemologies. We will aim at integrating paradigms from different disciplines, while taking into account the limits imposed by the modeling principles that we will expose in~\ref{sec:computation} (for example, the parsimony of models necessarily limits the integration of heterogenous ontologies).
	\item An important gap between this synthesis and the one of processes is the quasi absence here of models integrating governance processes. It will be a direction to be explored.
	\item On the contrary, a very good correspondance can be established between geographical models of urban systems and the theoretical positioning of the evolutive urban theory. This correspondance, more difficult to exhibit for all the other approaches reviewed, also suggests us to follow this direction.
\end{enumerate}
}{
Nous retrouvons les correspondances entre disciplines, échelles et types de modèles obtenues dans la modélographie en~\ref{sec:modelography}. Nous retirons les enseignement principaux suivants, en écho au tableau de synthèse obtenu en fin du Chapitre~\ref{ch:thematic} (Table~\ref{tab:thematic:processes}) :
\begin{enumerate}
	\item La dichotomie des ontologies et des processus pris en compte entre les échelles et entre les types est autant manifeste ici dans les modèles que dans les processus en eux-même\footnote{Puisqu'on a plus détaillé cette étude, elle parait même plus forte aussi, une plus grande précision permettant alors de séparer des catégories abstraites.}. Nous postulons qu'il existe bien des processus différents aux différentes échelles, et nous prendrons le parti d'étudier différentes échelles.
	\item Le cloisonnement des disciplines démontré en~\ref{sec:quantepistemo} se retrouve qualitativement dans cette synthèse : il est évident qu'elles divergent originellement dans leurs différentes épistémologies fondatrices. Nous tacherons d'intégrer des paradigmes de différentes disciplines, tout en prenant en compte les limites imposées par les principes de modélisation que nous présenterons en~\ref{sec:computation} (par exemple, la parcimonie des modèles limite nécessairement l'intégration d'ontologies hétérogènes).
	\item Un décalage important entre cette synthèse et celle des processus est la quasi absence ici de modèles intégrant des processus de gouvernance. Il s'agira d'une piste à explorer.
	\item Au contraire, une très bonne correspondance s'établit entre les modèles géographiques des systèmes urbains et les positionnements théoriques de la théorie évolutive des villes. Cette adéquation, plus difficile à retrouver pour l'ensemble des autres approches revues, nous suggère également de suivre cette piste.
\end{enumerate}
}


\newpage

\section*{Chapter Conclusion}{Conclusion du Chapitre}


\bpar{
The processes that we aim at modeling being multi-scalar, hybrid and heterogenous, the possible points of view and questionings are necessarily highly varied, complementary and rich. This could be a fundamental characteristic of socio-technical systems, that \noun{Pumain} formulates in~\cite{pumain2005cumulativite} as ``a new measure of complexity'', which would be linked to the number of viewpoints necessary to grasp a system at a given level of exhaustivity. This idea rejoins the position of \emph{applied perspectivism} that~\ref{app:sec:csframework} formalizes and which is implicitly present in the investigation of relations between economics and geography developed in~\ref{app:sec:ecogeo}. Thus, the modeling of interactions between networks and territories can be related to a very broad set of disciplines and approaches reviewed in section~\ref{sec:modelingsa}.
}{
Les processus que nous cherchons à modéliser étant multi-scalaires, hybrides et hétérogènes, les angles d'approches et questionnements possibles sont nécessairement extrêmement variés, complémentaires et riches. Il pourrait s'agir d'une caractéristique fondamentale des systèmes socio-techniques, que \noun{Pumain} formule dans~\cite{pumain2005cumulativite} comme ``une nouvelle mesure de complexité'', qui serait liée au nombre de point de vue nécessaires pour appréhender un système à un niveau donné d'exhaustivité. Cette idée rejoint la position de \emph{perspectivisme appliqué} que~\ref{app:sec:csframework} formalise et qui est implicitement présente dans l'investigation des relations entre économie et géographie développée en~\ref{app:sec:ecogeo}. Ainsi, la modélisation des interactions entre réseaux et territoires peut être reliée à un ensemble très large de disciplines et d'approches revues en section~\ref{sec:modelingsa}.
}

\bpar{
In order to better understand the neighboring scientific landscape, and quantify the roles or relative weights of each, we have lead several analysis in quantitative epistemology in~\ref{sec:quantepistemo}. A first premiliminary analysis based on an algorithmic systematic review suggests a certain compartmentalization of domains. This conclusion is confirmed by the hypernetwork analysis coupling citation network and semantic network, which also allowed to draw the disciplinary boundaries more precisely, both for the direct relations (citations) but also their scientific proximity for the terms and methods used. We can then use the constituted corpus and this knowledge of domains to achieve a semi-automatic systematic review in~\ref{sec:modelography}, which allows to constitute a corpus of works directly dealing with the subject, which is then fully screened, allowing to link characteristics of models to the different domains. We have thus at this stage a rather clear idea of what is done, why and how.
}{
Afin de mieux comprendre le paysage scientifique environnant, et quantifier les rôles ou poids relatifs de chacune, nous avons procédé à une série d'analyse en épistémologie quantitative en~\ref{sec:quantepistemo}. Une première analyse préliminaire basée sur une revue systématique algorithmique suggère un certain cloisonnement des domaines. Cette conclusion est confirmée par l'analyse d'hyperréseau couplant réseau de citations et réseau sémantique, qui permet également de dessiner plus finement les contours disciplinaires, à la fois sur leur relations directes (citations) mais aussi leur proximité scientifique pour les termes et méthodes utilisées. On peut alors utiliser le corpus constitué et cette connaissance des domaines pour une revue systématique semi-automatique en~\ref{sec:modelography}, qui permet de constituer un corpus de travaux traitant directement du sujet, qui est ensuite inspecté intégralement, permettant de lier caractéristique des modèles au différents domaines. Nous avons alors à ce stade une idée assez précise de ce qui se fait, pourquoi et comment.
}

\bpar{
The issue remains to determine the relative relevance of some approaches or ontologies, what will be the aim of the two chapters of the second part. We first conclude this first part with a discussion chapter~\ref{ch:positioning}, shedding a light on points which are necessary to be clarified before entering the core of the subject.
}{
L'enjeu reste de déterminer les pertinences relatives de certaines approches ou ontologies, ce qui sera le but des deux chapitres de la deuxième partie. Nous concluons d'abord cette première partie par un chapitre de discussion~\ref{ch:positioning}, éclairant des points nécessaires à clarifier avant une entrée dans le vif du sujet.
}

\stars




\bpar{
\chapter{Positioning}
}{
\chapter{Positionnements}
} 

\label{ch:positioning} 



\bigskip

\bpar{
Any research activity would be, according to some actors of it, necessarily political, starting from the choice of its objects. Thus, \noun{Ripoll} warns against the illusion of an objective research and the dangers of technocracy~\cite{ripoll2017jig}. We will not enter debates which are much too broad to be tackled even in one chapter, since they connect to themes in political sciences, ethics, philosophy, linked for example to scientific governance, to the insertion of science in society, to scientific responsibility.
}{
Toute activité de recherche serait, selon certains acteurs de celle-ci, nécessairement politisée, de par pour commencer le choix de ses objets. Ainsi, \noun{Ripoll} alerte contre l'illusion d'une recherche objective et les dangers de la technocratie~\cite{ripoll2017jig}. Nous ne rentrerons pas dans ces débats bien trop vastes pour être traités même en un chapitre, puisqu'il rejoignent des thèmes de sciences politiques, d'éthique, de philosophie, liés par exemple à la gouvernance scientifique, à l'insertion de la science dans la société, à la responsabilité scientifique.
}

\bpar{
It is clear that even subjects that are a priori intrinsically objective, such as particle and high energy physics, have implications regarding on one hand the choice of their financing and the externalities associated (for example, the existence of CERN has largely contributed to the development of distributed computing), but on the other hand also the potential applications of discoveries which can have considerable social impacts. In biology, ethics is at the heart of funding principles of disciplines, as witness the debates raised by the emergence of synthetic biology~\cite{gutmann2011ethics}. The advocates of prudent approches in it overlap with integrative biology, as integrative sciences defended by \noun{Paul Bourgine}, put into practice by the intermediate of the Unesco digital campus CS-DC\footnote{\url{https://www.cs-dc.org/}}, typically has social responsibility and citizen implication in the center of their virtuous circle. In social sciences and humanities, since researches interact with the object studied (to some extent the idea of \emph{interactive kind} from \noun{Hacking}~\cite{hacking1999social}), the political and social implications of research are naturally never questioned.
}{
Il est clair que même des sujets a priori intrinsèquement objectifs, comme la physique des particules et des hautes énergies, ont des implications regardant d'une part les choix de leur financements et les externalités associées (par exemple, l'existence du CERN a largement contribué au développement du calcul distribué), mais d'autre part aussi les applications potentielles des découvertes qui peuvent avoir des répercussions sociales considérables. En biologie, l'éthique est au coeur des principes fondateurs des disciplines, comme en témoignent les débats soulevés par l'émergence de la biologie synthétique~\cite{gutmann2011ethics}. Les tenants d'approches prudentes dans celle-ci se recoupent avec la biologie intégrative, or les sciences intégratives défendues par \noun{Paul Bourgine}, mises en oeuvre par l'intermédiaire du campus digital Unesco CS-DC\footnote{\url{https://www.cs-dc.org/}}, ont typiquement la responsabilité sociale et l'implication citoyenne au coeur de leur cercle vertueux. En sciences humaines et sociales, comme les recherches interagissent avec les objets étudiés (en quelque sorte l'idée des \emph{interactive kind} de \noun{Hacking}~\cite{hacking1999social})
 , les implications politiques et sociales de la recherche sont bien évidemment indiscutables.
}
 

\bpar{
We will be here positioned at an epistemological level, i.e. reflexions on the nature and content of scientific knowledge in the broad sense, i.e. co-constructed and validated within a community imposing certain criteria to be scientific~\cite{morin1991methode}, of course evolving since we will take position for the systematisation of some. But still, even staying at this level, some positioning is necessary, on several dimensions such as epistemological, methodological, thematic. The last two have already been sketched in the two previous chapters by the choice of objects of study, of problematics, and will be reinforced as long as we make progress.
}{
Nous nous placerons ici à un niveau épistémologique, c'est-à-dire à des réflexions sur la nature et le contenu des connaissances scientifiques au sens large, c'est-à-dire co-construites et validées au sein d'une communauté imposant certains critères de scientificité~\cite{morin1991methode}, bien sûr évolutifs puisque nous nous positionnerons pour la systématisation de certains. Mais donc, même en restant à ce niveau, des prises de positions sont nécessaires, celles-ci pouvant être épistémologiques, méthodologiques, thématiques. Ces dernières ont déjà été ébauchées dans les deux chapitres précédents par les choix des objets d'étude, des problématiques, et seront renforcées à mesure de la progression.
}
 
\bpar{
We thus propose here a relatively original exercise but that we judge necessary for a more fluent reading of the following. It consists in the precise development of certain positions which have a particular influence in our research approach.
}{
Nous proposons ainsi ici un exercice relativement original mais que nous jugeons nécessaire pour une lecture plus fluide de la suite. Il consiste en le développement précis de certains positionnements qui ont une influence particulière dans notre démarche de recherche.
}

\bpar{
In a first section (\ref{sec:computation}), we precise our position regarding simulation models. After having detailed the functions we will give to the models, we argue in an essay form for a cautious use of big data and intensive computation, and illustrate our positioning regarding model exploration through a methodological case study for the exploration of the sensitivity of models to spatial initial conditions.
}{
Dans une première section (\ref{sec:computation}), nous précisons notre position au regard des modèles de simulation. Après avoir détaillé les fonctions que nous prêterons aux modèles, nous argumentons sous forme d'essai pour un usage raisonné des données massives et du calcul intensif, et illustrons notre positionnement par rapport à l'exploration des modèles par une étude de cas méthodologique pour l'exploration de la sensibilité des modèles aux conditions initiales.
}

\bpar{
In a second section (\ref{sec:reproducibility}), we develop some examples to illustrate the need and the difficulty of reproducibility, and also links with new tools that can foster but also endanger it. We illustrate the question of data opening and interactive exploration through an empirical case study of traffic flows in Ile-de-France.
}{
Dans une deuxième section (\ref{sec:reproducibility}), nous développons des exemples pour illustrer le besoin et la difficulté de reproductibilité, ainsi que les liens avec des nouveaux outils pouvant la favoriser mais aussi la mettre en danger. Nous illustrons la question d'ouverture des données et d'exploration interactive par une étude de cas empirique des flux de trafic en Ile-de-France.
}

\bpar{
Finally, the last section (\ref{sec:epistemology}) modestly explicits epistemological positions, in particular regarding the field in which we are placed, the complexity of objects in social sciences, and the nature of complexity in a broad sense.
}{
Enfin, la dernière section (\ref{sec:epistemology}) explicite modestement des positions épistémologiques, notamment concernant le courant dans lequel nous nous plaçons, la complexité des objets en sciences sociales, et la nature de la complexité de manière générale.
}

\bpar{
The reader very familiar with \noun{Banos} ``commandments''~\cite{banos2013pour} will find in the first two sections original practical illustrations of these, our positioning being mainly in their direction.
}{
Le lecteur très familier avec les ``commandements'' de \noun{Banos}~\cite{banos2013pour} pourra trouver dans les deux premières sections des illustrations pratiques originales de ceux-ci, notre positionnement étant principalement dans leur lignée.
}


\stars

\bpar{
\textit{This chapter is composed of various works. The first section is novel for its two first parts, and for its last part describes ideas presented as \cite{cottineau2017initial}. The second section relates in its first part the theoretical content of \cite{raimbault2016cautious}, and corresponds to \cite{raimbault2017investigating} for the empirical illustration. The third section follows for its first part the epistemological foundations of \cite{raimbault:halshs-01505084} which were then deepthen by \cite{raimbault2017applied}, follows a part of \cite{raimbault2018coevolution} for its second part and uses \cite{raimbault2017complex} for its last part.}
}{
\textit{Ce chapitre est composé de divers travaux. La première section est inédite pour ses deux premières parties, et pour sa dernière partie reprend des idées présentées dans \cite{cottineau2017initial}. La deuxième section rend compte pour sa première partie du contenu théorique de \cite{raimbault2016cautious}, et reprend \cite{raimbault2017investigating} pour l'illustration empirique. La troisième section reprend dans sa première partie les bases épistémologiques de \cite{raimbault:halshs-01505084} approfondies par \cite{raimbault2017applied}, reprend une partie de \cite{raimbault2018coevolution} pour sa deuxième partie et rend compte de \cite{raimbault2017complex} pour sa dernière partie.}
}

%


\newpage

\section{Modeling, big data and intensive computing}{Modélisation, données massives et calcul intensif}

\label{sec:computation}


\bpar{
We now develop our positioning regarding issues linked to the use of modeling, of massive data and of intensive computing, what also induces by extension some comments on model exploration methods. It is not evident to what extent these new possibilities are necessarily accompanied of deep epistemological mutations, and we show on the contrary that their use necessitates more than ever a dialog with theory. Implicitly, this position foreshadows the epistemological frame for the study of complex systems of which we give the context in~\ref{sec:epistemology} and that we formalize in opening~\ref{sec:knowledgeframework}.
}{
Nous nous positionnons à présent sur les questions liées à l'utilisation de la modélisation, des données massives et du calcul intensif, ce qui induit aussi par extension une réflexion sur les méthodes d'exploration de modèles. Il n'est pas évident que ces nouvelles possibilités soient nécessairement accompagnées de mutations épistémologiques profondes, et nous montrons au contraire que leur utilisation nécessite plus que jamais un dialogue avec la théorie. Implicitement, cette position préfigure le cadre épistémologique pour l'étude des systèmes complexes dont nous donnons le contexte en~\ref{sec:epistemology} et que nous formalisons en ouverture~\ref{sec:knowledgeframework}.
}

\bpar{
The points developed here cover some crucial issues linked to modeling enterprises, and can be of an epistemological, theoretical or practical nature. We will first try to answer the question of why modeling. We will then give our position on more technical issues linked to the use of emerging computing resources and new data. Finally, the last point is methodological, and both illustrates the first two points and introduces a new method to explore models.
}{
Les points développés ici couvrent certains enjeux cruciaux liés aux entreprises de modélisation, et peuvent être de nature épistémologique, théorique, ou pratique. Nous tenterons tout d'abord de répondre à la question du pourquoi de la modélisation. Nous nous positionnerons ensuite sur des questions plus techniques liées à l'utilisation des ressources de calcul émergentes et des nouvelles données. Enfin, le dernier point est méthodologique, et illustre à la fois les deux premiers tout en introduisant une nouvelle méthode d'exploration de modèles.
}


\subsection{Why modeling ?}{Pourquoi modéliser ?}

\bpar{
We first develop the role of modeling in our process of scientific production. Models have in appearance diverse roles depending on disciplines: a model in physics results of a theory, allows to confront it with experiments and has to be validated through its predictive powers with strong requirements, whereas in computational social science one often settles for the reproduction of general stylized facts. A statistical model will be composed of assumptions on relations between variables and on the statistical distribution of an error term, and values of coefficients obtained will be interpreted even if the goodness-of-fit measure is very low. The aim here is thus to precise in which spirit our modeling approaches will be placed\footnote{If this work may appear as redundant, laborious and superficial to someone used to geosimulation models, it is crucial in our logic of disciplinary opening, in order first to avoid any misunderstanding on the status of results, and secondly to foster a dialog in the case of very different uses of models.}, what are their mechanisms and objectives.
}{
Nous développons dans un premier temps le rôle de la modélisation dans notre démarche de production scientifique. Les modèles ont en apparence des rôles divers selon les disciplines : un modèle en physique découle d'une théorie, permet de la confronter à l'expérimentation et devra être validé par ses pouvoirs prédictifs avec de fortes exigences, tandis qu'en science sociales computationnelles on se contentera souvent de la reproduction de faits stylisés généraux. Un modèle statistique sera composé d'hypothèses sur des relations entre variables et sur la distribution statistique d'un terme d'erreur, et les valeurs des coefficients obtenus seront interprétées même si la mesure d'ajustement est très faible. Il s'agit donc ici de préciser dans quelle logique nos travaux de modélisation se placeront\footnote{Si ce travail pourra paraître redondant, laborieux et superflu aux habitués des modèles de géosimulation, il est crucial dans notre logique d'ouverture disciplinaire, afin d'une part d'éviter tout malentendu sur le statut des résultats, d'autre part d'encourager un dialogue dans le cas d'utilisations très différentes des modèles.}, quels seront leurs ressorts et objectifs.
}


\subsubsection{Functions of models}{Fonctions des modèles}

\bpar{
As we just saw, the term of \emph{model} has multiple meanings, and implies different realities, practices, uses (we can assume a proper ontology to models which become real objects, at least when they are implemented). A way to propose a sort of typology for models is to proceed to a typology of their functions, as does~\cite{varenne2017theories}, based on the study of diverse disciplines (biology, geography, social sciences). This classification is to the best of our knowledge the most exhaustive existing. \noun{Varenne} thus distinguishes five broad classes of model functions\footnote{The broad classes of functions are declined into precise classes which form 21 classes. We do not detail them here, but give a synthesis describing the broad classes.}, which are in an increasing order according to their integration to a social practice :
}{
Comme nous venons de l'évoquer, le terme de \emph{modèle} a de multiples sens, et implique différentes réalités, pratiques, utilisations (on peut supposer une ontologie propre aux modèles qui deviennent des objets réels, au moins lorsque ceux-ci sont implémentés). Une façon d'en proposer une sorte de typologie est de procéder à celle de leurs fonctions, comme le fait~\cite{varenne2017theories}, en se basant sur l'étude de diverses disciplines (biologie, géographie, sciences sociales). Cette classification est à notre connaissance la plus exhaustive qui existe. \noun{Varenne} distingue donc cinq grandes classes de fonctions des modèles\footnote{Les grandes classes de fonctions sont déclinées en classes précises qui sont au nombre de 21. Nous ne les détaillons pas ici, mais donnons une synthèse décrivant les grandes classes.}, qui vont de manière croissante dans leur intégration à une pratique sociale :
}

\bpar{
\begin{enumerate}
	\item Function of perception and observation: make accessible an object which can not be observed through perception (physical model of a molecule), allow experiments, a memorization, the reading and visualization of data.
	\item Function of intelligibility: description of patterns, precision of ontologies, conception through prediction, explanation and comprehension of processes\footnote{The comprehension is more general than explanation, since it assumes a reconstruction of the system structure and a deductive use, i.e. a projection and generation of the system considered in the psychological structure considering it~\cite{morin1980methode}.}.
	\item Function of assistance to theorization: formulation, interpretation, illustration of a theory, internal consistence test (do deductive schemes induce model simulation results that are contradictory or consistent ?), applicability, computability (in the case of numerical schemes allowing to approximate the solutions of equations), co-computability (coupling of theories and models).
	\item Function of social communication: scientific communication, consultation, action with actors (\emph{stakeholders}\footnote{We do not develop this aspect at all, but we recall that \emph{stakeholder workshops} are one of the structuring axis of the Medium project we described in~\ref{sec:qualitative}. Even if the percolation with the axis focusing on the analysis and modeling of urban systems dynamics in which our work is situated is not explicit, they implicitly operate in exchanges between perspectives, and the cohabitation within a project foreshadows more integrated future perspectives.}).
	\item Function of decision-making: informing decision-making, action, self-fulfilling action in an abstract system (pricing models in finance).
\end{enumerate}
}{
\begin{enumerate}
	\item Fonction de perception et d'observation : rendre accessible un objet inobservable à la perception (modèle physique d'une molécule), permettre des expérimentations, une mémorisation, la lecture et la visualisation de données.
	\item Fonction d'intelligibilité : description de motifs, précision des ontologies, conception par la prédiction, explication et compréhension de processus\footnote{La compréhension est plus générale que l'explication, car suppose une reconstruction de la structure du système et un usage déductif, c'est-à-dire une projection et génération du système considéré dans la structure psychique le considérant~\cite{morin1980methode}.}.
	\item Fonction d'aide à la théorisation : formulation, interprétation, illustration d'une théorie, test de cohérence interne (les schémas déductifs induisent-ils des résultats de simulation de modèles contradictoires ou cohérents ?), applicabilité, calculabilité (dans le cas de schémas numériques permettant d'approcher les solutions d'équations), co-calculabilité (couplage de théories et modèles).
	\item Fonction de communication sociale : communication scientifique, concertation, action avec les acteurs (\emph{stakeholders}\footnote{Nous ne développerons pas du tout cet aspect, mais tenons à préciser que les \emph{stakeholder workshops} sont l'un des axes structurants du projet Medium que nous avons décrit en~\ref{sec:qualitative}. Même si la percolation avec l'axe d'analyse et de modélisation des dynamiques des systèmes urbains dans lequel notre travail s'inscrit n'est pas explicite, celle-ci s'opèrent implicitement dans les échanges entre perspectives, et la cohabitation au sein d'un projet laisse supposer des perspectives futures plus intégrées.}).
	\item Fonction de prise de décision : aide à la décision, action, action auto-réalisatrice dans un système abstrait (modèles de pricing en finance).
\end{enumerate}
}

\bpar{
It is clear that each discipline will have its own relation to these different functions, that some will be privileged, and others not accessible or without relevance for the object studied or the questions asked. In physics for example, the aspects of theory validation and of the existence of predictive models with a very high precision are at the heart of the discipline; whereas entire branches of social science such as urban planning for example are focused on models for communication and decision-making. Regarding this, we must not neglect the nature of social science for economics and stay doubtful of predictive aims of some modeling experiments\footnote{Even in finance at high frequencies, at which signals would be reasonably closer to physical systems than macro-econometric series for example as witnesses the appropriation of these problems by physicists, the predictability remains questionable and in any case limited~\cite{campbell2007predicting}.}.
}{
Il est clair que chaque discipline va avoir sa propre relation à ces différentes fonctions, que certaines seront privilégiées, d'autre non accessibles ou sans pertinence pour les objets étudiés ou questions posées. En physique par exemple, les aspects de validation des théories et l'existence de modèles prédictifs d'une très grande précision sont au coeur de la discipline, tandis que des branches entières des sciences sociales comme par exemple la planification urbaine sont axées sur des modèles pour la communication et la prise de décision. A cet égard, il ne faut pas négliger la nature de science sociale de l'économie et douter des visées prédictives de certaines expériences de modélisation\footnote{Même en finance à des fréquences élevées, où les signaux seraient plus raisonnablement assimilables à des systèmes physiques que des séries macro-économétriques par exemple comme en témoignent l'appropriation de ces problématiques par les physiciens, la prédictibilité reste questionnable et en tout cas limitée~\cite{campbell2007predicting}.}. 
}

\bpar{
This classification of functions can be found implicitly in modeling reasons developed outside any typology by~\cite{epstein2008model}: he insists on refuting the preconceived idea that models would only be used for prediction, and introduces diverse reasons, among which we can find intelligibility functions (explanation, uncover dynamics, reveal complexity or simplicity), of sustaining a theory (discover new questions, highlight uncertainties, suggest analogies), of informing decision-making (real-time crisis solutions, finding optimization compromises), and of communication (educate the public, train practitioners). 
}{
Cette classification des fonctions se retrouve en filigrane dans les raisons de modéliser développées en dehors de toute typologie par~\cite{epstein2008model} : celui-ci insiste pour tordre le cou à l'idée préconçue que les modèles servent uniquement à la prédiction, et introduit diverses raisons, parmi lesquelles on retrouve des fonctions d'intelligibilité (explication, mise en évidence de dynamiques, révéler la complexité ou la simplicité), de soutien à la théorie (découverte de nouvelles questions, mettre en valeur des incertitudes, suggérer des analogies), d'aide à la décision (solutions de crise en temps réel, trouver des compromis d'optimisation), et de communication (éduquer le public, entrainer les praticiens).
}


\bpar{
Within this frame of functional classification of models, our work will mainly use the following functions:
}{
Dans ce cadre de classification fonctionnelle des modèles, notre travail utilisera principalement les fonctions suivantes :
}

\bpar{
\begin{itemize}
	\item Descriptive models and pattern extraction: these will be the diverse empirical analyses aiming at establishing stylized facts on co-evolution processes for given case studies.
	\item Models with an explanation and comprehension goal: models simulating territorial dynamics that we will construct, with the objective of integrating co-evolution processes, will have the principal goal of explaining stylized facts linked to some processes (for example: variations of a given parameter corresponding to a given process explain a given stylized fact), and ideally the \emph{comprehension} of systems\footnote{Indeed, the boundary between explanation and comprehension is fuzzy and subjective. It is possible to consider that there already exists a certain level of comprehension when a model with a certain level of internal and ontological consistence, in relation with reasonable and relatively autonomous theoretical assumptions, allows to draw conclusions on global dynamics of the system considered.}.
	\item Models to test a theory: internal validation, i.e. consistence of model behavior regarding stylized facts implied by the theory, and external, in the sense of a more or less performant reproduction of dynamics for case studies considered in the frame of a theory; or more generally to answer a precise question or assumption.
\end{itemize}
}{
\begin{itemize}
	\item Modèles descriptif et extraction de motifs : il s'agira des diverses analyses empiriques visant à établir des faits stylisés sur les processus de co-évolution dans des cas d'étude donnés.
	\item Modèles à visée explicative et de compréhension : les modèles simulant des dynamiques territoriales que nous construirons, avec comme objectif l'intégration des processus de co-évolution, auront pour objectif principal l'explication de faits stylisés en lien avec des processus (par exemple : les variations de tel paramètre correspondant à tel processus expliquent tel fait stylisé), et dans l'idéal la \emph{compréhension} des systèmes\footnote{En fait, la frontière entre explication et compréhension est floue et subjective. Il est possible de considérer qu'il existe déjà un certain niveau de compréhension lorsqu'un modèle avec un certain niveau de cohérence interne et ontologique, en lien avec des hypothèses théoriques raisonnables et relativement autonomes, permet de tirer des conclusions sur les dynamiques globales du système considéré.}.
	\item Modèles pour éprouver la théorie : validation interne, c'est-à-dire cohérence du comportement du modèle par rapport aux faits stylisés impliqués par la théorie, et externe, au sens de reproduction plus ou moins performante de dynamiques de cas d'études précis considérés dans le cadre d'une théorie ; ou plus généralement pour répondre à une question ou hypothèse précise.
\end{itemize}
}

\subsubsection{Generative modeling}{Modélisation générative}

\bpar{
The \emph{type}\footnote{In the functional perspective, structures, contents and processes, i.e. the nature of models in themselves (what corresponds to the nature and principles of models evoked but not classified by \noun{Varenne}), are given as illustrating examples, but a given function is not restricted to a given model (although reciprocally some models are not able to fulfill some functions). There does not exist to the best of our knowledge a general typology of models by \emph{type}, that we could then define in terms of a typology of relations with other knowledge domains (see~\ref{sec:knowledgeframework}): for example a model using a given methodology, privileging a given tool, a particular or privileged use of data, etc. In any case, existing typologies or classifications of models are associated to literature reviews and synthesis that are proper to each discipline: for example, \cite{harvey1969explanation} (p.~157) proposes a general typology which remains however inspired from and limited to geography. Conditions for interdisciplinary typologies are an open question, which exploration is largely out of reach of our work.} of models that we will mainly use in our work is related to \emph{generative modeling}, in the sense given by~\cite{epstein2006generative} in its manifest for \emph{generative social sciences}. The fundamental principle is to propose to explain macroscopic regularities as emerging from interactions between microscopic entities, by simulating the evolution of the system in a generative way\footnote{Keeping in mind that the ability to generate is of course a necessary but not sufficient component of explanation, as illustrate the debate on this subject around the works of \noun{Epstein} synthesized by~\cite{rey2015plateforme} (p.~154).}. This paradigm can be linked to the paradigm of \emph{Pattern Oriented Modeling} in Ecology~\cite{grimm2005pattern}, which aims at explaining through the bottom-up production of patterns\footnote{Indeed, POM aims at reproducing by the model through simulation, i.e. \emph{generates}, patterns expected at several scales, constituting a virtual laboratory in which assumptions can be tested. Furthermore, \noun{Epstein}'s generativity is based on similar paradigms for explanation, implying models with a progressive complexity and which allow the test of assumptions, by isolating mechanisms sufficient to reproduce macroscopic patterns.}. Agent-based models, i.e. models implying a certain number of heterogeneous agents that are relatively autonomous and simulating their interactions, are a way to achieve it.
}{
Le \emph{type}\footnote{Dans la perspective fonctionnelle, les structures, contenus et processus, c'est-à-dire la nature des modèles en eux-mêmes (ce qui correspond à la nature et aux principes des modèles évoqués mais non classifiés par \noun{Varenne}), sont donnés comme exemples en illustration, mais une fonction donnée n'est pas restreinte à un modèle donné (bien que réciproquement certains modèles ne puissent remplir certaines fonctions). Il n'existe à notre connaissance pas de typologie générale des modèles par \emph{type}, qu'on pourrait alors définir en termes d'une typologie des relations avec les autres domaines de connaissance (voir~\ref{sec:knowledgeframework}) : par exemple un modèle utilisant telle méthodologie, privilégiant tel outil, un usage particulier ou privilégié de données, etc. Dans tous les cas, les typologies ou classification existantes de modèles sont associées aux revues de littérature et synthèses propres à chaque discipline : par exemple, \cite{harvey1969explanation} (p.~157) propose une typologie générale qui reste toutefois inspirée de et limitée à la géographie. Les conditions de typologies interdisciplinaires sont une question ouverte, dont l'exploration dépasse largement le propos de notre travail.} de modèles que nous utiliserons majoritairement dans notre travail s'apparente à de la \emph{modélisation générative}, au sens donné par~\cite{epstein2006generative} dans son manifeste pour des \emph{sciences sociales génératives}. Le principe fondamental est de proposer d'expliquer des régularités macroscopiques comme émergentes des interactions entre entités microscopiques, en simulant l'évolution du système de manière générative\footnote{En gardant à l'esprit que la capacité à générer est bien sûr une composante nécessaire mais pas suffisante à l'explication, comme l'illustre le débat à ce sujet autour des travaux d'\noun{Epstein} synthétisé par~\cite{rey2015plateforme} (p.~154).}. Ce paradigme peut être rapproché de celui du \emph{Pattern Oriented Modeling} en Ecologie~\cite{grimm2005pattern}, qui vise à expliquer par production de motifs par le bas\footnote{En effet, le POM vise à ce que le modèle reproduise par la simulation, c'est-à-dire \emph{génère}, des motifs (\emph{patterns}) attendus à différentes échelles, constituant un laboratoire virtuel dans lequel des hypothèses peuvent être testées. Par ailleurs, la générativité d'\noun{Epstein} se base sur des paradigmes similaires pour l'explication, impliquant des modèles à la complexité progressive et qui permettent le test d'hypothèses, en isolant des mécanismes suffisant pour reproduire des motifs macroscopiques.}. Les modèles basés-agents, c'est-à-dire des modèles impliquant un certain nombre d'agents hétérogènes relativement autonomes et simulant leur interactions, sont une façon d'y parvenir.
}

\bpar{
The use of generative modeling can be strongly linked to the notion of weak emergence introduced by~\cite{bedau2002downward}\footnote{We recall that weak emergence corresponds to the emergence of properties at an upper level that must effectively be computed by the system to be known.}. A system exhibiting emergent properties in the weak sense assumes that properties of the upper level (macro) must entirely be derived through simulation, while remaining reducible on the causal and ontological aspects. In other terms, the macro level does not posses irreducible causal powers, this being not incompatible with the existence of \emph{downward causation} and its autonomy. Some systems\footnote{As show conscience in neuroscience and psychology, or debates on the existence and autonomy of ``societal beings'' in sociology~\cite{angeletti2015etres}, for which we could only know a part of causal microscopic elements, namely individuals.} do not fall within this category in the current state of our knowledge since we are not able to exhibit microscopic causal elements. On the contrary, systems that we do not understand well but which simulate themselves and for which we are certain that the macro state emerges from microscopic interactions (as traffic and congestion for example), are perfect illustrations of this notion. The examples given by \noun{Bedau} to illustrate his demonstration are two dimensional cellular automatons, for which the role of computation is obvious and \emph{downward causation} can be illustrated by the behavior of macroscopic structures of the game of life which retroactively act on cells. Knowing the dynamics of weakly emergent systems by definition necessitates to simulate them, and thus to model them\footnote{On the difference between simulation of a model and model of simulation, \cite{phan2010agent} explains to what extent these two notions can be distinguished, but that it does not imply a fundamental difference for the concrete application: the simulation of a model consists in the operation of computing the successive states of a model in a given configuration, whereas a simulation model is a model conceived for the simulation of a system (for example a generative model) or the simulation of an other model (for example the numerical schemes to approximate equations). In all cases, the use of the model will imply simulation of a model. These remarks are verified in particular for the case of generative modeling. We will use the two without distinction in the following.}, this approach is thus natural to understand the structure or processes in a complex system.
}{
L'utilisation de modélisation générative peut être mise en correspondance forte avec la notion d'émergence faible introduite par~\cite{bedau2002downward}\footnote{On rappelle que l'émergence faible correspond à l'émergence de propriété à un niveau supérieur qui doivent être effectivement computées par le système pour être connues.}. Un système qui présente des propriétés émergentes au sens faible suppose que les propriétés du niveau supérieur (macro) doivent être entièrement dérivées par simulation, tout en restant réductible sur les plans causaux et ontologiques. En d'autre termes, le niveau macro ne possède pas de pouvoirs causaux irréductibles, cela n'étant pas incompatible avec l'existence de \emph{downward causation} et son autonomie. Certains systèmes\footnote{Comme le montrent la conscience en neuroscience et psychologie, ou les débats sur l'existence et l'autonomie ``d'êtres sociétaux'' en sociologie~\cite{angeletti2015etres}, pour lesquels nous pourrions ne connaitre que partie seulement des éléments causaux microscopiques à savoir les individus.} ne tombent pas dans cette catégorie à notre connaissance actuelle puisqu'on n'est pas capable de désigner des éléments microscopiques causaux. En revanche, des systèmes qu'on comprend mal mais qui se simulent eux-mêmes et dont on est certain que l'état macro émerge des interactions microscopiques (prenons le trafic et la congestion par exemple), sont des parfaites illustrations de cette notion. Les exemples donnés par \noun{Bedau} pour démontrer son propos sont des automates cellulaires en deux dimensions, pour lesquels le rôle de la computation est évident et la \emph{downward causation} peut être illustrée par le comportement des structures macroscopiques du jeu de la vie qui agissent rétroactivement sur les cellules. Connaitre les dynamiques de systèmes faiblement émergents nécessite par définition de les simuler, et donc de les modéliser\footnote{Sur la différence entre simulation de modèle et modèle de simulation, \cite{phan2010agent} explique dans quelle mesure ces deux notions peuvent être distinguées, mais que cela n'implique pas de différence fondamentale pour l'application concrète : la simulation d'un modèle consiste en l'opération de computation des états successifs d'un modèle dans une configuration donnée, tandis qu'un modèle de simulation est un modèle conçu pour la simulation d'un système (par exemple un modèle génératif) ou la simulation d'un autre modèle (par exemple les schémas numériques pour approcher des équations). Dans tous les cas, l'utilisation du modèle impliquera simulation d'un modèle. Ces remarques se vérifient particulièrement dans le cas de la modélisation générative. Nous utiliserons les deux de manière interchangeable par la suite.}, cette approche est ainsi naturelle pour connaître la structure ou les processus dans un système complexe.
}

\subsubsection{The model as a tool for indirect knowledge}{Le modèle comme outil de connaissance indirecte}

\bpar{
Thus, our models will principally be with a comprehension function (even if they do not reach their objective and remain at the level of an explanation). We will proceed in some cases to refined calibration on observed data, but these will never have the objective to predict. These calibrations will aim at extrapolating parameters and learn indirectly on modeled processed, and the model is thus indeed a tool of \emph{indirect knowledge}.
}{
Ainsi, nos modèles seront principalement à visée de comprehension (même s'ils n'atteignent pas l'objectif et restent au niveau d'une explication). Nous procéderons dans certains cas à des calibrations fines sur données observées, mais celles-ci n'auront à aucun moment l'objectif de prédiction. Ces calibrations serviront à extrapoler des paramètres et apprendre indirectement sur les processus modélisés, et le modèle est ainsi bien un instrument de \emph{connaissance indirecte}.
}

\bpar{
This knowledge of processes is allowed by the use of simulation as a virtual laboratory allowing the test of assumptions formulated from a theory or obtained from empirical stylized facts: this is exactly such a paradigm that is constructed by \cite{pumain2017urban}, which insist on (i) the need for parsimony in models; (ii) the need for multiple models (multi-modeling); and (iii) the role of extensive model exploration, to achieve it without falling into the trap of equifinality\footnote{Equifinality corresponds to the possibility for a system to reach a point in its phase space through different trajectories, i.e. in our case macroscopic patterns which can be generated by different microscopic processes. This concept was already formulated in the general systems theory~\cite{von1972history}. It challenges the notions of causality, et invalidates explanations of ``direct'' causality at the macroscopic level - we will come back to it more particularly in \ref{sec:causalityregimes}.}. Thus, the computation of \emph{Calibration Profiles} of the SimpopLocal model~\cite{reuillon2015} allow to establish necessary and sufficient conditions to reproduce a given pattern, and thus for example to indirectly declare a process necessary or not to produce a stylized fact.
}{
Cette connaissance des processus est permise par l'utilisation de la simulation comme un laboratoire virtuel permettant le test d'hypothèses formulées à partir d'une théorie ou issues de faits stylisés empiriques : c'est exactement ce type de paradigme que construit \cite{pumain2017urban}, qui insiste sur (i) le besoin de parcimonie dans les modèles ; (ii) le besoin de multiples modèles (multi-modélisation) ; et (iii) le rôle de l'exploration extensive des modèles, pour y parvenir sans tomber dans le piège de l'équifinalité\footnote{L'équifinalité correspond à la possibilité pour un système d'atteindre un point de son espace des phases par des trajectoires différentes, c'est-à-dire dans notre cas des motifs macroscopiques pouvant être générés par différents processus microscopiques. Ce concept était déjà formulé dans la théorie générale des systèmes~\cite{von1972history}. Il pose problème aux notions de causalité, et remet en cause des explications de causalité ``directe'' au niveau macroscopique - nous y reviendrons plus particulièrement en \ref{sec:causalityregimes}.}. Ainsi, l'établissement des \emph{Calibration Profiles} du modèle SimpopLocal~\cite{reuillon2015} permettent d'établir des conditions nécessaires et suffisantes pour reproduire un motif donné, et donc par exemple de déclarer indirectement un processus nécessaire ou non pour produire un fait stylisé.
}

\bpar{
Thus, we will here follow this approach of using models (of simulation principally), while keeping in mind that it does only partly answer to the fundamental challenges of urban modeling described by~\cite{perez2016agent}, in particular the capture of complexity and of multi-dimensionality of urban systems and also the possibility to generate future feasible scenarios (what is different from prediction), but not the issue of urban planning models, that could for example be participative and imply \emph{stakeholders}\footnote{The role of the applicative aim of models is linked first to a disciplinary sensitivity, such as the field of Luti models~\cite{wegener2004land} which is much more applicative than the one of theoretical and quantitative geography, but also to a ``cultural'' sensitivity, as illustrates~\cite{batty2013new} which shows an anglo-saxon branch of geography much closer to concrete applications.}.
}{
Ainsi, nous prendrons ici ce parti de l'utilisation des modèles (de simulation principalement), tout en gardant à l'esprit que celui-ci ne répond que partiellement aux challenges fondamentaux de la modélisation urbaine donnés par~\cite{perez2016agent}, notamment la capture de la complexité et de la multidimensionalité des systèmes urbains ainsi que la possibilité de générer des scenarii futurs possibles (ce qui est différent de la prédiction), mais pas la question des modèles de planification urbaine, pouvant par exemple être participatifs et impliquant les \emph{stakeholders}\footnote{Le rôle de la visée d'application des modèles est lié à la fois à une sensibilité disciplinaire, comme le domaine des Luti~\cite{wegener2004land} qui l'est bien plus que celui de la géographie théorique et quantitative, mais aussi à une sensibilité ``culturelle'', comme l'illustre~\cite{batty2013new} qui montre une branche de la géographie anglo-saxonne plus proche des applications concrètes.}.
}




\subsubsection{How to explore a model of simulation}{Comment explorer un modèle de simulation}

\bpar{
In order to maximally avoid the ``bricolage'' regarding all stages of a model genesis, from its specification, its conception, its use to its exploration, described by~\cite{bonhomme2017dictionnaire}, we propose to fix a protocol for the parts implying model exploration. More generally, there exists generic protocols such as the one introduced by~\cite{grimm2014towards} to accompany the full modeling approach. We consider the exploration stage and go into it with more details. We are in the context fixed above of a simulation model, mostly with a comprehension function.
}{
Afin d'éviter au maximum le ``bricolage'' concernant l'ensemble des étapes de la genèse d'un modèle, de sa spécification, sa conception, son utilisation à son exploration, décrit par~\cite{bonhomme2017dictionnaire}, nous proposons de nous fixer un protocole pour la partie d'exploration des modèles. Plus généralement, il existe des protocoles généraux comme celui introduit par~\cite{grimm2014towards} pour accompagner l'ensemble de la démarche de modélisation. Nous considérons l'étape d'exploration et creusons celle-ci plus en détails. Nous nous plaçons dans le cadre fixé ci-dessus d'un modèle de simulation, majoritairement à visée de compréhension.
}

\bpar{
The simplified protocol is directly obtained from the philosophy and structure of OpenMole. We can refer for example to~\cite{reuillon2013openmole} for the fundamental principles, the online documentation\footnote{Available at \url{https://next.openmole.org/Models.html}.} for a broad overview of available methods and their articulation within a standard frame, and \cite{pumain2017innovative} for a contextualization of the different methods. These works\footnote{Most have been realized in the interdisciplinary context of the Geodivercity ERC.} brought a considerable number of innovations simultaneously methodological, technical, thematic and theoretical. The OpenMole philosophy is articulated around three axes (see interview with \noun{R. Reuillon}, Appendix~\ref{app:sec:interviews}): the model as a ``black box'' to be explored (i.e. methods are independent of the model), use of advanced exploration methods, transparent access to high performance computing environments. These different components are in strong interdependency, and allow a paradigm shift in the use of simulation models: use of multi-modeling, i.e. variable structure of the model~\cite{cottineau2015modular}, change in the nature of questions asked to the model (e.g. complete determination of the feasible space~\cite{10.1371/journal.pone.0138212}), all this being allowed by the use of high performance computing~\cite{schmitt2014half}.
}{
Le protocole simplifié est issu directement de la philosophie et de la structure d'OpenMole. On peut se référer par exemple à~\cite{reuillon2013openmole} pour les principes fondamentaux, la documentation en ligne\footnote{Disponible à \url{https://next.openmole.org/Models.html}.} pour un aperçu global des méthodes disponibles et de leur articulation dans un cadre standard, et \cite{pumain2017innovative} pour une contextualisation des différentes méthodes. Ces travaux\footnote{La majorité ayant été réalisés dans le cadre interdisciplinaire de l'ERC Geodivercity.} ont apporté un nombre considérable d'innovations à la fois méthodologiques, techniques, thématiques et théoriques. La philosophie d'OpenMole s'articule autour de trois axes (voir entretien avec \noun{R. Reuillon}, Annexe~\ref{app:sec:interviews}) : le modèle comme ``boite noire'' à explorer (i.e. méthodes indépendantes du modèle), utilisation de méthodes avancées d'exploration, accès transparent aux environnements de calcul intensif. Ces différentes composantes sont en interdépendance forte, et permettent un changement de paradigme dans l'utilisation des modèles de simulation : utilisation de multi-modélisation, c'est-à-dire structure variable du modèle~\cite{cottineau2015modular}, changement de la nature des questions posées au modèle (par exemple détermination complète de l'espace faisable~\cite{10.1371/journal.pone.0138212}), tout cela permis par l'utilisation du calcul intensif~\cite{schmitt2014half}.
}

\bpar{
We consider a simulation model as an algorithm producing outputs from data and parameters as inputs. In this context, we propose in an ideal case all the following stages which should be necessary for a robust use of simulation models.
}{
Nous considérons un modèle de simulation comme un algorithme produisant des sorties à partir de données et de paramètres en entrée. Dans ce cadre, nous proposons dans un cas idéal l'ensemble des étapes suivantes qui devraient être nécessaire pour une utilisation robuste des modèles de simulation.
}

\bpar{
\begin{enumerate}
	\item Identification of main mechanisms and associated crucial parameters, possibly meta-parameters (here understood as a parameter generating the initial configuration of the model), and also their thematic domain; identification of indicators to evaluate the performance or the behavior of the model.
	\item Evaluation of stochastic variations: large number of repetitions of a reasonable number of parameters, establishment of the number of repetitions necessary to reach a certain level of statistical convergence.
	\item Evaluation of the sensitivity to meta-parameters, following the innovative methodology developed in the following\footnote{An example of this methodology consisting in the generation of synthetic data will be used in the remainder of this section; a formal description of the method is given in~\ref{app:sec:syntheticdata} and an other example of application in~\ref{app:sec:syntheticdata-finance}.}.
	\item Brutal exploration for a first sensitivity analysis, if possible statistical evaluation of the relations between parameters and outputs indicators.
	\item Calibration, targeted algorithmic exploration with the use of specific algorithms (\emph{Calibration Profile}, \emph{Pattern Space Exploration})\footnote{We will mostly not practice this last point, finding already enough responses to our questions with the previous points.}.
	\item Feedback on the model, extension and new bricks of multi-modeling, feedback on the stylized facts and the theory.
\end{enumerate}
}{
\begin{enumerate}
	\item Identification des mécanismes principaux et des paramètres cruciaux associés, possiblement des méta-paramètres (ici compris comme paramètre générant la configuration initiale du modèle), ansi que de leur domaine thématique ; identification des indicateurs pour évaluer la performance ou le comportement du modèle.
	\item Évaluation des variations stochastiques : grand nombre de répétitions pour un nombre raisonnable de paramètres, établissement du nombre de répétitions nécessaire pour atteindre un certain niveau de convergence statistique.
	\item Évaluation de la sensibilité aux meta-paramètres, suivant la méthodologie innovante développée par la suite\footnote{Un exemple de cette méthodologie consistant à la génération de données synthétiques sera utilisé dans la suite de cette section ; une description formelle de la méthode est donnée en~\ref{app:sec:syntheticdata} et un autre exemple d'application en~\ref{app:sec:syntheticdata-finance}.}.
	\item Exploration brutale pour une première analyse de sensibilité, si possible evaluation statistique des relations entre paramètres et indicateurs de sortie.
	\item Calibration, exploration algorithmique ciblée par l'utilisation d'algorithmes spécifiques (\emph{Calibration Profile}, \emph{Pattern Space Exploration})\footnote{Nous ne pratiquerons quasiment pas ce dernier point, trouvant suffisamment de réponses à nos questions avec les points précédents.}.
	\item Retours sur le modèle, extension et nouvelles briques de multi-modélisation, retours sur les faits stylisés et la théorie.
\end{enumerate}
}

\bpar{
In the corresponding case, some steps have no reason to be, for example the evaluation of stochasticity in the case of a deterministic model. Similarly, steps will take more or less importance depending on the nature of the question asked: calibration will not be relevant in the vase of fully synthetic models, whereas a systematic exploration of a large number of parameters will not systematically be necessary in the case of a model which aims at be calibrating on data. 
}{
Le cas échéant, certaines étapes n'ont pas lieu d'être, par exemple l'évaluation de la stochasticité dans le cas d'un modèle déterministe. De même, les étapes prendront plus ou moins d'importance selon la nature de la question posée : la calibration ne sera pas pertinente dans le cas de modèles complètement synthétiques, tandis qu'une exploration systématique d'un grand nombre de paramètres ne sera pas forcément nécessaire dans le cas d'un modèle qui a pour but d'être calibré sur des données.
}

\subsubsection{Link between modeling and open science}{Lien entre modélisation et science ouverte}

\bpar{
Finally, it is important to briefly highlight the links between modeling practices and open science, in parallel of the link between reproductibility and open science that we will do at the end of~\ref{sec:reproducibility}. In fact, open science is composed by several practices declined on different domains, thus this logical distribution in our positioning. To illustrate the issues, we propose to describe the example of model exploration workflows as a method for meta-sensitivity analysis, i.e. an aspect of the methodology applied below.
}{
Enfin, il est important de souligner brièvement les liens entre pratiques de modélisation et science ouverte, en parallèle du lien entre reproductibilité et science ouverte que nous ferons à la fin de~\ref{sec:reproducibility}. En fait, la science ouverte est composée d'un ensemble de pratiques se déclinant sur différents domaines, d'où sa ventilation logique dans nos positionnements. Pour illustrer les enjeux, nous proposons de décrire l'exemple des workflows d'exploration de modèle comme une méthode de méta-analyse de sensibilité, c'est-à-dire un aspect de la méthodologie appliquée ci-dessous.
}

\bpar{
The ideas of multi-modeling and extensive model exploration are nothing from new as \noun{Openshaw} already advocated for ``model-crunching'' in~\cite{openshaw1983data}, but their effective use only begins to emerge thanks to the apparition of new methods and tools together with an explosion of computation capabilities: \cite{cottineau2016back} claims for a renewed approach on multi-modeling. Coupling models as we do answers to similar questions. In that stream of research, the model exploration platform OpenMole~\cite{reuillon2013openmole} allows to embed any model as a blackbox, to write modulable exploration workflow using advanced methodologies such as genetic algorithms and to distribute transparently the computation on large scale computation infrastructures such as clusters or computation grids. In the case developed below, the workflow tool is a powerful way to embed both the sensitivity analysis and the meta-sensitivity analysis, and allow to couple any generator with any model in a straightforward way as soon as the model can be parametrized on its spatial initial configuration, given meta-parameters or an entire configuration.
}{
Les idées de multi-modélisation et d'exploration intensive de modèle sont tout sauf nouvelles puisque \noun{Openshaw} défendait déjà le ``model-crunching'' dans~\cite{openshaw1983data}, mais leur utilisation effective commence seulement à émerger grâce à l'apparition de nouvelles méthodes et outils en même temps qu'une explosion des capacités de calcul : \cite{cottineau2016back} propose une approche renouvelée de la multi-modélisation. Le couplage de modèles tel que nous l'opérons répond à des questions similaires. Dans cette lignée de recherche, la plateforme d'exploration de modèle OpenMole~\cite{reuillon2013openmole} permet d'embarquer n'importe quel modèle comme une boîte noire, d'écrire des workflow d'exploration modulables qui utilisent des méthodologies d'exploration avancées comme des algorithmes génétiques, et de distribuer de manière transparente les calculs sur des infrastructures de calcul à grande échelle comme des clusters ou grilles de calcul. Dans le cas développé par la suite, l'outil du workflow est un outil puissant pour intégrer à la fois l'analyse de sensibilité et la méta-analyse de sensibilité, et permet de coupler n'importe quel générateur avec n'importe quel modèle de façon très directe, à la condition minimale que le modèle puisse être paramétré sur sa configuration spatiale initiale, par la donnée de méta-paramètres ou d'une configuration entière.
}

\bpar{
Furthermore, an idea of workflows is to favor open and collaborative constructions, since OpenMole's ``marketplace'', directly integrated to the software\footnote{otherwise accessible at \url{https://github.com/openmole/openmole-market}}, allows to directly benefit from examples which have been shared on the collaborative repository. This is similar to model sharing platforms, which are numerous for agent-based models for example, but in an even more modular and participative spirit. Thus, some epistemological and methodological choices regarding modeling imply directly a positioning regarding open science: multi-modeling and model families, which go together with the coupling of heterogenous and multi-scalar models, can difficultly be reliable without opening, sharing and collaborative construction of models practices, as recalls~\cite{banos2013pour}.
}{
D'autre part, une idée des workflow est de favoriser des constructions ouvertes et collaboratives, puisque le ``marketplace'' d'OpenMole, directement intégré au logiciel\footnote{autrement accessible à \url{https://github.com/openmole/openmole-market}}, permet de bénéficier directement des exemples qui auront été partagés sur le dépôt collaboratif. Cela ressemble aux plateformes de partage de modèles, qui sont nombreuses pour les modèles agents par exemple, mais dans un esprit encore plus modulaire et participatif. Ainsi, certains choix épistémologiques et méthodologiques au regard de la modélisation impliquent directement un positionnement au regard de la science ouverte : la multi-modélisation et les familles de modèles, qui vont de pair avec le couplage de modèles hétérogènes et multi-échelles, ne peuvent guère être viables sans des pratiques d'ouverture, de partage et de construction collaborative des modèles, comme le rappelle~\cite{banos2013pour}.
}

\bpar{
Finally, one of the faces of the construction of open knowledge is pedagogy. \cite{chen2006effectiveness} proposes simulation as a tool to teach engineering students the processes underlying the systems that they will be brought to design and manage. This aspect is also to keep in mind for its performative character: models have then a retroaction on real situations, what complexifies even more the system considered.
}{
Enfin, l'un des visages de la construction de connaissances ouvertes est la pédagogie. \cite{chen2006effectiveness} propose la simulation comme outil pour apprendre aux élèves ingénieurs les processus sous-jacents aux systèmes qu'ils seront amenés à concevoir et gérer. Cet aspect est également à garder en tête de par son caractère performatif : les modèles ont alors une retroaction sur les situations réelles, ce qui complexifie encore le système considéré.
}


\subsubsection{Synthesis}{Synthèse}

\bpar{
We can briefly synthesize the ideas to keep in mind after this quick overview of crucial issues linked to modeling.
\begin{enumerate}
	\item Models can have a large number of functions \cite{varenne2017theories}, among which we will mostly use: information and patterns extraction, explanation and comprehension, verification and construction of theories.
	\item We will mostly be positioned within the paradigm of \emph{generative modeling}, with an aim of parsimony and multiple models with appropriated extensive exploration protocols~\cite{pumain2017urban}.
	\item This way to model both assumes and participates to an approach of open science~\cite{fecher2014open}.
\end{enumerate}
}{
Résumons brièvement les idées à garder à l'esprit à la suite de ce survol rapide d'enjeux cruciaux liés à la modélisation.
\begin{enumerate}
	\item Les modèles peuvent avoir un grand nombre de fonctions \cite{varenne2017theories}, parmi lesquelles nous utiliserons fondamentalement : extraction d'information et de motifs, explication et compréhension, vérification et construction des théories.
	\item Nous nous placerons majoritairement dans le paradigme de la \emph{modélisation générative}, dans un souci de parcimonie et de modèles multiples avec des protocoles d'exploration extensive appropriés~\cite{pumain2017urban}.
	\item Cette façon de modéliser à la fois suppose et participe à une démarche de science ouverte~\cite{fecher2014open}.
\end{enumerate}
}

\bpar{
In this context, we propose to now develop some issues particularly important for our question in a more precise way.
}{
Dans ce contexte, nous proposons de développer à présent certains enjeux particulièrement importants pour notre question de manière plus précise.
}


\subsection{For a cautious use of big data and computation}{Pour un usage raisonné des données massives et de la computation}

\bpar{
The so-called \emph{big data revolution} resides as much in the availability of large datasets of novel and various types as in the always increasing available computational power. Although the \emph{computational shift} (\cite{arthur2015complexity}) is central for a science aware of complexity and is undeniably the basis of future modeling practices in geography as \cite{banos2013pour} points out, we argue that both \emph{data deluge} and \emph{computational potentialities} are dangerous if not framed into a proper theoretical and formal framework. The first may bias research directions towards available datasets with the risk to disconnect from a theoretical background, whereas the second may overshadow preliminaries analytical resolutions essential for a consistent use of simulations. We argue that the conditions for most of results in this thesis are indeed the ones endangered by incautious big-data enthusiasm, concluding that a main challenge for future geocomputation is a wise integration of novel practices within the existing body of knowledge.
}{
La \emph{révolution des données massives} réside autant dans la disponibilité de grands jeux de données de nouveaux types variés, que dans la puissance de calcul potentielle toujours en augmentation. Même si le \emph{tournant computationnel} (\cite{arthur2015complexity}) est central pour une science consciente de la complexité et est sans douter la base des pratiques de modélisation futures en géographie comme \cite{banos2013pour} souligne, nous soutenons que à la fois le \emph{déluge de données} et les \emph{capacités de calcul} sont dangereuses si non cadrées dans un cadre théorique et formel propre. Le premier peut biaiser les directions de recherche vers les jeux de données disponibles avec le risque de se déconnecter d'un fond théorique, tandis que le second peut occulter des résolutions analytiques préliminaires essentielles pour un usage cohérent des simulations. Nous avançons que les conditions pour la majorité des résultats dans cette thèse sont en effet ceux mis en danger par un enthousiasme inconsidéré pour les données massives, tirant la conclusion qu'un challenge majeur pour la géocomputation future est une intégration sage des nouvelles pratiques au sein du corpus existant de connaissances.
}

\subsubsection{Increase in computing power}{Accroissement de la puissance de calcul}

\bpar{
The computational power available seems to follow an exponential trend, as some kind of Moore's law. Both effective Moore's law for hardware, and improvement of softwares and algorithms, combined with a democratization of access to large scale simulation facilities, makes always more and more CPU time available for the social scientist (and to the scientist in general but this shift happened quite before in other fields). About ten years ago, \cite{gleyze2005vulnerabilite} concluded that network analysis, for the case of Parisian public transportation network, was ``limited by computation''. Today most of these analyses would be quickly done on a personal computer with appropriated software and coding: \cite{2015arXiv151201268L} is a witness of such a progress, introducing new indicators with a higher computational complexity, computed on larger networks. The same parallel can be done for the Simpop models: the first Simpop models at the beginning of the millenium~\cite{sanders1997simpop} were ``calibrated'' by hand, whereas \cite{cottineau2015modular} calibrates the multi-modeling Marius model and~\cite{schmitt2014half} calibrates very precisely the SimpopLocal model, both on grid with billions of simulations. A last example, the field of Space Syntax, witnessed a long path and tremendous progresses from its theoretical origins~\cite{hillier1989social} to recent large-scale applications~\cite{hillier2016fourth}.
}{
La puissance de calcul disponible semble suivre une tendance exponentielle, comme une sorte de loi de Moore. Grace à d'une part la loi de Moore effective pour le matériel, d'autre part l'amélioration des logiciels et algorithmes, conjointement avec une démocratisation de l'accès au infrastructures de simulation à grande échelle, permet à toujours plus de temps processeur d'être disponible pour le chercheur en sciences sociales (et pour le scientifique en général, mais cette mutation a déjà été opérée depuis plus longtemps dans d'autres domaines). Il y a environ une dizaine d'années, \cite{gleyze2005vulnerabilite} était forcé de conclure que les analyses de réseau, pour les transports publics parisiens, étaient ``limitées par le calcul''. Aujourd'hui la plupart des mêmes analyses seraient rapidement réglée sur un ordinateur personnel avec les logiciels et programmes appropriés : \cite{2015arXiv151201268L} est un témoin d'un tel progrès, introduisant des nouveaux indicateurs avec une plus grande complexité de calcul, qui sont calculés sur des réseaux à grande échelle. Le même parallèle peut être fait pour les modèles Simpop : les premiers modèles Simpop au début du millénaire~\cite{sanders1997simpop} étaient ``calibrés'' à la main, tandis que \cite{cottineau2015modular} calibre le modèle Marius en multi-modélisation et~\cite{schmitt2014half} calibre très précisément le modèle SimpopLocal, chacun sur la grille avec des milliards de simulations. Un dernier exemple, le champ de la \emph{Space Syntax}, a témoigné d'une longue route et de progrès considérables depuis ses origines théoriques~\cite{hillier1989social} jusqu'à ses récentes applications à grande échelle~\cite{hillier2016fourth}.
}

\subsubsection{A data deluge?}{Un déluge de données ?}

\bpar{
Concerning the new and ``big'' data available, it is clear that always larger dataset are available and always newer type of data are available. Numerous examples of fields of application can be given. For example, mobility can now be studied from various entries, such as new data from smart transportation systems~\cite{o2014mining}, from social networks~\cite{frank2014constructing}, or other more exotic data such as mobile phone data~\cite{de2016death}. In an other spirit, the opening of ``classic'' datasets (such as city dashboards, open data government initiatives) should allow ever more meta-analyses. New ways to do research and produce data are also raising, towards more interactive and crowd-sourced initiatives. For example, \cite{10.1371/journal.pone.0183919} describes a web-application aimed at presenting a meta-analysis of Zipf's law across numerous datasets, but in particular features an upload option, where the user can upload its own dataset and add it to the meta-analysis. Other applications allow interactive exploration of scientific literature for a better knowledge of a complex scientific landscape, as~\cite{cybergeo20} does.
}{
Concernant les nouvelles données ``massives'' qui sont disponibles, il est clair que des quantités toujours plus grandes et des types toujours nouveaux sont disponibles. De nombreux exemples de champs d'application peuvent être donnés. La mobilité en est typique, puisque étudiée selon divers points de vue, comme les nouvelles données issues des systèmes de transport intelligents~\cite{o2014mining}, des réseaux sociaux~\cite{frank2014constructing}, ou des données plus exotiques comme des données de téléphonie mobile~\cite{de2016death}. Dans un autre esprit, l'ouverture de jeux de données ``classiques'' (comme les applications synthétiques urbaines, les initiatives gouvernementales pour les données ouvertes) devrait permettre toujours plus de méta-analyses. De nouvelles façons de pratiquer la recherche et produire des données sont également en train d'émerger, vers des initiatives plus interactives et venant de l'utilisateur. Ainsi, \cite{10.1371/journal.pone.0183919} décrit une application web ayant pour but de présenter une méta-analyse de la loi de Zipf sur de nombreux jeux de données, mais en particulier inclut une option de dépôt, à travers laquelle l'utilisateur peut télécharger son propre jeu de données et l'inclure dans la méta-analyse. D'autres applications permettent l'exploration interactive de la littérature scientifique pour une meilleure connaissance d'un horizon scientifique complexe, comme~\cite{cybergeo20} fait.
}

\subsubsection{On induced dangers}{Des dangers induits}

\bpar{
As always the picture is naturally not as bright as it seems to be at first sight, and the green grass that we try to go eating in the neighbor's field quickly turns into a sad reality. Indeed, the purpose and motivation are fuzzy and one can get lost. Some examples speak for themselves. 
}{
Comme toujours la situation n'est naturellement pas aussi idyllique qu'elle semble être au premier abord, et l'herbe verte du pré du voisin que nous pouvons être tentés d'aller brouter se transforme rapidement en un triste fumier. En effet, les objectifs et motivations d'un grand nombre d'approches restent flous et on peut facilement s'y perdre. Des illustrations parleront d'elles-mêmes.
}

\bpar{
\cite{barthelemy2013self} introduces a new dataset and rather new methods to quantify road network evolution, but the results, on which the authors seem to be astonished, are that a transition occurred in Paris at the Haussmann period. Any historian of urbanism would be puzzled by the exact purpose of the paper, as in the end a vague and bizarre feeling of reinventing the wheel floats in the air. The use of computation can also be exaggerated, and in the case of agent-based modeling it can be illustrated by the example of~\cite{axtell2016120}, for which the aim at simulating the system at scale 1:1 seems to be far from initial motivations and justifications for agent-based modeling, and may even give arguments to mainstream economists who easily detract ABMS.
}{
\cite{barthelemy2013self} introduit un nouveau jeu de données et des méthodes relativement nouvelles pour quantifier l'évolution du réseau de rues, mais les résultats, sur lesquels les auteurs semblent s'étonner, sont qu'une transition a eu lieu à Paris à l'époque d'Haussmann. Tout historien de l'urbanisme s'interrogerait sur le but exact de l'étude, puisqu'il reste à la fin un sentiment de réinvention de la roue. L'utilisation des ressources de calcul peut également être exagéré, et dans le cas de la modélisation multi-agents, on peut citer~\cite{axtell2016120}, pour lequel l'objectif de simuler le système à l'échelle 1:1 semble être loin des motivations et justifications originelles de la modélisation multi-agents, et pourrait même donner des arguments aux économistes \emph{mainstream} qui dénigrent facilement les ABMS.
}

\bpar{
Other anecdotes raise worries: there exist online curious examples, such as a web application\footnote{Voir \url{http://shiny.parisgeo.cnrs.fr/gibratsim/}.} that uses computational ressources to simulate Gaussian distributions for a Gibrat model in order to compute their mean and variance, that are input parameters of the model. It basically checks the Central Limit Theorem. Otherwise, the full distribution given by a Gibrat model is theoretically known as it was fully solved e.g. by \cite{gabaix1999zipf}. 
}{
D'autres anecdotes peuvent inquiéter :  il existe en ligne des exemples étonnants, comme une application web\footnote{Voir \url{http://shiny.parisgeo.cnrs.fr/gibratsim/}.} qui utilise des ressources de calcul pour simuler des distributions Gaussiennes afin de calculer pour un modèle de Gibrat les moyenne et variance, qui sont des paramètres d'entrée du modèle. En résumé, cela revient à vérifier le Théorème de la Limite Centrale. D'autre part, la distribution complète donnée par un modèle de Gibrat est entièrement connue théoriquement comme résolu e.g. par~\cite{gabaix1999zipf}.
}

\bpar{
On this point, we must partly disagree with the ninth commandment of \noun{Banos}, which recalls that ``mathematics are not the universal language of models'', or more precisely highlight the dangers of a misinterpretation of that principle\footnote{Generally, the \noun{Banos} commandments appear simple in their formulation, but are of a baffling depth and complexity when one tries to extract the implications and the global underlying philosophy, and must never be taken lightly.}: it postulates that alternative means to mathematics exist to help understand processes or methods, but insists that these are only an entry point and never pretends that it is possible to get rid of mathematics, drift that the previous example perfectly illustrates. Furthermore, it is possible to exhibit very simple mathematic structures, such as a simplex in any dimension, for which a ``simple'' visualization is an open problem.
}{
Sur ce point, nous devons partiellement être en désaccord avec le neuvième commandement de \noun{Banos}, qui rappelle que ``les mathématiques ne sont pas le language universel des modèles'', ou plutôt souligner les dangers d'une mauvaise interprétation de ce principe\footnote{De manière générale, les commandements de \noun{Banos} paraissent simples dans leur formulation, mais sont d'une profondeur et d'une complexité déconcertante lorsqu'on essaye d'en tirer les implications et la philosophie globale sous-jacente, et ne doivent jamais être pris à la légère.} : il postule que des moyens alternatifs aux mathématiques existent pour faire comprendre des processus ou des méthodes, mais précise que ceux-ci sont une porte d'entrée et ne prétend jamais qu'il est possible de se passer des mathématiques, dérive que l'exemple précédent illustre parfaitement. D'ailleurs, il est possible d'exhiber des structures mathématiques très simples, comme un simplexe en dimension quelconque, dont la visualisation ``simple'' est un problème ouvert.
}

\begin{figure}
\includegraphics[width=\linewidth]{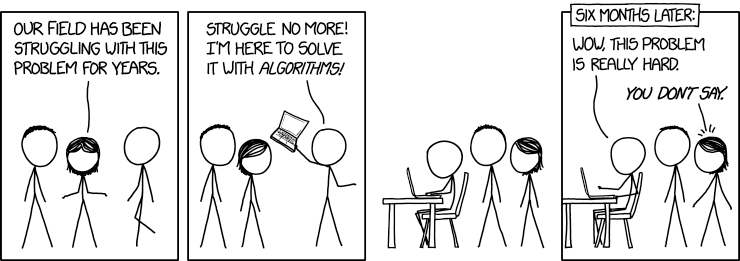}
\caption[Naive use of data mining]{On naive use of data mining and intensive computation. Source: \texttt{xkcd}\label{fig:computation:xkcd}}
\end{figure}

\bpar{
Data also provide their collections of misunderstandings. Recently on the French speaking diffusion list \emph{Geotamtam}, a sudden rush around \emph{Pokemon Go} data seemed to answer more to an urgent unexplained need to exploit this new data source before anyone else rather than an elaborated theoretical construction. Simple existing accurate datasets, such as historical cities population (for France the Pumain-INED database for example), are far from being fully exploited and it may be more important to focus on these already existing classic data. One must also be aware of the possible misleading applications of some results: \cite{louail2016crowdsourcing} makes a very good analysis of potential redistribution of bank card transactions within a city, but pushes the results as possible basis for social equity policy recommandation by acting on mobility, forgetting that urban form and function are coupled in a complex way and that moving transactions from one place to the other involves far more complex processes than policies, which furthermore never apply the way they were planned and lead to results different from the ones expected. Such an attitude, often observed for physicists, is well translated as an allegory by the figure~\ref{fig:computation:xkcd} which is only partly an exaggeration of some situations.
}{
Les données fournissent aussi leur collection de dérives. Récemment, sur la liste de diffusion de géographie francophone \emph{Geotamtam}, un soudain engouement autour des données issues de \emph{Pokemon Go} a semblé répondre plus à un besoin urgent et inexpliqué d'exploiter cette source de données avant tous les autres, plutôt qu'à des considérations théoriques élaborées. Des jeux de données existant et précis, comme la population historiques des villes (pour la France la base Pumain-INED par exemple), sont loin d'être entièrement exploités et il pourrait être plus pertinent de se concentrer sur ces jeux de données classiques qui existent déjà. De même, il faut être conscient des possibles applications de résultats basée sur des malentendus : \cite{louail2016crowdsourcing} analyse la redistribution potentielle des transactions de carte bancaire au sein d'une ville, mais présente les résultats comme la base possible de recommandations de politiques pour une équité sociale en agissant sur la mobilité, oubliant que la forme et les fonctions urbaines sont fortement couplées et que déplacer des transactions d'un endroit à un autre implique des processus bien plus complexes que des régulations directes, qui d'autant plus ne s'appliquent jamais de la façon prévue et conduisent à des résultats différents de ceux attendus. Une telle attitude, souvent observée de la part de physiciens, est très bien mise en allégorie par la figure~\ref{fig:computation:xkcd} qui n'est qu'à moitié une exagération de certaines situations.
}

\subsubsection{For a cautious use}{Pour un usage raisonné}

\bpar{
Our main claim here is that the computational shift and simulation practices will be central in geography, but may also be dangerous, for the reasons illustrated above, i.e. that data deluge may impose research subjects and elude theory, and that computation may elude model construction and solving. A stronger link is required between computational practices, computer science, mathematics, statistics and theoretical geography.
}{
Notre principal argument est que le tournant computationnel et les pratiques de simulation seront centrales en géographie, mais peuvent également être dangereux, pour les raisons illustrées ci-dessus, i.e. que le déluge de données peut imposer les sujets de recherche et occulter la théorie, et que la computation peut éluder la construction et la résolution de modèles. Un lien plus fort est nécessaire entre les pratiques de calcul, l'informatique, les mathématiques, les statistiques et la géographie théorique.
}

\bpar{
Theoretical and Quantitative Geography is at the center of this dynamic, as it was its initial purpose that seems forgotten in some cases. It implies the need for elaborated theories integrated with conscious simulation practices. In other words we can answer complementary naive questions that however need to be tackled once and for all. If a theory-free quantitative geography would be possible, the answer if naturally no as it is close to the trap of black-box data-mining analysis. Whatever is done in that case, the results will have a very poor explanatory power, as they can exhibit relations but not reconstruct processes. On an other hand, the possibility of a purely computational quantitative geography is a dangerous vision: even gaining three orders of magnitudes in computational power does not solve the dimensionality curse.
}{
La géographie théorique et quantitative est au centre de cette dynamique, puisqu'il s'agit de sa motivation initiale principale qui semble oubliée dans certains cas. Cela implique un besoin de recherche de théories élaborées intégrées avec des pratiques de simulation conscientes. En d'autres mots, on peut répondre à des questions naïves complémentaires qui ont toutefois besoin d'être traitées une bonne fois pour toutes. Quant à la question de la possibilité d'une géographie quantitative libérée de la théorie, la réponse est naturellement négative puisque cela se rapproche du piège de la fouille de données par boîte noire. Quoi qu'il soit fait par cette approche, les résultats auront un pouvoir explicatif très faible, puisqu'ils pourront mettre en valeur des relations mais pas reconstruire des processus. D'autre part, la possibilité d'une géographie quantitative purement basée sur le calcul est une vision dangereuse : même le gain de trois ordres de grandeur dans la puissance de calcul disponible ne résout pas le sort de la dimension.
}

\bpar{
We can take the example of non-stationarity results obtained in~\ref{sec:staticcorrelations}. The use of relatively massive data, because of the algorithms specifically designed to be able to tackle the processing, is a necessary condition to the results obtained, but both the scale and objects (i.e. the indicators computed) are co-determined by the theoretical constructs. Indeed the absence of theory would imply to not know the objects, measures and properties to study (e.g. the multi-scalar or dynamical character of processes), and without analytical resolutions, it would often be difficult to draw conclusions starting only from the empirical analysis, in particular for the multi-scalar aspect.
}{
Prenons l'exemple des résultats de non-stationnarité obtenus en~\ref{sec:staticcorrelations}. L'utilisation de données relativement massives, de par les algorithmes spécialement conçus pour être capable de faire les traitements, est une condition nécessaire aux résultats obtenus, mais à la fois l'échelle et les objets (c'est-à-dire les indicateurs calculés) sont co-déterminés par les constructions théoriques. En effet l'absence de théorie impliquerait de ne pas connaitre les objets, mesures et propriétés à étudier (e.g. le caractère multi-scalaire ou dynamique des processus), et sans résolutions analytiques, il serait souvent difficile de tirer des conclusions à partir des analyses empiriques seules, notamment pour l'aspect multi-scalaire.
}

\bpar{
Nothing is really new here but this position has to be stated and stood up, precisely because our work will use this kind of tools, trying to advance on a thin and fragile edge, with the void of the unfunded theoretical charlatanism on one side and the abyss of the technocratic blind drowning in foolish amounts of data. More than ever we need simple but powerful and funded theories {\`a}-la-Occam~\cite{batty2016theoretical}, to allow a wise integration of new techniques into existing knowledge.
}{
Rien n'est vraiment nouveau ici mais cette position doit être affirmée et tenue, précisément car notre travail se base sur ce type d'outils, essayant d'avancer sur une arête fine et fragile, avec d'un côté le vide du charlatanisme théorique infondé et de l'autre l'abîme de l'overdose technocratique dans des quantités de données folles. Plus que jamais on a besoin de théories simples mais fondées et puissantes {\`a}-la-Occam~\cite{batty2016theoretical}, pour permettre une intégration saine des nouvelles techniques au sein des connaissances existantes.
}

\subsection{Extend sensitivity analyses}{Étendre les analyses de sensibilité}

\subsubsection{Context}{Contexte}

\bpar{
When evaluating data-driven models, or even more simple partially data-driven models involving simplified parametrization, an unavoidable issue is the lack of control on ``underlying system parameters'' (what is a ill-defined notion but should be seen in our sense as parameters governing system dynamics). Indeed, a statistics extracted from running the model on enough different datasets can become strongly biased by the presence of confounding in the underlying real data, as it is impossible to know if result is due to processes the model tries to translate or to a hidden structure common to all data. The fundamental methodological question that we will study in the following is to be able to isolate effects which are proper to the model from the ones due to geography.
}{
Lors de l'évaluation de modèles basés sur les données, ou même de modèles plus simples partiellement basés sur les données impliquant une paramétrisation simplifiée, une issue inévitable est le manque de contrôle sur les ``paramètres implicites du systèmes'' (ce qui n'est pas une notion stricte mais doit être compris dans notre sens comme les paramètres régissant la dynamique). En effet, une statistique issue d'executions du modèle sur un nombre suffisant d'executions peut toutefois rester biaisée, au sens où il est impossible de savoir si les résultats sont dus aux processus que le modèle cherche à traduire ou à une structure présente dans les données initiales. La question méthodologique fondamentale qui nous intéressera pour la suite est d'être capable d'isoler les effets propres aux processus du modèles de ceux liés à la géographie.
}

\paragraph{Context}{Contexte}

\bpar{
Although simulation models of geographical systems in general and agent-based models in particular represent a fantastic opportunity to explore socio-spatial behaviours and to test a variety of scenarios for public policy, the validity of generative models is uncertain until their results are proven robust. Sensitivity analysis usually include the analysis of the effect of stochasticity on the variability of results, as well as the effects of small parameter changes. However, initial spatial conditions are usually taken for granted in geographical models, thus leaving completely unexplored the effect of spatial arrangements on the interaction of agents and of their interactions with the environment. In this part, we present a method to assess the effect of initial spatial conditions on simulation models, using a systematic generator controlled by meta-parameter to create density grids used in spatial simulation models. We show, with the example of a very classical agent-based model (Sugarscape model of ressource allocation) that the effect of space in simulation is significant, and sometimes even larger than parameters themselves. We do so using high performance computing in a very simple and straightforward open-source workflow. The benefits of this approach are various but include for example the knowledge of model behavior in an extended frame, the possibility of statistical control when regressing model outputs, or a finer exploration of model derivatives than with a direct approach.
}{
Bien que les modèles de simulation des systèmes géographiques en général et les modèles basés-agent en particulier représentent une opportunité considérable d'explorer les comportements socio-spatiaux et de tester une variété de scenarios pour les politiques publiques, la validité des modèles génératifs est incertaine tant que la robustesse des résultats n'a pas été établie. Les analyses de sensibilité incluent généralement l'analyse des effets de la stochasticité sur la variabilité des résultats, ainsi que les effets de variations locales des paramètres. Cependant, les conditions spatiales initiales sont généralement prise pour données dans les modèles géographiques, laissant ainsi totalement inexploré l'effet des motifs spatiaux sur les interactions des agents et sur leur interaction avec l'environnement. Dans cette partie, nous présentons une méthode pour établir l'effet des conditions spatiales initiales sur les modèles de simulation, utilisant un générateur systématique contrôlé par des meta-paramètres pour créer des grilles de densité utilisées dans les modèles de simulation spatiaux. Nous montrons, avec l'exemple d'un modèle agent très classique (le modèle Sugarscape d'extraction de ressources) que l'effet de l'espace dans les simulations est significatif, et parfois plus grand que l'effet des paramètres eux-mêmes. Nous y arrivons en utilisant le calcul haute performance en un workflow très simple et open source. Les bénéfices de notre approche sont variés mais incluent par exemple la connaissance du comportement du modèle dans un contexte plus large, la possibilité de contrôle statistique pour régresser les sorties du modèles, ou une exploration plus fine des dérivées du modèle que par rapport à une approche directe.
}

\paragraph{Role of spatio-temporal path dependency}{Role de la dépendance au chemin spatio-temporelle}

\bpar{
Spatio-temporal path dependancy is one of the main reasons making our approach relevant. Indeed, a crucial aspect of most spatio-temporal complex systems is their non-ergodicity~\cite{pumain2012urban} (the property that cross-sectional samples in space are not equivalent to samples in time to compute statistics such as averages), what witnesses generally strong spatio-temporal path-dependencies in their trajectories. Similar to what \noun{Gell-Mann} calls \emph{frozen accidents} in any complex system~\cite{gell1995quark}, a given configuration contains clues on past bifurcations, that can have had dramatic effects on the state of the system. Temporal and cumulative effects have been considered in various geographical subfields and at various geographical scales, for example in regional systems~\cite{Wilson1981} or the intra-urban scale~\cite{AllenSanglier1979}. Less studied is the impact of the spatial setting on models dynamics and potential bifurcations.
}{
La dépendance au chemin spatio-temporelle est une des raisons principales rendant notre approche pertinente. En effet, un aspect crucial de la plupart des systèmes complexes spatio-temporels est leur non-ergodicité~\cite{pumain2012urban} (la propriété que les échantillons dans l'espace ne sont pas équivalent aux échantillons dans le temps pour calculer des statistiques comme la moyenne), qui témoigne généralement de forte dépendances au chemin spatio-temporelles dans les trajectoires. De manière similaire à ce que \noun{Gell-Mann} appelle \emph{frozen accidents} dans tout système complexe~\cite{gell1995quark}, une configuration donnée contient des indices sur les bifurcations passées, qui peuvent avoir eu des effets considérables sur l'état du système. Les effets temporels et cumulatifs ont été considérés dans de nombreux sous-champs géographiques et à différentes échelles géographiques, par exemple les systèmes régionaux~\cite{Wilson1981} ou l'échelle intra-urbaine~\cite{AllenSanglier1979}. L'impact de la configuration spatiale sur les dynamiques du modèle et les bifurcations spatiales a été moins étudié.
}

\bpar{
The example of transportation networks is a good illustration, as their spatial shape and hierarchy is strongly influenced by past investment decisions, technical choices, or political decisions sometimes not rational~\cite{zembri2010new}. Some aggregated indicators will not take into account positions and trajectories of each agent (such as segregation in the Schelling model) but others, as in the case of spatial patterns of accessibility in a system of cities, fully capture the path-dependency and may therefore be highly dependent of the initial spatial configuration. It is not clear for example what shifted the economical and political capital of France from Lyon to Paris in the early Middle Age, some assumptions being the reconfiguration of trade patterns from South to North of Europe and thus an increased centrality for Paris due to its spatial position, while keeping in mind that geographical and political centralities are not equivalent and are in a complex relationship~\cite{guenee1968espace}. The bifurcation induced by socio-economic and political factors took a deep significance with worldwide repercussions until today when magnified by the spatial configuration.
}{
L'exemple des réseaux de transport est une bonne illustration, car leur forme spatiale et leur hiérarchie est fortement influencée par les décisions d'investissement du passé, les choix techniques, ou des décisions politiques qui ne sont parfois pas rationnelles~\cite{zembri2010new}. Certains indicateurs agrégés ne prendront pas en compte les positions et trajectoires de chaque agent (comme les inégalités totales dans le modèle Sugarscape) mais d'autres, comme dans le cas des motifs d'accessibilité spatiale dans un système de villes, capture entièrement la dépendance au chemin et peuvent ainsi être fortement dépendants à la configuration spatiale initiale. Il n'est pas clair par exemple ce qui a causé la transition de la capitale française de Lyon à Paris dans le bas Moyen-Age, certaines hypothèses étant la reconfiguration des motifs commerciaux du Sud au Nord de l'Europe et donc une centralité accrue pour Paris due à sa position spatiale, tout en gardant à l'esprit que les centralités géographique et politique ne sont pas équivalentes et entretiennent une relation complexe~\cite{guenee1968espace}. La bifurcation induite par des facteurs socio-économiques et politiques a pris une signification profonde avec des répercussions mondiales encore aujourd'hui quand elle a été concrétisée par la configuration spatiale.
}

\paragraph{Existing works}{Travaux existants}

\bpar{
The effect of the spatial configuration on area-based attributes of human behaviours has been largely discussed in geostatistics, meanly since the exposure of the Modifiable Areal Unit Problem (MAUP) \cite{Openshaw1984},\cite{FotheringhamWong1991}. Recently, \cite{Kwan2012} claims for a careful examination of what she coins the uncertain geographic context problem (UGCoP), that is of the spatial configuration of geographical units even if the size and delineation of the area are the same. On the contrary, the scarcity of these considerations in the geographic simulation model literature questions the generalisation of their results, as it has for instance been showed in the case of LUTI models \cite{Thomasetal2017}, or of diffusion processes using ABM \cite{LeTexierCaruso2017}. 
}{
L'effet de la configuration spatiale sur les attributs agrégés à la zone des comportements humains a été largement discuté en géostatistiques, approximativement depuis l'introduction du \emph{Modifiable Areal Unit Problem} (MAUP)~\cite{Openshaw1984}. Plus récemment, \cite{Kwan2012} plaide pour un examen plus attentif de ce qui serait un \emph{Uncertain Geographic Context Problem} (UGCoP), qui est la configuration spatiale des unités géographiques même si la taille et la délimitation des zones est la même. Au contraire, le faible nombre de considérations similaires dans la littérature traitant des modèles de simulation géographiques remet en question la généralisation de leur résultats, comme cela a été montré par exemple dans le cas des modèles LUTI~\cite{Thomasetal2017}, ou des processus de diffusion étudiés par modèles multi-agents~\cite{LeTexierCaruso2017}.
}

\subsubsection{Methods}{Méthodes}

\bpar{
We detail now the method developed to analyse the sensitivity of simulation models to initial spatial conditions. In addition to the usual protocol, which consists of running a model $\mu$ with various values of its parameters and relating these variations of values to the variations in the simulation results, we here introduce a spatial generator, which itself is determined by parameters and produces sets of spatial initial conditions. Initial spatial conditions are categorized to represent types of spaces ex-ante (for example: monocentric or polycentric density grids), and the sensitivity analysis of the model is now run against $\mu$ parameters as well as spatial parameters or spatial types. It allows the sensitivity analysis to produce qualitative conclusions regarding the influence of spatial distribution on the outputs of simulation models, alongside the classic variation of parameter values.
}{
Nous détaillons à présent la méthode développée pour analyser la sensibilité des modèles de simulation aux conditions spatiales initiales. S'ajoutant au protocole usuel, qui consiste à simuler un modèle $\mu$ pour différentes valeurs de ses paramètres et faire le lien entre ces variations aux variations des résultats de simulation, nous introduisons ici un générateur spatial, qui est lui-même déterminé par des paramètres et produit des ensembles de configurations spatiales initiales. Les configurations spatiales initiales sont catégorisées pour représenter des types d'espace typiques (par exemple des grilles de densité monocentriques ou polycentriques), et la sensibilité du modèle est à présent testée sur les paramètres de $\mu$ mais aussi sur les paramètres spatiaux ou les types spatiaux. Cela permet à l'analyse de sensibilité de fournir des conclusions qualitatives au regard de l'influence de la distribution spatiale sur les sorties des modèles de simulation, en parallèle des variation classiques des paramètres.
}

\paragraph{Spatial Generator}{Générateur spatial}

\bpar{
Our spatial generator applies an urban morphogenesis model developed and explored in~\ref{sec:densitygeneration}. To present it in a nutshell, grids are generated through an iterative process which adds a quantity $N_G$ of population at each time step, allocating it through preferential attachment characterised by its strength of attraction $\alpha$. This first growth process is then smoothed $n_d$ times using a diffusion process of strength $\beta$. Grids are thus generated from the combination of the values of these four meta-parameters $\alpha$, $\beta$, $n_d$ and $N_G$. To ease our exploration, only the distribution of density is allowed to vary rather than the size of the grid, which we fix to a 50x50 square environment of 100,000 units.
}{
Le générateur spatial applique un modèle de morphogenèse urbaine développé et exploré en~\ref{sec:densitygeneration}. Pour le présenter rapidement, les grilles sont générées par un processus itératif qui ajoute une quantité de population $N_G$ à chaque pas de temps, l'allouant selon un attachement préférentiel caractérisé par sa force d'attraction $\alpha$. Le premier processus est ensuite lissé $n_d$ fois par un processus de diffusion de force $\beta$. Les grilles sont donc générées aléatoirement par la combinaison des valeurs de ces quatre meta-paramètres $\alpha$, $\beta$, $n_d$ and $N_G$. Pour faciliter l'exploration, seule la distribution de densité est autorisée à varier plutôt que la taille de la grille, qui est fixée à un environnement carré 50x50 de population 100,000 unités.
}

\paragraph{Comparing phase diagrams}{Comparer les diagrammes de phase}

\bpar{
In order to test for the influence of spatial initial conditions, we need a systematic method to compare phase diagrams. Indeed, we have as many phase diagrams than we have spatial grids, what makes a qualitative visual comparison not realistic. A solution is to use systematic quantitative procedures. Several potential methods could be used: for example in the case of the Schelling model, an anisotropic spatial segregation index (giving the number of clusters found and in which region in the parameter spaces they are roughly situated) would differentiate strong \emph{meta phase transitions} (phase transitions in the space of meta parameters). The use of metrics comparing spatial distributions, such as the Earth Movers Distance which is used for example in Computer Vision to compare probability distributions~\cite{rubner2000earth}, or the comparison of aggregated transition matrices of the dynamic associated to the potential described by each distribution, would also be potential tools. Map comparison methods, popular in environmental sciences, provide numeral tools to compare two dimensional fields~\cite{visser2006map}. To compare a spatial field evolving in time, elaborated methods such as Empirical Orthogonal Functions that isolates temporal from spatial variations, would be applicable in our case by taking time as a parameter dimension, but these have been shown to perform similarly to direct visual inspection when averaged over a crowdsourcing~\cite{10.1371/journal.pone.0178165}. To keep it simple and as such methodological considerations are auxiliary to the main purpose of this paper, we propose an intuitive measure corresponding to the share of between-diagrams variability relative to their internal variability. More formally, the distance is given by
}{
Afin de tester l'influence des conditions spatiales initiales, nous avons besoin d'une méthode systématique pour comparer des diagrammes de phase. En effet, nous avons autant de diagramme de phase que de grilles spatiales, ce qui rend une comparaison visuelle qualitative non réaliste. Une solution est d'utiliser des procédures quantitatives systématiques. De nombreuses méthodes pourraient potentiellement être utilisées : par exemple, des indicators anisotropes comme la donnée de clusters et leur position dans le diagramme de phase, peuvent permettre de révéler des \emph{meta-transitions de phase} (transition de phase dans l'espace des meta-paramètres). L'utilisation de métriques comparant des distributions spatiales, comme la \emph{Earth Movers Distance} qui est utilisée en vision par ordinateur pour comparer des distributions de probabilité~\cite{rubner2000earth}, ou la comparaison de matrices de transition agrégées de la dynamique associée au potentiel décrit par chaque distribution, est également possible. Les méthodes de comparaison de cartes, répandues en sciences environnementales, fournissent de nombreux outils pour comparer des champs en deux dimensions~\cite{visser2006map}. Pour comparer un champ spatial évoluant dans le temps, des méthodes élaborées comme les Fonctions Orthogonales Empiriques qui isolent les variations temporelles des variations spatiales, seraient applicables dans notre cas en prenant le temps comme une dimension de paramètre, mais celles-ci ont été montrées ayant une performance similaire à la comparaison visuelle directe lorsqu'on prend la moyenne sur un ensemble de contributions crowdsourcées~\cite{10.1371/journal.pone.0178165}. Pour rester simple et car de telles considérations méthodologiques sont auxiliaires pour le propos principal de cette partie, nous proposons une mesure intuitive correspondant à la part de la variabilité inter-diagrammes relativement à leur variabilité interne. Plus formellement, cette distance est donnée par
}

\begin{equation}\label{eq:phase-distance}
d_r\left(\alpha_1,\alpha_2\right) = 2 \cdot \frac{d(f_{\vec{\alpha_1}},f_{\vec{\alpha_2}})^2}{Var\left[f_{\vec{\alpha_1}}\right] + Var\left[f_{\vec{\alpha_2}}\right]}
\end{equation}

\bpar{
where $\alpha \mapsto \left[\vec{x} \mapsto f_{\vec{\alpha}}\left(\vec{x}\right)\right]$ is the operator giving phase diagrams with $\vec{x}$ parameters and $\vec{\alpha}$ meta-parameters, and $d$ is a distance between probability distributions that can be taken for example as basic L2 distance or the Earth's Mover Distance. For each values $\vec{\alpha_i}$, the phase diagram is seen as a random spatial field, facilitating the definition of variances and distance.
}{
où $\alpha \mapsto \left[\vec{x} \mapsto f_{\vec{\alpha}}\left(\vec{x}\right)\right]$ est l'opérateur donnant les diagrammes de phase avec $\vec{x}$ paramètres et $\vec{\alpha}$ meta-paramètres, et $d$ une distance entre distributions de probabilité qui peut être prise par exemple comme la distance L2 basique ou la \emph{Earth Movers Distance}. Pour chaque valeur $\vec{\alpha_i}$, le diagramme de phase est vu comme un champ spatial aléatoire, ce qui facilite la définition des variances et de la distance.
}

\subsubsection{Results}{Résultats}

\bpar{
Sugarscape is a model of resource extraction which simulates the unequal distribution of wealth within a heterogenous population \cite{EpsteinAxtell1996}. Agents of different vision scopes and different metabolisms harvest a self-regenerating resource available heterogeneously in the initial landscape, they settle and collect this resource, which leads some of them to survive and others to perish. The main parameters of this model are the number of agents, their minimal and maximal resource. In addition, we are interested in testing the impact of the spatial distribution of the resource in this project, using the spatial generator. The outcome of the model is measured as a phase diagram of an index of inequality for ressource distribution (Gini index). We extend the implementation with agents wealth distribution, given by~\cite{li2009netlogo}.
}{
Sugarscape est un modèle d'extraction de ressources qui simule la distribution inégale des richesses dans une population hétérogène~\cite{EpsteinAxtell1996}. Des agents ayant différentes portées de vision et différents métabolismes collectent une ressource qui se régénère automatiquement et disponible de manière hétérogène dans le paysage initial. Ceux-ci s'établissent et collectent la ressource, ce qui mène certains d'entre eux à survivre et d'autres à périr. Les paramètres principaux du modèle sont le nombre d'agents, leur ressources minimale et maximale. Nous nous intéressons principalement à tester l'impact de la distribution spatiale, en utilisant le générateur spatial. La sortie du modèle est mesurée comme le diagramme de phase d'un index d'inégalité pour la distribution de la ressource (index de Gini). Nous étendons l'implémentation ayant initialement une distribution de richesse des agents, donnée par~\cite{li2009netlogo}.
}

\begin{figure}
\includegraphics[width=\linewidth]{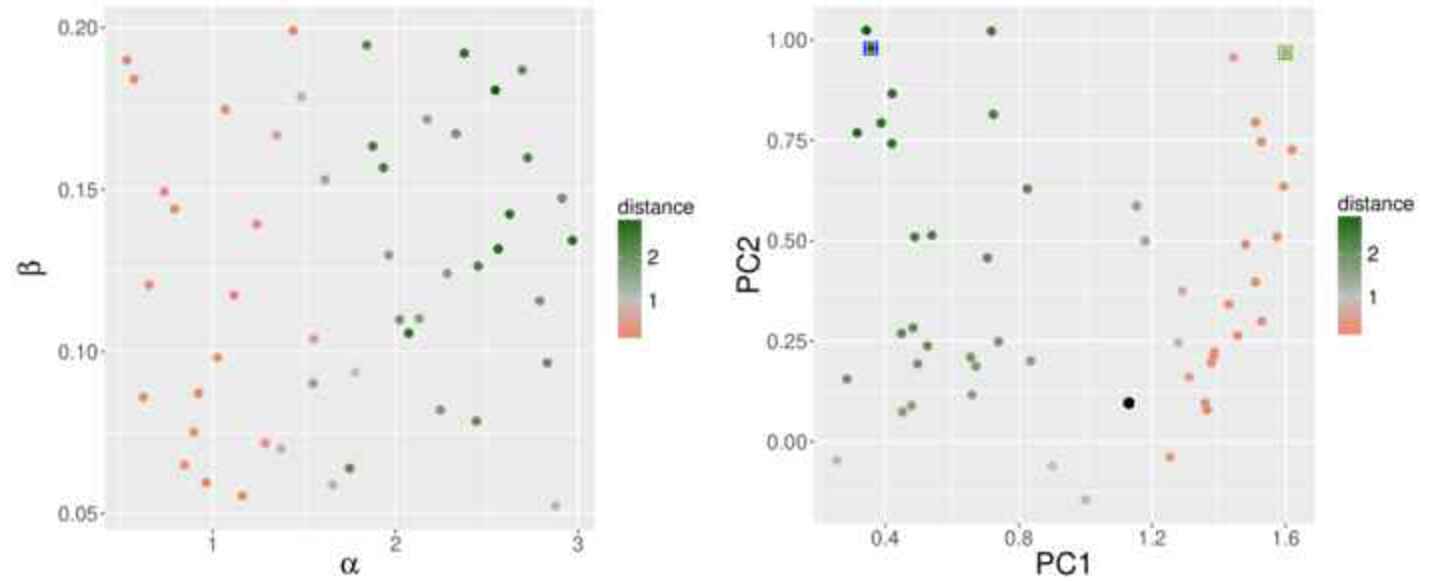}
\caption[Distances of phase diagrams to the reference]{\textbf{Relative distances of phase diagrams to the reference across grids.} (Left) Relative distance as a function of meta-parameters $\alpha$ (strength of preferential attachment) and diffusion ($\beta$, strength of diffusion process). (Right) Relative distance as a function of two first principal components of the morphological space (see text). Red point correspond to the reference spatial configuration. Green frame and blue frame give respectively the first and second particular phase diagrams shown in Fig.~\ref{fig:computation:sugarscape-phasediagrams}.\label{fig:computation:sugarscape-distance}}
\end{figure}

\bpar{
For the exploration, $2.5\cdot 10^6$ simulations (1000 parameter points x 50 density grids x 50 replications) allow us to show that the model is more sensitive to space than to its other parameters, both qualitatively and quantitatively: the amplitude of variations across density grids is larger than the amplitude in each phase diagram, and the behavior of phase diagram is qualitatively different in different regions of the morphological space. More precisely, we explore a grid of a basic parameter space of the model, which three dimensions are the population of agents $P\in \left[10;510\right]$, the minimal initial agent ressource $s_{-}\in \left[10;100\right]$ and the maximal initial agent ressource $s_{+}\in \left[110;200\right]$. Each parameter is binned into 10 values, giving 1000 parameter points. We run 50 repetitions for each configuration, what yield reasonable convergence properties. The initial spatial configuration varies across 50 different grids, generated by sampling meta-parameters for the generator in a LHS. We demonstrate the flexibility of our framework, by a direct sequential coupling of the generator and the model. We mesure the distance of all 3-dimensional phase diagrams to the reference phase diagram computed on the default model setup (see Fig.~\ref{fig:sugarscape-distance} for its morphological positioning regarded generated grids), using equation~\ref{eq:phase-distance} with the L2 distance to ensure direct interpretability. Indeed, it gives in that case the average squared distance between corresponding points of the phase diagrams, relative to the average of the variance of each. Therefore, values greater than 1 will mean that inter-diagram variability is more important than intra-diagram variability.
}{
Pour l'exploration, $2.5\cdot 10^6$ simulations (1000 points de paramètres x 50 grilles de densité x 50 réplications) nous permettent de montrer que le modèle est bien plus sensible à l'espace qu'à ses autres paramètres, à la fois quantitativement et qualitativement : l'amplitude des variations entre les grilles de densité est plus grande que l'amplitude dans chaque diagramme de phase, et le comportement de ces diagrammes de phase est qualitativement différent dans diverses régions de l'espace morphologique. Plus précisément, nous explorons une grille d'un espace de paramètre basique du modèle, dont les trois dimensions sont la population des agents $P\in \left[10;510\right]$, la ressource minimale initiale par agent $s_{-}\in \left[10;100\right]$ et la ressource initiale maximale par agent $s_{+}\in \left[110;200\right]$. Chaque paramètre est discrétisé en 10 valeurs, donnant 1000 points de paramètres. Nous procédons à 50 répétitions pour chaque configuration, ce qui donne des propriétés de convergence raisonnables. La distribution spatiale initiale varie parmi 50 grilles initiales, générée en échantillonnant les méta-paramètres du générateur dans un Hypercube Latin. Nous démontrons ainsi la flexibilité de notre cadre, par le couplage séquentiel direct du générateur avec le modèle. Nous mesurons la distance de l'ensemble des diagrammes de phase à 3 dimensions à un diagramme de phase de référence calculé sur l'initialisation du modèle par défaut (voir Fig.~\ref{fig:computation:sugarscape-distance} pour sa position morphologique au regard des grilles générées), en utilisant l'équation~\ref{eq:phase-distance} avec la distance L2 pour assurer une interprétation directe. En effet, cela donne dans ce cas la distance au carré moyenne entre chaque point en correspondance des diagrammes, relative à la moyenne des variances de chaque. Pour cela, des valeurs plus grandes que 1 signifient que la variabilité inter-diagramme est plus importante que la variabilité intra-diagramme.
}


\bpar{
We obtain a very strong sensitivity to initial conditions, as the distribution of the relative distance to reference across grids ranges from 0.09 to 2.98 with a median of 1.52 and an average of 1.30. It means that in average, the model is more sensitive to meta-parameters than to parameters, and the relation variation can reach a factor of 3. We plot in Fig.~\ref{fig:computation:sugarscape-distance} their distribution in a morphological space. The reduced morphological space is obtained by computing 4 raw indicators of urban form, namely Moran index, average distance, rank-size slope and entropy (see the section~\ref{sec:staticcorrelations} for precise definition and contextualization), and by reducing the dimension with a principal component analysis for which we keep the first two components (92\% of cumulated variance). The first measures a ``level of sprawl'' and of scattering, whereas the second measures aggregation.\footnote{We have $PC1 = 0.76\cdot distance + 0.60\cdot entropy + 0.03\cdot moran + 0.24\cdot slope$ and $PC2 = -0.26\cdot distance + 0.18\cdot entropy + 0.91\cdot moran + 0.26\cdot slope$.} We find that grids producing the highest deviations are the ones with a low level of sprawl and a high aggregation. It is confirmed by the behavior as a function of meta-parameters, as high values of $\alpha$ also yield high distance. In terms of model processes, it shows that congestion mechanisms induce rapidly higher levels of inequality.
}{
Nous obtenons une sensibilité très forte aux conditions initiales, puisque la distribution de la distance relative à la référence s'étend sur l'ensemble des grilles de 0.09 à 2.98, avec un médiane de 1.52 et une moyenne de 1.30. Cela signifie qu'en moyenne, le modèle est plus sensible aux méta-paramètres qu'aux paramètres, et que la variation relative peut atteindre jusqu'à un facteur 3. Nous montrons en Fig.~\ref{fig:computation:sugarscape-distance} leur distribution dans un espace morphologique. L'espace morphologique réduit est obtenu en calculant 4 indicateurs bruts de forme urbaine, qui sont l'index de Moran, la distance moyenne, le niveau de hiérarchie et l'entropie (voir la section~\ref{sec:staticcorrelations} pour une définition précise et une mise en contexte), et en réduisant la dimension avec une analyse par composantes principales pour laquelle nous gardons les deux premières composantes (92\% de variance cumulée). La première mesure un ``niveau d'étalement'' et d'éclatement, tandis que la seconde mesure l'agrégation.\footnote{Nous avons $PC1 = 0.76\cdot distance + 0.60\cdot entropy + 0.03\cdot moran + 0.24\cdot slope$ et $PC2 = -0.26\cdot distance + 0.18\cdot entropy + 0.91\cdot moran + 0.26\cdot slope$.} Nous trouvons que les grilles produisant les déviations les plus grandes sont celles avec un faible niveau d'étalement et une forte agrégation. Cela est confirmé par le comportement comme fonction des meta-paramètres, puisque des fortes valeurs de $\alpha$ donnent aussi une forte distance. En terme de processus du modèle, cela montre que les mécanismes de congestion induisent rapidement de plus haut niveaux d'inégalités.
}


\begin{figure}
\includegraphics[width=\linewidth]{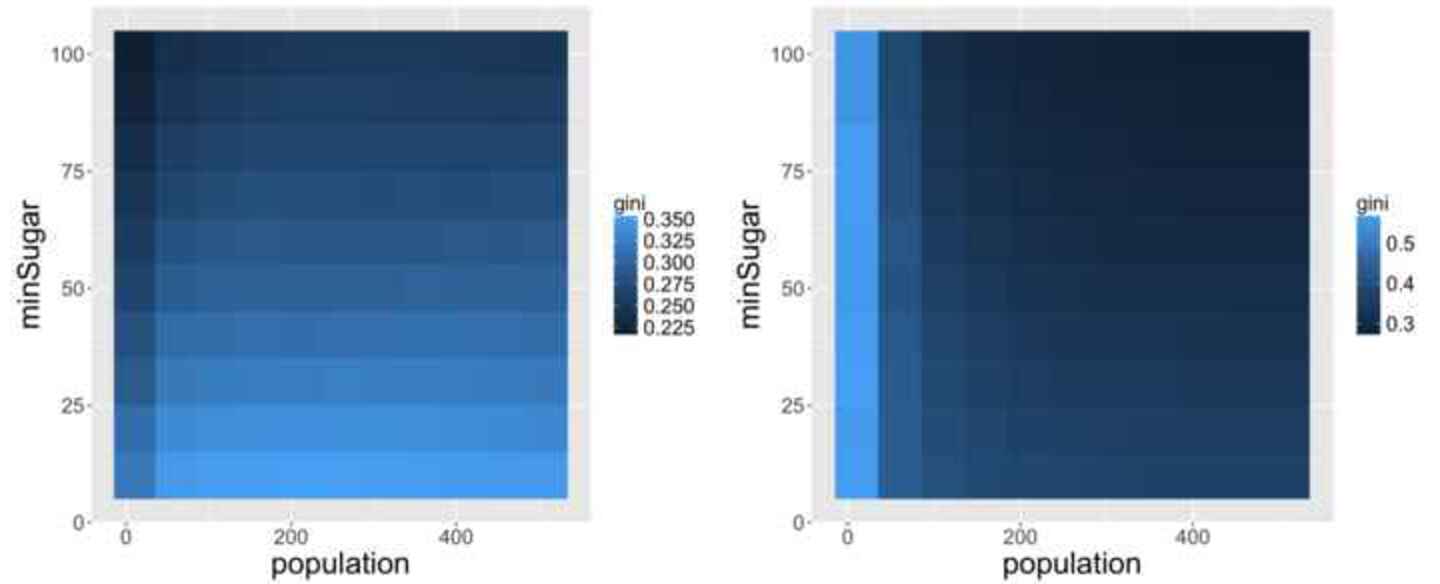}
\caption[Examples of phase diagrams]{\textbf{Examples of phase diagrams.} We show two dimensional phase diagrams on $(P,s_-)$, both at fixed $s_+ = 110$. (Left) Green frame, obtained with $\alpha = 0.79$, $n=2$, $\beta = 0.14$, $N=157$; (Right) Blue frame, obtained with $\alpha = 2.56$, $n=3$, $\beta = 0.13$, $N=128$.\label{fig:computation:sugarscape-phasediagrams}}
\end{figure}

\bpar{
We now check the sensitivity in terms of qualitative behavior of phase diagrams. We show in Fig.~\ref{fig:computation:sugarscape-distance} the phase diagrams for two very opposite morphologies in term of sprawling, but controlling for aggregation with the same $PC2$ value. These correspond to the green and blue frames in Fig.~\ref{fig:computation:sugarscape-distance}. The behaviors are rather stable for varying $s_+$, what means that the poorest agents have a determinant role in trajectories. The two examples have not only a very distant baseline inequality (the ceil of the first 0.35 is roughly the floor of the second 0.3), but their qualitative behavior is also radically opposite: the sprawled configuration gives inequalities decreasing as population decreases and decreasing as minimal wealth increases, whereas the concentrated one gives inequalities strongly increasing as population decreases and also decreasing with minimal wealth but significantly only for large population values. The process is thus completely inverted, what would have significant impacts if one tried to schematize policies from this model. This second example confirms thus the importance of sensitivity of simulation models to the initial spatial conditions.
}{
Nous contrôlons à présent la sensibilité en terme de comportement qualitatif des diagrammes de phase. Nous montrons en Fig.~\ref{fig:computation:sugarscape-phasediagrams} les diagrammes pour deux morphologies très opposées en terme d'étalement, mais en contrôlant l'agrégation par la même valeur de $PC2$. Ceux-ci correspondent au cadres vert et bleu en Fig.~\ref{fig:computation:sugarscape-distance}. Les comportements sont relativement stables pour $s_+$ variant, ce qui signifie que les agents les plus pauvres ont un rôle déterminant dans les trajectoires. Les deux examples ont non seulement une inégalité de base très distante (le plafond du premier 0.35 est environ le plancher du second 0.3), mais leur comportement qualitatif est également radicalement opposé : la configuration étalée donne des inégalités qui décroissent quand la population décroît et qui décroissent quand la richesse minimale augmente, tandis que la concentrée donne des inégalités augmentant fortement quand la population décroît et aussi décroissantes avec la richesse minimale mais significativement seulement pour des grandes valeurs de population. Le processus est ainsi complètement inversé, ce qui aurait un impact déterminant si l'on essayait de schématiser des politiques à partir du modèle. Cet exemple confirme ainsi l'importance de la sensibilité des modèles de simulation aux conditions spatiales initiales.
}

\stars

\bpar{
We saw in this section how to position ourselves regarding the use of simulation models, and more generally regarding intensive computation. We have seen in a recurrent way in problematics we studied the question of opening of scientific practices.
}{
Nous avons vu dans cette section comment nous positionner par rapport à l'usage des modèles de simulation, et plus généralement par rapport au calcul intensif. Nous avons vu revenir de manière récurrente dans les problématiques abordées la question de l'ouverture des pratiques scientifiques.
}

\bpar{
We propose in the next section to detail one aspect of it, the one of reproducibility, which is both a component but also a product: simultaneously product and producer, it allows a broader opening and is reciprocally encouraged by opening practices.
}{
Nous proposons dans la section suivante d'en détailler un aspect, celle de la reproductibilité, qui en est à la fois une composante mais aussi un produit : simultanément produit et producteur, elle permet une plus grande ouverture et est réciproquement encouragée par les pratiques d'ouverture.
}

\stars

%


\newpage


\section{Reproducibility and opening}{Reproductibilité et ouverture}

\label{sec:reproducibility}


\bpar{
The production of scientific knowledge finds its roots in the cumulative and collective nature of research, since progresses are made when, as \noun{Newton} put it, we ``stand on shoulder of giants'', in the sense that the scientific enterprise at a given time relies on all the previous work and that no advance would be possible without building on it. It includes the development of new theories, but also the extension, the test and the falsification of previous ones: the advance in the construction of the tower also means the deconstruction of some obsolete building bricks. This peer validation aspect and of constant questioning is also what legitimates science for a more robust knowledge and a societal progress based on a knowledge of an objective universe, compared to dogmatic systems wether they are political or religious~\cite{bais2010praise}.
}{
La production de connaissance scientifique trouve ses fondements dans la nature cumulative et collective de la recherche, puisque les progrès sont faits lorsque, comme \noun{Newton} l'a bien dit, on ``se tient sur les épaules de géants'', au sens que l'entreprise scientifique à un temps donné repose sur l'ensemble du travail précédent et qu'aucune avancée ne serait possible sans construire dessus. Cela inclut le développement de nouvelles théories, mais aussi l'extension, le test et la falsification de précédentes : l'avancée dans la construction de la tour signifie aussi la déconstruction de certaines briques obsolètes. Cet aspect de validation par les pairs et de remise en question constante est aussi ce qui légitime la science pour une connaissance plus robuste et un progrès sociétal basés sur une connaissance d'un univers objectif, par rapport aux systèmes dogmatiques qu'ils soient politiques ou religieux~\cite{bais2010praise}. 
}

\bpar{
The effective practice of reproducibility seems to be increasing~\cite{stodden2010scientific} and technical means to achieve it are always more developed (as e.g. ways to make data openly available, or to be transparent on the research process such as \texttt{git}~\cite{ram2013git}, or to integrate document creation and data analysis such as \texttt{knitr}~\cite{xie2013knitr}), at least in the field of modeling and simulation. However, the devil is indeed in the details and obstacles judged at first sight as minor become rapidly a burden for reproducing and using results obtained in some previous researches. We describe two cases studies where models of simulation are apparently highly reproducible but unveil as puzzles on which research-time balance is significantly under zero, in the sense that trying to exploit their results may cost more time than developing from scratch similar models.
}{
La reproductibilité semble être de plus en plus pratiquée de manière effective~\cite{stodden2010scientific} et les moyens techniques pour l'achever sont toujours plus développés (comme par exemple les outils pour déposer les données ouvertes, ou pour être transparent dans le processus de recherche comme \texttt{git}~\cite{ram2013git}, ou pour intégrer la création de document et l'analyse de données comme  \texttt{knitr}~\cite{xie2013knitr}), au moins dans le champ de la modélisation et de la simulation. Cependant le diable est bien dans les détails et des obstacles jugés dans un premier temps comme mineurs peuvent rapidement devenir un fardeau pour reproduire et utiliser des résultats obtenus dans des recherches précédentes. Nous décrivons deux études de cas où les modèles de simulation sont en apparence hautement reproductibles mais se révèlent vite des puzzles pour lesquels l'équilibre de temps de recherche passe rapidement sous zéro, au sens où essayer d'exploiter leur résultats coûtera plus en temps que de développer entièrement des modèles similaires.
}

\subsection{Explicitation, documentation and implementation of models}{Explicitation, documentation et implémentation des modèles}

\subsubsection{On the need to explicit the model}{Sur le besoin d'expliciter le modèle}

\bpar{
A current myth (to which we ourselves struggle to escape indeed) is that providing entire source code and data will be a sufficient condition for reproducibility, since complete computational reproducibility implies a similar environment what becomes quickly complicated to produce as show \cite{2016arXiv160806897H}. To solve this problem, \cite{10.1371/journal.pone.0152686} propose the use of Docker containers which allow to reproduce even the behavior of softwares with a graphic user interface independently of the environment. It is indeed one of the current direction of development of OpenMole, to simplify the packaging of libraries and of binary models (see the interview with \noun{R. Reuillon}). In any case, reproducibility has supplementary dimensions, the objective is not only to exactly produce the same plots and scientific analyses, assuming that code provided is the one which was indeed used to produce the given results. First, results must be as much implementation-independent as possible~\cite{crick2017reproducibility} (i.e. of language, libraries, of choice of data structure and of type of programming) for clear robustness purposes. Then, in relation with the previous point, one of the purposes of reproducibility is the reuse of methods or results as basis or modules for further research (what includes implementation in another language or adaptation of the method), in the sense that reproducibility is not replicability as it must be adaptable~\cite{drummond2009replicability}.
}{
Un mythe à la vie dure (auquel nous essayons en fait nous-même d'échapper) est que fournir le code source complet et les données seront une condition suffisante pour la reproductibilité, puisque la reproductibilité computationnelle complète implique un environnement similaire ce qui devient vite ardu à produire comme le montrent \cite{2016arXiv160806897H}. Pour résoudre ce problème, \cite{10.1371/journal.pone.0152686} proposent l'utilisation de conteneurs Dockers qui permet de reproduire même le comportement de logiciels avec interface graphique indépendamment de l'environnement. C'est d'ailleurs l'une des direction courantes de développement d'OpenMole, pour simplifier le packaging des bibliothèques et des modèles en binaire (voir l'entretien avec \noun{R. Reuillon}). Dans tous les cas, la reproductibilité a des dimensions supplémentaires, il ne s'agit pas de l'objectif unique qui serait de produire exactement les mêmes graphes et analyses statistiques, en supposant que le code fournit est celui qui a été effectivement utilisé pour produire les résultats donnés. Tout d'abord, ceux-ci doivent être autant que possible indépendants de l'implémentation~\cite{crick2017reproducibility} (c'est-à-dire du langage, des bibliothèques, des choix de structures de données et de type de programmation) pour des motifs clairs de robustesse. Ensuite, en relation avec le point précédent, un des buts de la reproductibilité est la réutilisation des méthodes ou résultats comme base ou modules pour une recherche future (ce qui comprend une implémentation dans un autre langage ou une adaptation de la méthode), au sens que la reproductibilité n'est pas la possibilité stricte de répliquer car elle doit être adaptable~\cite{drummond2009replicability}.
}


\bpar{
Our first case study fits exactly that scheme, as it was undoubtedly aimed to be shared with and used by the community since it is a model of simulation provided with the Agent-Based simulation platform NetLogo~\cite{wilensky1999netlogo}. The model is also available online~\cite{de2007netlogo} and is presented as a tool to simulate socio-economic dynamics of low-income residents in a city based on a synthetic urban environment, generated to be close in stylized facts from the real town of Tijuana, Mexico. Globally, the model works in the following way: (i) starting from urban centers, and land-use distribution is generated through a procedural modeling similar to \cite{lechner2006procedural}, i.e. roads are generated locally according to geometric rule and of local hierarchy, and a land-use with also a value are attributed as a function of the characteristics of the cell (distance to the center, to the road); (ii) in this urban environment are simulated residential dynamics of migrants, which aim at optimizing a utility function depending on the cost of life and the configuration of other migrants. Beside providing the source code, the model appears to be poorly documented in the literature or in comments and description of the implementation. Comments made thereafter are based on the study of the urban morphogenesis part of the model (setup for the ``residential dynamics'' component) as it is our global context of study. In the frame of that study, source code was modified and commented, which last version is available on the repository of the project\footnote{At \texttt{https://github.com/JusteRaimbault/CityNetwork/tree/master/Models/Reproduction/UrbanSuite}.}.
}{
Notre premier cas d'étude suit exactement ce schéma, puisqu'il a sans aucun doute été conçu pour être partagé avec la communauté et utilisé, s'agissant d'un modèle de simulation fournit avec la plateforme de modélisation agent NetLogo~\cite{wilensky1999netlogo}. Le modèle est également disponible en ligne~\cite{de2007netlogo} et est présenté comme un outil pour simuler les dynamiques socio-économiques des résidents à bas revenus d'une ville au sein d'un environnement urbain synthétique, généré pour ressembler en terme de faits stylisés à la ville réelle de Tijuana au Mexique. Globalement, le modèle fonctionne de la façon suivante : (i) à partir de centres urbains, une distribution d'usage du sol est générée par modélisation procédurale similaire à \cite{lechner2006procedural}, c'est-à-dire des routes sont générées de proche en proche selon des règles géométriques et de hiérarchie locales, et un usage du sol ainsi qu'une valeur est attribué en fonction des caractéristique de la cellule (distance au centre, à la route) ; (ii) dans cet environnement urbain sont simulées des dynamiques résidentielles de migrants, qui cherchent à optimiser une fonction d'utilité dépendant du coût de la vie et de la configuration des autres migrants. À part fournir le code source, le modèle n'est que peu documenté dans la littérature ou dans les commentaires et la description de l'implémentation. Les commentaires qui suivent sont basés sur l'étude de la partie du modèle simulant la morphogenèse urbaine (initialisation pour la composante ``dynamiques résidentielles'') comme il s'agit de notre contexte global d'étude. Dans le cadre de cette étude, le code source a été modifié et commenté, dont la dernière version est disponible sur le dépôt du projet\footnote{À \url{https://github.com/JusteRaimbault/CityNetwork/tree/master/Models/Reproduction/UrbanSuite}.}.
}

\paragraph{Rigorous formalization}{Formalisation rigoureuse}

\bpar{
An obvious part of model construction is its rigorous formalization in a formal framework distinct from source code. There is of course no universal language to formulate it~\cite{banos2013pour}, and many possibilities are offered by various fields (e.g. UML, DEVS, pure mathematical formulation), but the stage of precise formalization, which generally follows a more intuitive description giving the ideas and the main processes, can not be avoided. No paper nor documentation is provided with the model, apart from the embedded NetLogo documentation, that only thematically describes in natural language the ideas behind each step without developing more and provides information about role of different elements of the interface. As these elements lack here, the model is difficult to use as is. It could be objected here that the part we study is only an initialization procedure and not the core of the model: we maintain that all procedures must be equally documented and implemented with a similar care, or point towards an external reference in the case of the use of a third party model, as we indeed do for the coupling done in~\ref{sec:computation}.
}{
Une partie évidente de la construction d'un modèle est sa formalisation rigoureuse dans un cadre formel distinct du code source. Il n'y a bien sûr aucun langage universel pour le formuler~\cite{banos2013pour}, et de nombreuses possibilités sont offertes par de nombreux champs (e.g. UML, DEVS, formulation mathématique pure), mais l'étape de formalisation précise, qui suit généralement une description plus intuitive donnant les idées et processus dominants, ne peut pas être sautée. On pourrait se dire que le code source y est équivalent, mais ce n'est pas exactement vrai car on pourrait alors ne plus distinguer certains choix d'implémentation de la structure du modèle. Aucun article ni documentation n'accompagne le modèle ici, au delà de la documentation embarquée NetLogo, qui ne décrit que de manière thématique en langage naturel les idées derrière chaque étape sans plus développer et fournit de l'information sur le rôle des différents éléments de l'interface. Comme ces éléments manquent ici, le modèle n'est guère utilisable tel quel. On pourrait nous objecter ici que la partie que nous étudions est une procédure d'initialisation et non le coeur du modèle : nous maintenons que l'ensemble des procédures doit être également documenté et implémenté avec un soin équivalent, ou pointer vers une référence extérieure dans le cas d'utilisation d'un modèle tiers, comme nous le faisons d'ailleurs pour le couplage effectué en~\ref{sec:computation}.
}

\bpar{
This formulation is a key for it to be understood, reproduced and adapted; but it also avoids implementation biases such as:
\begin{itemize}
\item Architecturally dangerous elements: the world context is a sphere, what is not reasonable for this model at the scale of a city, proximity measures playing an important role in the production processes of the urban form. The agents may go from one side of the world to the other in the euclidian representation, what is not acceptable for a two dimensional projection of real world. To avoid that, many tricky tests and functions were used, including unadvised practices (e.g. death of agents based on position to avoid them jumping).
\item Lack of internal consistence: for example the patch variable \texttt{land-value} (undocumented but which use can be reconstructed from code analysis) used to represent different geographical quantities at different steps of the model (morphogenesis and residential dynamics), what becomes an internal inconsistence when both steps are coupled when the option allowing the city to grow is activated.
\item Coding errors: in an untyped language such as NetLogo, mixing types may conduct to unexpected runtime errors, or even \emph{bugs} not directly detectable and then more dangerous. This is the case of the patch variable \texttt{transport} in the model (although no error occurs in most of run configurations from the interface, what is more dangerous as the developer thinks implementation is secure). Such problems should be avoided if implementation is done from an exact formal description of the model.
\end{itemize}
}{
Une telle formulation est essentielle pour que le modèle soit compris, reproduit et adapté ; mais elle évite également des biais d'implémentation comme :
\begin{itemize}
\item Des éléments architecturaux dangereux : le contexte du monde est une sphère, ce qui n'est pas raisonnable pour ce modèle à l'échelle d'une ville, les mesures de proximité jouant un rôle important dans les processus de production de la forme urbaine. Les agents peuvent passer d'un côté du monde à l'autre dans la représentation euclidienne, ce qui n'est pas acceptable pour une projection en deux dimensions du monde réel. Pour éviter cela, de nombreux tests et fonctions subtiles sont utilisés, incluant des pratiques déconseillées (e.g. mort d'agents basée sur leur position pour les empêcher de sauter).
\item Manque de cohérence interne : par exemple la variable de cellule \texttt{land-value} (non documentée mais dont l'utilisation se reconstruit par analyse du code) utilisée pour représenter différentes quantités géographiques à différentes étapes du modèle (morphogenèse et dynamiques résidentielles), ce qui devient une incohérence interne quand les deux étapes sont couplées lorsque l'option permettant de faire croître la ville est activée.
\item Erreur de code : dans un langage non typé comme NetLogo, le mélange des types peut conduire à des erreurs inattendues à l'execution, ou même des \emph{bugs} non détectables directement et alors plus dangereux. C'est le cas de la variable de patch \texttt{transport} dans le modèle (même si aucune erreur ne survient dans la majorité des configurations depuis l'interface, ce qui est plus dangereux comme le développeur pense que l'implémentation est sûre). De tels problèmes devraient être évités si l'implémentation est faite à partir d'une description exacte du modèle.
\end{itemize}
}

\paragraph{Transparent implementation}{Implémentation transparente}

\bpar{
A totally transparent implementation must be expected, including ergonomics in architecture and coding, but also in the interface and the description of the expected behavior of the model.
}{
Une implémentation totalement transparente doit être attendue, incluant une certaine ergonomie dans l'architecture et le code, mais aussi dans l'interface et la description du comportement attendu du modèle.
}

\paragraph{Expected model behavior}{Comportement attendu du modèle}

\begin{figure}
\includegraphics[width=\linewidth]{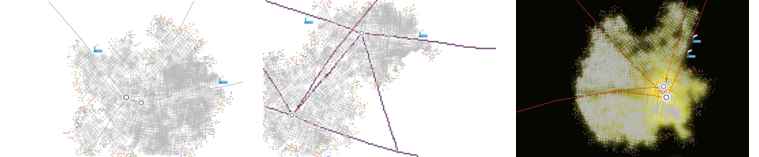}
\caption[Reproducibility and visualization]{\textbf{Example of simple improvement in visualization that can help understanding mechanisms implied in the model.} (Left) Example of original output; (Center) Visualization of main roads (in red) and underlying patches attribution, suggesting possible implementation bias in the use of discretized trace of roads to track their positions; (Right) Visualization of land values using a more readable color gradient. This step confirms the hypothesis, through the form of value distribution, that the morphogenesis step is an unnecessary detour to generate a random field for which simple diffusion method should provide similar results, as detailed in the paragraph on implementation. Initially, the interface of the model does not allow these visualization options, i.e. is limited to the first image. One can not understand the processes in play for morphogenesis, linked to the roads patches and to land values diffusing.\label{fig:reproducibility:tijuana}}
\end{figure}

\bpar{
Whatever the definition, a model can not be reduced to its formulation and/or implementation, as expected model behavior or model usage can be viewed as being part of the model itself. In the frame of \noun{Giere}'s perspectivism~\cite{giere2010scientific}, the definition of model includes the purpose of use but also the agent who aims to use it. Therefore a minimal explication of model behavior and exploration of the role of parameters is highly advised to decrease chances of misuses or misinterpretations of it. It includes simple runtime charts that are immediate on the NetLogo platform, but also indicators computations to evaluate outputs of the model. It can also be improved visualizations during runtime and model exploration, such as showed in Fig.~\ref{fig:reproducibility:tijuana}.
}{
Quelle que soit la définition, un modèle ne peut pas être réduit à sa formulation et/ou implémentation, comme le comportement attendu ou l'utilisation du modèle peuvent être vu comme des parties du modèle lui-même. Dans le cadre du perspectivisme de \noun{Giere}~\cite{giere2010scientific}, la définition du modèle inclut le motif de l'utilisation mais aussi l'agent qui vise à l'utiliser. Pour cela une explication minimale du comportement du modèle et une exploration du rôle des paramètres sont fortement recommandés pour diminuer les chances de mauvais usage ou mauvaises interprétations de celui-ci. Cela inclut des graphes simples obtenus immédiatement à l'exécution sur la plateforme NetLogo, mais aussi un calcul d'indicateurs pour évaluer les sorties du modèle. Il peut aussi s'agir de visualisations améliorées pendant l'execution et l'exploration du modèle, comme le montre la figure~\ref{fig:reproducibility:tijuana}.
}

\subsubsection{On the need of exactitude in model implementation}{Sur le besoin d'exactitude dans l'implémentation du modèle}

\bpar{
Possible divergences between model description in a paper and the effectively implemented processes may have grave consequences on the final reproducibility. The road network growth model given in~\cite{barthelemy2008modeling} is one example of such a discrepancy. A strict implementation of model mechanisms\footnote{Our implementation in NetLogo is available at \url{https://github.com/JusteRaimbault/CityNetwork/tree/master/Models/Reproduction/NWGrowth/LocalDistanceMin}.} provide slightly different results than the one presented in the paper, and as source code is not provided we need to test different hypotheses on possible mechanisms added by the programmer (that seems to be a connexion rule to intersections under a certain distance threshold). Lessons that could be possibly drawn from this example, which partly rejoin the ones drawn from the previous case study, are:
\begin{itemize}
\item the necessity of providing source code;
\item the necessity of providing architecture description along with code (if model description is in a langage too far from architectural specifications) in order to identify possible implementation biaises;
\item the necessity of performing and detailing explicitly model explorations, that would in that case have helped to identify the implementation bias.
\end{itemize}
}{
Des divergences potentielles entre la description du modèle dans un article et les processus effectivement implémentés peut avoir des conséquences graves sur la reproductibilité finale. Le modèle de croissance du réseau routier donné dans~\cite{barthelemy2008modeling} est un exemple d'un tel décalage. Une implémentation stricte des mécanismes du modèle\footnote{Notre implémentation en NetLogo est disponible à \url{https://github.com/JusteRaimbault/CityNetwork/tree/master/Models/Reproduction/NWGrowth/LocalDistanceMin}.} produit des résultats légèrement différents de ceux présentés dans l'article, et comme le code source n'est pas fourni nous devrions tester différentes hypothèses sur des mécanismes possibles ajoutés par le programmeur (qui semble être une règle de connexion aux intersections sous un certain seuil de distance). Des leçons qui peuvent éventuellement être tirées de cet exemple, qui rejoignent partiellement mais complètent celles tirées dans l'étude de cas précédente, sont :
\begin{itemize}
\item la nécessité de fournir le code source ;
\item la nécessité de fournir une description de l'architecture en même temps que le code (si la description du modèle est faite dans un langage trop loin de spécification architecturales) afin d'identifier des biais possibles d'implémentation ;
\item la nécessité de procéder à des explorations explicites du modèle et de les détailler, ce qui dans ce cas aurait permis d'identifier de possibles biais d'implémentation.
\end{itemize}
}

\bpar{
Making the last point mandatory may ensure a limited risk of scientific falsification as it is generally more complicated to fake false exploration results than to effectively explore the model. One could imagine an experiment to test the general behavior of a subset of the scientific community regarding reproducibility, that would consist in the writing of a false modeling paper in the spirit of~\cite{zilsel2015canular}, in which opposite results to the effective results of a given model are provided, without providing model implementation. A first bunch of test would be to test the acceptance of a clearly non-reproducible paper in diverse journals, possibly with a control on textual elements (using or not ``buzz-words'' associated to the journal, etc.). Depending on results, a second experiment may be tested with providing open source code for model implementation but still with false results, to verify if reviewers effectively try to reproduce results when they ask for the code (in reasonable computational power limits of course, HPC being not currently broadly available in social sciences). Our intuition is that results we would obtain would be strongly negative, given the difficulties met through an exigence of independent reproduction during the numerous reviews, even for journals making reproducibility a \emph{sine qua non} condition for publication, authors finding tricks to avoid constraints (postulate that simulation data are not data, publish only an wasteful aggregated version of the dataset used, etc.; we will come back later on the role of data). 
}{
Rendre le dernier point obligatoire pourrait assurer un risque limité de falsification puisqu'il est généralement plus compliqué de falsifier des résultats d'exploration plutôt que d'explorer effectivement le modèle. On pourrait imaginer une expérience pour tester le comportement général d'un sous-ensemble de la communauté scientifique au regard de la reproductibilité, qui consisterait en l'écriture d'un faux article de modélisation dans l'esprit de \cite{zilsel2015canular}, dans lesquels des résultats opposés aux résultats effectifs d'un modèle donné seraient fournis, sans fournir l'implémentation du modèle. Un premier test serait de tester l'acceptation d'un article clairement non reproductible dans divers journaux, si possible avec un contrôle sur les éléments textuels (par exemple en utilisant ou non des ``buzz-words'' chers au journal). Selon les résultats, une expérience plus poussée serait de fournir l'implémentation open source mais toujours avec des résultats modifiés plus ou moins fortement, afin de tester si les reviewers essayent effectivement de reproduire les résultats quand ils demandent le code (dans des capacités de calcul limitées bien sûr, le calcul intensif n'étant pas encore largement disponibles en sciences sociales). Notre intuition est que les résultats obtenus seraient fortement négatifs, vu les difficultés rencontrées par une exigence de discipline de reproduction indépendante lors de nombreuses relectures, même pour des revues faisant de la reproductibilité une condition \emph{sine qua non} de la publication, les auteurs trouvant des astuces pour se dérober aux contraintes (postuler que des données de simulation ne sont pas des données, ne fournir qu'une version agrégée inutile du jeu de données utilisées, etc. ; nous reviendrons sur le rôle des données plus loin).
}

\subsubsection{Interactive exploration and production of results}{Exploration interactive et production des résultats}

\bpar{
The use of interactive applications for data mining has non discussable advantages, such as a familiarization with the data structure through a global view which would be much more difficult to obtain or even impossible otherwise. It is the same underlying idea which justifies the interactivity for the preliminary exploration of agent-based models which is integrated to platforms such as NetLogo~\cite{wilensky1999netlogo} or Gamma~\cite{drogoul2013gama}. A similar objective is implicit in~\cite{rey2015plateforme}, i.e. a complete integration of the fine exploration of models and of the production of output plots and also their interactive exploration. As recalls \noun{Romain Reuillon} (Interview on the 11/04/2017, see \ref{app:sec:interviews}), the OpenMole platform which had to integrate this additional layer was at its beginning at this stage, and is still not mature enough today, since the state of the art of such practices is in full construction and changes significantly regularly~\cite{holzinger2014knowledge}.
}{
L'usage d'applications interactives pour la fouille de données a des avantages non discutables, tel qu'une familiarisation avec la structure des données par une vue d'ensemble qui serait beaucoup plus laborieuse voire impossible autrement. C'est la même idée sous-jacente qui justifie l'interactivité pour l'exploration préliminaire des modèles multi-agents intégrée à des plateformes comme Netlogo~\cite{wilensky1999netlogo} ou Gamma~\cite{drogoul2013gama}. Un objectif similaire est implicite dans~\cite{rey2015plateforme}, c'est-à-dire une intégration complète de l'exploration fine des modèles et de la production des graphes de sortie ainsi que leur exploration interactive. Comme le rappelle \noun{Romain Reuillon} (Entretien du 11/04/2017, voir \ref{app:sec:interviews}), la plateforme OpenMole qui devait accueillir cette couche supplémentaire était à ses débuts à l'époque et n'est toujours pas suffisamment mature aujourd'hui, puisque l'état de l'art de telles pratiques est en pleine construction et bouleversements réguliers~\cite{holzinger2014knowledge}.
}

\bpar{
Difficulties regarding reproducibility, which are particularly important for us here, are recurrent and far from being solved. Indeed, we must clearly situate the position of these tools and methods as a preliminary cognitive help\footnote{That we do not judge as superficial since we use them at least twice in the following, see below and also \ref{sec:energyprice}.}, but not often as allowing the production of final results: when parameters or dimensions are multiplied, the export of a plot is most often disconnected from the complete information that lead to its production. In a similar way, the use of integrated notebooks such as Jupyter, allowing to integrate analyses and redaction of the reporting, can become dangerous as we indeed can come back on a script, test different values of a parameter, and loose the values which produced a given plot. The use of versioning can partly be a solution but which is often heavy.
}{
Des difficultés au regard de la reproductibilité, qui nous concernent particulièrement ici, sont récurrentes et loin d'être résolues. En effet, il faut bien situer la position de ces outils et méthodes comme une aide cognitive préliminaire\footnote{Que nous ne jugeons pas superficielle puisque nous les mobilisons au moins deux fois par la suite, voir ci-dessous ainsi que \ref{sec:energyprice}.}, mais peu souvent comme permettant la production de résultats finaux : lorsque les paramètres ou dimensions se multiplient, l'export d'un graphe est bien souvent déconnecté de l'information complète ayant conduit à sa production. De la même manière, l'utilisation de notebooks intégrés tel Jupyter, permettant d'intégrer analyses et rédaction du compte-rendu, peut devenir dangereux car on peut justement revenir sur un script, tester différentes valeurs d'un paramètre, et perdre les valeurs qui avaient produit un graphe donné. L'utilisation de versioning peut être une solution partielle mais souvent lourde.
}

\bpar{
Ideally, any interactive software allowing to export results should simultaneously export a script or an exact and usable description which allows to exactly arrive at this point starting from raw data. Most applications for interactive exploration of spatio-temporal data are from this point of view relatively immature scientifically, since even in the case where they are totally honest and transparent on the analyses presented to the user, what is unfortunately not the rule, the progressive exploration steps are not reproducible and the method to extract characteristics is thus relatively random. By pushing the reasoning, their use would more reveal the acceptance of a weakness and a lack of systematic methods accompanying the discovery of patterns in complex spatio-temporal data in an efficient way.
}{
Dans l'idéal, tout logiciel interactif permettant l'export de résultats devrait en même temps exporter un script ou une description exacte et utilisable permettant d'arriver exactement à ce point à partir des données brutes. La plupart des applications d'exploration interactives de données spatio-temporelles sont à ce regard relativement immatures scientifiquement, car même dans le cas où elles sont totalement honnêtes et transparentes sur les analyses présentées à l'utilisateur, ce qui n'est malheureusement pas la règle, les tâtonnements d'exploration progressive ne sont pas reproductibles et la méthode d'extraction de caractéristiques est ainsi relativement aléatoire. En poussant le raisonnement, leur utilisation révélerait plutôt l'aveu d'une faiblesse d'un manque de méthodes systématiques accompagnant la découverte de motifs dans des données spatio-temporelles complexes de manière efficace.
}

\bpar{
Through a visionary plea, \noun{Banos} had already warned against ``the dangers of the jungle'' of data in~\cite{banos2001propos}, when he very appropriately highlights that interactive exploration must be accompanied of adapted local indicators, but even more of automatized exploration tools and of criteria for evaluating the choices made and the patterns discovered by the user. We come back again to the idea of an integrated platform of which OpenMole may be a precursor. The combination of human cognitive abilities to machine processing, in particular for computer vision problems, opens opportunities for novel discoveries, even more through a collective use as witnesses the Galaxy Zoo~\cite{2010AEdRv...9a0103R}\footnote{The principle rejoins the one of \emph{citizen science}, by making voluntaries from outside of the scientific community participate to tasks requiring cognition but no scientific knowledge: image classification, in order to train supervised algorithm, is the initial example of the Galaxi Zoo for the form of galaxies.}. Results of a crowdsourcing of human cognition can enter a competition with most advanced automated techniques \cite{10.1371/journal.pone.0178165} for the example of comparing spatial maps.
}{
Par un plaidoyer visionnaire, \noun{Banos} avait déjà mis en garde contre ``les dangers de la jungle'' des données dans~\cite{banos2001propos}, quand il souligne très justement que l'exploration interactive doit nécessairement se doubler d'indicateurs locaux adaptés, mais surtout d'outils d'exploration automatisés et de critère d'évaluation des choix faits et des motifs découverts par l'utilisateur. Nous revenons encore à l'idée d'une plateforme intégrée dont OpenMole pourrait être un précurseur. La combinaison des capacités cognitives humaines au traitement machine, notamment pour des problèmes de vision par ordinateur, ouvre des possibilités de découvertes inédites, encore plus via une utilisation collective comme en témoigne le Galaxy Zoo~\cite{2010AEdRv...9a0103R}\footnote{Le principe rejoint celui de \emph{citizen science}, en faisant participer des volontaires hors de la communauté scientifique à des tâches requérant cognition mais pas de connaissances scientifiques : la classification d'images, dans le but d'entraîner des algorithmes supervisés, est l'exemple initial du Galaxy Zoo pour la forme des galaxies.}. Les résultats d'un crowdsourcing de la cognition humaine peuvent rivaliser avec les techniques automatiques les plus avancées comme le montre~\cite{10.1371/journal.pone.0178165} pour l'exemple de la comparaison de cartes spatiales.
}

\bpar{
These possibilities must however not be overrated of used in the wrong context, and the questions of an efficient human-machine integration are indeed totally open. In the domain of geographic information visualization, \cite{pfaender2009spatialisation} introduces a specific semiology aiming at helping for the exploration of large heterogenous datasets, and experiments it on a specific application: it is a considerable advance towards an integrated platform and a sane and reproducible interactive exploration, the exploration directions answering to models based on cognitive sciences.
}{
Ces possibilités ne doivent cependant pas être sur-estimées ou utilisées à mauvais escient, et les questions d'intégration efficiente homme-machine sont d'ailleurs totalement ouvertes. Dans le domaine de la visualisation de l'information géographique, \cite{pfaender2009spatialisation} introduit une sémiologie spécifique visant à favoriser l'exploration de grands jeux de données hétérogènes, et l'expérimente sur une application spécifique : il s'agit d'une avancée considérable vers une plateforme intégrée et une exploration interactive saine et reproductible, les directions d'exploration répondant à des modèles basés sur les sciences cognitives. 
}

\bpar{
Finally, the role of interactivity in scientific communication and vulgarization is explored by the Appendix~\ref{app:sec:mediationecotox}, which suggests the elaboration of games, including an interactive computer game, to facilitate the transmission of scientific concepts to the public. This shows that the development of these innovative practices goes beyond the only frame of data analysis.
}{
Enfin, le rôle de l'interactivité dans la communication et la vulgarisation scientifiques est exploré par l'Annexe~\ref{app:sec:mediationecotox}, qui suggère la mise en place de jeux, notamment un jeu informatique interactif, pour faciliter la transmission de concepts scientifiques au public. Cela nous montre que le développement de ces pratiques innovantes dépasse largement le seul cadre de l'analyse de données.
}

\subsubsection{How to put into practice}{Mise en application}

\bpar{
Again, reproducibility and transparency are central and crucial elements of contemporaneous science, linked to the practices of open science and open access. Several examples (see a recent in experimental economics in~\cite{camerer2016evaluating}) in different disciplines show the lack of reproducibility of experiments results, whereas it should lead to a falsification or to a confirmation of results. Falsification is a costly practice since it requires a certain investment to the detriment of one's own research~\cite{chavalarias2005nobel}. It could thus be made more efficient through an increased transparency. Tools specifically dedicated to a direct reproducibility, often allowed by opening, should increase the global performance of science. But open access has much broader impacts than on science itself: \cite{teplitskiy2017amplifying} show an increased transfer of scientific knowledge towards society in the case of open articles, in particular through intermediaries such as Wikipedia. 
}{
Encore une fois, la reproductibilité et la transparence sont des éléments essentiels incontournables de la science contemporaine, liés aux pratiques de science ouverte et d'accès ouvert. Beaucoup d'exemples (voir un récent en économie expérimentale dans~\cite{camerer2016evaluating}) dans diverses disciplines montrent le manque de reproductibilité des résultats des expériences, alors que celle-ci doit pouvoir conduire à une falsification ou à une confirmation de ces résultats. La falsification est une pratique coûteuse car demandant un certain investissement au détriment de sa propre recherche~\cite{chavalarias2005nobel}. Elle pourrait ainsi être rendue plus efficiente grâce à une transparence augmentée. Des outils spécialement dédiés à une reproductibilité directe, souvent permise par l'ouverture, devraient accroître la performance globale de la science. Mais l'accès ouvert a des impacts bien plus larges que la science elle-même : \cite{teplitskiy2017amplifying} montrent un transfert des connaissances scientifiques accru vers la société dans le cas d'articles ouverts, notamment par des intermédiaires comme Wikipedia.
}

\bpar{
The development and systematisation of standards and good practices, in a joint way on the different issues raised, is a necessary condition for a scientific rigor which should be uniform for all the spectrum of existing disciplines. We construct for example tools facilitating the flow of scientific production, these being detailed in Appendix~\ref{app:workflow}. For example, for computational sciences, we already evoked the potentialities of using \texttt{git} which in fact extend without any constraint of discipline or of type of research of the good adaptations are introduced. The precise following of all the stages of a project, kept as a history which offers the possibility to come back to any step at any time, but also to work in a collaborative way, more or less in parallel depending on needs by using branches, is an example of service given by this tool. An example of good practices of use is given by~\cite{10.1371/journal.pcbi.1004947}.
}{
Le développement et la systématisation de standards et de bonnes pratiques, de manière conjointe sur les différentes problématiques évoquées, est une condition nécessaire à une rigueur scientifique qui devrait être uniforme au travers de l'ensemble des disciplines existantes. Nous construisons par exemple des outils facilitant le flot de production scientifique, ceux-ci étant détaillés en Appendice~\ref{app:workflow}. Par exemple, pour les sciences computationnelles, on a déjà évoqué les potentialités de l'utilisation de \texttt{git} qui s'étendent en fait sans contrainte de disciplines ni de types de recherche si les bonnes adaptations sont introduites. Le suivi précis de l'ensemble des étapes d'un projet, gardé en historique offrant la possibilité de revenir à n'importe laquelle à tout moment, mais aussi de travailler de façon collaborative, plus ou moins parallèlement selon les besoins en utilisant les branches, est un exemple de service fourni par cet outil. Un exemple de bonnes pratiques d'utilisation est donné par~\cite{10.1371/journal.pcbi.1004947}.
}

\bpar{
More generally, computational sciences necessitate the adoption of certain standards and practices to ensure a good reproducibility, and these remain mostly to be developed: \cite{wilson2017good} give first directions. Regarding the quality of data, numerous efforts are done to introduce frameworks to standardize data: for example~\cite{10.1371/journal.pone.0178731} describe a conceptual framework aiming at solving recurrent problems linked to the quality of biodiversity data (such as for example evaluating measures judging the possible use of a dataset for a given problem). New perspectives are opened for future frameworks for data processing which would be intrinsically open and reproducible, with the development of new techniques such as \emph{blockchain}\footnote{The \emph{blockchain} consists in the distribution of a transaction graph between users, these being validated (in the classical historical frame of type \emph{proof-of-work}) by the resolution of inverse cryptographic problems through brute force, by agents called miners, essential to the robustness of the ecosystem.}, as proposed by~\cite{2017arXiv170706552}.
}{
Plus généralement, les sciences computationnelles nécessitent l'adoption de certains standards et pratiques pour assurer une bonne reproductibilité, et ceux-ci restent majoritairement à développer : \cite{wilson2017good} donnent des premières pistes. Concernant la qualité des données, de nombreux efforts sont faits pour introduire des cadres de standardisation des données : par exemple~\cite{10.1371/journal.pone.0178731} décrivent un cadre conceptuel visant à guider la résolution de problèmes récurrents liés à la qualité des données de biodiversité (comme par exemple évaluer des mesures jugeant de l'usage possible d'un jeu de données pour un problème donné). De nouvelles perspectives s'ouvrent pour des futurs cadres de traitement de données intrinsèquement ouverts et reproductibles, avec le développement de nouvelles techniques comme le \emph{blockchain}\footnote{Le \emph{blockchain} consiste en la distribution d'un graphe de transactions entre utilisateurs, celles-ci étant validées (dans le cadre historique classique de type \emph{proof-of-work}) par la résolution de problèmes cryptographiques inverses par force brute, par des agents appelés mineurs, essentiels à la robustesse de l'écosystème.}, comme proposé par~\cite{2017arXiv170706552}.
}

\subsection{Opening data}{Ouverture des données}


\bpar{
The access to data is also a crucial point for reproducibility, and without taking too much time since it would imply developments on the definition, the philosophy, legal issues etc. which are research subjects in themselves, we give perspectives on opportunities offered by a systematic opening of research data. In geography, \emph{data papers} are a totally inexistent practice, and the rule is more to jealously keep the hand on a dataset produced, capitalizing on the fact to be the only one to have access to it\footnote{To the best of our knowledge there is no work quantifying the proportion of open data on all the data produced in geography. It could be the object of a work in quantitative epistemology applying similar techniques to the ones developed in chapter~\ref{ch:modelinginteractions}. The difficulty to find open data, compared to the frequency of publications in the domaines concerned, suggests a validity at least qualitative of this fact.}.
}{
L'accès aux données est également un point crucial pour la reproductibilité, et sans nous y attarder car cela impliquerait des développements sur la définition, la philosophie, le droit des données etc. qui sont des sujets de recherche en eux-mêmes, nous donnons des perspectives sur les opportunités offertes par une ouverture systématique des données en recherche. En géographie, les \emph{data paper} sont une pratique inexistante, et la règle est plutôt de garder la main jalousement sur un jeu produit, capitalisant sur le fait d'être le seul à y avoir accès\footnote{Il n'existe à notre connaissance pas de travail quantifiant la proportion de données ouvertes sur l'ensemble des données produites en géographie. Cela pourrait être l'objet d'un travail d'épistémologie quantitative appliquant des techniques similaires à celles développées en chapitre~\ref{ch:modelinginteractions}. La difficulté à trouver des données ouvertes, comparée à la fréquence des publications dans les domaines concernés, suggère une validité au moins qualitative de ce fait.}.
}

\bpar{
It is evident that the quality and quantity of knowledge produced will necessarily be greater if a dataset if publicly open, since at least the same will be obtained, and we can expect a use by other domains, other methods, and thus a higher richness\footnote{It is possible to argue that the system of scientific production is complex, and that an increased monetization, competition or privatization of research could be part of a research ecosystem which outputs could be judged of quality depending on the indicators chosen. These considerations are relevant, but out of our reach since corresponding to a work in anthropology and sociology of sciences. We postulate here this principle, and consider it as a subjective scientific positioning.}.
}{
Il est évident que la qualité et quantité des connaissances produites sera nécessairement plus grande si un jeu de données est publiquement ouvert, puisqu'au moins la même chose sera obtenue, et on peut s'attendre à une prise en main par d'autres domaines, d'autres méthodes, et donc à une plus grande richesse\footnote{Il est possible d'argumenter que le système de production scientifique est complexe, et qu'une monétarisation, compétition ou privatisation accrue de la recherche peut faire partie d'un écosystème de recherche dont les sorties pourront être jugées de qualité selon les indicateurs choisis. Ces considérations sont pertinentes, mais hors de notre portée puisque relevant d'un travail en anthropologie et sociologie des sciences. Nous postulons ici ce principe, et le considérons comme une position scientifique subjective.}.
}

\bpar{
Closing data will more induce negative effects, such as for example lost time to recode a vectorial database given only as a map in a paper. The argument of the time spent to justify closing is absurd, since on the contrary, seeing data as a component of knowledge in itself (see the knowledge framework in~\ref{sec:knowledgeframework}), the time spent should imply more citations, thus more use, what necessarily goes through the opening for data. Similarly, what king of logic, even the same absurd logic of a property of knowledge, leads geographers to insert a copyright on all their maps but also their figures, up to a copyright for a simple histogram, which would have naturally declined the offer if we could have interrogated it, honest of simplicity ?
}{
La fermeture induira plutôt des effets négatifs, comme par exemple du temps perdu à recoder un base vectorielle donnée uniquement sous forme de carte dans un article. L'argument du temps passé comme justification à la fermeture est absurde, puisqu'au contraire, en voyant les données comme une composante à part entière de la connaissance (voir le cadre de connaissances en~\ref{sec:knowledgeframework}), le temps passé doit impliquer plus de citations, donc plus d'utilisation, ce qui passe nécessairement par l'ouverture pour des données. De même, quelle logique, sinon la même absurde de propriété des connaissances, pousse les géographes à insérer un copyright sur l'ensemble de leurs cartes mais aussi leurs figures, jusqu'à un copyright pour un simple histogramme qui s'en serait bien passé si on avait pu l'interroger, honnête de simplicité ?
}

\bpar{
Our experience in reviewing papers leads us to truly worry on the value given to data opening by the authors: after a tenth of articles, including journals claiming as a priority and requirement the total opening of data and models, articles among which a single one is only partly open and all the others imply to believe on word the results presented (whereas one of the aim of the review is to avoid the cognitive biases which one or more humans necessarily have, through a cross-validation which must be done on raw results and not on interpretations containing these biases), it is difficult to believe that profound mutations of practices are not necessary.
}{
L'expérience d'évaluation d'articles nous induit à réellement nous inquiéter sur la valeur donnée à l'ouverture des données par les auteurs : au bout d'une dizaine d'articles, incluant des journaux affichant comme priorité et pré-requis l'ouverture totale des données et modèles, dont un seul est seulement partiellement ouvert et l'ensemble des autres implique de croire sur parole les résultats présentés (alors qu'un des but de la revue est de contourner les biais cognitifs qu'un ou des humains ont forcément par une validation croisée qui doit se faire sur les résultats bruts et non des interprétations contenant ces biais), il est difficile de croire que des mutations profondes des pratiques ne sont pas nécessaires.
}

\bpar{
But following the word of Framasoft\footnote{Network for the promotion of the free software, \url{https://framasoft.org/}}, ``the road is long but the path is free'', perspectives are numerous for an evolution which slowness is not inevitable. The Cybergeo journal, pioneer in the opening practices in social sciences (first journal fully online, first journal to launch a \emph{model paper} section), launches in 2017 a section \emph{data papers}\footnote{Which index is available at \url{https://cybergeo.revues.org/28545}. The first paper is~\cite{swerts2017database}, that we indeed use in~\ref{sec:lutecia}.} aiming at fostering the development of data sharing and opening in geography.
}{
Mais en suivant l'adage de Framasoft\footnote{Réseau pour la promotion du logiciel libre, \url{https://framasoft.org/}}, ``la route est longue mais la voie est libre'', les perspectives sont nombreuses pour une évolution dont la lenteur n'est pas inéluctable. Le journal Cybergéo, pionnier des pratiques d'ouverture en sciences sociales (première revue entièrement électronique, première revue à lancer une rubrique de \emph{model papers}), lance en 2017 une rubrique \emph{data papers}\footnote{Dont l'index est disponible à \url{https://cybergeo.revues.org/28545}. Le premier article est~\cite{swerts2017database}, que nous utilisons d'ailleurs en~\ref{sec:lutecia}.} visant à inciter le développement du partage de données et de l'ouverture en géographie.
}

\bpar{
There are still grey areas on which it is impossible today to have perspectives, in particular legal issues related to data. We have an example in the analyses we develop: bibliographical data are obtained at the price of a blocking war with Google and a considerable technical effort to win it (see~\ref{sec:quantepistemo} and \ref{app:sec:cybergeo}).
}{
Il reste des zones grises sur lesquelles il est impossible aujourd'hui d'avoir des perspectives, notamment le droit des données. Nous avons un exemple dans les analyses que nous développerons : les données bibliographiques sont obtenues au prix d'une guerre de blocage par Google et un effort technique considérable pour la gagner (voir~\ref{sec:quantepistemo} et \ref{app:sec:cybergeo}).
}

\bpar{
The opening implies an engagement which is definitively a part of our positioning. It is the same idea which underlies the construction of the application \texttt{CybergeoNetworks}\footnote{Which approach and context are detailed in Appendix~\ref{app:sec:cybergeonetworks}. It is available online at \url{http://shiny.parisgeo.cnrs.fr/CybergeoNetworks}.}, which couples the tools presented in~\ref{sec:quantepistemo} with other complementary approaches of corpus analysis, with the aim to foster scientific reflexivity, and to put this open tool at the disposal of independent editors, to emancipate from the new total control of big publishers which, searching for a new model to secure their profits, bet on the selling of meta-content and its analysis. Fortunately, the recent numeric law in France has won the conflict against their revendication of an exclusive right on full text mining.
}{
L'ouverture implique un engagement qui fait résolument partie de nos positionnements. C'est la même idée qui soutient la construction de l'application \texttt{CybergeoNetworks}\footnote{Dont la démarche et le contexte sont détaillés en Annexe~\ref{app:sec:cybergeonetworks}. Elle est disponible en ligne à \url{http://shiny.parisgeo.cnrs.fr/CybergeoNetworks}.}, qui couple les outils présentés en~\ref{sec:quantepistemo} avec d'autres approches complémentaires d'analyse de corpus, dans le but d'encourager la réflexivité scientifique, et de mettre cet outil ouvert à la disposition d'éditeurs indépendants, pour s'émanciper de la nouvelle main mise des géants de l'édition qui à la recherche d'un nouveau modèle pour sécuriser leur profits parient sur la vente de méta-contenu et de son analyse. Heureusement, la récente loi numérique en France a gagné le bras de fer contre leur revendication d'un droit exclusif sur la fouille de texte complets.
}


\subsection{Illustration by an empirical study}{Illustration par une étude empirique}

\bpar{
We propose now to develop a concrete example of an empirical study illustrating the last points shown above and allowing us a progressive entry into our problematic. In the case of road traffic in Ile-de-France, we proceed to a data collection where there is no open source. We also construct an application allowing its interactive exploration.
}{
Nous proposons à présent de développer un exemple concret d'étude empirique illustrant les derniers points relevés ci-dessus et nous permettant une entrée progressive dans notre problématique. Dans le cas du trafic routier en Ile-de-France, nous menons une collecte d'un jeu de données là où il n'existe pas de source ouverte. Nous mettons également en place une application permettant son exploration interactive.
}

\bpar{
We have developed in~\ref{sec:networkterritories} the concept of daily mobility as playing a key role in the interaction processes between transportation networks and territories, at a scale that we designated as microscopic. It is furthermore candidate to the mobilization of co-evolutive dynamics, as suggest the effect of localizations on congestion and reciprocally.
}{
Nous avons développé en~\ref{sec:networkterritories} le concept de mobilité quotidienne comme jouant un rôle clé dans les processus d'interaction entre réseaux de transport et territoires, à une échelle que nous avons désignée par microscopique. Il est de plus candidat à la mobilisation de dynamiques co-évolutives, comme le suggère l'effet des localisations sur la congestion et réciproquement.
}

\bpar{
Here, mobility will be captured by the traffic flow, and the co-evolution operates between network properties (congestion) and localization of agents. We study more precisely the hypothetic equilibrium of traffic flows, answering indirectly to issues that we detail below.
}{
Ici, la mobilité sera captée par le flux de trafic, et la co-évolution s'opère entre propriétés du réseau (congestion) et localisation des agents. Nous nous intéresserons plus particulièrement à l'équilibre hypothétique des flux de trafic, répondant indirectement à des problématiques que nous détaillons ci-dessous. 
}

\subsubsection{Context}{Contexte}

\bpar{
Traffic Modeling has been extensively studied since seminal work by~\cite{wardrop1952road} : economical and technical elements at stake justify the need for a fine understanding of mechanisms ruling traffic flows at different scales. Many approaches with different purposes coexist today, of which we can cite dynamical micro-simulation models, generally opposed to equilibrium-based techniques.
}{
La modélisation du trafic a été largement étudiée depuis les travaux séminaux de Wardrop (\cite{wardrop1952road}) : les enjeux économiques et techniques justifient le besoin d'une compréhension fine des mécanismes régissant les flux de trafic à différentes échelles. Des approches aux objectifs différents coexistent aujourd'hui, parmi lesquels on trouve par exemple les modèles dynamiques de micro-simulation, généralement opposés aux techniques se basant sur l'équilibre.
}

\bpar{
Whereas the validity of micro-based models has been largely discussed and their application often questioned, the literature is relatively poor on empirical studies assessing the stationary equilibrium assumption in the Static User Equilibrium (SUE) framework.
}{
Tandis que la validité des modèles microscopiques a été étudiée de façon conséquente et leur application souvent questionnée, la littérature est relativement pauvre en études empiriques testant l'hypothèse d'équilibre stationnaire du cadre de l'Equilibre Utilisateur Statique (EUS).
}

\bpar{
Various more realistic developments have been documented in the literature, such as Dynamic Stochastic User Equilibrium (DSUE) (see e.g. a description by~\cite{han2003dynamic}). An intermediate between static and stochastic frameworks is the Restricted Stochastic User Equilibrium, for which route choice sets are constrained to be realistic (\cite{rasmussen2015stochastic}).
}{
De nombreux développements plus précis dans les hypothèses de modélisation ont été documentés dans la littérature, tels l'Equilibre Utilisateur Dynamique Stochastique (EUDS) (voir pour une description par example~\cite{han2003dynamic}). À un niveau intermédiaire entre les cadres statiques et stochastiques se trouve l'Equilibre Utilisateur Stochastique Restreint, pour lequel les choix d'itinéraire des utilisateurs sont contraints à un ensemble d'alternatives réalistes (\cite{rasmussen2015stochastic}).
}

\bpar{
Extensions that incorporate user behavior with choice models have more recently been proposed, such as~\cite{zhang2013dynamic} taking into account both the influence of road pricing and congestion on user choice with a Probit model. Relaxations of other restricting assumptions such as pure user utility maximization have been also introduced, such as the Boundedly Rational User Equilibrium described by~\cite{mahmassani1987boundedly}. In this framework, user have a range of satisfying utilities and equilibrium is achieved when all users are satisfied. It produces more complex features such as the existence of multiple equilibria, and allows to account for specific stylized facts such as irreversible network change as developed by~\cite{guo2011bounded}.
}{
D'autres extensions prenant en compte le comportement de l'utilisateur via des modèles de choix ont été proposé plus récemment, comme~\cite{zhang2013dynamic} qui inclut à la fois l'influence de la tarification routière et de la congestion sur le choix avec un modèle Probit. La relaxation d'autres hypothèses restrictives comme la maximisation pure de l'utilité par l'utilisateur ont aussi été introduites, tels l'Equilibre Utilisateur Borné décrit par~\cite{mahmassani1987boundedly}. Dans ce cadre, l'utilisateur est satisfait si sa fonction d'utilité rentre dans une plage de valeurs tolérables, et l'équilibre est achevé lorsque chaque utilisateur est satisfait. Les dynamiques résultantes sont plus complexes comme révélé par l'existence d'équilibres multiples, et permet de rendre compte de faits stylisés spécifiques comme des évolutions irréversibles du réseau comme développé par~\cite{guo2011bounded}.
}

\bpar{
Other models for traffic assignment, inspired from other fields have also recently been proposed : in~\cite{puzis2013augmented}, an extended definition of betweenness centrality combining linearly free-flow betweenness with travel-time weighted betweenness yield a high correlation with effective traffic flows, acting thus as a traffic assignment model. It provides direct practical applications such as the optimization of traffic monitors spatial distribution.
}{
D'autres modèles d'attribution de trafic inspirés d'autres domaines ont également été plus récemment proposés : dans~\cite{puzis2013augmented}, une définition étendue de la centralité de chemin qui combine linéairement le centralité des flots non-contraints avec une centralité pondérée par le temps de parcours permet d'obtenir une forte corrélation avec les flux de trafic effectifs, fournissant ainsi un modèle d'attribution de trafic. Cela fournit également des applications pratiques comme l'optimisation de la distribution spatiale des capteurs de trafic.
}

\bpar{
Despite all these developments, some studies and real-world applications still rely on Static User Equilibrium. Parisian region e.g. uses a static model (MODUS) for traffic management and planning purposes. \cite{leurent2014user} introduce a static model of traffic flow including parking cruising and parking lot choice: it is legitimate to ask, specifically at such small scales, if the stationary distribution of flows is a reality. An example of empirical investigation of classical assumptions is given in~\cite{zhu2010people}, in which revealed route choices are studied. Their conclusions question ``Wardrop’s first principle'' implying that users choose among a well-known set of alternatives.
}{
Malgré ces nombreux développements, de nombreuses études et applications concrètes se basent sur l'Equilibre Utilisateur Statique. La région parisienne utilise par exemple un modèle statique (MODUS) pour gérer et planifier le trafic. \cite{leurent2014user} introduit un modèle statique de flots qui inclut les recherches locales et le choix du parking : dans ce cas particulier à de si faibles échelles, la stationnarité de la distribution des flux a encore moins de chances d'être une réalité. Un example d'exploration empirique des hypothèses classiques est donné par~\cite{zhu2010people}, pour lequel les choix d'itinéraires révélés sont étudiés. Les conclusions questionnent le ``premier principe de Wardrop'' qui postule que les utilisateurs choisissent parmi un ensemble d'alternatives parfaitement connu.
}

\bpar{
In the same spirit, we investigate the possible existence of the equilibrium in practice. More precisely, SUE assumes a stationary distribution of flows over the whole network. This assumption stays valid in the case of local stationarity, as soon as time scale for parameter evolution is considerably greater than typical time scales for travel. The second case which is more plausible and furthermore compatible with dynamical theoretical frameworks, is here tested empirically. The objective of this development is thus to study at a large scale the relations between networks and territories, through the intermediate of traffic flows which will be carried by the network but generated by territorial patterns.
}{
Dans le même esprit, nous proposons d'étudier l'existence empirique de l'équilibre statique. Plus précisément, l'EUS suppose une distribution stationnaire des flux sur l'ensemble du réseau. Cette hypothèse reste valable dans le cas d'une stationnarité locale, tant que l'échelle temporelle d'évolution des paramètres est considérablement plus grande que les échelles typiques de voyage. Le second cas qui est plus plausible et de plus compatible avec les cadres théoriques dynamiques est testé ici. L'objectif de ce développement est ainsi d'étudier à une grande échelle les relations entre réseaux et territoires, par l'intermédiaire des flux de trafic qui sont portés par le réseau mais générés par les motifs territoriaux.
}

\bpar{
In a first time, data collection procedure and dataset are described; we present then an interactive application for the interactive exploration of the dataset aimed to give intuitive insights into data patterns; we present then results of various quantitative analyses that give convergent evidence for the non-stationarity of traffic flows.
}{
Dans un premier temps, la procédure de collection de données ainsi que le jeu de données sont décrits ; nous présentons ensuite une application interactive pour l'exploration du jeu de données, dans le but de fournir une intuition sur les motifs présents ; puis nous donnons divers résultats d'analyses quantitatives allant dans le sens d'indices convergents pour une non-stationnarité des flux de trafic.
}

\subsubsection{Dataset}{Jeu de données}

\paragraph{Dataset construction}{Construction du jeu de données}

\bpar{
We propose to work on the case study of Parisian Metropolitan Region. An open dataset was constructed for highway links within the dense urban core\footnote{Mostly Paris and the \emph{Petite Couronne départements}.}, collecting public real-time open data for travel times (available at \url{www.sytadin.fr}). As stated by~\cite{bouteiller2013open}, the availability of open datasets for transportation is far to be the rule, and we contribute thus to a data opening by the construction of our dataset. Our data collection procedure consists in the following simple steps, executed each two minutes by a \texttt{python} script :
\begin{itemize}
\item fetch raw webpage giving traffic information
\item parse html code to retrieve traffic links id and their corresponding travel time
\item insert all links in a \texttt{sqlite} database with the current timestamp.
\end{itemize}
}{
Nous proposons de travailler sur l'étude de cas de la métropole parisienne. Un jeu de données ouvert a été construit, comprenant les liens autoroutiers du coeur urbain dense\footnote{Majoritairement Paris et les départements de la petite couronne.}, par collecte des données publiques en temps réel des temps de parcours (disponible sur \url{www.sytadin.fr}). Comme rappelé par~\cite{bouteiller2013open}, la disponibilité de jeux de données ouverts pour les transports est loin d'être la règle, et nous contribuons ainsi à une ouverture par la construction de notre jeu de données. La procédure de collecte de données consiste en les points suivants, exécutés toutes les deux minutes par un script \texttt{python} :
\begin{itemize}
\item récupération de la page web brute donnant les informations de trafic
\item parsing du code html afin de récupérer les identifiants des liens de trafic et les temps de parcours correspondants
\item insertion des liens dans une base \texttt{sqlite} avec le temps courant.
\end{itemize}
}

\bpar{
The automatized data collection script continues to enrich the database as time passes, allowing future extensions of this work on a larger dataset and a potential reuse by scientists or planners. The latest version of the dataset is available online as a sqlite format under a Creative Commons License\footnote{On the dataverse at the link \url{http://dx.doi.org/10.7910/DVN/X22ODA}.}.
}{
Le script automatisé de collection des données continue d'enrichir la base au fur et à mesure du temps, permettant des développements futurs de ce travail sur un jeu de données plus large, et une réutilisation potentielle pour des travaux scientifiques ou opérationnels. La dernière version du jeu de données au format sqlite est disponible en ligne sous une Licence \emph{Creative Commons}\footnote{Sur le dataverse au lien \url{http://dx.doi.org/10.7910/DVN/X22ODA}.}.
}

\paragraph{Data description}{Description des données}

\bpar{
A time granularity of 2 minutes was obtained for a three months period (February 2016 to April 2016 included)\footnote{As we will work at the intra-day temporal scale, we do not need a longer dataset in time to obtain significant conclusions as we will see in the following}. Spatial granularity is in average 10km, as travel times are provided for major links. The dataset contains 101 links. Raw data we use is effective travel time, from which we can construct travel speed and relative travel speed, defined as the ratio between optimal travel time (travel time without congestion, taken as minimal travel times on all time steps) and effective travel time. Congestion is constructed by inversion of a simple BPR function\footnote{It is a function linking speed to congestion in the link, largely used in transportation engineering~\cite{branston1976link}.} with exponent 1 as done by~\cite{barthelemy2016global}, i.e. we take $c_i = 1 - \frac{t_{i,min}}{t_i}$ with $t_i$ travel time in link $i$ and $t_{i,min}$ minimal travel time.
}{
Une granularité de deux minutes a été obtenue pour une période de trois mois (de février 2016 à avril 2016 inclus)\footnote{Comme nous allons travailler à l'échelle temporelle intra-journalière, nous n'avons pas besoin d'un jeu de données plus étendu dans le temps pour avoir des conclusions significatives comme nous le verrons par la suite.}. La granularité spatiale (la distance moyenne entre les centroïdes des liens) est en moyenne de 10km, les temps de trajet étant fournis pour les liens majeurs. Le jeu de données contient 101 liens. La variable brute utilisée est le temps de trajet effectif, à partir duquel il est possible de construire la vitesse de trajet et la vitesse relative de trajet, définie comme le rapport entre temps de trajet optimal (temps de trajet sans congestion, pris comme le temps minimal sur l'ensemble des pas de temps) et le temps de trajet effectif. La congestion est calculée par inversion d'une fonction BPR\footnote{Il s'agit d'une fonction permettant de relier vitesse à congestion dans un lien, largement utilisée en ingénierie des transports~\cite{branston1976link}.} simple avec exposant 1 comme il est fait par~\cite{barthelemy2016global}, i.e. en prenant $c_i = 1 - \frac{t_{i,min}}{t_i}$ avec $t_i$ temps de trajet effectif dans le lien $i$ et $t_{i,min}$ temps de trajet minimal.
}

\subsubsection{Analysis of traffic patterns}{Analyse des motifs de trafic}

\paragraph{Visualization of spatio-temporal congestion patterns}{Visualisation des motifs spatio-temporels de congestion}

\bpar{
As our approach is fully empirical, a good knowledge of existing patterns for traffic variables, and in particular of their spatio-temporal variations, is essential to guide any quantitative analysis. Taking inspiration from an empirical model validation literature, more precisely \emph{Pattern-oriented Modeling} techniques introduced by~\cite{grimm2005pattern}, we are interested in macroscopic patterns at given temporal and spatial scales: the same way stylized facts are in that approach extracted from a system before trying to model it, we need to explore interactively data in space and time to find relevant patterns and associated scales.
}{
Notre approche étant entièrement empirique, une bonne connaissance des motifs existants pour les variables de trafic, en particulier de leur variations spatio-temporelles, est crucial pour guider toute analyse quantitative. En s'inspirant de la littérature étudiant la validation empirique de modèles, plus précisément les techniques de \emph{modélisation orientée-motifs} introduites par~\cite{grimm2005pattern}, nous nous intéressons aux motifs macroscopiques, par exemple les corrélations, à des échelles temporelles et spatiales données : d'une manière équivalente aux faits stylisés qui sont dans cette approche extraits d'un système avant de tenter de le modéliser, nous devons explorer les données de manière interactive dans le temps et l'espace afin d'identifier des motifs pertinents et les échelles associées. 
}

\begin{figure}
\includegraphics[width=\linewidth]{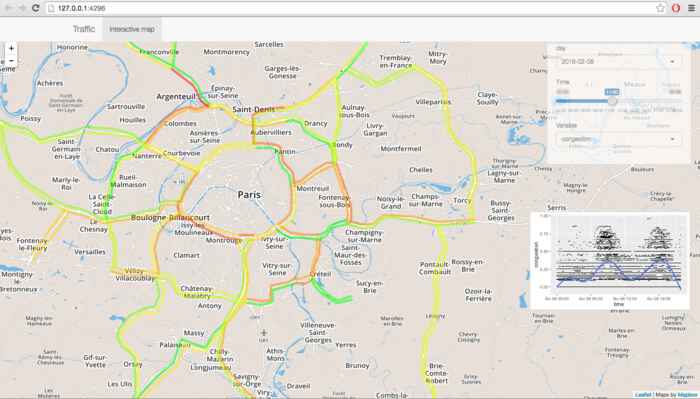}
\caption[Web-application for traffic data]{\textbf{Capture of the web-application.} It was developed to explore spatio-temporal traffic data for Parisian region. It is possible to select date and time (precision of 15min on one month, reduced from initial dataset for performance purposes). The inserted plot summarizes congestion patterns on the current day, by giving as a function of time all values (black dots) and their smoothing (blue curve).\label{fig:transportationequilibrium:fig-1}}
\end{figure}

\bpar{
We implemented therefore an interactive web-application for data exploration using \texttt{R} packages \texttt{shiny} and \texttt{leaflet}\footnote{Source code for the application and analyses is available on project open repository at \url{https://github.com/JusteRaimbault/TransportationEquilibrium}.}. It allows dynamical visualization of congestion among the whole network or in a particular area when zoomed in. The application is accessible online at \texttt{http://shiny.parisgeo.cnrs.fr/transportation}. A screenshot of the interface is presented in Figure~\ref{fig:transportationequilibrium:fig-1}.
}{
Une application web interactive a ainsi été implémentée pour explorer les données, à l'aide des packages \texttt{R} \texttt{shiny} et \texttt{leaflet}\footnote{Le code source de l'application et des analyses est disponible sur le dépôt ouvert du projet à \url{https://github.com/JusteRaimbault/TransportationEquilibrium}.}. L'application permet une visualisation dynamique des motifs de congestion sur l'ensemble du réseau ou dans une zone particulière grace au zoom. L'application est accessible en ligne à l'adresse \url{http://shiny.parisgeo.cnrs.fr/transportation}. La Figure~\ref{fig:transportationequilibrium:fig-1} présente une capture d'écran de l'interface.
}

\bpar{
Main conclusion from interactive data exploration is that strong spatial and temporal heterogeneity is the rule. The temporal pattern recurring most often, peak and off-peak hours is on a non-negligible proportion of days perturbed. In a first approximation, non-peak hours may be approximated by a local stationary distribution of flows, whereas peaks are too narrow to allow the validation of the equilibrium assumption. Spatially we can observe that no spatial pattern is clearly emerging. It means that in case of a validity of static user equilibrium, meta-parameters ruling its establishment must vary at time scales smaller than one day.
}{
La conclusion majeure de l'exploration interactive des données est qu'une grande hétérogénéité spatiale et temporelle est la règle. Le motif temporel le plus récurrent, la périodicité journalière des heures de pointe, est perturbée pour une proportion non négligeable de jours. En première approximation, les heures creuses peuvent être approchées par une distribution localement stationnaire des flux, tandis que la courte durée des heures de pointe suggère un système non-stationnaire sur ces périodes. Concernant l'espace, aucun motif spatial particulier n'émerge clairement. Cela signifie que dans le cas d'une validité de l'équilibre utilisateur statique, les méta-paramètres régissant son établissement doivent varier à des échelles temporelles plus courtes qu'un jour.
}

\bpar{
We postulate that traffic system must in contrary be far-from-equilibrium, especially during peak hours when critical phase transitions occur at the origin of traffic jams. 
}{
Nous postulons au contraire que le système de trafic est loin de l'équilibre, en particulier pendant les heures de pointe pendant lesquelles des transitions de phase critiques à l'origine des embouteillages émergent.
}

\paragraph{Spatio-temporal variability of travel paths}{Variabilité spatio-temporelle des trajets}

\begin{figure}
\includegraphics[width=\linewidth]{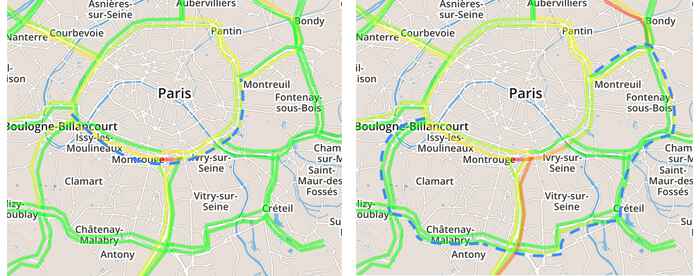}
\caption[Spatial variability of shortest paths]{\textbf{Spatial variability of a travel-time shortest path.} The trajectory of the shortest path is given in dotted blue. In an interval of only 10 minutes, between 11/02/2016 00:06 (left) and 11/02/2016 00:16 (right), the shortest path between \emph{Porte d'Auteuil} (West) and \emph{Porte de Bagnolet} (East), increases in effective distance of $\simeq 37$km (with an increase in travel time of only 6min), due to a strong disruption on the ring of Paris.\label{fig:transportationequilibrium:fig-2}}
\end{figure}

\bpar{
Following interactive exploration of data, we propose to quantify the spatial variability of congestion patterns to validate or invalidate the intuition that if equilibrium does exist in time, it is strongly dependent on space and localized. The variability in time and space of travel-time shortest paths is a first way to investigate flow stationarity from a game-theoretic point of view. Indeed, the static User Equilibrium is the stationary distribution of flows under which no user can improve its travel time by changing its route. A strong spatial variability of shortest paths at short time scales is thus evidence of non-stationarity, since a similar user will take a few time after a totally different route and not contribute to the same flow as a previous user. Such a variability is indeed observed on a non-negligible number of paths on each day of the dataset. We show in Figure~\ref{fig:transportationequilibrium:fig-2} an example of extreme spatial variation of shortest path for a particular Origin-Destination pair.
}{
A la suite de l'exploration interactive des données, nous proposons de quantifier la variabilité spatiale des motifs de congestion pour valider ou invalider l'intuition que si l'équilibre existe par rapport au temps, il est fortement dépendant de l'espace et localisé. La variabilité spatio-temporelle des plus courts chemins de trajet est une première façon d'étudier la stationnarité des flux d'un point de vue de théorie des jeux. En effet, l'Equilibre Utilisateur Statique est la distribution stationnaire des flux sous laquelle aucun utilisateur ne peut augmenter son temps de trajet en changeant son itinéraire. Une forte variabilité spatiale des plus courts chemins sur de courtes échelles spatiales révèle ainsi une non-stationnarité, puisqu'un même utilisateur prendra un chemin complètement différent après un court laps de temps et ne contribuera plus au même flux que précédemment. Une telle variabilité est en effet observée sur un nombre non-négligeable de chemins pour chaque jour du jeu de données. La figure~\ref{fig:transportationequilibrium:fig-2} montre un exemple de variation spatiale extrême d'un trajet pour une paire Origine-Destination particulière.
}

\bpar{
The systematic exploration of travel time variability across the whole dataset, and associated travel distance, confirms, as described in figure~\label{fig:transportationequilibrium:fig-3}, that travel time absolute variability has often high values of its maximum across OD pairs, up to 25 minutes with a temporal local mean around 10min. Corresponding spatial variability produces detours up to 35km.
}{
L'exploration systématique de la variabilité du temps de trajet sur l'ensemble du jeu de données, et des distances de trajet associées, confirme, comme présenté en figure~\label{fig:transportationequilibrium:fig-3}, que la variation absolue du temps de trajet présente fréquemment une forte variation de son maximum sur l'ensemble des paires O-D, jusqu'à 25 minutes avec une moyenne temporelle locale autour de 10 minutes. La variabilité spatiale correspondante entraine des détours allant jusqu'à 35km.
}

\begin{figure}
\includegraphics[width=\linewidth]{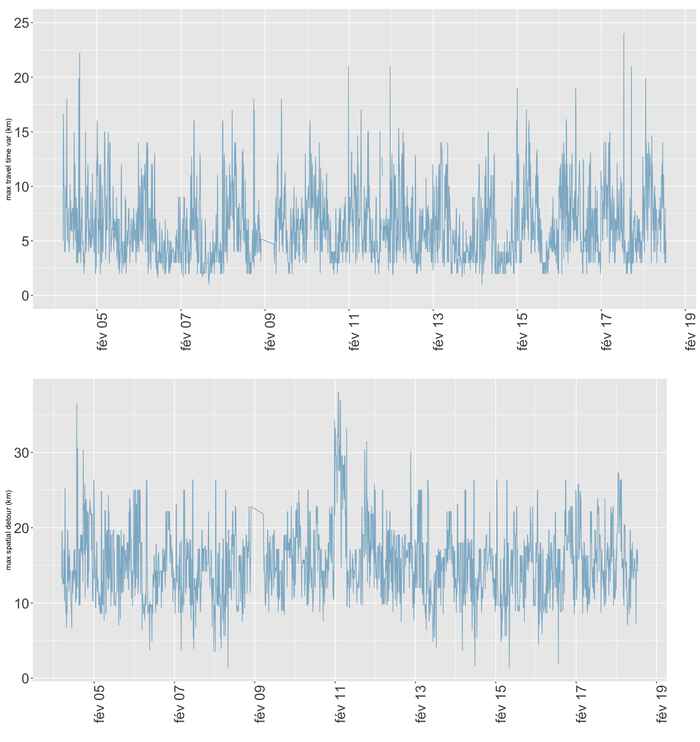}
\caption[Variability of travel times]{\textbf{Variability of travel times.} Maximal variability of travel time (\textit{Top}) in minutes and corresponding travel distance (\textit{Bottom}) maximal variability on a two weeks sample. We plot the maximal on all OD pairs of the absolute variability between two consecutive time steps. Peak hours imply a high time travel variability up to 25 minutes and a path length variability up to 35km.\label{fig:transportationequilibrium:fig-3}}
\end{figure}

\paragraph{Stability of network measures}{Stabilité des mesures de réseau}

\bpar{
The variability of potential trajectories observed in the previous section can be confirmed by studying the variability of network properties. In particular, network topological measures capture global patterns of a transportation network. Centrality and node connectivity measures are classical indicators in transportation network description as recalled in~\cite{bavoux2005geographie}. The transportation literature has developed elaborated and operational network measures, such as network robustness measures to identify critical links and measure overall network resilience to disruptions (an example among many is the Network Trip Robustness index introduced in~\cite{sullivan2010identifying}).
}{
La variabilité des trajectoires potentielles observée dans la section précédente peut être confirmée par l'étude de la variabilité des propriétés du réseau. En particulier, les mesures topologiques de réseau capturent les motifs globaux dans un réseau de transport. Les mesures de centralité et de connectivité des noeuds sont des indicateurs classiques pour la description des réseaux de transport comme rappelé par~\cite{bavoux2005geographie}. La littérature en transports a développé des mesures de réseau élaborées et opérationnelles, comme des mesures de robustesse pour identifier les liens critiques et mesurer la résilience globale du réseau aux perturbations (un exemple parmi d'autres est l'indice de \emph{Robustesse du Réseau Effective} introduit dans ~\cite{sullivan2010identifying}).
}

\bpar{
More precisely, we study the betweenness centrality of the transportation network, defined for a node as the number of shortest paths going through the node, i.e. by the equation
}{
Plus précisément, nous étudions la centralité de chemin du réseau de transport, défini pour un noeud comme le nombre de plus courts chemins passant par celui-ci, i.e. par l'équation
}

\begin{equation}
b_i = \frac{1}{N(N-1)}\cdot \sum_{o\neq d \in V}\mathbbm{1}_{i\in p(o\rightarrow d)}
\end{equation}

\bpar{
where $V$ is the set of network vertices of size $N$, and $p(o\rightarrow d)$ is the set of nodes on the shortest path between vertices o and d (the shortest path being computed with effective travel times). This index is more relevant to our purpose than other measures of centrality such as closeness centrality that does not include potential congestion as betweenness centrality does.
}{
où $V$ est l'ensemble des sommets du réseau de taille $N$, et $p(o\rightarrow d)$ est l'ensemble des noeuds sur le plus court chemin entre les sommets $o$ et $d$ (le plus court chemin étant calculé avec le temps de trajet effectif). Cette mesure de centralité est plus adaptée que d'autres dans notre cas, comme la centralité de proximité qui n'inclut pas la congestion potentielle comme la centralité de chemin.
}

\begin{figure}
\includegraphics[width=\linewidth]{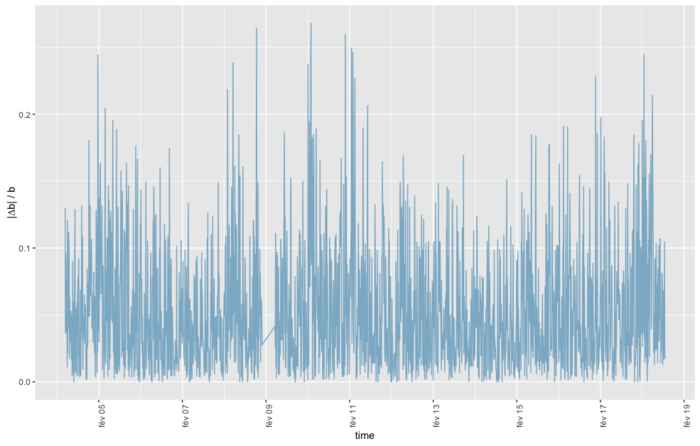}
\caption[Temporal stability of centrality]{\textbf{Temporal stability of maximal betweenness centrality.} We plot in time the normalized derivative of maximal betweenness centrality, that expresses its relative variations at each time step. The maximal value up to 25\% correspond to very strong network disruption on the concerned link, as it means that at least this proportion of travelers assumed to take this link in previous conditions should take a totally different path.\label{fig:transportationequilibrium:fig-4}}
\end{figure}

\bpar{
We show in Figure 4 the relative absolute variation of maximal betweenness centrality for the same time window than previous empirical indicators. More precisely it is defined by:
}{
Nous montrons en Fig.~\ref{fig:transportationequilibrium:fig-4} la variation relative absolue du maximum de la centralité de chemin, pour la même fenêtre temporelle que les indicateurs empiriques précédents. Plus précisément, elle est définie par :
}

\begin{equation}
\Delta b(t) = \frac{\left|\max_i (b_i(t + \Delta t)) - \max_i (b_i(t))\right|}{\max_i (b_i(t))}
\end{equation}

\bpar{
where $\Delta t$ is the time step of the dataset (the smallest time window on which we can capture variability). This absolute relative variation has a direct meaning : a variation of 20\% (which is attained a significant number of times as shown in Figure~\ref{fig:transportationequilibrium:fig-4}) means that in case of a negative variation, at least this proportion of potential travels have changed route and the local potential congestion has decrease of the same proportion. In the case of a positive variation, a single node has captured at least 20\% of travels.
}{
où $\Delta t$ est le pas de temps du jeu de données (la plus petite fenêtre temporelle sur laquelle une variabilité peut être capturée). Cette variation relative absolue a une signification directe : une variation de 20\% (qui est atteinte un nombre significatif de fois comme montré en Figure~\ref{fig:transportationequilibrium:fig-4}) implique dans le cas d'une variation négative, qu'au moins cette proportion de trajectoires potentielles ont changé et que la potentielle congestion locale a décru de la même proportion. Dans le cas d'une variation positive, un seul noeud a capturé au moins 20\% des trajets.
}

\bpar{
Under the assumption (that we do not try to verify in this work and assume to be also not verified as shown by~\cite{zhu2010people}, but that we use as a tool to give an idea of the concrete meaning of betweenness variability) that users rationally take the shortest path and assuming that a majority of travels are realized such a variation in centrality imply a similar variation in effective flows, leading to the conclusion that they can not be stationary in time (at least at a scale larger than $\Delta t$) nor in space.
}{
Sous l'hypothèse (qu'on ne tente pas de vérifier ici et qu'on peut également supposer non vérifiée comme montré par~\cite{zhu2010people}, mais que l'on utilise comme un outil pour donner une intuition sur la signification concrète de la variabilité de la centralité) que les utilisateurs choisissent rationnellement le plus court chemin, et supposant que la majorité des trajets est réalisée, une telle variation de la centralité implique une variation similaire dans les flux effectifs, conduisant à la conclusion qu'ils ne peuvent être stationnaires ni dans le temps (au moins sur une échelle plus grande que $\Delta t$) ni dans l'espace.
}

\paragraph{Spatial heterogeneity of equilibrium}{Hétérogénéité spatiale de l'équilibre}

\bpar{
To obtain a different insight into spatial variability of congestion patterns, we propose to use an index of spatial autocorrelation, the Moran index (defined e.g. in~\cite{tsai2005quantifying}). More generally used in spatial analysis with diverse applications from the study of urban form to the quantification of segregation, it can be applied to any spatial variable. It allows to establish neighborhood relations and unveils spatial local consistence of an equilibrium if applied on localized traffic variable. At a given point in space, local autocorrelation for the variable $c$ is computed by
}{
Afin d'obtenir un point de vue différent sur la variabilité spatiale des motifs de congestion, nous proposons d'utiliser un indice d'auto-corrélation spatiale, l'indice de Moran (défini par exemple dans~\cite{tsai2005quantifying}). Utilisé plus généralement en analyse spatiale, avec diverses applications allant de l'étude de la forme urbaine à la quantification de la ségrégation, il peut être appliqué à toute variable spatiale. Il permet d'établir des relations de voisinage et révèle la consistence spatiale locale d'un équilibre s'il est appliqué à une variable de trafic localisée. À un point donné de l'espace, l'auto-corrélation locale pour la variable $c$ est calculée par
}

\begin{equation}
\rho_i = \frac{1}{K}\cdot \sum_{i\neq j}{w_{ij}\cdot (c_i - \bar{c})(c_j - \bar{c})}
\end{equation}

\bpar{
where $K$ is a normalization constant equal to the sum of spatial weights times variable variance and $\bar{c}$ is variable mean. In our case, we take spatial weights of the form $w_{ij} = \exp{\left(\frac{-d_{ij}}{d_0}\right)}$ with $d_0$ typical decay distance and compute the autocorrelation of link congestion localized at link center. We capture therefore spatial correlations within a radius of same order than decay distance around the point $i$. The mean on all points yields spatial autocorrelation index $I$. A stationarity in flows should yield some temporal stability of the index.
}{
où $K$ est une constante de normalisation égale à la somme des poids spatiaux fois la variance de la variable et $\bar{c}$ est la moyenne de la variable. Dans notre cas, nous choisissons des poids spatiaux de la forme $w_{ij} = \exp{\left(\frac{-d_{ij}}{d_0}\right)}$ avec $d_0$ distance typique de décroissance. L'auto-corrélation est calculée sur la congestion des liens, localisée au centre du lien. Elle capture ainsi les corrélations spatiales dans un rayon du même ordre que la distance de décroissance autour du point $i$. La moyenne sur l'ensemble des points fournit l'indice d'auto-corrélation spatiale $I$. Une stationnarité des flots devrait impliquer une stabilité temporelle de l'index. 
}

\begin{figure}
\includegraphics[width=\linewidth]{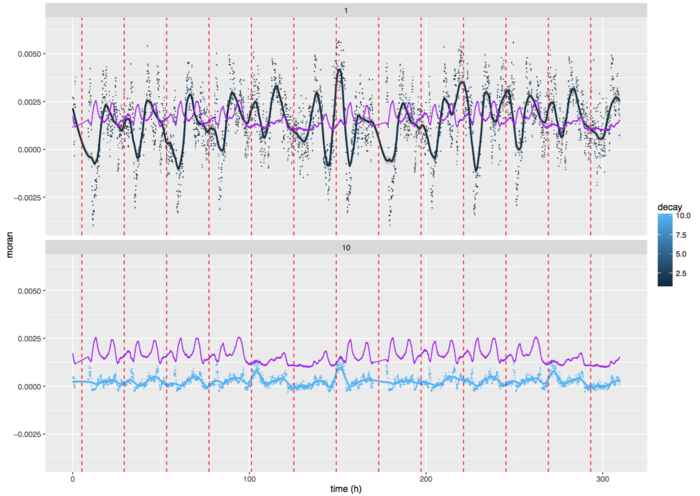}
\caption[Spatial auto-correlation for congestion]{\textbf{Spatial auto-correlations for relative travel speed on two weeks.} We plot values of auto-correlation index in time, for varying value of decay parameter (1km, 10km, given in color and title of plots). Intermediate values of decay parameter yield a rather continuous deformation between the two curves. Points are smoothed with a 2h span to ease reading. Vertical dotted lines correspond to midnight each day. Purple curve is relative speed fitted at scale to have a correspondence between auto-correlation variations and peak hours.\label{fig:transportationequilibrium:fig-5}}
\end{figure}

\bpar{
Figure~\ref{fig:transportationequilibrium:fig-5} presents temporal evolution of spatial autocorrelation for congestion. As expected, we have a strong decrease of autocorrelation with distance decay parameter, for both amplitude and temporal average. The high temporal variability implies short time scales for potential stationarity windows. When comparing with congestion (fitted to plot scale for readability) for 1km decay, we observe that high correlations coincide with off-peak hours, whereas peaks involve vanishing correlations.
}{
La figure~\ref{fig:transportationequilibrium:fig-5} présente l'évolution temporelle de l'auto-corrélation spatiale pour la congestion. Comme attendu, on observe une forte décroissance de l'auto-corrélation avec la distance de décroissance, à la fois sur l'amplitude et les moyennes temporelles. La forte variabilité temporelle implique de courtes échelles temporelles pour des fenêtres potentielles de stationnarité. Pour une distance de décroissance de 1km, en comparant l'auto-corrélation à la congestion (ajustée à l'échelle du graphe pour lisibilité), on observe que les fortes corrélations coincident avec les heures creuses, tandis que les heures de pointe correspondent à une décroissance des corrélations.
}

\bpar{
Our interpretation, combined with the observed variability of spatial patterns, is that peak hours correspond to chaotic behaviour of the system, as jams can emerge in any link: correlation thus vanishes as feasible phase space for a chaotic dynamical system is filled by trajectories in an uniform way what is equivalent to apparently independent random relative speeds.
}{
Notre interprétation, combinée avec la variabilité observée des motifs spatiaux, est que les heures de pointe correspondent à un comportement chaotique du système, puisque les bouchons peuvent émerger dans n'importe quel lien du réseau : la corrélation disparait alors puisque l'espace des phases atteignables pour un système dynamique chaotique est rempli uniformément par les trajectoires, de façon équivalente à des vitesses relatives qui apparaitraient comme aléatoires et indépendantes.
}

\bpar{
We have described an empirical study aimed at a simple but from our point of view necessary investigation of the existence of the static user equilibrium, more precisely of its stationarity in space and time on a metropolitan highway network. We constructed by data collection a traffic congestion dataset for the highway network of Greater Paris on 3 months with two minutes temporal granularity. The interactive exploration of the dataset with a web application allowing spatio-temporal data visualization helped to guide quantitative studies. Spatio-temporal variability of shortest paths and of network topology, in particular betweenness centrality, revealed that stationarity assumptions do not hold in general, what was confirmed by the study of spatial autocorrelation of network congestion.
}{
Nous avons décrit une étude empirique ayant pour but une approche simple, mais selon notre point de vue nécéssaire, de l'existence de l'équilibre utilisateur statique, plus précisément de sa stationnarité dans le temps et l'espace pour un réseau routier métropolitain principal. Un jeu de données de congestion du trafic est construite par collection de données, pour la métropole parisienne sur 3 mois avec une granularité temporelle de 2 minutes. L'exploration interactive du jeu de données via une application web permettant la visualisation spatio-temporelle aide à guider les analyses quantitatives. La variabilité spatio-temporelle des plus courts chemins et de la topologie du réseau, en particulier la centralité de chemin, révèle que l'hypothèse de stationnarité ne tient généralement pas, ce qui est confirmé par l'étude de l'auto-corrélation spatiale de la congestion du réseau.
}

\subsubsection{Perspective}{Mise en perspective}

\bpar{
We can now propose a perspective on this work regarding our general problematic of co-evolution. Traffic flows, which are representative of mechanisms of the transportation network, are generated by the spatial distribution of activities and the behaviors of microscopic agents. We just showed that the temporal evolution of these flows is complex, evoking chaotic dynamics, what can be also understood as a crucial role of non-linearity in the emergence of congestion.
}{
Nous pouvons proposer une mise en perspective de ce travail au regard de notre problématique générale de la co-évolution. Les flux de trafic, qui sont représentatifs du fonctionnement du réseau de transport, sont générés par la distribution spatiale des activités et les comportements des agents microscopiques. Or nous venons de montrer que l'évolution temporelle de ces flux est complexe, rappelant des dynamiques chaotiques, ce qui peut être également compris comme un rôle essentiel de la non-linéarité dans l'émergence de la congestion.
}

\bpar{
As we showed in chapter~\ref{ch:thematic}, these processes linked to daily mobility are probably linked to a proper level of co-evolution between networks and territories (for example the congestion inducing a network evolution, but also possibly some relocalizations), that we will not study in deep in our work. This illustration thus shows (i) an illustration of interactions between networks and territories at this microscopic scale, suggesting the existence of complex effects at this scale; (ii) the possible difficulty of modeling co-evolution at this microscopic scale, given the chaotic trajectories of the system studied.
}{
Comme nous l'avons montré au chapitre~\ref{ch:thematic}, ces processus liés à la mobilité quotidienne sont probablement liés à un niveau propre de co-évolution entre réseaux et territoires (par exemple la congestion induisant une évolution du réseau, mais aussi éventuellement des relocalisations), que nous n'abordons pas en profondeur dans notre travail. Cette illustration montre ainsi (i) une illustration des interactions entre réseaux et territoires à cette échelle microscopique, suggérant l'existence d'effets complexes à cette échelle ; (ii) la difficulté éventuelle d'une modélisation de la co-évolution à cette échelle microscopique, vu les trajectoires chaotiques du système étudié.
}

\bpar{
The empirical exploration allowed us to illustrate on the one hand the construction of an open dataset to counter the absence of data, and on the other hand the crucial role of interactive exploration, which must remain combined to more advanced analyses which are guided by it.
}{
Cette exploration empirique nous a permis d'illustrer d'une part la construction d'un jeu de données ouvertes pour combler l'absence de données, et d'autre part le rôle crucial de l'exploration interactive, qui doit rester combinée à des analyses plus poussées guidées par celle-ci.
}

\stars

\bpar{
We have thus detailed in this section some issues linked to reproducibility and open science, completing our specific positioning in terms of modeling with a more general positioning corresponding to the scientific practice.
}{
Nous avons ainsi détaillé dans cette section certains enjeux liés à la reproductibilité et à la science ouverte, complétant notre positionnement spécifiques en termes de modélisation avec un positionnement plus général correspondant à la pratique scientifique.
}

\bpar{
We will finally in the last following section again gain some generality and precise our epistemological positioning, i.e. concerning disciplines themselves and the production of knowledge. This step will be crucial, since our positioning regarding social systems and biological systems will allow us to introduce the fundamental elements for a more complete definition of co-evolution.
}{
Nous allons finalement dans la dernière section qui suit encore monter en généralité et préciser nos positionnement épistémologiques, c'est-à-dire concernant les disciplines elles-mêmes et la production de connaissance. Cette étape sera cruciale, puisque notre positionnement au regard des systèmes sociaux et des systèmes biologiques permettra d'introduire les éléments fondamentaux pour une définition plus complète de la co-évolution.
}

\stars

%


\newpage

\section{Epistemological positioning}{Positionnement épistémologique}\label{sec:epistemology}


\bpar{
The last section of this chapter aims at clarifying our epistemological positioning, since it has only been sketched at different points previously. Such a positioning is never harmless, since it strongly conditions the approaches, experiments and the interpretation of results: as \cite{morin1980methode} recalls, a positioning that pretends to be objective by rejecting any subjective component is much more biased than a conscious subjective approach.
}{
La dernière section de ce chapitre vise à clarifier notre positionnement épistémologique, celui-ci ayant déjà été ébauché à plusieurs occasions précédemment. Un tel positionnement n'est jamais anodin, puisqu'il conditionne fortement les démarches, les expériences et l'interprétation des résultats : comme le souligne~\cite{morin1980methode}, un positionnement qui se dit objectif en rejetant toute subjectivité est bien plus biaisé qu'une approche subjective consciente.
}

\bpar{
The points we wish to develop can be put into both a vertical perspective in terms of levels of abstraction and in a perspective of scientific domains: linearly, we first give the general epistemological context (typical to history of science, at a medium abstraction level), then switch at a less generic level to conceptually precise our particular objects (epistemology of the living and of the social), and finally take a broader perspective at the level of knowledge production itself (epistemology of complexity).
}{
Les points que nous souhaitons développer se placent dans une logique à la fois verticale de niveau d'abstraction et dans une logique de domaines scientifiques : dans l'ordre, nous posons d'abord le contexte épistémologique général (propre à l'histoire des sciences, à un niveau d'abstraction moyen), pour descendre en généralité et préciser conceptuellement nos objets particuliers (épistémologie du vivant et du social), pour finalement tout remettre en perspective au niveau de la production de connaissance elle-même (épistémologie de la complexité).
}

\subsection{Cognitive approach and perspectivism}{Approche cognitive et perspectivisme}

\bpar{
Our epistemological positioning relies on a cognitive approach to science, given by Giere in~\cite{giere2010explaining}. The approach focuses on the role of cognitive agents as carriers and producers of knowledge. It has been shown to be operational by \cite{giere2010agent} that studies an agent-based model of science. These ideas converge with \noun{Chavalarias}' Nobel Game~\cite{chavalarias2016s} which tests through a stylized model the balance between exploration and falsification in the collective scientific enterprise.
}{
Notre positionnement épistémologique se fonde sur une approche cognitive de la science, introduite par \noun{Giere} dans~\cite{giere2010explaining}. L'approche se concentre sur le rôle des agents cognitifs comme porteurs et producteurs de la connaissance. Son caractère opérationnel a été montré par \cite{giere2010agent} qui étudie un modèle multi-agents de la science. Ces idées convergent avec le jeu Nobel de \noun{Chavalarias}~\cite{chavalarias2016s} qui teste de manière stylisée l'équilibre entre production de nouvelles théories et tentative de falsification de théories existantes dans l'entreprise scientifique collective.
}

\bpar{
This epistemological positioning has been presented by \noun{Giere} as \emph{scientific perspectivism}~\cite{giere2010scientific}, which main feature is to consider any scientific entreprise as a \emph{perspective} in which \emph{agents} use \emph{media} (models) to represent something with a certain purpose. To make it more concrete, we can position it within Hacking's ``check-list'' of constructivism~\cite{hacking1999social}, a practical tool to position an epistemological position within a simplified three dimensional space which dimensions are different aspects on which realist approaches and constructivist approach generally diverge: first the contingency (path-dependency of the knowledge construction process) is necessary in the pluralist perspectivist approach which assumes parallel paths of knowledge construction. Secondly the ``degree of constructivism'' is quite high because agents produce knowledge. Finally, concerning the endogenous or exogenous explanation of the stability of theories, this stability depends on the complex interaction between the agents and their perspectives, and is thus strongly endogenous, close to the positioning of constructivism. It was presented for these reasons as an intermediate and alternative way between absolute realism and skeptical constructivism~\cite{brown2009models}. The concept of \emph{perspective} will play thus a central role in the framework developed in~\ref{sec:knowledgeframework}.

}{
Ce positionnement épistémologique a été présenté par \noun{Giere} comme \emph{perspectivisme scientifique}~\cite{giere2010scientific}, dont la caractéristique principale est de considérer toute entreprise scientifique comme une \emph{perspective} dans laquelle des \emph{agents} utilisent des \emph{media} (modèles) pour représenter quelque chose dans un certain but. Pour comprendre ses principes de manière plus concrète, nous pouvons le positionner sur la \emph{check-list} du constructivisme de \noun{Hacking}~\cite{hacking1999social}, un outil pratique pour situer une position épistémologique. Celle-ci suppose un espace simplifié tri-dimensionnel dans lequel les dimensions sont différents aspects sur lesquels les approches réalistes et constructivistes généralement divergent. Le premier point est le niveau de contingence (dépendance au chemin du processus de construction de connaissances) : celle-ci est nécessaire dans l'approche perspectiviste qui est pluraliste et suppose des chemins parallèles de construction de connaissance. Le deuxième point mesure un ``degré de constructivisme'', qui est assez haut en perspectivisme car les agents produisent la connaissance. Enfin, le dernier point qui concerne l'explication endogène ou exogène de la stabilité des théories, est fortement du côté du constructivisme, puisque cette stabilité dépend des interactions complexes entre les agents et leur perspectives et donc totalement endogène. Le perspectivisme a pour ces raisons été présenté comme un chemin intermédiaire et alternatif entre le réalisme absolu et le constructivisme sceptique~\cite{brown2009models}. Le concept de \emph{perspective} jouera pour nous un rôle fondamental dans le cadre développé en~\ref{sec:knowledgeframework}.
}

\bpar{
Since this approach puts the emphasis on auto-organization, we consider it to be fully compatible with an anarchist view of science as advocated by~\cite{feyerabend1993against}. He formulates doubts on the relevance of political anarchism but introduces \emph{scientific anarchism}, which must not be understood as a full refusal of any ``objective'' method, but of an artificial authority and legitimacy that some scientific methods or currents would like to impose. He demonstrates through a precise analysis of Galileo's work that most of his results were based on beliefs and that most were not accessible with the current tools and methods at that time, and postulates that a similar logic should apply to contemporary works. There is thus no \emph{perspective} that is objectively more legitimate than others as soon as they are evidence-based and peer review validated - and even in this case legitimacy should be questionable, since questioning is one foundation of knowledge. It corresponds exactly to the plurality of perspectives we defend.
}{
Cette approche mettant l'emphase sur l'auto-organisation, nous la voyons totalement compatible avec une vision anarchiste de la science comme défendue par~\cite{feyerabend1993against}. Celui-ci émet des doutes sur l'intérêt de l'anarchisme politique mais introduit l'\emph{anarchisme scientifique}, qu'il ne faut pas comprendre comme un refus total de toute méthode ``objective'', mais d'une autorité et légitimité artificielle que certaines méthodes ou courants scientifiques pourraient vouloir prendre. Il démontre par une analyse précise des travaux de Galilée que la plupart de ses résultats étaient basés sur des croyances et que la plupart n'étaient pas accessibles avec les outils et méthodes de l'époque, et postule qu'il devrait en être de même pour certains travaux contemporains. Il n'y a donc pas de \emph{perspective} objectivement plus légitime que d'autres dans la mesure de leurs validation par des faits et des pairs - et même dans ces cas la légitimité doit pouvoir être discutée, car la remise en question est un fondement de la connaissance. Cela correspond exactement à la pluralité des perspectives que nous défendons.
}

\bpar{
Assuming an auto-organization and emergence of knowledge can be interpreted as a priority given to the \emph{bottom-up} construction of paradigms, trying to take some distance with preconceptions or dogmas that impose a top-down view. In other words, it is similar to practicing the scientific anarchism proposed by \noun{Feyerabend}. Indeed, anarchist positioning have found a very relevant echo in the different currents of complexity, from cybernetics to self-organization during the 20th century~\cite{duda2013cybernetics}. Our knowledge framework developed in~\ref{sec:knowledgeframework} illustrates this emergence of knowledge. Moreover, our will for reflexivity and to give to this work diverse reading paths beyond linearity (see Appendix~\ref{app:reflexivity}), shows the application of these principles. Methodological recommendations and positioning given previously in this chapter could sound as totalitarian if they were given roughly out of context, but these are indeed exactly the contrary since they sprout from a recent dynamic of open science which is well bottom-up founded, and in part a consequence of opening and plurality.
}{
Supposer auto-organisation et émergence des connaissances peut être interprété comme une priorité donnée à la construction des paradigmes \emph{par le bas} (\emph{bottom-up}), en tentant de se distancer des préconceptions ou dogmes cadrant par le haut. En d'autres termes, il s'agit de pratiquer l'anarchisme scientifique prôné par \noun{Feyerabend}. En effet, les positions anarchistes ont trouvé un écho très cohérent dans les différents courants de la complexité, de la cybernétique à l'auto-organisation au cours du 20ème siècle~\cite{duda2013cybernetics}. Notre cadre de connaissances développé en~\ref{sec:knowledgeframework} illustre cette émergence de la connaissance. De plus, notre volonté de réflexivité et de donner à notre travail des pistes de lecture diverses au delà de la linéarité (voir Annexe~\ref{app:reflexivity}), montre l'application de ces principes. Les recommandations méthodologiques et les positionnements donnés précédemment dans ce chapitre pourraient sonner comme totalitaires s'ils étaient assénés de manière sèche sans contexte, mais ceux-ci sont en fait tout le contraire puisqu'ils découlent d'une dynamique récente de science ouverte qui a bien émergé par le bas, conséquence en partie de l'ouverture et de la pluralité.
}


\subsection{From life to culture}{De la Vie à la Culture}

\subsubsection{Biological systems and social systems}{Systèmes biologiques et systèmes sociaux}

\bpar{
The parallel between social and biological systems is not rare, sometimes more from an analogy perspective as for example in \noun{West}'s \emph{Scaling} theory which applies similar growth equations starting from scaling laws, with however inverse conclusions concerning the relation between size and pace of life~\cite{bettencourt2007growth}. Scaling relations do not hold when we try to apply them to a single ant, and they must be applied to the whole ant colony which is then the organism studied. When adding the property of cognition, we confirm that it is the relevant level, since the colony shows advanced cognitive properties, such as the resolution of spatial optimization problems, or the quick answer to an external perturbation. Human social organizations, cities, could be seen as organisms ? \cite{banos2013pour} extends the metaphor of the \emph{urban anthill} but recalls that the parallel stops quickly. We will however see to what extent some concepts from the epistemology of biology can be useful to understand social systems that we propose to study.
}{
Le parallèle entre les systèmes sociaux et les systèmes biologiques est souvent fait, parfois de manière plus qu'imagée comme par exemple pour la théorie du \emph{Scaling} de \noun{West} qui applique des équations de croissance similaires à partir des lois d'échelle, avec des conclusions inverses tout de même concernant la relation entre taille et rythme de vie~\cite{bettencourt2007growth}. Les relations d'échelle ne tiennent plus lorsqu'on essaye de les appliquer à une fourmi seule, et il faut alors l'appliquer à la fourmilière entière qui est l'organisme en question. En ajoutant la propriété de cognition, on confirme qu'il s'agit du niveau pertinent, puisque celle-ci possède des propriétés cognitives avancées, comme la résolution de problèmes d'optimisation spatiaux, ou la réponse rapide à une perturbation extérieure. Les organisations sociales humaines, les villes, peuvent-elles être vues comme des organismes ? \cite{banos2013pour} file la métaphore de la \emph{fourmilière urbaine} mais rappelle que le parallèle s'arrête assez vite. Nous allons voir cependant dans quelle mesure certains concepts de l'épistémologie de la biologie peuvent être utiles pour comprendre les systèmes sociaux que nous nous proposons d'étudier.
}

\bpar{
We start from the fundamental contribution of \noun{Monod} in~\cite{monod1970hasard}, which aims at developing crucial epistemological principles for the study of life. Thus, living organisms answer to three essential properties that differentiate them from other systems: (i) the teleonomy , i.e. the property that these are ``objects with a project'', project that is reflected in their structure and the structure of artifacts they produce\footnote{That must not be mistaken with teleology, typical of animist thoughts, that consists in giving a project or a meaning to the universe.}; (ii) the importance of morphogenetic processes in their constitution (see~\ref{sec:interdiscmorphogenesis}); (iii) the property of the invariant reproduction of information defining their structure. \noun{Monod} furthermore sketches in conclusion some paths towards a theory of cultural evolution. Teleonomy is crucial in social structures, since any organization aims at satisfying a set of objectives, even if in general it will not succeed and the objectives will co-evolve with the organization. This notion of multi-objective optimization is typical of complex socio-technical systems, and will be more crucial than for biological systems.
}{
Nous nous basons sur la contribution fondamentale de \noun{Monod} dans~\cite{monod1970hasard}, qui tente de développer les principes épistémologiques cruciaux pour l'étude du vivant. Ainsi, les organismes vivants répondent à trois propriétés essentielles qui permettent des les différencier d'autres systèmes : (i) la téléonomie, c'est-à-dire qu'il s'agit ``d'objets doués d'un projet'', projet qui se reflète dans leur structure et dans celles des artefacts qu'ils produisent\footnote{Qu'il ne faut pas confondre avec la téléologie, propres aux animismes, qui consiste à prêter un projet ou un sens à l'univers.} ; (ii) l'importance des processus morphogénétiques dans leur constitution (voir~\ref{sec:interdiscmorphogenesis}) ; (iii) la propriété de reproduction invariante de l'information définissant leur structure. \noun{Monod} esquisse de plus en conclusion des pistes pour une théorie de l'évolution culturelle. La téléonomie est essentielle dans les structures sociales, puisque toute organisation essaye de satisfaire un ensemble d'objectifs, même si en général elle n'y parviendra pas et que ceux-ci co-évolueront avec l'organisation. Cette notion d'optimisation multi-objectif est typique des systèmes complexes socio-techniques, et y sera plus cruciale que pour les systèmes biologiques.
}

\bpar{
Moreover, we postulate that the concept of morphogenesis is an essential tool to understand these systems, with a definition very similar to the one used in biology. A more thorough work to build this definition is done in~\ref{sec:interdiscmorphogenesis}, that we will sum up as the existence of relatively autonomous processes guiding the growth of the system and implying causal circular relations between form and function, that witness an emergent architecture. For social systems, isolating the system is more difficult and the notion of boundary will be less struct than for a biological system, but we will indeed find this link between form and function, such as for example the structure of an organization that impacts its functionalities.
}{
Ensuite, nous postulons que le concept de morphogenèse est un outil essentiel pour comprendre ces systèmes, avec une définition très proche de celle utilisée en biologie. Un travail approfondi pour donner cette définition est fait en~\ref{sec:interdiscmorphogenesis}, que nous résumerons en l'existence de processus relativement autonomes guidant la croissance du système et impliquant des relations causales circulaires entre forme et fonction qui témoignent d'une architecture émergente. Pour des systèmes sociaux, isoler le système est plus difficile et la notion de frontière sera moins stricte que pour un système biologique, mais on retrouvera bien ce lien entre forme et fonction, comme par exemple la structure d'une organisation ayant un impact sur ses fonctionnalités.
}

\bpar{
Finally, the reproduction of information is at the core of cultural evolution, through the transmission of culture and \emph{memetics}, the difference being that the ratio of scales between the frequency of transmission and mutation and cross-over processes or other non-memetic processes of cultural production is relatively low, whereas is many orders of magnitude in biology.
}{
Enfin, la reproduction de l'information est au coeur de l'évolution culturelle, par la transmission de la culture et la \emph{mémétique}, la différence étant que le rapport d'échelles de temps entre la fréquence de transmission et les processus de croisement et de mutation ou d'autres processus non mémétiques de production culturelle est relativement faible, alors qu'elle est de plusieurs ordres de grandeur en biologie.
}

\bpar{
An example shows that the parallel is not always absurd : \cite{2017arXiv170305917G} proposes an auto-catalytic network model for cognition, that would explain the apparition of cultural evolution through processes that are analogous to the ones that occurred at the apparition of life, i.e. a transition allowing the molecules to be self-sustained and to self-reproduce, mental representations being the analogous of molecules.
}{
Un exemple illustre que le parallèle n'est pas toujours absurde : \cite{2017arXiv170305917G} propose un modèle de réseau auto-catalytique pour la cognition, qui expliquerait l'apparition de l'évolution culturelle par des processus analogues à ceux s'étant produit à l'apparition de la vie, c'est-à-dire une transition permettant au molécules de s'auto-entretenir et s'auto-reproduire, les représentations mentales faisant office de molécules.
}

\bpar{
But even if processes are at the origin analogous, the nature of evolution is then quite different, as show \cite{vanderLeeuw2009}, darwinian criteria for evolution being not sufficient to explain the evolution of our organized societies. This is a complexity of a different nature in which the role of information flows is crucial (see the role of informational complexity in the next subsection).
}{
Mais si les processus à l'origine sont analogues, la nature de l'évolution est bien différente par la suite, comme le montrent \cite{vanderLeeuw2009}, les critères darwiniens d'évolution n'étant pas suffisant pour expliquer l'évolution de nos sociétés organisées. Il s'agit d'une complexité de nature différente dans laquelle le rôle des flux d'information est crucial (voir le rôle de la complexité informationnelle dans la sous-section suivante). 
}

\bpar{
One point that also must retain our attention is the greater difficulty to define levels of emergence for social systems: \cite{roth2009reconstruction} underlines the risk to fall into ontological dead-ends if levels were badly defined. He argues that more generally we must go past the single dichotomy micro-macro that is used as a caricature of the concepts of weak emergence, and that ontologies must often be multi-level and imply multiple intermediate levels.
}{
L'un des points sur lequel il s'agit également d'être attentif est la plus grande difficulté de définir les niveaux d'émergence pour les systèmes sociaux : \cite{roth2009reconstruction} souligne le risque de tomber dans des cul-de-sac ontologiques car les niveaux ont été mal définis. Il soutient qu'il faut d'une manière générale penser au-delà de la seule dichotomie micro-macro qui est utilisée pour caricaturer les concepts d'émergence faible, et que les ontologies doivent souvent être multi-niveaux et impliquant de multiples niveaux intermédiaires.
}

\bpar{
This last question must also be put into perspective with the problem of the existence of strong emergence in social structures, that in sociological terms corresponds to the idea of the existence of ``collective beings''~\cite{angeletti2015etres}. \noun{Morin} indeed distinguishes living systems of the second type (multi-cellular) and of the third type (social structures), but precises that the \emph{subjects} of the latest are necessarily unachieved\cite{morin1980methode} (p.~852). Thus, emergences from the biological to the social are analogous by stay fundamentally different.
}{
Cette dernière question est aussi à mettre en perspective avec le problème de l'existence d'émergence forte dans les structures sociales, qui en termes sociologiques correspond à l'idée de l'existence ``d'êtres collectifs''~\cite{angeletti2015etres}. \noun{Morin} distingue d'ailleurs les systèmes vivants du second type (multi-cellulaire) et du troisième type (structures sociales), mais précise que les \emph{sujets} de ces derniers sont nécessairement inachevés~\cite{morin1980methode} (p.~852). Ainsi, les émergences du biologique au social sont analogues mais restent fondamentalement différentes.
}

\subsubsection{Co-evolution}{Co-évolution}

\bpar{
This positioning on biological and social systems finds a direct echo for the concept of co-evolution. It indeed comes from biology, where it was developed following the concept of evolution, to be used more recently in social sciences and humanities. To what extent the concept was transfered ? Is there a parallel similar to the one between biological evolution and cultural evolution ? We propose, in order to answer these questions, to develop a brief multidisciplinary point of view on co-evolution\footnote{The approach here is slightly different from the one lead in~\ref{sec:interdiscmorphogenesis} in the case of morphogenesis, that will be \emph{interdisciplinary} in the sens that it aims at integrating approaches, whereas we stay here in an overview of concepts and thus more in a \emph{multidisciplinary} approach. The concept of \emph{co-evolution} being key for our empirical work in the following, we will therefore give an original characterization to it, and make the choice to not go into an integrative syncretism for this concept, but indeed to approach it from a \emph{geographical point of view}, and even more precisely in the frame of territorial systems. We could postulate a congruence between the empirical and modeling specialization and the one for theory, reading our process of knowledge production in a particular profile of knowledge domains dynamics (see~\ref{sec:knowledgeframework}).}. We will in the following review a broad spectrum of disciplines, starting from biology where the concept originated to progressively come to disciplines closer to territorial sciences.
}{
Ce positionnement sur les systèmes biologiques et sociaux trouve un écho immédiat pour le concept de co-évolution. Il provient en effet de la biologie, où il a été développé à la suite de celui d'évolution, pour être utilisé plus récemment en sciences humaines et sociales. Dans quelle mesure le concept a-t-il été transféré ? Retrouve-t-on un parallèle similaire à celui entre évolution biologique et évolution culturelle ? Nous proposons pour répondre à ces questions d'apporter un bref point de vue multidisciplinaire sur la co-évolution\footnote{La démarche ici est légèrement différente de celle que nous mènerons en~\ref{sec:interdiscmorphogenesis} dans le cas de la morphogenèse, qui sera \emph{interdisciplinaire} au sens où elle cherchera à intégrer les approches, tandis que nous restons ici dans un aperçu des concepts et donc plutôt dans du \emph{multidisciplinaire}. Le concept de \emph{co-évolution} étant clé pour notre travail empirique par la suite, nous en donnerons alors une caractérisation originale et prenons le parti de ne pas tomber dans le syncrétisme intégrateur pour ce concept, mais bien de l'approcher d'un \emph{point de vue géographique}, et même plus précisément dans le cadre des systèmes territoriaux. On pourrait postuler une congruence entre la spécialisation empirique/de modélisation et celle théorique, plaçant notre processus de production de connaissance dans un profil particulier de dynamiques de domaines de connaissance (voir~\ref{sec:knowledgeframework}).}. Nous passons par la suite en revue un large spectre de disciplines, partant de la biologie où le concept a initialement trouvé son origine pour arriver progressivement à des disciplines en relation avec les sciences du territoire.
}

\subsubsection{Biology}{Biologie}

\bpar{
The concept of co-evolution in biology is an extension of the well-known concept of \emph{evolution}, that can be tracked back to \noun{Darwin}. \cite{durham1991coevolution} (p.~22) recalls the components and systemic structures that are necessary to have evolution\footnote{And in that general context, evolution is not restricted to the biology of life and the presence of genes, but also to physical systems verifying these conditions. We will come back to that later.}.
}{
Le concept de co-évolution en biologie est une extension de celui bien connu d'\emph{évolution}, qui remonte à \noun{Darwin}. \cite{durham1991coevolution} (p.~22) rappelle les composantes et structure systémiques nécessaires pour qu'il y ait évolution\footnote{Et dans ce contexte général l'évolution n'est pas réservée à la biologie du vivant et la présence de gènes, mais aussi à des systèmes physiques vérifiant ces conditions. Nous y reviendrons plus loin.}.
}

\bpar{
\begin{enumerate}
\item Process of \emph{transmission}, implying transmission units and transmission mechanisms.
\item Process of \emph{transformation}, that necessitates sources of variation.
\item Isolation of sub-systems such that the effects of previous processes are observable in differentiations.
\end{enumerate}
}{
\begin{enumerate}
	\item Processus de \emph{transmission}, impliquant des unités de transmission et des mécanismes de transmission.
	\item Processus de \emph{transformation}, nécessitant des sources de variation.
	\item Isolation de sous-systèmes pour que les effets des processus précédents soient observables dans des différentiations.
\end{enumerate}
}

\bpar{
This way, a population submitted to constraints (often conceptually synthesized as a \emph{fitness}) that condition the transmission of the genetic heritage of individuals (transmission), and to random genetic mutations (transformation), will indeed be in evolution in the spatial territories it populates (isolation), and by extension the species to which it can be associated. 
}{
Ainsi, une population soumise à des contraintes (souvent synthétisées conceptuellement comme une \emph{fitness}) qui conditionnent la transmission du patrimoine génétique des individus (transmission), et à des mutation génétiques aléatoires (transformation), sera bien en évolution dans les territoires spatiaux qu'elle occupe (isolation), et par extension l'espèce à laquelle on peut l'associer.
}

\bpar{
Co-evolution is then defined as an evolutionary change in a characteristic of individuals of a population, in response to a change in a second population, which in turn responds by evolution to the change in the first, as synthesized by~\cite{janzen1980coevolution}. This author furthermore highlights the subtlety of the concept and warns against its unjustified uses: the presence of a congruence between two characteristics that seem adapted one to the other does not necessarily imply a co-evolution, since one species could have adapted alone to one characteristic already present in the other.
}{
La co-évolution est alors définie comme un changement évolutionnaire dans une caractéristique des individus d'une population, en réponse à un changement dans une deuxième population qui à son tour répond évolutionnairement au changement de la première, comme synthétisé par~\cite{janzen1980coevolution}. Cet auteur appuie par ailleurs la subtilité du concept et alerte contre ses utilisations injustifiées : la présence d'une congruence de deux caractéristiques qui semblent adaptées l'une à l'autre n'implique pas l'existence d'une co-évolution, l'une des deux espèces ayant pu s'adapter seule à une caractéristique déjà présente de l'autre.
}

\bpar{
This rough presentation partly hides the real complexity of ecosystems: populations are embedded in trophic networks and environments, and co-evolutionary interactions would imply communities of populations from diverse species, as presented by \cite{strauss2005toward} under the appellation of diffuse co-evolution. Similarly, spatio-temporal dynamics are crucial in the realization of these processes: \cite{dybdahl1996geography} study for example the influence of the spatial distribution on patterns of co-evolution for a snail and its parasite, and show that a higher speed of genetic diffusion in space for the parasite drive the co-evolutionary dynamics.
}{
Cette présentation brute de décoffrage mutile dans une certaine mesure la complexité réelle des écosystèmes : les populations s'insèrent dans des réseaux trophiques et des environnements, et les interactions co-évolutionnaires impliqueraient des communautés de populations d'espèces diverses, comme présenté par \cite{strauss2005toward} sous l'appellation de co-évolution diffuse. De même, les dynamiques spatio-temporelles sont cruciales dans la réalisation de ces processus : \cite{dybdahl1996geography} étudient par exemple l'influence de la distribution spatiale sur les motifs de co-évolution pour un escargot et son parasite, et montrent qu'une vitesse de diffusion génétique dans l'espace plus grande pour le parasite conduit les dynamiques de co-évolution.
}

\bpar{
The essential concepts to retain from the biological point of view are thus: (i) existence of evolution processes, in particular transmission and transformation; (ii) in circular schemas between populations in the case of co-evolution; and (iii) in a complex territorial frame (spatio-temporal and environmental in the sense of the rest of the ecosystem).
}{
Les concepts essentiels à retenir du point de vue biologique sont ainsi : (i) existence de processus d'évolution, en particulier transmission et transformation ; (ii) dans des schémas circulaires entre populations dans le cas de la co-évolution ; et (iii) dans un cadre territorial (spatio-temporel et environnemental au sens du reste de l'éco-système) complexe.
}

\subsubsection{Cultural evolution}{Evolution culturelle}

\bpar{
This development on co-evolution was brought by the parallel between biological and social systems. The evolution of culture is theorized within a proper field, and witnesses many co-evolutive dynamics. \cite{Mesoudi25072017} recalls the state of knowledge on the subject and future issues, such as the relation with the cumulative nature of culture, the influence of demography in evolution processes, or the construction of phylogenetic methods allowing to reconstruct branches of past evolutionary trees.
}{
Ce développement sur la co-évolution nous a été amené par le parallèle entre systèmes biologiques et systèmes sociaux. L'évolution de la culture est théorisée est explorée par un champ propre, et n'est pas en reste de dynamiques co-évolutives. \cite{Mesoudi25072017} rappelle l'état des connaissances sur le sujet et les défis à venir, comme la relation avec la nature cumulative de la culture, l'influence de la démographie dans les processus d'évolution, ou la construction de méthodes phylogénétiques permettant de reconstruire des arbres des branchements passés.
}

\bpar{
To give an example, \cite{carrignon2015modelling} introduces a conceptual frame for the co-evolution of culture and commerce in the case of ancient societies for which there are archeological data, and proposes its implementation with a multi-agent model which dynamics are partly validated by the study of stylized facts produced by the model. The co-evolution is here indeed taken in the sense of a mutual adaptation of socio-spatial structures, at comparable time scales, in this more general frame of cultural evolution.
}{
Pour donner un exemple, \cite{carrignon2015modelling} introduit un cadre conceptuel pour la co-évolution de la culture et du commerce dans le cas de sociétés anciennes sur lesquelles on dispose de données archéologiques, et propose son implémentation par un modèle multi-agents dont les dynamiques sont partiellement validées par l'étude des faits stylisés produits par le modèle. La co-évolution est bien prise ici au sens d'adaptation mutuelle de structures socio-spatiales, à des échelles de temps comparables, dans ce cadre plus général d'évolution culturelle.
}

\bpar{
Cultural evolution would even be indissociable from genetic evolution, since \cite{durham1991coevolution} postulates and illustrates a strong link between the two, that would themselves be in co-evolution. \cite{bull2000meme} explores a stylized model including two types of replicant populations (genes and memes) and shows the existence of phase transitions for the results of the genetic evolution process when the interaction with the cultural replicant is strong.
}{
L'évolution culturelle serait même indissociable de l'évolution génétique, puisque \cite{durham1991coevolution} postule et illustre un lien fort entre les deux, qui seraient eux-mêmes en co-évolution. \cite{bull2000meme} explore un modèle stylisé impliquant deux populations de répliquants (les gènes et les memes) et montre l'existence de transitions de phase pour les résultats du processus d'évolution génétique lorsque l'interaction avec le répliquant culturel est forte.
}


\subsubsection{Sociology}{Sociologie}

\bpar{
The concept was used in sociology and related disciplines such as organisation studies, following the parallel done before the same way as cultural evolution. In the field of the study of organisations, \cite{volberda2003co} develop a conceptual frame of inter-organisational co-evolution in relation with internal management processes, but deplore the absence of empirical studies aiming at quantifying this co-evolution. In the context of production systems management, \cite{tolio2010species} conceptualize an intelligent production chain where product, process and the production system must be in co-evolution.
}{
Le concept a été utilisé en sociologie et disciplines apparentées comme les études de l'organisation, suivant le parallèle effectué ci-dessus de la même manière que pour l'évolution culturelle. Dans le domaine de l'étude des organisations, \cite{volberda2003co} développent un cadre conceptuel de la co-évolution inter-organisationnelle en relations avec les processus de management internes, mais déplore l'absence d'études empiriques cherchant à quantifier cette co-évolution. Dans le cadre de la gestion des systèmes de production, \cite{tolio2010species} conceptualisent un chaine de production intelligente où produit, processus et système de production doivent être en co-évolution.
}

\subsubsection{Economic geography}{Economie géographique}

\bpar{
In economic geography, the concept of co-evolution has also largely been used. The idea of evolutionary entities in economy comes in opposition to the neo-classical current which remains a majority, but finds a more and more relevant echo~\cite{nelson2009evolutionary}. \cite{schamp201020} proceeds to an epistemological analysis of the use of co-evolution, and opposes the view of a neo-schumpeterian approach to economy which considers the emergence of populations that evolve from micro-economic rules (what would correspond to a direct and relatively isolationist reading of biological evolution) to a systemic approach that would consider the economy as an evolutive system in a global perspective (what would correspond to diffuse co-evolution that we previously developed), to propose a precise characterization that would correspond to the first case, assuming co-evolving \emph{institutions}. The most important for our purpose is that he underlines the crucial aspect of the choice of populations and of considered entities, of the geographical area, and highlights the importance of the existence of causal circular relations.
}{
En économie géographique, le concept de co-évolution a également largement été mobilisée. L'idée d'entités évolutionnaires en économie vient à contre-courant du courant néoclassique qui reste majoritaire, mais trouve un écho de plus en plus pertinent~\cite{nelson2009evolutionary}. \cite{schamp201020} procède à une analyse épistémologique de l'utilisation de la co-évolution, et oppose une approche néo-schumpeterienne de l'économie qui considère l'émergence de populations qui évoluent à partir de règles micro-économiques (qui correspondrait à une lecture directe et relativement isolationniste de l'évolution biologique) à une approche systémique qui considérerait l'économie comme un système évolutif de manière globale (qui correspondrait à l'évolution diffuse que nous avons développé précédemment), pour proposer une caractérisation précise tombant dans le premier cas, qui suppose des \emph{institutions} qui co-évoluent. Le plus important pour notre propos est qu'il souligne l'aspect crucial du choix des population et des entités considérées, de la zone géographique, et appuie l'importance de l'existence de relations causales circulaires.
}

\bpar{
Diverse examples of application can be given. \cite{doi:10.1080/00343400802662658} introduce a conceptual frame to allow to conciliate the evolutionary nature of companies, the theory of clusters and knowledge networks, in which the co-evolution between networks and companies  is central, and which is defined as a circular causality between different characteristics of these subsystems. \cite{colletis2010co} introduces a framework for the co-evolution of territories and technology (questioning for example the role of proximity on innovations), that reveals again the importance of the institutional aspect. The framework proposed by \cite{ter2011co} couples the evolutionary approach to companies, the literature on industries and innovation in clusters, and the approach through complex networks of connexions between the latest in the territorial system.
}{
Il est possible de donner divers exemples d'application. \cite{doi:10.1080/00343400802662658} introduisent un cadre conceptuel pour permettre de concilier nature évolutionnaire des entreprises, théorie des clusters et réseaux de connaissance, dans lequel la co-évolution entre réseaux et entreprises est centrale, et qui est définie comme une causalité circulaire entre différentes caractéristiques de ces sous-systèmes. \cite{colletis2010co} introduit un cadre de co-évolution des territoires et de la technologie (questionnant par exemple le rôle de la proximité pour les innovations), qui révèle l'importance à nouveau de l'aspect institutionnel. Le cadre proposé par \cite{ter2011co} couple la vision évolutionnaire des entreprises, la littérature sur les industries et l'innovation dans les clusters, et l'approche par réseau complexe des connexions entre ces premiers dans le système territorial.
}

\bpar{
In environmental economics, \cite{kallis2007coevolution} show that ``broad'' approaches (that can consider most of co-dynamics as co-evolutive) are opposed to stricter approaches (in the spirit of the definition given by \cite{schamp201020}), and that in any case a precise definition, not necessarily coming from biology, must be given, in particular for the search of an empirical characterization.
}{
En économie environnementale, \cite{kallis2007coevolution} montre que des approches ``larges'' (pouvant considérer la majorité des co-dynamiques comme co-évolutives) s'opposent à des approches plus strictes (dans l'esprit de la définition donnée par \cite{schamp201020}), et que dans tous les cas une définition précise, ne venant pas forcément de la biologie, doit être donnée, en particulier pour la recherche d'une caractérisation empirique.
}

\subsubsection{Geography}{Géographie}

\bpar{
For geography, as we already presented in introduction, the works that are the closest to notions of co-evolution empirically and theoretically are closely linked to the evolutive urban theory. It is not easy to track back in the literature at what time the notion was clearly formalized, but it is clear that it was present since the foundations of the theory as recalls \noun{Denise Pumain} (see~\ref{app:sec:interviews}): the complex adaptive system is composed of subsystems that are interdependent in a complex way, often with circular causalities. The first models indeed include this vision in an implicit way, but co-evolution is not explicitly highlighted of precisely defined, in terms that would be quantifiable or structurally identifiable. \cite{paulus2004coevolution} brings empirical proofs of mechanisms of co-evolution through the study of the evolution of economic profiles of French cities. The interpretation used by~\cite{schmitt2014modelisation} is based on an entry by the evolutive urban theory, and fundamentally consists in a reading of systems of cities as highly interdependent entities.
}{
Pour la géographie, comme nous l'avons déjà présenté en introduction, les travaux les plus proches empiriquement et théoriquement des notions de co-évolution sont étroitement liés à la théorie évolutive des villes. Il n'est pas évident de tracer dans la littérature à quel moment la notion a été clairement formalisée, mais il est évident qu'elle était présente dès les fondements de la théorie comme le rappelle \noun{Denise Pumain} (voir~\ref{app:sec:interviews}) : le système complexe adaptatif est composé de sous-systèmes en interdépendances complexes, souvent circulairement causales. Les premiers modèles incluent bien cette vision de manière implicite, mais la co-évolution n'est pas appuyée explicitement ou définie précisément, en termes qui seraient quantifiables ou identifiables structurellement. \cite{paulus2004coevolution} amène des preuves empiriques de mécanismes de co-évolution par l'étude de l'évolution des profils économiques des villes françaises. L'interprétation utilisée par~\cite{schmitt2014modelisation} repose sur une entrée par la théorie évolutive des villes, et consiste fondamentalement en une lecture des systèmes de villes comme entités fortement interdépendantes.
}

\subsubsection{Physical geography}{Géographie physique}

\bpar{
In the study of landscapes, \cite{sheeren2015coevolution} evoke the co-evolution of landscape and agricultural activities, but in fact do not consider any circular effect of one on the other. Their result show a priori that the evolution of agricultural practices yield an evolution of the landscape, and it is not clear to what extent the conceptual frame of co-evolution, evoked without any more details, is used.
}{
En étude des paysages, \cite{sheeren2015coevolution} parlent de co-évolution du paysage et des activités agricoles, mais ne considèrent en fait pas d'effet circulaires de l'un sur l'autre. A priori, leurs résultats montrent que l'évolution des pratiques agricoles entraine une évolution du paysage, et il n'est ainsi pas clair dans quelle mesure le cadre conceptuel de la co-évolution, mentionné sans plus de détails, est mobilisé.
}

\subsubsection{Physics}{Physique}

\bpar{
Finally, we can mention in an anecdotical way that the term of co-evolution has also been used by physics. Its use for physical systems may induce some debates, depending if we suppose or not that the transmission assumes a transmission of \emph{information}\footnote{Information is defined within the shanonian theory as an occurence probability for a chain of characters. \cite{morin1976methode} shows that the concept of information is indeed far more complex, and that it must be thought conjointly to a given context of the generation of a self-organizing negentropic system, i.e. realizing local decreases in entropy in particular thanks to this information. This type of system is necessarily alive. We will follow here this complex approach to information.}. In the case of a purely physical ontological transmission (\emph{physical beings}), then a large part of physical systems are evolutive. \cite{hopkins2008cosmological} develop a cosmological frame for the co-evolution of cosmic heterogenous objects which presence and dynamics are difficultly explained by more classical theories (some types of galaxies, quasars, supermassive black holes). \cite{antonioni2017coevolution} study the co-evolution between synchronisation and cooperation properties within a Kuramoto oscillators network\footnote{The Kuramoto model studies synchronization within complex systems, by studying the evolution of phases $\theta_i$ coupled by interaction equations $\dot{\vec{\theta}} = \vec{\omega} + \vec{W}\left[\vec{\theta}\right] + \mathbf{B}$ where $\vec{\omega}$ are proper forcing phases and the coupling strength between $i$ and $j$ is given by $\vec{W}_{i} = \sum_j w_{ij} \sin\left(\theta_i - \theta_j\right)$ and $\vec{B}$ is noise.}, showing on the one hand that the concept can be applied to abstract objects, and on the other hand that a complex network of relations between variables can be at the origin of dynamics witnessing circular causalities, i.e. a co-evolution in that sense.
}{
Enfin, on peut noter de manière anecdotique que le terme de co-évolution a également été utilisé par la physique. L'utilisation pour des systèmes physiques peut porter à débat, selon que l'on suppose ou non que la transmission suppose une transmission d'\emph{information}\footnote{L'information est définie dans la théorie shanonienne comme une probabilité d'occurrence d'une chaîne de caractère. \cite{morin1976methode} montre que le concept d'information est en fait bien plus complexe, et qu'il doit être pensé conjointement à un contexte donné de génération d'un système auto-organisateur néguentropique, i.e. réalisant des diminutions locales d'entropie notamment grâce à cette information. Ce type de système est nécessairement vivant. Nous prendrons ici cette vision complexe de l'information.}. Dans le cas d'une transmission ontologique uniquement physique (\emph{êtres physiques}), alors une grande partie des systèmes physiques sont évolutifs. \cite{hopkins2008cosmological} développent un cadre cosmologique pour la co-évolution d'objets cosmiques hétérogènes dont la présence et les dynamiques sont difficilement expliquées par des théories plus classiques (certains types de galaxies, quasars, trous noirs supermassifs). \cite{antonioni2017coevolution} étudient la co-évolution entre des propriétés de synchronisation et de coopération au sein d'un réseau d'oscillateurs de Kuramoto\footnote{Le modèle de Kuramoto s'intéresse à la synchronisation au sein de systèmes complexes, en étudiant l'évolution de phases $\theta_i$ couplée par les équations d'interaction $\dot{\vec{\theta}} = \vec{\omega} + \vec{W}\left[\vec{\theta}\right] + \mathbf{B}$ où $\vec{\omega}$ sont les phases propres de forçage et la force de couplage entre $i$ et $j$ est donnée par $\vec{W}_{i} = \sum_j w_{ij} \sin\left(\theta_i - \theta_j\right)$ et $\vec{B}$ du bruit.}, montrant d'une part que le concept peut être appliqué à des objets abstraits, et d'autre part qu'un réseau de relations complexes entre variables peut être à l'origine de dynamiques présentant des causalités circulaires, c'est-à-dire d'une co-évolution en ce sens.
}

\subsubsection{Synthesis}{Synthèse}

\bpar{
Most of these approaches fit in the theory of complex adaptive systems developed by \noun{Holland}, in particular in~\cite{holland2012signals}: it takes any system as an imbrication of systems of boundaries, that filter signals or objects. Within a given limit, the corresponding subsystem is relatively autonomous from the outside, and is called an \emph{ecological niche}, in a direct correspondence with highly connected communities within trophic or ecological networks. This way, interdependent entities within a niche are said to be co-evolving. We will come back on that approach in our theoretical construction in~\ref{sec:theory} when we will have developed other concepts that are necessary for it.
}{
La plupart de ces approches rentrent dans la théorie des systèmes complexes adaptatifs développée par \noun{Holland}, notamment dans~\cite{holland2012signals} : il voit tout système comme une imbrication de systèmes de limites, filtrant des signaux ou des objets. Au sein d'une limite donnée, le sous-système correspondant est relativement autonome de l'extérieur, est est appelé \emph{niche écologique}, en correspondance directe avec les communautés fortement connectées au sein des réseaux trophiques ou écologiques. Ainsi, des entités interdépendantes au sein d'une niche sont dites en co-évolution. Nous reviendrons sur cette entrée lors de la construction théorique en~\ref{sec:theory} lorsque nous aurons développé d'autres concepts qui lui sont nécessaire.
}

\bpar{
We retain from this multidisciplinary view of co-evolution the fundamental following points, that are precursors of a proper definition of co-evolution that will be given further, concluding the first part.
}{
Nous retenons de cet aperçu multidisciplinaire de la co-évolution les points fondamentaux suivants précurseurs à une définition propre de la co-évolution que nous donnerons plus loin, en conclusion de la première partie.
}

\bpar{
\begin{enumerate}
	\item The presence of \emph{evolution processes} is primary, and their definition is almost always based on the existence of transmission and transformation processes.
	\item Co-evolution assumes entities or systems, belonging to distinct classes, which evolutive dynamics are coupled in a circular causal way. Approaches can differ depending on the assumptions of populations of these entities, singular objects, or components of a global system then in mutual interdependency without a direct circularity.
	\item The delineation of systems and subsystems, both in the ontological space (definition of studied objects), but also in space and time, and their distribution in these spaces, is fundamental for the existence of co-evolutionary dynamics, and it seems in a large number of cases, of their empirical characterization.
\end{enumerate}
}{
\begin{enumerate}
	\item La présence de \emph{processus d'évolution} est primaire, et leur définition se ramène presque toujours à l'existence de processus de transmission et de transformation.
	\item La co-évolution suppose des entités ou systèmes, appartenant à des classes distinctes, dont les dynamiques évolutives sont couplées de manière circulaire causale. Les approches peuvent différer selon l'hypothèse de populations de ces entités, d'objets singuliers, ou de composantes d'un système global alors en interdépendance mutuelle sans qu'il y ait circularité directe. 
	\item La délimitation des systèmes ou des sous-systèmes, à la fois dans l'espace ontologique (définition des objets étudiés), mais aussi dans l'espace et le temps, ainsi que leur distribution dans ces espaces, est fondamental pour l'existence de dynamiques co-évolutives, et a priori dans un grand nombre de cas, pour leur caractérisation empirique.
\end{enumerate}
}


\subsection{Nature of complexity and knowledge production}{Nature de la complexité et production de connaissances}

\bpar{
The two previous epistemological points that we just developed were related respectively related first to the positioning in itself, i.e. the framework to read processes of production of scientific knowledge, and then to the nature of the concepts considered. We propose to again gain in generality compared to the first one and to introduce a development modestly contributing (i.e. in our context) to the \emph{knowledge of knowledge}. The aim is to interrogate the links between complexity and processes of knowledge production.
}{
Les deux premiers points épistémologiques que nous venons de traiter relevaient respectivement du positionnement en lui-même, c'est-à-dire du cadre de lecture des processus de production de connaissance scientifique, puis de la nature des concepts considérés. Nous proposons de monter encore en généralité par rapport au premier et d'introduire un développement contribuant modestement (c'est-à-dire dans notre contexte) à \emph{la connaissance de la connaissance}. Il s'agit d'interroger les liens entre complexité et processus de production de connaissances.
}

\bpar{
One aspect of knowledge production on complex systems, that we encounter several times here (see chapter~\ref{ch:theory}), and that seems to be recurrent and even inevitable, is a certain level of reflexivity (and that would be inherent to complex system in comparison to simple systems, as we will develop further). We mean by this term both a practical reflexivity, i.e. a necessity to increase the level of abstraction, such as the need to reconstruct in an endogenous way the disciplines in which a reflexion aims at positioning as proposed in \ref{sec:quantepistemo}, or to reflect on the epistemological nature of modeling when constructing a model such as in \ref{app:sec:csframework}, but also a theoretical reflexivity in the sense that theoretical apparels or produced concepts can apply recursively to themselves. This practical observation can be related to old epistemological debates questioning the possibility of an objective knowledge of the universe that would be independent of our cognitive structure, somehow opposed to the necessity of an ``evolutive rationality'' implying that our cognitive system, product of the evolution, mirrors the complex processes that led to its emergence, and that any knowledge structure will be consequently reflexive\footnote{We thank here \noun{D. Pumain} to have formulated this alternative view on the problem that we will develop in the following.}. We do not pretend here to bring a response to such a broad and vague question as such, but we propose a potential link between this reflexivity and the nature of complexity. 
}{
Un aspect de la production de connaissance sur des systèmes complexes, auquel nous nous heurtons plusieurs fois ici (voir chapitre~\ref{ch:theory}), et qui semble être récurrent voire inévitable, est un certain niveau de réflexivité (et qui serait inhérent aux systèmes complexes en comparaison aux systèmes simples, comme nous le développerons plus loin). Nous entendons par là à la fois une réflexivité pratique, c'est-à-dire la nécessité d'élever le niveau d'abstraction, comme le besoin de reconstruire de manière endogène les disciplines dans lesquelles une réflexion cherche à se positionner comme proposé en \ref{sec:quantepistemo}, ou de réfléchir à la nature épistémologique de la modélisation lors de l'élaboration d'un modèle comme en \ref{app:sec:csframework}, mais également une réflexivité théorique en le sens que les appareils théoriques ou les concepts produits peuvent s'appliquer de manière récursive à eux-mêmes. Cette constatation pratique fait écho à des débats épistémologiques anciens questionnant la possibilité d'une connaissance objective de l'univers qui serait indépendante de notre structure cognitive, ou bien la nécessité d'une ``rationalité évolutive'' impliquant que notre système cognitif, produit de l'évolution, reflète les processus complexes ayant conduit à son émergence, et que toute structure de connaissance sera par conséquent réflexive\footnote{Nous remercions \noun{D. Pumain} d'avoir pointé cette vue alternative du problème que nous allons développer par la suite.}. Nous ne prétendons pas ici apporter une réponse à une question aussi vaste et vague telle quelle, mais proposons un lien potentiel entre cette réflexivité et la nature de la complexité.
}

\subsubsection{Complexity and complexities}{Complexité et complexités}

\bpar{
What is meant by complexity of a system often leads to misunderstandings since it can be qualified according to different dimensions and visions. We distinguish first the complexity in the sense of weak emergence and autonomy between the different levels of a system, and on which different positions can be developed as in \cite{deffuant2015visions}. We will not enter a finer granularity, the vision of social complexity giving even more nightmares to the Laplace daemon, and since it can be understood as a stronger emergence (in the sense of weak and strong emergence as developed before in~\ref{sec:computation}). We thus simplify and assume that the nature of systems plays a secondary role in our reflexion, and therefore consider complexity in the sense of an emergence.
}{
Ce qui est entendu par complexité d'un système mène souvent à des malentendus car celle-ci peut être qualifiée selon différentes dimensions et visions. Nous distinguons dans un premier temps la complexité au sens d'émergence faible et d'autonomie entre les différents niveaux d'un système, et sur laquelle différentes positions peuvent être développées comme dans \cite{deffuant2015visions}. Nous ne rentrerons pas dans une granularité plus fine, la vision de la complexité sociale donnant encore plus de fil à retordre au démon de Laplace, et pouvant être par exemple comprise par une émergence plus forte (au sens d'émergence faible et forte développée précédemment en~\ref{sec:computation}). Nous simplifions ainsi et supposons que la nature des systèmes joue un rôle secondaire dans notre reflexion, et considérons la complexité au sens d'une émergence.
}

\bpar{
Moreover, we distinguish two other ``types'' of complexity, namely computational complexity and informational complexity, that can be seen as measures of complexity, but that are not directly equivalent to emergence, since there exists no systematic link between the three. We can for example consider the use of a simulation model, for which interactions between elementary agents translate as a coded message at the upper level: it is then possible by exploiting the degrees of freedom to minimize the quantity of information contained in the message. The different languages require different cognitive efforts and compress the information in a different way, having different levels of measurable complexity~\cite{febres2013complexity}. In a similar way, architectural artefacts are the result of a process of natural and cultural evolution, and witness more or less this trajectory.
}{
D'autre part, nous distinguons deux autres ``types'' de complexité, la complexité computationnelle et la complexité informationnelle, qui peuvent être vues comme des mesures de complexité, mais qui ne sont pas directement équivalentes à l'émergence, puisqu'il n'existe pas de lien systématique entre les trois. On peut par exemple imaginer utiliser un modèle de simulation, pour lequel les interactions entre agents élémentaires se traduisent par un message codé au niveau supérieur : il est alors possible en exploitant les degrés de liberté de minimiser la quantité d'information contenue dans le message. Les différentes langues demandent des efforts cognitifs différents et compressent différemment l'information, ayant différents niveau de complexité mesurables~\cite{febres2013complexity}. De même, des artefacts architecturaux sont le résultat d'un processus d'évolution naturelle puis culturelle et peuvent témoigner plus ou moins de cette trajectoire.
}

\bpar{
Numerous other conceptual or operational characterizations of complexity exist, and it is clear that the scientific community has not converged on a unique definition~\cite{chu2008criteria}\footnote{In an approach that is in a way reflexive, \cite{chu2008criteria} proposes to continue exploring the different existing approaches, as proxies of complexity in the case of an essentialism, or as concepts in themselves. The complexity should emerge naturally from the interaction between these different approaches studying complexity, hence the reflexivity.}. We propose to focus on these three concepts in particular, for which the relations are already not evident.
}{
De nombreuses autres caractérisations conceptuelles ou opérationnelles de la complexité existent, et il est clair que la communauté scientifique n'a pas convergé sur une définition unique~\cite{chu2008criteria}\footnote{Dans une approche en un sens réflexive, \cite{chu2008criteria} propose de continuer d'explorer les différentes approches existantes, comme des proxys de la complexité dans le cas d'un essentialisme, ou comme des concepts à part entière. La complexité devrait émerger d'elle même de l'interaction entre ces différentes approches étudiant la complexité, d'où la réflexivité.}. Nous proposons de nous concentrer sur ces trois concepts en particulier, pour lesquels les relations ne sont déjà pas évidentes.
}

\bpar{
Indeed, links between these three types of complexity are not systematic, and depend on the type of system. Epistemological links can however be introduced. We will develop the links between emergence and the two other complexities, since the link between computational complexity and informational complexity is relatively well explored, and corresponds to issues in the compression of information and signal processing, or moreover in cryptography.
}{
En effet, les liens entre ces trois types de complexité ne sont pas systématiques, et dépendent du type de système. Des liens épistémologiques peuvent néanmoins être introduits. Nous traitons ceux entre émergence et les deux autres complexités, étant donné que le lien entre complexité computationnelle et complexité informationnelle est assez bien compris et relève de problématiques de compression de l'information et de traitement du signal, ou encore de cryptographie.
}

\subsubsection{Computational complexity and emergence}{Complexité computationnelle et émergence}

\bpar{
Different clues suggest a certain necessity of computational complexity to have emergence in complex systems, whereas reciprocally a certain number of adaptive complex systems have high computational capabilities.
}{
Différents indices suggèrent une certaine nécessité de complexité computationnelle pour avoir émergence dans des systèmes complexes, tandis que réciproquement un certain nombre de systèmes complexes adaptatifs sont dotés de capacités de calcul élevées.
}

\bpar{
A first link where computational complexity implies emergence is suggested by an algorithmic study of fundamental problems in quantum physics. Indeed, \cite{2014arXiv1403.7686B} shows that the resolution of the Schrödinger equation with any Hamiltonian is a NP-hard and NP-complete problem, and thus that the acceptation of $\mathbf{P}\neq\mathbf{NP}$ implies a qualitative separation between the microscopic quantum level and the macroscopic level of the observation. Therefore, it is indeed the complexity (here in the sense of their computation) of interactions in a system and its environment that implies the apparent collapse of the wave function, what rejoins the approach of \noun{Gell-Mann} by quantum decoherence~\cite{gell1996quantum}, which explains that probabilities can only be associated to decoherent histories (in which correlations have led the system to follow a trajectory at the macroscopic scale)\footnote{The \emph{Quantum Measurement Problem} arises when we consider a microscopic wave function giving the state of a system that can be the superposition of several states, and consists in a theoretical paradox, on the one hand the measures being always deterministic whereas the system has probabilities for states, and on the other hand the issue of the non-existence of superposed macroscopic states (collapse of the wave function). As reviewed by~\cite{schlosshauer2005decoherence}, different epistemological interpretations of quantum physics are linked to different explanations of this paradox, including the ``classical'' Copenhagen one which attributes to the act of observation the role of collapsing the wave function. \noun{Gell-Mann} recalls that this interpretation is not absurd since it is indeed the correlations between the quantum object and the world that product the decoherent history, but that it is far more specific, and that the collapse happens in the emergence itself: the cat is either dead or living, but not both, before we open the box.}. The paradox of the Schrödinger cat appears then as a fundamentally reductionist perspective, since it assumes that the superposition of states can propagate through the successive levels and that there would be no emergence, in the sense of the constitution of an autonomous upper level. In other terms, the work of \cite{2014arXiv1403.7686B} suggests that computational complexity is sufficient for the presence of emergence\footnote{This effective separation of scales does not a priori imply that the lower level does not play a crucial role, since \cite{vattay2015quantum} proves that the properties of quantum criticality are typical of molecules of the living, without a priori any specificity for life in this complex determination by lower scales: \cite{2016arXiv161102269V} has recently introduced a new approach linking quantum theories and general relativity in which it is shown that gravity is an emergent phenomenon and that path-dependency in the deformation of the original space introduces a supplementary term at the macroscopic level, that allows to explain deviations attributed up to now to \emph{dark matter}.}.
}{
Un premier lien où complexité computationnelle implique émergence est suggéré par un examen algorithmique des problèmes fondamentaux de la physique quantique. En effet, \cite{2014arXiv1403.7686B} démontre que la résolution de l'équation de Schrödinger avec Hamiltonien quelconque est un problème NP-difficile et NP-complet, et donc que l'acceptation de $\mathbf{P}\neq\mathbf{NP}$ implique une séparation qualitative entre le niveau quantique microscopique et le niveau d'observation macroscopique. Ainsi, c'est bien la complexité (ici au sens de leur calcul) des interactions au sein du système et de son environnement qui explique l'apparente réduction du paquet d'onde, ce qui rejoint l'approche de \noun{Gell-Mann} par la décohérence quantique~\cite{gell1996quantum}, qui explique que des probabilités ne peuvent être associées qu'aux histoires décohérentes (dans lesquelles les corrélations ont fait prendre une trajectoire au système à l'échelle macroscopique)\footnote{Le \emph{Problème de la Mesure Quantique} se pose lorsqu'on considère une fonction d'onde microscopique donnant l'état d'un système pouvant être superposition de plusieurs états, et consiste en un paradoxe théorique, les mesures étant toujours déterministes alors que le système a des probabilité d'états d'une part, et le problème de la non-existence d'états macroscopiques superposés (réduction du paquet d'onde) d'autre part. Comme revu par~\cite{schlosshauer2005decoherence}, différentes interprétations épistémologiques de la physique quantiques sont liées à différentes explications de ce paradoxe, dont celle ``classique'' de Copenhague qui donne à l'acte d'observation le rôle de reduction du paquet d'onde. \noun{Gell-Mann} précise que cette interprétation n'est pas absurde puisque c'est bien les corrélations entre l'objet quantique et le monde qui produisent l'histoire décohérente, mais qu'elle est bien trop spécifique, et que la réduction a lieu dans l'émergence elle-même : le chat est bien mort ou vivant, mais pas les deux, avant que l'on ouvre la boîte.}. Le paradoxe du chat de Schrödinger nous apparait ainsi comme une perspective fondamentalement réductionniste, puisqu'il suppose que la superposition d'états peut se propager à travers les niveaux successifs et qu'il n'y aurait pas émergence, au sens de constitution d'un niveau supérieur autonome. En d'autres termes, le travail de \cite{2014arXiv1403.7686B} suggère que la complexité computationnelle est suffisante pour la présence d'émergence\footnote{A priori, cette séparation effective des échelles n'implique pas que le niveau inférieur ne joue pas un rôle crucial, puisque \cite{vattay2015quantum} prouve que les propriétés de criticalité quantiques sont typiques des molécules du vivant, sans qu'il n'y ait a priori de spécificité pour la vie dans cette détermination complexe par les échelles inférieures : \cite{2016arXiv161102269V} a introduit une nouvelle approche liant théories quantiques et relativité générale dans laquelle il est montré que la gravité est un phénomène émergent et que la dépendance au chemin dans la déformation de l'espace de base introduit un terme supplémentaire au niveau macroscopique, qui permet d'expliquer les déviations attribuées jusqu'alors à la \emph{matière noire}.}.
}


\bpar{
Reciprocally, the link between computational complexity and emergence is revealed by questions linked to the nature of computation~\cite{moore2011nature}. Cellular automatons, that are moreover crucial for the understanding of several complex systems, have been shown as Turing-complete\footnote{A system is said to be Turing-complete if it is able to compute the same functions than a Turing machine, commonly accepted as all what is ``computable'' (\noun{Church}'s thesis). We recall that a Turing machine is a finite automaton with an infinite writing band~\cite{moore2011nature}.}, such as the Game of Life~\cite{beer2004autopoiesis}\footnote{There even exists a programming language allowing to code in the \emph{Game of Life}, available at \url{https://github.com/QuestForTetris}. Its genesis finds its origin in a challenge posted on \emph{codegolf} aiming at the conception of a Tetris, and ended in an extremely advanced collaborative project.}. Some organisms without a central nervous system are capable of solving difficult decisional problems~\cite{reid2016decision}. An ant-based algorithm is shown by~\cite{Pintea2017} as solving a Generalized Travelling Salesman Problem (GTSP), problem which is NP-difficult. This fundamental link had already been conceived by \noun{Turing}, since beyond his fundamental contributions to contemporary computer science, he studied morphogenesis and tried to produce chemical models to explain it~\cite{turing1952chemical} (that were far from actually explaining it - it is still not well understood today, see~\ref{sec:interdiscmorphogenesis} - but which conceptual contributions were fundamental, in particular for the notion of reaction-diffusion). We moreover know that a minimum of complexity in terms of constituting interactions in a particular case of agent-based system (models of boolean networks), and thus in terms of possible emergences, implies a lower bound on computational complexity, which becomes significant as soon as interactions with the environment are added~\cite{tovsic2017boolean}.
}{
Dans le sens inverse, le lien entre complexité computationnelle et émergence est mis en valeur par les questions liées à la nature de la computation~\cite{moore2011nature}. Des automates cellulaires, qui sont par ailleurs cruciaux pour la compréhension de divers systèmes complexes, ont été montrés Turing-complets\footnote{Un système est Turing-complet s'il est capable de calculer les mêmes fonctions qu'une machine de Turing, communément accepté comme l'ensemble du ``calculable'' (thèse de \noun{Church}). Pour mémoire, une machine de Turing est un automate fini à bande d'écriture infinie~\cite{moore2011nature}.}, comme le Jeu de la Vie~\cite{beer2004autopoiesis}\footnote{Il existe même un langage de programmation permettant de programmer en \emph{Game of Life}, disponible à \url{https://github.com/QuestForTetris}. Sa genèse trouve son origine dans un défi posté sur \emph{codegolf} ayant pour but la conception d'un Tetris, et a abouti à un projet collaboratif extrêmement avancé.}. Des organismes sans système nerveux central sont capables de résoudre des problèmes décisionnels difficiles~\cite{reid2016decision}. Un algorithme à base de fourmis est montré par~\cite{Pintea2017} comme résolvant un Problème du Voyageur de Commerce Généralisé (GTSP), problème NP-difficile. Ce lien fondamental avait déjà été envisagé par \noun{Turing}, puisqu'au delà de ses contributions fondamentales à l'informatique moderne, il s'était intéressé à la morphogenèse et a tenté de produire des modèles chimiques d'explication de celle-ci~\cite{turing1952chemical} (qui étaient très loin de effectivement l'expliquer - elle n'est toujours pas bien comprise aujourd'hui, voir~\ref{sec:interdiscmorphogenesis} - mais dont les contributions conceptuelles ont été fondamentales, notamment pour la notion de réaction-diffusion). On sait par ailleurs qu'un minimum de complexité en termes d'interactions constituantes dans un cas particulier de système basé sur les agents (modèles de réseaux booléens), et donc d'émergences possibles, implique une borne inférieure sur la complexité computationnelle, qui devient conséquente dès que les interactions avec l'environnement sont ajoutées~\cite{tovsic2017boolean}.
}


\subsubsection{Informational complexity and emergence}{Complexité informationnelle et émergence}

\bpar{
Informational complexity, or the quantity of information contained in a system and the way it is stored, also bears some fundamental links with emergence. Information is equivalent to the entropy of a system and thus to its degree of organisation - this what allows to solve the apparent paradox of the Maxwell Daemon that would be able to diminish the entropy of an isolated system and thus contradict the second law of thermodynamics: it indeed uses the information on positions and velocities of molecules of the system, and its action balances to loss of entropy through its captation of information\footnote{The Maxwell Daemon is more than an intellectual construction: \cite{cottet2017observing} implements experimentally a daemon at the quantic level.}.
}{
La complexité informationnelle, ou la quantité d'information contenue dans un système et la manière dont celle-ci est stockée, entretient également des liens fondamentaux avec l'émergence. L'information est équivalente à l'entropie d'un système et donc à son degré d'organisation - c'est ce qui a permis de résoudre le paradoxe apparent du Démon de Maxwell qui serait capable de diminuer l'entropie d'un système isolé et donc contredire la deuxième loi de la thermodynamique : celui-ci utilise en fait l'information sur les positions et vitesses des molécules du système, et son action compense la perte d'entropie par sa captation d'information\footnote{Le démon de Maxwell est plus qu'une construction intellectuelle : \cite{cottet2017observing} implémente un démon expérimentalement au niveau quantique.}.
}

\bpar{
This notion of local increase in entropy has been largely studied by \noun{Chua} under the form of the \emph{Local Activity Principle}, which is introduced as a third principle of thermodynamics, allowing to explain with mathematical arguments the self-organization for a certain class of complex systems that typically involve reaction-diffusion equations~\cite{mainzer2013local}.
}{
Cette notion d'accroissement local de l'entropie a été étudiée largement par \noun{Chua} sous la forme du \emph{Local Activity Principle}, qui est introduit comme un troisième principe de la thermodynamique, permettant d'expliquer par des arguments mathématiques l'auto-organisation pour une certaine classe de systèmes complexes typiquement impliquant des équations de réaction-diffusion~\cite{mainzer2013local}.
}

\bpar{
The way information is stored and compressed is essential for life, since the ADN is indeed an information storage system, which role at different levels is far from being fully understood. Cultural complexity also witnesses of an information storage at different levels, for example within individuals but also within artefacts and institutions, and information flows that necessarily deal with the two other types of complexities. Information flows are essential for self-organization in a multi-agent system. Collective behaviors of fishes or birds are typical examples used to illustrate emergence and belong to the canonic examples of complex systems. We only begin to understand how these flows structure the system, and what are the spatial patterns of information transfer within a \emph{flock} for example: \cite{crosato2017informative} introduce first empirical results with transfer entropy for fishes and lay the methodological basis of this kind of studies. 
}{
La manière dont l'information est stockée et compressée est essentielle pour la vie, puisque l'ADN est bien un système de stockage d'information, dont le rôle à différents niveaux est bien loin d'être compris complètement. La complexité culturelle témoigne également d'un stockage de l'information à différents niveaux, par exemple au sein des individus mais aussi des artefacts et des institutions, et des flux d'information relevant nécessairement des deux autres types de complexité. Les flux d'information sont essentiels pour l'auto-organisation dans un système multi-agents. Les comportements collectifs de poissons ou d'oiseaux sont des exemples typiques utilisés pour illustrer l'émergence et font partie des cas d'école de systèmes complexes. On commence cependant seulement à comprendre comment ces flux structurent le système, et quels sont les motifs spatiaux de transfert d'information au sein d'un \emph{flock} par exemple : \cite{crosato2017informative} introduisent des premiers résultats empiriques avec l'entropie de transfert pour des poissons et posent les bases méthodologiques de ce type d'étude.
}

\subsubsection{Knowledge production}{Production de connaissances}

\bpar{
We know have enough material to come to reflexivity. It is possible to position knowledge production at the intersection of interactions between types of complexity developed above. First of all, knowledge as we consider it can not be dissociated from a collective construction, and implies thus an encoding and a transmission of information: it is at an other level all problematics linked to scientific communication. The production of knowledge thus necessitates this first interaction between computational complexity and informational complexity. The link between informational complexity and emergence is introduced if we consider the establishment of knowledge as a morphogenetic process. It is shown in~\ref{sec:interdiscmorphogenesis} that the link between form and function is fundamental in psychology: we can interpret it as a link between information and meaning, since semantics of a cognitive object can not be considered without a function. \noun{Hofstader} recalls in~\cite{hofstadter1980godel} the importance of symbols at different levels for the emergence of a thought, that consist in signals at an intermediate level. Finally, the last relation between computational complexity and emergence is the one allowing us a positioning in particular on knowledge production on complex systems, the previous links being applicable to any type of knowledge.
}{
Nous avons à présent la matière suffisante pour en venir à la réflexivité. Il est possible de positionner la production de connaissances à l'intersection des interactions entre types de complexité développées ci-dessus. Tout d'abord, la connaissance telle que nous l'envisageons ne peut se passer d'une construction collective, et implique donc un encodage et une transmission de l'information : il s'agit à un autre niveau de toutes les problématiques liées à la communication scientifique. La production de connaissances nécessite donc cette première interaction entre complexité computationnelle et complexité informationnelle. Le lien entre complexité informationnelle et émergence est mobilisé si on considère l'établissement de connaissances comme un processus morphogénétique. Il est montré en~\ref{sec:interdiscmorphogenesis} que le lien entre forme et fonction est fondamental en psychologie : nous pouvons l'interpréter comme un lien entre information et sens, puisque la sémantique d'un objet cognitif ne peut se passer d'une fonction. \noun{Hofstader} rappelle dans~\cite{hofstadter1980godel} l'importance des symboles à différents niveaux pour l'émergence d'une pensée, qui consistent à un niveau intermédiaire en des signaux. Enfin, la dernière relation entre complexité computationnelle et émergence est celle qui nous permet d'affirmer qu'on s'intéresse particulièrement à une production de connaissance sur des systèmes complexes, les deux premiers pouvant s'appliquer à tout type de connaissance.
}

\bpar{
Therefore, any \emph{knowledge of the complex} embraces not only all complexities and their relations in its content, but also in its nature as we just showed. The structure of knowledge in terms of complexity is analog to the structure of systems its studies. We postulate that this structural correspondence implies a certain recursivity, and thus a certain level of \emph{reflexivity} (in the sens of knowledge of itself and its own conditions).
}{
Ainsi, toute \emph{connaissance du complexe} embrasse non seulement toutes les complexités et leur relations dans son contenu, mais aussi dans sa nature comme nous venons de montrer. La structure de la connaissance en termes de complexité est analogue à la structure des systèmes qu'elle étudie. Nous postulons que cette correspondance structurelle implique une certaine récursivité, et donc un certain niveau de \emph{réflexivité} (au sens de connaissance d'elle-même et de ses propres conditions). 
}


\bpar{
We can try to extend to reflexivity in terms of a reflexion on the disciplinary positioning: following \cite{pumain2005cumulativite}, the complexity of an approach is also linked to the diversity of viewpoints that are necessary to construct it. To reach this new type of complexity\footnote{For which links with the previous types naturally appear: for example, \cite{gell1995quark} considers the effective complexity as an \emph{Algorithmic Information Content} (close to Kolmogorov complexity) of a Complex Adaptive System \emph{which is observing an other} Complex Adaptive System, what gives their importance to informational and computational complexities and suggests the importance of the observational viewpoint, and by extension of their combination - what furthermore must be related to the perspectivist approach of complex sciences presented above.}, that would be a supplementary dimension linked to the knowledge of complex systems, reflexivity must be at the core of the approach. \cite{read2009innovation} recall that innovation has been made possible when societies reached the ability to produce and diffuse innovation on their own structure, i.e when they were able to reach a certain level of reflexivity. The \emph{knowledge of the complex} would thus be the product and the support of its own evolution thanks to reflexivity which played a fundamental role in the evolution of the cognitive system: we could thus suggest to gather these considerations, as proposed by \noun{Pumain}, as a new epistemological notion of \emph{evolutive rationality}.
}{
On peut tenter d'étendre à la réflexivité en tant que réflexion sur le positionnement disciplinaire : suivant \cite{pumain2005cumulativite}, la complexité d'une approche est également liée à la diversité des points de vue nécessaires pour la construire. Pour atteindre ce nouveau type de complexité\footnote{Pour laquelle des liens avec les types précédents apparaissent naturellement : par exemple, \cite{gell1995quark} considère la complexité effective comme le \emph{Contenu d'Information Algorithmique} (proche de la complexité de Kolmogorov) d'un Système Complexe Adaptatif \emph{observant un autre} Système Complexe Adaptatif, ce qui donne son importance aux complexités informationnelle et computationnelle et suggère l'importance du point de vue d'observation, et par extension de la combinaison de ceux-ci - ce qui est par ailleurs à mettre en relation avec l'approche perspectiviste des sciences complexes présentée précédemment.}, qui serait une dimension supplémentaire liée à la connaissance des systèmes complexes, la réflexivité doit être au coeur de la démarche. \cite{read2009innovation} rappellent que l'innovation a été rendue possible quand les sociétés ont été capables de produire et diffuser de l'information sur leur propre structure, c'est-à-dire quand elles ont pu atteindre un certain niveau de réflexivité. La \emph{connaissance du complexe} serait donc le produit et le support de sa propre évolution grâce à la réflexivité qui a joué un rôle fondamental dans l'évolution du système cognitif : on pourrait ainsi suggérer de rassembler ces considérations, comme proposé par \noun{Pumain}, sous une nouvelle notion épistémologique de \emph{rationalité évolutive}.
}

\bpar{
To conclude, we can remark that given the law of \emph{requisite complexity}, proposed by \cite{gershenson2015requisite} as an extension of \emph{requisite variety}~\cite{ashby1991requisite}\footnote{One of the crucial principles of cybernetics, the \emph{requisite variety}, postulates that to control a system having a certain number of states, the controller must have at least as much states. \noun{Gershenson} proposes a conceptual extension of complexity, which can be justified for example by \cite{allen2017multiscale} which introduce the multi-scale \emph{requisite variety}, showing the compatibility with a theory of complexity based on information theory.}, the \emph{knowledge of the complex} will necessarily have to be a \emph{complex knowledge}. This other point of view reinforces the necessity of reflexivity, since following \noun{Morin} (see for example \cite{morin1991methode} on the production of knowledge), the \emph{knowledge of knowledge} is central in the construction of a complex thinking.
}{
Pour conclure, notons qu'étant donné la loi de la \emph{requisite complexity}, proposée par \cite{gershenson2015requisite} comme extension de la \emph{requisite variety}~\cite{ashby1991requisite}\footnote{L'un des principes cruciaux de la cybernétique, la \emph{requisite variety}, postule que pour contrôler un système ayant un certain nombre d'états, le contrôleur doit avoir au moins autant d'états. \noun{Gershenson} propose une extension conceptuelle à la complexité, qui peut être justifiée par exemple par \cite{allen2017multiscale} qui introduisent la \emph{requisite variety} multi-échelle, démontrant la compatibilité avec une théorie de la complexité basée sur la théorie de l'information.}, la \emph{connaissance du complexe} devra nécessairement être \emph{connaissance complexe}. Cet autre point de vue renforce la nécessité de la réflexivité, puisque suivant \noun{Morin} (voir par exemple \cite{morin1991methode} sur la production de connaissance), la \emph{connaissance de la connaissance} est centrale dans l'établissement d'une pensée complexe.
}



\subsubsection*{Practical implications}{Conséquences pratiques}

\bpar{
To conclude this epistemological section, we propose to synthesize all the ideas introduced as concrete manifestations that directly yield from them, and that strongly condition all the forms and semantics of knowledge introduced in the following. These directions (that we will not go up to name principles since they are only at the state of sketch) can be grouped into three large families: modeling practices, Open Science practices, and epistemology. On the domain of modeling practices, in each section emerge different axis that are more or less complementary:
}{
Pour conclure cette section épistémologique, nous proposons de synthétiser l'ensemble des idées introduites sous forme de manifestations concrètes en découlant directement, et qui conditionneront fortement l'ensemble de la forme et de la sémantique de la connaissance introduite par la suite. Ces directions (que nous n'irons pas jusqu'à nommer principes car seulement à l'état d'ébauche) peuvent être regroupées en trois grandes familles : pratiques de modélisation, pratique de la science ouverte, et épistémologie. Sur le plan des pratiques de modélisation, dans chaque section se dégagent différents axes plus ou moins complémentaires :
}

\bpar{
\begin{itemize}
	\item Modeling, which will be in most cases equivalent to simulation, must be understood as an indirect tool of knowledge on processes within a complex system or on its structure (according to the section on ``why modeling''), and models will necessarily have to be complex (following the reflexion on the different types of complexity) in the sense that they capture a phenomenon of weak emergence, but still respecting constraints of parsimony.
	\item The exploration of models is fully contained in the modeling enterprise (see reproducibility), and intensive computation is a cornerstone to efficiently explore simulation models (see intensive computation). Sensitivity analysis methods must be questioned and extended if needed (as illustrates the example of the sensitivity to space).
	\item As suggested by the perspectivist positioning, the coupling of models will have to play a crucial role in the capture of complexity.
\end{itemize}
}{
\begin{itemize}
	\item La modélisation, qui sera dans la majorité des cas équivalente à la simulation, doit être comprise comme un instrument de connaissance indirect sur des processus au sein d'un système complexe ou sur la structure de celui-ci (d'après la sous-section sur ``pourquoi modéliser''), et les modèles devront nécessairement être complexes (d'après la réflexion sur les différents types de complexité) au sens qu'il capturent un phénomène d'émergence faible, tout en respectant des exigences de parcimonie.
	\item L'exploration des modèles est partie intégrante de l'entreprise de modélisation (voir reproductibilité), et le calcul intensif est un élément clé pour explorer efficacement les modèles de simulation (voir calcul intensif). Les méthodes d'analyse de sensibilité doivent être questionnées et étendues si besoin (comme l'illustre l'exemple de la sensibilité à l'espace).
	\item Comme suggéré par le positionnement perspectiviste, le couplage de modèles devra jouer un rôle crucial dans la capture de la complexité.
\end{itemize}
}

\bpar{
Concerning open science, we can extract the following points:
}{
Pour la science ouverte, on peut extraire les points suivants :
}

\bpar{
\begin{itemize}
	\item The necessity of all measures linked to open science to allow the construction always more complex models, towards the co-construction of models by different disciplines.
	\item In this frame, the full opening of source code, together with its readability are crucial. The complete explicitation of the model in the scientific reporting, and a self-sustaining code documentation, are two aspect of it.
	\item The question of open data is not negotiable in that frame. The quasi-totality of our treatments is based on initially open data, and when it is not the case we work at an aggregated level for which data can be opened. Constructed open data are open.
	\item Concerning the methods of interactive exploration, which are an aspect of opening science, we develop some, but stay limited compared to the ideal requirement that these should be fully compatible with a reproducible approach.
\end{itemize}
}{
\begin{itemize}
	\item La nécessité de l'ensemble des démarches liées à la science ouverte pour parvenir à la construction de modèles toujours plus complexes, vers la co-construction de modèles par différentes disciplines.
	\item Dans ce cadre, l'ouverture complète du code source, ainsi que sa lisibilité sont cruciaux. L'explicitation complète du modèle dans le compte-rendu scientifique, ainsi qu'une documentation du code auto-suffisante, sont deux aspects de celle-ci.
	\item La question des données ouvertes n'est pas négociable dans ce cadre. La quasi-totalité de nos traitements est basée sur des données initialement ouvertes, et lorsque ce n'est pas le cas nous travaillons à un niveau agrégé auquel on peut fournir les données. Les jeux de données construits sont ouverts.
	\item Concernant les méthodes d'exploration interactive, qui sont un pendant de l'ouverture de la science, nous en développons un certain nombre, mais restons limités par rapport au pré-requis idéal qui devrait rendre celles-ci totalement compatibles avec une démarche reproductible.
\end{itemize}
}

\bpar{
Finally, from the epistemological point of view, we can also find ``practical'' implications that will naturally be more implicit in our approach, but not less structuring:
}{
Enfin, sur le point épistémologique, on peut également tirer des implications ``pratiques'' qui seront bien évidemment plus implicites dans notre démarche, mais pas moins structurantes :
}

\bpar{
\begin{itemize}
	\item Our inspiration will essentially be interdisciplinary and will aim at combining different points of view.
	\item Different knowledge domains (notion that we will precise in~\ref{sec:knowledgeframework}, but that we can understand for now in the sense of theoretical, modeling and empirical domains introduced by~\cite{livet2010}) can not be dissociated for any approach of scientific production, and we will use them in a strongly dependent way.
	\item Our approach will have to imply a certain level of reflexivity.
	\item The construction of a complex knowledge (\cite{morin1991methode}) is neither inductive nor deductive, but constructive in the idea of a morphogenesis of knowledge: it can be for example difficult to clearly identify precise ``scientific deadlocks'' since this metaphor assumes that an already constructed problem has to be unlocked, and even to constrain notions, concepts, objects or models in strict analytical frameworks, by categorizing them following a fixed classification, whereas the issue is to understand if the construction of categories is relevant. Doing it a posteriori is similar to a negation of the circularity and recursivity of knowledge production. The elaboration of ways to report that translate the diachronic character and the evolutive properties of it is an open problem.
\end{itemize}
}{
\begin{itemize}
	\item Notre inspiration sera essentiellement interdisciplinaire et cherchera à croiser les différents points de vue.
	\item Les différents domaines de connaissance (notion que nous préciserons en~\ref{sec:knowledgeframework}, mais qu'on peut comprendre pour l'instant au sens des domaines théorique, empirique et de la modélisation introduits par~\cite{livet2010}) sont indissociables pour toute démarche de production scientifique, et nous les mobiliserons de manière fortement dépendante.
	\item Notre démarche devra comprendre un certain niveau de réflexivité.
	\item La construction d'une connaissance complexe (\cite{morin1991methode}) est ni inductive ni déductive, mais constructive dans l'idée d'une morphogenèse de la connaissance : il peut par exemple être délicat d'identifier clairement des ``verrous scientifiques'' précis puisque cette métaphore suppose qu'il faut débloquer un problème déjà construit, et de même de faire rentrer notions, concepts, objet ou modèles dans des cadres analytiques stricts, en les catégorisant selon une classification fixe, alors que l'enjeu est de comprendre si la construction des catégories est pertinente. Le faire a posteriori relève d'une négation de la circularité et de la récursivité de la production de connaissance. L'élaboration de modes de compte-rendu rendant compte du caractère diachronique et des propriétés évolutives de celle-ci est un problème ouvert.
\end{itemize}
}

\stars

%


\newpage

\section*{Chapter Conclusion}{Conclusion du Chapitre}

\bpar{
Reading an article or a book is always more enlightening when we personally know the author, first because we can understand the \emph{private jokes} and extrapolate some developments of narrations which must be synthetic (even if the art of writing is indeed to try to transmit most of these elements, the ambiance in other words), and secondly because personality has complex implications on the way to apprehend the nature of knowledge and a certain a priori structure of the world. Therefore, scientific knowledge would highly probably be less rich if it was produced by machines with equivalent cognitive capacities, with equivalent empirical and subjective knowledge and experience and as diverse as the human ones, but that would have been programmed to minimize the impact of their personality and their convictions on writing and communication (still assuming that they would have a certain form of data and functions more or less equivalent). In these research laboratories typical from \emph{Blade Runner}, we doubt that the production of a knowledge of the complex would effectively be possible, since these machines would actually miss the \emph{evolutive rationality} developed in~\ref{sec:epistemology}, and we strongly doubt that is could be produced, at least given the current state of knowledge in artificial intelligence.
}{
La lecture d'un article ou d'un ouvrage est toujours bien plus éclairante lorsqu'on connait personnellement l'auteur, d'une part car on peut profiter des \emph{private joke} et extrapoler certains développements des narrations qui se doivent synthétiques (même si l'art de l'écriture est justement d'essayer de transmettre la majorité de ces éléments, l'ambiance en quelque sorte), et d'autre part car la personnalité a des implications complexes sur la manière d'appréhender la nature de la connaissance et une certaine structure a priori du monde. Pour cela, la connaissance scientifique serait très probablement moins riche si elle était produite par des machines aux capacités cognitives équivalentes, aux connaissances et experiences empiriques subjectives équivalentes et aussi diverses que celles humaines, mais qui auraient été programmées pour minimiser l'impact de leur personnalité et de leur convictions sur l'écriture et la communication (toujours en supposant qu'elles aient une certaine forme de données et fonctions plus ou moins équivalentes). Dans ces laboratoires de recherche dignes de \emph{Blade Runner}, nous doutons que la production d'une connaissance du complexe serait effectivement possible, puisqu'il manquerait à ces machines justement la \emph{rationalité évolutive} développée en~\ref{sec:epistemology}, et nous doutons fortement que celle-ci puisse être produite du moins dans l'état des connaissances actuelles en intelligence artificielle.
}

\bpar{
The aim of this chapter was thus to ``get to know each other'' on the positioning points which are inevitable for all our reflexion. These are furthermore crucial since they strongly condition some research directions.
}{
Le but de ce chapitre était donc ``de faire connaissance'' sur les points de positionnements incontournables pour l'ensemble de notre réflexion. Ceux-ci en sont d'autant plus cruciaux car conditionnent très fortement certaines directions de recherche.
}

\bpar{
Our positioning on reproducibility developed in~\ref{sec:reproducibility} implies some modeling choices, in particular the univocal use of open platforms, of open workflows and open implementations; it also implies a choice of data which must be accessible at the maximum, or made accessible, and thus some choices of objects and ontologies, or rather the non-choice of some: our problematic could be studied on fine company data while still keeping a consistence with the theoretical and thematic approach (the evolutive urban theory has largely made similar studies such as for example~\cite{paulus2004coevolution}), but the relative closing of this type of data does not make them usable in our approach.
}{
Notre positionnement sur la reproductibilité développé en~\ref{sec:reproducibility} implique certains choix de modélisation, notamment l'utilisation univoque de plateformes ouvertes, de workflow et d'implémentations ouverts ; il implique aussi un choix de données qui se doivent au maximum d'être accessibles ou rendues accessibles, et donc certains choix d'objets et d'ontologie, ou plutôt le non-choix de certains : nos problématiques pourraient être mobilisées sur des données d'entreprise fines tout en gardant une cohérence avec l'approche théorique et thématique (la théorie évolutive des villes a largement mobilisé ce type d'étude comme par exemple~\cite{paulus2004coevolution}), mais la relative fermeture de ce type de données ne les rend pas utilisables dans notre démarche.
}

\bpar{
Then, our positioning on the role of intensive computation and the need of model exploration~\ref{sec:computation} is source of all the numerical experiments and the methodologies used or developed.
}{
Ensuite, notre positionnement sur le rôle du calcul intensif et les besoins d'exploration des modèles~\ref{sec:computation} est source de l'ensemble des expériences numériques et des méthodologies utilisées ou développées.
}

\bpar{
Finally, our epistemological positioning~\ref{sec:epistemology} percolates in all our work, and allows to build the first bricks for more systematic theoretical formalizations which will be developed in chapter~\ref{ch:theory}.
}{
Enfin, notre positionnement épistémologique~\ref{sec:epistemology} percole dans l'ensemble de notre travail, et permet de poser les premières briques pour des formalisations théoriques plus systématiques qui seront développées en chapitre~\ref{ch:theory}.
}




\bpar{\chapter*{Conclusion of Part I: a definition of co-evolution}}{\chapter*{Conclusion de la Partie I : une définition de la co-évolution}}


\bpar{
\markboth{Conclusion of Part I}{Conclusion of Part I}
}{
\markboth{Conclusion de la Partie I}{Conclusion de la Partie I}
}




\bpar{
This first part allows us to formulate much more precisely our research question. Indeed, the first chapter allowed us to draw a sketch of the diversity of processes involved and of the temporal and spatial scales concerned. The second chapter gave us a very general view of existing modeling approaches and of their precise scientific context. Finally, the third chapter positions the question in an epistemological way, shed some light on co-evolution through a multi-disciplinary perspective, and clarifies the complexity with which we are dealing. It allows us to open on the directions to take in order to lead successfully the project of modeling the co-evolution.
}{
Cette première partie nous permet de cerner bien plus précisément notre question de recherche. En effet, le premier chapitre nous a permis de dresser un portrait de la diversité des processus impliqués et des échelles temporelles et spatiales concernées. Le deuxième chapitre nous a donné une vue très générale des modélisations existantes et de leur contexte scientifique précis. Enfin, le troisième chapitre positionne la question de manière épistémologique, apporte un éclairage multi-disciplinaire sur la co-évolution, et clarifie la complexité dans laquelle nous nous situons. Cela nous permet d'ouvrir sur les directions à prendre par la suite pour mener à bien l'entreprise de modélisation de la co-évolution.
}

\subsection*{Defining co-evolution}{Définir la co-évolution}

\bpar{
After the literature review given in~\ref{sec:modelingsa}, that includes different degrees of coupling between components of networks and territories, we are first able to precise what will be meant by \emph{modeling co-evolution}, by giving a definition of co-evolution in view of the multidisciplinary overview given in~\ref{sec:epistemology}.
}{
Après l'aperçu de la littérature donné en~\ref{sec:modelingsa}, incluant différents degrés de couplage entre les composantes des réseaux et territoires, nous sommes tout d'abord en mesure de préciser ce que nous entendrons par \emph{modéliser la co-évolution}, en fixant une définition de la co-évolution au regard de l'aperçu multidisciplinaire mené en~\ref{sec:epistemology}.
}

\bpar{
We propose the following entry for the specific case of transportation networks and territories, which echoes to the three main points (existence of evolutive processes, definition of entities or populations, isolation of subsystems in space and time) that we gave in~\ref{sec:epistemology}. It verifies the three following specifications.
}{
Nous proposons l'entrée suivante pour le cas spécifique des réseaux de transport et des territoires, qui fait écho au trois points essentiels (existence de processus évolutifs, définition des entités ou des populations, isolation de sous-systèmes dans le temps et l'espace) que nous avons dégagé en~\ref{sec:epistemology}. Celle-ci vérifie les trois spécifications suivantes.
}

\bpar{
First of all, evolutive processes correspond to transformations of components of the territorial system at the different scales: transformation of cities on the long time, of their networks, transmission between cities of socio-economic characteristics carries by microscopic agents but also cultural transmission, reproduction and transformation of agents themselves (firms, households, operators)\footnote{This list is based on assumptions of the evolutive urban theory that we already briefly introduced and that we will develop in itself in Chapter~\ref{ch:evolutiveurban}. It can not be exhaustive, since what would be the ``ADN of a city'' remains an open question as recalls \noun{Denise Pumain} in a dedicated interview~\ref{app:sec:interviews}.}.
}{
Dans un premier temps, les processus évolutifs correspondent aux transformations des composantes du système territorial aux différentes échelles : transformation sur le temps long des villes, de leurs réseaux, transmission entre villes des caractéristiques socio-économiques portées par les agents microscopiques mais aussi transmission culturelle, reproduction et transformation des agents eux-mêmes (firmes, ménages, opérateurs)\footnote{Cette liste s'appuie sur les hypothèses de la théorie évolutive des villes que nous avons déjà introduite brièvement et que nous développerons à part entière en Chapitre~\ref{ch:evolutiveurban}. Elle ne peut être exhaustive, puisque ce qui ferait ``l'ADN d'une ville'' reste une question ouverte comme nous le rappelle \noun{Denise Pumain} dans un entretien dédié~\ref{app:sec:interviews}.}.
}

\bpar{
These evolutive processes may imply a co-evolution. Within a territorial system, can simultaneously co-evolve: (i) given entities (a given infrastructure and given characteristics of a given territory for example, i.e. individuals), when their mutual influence will be circularly causal (at the corresponding scale); (ii) populations of entities, what will be translated for example as such type of infrastructure and given territorial components co-evolve at a statistical level in a given geographical region; (iii) all the components of a system at a small geographical scale when there exists strong global interdependencies. Our approach is thus fundamentally \emph{multi-scale} and articulates different significations at different scales.
}{
Ces processus évolutifs peuvent impliquer une co-évolution. Au sein d'un système territorial, pourront être en co-évolution à la fois : (i) des entités données (telle infrastructure et telles caractéristiques de tel territoire par exemple, c'est-à-dire des individus), lorsque leur influence mutuelle sera circulairement causale (à l'échelle leur correspondant) ; (ii) des populations d'entités, ce qui se traduira par exemple par tel type d'infrastructure et telle composante territoriale co-évoluent au niveau statistique dans une région géographique donnée ; (iii) l'ensemble des composantes d'un système à petite échelle géographique lorsqu'il existe de fortes interdépendances globales. Notre vision est donc fondamentalement \emph{multi-échelles} et articule différentes significations à différentes échelles.
}

\bpar{
Finally, the constraint of an isolation implies, in relation with the previous point, that co-evolution and the articulation of significations will have a meaning if there exists spatio-temporal isolations of subsystems in which differente co-evolutions operate, what is directly in accordance with a vision in \emph{Multi-scalar systems of systems}.
}{
Enfin, la contrainte d'une isolation implique, en lien avec le point précédent, que la co-évolution et l'articulation des significations auront un sens s'il existe des isolations spatio-temporelles de sous-systèmes où s'effectuent les différentes co-évolutions, ce qui est en accord direct avec un vision en \emph{Systèmes de systèmes multi-échelles}.
}


\bpar{
This extended definition will constitute our reference in the following when we will evoke the co-evolution of transportation networks and territories.
}{
Cette définition élargie constituera notre référence par la suite lorsqu'on parlera de co-évolution des réseaux de transport et des territoires.
}


\bpar{
We can then synthesize the fundamental results of this first part in the two following significant facts:
\begin{enumerate}
	\item The hypothesis of co-evolution of transportation networks and territories is supported from a theoretical and thematic point of view, et we construct a precise definition for it.
	\item Co-evolution remains relatively poorly explored in the literature of urban modeling, the characteristic of concerned disciplines and their interactions being a potential cause for it.
\end{enumerate}
}{
Nous pouvons alors synthétiser les résultats fondamentaux de cette première partie dans les deux faits marquants suivants :
\begin{enumerate}
	\item L'hypothèse de la co-évolution des réseaux de transport et des territoires est supportée d'un point de vue théorique et thématique, et nous en construisons une définition précise.
	\item La co-évolution reste très peu explorée dans la littérature de modélisation urbaine, les caractéristiques des disciplines concernées et leurs interactions pouvant en être une cause.
\end{enumerate}
}

\bpar{
We develop now the perspective that open at this stage.
}{
Développons à présent les perspectives qui s'ouvrent à ce stade.
}

\subsection*{On the need of an empirical characterization}{Du besoin d'une caractérisation empirique}

\bpar{
The broadest signification, i.e. generalized interdependency, is rapidly limited if its patterns are not finely characterized. It allows as an epistemological premise to consider certain ontologies and certain modeling approaches, but allows difficultly to finely understand the structure and processes of a system. The object will be then to decrease in generality and consider subsystems, in which we can consider the co-evolution of entities and of population. An understanding at this level necessitates a fine empirical characterization, without which our distinction would have no sense. A question that opens, and that we will tackle in the following, is then which are the possible empirical methods to characterize a co-evolution between entities or populations of entities.
}{
La signification la plus large, c'est-à-dire l'interdépendance généralisée, trouve vite ses limites si les motifs ne sont pas finement caractérisés. Elle permet comme prémisse épistémologique de considérer certaines ontologies et certaines démarches de modélisation, mais permet difficilement de comprendre finement la structure et les processus d'un système. Il s'agira alors de descendre en généralité et de considérer des sous-systèmes, au sein desquels on peut s'intéresser à la co-évolution d'entités et de population. Une compréhension à ce niveau nécessite une caractérisation empirique fine, sans quoi notre distinction n'aurait pas de sens. Une question qui s'ouvre, et que nous devrons traiter par la suite, est alors quelles sont les méthodes empiriques possibles pour caractériser une co-évolution entre entités ou populations d'entités.
}

\subsection*{Two complementary tracks}{Deux pistes complémentaires}


\bpar{
The state of the art done in~\ref{sec:modelingsa} above witnesses a weakness in the literature in the domain of strong coupling between the evolution of territories and network growth, given the restricted range and the disparity of reviewed works. The gap to fill on this point would thus be linked to the introduction of models strongly coupled in time more or less multi-processes and multi-scale, for which a part of the models described in~\ref{sec:modelingsa} then in~\ref{sec:modelography} are precursors.
}{
L'état de l'art fait en~\ref{sec:modelingsa} ci-dessus témoigne d'une faiblesse de la littérature dans le domaine du couplage fort entre évolution des territoires et croissance des réseaux, vu la portée restreinte et la disparité des travaux revus. Les lacunes à combler sur ce point seraient donc liées à l'introduction de modèles fortement couplés dans le temps plus ou moins multi-processus et multi-échelles, pour lesquels une partie des modèles décrits en~\ref{sec:modelingsa} puis en~\ref{sec:modelography} sont précurseurs.
}

\bpar{
The first exploratory research we will lead will have to answer to different conceptual tensions that result from the conclusions we just obtained:
\begin{itemize}
	\item allowing both an empirical approach, and more particularly a characterization method, and a modeling approach;
	\item allowing to take into account different scales;
	\item allowing the inclusion of ontologies for territories and for networks that are not always directly compatible.
\end{itemize}
}{
Les premières recherches exploratoires que nous allons mener doivent répondre à différentes tensions conceptuelles qui découlent des conclusions que nous venons de tirer :
\begin{itemize}
	\item permettre à la fois une approche empirique, et en particulier un méthode de caractérisation, ainsi qu'une approche de modélisation ;
	\item permettre la prise en compte de différentes échelles ;
	\item permettre l'inclusion d'ontologies pour les territoires et pour les réseaux qui ne sont pas toujours directement compatibles.
\end{itemize}
}

\bpar{
The scales will especially be a mesoscopic and a macroscopic scale since as we suggested in~\ref{sec:reproducibility} with the study of traffic flows, and as shows \cite{yasmin2017macro} for the validation of an activity model, the microscopic scale witnesses complex trajectories that are difficult to reproduce.
}{
Les échelles seront notamment une échelle mesoscopique et une échelle macroscopique puisque comme nous l'avons suggéré en~\ref{sec:reproducibility} avec l'étude des flux de trafic, et comme le montre \cite{yasmin2017macro} pour la validation d'un modèle d'activités, l'échelle microscopique présente des trajectoires complexes difficiles à reproduire.
}

\bpar{
We will choose to answer simultaneously to these different problematics with an original strategy of a double thematic entry.
}{
Nous choisirons pour répondre simultanément à ces différentes problématiques une stratégie originale de double entrée thématique.
}



\cleardoublepage 


%
\ctparttext{This part provides building blocks for the final objective of constructing models of co-evolution. These contain both stylized facts from empirical analyses and toy and hybrid modeling. They correspond to three distinct components of our overall construction: first analyses at the micro-scale confirming the chaotic and non-stationary nature of interactions between networks and territories, secondly a morphogenetic vision of these that corresponds roughly to a meso-scale, and finally an application of the evolutive urban theory at the macro-scale.}
%

\part{Building bricks}
%

\chapter*{Introduction of Part II}

\markboth{Introduction of Part II}{Introduction of Part II}



\bigskip


\bpar{
\textit{He finally would have realized his trip. No cities, or a very few. Which soul in these perpendicular \emph{streets} and \emph{avenues}, that we necessarily discover by car. Fill the tank again, maybe it's on purpose, just for the charm of the fuel smell. Anyway it would be funny to look at what these stations have to say, to keep in mind. A running return journey to Mount Elbert, then Longs Peak. Soon out of Colorado, goodbye my gummy bears. Damn it, so close to Denver, it could be worth it. Never mind, the mountains are calling and I must go, as someone used to say. What do we finally know of a territory as a consequences of our so selective discoveries ? A very narrow band on the spectrum of scales ? A tiny spatial extent: one supplementary dimension is not invented so easily. Maybe at least an awakening of conscience for antagonisms, of dualities. And the conscience to necessarily choose one of the aspects each time. To build bridges one must be prepared. To see the world with an eye catching multiple views, one must already have \emph{understood}, i.e. subjectively integrated, the corresponding processes. Memory of one of the first serious routes: Meije ridge traverse, 23 hours without interruption to end with hallucinations on the path to follow whereas the sparkles of crampons on the scree were not anymore enough to light the fallen night. This concrete feeling of the void on each side which imposes to grope, anchors in the subconscious even before reaching the stage of hallucinations: we travel at each moment a fine line, which is as much that of the arbitrariness of road trips as that of the bridges which difficultly resist the flood. On this ridge, anchors that are of course strong but also heterogenous are a pledge of life: diversity overcomes adversity.}
}{
\textit{Il aura finalement pu le faire, ce voyage. Pas, ou très peu de villes. Quelle âme dans ces \emph{streets} et \emph{avenues} perpendiculaires, qu'on traverse nécessairement en bagnole. Encore un plein, à croire que c'est fait exprès, pour le charme de l'odeur d'essence. Tiens ça serait amusant de regarder ce que racontent ces stations d'ailleurs, à garder en tête. Un aller-retour au pas de course au Mont Elbert, puis à Longs Peak. On sort bientôt du Colorado, faudra dire au revoir au gummy bears. Damn it, Denver est si proche, ça vaudrait la peine. Tant pis, the mountains are calling and I must go, comme dirait l'autre. Que connait-on finalement d'un territoire en conséquence de nos découvertes si sélectives ? Une infime partie du spectre des échelles ? Une infime étendue spatiale : on ne s'invente pas une dimension supplémentaire si facilement. Peut être au moins la prise de conscience des antagonismes, des dualités. Et la conscience d'avoir à chaque fois dû privilégier l'un des aspects. Pour faire des ponts il faut être préparé. Pour voir le monde par un regard qui en capture plusieurs, il faut déjà avoir \emph{compris}, c'est-à-dire intégré subjectivement, les processus correspondants. Souvenir d'une des premières courses sérieuses : les arêtes de la Meije, 23h consécutives pour terminer par des hallucinations sur le chemin à prendre que les étincelles des crampons sur les éboulis ne suffisaient plus à éclairer dans la nuit qui était retombée. Ce ressenti concret du gouffre de part et d'autre qui implique le tâtonnement, s'ancre dans l'inconscient avant même d'avoir atteint le stade des hallucinations : nous parcourons à chaque instant une fine arête, qui est autant celle de l'arbitraire du road trip que celle des ponts qui résistent difficilement quand vient la crue. Sur cette arête, les points d'ancrage bien évidemment solides mais aussi hétérogènes sont gages de vie : la diversité combat l'adversité.}
}


\bigskip


\bpar{
A paradox which is intrinsic to numerous knowledge production approaches is a need of an intrinsic consistence and of a reasonable reach of explication for concerned phenomena, which is opposed to an inevitable reduction of explored dimensions, but also to the fragility of bridges that it aims at creating towards other corpora of knowledge. The image we took above suggests that groping, i.e. a step-by-step progression without precipitation, and also the solidity of anchors, are solid assets to tackle this paradox.
}{
Un paradoxe intrinsèque à nombre de démarches de production de connaissance est un besoin de consistence intrinsèque et d'une portée satisfaisante d'explication des phénomènes concernés, qui s'oppose à une inévitable réduction des dimensions explorées mais également à la fragilité des ponts qu'elle tente de former vers d'autres corpus de connaissances. L'image prise ci-dessus suggère que le tâtonnement, c'est-à-dire une progression pas à pas sans précipitations, ainsi que la solidité des ancrages, sont des atouts solides pour affronter ce paradoxe.
}


\bpar{
This part directly opens thematic directions of answer for modeling co-evolution that we mentioned when concluding the first part, and thus builds these strong anchors. It however constructs basements without going into the heart of the problem in a spirit of robustness through progressive entries, and constructs therefore the \emph{elementary bricks} of our approach. Two chapters deal thus successively with the following thematics:
\begin{enumerate}
	\item A first chapter focuses on the evolutive urban theory, which is a privileged entry on urban systems from an evolutive point of view, and integrates in its core a multi-scalar approach to these systems. It unveils fundamental properties of territorial systems implied by the evolutive theory, by introducing a first empirical analysis of the spatial variability of interactions between urban form and network topology, then by developing a methodology to statistically characterize co-evolution (in the intermediate sense of population). It introduces then a first model of interaction between urban system and flows of the transportation network, with a static network.
	\item A second chapter explores the concept of morphogenesis, which allows a conceptual entry to the characteristic of modularity necessary to have co-evolution. After having developed an interdisciplinary definition of morphogenesis, it introduces a model of urban morphogenesis based on aggregation-diffusion processes for population density, and is then sequentially coupled to a network generation model.
\end{enumerate}
}{
Cette partie ouvre directement les pistes de réponse thématiques pour la modélisation de la co-évolution que nous avons évoqué en conclusion de la première partie, et pose ainsi ces ancrages forts. Elle pose toutefois des bases sans entrer dans le coeur du sujet par ce souci de robustesse par entrée progressive, et construit donc les \emph{briques élémentaires} de notre démarche. Deux chapitres traitent ainsi successivement les thématiques suivantes :
\begin{enumerate}
	\item Un premier chapitre s'intéresse à la théorie évolutive des villes, qui est une entrée privilégiée sur les systèmes urbains d'un point de vue évolutif, et intègre en son coeur un point de vue multi-scalaire de ces systèmes. Il éclaire des propriétés fondamentale des systèmes territoriaux impliquées par la théorie évolutive, en introduisant une première analyse empirique de la variabilité spatiale des interactions entre forme urbaine et forme de réseau, puis en développant une méthodologie de caractérisation statistique de la co-évolution (au sens intermédiaire de la population). Il introduit ensuite un premier modèle d'interaction entre système de ville et flux du réseau de transport, avec réseau statique.
	\item Un second chapitre explore le concept de morphogenèse, qui permet une entrée conceptuelle à la caractéristique de modularité nécessaire pour avoir co-évolution. Après avoir développé une définition interdisciplinaire de la morphogenèse, il introduit un modèle de morphogenèse urbaine basé sur des processus d'agrégation-diffusion pour la densité de population, et est ensuite couplé séquentiellement à un modèle de génération de réseau.
\end{enumerate}
}

\stars


\bpar{
\chapter*{Mathematical preliminaries}
}{
\chapter*{Préliminaires mathématiques}
}

\markboth{Mathematical preliminaries}{Mathematical preliminaries}

\label{ch:preliminarymath}

\bpar{
In order to be readable by the largest audience possible, we propose to precise in the preliminary interlude the definitions of notions or key methods that will be regularly used in the following, often out of a mathematical framework. This choice allows to keep a rigorous frame without making indigestible the reading of this manuscript to a large part of its legitimate audience. Without noted otherwise, the specifications given here will be the reference during the use of the corresponding terms.
}{
Afin de toucher l'audience la plus large possible, nous proposons de preciser dans cet intermède préliminaire les definitions de notions ou méthodes clés qui seront utilisées de manière régulière par la suite, souvent hors d'un cadre mathématique. Ce choix permet de garder un cadre rigoureux sans rendre indigeste la lecture du manuscrit à une grande partie de son public légitime. Sauf indication contraire, les specifications données ici feront référence lors de l'utilisation des termes correspondants.
}

\subsection*{Statistics}{Statistiques}

\bpar{
We will denote by $\Pb{\cdot}$ a probability, $\Eb{\cdot}$ an expectation, $\hat{\mathbb{E}}\left[\cdot\right]$ an associated estimator, and $\Covb{\cdot}{\cdot}$ a covariance.
}{
Nous noterons $\Pb{\cdot}$ une probabilité, $\Eb{\cdot}$ une espérance, $\hat{\mathbb{E}}\left[\cdot\right]$ un estimateur associé, et $\Covb{\cdot}{\cdot}$ une covariance.
}

\paragraph*{Correlation}{Corrélation}

\bpar{
Unless otherwise stated, we will estimate the covariance between two processes with a Pearson estimator, i.e. if $(X_i,Y_i)_i$ is a set of observations of processes $X,Y$, the correlation is estimated by
}{
Sauf indication contraire, nous estimerons la covariance entre deux processus par estimateur de Pearson, c'est-à-dire si $(X_i,Y_i)_i$ est un jeu d'observations des processus $X,Y$, la correlation est estimée par
}

\[
\hat{\rho} = \frac{\hat{\Cov}\left[X,Y\right]}{\sqrt{\hat{\Cov} \left[X\right] \cdot \hat{\Cov}\left[Y\right]}}
\]

\bpar{
where the covariance is estimated with the unbiased estimator $\hat{\Cov}$.
}{
où la covariance est estimée par l'estimateur non biaisé $\hat{\Cov}$.
}

\paragraph*{Granger causality}{Causalité de Granger}

\bpar{
A multi-dimensional time-series $\vec{X}(t)$ exhibits a Granger causality if with
\[
\vec{X}(t) = \mathbf{A}\cdot \left(\vec{X}(t-\tau)\right)_{\tau >0} + \varepsilon
\]
there exists $\tau,i$ such that $a_{i\tau}>0$ significantly. We will use a weak version of Granger causality, i.e. a test on lagged correlations defined by
\[
\rho_{\tau}\left[X_i,X_j\right] = \hat{\rho}\left[X_i(t-\tau),X_j(t)\right]
\]
with $\tau$ lag or advance. This will allow us to quantify relations between random variables defined in space and time.
}{
Une série temporelle multi-dimensionnelle $\vec{X}(t)$ présente une causalité de Granger si avec
\[
\vec{X}(t) = \mathbf{A}\cdot \left(\vec{X}(t-\tau)\right)_{\tau >0} + \varepsilon
\]
il existe $\tau,i$ tels que $a_{i\tau}>0$ significativement. Nous utiliserons une version faible de la causalité de Granger, c'est-à-dire un test sur les correlations retardées définies par
\[
\rho_{\tau}\left[X_i,X_j\right] = \hat{\rho}\left[X_i(t-\tau),X_j(t)\right]
\]
avec $\tau$ retard ou avance. Cela nous permettra de quantifier des relations entre variables aléatoires définies dans l'espace et dans le temps.
}

\paragraph*{Geographically Weighted Regression}{Regression Géographique Pondérée}

\bpar{
The Geographically Weighted Regression is an estimation technique for statistical models that allow to take into account the spatial non-stationarity of processes. If $Y_i$ is an explicated variable and $X_i$ a set of explicative variables, measures in the same points in space, we estimate a model $Y_i = f(X_i,\vec{x}_i)$ at each point $\vec{x}_i$, by taking into account the observations with spatial weighting around the point, where weights are fixed by a kernel that can take several forms, for example an exponential kernel is of the form
\[
w_i(\vec{x}) = \exp\left(- \norm{\vec{x} - \vec{x_i}} / d_0\right)
\]
The stationarity scale assumed by the model is then of the same order as $d_0$. It can be adjusted by cross-validation for example.
}{
La Régression Géographique Pondérée est une technique d'estimation de modèles statistiques permettant de prendre en compte la non-stationnarité spatiale des processus. Si $Y_i$ est une variable à expliquer et $X_i$ un jeu de variables explicatives, mesurés en des mêmes points de l'espace, on estime un modèle $Y_i = f(X_i,\vec{x}_i)$ à chaque point $\vec{x}_i$, en prenant en compte les observations par pondération spatiale autour du point, où les poids sont fixés par un noyau pouvant prendre plusieurs formes, par exemple un noyau exponentiel est de la forme
\[
w_i(\vec{x}) = \exp\left(- \norm{\vec{x} - \vec{x_i}} / d_0\right)
\]
L'échelle de stationnarité spatiale supposée par le modèle est alors de l'ordre de $d_0$. Celle-ci peut être ajustée par validation croisée par exemple.
}

\paragraph*{Machine learning}{Apprentissage statistique}

\bpar{
We will designate by \emph{Supervised Learning} any method to estimate a relation between variables $Y=f(X)$ where the value of $Y$ is known on a data sample. This will be a classification when the variable is discrete. Non-supervised classification consists in constructing $Y$ when only $X$ is given. In order to classify data, we will use a basic technique which gives good results on data without an exotic structure: the method of \emph{k-means}, repeated a sufficient number of times to take into account its stochastic character. The complexity of \emph{k-means} is in average polynomial, even if the exact solution of the partitioning problem is NP-hard.
}{
On désignera par \emph{Apprentissage supervisé} toute méthode d'estimation d'une relation entre variables $Y=f(X)$ où la valeur de $Y$ est connue sur un échantillon de données. On parlera de classification si la variable est discrete. La classification non-supervisée consiste à construire $Y$ lorsque seul $X$ est donné. On utilisera pour classifier une technique basique qui donne de bons résultats sur des données qui n'ont pas une structure exotique : la méthode des \emph{k-means}, répétée un nombre suffisant de fois pour prendre en compte son caractère stochastique. Le complexité du \emph{k-means} est polynomiale en moyenne, bien que la résolution exacte du problème de partition soit NP-difficile.
}

\paragraph*{Overfitting}{Overfitting}


\bpar{
The issue of \emph{overfitting} is particularly important during the estimation of models, since a too large number of parameters can lead to the capture of the realization noise as a structure. During the estimation of statistical models, information criteria can be used to quantify the gain in information produced by the addition of a parameter, and obtain a compromise between performance and parcimony.
}{
La question du sur-ajustement (\emph{overfitting}) est particulièrement importante lors de l'estimation de modèles, puisque un nombre trop important de paramètres pourra conduire a capturer le bruit de realisation comme structure. Lors de l'estimation de modèles statistiques, des critères d'information sont mobilisables pour quantifier le gain d'information produit par l'ajout d'un paramètre, et obtenir un compromis entre performance et parcimonie.
}

\bpar{
The \emph{Akaike Information Criteria} (AIC) allows to quantify the gain in information allowed by the addition of parameters in a model. For a statistical model which has a Likelihood function, the AIC is then defined by
\[
AIC = 2k - 2 \ln{\mathcal{L}}
\]
if $k$ is the number of parameters in the model and $\mathcal{L}$ the maximal value of the likelihood function. \cite{akaike1998information} shows that this expression corresponds to an estimation of the gain in Kullback-Leibler information. A correction for small samples of size $n$ is given by
\[
AICc = 2\cdot\left(k + \frac{k^2 + k}{n-k -1} - \ln{\mathcal{L}}\right)
\]
}{
Le \emph{Critère d'Information d'Akaike} (AIC) permet de quantifier le gain d'information permis par l'ajout de paramètres dans un modèle. Pour un modèle statistique qui dispose d'une Fonction de Vraisemblance (\emph{Likelihood}), l'AIC est alors défini par
\[
AIC = 2k - 2 \ln{\mathcal{L}}
\]
si $k$ est le nombre de paramètres du modèle et $\mathcal{L}$ la valeur maximale de la fonction de vraisemblance. \cite{akaike1998information} montre que cette expression correspond à une estimation du gain d'information de Kullback-Leibler. Une correction pour les petits échantillons de taille $n$ est donnée par
\[
AICc = 2\cdot\left(k + \frac{k^2 + k}{n-k -1} - \ln{\mathcal{L}}\right)
\]
}

\bpar{
A similar criteria but derived within a bayesian framework is the \emph{Bayesian Information Criterion} (BIC)~\cite{burnham2003model}, which leads to a stronger penalty for the number of parameters: $BIC = \ln n \cdot k - 2 \ln{\mathcal{L}}$.
}{
Un critère similaire mais dérivé dans un cadre bayésien est le \emph{Critère d'Information Bayésien} (BIC)~\cite{burnham2003model}, qui conduit à une pénalisation plus forte du nombre de paramètres : $BIC = \ln n \cdot k - 2 \ln{\mathcal{L}}$.
}

\bpar{
These criteria are applied to model selection by studying their differences between models (only differences have a meaning, since they are defined with an arbitrary constant): the ``best'' model is the one having the lowest criterion. In the case of models with comparable performances, it can be more relevant to combine models with the Akaike weights $w_i = \exp (- \Delta AIC / 2)$.
}{
Ces critères sont appliqués pour la sélection de modèles en étudiant leur différences entre modèles (seules les différences ont un sens, ceux-ci étant définis à une constante près) : le ``meilleur'' modèle est celui ayant le critère le plus faible. Dans le cas de modèles de performance comparables, il peut être pertinent de combiner les modèles par les poids d'Akaike $w_i = \exp (- \Delta AIC / 2)$.
}

\bpar{
This issue of overfitting is also implicit in the case of simulation models, but to the best of our knowledge there does not exist an established method allowing to tackle it.
}{
Cette question du sur-ajustement est également sous-jacente dans le cas des modèles de simulation, mais il n'existe à notre connaissance pas de méthode établie permettant de la traiter.
}

\subsection*{Stochastic processes: stationarity}{Processus stochastiques : stationnarité}


\bpar{
Stationarity properties inform on the variability of the distribution of a stochastic process. Let $(\vec{X}_i)_{i\in I}$ a multidimensional stochastic process. It will be said to be strongly stationary if its law does not depends on $i$, i.e. if $\Pb{\vec{X}_i} = \Pb{\vec{X}_{i+1}}$. Strong stationarity implies the equality of all moments for all $i$.
}{
Les propriétés de stationnarité informent sur la variabilité de la distribution d'un processus stochastique. Soit $(\vec{X}_i)_{i\in I}$ un processus stochastique multidimensionnel. Il sera dit fortement stationnaire si sa loi ne dépend pas de $i$, c'est-à-dire si $\Pb{\vec{X}_i} = \Pb{\vec{X}_{i+1}}$. La stationnarité forte implique l'égalité de tous les moments pour tout $i$.
}

\bpar{
We will use a wealer notion of stationarity for stochastic processes, or \emph{Weak Stationarity}, which uses the first two moments: $(\vec{X}_i)_{i\in I}$ is weakly stationary if
\begin{enumerate}
	\item $\Eb{\vec{X}_i} = \Eb{\vec{X}_0}$ for all $i$
	\item $\Covb{\vec{X}_i}{\vec{X}_j}$ depends only on $i-j$
\end{enumerate}
We can have a weak stationarity at the first order if only the condition on the expectation is verified, and at the second order if there is also the condition on autocovariance~\cite{zhang2014test}.
}{
Nous utiliserons une notion plus faible de la stationnarité des processus stochastiques, ou \emph{Weak Stationarity}, qui utilise les deux premiers moments : $(\vec{X}_i)_{i\in I}$ est faiblement stationnaire si
\begin{enumerate}
	\item $\Eb{\vec{X}_i} = \Eb{\vec{X}_0}$ pour tout $i$
	\item $\Covb{\vec{X}_i}{\vec{X}_j}$ ne dépend que de $i-j$
\end{enumerate}
On peut parler de stationnarité faible du premier ordre si seulement la condition sur l'espérance est vérifiée, et du second ordre si on a aussi la condition sur l'autocovariance~\cite{zhang2014test}.
}









\subsection*{Exploration of simulation models}{Exploration de modèles de simulation}

\bpar{
We will designate by simulation model any algorithm that associates a realisation $\mathcal{M}\left[\vec{x},\vec{\alpha}\right]$ to data $\vec{x}$ given parameters $\vec{\alpha}$. The question is then to understand the behavior of the model in an empirical way, by simulating it, possibly with several repetitions for the same parameters if it is stochastic. It is then for example possible to calibrate the model, i.e. find a set of parameters allowing to fulfil given objectives (that can be distances to observed data). 
}{
Nous désignerons par modèle de simulation tout algorithme associant une réalisation $\mathcal{M}\left[\vec{x},\vec{\alpha}\right]$ à des données $\vec{x}$ étant donnés des paramètres $\vec{\alpha}$. L'enjeu est alors de comprendre le comportement du modèle de manière empirique, en le simulant, possiblement avec plusieurs répétitions pour des mêmes paramètres si celui-ci est stochastique. Il est alors par exemple possible de calibrer le modèle, c'est-à-dire trouver un jeu de paramètres permettant de remplir des objectifs donnés (qui peuvent être des distances à des données observées).
}

\subsubsection*{Experience plan by sampling}{Plan d'expérience par échantillonnage}

\bpar{
The dimensionality curse corresponds to the fact that the size of the parameter space is exponential in the number of parameters. When it increases but we want to keep an overview of the behavior of a model on a large variety of input parameters, we can sample the space with a given number of points.
}{
Le sort de la dimension (\emph{Dimensionality Curse}) correspond simplement au fait que la taille de l'espace des paramètres est exponentielle en le nombre de paramètres. Lorsque celle-ci grandit mais qu'on veut garder un aperçu du comportement d'un modèle sur des valeurs très variées des paramètres d'entrée, on peut alors échantillonner l'espace par un nombre donné de points.
}

\bpar{
The Latin Hypercube Sampling (LHS) allows to ensure that for each dimension, the full range of values is covered when generated points are projected on the dimension. The Sobol sampling allows to generate point clouds with a weak discrepancy (see~\ref{app:sec:robustness} for a precise definition of discrepancy, that should be understood as a covering of space), and is particularly suited for the computation of integrals.
}{
L'échantillonnage par Hypercube Latin (LHS) permet d'assurer que pour chaque dimension, l'ensemble de la plage des valeurs est couverte lorsqu'on projette les points générés sur la dimension. L'échantillonnage par suite de Sobol permet de générer des points de discrépance faible (voir~\ref{app:sec:robustness} pour une définition précise de la discrépance, qu'il faut comprendre comme une couverture de l'espace), et est particulièrement adapté au calcul d'intégrales.
}

\bpar{
Sampling can become cumbersome is the model is very irregular, or for a precise calibration objective. Therefore, there exists specific algorithms for exploration and calibration, for which we can give some examples.
}{
L'échantillonnage peut devenir laborieux si le modèle est très irrégulier, ou pour un objectif précis de calibration. Pour cela, il existe des algorithmes spécifiques d'exploration et de calibration pour lesquels nous pouvons donner des exemples.
}

\subsubsection*{Genetic algorithm calibration}{Calibration par algorithme génétique}

\bpar{
Genetic algorithms are an alternative largely used in optimization, and are more generally a case of evolutionary computation meta-heuristics~\cite{rey2015plateforme}. We will generally use for the calibration of models the standard algorithm implemented in OpenMole, described in details by~\cite{pumain2017evaluation}. It is a stochastic extension of the NSGA2 algorithm for multi-objective optimization. It has the following main characteristics:
\begin{itemize}
	\item given a population of parameters that are candidates as solutions of the multi-objective problem, the Pareto front is determined as the non-dominated points;
	\item a set is constructed from this front by taking a constraint of diversity into account;
	\item an offspring is generated from this set by crossovers and mutations, and evaluated for its performance;
	\item the algorithm iterates on the new population. 
\end{itemize}
}{
Les algorithmes génétiques sont une alternative largement utilisée en optimisation, et font plus généralement partie des méta-heuristiques de computation évolutionnaire~\cite{rey2015plateforme}. Nous utiliserons généralement pour la calibration des modèles l'algorithme standard implémenté dans OpenMole, décrit en détails par~\cite{pumain2017evaluation}. Il s'agit d'une extension stochastique de l'algorithme NSGA2 pour l'optimisation multi-objectif. Il possède les caractéristiques principales suivantes :
\begin{itemize}
	\item étant donné une population de paramètres candidats comme solution au problème multi-objectif, le front de Pareto est déterminé comme les points non-dominés ;
	\item un ensemble est construit à partir de ce front en prenant en compte une contrainte de diversité ;
	\item une descendance est générée à partir de cet ensemble par croisements et mutations et évaluée pour sa performance ;
	\item l'algorithme itère sur la nouvelle population.
\end{itemize}
}

\bpar{
\cite{pumain2017evaluation} adds the objective of the number of replications to the objectives of the algorithm, in order to take into account stochasticity and find a compromise between optimality and robustness of solutions.
}{
\cite{pumain2017evaluation} ajoute l'objectif du nombre de réplications aux objectifs de l'algorithme, afin de prendre en compte la stochasticité et trouver un compromis entre optimalité et robustesse des solutions.
}

\subsubsection*{Specific algorithms}{Algorithmes spécifiques}

\bpar{
Based on genetic algorithms, various algorithms have been proposed to refine model exploration. We can mention two examples developed in the frame of OpenMole: the \emph{Pattern Space Exploration} algorithm (PSE)~\cite{10.1371/journal.pone.0138212} aims at discovering the set of outputs of a model, in the idea of a search for all feasible behaviors. The \emph{Calibration Profile} algorithm~\cite{reuillon2015} aims on the other hand at establishing the necessary character of a parameter to fulfill an objective, independently of other parameters.
}{
Se basant sur des algorithmes génétiques, divers algorithmes ont été proposés pour raffiner l'exploration des modèles. Mentionnons deux exemples développés dans le cadre d'OpenMole : l'algorithme PSE (\emph{Pattern Space Exploration})~\cite{10.1371/journal.pone.0138212} vise à découvrir l'ensemble de l'espace des sorties d'un modèle, dans l'idée d'une recherche de l'ensemble des comportements possibles. L'algorithme \emph{Calibration Profile}~\cite{reuillon2015} vise quant à lui à établir le caractère nécessaire d'un paramètre pour arriver à un objectif, indépendamment des autres paramètres.
}

\stars




\chapter[Evolutionary Urban Theory]{Co-evolution: an evolutionary urban theory approach} 

\label{ch:evolutiveurban} 

\bpar{
The study of interactions between transportation networks and territories can be studied from the standpoints of urban systems. Did the opening of the first High Speed Line in France between Paris and Lyon have an impact on the concerned territorial dynamics ? \cite{bonnafous1987regional} shows that it could have had some at the regional scale, in particular areas, as for example tourism in Burgundy. Did it have effects on the long time, beyond the decade ? At which scales, following which processes ? We rejoin the question of \emph{structuring effects}, that we evoked in chapter~\ref{ch:thematic} through a multi-scalar entry (micro, meso and macro), and also through the progressive development of the idea of co-evolution. These characteristics are indeed at the core of the evolutive urban theory, of which we propose therefore here to detail implications for our problematic.
}{
L'étude des interactions entre réseaux de transport et territoires peut s'aborder sous l'angle des systèmes urbains. L'ouverture de la première Ligne à Grande Vitesse en France entre Paris et Lyon a-t-elle eu un impact sur les dynamiques territoriales concernées ? \cite{bonnafous1987regional} montre qu'elle pourrait en avoir eu un à l'échelle régionale, dans des secteurs particuliers, comme par exemple le tourisme en Bourgogne. En-a-t-elle eu sur le temps long, au delà de la décade ? À quelles échelles, selon quels processus ? Nous retrouvons la question des \emph{effets structurants}, que nous avons abordée en chapitre~\ref{ch:thematic} par une entrée à plusieurs échelles (micro, meso et macro), ainsi que par le développement progressif de l'idée de co-évolution. Ces caractéristiques sont en fait au coeur de la théorie évolutive des villes, dont nous proposons donc ici d'approfondir les implications pour notre problématique.
}

\bpar{
After having recalled in preliminary the essential characteristics of the evolutive urban theory, we study in a first section at the mesoscopic scale the interactions between territories and networks, that we capture in morphological indicators for each, and for which we study the spatial correlations.
}{
Après avoir rappelé en préliminaire les caractéristiques essentielles de la théorie évolutive des villes, nous étudions dans une première section à l'échelle mesoscopique les interactions entre territoires et réseaux, que nous capturons dans des indicateurs morphologiques pour chacun, et pour lesquels nous étudions les correlations spatiales.
}

\bpar{
We then introduce the dynamical aspect by studying the notion of spatio-temporal causality in section~\ref{sec:causalityregimes}. The multiple configurations highlighted for a simple urban growth model that strongly couples network growth and density, that we will designate as \emph{causality regimes}, witness of circular causalities which are indeed markers of a co-evolution. The application to the case of rail network growth and urban populations in South Africa shows that this method empirically allows to reveal different regimes. This method is crucial on the one hand from a methodological point of view through the introduction of an original method allowing in some cases to better understand the respective influences between territories and networks, but also from a thematic point of view concerning the empirical presence of a co-evolution.
}{
Nous introduisons ensuite l'aspect dynamique en étudiant la notion de causalité spatio-temporelle dans la section~\ref{sec:causalityregimes}. Les multiples configurations mises en évidence pour un modèle simple de croissance urbaine couplant fortement croissance du réseau et densité, qu'on désignera comme \emph{régimes de causalité}, témoignent de causalités circulaires qui sont bien des marques d'une co-évolution. L'application au cas de la croissance du réseau ferroviaire et des populations urbaines en Afrique du Sud montre que cette méthode permet empiriquement de révéler différents régimes. Cette méthode est essentielle d'une part d'un point de vue méthodologique par l'introduction d'une méthode originale permettant dans certains cas de mieux cerner les influences respectives entre réseaux et territoires, mais également d'un point de vue thématique concernant la présence empirique d'une co-évolution.
}

\bpar{
We finally explore in a last section~\ref{sec:interactiongibrat} the possibilities offered by interaction models coming from the evolutive urban theory, at a small spatial scale and a long time scale, what suggests the existence of network effects in an indirect way, without even introducing co-evolution aspects in a first time.
}{
Nous explorons enfin dans une dernière section~\ref{sec:interactiongibrat} les possibilités offertes par les modèles d'interaction issus de la théorie évolutive des villes, à une petite échelle spatiale et longue échelle de temps, ce qui suggère l'existence d'effets de réseau de manière indirecte, sans même introduire d'aspects de co-évolution dans un premier temps.
}

\bpar{
This way, we build the first building bricks for different aspects of interactions and of co-evolution between networks and territories, in particular in the empirical domain for the characterization of co-evolution, and in the modeling domain by the introduction of a first model relating territories and networks.
}{
Ainsi, nous façonnons les premières briques pour différents aspects des interactions et de la co-évolution entre réseaux et territoires, en particulier sur le plan empirique pour la caractérisation de la co-évolution, et sur le plan de la modélisation par l'introduction d'un premier modèle mettant en relation territoires et réseaux.
}



\stars

\bpar{
\textit{This chapter is composed by various works. The first section includes a part from~\cite{raimbault2017calibration} for the morphological analysis, and the results presented by~\cite{raimbault2016cautious} for the analysis of correlations; the second section corresponds to the majority of~\cite{raimbault2017identification} for the theoretical formulation and the illustration on synthetic data, and then presents results of~\cite{raimbault:halshs-01584914} for the application. Finally the last section corresponds entirely to~\cite{raimbault2017indirect}.}
}{
\textit{Ce chapitre est composé de divers travaux. La première section reprend une partie traduite de~\cite{raimbault2017calibration} pour l'analyse morphologique, puis les résultats présentés par~\cite{raimbault2016cautious} pour l'analyse des correlations ; la deuxième section correspond à la majorité de~\cite{raimbault2017identification} pour la formulation théorique et l'illustration sur données synthétiques, puis présente les résultats de~\cite{raimbault:halshs-01584914} pour l'application. Enfin la dernière section est une traduction de~\cite{raimbault2017indirect}.}
}


\newpage

\section*{Evolutionary urban theory}{Théorie évolutive urbaine}

\bpar{
We have already evoked various aspects of the evolutionary urban theory, in relation to complexity in geography, and then to some models of urban systems it produced. A synthesis is here necessary to precisely draw the frame in which our developments will take place. This theory has initially been introduced in~\cite{pumain1997pour} which argues for a dynamical vision of systems of cities, in which self-organisation is crucial.
}{
Nous avons déjà évoqué divers aspects de la théorie évolutive des villes, en relation à la complexité en géographie, puis à certains modèles de systèmes urbains auxquels elle a conduit. Une synthèse est ici nécessaire pour poser précisément le cadre dans lequel nos développements s'inscriront. Cette théorie a été introduite initialement dans~\cite{pumain1997pour} qui argumente pour une vision dynamique des systèmes de villes, au sein desquels l'auto-organisation est essentielle.
}

\bpar{
The core of the evolutionary urban theory is perfectly synthesized by \noun{Denise Pumain} herself (interview in~\ref{app:sec:interviews}): it is ``\textit{a geographical theory with the ambition to gather most of stylized facts known on cities and their organisation within territories, in an out-of-equilibrium and non-static perspective, by following them on long time periods and putting an emphasis on structuring factors and bifurcations.}''
}{
Le coeur de la théorie évolutive urbaine est parfaitement synthétisé par \noun{Denise Pumain} elle-même (entretien en~\ref{app:sec:interviews}) : Il s'agit d'``\textit{une Théorie Géographique ayant pour ambition de rassembler la plupart des faits stylisés connus sur les villes et leur organisation dans les territoires, dans une perspective hors-équilibre et non statique, en les suivant sur de longues périodes de temps et mettant une emphase sur les facteurs structurants et les bifurcations.}''
}

\bpar{
Cities are interdependent evolutive spatial entities whose interrelations lead to the emergence of some macroscopic behavior at the system of cities scale. The system of cities is also seen as a network of cities, in correspondance with an approach through complex systems. Each city is itself a complex system in the spirit of~\cite{berry1964cities}, the multi-scalar aspect, in the sense of autonomous scales but that each have a specific role in the dynamics of the system, being essential in this theory, since microscopic agents carry processes of evolution of the system through complex retroactions between scales. The positioning of this theory within complexity approaches has later been confirmed~\cite{pumain2003approche}.
}{
Les villes sont des entités spatiales évolutives interdépendantes dont les interrelations font émerger le comportement macroscopique à l'échelle du système de villes. Le système de villes est aussi vu comme un réseau de villes, en correspondance avec une approche par les systèmes complexes. Chaque ville est elle-même un système complexe dans l'esprit de~\cite{berry1964cities}, l'aspect multi-scalaire, au sens d'échelles autonomes mais ayant chacune un rôle spécifique dans les dynamiques du système, étant essentiel dans cette théorie, puisque les agents microscopiques véhiculent les processus d'évolution du système à travers des rétroactions complexes entre les échelles. Le positionnement de cette théorie au regard des sciences des systèmes complexes a plus tard été confirmé~\cite{pumain2003approche}.
}

\bpar{
It has been shown that the evolutionary urban theory yields a candidate explanation to scaling laws, which are pervasive in urban systems\footnote{We recall that a scaling laws allows to link the size of cities in terms of population $P_i$ and an aggregated quantity $Z_i$, under the form $Z_i = Z_0\cdot \left(P_i/P_0\right)^{\alpha}$.}, which would be a consequence of the diffusion of innovation cycles between cities~\cite{pumain2006evolutionary}. These have furthermore been exhibited empirically for several urban systems~\cite{pumain2009innovation}. The notion of resilience of a system of cities, inducted by the adaptive character of these complex systems, implies that cities are drivers and incubators of social change~\cite{pumain2010theorie}. Finally, the path-dependancy is a source of non-ergodicity within these systems, making the ``universal'' interpretations of scaling laws developed by physicist not compatible with the evolutive urban theory~\cite{pumain2010theorie}.
}{
Il a été montré que la théorie évolutive des villes fournit une interprétation des lois d'échelle, omniprésentes dans les systèmes urbains\footnote{Nous rappelons qu'une loi d'échelle permet de relier taille des villes en termes de population $P_i$ et une quantité agrégée $Z_i$, sous la forme $Z_i = Z_0\cdot \left(P_i/P_0\right)^{\alpha}$.}, qui découleraient de la diffusion des cycles d'innovation entre les villes~\cite{pumain2006evolutionary}. Celles-ci ont par ailleurs été mises en évidence de manière empirique pour plusieurs systèmes urbains~\cite{pumain2009innovation}. La notion de résilience d'un système de villes, induit par le caractère adaptatif des ces systèmes complexes, implique que les villes sont les moteurs et les incubateurs du changement social~\cite{pumain2010theorie}. Enfin, la dépendance au chemin est source de non-ergodicité au sein de ces systèmes, rendant les interprétations ``universelles'' des lois d'échelle développées par les physiciens incompatibles avec la théorie évolutive~\cite{pumain2010theorie}.
}



\bpar{
The evolutionary urban theory has been conjointly elaborated with models of urban systems. For example the first Simpop model, described by~\cite{sanders1997simpop}, is a multi-agent model which works with the following rules: (i) settlements are initially villages with a uniquely agricultural production, and can in time transform into commercial cities, then administrative, then eventually industrial, the transition rules depending on threshold parameters in terms of population and neighborhood resources for the industrialization; (ii) settlements produce different types of goods depending on their functions and populations; (iii) these are exchanged through the intermediary of spatial interactions (depending on distance) in order to satisfy demands; (iv) populations evolve acoording to the size of the city and the level of demand satisfaction. This first model allows to simulate the evolution of an urban system in a stylized way.
}{
La théorie évolutive des villes a été élaborée conjointement avec des modèles de systèmes urbains. Par exemple le premier modèle Simpop, décrit par~\cite{sanders1997simpop}, est un modèle multi-agents qui fonctionne selon les principes suivants : (i) les établissement sont initialement des villages à la production uniquement agricole, et peuvent au cours du temps se transformer en villes commerciales, puis administratives, puis éventuellement industrielles, les règles de transition dépendant de paramètres de seuil en termes de population et des ressources environnantes pour l'industrialisation ; (ii) les établissement produisent différents types de biens selon leur fonctions et populations ; (iii) ceux-ci sont échangés par l'intermédiaire d'interactions spatiales (dépendant de la distance) afin de satisfaire les demandes ; (iv) les populations évoluent selon la taille de la ville et le niveau de satisfaction de la demande. Ce premier modèle permet de simuler l'évolution d'un système urbain de manière stylisée.
}

\bpar{
The Simpop2 model introduced by~\cite{bretagnolle2006theory} extends this model, allowing to include for example innovation cycles and the role of administrative boundaries in exchanges. It is applied on long time scales to urban growth patterns for Europe and the United States~\cite{doi:10.1177/0042098010377366}.
}{
Le modèle Simpop2 introduit par~\cite{bretagnolle2006theory} reprend et précise ce modèle, permettant d'inclure par exemple les cycles d'innovation et le rôle des limites administratives dans les échanges. Il est appliqué sur de longues échelles de temps aux motifs de croissance urbaine pour l'Europe et les Etats-Unis~\cite{doi:10.1177/0042098010377366}.
}

\bpar{
The most recent accomplishments of evolutive urban theory rely on the production of the ERC project GeoDivercity, presented in \cite{pumain2017urban}, which include considerable progresses from the technical point of view (OpenMole software\footnote{\url{http://openmole.org/}}~\cite{reuillon2013openmole}), from the thematic point of view (knowledge issued from the SimpopLocal model~\cite{schmitt2014modelisation} and the Marius model~\cite{cottineau2014evolution}), and from the methodological point of view (incremental modeling~\cite{cottineau2015incremental}). For an epistemological analysis through mixed methods of the evolutive theory, which allows to reinforce this bibliographical picture by a study of its genesis, in a sense of its \emph{form}, refer to~\ref{sec:knowledgeframework} which uses it as a case study to build a knowledge framework. In particular, an analysis of interviews with \noun{Denise Pumain} and \noun{Romain Reuillon}, reveals the cross-fertilisation between geographical knowledge and computer science knowledge, allowed by the interdisciplinary effort of model development and of their exploration methods.
}{
Les accomplissements les plus récents de la théorie évolutive reposent sur les productions du projet ERC GeoDivercity, présentées dans \cite{pumain2017urban}, qui incluent des progrès considérables à la fois techniques (logiciel OpenMole\footnote{\url{http://openmole.org/}}~\cite{reuillon2013openmole}), thématiques (connaissance issue des modèles SimpopLocal~\cite{schmitt2014modelisation} et Marius~\cite{cottineau2014evolution}) et méthodologiques (modélisation incrémentale~\cite{cottineau2015incremental}). Pour une analyse épistémologique par méthode mixtes de la théorie évolutive, qui permet de renforcer cet aperçu bibliographique par une étude de sa genèse, en quelque sorte de sa \emph{forme}, se référer à~\ref{sec:knowledgeframework} qui l'utilise comme cas d'étude pour construire un cadre de connaissance. En particulier, une analyse des entretiens avec \noun{Denise Pumain} et \noun{Romain Reuillon}, révèle la fertilisation croisée entre connaissances géographiques et connaissances informatiques, permise par l'effort interdisciplinaire de développement des modèles et de leurs méthodes d'exploration.
}

\subsubsection*{Implications}{Implications}

\bpar{
We can therefore consider the complexity of systems of cities in the sense of the evolutive urban theory as a morinian macro-concept~\cite{morin1976methode}, i.e. the complex combination of multiple concepts each necessary to the construction. The following concepts are thus necessary:
\begin{itemize}
	\item Out-of-equilibrium aspect of urban systems. The spatial character of systems often leads to complex spatio-temporal dynamics, and thus properties of non-stationarity for spatio-temporal associated processes.
	\item Systemic dynamics, i.e. existence of a strong interdependency between cities that can be interpreted as a co-evolution (in the last sense in the definition we gave).
	\item Central role of interactions between cities as drivers of growth processes, existence of structure effects on the long time.
\end{itemize}
}{
Nous pouvons ainsi considérer la complexité des systèmes de villes au sens de la théorie évolutive des villes comme un macro-concept morinien~\cite{morin1976methode}, c'est-à-dire la combinaison complexe de multiples concepts chacun nécessaires à la construction. Les concepts suivants sont ainsi nécessaires :
\begin{itemize}
	\item Aspect hors-équilibre des systèmes urbains. La spatialisation des systèmes conduit souvent à des dynamiques spatio-temporelles complexes, et donc des propriétés de non-stationnarité pour les processus spatio-temporels associés. 
	\item Dynamique systémique, c'est-à-dire existence d'une forte interdépendance entre villes pouvant être interprété comme une co-évolution (au dernier niveau dans la définition que nous en avons donné).
	\item Rôle central des interactions entre villes comme moteurs des processus de croissance, existence d'effets de structure sur le temps long.
\end{itemize}
}

\bpar{
These concepts will be thus explored following different perspectives in this chapter, in the following sections:
\begin{enumerate}
	\item From am empirical point of view, we will first study an exemple of non-stationarity properties of characteristics for territories and networks, and also of their interactions.
	\item We introduce then from a methodological point of view an approach allowing to better understand patterns of spatio-temporal interdependency, and thus \emph{co-evolution} that we will link to its intermediate statistical sense we gave.
	\item Finally, a modeling approach allows to explore interactions between cities on the long time, in particular in relation with the network in the context of our questionings.
\end{enumerate}
}{
Ces concepts seront ainsi explorés selon différentes perspectives dans ce chapitre, dans les sections suivantes :
\begin{enumerate}
	\item D'un point de vue empirique, nous étudierons d'abord un exemple de propriétés de non-stationnarité de caractéristiques pour les territoires et les réseaux, ainsi que de leur interactions.
	\item Nous introduisons ensuite d'un point de vue méthodologique une approche permettant de mieux comprendre les motifs d'interdépendance spatio-temporels, et donc la \emph{co-évolution} que nous rattacherons à son sens statistique intermédiaire que nous avons donné.
	\item Enfin, une approche de modélisation permet d'explorer les interactions entre villes sur le temps long, en particulier en lien avec le réseau dans le cadre de nos questionnements.
\end{enumerate}
}

\stars

%

\newpage


\section{Correlations between form of territories and network topology}{Corrélations entre forme des territoires et forme des réseaux}

\label{sec:staticcorrelations}


\bpar{
Through relocation processes, sometimes induced by networks, we can expect the latest to influence the distribution of populations in space. Reciprocally, network characteristics can be influenced by this distribution. We propose here to study these potential links by the intermediate of characterizations given by synthetic indicators for these two subsystems, and by correlations between these indicators.
}{
Au travers des processus de relocalisation, parfois induits par les réseaux, on peut s'attendre à ce que ces derniers influencent la distribution des populations dans l'espace. Réciproquement, les caractéristiques du réseau peuvent être influencées par celle-ci. Nous proposons ici d'étudier ces liens potentiels par l'intermédiaire de caractérisations issues d'indicateurs synthétiques pour ces deux sous-systèmes, et des corrélations entre ces indicateurs.
}

\bpar{
At the scale of the system of cities, the spatial nature of the urban system is captured by cities position, associated with aggregated city variables. We will work here at the mesoscopic scale, at which the precise spatial distribution of activities is necessary to understand the spatial structure of the territorial system. We will therefore use the term of morphological characteristics for population density and the road network.
}{
A l'échelle du système de villes, le caractère spatial du système urbain peut être synthétisé par les positions des villes, associées aux variables agrégées au niveau de la ville. Nous nous placerons ici à l'échelle mesoscopique, à laquelle la distribution spatiale fine des activités est nécessaire pour comprendre la structure spatiale du système territorial. Nous parlerons ainsi de caractéristiques morphologiques pour la densité de population et le réseau routier.
}

\bpar{
The choice of ``relevant'' boundaries for the territory or the city is a relatively open problem which will often depend on the question we are trying to answer \cite{paez2005spatial}. This way, \cite{guerois2002commune} show that the entities obtained are different if we consider an entry by the continuity of the built environment (morphological), by urban functions (employment area for example) or by administrative boundaries. We choose here the mesoscopic scale of a metropolitan center, of an order of one hundred kilometers, first for the relevance of the spatial field computed, and secondly because smaller scales become less relevant for the notion of urban form, whereas larger scales induce a too large variability.
}{
Le choix de limites ``pertinentes'' pour le territoire ou la ville est un problème relativement ouvert qui dépendra souvent de la question à laquelle on cherche à répondre \cite{paez2005spatial}. Ainsi, \cite{guerois2002commune} montrent que les entités obtenues sont différentes si on considère une entrée par la continuité du bâti (morphologique), par les fonctions urbaines (zones d'emploi par exemple) ou par les limites administratives. Nous choisissons ici l'échelle mesoscopique d'un centre métropolitain, de l'ordre de la centaine de kilomètres, d'une part pour la cohérence du champ spatial calculé, et d'autre part parce que des échelles plus petites deviennent moins pertinentes pour la notion de forme urbaine, tandis que des échelles plus grandes induisent une trop grande variabilité.
}


\bpar{
At this scale, we can assume that territorial characteristics, for population and network, are locally defined et vary in an approximatively continuous way in space. Thus, the construction of fields of morphological indicators will allow to endogenously reconstruct territorial entities through the emergent spatial structure of indicators at larger scales. For examples, cities should be distinguishable within non-urban spaces. The aim of this section is thus to study properties of these indicators and their interactions, and thus indirectly interactions between the territory and the network.  
}{
A cette échelle, on peut supposer les caractéristiques du territoire, pour la population et le réseau, comme étant définies localement et variant de manière relativement continue dans l'espace. Ainsi, la construction de champs d'indicateurs morphologiques permettra de reconstruire de manière endogène des ensembles territoriaux par la structure spatiale émergente des indicateurs aux échelles plus grandes. Par exemple, les villes devraient se distinguer au sein des espaces non urbains. L'enjeu de cette section est ainsi d'étudier les propriétés de ces indicateurs et de leurs interactions, et donc indirectement les interactions entre territoire et réseau.
}

\subsection{Morphological measures}{Mesures morphologiques}

\subsubsection{Urban morphology}{Morphologie Urbaine}

\bpar{
The approaches to quantify and qualify \emph{urban form} at the considered scale, and by extension to any population distribution in space what we can call \emph{territorial form}, are numerous.
}{
Les manières de quantifier et qualifier la \emph{forme urbaine} à l'échelle considérée, et par extension à toute distribution de population dans l'espace ce qu'on peut appeler \emph{forme territoriale}, sont nombreuses.
}

\bpar{
We need however quantities having a certain level of invariance to extract typical shapes. For example, two monocentric cities, i.e. concentrated around a given point, should be measured as morphologically close by a monocentricity indicator, whereas a direct comparison of population distributions can give a very high distance\footnote{Spatial distributions can be compared by an euclidian distance between corresponding matrices, or by more elaborated distances such as the Monge distance which solves a minimal transport problem and gives the quantity of displacements necessary to go from one distribution to the other.} between configurations depending on the position of centers.
}{
Nous avons cependant besoin de mesures ayant un certain niveau d'invariance pour extraire des formes typiques. Par exemple, deux villes monocentriques, c'est-à-dire concentrées autour d'un point donné, devraient être mesurées comme morphologiquement proches par un indicateur de monocentrisme tandis qu'une comparaison directe des distributions de population pourra donner une distance très élevée\footnote{On peut comparer des distributions spatiales par une distance euclidienne entre les matrices correspondantes, ou par des distances plus élaborées comme la distance de Monge qui résout un problème de transport minimal et donne la quantité de déplacements nécessaires pour passer d'une distribution à l'autre.} entre les configurations selon la position des centres.
}

\bpar{
We choose here to refer to the literature in urban morphology which proposes various set of indicators to describe urban form~\cite{tsai2005quantifying}. \cite{le2009quantifier} recalls the necessity of a multi-dimensional measure of the urban form. It is possible to obtain a robust description with a small number of independant indicators by a reduction of the dimension~\cite{Schwarz201029}.
}{
Nous choisissons pour notre étude de nous référer à la littérature en morphologie urbaine qui propose des jeux d'indicateurs variés pour décrire la forme urbaine~\cite{tsai2005quantifying}. \cite{le2009quantifier} rappelle la nécessité d'une mesure multi-dimensionnelle de la forme urbaine. Il est possible d'obtenir une description robuste avec un petit nombre d'indicateurs indépendants par une réduction de la dimension~\cite{Schwarz201029}.
}

\bpar{
Other solutions exist to quantify urban form\footnote{In operational urbanism, urban morphology is defined as ``the characteristics of the material form of cities and fabrics''~\cite{paquot2010abc}. We use this term here for fabrics at a mesoscopic scale, seen through the spatial distribution of populations.}. \cite{guerois2008built} study the form of European cities using a simple measure of density slopes from the center to the periphery. It is also possible to use indexes from fractal analysis, such as for example systematically applied by~\cite{2016arXiv160808839C} to classify urban forms. The link between urban morphology and topology of the underlying relational network has been suggested in a theoretical approach by~\cite{badariotti2007conception}. Other more original indexes can be proposed, such as by~\cite{lee2017morphology} which use the variations of trajectories for routes going through a city to establish a classification and show that it is strongly correlated with socio-economic variables.
}{
D'autres solutions existent pour quantifier la morphologie urbaine\footnote{Dans l'urbanisme opérationnel, la morphologie urbaine est définie comme ``les caractéristiques de la forme matérielle des villes et des tissus''~\cite{paquot2010abc}. Nous utilisons ce terme ici pour les tissus à une échelle mesoscopique, vus par la distribution spatiale des populations.}. \cite{guerois2008built} étudient la forme des villes européennes par l'utilisation d'une mesure simple des gradients de densité du centre vers la périphérie. Il est aussi possible d'utiliser des indices issus de l'analyse fractale, comme par exemple appliquée systématiquement par~\cite{2016arXiv160808839C} pour classifier les formes urbaines. Le lien entre morphologie urbaine et topologie du graphe de relations sous-jacent a été suggéré dans une approche théorique par~\cite{badariotti2007conception}. D'autre indices plus originaux peuvent être proposés, comme par~\cite{lee2017morphology} qui utilisent les variations de trajectoire d'itinéraires traversant une ville pour établir une classification et montrer que celle-ci est fortement corrélée aux variables socio-économiques.
}

\bpar{
Note that we consider here indicators on the spatial distribution of population density only, and that more elaborated considerations on urban form can include for example the distribution of economic opportunities and the combination of these two fields through accessibility measures. For the choice of indicators, we follow the analysis done in~\cite{le2015forme} where a morphological typology of large European cities is obtained. Its consistence suggests the ability of the indicator set used to capture urban form at this scale. We work at a comparable scale and must capture diverse aspects such as hierarchy, concentration, level of acentrism of the population distribution, hence the use of similar indicators.
}{
Il faut noter que nous considérons ici des indicateurs sur la distribution spatiale de la densité de population uniquement, et que des considérations plus élaborées sur la forme urbaine peuvent inclure par exemple la distribution des opportunités économiques et la combinaison de ces deux champs par des mesures d'accessibilité. Pour le choix des indicateurs, nous suivons l'analyse faite par~\cite{le2015forme} où une typologie morphologique des grandes villes européennes est obtenue. La cohérence de celle-ci suggère la capacité du jeu d'indicateurs utilisé à capturer la forme urbaine à cette échelle. Nous nous trouvons à une échelle comparable et devons capturer des aspects divers comme la hiérarchie, la concentration, le degré d'acentrisme de la distribution de population, d'où l'utilisation d'indicateurs similaires.
}

\subsubsection{Indicators}{Indicateurs}

\bpar{
We give now the formal definition of morphological indicators. We consider gridded population data $(P_i)_{1\leq i \leq N^2}$, write $M=N^2$ the number of cells, $d_{ij}$ the distance between cells $i,j$, and $P=\sum_{i=1}^{M} P_i$ total population. We measure urban form using:

\begin{enumerate}
\item Rank-size slope $\gamma$, expressing the degree of hierarchy in the distribution, computed by fitting with Ordinary Least Squares a power law distribution by $\ln \left( P_{\tilde{i}}/P_0\right) \sim k + \gamma\cdot \ln \left(\tilde{i}/i_0\right)$ where $\tilde{i}$ are the indexes of the distribution sorted in decreasing order (the constant $k$ of the adjustment does not play a role in hierarchy). It is always negative, and values close to zero mean a flat distribution.
\item Entropy of the distribution~\cite{le2015forme}, which expresses how uniform the distribution is, what is a way to capture a level of concentration:
\begin{equation}
\mathcal{E} = \sum_{i=1}^{M}\frac{P_i}{P}\cdot \ln{\frac{P_i}{P}}
\end{equation}
$\mathcal{E}=0$ means that all the population is in one cell whereas $\mathcal{E}=1$ means that the population is uniformly distributed.
\item Spatial-autocorrelation given by Moran index~\cite{tsai2005quantifying}, with simple spatial weights given by $w_{ij} = 1/d_{ij}$
\[
I = M \cdot \frac{\sum_{i\neq j} w_{ij} \left(P_i - \bar{P}\right)\cdot\left(P_j - \bar{P}\right)}{\sum_{i\neq j} w_{ij} \sum_{i}{\left( P_i - \bar{P}\right)}^2}
\]
Its theoretical bounds are -1 and 1, and positive values will imply aggregation spots (``density centers''), negative values strong local variations, whereas $I=0$ corresponds to totally random population values.
\item Average distance between individuals~\cite{le2009quantifier}, which captures a spatial dispersion of population and quantifies a level of acentrism (distance to a monocentric model):
\[
\bar{d} = \frac{1}{d_M}\cdot \sum_{i<j} \frac{P_i P_j}{P^2} \cdot d_{ij}
\]
where $d_M$ is a normalisation constant taken as the diagonal of the area on which the indicator is computed in our case.
\end{enumerate}
}{
Nous donnons à présent une définition formelle des indicateurs morphologiques utilisés ici. Nous considérons des données de population en grille $(P_i)_{1\leq i \leq N^2}$, écrivons $M=N^2$ le nombre de cellules, $d_{ij}$ la distance euclidienne entre les cellules $i$ et $j$, $P=\sum_{i=1}^{M} P_i$ la population totale, et $\bar{P}=\frac{1}{M}\cdot\sum_{i=1}^{M} P_i$ la population moyenne. La forme urbaine est mesurée par : 

\begin{enumerate}
\item Pente de la loi rang-taille $\gamma$, qui exprime le degré de hiérarchie de la distribution, calculé en ajustant une loi de puissance par Moindres Carrés Ordinaires par $\ln \left( P_{\tilde{i}}/P_0\right) \sim k + \gamma\cdot \ln \left(\tilde{i}/i_0\right)$ où $\tilde{i}$ sont les indices de la distribution triée de manière décroissante (la constante $k$ de l'ajustement ne joue pas de rôle dans la hiérarchie). Elle est toujours négative ou nulle, et des valeurs proches de zéro signifient une distribution complètement homogène. 
\item Entropie de la distribution~\cite{le2015forme}, qui exprime l'uniformité de la distribution, ce qui est une façon de capturer un niveau de concentration :
\begin{equation}
\mathcal{E} = \sum_{P_i\neq 0}\frac{P_i}{P}\cdot \ln{\frac{P_i}{P}}
\end{equation}
$\mathcal{E}=0$ signifie que toute la population est dans une cellule tandis que $\mathcal{E}=1$ signifie que la population est distribuée uniformément.
\item L'auto-corrélation spatiale donnée par l'indice de Moran~\cite{tsai2005quantifying}, avec des poids spatiaux simples donnés par $w_{ij} = 1/d_{ij}$ :
\[
I = M \cdot \frac{\sum_{i\neq j} w_{ij} \left(P_i - \bar{P}\right)\cdot\left(P_j - \bar{P}\right)}{\sum_{i\neq j} w_{ij} \sum_{i}{\left( P_i - \bar{P}\right)}^2}
\]
Celui-ci varie théoriquement entre -1 et 1, des valeurs positives impliquent des lieux d'agrégation (``centres de densité''), des valeurs négatives des fortes variations locales, tandis que $I=0$ correspond à des valeurs de population totalement aléatoires.
\item Distance moyenne entre individus~\cite{le2009quantifier}, qui témoigne de la dispersion spatiale de la population et quantifie un degré d'acentrisme (éloignement à un modèle monocentrique) :
\[
\bar{d} = \frac{1}{d_M}\cdot \sum_{i<j} \frac{P_i P_j}{P^2} \cdot d_{ij}
\]
où $d_M$ est une constante de normalisation que nous prenons comme la diagonale de la zone sur laquelle l'indicateur est calculé dans notre cas.
\end{enumerate}
}

\bpar{
The first two indexes are not spatial, and are completed by the last two that take space into account. Following \cite{Schwarz201029}, the effective dimension of the urban form justifies the use of all.
}{
Les deux premiers indices ne sont pas spatiaux, et sont complétés par les deux derniers prenant en compte l'espace. Suivant \cite{Schwarz201029}, la dimension effective de la forme urbaine justifie l'usage de l'ensemble de ceux-ci.
}


\paragraph{Results}{Résultats}


\begin{figure}
\includegraphics[width=0.9\linewidth]{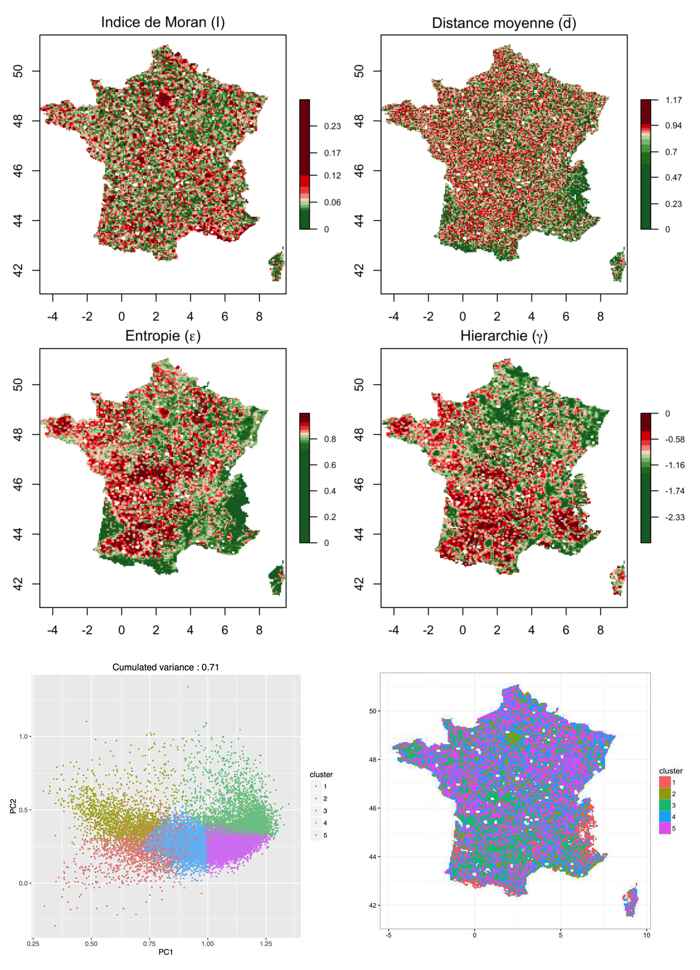}
\caption[Spatial distribution of morphologies]{\textbf{Empirical values of morphological indicators.} \textit{(Top four maps)} Spatial distribution of the morphological indicators for France. Scale color discretization is done using quantiles to ease map readability. \textit{(Bottom Left)} Projection of morphological values on the two first components on a Principal Component analysis. Color gives cluster in an unsupervised classification (see text). \textit{(Bottom right)} Spatial distribution of clusters. See text for details on the process to estimate spatial indicators and for the classification.\label{fig:staticcorrelations:empirical}}
\end{figure}


\bpar{
We compute the morphological measures given above on real urban density data, using the population density grid of the European Union at 100m resolution provided openly by Eurostat~\cite{eurostat}\footnote{This database has some precision issues that have been recognized~\cite{bretagnolle2016ville} but the aggregation at a larger resolution should allow to remove possible bias.}. The choice of the resolution, the spatial range, and the shape of the window on which indicators are computed, is made according to the thematic specifications given before. We consider 50km wide square windows. As it also does not make sense to have a too detailed resolution because of data quality\footnote{According to \cite{batista2013high} which details the construction of the dataset, good results were obtained after validation for seven countries on samples with a grid of resolution 1km. We are thus closer of this resolution with a resolution of 500m.}, we take $N=100$ and aggregate the initial raster data at a 500m resolution to meet this size on real windows of size 50km. To have a rather continuous distribution of indicators in space, we overlap windows by setting an offset of 10km between each, what induces a smoothing of values and allows to limit bord effects due to the shape. We have furthermore tested the sensitivity to window size by computing samples with 30km and 100km window sizes and obtained rather similar spatial distributions, and also strong correlations between the fields and their smoothing at a finer resolution, as detailed in Appendix~\ref{app:sec:staticcorrelations}.
}{
Nous calculons les mesures morphologiques données ci-dessus sur des données réelles de densité, en utilisant la grille de population de l'Union Européenne à la résolution de 100m fournie de manière ouverte par Eurostat~\cite{eurostat}\footnote{Cette base a certains défauts de précision qui ont été reconnus~\cite{bretagnolle2016ville} mais l'agrégation à une résolution supérieure devrait permettre de diminuer d'éventuels biais.}. Le choix de la résolution, de la portée spatiale, et de la forme de la fenêtre sur laquelle les indicateurs sont calculés, sont faits suivant les spécifications thématiques précédentes. Nous considérons des fenêtres carrées de largeur 50km. Comme une résolution trop détaillée n'est pas désirable à cause de la qualité des données\footnote{Selon \cite{batista2013high} qui détaille la construction du jeu de données, des bons résultats ont été obtenus après validation pour sept pays sur des échantillons avec une grille de résolution 1km. Nous nous rapprochons ainsi de cette résolution avec une grille de résolution 500m.}, nous agrégeons les données de la grille initiale à une résolution de 500m pour avoir des fenêtres de taille $N=100$ correspondant à 50km de côté. Pour obtenir une distribution des indicateurs relativement continue dans l'espace, nous superposons les fenêtres en posant un décalage de 10km entre chaque, ce qui induit un lissage des valeurs et permet de limiter les effets de bord dus à la forme. Nous avons par ailleurs testé la sensibilité à la taille de la fenêtre en calculant des échantillons avec des tailles de 30km et 100km et avons obtenu des distributions spatiales similaires, ainsi que de fortes corrélations entre les champs et leur lissage à une résolution plus fine, comme détaillé en Annexe~\ref{app:sec:staticcorrelations}.
}

\bpar{
The implementation of indicators must be done carefully, since computational complexities can reach $O(N^4)$ for the Moran index for example: we use convolution through Fast Fourier Transform, which is a technique allowing the computation of the Moran index with a complexity in $O(\log^2 N \cdot N^2)$\footnote{I.e. having an execution time bounded by $\log^2 N \cdot N^2$ if $N$ is the data size, what is a considerable gain compared to $N^4$: to process a grid of width 100, the asymptotic gain factor will be approximatively $10000$.}.
}{
L'implémentation des indicateurs doit être faite avec attention, puisque les complexités computationnelles peuvent atteindre $O(N^4)$ pour l'indice de Moran par exemple : nous utilisons la convolution par Transformée de Fourier Rapide, qui est une technique permettant de calculer l'indice de Moran avec une complexité en $O(\log^2 N \cdot N^2)$\footnote{C'est-à-dire ayant un temps d'execution borné par $\log^2 N \cdot N^2$ si $N$ est la taille des données, ce qui est un gain considérable par rapport à $N^4$ : pour le traitement d'une grille de côté 100, le facteur de gain asymptotique sera d'environ $10000$.}.
}


\bpar{
We show in Fig.~\ref{fig:staticcorrelations:empirical} maps giving values of indicators, for France only to ease maps readability. The first striking feature is the diversity of morphological patterns across the full territory. The auto-correlation is naturally high in Metropolitan areas (Paris, Lyon, Marseille for example), with the Parisian surroundings clearly detached. When looking at other indicators, it is interesting, regarding possible areas in which a co-evolution could happen, to denote regional regimes: rural areas have much less hierarchy in the South than in the North, whereas the average distance is rather uniformly distributed except for mountain areas. Regions of very high entropy are observed in the Center and South-West.
}{
Nous montrons en Fig.~\ref{fig:staticcorrelations:empirical} des cartes donnant les valeurs des indicateurs, pour la France seulement afin de permettre une lisibilité. La première caractéristique frappante est la diversité des motifs morphologiques au travers de l'ensemble du territoire. L'indice de Moran est relativement haut dans les zones englobant les métropoles (Paris, Lyon, Marseille par exemple), avec les environs de Paris qui se détachent clairement. Lorsqu'on s'intéresse aux autres indicateurs, il est intéressant, au regard de zones possibles dans laquelle une co-évolution peut s'opérer, de constater des régimes régionaux: les zones rurales ont beaucoup moins de hiérarchie dans le Sud que dans le Nord, tandis que la distance moyenne est plutôt distribuée uniformément sauf dans les zones montagneuses. Des régions qui présentent de fortes valeurs de l'entropie sont observées dans le centre et le Sud-ouest.
}

\bpar{
To have a better insight into existing morphological classes, we use unsupervised classification\footnote{Which consists in partitioning the data space according to their endogenous structure.} with a simple k-means algorithm\footnote{Given the distribution of points which have a relatively homogenous density, alternative methods such as the DBScan algorithm are relatively equivalent. We take here a number of repetitions $b=100$ of the algorithm to have a result robust to stochasticity.}. The number of clusters $k=5$ witnesses a transition in inter-cluster variance, what means that a variation of structure occurs at this number, that we then choose as the number of clusters. The split between classes is plotted in Fig.~\ref{fig:staticcorrelations:empirical}, bottom-left panel, where we show measures projected on the two first components of a Principal Component Analysis (explaining 71\% of variance, what is relatively large). The map of morphological classes confirms a North-South opposition in a background rural regime (clear green against blue), the existence of mountainous (red) and metropolitan (dark green) regimes. Such a variety of settlements forms will be the target for the model in~\ref{sec:densitygeneration}. A similar computation of morphological indicators was done for China using the gridded population data from~\cite{fu1km}. Maps are available in Appendix~\ref{app:sec:staticcorrelations}.
}{
Pour avoir une meilleure compréhension des classes morphologiques existantes, nous utilisons une classification non-supervisée\footnote{Qui consiste à partitioner l'espace des données selon leur structure endogène.} avec un algorithme des k-means\footnote{Vu la distribution des points qui ont une densité relativement homogène, des méthodes alternatives comme l'algorithme DBScan sont relativement équivalentes. Nous prenons ici un nombre de répétitions $b=100$ de l'algorithme pour avoir un résultat robuste à la stochasticité.}. Le nombre de clusters $k=5$ induit une transition dans la variance inter-cluster, ce qui signifie qu'une variation de structure s'opère à ce nombre, que nous choisissons alors comme nombre de clusters. La séparation entre les classes est montrée en Fig.~\ref{fig:staticcorrelations:empirical}, panneau bas gauche, où nous représentons les mesures projetées sur les deux premières composantes d'une Analyse en Composantes Principales (expliquant 71\% de la variance, ce qui est relativement conséquent). La carte des classes morphologiques confirme une opposition Nord-Sud dans le régime rural de fond (vert clair contre bleu), l'existence d'un régime de montagne (rouge) et d'un régime métropolitain (vert sombre). Une telle variété d'établissements sera l'un des objectifs du modèle en~\ref{sec:densitygeneration}. Un calcul similaire des indicateurs morphologiques a été effectué pour la Chine en utilisant la grille de population à 1km fournie par~\cite{fu1km}. Les cartes sont disponibles en Annexe~\ref{app:sec:staticcorrelations}.
}



\subsection{Network Measures}{Mesures de Réseau}

\bpar{
We consider network aggregated indicators as a way to characterize transportation network properties on a given territory, the same way morphological indicators yielded information on urban structure. We propose to compute some simple indicators on same extents as for morphology, to be able to explore relations between these static measures.
}{
Nous considérons d'autre part les mesures agrégées de réseau comme un moyen de caractériser les propriétés des réseaux de transport sur un territoire donné, de la même façon que les indicateurs morphologiques informent sur la structure urbaine. Nous proposons de calculer des indicateurs simples sur des étendues spatiales similaires à celles retenues pour la mesure de la morphologie, pour être en mesure d'explorer les relations entre ces mesures statiques.
}

\bpar{
Static network analysis has been extensively documented in the literature, such as for example \cite{louf2014typology} for a cross-sectional study of cities or \cite{2015arXiv151201268L} for the exploration of new measures for the road network. \cite{2017arXiv170902939M} uses techniques from deep learning to establish a typology of urban road networks for a large number of cities across the world.
}{
L'analyse statique de réseau a été intensément documentée dans la littérature, comme par example \cite{louf2014typology} pour une étude comparative des villes ou \cite{2015arXiv151201268L} pour l'exploration de nouvelles mesures pour le réseau de rues. \cite{2017arXiv170902939M} utilise des techniques issues de l'apprentissage profond pour établir une typologie des réseaux viaires urbains pour un grand nombre de villes dans le monde.
}

\bpar{
The questions behind such approaches are multiple: they can aim at finding typologies or at characterizing spatial networks, at understanding underlying dynamical processes in order to model morphogenesis, or even at being applied in urban planning such as \emph{Space Syntax} approaches~\cite{hillier1989social}. We are positioned here more within the two first logics since we aim at characterizing the shape of networks in a first step, and then to include their dynamics in models in a second step. Our significant contribution is the characterization of the road network on large spatial extents, covering Europe and China.
}{
Les enjeux derrière ce genre d'approches sont multiples : elles peuvent viser à des typologies ou caractérisations de réseaux spatiaux, à des compréhensions des processus dynamiques sous-jacents dans un but de modélisation de la morphogenèse, ou même de planification urbaine comme sont appliquées parfois les approches par \emph{Space Syntax}~\cite{hillier1989social}. Nous nous plaçons ici plutôt dans les deux premières logiques puisque nous cherchons à caractériser la forme des réseaux dans un premier temps, puis d'inclure leur dynamique dans des modèles dans un second temps. Notre contribution significative est la caractérisation du réseau routier sur de grandes étendues spatiales, couvrant l'Europe et la Chine.
}

\subsubsection{Indicators}{Indicateurs}

\bpar{
We introduce indicators to have a broad idea of the form of the network, using a certain number of indicators to capture the maximum of dimensions of properties of networks, more or less linked to their use. These indicators summarize the mesoscopic structure of the network and are computed on topological networks obtained through simplification steps that will be detailed later. If we denote the network with $N=(V,E)$, nodes have spatial positions $\vec{x}(V)$ and populations $p(v)$ obtained through an aggregation of population in the corresponding Voronoï polygon\footnote{A Voronoï diagram is a partition of the plan, constructed from a point cloud. The cell associated to each point is composed by the set of points closer to it than other points of the cloud. The graph of a Voronoï diagram is the dual of the associated Delaunay triangulation.}, and edges $E$ have \emph{effective distances} $l(E)$ taking into account impedances and real distances (to include the primary network hierarchy). We then use:
}{
Nous introduisons des indicateurs pour avoir une idée large de la forme du réseau, utilisant un certain nombre d'indicateurs pour capturer le maximum de dimensions des propriétés des réseaux, plus ou moins liées à l'utilisation de ceux-ci. Ces indicateurs résumant la structure mesoscopique du réseau sont calculés sur les réseau topologiques obtenus par des étapes de simplification détaillées plus loin. Notant le réseau $N=(V,E)$, les noeuds $V$ ont des positions spatiales $\vec{x}(v)$ et des populations $p(v)$ obtenues par agrégation de la population dans le polygone de Voronoï correspondant\footnote{Un diagramme de Voronoï est une partition du plan, formée à partir d'un nuage de point. La cellule associée à chaque point est constituée de l'ensemble des points plus proches de celui-ci que des autres points du nuage. Le graphe d'un diagramme de Voronoï est le dual de celui de la triangulation de Delaunay associée.}, les liens $E$ ont des \emph{distances effectives} $l(E)$ qui prennent en compte les impédances et les distances réelles (pour inclure la hiérarchie primaire du réseau). Nous utilisons alors :
}


\bpar{
\begin{itemize}
\item Characteristics of the graph, obtained from graph theory, as defined by~\cite{haggett1970network}: number of nodes $\left|V\right|$, number of links $\left|E\right|$, density $d$, average length of links $\bar{d_l}$, average clustering coefficient $\bar{c}$, number of components $c_0$.
\item Measures linked to shortest paths: diameter $r$, euclidian performance $v_0$ (defined by~\cite{banos2012towards}), average length of shortest paths $\bar{l}$.
\item Centrality measures: these are aggregated at the level of the network by taking their average and their level of hierarchy, computed by an ordinary least squares of a rank-size law, for the following centrality measures:
\begin{itemize}
\item Betweenness centrality~\cite{crucitti2006centrality}, average $\bar{bw}$ and hierarchy $\alpha_{bw}$: given the distribution of centrality on all nodes, we take the slope of a rank-size adjustment and the average of the distribution.
\item Closeness centrality~\cite{crucitti2006centrality}, average $\bar{cl}$ and hierarchy $\alpha_{cl}$.
\item Accessibility~\cite{hansen1959accessibility}, which is in our case computed as a closeness centrality weighted by populations: average $\bar{a}$ and hierarchy $\alpha_{a}$.
\end{itemize}
\end{itemize}
}{
\begin{itemize}
\item Caractéristiques du graphe, issues de la théorie des graphes, comme définies par~\cite{haggett1970network} : nombre de noeuds $\left|V\right|$, nombre de liens $\left|E\right|$, densité $d$, longueur moyenne des liens $\bar{d_l}$, coefficient de clustering moyen $\bar{c}$, nombre de composantes $c_0$.
\item Mesures liées au plus courts chemins : diamètre $r$, performance euclidienne $v_0$ (définie par~\cite{banos2012towards}), longueur moyenne des plus courts chemins $\bar{l}$.
\item Mesures de centralité : celles-ci sont agrégées au niveau du réseau en prenant leur moyenne et leur niveau de hiérarchie, calculé par un ajustement des moindres carrés d'une loi rang taille, pour les mesures de centralité suivantes :
\begin{itemize}
\item Centralité d'intermédiarité~\cite{crucitti2006centrality}, moyenne $\bar{bw}$ et hiérarchie $\alpha_{bw}$ : étant donné la distribution de la centralité sur l'ensemble des noeuds, on prend la pente d'un ajustement rang-taille ainsi que la moyenne de la distribution.
\item Centralité de proximité~\cite{crucitti2006centrality}, moyenne $\bar{cl}$ et hiérarchie $\alpha_{cl}$.
\item Accessibilité~\cite{hansen1959accessibility}, qui est dans notre cas calculée comme une centralité de proximité pondérée par les populations : moyenne $\bar{a}$ et hiérarchie $\alpha_{a}$.
\end{itemize}
\end{itemize}
}

\bpar{
The concept of accessibility is measured here by a network indicator, since its computation implies to attribute weights to the nodes with a corresponding population, and can be interpreted than as a potential of access to the rest of the population (as we did in chapter~\ref{ch:thematic}). This indicator is interesting a priori since it lies at the interface between the urban form and network topology, since the distribution of population on nodes is taken into account.
}{
Le concept d'accessibilité est mesuré ici par un indicateur de réseau, puisque son calcul implique d'attribuer des poids aux noeuds avec une population correspondante, et peut être interprété ensuite comme un potentiel d'accès au reste de la population (comme nous l'avons fait en chapitre~\ref{ch:thematic}). Cet indicateur est a priori intéressant car à l'interface entre forme urbaine et forme du réseau, puisque la distribution de population sur les noeuds est prise en compte.
}

\bpar{
Network performance is close to the rectilinearity measure (\emph{straightness}) proposed by \cite{josselin2016straightness}, which show that it efficiently differentiate rectilinear networks and radio-concentric networks, that are both recurring urban networks.
}{
La performance du réseau est proche de la mesure de rectilinéarité (\emph{straigthness}) proposée par \cite{josselin2016straightness}, qui montrent qu'elle différencie efficacement réseaux rectilinéaires et réseaux radio-concentriques, réseaux urbains récurrents.
}

\bpar{
Our indicators are conceived around network topology but not its use: developments with suited data could extend these analyses to the functional aspect of networks, such as for example performance measures computed by~\cite{trepanier2009calculation} using massive data for a public transportation network.
}{
Nos indicateurs sont conçus autour de la topologie du réseau mais pas son usage : des développements avec les données appropriées peuvent étendre ces analyses à l'aspect fonctionnel des réseaux, comme par exemple des mesures de performance calculées par~\cite{trepanier2009calculation} par l'intermédiaire de données massives pour un réseau de transports en commun.
}



\subsubsection{Data preprocessing}{Pré-traitement des données}

\bpar{
We work here with the road network, which structure is finely conditioned to territorial configuration of population densities. Furthermore, data for the current road network is openly available through the OpenStreetMap (OSM) project~\cite{openstreetmap}. Its quality was investigated for different countries such as England~\cite{haklay2010good} and France~\cite{girres2010quality}. It was found to be of a quality equivalent to official surveys for the primary road network. Concerning China, although \cite{zheng2014assessing} underlined a quick acceleration of OSM road data completeness and accuracy, its use for computation of network indicators may be questioned at a very fine scale. \cite{zhang2015density} highlights different regimes of data quality, partitioning China into regions among which qualitative behavior of OSM data varies. We will have to keep in mind this variability, and to ensure the robustness of results, we will simplify the network at a sufficient level of aggregation.
}{
Nous travaillons ici avec le réseau de rues, dont la structure est finement conditionnée aux configurations territoriales des densités de population. De plus, les données du réseau de routes actuel est disponible ouvertement par l'intermédiaire du projet OpenStreetMap (OSM)~\cite{openstreetmap}. Sa qualité a été étudiée pour différents pays comme l'Angleterre~\cite{haklay2010good} et la France~\cite{girres2010quality}. Il a été établi pour ces pays une qualité équivalente aux données officielles pour le réseau de rues primaire, au sens à la fois de la couverture spatiale et de la précision locale. Dans le cas de la Chine, bien que \cite{zheng2014assessing} soulève une récente accélération de la couverture et de la précision des données OSM pour les routes, leur usage pour le calcul d'indicateurs de réseau peut être questionné à une échelle très fine. \cite{zhang2015density} fournit une partition de la Chine en régions entre lesquelles le comportement qualitatif des données OSM varie. Nous devrons garder à l'esprit cette variabilité, et pour être assuré de la fiabilité des résultats, nous simplifierons le réseau à un niveau d'agrégation suffisant.
}

\bpar{
The network constituted by primary road segments is aggregated at the fixed granularity of the density grid to create a graph. It is then simplified to keep only the topological structure of the network, normalized indicators being relatively robust to this operation. This step is necessary for a simple computation of indicators and a thematic consistence with the density layer. We keep only the nodes with a degree strictly greater or smaller than two, and corresponding links, by taking care to aggregate the real geographical distance when constructing the corresponding topological link. Given the order of magnitude of data size (for Europe, the initial database has $\simeq 44.7\cdot 10^6$ links, and the final simplified database $\simeq 20.4\cdot 10^6$), a specific parallel algorithm is used, with a \emph{split-merge} structure. It separates the space into areas that can be independently processed and then merged. It is detailed in Appendix~\ref{app:sec:staticcorrelations}.
}{
Le réseau constitué des segments de rue primaires est agrégé à la granularité fixe de la grille de densité pour créer un graphe. Celui-ci est ensuite simplifié pour garder uniquement la structure topologique du réseau, les indicateurs normalisés étant relativement robustes à cette opération. Celle-ci est nécessaire pour un calcul simple des indicateurs et une cohérence thématique avec la couche de densité. On garde uniquement les noeuds ayant un degré strictement supérieur ou inférieur à deux, et les liaisons correspondantes, en prenant soin d'agréger la distance géographique réelle en construisant le lien topologique correspondant. Vu l'ordre de grandeur de taille des données (pour l'Europe, la base initiale a $\simeq 44.7\cdot 10^6$ liens, et la base finale simplifiée $\simeq 20.4\cdot 10^6$), un algorithme spécifique parallèle est mis en place, de structure \emph{split-merge}. Celui-ci découpe l'espace en zones qui peuvent être traitées indépendamment puis fusionnées. Il est détaillé en Annexe~\ref{app:sec:staticcorrelations}.
}


\subsubsection{Results}{Résultats}

\begin{figure}
\includegraphics[width=\linewidth]{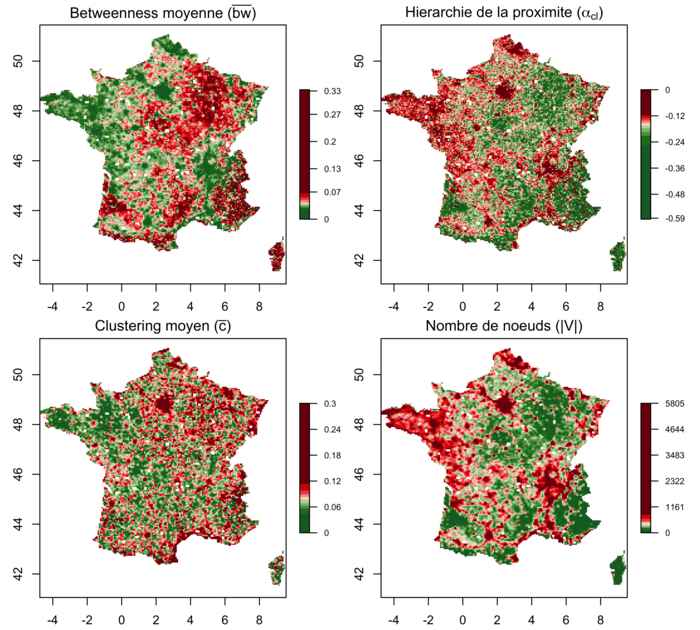}
\caption[Spatial distribution of network indicators]{\textbf{Spatial distribution of network indicators.} We show indicators for France, in correspondance with morphological indicators described previously. We give here the average betweenness centrality $\bar{bw}$, the hierarchy of closeness centrality $\alpha_cl$, the average clustering coefficient $\bar{c}$ and the number of nodes $\left|V\right|$.\label{fig:staticcorrs:network}}
\end{figure}

\bpar{
Network indicators have been computed on the same areas than urban form indicators, in order to put them in direct correspondance and later compute the correlations. We show in Fig.~\ref{fig:staticcorrs:network} a sample for France.
}{
Les indicateurs de réseau ont été calculés sur les mêmes zones que les indicateurs de forme urbaine, pour pouvoir les mettre en correspondance directe et calculer les correlations par la suite. Nous montrons en Fig.~\ref{fig:staticcorrs:network} un échantillon pour la France.
}

\bpar{
The spatial behavior of indicators unveils local regimes as for the urban form (urban, rural, metropolitan), but also strong regional regimes. They can be due to the different agricultural practices depending on the region for the rural for example, implying a different partition of parcels and also a particular organization of their serving. For network size, Brittany is a clear outlier and rejoins urban regions, witnessing very fragmented parcels (and a fortiori also of a land property fragmentation in the simplifying assumption of corresponding parcels and properties). This is partly correlated to a low hierarchy of accessibility. The South and the East of the extended \emph{Bassin Parisien} are distinguishable by a strong average betweenness centrality, in accordance with a strong hierarchy of the network.
}{
Le comportement spatial des indicateurs révèle comme pour la forme urbaine des régimes locaux (urbain, rural, métropolitain), mais aussi des régimes régionaux très marqués. Ceux-ci peuvent être dus aux différentes pratiques agricoles selon les régions dans le cas du rural par exemple, impliquant une partition différente des parcelles ainsi qu'une organisation particulière de leur desserte. En taille du réseau, la Bretagne se détache nettement et rejoint les régions urbaines, témoignant de parcelles très fragmentées (et a fortiori d'une découpe foncière fragmentée également dans l'hypothèse simplificatrice d'une coincidence des parcelles et du foncier). Cela est partiellement corrélé à une faible hiérarchie dans l'accessibilité. Le Sud et l'Est du Bassin Parisien étendu se distinguent par une forte centralité d'intermédiarité moyenne, en accord avec une forte hiérarchisation du réseau.
}

\bpar{
The same way as for urban form, this spatial variability suggests the search of variables regimes of interactions between indicators, as we will do for later through their correlations.
}{
De la même manière que pour la forme urbaine, cette variabilité spatiale suggère la recherche de régimes variables d'interactions entre indicateurs, comme nous le ferons plus loin par l'intermédiaire de leur corrélations.
}

\bpar{
For China, for which a selection of indicators is also given in~\ref{app:sec:staticcorrelations}, we observe even stronger local and regional variations. Highly populated urban areas detach themselves, corresponding to a particular regime.
}{
Pour la Chine, pour laquelle une selection d'indicateurs est également donnée en~\ref{app:sec:staticcorrelations}, on observe des variations locales et régionales encore plus marquées. Les zones urbaines fortement peuplées se détachent, correspondant à un régime bien particulier.
}

\bpar{
The accessibility indicator is finally strongly correlated with the same unweighted indicator, i.e. closeness centrality: we obtain a correlation of $\rho = 0.86$ estimated on all measure points for China.
}{
L'indicateur d'accessibilité est finalement fortement corrélé au même non-pondéré, c'est-à-dire la centralité de proximité : nous obtenons une corrélation de $\rho = 0.86$ estimée sur l'ensemble des points de mesure pour la Chine.
}


\subsection{Effective static correlations and non-stationarity}{Correlations statiques effectives et non-stationnarité}

\subsubsection{Spatial correlations}{Corrélations spatiales}

\bpar{
Local spatial correlations are computed on windows gathering a certain number of observations, and thus of windows on which indicators have been computed. We denote by $l_0$ (which is equal to 10km in preceding results) the resolution of the distribution of indicators. The estimation of correlations in then done on squares of size $\delta\cdot l_0$ (with $\delta$ which can vary typically from 4 to 100). $\delta$ gives simultaneously the number of observations used for the local estimation of correlation, and the spatial range of the corresponding window. Its value thus directly influences the confidence of the estimation.
}{
Les corrélations spatiales locales sont calculées sur des fenêtres regroupant un certain nombre d'observations, et donc de fenêtres sur lesquelles les indicateurs ont été calculés. Notons $l_0$ (qui vaut 10km dans les résultats précédents) la résolution des distributions des indicateurs. L'estimation des corrélations s'effectue alors sur des carrés de taille $\delta\cdot l_0$ (avec $\delta$ pouvant varier typiquement de 4 à 100). $\delta$ donne à la fois le nombre d'observations utilisées pour l'estimation locale de la corrélation, et la portée spatiale de la fenêtre correspondante. Sa valeur influe donc directement la fiabilité de l'estimation.
}

\begin{figure}[h!]
\includegraphics[width=\linewidth]{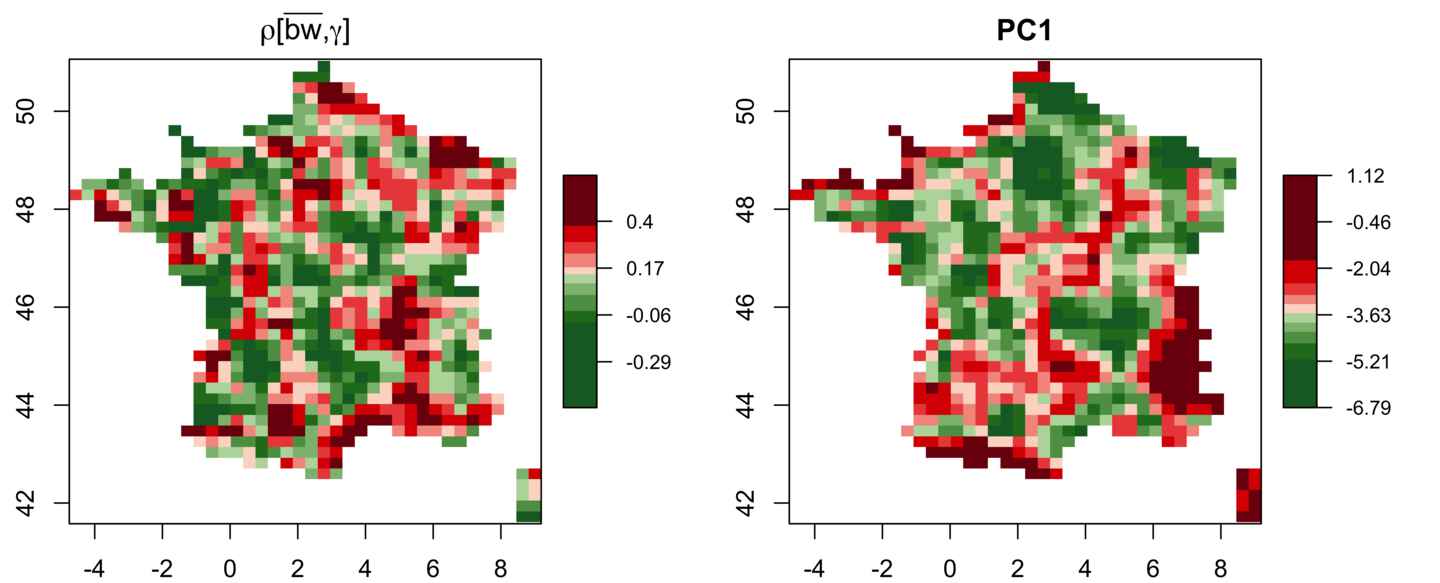}
\caption[Spatial correlations between morphological indicators and network indicators]{\textbf{Examples of spatial correlations.} For France, the maps give $\rho\left[\bar{bw},\gamma\right]$, correlation between the average betweenness centrality and the hierarchy of population (\textit{Left}) and the first component of the reduced matrix (\textit{Right}).\label{fig:staticcorrs:mapscorrs}}
\end{figure}

\bpar{
We show in Fig.~\ref{fig:staticcorrs:mapscorrs} examples of correlations estimated with $\delta = 12$ in the case of France. With 20 indicators, the correlation matrix is significantly large in size, but the effective dimension (the number of components required to reach the majority of variance) is reduced:  principal components analysis shows that 10 components already capture 62\% of variance, and the first component already captures 17\%, what is considerable in a space where the dimension is 190\footnote{This corresponds to the dimension of the correlation matrix between 20 indicators, i.e. the number of elements of its half without the diagonal. If correlations were randomly distributed, the first component would capture $1/190 = 0.5\%$ only, and the 10 first 5\%, since the variance is equally shared between independent dimensions.}.
}{
Nous montrons en Fig.~\ref{fig:staticcorrs:mapscorrs} des exemples de corrélations estimées avec $\delta = 12$ dans le cas de la France. Avec 20 indicateurs, la matrice de corrélation est assez conséquente en taille, mais la dimension effective (le nombre de composantes nécessaires pour atteindre la majorité de la variance) est réduite : une analyse en composante principale montre que 10 composantes capturent 62\% de la variance, et la première composante capture déjà 17\%, ce qui est considérable dans un espace où la dimension est de 190\footnote{Il s'agit de la dimension de la matrice de correlation entre 20 indicateurs, c'est-à-dire le nombre d'éléments de sa moitié moins sa diagonale. Si les corrélations étaient distribués aléatoirement, la première composante capturerait $1/190 = 0.5\%$ seulement, et les 10 premières 5\%, puisque la variance se répartit équitablement entre des dimensions indépendantes.}.
}

\bpar{
It is possible to examine the bloc for urban form, for the network, or for crossed correlations, which directly express a link between properties of the urban form and of the network. For example, the relation between average betweenness centrality and morphological hierarchy that we visualize allows to understand the process corresponding to the correspondance of hierarchies: a hierarchical population can induce a hierarchical network or the opposite direction, but it can also induce a distributed network or such a network create a population hierarchy - this must be well understood in terms of correspondence and not causality, but this correspondance informs on different urban regimes. Metropolitan areas seem to exhibit a positive correlation for these two indicators, as shows the Fig.~\ref{fig:staticcorrs:mapscorrs}, and rural spaces a negative correlation.
}{
Il est possible d'examiner le sous-bloc pour la forme urbaine, pour le réseau, ou les corrélations croisées, qui expriment directement un lien entre les propriétés de la forme urbaine et celles du réseau. Par exemple, la relation entre centralité de chemin moyenne et hiérarchie morphologique que l'on visualise permet de comprendre le processus correspondant à la correspondance des hiérarchies : une population hiérarchisée peut induire un réseau hiérarchisé ou le sens inverse, mais elle peut également induire un réseau distribué ou celui-ci peut créer une hiérarchie de population - il faut bien comprendre en terme de correspondance et non de causalité, mais cette correspondance informe sur différents régimes urbains. Les métropoles semblent présenter une corrélation positive pour ces deux indicateurs, comme le montre la Fig.~\ref{fig:staticcorrs:mapscorrs}, et des espaces ruraux une corrélation négative.
}

\bpar{
In order to give a picture of global relations between indicators, we can refer to the full correlation matrix in Fig.~\ref{fig:app:staticcorrelations:overallcorrs} (Appendix~\ref{app:sec:staticcorrelations}): for example, a strong population hierarchy is linked to a high and hierarchical betweenness centrality, but is negatively correlated to the number of edges (a diffuse population requires a more spread network to serve all the population). However, it is not possible this way to systematically link indicators, since they especially strongly vary in space. We give also in Appendix~\ref{app:sec:staticcorrelations}, Fig.~\ref{fig:app:staticcorrelations:europe-correlations}, maps for different correlation coefficients for all Europe.
}{
Afin de se donner une idée des relations globales entre indicateurs, on peut se référer à la matrice de corrélation complète en Fig.~\ref{fig:app:staticcorrelations:overallcorrs} (Annexe~\ref{app:sec:staticcorrelations}) : par exemple, une forte hiérarchie de population est liée à une centralité de chemin forte et hiérarchisée, mais est corrélée négativement au nombre de sommets (une population diffuse nécessite un réseau plus maillé pour la desserte). Toutefois, il n'est pas possible ainsi de systématiquement lier les indicateurs, puisque ceux-ci varient justement fortement dans l'espace. Nous donnons également en Annexe~\ref{app:sec:staticcorrelations}, Fig.~\ref{fig:app:staticcorrelations:europe-correlations}, des cartes de différents coefficients de corrélation pour l'ensemble de l'Europe. 
}

\bpar{
This suggests a very high variety of interaction regimes. The spatial variation of the first component of the reduced matrix confirms it, what clearly reveals the spatial non-stationarity of interaction processes between forms, since the first and second moments vary in space. The statistical significance of stationarity can be verified in different ways\footnote{There does not exist to the best of our knowledge a generic test for spatial non-stationarity. \cite{zhang2014test} develops for example a test for rectangular regions of any dimension, but in the specific case of \emph{point processes}.}. We use here the method of \cite{leung2000statistical} which consists in estimating through bootstrap the robustness of Geographically Weighted Regression models. These will be developed below, but we obtain for all tested models a significant non-stationarity without doubt ($p<10^{-3}$).
}{
Cela suggère une très grande variété de régimes d'interaction. La variation spatiale de la première composante de la matrice réduite confirme celle-ci, ce qui révèle clairement la non-stationnarité spatiale des processus d'interaction entre formes, puisque les premiers et second moments varient dans l'espace. La significativité statistique de la non-stationnarité spatiale peut être vérifiée de différentes façons\footnote{Il n'existe à notre connaissance pas de test générique de non-stationnarité spatiale. \cite{zhang2014test} développe par exemple un test pour des régions rectangulaires de dimensions quelconques, mais dans le cas spécifiques des \emph{point processes}.}. Nous utilisons ici la méthode de \cite{leung2000statistical} qui consiste à estimer par bootstrap la robustesse de modèles de Régression Géographique Pondérée. Ceux-ci seront développés ci-dessous, mais on obtient pour l'ensemble des modèles testés une non-stationnarité significative sans équivoque ($p<10^{-3}$).
}


\bpar{
Furthermore, the statistical distribution of correlations given in Fig.~\ref{fig:app:staticcorrelations:corr-distribs} in Appendix~\ref{app:sec:staticcorrelations} follows an asymmetric law for the morphology alone, and rather symmetric for the network and the cross-correlations, what would mean that some areas have rather strong morphological constraints whereas the shape of the network is rather free. Finally, we observe on the point clouds of the same figure, relating the values of correlations in the different blocs, that configurations for which cross-correlations are the strongest correspond to the ones for which morphological and network correlations are also strong, confirming the intrication of processes in that case.
}{
Par ailleurs, la distribution statistique des corrélations donnée en Fig.~\ref{fig:app:staticcorrelations:corr-distribs} en Annexe~\ref{app:sec:staticcorrelations} suit une loi asymétrique pour la morphologie seule, et plutôt symétrique pour le réseau et le croisement, ce qui voudrait dire que certaines zones ont des contraintes morphologiques assez fortes tandis que la forme du réseau est plutôt libre. Enfin, on constate sur les nuages de points de la même figure, croisant les valeurs des corrélations dans les différents blocs, que les configurations où les corrélations croisées sont les plus fortes correspondent à celles où les corrélations morphologiques et de réseau sont également fortes, confirmant l'imbrication des processus dans ce cas.
}

\subsubsection{Variations of the estimated correlations}{Variations des corrélations estimées}

\begin{figure}
\includegraphics[width=\linewidth,height=0.85\textheight]{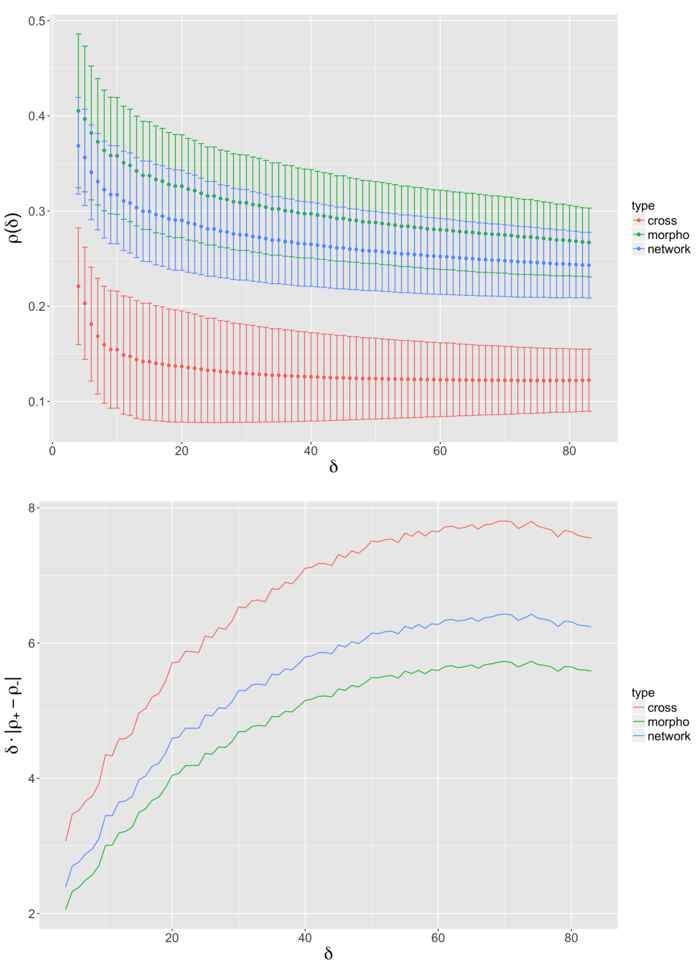}
\caption[Variation of correlations with scale]{\textbf{Variation of correlations with scale, for correlations computed on Europe.} (\textit{Top}) Average absolute correlations and their standard deviations, for the different blocs, as a function of $\delta$; (\textit{Bottom}) Normalized size of the confidence interval $\delta\cdot \left|\rho_{+} - \rho_{-}\right|$ (confidence interval $\left[\rho_{-} , \rho_{+}\right]$ estimated by the Fisher method) as a function of $\delta$.\label{fig:staticcorrs:corrsdistrib}}
\end{figure}

\bpar{
We show in Fig.~\ref{fig:staticcorrs:corrsdistrib} the variation of the estimation of correlation as a function of window size. More precisely, we observe a strong variation of correlations as a function of $\delta$, what is reflected in the average value of the matrix given here (which extends for example from $\rho(4)=0.22$ to $\rho(80)=0.12$ for average absolute cross-correlations). An increase of $\delta$ leads for all measures a shift towards positive values, but also a narrowing of the distribution, these two effects resulting in a decrease of average absolute correlations, which approximatively stabilize for large values of $\delta$. Such a variation could be a clue of a multi-scalar behavior: a change in window size should not influence the estimation if a single process would be implied, it should only change the robustness of the estimation. The development in Appendix~\ref{app:sec:staticcorrelations} illustrates this link in the case of processes superposed at two scales, and demonstrates that this structure of process implies a variation of the estimated correlation as a function of $\delta$, at least in low values, which is what we observe here in Fig.~\ref{fig:staticcorrs:corrsdistrib}.
}{
Nous montrons en Fig.~\ref{fig:staticcorrs:corrsdistrib} la variation de l'estimation des corrélations en fonction de la taille de la fenêtre. Plus précisément, on observe une forte variation des correlations en fonction de $\delta$, qui se reflète dans la valeur moyenne de la matrice donnée ici (s'étendant par exemple de $\rho(4)=0.22$ à $\rho(80)=0.12$ pour les corrélations croisées absolues moyennes). L'augmentation de $\delta$ cause pour l'ensemble un décalage dans le positif, mais également un rétrécissement de la distribution, ces deux effets se traduisant par une décroissance des corrélations absolues moyennes, qui se stabilisent approximativement pour les grandes valeurs de $\delta$. Cette variation pourrait être révélatrice d'un comportement multi-échelle : le changement de la taille de la fenêtre ne devrait pas influer l'estimateur si un seul processus était sous-jacent, elle devrait seulement changer la robustesse de l'estimation. Le développement en Annexe~\ref{app:sec:staticcorrelations} illustre ce lien dans le cas de processus superposés à deux échelles, et démontre que cette structure de processus implique une variation de la corrélation estimée en fonction de $\delta$, au moins dans les faibles valeurs, ce que nous observons ici en Fig.~\ref{fig:staticcorrs:corrsdistrib}.
}

\bpar{
Furthermore, the variation of the normalized size of the confidence interval for correlations, which in theory under an assumption of normality should lead $\delta\cdot \left|\rho_+ - \rho -\right|$ to remain constant, since bounds vary asymptotically as $1/\sqrt{N}\sim 1/\sqrt{\delta^2}$ (the demonstration is given in Appendix~\ref{app:sec:staticcorrelations}), follows the direction of this hypothesis of processes superposed at different scales as proposed previously.
}{
Par ailleurs, la variation de la taille normalisée de l'intervalle de confiance pour les corrélations, qui en théorie sous hypothèse de normalité devrait conduire $\delta\cdot \left|\rho_+ - \rho -\right|$ à être constant, puisque les bornes varient asymptotiquement comme $1/\sqrt{N}\sim 1/\sqrt{\delta^2}$ (la démonstration est donnée en Annexe~\ref{app:sec:staticcorrelations}), va dans la direction de cette hypothèse de processus superposés à plusieurs échelles comme proposé précédemment.
}

\bpar{
Thus, processes are both non-stationary, and clues suggest that they result of the superposition of processes at different scales\footnote{The notion of multi-scalar process is otherwise very broad, and can manifest itself in scaling laws for example~\cite{west2017scale}. An approach closer to the one we took is given by~\cite{Chodrow31102017} which measures intrinsic scales to segregation phenomenons by using measures from Information Theory.}.
}{
Ainsi, les processus sont à la fois non-stationnaires, et des indices poussent à laisser penser qu'ils résultent de la superposition de processus à différentes échelles\footnote{La notion de processus multi-scalaire est par ailleurs très diverse, et peut s'expliciter par exemple par la manifestation de lois d'échelles~\cite{west2017scale}. Une démarche plus proche de la notre est donnée par~\cite{Chodrow31102017} qui mesure les échelles intrinsèques aux phénomènes de ségrégation en utilisant des mesures issues de la Théorie de l'Information.}.
}

\subsubsection{Typical scales}{Echelles typiques}

\bpar{
We also propose to explore the possible property of multi-scalar processes by the extraction of endogenous scales which are present in the data. A Geographically Weighted Principal Component Analysis (GWRPCA)~\cite{harris2011geographically} in exploration suggests weights and importances that vary in space, what is in consistence with the non-stationarity of correlation structures obtained above. There is no reason a priori that the scales of variation of the different indicators are strictly the same. We propose thus to extract typical scales for crossed relations between the urban form and network topology.
}{
Nous proposons d'autre part d'explorer la propriété éventuelle de multi-scalarité par extraction d'échelles endogènes présentes dans les données. Une Analyse en Composantes Principales Géographique Pondérée (GWRPCA)~\cite{harris2011geographically} exploratoire suggère des poids et importances variables dans l'espace, ce qui est cohérent avec la non-stationnarité des structures de corrélation obtenue ci-dessus. Il n'y a a priori pas de raison pour que les échelles de variation des différents indicateurs soient strictement identiques. Nous proposons donc d'extraire les échelles typiques pour les relations croisées entre forme urbaine et forme de réseau.
}

\bpar{
We implement therefore the following method: we consider a typical sample of indicators (four for each aspect, see the list in Table~\ref{tab:staticcorrelations:gwr}), and for each indicator we formulate all the possible linear models as a function of opposite indicators (network for a morphological indicator, morphological for a network indicator), aiming at directly capturing the interaction without controlling on the type of form or of network. These models are then adjusted by a Geographically Weighted Regression (GWR) with an optimal range determined by a corrected information criteria (AICc)\footnote{By using the R package GWModel~\cite{gollini2013gwmodel}.}. For each indicator, we keep the model with the best value of the information criteria. We adjust the models on data for France, with a \emph{bisquare} kernel and an adaptative bandwidth in number of neighbors.
}{
Nous implémentons pour cela la méthode suivante : nous considérons un échantillon typique d'indicateurs (quatre pour chaque aspect, voir la liste en Table~\ref{tab:staticcorrelations:gwr}), et pour chaque indicateur nous formulons l'ensemble des modèles linéaires possibles en fonction des indicateurs opposés (réseau pour un indicateur morphologique, morphologique pour un indicateur de réseau), visant à capturer directement l'interaction sans contrôle sur le type de forme ou de réseau. Ces modèles sont alors ajustés par une Régression Géographique Pondérée (GWR) à portée optimale déterminée par critère d'information corrigé (AICc)\footnote{En utilisant le package R GWModel~\cite{gollini2013gwmodel}.}. Pour chaque indicateur, on retient le modèle ayant la meilleure valeur du critère d'information. Nous ajustons les modèles sur les données de la France, avec un noyau \emph{bisquare} et une portée adaptable en nombre de voisins.
}


\bpar{
Results are presented in Table~\ref{tab:staticcorrelations:gwr}. It is first interesting to note that all models have only one variable, suggesting relatively direct correspondances between topology and morphology. All morphological indicators are explained by network performance, i.e. the quantity of detours it includes. On the contrary, network topology is explained by Moran index for centralities, and by entropy for performance and the number of vertices. There is thus a dissymmetry in relations, the network being conditioned in a more complex way to the morphology than the morphology to the network. The adjustments are rather good ($R^2 > 0.5$) for most indicators, and \emph{p-values} obtained for all models (for the constant and the coefficient) are lower than $10^{-3}$. Concerning the scales corresponding to the optimal model, they are very localized, of the order of magnitude of ten kilometers, i.e a larger variation than the one obtained the correlations. This analysis confirms thus statistically on the one hand the non-stationarity, and on the other hand give a complementary point of view on the question of endogenous scales.
}{
Les résultats sont présentés en Table~\ref{tab:staticcorrelations:gwr}. Il est intéressant de noter dans un premier temps que l'ensemble des modèles ne comprend qu'une seule variable, suggérant des correspondances relativement directes entre topologie et morphologie. L'ensemble des indicateurs morphologiques est expliqué par la performance du réseau, c'est-à-dire la quantité de détour qu'il comprend. Au contraire, la topologie est expliquée par le Moran pour les centralités, et par l'entropie pour la performance et le nombre de sommets. On a ainsi une dissymétrie des relations, le réseau étant conditionné de manière plus complexe à la morphologie que la morphologie au réseau. Les ajustements sont bons ($R^2 > 0.5$) pour une majorité d'indicateurs, et les \emph{p-values} obtenues pour l'ensemble des modèles (constante et coefficient) sont inférieures à $10^{-3}$. Les échelles correspondant au modèle optimal sont quant à elles très localisées, de l'ordre de la dizaine de kilomètres, c'est-à-dire une plus grande variation que celle obtenue par les corrélations. Cette analyse confirme ainsi statistiquement d'une part la non-stationnarité, et d'autre part donne un point de vue complémentaire sur la question des échelles endogènes.
}



\begin{table}
\caption[Interrelations between morphological indicators and network indicators]{\textbf{Interrelations between network indicators and morphological indicators.} Each relation is adjusted by a Geographically Weighted Regression, for the optimal range adjusted by AICc.\label{tab:staticcorrelations:gwr}}
\bpar{
\begin{tabular}{|l|l|l|l|}
\hline
Indicator & Model & Range (km) & Adjustment ($R^2$) \\ \hline
Average distance $\bar{d}$ & $\bar{d} \sim v_0$ & 11.6 & 0.31 \\
Entropy $\mathcal{E}$  & $\mathcal{E} \sim v_0$ &  8.8  &0.75 \\
Moran $I$ & $I \sim v_0$ & 8.8 & 0.49 \\
Hierarchy $\gamma$ & $\gamma \sim v_0$ & 8.8  & 0.68 \\\hline
Average betweenness $\bar{bw}$ & $\bar{bw} \sim I$ & 12.3 & 0.58 \\
Average closeness $\bar{cl}$ & $\bar{cl}\sim I$ & 13.9 & 0.26 \\
Performance $v_0$ & $v_0 \sim \mathcal{E}$ & 8.6  & 0.86 \\
Number of nodes $\left|V\right|$ & $\left|V\right| \sim \mathcal{E}$ & 8.6  & 0.88 \\\hline
\end{tabular}
}{
\begin{tabular}{|l|l|l|l|}
\hline
Indicateur & Modèle & Portée (km) & Ajustement ($R^2$) \\ \hline
Distance moyenne $\bar{d}$ & $\bar{d} \sim v_0$ & 11.6 & 0.31 \\
Entropie $\mathcal{E}$  & $\mathcal{E} \sim v_0$ &  8.8  &0.75 \\
Moran $I$ & $I \sim v_0$ & 8.8 & 0.49 \\
Hiérarchie $\gamma$ & $\gamma \sim v_0$ & 8.8  & 0.68 \\\hline
\emph{Betwenness} moyenne $\bar{bw}$ & $\bar{bw} \sim I$ & 12.3 & 0.58 \\
\emph{Closeness} moyenne $\bar{cl}$ & $\bar{cl}\sim I$ & 13.9 & 0.26 \\
Performance $v_0$ & $v_0 \sim \mathcal{E}$ & 8.6  & 0.86 \\
Nombre de noeuds $\left|V\right|$ & $\left|V\right| \sim \mathcal{E}$ & 8.6  & 0.88 \\\hline
\end{tabular}
}
\end{table}
\subsubsection{Developments}{Développements}

\bpar{
We have thus shown empirically the non-stationarity of interactions between the morphology of the distribution of populations and the topology of the road network. Various developments of this analysis are possible.
}{
Nous avons ainsi montré empiriquement la non-stationnarité des interactions entre morphologie de la distribution des populations et topologie du réseau routier. Divers développements de cette analyse sont possibles.
}


\bpar{
Population density grids exist for all regions of the workd, such as for example the ones provided by~\cite{10.1371/journal.pone.0107042}\footnote{Available at \url{http://www.worldpop.org.uk/}. The potential variability of data quality depending on the areas should however lead to stay cautious on their use.}. The analysis may be repeated with other regions of the world, to compare the correlation regimes and test if urban system properties stay the same, keeping in mind the difficulties linked to the differences in data quality.
}{
Des grilles de densité de population existent pour l'ensemble des régions du monde, comme par exemple celles fournies par~\cite{10.1371/journal.pone.0107042}\footnote{Disponibles à \url{http://www.worldpop.org.uk/}. La variabilité potentielle de la qualité des données selon les zones doit toutefois amener une prudence dans leur utilisation.}. L'analyse peut être répétée pour d'autres régions, pour comparer les régimes de corrélations et tester si les propriétés des systèmes urbains restent les mêmes, en gardant à l'esprit les difficultés liées aux différences de qualité dans les données.
}



\bpar{
The research of local scales, i.e. with an adaptative estimation window in terms of size and shape for correlations, would allow to better understand the way processes locally influence their neighborhood. The validation criteria for window size would still be to determine: it can be as above an optimal range for explicative models that are locally adjusted.
}{
La recherche d'échelles locales, c'est-à-dire avec une fenêtre d'estimation adaptative en taille et forme pour les corrélations, permettrait de mieux comprendre la façon dont les processus influent localement sur leur voisinage. Le critère de validation de la taille de la fenêtre resterait à déterminer : il peut s'agir comme ci-dessus de portée optimale pour des modèles explicatifs ajustés localement.
}

\bpar{
The question of ergodicity should also be explored from a dynamical point of view, by comparing time and spatial scales of the evolution of processes, or more precisely the correlations between variations in time and variations in space, but the issue of the existence of databases precise enough in time appears to be problematic. The study of a link between the derivative of the correlation as a function of window size and of the derivatives of the processes is also a direction to obtain indirect informations on dynamics from static data.
}{
La question de l'ergodicité devrait également être explorée sur des bases dynamiques, en comparant les échelles de temps et d'espace d'évolution des processus, ou plus précisément les corrélations entre les variations dans le temps et celles dans l'espace, mais la question de l'existence de bases de données assez fines dans le temps paraît problématique. L'étude d'un lien entre la dérivée de la corrélation en fonction de la taille de la fenêtre et les dérivées des processus est également une piste pour obtenir des informations indirectes sur la dynamique à partir des données statiques.
}


\bpar{
Finally, the search of classes of processes on which it is possible to directly establish the relation between spatial correlations and temporal correlations, is a possible research direction. It stays out of the scope of this present work, but would open relevant perspectives on co-evolution, since it implies evolution in time and an isolation in space, and therefore a complex relation between spatial and temporal covariances.
}{
Enfin, la recherche de classes de processus sur lesquels il est possible d'établir directement la relation entre corrélations spatiales et corrélations temporelles, est une direction possible de recherche. Celle-ci est hors de portée de ce présent travail, mais ouvrirait des perspectives pertinentes sur la co-évolution, puisque celle-ci implique évolution dans le temps et une isolation dans l'espace, et donc une relation complexe entre covariances spatiales et temporelles. 
}



\stars

\bpar{
This section allowed us thus to study non-stationarity properties of morphological characteristics of territories and networks, and of their interactions in terms of static correlations. The indicators we computed will also be useful in the following.
}{
Cette section nous a donc permis d'étudier les propriétés de non-stationnarité des caractéristiques morphologiques des territoires et des réseaux, et de leurs interactions en termes de corrélations statiques. Les indicateurs calculés seront également utiles par la suite.
}

\bpar{
We propose in the next section to tackle a statistical approach to co-evolution, corresponding to the preliminary definition we gave. On the contrary to the previous approach, it will be based on dynamics.
}{
Nous proposons dans la section suivante de nous attaquer à une approche statistique de la co-évolution, correspondant à la définition intermédiaire que nous en avons donnée. Contrairement à l'approche précédente, celle-ci sera basée sur les dynamiques.
}

\stars

%

\newpage


\section{Spatio-temporal causalities}{Causalités spatio-temporelles}

\label{sec:causalityregimes}



\bpar{This section contributes to the understanding of strongly coupled spatio-temporal processes by describing a generic method based on Granger causality, which is a method introduced in economics to characterize possible causal relationships from correlation relations between variables lagged in time. We indeed introduce here a method allowing to characterize co-evolution at the statistical level.
}{
Cette section contribue à la compréhension des processus spatiotemporels fortement couplés, en proposant une méthode générique basée sur la causalité de Granger, qui est une méthode introduite en économie pour caractériser des possibles relations causales à partir de relations de corrélations entre variables décalées dans le temps. Il s'agit bien ici de l'introduction d'une méthode permettant de caractériser la co-évolution au niveau statistique.
}

\bpar{
The method is validated by the robust identification of causality regimes and of their phase diagram for an urban morphogenesis model that couples network growth with density. The application to the real case of South Africa unveils interactions that change in time, witnessing historical events between territorial demographic dynamics and network growth. 
}{
Notre méthode est validée par l'identification robuste de régimes de causalité et de leur diagramme de phase pour un modèle de morphogenèse urbaine couplant croissance du réseau et de la densité. L'application au cas réel de l'Afrique du Sud démontre des interactions qui changent dans le temps, témoins des évènements historiques entre les dynamiques démographiques territoriales et la croissance du réseau.
}

\bpar{
The exists in literature a small number of examples using statistical relationships on dynamical relations between network and territories, i.e. trying to establish a causal relationship between the two. For example, \cite{levinson2008density} explains for the case of London population and connectivity to network variables by these same variables lagged in time, unveiling circular causal effects. \cite{doi:10.1068/b39089} use similar techniques for a region in Italy with historical data on long time, but stays moderate on possible conclusions of systematic effects by recalling the importance of historical events on the estimated relations. \cite{cuthbert2005empirical} proceed to econometric estimations of reciprocal influence, and concludes that in their case study (in Canada at a sub-regional scale), the development of the network induces the development of land-use but not the opposite. Space and time scales influence thus significantly the results of such analysis. \cite{koning:hal-00962384} propose an estimation of relations between the existence of a High Speed Rail connection and economic variables on French Urban Units, and shows a negative effect of the connection itself, after controlling on the endogenous nature of the connection by a selection model, and a significant effect of the characteristics of Urban Units: for example, for urban units benefiting from a TGV connection without LGV, the effect is of -1\% on employments between 1982 and 2006. This study remains however limited as it takes neither a time lag larger than one time step nor spatial relations between entities. \cite{MANC:MANC1073} show on long time a causality link between infrastructure stock and economic growth on a global panel, but that these effects are moderated locally by under or over-investments: in that case, macro-economic effects are revealed.
}{
Il existe dans la littérature un petit nombre d'exemples d'utilisation de statistiques spatiales sur les relations dynamiques entre réseaux et territoires, c'est-à-dire cherchant à exhiber des relations de causalité entre les deux. Par exemple, \cite{levinson2008density} explique pour Londres les variations de population et de connectivité au réseau par ces mêmes variables décalées dans le temps, démontrant des effets causaux réciproques. \cite{doi:10.1068/b39089} utilisent des techniques similaires sur une région d'Italie sur des données historiques sur le temps long, mais modère les conclusions en rappelant l'importance des évènements historiques sur les relations estimées. \cite{cuthbert2005empirical} effectuent des estimations économétriques des influences réciproques, et conclut que dans le cas d'étude (au Canada à une échelle régionale) le développement du réseau induit le développement de l'usage du sol, mais pas l'inverse. L'échelle de temps et d'espace devrait logiquement être responsable de cette non-circularité. \cite{koning:hal-00962384} procèdent à une analyse économétrique de la relation entre existence d'une desserte TGV et variables économiques sur les unités urbaines Françaises, et conclut à un effet en propre négatif pour la desserte, après contrôle de l'endogénéité de la desserte par un modèle de sélection, et un effet significatif des caractéristiques propres des unités urbaines : par exemple, pour les unités urbaines desservies par TGV hors LGV, l'effet de la desserte est de -1\% sur les emplois entre 1982 et 2006. Cette étude reste cependant limitée car non spatialisée et prenant en compte un décalage d'une unité de temps seulement. \cite{MANC:MANC1073} montrent sur le temps long un lien de causalité entre stock d'infrastructure et croissance économique sur un panel mondial, mais que ces effets sont atténués localement par des sous ou sur-investissements : dans ce cas, des effets macroéconomiques sont révélés.
}


\subsection{Spatio-temporal causalities}{Causalités spatio-temporelles}

\bpar{The study of strongly coupled spatio-temporal processes implies to understand tangled intrications generally highly difficult to isolate. These interactions are the essence of complexity approaches, and are indeed at the origin of the emergent behavior of the system. They make sense as an object of study in itself and a separation of processes appears then contradictory with an integrated view of the system. In the case of territorial systems, the example of interactions between transportation networks and territories is a good illustration of this phenomenon, as shows the debate on structuring effects developed in chapter~\ref{ch:thematic}. We recall that we have suggested that the reality of territorial processes in in fact much more complicated that a simple causal relationship between the construction of an infrastructure and spillovers on local development, but indeed corresponds to a \emph{co-evolution}. 
}{
L'étude des processus spatio-temporels fortement couplés implique la prise en compte d'imbrications entre ceux-ci généralement difficiles à isoler. Essence même des approches par la complexité, ces interactions qui sont à l'origine du comportement émergent d'un système font sens comme objet d'étude en lui-même, et une séparation des processus paraît alors contradictoire avec une vision intégrée du système. Dans le cas des systèmes territoriaux, l'exemple des interactions entre réseaux de transport et territoires est une bonne illustration de ce phénomène, comme le montre le débat sur les effets structurants développé en chapitre~\ref{ch:thematic}. Nous rappelons que nous avons suggéré que la réalité des processus territoriaux est en fait bien plus compliquée qu'une simple relation causale entre la mise en place d'une infrastructure et les retombées sur le développement local, mais correspond bien à une \emph{co-évolution}.
}


\bpar{
At an other scale, still for relations between networks and territories, we can point at the relations between mobility practices, urban sprawl et ressource localisation in a metropolitan framework that are as much complex: \cite{cerqueira2017inegalites} shows for example a strong correspondence between conditioning of mobility practices by the accessibility and socio-professional category.
}{
À une autre échelle, toujours concernant les relations entre réseaux et territoires, on peut citer les liens entre pratiques de mobilité, étalement urbain et localisation des ressources dans un cadre métropolitain qui s'avèrent tout autant complexes : \cite{cerqueira2017inegalites} montre par exemple une forte correspondance entre conditionnement des pratiques de mobilité par l'accessibilité et classe socio-professionnelle.
}

\bpar{
This kind of issue is naturally present in other fields: in Economic Geography, the example of links between innovation, local spillovers of knowledge and aggregation of economic agents is a typical illustration of spatio-temporal economic processes exhibiting circular causalities difficult to disentangle~\cite{audretsch1996r}. Specific methods are introduced, as the use of statistical instruments: \cite{aghion2015innovation} shows that the geographical origin of US Congress members that attribute local subsidies is a powerful instrumental variable to link innovation and income inequalities for higher incomes, what confirms that the significant correlation between the two is indeed a causality of innovation on inequalities\footnote{This example is important from the methodological point of view, but not only since it implicitly links to the thematic of the diffusion of innovation which is crucial in the evolutive urban theory.}.
}{
Ce type de problématique est bien sûr présent dans d'autres domaines : en économie, l'exemple des liens entre innovation, impacts locaux de la connaissance et agrégation des agents économiques est une illustration typiques de processus économiques spatio-temporels présentant des causalités circulaires difficiles à démêler~\cite{audretsch1996r}. Des méthodes spécifiques sont introduites, comme l'utilisation d'instruments statistiques comme par~\cite{aghion2015innovation} dans lequel l'origine géographique des membres du Bureau du Congrès américain attribuant les subventions locales est une bonne variable instrumentale pour lier caractère innovant et inégalités des plus hauts salaires, et permet de montrer que la corrélation significative entre les deux est en fait un lien de causalité dirigé de l'innovation sur les inégalités\footnote{Cet exemple est important sur le plan méthodologique, mais pas seulement puisqu'il se lie en filigrane au thème de la diffusion de l'innovation qui est crucial dans la théorie évolutive des villes.}.
}

\subsubsection{Causality in geography}{Causalité en géographie}

\bpar{
Strong coupling in space and time generally implies a notion of causality, that geography has always studied: \cite{loi1985etude} shows that fundamental issues tackled by contemporary theoretical geography (isolation of objects, link between space and causal structures, etc.) were already implicit in \noun{Vidal}'s classical geography.
}{
Le couplage fort spatio-temporel implique généralement l'introduction de la notion de causalité, à laquelle la géographie s'est toujours intéressée : \cite{loi1985etude} montre que les questions fondamentales que se pose la géographie théorique récente (isolation des objects, lien entre espace et structures causales, etc.) étaient déjà présentes dans la géographie classique de \noun{Vidal}.
}

\bpar{
Beside, \cite{claval1985causalite} criticizes the new determinisms having emerged, in particular the one advocated by some scholars of systemic analysis\footnote{See~\cite{chamussy1984dynamique} for an example of model with a planning purpose positioned within that research stream.}: in its beginning, this approach inherited from cybernetics and thus of a reductionist vision implying a determinism even for a probabilistic formulation. \noun{Claval} observes that works contemporary to his writings could allow to capture the complexity that characterizes human decisions: the Prigogine School and the Theory of Catastrophes by René Thom.
}{
\cite{claval1985causalite} critique d'ailleurs les nouveaux déterminismes ayant émergé, notamment celui proposé par certains tenants de l'analyse systémique\footnote{Voir~\cite{chamussy1984dynamique} pour un exemple de modèle à but de planification se plaçant dans ce courant.} : dans ses débuts, cette approche héritait de la cybernétique et donc d'une vision réductionniste impliquant un déterminisme même dans une formulation probabiliste. \noun{Claval} note que des travaux contemporains à son écriture (l'école de Prigogine et la Théorie des Catastrophes de Thom) devraient permettre de capturer la complexité qui fait la particularité des décisions humaines.
}

\bpar{
This viewpoint has anticipated posterior developments, since as Pumain recalls in~\cite{pumain2003approche}, the shift from system analysis to self-organisation and complexity has been long and progressive, and these works have played a fundamental role for it. \noun{François Durand-Dastès} sums up this picture more recently in~\cite{durand2003geographes}, by focusing on the importance of bifurcations and path-dependency in the initial moments of the constitution of a system that he defines as \emph{systemogenesis}\footnote{This notion can be put closer to the one of \emph{morphogenesis} that we study more deeply in chapter~\ref{ch:morphogenesis}.}. This type of complex dynamics generally implies a co-evolution of system components, that can be understood as circular causalities between processes: the issue of identifying them is thus crucial regarding the notion of causality for contemporary complex geography.
}{
Ce point de vue a anticipé les développements antérieurs, puisque comme le rappelle~\cite{pumain2003approche}, le glissement de l'analyse des systèmes à l'auto-organisation puis à la complexité a été long et progressif, et ces travaux ont été fondamentaux pour le permettre. \noun{François Durand-Dastès} résume cette situation plus récemment dans \cite{durand2003geographes}, en appuyant l'importance des bifurcations et de la dépendance au chemin lors des instants initiaux de la constitution du système qu'il désigne par \emph{systèmogenèse}\footnote{Cette notion peut être rapprochée de celle de \emph{morphogenèse} que nous approfondissons en chapitre~\ref{ch:morphogenesis}.}. Ce type de dynamique complexe implique généralement une co-évolution des composantes du système, qu'on peut interpréter comme des causalités circulaires entre processus : la question de pouvoir les identifier est donc cruciale au regard de la notion de causalité pour la géographie complexe contemporaine.
}

\bpar{
This view of a complex causality~\cite{morin1976methode} can also be put into perspective with the concept of \emph{cumulative causality} in economics~\cite{skott1995cumulative}, which insists on the role of path-dependency and the possibility for small perturbations to cause significant effects by negative feedback: it is then impossible to separate the effects from their causes in infinitesimal perturbations.
}{
Cette vision d'une causalité complexe~\cite{morin1976methode} peut être aussi mise en perspective avec le concept de \emph{causalité cumulative} en économie~\cite{skott1995cumulative}, qui insiste sur le rôle de la dépendance au chemin et la possibilité pour de petites perturbations de causer des effets conséquents par rétroaction négative : il est alors impossible de séparer les effets des causes dans les perturbations infinitésimales.
}





\subsubsection{Identification of causalities}{Identification de causalités}

\bpar{
The operational character of the identification of causalities can take diverse forms, in different domains. It will depend on the definitions used, the same way than available methods for which we can give a few illustrations, by trying to cover diverse fields to highlight the different methodological issues and possibilities. \cite{goudet2017learning} use neural networks to infer causality relations between variables in the sense of conditional probabilities. \cite{liu2011discovering} propose to detect spatio-temporal relations between perturbations of trafic flows, introducing a particular definition of causality based on correspondance of extreme points. Associated algorithms are however specific and difficult to apply to other kind of systems. The use of spatio-temporal correlations has been shown to have in some cases a strong predictive power for trafic flows~\cite{min2011real}. Also in the field of transportation and land-use, \cite{xie2009streetcars} applies a Granger causality analysis, that can be interpreted as a lagged correlation, to show for a case study that network growth inducts urban development and is itself driven by externalities such as mobility habits.
}{
Le caractère opérationnel de l'identification des causalités peut prendre des formes très diverses, dans différents domaines. Celui-ci dépendra des définitions utilisées, de la même manière que des méthodes à disposition pour lesquelles nous pouvons donner quelques illustrations, en essayant de s'intéresser à des champs divers pour mettre en valeur les différents enjeux et possibilités méthodologiques. \cite{goudet2017learning} utilisent des réseaux de neurones pour inférer des relations de causalité entre variables au sens des probabilités conditionnelles. \cite{liu2011discovering} proposent la détection de relations spatio-temporelles entre perturbations des flux de trafic, introduisant une définition particulière de la causalité basée sur une correspondance de points extrêmes. Les algorithmes associés sont toutefois spécifiques et difficilement applicables à des types de systèmes différents. L'utilisation des corrélations spatio-temporelles a été démontrée comme ayant dans certains cas un fort pouvoir prédictif pour les flux de traffic~\cite{min2011real}. Également dans le domaine des transports et de l'usage du sol, \cite{xie2009streetcars} appliquent une analyse par causalité de Granger, qu'on pourra interpréter comme une corrélation retardée, pour montrer dans un cas particulier que la croissance du réseau induit le développement urbain et est elle-même tirée par des externalités comme les habitudes de mobilité.
}

\bpar{
Neuroscience has developed numerous methods answering similar issues. \cite{luo2013spatio} define a generalized Granger causality that takes into account non-stationarity and applies to abstracts regions produced by functional imaging. This kind of method is also developed in Computer Vision, as illustrated by \cite{ke2007spatio} which exploit spatio-temporal correlations of forms and flows between successive images to classify and recognize actions. Applications can be quite concrete such as compression of video files by extrapolation of motion vectors~\cite{chalidabhongse1997fast}. In all these cases, the study of spatio-temporal correlations meets the weak notions of causality described above, in the sense of a relation of correlation between variables in sace and time. These measures of causality are closer of ``predictive causality'' in opposition to the ``stimulus-response'' causality as recalled by \cite{bonnafous:halshs-00291521} (p.~90), but allow a large flexibility to be put into practice.
}{
Les neurosciences ont développé de nombreuses méthodes répondant à des problématiques similaires. \cite{luo2013spatio} définissent une causalité de Granger généralisée prenant en compte la non-stationnarité et s'appliquant à des régions abstraites issues d'imagerie fonctionnelle. Ce genre de méthode est également développée en Vision par Ordinateur, comme l'illustrent \cite{ke2007spatio} qui exploitent les corrélations spatio-temporelles de formes et de flux dans des successions d'images pour classifier et reconnaître des actions. Les applications peuvent être très concrètes comme la compression de fichiers videos par extrapolation des vecteurs de mouvement~\cite{chalidabhongse1997fast}. Dans l'ensemble de ces cas, l'étude des corrélations spatio-temporelles rejoint la plupart des notions faibles de causalité vues précédemment, au sens d'une relation de corrélation entre variables dans le temps et l'espace. Ces mesures de causalité sont plus proches de la ``causalité prédicative'' en opposition à la causalité ``stimulus-réponse'' comme indiqué par \cite{bonnafous:halshs-00291521} (p.~90), mais permettent une grande flexibilité de mise en pratique.
}

\bpar{
We aim here at exploring the possibility of an analog method for spatio-temporal data exhibiting a priori complex circular causalities, and thus to realize the difficult exercise to couple a certain level of simplicity with a grasping of complexity. We introduce therefore a method to analyse spatio-temporal correlations, similar to a Granger causality estimated in space and time. The robustness of the method is demonstrated in a systematic way by the application to a complex model of simulation of urban morphogenesis, what leads to the unveiling of distinct causality regimes in the phase space of the model. We also include the application to an empirical case study, what positions this work at the interface between knowledge domains of methodology, modeling and empirical.
}{
Nous cherchons ici à explorer la possibilité d'une méthode analogue pour des données spatio-temporelles présentant a priori des causalités circulaires complexes, et donc de tenter de concilier un certain niveau de simplicité et de caractère opérationnel à une prise en compte de la complexité. Nous introduisons ainsi une méthode d'analyse des corrélations spatio-temporelles similaire à une causalité de Granger estimée dans le temps et l'espace, dont la robustesse est démontrée par l'application systématique à un modèle de simulation complexe de morphogenèse urbaine et par l'isolation de régimes de causalités distincts dans l'espace des phases du modèle. Notre contribution inclut également l'application à un cas d'étude empirique, ce qui la positionne à l'interface des domaines de la méthodologie, de la modélisation et de l'empirique.
}

\bpar{
The rest of this section is organized as follows: the generic framework of the method is described in the next section. We then apply it to a synthetic dataset to partially validate it and test its potentialities, what allows us to apply it then on the South African urban system on long time. We finally discuss to proximity with existing methods and possible developments.
}{
La suite de cette section est organisée de la façon suivante : le cadre générique de la méthode proposée est décrit. Nous l'appliquons ensuite à un jeu de données synthétiques afin de la valider partiellement et de tester ses potentialités, ce qui permet de l'appliquer ensuite au système urbain sud-Africain sur le temps long. Nous discutons finalement la proximité avec d'autres méthodes existantes et des développements possibles.
}

\subsubsection{Method}{Méthode}

\bpar{
We formalize here the method in a generic way, based in a test similar to Granger causality\footnote{We recall that Granger causality corresponds to the existence of a significant relation between components of a vector lagged in time and itself.}, to try to identify causal relations in spatial systems. Let $X_j(\vec{x},t)$ spatio-temporal unidimensional random processes, which realizations occur in space and time. We give a set of fundamental spatial units  $(u_i)$ that can be for example raster cells or any paving of the geographical space\footnote{But which choice must be done with caution, in relation with the thematic studied, since our method does not escape a priori from the \emph{MAUP} problem~\cite{paez2005spatial}.}. We assume the existence of functions $\Phi_{i,j}$ allowing to make the correspondance between the realization of each components and spatial units, possibly through a first spatial aggregation. A realization of a system is given by a set of trajectories for each process $x_{i,j,t}$, and we write a set of realizations $x^{(k)}_{i,j,t}$ (accessible by stochastic repetitions in the case of a model of simulation for example, or by assumption of comparability of territorial sub-systems in real cases). We assume to have a correlation estimator $\hat{\rho}$ applying in time, space and repetitions, i.e. that covariance is estimated\footnote{The estimator $\hat{\mathbb{E}}$ spans here on time $t$, on spatial units $i$ and on repetitions $k$.} by
}{
Nous formalisons ici de manière générique la méthode, basée sur un test similaire à la causalité de Granger\footnote{On rappelle que la causalité de Granger correspond à l'existence d'une relation significative entre les composantes retardées dans le temps d'un vecteur et celui-ci.}, pour tenter d'identifier des relations causales dans des systèmes spatiaux. Soit $X_j(\vec{x},t)$ des processus aléatoires spatiaux unidimensionnels, se réalisant dans le temps et l'espace. On se donne un ensemble d'unités spatiales fondamentales $(u_i)$ qui peuvent être par exemple les cellules d'une image raster ou un pavage quelconque de l'espace géographique\footnote{Mais dont le choix sera à faire soigneusement, en lien avec la thématique étudiée, notre méthode n'échappant a priori pas au problème du \emph{MAUP}~\cite{paez2005spatial}.}. On suppose l'existence de fonctions $\Phi_{i,j}$ permettant de faire correspondre les réalisations de chaque composante aux unités spatiales, possiblement par une première agrégation locale. Une réalisation d'un système est donnée par un ensemble de trajectoires pour chaque processus $x_{i,j,t}$, et on pourra noter un ensemble de réalisations $x^{(k)}_{i,j,t}$ (accessibles dans le cas d'un modèle de simulation par exemple, ou par hypothèse de comparabilité de sous-systèmes territoriaux dans des cas réels). On suppose disposer d'un estimateur de corrélation $\hat{\rho}$ s'exerçant dans le temps, l'espace et les répétitions, c'est-à-dire que la covariance est estimée\footnote{L'estimateur $\hat{\mathbb{E}}$ portant ici sur le temps $t$, les unités spatiales $i$ et les répétitions $k$.} par
}

\[
\hat{\Cov}\left[X,Y\right] = \hat{\mathbb{E}}_{i,t,k}\left[XY\right] - \hat{\mathbb{E}}_{i,t,k}\left[X\right]\hat{\mathbb{E}}_{i,t,k}\left[Y\right]
\]

\bpar{
It is important to note here the hypothesis of spatial and temporal stationarity, that can however easily be relaxed in the case of local stationarity: we will have in such a case to estimate on sliding temporal or spatial windows. 
}{
Il est important de noter ici l'hypothèse de stationnarité spatiale et temporelle, qui peut toutefois aisément être relaxée dans le cas d'une stationnarité locale : il s'agira dans un tel cas d'estimer sur des fenêtres spatiales ou temporelles glissantes.
}

\bpar{
Furthermore, spatial auto-correlation is not explicitly included, but is taken into account either by the initial spatial aggregation is the characteristic scale of units is larger than the one of neighborhood effects, or by an adequate spatial estimator (weighted spatial statistics of type \emph{GWR}\footnote{We recall that Geographically Weighted Regression consists in estimating statistical models at different points in space, by weighting informations by distance, i.e. in other terms to take into account spatial non-stationarity.}~\cite{brunsdon1998geographically} for example). It allows us to define the lagged correlation between components $X_{j_1}$ and $X_{j_2}$ for a delay $\tau$ by
}{
D'autre part, l'auto-corrélation spatiale n'est pas explicitement incluse, mais est prise en compte soit par l'agrégation initiale si l'échelle caractéristique des unités est plus grande que celle des effets de voisinage, soit par un estimateur spatial adéquat (statistiques spatiales pondérées de type \emph{GWR}\footnote{On rappelle que la Regression Géographique Pondérée consiste à estimer des modèles statistiques à différents endroits de l'espace, en pondérant les informations par la distance, c'est-à-dire en d'autre termes de prendre en compte la non-stationnarité spatiale.}~\cite{brunsdon1998geographically} par exemple). Cela nous permet de définir la correlation retardée entre les composantes $X_{j_1}$ et $X_{j_2}$ pour le délai $\tau$ par
}


\begin{equation}
\rho_{\tau}\left[X_{j_1},X_{j_2}\right] = \hat{\rho}\left[x^{(k)}_{i,j_1,t - \tau},x^{(k)}_{i,j_2,t}\right]
\end{equation}

\bpar{
The lagged correlation is not directly symmetric, but we have evidently $\rho_{\tau}\left[X_{j_1},X_{j_2}\right] = \rho_{-\tau}\left[X_{j_2},X_{j_1}\right]$. This measure is then applied in a simple way: if $\textrm{argmax}_{\tau} \rho_{\tau}\left[X_{j_1},X_{j_2}\right]$ or $\textrm{argmin}_{\tau} \rho_{\tau}\left[X_{j_1},X_{j_2}\right]$ are ``clearly defined'' (both could be simultaneously), their sign will give the direction of causality between components $j_1$ and $j_2$ and their absolute value the propagation lag.
}{
La corrélation retardée n'est pas directement symétrique, mais on a de manière évidente $\rho_{\tau}\left[X_{j_1},X_{j_2}\right] = \rho_{-\tau}\left[X_{j_2},X_{j_1}\right]$. On applique alors cette mesure de manière simple : si $\textrm{argmax}_{\tau} \rho_{\tau}\left[X_{j_1},X_{j_2}\right]$ ou $\textrm{argmin}_{\tau} \rho_{\tau}\left[X_{j_1},X_{j_2}\right]$ sont ``clairement définis'' (les deux pouvant l'être simultanément), leur signe donnera alors le sens de la causalité entre les composantes $j_1$ et $j_2$ et leur valeur absolue le retard de propagation.
}

\bpar{
For example, $X_{j_1}$ can be a property linked to the network such as closeness centrality, and $X_{j_2}$ a property linked to territories, such as population density. This measure will allow then to define a direction of causality (possibly reciprocal) between these properties. The lag $\tau$ will typically be a number of years, in association with the spatial scale of estimation units which can vary from the scale of the district to urban areas, as we will see in the different cases of application in the following.
}{
Par exemple, $X_{j_1}$ pourra être une propriété liée au réseau comme la centralité de proximité, et $X_{j_2}$ une propriété liée aux territoires, comme la densité de population. Cette mesure permettra alors de déterminer un sens de causalité (éventuellement réciproque) entre ces propriétés. Le retard $\tau$ sera typiquement un certain nombre d'années, en association avec l'échelle spatiale des unités d'estimation qui pourra varier de l'échelle du quartier à celle des aires urbaines, comme nous le verrons dans les différents cas d'application par la suite.
}

\bpar{
The criteria for significance will depend on the case of application and of the estimator used. They can take into account different aspects of the robustness of the estimation. For example, a filtering on the significance of the statistical test (Fisher test in the case of a Pearson estimator) allows to ensure to isolate relations that are statistically significant. We could also want to ensure the significance of a minimal correlation, ans study the position of boundaries of a confidence interval of a given level. Finally, we can also fix an exogenous threshold $\theta$ on $\left|\rho_{\tau}\right|$ to ensure a certain level of correlation.
}{
Les critères de significativité dépendront du cas d'application et de l'estimateur utilisé. Ils peuvent prendre en compte différents aspects de la robustesse de l'estimation. Par exemple, un filtrage sur la significativité du test statistique (test de Fisher dans le cas d'un estimateur de Pearson) permet de s'assurer d'isoler des relations qui sont statistiquement significatives. On peut aussi vouloir s'assurer de la significativité d'une corrélation minimale, et regarder la position des bornes d'un intervalle de confiance à un niveau donné. Enfin, on peut aussi fixer un seuil exogène $\theta$ sur $\left|\rho_{\tau}\right|$ pour forcer un certain degré de corrélation.
}

\bpar{
To summarize the structure of the method and the articulation of the operations processed, we propose the scheme in Frame~\ref{frame:causalityregimes:regimes} below. The method we propose is not new in the elements used, but the combination of the different stages is original.
}{
Pour résumer la structure de la méthode et l'enchainement des traitements effectués, nous proposons le schéma dans l'Encadré~\ref{frame:causalityregimes:regimes} ci-dessous. La méthode que nous proposons n'est pas nouvelle dans les éléments utilisés, mais l'enchaînement des différentes étapes est originale.
}

\begin{figure}[h!]
\begin{mdframed}
\includegraphics[width=\textwidth]{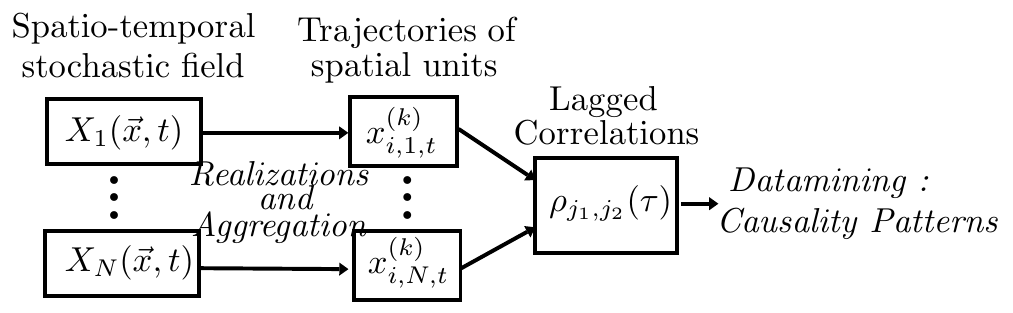}
\medskip
\framecaption{\textbf{Structure of the methodology.} We start from a stochastic field in time and space $X_j(\vec{x},t)$. A certain number of its realizations are captured, and measured on spatial units. We obtain trajectories $k$ for each unit $i$ in time $t$, denoted by $x_{i,j,t}^{(k)}$, on which the lagged correlation matrix $\rho_{j_1,j_2}(\tau)$ is estimated. The datamining on these allows to establish different \emph{regimes of causality}.\label{frame:causalityregimes:regimes}}{\textbf{Structure de la méthodologie.} Nous partons d'un champ stochastique dans le temps et l'espace $X_j(\vec{x},t)$. Un certain nombre de ses réalisations sont capturées, et mesurées sur des unités spatiales. Nous obtenons des trajectoires $k$ par unité $i$ dans le temps $t$, notées $x_{i,j,t}^{(k)}$, sur lesquelles la matrice des corrélations retardées $\rho_{j_1,j_2}(\tau)$ est estimée. Le datamining sur celles-ci permet d'établir différents \emph{régimes de causalité}. \label{frame:causalityregimes:regimes}}
\end{mdframed}
\end{figure}

\bpar{
Before diving into the empirical exploration of the method, we can give of it an intuitive vision to better understand its link with co-evolution. The Frame~\ref{frame:causalityregimes:twovars} synthesizes stylized situations that can occur in the case of two variables. In a caricatural way, with two variables $X,Y$, the profile of $\rho_{\tau}\left[X,Y\right]$ is summarized by the following characteristics: existence or not of an extremum for $\tau < 0$ and existence or not of an extremum for $\tau > 0$, i.e. possibilities of causality from $X$ to $Y$ and/or of causality from $Y$ to $X$. We illustrate four examples of profiles and represent the interactions between variables in a graphical way, in time and in a synthetic manner.
}{
Avant de nous plonger dans l'exploration empirique de la méthode, donnons-en une vision intuitive pour mieux comprendre son lien avec la co-évolution. L'Encadré~\ref{frame:causalityregimes:twovars} synthétise des situations stylisées pouvant se produire dans le cas de deux variables. De manière caricaturale, avec deux variables $X,Y$, le profil de $\rho_{\tau}\left[X,Y\right]$ est traduit selon les caractéristiques suivantes : existence d'un extremum ou non pour $\tau < 0$ et existence d'un extremum ou non pour $\tau > 0$, c'est-à-dire possibilités de causalité de $X$ vers $Y$ et/ou de causalité de $Y$ vers $X$. Nous illustrons quatre exemples de profils et représentons les interactions entre variables sous forme graphique, dans le temps et de manière synthétique. 
}

\begin{figure}[h!]
\begin{mdframed}
\bigskip
\includegraphics[width=\textwidth]{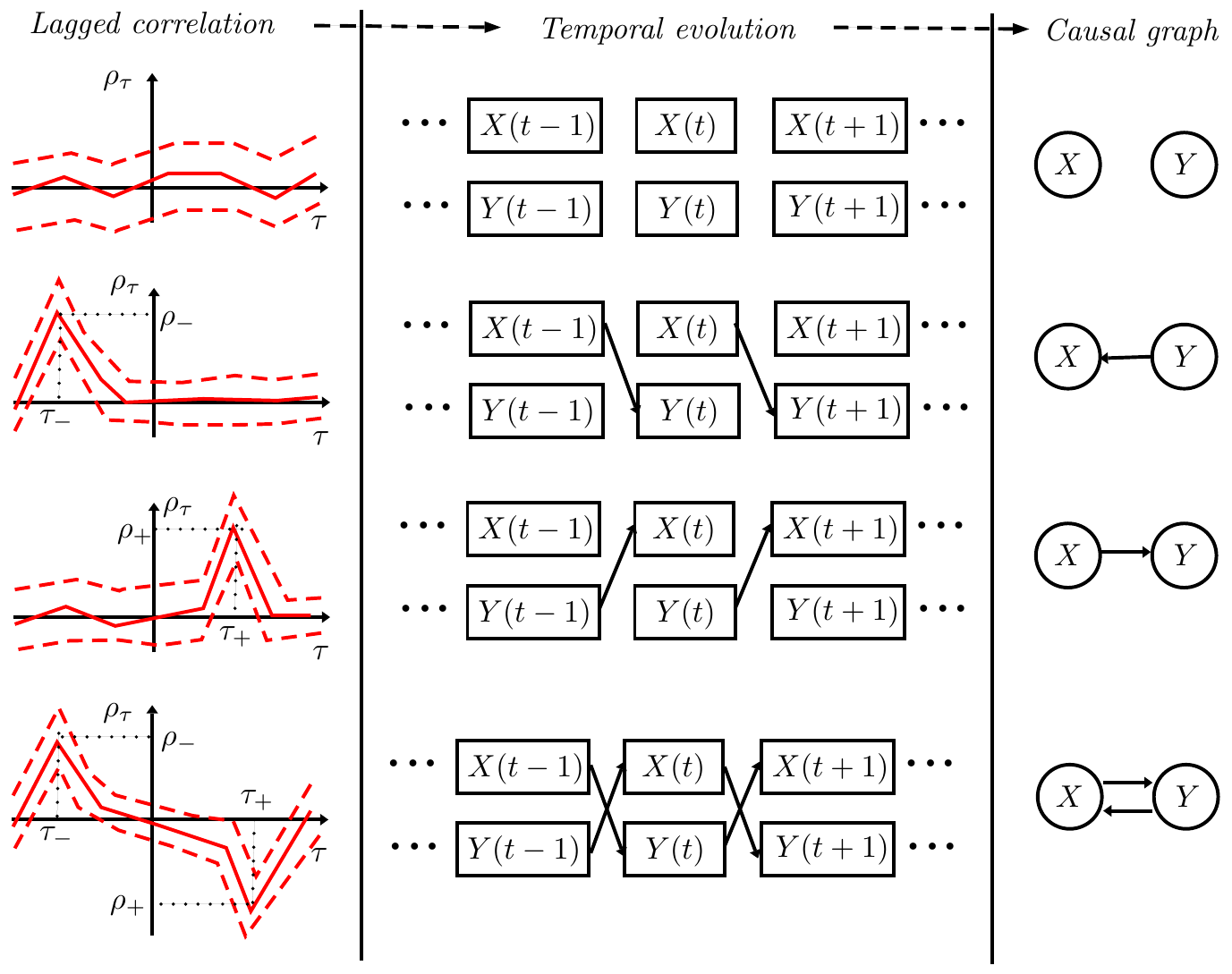}
\bigskip
\framecaption{\textbf{Illustration of possible situations in the case of two variables.} To simplify, we only differentiate situations through the existence or not of an extremum pour positive and negative values of the lag $\tau$ (and do not take into account the sign of the corresponding correlation). Dashed lines illustrate a significance threshold, for example a confidence interval on the correlation estimated. We have thus four situations: no significant extremum, existence of $\tau_-$, existence of $\tau_+$, existence of $\tau_-$ and of $\tau_+$. In the first case, there is no diachronic link between the variables (but possibly simultaneous correlations, specified by the double vertical arrows). In the two following cases, one variable ``causes'' the other (we will sometimes use this semantic shortcut to comment the results of analyses). Finally in the last case, we have circular causalities: such patterns will correspond to what we conceptually described as co-evolution.\label{frame:causalityregimes:twovars}}{\textbf{Illustration de situations possibles dans le cas de deux variables.} Pour simplifier, nous ne différencions les situations que par l'existence ou non d'un extremum pour les valeurs positives et négatives du délai $\tau$ (et ne prenons pas en compte le signe de la corrélation correspondante). Les lignes pointillées illustrent un seuil de significativité, par exemple un intervalle de confiance sur la corrélation estimée. On a donc quatre situations : aucun extremum significatif, existence de $\tau_-$, existence de $\tau_+$ et existence de $\tau_-$ et de $\tau_+$. Dans le premier, il n'y a pas de lien diachronique entre les variables (mais possiblement des corrélations simultanées, spécifiées par les doubles flèches verticales). Dans les deux suivants, l'une ``cause'' l'autre variable (nous utiliserons parfois ce raccourci sémantique pour commenter les résultats des analyses). Enfin dans la dernière, on a causalités circulaires : de tels motifs s'apparenteront à ce que l'on désignait conceptuellement par co-évolution.\label{frame:causalityregimes:twovars}}
\end{mdframed}
\end{figure}


\subsubsection{Emergence and measuring co-evolution ?}{Emergence et mesure de la co-évolution ?}


\bpar{
Let also take a brief instant to clarify the epistemological and ontological status expected through the application of this method, and to what extent we can expect to use it as an indirect measure of co-evolution. The Granger causality is estimated both \emph{in time}, \emph{in space} and \emph{between repetitions}. In the case where an historical phenomenon is observed, we have an unique trajectory and the estimation is done in time and space uniquely, but in any case we go from characteristics at the microscopic scale to a macroscopic measure\footnote{We use here these terms to simplify, it corresponds indeed to a given scale at an above scale which depends on the total temporal and spatial extent.}. Thus, we can have circular microscopic interactions, but emergence of a direction of the causality at the macroscopic level, or the opposite. Rejoining the question of populations and individuals for the definition of co-evolution in biology (see~\ref{sec:epistemology}), for which mutual adaptations emerge at the level of species, we postulate that the characterization of causality patterns is a way to characterize co-evolutive dynamics for territorial systems, corresponding then to our intermediate definition of co-evolution at the population level.
}{
Prenons également un court instant pour clarifier le statut épistémologique et ontologique attendu par l'application de cette méthode, et dans quelle mesure on peut espérer l'utiliser comme mesure indirecte de la co-évolution. La causalité de Granger est estimée à la fois \emph{dans le temps}, \emph{dans l'espace} et \emph{entre les répétitions}. Dans le cas où l'on observe un phénomène historique, on a une unique trajectoire et l'estimation est faite dans le temps et l'espace uniquement, mais dans tous les cas on passe de caractéristiques à l'échelle microscopique à une mesure macroscopique\footnote{Nous utilisons ici ces termes pour simplifier, il s'agit en fait d'une échelle donnée à une échelle supérieure qui dépend de l'étendue temporelle et spatiale totale.}. Ainsi, on peut avoir des interactions microscopiques circulaires, mais émergence d'un sens de la causalité au niveau macroscopique, ou l'inverse. Rejoignant la question des populations et individus pour la définition de la co-évolution en biologie (voir~\ref{sec:epistemology}), pour laquelle les adaptations mutuelles émergent au niveau des espèces, nous postulons que la caractérisation des motifs de causalité est une manière de caractériser des dynamiques co-évolutives pour les systèmes territoriaux, correspondant alors à notre définition intermédiaire de la co-évolution au niveau d'une population.
}

\bpar{
Is it then possible to answer in an univocal way to the question ``\textit{Is there co-evolution in a particular case}''\footnote{To which we add: for these components, on this spatial and temporal extent and on these temporal and spatial scales.} ? It would be known if we could reinvent the wheel but which is self-propelled. We mean here, and we will see it in the multiple developments, that numerous problems that are intrinsic to the study of geographical systems (the question of scales, of the definition of the system, of variables taken into account, the issue of observing unique trajectories, of sparse and noisy data, the MAUP, etc.) will still be present, and that the question above which is naturally subject to these issues appears to be naive. But we will see that it will be indeed possible to isolate clear signals, and will exhibit cases in which there exists a causal direction and others in which there is causality at the macroscopic level.
}{
Est-il alors possible de répondre de manière équivoque à la question ``\textit{y a-t-il co-évolution dans un cas particulier}''\footnote{A laquelle nous ajoutons : pour ces composantes, sur cette portée spatiale et temporelle et sur ces échelles spatiale et temporelle.} ? Cela se saurait si nous pouvions réinventer l'eau chaude mais qui se chauffe elle-même. Nous voulons dire par là, et nous le verrons dans les multiples développements, que de nombreux problèmes fondamentaux intrinsèques à l'étude des systèmes géographiques (la question des échelles, de la définition du système, des variables prises en compte, le problème de l'observation de trajectoire uniques, de données bruitées et éparses, le problème du MAUP, etc.) seront bien toujours présents, et que la question ci-dessus qui y est naturellement soumise s'avère naïve. Mais nous verrons qu'il sera bien possible d'isoler des signaux clairs, et mettrons en évidence des cas où il existe un sens causal et d'autres où il y a circularité au niveau macroscopique.
}



\subsection{Synthetic data}{Données synthétiques}

\bpar{
In a first time, we explore and validate the method on synthetic data, i.e. generated by the intermediary of a model with a certain level of control.
}{
Nous explorons et validons la méthode dans un premier temps sur données synthétiques, c'est-à-dire générées par l'intermédiaire d'un modèle avec un certain niveau de contrôle.
}

\subsubsection{Auto-regressive time series}{Séries temporelles auto-régressives}

\bpar{
Let illustrate the patterns that can be expected, in particular the stylized ones given previously in Frame~\ref{frame:causalityregimes:twovars}, on synthetic data with a simple structure. The idea is to generate time series on which the lag and the level of correlation are controlled, and also on which theoretical results are known.
}{
Illustrons les motifs qui peuvent être attendus, notamment ceux stylisés donnés précédemment en Encadré~\ref{frame:causalityregimes:twovars}, sur des données synthétiques avec une structure simple. L'idée est de générer des séries temporelles sur lesquelles le retard et le niveau de corrélation sont contrôlés, ainsi que les résultats théoriques connus.
}

\bpar{
Let $\vec{X}(t)$ a stochastic process verifying the auto-regression equation $\vec{X}(t) = \sum_{\tau > 0} \mathbf{A}(\tau) \cdot \vec{X}(t - \tau ) + \vec{\epsilon}(t)$. In the case where $\mathbf{A}(\tau) = 0$ for $\tau \neq \tau_0$ and $\mathbf{A}(\tau_0) = \left( {\begin{array}{cc} 0 & a \\ a & 0 \\ \end{array}} \right)$ for $-1<a<1$, the computation of theoretical correlations is possible (see Appendix~\ref{app:sec:causalityregimes}), and we obtain, by writing $\mathbf{X} = (X,Y)$, for $\tau > 0$
}{
Soit $\vec{X}(t)$ un processus stochastique suivant l'équation d'auto-régression $\vec{X}(t) = \sum_{\tau > 0} \mathbf{A}(\tau) \cdot \vec{X}(t - \tau ) + \vec{\epsilon}(t)$. Dans le cas où $\mathbf{A}(\tau) = 0$ pour $\tau \neq \tau_0$ et $\mathbf{A}(\tau_0) = \left( {\begin{array}{cc} 0 & a \\ a & 0 \\ \end{array}} \right)$ pour $-1<a<1$, le calcul des corrélations théoriques est possible (voir Annexe~\ref{app:sec:causalityregimes}), et on obtient, en notant $\mathbf{X} = (X,Y)$, pour $\tau > 0$
}

\bpar{
\[
\rho\left[X(t),Y(t-\tau)\right] = \begin{cases}
	a^{2k+1} \textrm{if } \tau = (2k+1)\tau_0\textrm{ for all }k\in \mathbb{Z} \\
	0 \textrm{ otherwise} 
\end{cases}
\]
}{
\[
\rho\left[X(t),Y(t-\tau)\right] = \begin{cases}
	a^{2k+1} \textrm{si } \tau = (2k+1)\tau_0\textrm{ pour tout }k\in \mathbb{Z} \\
	0 \textrm{ sinon} 
\end{cases}
\]
}

\bpar{
The expression is the same for $\tau<0$ by exchanging $X$ and $Y$. Thus, we control the lagged correlation at a fixed lag and at lags which are multiple of it with an odd factor. By changing one of the coefficients in 0 or in its opposite, we obtain for the first three maximums the three stylized profiles given in Frame~\ref{frame:causalityregimes:twovars}.
}{
L'expression est la même pour $\tau<0$ en échangeant $X$ et $Y$. Ainsi, on contrôle la corrélation retardée au retard voulu et aux retards qui en sont multiples avec un facteur impair. En changeant l'un des coefficients en 0 ou en son opposé, on obtient pour les premiers maximums les trois profils stylisés donnés en Encadré~\ref{frame:causalityregimes:twovars}.
}

\begin{figure}
	\includegraphics[width=\linewidth]{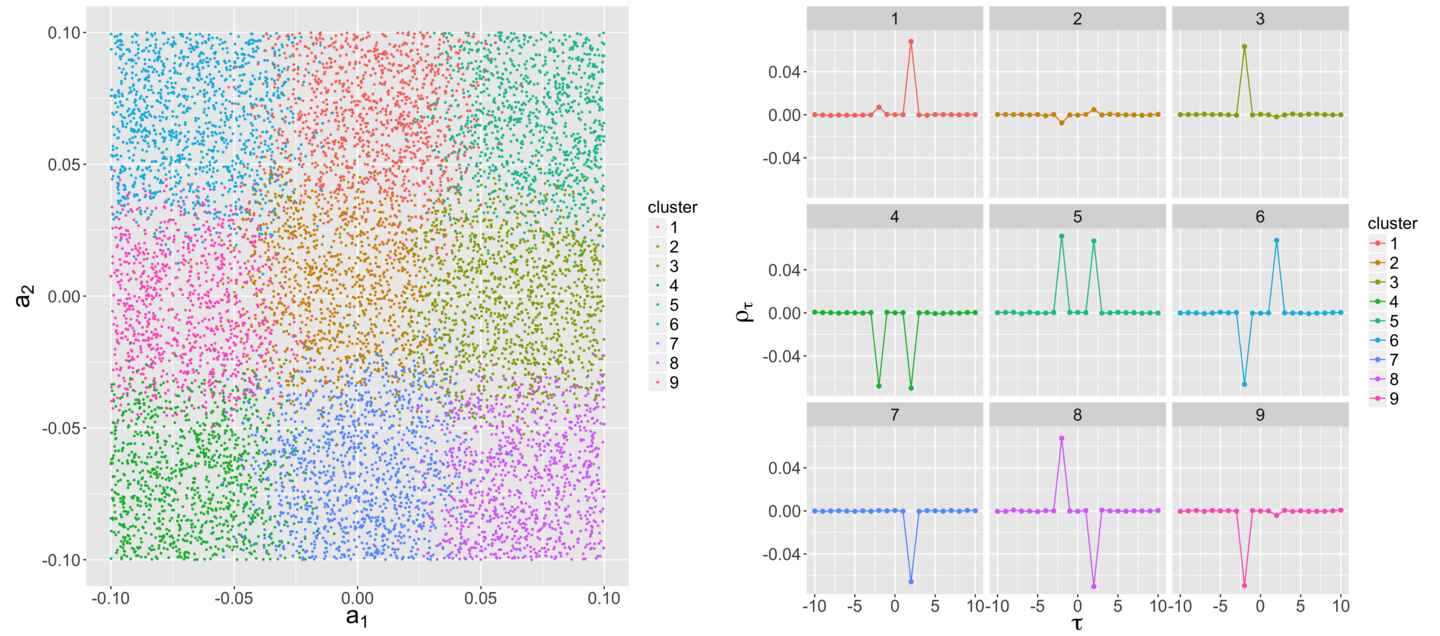}
	\caption[Auto-regressive time-series]{\textbf{Estimation of correlation regimes in the case of auto-regressive time series.} Results of the regimes classification method for simple AR processes. We simulate $b=10000$ time-series of length $t_f=10000$, with random coefficients $(a_1,a_2) \in [-0.1,0.1]$ and a lag $\tau_0 = 2$. (\textit{Left}) Values of coefficients $(a_1,a_2)$, the color giving the cluster obtained; (\textit{Right}) Trajectories of corresponding centroids. We obtain the expected stylized profiles, which correspond to the relative values of parameters: for example, cluster 1 is for a low $a_1$ and a high $a_2$, and indeed corresponds to a situation where $\rho_+$ exists i.e. a configuration $X\rightarrow Y$, and the sign of $\rho_+$ corresponds to $a_2 >0$.\label{fig:causalityregimes:arma}}
\end{figure}

\bpar{
Let use this example to numerically explore the possibility to classify lagged correlations profiles. We consider the same process for $\tau_0 = 2$ and $\mathbf{A}(\tau_0) = \left( {\begin{array}{cc} 0 & a_1 \\ a_2 & 0 \\ \end{array}} \right)$, with $-1<a_1,a_2<1$. We simulate with this model time series of length $t_f=10000$ by drawing $b=10000$ random values for parameters $(a_1,a_2)$. On each sample lagged correlations are estimated, and we proceed to a non-supervised classification\footnote{Using the \emph{k-means} algorithm with $k=9$ and $b_c = 1000$ repetitions.} on time-series $\left[\rho(\tau)\right]_{a_1,a_2}$. We show in Fig.~\ref{fig:causalityregimes:arma} typical profiles we obtain in correspondance with their position in the parameter space $(a_1,a_2)$> We exactly obtain the nine possible stylized profiles, in correspondance with the relative values of parameters as expected. Starting from very different profiles of lagged correlations, we are thus able to extract typical profiles of interaction between the variables. This conforts us with the idea to apply this method on more complex data in the following.
}{
Utilisons cet exemple pour explorer numériquement la possibilité de classifier les profils de corrélations retardées. Nous considérons le même processus pour $\tau_0 = 2$ et $\mathbf{A}(\tau_0) = \left( {\begin{array}{cc} 0 & a_1 \\ a_2 & 0 \\ \end{array}} \right)$, avec $-1<a_1,a_2<1$. Nous simulons avec ce modèle des séries temporelles de longueur $t_f=10000$ en tirant $b=10000$ valeurs aléatoires pour les paramètres $(a_1,a_2)$. Sur chaque série les corrélations retardées sont estimées, et nous procédons à une classification non-supervisée\footnote{Par algorithme des \emph{k-means} avec $k=9$ et $b_c = 1000$ répétitions.} sur les séries temporelles $\left[\rho(\tau)\right]_{a_1,a_2}$. Nous montrons en Fig.~\ref{fig:causalityregimes:arma} les profils typiques obtenus en correspondance avec leur position dans l'espace des paramètres $(a_1,a_2)$. Nous obtenons exactement les neuf profils stylisés possibles, en correspondance avec les valeurs relatives des paramètres comme attendu. À partir de profils très variés de corrélations retardées, nous sommes ainsi capable d'extraire des profils typiques d'interaction entre les variables. Cela nous renforce dans l'idée d'appliquer cette méthode sur des données plus complexes par la suite.
}


\subsubsection{Urban growth model}{Modèle de croissance urbaine}

\bpar{
This method must first be tested and partially validated, what we propose to do again on synthetic data. Echoing the example of relations between transportation networks and territories that introduced the research question before, we propose to generate stylized urban configurations in which network and density mutually interact, and for which causalities are not obvious \emph{a priori} knowing the parameters of the generative model.
}{
Cette méthode doit être testée et validée sur un système plus proche de nos préoccupations, ce que nous faisons à nouveau sur des données synthétiques.  En écho à l'exemple des relations entre réseaux de transport et territoires qui a permis d'introduire notre problématique précédemment, nous proposons de générer des configurations urbaines stylisées dans lesquelles réseau et densité s'influencent mutuellement, et pour lesquelles les causalités ne sont pas évidentes \emph{a priori} étant donné les paramètres du modèle génératif.
}

\bpar{
\cite{raimbault2014hybrid} describes and explores a simple model of urban morphogenesis\footnote{We do not explore here the concept of morphogenesis, which will be the object of chapter~\ref{ch:morphogenesis}, but use this model as a provider of synthetic data.} (the RBD model) which fulfils these constraints. This modèle is described in details for the configuration in which we use it in Frame~\ref{frame:causalityregimes:rbd}. Explicative variables of urban growth, processes of network extension and the coupling between urban density and the network are relatively simple. However, except for extreme cases (for example when distance to the center solely determines land value, the network will depend on density in a causal way; when only the distance to the network counts, the causality will be inverted), mixed regimes do not exhibit obvious causalities. It is for this reason an ideal case to test if the method is able to detect some. Synthetic data allow us to control the consistence in the cases where the relation is expected.
}{
\cite{raimbault2014hybrid} décrivent et explorent un modèle simple de morphogenèse urbaine\footnote{Nous n'explorons pas ici le concept de morphogenèse, qui fera l'objet du chapitre~\ref{ch:morphogenesis}, mais utilisons ce modèle comme producteur de données synthétiques.} (modèle RBD) répondant à ces contraintes. Ce modèle est décrit en détails pour la configuration dans laquelle nous l'utilisons en Encadré~\ref{frame:causalityregimes:rbd}. Les variables explicatives de la croissance urbaine, les processus d'extension du réseau et le couplage entre densité urbaine et réseau ne sont pas trop complexes. Cependant, hormis dans des cas particuliers (par exemple lorsque la distance au centre détermine la valeur foncière uniquement, le réseau dépendra de manière causale de la densité, ou lorsque la distance au réseau seule compte, la causalité sera inversée), les régimes mixtes ne présentent pas de causalités évidentes : c'est donc un cas adapté pour tester si la méthode est capable d'en détecter. Les données synthétiques nous permettent de contrôler la cohérence dans les cas où la relation est attendue.
}


\begin{figure}[h!]
\begin{mdframed}
\bpar{
The RBD model assumes a grid of size $N$, which cells have a binary state (occupied or not). In the version we use, there exists a unique urban center (particular node of the network) and the transportation network is initially empty. Each cell $i$ is characterized by variables $x_d (i)$ (population density within a fixed radius $r=5$), $x_r (i)$ (euclidian distance to the closest road) and $x_c (i)$ (distance to the center via the network). These variables allow to compute a value of potential for each cell $U_i = \sum w_k \tilde{x}_k (i)$, where the $w_k$ are model parameters allowing to influence the urban forms produced and $\tilde{x}_k(i)$ the variables normalized on all cells by $\tilde{x}_k(i) = \frac{\max_i x_k (i) - x_k (i)}{\max_i x_k (i) - \min_i x_k (i)}$.
}{
Le modèle RBD suppose une grille de côté $N$, dont les cellules ont un état binaire (occupée ou non). Dans la version utilisée, il existe un unique centre urbain (noeud particulier du réseau) et le réseau de transport est initialement nul. Chaque cellule $i$ est caractérisée par les variables $x_d (i)$ (densité de population dans un rayon fixé $r=5$), $x_r (i)$ (distance euclidienne à la route la plus proche) et $x_c (i)$ (distance au centre via le réseau). Ces variables permettent de calculer une valeur de potentiel pour chaque cellule $U_i = \sum w_k \tilde{x}_k (i)$, où les $w_k$ sont des paramètres du modèle permettant d'influencer les formes urbaines produites et $\tilde{x}_k(i)$ les variables normalisées sur l'ensemble des cellules par $\tilde{x}_k(i) = \frac{\max_i x_k (i) - x_k (i)}{\max_i x_k (i) - \min_i x_k (i)}$.
}

\bpar{
The potential can be interpreted as a utility aggregating preferences of agents which have to localize. A repulsion to density will for example yield very dispersed urban forms.
}{
Le potentiel peut être interprété comme une utilité agrégeant les préférences des agents devant se localiser. Une répulsion à la densité donnera par exemple des formes urbaines très dispersées.
}

\bpar{
The model evolves sequentially by progressively populating the grid. At each time step:
\begin{itemize}
	\item the $N_G$ cells with the largest $U_i$ value are simulatneously occupied;
	\item if a newly populated cell is at a distance to the network which is larger than a threshold $\theta_d$ (that we will fix here at $\theta_d = 5$), it is connected to the network by a new road taking the shortest path.
\end{itemize}
}{
Le modèle évolue séquentiellement en peuplant progressivement la grille. À chaque pas de temps :
\begin{itemize}
	\item les $N_G$ cellules avec plus grande valeur $U_i$ sont occupées de manière simultanée ;
	\item si une cellule nouvellement peuplée est à une distance au réseau supérieure à un seuil $\theta_d$ (que nous fixerons ici à $\theta_d = 5$), celle-ci est connectée au réseau par une nouvelle route prenant le chemin le plus court.
\end{itemize}
}


\bpar{
The growth stops at a fixed final time $t_f$.
}{
La croissance s'arrête à un temps final fixé $t_f$.
}

\medskip

\includegraphics[width=\linewidth]{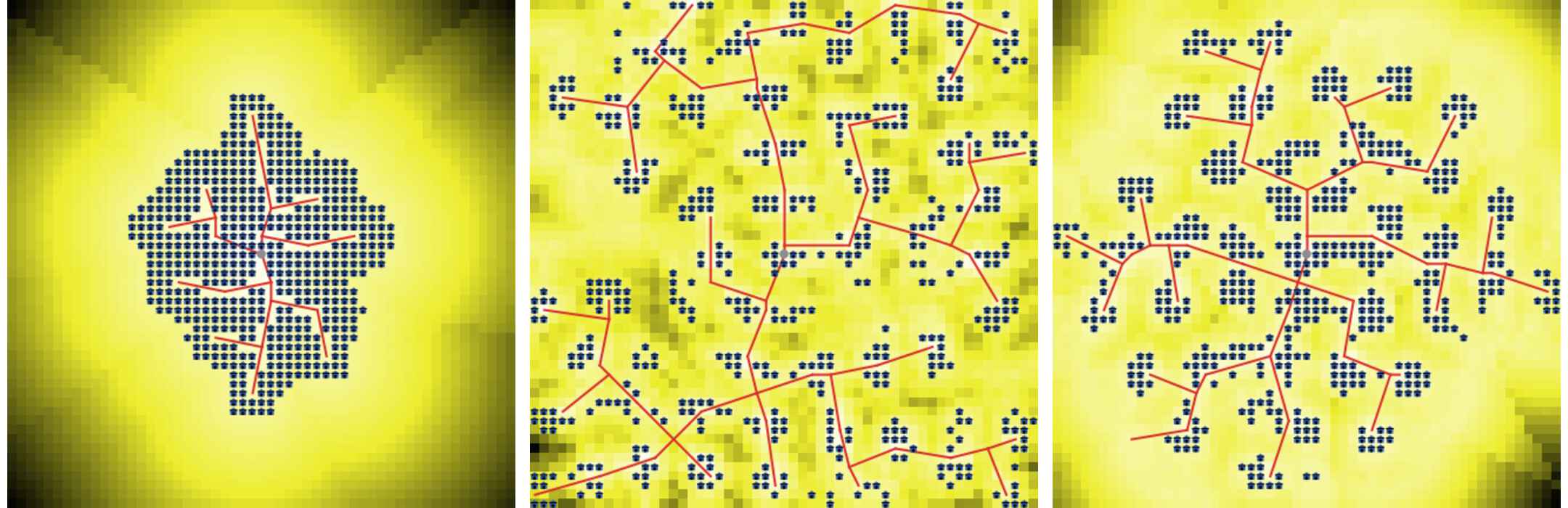}
\bpar{
\textit{Examples of various final configurations, obtained with weight parameters $(w_{d},w_{c},w_{r})$ respectively being equal to $(0,1,1)$,$(1,0,1)$, and $(1,1,1)$.}
}{
\textit{Exemples de configurations finales variées, obtenues avec les paramètres de poids $(w_{d},w_{c},w_{r})$ valant respectivement $(0,1,1)$,$(1,0,1)$, et $(1,1,1)$.}
}

\medskip

\framecaption{\textbf{Description of the RBD model.}\label{frame:causalityregimes:rbd}}{\textbf{Description du modèle RBD.}\label{frame:causalityregimes:rbd}}
\end{mdframed}
\end{figure}

\bpar{
We use an adapted implementation\footnote{The model is available on the open repository of the project at \\\texttt{https://github.com/JusteRaimbault/CityNetwork/tree/master/Models/Simple/ModelCA}.} of the original model, allowing to capture the values of studied variables for each cell of the cellular automaton and for each time step, and to calculate the lagged correlations in the sense described before, between variables of the model. We explore a grid of the parameter space of the RBD model, making the weight parameters for density $w_d$, distance to center $w_c$ and distance to network $w_r$ vary (see Frame~\ref{frame:causalityregimes:rbd} of model description), in $\left[0;1\right]$ with a step of $0.1$. Other parameters are fixed to their default values given by~\cite{raimbault2014hybrid}. For each parameter value, we proceed to $N=100$ repetitions, what is enough for a good convergence of indicators. Explorations are done with the OpenMole software~\cite{reuillon2013openmole}, the large number of simulations (1,330,000) implying the use of a computation grid\footnote{Simulation results are available at \url{http://dx.doi.org/10.7910/DVN/KGHZZB}.}.
}{
Nous utilisons une implémentation adaptée\footnote{Le modèle est disponible sur le dépôt ouvert du projet à \url{https://github.com/JusteRaimbault/CityNetwork/tree/master/Models/Simple/ModelCA}.} du modèle initial, permettant de capturer les valeurs des variables étudiées pour chaque cellule et à chaque pas de temps et de calculer les correlations retardées entre variables au sein du modèle. Nous explorons une grille de l'espace des paramètres du modèle RBD, faisant varier les paramètres de poids de la densité $w_d$, de la distance au centre $w_c$ et de la distance au réseau $w_r$ (voir Encadré~\ref{frame:causalityregimes:rbd} de description du modèle), dans $\left[0;1\right]$ avec un pas de $0.1$. Les autres paramètres sont fixés à leur valeurs par défaut données par \cite{raimbault2014hybrid}. Pour chaque valeur des paramètres, nous procédons à $N=100$ répétitions ce qui est suffisant pour une bonne convergence des indicateurs. Les explorations sont effectuées via le logiciel OpenMole~\cite{reuillon2013openmole}, le grand nombre de simulations (1330000) nécessitant l'utilisation d'une grille de calcul\footnote{Les résultats de simulation sont disponibles à \url{http://dx.doi.org/10.7910/DVN/KGHZZB}.}.
}

\bpar{
We compute for all cells the lagged correlations with the unbiased Pearson estimator between the variations of the following variables\footnote{Computing the correlations directly on the variables makes no sense since their value has no absolute meaning.}: local density, distance to center and distance to network. These are the explicative variables for model dynamics, and this the ones on which we can identify dynamical relations between local territorial characteristics.
}{
Nous calculons sur l'ensemble des cellules les corrélations retardées par estimateur de Pearson non biaisé entre les variations des variables suivantes\footnote{Calculer les corrélations sur les variables directement n'a pas de sens puisque leur valeur n'en a pas en absolu.} : densité locale, distance au centre et distance au réseau. Il s'agit des variables explicatives pour la dynamique du modèle, et donc celles sur lesquelles on peut identifier des relations dynamiques entre caractéristiques territoriales locales.
}

\begin{figure}
\vspace{-0.5cm}
\includegraphics[width=\linewidth,height=0.9\textheight]{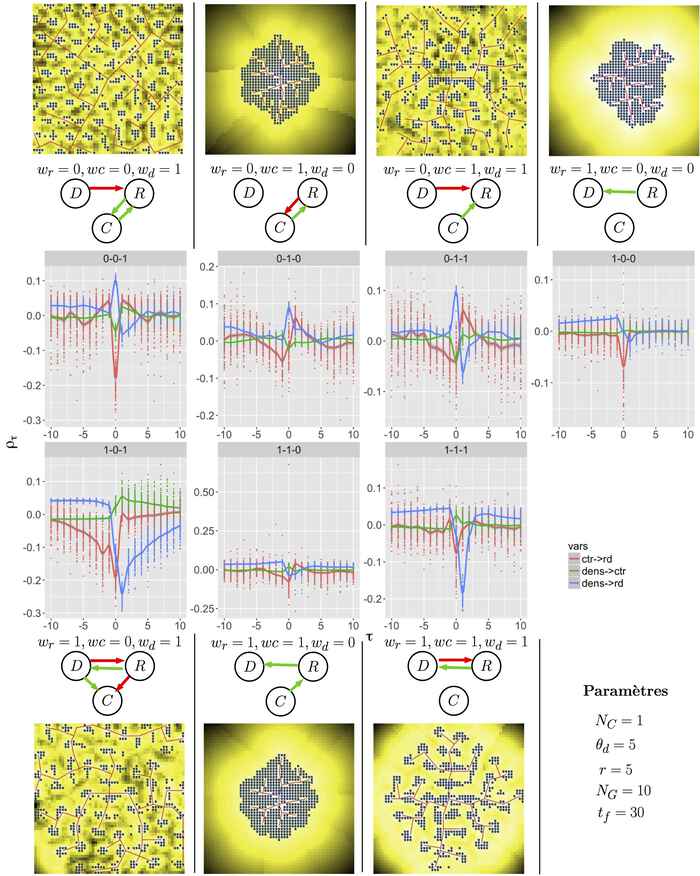}
\vspace{-0.8cm}
\caption[Correlation in the RBD model]{\textbf{Correlations in the RBD model.} Lagged correlations, for each combination of extreme values for all parameters $w_r,w_c,w_d$. Each graph gives the correlation $\rho_{\tau}$ as a function of the lag $\tau$. The different colors correspond to each couple of variables: distance to the center (\texttt{ctr}), density (\texttt{dens}) and distance to the network (\texttt{rd}). The dots correspond to individual correlations for each repetition of the model (estimators on $i$ and $t$), whereas the curves give the full estimator on all repetitions also. For each graph, we give in direct correspondance a final configuration, and the interpretation in terms of causality patterns as a graph between variables. The color of directed links gives the sign of the relation (red for a negative correlation, green for a positive correlation).\label{fig:causalityregimes:exrdb}}
\end{figure}

\bpar{
The figure~\ref{fig:causalityregimes:exrdb} shows the behavior of $\rho_{\tau}$ for each couple of variable (undirected, $\tau$ taking negative and positive values), for the combination of extreme values of parameters. We also give the interpretation in terms of a graph of relations between variables and an illustration of urban configuration generated for corresponding parameter values.
}{
La Fig.~\ref{fig:causalityregimes:exrdb} montre le comportement de $\rho_{\tau}$ pour chaque couple de variables (non dirigé, $\tau$ prenant des valeurs négatives et positives), pour les combinaisons des valeurs extrêmes des paramètres. Nous donnons également l'interprétation sous forme de graphe de relations entre variables et une illustration de configuration urbaine générée pour les valeurs de paramètres correspondantes.
}

\bpar{
We witness that a certain number of regimes emerge, and can draw the following synthetic interpretations:
\begin{itemize}
	\item The negative link $D\rightarrow R$ results of the simple mechanism of network extension: an increase in density leads to a diminution of network distance through the construction of a new road. Some configurations inhibit this link, through the complex interaction with other variables (for example $(0,1,0)$).
	\item Elaborated graphs of relation can emerge: $(1,0,1)$ leads for example to a circular relation between distance to the network and density, and a causality of these two variables on the distance to the center.
	\item Unexpected a priori behaviors emerge, such as for example the circular relation between distance to the network and distance to the center for $(0,0,1)$ where only density plays a role; on the contrary the model leads to the emergence of a relation which corresponds to the microscopic mechanism only when $w_r=1$.
\end{itemize}
}{
Nous voyons un certain nombre de régimes émerger, et pouvons tirer les interprétations synthétiques suivantes :
\begin{itemize}
	\item Le lien négatif $D\rightarrow R$ résulte du mécanisme simple d'extension du réseau : un accroissement de la densité conduit à une diminution de la distance au réseau par la construction d'une nouvelle route. Certaines configurations inhibent ce lien, par l'interaction complexe avec les autres variables (par exemple $(0,1,0)$).
	\item Des graphes de relations élaborés peuvent émerger : $(1,0,1)$ conduit par exemple à une relation circulaire entre distance au réseau et densité, et une causalité de ces deux variables sur la distance au centre.
	\item Des comportements non attendus a priori émergent, comme par exemple la relation circulaire entre distance au réseau et distance au centre pour $(0,0,1)$ où seule la densité joue ; au contraire le modèle fait émerger une relation qui correspond au mécanisme microscopique quand $w_r=1$ seulement.
\end{itemize}
}

\bpar{
The relevance of the method is clearer here, since it allows to unveil ``macroscopic'' causality patterns (i.e. that can effectively be measured at a statistical level), starting from ``microscopic'' patterns (for example the road connection rule), and in a non-linear manner. Links that can intuitively be expected such as $D\rightarrow R$ are in some cases inhibited. This confirms the relevance of the distinction between the first two levels of co-evolution, the ``processual'' co-evolution (at the level of entities or processes) and the statistical co-evolution at a population level.
}{
L'intérêt de la méthode se précise ici, puisqu'elle permet de dégager des motifs de causalité ``macroscopiques'' (c'est-à-dire effectivement mesurables à un niveau statistique), à partir de motifs ``microscopiques'' (par exemple la règle de connection de la route), et de manière non-linéaire. Des liens qu'on pourrait attendre intuitivement comme $D\rightarrow R$ sont dans certains cas inhibés. Cela confirme la pertinence de la distinction entre les deux premiers niveaux de co-évolution, la co-évolution ``processuelle'' (au niveau des entités ou des processus) et la co-évolution statistique au niveau d'une population.
}

\subsubsection{Causality regimes}{Régimes de causalité}

\begin{figure}
\includegraphics[width=\linewidth]{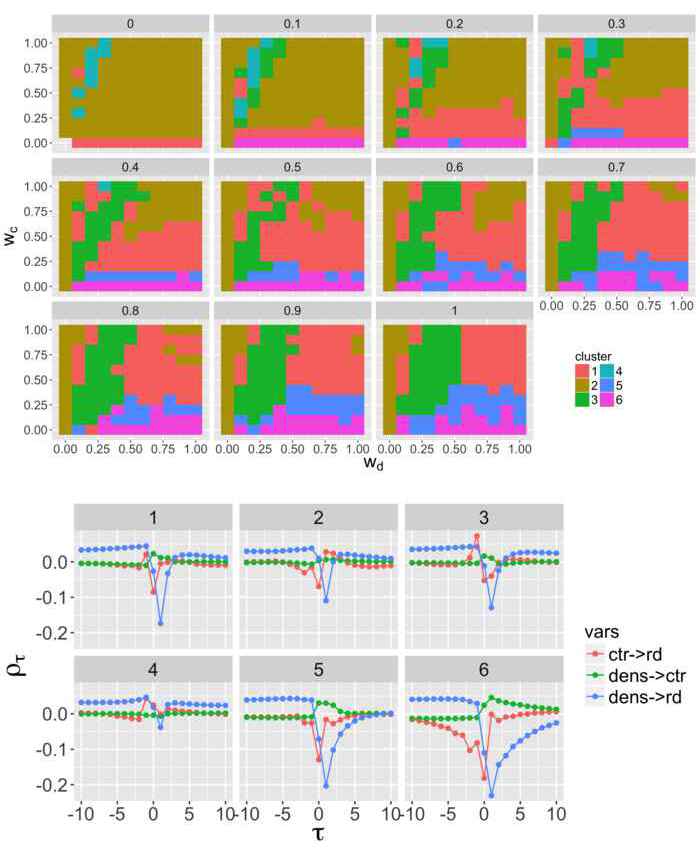}
\caption[Identification of interaction regimes]{\textbf{Causality regimes identified by a non-supervised classification.} \textbf{(Top)} Phase diagram of regimes (clusters) in the space $(w_{d},w_{c},w_{r})$, $w_r$ varying between the different sub-diagrams of $(w_{d},w_{c}$. \textbf{(Bottom)} Corresponding profiles of centroids, in terms of lagged correlations profiles $\rho_{\tau}$, for all variables couples (color) and for each cluster.\label{fig:causalityregimes:clustering}}
\end{figure}

\bpar{
We now demonstrate that it is possible to establish an endogenous typology of lagged correlations behaviors. To study these behaviors in a systematic way, we propose to identify regimes endogenously, by using non-supervised classification. We apply as previously a \emph{k-means} clustering, robust to stochasticity (5000 repetitions), with the following features: for each couple of variables, $\textrm{argmax}_{\tau} \rho_{\tau}$ and $\textrm{argmin}_{\tau} \rho_{\tau}$ if the corresponding value is such that $\frac{\rho_{\tau}-\bar{\rho}_{\tau}}{\left|\bar{\rho}_{\tau}\right|} > \theta$ with $\theta$ threshold parameter, 0 otherwise. The inclusion of supplementary features of values of $\rho_{\tau}$ does not significantly changes the results, these are therefore not taken into account to reduce the dimension. The choice of the number of clusters $k$ is generally a difficult problem in this kind of approach~\cite{hamerly2003learning}. In our case the system exhibits a structure which removes any ambiguity: the curves of inter-cluster variance proportion and its derivative (see  Fig.~\ref{fig:app:causalityregimes:clustering} in Appendix~\ref{app:sec:causalityregimes}), as a function of $k$ for different values of $\theta$, show a transition for $\theta = 2$, what gives for the corresponding curve a break around $k=5$. A visual screening of clusters in a principal plan confirms the good quality of the classification for these values. A class corresponds then to a \emph{causality regime}, for which we can represent the phase diagram as a function of model parameters, and also cluster centers profiles (computed as the barycenter in the full initial space) in figure~\ref{fig:causalityregimes:clustering}.
}{
Nous démontrons à présent qu'il est possible d'établir une typologie endogène des comportements des corrélations retardées. Afin d'étudier ces comportements de manière systématique, nous proposons d'identifier des régimes de manière endogène, en procédant à un apprentissage non-supervisé. Nous appliquons comme précédemment une classification des \emph{k-means}, robuste à la stochasticité (5000 répétitions), avec les points caractéristiques (\emph{features}) suivants : pour chaque couple de variable, $\textrm{argmax}_{\tau} \rho_{\tau}$ et $\textrm{argmin}_{\tau} \rho_{\tau}$ si la valeur correspondante est telle que $\frac{\rho_{\tau}-\bar{\rho}_{\tau}}{\left|\bar{\rho}_{\tau}\right|} > \theta$ avec $\theta$ paramètre de seuil, 0 sinon. L'inclusion des \emph{features} (variables caractéristiques) supplémentaires des valeurs de $\rho_{\tau}$ n'influence pas significativement les résultats, et celles-ci n'ont pas été prises en compte pour réduire la dimension. Le choix du nombre de clusters $k$ est en général épineux dans ce genre de problème~\cite{hamerly2003learning}, mais dans notre cas le système possède une structure qui lève l'ambiguïté : les courbes de la proportion de variance inter-cluster et de sa dérivée (voir  Fig.~\ref{fig:app:causalityregimes:clustering} en Annexe~\ref{app:sec:causalityregimes}), en fonction de $k$ pour différentes valeurs de $\theta$, présentent une transition pour $\theta = 2$, ce qui donne pour cette courbe une rupture à $k=5$. Un examen visuel des clusters dans un plan principal confirme la bonne qualité de la classification pour ces valeurs. Une classe correspond alors à un \emph{régime de causalité}, dont nous pouvons représenter le diagramme de phase en fonction des paramètres du modèle, ainsi que les trajectoires des centres des clusters (calculées comme barycentre dans l'espace complet initial) en Fig.~\ref{fig:causalityregimes:clustering}.
}

\subsubsection{Interpretation}{Interprétation}

\bpar{
We finally propose to interpret the regimes obtained, represented in Fig.~\ref{fig:causalityregimes:clustering}. The behavior obtained is particularly interesting: regions in the phase diagram depending on parameters corresponding to the different regimes are clearly delimited and connected. For example, we observe the emergence of regime (number 1) in which density strongly causes the distance to network in a negative way (in the sense of the existence of a negative $\rho_+$), but distance to the center causes the distance to the network, regime which maximal extent on $(w_d,w_r)$ is for an intermediate value $w_r=0.7$. Thus, to maximize the impact of network on density, the corresponding weight must not be maximized but take an intermediate value, what can be counter-intuitive at first sight: this illustrates the utility of the method in the case of circular causal relations difficult to entangle a priori. The regime 2, in which distance to network influences the density in the same way, but the relation between distance to center and to the network is inverted, predominates for low $w_r$ values: thus, the attenuation of the role of distance to the road leads the center to invert its relation with the distance to the road. The regime 6, extreme regarding its position within the parameter space since it is in a neighborhood of $w_c = 0$, corresponds to an isolated situation in which distance to the center does not play a role as an explicative variable, and we observe a causality of density on distance to the center ($\rho_+$ for $D\rightarrow R$) and of distance to the road on distance to the center ($\rho_-$ for $C\rightarrow R$), i.e. that this aspect is totally dominated by the others.
}{
Nous proposons finalement d'interpréter les régimes obtenus, représentés en Fig.~\ref{fig:causalityregimes:clustering}. Le comportement obtenu est particulièrement intéressant : les régions du diagramme de phase selon les paramètres correspondant aux régimes sont clairement délimitées et connexes. Par exemple, on observe l'émergence d'un régime (numéroté 1) où la densité cause fortement la distance au réseau de manière négative (au sens de l'existence d'un $\rho_+$ négatif), mais la distance au centre cause la distance au réseau, régime dont l'étendue maximale dans le plan $(w_d,w_c)$ est pour une valeur intermédiaire $w_r=0.7$. Ainsi, pour maximiser l'impact du réseau sur la densité, il ne faut pas maximiser le poids correspondant mais prendre une valeur intermédiaire, ce qui peut paraître contre-intuitif en premier abord : cela illustre l'intérêt de la méthode dans le cas de relations circulaires difficiles à démêler a priori. Le régime 2, où la distance au réseau influence la densité de la même manière, mais la relation entre distance au centre et route est inversée, est prédominant dans les faibles $w_r$ : ainsi, l'attenuation du rôle de la distance à la route conduit le centre à inverser sa relation avec la distance à la route. Le régime 6, extrême quant à sa position dans l'espace des paramètres car dans un voisinage de $w_c = 0$, correspond à une situation isolée dans laquelle la distance au centre n'importe pas comme variable explicative, et on observe une causalité de la densité sur la distance au centre ($\rho_+$ pour $D\rightarrow R$) et de la distance à la route sur la distance au centre ($\rho_-$ pour $C\rightarrow R$), c'est-à-dire que cet aspect est totalement dominé par les autres.
}

\bpar{
This application on synthetic data demonstrate thus on the one hand the robustness of the method given the consistence of obtained regimes, and realizes this way a much more finer qualification of model behavior than the one done in the original paper. In this precise case, it can be taken as an instrument of knowledge for relations between networks and territories in itself, allowing the test of assumption or the comparison of processes in the stylized model.
}{
Cette application sur données synthétiques démontre ainsi d'une part la robustesse de la méthode vu la cohérence des régimes obtenus, et constitue aussi une qualification beaucoup plus précise des comportements du modèle que celle réalisée dans l'article initial. Dans ce cas précis, il peut s'agir d'un instrument de connaissance des relations entre réseaux et territoires en lui-même, permettant le test d'hypothèses ou la comparaison de processus dans le modèle stylisé.
}


\subsection{Network-territory relations in South Africa}{Relations Réseaux-territoires en Afrique du Sud}

\bpar{
We demonstrate now the potentialities of our method to establish links between variables on geo-historical data on the long time, for the case of the railway network in South Africa during the 20th century. By making the assumption that territories and networks react differently to historical events, causality patterns should inform on their relation on the long time.
}{
Nous démontrons à présent les potentialités de notre méthode à établir des liens entre variables sur des données géo-historiques sur le temps long, pour le cas du réseau ferré en Afrique du Sud au cours du 20ème siècle. En faisant l'hypothèse que les territoires et les réseaux réagissent différemment aux événements historiques, les motifs de causalité devraient informer sur leur relations sur le temps long.
}

\subsubsection{Context}{Contexte}

\bpar{
Transportation Networks can be leveraged as a powerful tool to control populations, with even more significant outcomes when it percolates to their interaction with territories. The case of South Africa is an accurate illustration, as \cite{baffi:tel-01389347} shows that during apartheid railway network planning was used as a racial segregation tool by shaping strongly constrained mobility and accessibility patterns\footnote{The segregation policy is interpreted by~\cite{baffi:tel-01389347} (p.~189) as an intent of ``connecting without connectivity'', jointly with forced migrations of the segregated population in specific zones far from urban functions, called \emph{bantoustans}. The network has then been specifically developed to link these to the production areas without connecting them efficiently to urban centers.}. In particular, it is shown qualitatively that dynamics between territories and networks profoundly changed at the end of the apartheid, transforming a tool of planed segregation (network shaped was optimized to minimize unwanted accessibility) into an integration tool thanks to recent changes in network topology patterns. We investigate here the potential \emph{structural} properties of this historical process, by focusing on dynamical patterns of interactions between the railway network and city growth. More precisely, we try to establish if the segregative planning policies did actually modify the trajectory of the coupled system, what would correspond to deeper and wider impacts. 
}{
Les réseaux de transport peuvent être utilisés comme un puissant outil de contrôle des populations, avec des effets encore plus significatifs lorsque ceux-ci perturbent les relations avec les territoires. Le cas de l'Afrique du Sud est une illustration pertinente, puisque \cite{baffi:tel-01389347} montre que lors de l'apartheid la planification du réseau ferré était utilisée comme un outil de ségrégation raciale par l'établissements de motifs de mobilité et d'accessibilité fortement contraints\footnote{La politique de ségrégation est interprétée par~\cite{baffi:tel-01389347} (p.~189) comme une intention de ``connecter sans connectivité'', conjointement avec les migrations forcées de la population ségréguée dans des zones spécifiques à l'écart des fonctions urbaines, appelées \emph{bantoustans}. Le réseau a alors été spécifiquement développé pour relier ceux-ci aux zones de production sans les connecter efficacement aux centres urbains.}. En particulier, il est montré qualitativement que les dynamiques entre réseaux et territoires ont profondément changé à la fin de l'apartheid, transformant un outil de ségrégation planifiée (une forme de réseau conçue pour minimiser l'accessibilité des populations ségréguées) en un outil d'intégration grâce à des changement récents dans la topologie du réseau. Nous étudions ici les potentielles propriétés \emph{structurelles} de ce processus historique, en se concentrant sur les motifs dynamiques des interactions entre le réseau ferré et la croissance des villes. Plus précisément, nous essayons d'établir si les politiques de planification ségrégatives ont effectivement modifié la trajectoire du système couplé, ce qui correspondrait à des impacts plus larges et profonds que leurs effets immédiats.
}

\subsubsection{Data}{Données}

\bpar{
We use a comprehensive database covering the full South African railway network from 1880 to 2000 with opening and closing dates for each station and link, together with a city database spanning from 1911 to 1991 for which consistent ontologies for urban areas have been ensured. These database are described by~\cite{baffi:tel-01389347}, but they are not open. To ensure our opening constraint, we make available only the aggregated data we used in the analysis.
}{
Nous utilisons une base de données complète couvrant l'ensemble du réseau ferré Sud-Africain de 1880 à 2000 avec les dates d'ouverture et de fermeture pour chaque station et liaison, couplée à une base de données pour les villes s'étendant de 1911 à 1991 pour laquelle des ontologies consistantes pour les aires urbaines ont été assurées. Ces bases de données sont décrites par~\cite{baffi:tel-01389347}, mais ne sont pas ouvertes. Pour respecter notre exigence d'ouverture, nous ne mettons ainsi à disposition que les données agrégées utilisées dans l'analyse.
}

\subsubsection{Network measures}{Mesures de réseau}

\begin{figure}[h!]
\includegraphics[width=\linewidth]{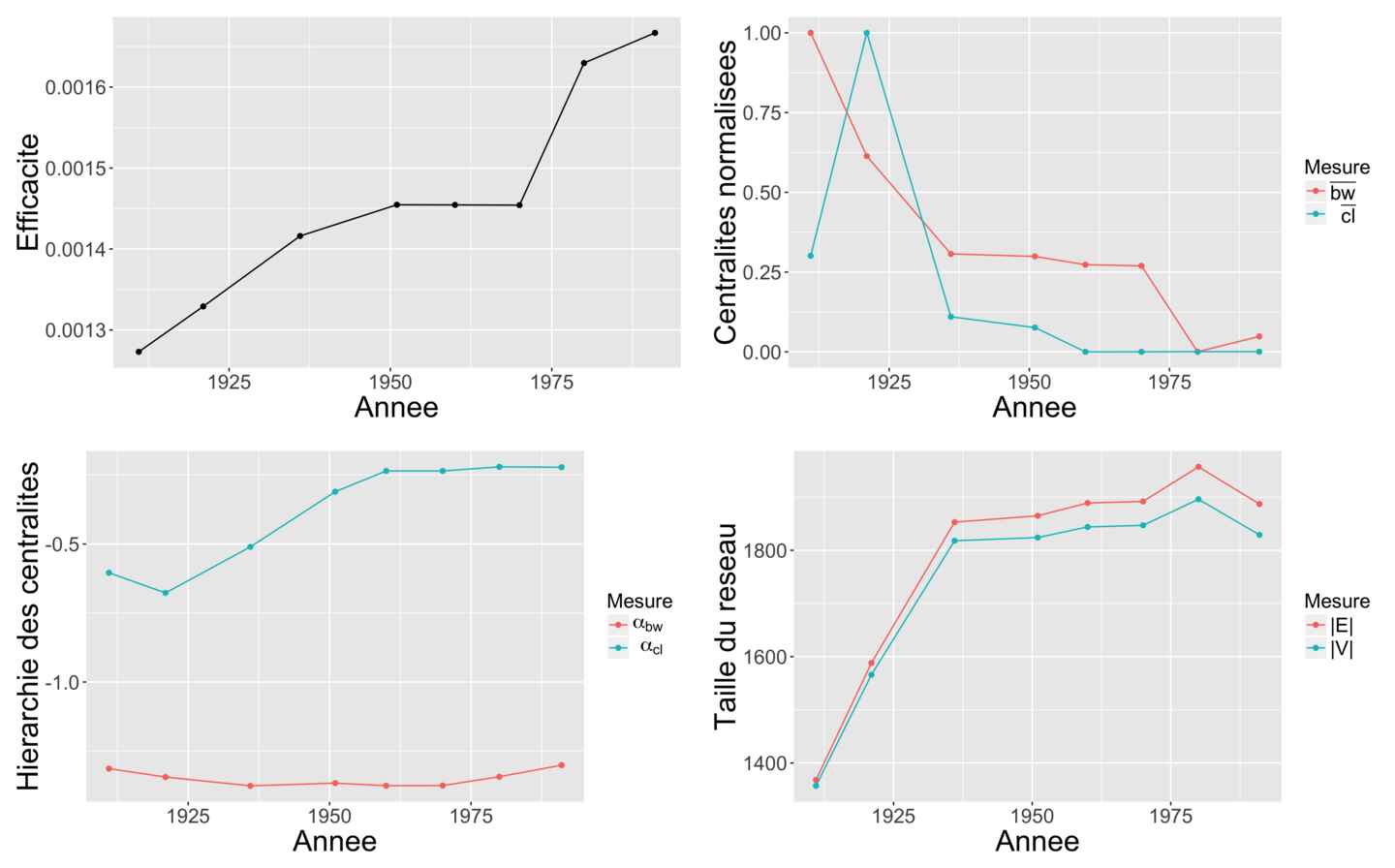}
\caption[Evolution of measures for the South African railway network]{\textbf{Evolution of measures for the South African railway network.} We compute for all dates the basic network measures: size, centralities summarized by their hierarchy and their average, efficiency. Centralities are normalized to comparison of their respective variation ($\max \bar{bw} = 0.07$, $\max \bar{cl} = 1.5e-4$).\label{fig:causalityregimes:network}}
\end{figure}

\bpar{
A preliminary analysis consists in studying the dynamical evolution of network measures, since these can witness of breaks in the structural properties of the network and thus of deep historical mutations. The evolution of some network properties, such as the distributions of centrality or of accessibility, can witness the existence of a planning having influenced them. We show in Fig.~\ref{fig:causalityregimes:network} the evolution of network measures in time\footnote{Globally, \cite{baffi:tel-01389347} (p.~154) shows that the South African railway network has developed with a tree-like structure and that segregation policies have fixed its structure, preventing it from meshing the territory.}, corresponding to some of the measures defined in~\ref{sec:staticcorrelations}. The closeness centrality, that we define as the average travel time towards the other nodes, has an interesting behavior. Indeed, network size and average values of centralities have a behavior in correspondence, which corresponds to the initial expansion of the network. On the contrary, the trend of the hierarchy of closeness centrality to decrease is suddenly broken at the date corresponding to the officialisation of segregation policies in 1951, whereas the size and the global geometrical shape of the network, translated by the efficiency, remain constant. Thus, in the best case the planning after this date is a coincidence with the variation of this property. It is highly probable that it is indeed responsible for this trend breaking, i.e. has had the expected effects on accessibility, with the aim to prevent the decrease of segregation, since the lowest the hierarchy is the highest the network equalitarian is.
}{
Une analyse préliminaire consiste à regarder l'évolution dynamique des mesures de réseau, celles-ci pouvant témoigner de ruptures dans les propriétés structurelles du réseau et donc de mutations historiques profondes. L'évolution de certaines propriétés du réseau, comme les distributions de la centralité ou de l'accessibilité, peut témoigner de l'existence d'une planification les ayant influencées. Nous montrons en Fig.~\ref{fig:causalityregimes:network} l'évolution des mesures de réseau dans le temps\footnote{Globalement, \cite{baffi:tel-01389347} (p.~154) montre que le réseau ferré sud-africain s'est développé en arborescence et que les politiques de ségrégation ont figé sa structure, l'empêchant de mailler le territoire.}, correspondant à certaines des mesures définies en~\ref{sec:staticcorrelations}. La centralité de proximité, que nous définissons comme le temps moyen de trajet vers les autres noeuds, présente un comportement intéressant. En effet, la taille du réseau et les valeurs moyennes des centralités présentent un comportement concordant, qui correspond à l'expansion initiale du réseau. Par contre, la tendance de la hiérarchie de la centralité de proximité à se réduire est soudainement rompue à la date correspondant à l'officialisation des politiques ségrégatives en 1951, alors que taille et forme géométrique globale du réseau, traduites par l'efficience, restent constants. Ainsi, dans le meilleur des cas la planification après cette date est une coincidence avec la variation de cette propriété. Il est très probable qu'elle soit en effet responsable de cette rupture de tendance, c'est-à-dire a eu les effets escomptés sur l'accessibilité, dans le but d'empêcher la diminution de la ségrégation, puisque plus la hiérarchie est faible plus le réseau est égalitaire.
}

\subsubsection{Causality patterns}{Motifs de causalité}

\bpar{
We then turn to dynamical interactions between the railway network and city growth. Therefore, we apply the method developed in the first part, which consists in the study of Granger causalities, in the broad sense of correlations between lagged variables, estimated between city population differentials and accessibility differentials due to network growth, for all cities or urban areas having a connection to the network. We test both the accessibility in terms of distance and weighted by population at the origin and at both extremities. If $P_i$ are populations, $d_{ij}$ the distance matrix within the network, the accessibility of $i$ will be given by $Z_i = w_i \sum_j w_j \exp \left(- d_{ij} / d_0 \right)$ where $d_0$ is a decay parameter and weights are $1/N$ or $P_i / \sum_j P_j$ depending on the modality. We vary the values of $d_0$ to take into account relations at different spatial scales. Furthermore lagged correlations are estimated on time windows with a variable size $T_W$, to test for different potential temporal stationarity scales.
}{
Nous examinons à présent les interactions dynamiques entre le réseau ferré et la croissance urbaine. Pour cela, nous appliquons la méthode développée dans la première partie, qui consiste à l'étude des causalités de Granger, au sens large des corrélations entre les variables retardées, estimées entre les différentiels de population des villes et les différentiels d'accessibilité dus à la croissance du réseau, pour toutes les villes ou aires urbaines ayant une connection au réseau. Nous testons à la fois l'accessibilité en termes de distance et pondérée par la population à l'origine et aux deux extrémités. Si $P_i$ sont les populations, $d_{ij}$ la matrice de distance dans le réseau, l'accessibilité de $i$ sera donnée par $Z_i = w_i \sum_j w_j \exp \left(- d_{ij} / d_0 \right)$ où $d_0$ est le paramètre de décroissance et les poids $w_i$ sont $1/N$ ou $P_i / \sum_j P_j$ selon la modalité. Nous faisons varier les valeurs de $d_0$ pour prendre en compte les relations à différentes échelles spatiales. De plus les corrélations retardées sont estimées sur des fenêtres temporelles de taille variable $T_W$, pour tester différentes échelles potentielles de stationnarité temporelle.
}

\begin{figure}
\includegraphics[width=\linewidth]{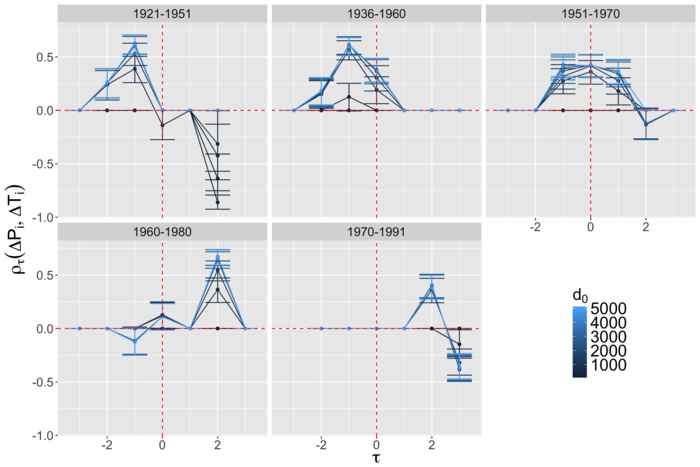}
\caption[Lagged correlations between population growth and accessibility gain in South Africa]{\textbf{Lagged correlations.} Lagged correlations as a function of lag $\tau$, for the window $T_W = 3$, on the different successive periods (columns), and for a variable $d_0$ (color). For the interpretation, we observe a maximum of lagged correlation which shifts in time going from a negative lag to a positive lag, what according to our definition corresponds to an inversion of the direction of causality.\label{fig:causalityregimes:sudafcorrs}}
\end{figure}

\bpar{
Results of estimations are shown in Figure~\ref{fig:causalityregimes:sudafcorrs}. We obtain results which are significant with travel-time accessibility only, that we show here\footnote{We give in Appendix~\ref{app:sec:causalityregimes}, Fig.~\ref{fig:app:causalityregimes:sudafcorrs} the lagged correlations profiles estimated for accessibilities weighted at the origin and at the destination. The auto-correlation a priori dominates the weighted accessibility: indeed, we have for the two weighted variables positive values for low values of $d_0$ only, the others being not significant.}. The best compromise for the time window appears to be around thirty years, if we seek to have both an important number of significant correlations (defined by $p<0.1$ for a Fisher test) and a high average level of absolute correlation for all lags and decay parameters. We interpret this value as approximatively the temporal stationarity scale of the system. It would be the duration on which a regime of the urban system is relatively stable, and is of the same order than the duration of the apartheid.
}{
Les résultats des estimations sont montrés en Figure~\ref{fig:causalityregimes:sudafcorrs}. Nous obtenons des résultats significatifs avec l'accessibilité non-pondérée seulement, que nous montrons ici\footnote{Nous donnons en Annexe~\ref{app:sec:causalityregimes}, Fig.~\ref{fig:app:causalityregimes:sudafcorrs} les profils de corrélations retardées estimées pour les accessibilités pondérées à l'origine et à l'origine et à la destination. L'auto-corrélation domine a priori l'accessibilité pondérée : en effet, on a pour les deux variables pondérées des valeurs positives pour les faibles valeurs de $d_0$ uniquement, les autres n'étant pas significatives.}. Le meilleur compromis pour la fenêtre temporelle apparaît être une trentaine d'années, si on cherche à avoir à la fois un bon nombre de corrélations significatives (définies par $p<0.1$ pour un test de Fisher) et un niveau moyen élevé de corrélation absolue sur l'ensemble des retards et des paramètres de décroissance. Nous interprétons cette valeur comme approximativement l'échelle temporelle de stationnarité du système. Il s'agirait de la durée sur laquelle un régime du système urbain est relativement stable, et est du même ordre de grandeur que la durée de l'apartheid.
}

\bpar{
Furthermore, the number of significant correlations clearly exhibits a phase transition in its intermediate values as a function of $d_0$ (Fig.~\ref{fig:app:causalityregimes:sudafcorrs}), what should correspond to the transition between the spatial scale of urban areas and the one of the country, and thus gives the local spatial stationarity scale, around $d_0 = 500km$. The cities of Cap Town and Johannesburg are at a distance of 1400km and correspond to two regions at the extremities of the country: this scale is thus a regional scale, smaller than the one of the urban system of the country.
}{
De plus, le nombre de corrélations significatives présente clairement une transition de phase dans ses valeurs intermédiaires en fonction de $d_0$ (Fig.~\ref{fig:app:causalityregimes:sudafcorrs}), ce qui devrait correspondre au passage entre l'échelle spatiale des aires urbaines et celle du pays, et donne ainsi l'échelle locale de stationnarité spatiale, autour de $d_0 = 500km$. Les villes du Cap et de Johannesburg sont à 1400km de distance et correspondent à deux régions aux extrémités du pays : cette échelle est donc une échelle régionale, inférieure à celle du système urbain du pays.
}

\bpar{
The study of the behavior of lagged correlations in Fig.~\ref{fig:causalityregimes:sudafcorrs} leads to the observation of relatively clear causality patterns, since the direction of Granger causality is inverted around 1950, this corresponding at each time to correlations up to 0.5 for some values of the decay parameter. We thus switch from an accessibility causing population growth with a delay between 10 and 20 years before the apartheid (1948), to the opposite, i.e. a population inducing the accessibility changes after the apartheid (with a lag of 20 years).
}{
L'examen du comportement des corrélations retardées en Fig.~\ref{fig:causalityregimes:sudafcorrs} conduit à l'observation de motifs de causalité assez clairs, puisque le sens de la causalité de Granger s'inverse autour de 1950, celle-ci étant à chaque fois marquée par des corrélations allant jusqu'à 0.5 pour certaines valeurs du paramètre de décroissance. On passe ainsi d'une accessibilité causant la croissance de la population avec un délai de 10 à 20 ans avant l'apartheid (1948), à l'opposé, c'est-à-dire une population induisant les changements d'accessibilité après l'apartheid (avec un délai de 20 ans).
}

\bpar{
This result is consistent with population relocations and the conception of the network following these. We interpret this phenomenon as a \emph{structural segregation}, i.e. a significant impact of planning policies on the dynamics of interactions between networks and territories. Indeed, the first regime can be interpreted as a direct effect of transportation on migration patterns within a context of liberty, in opposition to the second regime which would correspond to a control of the population and an adaptation of the network following. Therefore, the historical event had an effect at the second order on dynamical relations. These patterns rejoin the empirical conclusions obtained by~\cite{baffi:tel-01389347} on the subject of the apartheid, which for example shows a strong effect of measures on forced displacements of population, and also a decrease of accessibility for the target areas of segregation.
}{
Ce résultat est en cohérence avec les relocalisations de population et la conception du réseau en accord avec celles-ci. Nous interprétons ce phénomène comme une \emph{ségrégation structurelle}, c'est-à-dire un impact significatif des politiques de planification sur les dynamiques des interactions entre les réseaux et les territoires. En effet, le premier régime peut être interprété comme un effet direct du transport sur les motifs de migration dans un contexte de liberté, en opposition au second régime qui correspondrait à un contrôle de la population et d'une adaptation du réseau en fonction. Ainsi, l'évènement historique a eu un effet au second ordre sur les relations dynamiques. Ces motifs rejoignent les conclusions empiriques obtenues par~\cite{baffi:tel-01389347} sur le sujet de l'apartheid, qui montre par exemple un fort effet des mesures sur les déplacements forcés de population, ainsi qu'une baisse de l'accessibilité pour les zones cibles de la ségrégation. 

}

%


\subsubsection{Possible developments}{Développements possibles}

\bpar{
A first extension can consist in similar study with more precise socio-economic variables, for example quantifying directly segregation patterns. Furthermore, qualitative variables linked to historical events could be used as instrumental variables. The method of instrumental variables~\cite{angrist1996identification} is used to identify causal relationships between variables, in a complementary way to the one we introduced. We could try to render more robust conclusions, in particular check if the correlations are not spurious, by the application of this approach, that would however be difficult to be put into practice given the sparsity of data in our case.
}{
Une première extension pourra consister en une étude similaire avec des variables socio-économiques plus précises, pour quantifier par exemple directement les motifs de ségrégation. D'autre part, des variables qualitatives liées aux évènements historiques pourraient faire office de variable d'instrumentation. La méthode des variables instrumentales~\cite{angrist1996identification} est utilisée pour identifier des relations causales entre variables, d'une façon complémentaire à celle que nous avons mis en place. Nous pourrions chercher à rendre nos conclusions plus robustes, notamment vérifier si les corrélations ne sont pas fortuites, par l'application de cette approche, qui serait cependant difficile à réaliser vu la rareté des données dans notre cas.
}

\stars


\bpar{
We have up to here in this chapter explored two ingredients of the evolutive urban theory, that will be crucial to understand co-evolution between transportation networks and territories, namely non-stationarity properties of correlations, that will guide the construction of models at a similar scale (chapter~\ref{ch:morphogenesis} and then chapter~\ref{ch:mesocoevolution}) and the possibility to unveil causality regimes, that will be used as a tool to characterize co-evolution.
}{
Nous avons jusqu'ici dans ce chapitre exploré deux ingrédients de la théorie évolutive des villes, qui nous seront cruciaux pour comprendre la co-évolution entre réseaux de transport et territoires, à savoir les propriétés de non-stationnarité des corrélations, qui guideront la mise en place des modèles à une échelle similaire (chapitre~\ref{ch:morphogenesis} puis chapitre~\ref{ch:mesocoevolution}) et la possibilité de mise en évidence de régimes de causalité, qui nous servira d'outil de caractérisation de la co-évolution.
}

\bpar{
We propose now to introduce a last crucial element of the evolutive urban theory, which is the grasping of urban systems by the intermediate of interaction models between cities. These will not be co-evolutive in a first time but their ontology will aim at integrating the role of networks in the urban system.
}{
Nous proposons à présent d'introduire un dernier élément crucial de la théorie évolutive des villes, qui est l'appréhension des systèmes urbains par l'intermédiaire de modèles d'interaction entre villes. Ceux-ci ne seront pas co-évolutifs dans un premier temps mais leur ontologie visera à intégrer le rôle des réseaux dans le système urbain.
}


\stars

%

\newpage


\section{A macroscopic growth model}{Un modèle de croissance macroscopique}

\label{sec:interactiongibrat}


\bpar{
The last aspect of the evolutive urban theory that we propose to explore is positioned within the thematic and modeling perspectives: the study of cities themselves and of their interactions, through the intermediary of simulation models. As we will see, most models for systems of cities given by the evolutive urban theory are based on interactions between cities: this allows us to take a direct entry in our problematic since these operate through the intermediary of networks, that we will then be able to explicit in our models.
}{
Le dernier aspect de la théorie évolutive des villes que nous proposons d'explorer se positionne sur le plan thématique et sur le plan de la modélisation : l'étude des villes elles-mêmes et de leur interactions, par l'intermédiaire de modèles de simulation. Comme nous allons le voir, la plupart des modèles de systèmes de villes issus de la théorie évolutive des villes se basent sur les interactions entre villes : cela nous permet une entrée directe dans notre problématique puisque celles-ci se font par l'intermédiaire des réseaux, qu'on pourra alors expliciter dans nos modèles.
}

\bpar{
We describe thus a simple spatial model of urban growth for systems of cities at the macroscopic scale, which combines direct interaction between cities and an indirect effect of physical network flows as population growth drivers. The model is parametrized on population data for the French system of cities between 1831 and 1999.
}{
Nous décrivons ainsi un modèle spatial simple de croissance urbaine pour les systèmes de villes à l'échelle macroscopique, qui combine les interactions directes entre les villes et un effet indirect des flux du réseau physique comme moteurs de la croissance de population. Le modèle est paramétré et calibré sur les données de population pour le système de villes français entre 1831 et 1999.
}




\bpar{
The aim of this section is to explore further the assumption, central to \noun{Pumain}'s evolutive urban theory, according to which spatial interactions between cities are significant drivers of their growth. More precisely, we consider both abstract interactions and flow interactions mediated through the physical networks, mainly transportation network. We extend existing models accordingly.
}{
Le but de cette section est d'explorer plus en détail l'hypothèse, centrale à la théorie évolutive des villes de \noun{Pumain}, selon laquelle les interactions spatiales sont des moteurs significatifs de leur croissance. Plus précisément, nous considérons à la fois les interactions abstraites et les interactions avec les flux portés par les réseaux physiques, principalement les réseaux de transport. Nous étendons les modèles existants de manière correspondante.
}

\bpar{
Our contribution is twofold: (i) we show that very basic interaction models based on population only can be fitted to empirical data and that fitted parameter values are directly interpretable; and (ii) we introduce a novel methodology to quantify overfitting in models of simulation\footnote{Which as we already precised in~\ref{ch:preliminarymath} is a crucial issue for a correct adjustment of models.}, as an extension of information criteria for statistical models, which applied to our calibrated models confirms that the improvement in fit is not only due to additional parameters, but that the extended model effectively capture more information on system processes. This will unveil network effects in an indirect way. We first precise the thematic context in which the model will be situated and review modeling approaches to urban growth based on spatial interactions.
}{
Les apports de cette section consistent en deux points : (i) nous montrons que des modèles d'interaction très basiques basés uniquement sur la population peuvent être ajustés aux données empiriques et que les valeurs ajustées des paramètres sont directement interprétables ; et (ii) nous introduisons une nouvelle méthodologie pour quantifier le sur-ajustement dans les modèles de simulation\footnote{Qui comme nous l'avons déjà précisé en~\ref{ch:preliminarymath} est un problème crucial pour un ajustement correct des modèles.}, comme une extension de critères d'information pour les modèles statistiques, qui appliquée à nos modèles calibrés confirme que l'amélioration du fit n'est pas due seulement aux paramètres supplémentaires et donc artificielle, mais que le modèle étendu capture effectivement plus d'information sur les processus du système. Cela révèlera des effets de réseaux de manière indirecte. Nous précisons d'abord le contexte thématique dans lequel le modèle se placera et revoyons les approches de modélisation de la croissance urbaine basées sur les interactions spatiales.
}

\subsection{Modeling urban growth through spatial interaction}{Modéliser la croissance	urbaine par les interactions spatiales}

\subsubsection{Models for urban growth}{Modèles de croissance urbaine}

\bpar{
We replace the current approach within a more global context of modeling urban growth\footnote{That we understand here as the evolution in time of cities, typically seen through the evolution of their population or of economic activities, and of their spatial distribution.}, notion which is more general than our precise problematic. It however allows us here to construct a first model from the point of view of evolutive urban theory. A good knowledge of how cities differentiate, interact and grow is thus a relevant topic both for policy application and from a theoretical perspective. \cite{pumain2009innovation} suggests that cities are incubators of social change, their fate being closely linked to the one of societies, and thus as we developed in chapter~\ref{ch:thematic}, of their territories and their networks.
}{
Replaçons la présente démarche dans un contexte plus global de modélisation de la croissance urbaine\footnote{Qu'on comprend ici comme l'évolution dans le temps des villes, vue typiquement par l'évolution de leur population ou des activités économiques, et de leur distribution spatiale.}, notion plus générale que notre problématique précise. Elle nous permet ici toutefois de construire un premier modèle du point de vue de la théorie évolutive des villes. Une bonne connaissance de la façon dont les villes se différencient, interagissent et croissent est ainsi un sujet pertinent à la fois pour les applications en termes de politiques et d'un point de vue théorique. \cite{pumain2009innovation} suggère que les villes sont l'incubateur du changement social, leur destin étant étroitement lié à celui des sociétés, et donc comme nous l'avons développé en chapitre~\ref{ch:thematic}, de leurs territoires et de leurs réseaux.
}

\bpar{
Various disciplines have studied models of urban growth with different objectives and taking diverse aspects into account. For example, Economics are still cautious to include spatial interactions in their models~\citep{krugman1998space}, taking them into account in a very simplified way even in Economic Geograĥy, but producing models that are extremely detailed on market processes. On the contrary, Geography focuses more on territorial specificities and interactions in space but will produce general conclusion with more difficulty~\cite{marchionni2004geographical}. The example of this two disciplines shows how it is difficult to build bridges, as it needed exceptional efforts to translate from one to the other (as P. Hall did for Von Thunen work~\citep{taylor2016polymath}), and therefore how it is far from evident to grasp the complexity of Urban Systems in an integrated way.
}{
Diverses disciplines ont étudié des modèles de croissance urbaine avec différents objectifs et prenant en compte des aspects variés. Par exemple, l'économie est souvent prudente à inclure les interactions spatiales dans les modèles~\cite{krugman1998space}, les prenant en compte de façon très simplifiée même en économie géographique, mais produisant des modèles extrêmement détaillés en termes de processus de marché. Au contraire, la géographie se concentre plus sur les spécificités territoriales et les interactions dans l'espace mais produira des conclusions générales avec plus de difficulté~\cite{marchionni2004geographical}. L'exemple de ces deux disciplines montre comment il est difficile de créer des ponts, comme il a fallu des efforts exceptionnels pour effectuer des traductions de l'une à l'autre (comme \noun{P. Hall} le fit avec le travail de \noun{Von Thünen}~\cite{taylor2016polymath}), et ainsi comment il est loin d'être évident de capturer la complexité des systèmes urbains de manière intégrée.
}

\bpar{
The simplest model to explain\footnote{I.e. trying to translate and include its fundamental processes and to reproduce its stylized facts. A fundamental stylized fact is the hierachical distribution of the size of cities, following a scaling law, often close to a Zipf's law (rank-size law). It is considered as one of the most regular facts, at least in its generalized form as a scaling law~\cite{nitsch2005zipf}. It will give the population $P_i$ of cities as a function of their rank $i$ in the hierarchy, under the form $P_i = P_0 / i^{\alpha}$.} urban growth, the Gibrat model, assumes random independent growth rates. It has been shown by~\cite{gabaix1999zipf} to asymptotically produce the expected rank-size law for system of cities. Explaining urban scaling laws is closely related to the understanding of urban growth, as \cite{bettencourt2008large} suggests that these reflect underlying universal processes and that individual properties of cities can be explained by a change of scale. This approach however does not reflect the complex relation between economic agents for which~\cite{storper2009rethinking} is positioned, advocating for a new economic theory focusing on circular relations between the geography of production and the movement of jobs.
}{
Le modèle le plus simple pour expliquer\footnote{C'est-à-dire essayant de traduire et d'inclure ses processus fondamentaux et de reproduire ses faits stylisés. Un fait stylisé fondamental est la distribution hiérarchique de la taille des villes, en loi d'échelle, souvent proche d'une loi de Zipf (loi rang-taille). Elle est considérée comme l'un des faits les plus réguliers, au moins dans sa formulation généralisée sous forme de loi d'échelle~\cite{nitsch2005zipf}. Celle-ci donnera la population $P_i$ des villes en fonction de leur rang $i$ dans la hiérarchie, sous la forme $P_i = P_0 / i^{\alpha}$.} la croissance urbaine, le modèle de Gibrat, suppose des taux de croissance aléatoires indépendants. Il a été montré par~\cite{gabaix1999zipf} comme produisant asymptotiquement la loi rang-taille attendue pour les systèmes de villes. Expliquer les lois d'échelles urbaines est étroitement lié à la compréhension de la croissance urbaine, comme \cite{bettencourt2008large} suggèrent que celles-ci reflètent des processus universels sous-jacents et que les propriétés individuelles des villes s'expliquent par un passage à l'échelle. Cette approche reflète cependant peu les relations complexes entre agents économiques pour lesquelles~\cite{storper2009rethinking} se positionnent, plaidant pour une nouvelle théorie économique appuyant les relations circulaires entre la géographie de la production et les mouvements des emplois.
}

\bpar{
Using a bottom-up reconstruction of urban areas using dynamical microscopic population data, \cite{rozenfeld2008laws} shows indeed that positive deviations to the rank-size law systematically exist, these being systematically underestimated, and that these must be an effect of spatial interaction between urban areas. Complexity approaches are good candidates to integrate these into models. \cite{andersson2006complex} introduce for example a model of urban economy as a growing complex network of relations. The evolutive urban theory, introduced by~\cite{pumain1997pour}, focuses on cities as co-evolving entities and produces explanations for growth at the level of the system of cities.
}{
Par l'utilisation d'une reconstruction par le bas des zones de population contiguës via des données microscopiques dynamiques de population, \cite{rozenfeld2008laws} montrent en effet que des déviations aux tailles attendues par la loi rang-taille existent systématiquement, celles-ci étant sous-estimées, ce qui est probablement un effet des interactions spatiales entre les aires urbaines. Les approches par la complexité sont de bonnes candidates pour intégrer celles-ci dans les modèles. \cite{andersson2006complex} introduit par exemple un modèle d'économie urbaine comme un réseau croissant de relations. La théorie évolutive urbaine, introduite par~\cite{pumain1997pour}, se concentre sur les villes comme des entités en co-évolution et produit des explications pour la croissance au niveau du système de villes.
}

\bpar{
\cite{pumain2006evolutionary} show that scaling laws could be due to functional differentiation and diffusion of innovation between cities. The positioning regarding universality of laws is more moderate than \emph{Scaling} theories such as the one by~\cite{west2017scale}, as \cite{pumain2012urban} emphasizes that ergodicity can difficultly be assumed in the frame of complex territorial systems. One crucial feature of this paradigm is the importance of interactions between agents, generally the cities, to produce the emergent patterns at the scale of the system. \cite{pumain2013theoretical} has investigated the advantages of Agent-based models compared to more classical equation systems, and this methodological aspect is in accordance with the theoretical positioning, as it allows to take into account the heterogeneity of possible interactions, the geographical particularities, and to naturally translate emergence between levels and render multi-scale patterns.
}{
 \cite{pumain2006evolutionary} montrent que les lois d'échelles pourraient être dues à la différentiation fonctionnelle et la diffusion de l'innovation entre les villes. Les positionnement au regard de l'universalité des lois est plus modéré que les théories du \emph{Scaling} comme celle de~\cite{west2017scale}, puisque \cite{pumain2012urban} souligne que l'ergodicité peut difficilement être prise pour acquise dans le cadre des systèmes complexes territoriaux. Un aspect important de ce paradigme est l'importance des interactions entre agents constituants du système, généralement les villes, qui produisent les motifs émergents à l'échelle supérieure. \cite{pumain2013theoretical} ont investigué les avantages des modèles multi-agents en comparaison à des systèmes d'équations plus classiques, et cet aspect méthodologique est en accord avec le positionnement théorique, comme cette approche permet de prendre en compte l'hétérogénéité des interactions possibles, les particularités géographiques, et de traduire naturellement l'émergence entre les niveaux et rendre compte de motifs multi-échelles.
}


\subsubsection{Urban growth and spatial interactions}{Croissance urbaine et interactions spatiales}

\bpar{
We must recall that we consider in this section only models at the macroscopic scale, ruling out the numerous and rich approaches at the mesoscopic scale, that include for exemple cellular automatons models, models of urban morphogenesis that we will study in chapter~\ref{ch:morphogenesis} or land-use change models. We also rule out economics models that do not include explicitly spatial interactions. Several models for urban growth at the macroscopic scale have insisted on the role of space and spatial interactions, that we will illustrate in the following. \cite{bretagnolle2000long} propose a spatial extension of the Gibrat model. The gravity-based interaction model that~\cite{sanders1992systeme} use to apply concept of synergetics to cities is also close to this idea of interdependent urban growth, contained physically in the phenomenon of migration between cities. A more refined extension with economic cycles and innovation waves was developed by~\cite{favaro2011gibrat}, yielding a system dynamics version of the ontology of Simpop models~\citep{pumain2012multi} (that we already presented previously in chapter~\ref{ch:modelinginteractions} and in introduction of this chapter).
}{
Nous devons préciser que nous considérons dans cette section uniquement les modèles à l'échelle macroscopique, ne considérant pas les nombreuses approches très riches à l'échelle mesoscopique, qui incluent par exemple les modèles à automates cellulaires, les modèles de morphogenèse urbaine que nous aborderons en chapitre~\ref{ch:morphogenesis} ou les modèles de changement d'usage du sol. Nous excluons aussi les modèles économiques qui n'incluent pas explicitement les interactions spatiales. Un certain nombre de modèles de croissance urbaine à l'échelle macroscopique ont insisté sur le rôle de l'espace et des interactions spatiales, que nous allons illustrer par la suite. \cite{bretagnolle2000long} proposent une extension spatiale du modèle de Gibrat. Le modèle d'interaction basé sur la gravité que \cite{sanders1992systeme} utilise pour appliquer les concepts de la synergétique aux villes est également proche de cette idée de croissance urbaine interdépendante, contenue physiquement dans le phénomène de migration entre les villes. Une extension avec des cycles économiques et des vagues d'innovation a été développé par~\cite{favaro2011gibrat}, fournissant une version de l'ontologie des modèles Simpop~\cite{pumain2012multi} (que nous avons déjà présenté précédemment en chapitre~\ref{ch:modelinginteractions} et en introduction de ce chapitre) en termes de systèmes dynamiques.
}

\bpar{
The family of Simpop models has been developed in symbiosis with the evolutive urban theory, with the main characteristic of models based on agents taking into account spatial interaction. Models have been progressively refined, and specified for diverse case studies. We can give a chronological glimpse of a sample of these. This family of models have started with a toy-model based on economic interactions between cities as agents, that yield hierarchical patterns at the scale of the system~\citep{sanders1997simpop}. Later, the Simpop2 model, still based on distance interaction for commercial exchanges, including successive innovation waves, unveiled structural differences between the European and the US Urban Systems~\citep{bretagnolle2010comparer}. The SimpopLocal model~\citep{pumain2017simpoplocal} is used to show the emergence of initial settlement patterns. Finally, the most recent similar model, the Marius model~\citep{cottineau2014evolution} couples population and economic growth with cities interaction, allowing to accurately reproduce real trajectories (in the sense of the mean square error in time on all populations) on the former Soviet Union after calibration with multi-modeling of processes.
}{
La famille des modèles Simpop a été développée en symbiose avec la théorie évolutive des villes, avec la caractéristique principale de modèles basés sur les agents prenant en compte l'interaction spatiale. Les modèles ont été progressivement raffinés, et spécifiés pour divers cas d'étude. Nous pouvons  donner un aperçu chronologique d'un échantillon de ceux-ci. Cette famille de modèles a commencé avec un modèle stylisé basé sur les interactions économiques entre les villes comme agents, qui produit des motifs de hiérarchie à l'échelle du système~\cite{sanders1997simpop}. Plus tard, le modèle Simpop2, toujours basé sur l'interaction en fonction de la distance pour les échanges commerciaux, incluant les vagues successives d'innovation, a dévoilé des différences structurelles entre le système de villes européen et le système aux États-Unis~\cite{bretagnolle2010comparer}. Le modèle SimpopLocal~\cite{pumain2017simpoplocal} est utilisé pour montrer l'émergence des motifs initiaux d'établissement humains. Enfin, le dernier modèle similaire en date, le modèle Marius~\cite{cottineau2014evolution} couple la croissance de la population et économique avec les interactions entre les villes, permettant de reproduire assez fidèlement les trajectoires réelles (au sens de l'erreur carré moyenne dans le temps sur l'ensemble des populations) sur l'ancienne Union Soviétique après calibration avec multi-modélisation des processus.
}

\subsubsection{Urban growth and transportation networks}{Croissance urbaine et réseaux de transports}

\bpar{
We situate here the overview that we just give regarding models studying interactions between territories and networks that we largely reviewed in chapter~\ref{ch:modelinginteractions}.
}{
Nous situons ici l'aperçu que nous venons de donner en regard des modèles s'intéressant aux interactions entre territoires et réseaux que nous avons amplement revus en chapitre~\ref{ch:modelinginteractions}.
}

\bpar{
Under similar assumptions of previously reviewed models, the inclusion of transportation networks has been rarely pursued, contrary to the mesoscopic scale at which relations between networks and territories have been widely studied by Luti models (see for example~\cite{chang2006models}). Network growth models~\cite{xie2009modeling}, prolific in Economics and Physics, can not be used to explain urban growth.
}{
Dans des hypothèses similaires aux modèles précédemment revus, l'inclusion des réseaux de transports a été rarement poursuivie, contrairement à l'échelle mesoscopique à laquelle les relations entre réseaux et territoires ont été largement étudiées par les modèles Luti (voir par exemple~\cite{chang2006models}). Les modèles de croissance de réseau~\cite{xie2009modeling}, prolifiques en économie et physique, ne peuvent pas être utilisés pour expliquer la croissance urbaine.
}

\bpar{
\cite{bigotte2010integrated} studies an optimization model for network design combining the effects of urban hierarchy and of transportation network hierarchy. \cite{baptiste1999interactions} has modeled dynamical interplay between network links capacity and city growth on a subset of French city system. The SimpopNet model~\citep{schmitt2014modelisation} goes a step further in modeling the co-evolution between cities and transportation networks, as it allows new network links to be created in time. These examples shows the difficulty of coupling these two aspects of urban systems in models of growth, and we will for this reason take into account network effects in a simplified way as detailed further.

}{
\cite{bigotte2010integrated} étudie un modèle d'optimisation pour la conception du réseau combinant les effets de la hiérarchie urbaine et de la hiérarchie du réseau de transport. \cite{baptiste1999interactions} a modélisé l'intrication dynamique entre la capacité des liens du réseau et la croissance des villes sur un sous-ensemble du système de villes français. Le modèle SimpopNet~\cite{schmitt2014modelisation} va un pas plus loin dans la modélisation de la co-évolution entre les villes et les réseaux de transport, puisqu'il permet que de nouveaux liens soit créés dans le temps. Ces exemples montrent la difficulté de coupler ces deux aspects des systèmes urbains dans les modèles de croissance, et nous prendrons en compte pour cette raison les effets de réseau d'une manière simplifiée comme nous le détaillerons par la suite.
}

\bpar{
The rest of this section is organized as follows: our model is introduced and formally described; we then describe results obtained through exploration and calibration of the model on data for French cities, in particular the unveiling of network effects significantly influencing growth processes thanks to a novel methodology specifically introduced. We finally discuss the implications of these results.
}{
La suite de cette section est organisée de la manière suivante : nous introduisons notre modèle macroscopique et le décrivons de manière formelle ; puis nous donnons les résultats obtenus par l'exploration et la calibration du modèle sur les données pour les villes françaises, plus particulièrement la révélation d'effets de réseaux influençant de manière significative les processus de croissance, grâce à une nouvelle méthodologie spécifiquement introduite. Nous discutons finalement les implications de ces résultats.
}

\bpar{
The growth model at the macroscopic scale introduced and studied in details here will then be an elementary building brick for the construction of co-evolution models that we will propose in the following in chapter~\ref{ch:macrocoevolution}.
}{
Le modèle de croissance à l'échelle macroscopique introduit et étudié en détails ici servira alors de brique élémentaire pour la construction des modèles de co-évolution que nous proposerons par la suite au chapitre~\ref{ch:macrocoevolution}.
}

\subsection{Model and results}{Modèle et résultats}

\subsubsection{Model description}{Description du modèle}

\paragraph{Rationale}{Hypothèses}

\bpar{
A first fundamental issue to be clarified is the stochastic or deterministic character of the urban growth model. To what extent is a proposed model ``complex'' and is the simulation of stochasticity necessary ? Concerning the Gibrat model and most of its extensions, independence assumptions and linearity produce a totally predictable behavior and thus not complex in the sense of exhibiting emergence, in the sense of weak emergence~\citep{bedau2002downward}. In particular, the full distribution of random growth models can be analytically determined at any time~\citep{gabaix1999zipf}, and in the case of studying only first moment, a simple recurrence relation avoids to proceed to any Monte-Carlo simulation. Under these assumptions, it is reasonable to work with a deterministic model, as it is done for example for the Marius model~\citep{cottineau2014evolution}\footnote{The Marius model introduced by \cite{cottineau2014evolution} for the evolution of cities in former Soviet Union, links the economic and population variables at the scale of cities by taking their interactions into account. The model has been conceived in a perspective of incremental modeling and diverse processes can be considered.}. We will work under that hypothesis, capturing complexity through non-linearity.
}{
Un premier point fondamental à clarifier est le caractère stochastique ou déterministe du modèle de croissance urbaine. Dans quelle mesure un modèle proposé est-il ``complexe'' et la simulation de la stochasticité y est-elle nécessaire ? Concernant le modèle de Gibrat et la plupart de ses extensions, les hypothèses d'indépendance et la linéarité produisent un comportement pouvant être totalement prédit, ce qui ne les rend pas complexes au sens d'exhiber une émergence faible~\cite{bedau2002downward}. En particulier, la distribution complète des modèles de croissance aléatoire peut être déterminée analytiquement à tout instant~\cite{gabaix1999zipf}, et dans le cas de l'étude du premier moment seulement, une simple relation de récurrence évite de procéder à toute simulation de Monte-Carlo. Sous ces hypothèses, il est raisonnable de travailler avec un modèle déterministe, comme il est fait par exemple pour le modèle Marius~\cite{cottineau2014evolution}\footnote{Le modèle Marius introduit par \cite{cottineau2014evolution} pour l'évolution des villes en ex-Union soviétique, lie les variables économiques et de population à l'échelle de villes en prenant en compte leurs interactions. Le modèle a été conçu dans une perspective de modélisation incrémentale et divers processus peuvent être considérés.}. Nous travaillerons sous cette hypothèse, capturant la complexité par la non-linéarité.
}

\bpar{
We work on simple territorial systems assumed as regional city systems, in which cities are basic entities. We have thus around a hundred of cities, and intermediate territories are not taken into account. The time scale corresponds to the characteristic scale associated to this spatial scale, i.e. around one or two centuries. Spatial interactions are captured through gravity-type interactions, this formulation having the advantage of being simple and of capturing the ``first law of geography'' proposed by \noun{Tobler}~\cite{tobler2004first}\footnote{``\textit{Everything interacts with everything, but two closer things have more chance to interact}'', i.e. that interaction strength fades with distance.} This choice does not have a priori a fundamental influence on the behavior of the model, since other different approaches of the gravity paradigm such as the radiation model, more recently introduced, have a similar behavior in terms of flows prediction at this scale~\citep{masucci2013gravity}, and we will precisely base our approach on potential at their origin.

}{
Nous travaillons sur des systèmes territoriaux simples supposés comme des systèmes de villes régionaux, dans lesquels les villes sont les entités de base. Nous avons ainsi de l'ordre d'une centaine de villes, et les territoires intermédiaires ne sont pas pris en compte. L'échelle de temps correspond à l'échelle caractéristique associée à cette échelle spatiale, i.e. autour d'un ou deux siècles. Les interactions spatiales sont capturées par des interactions de type gravitaire, cette formulation ayant l'avantage de la simplicité et de capturer la ``première loi de la géographie'' de \noun{Tobler}~\cite{tobler2004first}\footnote{``\textit{Tout interagit avec tout, mais deux choses plus proches auront plus de chances d'interagir}'', c'est-à-dire que la force d'interaction décroit avec la distance.}. Ce choix n'a a priori pas une influence fondamentale sur le comportement du modèle, puisque d'autres approches différentes du paradigme gravitaire comme le modèle de radiation, introduites plus récemment, ont un comportement similaire en termes de prédiction des flux à cette échelle~\cite{masucci2013gravity}, et nous nous baserons précisément sur des potentiels à l'origine de ceux-ci.
}

\paragraph{Description}{Description}

\bpar{
Let $\vec{P}(t)=(P_i(t))_{1\leq i\leq n}$ be the population of cities in time. We consider on a deterministic extension of the Gibrat model. We recall that the Gibrat model assumes independent growth rates: $P_i (t+1) = r\cdot P_i(t)$ with $r$ random variable, such that $\Covb{P_i(t)}{P_j(t)}=0$, i.e. that cities do not influence themselves mutually in their growth process.
}{
Soit $\vec{P}(t)=(P_i(t))_{1\leq i\leq n}$ le vecteur des populations des villes dans le temps. Nous considérons une extension déterministe du modèle de Gibrat. Nous rappelons que le modèle de Gibrat suppose des taux de croissance aléatoires indépendants : $P_i (t+1) = r\cdot P_i(t)$ avec $r$ variable aléatoire, telle que $\Covb{P_i(t)}{P_j(t)}=0$, c'est-à-dire que les villes ne s'influencent pas mutuellement dans leur processus de croissance.
}

\bpar{
A linear extended version of the Gibrat model, taking into account interactions between cities would then write
}{
Une version étendue linéaire du modèle de Gibrat, prenant en compte les interactions entre villes s'écrirait alors
}
\[
\vec{P}(t+1)=\mathbf{R}\cdot \vec{P}(t)
\]
\bpar{
where $\mathbf{R}$ is a random matrix independent of growth rates (proportional to identity in the original case).
}{
où $\mathbf{R}$ est une matrice aléatoire indépendante des taux de croissance (l'identité à un scalaire près dans le cas original).
}

\bpar{
This directly leads thanks to the independence assumption to $\Eb{\vec{P}(t+1)}=\Eb{\mathbf{R}}\cdot\Eb{\vec{P}(t)}$, what reduces to a deterministic formulation of the Gibrat model which is equivalent to consider only expectancies of population in time and not simulate random trajectories anymore.
}{
Cela conduit directement grâce à l'hypothèse d'indépendance à $\Eb{\vec{P}(t+1)}=\Eb{\mathbf{R}}\cdot\Eb{\vec{P}(t)}$, ce qui revient à une formulation déterministe du modèle de Gibrat qui est équivalente à considérer seulement les espérances des populations dans le temps et ne plus simuler des trajectoires aléatoires.
}

\bpar{
We generalize this linear relation to a non-linear relation that allows to be more consistent in the interaction functions. Denoting $\vec{\mu}(t)=\Eb{\vec{P}(t)}$, we write this relation with a given function $f$ and an arbitrary time step $\Delta t$, under the form
}{
Nous généralisons cette relation linéaire à une relation non-linéaire qui permet d'être plus flexible dans les fonctions d'interaction. Notant $\vec{\mu}(t)=\Eb{\vec{P}(t)}$, nous écrivons cette relation avec une fonction donnée $f$ et un pas de temps quelconque $\Delta t$, sous la forme 
}
\[
\vec{\mu}(t+\Delta t)=\Delta t\cdot f(\vec{\mu}(t))
\]
\bpar{
Note that in that case, stochastic and deterministic versions are not equivalent anymore, precisely because of the non-linearity, but we stick to a deterministic version for the sake of simplicity. The specification of the interdependent growth rate is given by
}{
Il faut noter que dans ce cas, les versions stochastiques et déterministes ne sont plus équivalentes, précisément à cause de la non-linéarité, mais nous gardons une version déterministe pour rester simple. La spécification des taux de croissance interdépendants est donnée par
}

\begin{equation}
f(\vec{\mu}) = (1+r_0)\cdot \mathbf{Id}\cdot \vec{\mu} + \mathbf{G}\left(\vec{\mu}\right)\cdot \vec{1} + \vec{N}\left(\vec{\mu}\right)
\label{eq:interactiongibrat:model}
\end{equation}

\bpar{
where $\vec{1}$ is the column vector full of ones, and $\mathbf{G} = G_{ij} = w_G\cdot \frac{V_{ij}}{<V_{ij}>}$ is the direct interaction term, such that the interaction potential $V_{ij}$ follows a gravity-type expression given by, with $d_{ij}$ distance between $i$ and $j$ (euclidian or network distance),
}{
où  $\vec{1}$ est le vecteur colonne unité, et $\mathbf{G} = G_{ij} = w_G\cdot \frac{V_{ij}}{<V_{ij}>}$ est le terme d'interaction directe, de telle façon que le potentiel d'interaction $V_{ij}$ suit une expression de type gravitaire donnée par, avec $d_{ij}$ distance entre $i$ et $j$ (distance euclidienne ou distance de réseau),
}

\begin{equation}
V_{ij} = \left(\frac{\mu_i\mu_j}{\left(\sum_k{\mu_k}\right)^2}\right)^{\gamma_G}\cdot \exp{\left(-d_{ij}/d_G\right)}
\end{equation}

\bpar{
where $\gamma_G$ is a hierarchy exponent of interactions regarding populations, $d_G$ a decay parameter giving the typical interaction distance, and $w_G$ is the weight relative to direct interactions.
}{
où $\gamma_G$ est un exposant de hiérarchie des interactions par rapport aux populations, $d_G$ un paramètre de décroissance donnant la distance typique d'interaction, et $w_G$ est le poids relatif des interactions directes.
}

\bpar{
The last term captures a network effect: $\vec{N}$ is given by $N_{i} = w_N \cdot \frac{W_i}{<W_i>}$ where the network flow potential $W_i$ reads
}{
Le dernier terme capture un effet de réseau : $\vec{N}$ est donné par $N_{i} = w_N \cdot \frac{W_i}{<W_i>}$ où le potentiel du flux de réseau $W_i$ suit
}

\begin{equation}
W_{i} = \sum_{k < l} \left(\frac{\mu_k\mu_l}{\left(\sum_j\mu_j\right)^2}\right)^{\gamma_N} \cdot \exp{\left(-d_{kl,i}/d_N\right)}
\end{equation}

\bpar{
where $d_{kl,i}$ is the distance of city $i$ to the shortest path between $k,l$ computed in the geographical space, which can be through a transportation network or in an impedance field of the euclidian space. Parameters $\gamma_N$, $d_N$ and $w_N$ are analogous to the parameters for direct interaction. The seven model parameters are detailed below and summarized in Table~\ref{tab:interactiongibrat:parameters}. We first precise the rationale of this formulation.
}{
où $d_{kl,i}$ est la distance de la ville $i$ au plus court chemin entre $k,l$ calculé dans l'espace géographique, qui peut être par un réseau de transport ou dans un champ d'impédance dans l'espace euclidien. Les paramètres $\gamma_N$, $d_N$ et $w_N$ sont analogues à ceux pour l'interaction directe. Les septs paramètres du modèle sont détaillés ci-dessous et récapitulés en Table~\ref{tab:interactiongibrat:parameters}. Précisons d'abord la logique de cette formulation.
}

\bpar{
The first term of the equation is the pure Gibrat model, that we obtain by setting the weights $w_G = w_N = 0$. The second component captures direct interdependencies between cities, under the form of a separable gravity potential such as the one used in~\cite{sanders1992systeme}. The rationale for the third term, aimed at capturing network effects by expressing a feedback of network flow between cities $k,l$ on the city $i$. Intuitively, a demographic and economic flow physically transiting through a city or in its surroundings is expected to influence its development (through intermediate stops e.g.), this effect being of course dependent on the transportation mode since a high speed line with few stops will skip most of the traversed territories. Note that we don't use exactly gravity flows in the network term, since there is no decay of interactions generating flows with distance, but a decay of the effect of the flow as a distance to the network. This is equivalent to assuming long-range use of the network on average in time, since the attenuation term goes to 1 if the decay parameter goes to infinity, and is this way complementary to the first gravity term.
}{
Le premier terme de l'équation est le modèle de Gibrat seul, qui est obtenu en fixant les poids $w_G = w_N = 0$. La deuxième composante capture les interdépendances directes entre les villes, sous la forme d'un potentiel gravitaire séparable comme celui utilisé dans~\cite{sanders1992systeme}. La logique du troisième terme, qui a pour but de capturer l'effet de réseau en exprimant une rétroaction des flux du réseau entre les villes $k,l$ sur la ville $i$. Intuitivement, un flux démographique et économique transitant physiquement par une ville ou dans son voisinage est attendu d'avoir une influence sur son développement (par des arrêts intermédiaires e.g.), cet effet étant bien sûr dépendant du mode de transport puisqu'une ligne à grande vitesse avec peu d'arrêts ignorera la majorité des territoires traversés. Notons que nous n'utilisons pas exactement les flux gravitaires dans le terme de réseau, puisqu'il n'y a pas de décroissance des interactions générant les flux avec la distance, mais une décroissance de l'effet du flux en fonction de la distance au réseau. Cela est équivalent à supposer une utilisation du réseau sur de très longues portées en moyenne dans le temps, puisque le terme d'attenuation tend vers 1 si le paramètre de décroissance tend vers l'infini, ce qui est ainsi complémentaire au premier terme de gravité.
}

\paragraph{Parameter space}{Espace des paramètres}

\bpar{
We give in Table~\ref{tab:interactiongibrat:parameters} the description of model parameters, detailing the associated processes and parameter ranges. Both direct interaction and second order network flows effect have the same structure, namely separability between effect of distance and population influence, an exponential decay parameter and a hierarchy parameter expressing the inequality of contribution depending on cities relative sizes: the highest the exponent, the more contribution of smaller cities will be negligible regarding larger cities. The distance decay parameter can be interpreted as a characteristic attenuation distance for interactions\footnote{It is possible to formally obtain its exact expression. Let fix an arbitrary fraction $\alpha$ and typical spatial ranges for a local urban system $d_L$ and for a long range urban system $d_R$, consider a city $i$ and two neighbors $j,j'$ with same population $\mu_j=\mu_j'$, at distances $d_L$ and $d_R$ of $i$ respectively. If we want to answer the question to what distance difference is equivalent an attenuation of $\alpha$ of the interaction potential with $i$, we obtain $d_L - d_R = -d_G\cdot \ln \alpha$. Therefore, $d_G$ is exactly the proportionality coefficient answering this intuitive request.}. Finally, we will consider only positive weights $w_G$ and $w_N$, to follow empirical observations as detailed below. Numerical values for the weights will be given normalized by number of cities implied in the process, i.e. ${w'}_G = w_G / n$ and ${w'}_N = w_N / (n (n-1) / 2)$.
}{
Nous donnons en Table~\ref{tab:interactiongibrat:parameters} la description des paramètres du modèle, détaillant les processus associés et les bornes des paramètres. Les interactions directes et les effets au second ordre des flux du réseau ont tous deux la même structure, c'est-à-dire le caractère séparable de l'effet de la distance et de l'influence des populations, un paramètre de décroissance exponentielle et un paramètre de hiérarchie exprimant l'inégalité des contributions selon les tailles relatives des villes : plus l'exposant est grand, plus les contributions des petites villes seront négligeables au regard des grandes villes. Le paramètre de décroissance de la distance peut s'interpréter comme une distance caractéristique d'atténuation des interactions\footnote{Il est possible d'établir formellement son expression exacte. Fixons une fraction arbitraire $\alpha$ et des portées spatiales typiques pour un système urbain local $d_L$ et pour un système urbain à longue portée $d_R$, considérons une ville $i$ et deux voisines $j,j'$ de population égale $\mu_j=\mu_j'$, à des distances respectives $d_L$ et $d_R$ de $i$. Si on veut répondre à la question à quelle différence de distance est équivalent une atténuation de $\alpha$ du potentiel d'interaction avec $i$, nous obtenons $d_L - d_R = -d_G\cdot \ln \alpha$. Pour cela, $d_G$ est exactement le coefficient de proportionnalité répondant à ce questionnement intuitif.}. Finalement, nous ne considérerons que des poids positifs, pour suivre les observations empiriques comme détaillé ci-dessous. Les valeurs numériques pour les poids seront données normalisées par le nombre de villes impliquées dans le processus, i.e. ${w'}_G = w_G / n$ et ${w'}_N = w_N / (n (n-1) / 2)$.
}

\begin{table}[ht]
\caption[Parameter space of the interaction model]{\textbf{Parameter space of the interaction model.} We give the parameters names, notations, associated processes, possible interpretations, and typical variation ranges.\label{tab:interactiongibrat:parameters}}
\bpar{
\begin{tabular}{|l|l|l|l|l|}
\hline
Parameter & Notation & Process & Interpretation & Range\\
\hline
Growth rate & $r_0$ & Endogenous growth & Urban growth & $\left[ 0,1\right]$ \\
Gravity weight & $w_G$ & Direct interaction & Maximal growth & $\left[ 0,1\right]$ \\
Gravity gamma & $\gamma_G$ & Direct interaction & Level of hierarchy & $\left[ 0,+\infty\right]$ \\
Gravity decay & $d_G$ & Direct interaction & Interaction range & $\left[ 0,+\infty\right]$ \\
Feedback weight & $w_N$ & Effect of flows &  Maximal growth & $\left[ 0,1\right]$ \\
Feedback gamma & $\gamma_N$ & Effect of flows & Hierarchy level & $\left[ 0,+\infty\right]$ \\
Feedback decay& $d_N$ & Effect of flows & Range of the effect & $\left[ 0,+\infty\right]$ \\
\hline
\end{tabular}
}{
\begin{tabular}{|l|l|l|l|l|}
\hline
Paramètre & Notation & Processus & Interpretation & Domaine\\
\hline
Taux de Croissance & $r_0$ & Croissance Endogène & Croissance Urbaine & $\left[ 0,1\right]$ \\
Poids gravitaire & $w_G$ & Interaction directe & Croissance maximale & $\left[ 0,1\right]$ \\
Gamma gravitaire & $\gamma_G$ & Interaction directe & Niveau de hiérarchie & $\left[ 0,+\infty\right]$ \\
Décroissance gravitaire & $d_G$ & Interaction directe & Portée d'interaction & $\left[ 0,+\infty\right]$ \\
Poids de la rétroaction & $w_N$ & Effet des flux & Croissance maximale & $\left[ 0,1\right]$ \\
Gamma de la rétroaction & $\gamma_N$ & Effet des flux & Niveau de hiérarchie & $\left[ 0,+\infty\right]$ \\
Décroissance de la rétroaction & $d_N$ & Effet des flux & Portée de l'effet & $\left[ 0,+\infty\right]$ \\
\hline
\end{tabular}
}
\end{table}

\subsubsection{Data}{Données}

\bpar{
Our model is constructed to be hybrid, as we propose to study it on a semi-parametrization on empirical data. It could be possible to study it as a full toy-model, initial configuration and physical environment being constructed as synthetic data. We however aim at unveiling stylized facts on real data rather than on model behavior in itself, and setup therefore the model from the data we now describe.
}{
Le modèle est construit pour être hybride, car nous proposons de l'étudier sur une semi-paramétrisation par données empiriques. Il pourrait être possible de l'étudier comme un modèle entièrement stylisé, la configuration initiale et l'environnement physique étant construits comme données synthétiques. Nous visons cependant à révéler des faits stylisés sur des données réelles plutôt que sur le comportement du modèle en lui-même, et initialisons ainsi le modèle à partir des données que nous décrivons à présent.
}

\paragraph{Population data}{Données de population}

\bpar{
We work with the Pumain-INED historical database for French Cities~\citep{pumain1986fichier}, which give populations of \emph{Aires Urbaines} (INSEE definition) at time intervals of 5 years, from 1831 to 1999 (31 observations in time). The latest version of the database integrates Urban Areas, allowing to follow them on long time-period, according to Bretagnolle's long time cities ontology~\cite{bretagnolle:tel-00459720}, which constructs a functional definition of cities as entities with boundaries evolving in time. To simplify, we work on the 50 bigger cities in 1999\footnote{This choice has only a low influence on most of trajectories of cities since small cities do not have much influence in interaction process. It can have some on the adjustment for the cities added, but our objective is not to reproduce exactly all trajectories but to understand the role of the network, we fix that threshold.}. We furthermore isolate periods of similar length excluding wars, obtaining 9 periods\footnote{Which are pprecisely : 1831-1851, 1841-1861, 1851-1872, 1881-1901, 1891-1911, 1921-1936, 1946-1968, 1962-1982, 1975-1999.} of 20 years on which semi-stationary in time fit of the model will be done. 
}{
Nous travaillons avec la base de données historique Pumain-INED pour les villes françaises~\cite{pumain1986fichier}, qui donne les populations des aires urbaines (définition de l'INSEE) à des intervalles de temps de 5 ans, de 1831 à 1999 (31 observations temporelles). La version la plus récente de la base de données intègre les aires urbaines, permettant de les suivre sur de longues périodes de temps, suivant l'ontologie de \noun{Bretagnolle} pour les villes sur le temps long~\cite{bretagnolle:tel-00459720}, qui construit une définition fonctionnelle des villes comme entités dont les limites évoluent dans le temps. Pour simplifier, nous travaillons avec les 50 plus grandes villes en 1999\footnote{Ce choix n'ayant que peu d'influence sur la majorité des trajectoires des villes puisque les petites villes influent peu dans les processus d'interaction. Il peut en avoir sur l'ajustement des villes ajoutées, mais notre objectif n'étant pas de reproduire fidèlement l'ensemble des trajectoires mais de comprendre le rôle du réseau, nous fixons ce seuil.}. Nous isolons de plus des périodes de longueur similaires excluant les guerres, obtenant 9 périodes\footnote{Qui sont précisément : 1831-1851, 1841-1861, 1851-1872, 1881-1901, 1891-1911, 1921-1936, 1946-1968, 1962-1982, 1975-1999.} de 20 ans sur lesquelles l'ajustement du modèle non-stationnaire dans le temps sera exécuté.
}

\paragraph{Physical flows}{Flux physiques}

\bpar{
As stated before, this modeling exercise focuses on exploring the role of physical flows, whatever the effective shape of the network. We choose for this reason not to use real network data which is furthermore not easily available at different time periods, and physical flows are assumed to take the geographical shortest path taking into account terrain slope. It avoids geographical absurdities such as cities with a difficult access having an overestimated growth rate. Using the IGN 1km resolution Digital Elevation Model, we compute shortest paths in a standard way~\cite{collischonn2000direction}, by the construction of an impedance field of the form
}{
Comme rappelé précédemment, cet exercice de modélisation se concentre sur l'exploration du rôle des flux physiques, quelle que soit la forme effective du réseau. Nous choisissons pour cette raison de ne pas utiliser de vraies données de réseau qui sont de plus difficiles à obtenir à différentes périodes de temps, et nous supposons que les flux physiques prennent le plus court chemin géographique prenant en compte la pente du terrain. Cela évite des absurdités géographiques comme des villes difficilement accessibles ayant un taux de croissance surestimé. Utilisant le modèle d'élévation numérique de l'IGN à la résolution 1km, nous calculons les plus courts chemins de manière standard~\cite{collischonn2000direction}, par la construction d'un champ d'impédance de la forme
}

\[
Z = \left(1 + \frac{\alpha}{\alpha_0}\right)^{n_0}
\]

\bpar{
where $Z$ is the impedance of links of the 1km grid network in which each cell is connected to its eight neighbors. $\alpha$ is the terrain slope computed with elevation difference between the two cells. We take fixed parameter values $\alpha_0 = 3$ (corresponding to approximatively the real world value of a 5\% slope) and $n_0 = 3$ which yielded more realistic paths than smaller or larger values\footnote{More precisely, we ``eyeball'' validated, by inspecting visually the paths between typical destinations (including Paris-Lyon, Lyon-Marseille, Lyon-Bordeaux for example), for $\alpha_0. =2,3,4$. For $\alpha_0 = 4$, the path is generally too much rectilinear and goes through mountains avoided by main links; for $\alpha_0 = 2$ the path makes on the contrary too much detours. We tested $n_0 = 2,3$, the second being also more reasonable. A precise calibration of these parameters would necessitate the adjustment with the freeway network for example, but is out of the scope of this exercise here.}.
}{
où $Z$ est l'impédance des liens du réseau de la grille de 1km dans laquelle chaque cellule est connectée à ses huit voisins. $\alpha$ est la pente du terrain calculée avec la différence d'altitude entre les deux cellules. Nous prenons des valeurs des paramètres fixes $\alpha_0 = 3$ (correspondant approximativement à la valeur réelle d'une pente de 5\%) et $n_0 = 3$ ce qui donne des chemins plus réalistes que des valeurs significativement plus petites ou plus grandes\footnote{Plus précisément, on a validé ``à dire d'expert'', en inspectant visuellement les chemins entre quelques destinations typiques (incluant Paris-Lyon, Lyon-Marseille, Lyon-Bordeaux par exemple), pour $\alpha_0 = 2,3,4$. Pour $\alpha_0 = 4$, le chemin est généralement trop rectiligne et coupe à travers des reliefs évités par les grands axes ; pour $\alpha_0 = 2$ le chemin est trop sinueux au contraire. On a testé $n_0 = 2,3$, le deuxième étant également plus crédible. Une calibration précise de ces paramètres nécessiterait l'ajustement par rapport au réseau autoroutier par exemple, mais est hors de portée de cet exercice ici.}.
}

\subsubsection{Performance indicators}{Indicateurs de performance}

\bpar{
We work on an explanatory rather than an exploratory model. For this reason, indicators to evaluate model outputs are not directly linked to intrinsic properties of trajectories or obtained final states, but rather to a distance to the phenomenon we want to explain, i.e. the data. Given real population $p_i(t)$ (historical realizations of $P_i(t)$) and simulated expected populations $\mu_i(t)$ obtained with $\vec{\mu}(t_0) = \vec{p}(t_0)$ on a period of length $T$, we can evaluate two complementary aspects of model performance:
}{
Nous travaillons sur un modèle explicatif plutôt qu'un modèle exploratoire. Pour cette raison, les indicateurs pour évaluer les sorties du modèle ne sont pas directement liés aux propriétés intrinsèques des trajectoires ou des états finaux obtenus, mais plutôt à une distance au phénomène que l'on cherche à expliquer, i.e. les données. Étant donné des populations réelles $p_i(t)$ (réalisations historiques de $P_i(t)$) et les espérances simulées $\mu_i(t)$ obtenues par $\vec{\mu}(t_0) = \vec{p}(t_0)$ sur une période de longueur $T$, on peut évaluer deux aspects complémentaires de la performance du modèle :
}

\bpar{
\begin{itemize}
\item Overall model performance, given by logarithm of the mean-square error in space and time
\[
\varepsilon_G = \ln{\left(\frac{1}{T}\sum_t \frac{1}{n} \sum_i \left(p_i (t) - \mu_i (t) \right)^2\right)}
\]
\item Average local model performance, given by the mean-square error on logarithms, as proposed by~\cite{pumain2017incremental}
\[
\varepsilon_L = \frac{1}{T}\sum_t \frac{1}{n} \sum_i \left(\ln p_i(t) - \ln \mu_i (t)\right)^2
\]
\end{itemize}
}{
\begin{itemize}
\item Performance globale du modèle, donnée par le logarithme de l'erreur carrée moyenne dans l'espace et le temps
\[
\varepsilon_G = \ln{\left(\frac{1}{T}\sum_t \frac{1}{n} \sum_i \left(p_i (t) - \mu_i (t) \right)^2\right)}
\]
\item Performance locale moyenne, donnée par l'erreur carrée moyenne des logarithmes, comme proposé par~\cite{pumain2017incremental}
\[
\varepsilon_L = \frac{1}{T}\sum_t \frac{1}{n} \sum_i \left(\ln p_i(t) - \ln \mu_i (t)\right)^2
\]
\end{itemize}
}

\bpar{
Both are actually complementary, as using only $\varepsilon_G$ will focus only on larger cities and give mitigate results on medium-sized and small cities (for France only Paris will have reasonable fit as it strongly dominates other urban areas and cities). $\varepsilon_L$ allows therefore to take into account model performance in all cities simulated by the model.
}{
Les deux sont en fait complémentaires, puisque utiliser seulement $\varepsilon_G$ se concentrera seulement sur les plus grandes villes et donnera des résultats mitigés sur les villes de taille moyenne et les petites villes (pour la France seul Paris aura une estimation raisonnable comme il domine fortement les autres aires urbaines et villes en termes de population). $\varepsilon_L$ permet pour cela de prendre en compte la performance du modèle sur l'ensemble des villes simulées par le modèle.
}

\subsubsection{Results}{Résultats}

\paragraph{Stylized facts}{Faits stylisés}

\bpar{
Basic stylized facts can be extracted from such a database, as it has already been widely explored in the literature~\cite{guerin1990150}. We retrieve better fits of log-normal distributions of growth rates at all dates compared to normal fits, at the exception of growth rates for the interval 1886-1891. We also retrieve the fact that growth rates are mainly positive, on the cities we consider and when removing wars: when removing years 1872, 1921 and 1946, the quantile corresponding to a relative growth rate of zero as a median on all dates corresponding to the 13th quantile.
}{
Des faits stylisés typiques peuvent être extraits d'une telle base de données, comme il a déjà largement été exploré dans la littérature~\cite{guerin1990150}. Nous retrouvons les meilleurs ajustements de distributions log-normales des taux de croissance à toutes les dates en comparaison à des distributions normales, à l'exception des taux de croissance pour l'intervalle 1886-1891. Nous retrouvons également le fait que les taux de croissance sont essentiellement positifs, sur les villes que nous considérons et en enlevant les guerres : en ne considérant pas les années 1872, 1921 et 1946, le centile correspondant à un taux de variation relatif nul a une médiane sur l'ensemble des dates correspondant au 13ème centile.
}


\bpar{
An interesting feature to look at in relation with our considerations on spatial interactions are correlations between time-series, and more particularly their variation as a function of distance. We consider 50 years overlapping time-windows to have enough temporal observation, finishing respectively in (1881,1906,1931,1962,1999), and estimate on each, for each couple of cities $(i,j)$, the correlation between \emph{log-returns}, that we define by $\Delta X_i = X_i(t) - X_i(t-1)$ and $X_i(t) = \ln\left(\frac{P_i(t)}{P_i(t_0)}\right)$, is given by $\hat{\rho}_{ij}=\rho\left[\Delta X_i, \Delta X_j\right]$ with a classical Pearson estimator. This method allows to reveals dynamical interactions without being biased by sizes~\cite{mantegna1999introduction}.
}{
Un aspect intéressant à examiner en relation avec nos considérations sur les interactions spatiales sont les corrélations entre les séries temporelles, et plus particulièrement leur variation en fonction de la distance. Nous considérons des fenêtres temporelles de 50 ans se superposant pour avoir assez d'observations temporelles, finissant respectivement en (1881,1906,1931,1962,1999) et estimons sur chacune, pour chaque couple de villes $(i,j)$, la corrélation entre les \emph{log-returns}, que nous définissons par $\Delta X_i = X_i(t) - X_i(t-1)$ et $X_i(t) = \ln\left(\frac{P_i(t)}{P_i(t_0)}\right)$, est donnée par $\hat{\rho}_{ij}=\rho\left[\Delta X_i, \Delta X_j\right]$ avec un estimateur de Pearson classique. Cette méthode permet de révéler des interactions dynamiques sans être biaisée par les tailles~\cite{mantegna1999introduction}.
}

\begin{figure}
\includegraphics[width=\linewidth]{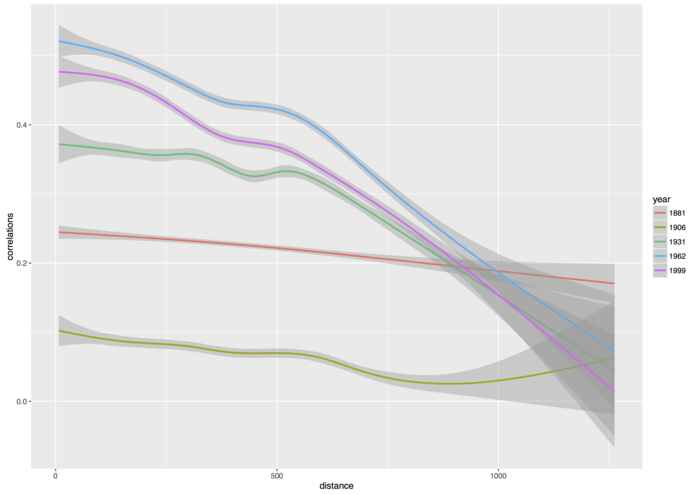}
\caption[Correlation of French population growth rates as a function of distance]{\textbf{Time-series correlations as a function of distance.} Solid line correspond to smoothed correlations, computed between each pairs $\Delta X_i,\Delta X_j$ for $i\neq j$, on successive periods given by the color of the curve.\label{fig:interactiongibrat:ts-correlations}}
\end{figure}

\bpar{
We show in Figure~\ref{fig:interactiongibrat:ts-correlations} the smoothed correlations curves as a function of distance, for each time period. These allow to establish the relation between the distance between cities and the effective interactions between these: two time-series strongly correlated will be interpreted as a strong interaction between the two cities.
}{
Nous montrons en Fig.~\ref{fig:interactiongibrat:ts-correlations} les courbes de corrélations lissées en fonction de la distance, pour chaque période temporelle. Celles-ci permettent d'établir la relation entre la distance entre villes et les interactions effectives entre celles-ci : deux séries temporelles fortement corrélées s'interpréteront comme une forte interaction entre les deux villes.
}

\bpar{
First of all, the strong differences between each confirms the non-stationarity of growth rates over the whole time period, and justifies the use of local fit in time for the model. We can also interpret these patterns in terms of historical events for the system of city and the transportation network. System dynamic begins with a flat correlation in 1881, around 0.2, that could be spurious due to simultaneous similar growth for all cities. It then stays flat but goes to zero, witnessing strong differentiations in growth patterns between 1856 and 1906. After 1931, the effect of the distance is clear with decreasing curves, starting between 0.4 and 0.5. We postulate that this evolution must be partly linked to transportation network evolution: considering railway network for example~\citep{thevenin2013mapping}, the initial overall development may have fostered long range interactions flattening thus the correlation curves, whereas its maturation over time has conducted to the return of more classical interactions decreasing quickly with distance.
}{
Tout d'abord, les fortes différences entre chaque courbe confirment la non-stationnarité des taux de croissance sur l'ensemble de la période, et justifient l'utilisation d'ajustements locaux dans le temps pour le modèle. Nous pouvons aussi interpréter ces motifs en termes d'événements historiques pour le système de villes et le réseau de transport. La dynamique du système commence par une corrélation plate en 1881, autour de 0.2, qui pourrait être fortuite à cause de croissance similaire simultanée pour toutes les villes. Elle reste ensuite plate mais tend vers 0, témoignant de fortes différentiations dans les motifs de croissance entre 1856 et 1906. Après 1931, l'effet de la distance se manifeste par des courbes décroissantes, commençant entre 0.4 et 0.5. Nous postulons que cette évolution doit être partiellement liée à l'évolution du réseau de transport : en considérant le réseau ferré par exemple~\cite{thevenin2013mapping}, le développement initial global a pu encourager des interactions à longue portée rendant ainsi les courbes de corrélation plates, tandis que sa maturation dans le temps a conduit au retour d'interactions plus classiques décroissant rapidement avec la distance.
}

\paragraph{Model exploration}{Exploration du modèle}


\bpar{
Data preprocessing, result processing and models profiling are implemented in \texttt{R}. For performances reasons and an easier integration into the OpenMole software for model exploration~\citep{reuillon2013openmole}, a \texttt{scala} version was also developed. The question of trade-off between implementation performance and interoperability is a typical issue in this kind of model, as a fully blind exploration and calibration can be misleading for further research directions or thematic interpretations. A NetLogo implementation, allowing interactive exploration and dynamical visualization, was also developed for this reason. Source code for models, cleaned raw data, simulation data, and results used here are available on the open repository of the project\footnote{At \url{https://github.com/JusteRaimbault/CityNetwork/tree/master/Models/InteractionGibrat}. The three versions of the model are aimed at being fetch, reused and extended, and each implementation has its own utility: the \texttt{R} version allow a direct integration with data analysis scripts, the \texttt{scala} version can be used as an OpenMole plugin, and the NetLogo version allows an interactive exploration and the direct visualization of trajectories.}.
}{
La pré-traitement des données, le traitement des résultats et le profilage des modèles sont implémentés en \texttt{R}. Pour des raisons de performance et une intégration plus facile dans le logiciel OpenMole pour l'exploration de modèles~\cite{reuillon2013openmole}, une version \texttt{scala} a également été développée. La question du compromis entre performance d'implémentation et interopérabilité est un problème typique de ce genre de modèle, puisque des explorations et calibrations totalement aveugles peuvent être trompeuses pour les directions de recherches futures ou les interprétations thématiques. Une implémentation NetLogo, permettant l'exploration interactive et la visualisation dynamique, a également été développée pour cette raison. Le code source des modèles, les données brutes nettoyées, les données de simulation, et les résultats utilisés ici sont disponibles sur le dépôt ouvert du projet\footnote{À \url{https://github.com/JusteRaimbault/CityNetwork/tree/master/Models/InteractionGibrat}. Les trois versions du modèle sont destinées à être reprises, réutilisées et étendues, et chaque implémentation a son utilité propre : la version \texttt{R} permet une intégration directe avec des scripts d'analyse de données, la version \texttt{scala} peut être utilisée comme plugin OpenMole, et la version NetLogo permet une exploration interactive et la visualisation directe des trajectoires.}.
}

\begin{figure}
\includegraphics[width=\linewidth]{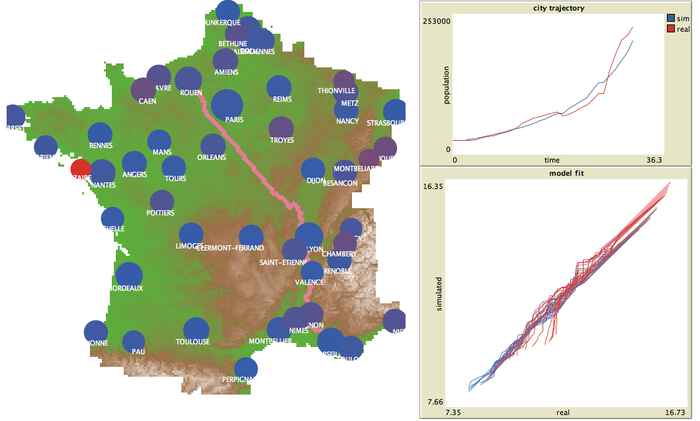}
\caption[Interface of the interaction model]{\textbf{Example of output of the model.} The graphical interface allows to explore interactively on which cities changes operate after a parameter change, what is necessary to interpret raw calibration results. The map gives adjustment errors by city (color) and their population (size). We illustrate in pink the geographical shortest path between Rouen and Marseille. The plot in top right panel gives in time the trajectory of a selected city, comparing simulated population with real population. The bottom right plot gives for each date the simulated data against real data: the closest the curve is to the diagonal, the better is the fit.\label{fig:interactiongibrat:interface}}
\end{figure}

\bpar{
 We show in Figure~\ref{fig:interactiongibrat:interface} an example of model output. Cities color give city-level fit error and their size the population. Outliers can therefore easily be spotted (as Saint-Nazaire having the worst fit in the example shown) and possible regional effects identified. We illustrate in pink an example of geographical shortest path, from Rouen to Marseille, which reasonably corresponds to the actual current shortest path by highway. Top right plot shows trajectory in time for a given city, whereas the bottom right plot gives overall fit quality in time, by plotting simulated data against real data. The closest the curve is from the diagonal, the better the fit.
}{
Nous montrons en Fig.~\ref{fig:interactiongibrat:interface} un exemple de sortie du modèle. Les couleurs des villes donnent l'écart à l'observation au niveau de la ville et leur taille la population. Les valeurs extrêmes peuvent ainsi être aisément repérées (comme Saint-Nazaire ayant le plus mauvais ajustement dans l'exemple montré) et des possibles effets régionaux identifiés. Nous illustrons en rose un exemple de plus court chemin géographique, de Rouen à Marseille, qui correspond raisonnablement au plus court chemin effectif actuel par autoroute. Le graphe du haut montre la trajectoire dans le temps pour une ville donnée, tandis que celui du bas donne la qualité globale de l'ajustement dans le temps, en traçant les données simulées en fonction des données réelles. Plus la courbe est proche de la diagonale, meilleur est l'ajustement.
}

\begin{figure}
\includegraphics[width=\linewidth]{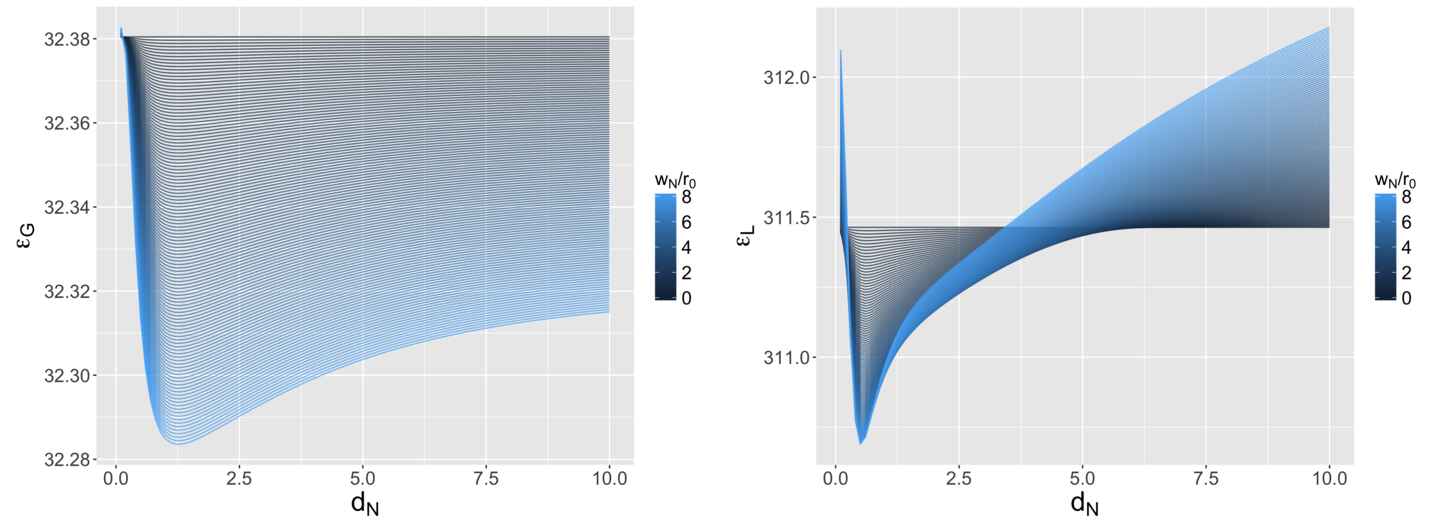}
\caption[Evidence of network effects]{\textbf{Evidence of network effects revealed by model exploration.} Left plot gives $\varepsilon_G$ as a function of $d_N$ for varying $r_0/w_N$, at fixed gravity effect and $\gamma_N=3$. Right plot is similar for $\varepsilon_L$. Starting from a pure gravity model (horizontal curve for $w_N$), progressively taking into account the network increases performances regarding the two objectives, in a restricted range for $d_N$. Values of $d_N$ giving the minima correspond to the typical distance of the network effect.\label{fig:interactiongibrat:networkeffects}}
\end{figure}

\bpar{
First model explorations, by simply sweeping fixed grids of the parameter space, already suggest the presence of network effects, in the sense that physical flow effectively have an influence on growth rates. We show in Figure~\ref{fig:interactiongibrat:networkeffects} a configuration of such a case. At fixed gravity parameters and growth rate, we study variations of the parameters $w_N, d_N$ and $\gamma_N$ and the corresponding response of $\varepsilon_G$ and $\varepsilon_L$. At fixed values of $\gamma_N$, we observe similar behaviors of the indicators when $w_N$ and $d_N$ change. The existence of a minimum of both as a function of $d_N$, that becomes stronger when $w_N$ increases, shows that introducing the network feedback terms improves local and global fits compared to the gravity model alone, i.e. that the associated process have potential explanatory power for growth patterns.
}{
Les premières explorations du modèle, en parcourant simplement des grilles fixées de l'espace des paramètres, suggèrent déjà la présence d'effets de réseau, au sens de flux physiques ayant effectivement une influence sur la reproduction des taux de croissance observés. Nous montrons en Fig.~\ref{fig:interactiongibrat:networkeffects} une configuration dans laquelle c'est le cas. À paramètres de gravité et taux de croissance fixés, nous étudions les variations des paramètres $w_N, d_N$ et $\gamma_N$ et la réponse correspondante de $\varepsilon_G$ et $\varepsilon_L$. À des valeurs fixes de $\gamma_N$, on observe un comportement similaire des indicateurs quand $w_N$ et $d_N$ varient. L'existence d'un minimum pour les deux comme fonction de $d_N$, qui devient plus marqué quand $w_N$ augmente, montre que l'introduction du terme de rétroaction du réseau améliore les ajustements locaux et globaux en comparaison du modèle de gravité seul, i.e. que les processus associés ont un pouvoir explicatif pour les motifs de croissance.
}

\paragraph{Calibrating the gravity model}{Calibration du modèle de gravité}

\begin{figure}
\includegraphics[width=\linewidth]{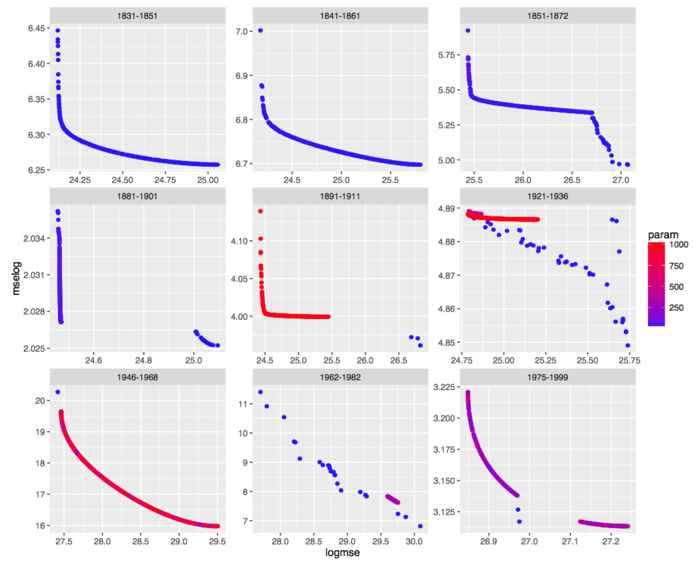}
\caption[Calibrating the gravity model]{\textbf{Calibrating the gravity model.} Pareto-front on successive periods. Color level gives the value of distance decay parameter.\label{fig:interactiongibrat:gravity-pareto}}
\end{figure}

\begin{figure}
\includegraphics[width=\linewidth]{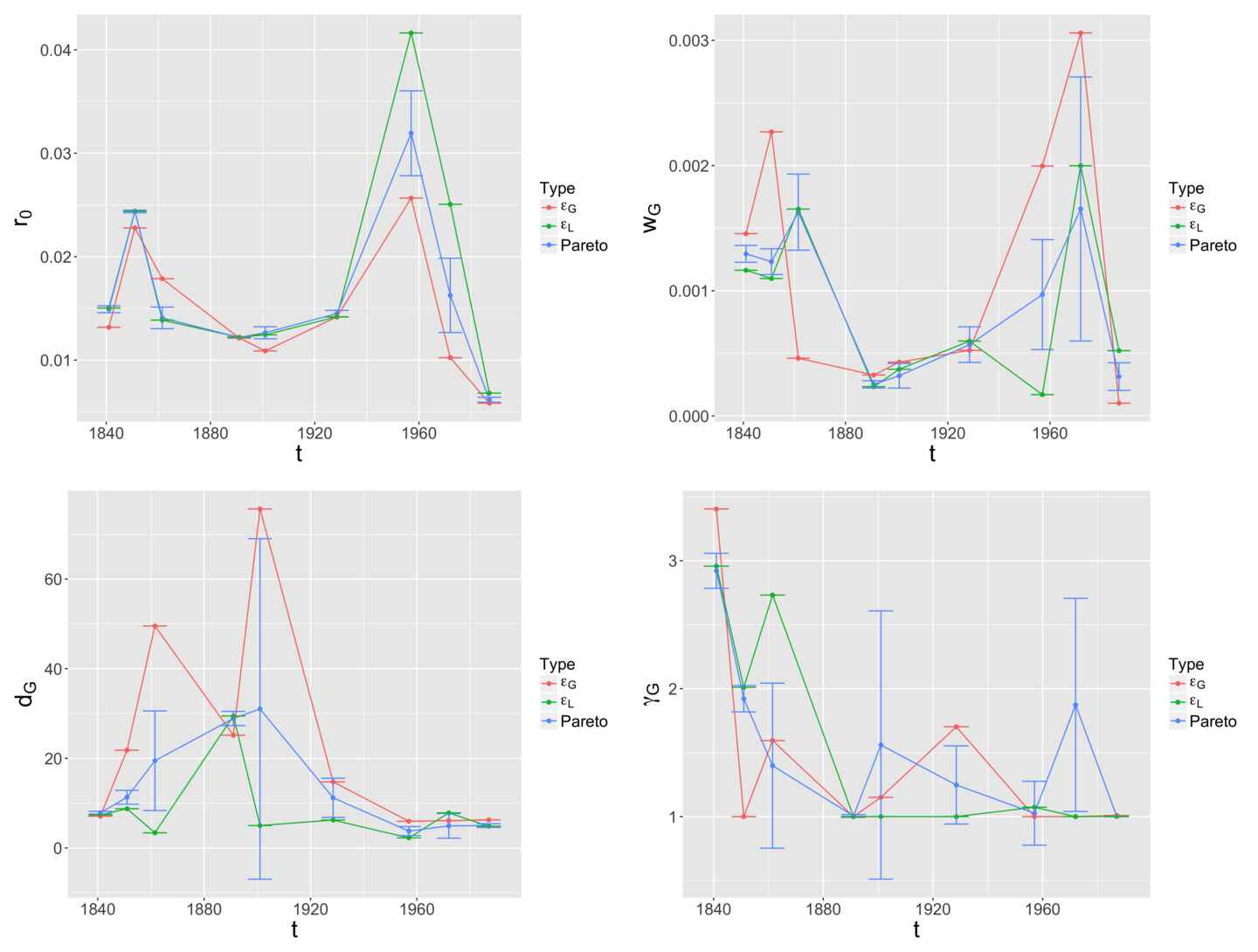}
\caption[Values of the calibrated parameters for the gravity model]{\textbf{Values of calibrated parameters for the gravity model only.} Each plot gives fitted values in time for each parameter. Red and Green curves correspond to best points for $\varepsilon_G$ (respectively $\varepsilon_L$), whereas the blue curves give the average value over the Pareto front with standard deviation. Depending if we are interested in the performance on small cities ($\varepsilon_L$), larges cities ($\varepsilon_G$), or compromises (\texttt{Pareto}), we will obtain an evolution of parameters that can differ. Large trends can be exhibited, such as the double peak of the endogenous growth rate ($r_0$) which is found also in the weight of interactions, and a decreasing hierarchy of interactions ($\gamma_G$).\label{fig:interactiongibrat:gravity-params}}
\end{figure}

\bpar{
We now use the model to indirectly extract information on processes in time. Indeed under assumption of non-stationarity, temporal evolution of locally fitted parameters show the evolution of the corresponding aspect of the processes. In a first experiment, we set $w_N=0$ and calibrate the model with four parameters on the 9 successive 20 years time windows. The optimization problem associated to model calibration does not present features allowing an easy solving (closed-form of a likelihood function, convexity or sparsity of the optimization problem), we must rely on alternative techniques to solve it. ``Brute force'' grid search\footnote{We mean by that a systematic exploration of a grid of the parameter space.} is rapidly limited by the dimensionality curse. Classical methods~\citep{batty1972calibration} such as gradient descent fail because of the rather complicated shape of the optimisation landscape.
}{
Nous utilisons à présent le modèle pour extraire de l'information de manière indirecte sur les processus dans le temps. En effet, sous l'hypothèse de non-stationnarité, l'évolution temporelle des paramètres ajustés localement montre l'évolution de l'aspect des processus correspondant. Dans une première expérience, nous fixons $w_N=0$ et calibrons le modèle avec quatre paramètres sur les neuf périodes temporelles successives de 20 ans. Le problème d'optimisation associé à la calibration du modèle ne présente pas de caractéristiques le rendant agréable à résoudre (expression fermée d'une fonction de likelihood, convexité ou caractère creux du problème d'optimisation), nous devons nous reposer sur des techniques alternatives pour le résoudre. Une exploration de grille par algorithme dit ``de force brute''\footnote{Nous entendons par là une exploration systématique d'une grille de l'espace des paramètres.} est rapidement limitée par le sort de la dimension. Les méthodes classiques~\cite{batty1972calibration} comme une descente du gradient échouent à cause de la forme assez compliquée du paysage d'optimisation.
}

\bpar{
Calibration using Genetic Algorithms (GA) are an efficient solution to find approximate solutions in a reasonable time. OpenMole embeds a collection of such meta-heuristics for different purposes: \cite{schmitt2014half} demonstrates the capabilities of these methods to calibrate models of simulation. In our case, it furthermore allow to do a bi-objective calibration on $(\varepsilon_G,\varepsilon_L)$. We use the standard steady state GA provided by OpenMole, distributed on 25 islands, with population of 200 and 100 generations\footnote{See~\cite{pumain2017urban} for the presentation of the most recent calibration method for urban models with Genetic Algorithms provided by OpenMole, as we also described with more details in~\ref{ch:preliminarymath}. The island distribution scheme make evolve independent instances of the algorithm on computation nodes, and regularly merges the results obtained on each.}.
}{
La calibration par Algorithme Génétique (GA) est une solution efficace pour trouver des solutions approximatives en un temps raisonnable. OpenMole inclut une collection de telles méta-heuristiques pour différents buts : \cite{schmitt2014half} démontre les potentialités de ces méthodes pour calibrer les modèles de simulation. Dans notre cas, cela permet de plus de procéder à une optimisation bi-objectif sur $(\varepsilon_G,\varepsilon_L)$. Nous utilisons le \emph{GA steady state standard} fournit par OpenMole, distribué sur 25 îles, avec une population de 200 individus et 100 générations\footnote{Voir~\cite{pumain2017urban} pour la présentation la plus récente des méthodes de calibration de modèles urbains par algorithme génétique fournit par OpenMole, comme nous l'avons également décrit plus en détails en~\ref{ch:preliminarymath}. Le schéma par distribution sur îles fait évoluer des instances indépendantes de l'algorithme sur des noeuds de calcul, puis fusionne régulièrement les résultats obtenus sur chaque.}.
}

\bpar{
We show in Figure~\ref{fig:interactiongibrat:gravity-pareto} the calibration results on successive periods, by plotting final population in the indicator space. As expected, Pareto fronts that corresponds to compromises between the two opposite objectives are the rule. It means that the model cannot be accurate both globally and locally, and an intermediate solution has to be found. This may due to the fact that interaction range changes with city size (i.e. that terms in the potential are no longer separable), that we keep as a possible model development. The shape of the Pareto front are revealing the chaotic optimisation landscape, as for some periods such as 1921-1936 or 1962-1982 fronts are not regular and sparse. The change in shapes also translates different dynamical regimes across the periods: for 1881-1901, the quasi-vertical shape followed by an isolated front at high $\varepsilon_G$ values reveals a quasi-binary model behavior in the optimal regimes, in the sense that improving $\varepsilon_L$ under the limit is only possible through a qualitative jump at a high price for $\varepsilon_G$.
}{
Nous montrons en Fig.~\ref{fig:interactiongibrat:gravity-pareto} les résultats de la calibration sur les périodes successives, en représentant la population finale dans l'espace des indicateurs. Comme attendu, des fronts de Pareto correspondant à des compromis entre les deux objectifs opposés sont la règle. Cela signifie que le modèle ne peut pas être précis à la fois globalement et localement, et qu'une solution intermédiaire doit être trouvée. Cela peut être dû au fait que la portée d'interaction change avec la taille de la ville (i.e. que les termes dans le potentiel ne sont plus séparables), que nous gardons comme un développement potentiel du modèle. La forme des fronts de Pareto révèle un paysage d'optimisation chaotique, puisque pour certaines périodes comme 1921-1936 ou 1962-1982 les fronts ne sont pas réguliers et éparpillés. Le changement dans les formes traduit également différents régimes dynamiques selon les périodes. Par exemple, pour la période 1881-1901, la forme quasi-verticale dans les faibles valeurs de $\varepsilon_G$, suivie par un front isolé à de fortes valeurs de $\varepsilon_G$ révèle un comportement quasi-binaire du modèle dans les régimes optimaux. En effet, l'amélioration de $\varepsilon_L$ sous la limite du premier sous-front est possible uniquement à travers un saut qualitatif à un fort coût pour $\varepsilon_G$.
}

\bpar{
The values taken by $d_G$ for periods 1892-1911 and 1921-1936 show that larger cities have longer interaction range, as high value give better values of $\varepsilon_G$. We show in Figure~\ref{fig:interactiongibrat:gravity-params} the values of fitted parameters in time, averaged over the Pareto front and for best single-objective solutions. First, the two peaks patterns for $r_0$ corresponds roughly to the patterns observed in average growth rates. The evolution of $w_G$ has a similar shape but lagged by 20 years: it can be interpreted as a repercussion of endogenous growth on interaction patterns in the following years, which is consistent with an interpretation of the interaction process in terms of migration. The values of $d_G$, with an increase until 1900 followed by a progressive decrease, is consistent with the behavior of empirical correlations commented above: the first 50 years windows have higher interaction range what corresponds to flat correlation curves. Finally, the level of hierarchy $\gamma_G$ has regularly decreased, corresponding to an attenuation of the power of large cities that can be understood in terms of progressive decentralization in France that has been fostered by the administration\footnote{Keeping in mind that reality is naturally much more complex and that such a trend could also be lead by a more global integration within a change in nature of urban structures, witnessed for example by the emergence of mega-city regions that we introduced in~\ref{sec:casestudies}.}.
}{
Les valeurs prises par $d_G$ pour les périodes 1892-1911 et 1921-1936 montrent que les grandes villes ont des portées d'interaction plus grandes, puisqu'une valeur plus grande donne des meilleurs valeurs pour $\varepsilon_G$. Nous montrons en Fig.~\ref{fig:interactiongibrat:gravity-params} les valeurs des paramètres ajustés dans le temps, par leur moyenne sur le front de Pareto et pour les deux meilleures solutions à objectif simple. Tout d'abord, les deux motifs en pic pour $r_0$ correspondent globalement au comportement observé sur les taux de croissance moyens. L'évolution de $w_G$ a une forme similaire mais décalée de 20 ans : cela peut être interprété comme une répercussion de la croissance endogène sur les motifs d'interaction les années suivantes, ce qui est cohérent avec une interprétation des processus d'interaction en termes de migration. Les valeurs de $d_G$, avec une augmentation jusqu'en 1900 suivie d'une décroissance progressive, sont cohérentes avec le comportement des corrélations empiriques commenté précédemment : les deux premières fenêtres de 50 ans ont des portées d'interaction plus grandes ce qui correspond à des courbes de corrélations plates. Enfin, le niveau de hiérarchie $\gamma_G$ a été régulièrement décroissant, que nous lisons comme une atténuation du pouvoir des grandes villes, qui peut être comprise en termes de la décentralisation progressive en France qui a été encouragée par l'administration\footnote{Sachant que la réalité est forcément plus complexe et qu'une telle tendance peut être tirée aussi par une inscription plus globale dans un changement de nature des structures urbaines, dont témoigne par exemple l'émergence des méga-régions urbaines que nous avons introduit en~\ref{sec:casestudies}.}.
}

\paragraph{Network Effects}{Effets de Réseau}

\begin{figure}
\includegraphics[width=\linewidth]{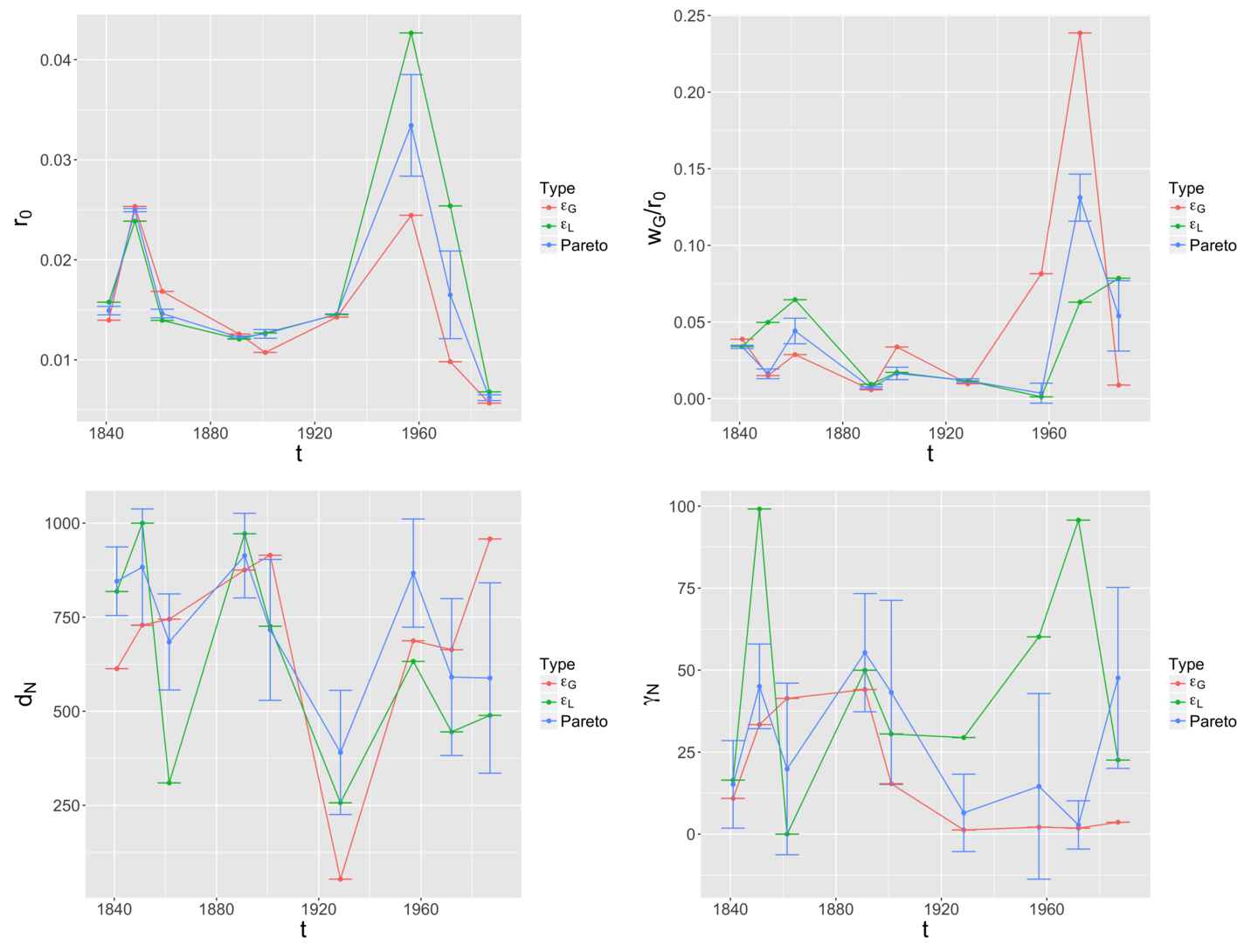}
\caption[Full model calibration]{\textbf{Calibrated parameters for the full model.} We plot values of $r_0, w_G/r_0, d_N$ and $\gamma_N$ in time, for single-objective optimal points (Red and Green curves) and averaged over the Pareto front (Blue).\label{fig:interactiongibrat:feedback}}
\end{figure}

\bpar{
We now turn to the calibration of the full model on successive periods, in order to interpret parameters linked to network flows and gain insight into network effects. The full calibration is done in a similar way with seven parameters being free. We plot in Figure~\ref{fig:interactiongibrat:feedback} the fitted values in time for some of these parameters. The behavior of growth rate and of the gravity weight relative to growth rate, that is similar to the gravity model only, confirms that network effects are well at the second order and that endogenous growth and direct interactions are main driver. Network effects are however not negligible, as they improve the fit as shown before in model exploration, capturing therein second order processes. The evolution of $d_N$, corresponding to the range on which network influences the territories it goes through, shows a minimum in 1921-1936 to stabilize again later, but at a value lower that past values. This could correspond to the ``tunnel effect'', when high-speed transportation do not stop much. Indeed, the evolution of railway has witnessed a high decrease in local lines at a date similar to the minimum, and later the emergence of specific High Speed lines, explaining this lower final value. Hierarchy of flows have slightly decreased as for gravity, but are extremely high. This means that only flows between larger cities have a significant effect. This way, the model gives indirect information on the processes linked to network effects.
}{
Nous nous intéressons à présent à la calibration du modèle complet sur des périodes successives, dans le but d'interpréter les paramètres liés aux flux de réseau et obtenir des informations sur les effets de réseau. La calibration complète est faite de manière similaire avec les sept paramètres libres. Nous montrons en Fig.~\ref{fig:interactiongibrat:feedback} les values ajustées dans le temps pour certains de ces paramètres. Le comportement du taux de croissance et du poids de la gravité relatif au taux de croissance, qui est similaire au modèle de gravité seul, confirme que les effets de réseau sont bien au second ordre et que la croissance endogène et les interactions directes sont les facteurs principaux. Les effets de réseaux sont cependant loin d'être négligeables, puisqu'ils améliorent l'ajustement comme montré précédemment lors de l'exploration du modèle, capturant ainsi des processus de second ordre. L'évolution de $d_N$, correspondant à la portée sur laquelle le réseau influence le territoire qu'il traverse, montre un minimum en 1921-1936 pour se stabiliser à nouveau plus tard, mais à une valeur plus basse que les valeurs du passé. Cela pourrait correspondre à l'effet tunnel, puisque les transports à grande vitesse s'arrêtent peu sur les territoires qu'ils traversent. En effet, l'évolution du réseau ferré a témoigné d'une forte décroissance des lignes locales à une date similaire au minimum, et plus tard l'émergence de lignes à grande vitesse spécifiques, ce qui expliquerait cette valeur finale plus basse. La hiérarchie de l'effet de réseau a été légèrement décroissante comme pour la gravité, mais est extrêmement haute. Cela signifie que seuls les flux entre les grandes villes ont un effet significatif. Ainsi, le modèle donne une information indirecte sur les processus liés aux effets de réseau.
}

\bpar{
We retain from the calibration of the full model the following stylized facts.
\begin{itemize}
	\item Network effects are captured at the second order by the model.
	\item Variations in the range of the network effect suggest the emergence of the tunnel effect.
	\item Main flows largely dominate in the network effect.
\end{itemize}
}{
Nous retenons de la calibration du modèle complet les faits stylisés suivants.
\begin{itemize}
	\item Des effets des réseaux sont capturés au second ordre par le modèle.
	\item Les variations de la portée de l'effet du réseau suggèrent l'émergence de l'effet tunnel.
	\item Les flux principaux dominent largement dans l'effet de réseau.
\end{itemize}
}

\paragraph{Estimating the compromise between fitting power and number of parameters}{Estimer le compromis entre puissance d'ajustement et nombre de paramètres}

\bpar{
We focus in this last experiment on quantifying the ``performance'' of the model, taking into account its predictive abilities, but also its structure. More precisely, we want to tackle the issue of overfitting, which has been for long recognized in Machine Learning for example~\cite{dietterich1995overfitting}, but for which there is a lack of methods for models of simulation\footnote{This question is indeed at the heart of the understanding of complexity: the issue is to find a dimension capturing the main part of system dynamics at a given level, what is similar to determine a level of simplification which captures the aggregated dynamics, and thus emergence. There could furthermore exist a link between this question and the problem of determining the \emph{embedding dimension} for time series, which is equivalent to finding an effective dimension of the phase space of a system.}. We need to introduce a tool to confirm that the improvement in model fit is not only artificially due to additional parameters. This stage is important at this stage of introduction of preliminary models: the question is indeed to construct relevant bricks but also simple.
}{
Nous visons dans cette dernière expérience à quantifier la ``performance'' du modèle, prenant en compte ses capacités prédictives, mais aussi sa structure. Plus précisément, nous voulons traiter la question du sur-ajustement (\textit{overfitting}), qui a été reconnue depuis un certain temps en apprentissage statistique par exemple~\cite{dietterich1995overfitting}, mais pour lequel il manque des méthodes applicables aux modèles de simulation\footnote{Cette question est en fait au coeur de la compréhension de la complexité : il s'agit de trouver une dimension qui capture la plus grande partie des dynamiques du système à un niveau donné, ce qui revient à déterminer un niveau de simplification qui capture les dynamiques agrégées, et donc l'émergence. Il pourrait d'ailleurs exister un lien entre cette question et le problème de la détermination de l'\emph{embedding dimension} pour les séries temporelles, qui revient à trouver une dimension effective de l'espace des phases d'un système.}. Nous avons besoin d'introduire un outil qui confirme que l'amélioration de l'ajustement n'est pas uniquement artificiellement due aux paramètres supplémentaires. Cette étape est importante à ce stade d'introduction de modèles préliminaires : il s'agit en effet de construire des briques pertinentes mais également relativement simples.
}

\bpar{
The Akaike Information Criterion (AIC) provides for statistical models for which a likelihood function is available the gain in information between two models~\citep{akaike1998information}, correcting fit improvement for number of parameters. Similar methods include the Bayesian Information Criterion (BIC), which relies on slightly different assumptions and corrects differently. \cite{biernacki2000assessing} proposes an integrated likelihood as a generalization of these criteria in unsupervised classification. \cite{2017arXiv170108673P} shows that in the case of selecting the number of states in Hidden Markov Models, real cases induces too much pitfalls for standard methods to work robustly, and suggest pragmatic selection based on their results and expert judgement. In our case, the problem is that it is not even possible to define these.
}{
Le critère d'information d'Akaike (AIC) fournit pour les modèles statistiques pour lesquels une fonction de vraisemblance est disponible le gain d'information entre deux modèles~\cite{akaike1998information}, corrigeant l'amélioration du fit par le nombre de paramètres. Des méthodes similaires incluent le critère d'information Bayesien (BIC), qui repose sur des hypothèses légèrement différentes et corrige d'une autre façon. \cite{biernacki2000assessing} propose une vraisemblance intégrée comme une généralisation de ces critères pour la classification non-supervisée. \cite{2017arXiv170108673P} montre que dans le cas de la selection du nombre d'états pour des modèles de Markov cachés, les cas réels induisent trop d'embûches pour que les méthodes standard fonctionnement de manière robuste, et suggèrent une sélection pragmatique basée sur leurs résultats et le jugement d'expert. Dans notre cas, le problème est qu'il n'est pas possible de les définir.
}

\bpar{
The method we propose is based on the intuitive idea of approaching models of simulation by statistical models and using the corresponding AIC under certain validity conditions. \cite{2017arXiv170609773B} uses a similar trick of considering the models as black boxes and approaching them to gain insights, in their case to extract interpretable structure as decision trees.
}{
La méthode que nous proposons est basée sur l'idée intuitive d'approcher les modèles de simulation par des modèles statistiques et d'utiliser l'AIC correspondant sous certaines conditions de validité. \cite{2017arXiv170609773B} utilise une astuce similaire de considérer les modèles comme des boites noires et de les approcher pour gagner de l'information, dans leur cas pour extraire une structure interprétable sous forme d'arbres de décision.
}

\bpar{
Let $(X,Y)$ be the data and observations. We consider computational models as functions $(X,\alpha_k) \mapsto M_{\alpha_k}^{(k)}(X)$ mapping data values to a random variable. What is seen as data and parameters is somehow arbitrary but is separated in the formulation as corresponding dimensions will have different roles.
}{
Soit $(X,Y)$ les données initiales et les observations des réalisation. Nous considérons un modèle computationnel $M_k$ comme une fonction $M_k : (X,\alpha_k) \mapsto M_{\alpha_k}^{(k)}(X)$ faisant correspondre les valeurs des données $X$ à une variable aléatoire, étant donné des paramètres $\alpha_k$. Ce qui est vu comme données et comme paramètres est dans une certaine mesure arbitraire mais est séparé dans la formulation puisque les dimensions correspondantes auront des rôles différents. 
}

\bpar{
We assume that the models have been fitted to data in the sense that an heuristic has been used to find an approximate optimal solution $\alpha^{\ast}_k = \argmin_{\alpha_k}\norm{M_{\alpha_k}^{(k)}(X) - Y}$, and we write $\varepsilon_k = \norm{M_{\alpha_k}^{(k)}(X) - Y}^2$ the corresponding mean-square error.
}{
Nous supposons que les modèles ont été ajustés aux données au sens où une heuristique a été utilisée pour trouver une solution optimale approximative $\alpha^{\ast}_k = \argmin_{\alpha_k}\norm{M_{\alpha_k}^{(k)}(X) - Y}$, et nous écrivons $\varepsilon_k = \norm{M_{\alpha_k}^{(k)}(X) - Y}^2$ l'erreur carrée moyenne correspondante.
}

\bpar{
For each optimized computational model, a statistical model $S^{(k)}$ with the same degree of freedom can be fitted on a set of realizations: $M^{(k)}_{\alpha^{\ast}_k}(X) = S^{(k)} (X)$, with an error $s_k = \norm{M_{\alpha^{\ast}_k}^{(k)}(X) - S^{(k)}(X)}^2$. We make the assumption that if the fit of statistical models is good, then the gain of information between the two should capture the gain of information between simulation models.
}{
Pour chaque modèle computationnel optimisé, un modèle statistique $S^{(k)}$ avec le même degré de liberté peut être ajusté sur un ensemble de réalisations : $M^{(k)}_{\alpha^{\ast}_k}(X) = S^{(k)} (X)$, avec une erreur $s_k = \norm{M_{\alpha^{\ast}_k}^{(k)}(X) - S^{(k)}(X)}^2$. Nous faisons alors l'hypothèse que si l'ajustement des modèles statistiques est bon, alors le gain d'information entre les deux devrait capturer le gain d'information entre les modèles de simulation.
}


\bpar{
We define therefore an \emph{Empirical AIC} measure between two simulation models by
}{
Nous définissons ainsi une mesure d'\emph{AIC empirique} entre deux modèles de simulation par
}

\begin{equation}
I\left( M^{(1)}, M^{(2)}\right) = \Delta AIC \left[S^{(1)},S^{(2)}\right]
\end{equation}


%
%

\bpar{
In practice we calibrate the gravity only model and the full model on the full time span, and choose two intermediate solutions giving $M^{(1)}$ at $r_0=0.0133, d_G = 4.02e12, w_G = 1.28e-4, \gamma_G = 3.82$ with $\varepsilon_G=31.2375,\varepsilon_L=302.89$ and the full model $M^{(2)}$ at $r_0=0.0128, d_G = 8.43e14, w_G = 1.230e-4, \gamma_G = 3.81, w_N=0.60, d_N=7.47e14, \gamma_N = 1.15$ with $\varepsilon_G=31.2366,\varepsilon_L=302.93$. It is not clear how the empirical method is sensitive to the type of statistical model used, we use therefore severals for robustness, each time with the corresponding number of parameters (4 for the first and 7 for the second model): a polynomial model of the form $a_0 + \sum_{i>0} a_i X^i$, a mixture of logarithm and polynomial as $a_0 + a_1 \ln X + \sum_{i>1} a_i X^i$ and a generalized polynomial with real power coefficients that are optimized for model fit using a genetic algorithm $a_0 + \sum_{i>0} a_i X^{\alpha_i}$. We fit the statistical models using successive years as different realizations. Results for each are shown in Table~\ref{tab:interactiongibrat:empiricalaic}. We give the value of $s_k / \varepsilon_k$ and the $\Delta AIC$. We also provide the $\Delta BIC$ to check the robustness regarding the information criterion used.
}{
En pratique, nous calibrons le modèle de gravité seul et le modèle complet sur la période temporelle complète, et choisissons deux solutions intermédiaires donnant $M^{(1)}$ à $r_0=0.0133, d_G = 4.02e12, w_G = 1.28e-4, \gamma_G = 3.82$ avec $\varepsilon_G=31.2375,\varepsilon_L=302.89$ et le modèle complet $M^{(2)}$ à $r_0=0.0128, d_G = 8.43e14, w_G = 1.230e-4, \gamma_G = 3.81, w_N=0.60, d_N=7.47e14, \gamma_N = 1.15$ avec $\varepsilon_G=31.2366,\varepsilon_L=302.93$. Il n'est pas clair dans quelle mesure la méthode empirique est sensible au type de modèle statistique utilisé, nous utilisons pour cela un certain nombre pour la robustesse, à chaque fois avec les nombre de paramètres correspondants (4 pour le premier et 7 pour le second modèle) : un modèle polynomial de la forme $a_0 + \sum_{i>0} a_i X^i$, une mixture de logarithme et polynôme comme $a_0 + a_1 \ln X + \sum_{i>1} a_i X^i$ et un polynôme généralisé avec des exposants réels qui ont été optimisés pour la performance du modèle par utilisation d'un algorithme génétique $a_0 + \sum_{i>0} a_i X^{\alpha_i}$. Nous ajustons les modèles statistiques en utilisant les années successives comme des réalisations différentes. Les résultats pour chaque sont donnés en Table~\ref{tab:interactiongibrat:empiricalaic}. Nous donnons les valeurs de $s_k / \varepsilon_k$ et le $\Delta AIC$. Nous donnons aussi le $\Delta BIC$ (défini de la même manière avec le critère d'information bayésien) pour vérifier la robustesse au regard du critère d'information utilisé.
}

\bpar{
We find a positive value for 5 criteria out of 6, what means that information gain is indeed positive. The gain decreases when statistical model fit improves, and only the BIC for the optimized model fails to show an improvement. The assumption of negligible errors is always verified as the rate is always around 1\%. This approach is of course preliminary and further work should be done for a more systematic testing and more robust justification of the method. It suggests however that fit improvement in the model of simulation are effective, and that the model reveals therefore network effects.
}{
Nous trouvons une valeur positive pour 5 critères sur 6, ce qui signifie que le gain d'information est effectivement positif. Le gain décroît quand la performance du modèle statistique augmente, et seul le BIC pour le modèle optimisé échoue à montrer une amélioration. L'hypothèse des erreurs négligeables est toujours vérifiée puisque le taux est toujours autour de 1\%. Cette approche est bien sûr préliminaire et des développements supplémentaires seraient nécessaires pour un test plus systématique et une exploration de la robustesse de la méthode. Cela suggère cependant que l'amélioration de l'ajustement dans le modèle de simulation sont effectifs, et que le modèle révèle par cela des effets de réseau.
}






\begin{table}[ht]
\caption[Values of empirical fit criteria]{\textbf{Values of empirical fit criteria.} We give the relative adjustments of models, and the empirical criteria for AIC and for BIC, for each type of statistical model.\label{tab:interactiongibrat:empiricalaic}}
\bpar{
\begin{tabular}{|l|l|l|l|l|l}
\hline
Statistical Model & Fit for $M^{(1)}$ & Fit for $M^{(2)}$ & $\Delta AIC$ & $\Delta BIC$\\
\hline
Polynomial & 0.01438 & 0.01415 & 19.59 & 3.65\\
Log-polynomial & 0.01565  & 0.01435 & 125.37 & 109.43\\
Generalized Polynomial & 0.01415  & 0.01399 & 11.70 & -4.23\\
\hline
\end{tabular}
}{
\begin{tabular}{|l|l|l|l|l|l}
\hline
Modèle Statistique & Ajustement pour $M^{(1)}$ & Ajustement pour $M^{(2)}$ & $\Delta AIC$ & $\Delta BIC$\\
\hline
Polynomial & 0.01438 & 0.01415 & 19.59 & 3.65\\
Log-polynomial & 0.01565  & 0.01435 & 125.37 & 109.43\\
Polynomial Généralisé & 0.01415  & 0.01399 & 11.70 & -4.23\\
\hline
\end{tabular}
}
\end{table}

\subsection{Towards co-evolutive models}{Vers des modèles co-évolutifs}

\bpar{
Let recall the positioning of the study we just lead within our global objective. Our focus on network effects remains quite limited since (i) we do not consider an effective infrastructure but abstract flows only, and (ii) we do not take into account the possible network evolution, due to technical progresses~\citep{bretagnolle2000long} and infrastructure growth in time. An interesting development would be first the application of our model with real network data, using effective distance matrices in time, computed e.g. with the train network used by~\cite{thevenin2013mapping}. Then, allowing the network to dynamically evolve in time, as a function of flows, would yield a model of co-evolution between cities and transportation networks for a system of cities, which has been suggested by the empirical analyses by~\cite{bretagnolle:tel-00459720} (in a different sense that the one we are constructing).
}{
Rappelons le positionnement de l'étude que nous venons de mener par rapport à notre objectif général. Notre compréhension des effets de réseau reste ici assez limitée puisque (i) nous ne considérons pas une infrastructure réelle mais uniquement des flux abstraits, et (ii) nous ne prenons pas en compte la possible évolution du réseau, due aux progrès techniques~\cite{bretagnolle2000long} et à la croissance de l'infrastructure dans le temps. Un développement intéressant sera d'abord l'application du modèle sur des données réelles de réseau, en utilisant les matrices de distance réelles dans le temps, calculées e.g. avec le réseau ferré utilisé par~\cite{thevenin2013mapping}. Ensuite, permettre au réseau d'évoluer de manière dynamique dans le temps, comme fonction des flux, produira un modèle de co-évolution entre les villes et les réseaux de transport pour un système de villes, qui a été suggérée par les analyses empiriques de~\cite{bretagnolle:tel-00459720} (en un sens différent de celui que nous construisons).
}

\bpar{
This kind of model is very rare, and \cite{schmitt2014modelisation} provides with SimpopNet one of the few examples. It is shown in section~\ref{sec:quantepistemo} that disciplinary compartmentalization may be at the origin of the relative absence of such type of models in the literature. Indeed, it would imply to include heterogenous processes such as economic rules to drive network growth, that are quite far from the approach taken. It would however allow to investigate to what extent the refinement of network spatial structure and network dynamics can improve the explanation of urban system dynamics. The relevance of such as development is confirmed by empirical approaches, as~\cite{dupuy1996cities} which shows the role of the position of cities in the European freeway network for their respective relations and their competitivity.
}{
Ce type de modèle est très rare, et \cite{schmitt2014modelisation} fournit avec SimpopNet l'un des exemples existants. Il est montré dans la section~\ref{sec:quantepistemo} que la séparation des disciplines pourrait être à l'origine de l'absence relative de tels types de modèles dans la littérature. En effet, cela impliquerait d'inclure des processus hétérogènes comme des règles économiques pour régir la croissance du réseau, qui sont assez loin de l'approche prise. Cela permettrait cependant d'investiguer dans quelle mesure le raffinement de la structure spatiale du réseau et des dynamiques de réseau peut améliorer l'explication des dynamiques des systèmes urbains. La pertinence d'un tel développement est confirmée par les approches empiriques, comme~\cite{dupuy1996cities} qui montre le rôle de la position des villes dans le réseau autoroutier européen pour leur relations respectives et leur compétitivité. 
}

\bpar{
We have introduced a spatial model of growth for a system of cities at the macroscopic scale, including second order network effects among endogenous growth and direct interaction growth drivers. The model is parametrized on real data for the French city system between 1831 and 1999. The calibration of the model in time provides interpretations for the evolution of processes of interaction within the system of cities. We furthermore show that the model effectively unveils network effects by controlling for overfitting. This work paves the way for more complicated models with dynamical networks, that would capture the co-evolution between transportation network and territories, which will be developed in chapter~\ref{ch:macrocoevolution}.
}{
Nous avons introduit un modèle spatial de croissance pour un système de villes à l'échelle macroscopique, incluant des effets de réseau au second ordre avec la croissance endogène et les interactions directes comme moteurs de la croissance. Le modèle est initialisé sur les données réelles du système de villes français entre 1831 et 1999. La calibration du modèle dans le temps fournit des interprétations pour l'évolution des processus d'interaction dans le système de villes. Nous montrons de plus que le modèle révèle effectivement des effets de réseau en contrôlant le sur-ajustement. Ce travail ouvre la voie pour des modèles plus compliqués avec des réseaux dynamiques, qui captureraient la co-évolution entre les réseaux de transport et les territoires, qui seront développés au chapitre~\ref{ch:macrocoevolution}.
}

\stars

%


\newpage

\section*{Chapter Conclusion}{Conclusion du Chapitre}

\bpar{
The notion of co-evolution, which was until here in our work relatively conceptual, appears under multiple new complementary angles. This chapter allows to clarify its role within the evolutive urban theory. It will also be central for the theoretical construction that we will elaborate in~\ref{sec:theory}.
}{
La notion de co-évolution, qui était jusqu'ici dans notre travail relativement conceptuelle, apparaît sous de multiples angles nouveaux complémentaires. Ce chapitre permet d'éclairer son rôle au sein de la théorie évolutive des villes. Celle-ci sera également centrale pour la construction théorique que nous élaborerons en~\ref{sec:theory}.
}

\bpar{
Indeed, strong interdependencies can translate as variable local correlations, i.e. a spatial non-stationarity, induced on the one hand by the local patterns corresponding to a given interaction regime, of which we managed to capture the static manifestations in section~\ref{sec:staticcorrelations}, on the other hand by the multi-scalar nature of implied processes that we also showed, and thus by interaction at a small scale and long range between the different territorial entities, that we illustrated on a simple case with the interaction model studied in~\ref{sec:interactiongibrat}, which already allowed to indirectly reveal network effects in systems of cities.
}{
En effet, des interdépendances fortes peuvent se traduire par des corrélations locales variables, c'est-à-dire une non-stationnarité spatiale, induite d'une part par les motifs locaux correspondant à un régime d'interaction donné, dont nous avons pu capturer les manifestations statiques en section~\ref{sec:staticcorrelations}, d'autre part par le caractère multi-scalaire des processus impliqués que nous avons également montré, et donc par les interactions à petite échelle et longue portée entre les différentes entités territoriales, que nous avons illustré sur un cas simple par le modèle d'interaction étudié en~\ref{sec:interactiongibrat}, qui a déjà permis de révéler indirectement des effets de réseaux dans les systèmes de villes. 
}

\bpar{
We also shed light on a dynamical approach of co-evolution, by showing the potential complexity of the structure of causal relationships in the cas of a simple model of urban morphogenesis. The methodology developed was also shown efficient on real data for South Africa on long time, allowing to show an effect of segregation policies at the second order on the co-evolution itself. This method will be used as an empirical characterization of co-evolution in the following.
}{
On a également éclairé une approche dynamique de la co-évolution, en montrant la complexité potentielle de la structure des relations causales dans le cas d'un modèle de morphogenèse urbaine simple. La méthodologie développée s'est montrée également efficace sur les données réelles de l'Afrique du Sud sur le temps long, permettant de découvrir un effet des politiques de ségrégation au second ordre sur la co-évolution elle-même. Cette méthode nous servira de caractérisation empirique de la co-évolution par la suite.
}


\stars





\bpar{
\chapter{Urban Morphogenesis}
}{
\chapter{Morphogenèse urbaine}
}

\label{ch:morphogenesis} 




\bpar{
Geography gives a great importance to spatial relations and to the establishment of networks, as shows for example the first law of \noun{Tobler} combined to the fact that networks are the support of interactions. We unveiled it for the relations between networks and territories for example in section~\ref{sec:interactiongibrat}. However, our results on non-stationarity, together with the evidence of endogenous spatial scales, suggest a certain relevance of the idea of relatively independent sub-systems. It would be then possible to isolate some local rules ruling a sub-system, after some exogenous parameters have been fixed which in particular capture the relations with other sub-systems. This question is related simultaneously to the spatial scale, the temporal scale, but also the elements concerned.
}{
La géographie accorde une grande importance aux relations spatiales et à la mise en réseau, comme l'atteste par exemple la première loi de \noun{Tobler} combinée au fait que les réseaux sont vecteurs des interactions. Nous l'avons mis en évidence pour les relations entre réseaux et territoires par exemple en section~\ref{sec:interactiongibrat}. Toutefois, nos résultats sur la non-stationnarité, ainsi que la mise en valeur d'échelles locales endogènes, suggèrent une certaine pertinence de l'idée de sous-systèmes relativement indépendants. Il serait alors possible d'isoler certaines règles locales régissant un sous-système, un fois fixés certains paramètres exogènes capturant justement les relations avec d'autres sous-systèmes. Cette question porte à la fois sur l'échelle d'espace, de temps, mais aussi sur les éléments concernés.
}

\bpar{
We can go back to a concrete fieldwork example already evoked in chapter~\ref{ch:thematic}: the difficult launch of Zhuhai tramway. The impact of the delay in the operationalization and the questioning of future lines (due to an unexpected technical problem linked to an electrical current transfer technology with a third rail, imported from Europe but which had never been tested in the local climatic conditions which are rather exceptional in terms of humidity), will have a very different nature depending on the scale and the urban actors considered. The lack of general coordination between transportation and urbanism leads to assume that urban dynamics in terms of population and employment are relatively insensitive to it for the moment. The Transportation Bureau of Zhuhai Municipality and also the European technical company having conceived the failing technology may have been the subject of much more significant political and economical impacts, whereas otherwise, let it be in Zhongshan, Macao or Hong-Kong, we can assume that the issue has a nearly inexistant consequence, since the project has a fully local role. There therefore exist some complex interplays of relative independencies and interdependencies in territorial systems.
}{
Reprenons un exemple concret de terrain déjà évoqué au chapitre~\ref{ch:thematic} : la laborieuse mise en place du tramway de Zhuhai. L'impact du retard de la mise en place et la remise en question de futures lignes (dus à un problème technique inattendu lié à une technologie de transfert de courant par troisième rail importée d'Europe qui n'avait jamais été testée dans les conditions climatiques locales assez exceptionnelles en termes d'humidité), aura une nature très différente selon l'échelle et les acteurs urbains considérés. Le manque de coordination générale entre transports et urbanisme laisse supposer que les dynamiques urbaines en termes de populations et d'emplois y sont relativement insensibles dans l'immédiat. Le Bureau des Transports de la Municipalité de Zhuhai ainsi que le bureau technique européen ayant conçu la technologie défectueuse ont pu subir des répercussions politiques et économiques bien plus conséquentes, tandis que par ailleurs, que ce soit à Zhongshan, Macao ou Hong-Kong, nous pouvons supposer que le problème a une repercussion quasi-nulle, le projet ayant un rôle uniquement local. Il existe ainsi des jeux complexes d'indépendances et d'interdépendances relatives dans les systèmes territoriaux.
}


\bpar{
Generalizing to the local transport system, it can be relatively well isolated from neighbor systems, and thus its relations with the city considered in a local context. It is possible to assume both a certain form of local stationarity but also a certain independency with the exterior. The type of reasoning we just sketched implies the crucial elements which are proper to the idea of \emph{urban morphogenesis}.
}{
Généralisant au système de transport local, celui-ci peut être relativement bien isolé des systèmes voisins, et donc ses relations avec la ville considérée dans un contexte local. Il est possible de supposer à la fois une certaine forme de stationnarité locale mais aussi une certaine indépendance avec l'extérieur. Le type de raisonnement que nous avons esquissé mobilise les éléments essentiels propres à l'idée de \emph{morphogenèse urbaine}.
}



\bpar{
We will in this chapter clarify its definition and show the potentialities it gives to shed light on the relations between networks and territories. First of all, an epistemological effort through complementary viewpoints from different disciplines allow to shed light on the nature of morphogenesis in section~\ref{sec:interdiscmorphogenesis}. This allows to clarify the concept by giving it a very precise definition, distinct from self-organisation, which insists on the causal circular relations between form and function.
}{
Nous allons dans ce chapitre clarifier sa définition et montrer les potentialités qu'elle donne pour éclairer les relations entre réseaux et territoires. Dans un premier temps, un effort d'épistémologie par des points de vue complémentaires de plusieurs disciplines permet d'éclairer la nature de la morphogenèse dans la section~\ref{sec:interdiscmorphogenesis}. Cela permet de clarifier le concept en lui donnant une définition bien précise, distincte de celle de l'auto-organisation, qui appuie les relations causales circulaires entre forme et fonction.
}

\bpar{
We then explore a simple model of urban morphogenesis, based on population density only, at the mesoscopic scale, in section~\ref{sec:densitygeneration}. The demonstration that abstract processes of aggregation and diffusion are sufficient to reproduce a large diversity of forms of human settlements in Europe, by using the results of section~\ref{sec:staticcorrelations}, confirms the relevance of the idea of morphogenesis for modeling at certain scales and for the morphological dimensions.
}{
Nous explorons ensuite un modèle simple de morphogenèse urbaine, basé sur la densité de population seule, à l'échelle mesoscopique, dans la section~\ref{sec:densitygeneration}. La démonstration que les processus abstraits d'agrégation et de diffusion sont suffisants pour reproduire une grande diversité de formes d'établissements humains en Europe, en utilisant les résultats de la section~\ref{sec:staticcorrelations}, confirme la pertinence de l'idée de morphogenèse pour la modélisation à certaines échelles et pour les dimensions morphologiques.
}

\bpar{
This model is then coupled in a sequential way to a network morphogenesis model in section~\ref{sec:correlatedsyntheticdata}, in order to establish a possible space of static correlations between indicators of urban form and network indicators, which are as we previously saw a witness of local relations between networks and territories.
}{
Ce modèle est ensuite couplé de manière séquentielle à un modèle de morphogenèse de réseau dans la section~\ref{sec:correlatedsyntheticdata}, afin d'établir un espace possible des correlations statiques entre indicateurs de forme urbaine et indicateurs de réseau, qui sont comme on l'a vu précédemment un témoin des relations locales entre réseaux et territoires.
}

\bpar{
We thus introduce other building bricks for modeling co-evolution, at the mesoscopic scale through the entry of urban morphogenesis.
}{
Nous posons ainsi d'autres briques de modélisation de la co-évolution, à l'échelle mesoscopique par l'entrée de la morphogenèse urbaine.
}

\stars

\bpar{
\textit{This chapter is composed by various works. The first section is adapted from a work in collaboration with \noun{C. Antelope} (University of California), \noun{L. Hubatsch} (Francis Crick Institute) and \noun{J.M. Serna} (Université Paris VII) following the Santa Fe Institute 2016 summer school~\cite{antelope2016interdisciplinary}; the second section corresponds to~\cite{raimbault2017calibration}; and finally the third section has been written for \emph{Actes des Journées de Rochebrune 2016}~\cite{raimbault2016generation}.}
}{
\textit{Ce chapitre est composé de divers travaux. La première section est adaptée d'un travail en anglais en collaboration avec \noun{C. Antelope} (University of California), \noun{L. Hubatsch} (Francis Crick Institute) et \noun{J.M. Serna} (Université Paris VII) à la suite de l'école d'été 2016 du Santa Fe Institute~\cite{antelope2016interdisciplinary} ; la deuxième section est traduite de~\cite{raimbault2017calibration} ; et enfin la troisième section a été écrite pour les Actes des Journées de Rochebrune 2016~\cite{raimbault2016generation}.}
}



%

\newpage


\section{An interdisciplinary approach to morphogenesis}{Une approche interdisciplinaire de la morphogenèse}

\label{sec:interdiscmorphogenesis}


\bpar{
A first crucial step is a clarification of what is meant by the term of morphogenesis. Central brick in our constructions, it is indeed crucial to give it a rigorous and clear structure. The approach taken here is aimed at being \emph{interdisciplinary}, in the sense that it takes as an objective to construct a synthetic knowledge implying the different disciplines included\footnote{The approach is slightly different from the effort regarding co-evolution done in~\ref{sec:epistemology}, which proposed a multidisciplinary review but constructed a definition tailored for territorial systems. The work here is the consequence of an interdisciplinary collaboration and is aimed at being more integrative.}, and is therefore aimed as being broad (covered disciplines), deep (depth in each discipline\footnote{Naturally not in the sense of the current knowledge fronts, but in the sense of a non-vulgarized description. The issue is to capture the depth of the discipline while remaining accessible to an interdisciplinary audience.}) and synthetic (integration resulting from it). 
}{
Une première étape essentielle est la clarification de ce qui est entendu par le terme de morphogenèse. Brique essentielle de nos constructions, il est en effet crucial de lui donner une armature rigoureuse et claire. La démarche prise ici est voulue \emph{interdisciplinaire}, au sens où elle se pose comme objectif de construire une connaissance synthétique à partir des disciplines abordées\footnote{Nous nous différentions ici de l'effort concernant la co-évolution mené en~\ref{sec:epistemology}, qui proposait une revue multidisciplinaire mais construisait une définition propre aux systèmes territoriaux. Le travail ici est issu d'une collaboration interdisciplinaire et s'inscrit dans une logique plus intégrative.}, et se veut donc à la fois large (disciplines couvertes), profonde (profondeur dans chaque discipline\footnote{Bien sûr pas au sens d'un compte rendu des fronts de connaissance actuels, mais d'une présentation non vulgarisée. Il s'agit de capter la profondeur de la discipline tout en restant accessible à une audience interdisciplinaire.}) et synthétique (intégration en résultant).
}

\bpar{
The notion of morphogenesis seems to play an important role in the study of a broad range of complex systems. If the concept was introduced in embryology to design growth of organisms, it was rapidly used in various fields, e.g. urbanism, geomorphology and even psychology. However, the use of the concept seems generally fuzzy and to have a field-specific definition for each use. We propose in this section an epistemological study, starting with a broad interdisciplinary review and extracting essential notions linked to morphogenesis across fields. We propose a broad overview, within the spirit of an applied perspectivism such as described in section~\ref{sec:epistemology}, to produce concepts as much generic and broad as possible. This allows to build a consistent general meta-framework for morphogenesis. Further work may include concrete application of the framework on particular cases to operate interdisciplinary transfers of concepts, and quantitative text analysis to strengthen qualitative results.
}{
Le concept de morphogenèse semble jouer un rôle important dans l'étude d'une large gamme de systèmes complexes. S'il a été introduit initialement en embryologie pour désigner la croissance des organismes, il a été rapidement utilisé dans différentes disciplines, e.g. l'urbanisme, la géomorphologie, et même la psychologie. Toutefois, l'utilisation du concept semble généralement floue et avoir une définition spécifique à chaque champ pour chacune de ses utilisations. Nous menons dans cette section une étude épistémologique, commençant par une revue interdisciplinaire large puis en extrayant les notions essentielles liées à la morphogenèse dans chaque champ. Nous prenons le parti d'une vision croisée, dans l'idée d'un perspectivisme appliqué comme introduit en section~\ref{sec:epistemology}, pour obtenir des concepts aussi génériques et larges que possible. Cela permet de construire un méta-cadre général consistent pour la morphogenèse. Des applications peuvent inclure une application concrète du cadre sur des cas particuliers pour opérer un transfert interdisciplinaire de concepts entre disciplines.
}

\paragraph{Context}{Contexte}

\bpar{
During every historical period, people use the main technological advance as a metaphor to explain other phenomena in nature. First, nature was mechanical, then electrical, and now computational. Here, we suggest that taking an alternative metaphor might allow us to better study some properties of a system, and study how the concept of morphogenesis that originated in the study of developmental biology, can be used across systems. Morphogenesis is a very powerful metaphor that is distinct from the previous three that have been very popular in history. Unlike the mechanical, electrical or computational explanations of nature, morphogenesis is not a human designed process. Morphogenesis emphasizes the role of change and growth, rather than a static state. As \cite{thompson1942growth} already pointed out, ``\textit{natural history deals with ephemeral and accidental, not eternal nor universal things}''. The goal of our exercixse is to study three questions: (i) how is morphogenesis defined in different fields?; (ii) are there fields that use approaches and concepts that embody the notion of morphogenesis but do not use the word ?; (iii) to what extent an approache studying morphogenesis can be applied across different fields? A similar effort is described in~\cite{bourgine2010morphogenesis}, but it consists more of a collection of viewpoints from subjects that can be related to morphogenesis rather than an epistemological reconstruction of the notion as we propose to do. Furthermore, examples are far from exhausted and our review is thus complementary.
}{
Durant chaque période historique, l'avancée technologique principale a été utilisée comme une métaphore pour expliquer d'autres phénomènes de la nature. D'abord, la nature a été mécanique, puis électrique, et à présent computationnelle. Ici, nous suggérons qu'une métaphore alternative peut permettre de mieux étudier les propriétés d'un système, et ainsi comprendre comment le concept de morphogenèse qui a trouvé son origine en biologie du développement, peut être utilisé pour d'autres types de systèmes. La morphogenèse est une métaphore très puissante qui est bien distincte des trois précédentes qui ont été très populaires dans l'histoire. Ainsi, contrairement aux explications mécaniques, électriques ou computationnelles de la nature, la morphogenèse n'est pas un processus conçu par l'homme. La morphogenèse met l'accent sur le rôle du changement et de la croissance, plutôt qu'un état statique. Comme \cite{thompson1942growth} mentionnait déjà, ``\textit{l'histoire naturelle traite de l'éphémère et les accidents, pas par des choses éternelles ou universelles}''. Le but de notre exercice est de répondre à trois questions : (i) comment la morphogenèse est définie dans différents champs ; (ii) existe-t-il des champs qui utilisent des approches et concepts incluant la notion de morphogenèse mais sans utiliser le terme ; (iii) dans quelle mesure les approches étudiant la morphogenèse peuvent-elles être transférées entre les champs ? Un effort similaire a été mené par~\cite{bourgine2010morphogenesis} mais consiste plus en une collection de points de vue de sujets liés à la morphogenèse plutôt qu'une reconstruction épistémologique de la notion comme nous proposons de faire. De plus, les exemples sur ce sujet sont loin d'être épuisés et notre revue est pour cela complémentaire.
}

\bpar{
In the context of our global problematic, this work will allow us first to consider consistent territorial systems through the assumption of existing morphogenetic subsystems, and secondly to link territories and networks by the intermediary of the crucial link between form and function that we will develop below.
}{
Dans le cadre de notre problématique globale, cet effort nous permettra d'une part de considérer des systèmes territoriaux cohérents par l'hypothèse de l'existence de sous-système morphogénétiques, et d'autre part de lier territoires et réseaux par l'intermédiaire du lien crucial entre forme et fonction que nous allons développer ci-dessous.
}

\bpar{
The rest of this section is organized as follows: we first provide am autonomous review of the notion of morphogenesis across various fields, ranging from biology to social sciences, psychology and territorial sciences. A synthesis is then made and a framework as general as possible proposed. We finally discuss further developments and potential application of this epistemological analysis.
}{
La suite de cette section est organisée de la façon suivante : nous produisons d'abord une revue autonome de la notion de morphogenèse pour différents champs, s'étendant de la biologie aux sciences sociales, la psychologie et les sciences territoriales. Une synthèse est ensuite faite et un cadre aussi général que possible proposé. Nous discutons finalement des développements futurs et des applications potentielles de cette analyse épistémologique.
}

\subsection{Reviews}{Revues}

\bpar{
We propose a broad overview of the way the notion of morphogenesis if used in field that are a priori rather far from each other. Our review does not pretend to be exhaustive and we do not use any systematic method, the idea being to invoke and cross various relevant conceptions of the notion.
}{
Nous proposons un aperçu large de la manière dont est utilisée la notion de morphogenèse dans des domaines a priori très éloignés. Notre revue ne se prétend pas exhaustive et nous n'utilisons pas de méthode systématique, l'idée étant de mobiliser et de croiser différentes conceptions pertinentes de la notion.
}

\subsubsection{Developmental Biology}{Biologie du développement}

\bpar{
In developmental biology, morphogenesis refers to the mechanisms of how an organism acquires its shape and different functional units, starting from only one cell. Generally, these mechanisms need to work reliably in order to guarantee similar outcomes for every individual. This often requires cells to know their position relative to some reference frame, in order to differentiate, i.e, take a specific function, or to decide whether or not to divide\footnote{The growth of an organism is entirely done through cellular divisions, starting from an initial egg cell. A mature organism is continuously renewed through this uninterrupted division.}, what is a crucial stage in growth. The following part describes models that have been applied in developmental biology.
}{
En biologie du développement, la morphogenèse fait référence aux mécanismes conduisant un organisme à acquérir sa forme et différentes unités fonctionnelles, en partant d'une unique cellule. De manière générale, ces mécanismes doivent être fiables pour garantir une issue similaire pour chaque individu. Cela suppose que les cellules connaissent leur position par rapport à un cadre de référence afin de se différencier, c'est-à-dire prendre une fonction particulière, ou pour décider si elles doivent se diviser\footnote{La croissance d'un organisme a lieu entièrement par division cellulaires, à partir d'une cellule oeuf initiale. Un organisme mature se renouvelle en continu par cette division ininterrompue.} ou non, ce qui est une étape cruciale lors de la croissance. Nous décrivons par la suite les modèles qui ont été appliqués en biologie du développement.
}

\paragraph{Reaction-diffusion mechanisms}{Mécanismes de réaction-diffusion}

\bpar{
Alan Turing used the term reaction-diffusion system in his seminal 1952 paper 'The Chemical Basis of Morphogenesis' to describe simple patterning in a theoretical ring of cells~\cite{turing1952chemical}. Even though this work is now considered one of the most fundamental contributions to the field of pattern formation, it took many years until his work started getting recognition as an actual model for biological systems. \cite{gierer1972theory} has later suggested to use similar models also for intracellular polarity, a ubiquitous phenomenon in biology in which a cell establishes and maintains two different regions within itself. These reaction diffusion networks are one example of the emergence of patterns from a homogeneous state. Using this framework we can recapitulate many pattern formation mechanisms in development, such as coloration or segmentation. These larger scale patterns are generated by the interaction of a few species of chemicals, each chemical species also undergoing a diffusion, production and degradation. Thus it is possible to represent this model using a system of partial differential equations, and certain parameters will generate stable patterns from homogeneous initial condition, where random perturbations are amplified by the system. With only a few molecular species, very complex patterns can be formed~\cite{kondo2010reaction}. One of the most studied reaction-diffusion model capable of producing stable patterns comprises of two types of molecules, one activator and one repressor. The difference in diffusion rate between the two molecules is what amplifies random noise in the system~\cite{gierer1972theory}. The system the most studied at the origin of a coloration corresponds to the reactions responsible from the yellow-black stripes in the Zebrafish~\cite{nakamasu2009interactions}. The emergence of cell polarity is explained in some yeasts by a similar mechanism~\cite{goryachev2008dynamics}. Examples implying function such as body segmentation in \textit{Drosophila melanogaster} usually involves a more complex system than the two previously discussed examples to ensure the robustness of the emergence of these functions.
}{
Le terme de réaction-diffusion avait été utilisé par \noun{Alan Turing} dans son article séminal de 1952~\cite{turing1952chemical}, pour décrire l'émergence de motifs dans un anneau théorique de cellules. Bien que ce travail soit aujourd'hui reconnu comme l'une des contributions les plus fondamentales dans le champ de la formation de motifs, il a fallu des années pour qu'il trouve une reconnaissance comme modèle effectif pour les systèmes biologiques. \cite{gierer1972theory} a plus tard suggéré d'utiliser des modèles similaires pour expliquer la polarité intracellulaire, qui correspond à la capacité d'une cellule à différencier des zones dans son intérieur. Ces réseaux de réaction-diffusion sont un exemple de l'émergence de motifs à partir d'un état homogène, parmi d'autres comme la coloration ou la segmentation. Ces motifs à grande échelle sont générés par l'interaction entre un petit nombre d'espèces chimiques, chacune suivant une diffusion, une production et une dégradation. Il est ainsi possible d'utiliser des systèmes d'équations aux dérivées partielles, pour lesquelles certains paramètres généreront des motifs stables à partir de conditions initiales homogènes, où les perturbations aléatoires sont amplifiées par le système. Des motifs complexes peuvent être produits à partir d'un nombre très restreint d'espèces~\cite{kondo2010reaction}. L'une des réactions capables de produire des motifs stables les plus étudiées comporte deux types de molécules, un activateur et un répresseur. La différence dans le taux de diffusion entre les deux molécules est responsable de l'amplification du bruit dans le système~\cite{gierer1972theory}. Le système à l'origine d'une coloration le plus étudié correspond aux réactions responsables des rayures jaunes et noires du poisson zèbre~\cite{nakamasu2009interactions}. L'émergence de la polarité cellulaire est expliquée chez certaines levures par un mécanisme similaire~\cite{goryachev2008dynamics}. Des exemples impliquant des fonctions comme la segmentation du corps de \textit{Drosophila melanogaster} impliquent des réseaux d'espèces chimiques bien plus complexes pour assurer la robustesse de l'émergence de ces fonctions.
}

\paragraph{The French Flag Model}{Le modèle French Flag}

\bpar{
In a similar way, the French Flag model was initially conceived to explain differentiation of cells in a regular fashion~\cite{Wolpert1969}. The model assumes a graded concentration of a protein, generally called the morphogen, to which tissue cells will react differently depending on its level (thus the flag stripes). Such a gradient must be produced through a diffusion, originating from a source, completed by a stabilisation mechanism implying a sink or a local degradation within the tissue (mechanisms reviewed by~\cite{Rogers2011}). The gradient can then be used locally in a linear way (for example by increasing the expression of a gene linearly with morphogen concentration) or switch-like by local feedback mechanisms. According to~\cite{Wolpert2011}, none of these systems is actually well understood, but empirical evidences of their existence are clear at a large enough granularity, since gradients are indeed observed within a certain number of species. The experiments which are necessary for their exact confirmation are indeed very difficult and within the reach of the current state-of-the-art for most (as they require precise \emph{in vivo} measures of the mobility and decline of given molecules).
}{
De façon similaire, le modèle French Flag a été conçu initialement pour expliquer la différentiation des cellules de manière régulière~\cite{Wolpert1969}. Le modèle suppose un gradient de concentration d'une protéine, généralement appelée le morphogen, auquel les cellules d'un tissu réagiront différemment selon leur niveau (d'où les rayures du drapeau). Un tel gradient doit être produit par une diffusion, à partir d'une source, complété par un mécanisme de stabilisation impliquant un puits ou une dégradation locale dans le tissu (mécanismes qui sont passés en revue par \cite{Rogers2011}). Le gradient peut ensuite être utilisé localement de manière linéaire (l'expression d'un gène variant de manière linéaire par exemple) ou par seuils grâce à des boucles de rétroaction locales. D'après \cite{Wolpert2011}, aucun de ces systèmes n'est parfaitement bien compris, mais les preuves empiriques de leur existence sont claires à une granularité assez grande, puisque les gradients sont effectivement observés chez un certain nombre d'espèces. Les expériences nécessaires pour leur vérification exacte sont en effet très difficiles et encore hors de portée pour la plupart (car supposent des mesures précises \emph{in vivo} de la mobilité et du déclin de molécules données).
}

\paragraph{Intra-cellular forces}{Forces inter-cellulaires}

\bpar{
Cellular rearrangements are often driven by intracellular forces~\cite{Heisenberg2013}, which are then mediated between cells via modulable cell-cell junctions. This phenomenon can lead to a seemingly fluid behavior when an external stress is applied for a given period. On shorter timescales, however, cells exhibit an elastic response, assuming their previous shape if an intermittent external force is no longer present. For the tissue to change in form, occur dome divisions, cell deaths, cell extrusion, or intercalations~\cite{Guillot2013}. An example of a well-studied tissue shape change can be also found in \textit{Drosophila melanogaster}. In this case cells that initially form a flat layer become a long furrow by constricting their cell membrane area on one side~\cite{Lecuit2007}.
}{
Les réagencements cellulaires sont souvent conduits par des forces physiques intracellulaires~\cite{Heisenberg2013}, qui sont ensuite transmises entre cellules, par des jonctions intercellulaires modulables. Ce phénomène peut conduire à un comportement quasi-fluidique lorsqu'un stress extérieur est appliqué pour une certaine durée. À de plus petites échelles temporelles, les cellules gardent cependant un comportement élastique et gardent leur forme lorsqu'aucune force extérieure n'est appliquée. Pour que le tissu change de forme, ont lieu des divisions, morts, extrusions ou intercalages de cellules~\cite{Guillot2013}. Un exemple de dynamique de tissu bien étudiée s'observe chez \textit{Drosophila melanogaster} également. Dans ce cas, des cellules formant initialement une couche plate deviennent un long sillon en contractant la membrane cellulaire d'un coté~\cite{Lecuit2007}.
}

\bpar{
These considerations are both far and close to our general problematic: there exists for example bio-mimetic models applied to urban systems, such as for the generation of a transportation network~\cite{tero2010rules}\footnote{See also a recent initiative by biologists constructing an interdisciplinary project aiming at applying the complex principles of the complex spatial organisation of proteins to the urban space, which has been made concrete through the conference ``Optimized management of space: from cities to natural systems'' in December 2017:  : \url{https://gopro2017.sciencesconf.org/}.}.
}{
Ces considérations sont à la fois lointaines et proches de notre problématique générale : il existe par exemple des modèles bio-mimétiques appliqués aux systèmes urbains, comme pour la génération d'un réseau de transport~\cite{tero2010rules}\footnote{Voir aussi une initiative récente par des biologistes montant un projet interdisciplinaire visant à appliquer les principes complexes d'organisation spatiale complexe des protéines à l'espace urbain, qui s'est concrétisée dans la conférence ``Gestion optimisée de l'Espace : des villes aux systèmes naturels'' en décembre 2017 : \url{https://gopro2017.sciencesconf.org/}.}.
}

\subsubsection*{Artificial Life}{Artificial Life}

\bpar{
The notion of \emph{Programmable Self-assembly} seems in \emph{Artificial Life}\footnote{In the sense of the proper scientific field, leaded for example by the \emph{International Society for Artificial Life} (see \url{http://alife.org/}).} to be very close from the biological concept of morphogenesis: \cite{crosato2014self} notices in a broad review that ``\textit{the best example of \emph{Programmable Self-Assembly} in nature is probably the cell organisation in multicellular organisms, which is encoded by the DNA}''. An important approach in this field is the concept of \emph{Morphogenetic Engineering} introduced by \noun{Doursat}, which focuses on designing complex systems from the bottom-up. A review of the field is done in~\cite{doursat2013review}.
}{
La notion de \emph{Programmable Self-Assembly} semble être en \emph{Artificial Life}\footnote{Au sens du domaine scientifique propre, animé par exemple par la \emph{International Society for Artificial Life} (voir \url{http://alife.org/}).} très proche du concept biologique de morphogenèse : \cite{crosato2014self} note dans une large revue que ``\textit{le meilleur exemple de \emph{Programmable Self-Assembly} dans la nature est probablement l'organisation des cellules en organismes multi-cellulaires, qui est encodée par l'ADN}''. Une approche importante dans ce champ est le concept de \emph{Morphogenetic Engineering} introduit par \noun{Doursat}, qui se concentre sur la conception de systèmes complexes par le bas. Une revue du champ est faite dans~\cite{doursat2013review}.
}

\bpar{
An essential distinction between self-organization and morphogenesis which is introduced in that review is the presence of an \emph{architecture}, in the sense of a macroscopic structure that can be easily isolated and with functional properties~\cite{varenne2015programming} (but that we firstly do not consider as teleonomic~\cite{monod1970hasard} to keep a certain level of generality).
}{
Une distinction essentielle entre auto-organisation et morphogenèse qui y est introduite est la présence d'une \emph{architecture}, au sens d'une structure macroscopique bien discernable ayant des propriétés fonctionnelles~\cite{varenne2015programming} (mais que nous ne considérerons dans un premier temps pas nécessairement téléonomiques~\cite{monod1970hasard} pour garder un certain niveau de généralité).
}

\bpar{
An example of a heterogeneous swarm of particles, yielding complex architectures is described in~\cite{doursat2008programmable}. Processes of local interactions (corresponding in biology to local physical forces) and positional information through gradient propagation are both integrated in the swarm model and allow bottom-up assembly of complex patterns. \cite{sayama2009swarm} develops similar models by adding to it the possibility of an evolution of particle species, directed in real time by the modeler, what allows to effectively orientate the emergent architecture. To what extent these artificial systems are actually close to living systems remains an open question: \cite{Schmickl_2016} exhibit similar swarming rules which lead to the emergence of structures with reproduction properties, and with different functions within a proper ecosystem, which they qualify as ``imitating life''\footnote{We will develop further with more details the necessary concepts to dig further into that affirmation.}. The addition of an environment with its proper properties strongly influences the morphogenetic dynamics, as shown by~\cite{cussat2012synthesis} which combines a chemical reaction layer with an hydrodynamic layer in which the former takes place.
}{
Un exemple d'une nuée hétérogène de particules, produisant des architectures complexes, est décrit dans~\cite{doursat2008programmable}. Les processus d'interactions locales (correspondant en biologie aux forces physiques locales) et l'information de position par la propagation du gradient sont tous deux intégrés dans le modèle et permettent l'émergence de motifs complexes. \cite{sayama2009swarm} développe des modèles similaires en y ajoutant la possibilité d'évolution des espèces de particules, dirigée en direct par le modélisateur, ce qui permet effectivement d'orienter l'architecture émergente. Dans quelle mesure ces systèmes artificiels sont proches de systèmes vivants est une question ouverte : \cite{Schmickl_2016} exhibent des règles de mouvement similaires qui conduisent à l'émergence de structures aux propriétés de reproduction, avec différentes fonctions dans un écosystème propre, qu'ils qualifient comme ``imitant la vie''\footnote{Nous développerons plus en détails plus loin les concepts nécessaires pour creuser cette affirmation.}. L'ajout d'un environnement avec ses propriétés propres influence fortement les dynamiques morphogénétiques, comme montré par~\cite{cussat2012synthesis} qui combine une couche de réaction chimique avec une couche hydrodynamique dans laquelle cette première prend place.
}

\bpar{
The application of these methods to concrete engineering issue begins to be developed with promising results: \cite{Aage:2017aa} use a morphogenesis model for the design of the internal structure of a plane wing and obtains gains in mass up to 5\%, and a final structure very close to forms with similar functions in nature such as a bird wing. In this latest case, bio-mimetism is emergent since it is not assumed within microscopic morphogenesis processes, but is observed at the scale of the whole system.
}{
L'application de ces méthodes à des questions concrètes d'ingénierie commence à être développée avec des résultats prometteurs : \cite{Aage:2017aa} utilise un modèle de morphogenèse pour la conception de la structure interne d'une aile d'avion et obtient des gains de masse allant jusqu'à 5\%, et une structure finale très proche de formes aux fonctions similaires dans la nature comme une aile d'oiseau. Dans ce dernier cas, le bio-mimétisme est émergent puisqu'il n'est pas supposé dans les processus microscopiques de morphogenèse, mais se constate à l'échelle de l'ensemble du système.
}

\subsubsection*{Territorial sciences}{Sciences territoriales}

\bpar{
The concept is used in various disciplines dealing with territories and the built environment: geography, urban planning and design, urbanism, architecture. There does not seem to be a unified view nor theory within these fields, not even within each field itself.
}{
Le concept de morphogenèse est utilisé dans de nombreuses disciplines s'intéressant aux territoires et à l'environnement bâti: géographie, planification et \emph{urban design}, urbanisme, architecture. Il ne semble pas exister de vue unifiée ni de théorie entre les champs ni dans chaque champ lui-même.
}

\paragraph{Built environment}{Environnement bâti}

\bpar{
Architecture and urbanism are disciplines studying human settlements and the built environment at relatively large scale\footnote{We do not include territorial planning, but indeed consider the context of urban projects not broader than the metropolitan scale.}. The theory of Urban Metabolism by \cite{olsen1982urban} links city morphogenesis with urban metabolism and urban ecology. The city is seen as a living organism with different time scales of evolution (the life cycles). The study of urban morphology~\cite{moudon1997urban}, which focuses on morphogenetic processes, is presented as an emerging field in itself, at the interface of geography, architecture and urban planning: this view emphasizes the crucial role of the form in these kind of processes. \cite{burke1972dublin} studies the growth of a particular city (Dublin) during a given period of time, and attributes the evolution of urban morphology to \emph{morphogenetic agents}, i.e. people and developers.
}{
L'architecture et l'urbanisme sont des disciplines étudiant les établissements humains et l'environnement bâti à des échelles généralement grandes\footnote{Nous n'incluons pas l'aménagement du territoire, mais considérons bien le contexte de projets urbains qui ne dépassent jamais l'échelle métropolitaine.}. La théorie du Métabolisme Urbain de \cite{olsen1982urban} relie la morphogenèse de la ville à son métabolisme et à l'écologie urbaine. La ville est vue comme une organisme vivant avec différentes échelles de temps d'évolution (les cycles de vie). L'étude de la morphologie urbaine~\cite{moudon1997urban}, qui s'intéresse aux processus morphogénétiques, est présenté comme un champ émergent en lui-même, à l'interface de la géographie, l'architecture et la planification urbaine : cette vision appuie sur le rôle crucial de la forme dans ce genre de processus. \cite{burke1972dublin} étudie la croissance d'une ville particulière (Dublin) durant une période temporelle donnée, et attribue l'évolution de la morphologie urbaine aux \emph{agents morphogénétiques}, i.e. les habitants et les promoteurs.
}

\bpar{
At another scale, in architecture, a building can be seen as the results of microscopic processes having their own meaning, and a particular architectural style interpreted through the use of a generative shape grammar~\cite{ceccarini2001essai}. This methodology is not far from the work of \noun{C. Alexander}, an architect who produced a theory of design process \cite{mehaffy2007notes}, inspired from computer science and biology and linked in some aspects to complexity. The notion of morphogenesis is in that case however quite loose, as referring to the process of form generation in general, in the same way as~\cite{whitehand1999urban} which studies concrete changes in house forms as witnesses of urban morphogenesis, showing for example that higher density districts are more subject to contagion by minor adaptations by inhabitants.
}{
À une autre échelle, en architecture, un bâtiment peut être vu comme le résultat de processus microscopiques ayant un sens propre, et un style architectural particulier peut être interprété par l'utilisation d'une grammaire générative de formes~\cite{ceccarini2001essai}. Cette méthodologie se rapproche du travail de \noun{C. Alexander}, un architecte ayant produit une théorie des processus de design~\cite{mehaffy2007notes}, inspirée de l'informatique et de la biologie et liée par certains aspects à la complexité. La notion de morphogenèse est dans ce cas cependant assez floue, puisqu'elle se rapporte au processus de la génération de forme en général, de la même façon que~\cite{whitehand1999urban} étudie les changements concrets dans la forme des maisons comme un témoin de la morphogenèse urbaine, montrant par exemple que les quartiers de plus forte densité étaient plus susceptibles à la contagion des adaptations mineures par les habitants.
}

\bpar{
\noun{Dollens} refers to autopoiesis~\cite{dollens2014alan}\footnote{The concept of \emph{autopoiesis} on which we will come back with more details in the following, mainly consists in the ability of a system to be self-sustained as an autonomous network of processes.}, implying a particular case of morphogenesis, to advocate \noun{Turing}'s influence on contemporary design thinking, and to propose a more organic approach to architecture. \cite{desmarais1992premisses} sustains the thesis that human structures carry an abstract morphology, and that it is generated by processes carrying meaning, rejoining the conception by~\cite{ceccarini2001essai}. This echoes with the use of morphogenesis in psychology as we will see further: the elaboration of the concrete form of the artefact is then converging with the cognitive process of agents generating it, and this cognitive process can in itself be understood as a morphogenesis. \cite{levy2005formes} highlights the difficulty to give a proper definition of the term of urban form, and proposes to revisit it by linking the production of the form to the production of meaning within the full dynamic of the system. This positioning partly rejoins the one we will take later to define morphogenesis.
}{
\noun{Dollens} fait référence à l'autopoièse~\cite{dollens2014alan}\footnote{Le concept d'\emph{autopoièse} sur lequel nous reviendrons plus en détail par la suite, consiste principalement en la capacité d'un système de s'auto-entretenir comme réseau autonome de processus.}, impliquant un cas particulier de morphogenèse, pour défendre l'influence de \noun{Turing} sur la pensée contemporaine en design, et pour proposer une approche plus organique de l'architecture. \cite{desmarais1992premisses} soutient que les structures humaines sont porteuses d'une morphologie abstraite, et que celle-ci est générée par des processus porteurs de sens, rejoignant la conception de~\cite{ceccarini2001essai}. Cela fait écho aux usages de la morphogenèse en psychologie comme nous verrons plus loin : l'élaboration de la forme concrète de l'artefact va alors de pair avec le processus cognitif des agents la générant, et ce processus cognitif peut lui-même être compris comme une morphogenèse. \cite{levy2005formes} soulève la difficulté d'une définition propre du terme de forme urbaine, et propose de le revisiter en liant la production de la forme à celle du sens dans l'ensemble de la dynamique du système. Ce positionnement rejoint partiellement celui que nous prendrons plus loin pour définir la morphogenèse.
}

\paragraph{Urban modeling}{Modélisation urbaine}

\bpar{
The literature in modeling urban growth often refers to the growth process as morphogenesis when the scale implied allows to exhibit shape patterns. An example of the emergence of qualitatively different urban functions, based on the Alonso-Mills-Muth model is proposed in~\cite{bonin2012modele}. \cite{bonin2014modelisation} also extends this standard model of urban economics and directly refers to \emph{morphogens}, the substance which diffuse and react in the initial formulation of morphogenesis by \noun{Turing}. This model takes into account population, real estate prices, housing surfaces, employments and endogenous amenities, and simulates the evolution of their spatial distribution and of the resulting urban form.
}{
La littérature de modélisation de la croissance urbaine se réfère souvent au processus de croissance comme morphogenèse quand l'échelle impliquée permet de révéler des motifs de forme. Un exemple de l'émergence de fonctions urbaines qualitativement différenciées, dans un modèle basé sur le modèle d'Alonso-Mills-Muth, est proposé dans~\cite{bonin2012modele}. \cite{bonin2014modelisation} étend également ce modèle standard d'économie urbaine et fait directement référence aux \emph{morphogens}, substance qui diffusent et réagissent dans la formulation initiale de la morphogenèse par \noun{Turing}. Ce modèle prend en compte la population, les prix immobiliers, les surfaces de logement, les emplois et des aménités endogènes, et simule l'évolution de leur distribution spatiale et de la forme urbaine résultante.
}

\bpar{
\cite{makse1998modeling} studies a model of urban growth involving the local urban form. In this case the local spatial correlations induce urban structure when the cities gain new inhabitants. More heterogeneous models imply a coupling between city components and transportation networks. \cite{achibet2014model} describe a model of co-evolution between road network and urban blocks structure. At a smaller scale and involving more abstract functions, \cite{raimbault2014hybrid} couple city growth with network growth, including a local feedback of the form through a density constraint and a global feedback of position through network centrality and accessibility to amenities. These two mechanisms are analogous to local interactions and the diffusion of a global information flow in biology.
}{
\cite{makse1998modeling} étudie un modèle de croissance urbaine impliquant la forme urbaine locale. Dans ce cas les corrélations spatiales locales induisent la structure urbaine quand les villes gagnent de nouveaux habitants. Des modèles plus hétérogènes impliquent un couplage entre les composantes urbaines et les réseaux de transport. \cite{achibet2014model} décrit un modèle de co-évolution entre réseau de rues et la structure des blocs urbains. À une plus petite échelle et impliquant des fonctions plus abstraites, \cite{raimbault2014hybrid} couplent croissance urbaine et croissance de réseau, incluant une rétroaction locale de la forme par une contrainte de densité et une rétroaction globale de la position par la centralité de réseau et l'accessibilité aux aménités. Ces deux mécanismes sont analogues aux interactions locales et à la diffusion du flux d'information global en biologie. 
}



\paragraph{Archeology}{Archéologie}

\bpar{
The morphogenesis of past human settlements viewed from Thom's Catastrophe Theory point of view, is introduced by~\cite{renfrew1978trajectory}. Sudden changes (qualitative changes, or regime shifts) have occurred at any time and can be viewed as bifurcations during the morphogenesis process. Another simplified way to see this is to interpret the transition as a change of meta-parameters of a stationary dynamic.
}{
La morphogenèse des établissements humains du passé, vue du point de vue de la théorie des catastrophes de \noun{Thom}, est introduite dans~\cite{renfrew1978trajectory}. Des changement soudains (changement qualitatifs, ou changements de régime) se sont produits à toute époque et peuvent être vus comme des bifurcations durant le processus de morphogenèse. Une autre manière simplifiée de le comprendre est d'interpréter la transition comme un changement des meta-paramètres d'une dynamique stationnaire.
}

\subsubsection*{Social science and psychology}{Sciences sociales et psychologie}

\bpar{
Morphogenesis has been occasionally used as a suitable metaphor to understand different processes in social science and various psychological fields (we will in the following designate the corresponding systems as psychological systems). In developmental psychology for example, the influence of cultural learning processes on behavior are a suitable example~\cite{hart_held_2013}. Regarding clinical psychology, analogies are used for the self-organization of relations with the Self and the Other, and also for dynamics implying creative emergence, which must be fostered for a ``successful'' psychotherapy~\cite{piers_self-organizing_2007}. Moreover, in the field of neuroscience, the structure of brain in itself and the development of neural networks is typically the product of morphogenetic processes~\cite{_issues_2013}. In social psychology, the co-evolution between the individual and society can also be understood through. that approach~\cite{archer_margaret_1999}. The theory by \noun{René Thom} which we will detail later has certainly played a role in the use of this concept in psychology~\cite{de_luca_picione_processes_2016}. Nonetheless, more than a systematic and widespread unity throughout these different fields, we encounter multiple uses that are sometimes discontinuous, and one could argue that the utility of morphogenesis could be more tangible on an epistemological level. This would consist of a shared perception of morphogenesis’s descriptive power to further understand the emergence and structure of various phenomena.
}{
La morphogenèse a été occasionnellement utilisée comme une métaphore efficace pour comprendre différents processus en sciences sociales et dans divers champs de la psychologie (nous désignerons les systèmes concernés par le terme générique de systèmes psychologiques). En psychologie du développement par exemple, l'influence des processus d'apprentissage culturel sur le comportement sont une bonne illustration~\cite{hart_held_2013}. Pour la psychologie clinique, des analogies sont utilisées pour l'auto-organisation des relations avec le Moi et l'Autre, ainsi que pour les dynamiques impliquant l'émergence créative, qui doit être encouragée pour une psychothérapie ``aboutie''~\cite{piers_self-organizing_2007}. D'autre part, en neurosciences, la structure du cerveau en elle-même et la mise en place des réseaux de neurones est typiquement le produit de processus morphogénétiques~\cite{_issues_2013}. En psychologie sociale, la co-évolution de l'individu et de la société peut également être vu par ce prisme~\cite{archer_margaret_1999}. La théorie de \noun{René Thom} que nous détaillerons plus loin a certainement joué un rôle dans l'utilisation de ce concept en psychologie~\cite{de_luca_picione_processes_2016}. Toutefois, au delà d'une unité systématique au travers de ces différents champs, les usages sont plutôt discontinus, et nous pouvons supposer que l'utilité du concept de morphogenèse réside plutôt dans sa portée épistémologique. Celle-ci consisterait dans une perception partagée du pouvoir descriptif de la morphogenèse pour mieux comprendre l'émergence de la structure des divers phénomènes.
}

\subsubsection{Epistemology}{Epistémologie}

\bpar{
Morphogenesis is also used to study science itself: for example~\cite{gilbert2003morphogenesis} studies the evolution of evolutionary developmental biology through the metaphor of morphogenesis. He sees scientific ideas as interacting agents from which emerge new phenotype through differentiation processes, what is designed as the morphogenesis of the field.
}{
La morphogenèse peut aussi être utilisée pour étudier la science elle-même : par exemple~\cite{gilbert2003morphogenesis} étudie l'évolution de la biologie évolutionnaire du développement par la métaphore de la morphogenèse. Il voit les idées scientifiques comme des agents en interaction, desquels émergent de nouveaux phénotypes par des processus de différentiation, qui sont désignés comme la morphogenèse du champ.
}

\subsubsection{History of the notion}{Histoire de la notion}

\bpar{
The study of morphogenesis started with embryology between just before 1930's. This is about the same period at which cellular moves of bacteria have been discovered~\cite{abercrombie1977concepts}. Statistics obtained from Google Books\footnote{See the \emph{Google Books ngram viewer} which allows to visualize the evolution of terms use in time in the full Google Books corpus, available at \url{https://books.google.com/ngrams/graph?content=morphogenesis}.} give the first use of the word morphogenesis in a book is in 1871. The use then saw a large peak in usage between 1907 and1909, and continued to increase in usage until the 1990's before slowing decreasing in usage.
}{
L'étude de la morphogenèse a démarré avec l'embryologie juste avant les années 30. Il s'agit environ de la même période à laquelle les mouvement cellulaires de bactéries ont été découverts~\cite{abercrombie1977concepts}. Les statistiques issues de Google Books\footnote{Voir le \emph{Google Books ngram viewer} qui permet de visualiser l'évolution de l'usage de termes dans le temps dans l'ensemble du corpus Google Books, disponible à \url{https://books.google.com/ngrams/graph?content=morphogenesis}.} donne le premier usage du mot dans un livre en 1871. L'usage montre ensuite un pic d'utilisation entre 1907 et 1909, pour continuer d'augmenter jusqu'en 1990 avant de décroître progressivement.
}

\subsubsection{Putting into perspective}{Mise en perspective}

\bpar{
These journeys through diverse disciplines have already allowed us to unveil key ideas and concepts similar to morphogenesis. We conclude this review with a broader perspective to be more general.
}{
Ces voyages par diverses disciplines nous ont permis déjà de dégager des idées clés et des concepts voisins à la morphogenèse. Nous concluons cette revue par une mise en perspective pour gagner en généralité.
}

\paragraph{A mathematical approach}{Une approche mathématique}

\bpar{
Ren{\'e} Thom developed in \emph{Structural stability and Morphogenesis}~\cite{thom1974stabilite} a theory of system dynamics, the catastrophe theory, which can be specified as understanding the impact of the topological structure of the phase space on system dynamics. We detail the synthesis done by~\cite{petitot2011morphogenetic}, in particular in the case of the specification for dynamical systems. Let $M$ be a differentiable manifold, in which the system state $(\vec{x},\dot{\vec{x}})_w$ is embedded, parametrized by an external control $w\in W$. There exists than a closed set $K\subset W$ called \emph{catastrophe set}. The topological type of $K$ is indeed endogenously determined by system dynamics (in simple cases, it refers to the ``classical'' type of attractors and fixed points usually known: points and limit cycles). When $w$ encounters $K$, the system undergoes a \emph{qualitative} change in its form, what constitutes the basis of \emph{morphogenesis}.
}{
Ren{\'e} Thom a développé dans \emph{Stabilité Structurelle et Morphogenèse}~\cite{thom1974stabilite} une théorie de la dynamique des systèmes, la théorie des catastrophes, qui peut être spécifiée pour comprendre l'impact de la structure topologique de l'espace des phases sur les dynamiques du système. Nous précisons la synthèse qui en est faite par~\cite{petitot2011morphogenetic}, en particulier dans le cas de la spécification aux systèmes dynamiques. Soit $M$ une variété différentielle, dans laquelle l'état du système $(\vec{x},\dot{\vec{x}})_w$ est embarqué, paramétré par un contrôle extérieur $w\in W$. Il existe alors un ensemble fermé $K\subset W$ appelé \emph{ensemble de catastrophe}. Le type topologique de $K$ est en fait déterminé de manière endogène par la dynamique du système (dans les cas simples, il réfère au types ``classiques'' d'attracteurs et points fixes que l'on connait habituellement : points et cycles limites). Quand $w$ rencontre $K$, le système subit un changement \emph{qualitatif} dans sa forme, ce qui constitue la base de la \emph{morphogenèse}.
}

\bpar{
This abstract theory of morphogenesis is independent of the nature of the system studied, its main contribution being to classify local catastrophes that occur during morphogenesis. Differentiation and richness of patterns have thus a geometrical explanation through the topological types of catastrophes. \noun{Thom} notes that at the time of his writing, the study of form has mainly be the focus of biology, but that many applications could be done in physics and geomorphology for example. He formulates the hypothesis that because this implies discontinuities and self-organisation, to which mathematicians were repulsive, this was not applied easily to various fields. We can link this to the rise of complexity approaches, with complexity paradigms that slowly spread in various disciplines, and which the study of morphogenesis seem to have followed.
}{
Cette théorie abstraite de la morphogenèse est indépendante de la nature du système étudié, sa contribution principale étant de classifier les catastrophes locales qui surviennent lors de la morphogenèse. La différentiation et la richesse des motifs ont ainsi une explication géométrique à travers les types topologiques des catastrophes. \noun{Thom} note qu'à cette époque, l'étude de la forme a majoritairement été ciblée par la biologie, mais que de nombreuses applications pourraient être développées en physique et géomorphologie par exemple. Il formule l'hypothèse que parce que cela implique des discontinuités et de l'auto-organisation, à laquelle les mathématiciens étaient réticents, la théorie n'a pas été appliquée facilement à divers champs. Nous pouvons lier cela à l'émergence des approches complexes, avec des paradigmes de la complexité qui se sont progressivement répandus dans diverses disciplines, et l'étude de la morphogenèse semble avoir suivi. 
}


\bpar{
Mathematics, not mentioned that much in our review, are however concerned both as a tool but also as a discipline in itself, since mathematical constructions obtained from questions linked to morphogenesis are research subjects in themselves. As recently recalled \noun{Cedric Villani} in~\cite{villani2017chauvesouris}, ``\textit{morphogenesis is a discipline which is not yet clearly identified and still with many mysteries, at the intersection of mathematics, chemistry and biology, where mathematical models play a role to make the structures emerge}''.
}{
Les mathématiques, peu mentionnées dans notre revue, sont toutefois concernées à la fois comme outil mais aussi comme discipline à part entière, les constructions mathématiques obtenues à partir des questions liées à la morphogenèse sont des sujets de recherche propres. Comme l'a récemment rappelé \noun{Cedric Villani} dans~\cite{villani2017chauvesouris}, ``\textit{la morphogenèse est une discipline pas très bien identifiée ayant toujours un certain nombre de mystères, à l'intersection entre les mathématiques, la chimie et la biologie, où des modèles mathématiques jouent un rôle pour faire émerger les structures}''.
}

\paragraph{Autopoiesis}{Autopoièse}

\bpar{
The concept of autopoiesis, which initially originates from biology, is intimately linked to morphogenesis. In the case of psychological systems, it provides an interpretation of cognition and conscience which depends on the observer. This had impacts in psychology and sociology, such as some theories of systems~\cite{gershenson2015requisite}. Social and phychological systems are then understood as strongly coupled systems, as witnesses language which is a social phenomenon deeply anchored within cognitive manifestations~\cite{seidl_luhmanns_2004}. These approaches also rejoin the views of the self as dynamical and recursive~\cite{pichon_riviere_processus_2004}. The interpenetration of the social and the psychological have an echo in the psychanalytical anthropology of \noun{Freud} which highlights the relations between nevrotic symptoms and socio-cultural phenomena~\cite{freud_totem_1989}.
}{
Le concept d'autopoièse, qui provient initialement de la biologie, est intimement lié à la morphogenèse. Dans le cas des systèmes psychologiques, il fournit une interprétation dépendante de l'observateur de la cognition et de la conscience. Celle-ci a eu des impacts en psychologie et sociologie, comme certaines théories des systèmes~\cite{gershenson2015requisite}. Les systèmes sociaux et psychiques sont alors compris comme des systèmes fortement couplés, comme le témoigne le language qui est un phénomène social profondément ancré dans les manifestations cognitives~\cite{seidl_luhmanns_2004}. Ces approches rejoignent également les visions du sujet comme dynamique et récursif~\cite{pichon_riviere_processus_2004}. L'interpénétration du social et du psychologique trouvent écho chez l'anthropologie psychanalytique de \noun{Freud} qui appuie les relations entre les symptômes de névrose et les phénomènes socio-culturels~\cite{freud_totem_1989}.
}

\bpar{
In its biological sense, it expresses the ability for a system to reproduce itself. A basic characterization is the existence of a semi-permeable boundary produced within the system and the ability to reproduce its components. A more general definition is proposed by~\cite{bourgine2004autopoiesis}\footnote{\textit{``An autopoietic system is a network of processes that produces the components that reproduce the network, and that also regulates the boundary conditions necessary for its ongoing existence as a network''}.}. The notion of dynamical processes is key, and could be linked to \noun{Thom}'s theory of morphogenesis. They furthermore introduce a definition of cognition (trigger actions as function of sensory inputs to ensure viability), and of living organisms as autopoietic and cognitive, both notions being however distinct~\cite{bitbol2004autopoiesis}. In that frame for example, the arbotron~\cite{jun2005formation} is cognitive but not autopoietic. An example of link between autopoiesis and morphogenesis is shown in~\cite{niizato2010model}, where a type of Physarum organism has to play both on cell mobility and form evolution to be able to collect the food necessary for its survival. At this stage, we can postulate a strict inclusion from autopoietic systems, morphogenetic systems to self-organizing systems.
}{
Dans son approche biologique, il exprime la capacité d'un système à s'auto-reproduire. Une caractérisation rudimentaire est l'existence d'une frontière semi-perméable produite par le système et la capacité à reproduire ses composants. Une définition plus générale est proposée par~\cite{bourgine2004autopoiesis}\footnote{\textit{``Un système autopoiétique est un réseau de processus qui produit les composants permettant de reproduire le réseau, et qui régule également les conditions au bord nécessaire pour son existence continue en tant que réseau''}.}. La notion de processus dynamique est une notion clé, et pourrait être liée à la théorie de la morphogenèse de \noun{Thom}. Ils introduisent de plus une définition de la cognition (déclenchement d'actions en fonction d'entrées sensorielles pour assurer la viabilité), et d'un organisme vivant comme autopoiétique et cognitif, les deux concepts étant bien distincts~\cite{bitbol2004autopoiesis}. Dans ce cadre par exemple, l'arbotron~\cite{jun2005formation} est cognitif mais pas autopoiétique. Un exemple de lien entre autopoièse et morphogenèse est montré dans~\cite{niizato2010model}, où un type d'organisme Physarum doit jouer à la fois sur la mobilité des cellules et sur l'évolution de la forme pour être capable de collecter la nourriture nécessaire à sa survie. À cette étape, nous pouvons déjà postuler une inclusion stricte des systèmes autopoiétiques, aux systèmes morphogénétiques, aux systèmes auto-organisés.
}

\paragraph{Co-evolution}{Co-évolution}

\bpar{
Since morphogenesis can be transposed to ecosystem or societies, and the components of the system are co-evolving in those cases, the existence of co-evolution may be linked with morphogenesis, as an other way of understanding the system. Symbiosis in biology can lead to very strong causalities in organism evolution (co-evolution): this phenomenon has been described as \emph{symbiogenesis}. The symbiosis induces an change in morphogenetic patterns of symbiotic organisms as exemplified for different species in~\cite{chapman1998morphogenesis}. Thus a strong link between morphogenesis and co-evolution: in this case morphogenesis refers more to evolutionary paths of morphogenetic patterns, i.e. at a different time scale.
}{
La morphogenèse pouvant être transposée aux écosystèmes ou aux sociétés, dont les composantes sont en co-évolution dans ce cas, la présence d'une co-évolution pourrait être liée à la morphogenèse, comme une autre façon de voir le système. La symbiose en biologie peut mener à des causalités très fortes dans l'évolution de l'organisme (co-évolution) : ce phénomène a été désigné comme \emph{symbiogenesis}. La symbiose induit un changement dans les motifs morphogénétiques des organismes symbiotiques comme montré pour différentes espèces par~\cite{chapman1998morphogenesis}. D'où un lien potentiellement fort entre morphogenèse et co-évolution : dans ce cas la morphogenèse est plus utilisée pour désigner des trajectoires évolutionnaires de motifs morphogénétiques, i.e. sur une échelle de temps différente.
}

\paragraph{System definition and boundaries}{Definition et frontières du système}

\bpar{
The morphogenesis of a system must be considered conjointly with the definition of system boundaries, and the ability of boundaries to jointly open and close it\footnote{In chapter~\ref{ch:evolutiveurban}, the definition of these boundaries has been fixed in an exogenous way. Their role for morphogenetic systems suggests the possibility of an endogenous approach, as we finally did when unveiling endogenous territorial regimes in~\ref{sec:staticcorrelations}.}. The theory of complex adaptive systems by~\cite{holland2012signals} is based on a representation of these as systems of boundaries which can filter the signals exchanged between systems. This converges with the approach of an autopoietic system, and in the morphogenetic case, it is possible to assume fuzzy limits (the difficulty in modeling such systems being then the definition of the system and its boundaries). Such systems are however capable to maintain a complexity through the complex combination of opening and closing~\cite{morin1976methode}. In the case of urban systems, morphogenesis seems to be more relevant than a strict autopoiesis, since their properties change depending on the definition of boundaries \cite{2015arXiv150707878C} (see also Appendix~\ref{app:sec:scaling} which shows in a theoretical way the sensitivity of scaling laws to the definition of the system).
}{
La morphogenèse d'un système doit être considérée en même temps que la définition des limites d'un système, et la capacité des frontières à l'ouvrir et le fermer à la fois\footnote{Au chapitre~\ref{ch:evolutiveurban}, la définition de ces frontières avait été fixée de manière exogène. Leur rôle pour les systèmes morphogénétiques suggère la possibilité d'une approche endogène, comme finalement nous l'avons fait en révélant des régimes territoriaux endogènes en~\ref{sec:staticcorrelations}.}. La théorie des systèmes complexes adaptatifs de~\cite{holland2012signals} se base sur une représentation de ceux-ci par des systèmes de frontières pouvant filtrer des signaux échangés entre systèmes. Cela rejoint la vison d'un système autopoiétique, et dans le cas morphogénétique, il est possible de supposer des limites floues (la difficulté dans la modélisation de tels systèmes étant alors la définition du système et de ses limites). Ces systèmes sont toutefois capables de maintenir une complexité par la combinaison complexe de l'ouverture et de la fermeture~\cite{morin1976methode}. Dans le cas des systèmes urbains, la morphogenèse est plus crédible qu'une autopoièse stricte, puisque leur propriétés changent selon la définition des limites \cite{2015arXiv150707878C} (voir aussi l'Annexe~\ref{app:sec:scaling} qui montre de manière théorique la sensibilité des lois d'échelles à la définition du système).
}


\subsection{Synthesis}{Synthèse}

\subsubsection{Key notions}{Notions clés}

\bpar{
We list now important concepts that come out from this review to describe morphogenesis. Each may be domain-dependent, and underlying conceptions may vary from one field to the other.
}{
Nous listons à présent les concepts importants découlant de cette revue pour décrire la morphogenèse. Chacun peut être dépendant du domaine, et les conceptions sous-jacentes peuvent varier d'un champ à l'autre.
}

\bpar{
\begin{itemize}
\item \textbf{Self-organisation} : Morphogenesis implies self-organisation but the contrary is not necessarily true (the two concepts being sometimes used similarly in geomorphology~\cite{cholley1950morphologie}), some aspects are specific of morphogenesis, such as the presence of functions resulting from the form.
\item \textbf{Patterns and shape} : The ``formation of shapes'' seems to be common to all approaches to morphogenesis.
\item \textbf{Embryogenesis / tissue modeling} In biology, typical processes of morphogenesis are generally observed at early stages of life, during empryogenesis, including the initial formation of tissues.
\item \textbf{Apoptosis} Morphogenesis is often related to life (see section on autopoiesis), but also to death: the programmed death of cells, apostosis, can in some cases be a part of morphogenetic processes.
\item \textbf{Qualitative vs Quantitative} Qualitative bifurcations are a fundamental concept in morphogenesis : e.g. differentiation of organs in biology ; emergence of differentiated urban functions
\item \textbf{Symmetry} Symmetry breaking occurs, mostly at early stages, but also at all stages of morphogenesis.
\item \textbf{Unit and Scale} Are systems top-down or bottom-up designed, self-organized or exhibiting architecture ? Both are not necessarily incompatible, fundamental units and scales playing a crucial role in defining morphogenesis. Fractal-like systems, such as corals (collaborating tissues) or cities, but also the self and the society, can be studied from the point of view of morphogenetic processes at different levels.
\item \textbf{Boundaries} Boundaries are a major aspect in Complex Adaptive systems (see e.g. Holland's approach as \emph{Signals and Boundaries}~\cite{holland2012signals}). Morphogenesis can imply clear boundaries (of an embryo e.g.) but not necessarily (social organisms, cities for which the definition of boundaries is still an open question~\cite{2015arXiv150707878C}).
\item \textbf{Relation between Form and Function}: causal relations between form and function are at the center of emerging architecture.
\end{itemize}
}{
\begin{itemize}
\item \textbf{Auto-organisation} : la morphogenèse implique auto-organisation mais le contraire n'est pas nécessairement vrai (les deux concepts étant parfois confondus comme en géomorphologie~\cite{cholley1950morphologie}), certains aspects sont spécifiques à la morphogenèse, comme la présence de fonctions résultant de la forme.
\item \textbf{Motifs et Forme} : ``l'émergence de formes'' semble être commun à toutes les approches de la morphogenèse.
\item \textbf{Embryogenèse / modélisation des tissus} en biologie, les processus typiques de la morphogenèse sont généralement observés au stades initiaux de la vie, durant l'embryogenèse, incluant la formation initiale des tissus.
\item \textbf{Apostosis} la morphogenèse est souvent liée à la vie (voir la section sur l'autopoièse), mais aussi à la mort : la mort programmée de cellules, l'apostosis, peut dans certains cas faire partie de processus morphogénétiques.
\item \textbf{Qualitatif vs Quantitatif} Les bifurcations qualitatives sont un concept fondamental pour la morphogenèse : e.g. la différentiation des organes en biologie ; l'émergence de fonctions urbaines différenciées.
\item \textbf{Symétrie} Des ruptures de symétrie se produisent, majoritairement dans les étapes initiales, mais aussi à tous les stades de la morphogenèse.
\item \textbf{Unité et Echelle} : les systèmes sont-ils conçus par le haut ou par le bas, auto-organisés, ou présentant une architecture ? Les deux ne sont pas nécessairement incompatibles, les unités fondamentales et les échelles jouant un rôle crucial dans la définition de la morphogenèse. Les systèmes semblables à des fractales, comme les coraux (tissus collaboratifs) ou les villes, mais aussi le sujet et la société peuvent être étudiés du point de vue des processus morphogénétiques à différents niveaux.
\item \textbf{Frontières} : les frontières sont un aspect crucial pour l'étude des Systèmes Complexes Adaptatifs (voir par exemple l'approche de \noun{Holland} par \emph{Signals and Boundaries}~\cite{holland2012signals}). La morphogenèse peut impliquer des frontières claires (d'un embryon e.g.) mais pas nécessairement (organismes sociaux, villes pour lesquelles la définition des frontières est toujours une question ouverte~\cite{2015arXiv150707878C}).
\item \textbf{Relation entre forme et fonction} : les relations causales entre forme et fonction sont au centre de l'architecture émergente.
\end{itemize}
}

\subsubsection{Common processes and differences}{Processus communs et divergences}

\paragraph{From local interactions to global information flow}{Des interactions locales aux flux globaux d'information}

\bpar{
The interplay between agent-to-agent interactions, either through neighborhood effects such as mechanistic interactions and diffusion, or through network interactions such as signaling, and the feedback of a global information flow (i.e. a downward causation of the upper level) appears to be common to most use of morphogenesis. It highlights the fundamental multi-level nature of morphogenetic processes and the central role of emergence.
}{
Les imbrications des relations entre agents, soit par des effets de voisinage comme des interactions mécaniques et la diffusion, ou par des interactions de réseaux comme le signalement, et la retroaction d'un flux d'information global (i.e. une causalité descendante du niveau supérieur) apparaît être commun à la majorité des utilisations de la morphogenèse. Cela souligne la nature fondamentalement multi-niveaux des processus morphogénétiques et le rôle central de l'émergence.
}

\paragraph{From self-organization to morphogenesis: the notion of architecture}{De l'auto-organisation à la morphogenèse : la notion d'architecture}

\bpar{
Most system studied seem to have the particularity to exhibit an architecture, what would make the distinction between self-organization and morphogenesis. This idea comes from the field of morphogenetic engineering, which can be seen as a subfield of artificial intelligence~\cite{doursat2012morphogenetic}. This point may be a divergence point on some fields, as for example in physical science, where the ``morphogenesis'' of terrain patterns is a self-organization in our sense. The notion of architecture may be tricky to define. A way to do it is to consider the functions of macro-levels in the system: the emergence of function at an upper level implies an architecture, which is \emph{the link between the form and the function}. Here this last concept takes all its sense and importance in regard to morphogenesis.
}{
La plupart des systèmes étudiés semblent avoir la particularité de présenter une architecture, ce qui permettrait de faire la distinction entre auto-organisation et morphogenèse. Cette idée vient du champ du \emph{morphogenetic engineering}, qui peut être vu comme un sous-champ de l'intelligence artificielle \cite{doursat2012morphogenetic}. Ce point peut être une divergence pour certains champs, comme par exemple en géographie physique où la ``morphogenèse'' de motifs d'érosion est une auto-organisation en notre sens. La notion d'architecture peut être difficile à définir. Une façon d'y parvenir est de considérer les fonctions des niveaux macroscopiques du système : l'émergence d'une fonction à un niveau supérieur implique une architecture, qui est \emph{le lien entre la forme et la fonction}. Ici ce dernier concept prend tout son sens et son importance au regard de la morphogenèse.
}

\subsubsection{Proposition of a meta-epistemological framework}{Proposition d'un cadre méta-épistémologique}

\paragraph{Framework}{Cadre}

\bpar{
We propose a hierarchical organisation of concepts, that can be seen as a meta-epistemological framework, since definitions are built from synthesis of the many disciplines evoked here, and that their application in each particular discipline yields an epistemological frame. The concepts are organized the following way :

\begin{equation}
\textrm{Self-organization} \supsetneq \textrm{Morphogenesis} \supsetneq \textrm{Autopoiesis} \supsetneq \textrm{Life} 
\end{equation}

each having a generic definition, elaborated from the synthesis of disciplines. The strict inclusion denoted by $\supsetneq$ means that a concept implies the other but that they remain different. All concepts are necessary to be able to situate morphogenesis well.
}{
Nous proposons une imbrication hiérarchique des concepts, qui peut être vue comme un cadre méta-épistémologique, puisque les définitions sont construites de la synthèse des diverses disciplines évoquées ici, et que leur application dans chaque discipline particulière fournit un cadre épistémologique. Les concepts sont organisés de la façon suivante:

\begin{equation}
\textrm{Auto-organisation} \supsetneq \textrm{Morphogenèse} \supsetneq \textrm{Autopoïèse} \supsetneq \textrm{Vie}
\end{equation}

chacun ayant une définition générique, élaborée de la synthèse des disciplines. L'inclusion stricte dénotée par $\supsetneq$ signifie qu'un concept implique l'autre mais qu'ils sont différents. L'ensemble des concepts est nécessaire pour bien situer la morphogenèse.
}

\bpar{
\textbf{Definition : \textit{Self-organization}.} A system is self-organized if it exhibits weak emergence~\cite{bedau2002downward}.
}{
\textbf{Definition : \textit{Auto-organisation}.} Un système est dit auto-organisé s'il exhibe une émergence faible~\cite{bedau2002downward}.
}

\bpar{
\textbf{Definition : \textit{Morphogenesis}.} A self-organized system is the result of morphogenetic processes if it exhibits an emergent architecture, in the sens of causal relations between form and function at different levels.
}{
\textbf{Définition : \textit{Morphogenèse}.} Un système auto-organisé est le produit de processus morphogénétiques s'il présente une architecture émergente, au sens de relations causales circulaires entre forme et fonction à différents niveaux.
}

\bpar{
The \emph{form} is understood as \emph{topological or geometrical properties} of a system or one of its parts, whereas the \emph{function} is its role within the chains of processes, in a \emph{teleonomic} perspective\footnote{In the sense given by \noun{Monod} in~\cite{monod1970hasard}, i.e. aiming at answering a project, or a given purpose. Living organisms are teleonomic in the sense that all their functions in the end aim at reproducing their ADN. A non-\emph{teological} vision of the universe postulate that it does not have any project, and that most of physical objects do not enter this category. All the other case studies we reviewed in our construction are teleonomic at different levels: territorial systems are planned following the logic of actors which answer to some projects; robotic systems in \emph{morphogenetic engineering} answering to a purpose; ideas or thought are part of the ecosystem of the mind. We therefore postulate this teleonomic necessity of the function to have a morphogenesis, position which can be discussed, such as in geomorphology the river network will be supposed to have the function of draining rainfall. In any case a clear dichotomy between morphogenesis in our sense and self-organization can not be established, and a continuum surely corresponds more to reality (the same way that \noun{Bedau} imagines a continuum between weak emergence and strong emergence). Indeed, in a perspectivist vision (voir~\ref{sec:epistemology}), the observer plays an essential role in the definition of a function: the Game of Life used as a computer (through its properties of being Turing-complete) will be morphogenetic, whereas it will be self-organized if it is simulated without reason, rejoining the absurdity of defining an \emph{object} without \emph{subject} highlighted by \noun{Morin} in~\cite{morin1976methode}.}.
}{
La \emph{forme} est comprise comme \emph{propriétés topologiques ou géométriques} d'un système ou de l'une de ses parties, tandis que la \emph{fonction} est son rôle au sein des chaînes de processus, dans une perspective \emph{téléonomique}\footnote{Au sens donné par \noun{Monod} dans~\cite{monod1970hasard}, c'est-à-dire visant à répondre à un projet, à un but donné. Les êtres vivants sont téléonomiques au sens que l'ensemble de leur fonctions visent à finalement reproduire leur ADN. Une vision non \emph{téléologique} de l'univers postule que celui-ci n'a pas de projet, et que la plupart des objets physiques ne rentrent pas dans cette catégorie. L'ensemble des autres cas d'étude que nous avons revu dans notre construction sont téléonomiques à différents niveaux : les systèmes territoriaux sont aménagés selon des logiques d'acteurs qui répondent à des projets ; les systèmes de robots en \emph{morphogenetic engineering} répondent à un besoin ; les idées ou pensées participent à l'écosystème de l'esprit. Nous postulons ainsi cette nécessité téléonomique de la fonction pour avoir morphogenèse, position qui peut être discutée, comme en géomorphologie le réseau de rivières sera supposé avoir la fonction de drainer l'eau de pluie. Dans tous les cas une dichotomie claire entre morphogenèse en notre sens et auto-organisation ne pourra être distinctement établie, et un continuum correspond plus sûrement à la réalité (de la même manière que \noun{Bedau} imagine un continuum entre émergence faible et émergence forte). En effet, dans une vision perspectiviste (voir~\ref{sec:epistemology}), l'observateur joue un rôle essentiel dans la définition d'une fonction : le Jeu de la Vie utilisé comme ordinateur (par ses propriétés de Turing-complétude) sera morphogénétique, tandis qu'il sera auto-organisé s'il est simulé sans raison, rejoignant l'absurdité de la définition d'un \emph{objet} sans \emph{sujet} soulevée par \noun{Morin} dans~\cite{morin1976methode}.}.
}

\bpar{
\textbf{Definition : \textit{Autopoiesis and Life}.} We take the definition of Bourgine \cite{bourgine2004autopoiesis} for autopoiesis, that extends Bitbol's~\cite{bitbol2004autopoiesis}, who also define life as autopoiesis with cognition.
}{
\textbf{Définition : \textit{Autopoièse et Vie}.} Nous prenons la définition de \noun{Bourgine} pour l'autopoièse~\cite{bourgine2004autopoiesis}, qui étend celle de \noun{Bitbol}~\cite{bitbol2004autopoiesis}, qui définit également la vie comme autopoièse avec cognition.
}

\bpar{
The boundary between self-organization and morphogenesis is the existence of causal links between form and function, which can be defined as \emph{architecture}~\cite{doursat2013review}, generally emergent from the bottom-up. We observe that the complexity of systems increase with notion depth, what can be loosely translated in the fact that :
\begin{itemize}
\item Emergence strength~\cite{bedau2002downward} diminishes with depth, in the sense that the number of autonomous scales increases.
\item Number of bifurcations increases~\cite{thom1974stabilite}, i.e. path-dependancy increases.
\end{itemize}

These two properties can be interpreted as \emph{one of} caracterisations of complexity (see~\ref{sec:epistemology}).
}{
La frontière entre auto-organisation et morphogenèse est l'existence de liens causaux entre forme et fonction, qui peut être définie comme une \emph{architecture}~\cite{doursat2013review}, qui sera généralement émergente. Nous observons que la complexité du système augmente avec la profondeur de la notion, ce qui peut être traduit de façon simplifiée par :
\begin{itemize}
\item La force de l'émergence~\cite{bedau2002downward} diminue avec la profondeur, au sens que le nombre d'échelles autonomes, ainsi que le nombre de propriétés aux pouvoir causaux irréductibles, augmentent.
\item Le nombre de bifurcations augmente~\cite{thom1974stabilite}, i.e. la dépendance au chemin augmente.
\end{itemize}

Ce deux propriétés peuvent être interprétées comme \emph{l'une des} caractérisations de la complexité (voir~\ref{sec:epistemology}).
}

\paragraph{Application}{Application}

\bpar{
An ontological specification~\cite{livet2010}, i.e. the definition of entities to which the notion apply, yields an application to a particular field, each one developing its own properties and level of inclusion between concepts. There is a priori no reason for a direct correspondence or equivalence of projected concepts, thus transfer of knowledge between fields may be subject to caution.
}{
Une spécification ontologique~\cite{livet2010}, i.e. la définition des entités à laquelle la notion s'applique, fournit une application à un champ donné, chaque champ développant ses propres propriétés et niveaux d'inclusion entre les concepts. Il n'existe a priori pas de raison pour une correspondance directe ou une équivalence entre les concepts projetés, ainsi le transfert de connaissances entre les domaines doit rester sujet à précaution.
}

\bpar{
We illustrate in Frame~\ref{frame:interdiscmorphogenesis:examples} diverse examples of systems which can be qualified as morphogenetic or not, depending on the functional perspective which is taken. These are presented by disciplines. We see therein the generic character of the framework and also its flexibility.
}{
Nous illustrons en Encadré~\ref{frame:interdiscmorphogenesis:examples} divers exemples de systèmes pouvant être qualifiés de morphogénétiques ou non, selon la perspective fonctionnelle que l'on en prend. Ceux-ci sont présentés par disciplines. Nous voyons ainsi le caractère générique du cadre ainsi que sa flexibilité.
}

\begin{figure}[h!]
	\begin{mdframed}
		
		\includegraphics[width=\linewidth]{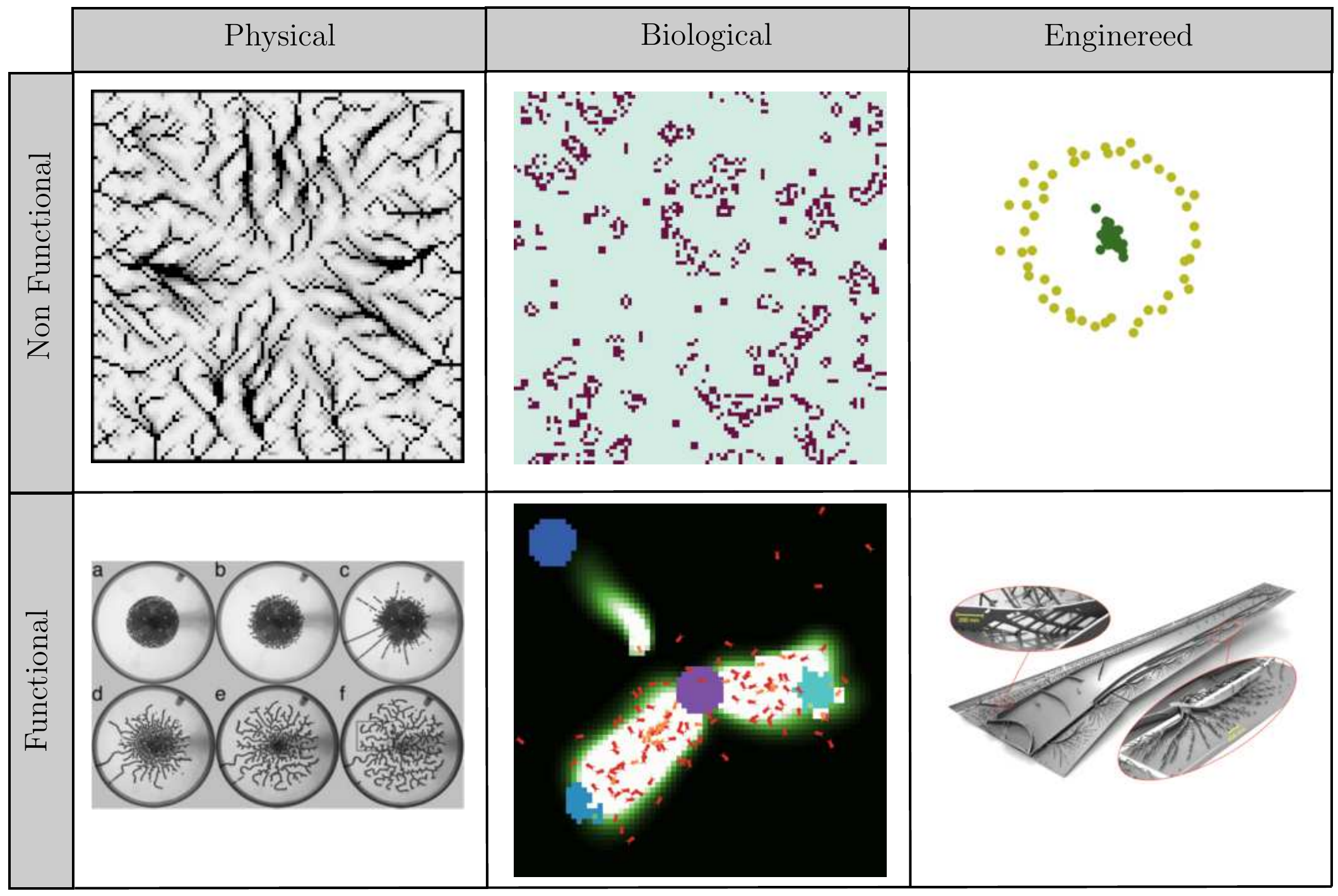}
		
		\medskip
		
		\bpar{
		We give illustrations in three typical disciplines, and propose an interpretation of the functional or not functional character of systems (making the systems in the first row non-morphogenetic in our sense). In the order from left to right and top to bottom: erosion model (NetLogo library); Game of Life (see~\ref{sec:epistemology}, NetLogo library); \emph{Swarm chemistry} (abstract particles with movement and interaction rules), implemented from \cite{sayama2007decentralized}; arbotron (metal balls under the influence of an electric potential) \cite{jun2005formation}; ant model (NetLogo library); industrial conception \cite{Aage:2017aa}. The Game of Life, if used as a computer, can have functional aspects. Similarly the arbotron or the particle cloud of \emph{Swarm chemistry} can be used as instruments. Everything must be seen within the perspective in which the system is placed.
		}{
		Nous donnons des illustrations dans trois disciplines typiques, et proposons une interprétation des caractères fonctionnels ou non (rendant les systèmes de la première ligne non-morphogénétiques en notre sens). Dans l'ordre de gauche à droite et de haut en bas : modèle d'érosion (bibliothèque NetLogo) ; jeu de la vie (voir~\ref{sec:epistemology}, bibliothèque NetLogo) ; \emph{Swarm chemistry} (particules abstraites ayant des règles de mouvement et d'interaction), implémenté à partir de \cite{sayama2007decentralized} ; arbotron (billes de métal sous l'influence d'un potentiel électrique) \cite{jun2005formation} ; modèle de fourmis (bibliothèque NetLogo) ; conception industrielle  \cite{Aage:2017aa}. Le jeu de la vie, s'il est utilisé comme ordinateur, peut avoir des aspects fonctionnels. De même l'arbotron ou le nuage de particules de la \emph{Swarm chemistry} peuvent être utilisés comme instruments. Tout est question de perspective dans laquelle le système est placé.
		}
		
		\framecaption{\textbf{Examples of morphogenetic systems.}\label{frame:interdiscmorphogenesis:examples}}{\textbf{Exemples de systèmes morphogénétiques.} \label{frame:interdiscmorphogenesis:examples}}
	\end{mdframed}
\end{figure}

\subsection{Discussion}{Discussion}

\bpar{
Before positioning the utility of constructing this concept regarding our general problematic, we can detail some potential developments which are proper to this interdisciplinary effort.
}{
Avant de positionner l'utilité de la construction de ce concept par rapport à notre problématique générale, détaillons quelque développements potentiels propres à cet effort interdisciplinaire.
}

\subsubsection{Towards a more systematic construction}{Vers une construction systématique}

\bpar{
Our work relies for now on a broad but not \emph{systematic} review, in the sense of the methodology used for example in therapeutic evaluation, and where they play a role as important as primary studies, new knowledge being created through systematic comparison of results and meta-analysis. It would imply in our case an iterative approach, by using in a coupled way the different tools and methods developed in~\ref{sec:quantepistemo}, with the following scheme.
}{
Ce travail repose pour l'instant sur une revue large mais non \emph{systématique}, au sens de la méthodologie utilisée en évaluation thérapeutique par exemple, et où elle joue un rôle aussi important que les études primaires, une nouvelle connaissance étant créée par la comparaison systématique des résultats et la meta-analyse. Cela impliquerait dans notre cas une approche itérative, en utilisant de manière couplée les différents outils et méthodes développés en~\ref{sec:quantepistemo}, selon le schéma suivant.
}

\bpar{
\begin{itemize}
\item Blind systematic review, without any a priori on the fields concerned and on the way to express the notion.
\item Extraction of main fields; extraction of synonyms and close notions (such as we did here with autopoiesis and self-assembly for example ; if needed iteration of the first general review.
\item Systematic reviews specific to each field, as each one has its own bibliographical databases, specific ways of communication, etc.
\item Confrontation of each notion from one field to other fields.
\end{itemize}
}{
\begin{itemize}
\item Une revue systématique aveugle, sans aucun a priori des champs concernés ou des moyens d'exprimer la notion.
\item Extraction des champs principaux ; extraction des synonymes et notions proches (comme il a été fait ici avec l'autopoièse et la \emph{self-assembly} par exemple) ; si besoin itération de la première revue générale.
\item Revue systématique spécifique à chaque champ, puisque chaque a ses propres bases bibliographiques, moyens spécifiques de communiquer, etc.
\item Confrontation de chaque notion depuis un champ vers les autres.
\end{itemize}
}
 
\bpar{
The objective in our case would be to enrich, as we already did in a preliminary way, but systematically, the concept of urban morphogenesis.
}{
L'objectif dans notre cas serait d'enrichir, comme nous l'avons déjà fait de manière préliminaire, mais systématiquement, le concept de morphogenèse urbaine.
}

\subsubsection{Quantitative Epistemology}{Epistemologie quantitative}

\bpar{
Our position may be also strengthen by quantitative approaches to literature analysis. With text-mining, keywords and concept extraction from abstracts (or even full texts) is possible, and would allow to confront our qualitative analysis to empirical data, by answering questions such as: is a concept indeed central, or what concept is used the same way in most disciplines. \cite{chavalarias2013phylomemetic} for example reconstructs scientific fields from the bottom-up through text-mining, and studies their lineage and dynamics in time.
}{
Notre position peut également être renforcée par des approches quantitatives à l'analyse de la littérature. Avec la fouille de texte, l'extraction de mots-clés et de concepts à partir des résumés (ou même des textes complets) est possible, et devrait permettre de confronter notre analyse qualitative à la réalité empirique, en répondant à des questions telles que : un concept est-il central, ou quel concept est utilisé de la même façon dans la plupart des disciplines. \cite{chavalarias2013phylomemetic} par exemple reconstruisent des champs scientifiques par le bas par une analyse textuelle, et étudie leur lignée et dynamique dans le temps.
}


\subsubsection{Transfer of knowledge}{Transfert de connaissances}

\bpar{
Concrete applications of our framework include potential transfer of knowledge between fields. As biological systems inspire system architecture in morphogenetic engineering, or as the use of gravity models inspired from physics have flourishing applications in geography, we think that trying to decline the general framework in specific disciplines may bring analogies or new models that would have been difficult to formulate otherwise.
}{
Les applications concrètes de ce cadre incluent un transfert potentiel de connaissances entre champs. Comme les systèmes biologiques inspirant l'architecture en \emph{morphogenetic engineering}, ou comme l'usage des modèles gravitaires inspirés par la physique a eu des applications riches en géographie, nous postulons que les tentatives de déclinaison du cadre dans des disciplines spécifiques peuvent favoriser des analogies ou d'autres modèles qui auraient été difficiles à formuler autrement.
}

\stars

%


\newpage

\bpar{
The exploration of the concept of morphogenesis realized in the previous section allows to guide the conception of urban growth models. Models bqsed on this concept will have to present the following properties:
\begin{enumerate}
	\item Crucial role of the \emph{form}, and thus inclusion both of a definition and a measure of the form, but also role of it in ontologies.
	\item Strong coupling between form and function. In a first time, the function will not be explicit in the ontology but indeed present in abstract processes.
	\item Autonomy of sub-systems, i.e. existence of a certain level of modularity in the global system. This property guides us both in the modeling scale, that we will take as ``intermediate'', or mesoscopic, and also in the search for simple models, i.e. that are parcimonious in processes taken into account and in the number of parameters.
\end{enumerate}
}{
L'exploration du concept de morphogenèse réalisée dans la section précédente permet de guider la conception de modèles de croissance urbaine. Des modèles qui se baseront sur ce concept devront avoir les propriétés suivantes :
\begin{enumerate}
	\item Rôle crucial de la \emph{forme}, et donc inclusion à la fois d'une définition et d'une mesure de la forme, mais également rôle de celle-ci dans les ontologies.
	\item Couplage fort de la forme avec la fonction. Dans un premier temps, la fonction ne sera pas explicite dans l'ontologie mais bien présente dans les processus abstraits.
	\item Autonomie des sous-systèmes, c'est-à-dire présence d'un certain niveau de modularité dans le système global. Cette propriété nous guide à la fois dans l'échelle de modélisation, que nous prendrons ``moyenne'', ou mesoscopique, ainsi que dans la recherche de modèles simples, c'est-à-dire parcimonieux dans les processus pris en compte et dans le nombre de paramètres.
\end{enumerate}
}

\bpar{
The strategy we follow to integrate these properties in morphogenesis models which will lead us to co-evolution models between transportation networks and territories, is progressive: progression in the span of ontologies (in terms of number of aspects taken into account) and progression in complexity, that we will interpret here as a coupling strength. The two following sections present thus first a morphogenesis model aimed at being minimalist for population density only, secondly the weak (sequential) coupling of it with a road network generation model. The strong coupling and the explicitation of functions through the network, producing the basis of a co-evolution model, will be the object of chapter~\ref{ch:mesocoevolution}.
}{
La stratégie que nous suivons pour intégrer ces propriétés dans des modèles de morphogenèse qui doivent nous conduire à des modèles de co-évolution entre réseaux de transport et territoires, est progressive : progression en largeur des ontologies (en nombre d'aspects pris en compte) et progression en complexité, que nous interpréterons ici comme force du couplage. Les deux sections suivantes présentent ainsi d'abord un modèle de morphogenèse à visée minimaliste pour la densité de population uniquement, puis le couplage faible (séquentiel) de celui-ci avec un modèle de génération du réseau routier. Le couplage fort et l'explicitation des fonctions par le réseau, fournissant les bases d'un modèle de co-évolution, feront l'objet du chapitre~\ref{ch:mesocoevolution}.
}

\stars


\newpage

\section{Urban morphogenesis by aggregation-diffusion}{Morphogenèse urbaine par agrégation-diffusion}

\label{sec:densitygeneration}

\bpar{
We study a stochastic model of urban growth generating spatial distributions of population densities at an intermediate mesoscopic scale. The model is based on the antagonist interplay between the two opposite abstract processes of aggregation (preferential attachment) and diffusion (urban sprawl). Introducing indicators to quantify precisely urban form, the model is first statistically validated and intensively explored to understand its complex behavior across the parameter space. We then compute real morphological measures on local areas of size 50km covering all European Union, and show that the model can reproduce most of existing urban morphologies in Europe. It implies that the morphological dimension of urban growth processes at this scale are sufficiently captured by the two abstract processes of aggregation and diffusion.
}{
Nous étudions ici un modèle stochastique de croissance urbaine générant des distributions spatiales de densité de population à une échelle mesoscopique. Le modèle se base sur le jeu antagoniste entre les deux processus abstraits opposés de l'agrégation (attachement préférentiel) et de la diffusion (étalement urbain). En utilisant des indicateurs pour quantifier précisément la forme urbaine, le modèle est d'abord validé statistiquement puis exploré intensivement pour comprendre son comportement complexe sur son espace de paramètres. Ayant calculé précédemment en~\ref{sec:staticcorrelations} les mesures morphologiques réelles sur des zones de taille 50km couvrant l'ensemble de l'Union Européenne, nous les utilisons pour montrer que le modèle peut reproduire la plupart des morphologies urbaines existantes en Europe. Cela implique que la dimension morphologique des processus de croissance urbaine à cette échelle est capturée de manière suffisante par les deux processus abstraits d'agrégation et de diffusion.
}

\subsection{Context}{Contexte}

\subsubsection{Urban Growth}{Croissance urbaine}

\bpar{
The study of urban growth, and more particularly its quantification, is more than ever a crucial issue in a context where most of the world population live in cities which expansion has significant environmental impacts~\cite{seto2012global} and that have therefore to ensure an increased sustainability and resilience to climate change. The understanding of drivers for urban growth can lead to better integrated policies.
}{
L'étude de la croissance urbaine, et plus particulièrement sa quantification, est plus que jamais un enjeu crucial dans un contexte où la majorité de la population mondiale vit dans des villes dont l'expansion a des impacts environnementaux significatifs~\cite{seto2012global} et qui doivent pour cela assurer une soutenabilité et une résilience au changement climatique accrues. La compréhension des moteurs de la croissance urbaine devrait conduire à l'élaboration de politiques mieux intégrées.
}

\bpar{
It is however a question far from being solved in the diverse related disciplines: Urban Systems are complex socio-technical systems that can be studied from a large variety of viewpoints. \noun{Batty} has advocated in that sense for the construction of a dedicated science defined by its objects of study more than the methods used~\cite{batty2013new}, what would allow easier coupling of approaches and therefore urban growth models taking into account heterogeneous processes. The processes that a model can grasp are also linked to the choice of the scale of study.
}{
Il s'agit cependant d'une question loin d'être résolue dans les diverses disciplines concernées : les systèmes urbains sont des systèmes socio-techniques complexes qui peuvent être étudiés depuis une grande variété de points de vue. \noun{Batty} défend en ce sens la construction d'une science dédiée définie par ses objets d'étude plus que par les méthodes utilisées~\cite{batty2013new}, ce qui devrait permettre des couplages plus faciles entre approches et donc des modèles de croissance urbaine prenant en compte des processus hétérogènes. Les processus qu'un modèle peut prendre en compte sont également liés au choix de l'échelle d'étude.
}

\bpar{
At a macroscopic scale, models of growth in system of cities are mainly the concern of economics and geography. We reviewed them in~\ref{sec:interactiongibrat}, and recall here that these can be more or less spatialized, and include interaction models to which belong for example Simpop models and their offspring.
}{
A une échelle macroscopique, les modèles de croissance pour les systèmes de villes sont majoritairement le sujet de l'économie et de la géographie. Nous les avons passé en revue en~\ref{sec:interactiongibrat}, et rappelons ici que ceux-ci peuvent être plus ou moins spatialisés, et incluent les modèles d'interaction dont font partie par exemple les modèles Simpop et leur descendants.
}

\subsubsection{Cellular automatons}{Automates cellulaires}

\bpar{
At larger scales, agents of models fundamentally differ. Space is generally taken into account in a finer way, through neighborhood effects for example. For example, \cite{andersson2002urban} propose a micro-based model of urban growth, with the purpose to replace non-interpretable physical mechanisms with agent mechanisms, including interactions forces and mobility choices. Local correlations are used in \cite{makse1998modeling}, which develops the model introduced in~\cite{makse1995modelling}, to modulate growth patterns to ressemble real configurations. The world of Cellular Automata (CA) models of Urban Growth~\cite{batty1994cells} also offers numerous examples. \cite{GEAN:GEAN940} introduced a generic framework for CA with multiple land use, based on local evolution rules. A model with simpler states (occupied or not) but taking into account global constraints is studied by \cite{ward2000stochastically}. The Sleuth model, initially introduced by \cite{clarke1998loose} for the San Francisco Bay area, and for which an overview of diverse applications is given in~\cite{clarke2007decade}, was calibrated on areas all over the world, yielding comparative measures through the calibrated parameters.
}{
À de plus grandes échelles, les agents des modèles diffèrent fondamentalement. L'espace est généralement pris en compte de manière plus fine, par les effets de voisinage par exemple. \cite{andersson2002urban} décrit un modèle de croissance urbaine basé sur le microscopique, dans le but de remplacer des mécanismes physiques non interprétables par des mécanismes d'agents, incluant des forces d'interaction et des choix de mobilité. Les corrélations locales sont utilisées par \cite{makse1998modeling}, qui développe le modèle introduit dans \cite{makse1995modelling}, pour moduler les motifs de croissance pour qu'ils ressemblent à des configurations réelles. Le monde des modèles de croissance urbaine à automates cellulaires (CA)~\cite{batty1994cells} offre aussi de nombreux exemples. \cite{GEAN:GEAN940} introduit un cadre générique pour les CA avec usage du sol multiple, basé sur des règles d'évolution locales. Un modèle avec des états plus simples (occupé ou non) mais prenant en compte des contraintes globales est étudié par~\cite{ward2000stochastically}. Le modèle Sleuth, introduit initialement par~\cite{clarke1998loose} pour la zone de la Baie de San Francisco, et pour lequel un aperçu des diverses applications est donné dans~\cite{clarke2007decade}, a été calibré sur diverses régions du monde, fournissant des mesures comparatives au travers des paramètres calibrés.
}

\subsubsection{Urban morphogenesis}{Morphogenèse urbaine}

\bpar{
Closely related to CA models but not exactly similar are Urban Morphogenesis models, which aim to simulate the growth of urban form from autonomous rules. We already described several in~\ref{sec:interdiscmorphogenesis}, and propose now to situate them regarding the models above. The link is clear, since for example \cite{frankhauser1998fractal} suggests that the fractal nature of cities is closely linked to the emergence of the form from the microscopic socio-economic interactions, namely urban morphogenesis. \cite{courtat2011mathematics} develops a morphogenesis model for urban roads alone, with growth rules based on geometrical considerations. These are shown sufficient to produce a broad range of patterns analog to existing ones. Similarly, \cite{raimbault2014hybrid} couples a CA with an evolving network to reproduce stylized urban form, from concentrated monocentric cities to sprawled suburbs. The Diffusion-Limited-Aggregation model, originating from physics, and which was first studied for cities by~\cite{batty1991generating}, can also be seen as a morphogenesis model. These typpe of models, that sometimes can be classified as CA, have generally the particularity of being parsimonious in their structure. Similar models have also been studied in biology for the diffusion of population as for example~\cite{bosch1990velocity}.
}{
Enfin, assez proches des modèles CA, mais au coeur de nos préoccupations ici, sont les modèles de morphogenèse urbaine, qui visent à simuler la croissance de la forme urbaine à partir de règles autonomes. Nous en avons déjà décrit un certain nombre en~\ref{sec:interdiscmorphogenesis}, et proposons maintenant de les situer par rapport aux modèles précédents. Le lien est clair, puisque par exemple \cite{frankhauser1998fractal} suggère que la nature fractale des villes est en relation étroite avec l'émergence de la forme urbaine à partir des interactions socio-économiques microscopiques, à savoir la morphogenèse urbaine. \cite{courtat2011mathematics} développe un modèle de morphogenèse pour les routes urbaines seules, avec des règles de croissance basées sur des considérations géométriques. Celles-ci sont montrées suffisantes pour produire un large spectre de motifs analogues à des motifs existants. De manière similaire, \cite{raimbault2014hybrid} couple un CA avec un réseau évolutif pour reproduire des formes urbaines stylisées, de villes monocentriques concentrées à des banlieues étalées. Le modèle \emph{Diffusion-Limited-Aggregation}, venant de la physique, et qui a été appliqué aux villes en premier par~\cite{batty1991generating}, peut aussi être vu comme un modèle de morphogenèse. Ce type de modèles, qui peuvent parfois être classifiés comme CA, ont généralement la particularité d'être parcimonieux dans leur structure. Des modèles similaires ont également été étudiés en biologie pour la diffusion de population par exemple~\cite{bosch1990velocity}.
}

\bpar{
The particularity of these models, compared to cellular automatons, is the crucial role of the form in their evolution rules, and for some of the function, such as for~\cite{bonin2012modele}. We will follow here a similar logic of rules based on form (in a first time) and function (in chapter~\ref{ch:mesocoevolution}) to construct interaction models between territories and networks.
}{
La particularité de ces modèles, en comparaison aux automates cellulaires, est la rôle crucial de la forme dans leur règles d'évolution, et pour certains de la fonction, comme~\cite{bonin2012modele}. Nous nous placerons ici dans cette même logique de règles basées sur la forme (dans un premier temps) et la fonction (en chapitre~\ref{ch:mesocoevolution}) pour construire des modèles d'interaction entre territoires et réseaux.
}

\subsubsection{Objective}{Objectif}

\bpar{
We study in this section a morphogenesis model, at the mesoscopic scale, aimed at being simplistic in its rules and variables, but trying to be accurate in the reproduction of existing patterns for the urban form (in the sense of~\ref{sec:staticcorrelations}). The underlying question is to explore the performance of simple mechanisms in reproducing complex urban patterns. We consider abstract processes, namely aggregation and diffusion, candidates as partially explanatory drivers of urban growth, based on population only, that will be detailed in model rationale below. An important aspect we introduce is the quantitative measure of urban form, based on a combination of morphological indicators, to quantify and compare model outputs and real urban patterns. Our contribution is significant on several points: (i) we compute local morphological characteristics on a large spatial extent (full European Union); (ii) we give significant insights into model behavior through extensive exploration of the parameter space; (iii) we show through calibration that the model is able to reproduce most of existing urban forms across Europe, and that these abstract processes are sufficient to explain urban form alone. The rest of this paper is organized as follows: we first describe formally the model and the morphological indicators. We then detail values of morphological measures on real data, study the behavior of the model by exploring its parameter space and through a semi-analytical approach to a simplified case, and we describe results of model calibration.
}{
Nous étudions dans cette section un modèle de morphogenèse, à l'échelle mesoscopique, dont le but est d'être performant pour la reproduction de motifs de forme urbaine existants (au sens de~\ref{sec:staticcorrelations}), sous contrainte de simplicité dans ses règles et variables. La question sous-jacente est l'exploration de la performance de mécanismes simples pour reproduire des formes urbaines complexes. Nous considérons des processus abstraits, précisément l'agrégation et la diffusion, comme facteurs potentiellement explicatifs de la croissance urbaine, basés sur la densité de population seule, qui seront détaillés ci-dessous. Un aspect important que nous utilisons est la mesure quantitative de la forme urbaine, basée sur une combinaison d'indicateurs morphologiques, pour quantifier et comparer les sorties de modèle et les formes urbaines réelles. Notre contribution est significative sur plusieurs points : (i) le calcul des caractéristiques morphologiques réelles sur une étendue spatiale conséquente (Union Européenne complète) ; (ii) nous apprenons le comportement du modèle par une exploration conséquente de l'espace des paramètres ; (iii) nous montrons par la calibration que le modèle est capable de reproduire la majorité des formes urbaines existantes en Europe, et que ces processus abstraits sont suffisants pour expliquer la forme urbaine seule. La suite de cette section est organisée de la façon suivante : nous décrivons d'abord formellement le modèle. Nous étudions ensuite le comportement du modèle par une exploration de l'espace des paramètres et par une approche semi-analytique d'un cas simplifié, puis nous décrivons les résultats de la calibration du modèle.
}

\subsection{Model and results}{Modèle et résultats}

\subsubsection{Urban growth model}{Modèle de croissance urbaine}

\paragraph{Rationale}{Description}

\bpar{
Our model is based on widely accepted ideas of diffusion-aggregation processes for Urban Processes. The combination of attraction forces with repulsion, due for example to congestion, already yield a complex outcome that has been shown under some simplifying assumptions to be representative of urban growth processes. A model capturing these processes was introduced by~\cite{batty2006hierarchy}, as a cell-based variation of the \emph{Diffusion-Limited-Aggregation} (DLA) model~\cite{batty1991generating}. Indeed, the tension between antagonist aggregation and sprawl mechanisms may be an important process in urban morphogenesis. For example \cite{fujita1996economics} opposes centrifugal forces with centripetal forces in the equilibrium view of urban spatial systems, what is easily transferable to non-equilibrium systems in the framework of self-organized complexity: a urban structure is a far-from-equilibrium system that has been driven to this point by these opposite forces. For example, concrete dispersion forces are congestion or the search for low density by residents, whereas aggregation forces can be the presence of amenities, of places of interest, of increased possibilities of social interactions \cite{krugman1992dynamic}.
}{
Notre modèle est basé sur des idées largement acceptées de processus d'agrégation-diffusion pour les processus urbains. La combinaison de forces d'attraction avec celles de répulsion, dues par exemple à la congestion, fournit déjà une issue complexe qui a été montrée représentative des processus de croissance urbaine sous certaines hypothèses simplificatrices. Un modèle capturant ces processus a été introduit dans~\cite{batty2006hierarchy}, comme une variante dans un formalisme d'automate cellulaire du modèle de \emph{Diffusion-limited Aggregation} (DLA)~\cite{batty1991generating}. En effet, la tension entre les mécanismes antagonistes d'agrégation et d'étalement peut être un processus important pour la morphogenèse urbaine. Par exemple, \cite{fujita1996economics} opposent les forces centrifuges aux forces centripètes dans une vision d'équilibre des systèmes urbains spatiaux, ce qui peut facilement être transféré aux systèmes hors équilibre dans le cadre de la complexité auto-organisée : une structure urbaine est un système hors-équilibre qui a été conduit à ce point par ces forces opposées. Par exemple, des forces concrètes de dispersion sont la congestion ou la recherche de faible densité par les habitants, tandis que des forces d'agrégation peuvent être la présence d'aménités, de lieux d'intérêts, de possibilités accrues d'interactions sociales \cite{krugman1992dynamic}.
}

\bpar{
The two contradictory processes of urban concentration and urban sprawl are captured by the model, what allows to reproduce with a good precision a large number of existing morphologies. We can expect aggregation mechanisms such as preferential attachment to be good candidates in urban growth explanation, as it was shown that the Simon model based on them generates power-laws typical of urban systems (scaling laws for example)~\cite{dodds2017simon}. The question at which scale is it possible and relevant to define and try to simulate urban form is rather open, and will in fact depend on which issues are being tackled. Working in a typical setting of morphogenesis, the processes considered are local and our model must have a resolution at the micro-level. We however want to quantify urban form on consistent urban entities, and will work therefore on spatial extents of order 50$\sim$100km. We sum up these two aspects by stating that the model is at the \emph{mesoscopic scale}.
}{
Les deux processus contradictoires de concentration urbaine et d'étalement urbain sont capturés par le modèle, ce qui permet de reproduire avec une bonne précision un grand nombre de morphologies existantes. Nous pouvons supposer que des mécanismes d'agrégation comme l'attachement préférentiel sont des bons candidats pour expliquer la croissance urbaine. En effet, il a été montré que le modèle de Simon, pour lequel l'attachement préférentiel est le principal mécanisme, génère des \emph{power-law} qui sont typiques des systèmes urbains (lois d'échelles par exemple)~\cite{dodds2017simon}. La question de l'échelle à laquelle il est possible et pertinent de définir et d'essayer de simuler la croissance urbaine est relativement ouverte, et dépendra en fait de quels problèmes sont considérés. Travaillant dans un cadre typique de la morphogenèse, les processus considérés sont locaux et notre modèle doit avoir une résolution au niveau microscopique. Nous voulons cependant quantifier la forme sur des unités urbaines cohérentes, et travaillerons ainsi sur des étendues spatiales d'ordre 50$\sim$100km. Nous résumons ces deux aspects en posant que le modèle est à l'échelle \emph{mesoscopique}.
}

\paragraph{Formalization}{Formalisation}

\bpar{
We formalize now the model and its parameters. The world is a square grid of width $N$, in which each cell is characterized by its population $(P_i(t))_{1\leq i\leq N^2}$. We consider the grid initially empty, i.e. $P_i(0)=0$, but the model can be easily generalized to any initial population distribution. The population distribution is updated in an iterative way. At each time step,
}{
Nous formalisons à présent le modèle et ses paramètres. Le monde du modèle est une grille carrée de côté $N$, dans lequel chaque cellule est caractérisée par sa population $(P_i(t))_{1\leq i\leq N^2}$. Nous considérons la grille initialement vide, i.e. $P_i(0)=0$, mais le modèle peut être facilement généralisé à n'importe quelle distribution initiale de population. La distribution de population est mise à jour de façon itérative. À chaque pas de temps,
}

\bpar{
\begin{enumerate}
\item Total population is increased by a fixed number $N_G$ (growth rate). Each population unit is attributed independently to a cell following a preferential attachment such that
\begin{equation}
\Pb{P_i(t+1)=P_i(t)+1|P(t+1)=P(t)+1}=\frac{(P_i(t)/P(t))^{\alpha}}{\sum(P_i(t)/P(t))^{\alpha}}
\end{equation}
The attribution being uniformly drawn if all population are equal to 0.
\item A fraction $\beta$ of population is diffused to cell neighborhood (8 closest neighbors receiving each the same fraction of the diffused population). This operation is repeated $n_d$ times.
\end{enumerate}
}{
\begin{enumerate}
\item La population totale est augmentée par un nombre fixe $N_G$ (de cette façon, $1 + N_G / P(t)$ est le taux de croissance). Chaque unité de population est attribuée indépendamment à une cellule suivant un attachement préférentiel tel que la probabilité qu'une cellule gagne une unité conditionnellement au fait qu'une unité soit ajoutée à la population totale est donnée par
\begin{equation}
\hspace{-1cm}
\Pb{P_i(t+1)=P_i(t)+1|P(t+1)=P(t)+1}=\frac{(P_i(t)/P(t))^{\alpha}}{\sum(P_i(t)/P(t))^{\alpha}}
\end{equation}
L'attribution est tirée de manière uniforme si toutes les populations sont égales à 0.
\item Une fraction $\beta$ de la population est diffusée au voisinage de chaque cellule (les 8 plus proches voisins recevant chacun la même fraction de la population diffusée). Cette opération est répétée $n_d$ fois, ce qui permet de lisser plus ou moins le processus de diffusion, et correspond de manière thématique à une diffusion de plus ou moins grande portée.
\end{enumerate}
}

\bpar{
The model stops when total population reaches a fixed parameter $P_m$. To avoid bord effects such as reflecting diffusion waves, border cells diffuse their due proportion outside of the world, implying that the total population at time $t$ is strictly smaller than $N_G\cdot t$.
}{
Le modèle s'arrête quand la population totale atteint un paramètre fixé $P_m$. Pour éviter les effets de bord comme des ondes de diffusion se réfléchissant, les cellules du bord diffusent aux voisins virtuels hors du monde, ce qui implique que la population totale à l'instant $t$ est strictement inférieure à $N_G\cdot t$.
}

\bpar{
We summarize model parameters in Table~\ref{tab:density:parameters}, giving the associated processes and values ranges we use in the simulations. The total population of the area $P_m$ is exogenous, in the sense that it is supposed to depend on macro-scale growth patterns on long times. Growth rate $N_G$ captures both endogenous growth rate and migration balance within the area. The aggregation rate $\alpha$ sets the differences in attraction between cells, what can be understood as an abstract attraction coefficient following a scaling law of population. Finally, the two diffusion parameters are complementary since diffusing with strength $n_d\cdot \beta$ is different of diffusing $n_d$ times with strength $\beta$, the later giving flatter configurations.
}{
Nous résumons les paramètres du modèle dans la Table~\ref{tab:density:parameters}, donnant les processus associés et les bornes des valeurs utilisées dans les simulations. La population totale de la zone $P_m$ est exogène, au sens qu'elle est supposée dépendre de processus de croissance à l'échelle macroscopique sur le temps long. Le taux de croissance $N_G$ capture à la fois la croissance endogène et la balance migratoire dans la zone. Le taux d'agrégation $\alpha$ fixe la différence d'attractivité entre cellules, qui peut être interprétée comme un coefficient abstrait d'attraction suivant une loi d'échelle de la population. Enfin, les deux paramètres de diffusion sont complémentaires puisque diffuser avec force $n_d\cdot \beta$ est différent de diffuser $n_d$ fois avec force $\beta$, le dernier cas donnant des configurations plus plates.
}

\begin{table}[!ht]
\caption[Summary of parameters of the morphogenesis model]{\textbf{Summary of parameters of the morphogenesis model.} We give the corresponding processes and the typical variation range within the configuration we use.\label{tab:density:parameters}}
\bpar{
\begin{tabular}{|l|l|l|l|}
\hline
Parameter & Notation & Processus & Range\\ \hline
Total population & $P_m$ & Macroscopic growth & $[1e4,1e6]$\\ \hline
Growth rate & $N_G$ & Mesoscopic growth & $[500,30000]$\\ \hline
Aggregation force & $\alpha$ & Aggregation & $[0.1,4]$\\ \hline
Diffusion force & $\beta$ & Diffusion & $[0,0.5]$\\ \hline
Number of diffusions & $n_d$ & Diffusion & $\{1,\ldots , 5\}$\\ \hline
\end{tabular}
}{
\begin{tabular}{|l|l|l|l|}
\hline
Paramètre & Notation & Processus & Domaine\\ \hline
Population totale & $P_m$ & Croissance macroscopique & $[1e4,1e6]$\\ \hline
Taux de croissance & $N_G$ & Croissance mesoscopique  & $[500,30000]$\\ \hline
Force d'agrégation & $\alpha$ & Agrégation & $[0.1,4]$\\ \hline
Force de diffusion & $\beta$ & Diffusion & $[0,0.5]$\\ \hline
Nombre de diffusions & $n_d$ & Diffusion & $\{1,\ldots , 5\}$\\ \hline
\end{tabular}
}
\end{table}

\paragraph{Measuring urban form}{Mesure de la forme urbaine}

\bpar{
As our model is only density-based, we propose to quantify its outputs through spatial morphology, i.e. properties of the spatial distribution of density. At the scale chosen, these will be expected to translate various functional properties of the urban landscape. The context and definition of indicators has already been given in section~\ref{sec:staticcorrelations}.
}{
Comme le modèle se base uniquement sur la densité, nous proposons de quantifier ses sorties par la morphologie spatiale, i.e. les propriétés de la distribution spatiale de la densité. A l'échelle choisie, on s'attend à ce qu'elle traduise diverses propriétés fonctionnelles de l'environnement urbain. Le contexte et la définition des indicateurs a déjà été donnée en section~\ref{sec:staticcorrelations}.
}

\subsubsection{Real data}{Données réelles}

\bpar{
We work with values of indicators computed in section~\ref{sec:staticcorrelations} for Europe, on windows of size 50km with a resolution of 100 cells. We set thus in the following $N=100$ for model simulations.
}{
Nous travaillons sur les valeurs des indicateurs calculées en section~\ref{sec:staticcorrelations} pour l'Europe, sur les fenêtres de côté 50km avec résolution de 100 cellules. Nous posons donc pour la suite $N=100$ pour les simulations du modèle.
}

\subsubsection*{Generation of urban patterns}{Génération de structures urbaines}

\paragraph{Implementation}{Implémentation}

\bpar{
The model is implemented both in NetLogo~\cite{wilensky1999netlogo} for exploration and visualization purposes, and in \texttt{Scala} for performance reasons and easy integration into OpenMole~\cite{reuillon2013openmole}, which allows a transparent access to High Performance Computing environments. Computation of indicator values on geographical data is done in \texttt{R} using the \texttt{raster} package~\cite{hijmans2015geographic}. Source code and results are available on the open repository of the project\footnote{At \texttt{https://github.com/JusteRaimbault/Density}.}. Raw datasets for real indicator values and simulation results are available on Dataverse\footnote{At \texttt{http://dx.doi.org/10.7910/DVN/WSUSBA}.}. We have in the context of the \texttt{scala} implementation implemented the convolution of distributions in two dimensions through fast Fourier tranform, allowing to capture a complexity of $O(N^4)$ in $O(N^2\log^2 N)$\footnote{We recall that a measure of complexity of an algorithm corresponds to the evaluation of time necessary to solve a problem as a function of data size, denoted $N$. An asymptotic order of magnitude is written $O(f(N))$. Therefore, a switch from a fourth order of magnitude to an order very close to a square is significant for computation time, making quasi-instantaneous a computation that would take around 10 seconds for the grid size we have. The fast Fourier transform uses a sparse decomposition to compute the discrete Fourier transform in $O(N\log N)$ instead of $O(N^2)$. The morphism of the transform of product to convolution, i.e. $\mathcal{F}\left[f\ast g \right] = \mathcal{F}\left[f \right]\cdot \mathcal{F}\left[f \right]$, allows to transfer this gain to the computation of a convolution.}, and implemented indicators which have been integrated to a dedicated NetLogo extension (it is detailed in~\ref{app:subsec:morphologyextension}).
}{
Le modèle est implémenté à la fois en NetLogo~\cite{wilensky1999netlogo} pour des raisons d'exploration et de visualisation, et en \texttt{scala} pour des raisons de performance et d'intégration plus aisée dans OpenMole~\cite{reuillon2013openmole}, qui permet un accès transparent aux environnements de calcul haute performance. Le calcul des valeurs des indicateurs sur les données géographiques est fait en \texttt{R} avec le package \texttt{raster}~\cite{hijmans2015geographic}. Le code source et les résultats sont disponibles sur le dépôt ouvert du projet\footnote{À \url{https://github.com/JusteRaimbault/Density}.}. Les données des valeurs réelles des indicateurs et des résultats de simulation sont disponibles sur Dataverse\footnote{À \url{http://dx.doi.org/10.7910/DVN/WSUSBA}.}. Nous avons dans le cadre de l'implémentation \texttt{scala} implémenté la convolution de distributions en deux dimensions par Transformée de Fourier rapide, permettant de transformer une complexité $O(N^4)$ en $O(N^2\log^2 N)$\footnote{On rappelle qu'une mesure de complexité d'un algorithme correspond à l'évaluation du temps nécessaire à la résolution d'un problème en fonction de la taille des données, notée $N$. Un ordre de grandeur asymptotique est noté $O(f(N))$. Ainsi, un passage d'un ordre puissance quatre à un ordre quasi puissance deux est significatif pour le temps de calcul, rendant quasi instantané un calcul prenant une dizaine de secondes pour nos tailles de grilles. La transformée de Fourier rapide utilise une décomposition creuse pour calculer la transformée de Fourier discrete en $O(N\log N)$ au lieu de $O(N^2)$. Le morphisme de la transformée du produit vers la convolution, c'est-à-dire $\mathcal{F}\left[f\ast g \right] = \mathcal{F}\left[f \right]\cdot \mathcal{F}\left[f \right]$, permet de transférer ce gain au calcul d'une convolution.}, puis implémenté les indicateurs qui ont pu être intégrés à une extension NetLogo dédiée (celle ci est détaillée en~\ref{app:subsec:morphologyextension}).
}

\paragraph{Generated shapes}{Formes générées}

\bpar{
The model has a relatively small number of parameters but is able to generate a large variety of shapes, extending beyond existing forms. We run the model for parameters varying in ranges given in Table~\ref{tab:parameters}, for a world size $N=100$.
}{
Le modèle a un nombre relativement faible de paramètres mais est capable de générer une grande variété de formes. Nous simulons le modèle pour des valeurs de paramètres variant dans les bornes données en Table~\ref{tab:density:parameters}, pour une taille de monde $N=100$.
}

\begin{figure}
\includegraphics[width=\linewidth]{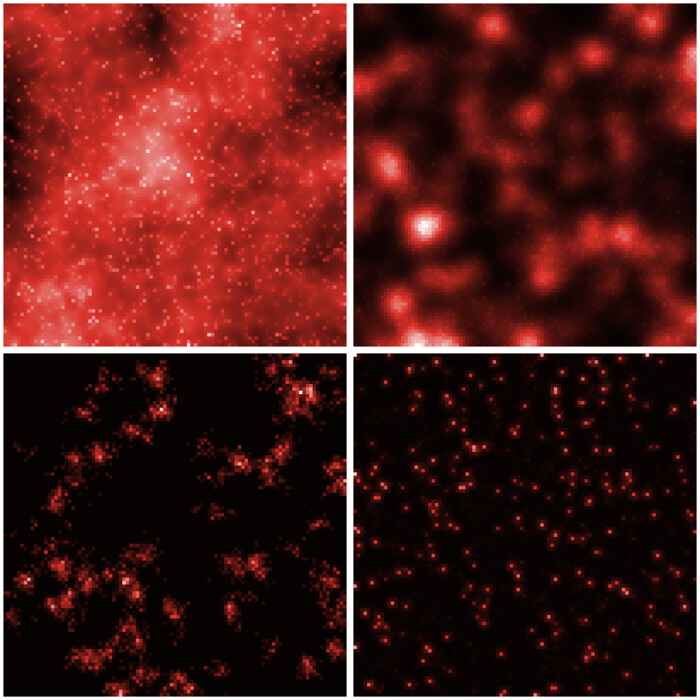}
\caption[Example of generated territorial forms]{\textbf{Example of the variety of generated urban shapes.} \textit{(Top left)} Very diffuse urban configuration, $\alpha = 0.4,\beta = 0.05, n_d = 2, N_G = 76, P_m = 75620$; \textit{(Top Right)} Semi-stationary polycentric urban configuration, $\alpha = 1.4,\beta = 0.047, n_d = 2, N_G = 274, P_m = 53977$; \textit{(Bottom Left)} Intermediate settlements (peri-urban or densely populated rural area), $\alpha = 0.4,\beta = 0.006, n_d = 1, N_G = 25, P_m = 4400$; \textit{(Bottom Right)} Rural area, $\alpha = 1.6,\beta = 0.006, n_d = 1, N_G = 268, P_m = 76376$.\label{fig:density:fig2}}
\end{figure}

\bpar{
Fig.~\ref{fig:fig2} shows examples of the variety of generated shapes for different parameter values, with corresponding interpretations. The four very different shapes can be obtained with variation of a single parameter sometimes: going from a peri-urban area from a rural area implies an increased aggregation at the same level of diffusion. Note that the model is density driven, and that the parameter $P_m/N_G$ is what really influences the dynamics: the values of $P_m$ are in some cases not directly corresponding to the interpretations we made (for the rural in particular) that are done on densities. A rescaling keeps the settlement form and solves this issue.
}{
La Fig.~\ref{fig:density:fig2} montre des exemples de la variété des formes territoriales générées pour différentes valeurs des paramètres, avec les interprétations correspondantes. Parmi les quatre formes très différentes, certaines peuvent être obtenues avec la variation d'un seul paramètre seulement : passer d'une zone péri-urbaine à une zone rurale implique une agrégation accrue au même niveau de diffusion. Il faut noter que le modèle est basé sur la densité, et que le paramètre $P_m/N_G$ est celui qui influence réellement la dynamique : les valeurs de $P_m$ ne correspondent dans certains cas pas directement aux interprétations qui sont faites sur les densités relatives (pour le rural en particulier). L'homothétie garde la forme des établissements et résout ce problème.
}

\bpar{
It appears that the dynamical nature of the model allows through the combination of parameters $P_m$ and $N_G$ to choose between configurations that can be non-stationary or semi-stationary, whereas the interaction between $\alpha$ and $\beta$ modulates the sprawl and the compact character of forms.
}{
Il apparait que la nature dynamique du modèle permet par la combinaison des paramètres $P_m$ et $N_G$ de choisir entre des configurations qui peuvent être non-stationnaires ou semi-stationnaires, tandis que l'interaction entre $\alpha$ et $\beta$ module l'étalement et le caractère compact des formes.
}

\bpar{
These examples show the potentiality of the model to produce diverse forms. We have then to systematically study its stochasticity and explore its parameter space.
}{
Ces exemples montrent la potentialité du modèle à produire des formes diverses. Nous devons ensuite étudier systématiquement sa stochasticité et explorer son espace des paramètres.
}

\subsubsection{Model behavior}{Comportement du modèle}

\bpar{
In the study of such a computational model of simulation, the lack of analytical tractability must be compensated by an extensive knowledge of model behavior in the parameter space~\cite{banos2013pour}. This type of approach is typical of what Arthur calls the \emph{Computational shift in modern science}~\cite{arthur2015complexity}: knowledge is less extracted through analytical exact resolution than through intensive computational experiments, even for ``simple'' models such as the one we study.
}{
Dans l'étude d'un tel modèle computationnel de simulation, l'impossibilité d'une résolution analytique doit être compensé par une connaissance extensive du comportement du modèle dans l'espace des paramètres~\cite{banos2013pour}. Ce type d'approche est typique de ce que \noun{Arthur} nomme le \emph{tournant computationnel dans la science moderne}~\cite{arthur2015complexity} : la connaissance est moins extraite de résolutions analytiques exactes que par des expériences de calcul intensif, même pour des modèles ``simples'' comme celui que nous étudions.
}

\paragraph{Convergence}{Convergence}

\bpar{
First of all we need to assess the convergence of the model and its behavior regarding stochasticity. We run for a sparse grid of the parameter space consisting of 81 points, with 100 repetitions for each point. Corresponding histograms are shown in~\nameref{S1_Text}. Indicators show good convergence properties: most of indicators are easily statistically discernable across parameter points: for example the Moran index, which is among the most dispersed, has a spread between 0 and 0.1 between parameters but a maximal variability of 0.01 between replications.
}{
Dans un premier temps il est important d'assurer la convergence du modèle et son comportement au regard de la stochasticité. Nous simulons le modèle pour une grille creuse de l'espace des paramètres contenant 81 points, avec 100 répétitions à chaque point. Les histogrammes correspondants sont montrés en Annexe~\ref{app:sec:density}. Les indicateurs présentent de bonnes propriétés de convergence: la plupart des indicateurs sont aisément discernables de manière statistique entre les points de paramètres : par exemple l'indice de Moran, parmi les plus dispersés, a une étendue de 0 à 0.1 entre paramètres mais une variabilité maximale de 0.01 entre réplications.
}

\bpar{
We use this experiment to find a reasonable number of repetitions needed in larger experiments. For each point, we estimate the Sharpe ratios for each indicators, i.e. mean normalized by standard deviation. The more variable indicator is Moran with a minimal Sharpe ratio of 0.93, but for which the first quartile is at 6.89. Other indicators have very high minimal values, all above 2. Its means than confidence intervals large as $1.5 \cdot \sigma$ are enough to differentiate between two different configurations. In the case of gaussian distribution, we know that the size of the 95\% confidence around the average is given by $2\cdot \sigma \cdot 1.96 / \sqrt{n}$, what gives $1.26Â \cdot \sigma$ for $n=10$. We run therefore this number of repetitions for each parameter point in the following, what is highly enough to have statistically significant differences between average as shown above. In the following, when referring to indicator values for the simulated model, we consider the ensemble averages on these stochastic runs.
}{
Nous utilisons cette expérience pour établir un nombre raisonnable de répétitions nécessaires pour des expériences plus volumineuses. Pour chaque point, nous estimons le ratio de Sharpe pour chaque indicateur, i.e. sa moyenne normalisée par la déviation standard. L'indicateur le plus variable est l'indice de Moran avec un Sharpe minimal de 0.93, mais pour lequel le premier quartile est à 6.89. Les autres indicateurs ont tous des valeurs minimales très hautes, toutes au dessus de 2. Cela signifie que des intervalles de confiance larges comme $1.5 \cdot \sigma$ ($\sigma$ écart-type estimé des distributions) sont suffisants pour différencier entre deux configurations différentes. Dans le cas d'une distribution Gaussienne, nous savons que la taille de l'intervalle de confiance à 95\% autour de la moyenne est donné par $2\cdot \sigma \cdot 1.96 / \sqrt{n}$, ce qui donne $1.26 \cdot \sigma$ pour $n=10$. Nous utilisons pour cela ce nombre de répétitions pour chaque point de paramètres par la suite, ce qui est largement suffisant pour avoir des différences entre les moyennes qui sont statistiquement significatives comme montré précédemment. Par la suite, lorsque nous considérons les valeurs des indicateurs pour le modèle simulé, nous considérons la moyenne d'ensemble sur ces répétitions stochastiques.
}

\paragraph{Exploration of parameter space}{Exploration de l'espace des paramètres}

\bpar{
We sample the Parameter space using a Latin Hypercube Sampling, with parameter as $\alpha \in [0.1,4],\beta \in [0,0.5],n_d \in \{1,\ldots , 5\}, N_G \in [500,30000], P_m \in [1e4,1e6]$. As we already explained, relative values of $P_m$ and $N_G$ have a stronger influence on the forms obtained than their values in absolute, and we thus set $P_m$ to obtain territories containing at most 1 million inhabitants, what is a strong but not extreme density (for comparison, the Parisian region concentrates around 8 millions inhabitants on an area of a similar size). Values of $N_G$ vary considerably to cover a large number of possible dynamical regimes. Values of $\alpha$ and $\beta$ have been obtained through successive experimentations.
}{
Nous échantillonnons l'espace des paramètres en utilisant un \emph{Latin Hypercube Sampling}, les paramètres variant dans $\alpha \in [0.1,4],\beta \in [0,0.5],n_d \in \{1,\ldots , 5\}, N_G \in [500,30000], P_m \in [1e4,1e6]$. Comme nous l'avons expliqué, les valeurs relatives de $P_m$ et $N_G$ jouent plus sur les formes obtenues que leur valeurs dans l'absolu, et nous fixons donc $P_m$ pour obtenir des territoires contenant au maximum 1 million d'habitants, ce qui est une forte densité mais pas extrême (pour comparaison, la métropole parisienne concentre autour de 8 millions sur une zone de taille équivalente). Les valeurs de $N_G$ varient considérablement pour couvrir un grand nombre de régimes dynamiques possibles. Les valeurs de $\alpha$ et $\beta$ ont été obtenues par expérimentations successives.
}

\bpar{
This type of cribbing is a good compromise to have a reasonable sampling without being subject to the dimensionality curse within normal computation capabilities. We sample around 80000 parameters points, with 10 repetitions each (as established in the previous experiment). We recall the protocol followed here to obtain the behavior of a simulation model, which can be put into perspective into the more general one presented in~\ref{sec:computation}:
\begin{itemize}
	\item sampling of parameter points;
	\item simulation of the models for each parameter point, repeated 10 times;
	\item computation for each model execution of urban form indicators;
	\item aggregation for each parameter point by computing averages on repetitions\footnote{Given the shape of distributions obtained for 100 repetitions, presented in~\ref{app:sec:density}, the use of the average or the median given equivalent results.}.
\end{itemize}
}{
Ce type de criblage est un bon compromis pour avoir un échantillonnage raisonnable sans être soumis au sort de la dimension dans des capacités de calcul normales. Nous échantillonnons autour de 80000 points, avec 10 répétitions chacun (comme établi avec l'expérience précédente). Rappelons le protocole suivi ici pour obtenir le comportement d'un modèle de simulation, qui est à placer dans celui plus général présenté en~\ref{sec:computation} :
\begin{itemize}
	\item échantillonnage des points de paramètres ;
	\item simulation du modèle pour chaque point de paramètre, répétée 10 fois ;
	\item calcul pour chaque execution du modèle des indicateurs de forme ;
	\item agrégation pour chaque point de paramètres par calcul des moyennes sur les répétitions\footnote{Vu la forme des distributions obtenues pour 100 répétitions, présentées en~\ref{app:sec:density}, l'utilisation de la moyenne ou de la médiane donnent des résultats équivalents.}.
\end{itemize}
}

\bpar{
Full plots of model behavior as a function of parameters are given in Appendix~\ref{app:sec:density}. We show in~\ref{fig:density:fig3} some particularly interesting behavior for slope $\gamma$ and average distance $\bar{d}$. First of all, the overall qualitative behavior depending on aggregation strength, namely that lower alpha giver less hierarchical and more spread configurations, confirms the expected intuitive behavior.
}{
Des graphes complets du comportement du modèle en fonction des paramètres sont donnés en Annexe~\ref{app:sec:density}. Nous montrons en Fig.~\ref{fig:density:fig3} des comportements particulièrement intéressants pour la pente $\gamma$ et la distance $\bar{d}$. Tout d'abord, le comportement qualitatif général en fonction de la force d'agrégation, c'est-à-dire que des valeurs faibles de $\alpha$ donnent des configurations moins hiérarchiques et plus étalées, confirme le comportement attendu intuitivement.
}

\bpar{
The effect of diffusion strength $\beta$ is more difficult to grasp: the effect is inverted for slope between high and low growth rates but not for distance, that shows an inversion when $\alpha$ varies. In the low $N_G$ case, low diffusion creates more sprawled configuration when aggregation is low, but less sprawled when aggregation is high. Furthermore, all indicators show a more or less smooth transition around $\alpha \simeq 1.5$. Slope stabilize over certain values, meaning that the hierarchy cannot be forced more and indeed depends of the diffusion value, at least for low $N_G$ (right column). In general, higher valued for $P_m/N_G$ increase the effect of diffusion what could have been expected.
}{
L'effet de la force de diffusion $\beta$ est plus difficile à cerner : l'effet est inversé pour la pente entre des haut et bas taux de croissance mais pas pour la distance, qui elle présente une inversion quand $\alpha$ varie. Dans le cas où $N_G$ est faible, une diffusion faible crée des configurations plus étalées quand l'agrégation est basse, mais moins étalées quand l'agrégation est forte. De plus, tous les indicateurs présentent une transition plus ou moins abrupte autour de $\alpha \simeq 1.5$. La pente se stabilise au dessus de certaines valeurs, ce qui veut dire que la hiérarchie ne peut pas être forcée plus et dépend alors de la valeur de la diffusion, au moins pour les faibles $N_G$ (colonne de droite). En général, des valeurs fortes pour $P_m/N_G$ augmentent les effets de la diffusion, ce à quoi on pouvait s'attendre.
}

\bpar{
The existence of a minimum for slope at $n_d=1,P_m/N_G\in\left[13,26\right]$ and lowest $\beta$ is unexpected and witnesses a complex interplay between aggregation and diffusion. The emergence of this ``optimal'' regime is associated with shifts of the transition points in other cases: for example, lowest diffusion imply a transition beginning at lower values of $\alpha$ for average distance. This exploration confirms that complex behavior, in the sense of unpredictable emerging forms, occurs in the model: one cannot predict in advance the final form given some parameters, without referring to the full exploration of which we give an overview here.
}{
L'existence d'un minimum pour la pente à $n_d=1,P_m/N_G\in\left[13,26\right]$ et les valeurs faibles de $\beta$ est inattendue et témoigne d'une interaction complexe entre agrégation et diffusion. L'émergence de ce régime ``optimal'' est associé avec un décalage des points de transition dans les autres cas : par exemple une diffusion plus faible implique une transition commençant à des valeurs plus faible de $\alpha$ pour la distance. Cette exploration confirme qu'un comportement complexe, au sens de formes émergentes qui ne peuvent être prédites, est présent dans le modèle : il n'est pas possible de donner à l'avance la forme finale étant donné un jeu de paramètres, sans se référer à l'exploration complète dont nous avons donné un aperçu ici.
}

\begin{figure}
\includegraphics[width=\linewidth]{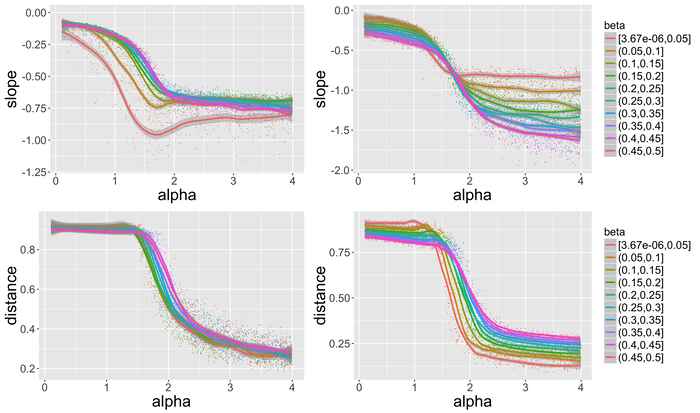}
\caption[Behavior of indicators]{\textbf{Behavior of indicators.} Slope $\gamma$ (top row) and average distance $\bar{d}$ (bottom row) as a function of $\alpha$, for different bins for $\beta$ given by curve color, for particular values $n_d=1,P_m/N_G\in\left[13,26\right]$ (left column) and $n_d=4,P_m/N_G\in\left[41,78\right]$ (right column). We observe in each case a transition as a function of $\alpha$, which properties are influenced by other parameters. For low values of $P_m/N_G$ and of $\beta$ emerges a counter-intuitive non-monotony.\label{fig:density:fig3}}
\end{figure}

\subsubsection{Semi-analytical analysis}{Analyse semi-analytique}


\bpar{
Our model can be understood as a type of reaction-diffusion model, that have been widely used in other fields such as biology as we synthesized in~\ref{sec:interdiscmorphogenesis}. An other way to formulate the model typical to these approaches is by using Partial Differential Equations (PDE). In the case of a firm growth model, which is a generalization of the Simon model with an arbitrary form of the attachment function, \cite{2017arXiv171007580R} show that a PDE and its general solution can be derived. We propose to gain insights into long-time dynamics by studying them on a simplified case. We consider the system in one dimension, such that $x\in \left[0;1\right]$ with $1/\delta x$ cells of size $\delta x$. A time step is given by $\delta t$. Each cell is characterized by its population as a random variable $P(x,t)$. We work on their expected values $p(x,t) = \Eb{P(x,t)}$, and assume that $n_d=1$. As developed in Appendix~\ref{app:sec:density}, we show that this simplified process verifies the following PDE:
}{
Notre modèle peut être compris comme un type de modèle de réaction-diffusion, qui ont été utilisés largement dans d'autres champs comme la biologie comme nous l'avons résumé en~\ref{sec:interdiscmorphogenesis}. Une autre façon de formuler les modèles typiques de ces approches est d'utiliser des équations aux dérivées partielles (EDP). Dans le cas d'un modèle de croissance de firmes, généralisation du modèle de Simon avec forme quelconque de la fonction d'attachement, \cite{2017arXiv171007580R} montrent qu'une EDP et sa solution générale peuvent être dérivées. Notre cas est plus délicat par l'ajout du processus de diffusion. Nous proposons d'éclairer des comportements des dynamiques de temps long en les étudiant sur un cas simplifié. Nous considérons le système en une dimension, tel que $x\in \left[0;1\right]$ avec $1/\delta x$ cellule de taille $\delta x$. Un pas de temps est donné par $\delta t$. Chaque cellule est caractérisée par sa population comme une variable aléatoire dépendant de la position $x$ et du temps $t$, que nous notons $P(x,t)$. Nous travaillons sur les espérances $p(x,t) = \Eb{P(x,t)}$, et supposons que $n_d=1$. Comme développé en Annexe~\ref{app:sec:density}, on peut montrer que ce processus simplifié obéit à l'EDP suivante :
}

\begin{equation}\label{eq:pde}
\hspace{-1.5cm}
\delta t \cdot \frac{\partial p}{\partial t} = \frac{N_G \cdot p^{\alpha}}{P_{\alpha}(t)} + \frac{\alpha \beta (\alpha - 1) \delta x^2}{2}\cdot \frac{N_G \cdot p^{\alpha-2}}{P_{\alpha}(t)} \cdot \left(\frac{\partial p}{\partial x}\right)^2 + \frac{\beta \delta x^2}{2} \cdot \frac{\partial^2 p}{\partial x^2} \cdot\left[ 1 + \alpha \frac{N_G p^{\alpha - 1}}{P_{\alpha(t)}} \right]
\end{equation}

\bpar{
where $P_{\alpha}(t) = \int_x p(x,t)^{\alpha} dx$. This non-linear equation can not be solved analytically, the presence of integral terms putting it out of standard methods, and numerical resolution must be used~\cite{tadmor2012review}.
}{
où $P_{\alpha}(t) = \int_x p(x,t)^{\alpha} dx$. Cette équation non-linéaire ne peut pas être résolue analytiquement, la présence de termes intégraux la mettant hors des méthodes standard, et la résolution numérique doit être utilisée comme le suggère~\cite{tadmor2012review}.
}

\bpar{
It is important to note that the simplified model can be expressed by a PDE analog to reaction-diffusion equations, as the one partially solved for a simpler model in~\cite{bosch1990velocity}. We show in \ref{app:sec:density} that because of the boundaries conditions, density (proportion of population) converges towards a stationary solution at long times, going through intermediate states in which the solution is partially stabilized, in the sense that its evolution speed becomes rather slow. These ``semi-stationary'' states are the ones used in two dimensions along with the dynamical ones. This study confirms that the variety of shapes obtained through the model is permitted both by the interplay of aggregation and diffusion as the equation couples them, but also by the values of $P_m / N_G$ that allow to set the convergence level. Indeed, the sensitivity of the stationary solution to parameters is very low compared to the shape of the world, and using the model in stationary mode would make no sense in our case.
}{
Il est important de noter que le modèle simplifié peut être exprimé comme une EDP analogue aux équations de réaction-diffusion, comme celle partiellement résolue pour un modèle plus simple dans \cite{bosch1990velocity}. Nous montrons en \ref{app:sec:density} qu'à cause des conditions au bord, la densité (au sens de la proportion de population) converge vers une solution stationnaire sur le temps long, en passant par des états intermédiaires pour lesquels la solution est partiellement stabilisée, au sens où sa vitesse d'évolution devient relativement lente. Ces états ``semi-stationnaires'' sont ceux utilisés en deux dimensions avec les états dynamiques. Cette étude confirme que la variété des formes obtenues par le modèle est permise à la fois par l'interactions entre l'agrégation et la diffusion puisque l'équation les couple, mais aussi par les valeurs de $P_m / N_G$ qui permet de fixer le niveau de convergence. En effet, la sensibilité de la solution stationnaire aux paramètres est très faible en comparaison de la forme du monde (en écho à notre étude sur la sensibilité aux conditions spatiales initiales en~\ref{sec:computation}), et utiliser le modèle en mode stationnaire n'aurait aucun sens dans notre cas.
}

\begin{figure}[h!]
\includegraphics[width=\linewidth]{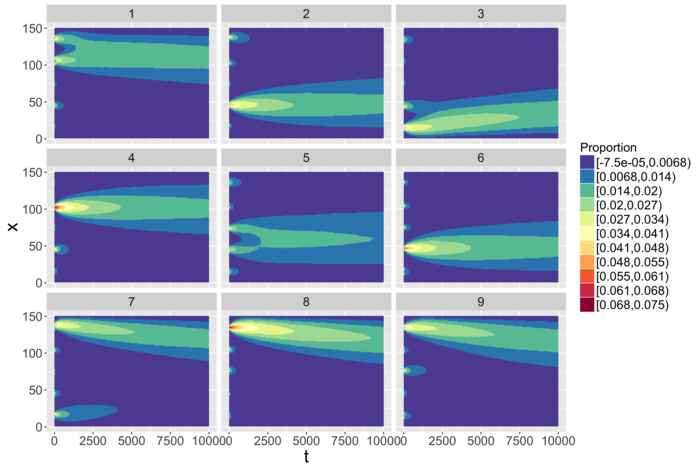}
\caption[Randomness and frozen accidents]{\textbf{Randomness and frozen accidents.} We show nine random realizations of the one dimensional system with similar initial conditions, namely five equidistant equally populated initial cells. Parameters are $\alpha = 1.4,\beta =0.1,N_G=10$. Each plot shows time against space, color level giving the proportion of population in each cell.\label{fig:density:fig4}}
\end{figure}

\bpar{
Finally, we use this toy case to demonstrate the importance of bifurcations in model dynamics. More precisely, we show that path-dependence is crucial for the final form. As illustrated in Fig.~\ref{fig:density:fig4}, using an initial condition making the choice ambiguous, corresponding to five equidistant equally populated cells, produces very different trajectories, as generally one of the spots will end dominating the others, but is totally random, witnessing dramatic bifurcations in the system at initial times. This aspect is typically expected in urban systems, since very precise characteristics will be included in the determinants of localization at the initial moments of system genesis: the existence of a very local resource, or the strategic advantage of the site (defensive or crossing site for example~\cite{hypergeo}), will determine on very long times the local territorial form. This aspect confirms the importance of robust indicators described before.
}{
Enfin, nous utilisons ce cas simplifié pour démontrer l'importance des bifurcations dans la dynamique du modèle. Plus précisément, nous montrons que la dépendance au chemin est cruciale pour la forme finale. Comme illustré en Fig.~\ref{fig:density:fig4}, l'utilisation d'une condition initiale rendant les choix ambigus, correspondant à 5 cellules équidistantes et de population égale, produit des trajectoires très différentes, puisqu'en général l'un des lieux finira par dominer les autres, mais est complètement aléatoire, témoignant de bifurcations cruciales dans le système aux instants initiaux. Cet aspect est typiquement attendu dans les systèmes urbains, puisque des caractéristiques très précises feront partie des déterminants de la localisation aux instants initiaux de la genèse du système : l'existence d'une ressource très locale, ou l'avantage stratégique du site (site défensif ou site de franchissement par exemple~\cite{hypergeo}), détermineront sur des temps très longs la forme territoriale locale. Cet aspect confirme l'importance d'indicateurs morphologiques robustes décrits précédemment.
}

\subsubsection{Model calibration}{Calibration du modèle}

\bpar{
We finally turn to the the calibration of the model, that is done on the morphological objectives. As a single calibration for each real cell is computationally out of reach, we use the previous model exploration and superpose the point clouds with real indicator values. Full scatterplots of all indicators against each other, for simulated and real configurations, are given in~\ref{app:sec:density}. We find that the real point cloud is mostly contained within the simulated, that extend in significantly larger areas. It means that for a large majority of real configuration, there exist model parameters producing in average exactly the same morphological configuration. The highest discrepancy is for the distance indicator, the model failing to reproduce configuration with high distance, low Moran and intermediate hierarchy. These could for example correspond to polycentric configurations with many consequent centers.
}{
Nous traitons finalement la calibration du modèle, qui est faite sur les objectifs morphologiques. Comme une calibration pour chaque cellule réelle est hors de portée en terme de calcul, nous utilisons l'exploration précédente du modèle et superposons le nuage de points avec les valeurs réelles des indicateurs. Les scatterplots pour chaque couple d'indicateurs, pour les configurations simulées et les réelles, sont donnés en~\ref{app:sec:density}. Nous constatons que le nuage de points réels est en majorité contenu dans le simulé, qui s'étend également sur de grandes surfaces non couvertes par le réel. Cela signifie que pour une grande majorité des configurations réelles, il existe des valeurs des paramètres qui produisent en moyenne exactement la même configuration telle que résumée par les indicateurs de forme. Les plus grands écarts sont pour l'indicateur de distance, le modèle échouant à produire des configurations avec une valeur élevée de la distance, un Moran faible et une hiérarchie intermédiaire. Cela peut par exemple correspondre à des configurations polycentriques avec de nombreux centres conséquents.
}

\begin{figure}
\includegraphics[width=\linewidth]{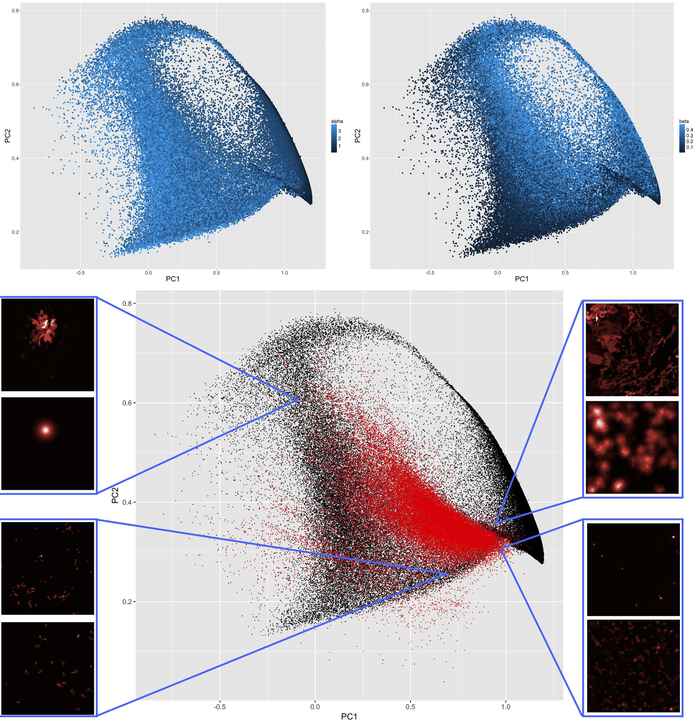}
\caption[Model calibration]{\textbf{Model calibration.} \textit{(Top)} Simulated configurations in the two first principal components plan, color level giving the influence of $\alpha$ (left) and of $\beta$ (right); \textit{(Bottom)} Simulated points in the same space (in black) with real configurations (in red). We show around the plot typical examples of real configurations and their simulated counterparts in different regions of the space, the first being the real and the second the simulated in each case: Top left geographical coordinates 25.7361,44.69989 - Romania, Bucharest - simulation parameters $\alpha=3.87,\beta=0.432,N_G=1273,nd=4,P_m=63024$ ; Top right geographical coordinates -2.561874,41.30203 - Spain, Castilla et Leon, Soria - simulation parameters $\alpha=1,\beta=0.166,N_G=100,nd=1,P_m=10017$; Bottom left geographical coordinates 27.16068,65.889 - Finland, Lapland - simulation parameters $\alpha=0.4,\beta=0.006,N_G=25,nd=1,P_m=849$; Bottom right geographical coordinates -2.607152,39.74274 - Spain, Castilla-La Mancha, Cuenca - simulation parameters $\alpha=1.14,\beta=0.108,N_G=637,nd=1,P_m=13235$.\label{fig:density:densitycalib}}
\end{figure}

\bpar{
We consider a more loose calibration constraint, by doing a Principal Component Analysis on synthetic and real morphological values, and consider the two first components only. These represent 85\% of cumulated variance. The rotated point clouds along these dimensions are shown in Fig.~\ref{fig:density:densitycalib}. Most of the real point cloud falls in the simulated one in this simplified setting. We illustrate particular points with real configurations and their simulated counterparts: for example Bucharest, Romania, corresponds to a monocentric semi-stationary configuration, with very high aggregation but also diffusion and a rather low growth rate. Other examples show less populated areas in Spain and Finland. From the plots giving parameter influence, we can show that most real situation fall in the region with intermediate $\alpha$ but quite varying $\beta$. It is consistent with real scaling urban exponents having a variation range rather small (between 0.8 and 1.3 generally~\cite{pumain2006evolutionary}) compared to the one we allowed in the simulations, whereas the diffusion processes may be much more diverse.
}{
 Nous considérons une contrainte de calibration plus faible, en procédant à une analyse en composantes principales sur les valeurs normalisées des indicateurs morphologiques pour les configurations synthétiques et réelles, et ne considérons que les deux premières composantes seulement. Celles-ci représentent 85\% de la variance cumulée. Les nuages de points projeté sur ces dimensions est montré en Fig.~\ref{fig:density:densitycalib}. La majorité du nuage réel tombe dans le simulé dans cette configuration simplifiée. Nous illustrons des points particuliers avec des configurations réelles et leur contrepartie simulée : par exemple Bucarest, Roumanie, correspond à une configuration monocentrique semi-stationnaire, avec une forte agrégation mais aussi diffusion et un taux de croissance plutôt bas. Les autres exemples montrent des zones moins peuplées en Espagne et en Finlande. À partir des graphes montrant l'influence des paramètres, on peut montrer que la plupart des situations réelles tombent dans la région avec des valeurs intermédiaires pour $\alpha$ mais $\beta$ assez variable. Cela est cohérent avec le fait que les exposants de lois d'échelles urbaines ont une plage de variation plutôt étroite (entre 0.8 et 1.3 généralement~\cite{pumain2006evolutionary}) comparée à celle que nous avons permis dans les simulations, tandis que les processus de diffusion peuvent être bien plus divers.
}

\bpar{
This way, we have shown that the model is able to reproduce most of existing urban density configuration in Europe, despite its rather simplicity. It confirms that in terms of urban form, most of drivers at this scale can be translated into these abstract processes of aggregation and diffusion. It also implies that urban functions, which can be quantified by similar indicators on their spatial distribution, play a reduced role in the location of populations, or that they must be quite correlated to the distribution of population (and are indeed taken into account in an abstract way in the aggregation function).
}{
Ainsi, nous avons montré que le modèle est capable de reproduire la majorité des configurations de densité en Europe, malgré sa relative simplicité. Cela confirme qu'en termes de forme urbaine, la plupart des facteurs à cette échelle peuvent être traduits dans ces processus abstraits d'agrégation et de diffusion. Cela implique également que les fonctions urbaines, qui pourraient être quantifiées par des indicateurs similaires sur leur distribution spatiale, ne jouent que peu de rôle dans la localisation des populations, ou alors que celles-ci sont fortement corrélées à la distribution de la population (et sont en fait prises en compte de manière abstraite dans la fonction d'agrégation).
}


\subsection{Discussion}{Discussion}

\paragraph{Calibration and model refinement}{Raffinement de la calibration et du modèle}

\bpar{
Further work on this simple model may consist in extracting the exact parameter space covering all real situations and provide interpretation of its shape, in particular through correlations between parameters and expressions of boundaries functions. Its volume in different directions should furthermore give the relative importance of parameters.
}{
Des développements futurs sur ce modèle simple peuvent consister en l'extraction de l'espace des paramètres exact couvrant l'ensemble des situations réelles et fournir une interprétation de sa forme, en particulier par le lien entre les valeurs des paramètres de génération et l'expression de la frontière de l'espace faisable en termes d'indicateurs. Son volume dans différentes directions devrait de plus donner l'importance relative des paramètres. 
}

\begin{figure}
\includegraphics[width=\linewidth]{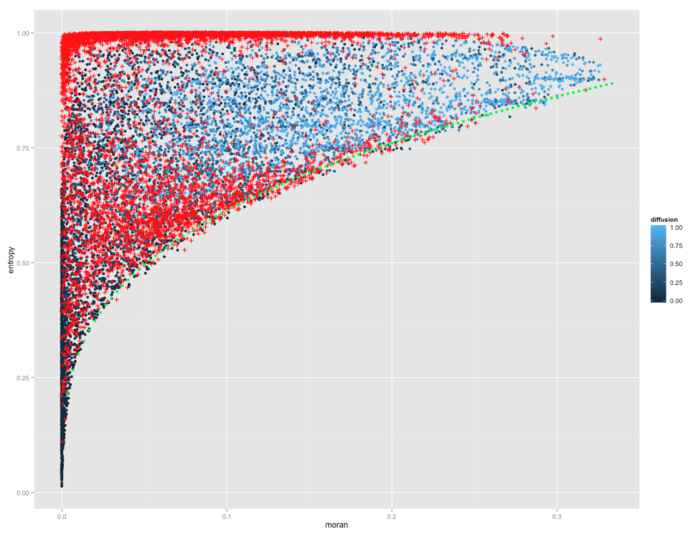}
\caption[PSE exploration]{\textbf{PSE exploration.} Scatterplot of Moran against Entropy, with blue points obtained with LHS and red with PSE exploration. Green dashed line gives feasible lower bound.\label{fig:density:fig6}}
\end{figure}

\bpar{
Concerning the feasible space for the model of simulation itself, we tested a targeted exploration algorithm, giving promising results. More precisely, the Parameter Space Exploration (PSE) algorithm~\cite{10.1371/journal.pone.0138212} which is implemented in OpenMole, is aimed at determining all the possible outputs of a simulation model, i.e. samples its output space rather than input space. We obtain promising results as shown in Fig.~\ref{fig:fig6}: we find that the lower bound in Moran-entropy plan, confirmed by the algorithm, unexpectedly exhibit a scaling relationship. It would mean that at a given level of auto-correlation, that one could want to attain for sustainability reasons for example (optimality through co-location), imposes a minimal disorder in the configuration of activities.
}{
Concernant l'espace faisable pour le modèle de simulation en lui-même, nous avons testé un algorithme d'exploration ciblée, qui donne des résultats prometteurs. Plus précisément, l'algorithme PSE~\cite{10.1371/journal.pone.0138212} qui est implémenté dans OpenMole, a pour but de déterminer toutes les sorties possibles d'un modèle de simulation, c'est-à-dire échantillonne son espace de sortie plutôt que d'entrée. Nous obtenons des résultats intéressants comme montré en Fig.~\ref{fig:density:fig6} : nous trouvons que la borne inférieure dans le plan Moran-entropie, confirmée par l'algorithme, exhibe une loi d'échelle de manière inattendue (puisqu'il est impossible a priori de déterminer cet espace non-faisable avec seulement les formules d'indicateurs, celui-ci étant témoin de la réalité de structures urbaines même simulées). Cela voudrait dire qu'un niveau fixé d'auto-corrélation, qu'on pourrait vouloir atteindre pour des raisons de soutenabilité par exemple (optimalité par co-localisation), impose un désordre minimal dans la configuration des activités.
}

\bpar{
Other relations between indicators and as a function of parameters can be the object of similar future developments. The question of doing a dynamical calibration of the model, i.e. trying to reproduce configurations at successive times, is conditioned to the availability of population data at this resolution in time.
}{
D'autres relations entre indicateurs et comme fonction des paramètres peuvent être l'objet de développements futurs similaires. La possibilité d'une calibration dynamique du modèle, i.e. essayer de reproduire des configurations à des dates successives, est conditionnée à la disponibilité des données de population à cette résolution dans le temps.
}


\bpar{
We aimed at using abstract processes rather than having a highly realistic model. Tuning some mechanisms is possible to have a model closer to reality in microscopic processes: for example thresholding the local population density, or stopping the diffusion at a given distance from the center if it is well defined. It is however far from clear if these would produce such a variety of forms and could be calibrated in a similar way, as being accurate locally does not mean being accurate at the mesoscopic level for morphological indicators. Allowing the parameters to locally vary, i.e. being non stationary in space, or adding randomness to the diffusion process, are also potential model refinements.
}{
Nous avons visé à utiliser des processus abstraits plutôt que d'avoir un modèle hautement réaliste. La modification de certains mécanismes est possible pour avoir un modèle plus proche de la réalité des processus microscopiques : par exemple plafonner la densité de population locale, ou stopper la diffusion à une distance donnée du centre s'il est bien défini. Il est cependant loin d'être clair dans quelle mesure ceux-ci produiraient une telle variété de formes et pourraient être calibrés de la même façon, puisqu'être précis localement n'implique pas d'être précis au niveau mesoscopique pour les indicateurs morphologiques. Permettre aux paramètres de varier localement, i.e. être non-stationnaires dans l'espace, ou ajouter de l'aléatoire au processus de diffusion, sont également des raffinements possibles du modèle.
}

\bpar{
In conclusion, we have provided a calibrated spatial urban morphogenesis model at the mesoscopic scale that can reproduce a significant proportion of European urban pattern in terms of morphology. We demonstrate that the abstract processes of aggregation and diffusion are sufficient to capture urban growth processes at this scale. It is meaningful in terms of policies based on urban form such as energy efficiency, but also means that issues out of this scope must be tackled at other scales or through other dimensions of urban systems.
}{
En conclusion, nous avons produit un modèle spatial de morphogenèse urbaine à l'échelle mesoscopique, dont la calibration permet de reproduire une proportion considérable des configurations territoriales européennes en termes de morphologie. Nous démontrons que les processus abstraits d'agrégation et diffusion sont suffisants pour capturer la dimension morphologique des processus de croissance urbaine à cette échelle. Cela a des implications par exemple en termes de politiques basées sur la forme urbaine comme l'efficacité énergétique, mais aussi signifie que les questions hors de ce cadre doivent être traitées à d'autres échelles ou par d'autres dimensions des systèmes urbains.
}

\stars

\bpar{
The first section of this chapter allowed to deepen the definition of morphogenesis, whereas the second lead to the construction of a simple model of urban morphogenesis, allowing to generate population distributions, that can be understood as territorial configurations.
}{
La première section de ce chapitre a permis de creuser la définition de la morphogenèse, tandis que la deuxième a mené à la construction d'un modèle simple de morphogenèse, permettant de générer des distributions de population, qu'on peut voir comme des configurations territoriales.
}

\bpar{
We will now use this framework to introduce the morphogenesis of networks conditioned to a territorial morphogenesis, to progressively shift towards co-evolution models.
}{
Nous allons à présent exploiter ce cadre pour introduire la morphogenèse des réseaux conditionnée à une morphogenèse territoriale, pour nous diriger progressivement vers des modèles de co-évolution.
}

\stars

%


\newpage

\section{Generation of correlated territorial configurations}{Génération de configurations territoriales corrélées}

\label{sec:correlatedsyntheticdata}

\bpar{
This section aims to explore the sequential coupling (or simple coupling) between previous model of density generation and an heuristic of network growth. We explore therein the feasible space of correlations between network measures and morphological measures. We first recall the issues linked to the notion of synthetic data and the role of correlation structures in these.
}{
Cette section vise à explorer un couplage séquentiel (ou couplage simple) du modèle de génération de densité précédent avec une heuristique de croissance de réseau. Nous explorons par là un espace faisable de corrélations entre les mesures de réseau et les mesures morphologiques. Rappelons dans un premier temps les enjeux de la notion de données synthétiques et du rôle des structures de corrélation dans celles-ci.
}

\subsection{Correlated geographical data of density and network}{Données géographiques corrélées de densité et de réseau}

\bpar{
One of the inspirations and applications of the current approach is the generation of synthetic data, for example to feed sensitivity analysis to the spatial configuration presented in section~\ref{sec:computation}. The use of synthetic data in geography is generally directed towards the generation of synthetic populations within agent-based models (mobility, \emph{LUTI} models)~\cite{pritchard2009advances}. We can make a link with some spatial analysis techniques. The extrapolation of a continuous spatial field from a discrete spatial sample through a kernel density estimation for example can be understood as the creation of a synthetic dataset (even if it is not generally the initial view, as in Geographically Weighted Regression~\cite{brunsdon1998geographically} in which variable size kernels do not interpolate data \emph{stricto sensu} but extrapolate abstract variables representing interaction between explicit variables).
}{
L'une des inspirations et applications de la présente démarche est la génération de données synthétiques, par exemple pour alimenter les analyses de sensibilité à la configuration spatiale présentées en section~\ref{sec:computation}. En géographie, l'utilisation de données synthétiques est plus généralement axée vers l'utilisation de populations synthétiques au sein de modèles multi-agents, comme par exemple des modèles de mobilité, des modèles \emph{LUTI}~\cite{pritchard2009advances}. On peut également citer des méthodes d'analyse spatiale qui s'en rapprochent : par exemple, l'extrapolation d'un champ spatial continu à partir d'un échantillon discret, par une estimation par noyaux par exemple, peut être compris comme la génération d'un jeu de données synthétiques, même si ce n'est pas le point de vue initial. Par exemple, dans le cas de la Regression Géographique Pondérée~\cite{brunsdon1998geographically}, les noyaux de tailles variables construisent des champs abstraits représentant un potentiel généré par des variables explicites définies en des points précis de l'espace.
}

\bpar{
In the field of modeling in quantitative geography, \emph{toy-models} or hybrid models require a consistent initial spatial configuration. A set of possible initial configurations becomes a synthetic dataset on which the model is tested. The first Simpop model~\cite{sanders1997simpop}, precursor of a large family of models later parametrized with real data, could enter that frame but was studied on an unique synthetic spatialization. Similarly underlined was the difficulty to generate an initial transportation infrastructure in the case of the SimpopNet model~\cite{schmitt2014modelisation} although it was admitted as a cornerstone of knowledge on the behavior of the model.
}{
Dans le domaine de la modélisation en géographie quantitative, dans le cas de \emph{modèles jouets} ou de modèles hybrides, une configuration initiale cohérente est souvent essentielle : un ensemble de configurations initiales possibles est alors un jeu de données synthétiques sur lesquelles le modèle est testé : le premier modèle Simpop~\cite{sanders1997simpop}, pionnier d'une famille de modèles par la suite configurés sur des données observées, pourrait rentrer dans ce cadre mais était utilisé sur une spatialisation synthétique unique. De même, il a été souligné la difficulté de générer une configuration initiale pour une infrastructure de transport dans le cas du modèle SimpopNet~\cite{schmitt2014modelisation}, alors qu'il s'agit d'un point essentiel dans la connaissance du comportement du modèle.
}

\bpar{
A systematic control of spatial configuration effects on the behavior of simulation models was only recently proposed~\cite{cottineau2015revisiting}, and as we developed in~\ref{sec:computation}, this approach that can be interpreted as a statistical control on spatial data. The aim is to be able to distinguish proper effects due to intrinsic model dynamics from particular effects due to the geographical structure of the case study. Such results are essential for the validation of conclusions obtained with modeling and simulation practices in quantitative geography. Indeed, as we reviewed in~\ref{sec:computation}, most modeling experiments systematically explore the influence of parameters, but not of spatial initial configurations, although they may have a stronger influence than parameters.
}{
Il a récemment été proposé de contrôler systématiquement les effets de la configuration spatiale sur le comportement de modèles de simulation spatialisés~\cite{cottineau2015revisiting}, et comme nous l'avons développé en~\ref{sec:computation}, cette méthodologie pouvant être interprétée comme un contrôle par données statistiques spatiales. L'enjeu est de pouvoir alors distinguer les effets propres dus à la dynamique intrinsèque du modèle, des effets particuliers dus à la structure géographique du cas d'application. Celui-ci est crucial pour la validation des conclusions issues des pratiques de modélisation et simulation en géographie quantitative. En effet, comme nous l'avons revu en~\ref{sec:computation}, la plupart des expériences de modélisation explorent systématiquement l'influence des paramètres, mais pas des configurations spatiales initiales, alors que celles-ci peuvent avoir une influence plus grande que les paramètres.
}

\subsection{Model and results}{Modèle et résultats}

\subsubsection{Formalization}{Formalisation}

\bpar{
We propose in our case to generate territorial systems summarized in a simplified way as a spatial population density $d(\vec{x})$ and a transportation network $n(\vec{x})$. Correlations we aim to control are correlations between urban morphological measures and network measures. The question of interactions between territories and networks is already well-studied~\cite{offner1996reseaux} but stays highly complex and difficult to quantify~\cite{offner1993effets}. A dynamical modeling of implied processes should shed light on these interactions \cite{bretagnolle:tel-00459720} (p. ~162). We develop in that frame a \emph{simple} coupling (i.e. without any feedback loop) between a density distribution model and a network morphogenesis model.
}{
Dans notre cas, nous proposons de générer des systèmes de villes représentés par une densité spatiale de population $d(\vec{x})$ et la donnée d'un réseau de transport $n(\vec{x})$, représenté de façon simplifiée, pour lesquels on serait capable de contrôler les corrélations entre mesures morphologiques de la densité urbaine et caractéristiques du réseau. Nous rappelons que la question des interactions entre territoires et réseaux de transport est un sujet d'étude classique~\cite{offner1996reseaux}, mais toujours majoritairement ouvert, extrêmement complexe et difficile à quantifier~\cite{offner1993effets}. Une modélisation dynamique des processus impliqués devrait apporter des connaissances sur ces interactions \cite{bretagnolle:tel-00459720} (p.~162). Dans ce cadre, nous développons un couplage \emph{simple} (c'est-à-dire sans boucle de rétroaction) entre un modèle de morphogenèse urbaine et un modèle de génération de réseau.
}

\paragraph{Density model}{Modèle de densité}

\bpar{
The density model is the model described and explored in the previous section~\ref{sec:densitygeneration}. We use it for the conditional generation of network.
}{
Les modèle de densité est celui décrit et exploré dans la section précédente~\ref{sec:densitygeneration}. Nous l'utilisons pour la génération conditionnelle du réseau.
}

\paragraph{Network model}{Modèle de réseau}

\bpar{
On the other hand, we are able to generate a planar transportation network by a model $N$, at a similar scale and given a density distribution. Because of the conditional nature to the density of the generation process, we will first have conditional estimators for network indicators, and secondly natural correlations between network and urban shapes should appear as processes are not independent. The nature and modularity of these correlations as a function of model parameters are still to determine by exploration of the coupled model.
}{
D'autre part, il est possible de générer par un modèle $N$ un réseau de transport planaire à une échelle équivalente, étant donné une distribution de densité. La génération du réseau étant conditionnée à la donnée de la densité, les estimateurs des indicateurs de réseau seront conditionnels d'une part, et d'autre part les formes urbaines et du réseau devraient nécessairement être corrélées, les processus n'étant pas indépendants. La nature et la modularité de ces correlations selon la variation des paramètres des modèles restent à déterminer par l'exploration du modèle couplé.
}

\bpar{
Concerning the choice of the heuristic to generate an infrastructure network, we have reviewed in~\ref{sec:modelingsa} several models allowing it. Furthermore, we will compare different models in an operational manner in~\ref{sec:networkgrowth}. The aim here being to demonstrate the feasibility of the coupling within a morphogenesis model and also to explore the feasible space of correlation, we propose a unique heuristic, which is inspired by the model of~\cite{schmitt2014modelisation}, and simplifies it by removing the stochastic character. This heuristic is detailed below.
}{
Concernant le choix de l'heuristique de génération d'un réseau d'infrastructure, nous avons revu en~\ref{sec:modelingsa} de nombreux modèles le permettant. D'autre part, nous comparerons différents modèles de manière opérationnelle en~\ref{sec:networkgrowth}. Le but ici étant de démontrer la faisabilité du couplage au sein d'un modèle de morphogenèse ainsi que d'explorer l'espace faisable des corrélations, nous proposons une heuristique unique, qui s'inspire du modèle de~\cite{schmitt2014modelisation}, tout en le simplifiant par suppression du caractère stochastique. Cette heuristique est détaillée ci-dessous.
}

\bpar{
The heuristic network generation procedure is the following :
\begin{enumerate}
\item A fixed number $N_c$ of centers that will be first nodes of the network si distributed given density distribution, following a similar law to the aggregation process, i.e. the probability to be distributed in a given patch is $\frac{(P_i/P)^{\alpha}}{\sum (P_i/P)^{\alpha}}$. Population is then attributed according to Voronoi areas of centers, such that a center cumulates population of patches within its extent.
\item Centers are connected deterministically by percolation \cite{callaway2000network} between closest clusters : as soon as network is not connected, two closest connected components in the sense of minimal distance between each vertices are connected by the link realizing this distance. It yields a tree-shaped network.
\item Network is modulated by potential breaking in order to be closer from real network shapes. More precisely, a generalized gravity potential between two centers $i$ and $j$ is defined by
\[
V_{ij}(d) = \left[ (1 - k_h) + k_h \cdot \left( \frac{P_i P_j}{P^2} \right)^{\gamma_G} \right]\cdot \exp{\left( -\frac{d}{r_g (1 + d/d_0)} \right)}
\]
where $d$ can be euclidian distance $d_{ij}=d(i,j)$ or network distance $d_N(i,j)$, $k_h \in [0,1]$ a weight to modulate role of populations in the potential, $\gamma$ giving shape of the hierarchy across population values, $r_g$ characteristic interaction distance and $d_0$ distance shape parameter (allowing to flatten the distribution in low values). This form of potential assumes on the one hand that the attenuation of interaction to distance is independent from the strength of interaction due to weights (standard assumption of gravity models); on the other hand that a constant term due to distance can wight more or less (weighting by $k_h$); and finally that the distance function take as parameter a characteristic distance, but also a shape parameter, allowing for example to control the decrease on low distances.
\item A fixed number $K\cdot N_L$ of potential new links is taken among couples having greatest euclidian distance potential ($K=5$ is fixed, this value producing experimentally reasonable length links): this stage allows to eliminate very short links with a small population and very long links.
\item Among potential links, $N_L$ are effectively realized, that are the one with smallest rate $\tilde{V}_{ij} = V_{ij}(d_N)/V_{ij}(d_{ij})$. At this stage only the gap between euclidian and network distance is taken into account : $\tilde{V}_{ij}$ does indeed not depend on populations and is increasing with $d_N$ at constant $d_{ij}$.
\item Planarity of the network is forced by creation of nodes at possible intersections created by new links (with the former network or between new links)\footnote{Our model is different from~\cite{schmitt2014modelisation} on that point, as we simplify and do not assume levels of hierarchy between links.}.
\end{enumerate}
}{
La procédure de génération heuristique de réseau est la suivante :
\begin{enumerate}
\item Un nombre fixé $N_c$ de centres qui seront les premiers noeuds du réseau est distribué selon la distribution de densité, suivant une loi similaire à celle d'agrégation, i.e. la probabilité d'être distribué sur une case est $\frac{(P_i/P)^{\alpha}}{\sum (P_i/P)^{\alpha}}$ en notant $P_i$ la population d'une case et $P$ la population totale. La population est ensuite répartie selon les zones de Voronoï des centres, un centre cumulant la population des cases dans son emprise.
\item Les centres sont connectés de façon déterministe par percolation \cite{callaway2000network} entre plus proches clusters : tant que le réseau n'est pas connexe, les deux composantes connexes les plus proches au sens de la distance minimale entre chacun de leurs sommets sont connectées par le lien réalisant cette distance. Nous obtenons alors un réseau arborescent.
\item Le réseau est alors modulé par ruptures de potentiels afin de se rapprocher de formes réelles. Plus précisément, un potentiel d'interaction gravitaire généralisé entre deux centres $i$ et $j$ de populations $P_i,P_j$ est défini par
\[
V_{ij}(d) = \left[ (1 - k_h) + k_h \cdot \left( \frac{P_i P_j}{P^2} \right)^{\gamma_G} \right]\cdot \exp{\left( -\frac{d}{d_G (1 + d/d_0)} \right)}
\]
où $d$ peut être la distance euclidienne ou la distance par le réseau, $k_h \in [0,1]$ est un poids permettant de changer le rôle des populations dans le potentiel, $\gamma_G$ régit la forme de la hiérarchie selon les valeurs des populations, $d_G$ est la distance caractéristique de décroissance et $d_0$ paramètre de forme (permettant d'aplatir la distribution dans les faibles valeurs). Cette forme de potentiel suppose d'une part que l'attenuation de l'interaction due à la distance est indépendante de la force de l'interaction due aux poids (hypothèse standard des modèles gravitaires) ; d'autre part qu'un terme constant dû à la distance peut prendre plus ou moins de poids (pondération par $k_h$) ; et enfin que la fonction de distance prend comme paramètre une distance caractéristique, mais aussi un paramètre de forme, permettant par exemple de contrôler la décroissance sur les faibles distances.
\item Un nombre $K\cdot N_L$ de nouveaux liens potentiels est pris comme les couples ayant le plus grand potentiel pour la distance euclidienne ($K=5$ est fixé, cette valeur produisant expérimentalement des liens pas trop courts) : cette étape permet d'éliminer les liens très courts à faible population et les liens trop long.
\item Parmi les liens potentiels, $N_L$ sont effectivement réalisés, qui sont ceux ayant le plus faible rapport $V_{ij}(d_N)/V_{ij}(d_{ij})$ où $d_{ij}=d(i,j)$ est la distance euclidienne et $d_N(i,j)$ la distance par le réseau. À cette étape seul l'écart entre distance euclidienne et distance par le réseau compte, ce rapport ne dépendant plus des populations et étant croissant en $d_N$ à $d_{ij}$ fixé.
\item Le réseau est planarisé par création de noeuds aux intersections induites par les nouveaux liens (avec l'ancien réseau ou entre nouveaux liens)\footnote{Notre modèle diffère également de~\cite{schmitt2014modelisation} sur ce point car nous simplifions et ne supposons pas différents niveaux de hiérarchie entre les liens.}.
\end{enumerate}
}

\bpar{
We insist on the fact that the network generation procedure is entirely heuristic and result of thematic assumptions (connected initial network, gravity-based link creation) combined with trial-and-error during first explorations. Other model types could be used as well, such biological self-generated networks~\cite{tero2010rules}, local network growth based on geometrical constraints optimization~\cite{barthelemy2008modeling}, or a more complex percolation model than the initial one that would allow the creation of loops for example. We could thus in the frame of a modular architecture, in which the choice between different implementations of a functional brick can be seen as a meta-parameter~\cite{cottineau2015incremental}, choose network generation function adapted to a specific need (as e.g. proximity to real data, constraints on output indicators, variety if generated forms).
}{
Notons que la construction du modèle de génération est heuristique, et que d'autres types de modèles comme un réseau biologique auto-généré~\cite{tero2010rules}, une génération par optimisation locale de contraintes géométriques \cite{barthelemy2008modeling} ou un modèle de percolation plus complexe que celui utilisé, peuvent le remplacer, et permettraient la création de boucles dans le réseau. Ainsi, dans le cadre d'une architecture modulaire où le choix entre différentes implémentations d'une brique fonctionnelle peut être vue comme méta-paramètre~\cite{cottineau2015incremental}, on pourrait choisir la fonction de génération adaptée à un besoin donné (par exemple proximité à des données réelles, contraintes sur les relations entre indicateurs de sortie, variété de formes générées).
}


\paragraph{Parameter space}{Espace des paramètres}

\bpar{
Parameter space for the coupled model\footnote{Weak coupling allows to limit the total number of parameters as a strong coupling would involve retroaction loops and consequently associated parameters to determine their structure and intensity. In order to diminish it, an integrated model would be preferable to a strong coupling, what is slightly different in the sense where it is not possible in the integrated model to freeze one of the subsystems to obtain a model of the other subsystem that would correspond to the non-coupled model.} is constituted by density generation parameters $\vec{\alpha}_D = (P_m/N_G , \alpha,\beta , n_d)$ (see section~\ref{sec:densitygeneration}; we study for the sake of simplicity the rate between population and growth rate instead of both varying, i.e. the number of steps needed to generate the distribution) and network generation parameters $\vec{\alpha}_N=(N_C,k_h,\gamma , r_g , d_0)$. We denote $\vec{\alpha} = (\vec{\alpha}_D,\vec{\alpha}_N)$. 
}{
L'espace des paramètres du modèle couplé\footnote{Le couplage faible permet de limiter le nombre total de paramètres puisqu'un couplage fort incluant des boucles de retroaction comprendrait nécessairement des paramètres supplémentaires pour régler la forme et l'intensité de celles-ci. Pour espérer le diminuer, il faudrait concevoir un modèle intégré, ce qui est différent d'un couplage fort dans le sens où il n'est pas possible de figer l'un des sous-systèmes pour obtenir un modèle de l'autre correspondant au modèle non-couplé.} est constitué des paramètres de génération de densité $\vec{\alpha}_D = (P_m/N_G , \alpha,\beta , n_d)$ (voir section~\ref{sec:densitygeneration} ; on s'intéresse pour simplifier au rapport entre population et taux de croissance, i.e. le nombre d'étapes nécessaires pour générer, et on fixe la population totale) et des paramètres de génération de réseau $\vec{\alpha}_N=(N_C,k_h,\gamma_G , d_G , d_0)$. Nous noterons $\vec{\alpha} = (\vec{\alpha}_D,\vec{\alpha}_N)$ l'ensemble des paramètres.
}

\paragraph{Indicators}{Indicateurs}

\bpar{
Urban form and network structure are quantified by numerical indicators in order to modulate correlations between these. Morphology is defined as a vector $\vec{M}=(r,\bar{d},\varepsilon,a)$ giving spatial auto-correlation (Moran index), mean distance, entropy and hierarchy (see~\ref{sec:staticcorrelations} for a precise definition of these indicators). Network measures $\vec{G} = (\bar{c},\bar{l},\bar{s},\delta)$ are with network denoted $(V,E)$
\begin{itemize}
\item Mean centrality $\bar{c}$ defined as average \emph{betweeness-centrality} (normalized in $[0,1]$) on all links.
\item Mean path length $\bar{l}$ given by
\[
\frac{1}{d_m}\frac{2}{|V|\cdot (|V|-1)}\sum_{i<j}d_N(i,j)
\]
\item with $d_m$ normalization distance taken here as world diagonal $d_m=\sqrt{2}N$.
\item Mean network speed~\cite{banos2012towards} which corresponds to network performance compared to direct travel, defined as $\bar{s} = \frac{2}{|V|\cdot (|V|-1)}\sum_{i<j}{\frac{d_{ij}}{d_N(i,j)}}$.
\item Network diameter $\delta = \max_{ij}d_N(i,j)$.
\end{itemize}
}{
On quantifie la forme urbaine et la forme du réseau, dans le but de moduler la corrélation entre ces indicateurs. La forme est définie par un vecteur $\vec{M}=(r,\bar{d},\varepsilon,a)$ donnant auto-corrélation spatiale (indice de Moran), distance moyenne, entropie, hiérarchie (voir~\ref{sec:staticcorrelations} pour une définition précise de ces indicateurs). Les mesures de la forme du réseau $\vec{G} = (\bar{c},\bar{l},\bar{s},\delta)$ sont, avec le réseau noté $(V,E)$,
\begin{itemize}
\item Centralité moyenne $\bar{c}$, définie comme la moyenne de la \emph{betweeness-centrality} (normalisée dans $[0,1]$) sur l'ensemble des liens.
\item Longueur moyenne des chemins $\bar{l}$ définie par
\[
\frac{1}{d_m}\frac{2}{|V|\cdot (|V|-1)}\sum_{i<j}d_N(i,j)
\]
avec $d_m$ distance de normalisation prise ici comme la diagonale du monde $d_m=\sqrt{2}N$.
\item Vitesse moyenne~\cite{banos2012towards}, qui correspond à la performance du réseau par rapport au trajet à vol d'oiseau, définie par $\bar{s} = \frac{2}{|V|\cdot (|V|-1)}\sum_{i<j}{\frac{d_{ij}}{d_N(i,j)}}$.
\item Diamètre du réseau $\delta = \max_{ij}d_N(i,j)$
\end{itemize}
}

\bpar{
We do not have at this stage any ``performance'' indicator for the network generation process, i.e. aiming at reproducing typical patterns or optimizing some criteria. These will come later in~\ref{sec:networkgrowth} when we will calibrate similar models on real data. We consider the examples shown in~\ref{fig:correlatedsyntheticdata:example} as elements of the feasible space, the question being if network shapes corresponding to realities or given stylized facts will be also the object of this calibration.
}{
Nous n'avons à ce stade pas d'indicateur de ``performance'' du processus de génération de réseau, c'est-à-dire visant à reproduire des motifs typiques ou optimisant certains critères. Ceux-ci viendront plus tard en~\ref{sec:networkgrowth} lorsqu'on calibrera des modèles similaires sur des données réelles. Nous considérons les exemples montrés en~\ref{fig:correlatedsyntheticdata:example} comme des éléments de l'espace faisable, la question de savoir si les formes de réseau correspondent à des réalités ou à des faits stylisés donnés sera également l'objet de cette calibration.
}

\paragraph{Covariance and correlation}{Covariance et corrélation}

\bpar{
We study the cross-correlation matrix $\Covb{\vec{M}}{\vec{G}}$ between morphology and network. We estimate it on a set of $n$ realizations at fixed parameter values $(\vec{M}\left[D(\vec{\alpha})\right],\vec{G}\left[N(\vec{\alpha})\right])_{1\leq i\leq n}$ with standard unbiased estimator. We will study the Pearson correlation associated to it, estimated in the same way. 
}{
On utilisera la matrice de covariance croisée $\Covb{\vec{M}}{\vec{G}}$ entre densité et réseau, estimée sur un jeu de $n$ réalisations à paramètres fixés $(\vec{M}\left[D(\vec{\alpha})\right],\vec{G}\left[N(\vec{\alpha})\right])_{1\leq i\leq n}$ par l'estimateur standard non-biaisé. On s'intéressera à la corrélation de Pearson qui y est associée, estimée de la même façon.
}

\subsubsection{Implementation}{Implémentation}

\bpar{
Coupling of generative models is done both at formal and operational levels. We interface therefore independent implementations. The OpenMole software~\cite{reuillon2013openmole} for intensive model exploration offers for that the ideal frame thanks to its modular language allowing to construct \emph{workflows} by task composition and interfacing with diverse experience plans and outputs. For operational reasons, density model is implemented in \texttt{scala} language as an OpenMole \texttt{plugin}, whereas network generation is implemented in agent-oriented language \texttt{NetLogo}~\cite{wilensky1999netlogo} because of its possibilities for interactive exploration and heuristic model construction. Source code is available for reproducibility on project repository\footnote{at \texttt{https://github.com/JusteRaimbault/CityNetwork/tree/master/Models/Synthetic}}.
}{
Le couplage des modèles génératifs est effectué à la fois au niveau formel et au niveau opérationnel, c'est-à-dire qu'on fait interagir des implémentations indépendantes. Pour cela, le logiciel OpenMole~\cite{reuillon2013openmole} utilisé pour l'exploration intensive, offre le cadre idéal de par son langage modulaire permettant de construire des \emph{workflows} par composition de tâches à loisir et de les brancher sur divers plans d'expérience et sorties. Pour des raisons opérationnelles, le modèle de densité est implémenté en langage \texttt{scala} comme un \texttt{plugin} d'OpenMole, tandis que la génération de réseau est implémentée en langage basé-agent \texttt{NetLogo}~\cite{wilensky1999netlogo}, ce qui facilite l'exploration interactive et la construction heuristique interactive. Le code source est disponible pour reproductibilité sur le dépôt du projet\footnote{À l'adresse \url{https://github.com/JusteRaimbault/CityNetwork/tree/master/Models/Synthetic}.}.
}

\begin{figure}
\includegraphics[width=0.9\linewidth,height=0.87\textheight]{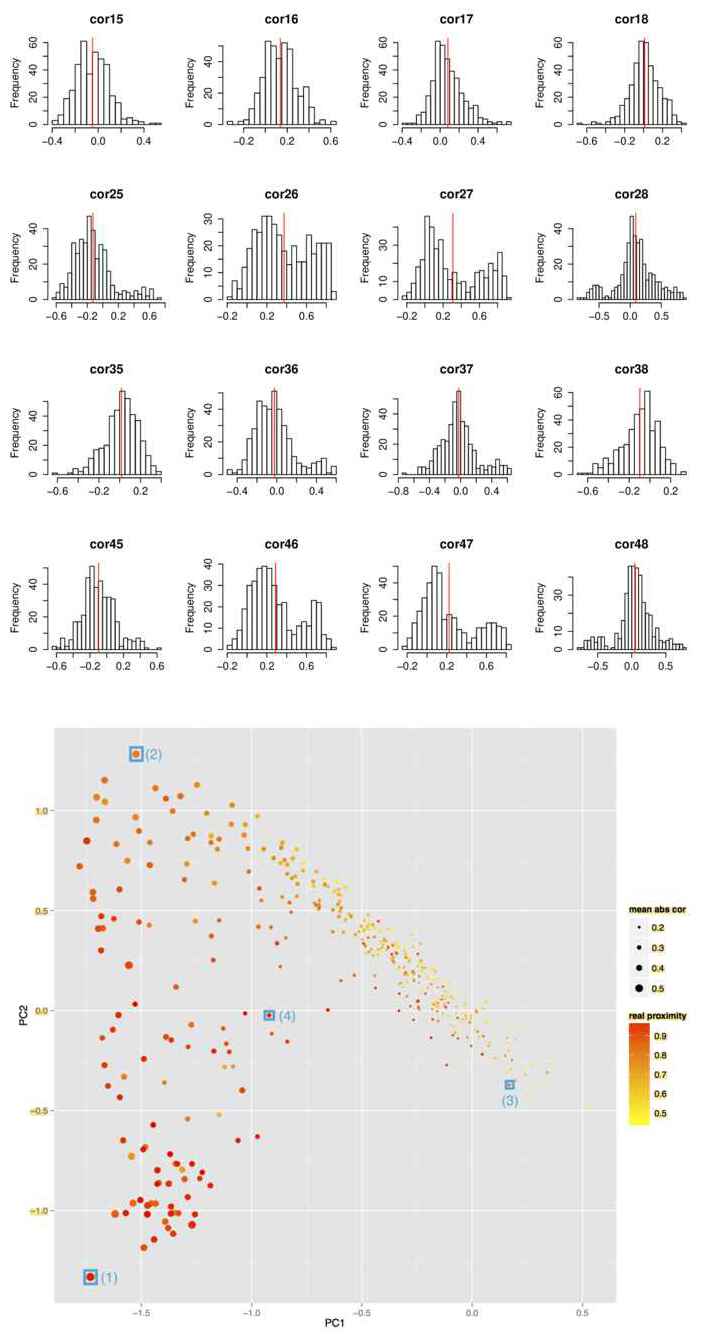}
\caption[Exploration of feasible space for correlations between urban morphology and network structure]{\textbf{Exploration of feasible space for correlations between urban morphology and network structure.} \textit{(a)} Distribution of crossed-correlations between vectors $\vec{M}$ of morphological indicators (in numbering order Moran index, mean distance, entropy, hierarchy) and $\vec{N}$ of network measures (centrality, mean path length, speed, diameter). \textit{(d)} Representation in the principal plan, scale color giving proximity to real data defined as $1 - \min_r \norm{\vec{M}-\vec{M}_r}$ where $\vec{M}_r$ is the set of real morphological measures, point size giving mean absolute correlation.\label{fig:correlatedsyntheticdata:densnwcor}}
\end{figure}

\begin{figure}
\includegraphics[width=\linewidth]{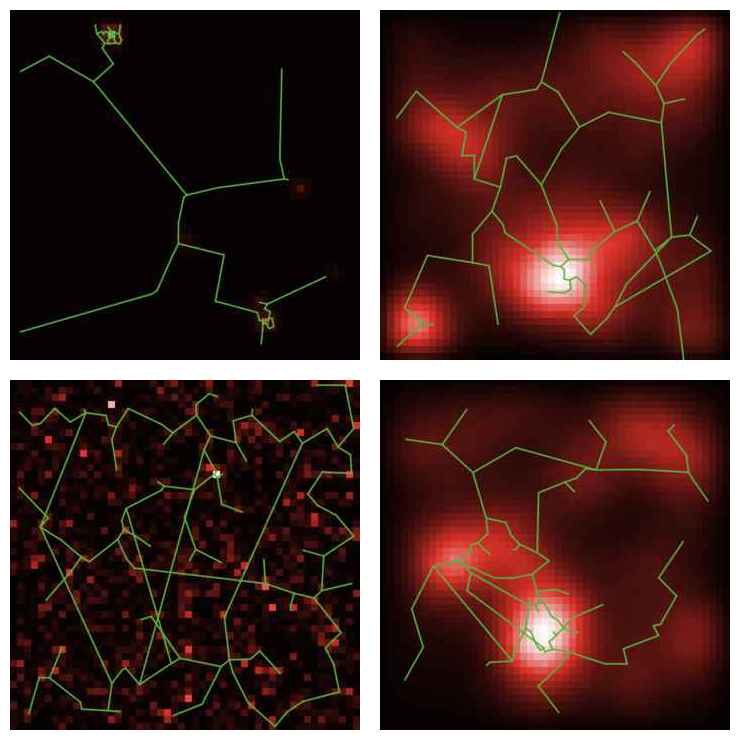}
\caption[Examples of generated coupled configurations]{Configurations obtained for parameters giving the four emphasized points in (d), in order from left to right and top to bottom. We recognize polycentric city configurations (2 and 4), diffuse rural settlements (3) and aggregated weak density area (1). See appendice for exhaustive parameter values, indicators and corresponding correlations. For example $\bar{d}$ is highly correlated with $\bar{l},\bar{s}$ ($\simeq$0.8) in (1) but not for (3) although both correspond to rural environments; in the urban case we observe also a broad variability: $\rho[\bar{d},\bar{c}]\simeq 0.34$ for (4) but $\simeq-0.41$ for (2), what is explained by a stronger role of gravitation hierarchy in (2) $\gamma=3.9,k_h=0.7$ (for (4), $\gamma=1.07,k_h=0.25$), whereas density parameters are similar.\label{fig:correlatedsyntheticdata:example}}
\end{figure}

\subsubsection{Results}{Résultats}

\bpar{
The study of density model alone is developed in the previous section. We recall that it is calibrated on European density grid data, on 50km width square areas with 500m resolution for which real indicator values have been computed on whole Europe. Furthermore, a grid exploration of model behavior yields feasible output space in reasonable parameters bounds (roughly $\alpha \in [0.5,2],N_G\in [500,3000], P_m \in [10^4,10^5],\beta\in [0,0.2], n_d \in \{ 1, \ldots , 4\}$). The reduction of indicators space to a two dimensional plan through a Principal Component Analysis (variance explained with two components $\simeq 80\%$) allows to isolate a set of output points that covers reasonably precisely real point cloud. It confirms the ability of the model to reproduce morphologically the set of real configurations.
}{
L'étude du modèle de densité seul est développée dans la section précédente. Pour rappel, il est notamment calibré sur les données de la grille européenne de densité, sur des zones de 50km de côté et de résolution 500m pour lesquelles les valeurs réelles des indicateurs ont été calculées pour l'ensemble de l'Europe. D'autre part, une exploration préliminaire du modèle permet d'estimer l'ensemble des sorties possibles dans des bornes raisonnables pour les paramètres (grossièrement $\alpha \in [0.5,2],N_G\in [500,3000], P_m \in [10^4,10^5],\beta\in [0,0.2], n_d \in \{ 1, \ldots , 4\}$). La réduction à un plan de l'espace des objectifs par une analyse en composantes principales (variance expliquée à deux composantes $\simeq 80\%$) permet d'isoler un nuage de points de sorties recouvrant assez fidèlement le nuage des points réels, ce qui veut dire que le modèle est capable de reproduire morphologiquement l'ensemble des configurations existantes.
}


\bpar{
Given the large relative dimension of parameter space, an exhaustive grid exploration is not possible. We use a Latin Hypercube sampling procedure with bounds given above for $\vec{\alpha}_D$ and for $\vec{\alpha}_N$, we take $N_C \in [50,120], r_g \in [1,100] , d_0 \in [0.1,10] , k_h \in [0,1] , \gamma \in [0.1,4],N_L\in [4,20]$. For number of model replications for each parameter point, less than 50 are enough to obtain confidence intervals at 95\% on indicators of width less than standard deviations. For correlations a hundred give confidence intervals (obtained with Fisher method) of size around 0.4, we take thus $n=80$ for experiments. 
}{
Dans le but d'illustrer la méthode de génération de données synthétiques, l'exploration a été orientée vers l'étude des corrélations. Étant donné la grande dimension relative de l'espace des paramètres, une exploration par grille exhaustive est impossible. Nous utilisons un plan d'expérience par criblage (hypercube latin), avec les bornes indiquées ci-dessus pour $\vec{\alpha}_D$ et pour $\vec{\alpha}_N$, on a $N_C \in [50,120], d_G \in [1,100] , d_0 \in [0.1,10] , k_h \in [0,1] , \gamma_G \in [0.1,4],N_L\in [4,20]$. Concernant le nombre de réplications du modèle pour chaque valeur des paramètres, moins de 50 sont nécessaires pour obtenir sur les indicateurs des intervalles de confiance à 95\% de taille inférieure aux déviations standard. Pour les corrélations, une centaine donne des intervalles de confiance (obtenus par méthode de Fisher) de taille moyenne 0.4, nous fixons donc $n=80$ pour l'expérience.
}

\bpar{
Figure~\ref{fig:correlatedsyntheticdata:densnwcor} gives details of experiment results. Regarding the subject of correlated synthetic data generation, we can sum up the main lines as following :
\begin{itemize}
\item Empirical distributions of correlation coefficients between morphology and network indicators are not simple and some are bimodal (for example $\rho[I,\bar{l}]$  between Moran index and mean path length which corresponds to \texttt{cor46} on Fig.~\ref{fig:correlatedsyntheticdata:densnwcor}).
\item it is possible to modulate up to a relatively high level of correlation for all indicators, maximal absolute correlation varying between 0.6 and 0.9. Amplitude of correlations varies between 0.9 and 1.6, allowing a broad spectrum of values. Point cloud in principal plan has a large extent but is not uniform : it is not possible to modulate at will any coefficient as they stay themselves correlated because of underlying generation processes. A more refined study at higher orders (correlation of correlations) would be necessary to precisely understand degrees of freedom in correlation generation.
\item Most correlated points are also the closest to real data, what confirms the intuition and stylized fact of a strong interdependence in reality.
\item Concrete examples taken on particular points in the principal plan show that similar density profiles can yield very different correlation profiles.
\end{itemize}
}{
La figure~\ref{fig:correlatedsyntheticdata:densnwcor} donne le détail des résultats de l'exploration. Nous retiendrons les résultats marquants suivants au regard de la génération de données synthétiques corrélées.
\begin{itemize}
\item les distributions empiriques des coefficients de corrélation entre indicateurs de forme et indicateurs de réseaux ne sont pas simples, pouvant être bimodales (par exemple $\rho[I,\bar{l}]$ entre l'index de Moran et le chemin moyen qui correspond à \texttt{cor46} sur la Fig.~\ref{fig:correlatedsyntheticdata:densnwcor}). 
\item Nous arrivons à générer un assez haut niveau de corrélation pour l'ensemble des indicateurs, la corrélation absolue maximale variant entre 0.6 et 0.9 ; l'amplitude varie quant à elle entre 0.9 et 1.6, ce qui permet un large spectre de valeurs. L'espace couvert dans un plan principal a une étendue certaine mais n'est pas uniforme : on ne peut pas moduler à loisir n'importe quel coefficient, ceux-ci étant liés par les processus de génération sous-jacents. Une étude plus fine aux ordres suivants (corrélation des corrélations) serait nécessaire pour cerner exactement la latitude dans la génération.
\item Les points les plus corrélés en moyenne sont également ceux les plus proches des données réelles, ce qui confirme l'intuition d'une forte interdépendance en réalité.
\item Des exemples concrets pris sur des points particuliers distants dans le plan principal montre que des configurations de densité proches peuvent présenter des profils de corrélations très différents.
\end{itemize}
}



\subsection{Discussion}{Discussion}

\subsubsection{Developments}{Développements}

\bpar{
This case study could be refined by extending correlation control method. A precise knowledge of $N$ behavior (statistical distributions on an exhaustive grid of parameter space) conditional to $D$ would allow to determine $N^{<-1>} | D$ and have more latitude in correlation generation. We could also apply specific exploration algorithms to reach exceptional configurations realizing an expected correlation level, or at least to obtain a better knowledge of the feasible space of correlations~\cite{10.1371/journal.pone.0138212}.
}{
Il est possible de raffiner cette étude en étendant la méthode de contrôle des corrélations. La connaissance très fine du comportement de $N$ (distribution statistique sur une grille fine de l'espace des paramètres) conditionnée à $D$ devrait permettre de déterminer exhaustivement $N^{<-1>} | D$ et avoir plus de latitude dans la génération des corrélations. Il est aussi possible d'appliquer des algorithmes spécifiques d'exploration pour essayer d'atteindre des configurations exceptionnelles réalisant un niveau de corrélation voulu, ou au moins pour découvrir l'espace complet des corrélations atteignables par la méthode de génération~\cite{10.1371/journal.pone.0138212}.
}

\subsubsection{Direct applications}{Applications directes}

\bpar{
Starting from the second example which was limited to data generation, we propose examples of direct applications that should give an overview of the range of possibilities.
}{
Le couplage que nous venons de présenter s'est arrêté à la génération des données synthétiques. Nous proposons des pistes d'applications directes qui donneront un aperçu de l'éventail des possibilités.
}

\bpar{
Calibration of network generation component at given density, on real data for transportation network\footnote{Typically road network given the shape of generated networks ; it should be straightforward to use OpenStreetMap open data that have a reasonable quality for Europe, at least for France~\cite{girres2010quality}, with however adjustments on generation procedure in order to avoid edge effects due its restrictive frame, for example by generating on an extended surface to keep only a central area on which calibration would be done.} should theoretically allow to unveil parameter sets reproducing accurately existing configurations both for urban morphology and network shape. It could be then possible to derive a ``theoretical correlation'' for these, as an empirical correlation is according to some theories of urban systems not computable as a unique realization of stochastic processes is observed. Because of non-ergodicity of urban systems~\cite{pumain2012urban}, there are strong chances that involved processes are different across different geographical areas (or from an other point of view that they are in an other state of meta-parameters, i.e. in an other regime) and that their interpretation as different realizations of the same stochastic process makes no sense, the impossibility of covariation estimation following, except under simplified assumptions as we did in\ref{sec:staticcorrelations}. By attributing a synthetic dataset similar to a given real configuration, we would be able to compute a sort of \emph{intrinsic correlation} proper to this configuration. As territorial configurations emerge from spatio-temporal interdependences between components of territorial systems, this intrinsic correlation emerges the same way, and its knowledge gives information on these interdependences and thus on relations between territories and networks.
}{
La calibration de la composante de génération de réseau, à densité donnée, sur des données réelles de réseau de transport\footnote{Typiquement routier vu les formes heuristiques obtenues, il devrait par exemple être aisé d'utiliser les données ouvertes d'OpenStreetMap qui sont de qualité raisonnable pour l'Europe, du moins pour la France~\cite{girres2010quality} et pour lesquelles nous avons déjà simplifié le réseau et calculé les indicateurs en~\ref{sec:staticcorrelations}. Il y a toutefois des ajustements à faire sur le modèle pour supprimer les effets de bord dus à sa structure, par exemple en le faisant générer sur une surface étendue pour ne garder qu'une zone centrale sur laquelle la calibration aurait lieu.} permettrait en théorie d'isoler un jeu de paramètres représentant fidèlement des situations existantes à la fois pour la forme urbaine et la forme du réseau. Il serait alors possible de dériver une ``corrélation théorique'' pour celles-ci, étant donné qu'une corrélation empirique n'est en théorie pas calculable puisqu'une seule instance des processus stochastiques est observée. Vu la non-ergodicité des système urbains~\cite{pumain2012urban}, il y a de fortes chances pour que ces processus soient différents d'une zone géographique à l'autre (ou selon un autre point de vue qu'ils soient dans un autre état des méta-paramètres, dans un autre régime) et que leur interprétation en tant que réalisations d'un même processus stochastique n'ait aucun sens, entrainant l'impossibilité du calcul des covariations, sauf sous des hypothèses simplifiées comme nous l'avons fait en \ref{sec:staticcorrelations}. Il s'agit alors de supposer une stationnarité locale, c'est-à-dire des processus dominants se manifestant selon des paramètres variables selon les régions de l'espace. En attribuant un jeu de données synthétiques similaire à une situation donnée, on serait capable de calculer une sorte de \emph{corrélation intrinsèque} propre à la situation, qui émerge en fait en réalité des interdépendances temporelles des composantes. Connaitre celle-ci renseigne alors sur ces interdépendances, et donc sur les relations entre réseaux et territoires.
}

\bpar{
As already mentioned, most of models of simulation need an initial state generated artificially as soon as model parametrization is not done completely on real data. An advanced model sensitivity analysis implies a control on parameters for synthetic dataset generation, seen as model meta-parameters~\cite{cottineau2015revisiting}. In the case of a statistical analysis of model outputs it provides a way to operate a second order statistical control.
}{
Comme déjà évoqué, la plupart des modèles de simulation nécessitent un état initial, généré artificiellement à partir du moment où la paramétrisation n'est pas effectuée totalement à partir de données réelles. Une analyse de sensibilité avancée du modèle implique alors un contrôle sur les paramètres de génération du jeu de données synthétique, vu comme méta-paramètre du modèle~\cite{cottineau2015revisiting}. Dans le cas d'une analyse statistique des sorties du modèle, on est alors capable d'effectuer un contrôle statistique au second ordre.
}

\bpar{
We study stochastic processes for the study of synthetic data in~\ref{app:sec:syntheticdata-finance}, in the sense of random time-series, whereas time did not have a role in the second case. We can suggest a strong coupling between the two model components (or the construction of an integrated model) and to observe indicators and correlations at different time steps during the generation. In a dynamical spatial models we have because of feedbacks necessarily propagation effects and therefore the existence of lagged interdependences in space and time~\cite{pigozzi1980interurban}. It would drive our field of study towards a better understanding of dynamical correlations. A co-evolution model in that spirit will be proposed in chapter~\ref{ch:mesocoevolution}.
}{
Nous étudions des processus stochastiques dans l'étude des données synthétiques en~\ref{app:sec:syntheticdata-finance}, au sens de séries temporelles aléatoires, alors que le temps ne jouait pas de rôle ici. Nous pouvons suggérer un couplage fort entre les deux composantes du modèle (ou la construction d'un modèle intégré) et l'observation les indicateurs et correlations à différents pas de temps de la génération. Dans le cas d'une dynamique, de par les rétroactions, on a nécessairement des effets de propagation et donc l'existence d'interdépendances décalées dans l'espace et le temps~\cite{pigozzi1980interurban}, étendant le domaine d'étude vers une meilleure compréhensions des corrélations dynamiques. Un modèle de co-évolution dans cet esprit sera proposé en chapitre~\ref{ch:mesocoevolution}.
}

\subsubsection{Generalization}{Généralisation}

\bpar{
We were limited to the control of first and second moments of generated data, but we could imagine a theoretical generalization allowing the control of moments at any order. However, as shown by the geographical example, the difficulty of generation in a concrete complex case questions the possibility of higher orders control when keeping a consistent structure model and a reasonable number of parameters. The study of non-linear dependence structures as proposed in~\cite{chicheportiche2013nested} is in an other perspective an interesting possible development.
}{
Nous nous sommes limité au contrôle des premier et second moments des données générées, mais il est possible d'imaginer une généralisation théorique permettant le contrôle des moments à un ordre arbitraire. Toutefois, la difficulté de génération dans un cas concret complexe, comme le montre l'exemple géographique ici, questionne la possibilité de contrôle aux ordres supérieurs tout en gardant un modèle à la structure cohérente et au nombre de paramètres relativement faible. Par contre, l'étude de structures de dépendances non-linéaires comme celles utilisées dans~\cite{chicheportiche2013nested} est une piste de développement intéressante.
}

\bpar{
We described a model allowing to generate synthetic datasets in which correlation structure is controlled, for which a generic formulation is given in Appendix~\ref{app:sec:syntheticdata}. Its partial implementation in two very different domains, in Apppendix~\ref{app:sec:syntheticdata-finance} and here, shows its flexibility and the broad range of possible applications. More generally, it is crucial to favorise such practices of systematic validation of computational models by statistical analysis, in particular for agent-based models for which the question of validation stays an open issue. 
}{
Nous avons ainsi proposé une méthode de génération de données synthétiques corrélées à un niveau contrôlé, pour laquelle une formulation générique est donnée en Annexe~\ref{app:sec:syntheticdata}. Son implémentation partielle dans deux domaines très différents, en Annexe~\ref{app:sec:syntheticdata-finance} et ici, montre sa flexibilité et l'éventail des applications potentielles. De manière générale, il est essentiel de généraliser de telles pratiques de validation systématique de modèles par étude statistique, en particulier pour les modèles agents pour lesquels la question de la validation reste encore relativement ouverte.
}

\bpar{
Regarding our general problematic, we have introduced a first coupling between transportation networks and territories at the mesoscopic scale, through a sequential coupling of morphogenesis models. The development of this model into a co-evolution model will be the object of chapter~\ref{ch:mesocoevolution}.
}{
Concernant notre problématique générale, nous avons introduit un premier couplage entre réseaux de transport et territoires à l'échelle mesoscopique, par un couplage séquentiel de modèles de morphogenèse. Le développement de ce modèle en un modèle de co-évolution fera l'objet du chapitre~\ref{ch:mesocoevolution}.
}

\stars

%


\newpage

\section*{Chapter Conclusion}{Conclusion du Chapitre}

\bpar{
A general question relatively open regarding urban systems is the one of the \emph{link between form and function}. Even if it is in some cases and at certain scales easily extricable, there does not seem to exist any general rule nor theory answering this fundamental problem. Will future \emph{smart cities} be able to totally disconnect the form from the function as hypothesizes~\cite{batty2017age} ?
}{
Une question générale relativement ouverte concernant les systèmes urbains est celle du \emph{lien entre forme et fonction}. Si, dans certains cas et à certaines échelles, celui-ci est aisément extricable, il ne semble pas exister de règle générale ni de théorie répondant à ce problème fondamental. Les futures \emph{smart cities} seront-elles capables de totalement déconnecter la forme de la fonction comme le suppose~\cite{batty2017age} ?
}

\bpar{
By situating oneself at the scale of a system of cities or a mega-urban region, for which the form will manifest in relative positions both geographically, but also according to multi-layer networks, of cities according to their specializations, or in the fine localization of the different types of activities within the region and the links formed by the transportation network, we can assume on the contrary that the new urban forms will be linked in ever more intricate and complex ways with their functions, at different scales and according to different dimensions.
}{
En se plaçant à l'échelle d'un système de villes ou d'une méga-région urbaine, pour lesquels la forme se manifestera dans les positions relatives à la fois géographiques, mais aussi selon des réseaux multi-couches, des villes selon leur spécialisations, ou dans la localisation fine des différents types d'activités dans la région et les liens formés par le réseau de transport, nous pouvons supposer au contraire que les nouvelles formes urbaines seront liées de manière toujours plus intriquées et complexes avec leurs fonctions, à différentes échelles et selon différentes dimensions.
}

\bpar{
The notion of morphogenesis, that we defined and partly explored, seems to be a good candidate to link form and function as we showed in~\ref{sec:interdiscmorphogenesis}. A simple model such as the one studied in~\ref{sec:densitygeneration} integrates this paradigm without providing any possible interpretation since functions are implicit in the processes considered. By coupling the model to the transportation network as done in~\ref{sec:correlatedsyntheticdata}, we explicitly introduce notions of functions since for example accessibility has now a role, but also because the network is a function in itself.
}{
La notion de morphogenèse, que nous avons définie et explorée partiellement, semble être bonne candidate pour lier forme et fonction comme nous l'avons montré en~\ref{sec:interdiscmorphogenesis}. Un modèle simple comme celui étudié en~\ref{sec:densitygeneration} intègre ce paradigme sans pouvoir offrir d'interprétation possible puisque les fonctions sont implicites dans les processus considérés. En couplant le modèle au réseau de transport comme fait en~\ref{sec:correlatedsyntheticdata}, nous introduisons explicitement des notions de fonctions puisque par exemple l'accessibilité se met à jouer un rôle, mais aussi parce que le réseau est une fonction en lui-même.
}

\bpar{
These paradigms will be used in the following to model co-evolution within a corresponding perspective in~\ref{sec:mesocoevolmodel}, i.e. at the mesoscopic scale with the same assumptions of autonomous processes and well defined sub-systems. We will deepen the reflexion on the role of functions with a multi-dimensional urban form in the study of the Lutecia model in~\ref{sec:lutecia}, which will integrate the governance of the transportation system and relations between actives and employments within a metropolitan region.
}{
Ces paradigmes seront utilisés par la suite pour modéliser la co-évolution dans une perspective correspondante en~\ref{sec:mesocoevolmodel}, c'est-à-dire à l'échelle mesoscopique avec les mêmes hypothèses de processus autonomes et de sous-système bien défini. Nous pousserons la reflexion sur le rôle des fonctions avec une forme urbaine multi-dimensionnelle dans l'étude du modèle Lutecia en~\ref{sec:lutecia}, qui intègrera la gouvernance du système de transport et les relations entre actifs et emplois dans une région métropolitaine.
}

\stars



\bpar{
\chapter*{Conclusion of Part II: Co-evolution, a complex concept with multiple faces}
}{
\chapter*{Conclusion de la Partie II : co-évolution, un concept complexe aux visages multiples}
}

\bpar{
\markboth{Conclusion of Part II}{Conclusion of Part II}
}{
\markboth{Conclusion de la partie II}{Conclusion de la partie II}
}


\bpar{
This part allowed us to bring diverse first elements of answer to our problematic of modeling co-evolution, by both constructing tools and opening particular perspectives.
}{
Cette partie nous a permis d'apporter divers premiers éléments de réponse à notre problématique de modélisation de la co-évolution, en construisant à la fois des outils et en ouvrant des perspectives particulières. 
}

\bpar{
The first chapter, with an heterogenous composition, digs into fundamental concepts spanning from the evolutive urban theory, which is thus confirmed as a consequent part of our conceptual skeleton. The study of static correlations confirms the non-stationarity and suggests the multi-scalarity of interactions between networks and territories, and allows us on the one hand to confirm the relevance of our approach at two distinct scales, and on the other hand provides an empirical analysis constructing observed data which will allow to calibrate models. Then, the construction of an operational characterization of co-evolution, in terms of causality regimes, is essential both from the empirical viewpoint and for the characterization of the behavior of models which will be introduced in the following. Finally, we explore the potentialities of interaction models in systems of cities, what allows us to confirm the existence of network effects.
}{
Le premier chapitre, à la composition hétérogène, creuse des concepts fondamentaux issus de la théorie évolutive des villes, qui s'affirme ainsi comme partie intégrante de notre squelette conceptuel. L'étude des corrélations statiques confirme la non-stationnarité et suggère la multi-scalarité des interactions entre réseaux et territoires, et nous permet d'une part de confirmer la pertinence de notre approche à deux échelles distinctes, et d'autre part fournit une analyse empirique construisant des données observées qui permettront de calibrer les modèles. Ensuite, la construction d'une caractérisation opérationnelle de la co-évolution, en termes de régimes de causalité, est essentielle à la fois du point de vue empirique et pour la caractérisation du comportement des modèles à venir. Enfin, nous explorons les potentialités des modèles d'interaction dans les systèmes de villes, ce qui nous permet de confirmer l'existence d'effets de réseau.
}

\bpar{
The second chapter explores the concept of morphogenesis, starting by constructing for it an interdisciplinary definition which suggests the modeling paradigms through form and function and introduces an implicit link with co-evolution. We then develop a simple model based on form only, through aggregation-diffusion principles, and show that it reproduces a large spectrum of territorial forms existing in Europe. We finally construct the first building brick of a coupling with a network growth model and explore the space of potential static correlations.
}{
Le second chapitre explore le concept de morphogenèse, en commençant par en construire une définition interdisciplinaire qui suggère les paradigmes de modélisation par la forme et la fonction et introduit un lien implicite avec la co-évolution. Nous développons alors un modèle simple se basant uniquement sur la forme, par des principes d'agrégation-diffusion, et montrons que celui-ci reproduit une large gamme de formes territoriales existantes en Europe. Nous posons alors la première brique d'un couplage avec un modèle de croissance de réseau et explorons l'espace des corrélations statiques potentielles.
}

\bpar{
We can at this stage make a conceptual summary of our progressive construction.
}{
Nous pouvons faire à ce stade un bilan conceptuel de notre construction progressive.
}

\subsection*{Conceptual definition}{Définition conceptuelle}

\bpar{
We recall the conceptual definition of co-evolution constructed in particular through multi-disciplinary transfer in the first part: evolutive territorial systems can exhibit co-evolution properties at three distinct levels: (i) local entities in reciprocal interactions; (ii) regional population of entities exhibiting circular causalities from a statistical viewpoint; (iii) global systemic interdependencies.
}{
Rappelons la définition conceptuelle de la co-évolution construite en particulier par transfert multidisciplinaire en première partie : des systèmes territoriaux évolutifs pourront présenter des propriétés de co-évolution à trois niveaux distincts : (i) entités locales en interactions réciproques ; (ii) population régionale d'entités présentant des causalités circulaires d'un point de vue statistique ; (iii) interdépendances systémiques globales.
}

\subsection*{An operational characterization}{Une caractérisation opérationnelle}

\bpar{
This part will also have been crucial since it allowed us to introduce an operational measure of complex causal relationships, that we propose to consider as a method to characterize co-evolution, i.e. a proxy for it. This characterization, introduced and explored in~\ref{sec:causalityregimes}, is based on the idea of \emph{causality regimes}, which correspond to causality patterns in the Granger sense between an ensemble of variables. In the case of reciprocal causalities between two populations of entities, we will speak indeed of a \emph{co-evolution} in the second sense given above. We therefore have an empirical and operational characterization of co-evolution.
}{
Cette partie aura également été cruciale puisqu'elle aura permis d'introduire une mesure opérationnelle de relations causales complexes, que nous proposons de considérer comme une méthode de caractérisation de la co-évolution, c'est-à-dire un proxy de celle-ci. Cette caractérisation, introduite et explorée en~\ref{sec:causalityregimes}, se base sur l'idée de \emph{régimes de causalité}, qui correspondent à des motifs de causalité au sens de Granger entre un ensemble de variables. Dans le cas de causalités réciproques entre deux populations d'entités, nous aurons bien \emph{co-évolution} au deuxième sens donné ci-dessus. Nous avons donc ainsi une caractérisation empirique et opérationnelle de la co-évolution.
}

\subsection*{The morphogenetic approach}{L'approche morphogénétique}

\bpar{
Morphogenesis highlights the question of autonomy and interdependency, of boundaries and the environment, the question of scales. We can precise to what extent it reinforces the construction of the concept of co-evolution. The idea of independent subsystem rejoins the one of ecological niche which is equivalent to a system of boundaries in the theory of \noun{Holland}~\cite{holland2012signals}. This theory indeed consider the entities of a given niche as co-evolving: we can see implicitly that this concept allows on the one hand a relevant entry for models at the mesoscopic scale, but on the other hand that it creates inevitable yet unexpected deep links with the conceptual context that we are progressively building.
}{
La morphogenèse appuie la question de l'autonomie et de l'interdépendance, des limites et de l'environnement, la question des échelles. Précisons dans quelle mesure celle-ci renforce la construction du concept de co-évolution. L'idée de sous-système indépendant rejoint celle de niche écologique équivalente à un système de frontières dans la théorie de \noun{Holland}~\cite{holland2012signals}. Or celui-ci suppose les entités d'une même niche en co-évolution : on voit ainsi en filigrane que ce concept nous permet d'une part une entrée pertinente pour des modèles à l'échelle mesoscopique, mais qu'il tisse d'autre part indubitablement bien que subrepticement des liens profonds avec la sphère conceptuelle que nous mettons progressivement en place.
}

\subsection*{Towards a modeling approach of co-evolution}{Vers une approche de modélisation de la co-évolution}

\bpar{
By recalling the three knowledge domains conceptual-empirical-models~\cite{livet2010}, we can consider to be equiped for the still missing component and which is our final objective: the one of models, since we extensively developed co-evolution from a conceptual and empirical point of view.
}{
En se raccrochant au triptyque des domaines de connaissance concepts-empirique-modèles~\cite{livet2010}, nous pouvons considérer être armé pour la composante encore manquante et qui est notre objectif final : celle des modèles, puisque nous avons amplement traité la co-évolution du point de vue conceptuel et du point de vue empirique. 
}


\bpar{
The aim of the next part will thus be to produce a synthesis of the bricks we introduced, and progressively construct co-evolution models at the two scales (macroscopic and mesoscopic), mostly by extending the models already studied.
}{
L'enjeu de la partie suivante va donc être de produire une synthèse des briques que nous avons introduites, et construire progressivement des modèles de co-évolution aux deux échelles (macroscopique et mesoscopique), principalement en étendant les modèles déjà étudiés.
}


\cleardoublepage 


%
\ctparttext{The buildings bricks, methods and tools are mainly set up for the culminating part of our work, which consists in the construction of models of co-evolution at different scales.}
%
\part{Synthesis}
%

\bpar{
\chapter*{Introduction of part III}
}{
\chapter*{Introduction de la Partie III}
}

\markboth{Introduction}{Introduction}


\bpar{
The contradictions felt within a constraining academic context can rapidly limit the possibilities to dig deeper but also of synthesis. The saturation threshold is easily reached and the resignation to bury idealist illusions of the past is rapidly the rule. But the receptivity of the public may provide an invisible way out of these constraints, and some media play a determining role in it. The most important has been the experimental modeling experience. The model as a communication tool. The model as a game for teaching. The model as a pretext to develop a personal thinking. The model to dig deeper into notions evoked. The model at the crossroad of viewpoints and sensitivities. The model at the convergence of concepts understood. The model as a complex synthesis. Pessimism should finally not be the rule, the most original ways to get out would also be the most efficient.
}{
\textit{Les contradictions ressenties au sein d'un contexte académique contraignant peuvent rapidement limiter les possibilités à la fois d'approfondissement mais aussi de synthèse. Le niveau de saturation est facilement atteint et la résignation à enterrer des illusions idéalistes passées est rapidement de mise. Mais la réceptivité du public permet d'échapper invisiblement à ces contraintes, et certains media y jouent un rôle déterminant. La plus marquante aura été celle par modélisation expérimentale. Le modèle comme outil de communication. Le modèle comme un jeu pour l'enseignement. Le modèle comme prétexte au développement d'une réflexion personnelle. Le modèle comme approfondissement de notions effleurées. Le modèle comme croisement des points de vue et des sensibilités. Le modèle à la convergence des concepts compris. Le modèle comme synthèse complexe. Le pessimisme ne doit finalement pas être de mise, les moyens les plus originaux de s'évader seraient aussi les plus efficaces.}
}


\bigskip

\bpar{
At the heart of our subject, we must both do the synthesis of conceptual and empirical entries on co-evolution, and to go deeper into thematic entries. Model will be together both products and producers of this synthesis and this deepening, and allow to extract us from the restrictive disciplinary framework previously highlighted, by exploring fuzzy boundaries of domains and of knowledge, at the image of the experiment in collective modeling for teaching which was described above\footnote{Which lead to a concrete realization, see \url{https://github.com/JusteRaimbault/ExperimentalModeling}.} in which the model both allowed to get out of the frame and to operate a synthesis and a deepening.
}{
Au coeur de notre sujet, nous devons à la fois faire la synthèse des entrées conceptuelle et empirique sur la co-évolution, et l'approfondissement des entrées thématiques. Les modèles vont être à la fois produits et producteurs de cette synthèse et de cet approfondissement, et permettre de nous extraire du cadre disciplinaire restrictif mis en valeur précédemment, en explorant des frontières floues des domaines et de la connaissance, à l'image de l'expérience de modélisation collective en enseignement imagée ci-dessus\footnote{Qui a conduit à une réalisation concrète, voir \url{https://github.com/JusteRaimbault/ExperimentalModeling}.} dans laquelle le modèle a à la fois permis de s'extraire du cadre et d'opérer une synthèse et un approfondissement.
}

\bpar{
This part thus aims at formulating and exploring co-evolution models, answering our second axis of the problematic, i.e. how to integrate co-evolution processes within models. The question of scales has been implicitly dealt with in the complementary thematic entries of the previous part: at a mesoscopic scale, it will be more relevant to focus on the precise form, whereas at a macroscopic scale the interaction between agents are fundamental. This complementarity of scales furthermore echoes two seminal models of urban growth, the Gibrat and the Simon models. We demonstrate in Appendix~\ref{app:sec:stochurbgrowth} that these are two specifications of a more global framework of stochastic models of urban growth, what suggests that our two approaches are not only complementary but also can be synthesized.
}{
Cette partie vise ainsi à formuler et explorer des modèles de co-évolution, répondant à notre deuxième axe de la problématique, c'est-à-dire comment intégrer les processus de co-évolution dans des modèles. La question des échelles a été traitée de manière sous-jacente par les entrées thématiques complémentaires de la partie précédente : à une échelle mesoscopique, il sera plus pertinent de s'intéresser à la forme précise, tandis qu'à une échelle macroscopique les interactions entre agents sont fondamentales. Cette complémentarité des échelles fait par ailleurs écho à deux modèles séminaux de croissance urbaine, le modèle de Gibrat et le modèle de Simon. Nous démontrons en Annexe~\ref{app:sec:stochurbgrowth} que ceux-ci sont deux spécifications d'un cadre plus global de modèles stochastique de croissance urbaine, ce qui suggère que nos deux approches sont non seulement complémentaires mais synthétisables.
}

\bpar{
We thus construct the models in two chapters, which order has been fixed to have a progressive degree of complexity of models. The chapter~\ref{ch:macrocoevolution} develops the models at the macroscopic scale. We first introduce the necessary indicators to qualify the behavior of such a type of models, which are tested by applying them to a model from the literature. A direct extension of the interaction model of~\ref{sec:interactiongibrat} is then proposed as a co-evolution model at the macroscopic scale.
}{
Nous construisons ainsi les modèles dans deux chapitres, dont l'ordre a été fixé pour avoir un degré progressif de complexité des modèles. Le chapitre~\ref{ch:macrocoevolution} développe les modèles à l'échelle macroscopique. Nous introduisons d'abord les indicateurs nécessaire pour qualifier le comportement de ce type de modèle, qui sont testés par application à un modèle de la littérature. Une extension directe du modèle d'interaction de~\ref{sec:interactiongibrat} est ensuite proposée comme modèle de co-évolution à l'échelle macroscopique.
}

\bpar{
The chapter~\ref{ch:mesocoevolution} begins with developing a string coupling of the morphogenesis model of~\ref{sec:densitygeneration} and of network growth models, in a multi-modeling approach. It is calibrated on static data computed in~\ref{sec:staticcorrelations}. We then introduce a model at the metropolitan scale that takes into account governance processes for the extension of the transportation network.
}{
Le chapitre~\ref{ch:mesocoevolution} développe pour commencer un couplage fort du modèle de morphogenèse de~\ref{sec:densitygeneration} et de modèles de croissance de réseau, dans une démarche de multi-modélisation. Celui-ci est calibré sur données statiques calculées en~\ref{sec:staticcorrelations}. Nous introduisons ensuite un modèle à l'échelle métropolitaine prenant en compte des processus de gouvernance pour l'extension du réseau de transport.
}

\stars

\bpar{
\chapter{Co-evolution at the macroscopic scale}
}{
\chapter{Co-évolution à l'échelle macroscopique}
}

\label{ch:macrocoevolution} 


\bpar{
Coupled dynamics between territories and networks can be grasped at the macroscopic scale through an approach by interactions, as we showed in chapter~\ref{ch:evolutiveurban}. The explicative power is then different to the one of classical economic models and concerns other types of processes, based on interactions at smaller spatial scales and longer time scales. In this frame, transportation networks and systems of cities co-evolve on long time.
}{
Les dynamiques couplées des territoires et des réseaux peuvent être appréhendées à l'échelle macroscopique au moyen d'une approche par les interactions, comme nous l'avons montré au chapitre~\ref{ch:evolutiveurban}. Le pouvoir explicatif est alors different de celui des modèles économiques classiques et concerne d'autres types de processus, basés sur les interactions à des échelles d'espace plus petites et des échelles de temps plus longues. Dans ce cadre, les réseaux de transports et les systèmes de villes co-évoluent sur le temps long.
}

\bpar{
To what extent the construction of the railway link through the Channel tunnel could have consolidated the economic power of London or reinforce its interactions with its close European neighbours, and to what extent the recent political events could lead to a modification of economic trajectories and then as a consequence to a modification of transportation patterns through a feedback of demand ? In a similar way, to what extent the projects of high speed lines on the East coast of the United States and in the California corridor are coordinated with regional dynamics, and if they are effectively realized, to what extent can they influence trajectories of the system of cities ?
}{
Dans quelle mesure la construction du lien ferroviaire par le tunnel sous la Manche a-t-elle pu conforter le pouvoir économique de Londres ou renforcer ses interactions avec ses proches voisins Européens, et dans quelle mesure les évènements politiques recents peuvent-ils conduire a une modification des trajectoires économiques puis par conséquent à une modification des motifs de transports par une rétroaction de la demande ? D'une façon similaire, dans quelle mesure les projets de lignes à grande vitesse sur la côte Est des Etats-Unis et dans le corridor Californien sont-ils coordonnés aux dynamiques régionales, et s'ils sont effectivement réalisés, dans quelle mesure peuvent-ils influencer les trajectoires du système de villes ?
}

\bpar{
We have already studied similar issues in the case of South Africa and with an empirical approach in~\ref{sec:causalityregimes}, and we propose in this chapter to reflect it from the point of view of modeling, by introducing co-evolution processes in interaction models already developed.
}{
Nous avons déjà étudié des questions analogues dans le cas de l'Afrique du Sud de manière empirique en~\ref{sec:causalityregimes}, et nous proposons dans ce chapitre d'y faire écho du point de vue de la modélisation, en introduisant les processus de co-evolution dans les modèles d'interactions déjà développés.
}

\bpar{
To give an idea of the nature of conclusions we can expect to draw from such an approach, we begin in~\ref{sec:macrocoevolexplo} by a systematic exploration of the SimpopNet model, approach which is the most advanced in terms of modeling the co-evolution of cities and transportation networks at this scale, as established in chapter~\ref{ch:modelinginteractions}. It also allows us to introduce the suited indicators for the evaluation of trajectories of systems of cities.
}{
Pour donner une idée de la nature des enseignements qu'il est possible de tirer d'une telle approche, nous commençons en~\ref{sec:macrocoevolexplo} par une exploration systématique du modèle SimpopNet, approche la plus avancée en termes de modélisation de la co-evolution des villes et des réseaux de transport à cette échelle, comme établi au chapitre~\ref{ch:modelinginteractions}. Cela nous permet également d'introduire les indicateurs adaptés pour l'évaluation des trajectoires des systèmes de villes.
}

\bpar{
We then describe in~\ref{sec:macrocoevol} the generic model of co-evolution, which is tested on synthetic data at two levels of detail for network representation, and then on the French system of cities.
}{
Nous décrivons ensuite en~\ref{sec:macrocoevol} le modèle générique de co-évolution, qui est testé sur des données synthétiques à deux niveaux de détail pour la représentation du réseau, puis sur le système de villes français.
}

\stars

\bpar{
\textit{This chapter will be published as a book chapter~\cite{raimbault2018unveiling} for its first section. The second section describes the results of~\cite{raimbault2017macro} for synthetic data, and will be published also as a book chapter~\cite{raimbault2018models}.}
}{
\textit{Ce chapitre va paraître prochainement comme chapitre d'ouvrage~\cite{raimbault2018unveiling} pour sa première section. La deuxième section reprend les résultats de~\cite{raimbault2017macro} pour les données synthétiques, et va paraître prochainement également comme chapitre d'ouvrage~\cite{raimbault2018models}.}
}



%


\newpage

\section{Exploring macroscopic models of co-evolution}{Explorer les modèles macroscopiques de co-évolution}

\label{sec:macrocoevolexplo}


\bpar{
We first propose to introduce co-evolution models at the macroscopic scale by exploring the results produced by an existing model, what will also allow to introduce methods and indicators that are necessary to the exploration, and to grasp the typical questions linked to this type of models. In particular, we proceed to a systematic exploration of the SimpopNet model \cite{schmitt2014modelisation}, which is to the best of our knowledge one of the rare initiatives to model co-evolution within a system of cities.
}{
Nous proposons dans un premier temps d'introduire les modèles de co-évolution à l'échelle macroscopique en explorant les résultats produits par un modèle existant, ce qui permettra également d'introduire les méthodes et indicateurs nécessaires à l'exploration, ainsi que d'appréhender les questionnements typiques liés à ce type de modèles. En particulier, nous procédons à une exploration systématique du modèle SimpopNet~\cite{schmitt2014modelisation}, à notre connaissance l'une des rares initiatives pour modéliser la co-évolution au sein d'un système de villes.
}

\subsection{Context}{Contexte}

\bpar{
A considerable gain in knowledge can be observed, from the conceptual or thematic description of a model, to its mathematical formalisation, its implementation, its systematic exploration, up to its exploration in deep with the help of specific meta-heuristics. Our postulate, that is a consequence of both our positioning (see \autoref{ch:positioning} on simulation) and experiments of which previously developed models are part, is that it is important, but furthermore of a \emph{qualitative} nature, in the sense that the nature of knowledge follows abrupt transitions during the advance of the investigation in this continuum.
}{
Un gain considérable de connaissances peut s'observer, de la description conceptuelle ou thématique d'un modèle, à sa formalisation mathématique, son implémentation, son exploration systématique, jusqu'à son exploration approfondie à l'aide de méta-heuristiques spécifiques. Notre postulat, qui découle à la fois de notre positionnement (voir le chapitre~\ref{ch:positioning} sur la simulation) et d'expériences dont les modèles déroulés précédemment font partie, est que celui-ci est important, mais surtout de nature \emph{qualitative}, c'est-à-dire que la nature même des connaissances subit des transitions abruptes lors de l'avancée de la démarche dans ce continuum.
}

\bpar{
The SimpopNet model introduced by~\cite{schmitt2014modelisation}, which is to our knowledge the only co-evolution model in the perspective of the evolutive urban theory. Its behavior was however not systematically explored, what makes it a good candidate for our approach.
}{
Le modèle SimpopNet introduit par~\cite{schmitt2014modelisation}, est à notre connaissance l'unique modèle de co-évolution selon la perspective de la théorie évolutive des villes. Son comportement n'a cependant pas été exploré systématiquement, ce qui en fait un bon candidat pour notre démarche.
}

\subsubsection{Studied model}{Modèle étudié}

\bpar{
We briefly reformulate the model, following the notations for the formalization of the interaction model in~\ref{sec:interactiongibrat}, since a certain number of parameters and processes are similar. Cities grow following a specification that rejoins equation~\ref{eq:interactiongibrat:model}, i.e. in this specific case
}{
Nous reformulons brièvement le modèle, suivant les notations de la formulation du modèle d'interaction en~\ref{sec:interactiongibrat}, un certain nombre de paramètres et de processus se recoupant. Les villes croissent suivant une spécification qui peut être rapprochée de l'équation~\ref{eq:interactiongibrat:model}, c'est-à-dire dans ce cas spécifique
}
\[
\mu_i(t+1) - \mu_i (t) = \mu_i (t) \cdot \frac{\lambda^{\beta}}{N} \sum_{j} \frac{V_{ij}}{<V_{ij}>}
\]
\bpar{
where the potential is of the form $V_{ij} = \mu_j / d_{ij}^\beta$ and $V_{ii}=0$, and $\beta$ is a parameter for the distance decay and $\lambda$ shape parameter for the decay function. We thus find our formulation, with $r_0 = 0$ and $w_G = \lambda^\beta \cdot N$. Since $\lambda$ gives the typical distance of interaction, it will be noted $d_G$ in the following, and $\beta$ will be noted $\gamma_G$ (it is indeed a level of hierarchy as a function of distance).
}{
où le potentiel est de la forme $V_{ij} = \mu_j / d_{ij}^\beta$ et $V_{ii}=0$, et $\beta$ est un paramètre de vitesse de décroissance en fonction de la distance et $\lambda$ paramètre de forme de la fonction de décroissance. Nous retrouvons ainsi notre formulation, avec $r_0 = 0$ et $w_G = \lambda^\beta \cdot N$. Comme $\lambda$ fixe la distance typique d'interaction, nous le noterons par la suite $d_G$, et $\beta$ sera noté $\gamma_G$ (il s'agit bien d'un niveau de hiérarchie en fonction de la distance).
}


\bpar{
The network growths at each time step through a process that can be seen as a potential breakdown (as described in chapter~\ref{ch:thematic}): a couple of cities is chosen, the first according to populations with a hierarchy $\gamma_N$ (i.e. with a probability proportional to ${\mu_i}^{\gamma_N}$) and the second following interaction forces $\mu_i \mu_j / d_{ij}^\beta$ with the same hierarchy $\gamma_N$. A link is then created if the network is not efficient enough, i.e. if $d_{ij}/d^{(N)}_{ij}> \theta_N$. The links created at a date $t$ have a speed $v(t)$, which will depend on current transportation technologies. The creation of new intersections to yield a planar graph is only done for links with a similar speed.
}{
Le réseau croît à chaque pas de temps par un processus que nous pouvons qualifier de rupture de potentiel (comme décrit au chapitre~\ref{ch:thematic}) : un couple de villes est choisi, la première selon les populations avec une hiérarchie $\gamma_N$ (c'est-à-dire avec une probabilité proportionnelle à ${\mu_i}^{\gamma_N}$) et la seconde selon les forces d'interactions $\mu_i \mu_j / d_{ij}^\beta$ avec la même hiérarchie $\gamma_N$. Un lien est alors créé si le réseau n'est pas assez efficace, c'est-à-dire si $d_{ij}/d^{(N)}_{ij}> \theta_N$. Les liens créés à une date $t$ ont une vitesse $v(t)$, qui dépendra des technologies de transport courantes. La création de nouvelles intersections pour produire un graphe planaire n'est effectuée que pour les liens de vitesse semblable.
}

\bpar{
In order to study a stylized version of the model, we consider a configuration such that $v(t > 0) = v_0$ and $v(0) = 1$ (the initial model considers three values for speed that correspond to the reality of transportation technologies between 1830 and 2000).
}{
Afin d'étudier une version stylisée du modèle, nous considérons une configuration telle que $v(t > 0) = v_0$ et $v(0) = 1$ (le modèle initial considère trois valeurs de la vitesse correspondant aux réalités des technologies de transport entre 1830 et 2000).
}

\subsubsection{Perspectives}{Perspectives}

\bpar{
We can put the structure of this model into perspective. Some modeling choices are not in direct consistency with the application it is used for: for example, such a precision in the parametrization of dates and speeds (historical dates from 1800 to 2000 and speed that approximatively corresponds to transportation technologies) makes it a hybrid model, and should correspond to an application on a real spatial configuration. In a synthetic configuration as used in the model, these parameters have a sense only if we know the behavior of simulated dynamics, and in particular the role of the spatial configuration, i.e. if we are able to differentiate effects linked to the dynamics from effects linked to the initial spatial configuration.
}{
Mettons en perspective la structure de ce modèle. Certains choix de modélisation ne sont pas en cohérence directe avec l'application qui en est faite : par exemple, une telle précision dans la paramétrisation des dates et des vitesses (dates historiques de 1800 à 2000 et vitesse correspondant approximativement aux technologiques de transport) en fait un modèle hybride, et devrait correspondre à une application sur une configuration spatiale réelle. Dans une configuration synthétique comme employée dans le modèle, ces paramètres n'ont de sens que si l'on connait le comportement des dynamiques simulées, et en particulier le rôle de la configuration spatiale, c'est-à-dire si l'on est capable de différencier effets liés à la dynamique et effets liés à la configuration spatiale initiale. 
}

\bpar{
Furthermore, the use of the interaction model without the endogenous Gibrat term would be difficultly adaptable to an application of the model on real data since the values we obtained in the precedent studies of interaction models, but stays relevant in a stylized model, in order to understand the interaction processes in an isolated way, as we will do later (keeping in mind that this knowledge does not necessarily describes the coupled behavior, since the interaction between processes can lead to the emergence of new behaviors).
}{
D'autre part, l'utilisation du modèle d'interaction sans le terme de Gibrat endogène serait difficilement adaptable pour une application du modèle sur données réelles vu les valeurs obtenues dans les études précédentes des modèles d'interaction, mais est bien cohérent dans un modèle stylisé, afin de comprendre les processus d'interaction de manière isolée, comme nous le ferons plus loin (tout en gardant à l'esprit que cette connaissance ne reflète pas nécessairement le comportement couplé, l'interaction entre les processus pouvant faire émerger de nouveaux comportements).
}

\bpar{
The formulation of the potential, given above, as $(\lambda / d_{ij})^\beta$, implies that $\lambda$ captures both the weight of the potential and the shape of the decreasing function, but imposes a dependence between these two effects, on the contrary to the specification we use previously. It furthermore does not allow an interpretation in terms of limit flows\footnote{The weight parameter in our model in~\ref{sec:interactiongibrat} gives indeed the value of the flow when the distance attenuation goes to infinity and for all the population.}.
}{
La formulation du potentiel, donnée ci-dessus, en $(\lambda / d_{ij})^\beta$, implique que $\lambda$ capture à la fois le poids du potentiel et la forme de la décroissance, mais contraint à une dépendance entre ces deux effets, contrairement à la spécification que nous avons utilisé précédemment. Elle ne permet par ailleurs pas d'interprétation en terme de flux limite\footnote{Le paramètre de poids dans notre modèle en~\ref{sec:interactiongibrat} donne en fait la valeur du flux lorsque l'atténuation par la distance tend vers l'infini et pour l'ensemble de la population.}.
}

\bpar{
Finally, rules allowing variable values for $v(t)$ and the non-planarity mechanism\footnote{When a new link is constructed, it does create intersections only with links of similar speed.}, allows the introduction of a tunnel effect, which is as we recall is the absence of interaction of an infrastructure traversing a territory with it. The effect is however exogenous since explicitly specified in model rules, on the contrary to the interaction model with feedback of flows, in which the variations of $w_N$ and $d_N$ should capture an endogenous tunnel effect. The introduction of specific indicators to measure it would be an interesting development direction, but we stay here at considering the hierarchy of centralities which is already a good indicator for it\footnote{Indeed, a highly hierarchical distribution of accessibilities means that there exists a small number of cities very accessible and a large number with a low accessibility. If main cities reasonably cover the space, then their links necessarily ignore the overflown cities with low accessibility, otherwise the distribution would be less hierarchical.}.
}{
Enfin, les règles permettant des valeurs variables de $v(t)$ et le mécanisme de non-planarité du réseau\footnote{Lorsqu'un nouveau lien est construit, celui-ci ne forme des intersections qu'avec les liens de vitesse similaire.}, permet l'introduction d'un effet tunnel, qui nous le rappelons est la non-interaction d'une infrastructure traversant un territoire avec celui-ci. L'effet est cependant exogène puisque spécifié explicitement dans les règles du modèle, contrairement au modèle d'interaction avec rétroaction des flux, dans lequel les variations de $w_N$ et $d_N$ doivent capturer un effet tunnel endogène. L'introduction d'indicateurs spécifiques pour le mesurer serait une piste intéressante de développement, mais nous nous contentons de regarder ici la hiérarchie des centralités qui en est déjà un bon indicateur\footnote{En effet, une distribution très hiérarchique des accessibilités signifie qu'il existe un petit nombre de villes très accessibles et un grand nombre peu accessibles. Si les villes importantes couvrent raisonnablement l'espace, alors leurs liens ignorent nécessairement les villes peu accessibles survolées, sinon la distribution serait moins hiérarchique.}.
}


\subsection{Methodology}{Méthode}

\subsubsection{Spatial configuration}{Configuration spatiale}

\bpar{
An important aspect for understanding co-evolution processes implied in this model is the role of the initial spatial configuration in emerging patterns observed. We therefore apply the methodology developed in~\ref{sec:computation}, which allows to extend the sensitivity analysis of a model to spatial meta-parameters\footnote{We recall that in our case a meta-parameter is a parameter allowing to generate an initial configuration upstream of the model.}.
}{
Un aspect important de la compréhension des processus de co-évolution impliqués dans ce modèle est le rôle de la configuration spatiale initiale dans les motifs émergents observés. Nous appliquons pour cela la méthodologie développée en~\ref{sec:computation}, qui permet d'étendre l'analyse de sensibilité d'un modèle à des méta-paramètres spatiaux\footnote{Nous rappelons que dans notre cas un méta-paramètre est un paramètre permettant de générer une configuration initiale en amont du modèle.}.
}

\paragraph{Generation of synthetic configurations}{Génération de configurations synthétiques}

\bpar{
A synthetic system of cities is constructed the following way (see Appendix~\ref{app:sec:syntheticdata} for the notion of synthetic data, calibrated at the first and the second order). A fixed number $N$ of cities is uniformly distributed in space, under the constraint of a minimal distance between each, and their population is attributed following a rank-size law which parameters $P_{m}$ and $\alpha$ can be adjusted (the distribution of city sizes in the initial model corresponds to $\alpha\simeq 0.68$ with $R^2=0.98$).
}{
Une système de villes synthétique est construit de la façon suivante (voir l'Annexe~\ref{app:sec:syntheticdata} pour la notion de données synthétiques, calibrées au premier et second ordre). Un nombre fixé de villes $N$ est réparti uniformément dans l'espace en respectant une distance minimale entre chaque, et leur population est attribuée suivant une loi rang-taille dont les paramètres $P_{m}$ et $\alpha$ peuvent être ajustés (la distribution de la taille des villes dans le modèle initial correspond à $\alpha\simeq 0.68$ avec $R^2=0.98$).
}

\bpar{
A skeleton of network is created by progressive connection: the algorithm connects cities two by two by closest neighbour in terms of euclidian distance, and then iteratively selects randomly a cluster and connects it perpendicularly to the closest link outside the cluster. The network is then extended by the creation of local shortcuts, through a repetition $n_s$ times of the random selection of a city according to populations, and its connection to a neighbour in a radius $r_s$ under conditions of a maximal degree $d_s$. The final network is then made planar.
}{
Un squelette de réseau est créé par connection progressive : l'algorithme connecte les villes deux à deux par plus proche voisin en distance euclidienne, puis itérativement sélectionne un cluster aléatoirement et le connecte perpendiculairement au lien le plus proche hors du cluster. Le réseau est ensuite étoffé par la création de raccourcis locaux, par répétitions $n_s$ fois de la sélection aléatoire d'une ville selon les populations, et sa connexion à un voisin dans un rayon $r_s$ sous conditions de degré maximal $d_s$. Le réseau final est ensuite planarisé.
}

\bpar{
This process creates networks that visually correspond (in terms of the order of magnitude of the number of loops, and their spatial range) to the initialization of the model, knowing that a single instance of the network does not allow to determine distributions of topological parameters for which a more precise calibration could be done.
}{
Cette procédure crée des réseaux correspondant visuellement (ordre de grandeur du nombre de boucles, portée spatiale de celles-ci) à l'initialisation du modèle, sachant qu'une instance du réseau ne permet pas de déterminer les distributions de paramètres topologiques sur lesquels une calibration plus fine pourrait être opérée.
}

\subsubsection{Indicators}{Indicateurs}

\bpar{
A crucial aspect of the study of simulation models is the definition of relevant indicators, particularly in the case of synthetic models where it is not possible to produce outputs that are directly linked to data for example. Very general stylized facts, as aiming at producing an urban hierarchy or a network hierarchy, are relatively limited. Moreover, the hierarchy is mechanically produced by most models including aggregation processes. We therefore need more elaborated indicators to understand the dynamics of the system. These indicators must in particular give elements of answer to the following questions:
\begin{itemize}
	\item types of systems of cities produced by the model;
	\item change in time of the organization of the system of cities;
	\item typical profiles of trajectories;
	\item ability to ``produce some co-evolution''.
\end{itemize}
}{
Un aspect crucial de l'étude des modèles de simulation est la définition d'indicateurs pertinents, surtout dans le cas de modèles synthétiques où il n'est pas possible de produire des sorties directement liées aux données par exemple. Des faits stylisés très généraux, comme vouloir produire une hiérarchie urbaine ou une hiérarchie de réseau, sont relativement limités. De plus, la hiérarchie est produite mécaniquement par la majorité des modèles incluant des processus d'agrégation. Il faut donc des indicateurs plus élaborés pour comprendre les dynamiques du système. Ces indicateurs doivent notamment apporter des éléments de réponse aux questions suivantes : 
 \begin{itemize}
 	\item types de systèmes de villes produits par le modèle ;
 	\item changement dans le temps de l'organisation du système de ville ;
 	\item profils typiques de trajectoires ;
 	\item capacité à ``produire de la co-évolution''.
 \end{itemize}
}

\bpar{
In order to concentrate on the ability of the model to produce trajectories that are both diverse and complex, and for example its ability to produce bifurcations that would manifest as inversions in ranks, and also its ability to capture different aspects of co-evolutive dynamics, we propose a set of indicators, including for example lagged correlation measures in the spirit of causality regimes exhibited in~\ref{sec:causalityregimes}, or a correlation measure as a function of distance, to understand the role of spatial interactions in the coupling of trajectories. Given a variable $X_i(t)$ defined for each city and in time (that will be the population or centrality measures for example), we define the following indicators.
}{
Pour se concentrer sur la capacité du modèle à produire des trajectoires à la fois diverses et complexes, et par exemple sa capacité à produire des bifurcations qui se traduiraient par inversions de rang, ainsi que sa capacité à capturer différents aspects des dynamiques co-évolutives, nous proposons un jeu d'indicateurs, incluant par exemple des mesures de corrélation retardée en écho aux régimes de causalité exhibés en~\ref{sec:causalityregimes}, ou une mesure de corrélation en fonction de la distance, pour comprendre le rôle des interactions spatiales dans les couplages de trajectoires. Étant donné une variable $X_i(t)$ définie sur chacune des villes et dans le temps (qui pourra être la population ou des mesures de centralité par exemple), nous définissons les indicateurs suivants.
}

\bpar{
\begin{itemize}
	\item Indicators characterizing the distribution of $X_i$ in time: hierarchy (slope of the least squares adjustment of $X_i$ as a function of rank) $\alpha (t)$, entropy of the distribution $\varepsilon (t)$, descriptive statistics (average $\hat{\Eb{X}} (t)$ and standard deviation $\hat{\sigma} (t)$).
	\item Rank correlation between the initial time and the final time, which translates the quantity of change in the hierarchy during the evolution of the system, and is defined by $\rho_r = \hat{\rho}\left[rg(X_i(t=0)),rg(X_i(t=t_f))\right]$, where $rg(X_i)$ is the rank of $X_i$ among all values.
	\item Diversity of trajectories $\mathcal{D}\left[X_i\right]$, which captures a diversity of time series profiles for the considered variable. With $\tilde{X}_i(t)\in \left[0;1\right]$ the trajectories that have been individually rescaled, it is defined by
\[
\mathcal{D}\left[X_i\right] = \frac{2}{N\cdot(N-1)}\sum_{i<j} \left(\frac{1}{T}\int_{t} \left(\tilde{X}_i(t) - \tilde{X}_j(t)\right)^2 \right)^{\frac{1}{2}}
\]
	\item Changes in direction of trajectories $\mathcal{C}\left[X_i\right]$, that we take as the number of inflexion points. In the context of such a type of model, which mainly produces monotonous trajectories, this indicator witnesses in a certain way of a ``complexity'' of trajectories.
	\item Correlations as a function of distance, to understand the way the effect of distance is translated at the macroscopic scale. The profile of this function, regarding interaction distance parameters included in the model, will translate the tendency of the model to lead to the emergence of one level of interaction or the other. It is computed as
\[
\rho_d = \hat{\rho}\left[(X(\vec{x}_k,Y(\vec{x}_{k'}))\right]
\]
where $X_i, Y_i$ are the two variables considered and $(k,k')$ the set of couples such that $\norm{\vec{x}_k-\vec{x}_{k'}} - d \leq \varepsilon$ with $\varepsilon$ a tolerance threshold (in practice taken to regroup couples by distance deciles).
	\item Lagged correlations between the variations of variables, to identify causality patterns between variables $X$ and $Y$. The patterns $\hat{\rho}_{\tau}$ for all variables, and for $\tau$ lag or anticipation, must be understood in the sense of potential regimes, explored in~\ref{sec:causalityregimes}.
\[
\rho_{\tau} = \hat{\rho}\left[\Delta X(t-\tau),\Delta Y(t)\right]
\]
\end{itemize}
}{
\begin{itemize}
  \item Indicateurs caractérisant la distribution de $X_i$ dans le temps : hiérarchie (pente de l'ajustement moindres carrés de $X_i$ en fonction du rang) $\alpha (t)$, entropie de la distribution $\varepsilon (t)$, statistiques descriptives (moyenne $\hat{\Eb{X}} (t)$ et écart-type $\hat{\sigma} (t)$).
  \item Corrélation de rang entre l'instant initial et l'instant final, qui traduit la quantité de changement dans la hiérarchie lors de l'évolution du système, et est définie par $\rho_r = \hat{\rho}\left[rg(X_i(t=0)),rg(X_i(t=t_f))\right]$, où $rg(X_i)$ est le rang de $X_i$ parmi l'ensemble des valeurs.
  \item Diversité des trajectoires $\mathcal{D}\left[X_i\right]$, qui capture une diversité de profil des séries temporelles pour la variable considérée. Avec $\tilde{X}_i(t)\in \left[0;1\right]$ les trajectoires mises à l'échelle individuellement, elle est définie par
\[
\mathcal{D}\left[X_i\right] = \frac{2}{N\cdot(N-1)}\sum_{i<j} \left(\frac{1}{T}\int_{t} \left(\tilde{X}_i(t) - \tilde{X}_j(t)\right)^2 \right)^{\frac{1}{2}}
\]
\item Changements de direction des trajectoires $\mathcal{C}\left[X_i\right]$, que nous prenons comme le nombre de points d'inflexion. Dans le cadre de ce type de modèle, qui produit majoritairement des trajectoires monotones, cet indicateur témoigne dans une certaine mesure d'une ``complexité'' des trajectoires.
\item Corrélations en fonction de la distance, pour comprendre la manière dont l'effet de la distance est traduit au niveau macroscopique. Le profil de cette fonction, au regard des valeurs des paramètres de distance d'interaction inclus dans le modèle, traduira la tendance du modèle à faire émerger tel ou tel niveau d'interaction. Elle est calculée comme
\[
\rho_d = \hat{\rho}\left[(X(\vec{x}_k,Y(\vec{x}_{k'}))\right]
\]
où $X_i, Y_i$ sont les deux variables considérées et $(k,k')$ l'ensemble des couples tels que $\norm{\vec{x}_k-\vec{x}_{k'}} - d \leq \varepsilon$ avec $\varepsilon$ seuil de tolérance (en pratique pris pour regrouper les couples par déciles de distance).
\item Corrélations retardées entre les variations des variables, pour identifier des motifs de causalité entre les variables $X$ et $Y$. Les motifs $\hat{\rho}_{\tau}$ pour l'ensemble des variables, et pour $\tau$ retard ou anticipation, sont à lire dans le sens des régimes potentiels, explorés en~\ref{sec:causalityregimes}.
\[
\rho_{\tau} = \hat{\rho}\left[\Delta X(t-\tau),\Delta Y(t)\right]
\]
\end{itemize}
}


\bpar{
These indicators are used on the following variables:
\begin{itemize}
	\item populations $\mu_i(t)$,
	\item closeness centralities
	\[c_i(t) = \frac{1}{N-1}\sum_{i\neq j} \frac{1}{d_{ij}(t)}\]
	which capture the position within the urban system,
	\item accessibilities \[X_i = \frac{1}{\sum_k \mu_k}\sum_{i\neq j} P_j \exp{\left(- d_{ij}(t)/d_G\right)}\] which capture the insertion within the urban system.
\end{itemize}
}{
Ces indicateurs sont utilisés sur les variables suivantes :
\begin{itemize}
	\item populations $\mu_i(t)$,
	\item centralités de proximité
	\[c_i(t) = \frac{1}{N-1}\sum_{i\neq j} \frac{1}{d_{ij}(t)}\]
	qui capturent la position dans le système urbain,
	\item accessibilités \[X_i = \frac{1}{\sum_k \mu_k}\sum_{i\neq j} P_j \exp{\left(- d_{ij}(t)/d_G\right)}\] qui capturent l'insertion dans le système urbain.
\end{itemize}
}

\bpar{
We furthermore introduce diverse indicators for network topology, to understand the final forms produced by the heuristic: diameter, average path length, average betweenness centrality and its level of hierarchy, average performance, total length, as they have been defined in~\ref{sec:staticcorrelations}.
}{
Nous introduisons de plus divers indicateurs de topologie du réseau, pour comprendre les formes finales produites par l'heuristique : diamètre, longueur moyenne de chemin, centralité de chemin moyenne et son niveau de hiérarchie, performance moyenne, longueur totale, comme ils ont été définis en~\ref{sec:staticcorrelations}.
}

\subsection{Results}{Résultats}

\subsubsection{Experience plan}{Plan d'expérience}

\bpar{
Given an initial spatial configuration (i.e. a value of meta-parameters), we establish the behavior of indicators by exploring a grid of the parameter space. The number of parameters being low and the objective being a first grasp of the model behavior, in particular if it is able to produce co-evolution dynamics, we do not use more elaborated exploration methods. The parameters are $(d_G,\gamma_G,\gamma_N,\theta_N,v_0)$ and meta-parameters $(N_S,\alpha_S,d_S,n_S)$. We take also the meta-parameters into account in order to understand the sensitivity of the model to space.
}{
Étant donné une configuration spatiale initiale (c'est-à-dire une valeur des méta-paramètres), nous établissons le comportement des indicateurs par l'exploration d'une grille de l'espace des paramètres. Le nombre de paramètres étant restreint et l'objectif étant un premier aperçu du comportement du modèle, notamment s'il est capable de produire des dynamiques de co-évolution, nous ne faisons pas appel à des méthodes d'exploration plus élaborées. Les paramètres sont $(d_G,\gamma_G,\gamma_N,\theta_N,v_0)$ et les méta-paramètres $(N_S,\alpha_S,d_S,n_S)$. Nous prenons en compte également les méta-paramètres pour comprendre la sensibilité du modèle à l'espace.
}

\bpar{
We explore a grid of 16 configurations of meta-parameters, 324 configurations of parameters, and 30 random replications, what corresponds to $155520$ simulations. They are executed on a computation grid with the intermediary of OpenMole\footnote{Simulation results are available at \url{http://dx.doi.org/10.7910/DVN/RW8S36}.}.
}{
Nous explorons une grille de 16 configurations des méta-paramètres, 324 configurations de paramètres, et 30 réplications aléatoires, ce qui correspond à $155520$ simulations. Celles-ci sont exécutées sur grille de calcul par l'intermédiaire d'OpenMole\footnote{Les résultats de simulation sont disponibles à \url{http://dx.doi.org/10.7910/DVN/RW8S36}.}.
}

\subsubsection{Convergence}{Convergence}

\bpar{
Since the model is stochastic, it is important to control the convergence of indicators, that will be more or less easy depending on their variability. To quantify the variability of an indicator $X$ regarding stochasticity, we use a measure similar to the one used in~\ref{sec:densitygeneration}, given by $v\left[X\right] = \hat{\mathbb{E}}\left[X\right]/\hat{\sigma}\left[X\right]$ with basic estimators for the expectance and the standard deviation. On the full set of replications, we obtain for all indicators given previously, a median for the ratio $v\left[X\right]$ estimated within replications, estimated on all parameter values, which takes a minimal value of $3.94$, for the average accessibility at final time, what witnesses a low stochastic variability. We can furthermore use this value to estimate the level of convergence: it corresponds to a 95\% confidence interval around the mean of relative size $0.18$ (under the assumption of a normal distribution of the average), i.e. a good convergence. This aspect is crucial for the robustness of results.
}{
Le modèle étant stochastique, il est important de contrôler la convergence des indicateurs, qui sera plus ou moins facile selon leur variabilité. Pour quantifier la variabilité d'un indicateur $X$ par rapport à la stochasticité, nous utilisons une mesure similaire à celle de~\ref{sec:densitygeneration}, donnée par $v\left[X\right] = \hat{\mathbb{E}}\left[X\right]/\hat{\sigma}\left[X\right]$ avec les estimateurs basiques pour l'espérance et l'écart-type. Sur l'ensemble des réplications, on obtient sur l'ensemble des indicateurs donnés précédemment, une médiane pour le ratio $v\left[X\right]$ estimé au sein des réplications, estimée sur toutes les valeurs des paramètres, qui prend une valeur minimale de $3.94$, pour la moyenne de l'accessibilité à l'instant final, ce qui témoigne d'une faible variabilité stochastique. On peut de plus utiliser cette valeur pour estimer le niveau de convergence : elle correspond à un intervalle de confiance à 95\% autour de la moyenne de taille relative $0.18$ (sous hypothèse de distribution normale de la moyenne), c'est-à-dire une bonne convergence. Cet aspect est essentiel pour la robustesse des résultats.
}

\subsubsection{Sensitivity to space}{Sensibilité à l'espace}

\bpar{
The Table~\ref{tab:macrocoevolexplo:spacematters} give values of $\tilde{d}$ for 16 configurations of meta-parameters\footnote{The definition of the relative measure of sensitivity, given in~\ref{sec:computation}, is for two phase diagrams $f_1,f_2$ and $d$ euclidian distance, $\tilde{d} = 2 d(f_1,f_2)/(\Varb{f_1}+\Varb{f_2})$.}, in comparison to an arbitrary reference configuration (first column). The hierarchy within the initial system of cities appears as the stronger determinant of variability, since all configurations with $\alpha_S = 1.5$ give values larger than $1.7$, what witnesses a very strong sensitivity relative to this hierarchy.
}{
La Table~\ref{tab:macrocoevolexplo:spacematters} donne les valeurs de $\tilde{d}$ pour 16 configurations des méta-paramètres\footnote{La definition de la mesure relative de sensibilité, donnée en~\ref{sec:computation}, est pour deux diagrammes de phase $f_1,f_2$ et $d$ distance euclidienne, $\tilde{d} = 2 d(f_1,f_2)/(\Varb{f_1}+\Varb{f_2})$.}, par rapport à une configuration de référence arbitraire (première colonne). La hiérarchie au sein du système de villes initial apparaît comme le plus fort déterminant de la variabilité, puisque l'ensemble des configurations avec $\alpha_S = 1.5$ donnent des valeurs supérieures à $1.7$, ce qui témoigne d'une très forte sensibilité relative à cette hiérarchie.
}

\bpar{
Then, the number of cities plays a non negligible secondary role , giving the stronger effects of space. Thus, it is crucial to keep in mind this role of the initial configuration during the analysis of phase diagrams. To stay within the same spirit than the model that was initially proposed, we will however comment a phase diagram for a given spatial configuration. The study of the extended model with integration of meta-parameters to which it is sensitive at their full extent is beyond the reach of this auxiliary analysis.
}{
Ensuite, le nombre de villes joue un rôle secondaire non négligeable, donnant les plus forts effets de l'espace. Ainsi, il est crucial de garder à l'esprit ce rôle de la configuration initiale lors de l'analyse des diagrammes de phase. Pour rester dans l'esprit du modèle initialement proposé, nous commenterons toutefois un diagramme de phase pour une configuration spatiale donnée. L'étude du modèle étendu avec intégration des méta-paramètres auxquels il est sensible comme paramètres à part entière est hors de portée de cette analyse auxiliaire.
}

\begin{table}[!ht]
\caption[Sensitivity to space of the SimpopNet model]{\textbf{Sensitivity to space of the SimpopNet model.} Each column corresponds to an instance of the phase diagram, for which meta-parameters are given, with the relative distance to an arbitrary reference diagram. As inputs we have the meta-parameters $N_S,\alpha_S,d_S,n_S$ and as outputs of simulations the distance $\tilde{d}$.\label{tab:macrocoevolexplo:spacematters}}
\begin{tabular}{|l|l|l|l|l|l|l|l|l|l|l|l|l|l|l|l|l|}
\hline
$N_S$ & 40 & 40 & 40 & 40 & 40 & 40 & 40 & 40 & 80 & 80 & 80 & 80 & 80 & 80 & 80 & 80\\
$\alpha_S$ & 0.5 & 0.5 & 0.5 & 0.5 & 1.5 & 1.5 & 1.5 & 1.5 & 0.5 & 0.5 & 0.5 & 0.5 & 1.5 & 1.5 & 1.5 & 1.5\\
$d_S$ & 5 & 5 & 10 & 10 & 5 & 5 & 10 & 10 & 5 & 5 & 10 & 10 & 5 & 5 & 10 & 10\\
$n_S$ & 10 & 30 & 10 & 30 & 10 & 30 & 10 & 30 & 10 & 30 & 10 & 30 & 10 & 30 & 10 & 30\\\hline
$\tilde{d}$ & 0 & 0.05 & 0.26 & 0.21 & 1.79 & 1.80 & 1.79 & 1.72 & 0.44 & 0.36 & 0.42 & 0.42 & 2.25 & 2.23 & 2.24 & 2.21\\\hline
\end{tabular}
\end{table}

\subsubsection{Model behavior}{Comportement du modèle}

\begin{figure}
	\includegraphics[width=\linewidth,height=0.85\textheight]{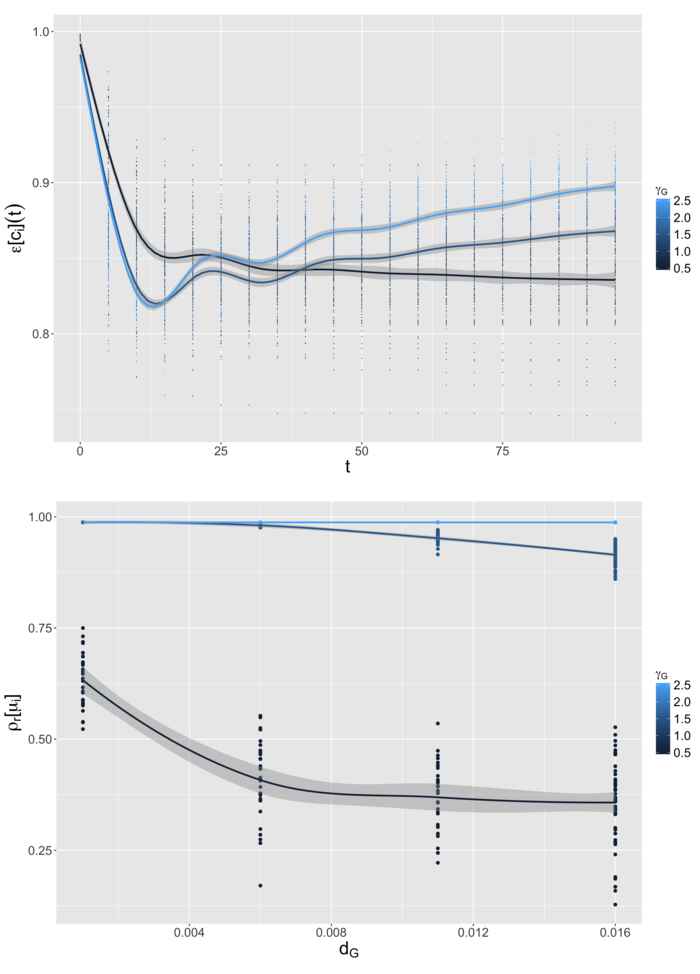}
	\caption[Behavior of the SimpopNet model]{\textbf{Model behavior for the spatial configuration $N_S=80,\alpha_S=0.5,d_S=10,n_S=30$.} (\textit{Top}) Temporal trajectories of the entropy for closeness centralities, for $\gamma_N = 2.5$, $v_0 = 110$, $d_G = 0.016$, $\theta_N = 11$, as a function of $\gamma_G$ (color); (\textit{Bottom}) Rank correlation for population, as a function of $d_G$ and of $\gamma_G$ (color), for $\theta_N = 11$, $\gamma_N = 2.5$.\label{fig:macrocoevolexplo:behavior}}
\end{figure}

\bpar{
The Fig.~\ref{fig:macrocoevolexplo:behavior} reports the behavior of the model according to a selection among the diverse indicators given above. We comment a particular spatial configuration which corresponds to a low hierarchical system with a network having only local shortcuts, given by meta-parameters $N_S=80,\alpha_S=0.5,d_S=10,n_S=30$, which are the values giving configurations that are the most similar to the one of the initial model. Complete plots are available in Appendix~\ref{app:sec:macrocoevolexplo}.
}{
La Fig.~\ref{fig:macrocoevolexplo:behavior} rend compte du comportement du modèle selon une selection parmi les divers indicateurs donnés ci-dessus. Nous commentons une configuration spatiale particulière qui correspond à un système peu hiérarchisé avec un réseau n'ayant que des raccourcis locaux, donnée par les méta-paramètres $N_S=80,\alpha_S=0.5,d_S=10,n_S=30$, qui sont les valeurs donnant des configurations les plus similaires à celle du modèle initial. Les graphes complets sont disponibles en Annexe~\ref{app:sec:macrocoevolexplo}.
}

\bpar{
The values taken by the entropy for centralities (first panel of Fig.~\ref{fig:macrocoevolexplo:behavior}), as a function of time, for $\gamma_N = 2.5$ and $v_0 = 110$, exhibit different regimes depending on $d_G$ and $\gamma_G$. A low hierarchy leads to an entropy stabilizing in time, what corresponds to a certain uniformization of distances. On the contrary, a strong hierarchy produces a regime with a minimum, and then an increase of disparities in time.
}{
Les valeurs prises par l'entropie pour les centralités (premier panel de la Fig.~\ref{fig:macrocoevolexplo:behavior}), en fonction du temps, pour $\gamma_N = 2.5$ et $v_0 = 110$, exhibent différents régimes en fonction de $d_G$ et $\gamma_G$. Une faible hiérarchie conduit à une entropie se stabilisant dans le temps, correspondant à une certaine uniformisation des distances. Au contraire, une forte hiérarchie produit un régime avec un minimum, puis une augmentation des disparités dans le temps.
}

\bpar{
This variety of behaviors can be found again with the rank correlation $\rho_R$, that we show here for the population variable, as a function of $d_G$. It has a low sensitivity to $\theta_N$ and $\gamma_N$ (see Appendix~\ref{app:sec:macrocoevolexplo}), but strongly varies as a function of $d_G$ and $\gamma_G$: interactions at a higher distance induce systematically a larger number of changes in the hierarchy of populations. These can occur when the hierarchy of distance is low. To summarize, the increase of the range of interactions will diminish the inertia of trajectories of the system of cities, whereas the increase of their hierarchy will increase it. This is relatively credible from a thematic point of view: longer and uniform interactions have more chances to make individual trajectories change.
}{
Cette variété de comportements se retrouve avec la corrélation de rang $\rho_R$, que nous montrons ici pour la variable de population, en fonction de $d_G$. Celle-ci est peu sensible à $\theta_N$ et $\gamma_N$ (voir Annexe~\ref{app:sec:macrocoevolexplo}), mais varie fortement en fonction de $d_G$ et $\gamma_G$ : des interactions à plus longue distance induisent systématiquement un plus grand nombre de changements dans la hiérarchie des populations. Celles-ci peuvent avoir lieu quand la hiérarchie de la distance est faible. En résumé, l'augmentation de la portée des interactions diminuera l'inertie des trajectoires du système de villes, tandis que l'augmentation de leur hiérarchie la fera croître. Cela est relativement crédible du point de vue thématique : des interactions plus lointaines et uniformes ont plus de chances de faire changer les trajectoires individuelles.
}

\bpar{
The behavior of correlation indicators is shown in Fig.~\ref{fig:macrocoevolexplo:correlations}. Concerning the effect of distance on correlations between variables, i.e. the evolution of $\rho_d$, it is interesting to note that an increase of $d_G$ systematically diminishes the levels of correlation, what corresponds to the complexification that we previously showed. As expected, $\rho_d\left[d\right]$ decreases as a function of distance, and exhibits non zero values for the correlation between population and centrality for a high hierarchy $\gamma_G$, what shows that simultaneous adaptation regimes are rare in this model.
}{
Le comportement des indicateurs de corrélation est montré en Fig.~\ref{fig:macrocoevolexplo:correlations}. Concernant l'effet de la distance sur les corrélations entre variables, c'est-à-dire l'évolution de $\rho_d$, il est intéressant de noter que l'augmentation de $d_G$ diminue systématiquement les niveaux de corrélation, ce qui correspond à la complexification mise en valeur précédemment. Comme attendu, $\rho_d\left[d\right]$ décroit en fonction de la distance, et montre des valeurs non nulles pour la corrélation entre population et centralité pour une forte hiérarchie $\gamma_G$, ce qui montre que les régimes d'adaptation simultanée sont rares dans ce modèle.
}

\subsubsection{Causality regimes}{Régimes de causalité}

\begin{figure}
\includegraphics[width=\linewidth,height=0.9\textheight]{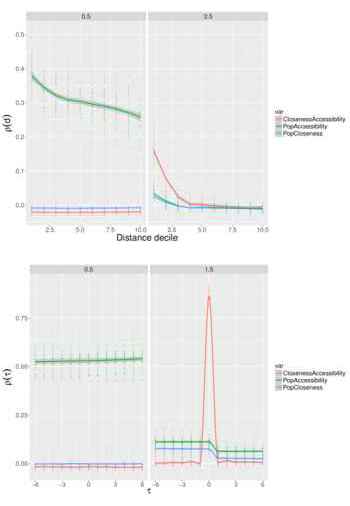}
	\caption[Correlations patterns in space and time]{\textbf{Correlations in the model for the spatial configuration $N_S=80,\alpha_S=0.5,d_S=10,n_S=30$.} (\textit{Top}) Correlations as a function of distance, for couples of variables (color), for $\gamma_N = 2.5$, $\theta_N = 21$, $v_0 = 10$, and for $d_G$ (columns) and $\gamma_G$ (rows) variables; (\textit{Bottom}) Lagged correlations for the same parameters. \label{fig:macrocoevolexplo:correlations}}
\end{figure}

\bpar{
Finally, by studying $\rho_{\tau}$ (Fig.~\ref{fig:macrocoevolexplo:correlations}, bottom panel), we observe that causality regimes in the sense of~\ref{sec:causalityregimes} are not very varied (as the Fig.~\ref{fig:app:macrocoevolexplo:laggedcorrs} in Appendix~\ref{app:sec:macrocoevolexplo} confirms it for a broader range of parameters). The population is systematically caused by the centrality, but there exists no regime in which we observe the contrary. This is a logic of an effect of reinforcement of hierarchy by centrality, but not a configuration with circular causalities, and thus not a co-evolution properly speaking as we defined in the statistical sense.
}{
Enfin, en étudiant $\rho_{\tau}$ (Fig.~\ref{fig:macrocoevolexplo:correlations}, panneau du bas), nous constatons que les régimes de causalité au sens de~\ref{sec:causalityregimes} ne sont pas très variés (comme le confirme la Fig.~\ref{fig:app:macrocoevolexplo:laggedcorrs} en Annexe~\ref{app:sec:macrocoevolexplo} pour une plage plus large de paramètres). La population est systématiquement causée par la centralité, mais il n'existe pas de régime où l'on observe le contraire. Il s'agit d'une logique d'effet de renforcement de la hiérarchie par la centralité, mais pas d'une configuration avec causalités circulaires, et donc pas d'une une co-évolution à proprement parler comme nous l'avons définie au sens statistique.
}



\bpar{
This brief exploration allows us to say that this model captures urban trajectories of a certain complexity, but that it does apparently not reproduces co-evolution regimes.
}{
Cette exploration brève nous permet d'affirmer que ce modèle capture des trajectoires urbaines d'une certaine complexité, mais qu'il ne reproduit apparemment pas de régimes de co-évolution.
}

\stars

\bpar{
We have thus in this section introduced the tools to understand trajectories produced by a co-evolution model, and tested these on the SimpopNet model.
}{
Nous avons ainsi dans cette section introduit les outils pour comprendre les trajectoires produites par un modèle de co-évolution, et testé ceux-ci sur le modèle SimpopNet.
}

\bpar{
In the following, we will explore in the same spirit a co-evolutive extension of the interaction model developed in~\ref{sec:interactiongibrat}, and will aim at establishing to what extent it is able to capture co-evolutive dynamics.
}{
Par la suite, nous explorerons dans le même esprit une extension co-évolutive du modèle d'interaction développé en~\ref{sec:interactiongibrat}, et chercherons à établir dans quelle mesure il est capable de capturer des dynamiques co-évolutives.
}

\stars

%


\newpage

\section{Dynamical extension of the interaction model}{Extension dynamique du modèle d'interaction}

\label{sec:macrocoevol}


\bpar{
This section extends the logic of integrating a system of cities with a transportation network, which has been pursued in a static way for network behavior in the interaction model developed and explored in section~\ref{sec:interactiongibrat}, to propose a \emph{macroscopic model of co-evolution for systems of cities}.
}{
Nous pouvons à présent étendre la logique d'intégration d'un système de ville et du réseau de transport, effectuée de manière statique pour le comportement du réseau dans le modèle d'interaction développé et exploré en section~\ref{sec:interactiongibrat}, pour proposer une formulation d'un \emph{modèle de co-évolution macroscopique pour les systèmes de villes}.
}

\subsection{Macroscopic Model of Co-evolution}{Modèle macroscopique de co-évolution}

\subsubsection{Rationale}{Hypothèses et choix de modélisation}

\bpar{
This first approach relies in a direct extension of the interaction model within a system of cities described in chapter~\ref{ch:evolutiveurban}, at a macroscopic scale with an ontology typical to systems of cities. For the sake of simplicity, we still stick to an unidimensional description of cities by their population.
}{
Cette première approche se place dans une logique d'extension directe du modèle d'interactions au sein d'un système de villes présenté en chapitre~\ref{ch:evolutiveurban}, c'est-à-dire à une échelle macroscopique et avec une ontologie typique aux systèmes de villes. Toujours dans un choix de simplicité, nous gardons ici une description unidimensionnelle des villes par leur population.
}

\bpar{
Concerning network growth, we propose also to stay at a relatively aggregated and simplified level, allowing to test growth heuristics at different levels of abstraction. In order to be flexible on model mechanisms, diverse processes can be taken into account, such as direct interactions between cities, intermediate interactions through the network, the feedback of network flows and a demand-induced network growth.
}{
Concernant la croissance du réseau, nous proposons de nous placer également à un niveau relativement agrégé et simplifié, en permettant de tester des heuristiques de croissance à différents niveaux d'abstraction. Dans une logique de flexibilité des mécanismes du modèle, il peut prendre en compte divers processus comme les interactions directes entre les villes, les interactions intermédiaires par le réseau, la rétroaction des flux de réseau et une croissance du réseau induite par la demande.
}

\bpar{
Empirical characteristics emphasized by~\cite{thevenin2013mapping} for the French railway network suggest the existence of feedbacks of network use, or of flows traversing it, on its persistence and its development, whose properties have evolved in time: a first phase of strong development would correspond to an answer to a high need of coverage, followed by a reinforcement of main link and the disappearance of weakest links.
}{
Les éléments empiriques mis en valeur pour le réseau ferré français par~\cite{thevenin2013mapping} suggèrent l'existence de rétroactions de l'utilisation du réseau, ou des flux le traversant, sur sa persistence et son développement, dont les propriétés ont évolué dans le temps : une première phase de développement fort correspondrait à la réponse du réseau à une demande de forte couverture, suivie d'une phase de renforcement des liens principaux et la disparition des liens les plus faibles.
}

\bpar{
The coupling between cities and the network will be achieved by the intermediate of flows between cities in the network: these capture the interactions between cities and have simultaneously an influence on the network in which they flow.
}{
Le couplage entre réseau et ville sera fait par l'intermédiaire des flux entre villes dans le réseau : ceux-ci portent les interactions entre villes et ont simultanément une influence sur le réseau dans lequel ils circulent.
}

\subsubsection{General Formulation}{Formulation générique}

\begin{figure}
\includegraphics[width=\linewidth]{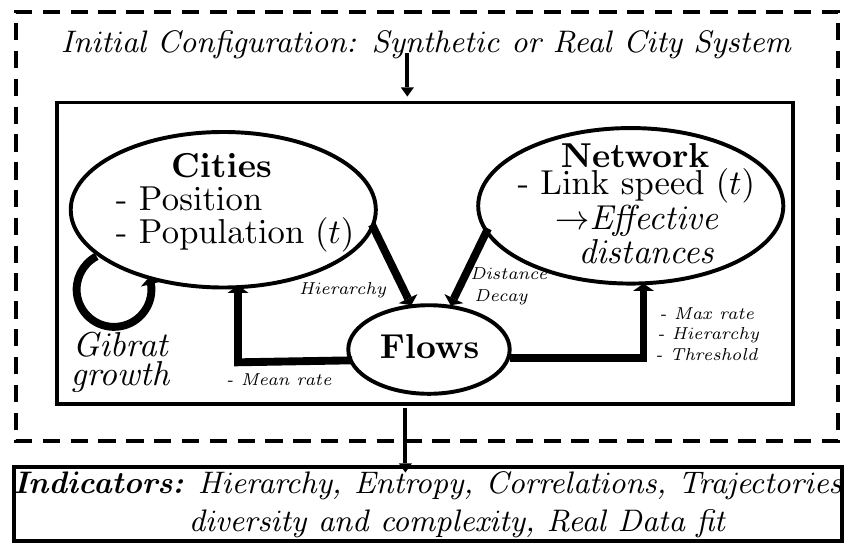}
\caption[Schematic model representation]{\textbf{Abstract representation of the model.} Ellipses correspond to main ontological elements (cities, network, flows), whereas arrows translate processes for which associated parameters are given. The model is described in its broader ecosystem of initialisation and output indicators.\label{fig:macrocoevol:model}}
\end{figure}

\bpar{
The urban system is characterized by populations $\mu_i(t)$ and the network $\mathbf{G}(t)$, to which can be associated a distance matrix $d^G_{ij}(t)$. Flows between cities $\phi_{ij}$ follow the expression given in~\ref{sec:interactiongibrat} with network distance. The same way, the evolution of populations follows the specifications of the base model. The Fig.~\ref{fig:macrocoevol:model} shows the structure of the model.
}{
Le système urbain est caractérisé par les populations $\mu_i(t)$ et le réseau $\mathbf{G}(t)$, auquel on peut associer une matrice de distance $d^G_{ij}(t)$. Les flux entre villes $\phi_{ij}$ suivent les expressions données en~\ref{sec:interactiongibrat} avec la distance réseau. De la même manière, la variation des populations suit les spécifications du modèle de base. La Fig.~\ref{fig:macrocoevol:model} exprime le modèle sous forme schématisée.
}

\paragraph{Network Growth}{Croissance du réseau}

\bpar{
Concerning the network, we assume that it evolves following the equation
\begin{equation}
\mathbf{G}(t + 1) = F(\mathbf{G}(t),\phi_{ij}(t))
\end{equation}
such that the assignment of flows within the network and a local variation of its elements is possible. We propose in a first time to consider patterns linked to distance only, and to specify a relation on an abstract network as
\begin{equation}
d^G_{ij}(t+1) = F(d^G_{ij}(t),\phi_{ij}(t))
\end{equation}
i.e. an evolution of the distance matrix only. In this spirit, we keep an interaction model strictly at a macroscopic scale, since a precise spatialization of the network would imply to take into account a finer scale that includes the local shape of the network which determines shortest paths.
}{
Concernant le réseau, nous faisons l'hypothèse que celui-ci évolue suivant
\begin{equation}
\mathbf{G}(t + 1) = F(\mathbf{G}(t),\phi_{ij}(t))
\end{equation}
de telle façon qu'une assignation des flux dans le réseau ainsi qu'une variation locale de ses éléments est possible. Nous proposons dans un premier temps de nous intéresser aux motifs liés à la distance uniquement, et de spécifier une relation sur un réseau abstrait par
\begin{equation}
d^G_{ij}(t+1) = F(d^G_{ij}(t),\phi_{ij}(t))
\end{equation}
c'est-à-dire une évolution de la matrice de distance uniquement. Dans cette logique, nous restons dans un modèle d'interaction à l'échelle strictement macroscopique, puisqu'une spatialisation précise du réseau impliquerait la prise en compte d'une échelle plus fine qui comporte la forme locale du réseau déterminante des plus court chemins.
}

\bpar{
Following a thresholded feedback heuristic, given a flow $\phi$ in a link, we assume its effective distance to be updated by:
}{
Suivant l'heuristique de rétroaction par seuil, étant donné un flux $\phi$ dans un lien, on suppose que sa distance effective est mise à jour par :
}

\begin{equation}
d(t+1) = d(t)\cdot \left( 1 + g_{max} \cdot \left[\frac{1 - \left(\frac{\phi}{\phi_0}\right)^{\gamma_s}}{1 + \left(\frac{\phi}{\phi_0}\right)^{\gamma_s}}\right]\right)
\end{equation}

\bpar{
with $\gamma_s$ a hierarchy parameter, $\phi_0$ the threshold parameter and $g_{max}$ the maximal growth rate at each step. This auto-reinforcement function can be interpreted the following way: above a limit flow $\phi_0$, the travel conditions improve, whereas they deteriorate below. The hierarchy of gain is given by $\gamma_s$, and since $\frac{1 - \left(\frac{\phi}{\phi_0}\right)^{\gamma_s}}{1 + \left(\frac{\phi}{\phi_0}\right)^{\gamma_s}} \rightarrow_{\phi\rightarrow \infty} -1$, $g_{max}$ is the maximal distance gain. This function is similar to the one used by \cite{tero2007mathematical}\footnote{Which uses $\Delta d = \Delta t \left[ \frac{\phi^\gamma}{1 + \phi^\gamma} - d\right]$. This function yield similarly a threshold effect, since the derivative vanishes at $\phi^{\ast} = \left(\frac{d}{1 - d}\right)^{1/\gamma}$, but it can not be adjusted.}.
}{
avec $\gamma_s$ un paramètre de hiérarchie, $\phi_0$ le paramètre de seuil et $g_{max}$ le taux de croissance maximal à chaque étape. Cette fonction d'auto-renforcement s'interprète de la façon suivante : au dessus d'un flux limite $\phi_0$, les conditions de trajet s'améliorent, tandis qu'elles diminuent en dessous. La hiérarchie du gain est donnée par $\gamma_s$ et comme $\frac{1 - \left(\frac{\phi}{\phi_0}\right)^{\gamma_s}}{1 + \left(\frac{\phi}{\phi_0}\right)^{\gamma_s}} \rightarrow_{\phi\rightarrow \infty} -1$, $g_{max}$ est le gain de distance maximal. Il s'agit d'une fonction similaire à celle utilisée par \cite{tero2007mathematical}\footnote{Qui utilise $\Delta d = \Delta t \left[ \frac{\phi^\gamma}{1 + \phi^\gamma} - d\right]$. Cette fonction permet également un effet de seuil, puisque la dérivée s'annule en $\phi^{\ast} = \left(\frac{d}{1 - d}\right)^{1/\gamma}$, mais celui-ci ne peut pas être ajusté.}.
}



\subsubsection{Implementation}{Implémentation}

\bpar{
The coupling of the interaction model to a finer representation of the network (for example an encoding of the whole network structure) makes the full integration into an OpenMole plugin more difficult, as it was done for the model studied in~\ref{sec:interactiongibrat}. We need here an \emph{ad hoc} implementation. The use of a workflow as a mediator for coupling is an interesting solution but which is realistic only for a weak coupling as in~\ref{sec:correlatedsyntheticdata}. One of the issues that the meta-modeling library for OpenMole that is currently being developed around OpenMole will have to tackle is the possibility to allow strong coupling (for example in the sense of a dynamical coupling during the evolution of the simulation) of heterogeneous components in a transparent way, in order to benefit from the advantages of different languages or of already existing implementations. 
}{
Le couplage du modèle d'interaction à une explicitation du réseau plus fine (par exemple encodage de l'ensemble de la structure du réseau) rend plus difficile l'intégration complète dans un plugin OpenMole comme c'était le cas pour le modèle étudié en~\ref{sec:interactiongibrat}, nécessitant une implémentation \emph{ad hoc}. L'utilisation d'un workflow comme médiateur pour le couplage est une solution intéressante mais réaliste uniquement dans le cas d'un couplage faible comme en~\ref{sec:correlatedsyntheticdata}. L'un des défis que devra relever la bibliothèque de méta-modélisation en cours de développement autour d'OpenMole, est la possibilité de coupler fortement (par exemple au sens de dynamiquement dans l'évolution de la simulation) des composantes hétérogènes de manière transparente, permettant de tirer parti des avantages de différents langages ou d'implémentations déjà existantes.
}


\bpar{
We choose here a full implementation with NetLogo, for the simplicity of coupling between components. A particular care is taken for the duality of network representation, both as a distance matrix and as a physical network, in order to facilitate the extension to physical network heuristics.
}{
Nous optons ici pour une implémentation complète en NetLogo pour une simplicité de couplage des composantes. Une attention particulière est portée à la dualité de la représentation du réseau, à la fois sous forme de matrice de distance et sous forme physique, pour permettre facilement l'extension à des heuristiques de réseau physique.
}

\subsection{Application to Synthetic Data}{Application à des données synthétiques}

\bpar{
The model is first tested and explored on synthetic city systems, in order to understand some of its intrinsic properties. In this case, we consider the model with an abstract network as specified above, i.e. without spatial description of the network and with evolution rules acting directly on $d^G_{ij}$ given the previous specifications. 
}{
Le modèle est d'abord testé et exploré sur des systèmes de villes synthétiques, afin de comprendre certaines de ses propriétés intrinsèques. Dans ce cas, nous considérons le modèle avec réseau abstrait comme spécifié ci-dessus, c'est-à-dire sans explicitation spatiale du réseau et avec les règles d'évolution agissant directement sur $d^G_{ij}$ selon les spécifications données précédemment.
}

\subsubsection{Synthetic data}{Données synthétiques}

\bpar{
A synthetic city system is generated following the heuristic used in the previous section: (i) $N_S$ cities are randomly distributed in the euclidian plan; (ii) populations are attributed to cities following an inverse power law, with a hierarchy parameter $\alpha_S$ and such that the largest city has a population equal to $P_{max}$, i.e. following $P_i = P_{max} \cdot i^{-\alpha_S}$.
}{
Un système de villes synthétiques est généré, en suivant l'heuristique utilisée dans la section précédente : (i) des villes en nombre $N_S$ sont placées aléatoirement dans le plan euclidien ; (ii) les populations sont attribuées aux villes selon une loi de puissance inverse, avec un paramètre de hiérarchie $\alpha_S$ et de telle façon que la plus grande ville ait une population $P_{max}$, c'est-à-dire suivant $P_i = P_{max} \cdot i^{-\alpha_S}$.
}

\bpar{
To simplify, several meta-parameters are fixed: the number of cities is fixed at $N_S = 30$, the maximal population at $P_{max} = 100000$ and the maximal network growth to $g_{max} = 0.005$. Final time is fixed at $t_f = 30$, what corresponds to distances divided approximatively by 5\footnote{Indeed, we can compute that the minimal multiplicative factor for distance is $(1 - g_{max})^{t_f}$, what gives for these values $(1 - 0.05)^{30} \simeq 0.214$, i.e. a division by 5 of the travel time.}, in order to comply to an empirical constraint: this corresponds to the evolution of the travel time between Paris and Lyon from around ten hours at the beginning of the century to two hours today, showed for example by~\cite{thevenin2013mapping}. We also neglect network effects at the second order by taking $w_N = 0$.
}{
Pour simplifier, nous fixons un certain nombre de méta-paramètres : le nombre de villes est fixé à $N_S = 30$, la population maximale à $P_{max} = 100000$ et la croissance maximale du réseau à $g_{max} = 0.005$. Le temps final est fixé à $t_f = 30$, ce qui correspond à des distances divisées par 5 environ\footnote{En effet, on peut calculer que le facteur multiplicatif minimal pour la distance est de $(1 - g_{max})^{t_f}$, ce qui donne pour ces valeurs $(1 - 0.05)^{30} \simeq 0.214$, c'est-à-dire une division par 5 du temps de trajet.}, afin de respecter un critère empirique : cela correspond à un passage du Paris-Lyon en une dizaine d'heures au début du 19ème siècle à deux heures aujourd'hui, mis en évidence par exemple par~\cite{thevenin2013mapping}. Nous négligeons aussi les effets de réseau au second ordre en fixant $w_N = 0$.
}


\bpar{
We explore a grid in the parameter space $\alpha_S$, $\phi_0$, $\gamma_s$, $w_G$, $d_G$, $\gamma_G$. We use the indicators introduced in~\ref{sec:macrocoevolexplo} to quantify model behavior in the parameter space. We describe the results for $\alpha_S = 1$, what is the closest to existing city systems (in comparison to 0.5 and 1.5, see the systematic review of the rank-size law estimations done by~\cite{10.1371/journal.pone.0183919}).
}{
Nous explorons une grille de l'espace des paramètres $\alpha_S$, $\phi_0$, $\gamma_s$, $w_G$, $d_G$, $\gamma_G$. Nous utilisons les indicateurs introduits en~\ref{sec:macrocoevolexplo} pour quantifier le comportement du modèle dans l'espace des paramètres. Nous donnons les résultats pour $\alpha_S = 1$, valeur la plus proche pour la plupart des systèmes de villes actuels (par rapport à 0.5 et 1.5, voir la revue systématique des estimations de la loi rang-taille faite par~\cite{10.1371/journal.pone.0183919}).
}


\subsubsection{Trajectories}{Trajectoires}

\begin{figure}
\includegraphics[width=\linewidth,height=0.9\textheight]{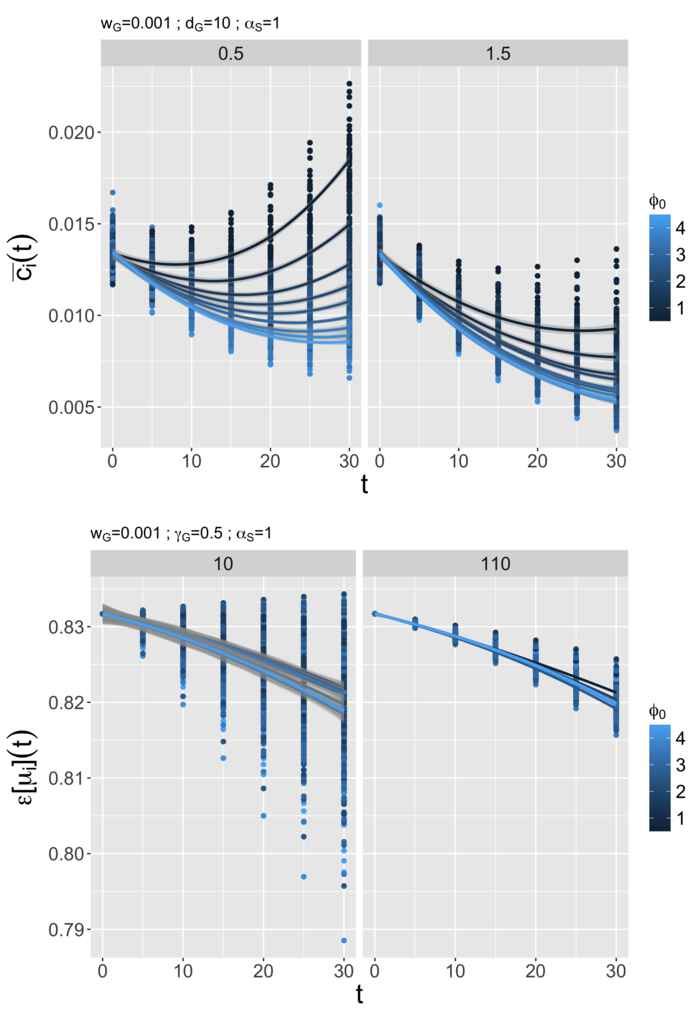}
\caption[Temporal behavior of the co-evolution model]{\textbf{Temporal behavior of the co-evolution model with abstract network on a synthetic system of cities.} \textit{(Top)} Average closeness centralities, as a function of time, for $\gamma_G$ (rows) and $\phi_0$ (color) variable, at fixed $w_G = 0.001$ and $d_G = 10$; \textit{(Bottom)} Entropy of populations, as a function of time, for $d_G$ (columns) and $\phi_0$ (color) variable, at fixed $w_G = 0.001$ and $\gamma_G = 0.5$. See main text for interpretation. Trajectories on the explored subspace of the parameter space are given in Fig.~\ref{fig:app:macrocoevol:behavior-time}, Appendix~\ref{app:sec:macrocoevol}.\label{fig:macrocoevol:behavior-time}}
\end{figure}

\bpar{
The evolution of the average closeness centrality in time is shown in Fig.~\ref{fig:macrocoevol:behavior-time} (top) for $w_G = 0.001$, and with variables $(\gamma_G,\phi_0)$. The behavior is not sensitive to $d_G$ (see the complete plots in~\ref{app:sec:macrocoevol}). This evolution witnesses a transition as a function of the level of hierarchy: when it decreases, we observe the emergence of trajectories for which the average centrality increases in time, what corresponds to configurations in which all cities profit in average from accessibility gains. 
}{
L'évolution de la centralité de proximité moyenne dans le temps est visualisée en Fig.~\ref{fig:macrocoevol:behavior-time} (haut) pour $w_G = 0.001$, et à $(\gamma_G,\phi_0)$ variables. Le comportement n'est pas sensible à $d_G$ (voir graphique complet en~\ref{app:sec:macrocoevol}). Cette évolution témoigne d'une transition en fonction du niveau de hiérarchie : lorsque celui-ci décroit, on observe l'émergence de trajectoires où la centralité moyenne croît dans le temps, ce qui correspond à des situations où l'ensemble des villes bénéficie en moyenne d'accroissements d'accessibilité.
}

\bpar{
Concerning the entropy of populations, for which the temporal trajectory is shown in Fig.~\ref{fig:macrocoevol:behavior-time} (bottom), all parameters give a decreasing entropy, i.e. a behavior of convergence of cities trajectories in time\footnote{Indeed, the entropy for the population variable gives the dispersion of the distribution of populations, and thus its decrease translate a trend to concentrate in time.}.
}{
En terme d'entropie des populations, dont nous traçons la trajectoire temporelle en Fig.~\ref{fig:macrocoevol:behavior-time} (bas), l'ensemble des paramètres donne une entropie décroissante, c'est-à-dire des comportements de convergence des trajectoires des villes dans le temps\footnote{En effet, l'entropie pour la variable de population exprime la dispersion de la distribution des populations, et donc une décroissance de celle-ci exprime une tendance à la concentration dans le temps.}.
}

\bpar{
Looking at the complexity of accessibility trajectories, we observe for values of $\phi_0 > 1.5$ a maximum of complexity as a function of interaction distance $d_G$, stable when $w_G$ and $\gamma_G$ vary (see also the exhaustive plots in Fig.~\ref{fig:app:macrocoevol:behavior-aggreg}, Appendix~\ref{app:sec:macrocoevol}). This intermediate scale can be interpreted as producing regional subsystems, large enough for each to develop a certain level of complexity, et isolated enough to avoid the convergence of trajectories over the whole system. We reconstruct therein a spatial non-stationarity, typically observed in~\ref{sec:staticcorrelations}, and rejoin the concept of the ecological niche\footnote{As it was already described in~\ref{sec:interdiscmorphogenesis}, an ecological niche in the sense of~\cite{holland2012signals} corresponds to the relatively independent ecosystem in which there is co-evolution between the species.} localized in space: the emergent subsystems that are relatively independent, are good candidates to contain processes of co-evolution. The emergence of this intermediate scale can be compared to the modularity of the French urban system showed by~\cite{berroir2017systemes}.
}{
Lorsqu'on s'intéresse à la complexité des trajectoires d'accessibilité, on note pour des valeurs de $\phi_0 > 1.5$ un maximum de la complexité en fonction de la distance d'interaction $d_G$, stable lorsque $w_G$ et $\gamma_G$ varient (voir également graphes exhaustifs en Fig.~\ref{fig:app:macrocoevol:behavior-aggreg}, Annexe~\ref{app:sec:macrocoevol}). Cette échelle intermédiaire peut être interprétée comme produisant des sous-systèmes régionaux, assez grands pour développer chacun un certain niveau de complexité, et assez isolés pour ne pas uniformiser les trajectoires sur l'ensemble de l'espace. Nous reconstruisons ainsi une non-stationnarité spatiale, typiquement observée en~\ref{sec:staticcorrelations}, et rejoignons le concept de niche écologique\footnote{Comme nous l'avons déjà présenté en~\ref{sec:interdiscmorphogenesis}, une niche écologique au sens de~\cite{holland2012signals} correspond à l'écosystème relativement indépendant au sein de laquelle il y a co-évolution entre les espèces.} localisée dans l'espace : les sous-systèmes émergents, relativement indépendants, sont de bons candidats pour être porteurs de processus de co-évolution. L'émergence de cette échelle intermédiaire peut être mise en parallèle avec la modularité du système urbain français montrée par~\cite{berroir2017systemes}.
}

\bpar{
Finally, the behavior of rank correlations for accessibility reveals that the interaction distance systematically increases the number of hierarchy inversions, what corresponds in a sense to an increase in overall system complexity. The hierarchy parameter diminishes this correlation, what means that a more hierarchical organization will impact a larger number of cities in the qualitative aspects of their trajectories. This effect is similar to the ``first mover advantage'' showed by \cite{levinson2011does}, which unveils a path dependency and an advantage to be rapidly connected to the network: in our case, the modifications in the hierarchy correspond to cities that benefit from their positioning in the network.
}{
Enfin, le comportement des corrélations de rang pour l'accessibilité révèle que la distance d'interaction augmente systématiquement le nombre d'inversions de hiérarchie, ce qui correspond en un sens à une augmentation de la complexité globale du système. Le paramètre de hiérarchie diminue quant à lui cette corrélation, ce qui veut dire qu'une évolution plus hiérarchique affectera un plus grand nombre de villes dans l'aspect qualitatif de leur trajectoires. Cet effet est similaire à celui du ``\textit{first mover advantage}'' montré par \cite{levinson2011does}, qui révèle une dépendance au chemin et un avantage à être connecté rapidement au réseau : dans notre cas, les modifications de hiérarchie correspondent à des villes qui tirent avantage de leur positionnement dans le réseau.
}

\begin{figure}
\includegraphics[width=\linewidth]{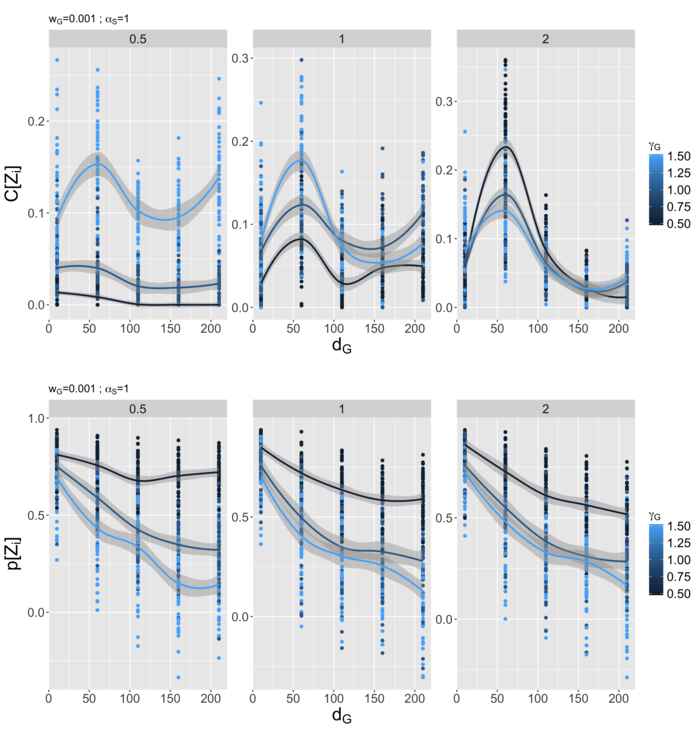}
\caption[Agregated behavior of the co-evolution model]{\textbf{Agregated behavior of the co-evolution model.} \textit{(Top)} Complexity of accessibilities, as a function of $d_G$, for $\phi_0$ (columns) and $\gamma_G$ (color) variable, at fixed $w_G = 0.001$; \textit{(Bottom)} Rank correlations of accessibilities as a function of $d_G$, for the same parameters. The behavior on the explored subspace of the parameter space are given in Fig.~\ref{fig:app:macrocoevol:behavior-aggreg}, Appendix~\ref{app:sec:macrocoevol}.\label{fig:macrocoevol:behavior-aggreg}}
\end{figure}

\subsubsection{Correlations}{Corrélations}

\bpar{
We can in a first time focus on the variations of correlations between variables as a function of distance. Profiles of $\rho_d$ for the three couples of variables show that intermediate and large values of the interaction distance ($d_G > 50$) induce populations totally uncorrelated with centralities and accessibilities (Fig.~\ref{fig:app:macrocoevol:distcorrs}, Appendix~\ref{app:sec:macrocoevol}). For small values of $d_G$, a decreasing then vanishing profile confirms the existence of strong local effects, where very close cities will have a strong reciprocal influence. The behavior of the correlation between accessibility and centrality is more difficult to interpret, and may be due to autocorrelation phenomenons\footnote{These can not be computed, as it implies to decompose $\rho\left[\sum_{i\neq j} \frac{1}{d_{ij}}; \sum_{i\neq j} P_j \exp{\left(-d_{ij}/d_G\right)}\right]$. It is for example possible to approximate $\rho\left[X+Y;Z\right]$ under the condition that $\varepsilon = \sigma_Y / \sigma_X \ll 1$ at the first order by $\rho\left[ X+Y;Z \right] \simeq \left(\rho\left[ X;Z \right]  + \varepsilon \rho\left[Y;Z\right]\right)\cdot\left(1 - \frac{1}{2}\rho\left[X;Y\right]\varepsilon - \frac{\varepsilon^2}{2})\right)$, by this assumption is too restrictive to be used for all terms in the sum.}. Its level does not depend on distance but on $d_G$, and decreases to end at a negative correlation.
}{
Nous pouvons dans un premier temps nous intéresser aux variations des corrélations entre variables en fonction de la distance. Les profils de $\rho_d$ pour les trois couples de variables montrent que des valeurs moyennes et grandes de la distance d'interaction ($d_G > 50$) induisent des populations totalement décorrélées aux centralités et accessibilités (Fig.~\ref{fig:app:macrocoevol:distcorrs}, Annexe~\ref{app:sec:macrocoevol}). Pour des petits $d_G$, un profil décroissant puis nul confirme l'existence d'effets locaux forts, où des villes très proches s'influenceront fortement. Le comportement de la corrélation entre accessibilité et centralité est plus difficile à interpréter, et peut être dû aux phénomènes d'auto-corrélation\footnote{Celles-ci ne sont pas calculables, car il s'agirait de décomposer $\rho\left[\sum_{i\neq j} \frac{1}{d_{ij}}; \sum_{i\neq j} P_j \exp{\left(-d_{ij}/d_G\right)}\right]$. Il est possible par exemple d'approximer $\rho\left[X+Y;Z\right]$ sous la condition que $\varepsilon = \sigma_Y / \sigma_X \ll 1$ au premier ordre par $\rho\left[ X+Y;Z \right] \simeq \left(\rho\left[ X;Z \right]  + \varepsilon \rho\left[Y;Z\right]\right)\cdot\left(1 - \frac{1}{2}\rho\left[X;Y\right]\varepsilon - \frac{\varepsilon^2}{2})\right)$, mais cette hypothèse est trop restrictive pour être valable sur l'ensemble de la somme.}. Son niveau ne dépend pas de la distance mais de $d_G$, et est décroissant pour finir à une corrélation négative.
}

\subsubsection{Causality regimes}{Régimes de causalité}

\bpar{
We can now study lagged correlation patterns produced by the model, i.e. its ability to effectively produce co-evolution in the sense we defined.
}{
Tournons nous finalement vers les motifs de corrélations retardées produits par le modèle, c'est-à-dire l'étude de sa capacité à effectivement produire de la co-évolution comme nous l'avons définie.
}

\begin{figure}
	\includegraphics[width=\linewidth]{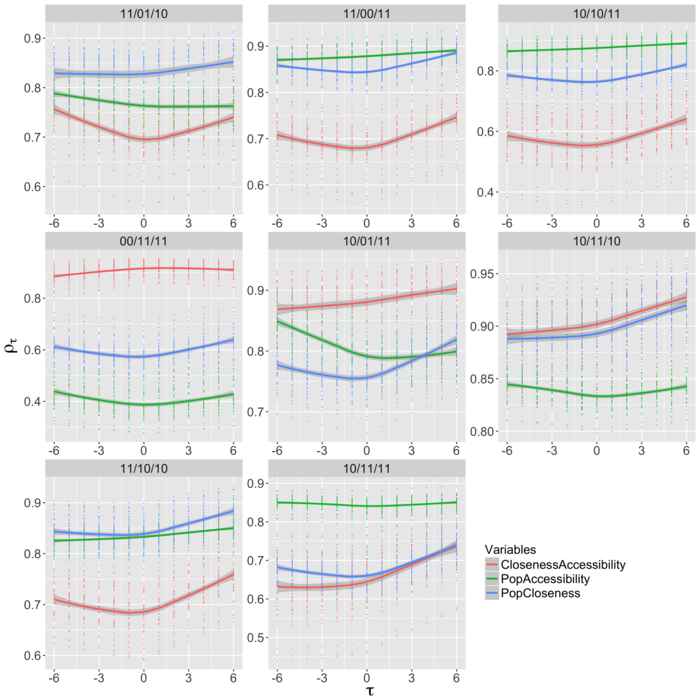}
\caption[Profiles of lagged correlations]{\textbf{Lagged correlations.} We give here for the 8 configurations showing at least 4 links between variables (coded in the order of couples, by the existence or not of a link for $\tau_+$ and for $\tau_-$), the lagged correlation profiles $\rho_{\tau}$ as a function of $\tau$, for all couples of variables (color).\label{fig:macrocoevol:correlations}}
\end{figure}

\bpar{
The exploration of profiles for $\rho_\tau$ for varying parameter values is illustrated in Appendix~\ref{app:sec:macrocoevol}, and suggests the existence of multiple causality regimes. The Fig.~\ref{fig:macrocoevol:correlations} give examples of such profiles. We however observe (i) the systematic existence of a constant correlation at $\tau = 0$ and (ii) the small variations of correlations that impose the need for a statistical test to ensure that we isolate a significant effect.
}{
L'exploration de profils de $\rho_{\tau}$ selon les paramètres est illustrée en Annexe~\ref{app:sec:macrocoevol}, et suggère l'existence de régimes de causalité variés. La Fig.~\ref{fig:macrocoevol:correlations} donne des exemples de profils. Nous constatons cependant (i) l'existence systématique d'une corrélation constante à $\tau = 0$ et (ii) les faibles variations des corrélations qui nécessitent la mise en place d'un test statistique pour être certains d'isoler un effet significatif.
}

\bpar{
We add here for this reason an additional criteria based on a statistical test: for $\tau_+ = \textrm{argmax}_{\tau>0} \left|\rho_{\tau} - \rho_0\right|$ and $\tau_- = \textrm{argmax}_{\tau<0} \left|\rho_{\tau} - \rho_0\right|$, a Kolmogorov-Smirnov test is used to compare the distributions of $\rho_{\tau_{\pm}}$ and of $\rho_0$. If they are declared different with a p-value smaller than $0.01$, and if $\left|\rho_{\tau_{\pm}}\right| > \left|\rho_0\right|$, we accept the causality link between variables in the corresponding direction.
}{
Nous ajoutons donc ici un critère basé sur un test statistique : pour $\tau_+ = \textrm{argmax}_{\tau>0} \left|\rho_{\tau} - \rho_0\right|$ et $\tau_- = \textrm{argmax}_{\tau<0} \left|\rho_{\tau} - \rho_0\right|$, un test de Kolmogorov-Smirnov est utilisé pour comparer les distributions de $\rho_{\tau_{\pm}}$ et de $\rho_0$. S'il les déclare différentes avec une \emph{p-value} inférieure à $0.01$, et si $\left|\rho_{\tau_{\pm}}\right| > \left|\rho_0\right|$, nous convenons d'un lien de causalité entre les variables dans le sens correspondant.
}

\bpar{
A configuration is then coded by a representation of its graph between variables, given by the six discrete variables equal to 0 if there is no link between the variables (within all directed couples between population, accessibility and centrality) and 1 or -1 depending on the sign of the correlation if there exists a statistically significant link (in practice we observe only positive correlations).
}{
Nous codons alors une configuration par une représentation de son graphe entre variables, donnée par les 6 variables discrètes valant 0 s'il n'y a pas de lien entre les variables (parmi l'ensemble des couples dirigés entre population, accessibilité et centralité) et 1 ou -1 selon le signe de la corrélation s'il existe un lien significatif statistiquement (en pratique toutes les corrélations sont positives).
}

\bpar{
We obtain overall 33 different configurations of links between variables, out of the 64 possible configurations ($2^6$ possible choices for positive correlations only). In comparison, the application of this method on the results of~\ref{sec:macrocoevolexplo} give only 8 distinct configurations\footnote{In which two configurations correspond to a negative circular causality between accessibility and centrality, what suggests that the SimpopNet model can produce a co-evolution between variables, but in a restricted number compared to the configurations obtained here and only between two network variables.}.
}{
Nous obtenons au total 33 configurations de liens entre variables, sur les 64 possibles ($2^6$ choix possibles pour des corrélations uniquement positives). En comparaison, l'application de cette méthode sur les résultats de~\ref{sec:macrocoevolexplo} ne donne que 8 configurations distinctes\footnote{Parmi lesquelles deux correspondent à une causalité circulaire négative entre accessibilité et centralité, ce qui fait dire que le modèle Simpopnet peut produire une co-évolution entre variables, mais en nombre restreint par rapport aux configurations obtenues ici et uniquement entre deux variables de réseau.}.
}

\bpar{
The type of relations we obtain are particularly interesting regarding co-evolution. We indeed observe:
\begin{itemize}
\item a configuration without any link between variables;
\item 13 configurations of type ``structuring effect'', i.e. for which the graph does not have any loop;
\item a configuration of type ``indirect co-evolution'', for which the graph has a loop of length three ($c_i \rightarrow X_i \rightarrow \mu_i \rightarrow c_i$) ;
\item 18 configurations of type ``co-evolution'', in which there exists at least a loop of length two (direct circular relation between two variables).
\end{itemize}
}{
Les types de relations obtenues sont particulièrement éclairantes au regard de la co-évolution. On observe en effet :
\begin{itemize}
	\item une configuration sans lien entre variables ;
	\item 13 configurations de type ``effet structurant'', c'est-à-dire dont le graphe ne possède aucune boucle ; 
	\item une configuration de type ``co-évolution indirecte'', dont le graphe possède une boucle de longueur trois ($c_i \rightarrow X_i \rightarrow \mu_i \rightarrow c_i$) ;
	\item 18 configurations de type ``co-évolution'', dans lesquelles il existe au moins une boucle de longueur deux (relation circulaire directe entre deux variables).
\end{itemize}
}







\bpar{
Among all these regimes, 8 correspond to a graph with at least 4 links (which are then necessarily co-evolutive): we show these profiles in Fig.~\ref{fig:macrocoevol:correlations}. Two regimes witness a positive deviation of the correlation between population and accessibility for positive delays, increasing up to the maximal delay, what could be a clue of a reinforcement of population dynamics through centrality, stylized fact shown for the French system of cities by~\cite{bretagnolle:tel-00459720}.
}{
Parmi tous ces régimes, 8 correspondent à un graphe avec au moins 4 liens (qui sont alors nécessairement co-évolutifs) : il s'agit des profils que nous représentons en Fig.~\ref{fig:macrocoevol:correlations}. Deux régimes présentent une déviation positive de la corrélation entre population et accessibilité pour les retards positifs, en croissance jusqu'au retard maximal, ce qui pourrait être un marqueur du renforcement des dynamiques de population par la centralité, fait stylisé exhibé pour le système de ville Français par~\cite{bretagnolle:tel-00459720}.
}

\bpar{
The regimes in which the centrality is co-evolving with population correspond to the ones where the co-evolution between the network and the territory is the strongest (since the accessibility depends on both), and are observed for large values of $d_G$ (average $d_G=183$ on 62 parameter points). This way, this co-evolution is favored by long interaction ranges. 
}{
Les régimes où la centralité est en co-évolution avec la population correspondent à ceux où la co-évolution entre réseau et territoire est la plus marquée (puisque l'accessibilité relève des deux), et sont observés pour des grandes valeurs de $d_G$ (moyenne $d_G=183$ sur 62 points de paramètres). Ainsi, cette co-évolution est favorisée par de grandes portées d'interaction.
}

\bpar{
Finally, the regime with the largest number of links\footnote{That corresponds to the regime coded by ``10/11/11'', with co-evolution of population and centrality and of population and accessibility, and a causality of centrality on accessibility.}, is obtained for a long interaction range $d_G = 160$, a strong interaction hierarchy $\gamma_G = 1.5$, but a low hierarchy of the initial system of cities $\alpha_S$: far-reaching but hierarchical interactions in an uniform system of cities lead to a maximum of entanglement between variables.  
}{
Enfin, le régime avec le plus fort nombre de liens\footnote{Correspondant au régime codé ``10/11/11'', avec co-évolution de la population et de la centralité ainsi que de la population et de l'accessibilité, et une causalité de la centralité sur l'accessibilité.}, est obtenu pour une grande portée d'interaction $d_G = 160$, une forte hiérarchie d'interaction $\gamma_G = 1.5$, mais une faible hiérarchie du système de ville initial $\alpha_S$ : des interactions lointaines et hiérarchisées dans un système uniforme conduisent à un maximum d'intrication entre variables.
}

\bpar{
We finally confirm these results of variety in causality regimes produced by the model by applying the \emph{Pattern Space Exploration} algorithm~\cite{10.1371/journal.pone.0138212} to the model, with objectives the six correlations studied above (evaluated as zero in the case of a non-significance). A graphical presentation of results is given in Appendix~\ref{app:sec:macrocoevol}. We mainly obtain a number of regimes produced by the model larger than the ones obtained before (with negative correlations, 260 realized regimes out of $3^6 = 729$ possible). This short complementary study confirms the ability of the model to produce a large number of co-evolution regimes. 
}{
Nous confirmons finalement ces résultats de variété des régimes de causalité produits par le modèle en appliquant l'algorithme \emph{Pattern Space Exploration}~\cite{10.1371/journal.pone.0138212} au modèle, avec comme objectifs les 6 corrélations étudiées précédemment (évaluées comme nulles dans le cas d'une non-significativité). Une présentation graphique des résultats est donnée en Annexe~\ref{app:sec:macrocoevol}. Nous obtenons principalement un nombre de régimes produits par le modèle encore plus important que ceux obtenus précédemment (avec des corrélations négatives, 260 régimes réalisés sur $3^6 = 729$ possibles). Cette brève étude complémentaire confirme la capacité du modèle à produire une grande variété de régimes de co-évolution.
}

\subsubsection{Synthesis}{Synthèse}

\bpar{
The important stylized facts that can be drawn from the exploration of the model on synthetic data are the following.
}{
Les faits stylisés marquants qui ressortent de l'exploration du modèle sur données synthétiques sont les suivants.
}

\bpar{
\begin{enumerate}
\item We observe the existence of an intermediate spatial scale allowing the evolution of relatively independent niches, corresponding to a maximal level of complexity for cities trajectories.
\item Lagged correlations unveil at least three different types of interaction regimes, that we interpret as an adaptation regime, a direct co-evolution regime, and an indirect co-evolution regime.
\end{enumerate}
}{
\begin{enumerate}
	\item On révèle l'existence d'une échelle spatiale intermédiaire permettant l'évolution de niches relativement indépendantes, correspondant à un niveau de complexité des trajectoires maximal.
	\item Les corrélations retardées mettent en évidence au moins trois types différents de régimes d'interaction, que l'on interprète comme un régime d'adaptation, un régime de co-évolution direct et un régime de co-évolution indirecte.
\end{enumerate}
}

\subsection{Applications to French City System}{Applications au système de villes français}

\bpar{
The model is then applied to the French system of cities on long time dynamical data: the Pumain-INED database for populations, spanning from 1831 to 1999 \cite{pumain1986fichier}, with the evolving railway network from 1840 to 2000 \cite{thevenin2013mapping}. Such a time span can be associated with structural effect on long time, as developed in~\ref{sec:networkterritories}. This application aims on the one hand at testing the ability of the model to reproduce a real dynamic of co-evolution, and on the other hand at extracting thematic information on processes through calibrated parameter values.
}{
Le modèle est ensuite appliqué au système de villes français sur des données dynamiques sur le temps long : la base Pumain-INED pour les populations, couvrant de 1831 à 1999 \cite{pumain1986fichier}, avec le réseau ferré dynamique de 1840 à 2000 \cite{thevenin2013mapping}. Cette durée temporelle découle de la logique des effets de structure sur le temps long, comme développé en~\ref{sec:networkterritories}. Cette application vise d'une part à tester la capacité du modèle à reproduire une dynamique de co-évolution réelle, et d'autre part à extraire une information thématique sur les processus via les valeurs calibrées des paramètres.
}

\subsubsection{Network Data}{Données de réseau}

\bpar{
We work on railway network data constructed by~\cite{thevenin2013mapping}. The French railway network is particularly interesting jointly with population data already presented, since the covered time span is relatively close, and as \cite{thevenin2013mapping} recalls, this transportation mode has at any times materialized the implication of public and private actors. It corresponds to different processes depending on the period, from a more decentralized management to a more centralized recently, and different technological materializations with for example the recent emergence of high speed trains \cite{zembri1997fondements}. For each date in the population database, we extract the simplified abstract network in which all stations and intersections with a degree larger than two are linked with abstract links which speed and length attributes correspond to real values, at a granularity of 1km\footnote{This processing is achieved thanks to the R package for transportation network analysis specifically developed for this thesis, see~\ref{app:sec:packages}.}. This yields the time-distance matrices between the cities included in the model.
}{
Nous travaillons sur les données de réseau ferré construites par~\cite{thevenin2013mapping}. Le réseau ferré français est particulièrement intéressant en conjonction avec les données de population déjà présentées, puisque la période couverte est relativement similaire, et que comme le rappelle \cite{thevenin2013mapping}, ce moyen de transport a à toute période concrétisé l'implication d'acteurs publiques et privés importants, tout en correspondant à différents processus selon les époques, d'une gestion plutôt décentralisée à une centralisation très forte plus récemment, et différentes concrétisations technologiques avec par exemple l'émergence récente de la grande vitesse~\cite{zembri1997fondements}. Pour chaque date de la base de donnée de population, nous extrayons le graphe abstrait simplifié où toutes les gares et intersections de degré supérieur à deux sont reliés par les liens abstrait avec attributs de vitesse et distance traduisant la valeur réelle, à une granularité de 1km\footnote{Ce traitement est effectué grâce au package R pour l'analyse des réseaux de transport développé spécifiquement dans le cadre de cette thèse, voir~\ref{app:sec:packages}.}. Cela permet également de construire les matrices de distance-temps entre les villes considérées dans le modèle.
}

\subsubsection{Stylized facts}{Faits stylisés}

\begin{figure}
	\includegraphics[width=\linewidth]{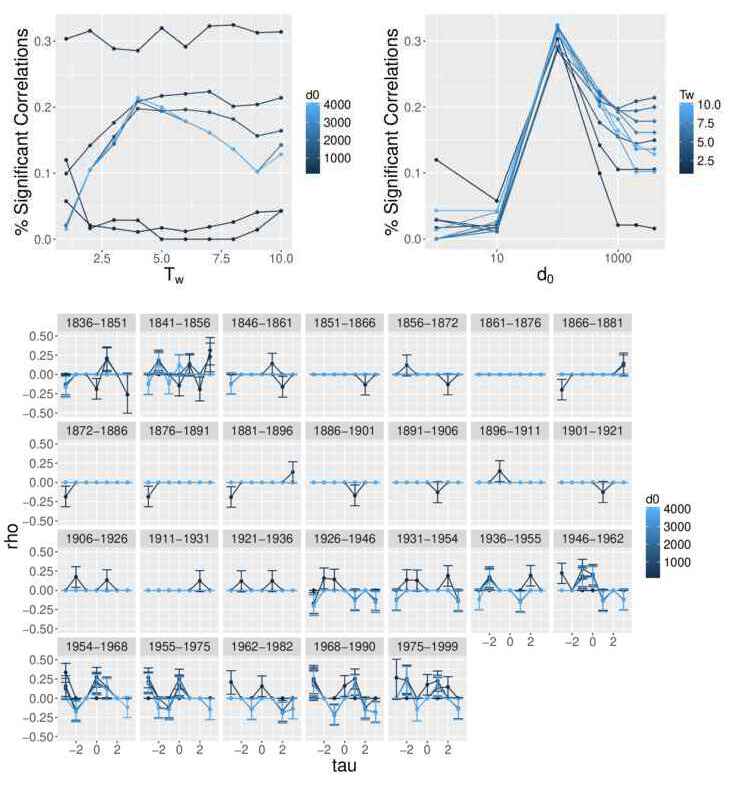}
	\caption[Empirical lagged correlations for the French system of cities]{\textbf{Empirical lagged correlations for the French system of cities.} Correlations are estimated on a window of duration $5\cdot T_w$, between population growth rates and the variations of closeness centrality with a decay parameter $d_0$ (see text). \textit{(Top left)} Number of significant correlations (taken such that $p<0.1$ at 95\%) as a function of $T_w$ for $d_0$ variable; \textit{(Top right)} Number of significant correlations as a function of $d_0$ for $T_w$ variable; \textit{(Bottom)} For the ``optimal'' window $T_w = 4$, value of $\rho_{\tau}$ as a function of $\tau$, for all successive periods.\label{fig:macrocoevol:empirical}}
\end{figure}

\bpar{
Before calibrating the model, we can observe the lagged correlation patterns in the dataset, by applying the causality regimes method. This empirical study should on the one hand allow us to verify well known stylized facts, and on the other hand to produce a preliminary knowledge of empirical system behavior. We compute as detailed above the closeness centrality through the network, given by $T_i = \sum_j \exp{-d_{ij}/d_0}$, and we study the lagged correlation between its derivative $\Delta T_i$ and the derivative of the population $\Delta P_i$, given by $\hat{\rho}_{\tau} = \hat{\rho}\left[\Delta P_i(t),\Delta T_i(t-\tau)\right]$ estimated on a moving window containing $T_w$ successive dates. We show in Fig.~\ref{fig:macrocoevol:empirical} the results obtained.
}{
Avant de calibrer le modèle, observons les motifs de corrélation présents dans les données, en appliquant la méthode des corrélations retardées. Cette étude empirique devrait permettre d'une part de vérifier des faits stylisés connus, d'autre part d'établir une connaissance préliminaire du comportement empirique du système. Nous calculons comme précisé ci-dessus la centralité de proximité via le réseau, donnée par $T_i = \sum_j \exp{-d_{ij}/d_0}$, et étudions la corrélation retardée entre sa dérivée $\Delta T_i$ et celle de la population $\Delta P_i$, donnée par $\hat{\rho}_{\tau} = \hat{\rho}\left[\Delta P_i(t),\Delta T_i(t-\tau)\right]$ estimée sur une fenêtre glissante comprenant $T_w$ dates successives. Nous montrons en Fig.~\ref{fig:macrocoevol:empirical} les résultats obtenus.
}

\bpar{
These results are important for at least two reasons. First, the behavior of the number of significant correlations as a function of $T_w$ and $d_0$ allows us to find stationarity scales in the system. We observe on the one hand a specific spatial scale that gives a maximum for all temporal windows, at $d_0 = 100km$, what suggests the existence of consistent regional subsystems, which existence is stable in time: indeed, this value corresponds to the interaction distance. It remarkably coincides with the intermediate scale isolated in the synthetic model. On the other hand, long spatial ranges induce an optimal temporal scale, for $T_w = 4$ what corresponds to around twenty years: we identify it as the overall temporal stationarity scale of the system and study the lagged correlations for this value.
}{
Ces résultats sont importants pour au moins deux raisons. Dans un premier temps, le comportement du nombre de corrélations significatives en fonction de $T_w$ et de $d_0$ permet la recherche d'échelles de stationnarité dans le système. Nous observons d'une part une échelle spatiale spécifique donnant un maximum pour l'ensemble des fenêtres temporelles, à $d_0 = 100km$, ce qui suggère l'existence de sous-systèmes régionaux cohérents, dont l'existence est stable dans le temps : en effet, cette valeur correspond à la distance d'interaction. Celle-ci coincide remarquablement avec l'échelle intermédiaire isolée dans le modèle synthétique. D'autre part, les grandes portées spatiales induisent une échelle temporelle optimale, pour $T_w = 4$ ce qui correspond à une vingtaine d'année : nous l'identifions comme l'échelle de stationnarité temporelle du système dans son ensemble et étudions les corrélations retardées pour cette valeur.
}

\bpar{
Secondly, the behavior of lagged correlations does not seem to comply to the existing literature. At the intermediate spatial scale, the values of $\rho_+,\rho_-$ exhibit no regularity. On the whole system, there is until 1946 close to no significant effect, then no causality between 1946 and 1975 (maximum at $\tau = 0$, non-significant minimum), and a 5 years shift of accessibility causing population after 1968 (the effect staying however doubtful). We do not reproduce the correlation effect between network centrality and place in the urban hierarchy advocated by~\cite{bretagnolle2003vitesse}\footnote{As~\cite{lemoy2017scaling} is not able to reproduce, for density profiles as a function of the distance to the center of European metropolis, the transition that allows~\cite{guerois2008built} to define the peri-urban. These more or less recent works are not reproducible, producing neither code nor data, and giving only a superficial description of the methods, and it is thus impossible to know the origin of the qualitative divergence obtained. A good reproducibility together with the construction of systematic comparisons (\emph{benchmarks}) of models, empirical analysis, that are recent but also to validate old studies, seems to be a reasonable solution to this kind of issue.}, what lead us to question the existence of the ``structural co-evolution'' on long time described by \noun{Bretagnolle} in~\cite{espacegeo2014effets}. What \cite{bretagnolle2003vitesse} obtains is a simultaneous correspondence between growth rate and level of connectivity to the network (and not with network dynamic), but not in our sense a co-evolution, since no statistical relation is furthermore exhibited.
}{
Dans un second temps, le comportement des corrélations retardées ne semble pas en accord avec la littérature existante. À l'échelle spatiale intermédiaire, les valeurs de $\rho_+,\rho_-$ n'exhibent aucune régularité. Sur l'ensemble du système, on a jusqu'en 1946 quasiment aucun effet significatif, puis aucune causalité entre 1946 et 1975 (maximum à $\tau = 0$, minimum non significatif), puis un décalage de 5 ans de l'accessibilité causant la population après 1968 (l'effet restant tout de même douteux). Nous ne reproduisons pas l'effet de corrélation entre centralité dans le réseau et place dans la hiérarchie urbaine défendu par~\cite{bretagnolle2003vitesse}\footnote{Tout comme~\cite{lemoy2017scaling} n'arrivent pas à reproduire, pour les profils de densité en fonction de la distance au centre des métropoles européennes, la transition permettant à \cite{guerois2008built} de définir le péri-urbain. Ces travaux plus ou moins anciens ne sont pas reproductibles, ne fournissant ni code, ni données et ne donnant qu'une description très succincte des méthodes, et il est ainsi impossible de connaitre l'origine de la divergence qualitative obtenue. Une bonne reproductibilité ainsi que la construction de comparaisons systématiques (\emph{benchmarks}) de modèles, analyses empiriques, récentes mais aussi en validation d'études passées, nous semble une bonne solution à ce genre de problèmes.}, ce qui amène à relativiser l'existence de la ``co-évolution structurelle'' sur le temps long décrite par \noun{Bretagnolle} dans~\cite{espacegeo2014effets}. Ce que trouve \cite{bretagnolle2003vitesse}, c'est une correspondance simultanée entre taux de croissance et niveau de connectivité aux réseaux (et non avec les dynamiques du réseau), mais pas à notre sens une co-évolution, d'autant plus qu'aucune relation statistique n'est exhibée.
}

\bpar{
We rejoin the recent results of~\cite{mimeur:hal-01616746} that show the statistical non-significance of the correlation between growth rate and evolution of network coverage and accessibility, at a zero delay. Our results are less precise on the class of cities studied (they differentiate large and small cities, and work on a larger panel), but more general as they study variable delays and accessibility ranges, and are thus complementary.
}{
Nous rejoignons les résultats récents de~\cite{mimeur:hal-01616746} qui montrent la non-significativité statistique de la corrélation entre taux de croissance et évolution de la couverture du réseau ainsi que l'évolution de l'accessibilité, à délai nul. Nos résultats sont moins précis pour les classes de villes étudiées (ils différencient grandes villes et petites, et travaillent sur un panel plus grand), mais plus généraux car pour un délai et une portée de l'accessibilité variables, et donc complémentaires.
}

\subsubsection{Calibration of the abstract model}{Calibration du modèle abstrait}

\bpar{
Expected results of the calibration on real data concern both the more or less accurate reproduction of real city population growth dynamics, i.e. to what extent the inclusion of a dynamical network can increase the explanatory power for trajectories, and also how realistic the evolution of network distance is. We still work with the abstract model.
}{
Les résultats attendus de la calibration sur données réelles concernent à la fois la reproduction plus ou moins précise des dynamiques réelles de croissance de population, c'est-à-dire dans quelle mesure la prise en compte d'un réseau dynamique peut augmenter le pouvoir explicatif pour les trajectoires, et aussi quel est le niveau de réalisme de l'évolution de la distance par le réseau. Nous travaillons toujours avec le modèle abstrait.
}

\paragraph{Model evaluation}{Evaluation du modèle}

\bpar{
We can add to the indicators used before a calibration indicator for distance. The particular property of adjustment for populations, that resides in the existence of a power law for the sizes of cities that made negligible the performance on medium and small cities in the case of a cumulated error, and suggested the addition of the indicator on the error on logarithms, is not present for distances that follow a distribution concentrated on a single order of magnitude. We use therefore a standard measure of fit, given by
\[
\varepsilon_D = \log \left[ \sum_t \sum_{i,j} \left(d_{ij}(t) - \tilde{d}_{ij}(t)\right)^2\right]
\]
where $d_{ij}(t)$ are observed distances and $\tilde{d}_{ij}(t)$ the simulated distances. It is simply a cumulated squared-error, as used for the comparison of origin-destination matrices in a similar case of simulation of a transportation network in~\cite{jacobs2016transport}.
}{
On peut ajouter aux indicateurs utilisés précédemment un indicateur de calibration pour la distance. L'aspect particulier de l'ajustement pour les populations, qui résidait dans la présence d'une loi de puissance pour les tailles de villes rendant négligeables les performances sur les villes moyennes et les petites villes dans le cas d'une erreur cumulée, et suggérait l'ajout de l'indicateur de l'erreur sur les logarithmes, n'est pas présent pour les distances qui suivent une distribution concentrée sur un ordre de grandeur unique. Nous utilisons ainsi une mesure standard d'ajustement, donnée par
\[
\varepsilon_D = \log \left[ \sum_t \sum_{i,j} \left(d_{ij}(t) - \tilde{d}_{ij}(t)\right)^2\right]
\]
où $d_{ij}(t)$ sont les distances observées et $\tilde{d}_{ij}(t)$ les distances simulées. Il s'agit simplement d'une erreur carré cumulée, comme utilisée pour la comparaison de matrices origine-destination dans un cas similaire de simulation d'un réseau de transport dans~\cite{jacobs2016transport}.
}




\paragraph{Results}{Résultats}

\begin{figure}
	\includegraphics[width=\linewidth]{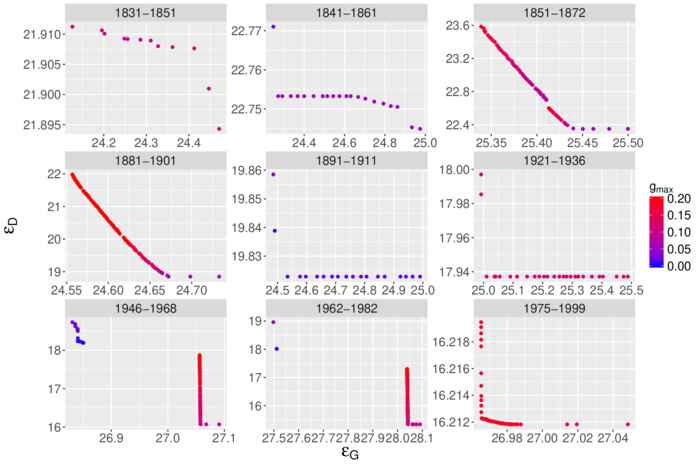}
	\caption[Pareto fronts for the calibration on population and distance]{\textbf{Pareto fronts for the bi-objective calibration between population and distance.} Fronts are given for each calibration period and are colored according to $g_{max}$.\label{fig:macrocoevol:pareto}}
\end{figure}

\begin{figure}
	\includegraphics[width=\linewidth]{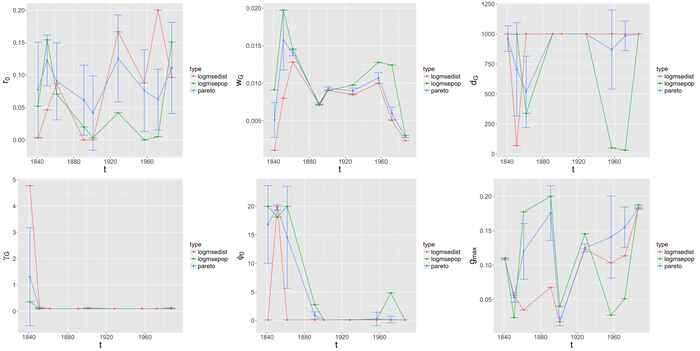}
	\caption[Evolution of calibrated parameters]{\textbf{Temporal evolution of optimal parameters.} From left to right and top to bottom, values of parameters $(r_0,w_G,d_G,\gamma_G,\phi_0,g_{max})$, respectively for the full Pareto front (blue), for the optimal point in the sense of the distance (red) and the optimal point in the sense of the population (green). \label{fig:macrocoevol:parameters}}
\end{figure}

\bpar{
We proceed to a non-stationary calibration, on the $(\varepsilon_P,\varepsilon_D)$ objectives, i.e. the squared-error on populations and on distances. The estimation is done with a moving window with the periods already used in~\ref{sec:interactiongibrat}. In order to have a limited dimension to explore, we take a fixed $w_N = 0$ to study the interactions only at the first order, knowing that the abstract network parameters $(g_{max},\gamma_S,\varphi_0)$ are taken into account in the calibration. The calibration is done with a genetic algorithm in a way similar as in~\ref{sec:interactiongibrat}. The Fig.~\ref{fig:macrocoevol:pareto} shows the obtained Pareto fronts, and the Fig.~\ref{fig:macrocoevol:parameters} the evolution in time of parameter values for the optimal solutions.
}{
Nous procédons à une calibration non-stationnaire, sur les objectifs $(\varepsilon_P,\varepsilon_D)$, c'est-à-dire l'erreur carrée sur les population et celle sur les distances. L'estimation est faite par fenêtre mobile sur les périodes déjà utilisées en~\ref{sec:interactiongibrat}. Pour limiter la dimension à explorer, nous fixons $w_N = 0$ pour n'étudier que les interactions au premier ordre, sachant que les paramètres de réseau abstrait $(g_{max},\gamma_S,\varphi_0)$ sont pris en compte dans la calibration. La calibration est effectuée par algorithme génétique de façon similaire à~\ref{sec:interactiongibrat}. La Fig.~\ref{fig:macrocoevol:pareto} montre les fronts de Pareto obtenus, et la Fig.~\ref{fig:macrocoevol:parameters} l'évolution dans le temps des valeurs des paramètres pour les solution optimales.
}

\bpar{
We observe a large variability of the shape of Pareto fronts for the bi-objective calibration on population and distance, what witnesses more or less difficulty to simultaneously adjust population and distance. Some periods, such as 1891-1911 and 1921-1936, are close to have a simultaneous objective point for the two objectives, what would correspond to a good correspondence of the model to both trajectories of cities and trajectory of the network on these periods.  
}{
Nous observons une forte variabilité des formes des fronts de Pareto pour la calibration bi-objectif population et distance, ce qui témoigne d'une plus ou moins grande difficulté à ajuster simultanément population et distance. Certaines périodes, comme 1891-1911 et 1921-1936, sont proches de présenter un point optimal simultané pour les deux objectifs, ce qui correspondrait à une bonne adéquation du modèle à la fois aux trajectoires de villes et à celle du réseau sur ces périodes.
}

\bpar{
In comparison with calibration results of the model with static network of~\ref{sec:interactiongibrat}, when comparing the performances for the objective $\varepsilon_G$, we find periods where the static is clearly better (1831 and 1841 for example) and others where the co-evolutive model is better (1946 and 1962): thus, taking into account the co-evolution helps in some cases to have a better reproduction of population trajectories.
}{
En comparaison aux résultats de calibration sur le modèle avec réseau statique de~\ref{sec:interactiongibrat}, en regardant les performances pour l'objectif $\varepsilon_G$, nous trouvons des périodes où le statique est clairement meilleur (1831 et 1841 par exemple) et d'autres où le modèle co-évolutif est meilleur (1946 et 1962) : ainsi, la prise en compte de la co-évolution permet dans certains cas une meilleure reproduction des trajectoires de population.
}

\bpar{
The values of optimal parameters in time, shown in Fig.~\ref{fig:macrocoevol:parameters}, seem to contain some signal. The evolution of $w_G$ and $\gamma_G$ are coherent with the evolutions observed for the static model. For $d_G$, the model principally saturates on the maximal distance and the evolution is difficult to interpret. 
}{
Les valeurs des paramètres optimaux dans le temps, présentées en Fig.~\ref{fig:macrocoevol:parameters}, contiennent un certain signal. L'évolution de $w_G$ et $\gamma_G$ sont consistantes avec celles observées pour le modèle statique. Pour $d_G$, le modèle sature principalement sur une distance maximale et l'évolution est difficile à interpréter. 
}

\bpar{
However, the evolution of $\phi_0$ could be a sign of a ``TGV effect'' in recent periods, through the secondary peak for population after 1960. Indeed, the construction of high speed lines has shortened distances between cities on top of the hierarchy, and an increase of the threshold $\phi_0$ corresponds to an increase of the selectivity for a potential diminution of distances.
}{
Par contre, l'évolution de $\phi_0$ pourrait témoigner d'un ``effet TGV'' dans les instants récents, par le pic secondaire pour la population après 1960. En effet, la construction des LGV a raccourci les distances entre les villes au plus haut de la hiérarchie, et une augmentation du seuil $\phi_0$ correspond à une augmentation de la sélectivité pour une diminution potentielle des distances.
}

\bpar{
The calibrated $g_{max}$ can finally be interpreted according to the history of the railway network (at least of all points in the Pareto front): a significant secondary peak in the first years, a minimum in the years corresponding to the stabilization of the network (1900), and an increase until today linked to the increase of train speeds and the opening of high speed lines. 
}{
Le $g_{max}$ calibré peut enfin être interprété selon l'histoire du réseau ferré (du moins pour l'ensemble des points du font de Pareto) : un pic significatif dans les premières années, un creux dans les années de stabilisation du réseau (1900), puis une augmentation jusqu'à aujourd'hui liée à l'augmentation de la vitesse des trains puis l'ouverture des lignes à grande vitesse.
}

\bpar{
We have this way in a certain extent indirectly quantify interaction processes through the network and the processes of network adaptation to flows, in the case of a real system.
}{
Nous avons pu ainsi dans une certaine mesure indirectement quantifier les processus d'interaction par le réseau et ceux d'adaptation du réseau au flux, dans le cas d'un système réel.
}

\subsubsection{Model with a physical network}{Modèle avec réseau physique}

\bpar{
We now sketch the outline of a specification of the model with a physical network, what would in a sense correspond to an hybrid model combining different scales. The objective of such a specification would be on the one hand to study the difference in trajectories compared to the abstract network, i.e. to quantify the importance of economies of scale (due to common links), of congestion and also the possible compromises to take in order to spatialize the network. On the other hand, it would help to understand to what extent it is possible to produce realistic networks in comparison to autonomous network growth models for example. These issues are tackled at an other scale and for other ontological specifications in chapter~\ref{ch:mesocoevolution}.
}{
Nous esquissons à présent les contours d'une spécification du modèle avec réseau physique, qui correspondrait en un sens à un modèle hybride combinant plusieurs échelles. L'idée d'une telle spécification serait d'une part d'étudier l'écart de trajectoire par rapport au réseau abstrait, c'est-à-dire quantifier l'importance des économies d'échelles (liées aux tronçons communs) et de la congestion, ainsi que les possibles compromis à effectuer liés à la spatialisation du réseau, et d'autre part d'étudier dans quelle mesure il est possible de reproduire des réseaux réalistes en comparaison à des modèles autonomes de croissance de réseau par exemple. Ces questions sont traitées à une autre échelle et pour d'autres spécifications ontologiques au chapitre~\ref{ch:mesocoevolution}. 
}

\bpar{
Such a specification follows the frame of \cite{li2014modeling}, which model the co-evolution between transportation corridors and the growth of main poles at a regional scale.
}{
Une telle spécification rejoint la logique de \cite{li2014modeling}, qui modélisent la co-évolution des couloirs de transport et de la croissance des principaux pôles à une échelle régionale.
}

\bpar{
The physical network we implement aims at satisfying a greedy criteria of local time gain. More precisely, we assume a self-reinforcement similar to~\cite{tero2010rules} A specification analog to the one used before assumes a growth for each link, given also in a logic of self-reinforcement by:
}{
Le réseau physique que nous implémentons cherche à satisfaire un critère de gain de temps local. Plus précisément, on suppose un auto-renforcement à la manière de~\cite{tero2010rules}. Une specification analogue à celle utilisée précédemment suppose une croissance pour chaque lien, donnée également dans une logique d'auto-renforcement par :
}

\[
d(t+1) = d(t)\cdot \left(1 + g_{max} \cdot \left[\frac{\phi}{\max \phi}\right]^{\gamma_s}\right)
\]

\bpar{
if $\phi$ is the flow in the link and $d(t)$ its effective distance. The threshold specification used before does indeed not allow a good convergence in time, in particular with the emergence of local oscillation phenomena.
}{
si $\phi$ est le flux dans le lien et $d(t)$ sa distance effective. La specification par seuil utilisée précédemment ne permet en effet pas une bonne convergence dans le temps, notamment par l'émergence de phénomènes d'oscillation locale.
}

\begin{figure}
	\includegraphics[width=\linewidth]{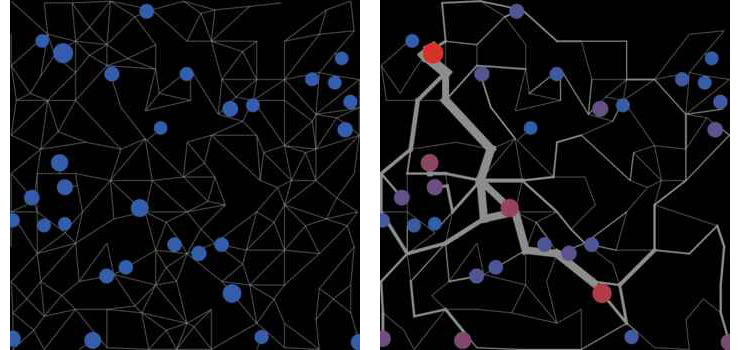}
	\caption[Example of application of the macroscopic model with a self-reinforcing network]{\textbf{Example of configuration obtained with a self-reinforcing network.} \textit{(Left)} Inital random configuration, with uniform impedances; \textit{(Right)} Final configuration obtained after 100 iterations.\label{fig:macrocoevolution:slimemould}}
\end{figure}

\bpar{
We generate a random initial network, by perturbing the position of vertices of a grid for which a fixed proportion of links has been removed (40\%) and by linking cities to the network through the shortest path. Links have all the same impedance, which then evolves according to the equation above. An example of a configuration obtained with this specification is given in Fig.~\ref{fig:macrocoevolution:slimemould}. The good convergence properties (visual stabilization of network structure during restricted experiments) suggest the potentialities offered by this specification, which systematic exploration is out of the scope of this work.
}{
Nous générons un réseau initial aléatoire, en perturbant la position des sommets d'une grille dont une proportion fixée de liens a été supprimée (40\%) et en y reliant les villes au plus court. Les liens ont tous même impédance, puis celle-ci évolue selon l'équation ci-dessus. Un exemple de configuration obtenue par cette spécification est donné en Fig.~\ref{fig:macrocoevolution:slimemould}. Les bonnes propriétés de convergence (stabilisation visuelle de la structure du réseau lors d'expériences restreintes) suggèrent les potentialités offertes par cette spécification, dont l'exploration systématique est hors de notre portée ici.
}

\subsubsection{Perspectives}{Perspectives}

\paragraph{Particular trajectories}{Trajectoires particulières}

\bpar{
The study of particular trajectories within a system of cities can allow to answer to specific thematic questions: for example, the influence of medium-sized cities on the global trajectory of the system, or the drivers of a more or less ``successful'' trajectory for this type of profile. In the case of the application to a real system, the mapping of deviation to the model in time can suggest regional particularities.
}{
L'étude de trajectoires particulières au sein du système de villes peut permettre de répondre à des questions thématiques spécifiques : par exemple, l'influence des villes moyennes sur la trajectoire globale du système, ou les déterminants d'une plus ou moins bonne ``réussite'' pour ce type de profil. Dans le cas de l'application à un système réel, la cartographie des déviations au modèle dans le temps peut suggérer des particularités régionales.
}

\paragraph{Comparison of urban systems}{Comparaison de systèmes urbains}

\bpar{
We also finally expect to be able through the model to compare urban systems in different geographical and political contexts, and at different scales. This should foster the understanding the implications of planning actions on the interactions between networks and territories. For example, French railway network has emerged through multiple operators, on the contrary to the Chinese high speed railway network, for which a more precise development could be considered.
}{
Nous nous attendons finalement également à pouvoir par l'intermédiaire de ce modèle comparer des systèmes urbains dans des contextes géographiques et politiques différents, ainsi qu'à différentes échelles. Cela devrait permettre de révéler les implications des actions de planification sur les interactions entre réseaux et territoires. Par exemple, le réseau ferré français a émergé par l'intermédiaire de multiples opérateurs, au contraire du réseau ferré à grande vitesse Chinois, pour lequel un développement précis pourrait être envisagé.
}

\stars

%


\newpage

\section*{Chapter Conclusion}{Conclusion du Chapitre}

\bpar{
This macroscopic entry into co-evolution processes aimed at understanding them (i) within a system of cities, i.e. in an aggregated way and at an abstract level; and (ii) on a long time scale, of the order of a century. The processes we considered are: growth of city as a consequence of interactions which depend on the network; effect of flows at the second order on these growths (that we did not explore here); effect of feedback of flows on distances in the network in a thresholded way (the latest being refined with an effect of network topology in the case of SimpopNet).
}{
Cette entrée macroscopique dans les processus de co-évolution visait à les comprendre (i) au sein d'un système de villes, c'est-à-dire de manière agrégée et à un niveau abstrait ; et (ii) sur une échelle temporelle longue, de l'ordre de la centaine d'années. Les processus considérés sont : croissance des villes entrainée par les interactions qui dépendent du réseau ; effet de flux au second ordre sur ces croissances (que nous n'avons pas exploré ici) ; effet de retroaction des flux sur les distances dans le réseau de manière seuillée (ce dernier étant raffiné avec un effet de la topologie du réseau dans le cas de SimpopNet).
}

\bpar{
We first show, through a systematic exploration of the SimpopNet model, that it is highly sensitive to the spatial configuration, suggesting that potential conclusions on processes will always have to be contextualized. We also show that it difficultly produces a co-evolution in the sense of circular causalities between network and cities, and that the dominating process is more an adaptation of cities to the network.
}{
Nous démontrons dans un premier temps, par exploration systématique du modèle SimpopNet, que celui-ci est très sensible à la configuration spatiale, suggérant que les conclusions potentielles sur des processus devront toujours être contextualisées. Nous montrons également que celui-ci produit difficilement une co-évolution au sens de circularités causales entre réseau et villes, et que le processus dominant est plutôt une adaptation des villes au réseau.
}

\bpar{
Our model we then explore allows on the other hand, at the price of an abstraction of the network, to reveal in a synthetic way first an intermediate scale of maximal complexity suggesting the emergence of regional subsystems, allowed by intermediate values of the interaction distance and high values of the feedback threshold for the network; secondly the existence of at least three regimes of causality, among which at least two can be qualified as co-evolutive. The study of real data for the French system of cities indeed confirms the existence of the regional scale, and also a short stationarity time scale of around twenty years, but very few significant interactions at this scale, in contradiction with the existing literature. The calibration of the model on real data reproduces well the known patterns of railway network growth, and suggest more recently a ``TGV effect''.
}{
Notre modèle exploré par la suite permet quant à lui, au prix d'une abstraction du réseau, de révéler de manière synthétique d'une part une échelle intermédiaire de complexité maximale suggérant l'émergence de sous-systèmes régionaux, permis par des valeurs intermédiaire de la distance d'interaction et des valeurs fortes du seuil de rétroaction pour le réseau ; d'autre part l'existence d'au moins trois régimes de causalité, dont deux peuvent être qualifiés de co-évolutifs. L'étude des données réelles pour le système de ville français confirme bien l'existence de l'échelle régionale, ainsi que d'une échelle temporelle de stationnarité courte d'une vingtaine d'années, mais très peu d'interactions significatives à cette échelle, en contradiction avec la littérature existante. La calibration du modèle sur données réelles reproduit bien les motifs connus de croissance du réseau ferré, et suggèrent un ``effet TGV'' plus récemment.
}

\bpar{
We introduce a development with physical network, which allows to make the link with ontologies we will explore in the following in chapter~\ref{ch:mesocoevolution}: the co-evolution at a mesoscopic scale, by insisting on the role of form and function, and thus of precise mechanisms of network development.
}{
Nous introduisons un développement avec réseau physique, qui permet de faire le lien avec les ontologies que nous allons explorer par la suite en chapitre~\ref{ch:mesocoevolution} : la co-évolution à l'échelle mesoscopique, en appuyant sur le rôle de la forme et de la fonction, et donc des mécanismes précis de développement du réseau.
}

\stars





\bpar{
\chapter{Co-evolution at the mesoscopic scale}
}{
\chapter{Co-évolution à l'échelle mesoscopique}
}

\label{ch:mesocoevolution} 


\bpar{
Processes underlying co-evolution are not exactly similar when switching from the macroscopic scale to the mesoscopic scale, as suggest our different empirical analysis: for exemple, causality regimes obtained at a small scale for South Africa in~\ref{sec:causalityregimes} are clearer than the ones for real estate transactions and the Grand Paris in~\ref{sec:casestudies}. At the metropolitan scale, relocation processes are crucial to explain the evolution of the urban form, and these can partly be attributed to accessibility differentials, knowing that the evolution of networks answers on the other hand to complex logics conditioned by territorial distributions. Centrality, density, accessibility, as much properties potentially implied in co-evolutive processes, and that are proper to the concept of urban form.
}{
Les processus sous-jacents à la co-évolution ne sont pas exactement similaires en passant de l'échelle macroscopique à l'échelle mesoscopique, comme le suggèrent nos différentes analyses empiriques : par exemple, les régimes de causalités obtenus à petite échelle pour l'Afrique du Sud en~\ref{sec:causalityregimes} sont plus clairs que ceux pour les transactions immobilières et le Grand Paris en~\ref{sec:casestudies}. À l'échelle métropolitaine, les processus de relocalisation sont essentiels pour expliquer l'évolution de la forme urbaine, et ceux-ci peuvent partiellement être attribués aux différentiels d'accessibilité, sachant que l'évolution des réseaux répond quant à elle à des logiques complexes conditionnées par les distributions territoriales. Centralité, densité, accessibilité, autant de propriétés potentiellement impliquées dans les processus co-évolutifs, et propres au concept de forme urbaine.
}

\bpar{
We make the choice to insist on the role of the urban form at the mesoscopic scale, and use urban morphogenesis as a modeling paradigm for co-evolution: the strong coupling of the urban form with the network through co-evolution allows to consider urban functions more explicitely. This chapter follows the chapter~\ref{ch:morphogenesis}, and extends the model that have been developed in it.
}{
Nous faisons le choix d'appuyer le rôle de la forme urbaine à l'échelle mesoscopique, et utilisons la morphogenèse urbaine comme paradigme de modélisation de la co-évolution : le couplage fort de la forme urbaine avec le réseau par la co-évolution permet de considérer les fonctions urbaines plus explicitement. Ce chapitre fait suite au chapitre~\ref{ch:morphogenesis}, et étend les modèles qui y ont été développés.
}


\bpar{
Different network generation heuristics are compared in a first section~\ref{sec:networkgrowth}, still in a weak coupling paradigm, in order to establish the topologies produced by different rules.
}{
Différentes heuristiques de génération de réseau sont comparées dans une première section~\ref{sec:networkgrowth}, toujours dans un paradigme de couplage simple, afin d'établir les topologies produites par différentes règles.
}

\bpar{
This step allows to introduce a co-evolution model through morphogenesis in~\ref{sec:mesocoevolmodel}, which is calibrated on coupled objectives of urban morphology and network topology.
}{
Cette étape permet d'introduire un modèle de co-évolution par morphogenèse en~\ref{sec:mesocoevolmodel}, qui est calibré sur les objectifs couplés de morphologie urbaine et de topologie de réseau.
}

\bpar{
Finally, we describe in~\ref{sec:lutecia} a model allowing the exploration of complex processes for network growth, in particular endogenous governance processes implying deciding agents at the metropolitan scale.
}{
Enfin, nous décrivons en~\ref{sec:lutecia} un modèle permettant l'exploration de processus complexes pour la croissance du réseau, notamment des processus endogènes de gouvernance impliquant des agents décideurs à l'échelle métropolitaine.
}

\vspace{-0.5cm}

\stars

\vspace{-0.5cm}

\bpar{
\textit{The results of the two first sections of this chapter have been presented at CCS 2017 as~\cite{raimbault:halshs-01590624}, and will be published in a synthetic way as a book chapter~\cite{raimbault2018urban}; the structure of the model and preliminary results for the third section have been presented at ECTQG 2015 as \cite{le2015modeling}.}
}{
\textit{Les résultats des deux premières sections de ce chapitre ont été présentés à CCS 2017 comme~\cite{raimbault:halshs-01590624}, et paraitront prochainement de façon synthétique comme chapitre d'ouvrage~\cite{raimbault2018urban} ; la structure du modèle et des résultats préliminaires pour la troisième section ont été présentés à ECTQG 2015 comme \cite{le2015modeling}.}
}


%


\newpage

\section{Network growth models}{Modèles de croissance de réseau}

\label{sec:networkgrowth}


\bpar{
We propose first to study with more details processes of network growth for the mesoscopic sclae. The idea is to understand intrinsic properties of different network growth heuristics. This exercise is interesting in itself since there is to the best of our knowledge no systematic comparison of spatial networks morphogenesis models: \cite{xie2009modeling} propose for example a review from the point of view of network economics, it does not include on the one hand some disciplines (see chapter~\ref{ch:modelinginteractions}), and on the other hand does not compare performances of models on dedicated comparable implementations.
}{
Nous proposons dans un premier temps d'étudier en détails les processus de croissance de réseau pour l'échelle mesoscopique. L'idée est de comprendre les propriétés intrinsèques des différentes heuristiques de croissance de réseau. Cet exercice est intéressant en lui-meme puisqu'il n'existe pas à notre connaissance de comparaison systématique de modèles de morphogenèse des réseaux spatiaux : si~\cite{xie2009modeling} proposent par exemple une revue du point de vue de l'économie des réseaux, celle-ci ne prend pas en compte certaines disciplines d'une part (voir chapitre~\ref{ch:modelinginteractions}), et ne compare d'autre part pas les performances des modèles sur des implémentations dédiées comparables. 
}


\subsection{Benchmarking network growth heuristics}{Comparer les heuristiques de croissance de réseau}

\bpar{
Considering network growth in itself, several heuristics exist in order to generate a network under some constraints. As already developed especially in~\ref{sec:modelingsa}, from economic network growth approaches to local optimization heuristics, geographical mechanisms or biological network growth, each has its own advantages and particularities. We already tested in~\ref{sec:correlatedsyntheticdata} an heuristic based on interaction potential breakdown. In order to be able to compare different network growth heuristics ``everything else being equal'', it is necessary to explore them at fixed density, although the thematic meaning of results will not have any value, neither on long times nor for co-evolution.
}{
Pour la croissance du réseau en tant que telle, de nombreuses heuristiques existent pour générer un réseau sous certaines contraintes. Comme déjà développé précédemment notamment en~\ref{sec:modelingsa}, des modèles économiques de croissance de réseau aux heuristiques d'optimisation locale, aux mécanismes géographiques ou à la croissance de réseau biologique, chacun a ses avantages et particularités propres. Nous avons déjà testé en~\ref{sec:correlatedsyntheticdata} une heuristique basée sur la rupture de potentiel d'interaction. Pour pouvoir comparer ``toutes choses égales par ailleurs'' les différentes heuristiques de génération de réseau, il est nécessaire de les explorer à densité fixée, même si le sens thématique des résultats ne peut avoir de valeur ni sur le temps long, ni pour la co-évolution.
}

\bpar{
The importance of heuristics capturing a topological structure allowing a certain compromise between performance, congestion and cost, is shown by empirical analyses such as \cite{2012arXiv1202.1747W} for metropolitan networks, which shows that patterns of evolution for correlations between degrees witness an evolution of networks towards such a topology.
}{
L'importance d'heuristiques pouvant capturer une structure topologique permettant un certain compromis entre performance, congestion et coût, est montrée par des analyses empiriques comme \cite{2012arXiv1202.1747W} pour les réseaux de métro, qui montre que les motifs d'évolution des corrélations entre degrés témoignent d'une évolution des réseaux vers une telle topologie.
}

\bpar{
We precise in the following the core of the network growth model together with several heuristics from diverse origins, compared in similar conditions through their integration within the common basis.
}{
Nous précisons par la suite le coeur du modèle de croissance de réseau ainsi qu'un certain nombre d'heuristiques aux origines variées, comparées dans des conditions similaires par leur intégration à la base commune.
}

\subsubsection{Core of the network growth model}{Base du modèle de croissance de réseau}

\bpar{
A common process to the different heuristics constitutes the core of the network growth model, and bridges population density distribution with the network. In concrete terms, the aim is to attribute new centers according to this density, and we make the choice of specifying this process exogenously to network growth itself\footnote{This intermediate stage is close in our case to the idea of procedural modeling, since the implemented rule aims at reproducing a shape without needing the actual processes. This raises the issue of equifinality and of the potential existence of equivalent models for this submodel or for the full model capturing a real process corresponding to it. The use of multi-modeling also at this stage could be a solution, but frameworks allowing to tackle an arbitrary number of stationarity levels or even allowing the model to be autonomous on these choices do not exist yet.}.
}{
Un processus commun aux différentes heuristiques constitue le coeur du modele de croissance de réseau, et fait le pont entre la distribution de densité de population et le réseau. Concrètement, il s'agit d'attribuer des nouveaux centres en fonction de cette densité, et nous faisons le choix de specifier ce processus de manière exogène à la croissance de réseau elle-même\footnote{Cette étape intermédiaire se rapproche dans notre cas d'un esprit de modélisation procédurale, puisque la règle implémentée cherche à reproduire une forme sans besoin des processus réels. Cela pose la question de l'équifinalité et de l'existence potentielle de modèles equivalents pour ce sous-modèle ou pour le modèle complet capturant un processus réel correspondant à celle-ci. L'utilisation de multi-modélisation également sur cette étape pourrait être une solution mais les cadres permettant de s'extraire d'un nombre arbitraire de niveaux de stationnarité ou même permettant une autonomie du modèle sur ces choix n'existent pas encore.}.
}

\bpar{
We recall the context used in~\ref{sec:correlatedsyntheticdata}, i.e. a grid of cells characterized by their population $P_i$, on which a network composed of nodes and links develops. The population distribution will here be fixed in time $P_i(t) = P_i(0)$, and the network evolves sequentially starting from an initial network.
}{
Reprenons le contexte utilisé en~\ref{sec:correlatedsyntheticdata}, c'est-à-dire une grille de cellules caractérisées par leur population $P_i$, sur laquelle un réseau composé de noeuds et de liens se développe. La distribution de la population sera ici fixe dans le temps $P_i(t) = P_i(0)$, et le réseau évolue séquentiellement à partir d'un réseau initial.
}

\bpar{
A step of network growth is realized at fixed time intervals $t_N$ (parameter which allows to adjust the respective evolution speeds for population and for the network). It corresponds to the following stages, of which the firt two refine the logic of~\cite{raimbault2014hybrid} (which stipulates that population centers must be connected to the existing network in a basic way).
}{
Une étape de croissance de réseau est réalisée à intervalles de temps $t_N$ (paramètre permettant d'ajuster les vitesses respectives d'évolution pour la population et pour le réseau). Elle correspond aux étapes suivantes, dont les deux premières raffinent la logique de~\cite{raimbault2014hybrid} (qui stipule que des centres de peuplement doivent être connectés au réseau existant de manière basique).
}

\bpar{
\begin{enumerate}
	\item A fixed number $n_N$ of new nodes is added. Sequentially, the probability to receive a new node is given by
\[
p_i = \frac{P_i}{P_{max}} \cdot \frac{\delta_M - \delta_i}{\delta_M}
\]
what means that an elementary node corresponds to the conjunction of events: (i) high density $P_i$ of population in cell compared to the maximal population for each cell $P_{max}$, (ii) density of nodes $\delta_i$ in a radius $r_n$ low compared to a maximal density $\delta_M$. Population of nodes is reattributed at each stage through triangulation the same way as in~\ref{sec:correlatedsyntheticdata}.
	\item New nodes are then connected by a new link, following the shortest path to the network (perpendicular connexion or towards the closest node).
	\item New links are added, until they reach a maximal number of added links $l_{m}$, following an heuristic that varies among: no heuristic (no supplementary links added), random, deterministic potential breakdown (see~\ref{sec:correlatedsyntheticdata}), random potential breakdown \cite{schmitt2014modelisation}, cost-benefits \cite{louf2013emergence}, biological network generation (heuristic based on \cite{tero2010rules}).
\end{enumerate}
}{
\begin{enumerate}
	\item Un nombre fixe $n_N$ de nouveaux noeuds est ajouté. Séquentiellement, la probabilité de recevoir un nouveau noeud est donnée par
\[
p_i = \frac{P_i}{P_{max}} \cdot \frac{\delta_M - \delta_i}{\delta_M}
\]
c'est-à-dire qu'un noeud élémentaire correspond à la conjonction des évènements : (i) densité élevée de population de la cellule $P_i$ par rapport à la population maximale par cellule $P_{max}$, (ii) densité de noeuds $\delta_i$ dans un rayon $r_n$ faible par rapport à une densité maximale de noeuds $\delta_M$. La population des noeuds est re-attribuée à chaque étape par triangulation comme en~\ref{sec:correlatedsyntheticdata}.
	\item Les nouveaux noeuds sont alors connectés par un nouveau lien, suivant le plus court chemin vers le réseau (raccord perpendiculaire ou avec le sommet le plus proche).
	\item Des nouveaux liens sont ajoutés, jusqu'à atteindre un nombre maximal de liens ajoutés $l_{m}$, suivant une heuristique pouvant varier parmi :  aucune (pas d'ajouts de liens), aléatoire, rupture de potentiel déterministe (voir~\ref{sec:correlatedsyntheticdata}), rupture de potentiel aléatoire \cite{schmitt2014modelisation}, coût-bénéfices \cite{louf2013emergence}, génération de réseau biologique (heuristique basée sur \cite{tero2010rules}).
\end{enumerate}
}


\bpar{
We fix to simplify the parameters $r_n = 5$, $\delta_M = 10$ and $n_N=20$, and the parameters $t_N$ and $l_m$ will be variable.
}{
Nous fixons pour simplifier $r_n = 5$, $\delta_M = 10$ et $n_N=20$, et les paramètres $t_N$ et $l_m$ seront variables.
}

\subsubsection{Baseline heuristics}{Heuristiques de référence}

\bpar{
We consider two baseline heuristics to better situate the ones we will explore in the following: the one composed uniquely by the base described previously, which produces tree networks; and random network generation, which consists in creating a fixed number $l_m$ of new links between randomly chosen nodes, and to make the final network planar\footnote{The algorithm to obtain a planar network consists in the creation of nodes at the possible intersections of new links (``flattening'' of the network).}.
}{
Nous considérons deux heuristiques de référence pour mieux situer celles explorées par la suite : celle composée uniquement de la base décrite précédemment, qui produit des réseaux arborescents ; et la generation de réseau aléatoire, qui consiste à créer un nombre fixe $l_m$ de nouveaux liens entre des sommets choisis aléatoirement, puis à planariser le réseau final\footnote{L'algorithme de planarisation consiste en la création de noeuds aux intersections éventuelles de nouveaux liens (``aplatissement'' du réseau).}.
}


\subsubsection{Euclidian heuristic}{Heuristique euclidienne}


\bpar{
This heuristic, which rationale relies on ideas of gravity potential breakdown, corresponds to the method developed in~\ref{sec:correlatedsyntheticdata}. It is a method close to the one introduced by~\cite{schmitt2014modelisation}, without the stochastic aspect and prone to miss path-dependency phenomena, but more refined in the mechanisms of gravity potentials.
}{
Cette heuristique, dont la logique repose sur des idées de rupture de potentiel gravitaire, correspond à la méthode développée en~\ref{sec:correlatedsyntheticdata}. Il s'agit d'une méthode proche de celle introduite par~\cite{schmitt2014modelisation}, sans l'aspect stochastique et pouvant passer à côté de phénomènes de dépendance au chemin, mais plus raffinée dans les mécanismes de potentiels gravitaires.
}

\subsubsection{Random potential breakdown}{Rupture de potentiel aléatoire}

\bpar{
Random potential breakdown is the heuristic used in the SimpopNet model \cite{schmitt2014modelisation}, which is inspired by the model introduced by \cite{blumenfeld2010network}. At each step, two cities are randomly drawn, the first following a probability proportional to $P_i^{\gamma_R}$ and the second following $V_{i_0j}^{\gamma_R}$ such that $i_0$ is the first city drawn and $V_{ij}$ are euclidian gravity potentials. If
\[
d_N(i_0,j_0) / d(i_0,j_0) > \theta_R
\]
i.e if the relative detour through the network is larger than a threshold parameter, a link is created between the two cities\footnote{To remain comparable to the other heuristics that do not include speeds in links, newly created links are of speed 1 and not $v_0$ as in the implementation of~\ref{sec:macrocoevolexplo}.}. At each time step, $l_m$ new links are created following this process. The final network is made planar.
}{
La rupture de potentiel aléatoire est celle utilisée par SimpopNet \cite{schmitt2014modelisation}, qui reprend le modèle introduit par \cite{blumenfeld2010network}. À chaque étape, deux villes sont tirées aléatoirement, la premiere selon une probabilité proportionnelle à $P_i^{\gamma_R}$ et la deuxième selon $V_{i_0j}^{\gamma_R}$ sachant que $i_0$ est la première ville tirée et $V_{ij}$ sont les potentiels gravitaires euclidiens. Si
\[
d_N(i_0,j_0) / d(i_0,j_0) > \theta_R
\]
c'est-à-dire si le détour relatif par le réseau est supérieur à un paramètre de seuil, un lien est créé entre les deux villes\footnote{Pour rester comparable aux autres heuristiques qui n'incluent pas de vitesse des liens, les nouveaux liens sont de vitesse 1 et non $v_0$ comme dans l'implémentation de~\ref{sec:macrocoevolexplo}.}. À chaque pas de temps, $l_m$ nouveaux liens sont créés selon ce processus. Le réseau final est planarisé.
}

\subsubsection{Biological heuristic}{Heuristique biologique}

\bpar{
\cite{raimbault2015labex} explores applications of biological network growth models (\emph{slime mould}), in particular their ability to produce from the bottom-up optimal solutions in the Pareto sense for contradictory objectives, such as cost and robustness. The considered model comes from~\cite{tero2010rules}.
}{
\cite{raimbault2015labex} explore des applications des modèles de croissance de réseau biologique (\emph{slime mould}), notamment leur capacité à produire de manière émergente des solutions optimales au sens de Pareto pour des indicateurs contradictoires, comme le coût et la robustesse. Le modèle considéré est issu de~\cite{tero2010rules}.
}

\bpar{
The advantage of such an heuristic is confirmed in some cases by the reality of multi-objective optimizations: \cite{padeiro:tel-00438092} (p.~72) illustrates in particular the extension of the Parisian metro in Bobigny in the seventies, and the consideration of indicators for cost, served population, expected rush hour traffic, and average travel time.
}{
L'intérêt d'une telle heuristique est confirmé dans certains cas par la réalité des optimisations multi-objectifs : \cite{padeiro:tel-00438092} (p.~72) illustre entre autres le prolongement du métro Parisien à Bobigny dans les années 1970, et la prise en compte des indicateurs de coût, de population desservie, de traffic attendu en heure de pointe, et de temps de trajet moyen.
}

\bpar{
The \emph{slime mould} model works the following way. Given an initial network with links of uniform capacities, a fluid is distributed in the network from a source to a sink, establishing a flow in each link. An equilibrium of fluid pressures at network nodes can be found, which corresponds to the stationary state for flows\footnote{More precisely, the problem is equivalent to an electrostatic linear equations system that we just have to solve.}. Given an equilibrium for pressures, capacities of links evolve according to the traversing flow. An iteration of equilibria and of tubes evolution allows then a convergence towards a stable hierarchical distribution of capacities. The detail of the procedure is described in Appendix~\ref{app:sec:networkgrowth}, following the mathematical details developed by \cite{tero2007mathematical}.
}{
Le modèle de \emph{slime mould} fonctionne de la façon suivante. Étant donné un réseau initial dont les liens ont des capacités uniformes, un fluide est distribué dans le réseau d'une source à un puit, établissant un flux dans chaque lien. Un équilibre des pressions du fluide aux noeuds du réseau peut être établi, qui correspond à l'état stationnaire pour les flux\footnote{Plus précisément, le problème est équivalent à un système d'équations linéaire électrostatiques qu'il suffit de résoudre.}. Étant donné un équilibre des pressions, les capacités des liens évoluent en fonction du flux traversant. Une itération des équilibres et de l'évolution des tubes permet alors une convergence vers une distribution hiérarchique stable des capacités. Le détail de la procédure est décrit en Annexe~\ref{app:sec:networkgrowth}, suivant les détails mathématiques développés par \cite{tero2007mathematical}.
}

\bpar{
Our logic if to use this mechanism to determine at a given time a given number of realized links. Advantages of the heuristic we are going to detail are especially that (i) it can be used in an iterative way to represent a sequential topological evolution of the network, in comparison to most investment models that evolve only capacities in time; and (ii) it translates processes of network self-organization, and moreover produces optimal networks in the Pareto sense for cost and robustness.
}{
Notre logique est d'utiliser ce mécanisme pour, à un instant donné, déterminer un certain nombre de liens réalisés. Les avantages de l'heuristique que nous allons détailler sont notamment que (i) elle peut être utilisée de manière itérative pour traduire une évolution topologique séquentielle du réseau, en comparaison à la plupart des modèles d'investissement qui font évoluer uniquement les capacités dans le temps ; et (ii) elle traduit des processus d'auto-organisation du réseau, et produit par ailleurs des réseaux optimaux au sens de Pareto pour le coût et la robustesse.
}

\bpar{
The application of the slime-mould model to network generation is done according to the following steps, within the global frame described previously.
}{
L'application du modèle de slime-mould à la generation de réseau s'effectue de la façon suivante, en s'insérant dans le cadre global décrit précédemment.
}

\bpar{
\begin{enumerate}
	\item Starting from the existing network to which we add a grid network (with diameters two times smaller to take into account the preponderance of the existing network) with diagonal connexions, and in which 20\% of links are randomly deleted to simulate perturbations linked to topology, we constitute the initial support in which slime-mould flows will be simulated.
	\item We proceed by iteration of successive generations, which consist in the following steps, for an increasing value of $k$ ($k\in \{ 1,2,4 \}$ in practice):
	\begin{itemize}
		\item given the distribution of population, the slime-mould model is iterated $k\cdot n_b$ times to obtain the emergent network through convergence of capacities;
		\item links with a capacity inferior to a threshold parameter $\theta_b$ are removed;
		\item the largest connected component is kept.
	\end{itemize}
	\item The final network is simplified\footnote{The simplification algorithm consists in the replacement of link sequences which extremities have all a degree of two, excepted the start and end nodes, by a unique link.} and made planar.
\end{enumerate}
}{
\begin{enumerate}
	\item À partir du réseau existant auquel on ajoute un réseau en grille (diamètres deux fois moindres pour prendre en compte la prépondérance du réseau existant) avec connexion diagonales, et dans lequel on supprime de manière aléatoire 20\% des liens pour simuler les perturbations liées à la topologie, nous constituons le support initial dans lequel les flux du slime-mould seront simulés.
	\item On procède par itération de générations successives, qui consistent pour $k$ croissant ($k\in \{ 1,2,4 \}$ en pratique) en les étapes suivantes :
	\begin{itemize}
		\item étant donné la distribution de la population, on itère $k\cdot n_b$ fois le modèle de slime mould pour obtenir le réseau emergent par convergence des capacités ;
		\item les liens de capacité inférieure à un paramètre de seuil $\theta_b$ sont supprimés ;
		\item la plus grande composante connexe est conservée.
	\end{itemize}
	\item Le réseau final est simplifié\footnote{L'algorithme de simplification consiste en un remplacement des séquences de liens dont les sommets hors extrémités ont tous degré 2 par un lien unique.} et planarisé.
\end{enumerate}
}

\begin{figure}
\includegraphics[width=\linewidth]{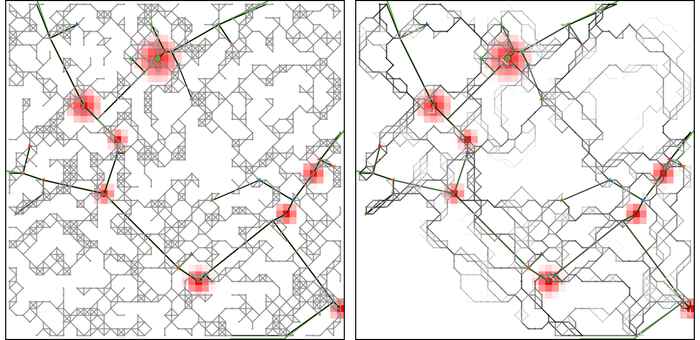}
\caption[Biological network generation example]{\textbf{Biological heuristic for network generation.} This visualization example illustrates the intermediate stages for the addition of links. \textit{(Left)} The initial semi-random network in which the slime-mould is launched; \textit{(Right)} same network after 80 iterations of the slime-mould, the thickness of links giving the capacity.\label{fig:networkgrowth:bioexample}}
\end{figure}

\bpar{
We illustrate in Fig.~\ref{fig:networkgrowth:bioexample} two stages of this generation process, showing the basis structure on which the self-reinforcement model is launched, and the convergence of link capacities after a certain number of steps.
}{
Nous illustrons en Fig.~\ref{fig:networkgrowth:bioexample} deux étapes de ce processus de génération, montrant la structure de base sur lequel le modèle d'auto-renforcement est lancé, et la convergence des capacité des liens après un certain nombre d'étapes.
}

\subsubsection{Cost-benefits evaluation}{Evaluation coûts-bénéfices}

\bpar{
The notion of cost is not explicitly included in all the growth heuristics presented up to here - it is implicitly in gravity potentials through the distance decay parameter, and also in the slime-mould since it generates networks exhibiting a compromise between cost and robustness. We therefore add a simple heuristic which is focused on the cost of network links during their extension. It is the heuristic studied by~\cite{louf2013emergence}, which relies on a rationale in transportation economics. Following a logic of cost-benefits analysis by network developments actors, links are sequentially realized for the couple of non-connected cities with a minimal cost, with a cost of the form $d_{ij} - \lambda / V_{ij}$, where the parameter $\lambda$ is the compromise between construction cost and gain in connected potential.
}{
La notion de coût n'est pas présente de manière explicite dans l'ensemble des heuristiques de croissance présentées jusqu'ici - elle l'est de manière implicite dans les potentiels de gravité par le paramètre d'attenuation de la distance, ainsi que dans le \emph{slime-mould} puisque celui-ci génère des réseau compromis entre robustesse et coût. Nous ajoutons donc une heuristique simple qui est centrée sur le coût des tronçons de réseau lors de leur extension. Il s'agit de celle étudiée par~\cite{louf2013emergence}, qui se base sur des arguments d'économie des transports. Suivant une logique d'analyse coûts-bénéfices par les acteurs du développement du réseau, les liens sont réalisés séquentiellement pour les couples de villes non connectées ayant un coût minimal, avec un coût de la forme $d_{ij} - \lambda / V_{ij}$, où le paramètre $\lambda$ est le compromis entre coût de construction et gain de potentiel connecté.
}


\subsubsection{Parameters}{Paramètres}

\bpar{
We summarize the parameters that will vary in the following in Table~\ref{tab:networkgrowth:parameters}. An additional ``parameter'', or more precisely a meta-parameter, is the choice of the heuristic to add links.
}{
Nous résumons les paramètres que nous ferons varier par la suite en Table~\ref{tab:networkgrowth:parameters}. Un ``paramètre'' supplémentaire, ou plutôt un méta-paramètre, est le choix de l'heuristique pour l'ajout des liens.
}

\begin{table}
\caption[Summary of network growth parameters]{\textbf{Summary of network growth parameters for all heuristics.} We also give the corresponding processes, typical variation ranges and their default values.\label{tab:networkgrowth:parameters}}
\bpar{
\begin{tabular}{|c|c|c|c|c|c|}
  \hline
Heuristic & Parameter & Name & Process & Domain & Default\\
  \hline
\multirow{5}{*}{Base}& $l_m$ & added links & growth & $[0;100]$ & $10$ \\\cline{2-6}
 & $d_G$ & gravity distance & potential & $]0;5000]$ & $500$ \\\cline{2-6}
 & $d_0$ & gravity shape & potential & $]0;10]$ & $2$ \\\cline{2-6}
 & $k_h$ & gravity weight & potential & $[0;1]$ & $0.5$ \\\cline{2-6}
 & $\gamma_G$ & gravity hierarchy & potential & $[0.1;4]$ & $1.5$ \\\hline
\multirow{2}{*}{Random breakdown}& $\gamma_R$ & random selection hierarchy & hierarchy & $[0.1;4]$ & $1.5$ \\\cline{2-6}
& $\theta_R$ & random threshold & breakdown & $[1;5]$ & $2$ \\\hline
Cost-benefits & $\lambda$ & compromise & compromise & $[0;0.1]$ & $0.05$ \\\hline
\multirow{2}{*}{Biological}& $n_b$ & iterations & convergence & $[40;100]$ & $50$ \\\cline{2-6}
& $\theta_b$ & biological threshold & threshold & $[0.1;1.0]$ & $0.5$ \\\hline
\end{tabular}
}{
\begin{tabular}{|c|c|c|c|c|c|}
  \hline
Heuristique & Paramètre & Nom & Processus & Domaine & Défaut\\
  \hline
\multirow{5}{*}{Base}& $l_m$ & liens ajoutés & croissance & $[0;100]$ & $10$ \\\cline{2-6}
 & $d_G$ & distance gravitaire & potentiel & $]0;5000]$ & $500$ \\\cline{2-6}
 & $d_0$ & forme gravitaire & potentiel & $]0;10]$ & $2$ \\\cline{2-6}
 & $k_h$ & poids gravitaire & potentiel & $[0;1]$ & $0.5$ \\\cline{2-6}
 & $\gamma_G$ & hiérarchie gravitaire & potentiel & $[0.1;4]$ & $1.5$ \\\hline
\multirow{2}{*}{Rupture aléatoire}& $\gamma_R$ & hiérarchie aléatoire & hiérarchie & $[0.1;4]$ & $1.5$ \\\cline{2-6}
& $\theta_R$ & seuil aléatoire & rupture & $[1;5]$ & $2$ \\\hline
Coût-Bénéfices& $\lambda$ & compromis & compromis & $[0;0.1]$ & $0.05$ \\\hline
\multirow{2}{*}{Biologique}& $n_b$ & itérations & convergence & $[40;100]$ & $50$ \\\cline{2-6}
& $\theta_b$ & seuil biologique & seuil & $[0.1;1.0]$ & $0.5$ \\\hline
\end{tabular}
}
\end{table}

\subsection{Results}{Résultats}

\subsubsection{Model setup}{Initialisation du modèle}

\bpar{
The model is initialized on synthetic or semi-synthetic configurations, with a grid of size $N=50$, with the following steps.
\begin{enumerate}
	\item Population density is initialized either with an exponential mixture, which centers (network nodes) follow the configuration of a synthetic city system as done in~\ref{sec:macrocoevolexplo}; or from a real configuration extracted from the density raster for France. We will use the second option here in systematic explorations.
	\item In the second case, a fixed number of network nodes are generated and located following a preferential attachment to density (see~\ref{sec:correlatedsyntheticdata})\footnote{To avoid bord effects of a network with no connection to the exterior, we add a fixed number $n_e$ of nodes (that we take as $n_e = 6$) at random locations on the border of the world.}. We do not initialize on real networks, since these will be the calibration target, but impose an initial synthetic skeleton that can be interpreted as an archaic network.
	\item An initial network is generated by connecting the nodes as detailed in~\ref{sec:correlatedsyntheticdata}.
\end{enumerate}
}{
Le modèle est initialisé sur configurations synthétiques ou semi-synthétiques, avec une grille de taille $N=50$, selon les étapes suivantes.
\begin{enumerate}
	\item La densité de population est initialisée soit par mélange d'exponentielles, dont les centres (noeuds du réseau) suivent la configuration d'un système de ville synthétique comme fait en~\ref{sec:macrocoevolexplo} ; soit à partir d'une configuration réelle extraite du raster de densité pour la France. Nous utiliserons la deuxième option dans les explorations systématiques ici.
	\item Dans le second cas, un nombre fixe de noeud du réseau sont générés et localisés de manière préférentielle selon la densité (voir~\ref{sec:correlatedsyntheticdata})\footnote{Pour éviter les effets de bord d'un réseau n'ayant aucune connexion avec l'extérieur, nous ajoutons un nombre fixe $n_e$ de noeuds (que nous prenons $n_e = 6$) à des points aléatoires sur le bord du monde.}. Nous n'initialisons pas sur réseau réel puisqu'il s'agira de la cible de calibration, mais imposons un squelette initial synthétique pouvant être interprété comme un réseau archaïque.
	\item Un réseau initial est créé par connection des noeuds comme détaillé en~\ref{sec:correlatedsyntheticdata}.
\end{enumerate}
}

\subsubsection{Generated networks}{Réseaux générés}

\begin{figure}
	\includegraphics[width=\linewidth]{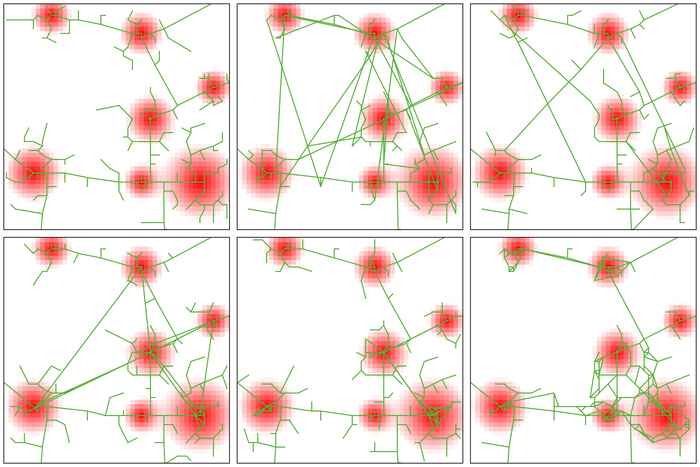}
\caption[Network examples for different generation heuristics]{\textbf{Examples of networks obtained with the different heuristics.} Networks are obtained for the same density configuration composed of 7 centers, and for the same initial network connecting them. We take $l_m = 10$ and fix the final size to 200 nodes. Gravity parameters are $d_G = 2000$, $d_0 = 3$, $\gamma_G = 0.3$, $k_h = 0.6$. In the order from left to right and top to bottom: network with connexion only; random network; random potential breakdown with $\gamma_R = 2$ and $\theta_R = 1.6$; deterministic potential breakdown; cost-benefits with $\lambda = 0.009$; biological with $n_b = 50$ and $\theta_b = 0.6$.\label{fig:networkgrowth:examples}}
\end{figure}

\bpar{
A visual illustration of the different generated topologies is given in Fig.~\ref{fig:networkgrowth:examples} for a synthetic density configuration. This allows us to compare the particularities of each heuristic. For example, links formed through random breakdown compared to deterministic breakdown witness the path-dependency and produce a less redundant network, whereas deterministic breakdown reinforces the strongest link between the two large cities that are close. The cost-based heuristic gives network that are dense in a very localized way, but avoids too long links. Finally, the biological heuristic produces a dense mesh in the sub-region where interactions are the strongest.
}{
Une illustration visuelle des différentes topologies générées est donnée en Fig.~\ref{fig:networkgrowth:examples} pour une configuration de densité synthétique. Cela permet de comparer les particularités de chacune des heuristiques. Par exemple, les liens formés par la rupture aléatoire en comparaison à la rupture déterministe témoignent de la dépendance au chemin et produisent un réseau moins redondant, tandis que la rupture déterministe renforce le lien le plus fort entre les deux grandes villes proches. L'heuristique basée sur le coût donne des réseaux denses de manière très localisée, mais évite les liens trop longs. Enfin, l'heuristique biologique produit un maillage dense dans la sous-région où les interactions sont les plus fortes.
}

\subsubsection{Experience plan}{Plan d'expérience}

\bpar{
We detail now an experience plan to explore the space of networks generated by the different heuristics. Network generation is done with constant population densities, on real configurations that have been morphologically classified in~\ref{sec:staticcorrelations}. We consider 50 real density grids, corresponding to areas in France, classified into 5 morphological classes. Their description is given in Appendix~\ref{app:sec:networkgrowth}, and show that they cover a set of morphologies spanning to very localized and sparse settlements to polycentric structures, and intermediate configurations.
}{
Détaillons un plan d'expérience pour explorer l'espace des réseaux générés par les différentes heuristiques. La génération de réseau est faite à densité de population constante, sur configurations réelles classifiées morphologiquement en~\ref{sec:staticcorrelations}. Nous considérons 50 grilles réelles de densité, correspondant à des zones en France, classées dans 5 classes morphologiques. La description de celles-ci est donnée en Annexe~\ref{app:sec:networkgrowth}, et montre qu'elle couvrent un ensemble de morphologies allant d'établissements très localisés et dispersés à des structures polycentriques, et des configurations intermédiaires.
}


\bpar{
Given the parameter ranges previously given for each heuristic, we compare the feasible space for a basic exploration with a Latin Hypercube Sampling of parameter space, for all density grids, with 5 repetitions for each parameter point\footnote{What corresponds to around 240000 repetitions of the model. The simulation data is available at \url{http://dx.doi.org/10.7910/DVN/OBQ4CS}.}.
}{
Étant donné les plages de paramètres données précédemment pour chacune des heuristiques, nous comparons l'espace faisable pour une exploration basique en criblage LHS de l'espace des paramètres, pour l'ensemble des grilles de densité, avec 5 répétitions par point de paramètre\footnote{Correspondant à environ 240000 répétitions du modèle. Le jeu de données issu des simulations est disponible à \url{http://dx.doi.org/10.7910/DVN/OBQ4CS}.}.
}


\subsubsection{Obtained topologies}{Topologies obtenues}

\bpar{
Networks are characterized here with the following indicators: average betweenness centrality $\bar{bw}$ and average closeness centrality $\bar{cl}$, diameter $r$, average path length $\bar{l}$, relative speed $v_0$. To visualize feasible spaces and then compare them to real networks, we reduce the space in a principal hyperplan, from points obtained in simulations. The first two components can be interpreted the following way\footnote{Their composition is given by: $PC1 = - 0.51 \bar{bw} - 0.45 \bar{l} + 0.57 v_0 - 0.43 r + 0.05 \bar{cl}$ and $PC2 = -0.45 \bar{bw} + 0.17 \bar{l} +0.33 v_0 + 0.8 r +0.1 \bar{cl}$.}: the first will characterize networks in which paths are shorter, whereas the second corresponds to networks with a higher average distance, thus more spread in space, but more efficient.
}{
Les réseaux sont caractérisés ici par les indicateurs suivants : centralité de chemin moyenne $\bar{bw}$ et centralité de proximité moyenne $\bar{cl}$, diamètre $r$, longueur moyenne de chemin $\bar{l}$, vitesse relative $v_0$. Pour visualiser les espaces faisables et les comparer aux réseaux réels par la suite, nous réduisons l'espace dans un hyperplan principal, à partir des points obtenus dans les simulations. Les deux premières composantes s'interprètent de la façon suivante\footnote{Leur composition est donnée par : $PC1 = - 0.51 \bar{bw} - 0.45 \bar{l} + 0.57 v_0 - 0.43 r + 0.05 \bar{cl}$ et $PC2 = -0.45 \bar{bw} + 0.17 \bar{l} +0.33 v_0 + 0.8 r +0.1 \bar{cl}$.} : la première va caractériser des réseaux où les chemins sont courts, tandis que la deuxième exprime des réseaux à distance moyenne plus grande, donc plus étalés, mais plus efficients. 
}
 
\begin{figure}
\includegraphics[width=\linewidth]{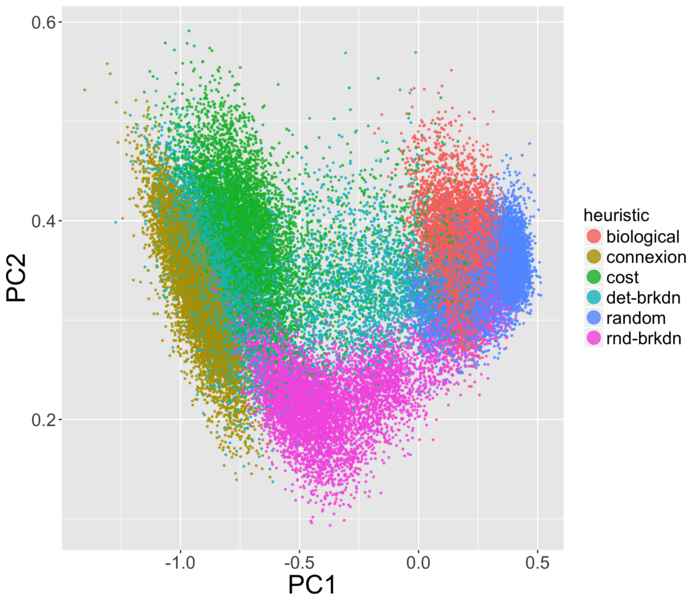}
\caption[Feasible topological space]{\textbf{Feasible topological space for the different generation heuristics.} Point clouds cover complementary regions of the topological space, the color giving the heuristic: biological (\texttt{biological}), reference (\texttt{connexion}), cost-benefits (\texttt{cost}), deterministic breakdown (\texttt{det-brkdn}), random (\texttt{random}) and random breakdown (\texttt{rnd-brkdn}). The same figure conditioned to the morphological class for density is given in Appendix~\ref{app:sec:networkgrowth}.\label{fig:networkgrowth:feasiblespace}}
\end{figure}

\bpar{
The point cloud of the topological feasible space, obtained with the experience plan described above, is given in Fig.~\ref{fig:networkgrowth:feasiblespace}. The coverage is allowed by the complementarity of different clouds for each heuristic. For example, the random heuristic is at the total opposite of the reference heuristic along the first component: the reference tree network logically induces a larger number of detours, and thus longer paths. Random breakdown allows to cover a large span of $PC1$ and corresponds more to low values of $PC2$. 
}{
Le nuage de points de l'espace topologique faisable, obtenu avec le plan d'expérience décrit ci-dessus, est donné en Fig.~\ref{fig:networkgrowth:feasiblespace}. La couverture est permise par la complémentarité des différents nuages pour chaque heuristique. Par exemple, l'heuristique aléatoire est à l'opposée complète de l'heuristique de référence le long de la première composante : le réseau arborescent de référence induit logiquement un plus grand nombre de détours, et donc des chemins plus longs. La rupture aléatoire permet de couvrir une grande plage sur $PC1$ et occupe une place privilégiée pour les faibles valeurs de $PC2$.
}


\bpar{
To better understand the complementarity of approaches, we can quantify the intersection of point clouds in Fig.~\ref{fig:networkgrowth:feasiblespace} with a simple method: by dividing the plan into a grid (that we take of size 20x20), the proportions $p_{ij}$ of points for each heuristic $j$ for each cell $i$ can be aggregated into a concentration index $h_i = \sum_j p_{ij}^2$ (Herfindhal index) which distribution describes the balance between heuristics in the different regions of space. We obtain for cells a first quartile at $0.54$, a median at $0.76$ and a third quartile at $1$. For comparison, in the case of two types of points only, a repartition 65-35\% gives an index of $0.55$ and a repartition 85-15\% an index of $0.75$, what means that at least half of cells have more than three quarters of points in a unique category. This confirms the conclusion of a strong complementarity of heuristics.
}{
Pour mieux comprendre la complémentarité des approches, nous pouvons quantifier l'intersection des nuages de points de la Fig.~\ref{fig:networkgrowth:feasiblespace} par une méthode simple : en divisant le plan en une grille (qu'on prend de taille 20x20), les proportions $p_{ij}$ de points de chaque heuristique $j$ pour chaque cellule $i$ peuvent être agrégées en un indice de concentration $h_i = \sum_j p_{ij}^2$ (indice de Herfindhal) dont la distribution décrit les équilibres entre heuristiques dans les régions de l'espace. Nous obtenons pour les cellules un premier quartile à $0.54$, une médiane à $0.76$ et un troisième quartile à $1$. Pour comparaison, dans le cas de deux types de points seulement, une répartition 65-35\% donne un indice de $0.55$ et une répartition 85-15\% un indice de $0.75$, ce qui veut dire qu'au moins la moitié des cellules ont plus de trois quarts de points dans une unique catégorie. Cela confirme la conclusion de forte complémentarité des heuristiques.
}

%

\subsubsection{Comparison to real networks}{Comparaison aux réseaux réels}

\begin{figure}
\includegraphics[width=\linewidth,height=0.85\textheight]{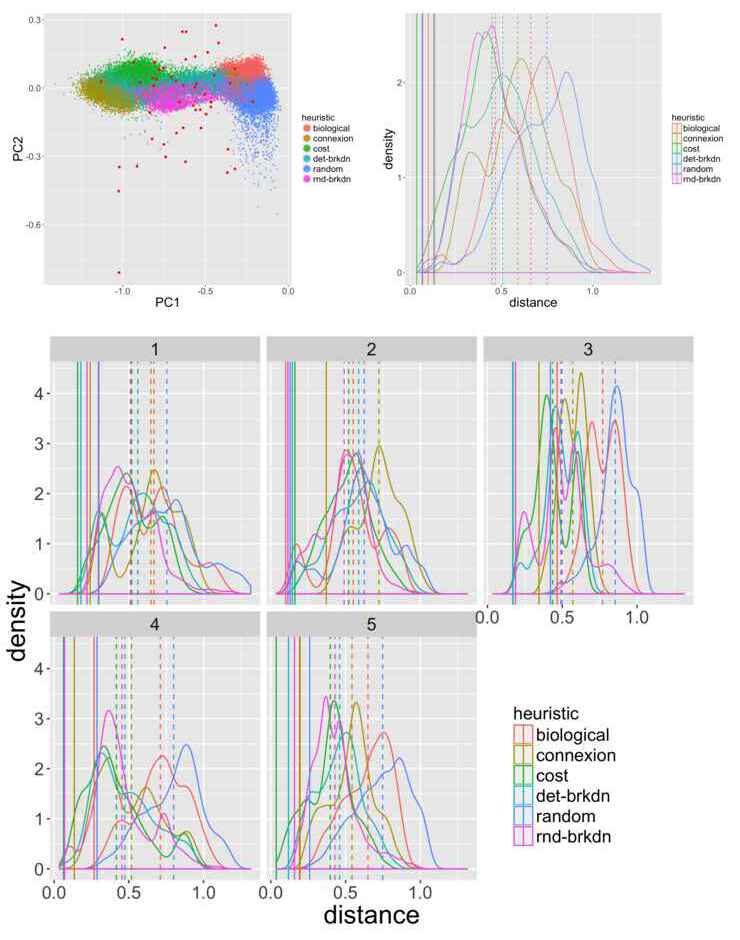}
\caption[Comparison to real networks]{\textbf{Comparison to real networks.} (\textit{Top Left}) Point clouds for simulated configurations (color in the legend) and for real configurations (in red), in a principal plan such that $PC1 = 0.12 \bar{bw} - 0.09 \bar{cl} + 0.98 \bar{l}$ and $PC2 = -0.20 \bar{bw} - 0.97 \bar{cl} - 0.06 \bar{l}$. (\textit{Top right}) Distribution of distances $d_{min}$ for all simulated points, for each heuristic (color). Dashed vertical lines give the average and solid lines the minimum for each distribution. (\textit{Bottom}) Same histograms, conditioned by morphological class for density distribution.\label{fig:networkgrowth:realdistance}}
\end{figure}

\bpar{
We use the measures on real road networks obtained in~\ref{sec:staticcorrelations} to compute a distance of generated configurations to observed configurations, by considering real networks corresponding to density configurations used for initialization. We take for a given parameter point the minimum of the euclidian distance on vectors of indicators for all real points\footnote{What means that if $d(1,2) = \sqrt{(\bar{bw}_1 - \bar{bw}_2)^2 + (\bar{cl}_1 - \bar{cl}_2)^2 + (\bar{l}_1 - \bar{l}_2)^2}$, we consider $d_{min} = \min_j d(S,R_j)$ if $S$ is the simulated point and $R_j$ the set of real points. We keep here only the indicators $\bar{bw}$, $\bar{cl}$ and $\bar{l}$, for normalization reasons.}. This comparison is made possible since indicators are normalized, and indicators on real networks are comparable to indicators on synthetic networks.
}{
Nous utilisons les mesures sur réseaux routiers réels obtenues en~\ref{sec:staticcorrelations} pour calculer une distance des configurations générées aux configurations observées, en considérant les réseaux réels correspondant aux configurations de densité utilisées pour l'initialisation. Nous prenons pour un point de paramètre le minimum de distance euclidienne sur les vecteurs d'indicateurs pour l'ensemble des points réels\footnote{C'est-à-dire si $d(1,2) = \sqrt{(\bar{bw}_1 - \bar{bw}_2)^2 + (\bar{cl}_1 - \bar{cl}_2)^2 + (\bar{l}_1 - \bar{l}_2)^2}$, on considère $d_{min} = \min_j d(S,R_j)$ si $S$ est le point simulé et $R_j$ l'ensemble des points réels. Nous conservons ici uniquement les indicateurs $\bar{bw}$, $\bar{cl}$ et $\bar{l}$, pour des raisons de normalisation.}. Cette comparaison est possible car les indicateurs sont normalisés, et les indicateurs sur réseaux réels sont comparables aux indicateurs sur réseaux synthétiques.
}


\bpar{
Comparison results to real points are given in Fig.~\ref{fig:networkgrowth:realdistance}. We give a representation as a point cloud and histograms for distributions of distances, on all grids and by morphological class. We observe that around ten real configuration (one fifth) fall far outside the point cloud. Once again, heuristics are complementary to approach a larger number of points. Concerning distances, the random heuristic is the worse in terms of mode and average, followed by the biological, the reference (connexion only), the deterministic breakdown and finally teh random-breakdown and the cost which are approximatively equivalent. All realize very low minimal distances.
}{
Les résultats de comparaison aux points réels sont donnés en Fig.~\ref{fig:networkgrowth:realdistance}. Nous donnons une représentation en nuage de points et les histogrammes de distribution des distances, sur l'ensemble des grilles et par classe morphologique. On constate qu'une dizaine de configurations réelles (1/5ème) se retrouvent à grande distance du nuage de points simulés, mais que les autres tombent à distance faible ou à l'intérieur du nuage de points. Encore une fois, les différentes heuristiques sont complémentaires pour approcher un plus grand nombre de points. Concernant les distances, l'aléatoire est le plus mauvais en termes de mode et de moyenne, suivi par le biologique, la référence (connection), la rupture déterministe puis la rupture aléatoire et le coût qui sont à peu près équivalentes. Elles réalisent toutes des distances minimales très faibles.
}

\bpar{
When conditioning by morphological classes, we see that classes 3, 4 and 5 give the most difficulties for all heuristics in terms of minima - they are indeed the configurations with very localized settlements or a diffuse population (see~\ref{app:sec:networkgrowth}): it is therefore easier to reproduce real network configurations in the case of polycentric structures. In all cases, the biological heuristic is not very efficient, but it is not directly possible to know if this is a consequence of its under-exploitation and its fixed parameters, or of its intrinsic dynamics.
}{
En conditionnant par les classes morphologiques, nous voyons que les classes 3, 4 et 5 donnent le plus de difficultés à l'ensemble des heuristiques en termes de minimum - or il s'agit des configurations avec établissements très localisés ou population diffuse (voir~\ref{app:sec:networkgrowth}) : il est donc plus facile de reproduire les configurations réelles de réseau dans le cas de structures polycentriques. Dans tous les cas, l'heuristique biologique est peu performante, mais il n'est pas directement possible de savoir si cela est dû à sa sous-exploration et aux paramètres fixés ou à sa dynamique intrinsèque.
}

%


%
%
%
%

\subsection{Discussion}{Discussion}

\bpar{
If the slime-mould model is able to generate robust networks in a simplified way, its use for planning has been questioned, in particular because it does not take into account external factors and the urban environment~\cite{adamatzky2010road}. Our results seem to confirm these analyses, since this heuristic is the least performing in terms of distance to real networks.
}{
Si le modèle slime-mould est capable de conduire de manière simplifiée à une génération de réseaux robustes, son utilisation pour la planification a été mise en question, notamment pour sa non prise en compte de facteurs extérieurs et de l'environnement urbain~\cite{adamatzky2010road}. Nos résultats semblent confirmer ces analyses, puisque cette heuristique est la moins performante au sens de la distance aux réseaux réels.
}

\bpar{
We have thus explored and compared different network generation heuristics, at a fixed density. We note the following points.
\begin{itemize}
	\item Different models produce networks that appear as complementary in an indicator space.
	\item Similarly, they are complementary to resemble configurations of real networks, while showing different performances. Very localized or diffuse density configurations correspond to networks that are more difficult to reproduce, in comparison to polycentric structures. 
\end{itemize}
}{
Nous avons donc exploré et comparé différentes heuristiques de génération de réseau, à densité fixée. Nous en retirons les enseignements suivants.
\begin{itemize}
	\item Les différents modèles produisent des réseaux qui apparaissent complémentaires dans un espace d'indicateurs.
	\item De même, ils sont complémentaires pour s'approcher des configurations des réseaux réels, tout en présentant des performances différentes. Des configurations de densité très localisées ou diffuses correspondent à des réseaux plus difficiles à reproduire, en comparaison aux structures polycentriques.
\end{itemize}
}

\stars

\bpar{
Armed with these network growth models, we will be able to couple them to a density model, in order to develop a co-evolution model at the mesoscopic scale, which will be the subject of the following section.
}{
Disposant de ces modèles de croissance de réseau, nous allons pouvoir les utiliser en couplage avec un modèle de densité, afin de développer un modèle de co-évolution à l'échelle mesoscopique, qui fera l'objet de la section suivante.
}

\stars

%


\newpage


\section{Co-evolution at the mesoscopic scale}{Co-évolution à l'échelle mesoscopique}

\label{sec:mesocoevolmodel}


\bpar{
Urban settlements and transportation networks have been shown to be co-evolving, in the different thematic, empirical  and modeling studies of territorial systems developed up to here. As we saw, modeling approaches of such dynamical interactions between networks and territories are poorly developed. We propose in this section to realize a first entry at an intermediate scale, focusing on morphological and functional properties of the territorial system in a stylized way. We introduce a stochastic dynamical model of urban morphogenesis which couples the evolution of population density within grid cells with a growing road network.
}{
Les établissements urbains et les réseaux de transport ont été montrés comme co-évolutifs, dans les différentes approches thématiques, empiriques, et de modélisation des systèmes territoriaux développées jusqu'ici. Comme on l'a vu, les approches modélisant ces interactions dynamiques entre réseaux et territoires sont peu développées. Nous proposons dans cette section de réaliser une première entrée à une échelle intermédiaire, en s'intéressant aux propriétés morphologiques et fonctionnelles des systèmes territoriaux de manière stylisée. Nous introduisons un modèle dynamique et stochastique de morphogenèse urbaine qui couple l'évolution de la densité de population dans les cellules d'une grille avec l'évolution d'un réseau routier.
}

\subsection{Model description}{Description du modèle}

\subsubsection{General structure}{Structure générale}

\begin{figure}
	\includegraphics[width=\linewidth]{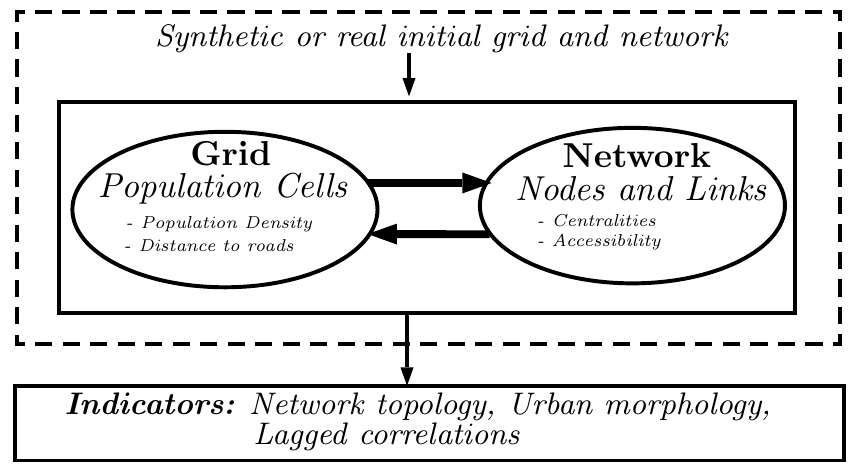}
	\caption[Morphogenesis at the mesoscopic scale]{\textbf{Structure of the co-evolution model at the mesoscopic scale.}\label{fig:mesocoevolmodel:workflow}}
\end{figure}

\bpar{
The general principles of the model are the following. With an overall fixed growth rate, new population aggregate preferentially to a local potential, for which parameters control the dependance to various explicative variables. These are in particular local density, distance to the network, centrality measures within the network and generalized accessibility. \cite{doi:10.1080/13658816.2014.893347} shows in the case of Stockholm the very strong correlation between centrality measures in the network and the type of land-use, what confirms the inportance to consider centralities as explicative variables for the model at this scale. We generalize thus the morphogenesis model studied in~\ref{sec:densitygeneration}, with aggregation mechanisms similar to the ones used by~\cite{raimbault2014hybrid}. A continuous diffusion of population completes the aggregation to translate repulsion processes generally due to congestion. Because of the different time scales of evolution for the urban environment and for networks, the network grows at fixed time steps, following the submodel developed in~\ref{sec:networkgrowth}: a first fixed rule ensures connectivity of newly populated patches to the existing network. The different network generation heuristics are then included in the model. We expect the different heuristics to be complementary since for example the gravity model would be more typical of planned top-down network evolution, whereas the biological model will translate bottom-up processes of network growth. The Fig.~\ref{fig:mesocoevolmodel:workflow} summarizes the general structure of the morphogenesis model.
}{
Les principes généraux du modèle sont les suivants. Avec un taux de croissance global fixé, une nouvelle population s'agrège préférentiellement à un potentiel local, dont la dépendance à diverses variables explicatives est contrôlé par des paramètres. Celles-ci sont la densité locale, la distance au réseau, les mesures de centralité dans le réseau et l'accessibilité généralisée. \cite{doi:10.1080/13658816.2014.893347} montre dans le cas de Stockholm la très forte corrélation entre mesures de centralité et le type d'usage du sol, ce qui confirme l'importance de considérer les centralités comme variables explicatives pour le modèle à cette échelle. Nous généralisons ainsi le modèle de morphogenèse étudié dans~\ref{sec:densitygeneration}, avec des mécanismes d'agrégation similaires à ceux utilisés par~\cite{raimbault2014hybrid}. Une diffusion continue de la population complète l'agrégation pour traduire les processus de répulsion généralement dus à la congestion. À cause des différentes échelles de temps impliquées dans l'évolution de l'environnement urbain et des réseaux, le réseau croit à pas de temps fixés, suivant le sous-modèle développé en~\ref{sec:networkgrowth} : une première règle fixe assure la connectivité des cellules nouvellement peuplés au réseau existant. Les différentes heuristiques de génération de réseau sont ensuite incluses dans le modèle. Nous nous attendons à une complémentarité de celles-ci, puisque par exemple le modèle gravitaire sera plus typique d'une évolution de réseau planifiée, tandis que le modèle biologique traduit des processus auto-organisés de croissance de réseau. La Fig.~\ref{fig:mesocoevolmodel:workflow} résume la structure générale du modèle de morphogenèse.
}

\subsubsection{Formalization}{Formalisation}

\bpar{
The model is based on a squared population grid of size $N$, which cells are defined by populations $(P_i)$. A road network is included in a way similar as in~\ref{sec:networkgrowth}. We assume at the initial state a given population distribution and a network.
}{
Le modèle est basé sur une grille carrée de population de côté $N$, dont les cellules sont définies par les populations $(P_i)$. Un réseau routier s'y superpose de la même manière qu'en~\ref{sec:networkgrowth}. Nous supposons une distribution de population à l'instant initial ainsi qu'un réseau.
}

\bpar{
The evolution of densities is based on a utility function, influenced by local characteristics of the urban form and function, that we call \emph{explicative variables}. Let $x_k(i)$ a local explicative variable for cell $i$, which will be among the following variables:
\begin{itemize}
	\item population $P_i$;
	\item proximity to roads\footnote{Taken as $\exp (-d / d_n)$ where $d$ is the distance by projection on the closest road, and $d_n=10$ is fixed.};
	\item betweenness centrality;
	\item closeness centrality;
	\item accessibility.
\end{itemize}
For the last three, they are defined as previously for network nodes, and then associated to cells by taking the value of the closest node, weighted by a decreasing function of the distance to it\footnote{I.e. of the form $x_k = x^{(n)}_k (\argmin_j d(i,j)) \cdot \exp \left( -  \min_j d(i,j) / d_0 \right)$, with $x^{(n)}_k$ the corresponding variable for nodes, the index $j$ being taken on all nodes, and the decay parameter $d_0$ is in our case fixed at $d_0=1$ to keep the property that network variables are essentially significant at close distances from the network.}. We consider then normalized explicative variables defined by $\tilde{x}_k(i) = x_k(i) - \min_j x_k(j) / (\max_j x_k(j) - \min_j x_k(j))$.
}{
L'évolution des densités se base sur une fonction d'utilité, influencée par des caractéristiques locales de la forme et de la fonction urbaine, que l'on appelle \emph{variables explicatives}. Soit $x_k(i)$ une variable explicative locale pour la cellule $i$, qui sera parmi les variables suivantes : 
\begin{itemize}
	\item population $P_i$ ;
	\item proximité aux routes\footnote{Prise sous la forme $\exp (-d / d_n)$ avec $d$ distance par projection sur la route la plus proche, et $d_n =10$ fixé.} ;
	\item centralité de chemin ;
	\item centralité de proximité ;
	\item accessibilité.
\end{itemize}
Pour les trois dernières, celles-ci sont définies comme précédemment pour les noeuds du réseau, puis associées aux cellules en prenant la valeur du noeud le plus proche, pondérée par une fonction décroissante en fonction de la distance à celui-ci\footnote{C'est-à-dire de la forme $x_k = x^{(n)}_k (\argmin_j d(i,j)) \cdot \exp \left( -  \min_j d(i,j) / d_0 \right)$, avec $x^{(n)}_k$ variable correspondante pour les noeuds, l'indice $j$ étant pris sur l'ensemble des noeuds, et le paramètre de décroissance $d_0$ étant dans notre cas fixé à $d_0 = 1$ pour garder la caractéristique que les variables de réseau sont essentiellement significatives à des distances proches de celui-ci.}. Nous considérons alors les variables explicatives normalisées définies par $\tilde{x}_k(i) = x_k(i) - \min_j x_k(j) / (\max_j x_k(j) - \min_j x_k(j))$. 
}

\bpar{
The utility of a cell is then given by a linear aggregation\footnote{An alternative could be for example a Cobb-Douglas function, which is equivalent to a linear aggregation on the logarithms of variables.}
\begin{equation}
U_i = \sum_k w_k \cdot \tilde{x}_k(i)
\end{equation}
where $\tilde{x}_k$ are the normalized local explicative variables, and $w_k$ are weight parameters, which allow to weight between the different influences.
}{
L'utilité d'une cellule est alors donnée par une agrégation linéaire\footnote{Une alternative étant par exemple une fonction de Cobb-Douglas, qui revient à une agrégation linéaire sur les logarithmes des variables.}
\begin{equation}
U_i = \sum_k w_k \cdot \tilde{x}_k(i)
\end{equation}
où les $\tilde{x}_k$ sont les variables explicatives locales normalisées, et $w_k$ des paramètres de poids, qui permettent de pondérer les différentes influences.
}


\bpar{
A time step of model evolution includes then the following stages.
\begin{enumerate}
	\item Evolution of the population following rules similar to the morphogenesis model developed in~\ref{sec:densitygeneration}. Given an exogenous growth rate $N_G$, individuals are added independently following an aggregation done with a probability $U_i^\alpha/\sum_k U_k^\alpha$, followed by a diffusion of strength $\beta$ to neighbor cells, done $n_d$ times.
	\item Network growth following the rules described in~\ref{sec:networkgrowth}, knowing that this takes place is the time step is a multiple of a parameter $t_N$, which allows to integrate a differential between temporal scales for population growth and for network growth.
\end{enumerate}
}{
Un pas de temps d'évolution du modèle comporte alors les étapes suivantes.
\begin{enumerate}
	\item Évolution de la population selon des règles similaires au modèle de morphogenèse développé en~\ref{sec:densitygeneration}. Étant donné un taux de croissance exogène $N_G$, les individus sont ajoutés de manière indépendante suivant une agrégation faite selon la probabilité $U_i^\alpha/\sum_k U_k^\alpha$, suivie d'une diffusion aux voisins de force $\beta$, effectuée $n_d$ fois.
	\item Croissance du réseau selon les règles décrites en~\ref{sec:networkgrowth}, sachant que celle-ci a lieu si le pas de temps est un multiple d'un paramètre $t_N$, qui permet d'intégrer un différentiel d'échelles temporelles entre la croissance de la population et celle du réseau.
\end{enumerate}
}

\bpar{
The aggregation following a power of the utility yields a flexibility in the underlying optimization problem, since as \cite{josselin2013revisiting} recall, the use of different norms in spatial optimal location problems corresponds to different logics of optimization.
}{
L'agrégation selon une puissance de l'utilité permet une flexibilité dans le problème d'optimisation sous-jacent, puisque comme le rappellent \cite{josselin2013revisiting} l'utilisation de différentes normes dans les problèmes de localisation spatiale optimale correspond à des logiques d'optimisation différentes.
}

\bpar{
The parameters of the model that we will make vary are then:
\begin{itemize}
	\item aggregation-diffusion parameters $\alpha,\beta,N_g,n_d$, summarized in Table~\ref{tab:density:parameters};
	\item the four weight parameters $w_k$ for the explicative variables, which vary in $[0;1]$;
	\item network growth parameters for the different heuristics, summarized in Table~\ref{tab:networkgrowth:parameters}.
\end{itemize}
}{
Les paramètres du modèle que nous ferons varier sont donc :
\begin{itemize}
	\item les paramètres d'agrégation-diffusion $\alpha,\beta,N_g,n_d$, résumés en Table~\ref{tab:density:parameters} ;
	\item les paramètres de poids des variables explicatives $w_k$, au nombre de 4, compris dans $[0;1]$ ;
	\item les paramètres de croissance de réseau des différentes heuristiques, résumés en Table~\ref{tab:networkgrowth:parameters}.
\end{itemize}
}

\bpar{
Output model indicators are the urban morphology indicators, topological network indicators, and lagged correlations between the different explicative variables.
}{
Les indicateurs de sortie du modèle sont les indicateurs de morphologie urbaine, les indicateurs topologiques du réseau, et les corrélations retardées entre les différentes variables explicatives.
}

\subsection{Results}{Résultats}

\subsubsection{Implementation}{Implémentation}

\bpar{
The model is implemented in NetLogo, given the heterogeneity of aspects that have to be taken into account, and this language being particularly suitable to couple a grid of cells with a network. Urban morphology indicators are computed thanks to a NetLogo extension specially developed (see Appendix~\ref{app:tools}).
}{
Le modèle est implémenté en NetLogo, vu l'hétérogénéité des aspects à prendre en compte, et ce langage se montrant particulièrement efficace pour coupler une grille de cellules à un réseau. Les indicateurs de morphologie urbaine sont calculés grâce à une extension NetLogo spécifiquement développée (voir Annexe~\ref{app:tools}).
}

\subsubsection{Experience plan}{Plan d'expérience}

\bpar{
We propose to focus on the ability of the model to capture relations between networks and territories, and more particularly the co-evolution. Therefore, we will try to establish if (i) the model is able to reproduce, beyond the form indicators, the static correlation matrices computed in~\ref{sec:staticcorrelations}; and (ii) the model produces a variety of dynamical relations in the sense of causality regimes developed in~\ref{sec:causalityregimes}.
}{
Nous proposons de nous concentrer sur la capacité du modèle à capturer les relations entre réseaux et territoires, et en particulier la co-évolution. Pour cela, nous chercherons si (i) le modèle est capable de reproduire, en plus des indicateurs de forme, les matrices de corrélations statiques calculées en~\ref{sec:staticcorrelations} ; et (ii) le modèle produit une variété de relations dynamiques au sens des régimes de causalité développés en~\ref{sec:causalityregimes}.
}

\bpar{
The model is initialized on fully synthetic configurations, with a grid of size $50$. Configurations are generated through an exponential mixture in a way similar to \cite{anas1998urban}: $N_c = 8$ centers are randomly located, to which a population is attributed following a scaling law $P_i = P_0\cdot (i+1)^{-\alpha_S}$ with $\alpha_S = 0.8$ and $P_0 = 200$. The population of each center is distributed to all cells with an exponential kernel of shape $d(r) = P_{max}\exp\left( - r / r_0\right)$ where the parameter $r_0$ is determined to fix the population at $P_i$, with $P_{max} = 20$ (density at the center)\footnote{We have indeed $P_i = \iint d(r) = \int_{\theta=0}^{2\pi} \int_{r=0}^{\infty} d(r) rdrd\theta = 2 \pi P_{max} \int_r r\cdot \exp\left( - r / r_0\right) = 2 \pi P_{max} r_0^2$, and therefore $r_0 = \sqrt{\frac{P_i}{2\pi P_{max}}}$.}. The initial network skeleton is generated as detailed in~\ref{sec:networkgrowth}.
}{
Le modèle est initialisé sur configurations entièrement synthétiques, avec une taille de grille $50$. Les configurations sont générées par mélange d'exponentielles d'une manière similaire à \cite{anas1998urban} : $N_c = 8$ centres sont localisés de manière aléatoire, et une population leur est attribuée selon une loi d'échelle $P_i = P_0\cdot (i+1)^{-\alpha_S}$ avec $\alpha_S = 0.8$ et $P_0 = 200$. La population de chaque centre est distribuée à l'ensemble des cellules avec un noyau exponentiel de forme $d(r) = P_{max}\exp\left( - r / r_0\right)$ où le paramètre $r_0$ est déterminé pour fixer la population à $P_i$, avec $P_{max} = 20$ (densité au centre)\footnote{On a en effet $P_i = \iint d(r) = \int_{\theta=0}^{2\pi} \int_{r=0}^{\infty} d(r) rdrd\theta = 2 \pi P_{max} \int_r r\cdot \exp\left( - r / r_0\right) = 2 \pi P_{max} r_0^2$, et donc $r_0 = \sqrt{\frac{P_i}{2\pi P_{max}}}$.}. Le squelette de réseau initial est généré comme détaillé en~\ref{sec:networkgrowth}.
}

\bpar{
We explore a Latin Hypercube Sampling of the parameter space, with 10 repetitions for around 7000 parameter points, corresponding to a total of around 70000 model repetitions\footnote{For which simulation results are also available at \url{http://dx.doi.org/10.7910/DVN/OBQ4CS}.}, realized on a computation grid by using OpenMole.
}{
Nous explorons un échantillonnage LHS de l'espace des paramètres, avec 10 répétitions pour environ 7000 points de paramètres, correspondant à un total autour de 70000 répétitions du modèle\footnote{Pour lesquelles les résultats de simulation sont disponibles également à \url{http://dx.doi.org/10.7910/DVN/OBQ4CS}.}, effectuées sur grille de calcul par l'intermédiaire d'OpenMole.
}

\subsubsection{Static and dynamical calibration}{Calibration statique et dynamique}

\begin{figure}
	\includegraphics[width=\linewidth]{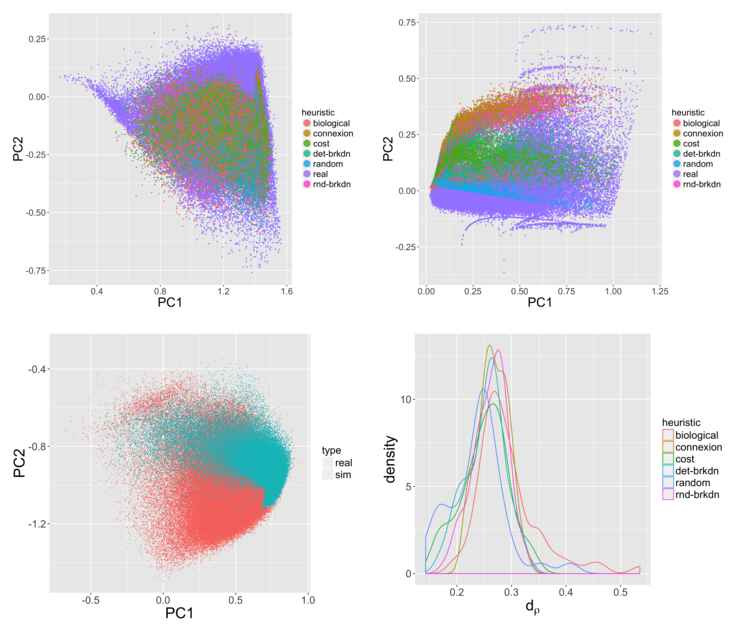}
	\caption[Calibration of the morphogenesis model]{\textbf{Calibration of the morphogenesis model at the first and second order.} (\textit{Top Left}) Simulated and observed point clouds in a principal plan for urban morphology indicators. (\textit{Top Right}) Simulated and observed could points in a principal plan for network indicators. (\textit{Bottom Left}) Simulated and observed point clouds in a principal plan for all indicators. (\textit{Bottom Right}) Distributions of distances on correlations $d_{\rho}$, for the different heuristics.\label{fig:mesocoevolmodel:calibration}}
\end{figure}

\bpar{
The model is calibrated at the first order, on indicators for the urban form and network measures, and at the second order on correlations between these. Real data used are still the same as introduced in~\ref{sec:staticcorrelations}, which as we recall it are based on Eurostat population grid and the road network from OpenStreetMap. We use here the full set of points from Europe.
}{
Le modèle est calibré au premier ordre, sur les indicateurs de forme urbaine et de mesure de réseau, ainsi qu'au second ordre sur les correlations entre ceux-ci. Les données réelles utilisées sont toujours celles introduites en~\ref{sec:staticcorrelations}, qui nous le rappelons sont basées sur les données de population raster Eurostat et le réseau routier issu d'OpenStreetMap. Nous utilisons ici l'ensemble des points de l'Europe.
}

\bpar{
We introduce an \emph{ad hoc} calibration procedure in order to take into account the first two moments, that we detail below. More elaborated procedures are used for example in economics, such as \cite{watson1993measures} which uses the noise of the difference between two variables to obtain the same covariance structure for the two corresponding models, or in finance, such as \cite{frey2001copulas} which define a notion on equivalence between latent variables models which incorporates the equality of the interdependence structure between variables. We avoid here to add supplementary models, and consider simply a distance on correlation matrices. The procedure is the following.
\begin{itemize}
	\item Simulated points are the ones obtained through the sampling, with average values on repetitions.
	\item In order to be able to estimate correlation matrices between indicators for simulated data, we make the assumption that second moments are continuous as a function of model parameters, and split for each heuristic the parameter space into areas to group parameter points\footnote{Each parameter being binned into $15 / k$ equal segments, where $k$ is the number of parameters: we empirically observed that this allowed to always have a minimal number of points in each area.}, what allows to estimate for each group indicators and the correlation matrix.
	\item For each estimation done this way, that we write $\bar{S}$ (indicators) and $\rho [S]$ (correlations), we can then compute the distance to real points on indicators $d_I (R_j) = d(\bar{S},R_j)$ and on correlation matrices $d_{\rho} (R_j) = d(\rho [S],\rho[R_j])$ where $R_j$ are the real points with their corresponding correlations\footnote{That are estimated in~\ref{sec:staticcorrelations} as we recall, with a square window centered around the point, that we take here for $\delta = 4$.}, and $d$ an euclidian distance normalized by the number of components.
	\item We consider then the aggregated distance defined as $d_A^2 (R_j) = d_I^2 (R_j) + d_{\rho}^2 (R_j)$. Indeed, as developed empirically and analytically in Appendix~\ref{app:sec:mesocoevolmodel}, the shape of Pareto fronts for the two distances considered suggests the relevance of this aggregation. The real point closest to a simulated point is then the one in the sense of this distance.
\end{itemize}
}{
Nous introduisons un processus \emph{ad hoc} de calibration pour pouvoir tenir compte des deux premiers moments, que nous détaillons ci-dessous. Des procédures plus élaborées sont utilisées par exemple en économie, comme \cite{watson1993measures} qui utilise le bruit de la différence entre deux variables pour obtenir la même structure de covariance pour les deux modèles correspondants, ou en finance, comme \cite{frey2001copulas} qui définissent une notion d'équivalence entre modèles à variables latentes qui incorpore l'égalité de la structure d'interdépendance entre variables. Nous évitons ici d'ajouter des modèles supplémentaires, et considérons simplement une distance sur les matrices de corrélation. La procédure est la suivante.
\begin{itemize}
	\item Les points simulés sont ceux issus de l'échantillonnage, avec les valeurs moyennes sur les répétitions.
	\item Afin de pouvoir estimer des matrices de corrélation entre indicateurs pour les données simulées, nous faisons l'hypothèse que les seconds moments sont continus en les paramètres du modèle, et découpons pour chaque heuristique l'espace des paramètres en zones pour grouper les points de paramètres\footnote{Chaque paramètre étant découpé en $15 / k$ segments égaux avec $k$ nombre de paramètres : nous avons constaté empiriquement que cela permettait d'avoir toujours un nombre minimal de mesures dans chaque zone.}, ce qui permet d'estimer pour chaque groupe les indicateurs et la matrice des corrélations.
	\item Pour chaque estimation ainsi menée, qu'on note $\bar{S}$ (indicateurs) et $\rho [S]$ (corrélations), on peut alors calculer la distance aux points réels sur les indicateurs $d_I (R_j) = d(\bar{S},R_j)$ et sur les matrices de corrélation $d_{\rho} (R_j) = d(\rho [S],\rho[R_j])$ où les $R_j$ sont les points réels avec leurs corrélations correspondantes\footnote{Estimées on le rappelle en~\ref{sec:staticcorrelations}, par fenêtre centrée sur le point, qu'on prend ici pour $\delta = 4$.}, et $d$ une distance euclidienne normalisée par le nombre de composantes.
	\item Nous considérons alors la distance agrégée définie comme $d_A^2 (R_j) = d_I^2 (R_j) + d_{\rho}^2 (R_j)$. En effet, comme développé empiriquement et analytiquement en Annexe~\ref{app:sec:mesocoevolmodel}, la forme des fronts de Pareto pour les deux distances considérées suggère la pertinence de cette agrégation. Le point réel le plus proche du point simulé est alors celui au sens de cette distance.
\end{itemize}
}

\bpar{
The Fig.~\ref{fig:mesocoevolmodel:calibration} summarizes calibration results. Morphological indicators are easier to approach than network indicators, for which a part of the simulated clouds does not superpose with observed points. We find again a certain complementarity between network heuristics. When considering the full set of indicators, few simulated points are situated far from the observed points, but a significant proportion of these is beyond the reach of simulation. Thus, the simultaneous capture of morphology and topology is obtained at the price of less precision.
}{
La Fig.~\ref{fig:mesocoevolmodel:calibration} résume les résultats de la calibration. Les indicateurs morphologiques sont plus aisément approchés que ceux de réseau, pour lesquels une partie des nuages simulés ne se superpose pas avec les points observés. Nous retrouvons une certaine complémentarité dans les heuristiques de réseau. En considérant l'ensemble des indicateurs, peu de points simulés tombent loin des points observés, mais une proportion significative de ceux-ci est hors d'atteinte de la simulation. Ainsi, la capture simultanée de la morphologie et de la topologie se fait au prix d'une moins grande précision.
}

\bpar{
We however obtain a good reproduction of correlation matrices as shown in Fig.~\ref{fig:mesocoevolmodel:calibration} (histogram for $d_{\rho}$, bottom right). The worse heuristic for correlations is the biological one in terms of maximum, whereas the random produces rather good results: this could be due for example to the reproduction of very low correlations, which accompany a structure effect due to the initial addition of nodes which imposes already a certain correlation. On the contrary, the biological heuristic introduces supplementary processes which can possibly be beneficial to the network in terms of independence (or following the opposed viewpoint be detrimental in terms of correlations). In any case, this application shows that our model is able to resemble real configurations both for indicators and their correlations.
}{
Nous obtenons toutefois une bonne reproduction des matrices de corrélation, comme présenté en Fig.~\ref{fig:mesocoevolmodel:calibration} (histogramme de $d_{\rho}$, bas droite). La moins bonne heuristique pour les corrélations est la biologique en termes de maximum, tandis que l'aléatoire produit d'assez bons résultats : cela pourrait par exemple être dû à la reproduction des corrélations quasi nulles, accompagnant un effet de structure dû à l'ajout initial des noeuds qui impose déjà une certaine corrélation. Au contraire, l'heuristique biologique introduit des processus supplémentaires qui peuvent éventuellement bénéficier au réseau en termes d'indépendance (ou selon le point de vue opposé être préjudiciable en termes de corrélations). En tout cas, cette application démontre que notre modèle est capable à la fois de s'approcher de configurations réelles pour les indicateurs et pour leurs corrélations.
}

\subsubsection{Causality regimes}{Régimes de causalité}

\begin{figure}
	\includegraphics[width=\linewidth,height=0.93\textheight]{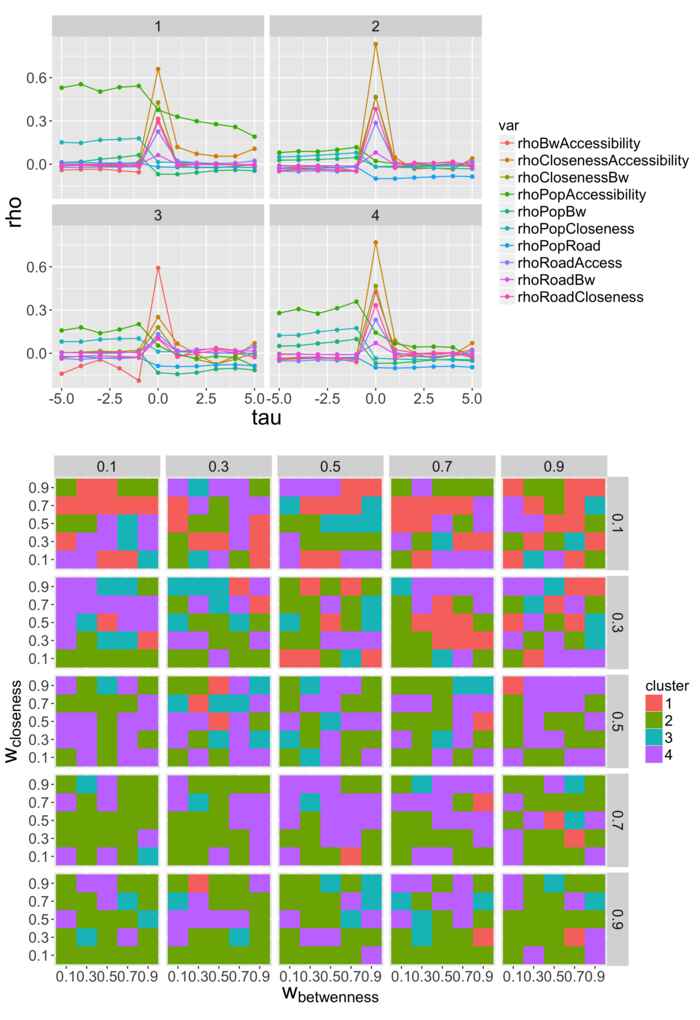}
	\caption[Causality regimes for the morphogenesis model]{\textbf{Causality regimes for the co-evolution model.} (\textit{Top}) Trajectories of classes centers in terms of $\rho[\tau]$ between the different explicative variables. (\textit{Bottom}) Phase diagram of regimes in the parameter space for $w_k$, represented here as the variation of diagrams for $(w_{bw},w_{cl})$, along the variations of $w_{road}$ (in rows) and of $w_{pop}$ (in columns).\label{fig:mesocoevolmodel:causality}}
\end{figure}

\bpar{
We furthermore study dynamical lagged correlations between the variations of the different explicative variables for cells (population, distance to the network, closeness centrality, betweenness centrality, accessibility). We apply the method of causality regimes introduced in~\ref{sec:causalityregimes}. The Fig.~\ref{fig:mesocoevolmodel:causality} summarizes the results obtained with the application of this method on simulation results of the co-evolution model. The number of classes inducing a transition is smaller than for the RDB model, translating a smaller degree of freedom, and we fix in that case $k=4$. Centroid profiles allow to understand to ability of the model to more or less capture a co-evolution.
}{
Nous étudions d'autre part les correlations retardées dynamiques entre les variations des différentes variables explicatives des cellules (population, distance au réseau, centralité de proximité, centralité de chemin, accessibilité). Nous appliquons la méthode des régimes de causalité introduite en~\ref{sec:causalityregimes}. La Fig.~\ref{fig:mesocoevolmodel:causality} résume les résultats obtenus par l'application de cette méthode sur les résultats de simulation du modèle de co-évolution. Le nombre de classes induisant une transition est plus faible que pour le modèle RDB, traduisant un plus faible degré de liberté, et nous fixons dans ce cas $k=4$. Les profils des centroïdes permettent de comprendre la capacité du modèle à capturer plus ou moins une co-évolution.
}


\bpar{
The regimes obtained appear to be less diverse than the ones obtained in~\ref{sec:causalityregimes} or for the macroscopic co-evolution in~\ref{sec:macrocoevol}. Some variables have naturally a strong simultaneous correlation, spurious from their definitions, such as closeness centrality and accessibility, or the distance to the road and the closeness centrality. For all regimes, population significantly determines the accessibility. The regime 1 corresponds to a full determination of the network by the population. The second is partly circular, through the effect of roads on populations. The regime 3 is more interesting, since closeness centrality negatively causes the accessibility: this means that in this configuration, the coupled evolution of the network and the population follow the direction of a diminution of congestion. Furthermore, as population causes the closeness centrality, there is also circularity and thus co-evolution in that case. When we locate it in the phase diagram, this regime is rather sparse and rare, contrary for example to the regime 1 which occupies a large portion of space for a low importance of the road ($w_{road} \leq 0.3$). This confirms that the co-evolution produced by the model is localized and not a characteristic always verified, but that it is however able to generate some in particular regimes.
}{
Les régimes obtenus apparaissent moins divers que ceux obtenus en~\ref{sec:causalityregimes} ou pour la co-évolution macroscopique en~\ref{sec:macrocoevol}. Certaines variables ont naturellement une forte corrélation simultanée, fortuite par leur définitions, comme la centralité de proximité et l'accessibilité, ou la distance à la route et la centralité de proximité. Dans l'ensemble des régimes, la population détermine significativement l'accessibilité. Le régime 1 correspond à une détermination entière du réseau par la population. Le second est partiellement circulaire, de par l'effet des routes sur la population. Le régime 3 est intéressant, la centralité de chemin causant négativement l'accessibilité : cela veut dire que dans cette configuration, l'évolution couplée du réseau et de la population vont dans le sens d'une diminution de la congestion. De plus, comme la population cause la centralité de proximité, il y a également circularité et donc co-évolution dans ce cas. En le localisant dans le diagramme de phase, ce régime est assez dispersé et rare, au contraire par exemple du régime 1 qui occupe une grande partie de l'espace pour une importance faible de la route ($w_{road} \leq 0.3$). Cela confirme que la co-évolution produite par le modèle est ponctuelle et non une caractéristique toujours vérifiée, mais qu'il est toutefois capable d'en générer dans des régimes particuliers.
}

\subsection{Discussion}{Discussion}


\bpar{
We have thus proposed a co-evolution model at the mesoscopic scale, based on a multi-modeling paradigm for the evolution of the network. The model is able to reproduce a certain number of observed situations at the first and second order, capturing thus a static representation of interactions between networks and territories. It also yields different dynamical causality regimes, being however less diverse than the simple model studied before: therefore, a more elaborated structure in terms of processes must be paid in flexibility of interaction between these. This suggests a tension between a ``static performance'' and a ``dynamical performance'' of models.
}{
Nous avons ainsi proposé un modèle de co-évolution à l'échelle mesoscopique, se basant sur un paradigme de multi-modélisation pour l'évolution du réseau. Le modèle est capable de reproduire un certain nombre de situations observées au premier et au second ordre, capturant une représentation statique des interactions entre réseaux et territoires. Il dégage également différents régimes dynamiques de causalité, en étant toutefois moins riche que le modèle simple étudié plus tôt : ainsi, une structure plus élaborée en termes de processus se paie en flexibilité d'interaction entre ceux-ci. Cela suggère une tension entre ``performance statique'' et ``performance dynamique'' des modèles.
}

%

\bpar{
An open question is to what extent a pure network model with preferential attachment for nodes would reproduce results close to what we obtained. The complex coupling between aggregation and diffusion (shown in~\ref{sec:densitygeneration}) could not be easily included, and the model could in any case not answer to questions on the coupling of the dynamics.
}{
Une question ouverte est dans quelle mesure un modèle de réseau pur avec attachement préférentiel des noeuds reproduirait des résultats proches des nôtres. Le couplage complexe entre agrégation et diffusion (montré en~\ref{sec:densitygeneration}) ne pourrait pas être inclut aisément, et le modèle ne pourrait dans tous les cas répondre à des problématiques concernant le couplage des dynamiques.
}


\stars

\bpar{
We have thus explored a co-evolution model based on morphogenesis that takes into account multiple processes for the evolution of the network. We studied its calibration on observed data at the first and the second order, and explored the causality regimes it produces.
}{
Nous avons ainsi exploré un modèle de co-évolution se reposant sur la morphogenèse et prenant en compte de multiples processus d'évolution du réseau. Nous avons étudié sa calibration sur données observées au premier et second ordre, et exploré les régimes de causalité qu'il produit.
}

\bpar{
We propose now a last entry into co-evolution at the mesoscopic scale, by developing a model that considerably complexifies the influence of the territory on the network, by taking into account governance processes.
}{
Nous proposons à présent une dernière incursion dans la co-évolution à l'échelle mesoscopique, en développant un modèle qui complexifie considérablement l'influence du territoire sur le réseau, en prenant en compte des processus de gouvernance. 
}

\stars

%

\section{Co-evolution and governance}{Co-évolution et gouvernance}

\label{sec:lutecia}



\bpar{
This section aims at giving directions for a more complex modeling of co-evolution, still at the mesoscopic scale. We have seen in~\ref{sec:networkterritories} that governance processes correspond to a level that intrinsically couples networks and territories: collective decisions concern jointly transportation, territories, and their articulation. We have moreover studied the particular case of a Mega-city Region (MCR) in~\ref{sec:casestudies}, and saw to what extent this context favoured a complexity of interactions. The emergence of MCR raises the question of the emergence of new modes of governance, more or less easy to implement as show the examples of Stuttgart and the Rhin-Rhur metropolitan areas according to~\cite{lenechet2017peupler}.
}{
Cette section se propose de donner des pistes vers une modélisation plus complexe de la co-évolution, toujours à l'échelle mesoscopique. Nous avons vu en~\ref{sec:networkterritories} que les processus de gouvernance relevaient d'un niveau qui couple intrinsèquement les réseaux et les territoires : les décisions collectives portent à la fois sur les transports, les territoires et leur articulation. Nous avons par ailleurs étudié le cas particulier d'une méga-région urbaine (MCR) en~\ref{sec:casestudies}, et vu dans quelle mesure ce contexte était propice à une complexité des interactions. L'émergence des MCR soulève la question de l'émergence de nouveaux modes de gouvernance, plus ou moins aisés à mettre en place comme en témoignent selon~\cite{lenechet2017peupler} les exemples de la métropole de Stuttgart et de la métropole Rhin-Rhur.
}

\bpar{
We develop therefore here a co-evolution model at the scale of a MCR, which aims in particular at endogenizing some processes of governance of the transportation network. This model extends in particular the one introduced by~\cite{le2010approche} which was then developed by~\cite{lenechet:halshs-00674059}.
}{
Nous développons donc ici un modèle de co-évolution à l'échelle d'une MCR, qui vise en particulier à endogénéiser certains processus de gouvernance du réseau de transport. Ce modèle étend en particulier celui introduit par~\cite{le2010approche} puis développé par~\cite{lenechet:halshs-00674059}.
}


\subsection{Context}{Contexte}

\subsubsection{Mega-city regions and Gouvernance}{Mega-régions urbaines et gouvernance}

\bpar{
We recall that a mega-city region is a network of highly connected cities in terms of economic and population flows, forming a polycentric region~\cite{hall2006polycentric}. It is the last ``urban regime'' which emerged within systems of cities, and it could be a more plausible trajectory for large urban agglomerates than always larger monocentric cities. \cite{neuman2009futures} point out that the future sustainability of these MCR will be closely linked to their ability to \emph{learn} new governance schemes, in the sense of an increased adaptability and flexibility of governance processes. \cite{innes2010strategies} suggest also that strategies implying self-organisation through the dialogue between stakeholders is a path to tackle the complexity of governing a MCR. We propose in the following to partly answer this question of the link between governance structure and evolution of the MCR, through the model we will develop.
}{
Nous rappelons qu'une méga-région urbaine est un réseau de villes fortement connecté en termes de flux économiques et de population, formant une région polycentrique~\cite{hall2006polycentric}. Il s'agit du dernier ``régime urbain'' qui a émergé au sein des systèmes de villes, et il pourrait s'agir d'une trajectoire plus plausible que des villes monocentriques toujours plus grandes pour les agglomérats urbains considérables. \cite{neuman2009futures} soulignent que la soutenabilité future de ces MCR sera intimement liée à leur capacité à \emph{apprendre} de nouveaux schémas de gouvernance, au sens d'une adaptabilité et flexibilité accrue des processus de gouvernance. \cite{innes2010strategies} suggèrent par ailleurs que des stratégies impliquant auto-organisation par le dialogue entre acteurs sont un moyen de répondre efficacement à la complexité de la gouvernance d'une MCR. Nous proposons par la suite de répondre partiellement à cette question du lien entre structure de gouvernance et évolution de la MCR, par l'intermédiaire du modèle que nous développons. 
}

\subsubsection{Modeling co-evolution with governance processes}{Modélisation de la co-évolution par des processus de gouvernance}

\bpar{
The role of governance processes in models coupling the evolution of transportation network with the evolution of land-use has already been investigated from different points of view in modeling approaches.
}{
Le rôle des processus de gouvernance dans les modèles couplant l'évolution des réseaux de transport à l'évolution de l'usage du sol a déjà été considéré selon différents points de vue dans les approches de modélisation.
}

\paragraph{Network growth}{Croissance du réseau}

\bpar{
\cite{li2016integrated} couples a network investment model with a traffic and localization model, and show that the obtained steady state configurations outperform an operational research approach to network design in terms of overall accessibility.
}{
\cite{li2016integrated} couple un modèle d'investissement de réseau avec un modèle de traffic et de localisation, et montre que les solutions stationnaires obtenues sont plus performantes qu'une approche en recherche opérationnelle pour la conception du réseau en termes d'accessibilité totale.
}

\bpar{
Concerning network growth only, \cite{jacobs2016transport} proposes a simulation model in which alternatives between plausible investments (by different investors) are evaluated with a discrete choice model which utility function takes into account returns on investment but also variables to optimize such as accessibility. It is applied to the growth of the Dutch railway in the 19th century, and shown to reproduce quite accurately the historical network.
}{
Concernant la croissance du réseau seule, \cite{jacobs2016transport} propose un modèle de simulation dans lequel les alternatives entre investissements plausibles (par des investisseurs différents) sont évalués avec un modèle de choix discrets dont la fonction d'utilité prend en compte les retours sur investissement mais également des variables à optimiser comme l'accessibilité. Il est appliqué à la croissance du réseau ferré néerlandais au 19ème siècle, et démontré capable de reproduire assez fidèlement le réseau historique. 
}

\paragraph{Modeling gouvernance}{Modéliser la gouvernance}

\bpar{
\cite{Xie2011} introduces a theoretical economic model of infrastructure investment. Two levels of governance, local and centralized are considered in the model. For the provision of new infrastructure that has to be split between two contiguous districts (space being one-dimensional), a game between governance agents determines both the level of decision and the attribution of the stock proportion to each district. Governments either want to maximize the aggregated utility (Pigovian government), or include explicit political strategies to satisfy a median voter. Numerical exploration of the model show that these processes are equivalent to compromises between cost and benefits, and that the level of governance depends on the state of the network.
}{
\cite{Xie2011} introduit un modèle économique théorique d'investissement dans les infrastructures. Deux niveaux de gouvernance, local et centralisé, sont considérés dans le modèle. Pour la provision d'une nouvelle infrastructure qui doit être partagée entre deux zones contiguës (l'espace étant à une dimension), un jeu entre des agents de gouvernance détermine à la fois le niveau de décision et l'attribution de proportion du stock à chaque zone. Les gouvernements veulent soit maximiser l'utilité agrégée (gouvernement Pigovien), ou bien inclure des stratégies politiques explicites pour satisfaire un électeur médian. L'exploration numérique du modèle montre que ces processus sont équivalents à des compromis entre coûts et bénéfices, et que le niveau de gouvernance effectif dépend de l'état du réseau.
}

\bpar{
\cite{xie2011governance} proposes a simpler version of this model on the governance side but coupled with a more realistic travel side : it couples on a synthetic growing network a traffic model with a pricing model and an investment model, and show that under the assumption of centralization, an equilibrium between demand and network performance can be reached, but that investments are not efficient on the long run, with a higher loss for decentralized investments.
}{
\cite{xie2011governance} propose une version plus simple de ce modèle du point de vue de la gouvernance mais couplé à un modèle de transport plus réaliste : il couple sur un réseau synthétique croissant un modèle de traffic avec un modèle de prix et un modèle d'investissement, et montre que sous l'hypothèse d'une centralisation, un équilibre entre la demande et la performance du réseau peut être atteint, mais que les investissements ne sont pas efficients sur le long terme, avec une perte plus importante pour les investissements décentralisés.
}

\bpar{
We will be positioned in a logic close to the first model for the role of the governance structure, and close to the second for the precision of the inclusion of space.
}{
Nous nous placerons dans une logique proche du premier modèle dans le rôle de la structure de gouvernance, et proche du second dans la précision de prise en compte de l'espace.
}

\paragraph{Game theory}{Théorie des jeux}

\bpar{
Some of these models, in particular \cite{Xie2011}, are based on game theory to model the behavior of stakeholders. It has already been widely applied for modeling in social and political sciences to questions dealing with cognitive interacting agents with individual interests~\cite{ordeshook1986game}. \cite{abler1977spatial} (p.~487) formulate a location decision problem for coffee farms on Kilimanjaro as a game combining a production strategy and a location strategy (fixing then the environmental conditions). This framework has furthermore already been used in transportation investment studies, such as e.g. in~\cite{Roumboutsos2008209} which use the notion of Nash equilibrium to understand choices of public or private operators concerning the integration of their system in the broader mobility system. We will use game theory paradigms to integrate governance in a simple way in our model.
}{
Certains de ces modèles, en particulier \cite{Xie2011}, se fondent sur la théorie des jeux pour modéliser le comportement des acteurs. Celle-ci a déjà été largement appliquée à des questions de modélisation en sciences sociales ou politiques pour des problèmes impliquant des agents cognitifs en interaction avec des intérêts individuels~\cite{ordeshook1986game}. \cite{abler1977spatial} (p.~487) formulent un problème de décision de localisation pour des fermes de café au Kilimanjaro comme un jeu combinant une stratégie de production et une stratégie de localisation (fixant alors les conditions environnementales). Ce cadre a par ailleurs été utilisé pour étudier les investissements en termes de transports, comme par exemple par \cite{Roumboutsos2008209} qui utilisent la notion d'équilibre de Nash pour comprendre les choix des opérateurs publics ou privés quant à l'intégration de leur système dans le système de mobilité plus global. Nous utiliserons des paradigmes de théorie des jeux pour intégrer la gouvernance de manière simple dans notre modèle.
}


\bpar{
The aim of this section is thus to follow these different models, and to propose a co-evolution model in which network growth is integrated in an endogenous way, through the modeling of implied governance processes.
}{
Le but de cette section est donc de se placer dans la lignée de ces différents modèles, et de proposer un modèle de co-évolution au sein duquel la croissance du réseau est intégrée de manière endogène, par la modélisation des processus de gouvernance impliqués.
}

\subsection{The Lutecia Model}{Le Modèle Lutecia}

\bpar{
We now describe the Lutecia model\footnote{The name comes from an acronym linked to its structure which is detailed in the following. Naming models is a delicate operation since it induces a kind of reification or even personification, in any case can be seen as a kind of fetichism. It can potentially perturb the role of the model within the knowledge production process and make the model an end in itself. We are convinced that an endogenous naming through the uses of the model by the community is more appropriate. We make here an exception given the particular story of its genesis.}, in its general structure, and then in the specification we will later develop.
}{
Nous décrivons à présent le modèle Lutecia\footnote{L'appellation est un acronyme lié à sa structure détaillée par la suite. Nommer les modèles est une opération délicate puisqu'elle induit une certaine réification voire personification, dans tous les cas relève d'un certain fétichisme. Celle-ci peut potentiellement perturber la place du modèle au sein du processus de production de connaissance et faire du modèle une fin en soi. Nous sommes convaincu qu'une dénomination endogène via les usages du modèles par la communauté est plus approprié. Nous faisons ici une exception vu l'histoire particulière de sa genèse.}, dans sa structure générale, puis dans la spécification que nous développerons par la suite.
}

\subsubsection{Global model structure}{Structure globale du modèle}

\bpar{
The model couples in a complex way a module for land-use evolution with a module for transportation network growth. Submodels (or submodules), detailed in the following, include in particular a governance module that rules processes of network evolution. The most important feature of the Lutecia model is the inclusion of an endogenous infrastructure provision submodel, based on iterative increases in accessibility, within a Luti model.
}{
Le modèle couple de manière complexe un module pour l'évolution de l'usage du sol à un module de croissance de réseau de transport. Les sous-modèles (ou modules), détaillés par la suite, incluent en particulier un modèle de gouvernance pour régir l'évolution du réseau. L'inclusion d'un modèle endogène de provision d'infrastructures basé sur les augmentation itératives de l'accessibilité, au sein d'un modèle Luti, consiste en la contribution principale du modèle Lutecia.
}

\bpar{
The accessibility, that we will take here as a potential of access of actives to employments, is a cornerstone of the model. Indeed, micro-economic agents will relocate in order to maximize their accessibility, whereas new transportation infrastructure decisions will be taken by governance agents based on a criteria of maximization of accessibility increase in their area.
}{
L'accessibilité, que nous prendrons ici comme un potentiel d'accès des actifs aux emplois, est au coeur du modèle. En effet, les agents micro-économiques se relocalisent afin de maximiser leur accessibilité, tandis que les décisions de nouvelles infrastructures de transport sont prises par des agents de gouvernance selon un critère de maximisation de l'augmentation d'accessibilité dans leur zone. 
}

\bpar{
In its more general structure, the Lutecia model is composed by five sub-models, of which only three will be studied here for simplicity reasons. The sub-models are the following :
\begin{itemize}
\item LU stands for Land Use module : it proceeds to the re-localization of actives and employments given current conditions of accessibility.
\item T stands for Transport module : it computes the transportation conditions such as flows and congestion in the urban region.
\item EC stands for Evaluation of Cooperation module : it evaluates the agent or agents that will proceed to build a new infrastructure. 
\item I stands for Infrastructure provision module : it determines the localization of the new transportation infrastructure, based on a criteria of accessibility maximization.
\item A stands for Agglomeration economies module : it evaluates the productivity of firms, depending on the accessibility to employments.
\end{itemize}
We will in the following study the coupling between the LU-EC-I sub-models: we assume at the first order no significant effect of congestion, and thus no role of transport modeling; and furthermore consider simple assumptions for economics and neglect agglomeration economies.
}{
Dans sa structure la plus générale, le modèle Lutecia est composé de cinq sous-modèles, parmi lesquels nous n'en développerons que trois ici pour des raisons de simplicité. Les sous-modèles sont les suivants :
\begin{itemize}
	\item LU correspond à l'usage du sol : il opère la relocalisation des actifs et des emplois étant donné les conditions courantes d'accessibilité.
	\item T correspond à Transport : il évalue les conditions de transport (flux, congestion) dans la région urbaine.
	\item EC correspond à l'évaluation de la coopération : il évalue le ou les agents qui procéderont à la construction d'une nouvelle infrastructure.
	\item I correspond à provision d'infrastructure : il détermine la localisation de la nouvelle infrastructure de transport en fonction d'un critère de maximisation d'accessibilité.
	\item A correspond aux agglomérations d'économie : il évalue la productivité des firmes, selon l'accessibilité aux emplois.
\end{itemize}
Nous étudierons par la suite le couplage entre les sous-modèles LU-EC-I : nous supposons au premier ordre pas d'effet significatif de la congestion, et donc pas de rôle de la modélisation du transport ; et par ailleurs prenons des hypothèses simples sur le plan économique et négligeons les agglomérations d'économie.
}

\bpar{
Different time scales are included in the model: a short scale, corresponding to daily mobility that yields flows in the transportation network and to firms productivity (modules T and A); an intermediate scale for residential and firms dynamics (module LU) ; and a long time scale for the evolution of the network (modules EC and I). Levels of stochasticity are considered accordingly: the smallest scales have deterministic dynamics whereas the longer exhibits randomness.
}{
Des échelles de temps imbriquées sont incluses dans le modèle : une échelle courte, correspondant à la mobilité quotidienne qui produit les flux dans le réseau de transport et aux productivités des entreprises (modules T et A) ; une échelle intermédiaire pour les dynamiques de localisation des actifs et emplois (module LU) ; et une longue échelle de temps pour l'évolution du réseau (modules EC et I). Les niveau de stochasticité sont pris en conséquence : les échelles les plus petites ont des dynamiques déterministes tandis que la plus longue présente un comportement aléatoire.
}

\subsubsection{Detailed description of the model}{Description détaillée du modèle}

\paragraph{Description of the environment}{Description de l'environnement}

\bpar{
The mega-city region is modeled with a two level spatial zoning. The world is composed by a lattice of patches, that are the basic units to quantify land use. We assume that each patch $k$ is characterized at time $t$ by its resident actives $A_k(t)$ and number of employments $E_k(t)$. At a higher level, the MCR is decomposed into administrative areas that correspond to the city governance levels, to which we attribute $M$ abstract agents called \emph{mayors}: $M_k$ gives thus the administrative area to which each patch belongs. We assume furthermore the existence of a global governance agent that correspond to a regional authority at the level of the MCR.
}{
La méga-région Urbaine est modélisée avec un zonage spatial à deux niveaux. L'environnement du modèle est composé par une grille, dont les cellules sont les unités élémentaires pour quantifier l'usage du sol. Nous supposons que chaque cellule $k$ est caractérisée au temps $t$ par le nombre d'actifs y résidant $A_k(t)$ et son nombre d'emplois $E_k(t)$. À un niveau supérieur, la MCR est décomposée en unités administratives qui correspondent au niveau de gouvernance des villes, auxquelles sont attribués $M$ agents abstraits appelés \emph{maires} : $M_k$ désigne ainsi la zone administrative à laquelle chaque cellule appartient. Nous supposons de plus l'existence d'un agent de gouvernance global qui correspond à une autorité typiquement régionale, au niveau de la MCR.
}

\bpar{
On top of this patch-level land-use and governance setup, we introduce a transportation network $G = (V,E)$ localized in space by its nodes coordinates $(x_v,y_v)$, and characterized by a speed $v_G$ relative to movements in the euclidian space. Assuming that the network can be taken anywhere on each link, it unequivocally induces a geographical travel-time distance that we describe by the shortest path distance matrix between each patch $D = (d_{k,k'}(t))$. The accessibility of actives to employments is then defined for each patch as a Hansen accessibility with a decay of distance $\lambda$ capturing typical commuting range, by
}{
De manière complémentaire à cette configuration d'usage du sol et de gouvernance, nous introduisons un réseau de transport $G = (V,E)$ localisé dans l'espace par les coordonnées de ses noeuds $(x_v,y_v)$, et caractérisé par une vitesse $v_G$ relative aux déplacements dans l'espace euclidien. Sous l'hypothèse que le réseau peut être rejoint à tout endroit sur les liens, il induit de manière univoque une distance-temps géographique, que nous décrivons par la matrice des plus courts temps entre chaque cellule $D = (d_{k,k'}(t))$. L'accessibilité des actifs aux emplois est alors définie pour chaque cellule comme une accessibilité de Hansen, avec un paramètre de décroissance de la distance $\lambda$ qui capture un potentiel d'accès des actifs aux emplois, par
}

\begin{equation}
X^{(A)}_k = A_k\cdot \sum_{k'} E_{k'} \exp{\left(-\lambda \cdot d_{k,k'}\right)}
\end{equation}

\bpar{
The accessibility of employments to actives is defined in a similar manner. Dynamics are taken in a discrete way: $t \in \{t_0 = 0 , \ldots , t_f\}$, with time ticks corresponding to a time scale at which land use typically evolves, i.e. 5 to 10 years. We take thus a slower speed for the evolution of the network which will be constructed by segments at each time step, whereas land-use will be considered as being in equilibrium at the scale of the decade, in consistence with the frame developed in chapter~\ref{ch:thematic}.
}{
L'accessibilité des emplois aux actifs est définie de manière similaire. La dynamique est considérée de façon discrete : $t \in \{t_0 = 0 , \ldots , t_f\}$, avec les pas de temps correspondant à une échelle à laquelle l'usage du sol évolue en moyenne, i.e. de 5 à 10 ans. Nous prenons ainsi une vitesse plus lente pour l'évolution du réseau qui se construira par tronçons à chaque pas de temps, tandis que l'usage du sol sera considéré comme en équilibre à l'échelle de la décade, en cohérence avec le cadre développé en chapitre~\ref{ch:thematic}.
}




\paragraph{Evolution of land-use}{Évolution de l'usage du sol}

\bpar{
For the land-use module, the model is based on the Lowry model~\cite{lowry1964model}. We assume that residential/employments relocations are at equilibrium at the time scale of a tick. In comparison, the evolution of transportation infrastructure is much slower~\citep{wegener2004land}\footnote{We do not consider land values, rents or transportation costs, that are the core of models in Urban Economics such as the Alonso and Fujita models for example (see \cite{lemoy2017exploring} for a recent agent-based approach to these).}. Actives and Employments relocate given some utilities that take into account both accessibility and the urban form. Indeed, one of the drivers of Urban Sprawl may be interpreted as a repulsion of residents by density. To aggregate both effects in a simple way, we take a Cobb-douglas function for utilities of actives and employments
}{
Pour le module d'usage du sol, le modèle s'inspire du modèle de Lowry~\cite{lowry1964model}. Les relocalisations d'une proportion fixe d'actifs et d'emplois sont supposées à l'équilibre à l'échelle d'un pas de temps. En comparaison, l'évolution de l'infrastructure de transport est largement plus lente~\cite{wegener2004land}\footnote{Nous ne considérons pas ici les valeurs foncières, les loyers ou les coûts de transport, qui sont au coeur des modèles d'économie urbaine comme le modèle d'Alonso ou de Fujita par exemple (voir \cite{lemoy2017exploring} pour une approche récente multi-agents de ceux-ci).}. Les actifs et les emplois se relocalisent selon des utilités qui prennent en compte à la fois l'accessibilité et la forme urbaine. En effet, l'un des moteurs de l'étalement urbain peut être interprété comme une répulsion des résidents par la densité. Pour agréger les deux effets, nous prenons une fonction de Cobb-Douglas pour l'utilité
}

\begin{equation}\label{eq:utility}
U_k^{(A)} = {X_k^{(A)}}^{\gamma_A}\cdot {F_k^{(A)}}^{1-\gamma_A}
\end{equation}

\bpar{
what is equivalent to have a linear aggregation of the logarithm of explicative variables. Employments follow an analog expression with a dedicated weight parameter $\gamma_E$. Here the utility is simply influenced only by accessibility and by an indicator of local urban form called \emph{form factor}, given in the case of actives by $F_k^{(A)} = \frac{1}{A_k \cdot E_k}$, meaning that population is repulsed by density. The combination of the positive effect of accessibility to the negative effect of density produces a tension between contradictory objectives allowing a certain level of complexity already in the land-use sub-model alone. The form factor for jobs is taken as $F_k^{(E)}=1$ for the sake of simplicity and following the fact that jobs can aggregate far more than dwellings. 
}{
ce qui est équivalent à une agrégation linéaire du logarithme des variables explicatives. Les emplois suivent une expression analogue avec un paramètre de poids spécifique $\gamma_E$. L'utilité est influencée ici uniquement par l'accessibilité et par un indicateur de forme urbaine locale nommé \emph{facteur de forme}. Nous le définissons dans le cas des actifs par $F_k^{(A)} = \frac{1}{A_k \cdot E_k}$, ce qui signifie que la population est repoussée par la densité. La combinaison de l'effet positif de l'accessibilité à celui négatif de la densité produit une tension entre des objectifs contradictoires, permettant un certain niveau de complexité déjà dans le sous-modèle d'usage du sol seul. Le facteur de forme pour les emplois est pris comme $F_k^{(E)}=1$ pour simplifier et suivant la logique que les emplois peuvent s'agréger bien plus que les logements.
}

\bpar{
Relocations are then done deterministically following a discrete choice model, which yields the value of actives at the next step as
}{
Les relocalisations sont ensuite faites de manière déterministe suivant un modèle de choix discret, qui donne les valeurs des actifs à l'étape suivante comme
}

\begin{equation}\label{eq:discretechoicereloc}
A_i(t+1) = (1 - \alpha) A_i(t) + \alpha \cdot \left(\sum_j{A_j(t)}\right)\cdot\frac{\exp{(\beta \tilde{U}_i(A))}}{\sum_j{\exp{(\beta \tilde{U}_j(A))}}}
\end{equation}

\bpar{
where $\beta$ is the Discrete Choice parameter that can be interpreted as a ``level of randomness''\footnote{When $\beta \rightarrow 0$, all destination patches have an equal probability from any origin patch, whereas $\beta \rightarrow \infty$ gives fully deterministic behavior towards the patch with the best utility.} and $\tilde{U}_i$ are the utilities normalized by the maximal utility. $\alpha$ is the fixed fraction of actives relocating. Employments follow again a similar expression.
}{
où $\beta$ est le paramètre de choix discrets qui peut être interprété comme un ``niveau d'aléatoire''\footnote{Quand $\beta \rightarrow 0$, toutes les cellules de destination ont une probabilité égale depuis l'ensemble des cellules d'origine, tandis que $\beta \rightarrow \infty$ donne un comportement totalement déterministe vers la cellule avec meilleure utilité.} et $\tilde{U}_i$ sont les utilités normalisées par l'utilité maximale. $\alpha$ est la fraction fixe d'actif se relocalisant. Les emplois suivent une expression similaire.
}

\subsubsection{Network evolution : governance process}{Évolution du réseau : processus de gouvernance}

\paragraph{Assumptions}{Hypothèses}

\bpar{
The governance part of the model has the following rationale :
\begin{itemize}
\item Three levels of governance are included, namely a central actor (the region, or regional government), local actors (municipalities) acting individually, and local actors cooperating what constitutes an intermediate level.
\item Assuming a new infrastructure is to be built, the planning can be either from top-down decision (region) or from the bottom-up (local actors). We make the assumption that the processes behind the determination of the level of decision are far too complex (since they are generally political processes) to be taken into account in the model. This step is thus determined exogenously following an uniform law given a parameter.
\item If the decision is taken at the local level, negotiations between actors occur. We assume that
\begin{itemize}
\item the initiator of the new infrastructure can be any of the local actors, but richer cities will have more chance to built;
\item negotiations for possible collaboration are only done between neighbor cities, what is related to the medium range of infrastructure segments considered;
\item for this reason, and as $n$-players games have been shown to exhibit a chaotic behavior~\cite{2016arXiv161208111S} when $n$ increases, we consider negotiations between two actors only. The probability of cooperation that are endogenously determined can be furthermore directly interpreted.
\end{itemize}
\item For the sake of simplicity, the total stock of infrastructure built at one governance time step is constant, and decision times are also fixed\footnote{See the discussion for the implications of that hypothesis and possible relaxations.}.
\end{itemize}
}{
Le sous-modèle pour la gouvernance suit les hypothèses suivantes.
\begin{itemize}
	\item Trois niveaux de gouvernance sont inclus, qui sont un acteur central (la région, ou le gouvernement régional), les acteurs locaux (municipalités) qui agissent seuls, et les acteurs locaux qui coopèrent.
	\item Sous l'hypothèse qu'une nouvelle infrastructure doit être construite, la planification peut être soit par le haut (région) soit par le bas (acteurs locaux). Nous supposons que les processus derrière la détermination du niveau de décision sont bien trop complexes (puisqu'il incluent généralement des processus politiques) pour être pris en compte par le modèle. Cette étape est donc déterminée de manière exogène selon une loi uniforme à paramètre fixe.
	\item Si la décision est prise au niveau local, des négociations entre les acteurs ont lieu. Les concernant, nous supposons que :
	\begin{itemize}
		\item l'initiateur de la nouvelle infrastructure peut être n'importe quel acteur local, mais les villes riches ont plus de chance de construire ;
		\item les négociations pour des possibles collaborations n'ont lieu qu'entre acteurs voisins, ce qui est en cohérence avec des segments d'infrastructure de longueur moyenne ;
		\item pour cette raison, et d'autant plus que les jeux à $n$ joueurs présentent des comportements chaotiques quand $n$ augmente~\cite{2016arXiv161208111S}, nous ne considérons des négociations qu'entre deux acteurs uniquement. De plus, la probabilité de coopération endogène peut alors être directement interprétée.
	\end{itemize}
	\item Pour rester simple, le stock total d'infrastructure construit à un pas de temps de gouvernance est constant, et les temps de décision sont également fixés\footnote{Voir également la discussion pour de possibles relaxations de ces hypothèses.}.
\end{itemize}
}



\paragraph{Network evolution}{Évolution du réseau}

\bpar{
The workflow for transportation network development is the following :
\begin{enumerate}
\item At each time step, 2 new road segments of length $l_r$ are built. The choice between local and global is done by a uniform draw with probability $\xi$. In the case of local building, roads are attributed successively to mayors (one road maximum per mayor) with probabilities $\xi_i$ which are proportional to the number of employments of each, what means that richer areas will get more roads.
\item Areas building a road will enter negotiations. Possible strategies for players (negotiating areas, $i=0,1$, the strategies being written $S_i$) are to not collaborate ($NC$), i.e. develop his road segment alone, and to collaborate ($C$), i.e. wanting to develop conjointly. Strategies are chosen simultaneously (non-cooperative game), in a random way according probabilities determined as detailed below. For $(C,NC)$ and $(NC,C)$ combinations, roads are built separately. For $(NC,NC)$ both act as alone, and for $(C,C)$ a common development is done.
\item Depending on the level of governance and the strategies chosen, the corresponding optimal infrastructures are build.
\end{enumerate}
}{
Les étapes pour le développement du réseau de transport sont les suivantes.
\begin{enumerate}
	\item À chaque pas de temps, 2 nouveaux segments de route de longueur $l_r$ sont construits. Le choix entre le niveau local et global est déterminé par un tirage uniforme avec une probabilité $\xi$. Dans le cas d'une construction locale, les routes sont attribuées successivement aux maires (une route par maire au maximum) avec des probabilités $\xi_i$ qui sont proportionnelles au nombre d'emplois de chacun, ce qui signifie que les zones plus riches auront plus de routes.
	\item Les zones devant construire chacune une route entrent en négociations. Les stratégies possibles pour les acteurs (zones en négociation, $i=0,1$, les stratégies étant notées $S_i$) sont de ne pas collaborer ($NC$), c'est-à-dire développer son tronçon de réseau seul, et de collaborer ($C$), c'est-à-dire vouloir développer conjointement. Les stratégies sont choisies simultanément (jeu non-coopératif), de manière aléatoire selon des probabilités déterminées comme détaillé ci-dessous. Pour les combinaisons $(C,NC)$ et $(NC,C)$, les routes sont construites séparément. Pour $(NC,NC)$ les deux agissent séparément et pour $(C,C)$ un développement commun est mené.
	\item Selon le niveau de gouvernance et les stratégies choisies, l'infrastructure optimale correspondante est construite.
\end{enumerate}
}

\paragraph{Evaluation of cooperation}{Évaluation de la coopération}

\bpar{
We detail now the way the cooperation probabilities are established. We denote $Z^{\ast}_i(S_0,S_1)$ the optimal infrastructure for area $i$ with $(S_0,S_1)\in \{(NC,C),(C,NC),(NC,NC)\}$ which are determined by an heuristic in each zone separately (see implementation details), and $Z^{\ast}_C$ the optimal common infrastructure computed with a 2 segments infrastructure on the union of both areas. It corresponds to the case where both strategies are $C$. Marginal accessibilities for area $i$ and infrastructure $Z$ is defined as $\Delta X_i(Z)=X^Z_i - X_i$. We introduce construction costs, noted $I$ for a road segment, assumed spatially uniform. We furthermore introduce a cost of collaboration $J$ that corresponds to a shared cost for building a larger infrastructure.
}{
Détaillons à présent la manière dont les probabilités de coopération sont établies. Nous notons $Z^{\ast}_i(S_0,S_1)$ l'infrastructure optimale en termes de gain d'accessibilité pour la zone $i$ avec $(S_0,S_1)\in \{(NC,C),(C,NC),(NC,NC)\}$ qui sont déterminées de manière heuristique pour chaque zone séparément (voir détails d'implémentation), et $Z^{\ast}_C$ l'infrastructure optimale commune calculée sur l'union des deux zones avec une infrastructure composée de deux segments élémentaires. Cette dernière correspond au cas où les deux stratégies sont $C$. Les accessibilités marginales pour la zone $i$ et l'infrastructure $Z$ sont définies comme $\Delta X_i(Z)=X^Z_i - X_i$. Nous introduisons des coûts de construction, notés $I$ pour un segment de route, supposés uniformes dans l'espace. Nous introduisons de plus un coût de collaboration $J$ qui correspond à un coût partagé pour construire une infrastructure plus grande.
}

\bpar{
The determination of probabilities defining mixed strategies is based on the payoff matrix, which gives is the value of utility gains for each players and each possible decision configuration. The payoff matrix of the game is the following, with $\kappa$ a normalization constant (``price of accessibility''), and the players being written $i\in \{ 0;1\}$ (such that $1-i$ denotes the player opposed to $i$)
}{
La détermination des probabilités donnant la composition des stratégies mixtes se base sur la matrice de gain, qui donne les gains d'utilité pour chaque joueur et chaque combinaison de décisions. La matrice de gain du jeu est la suivante, avec $\kappa$ une constante de normalisation (``prix de l'accessibilité''), et les joueurs étant notés $i\in \{ 0;1\}$ (tel que $1-i$ désigne le joueur opposé à $i$)
}

\begin{center}
\begin{tabular}{ |c|c|c| } 
 \hline
 0 $|$ 1  & C & NC \\ \hline
 C & $U_i = \kappa \cdot \Delta X_i(Z^{\ast}_C) - I - \frac{J}{2}$
   & $\begin{cases}U_0 = \kappa \cdot \Delta X_0(Z^{\ast}_0)-I \\U_1 = \kappa \cdot \Delta X_1(Z^{\ast}_1)-I - \frac{J}{2}\end{cases}$ \\ \hline
 NC & $\begin{cases}U_0 = \kappa \cdot \Delta X_0(Z^{\ast}_0)-I - \frac{J}{2}\\U_1 = \kappa \cdot \Delta X_1(Z^{\ast}_1)-I\end{cases}$
   & $U_i = \kappa \cdot \Delta X_i(Z^{\ast}_i) - I$ \\
 \hline
\end{tabular}
\end{center}

\bpar{
To simplify, we assume the cost parameters dimensioned as an accessibility what is equivalent to have $\kappa = 1$. We will furthermore see that since only accessibility differentials are determining, the construction cost $I$ does finally not play any role. This payoff matrix is used in two games corresponding to complementary processes:
\begin{itemize}
	\item the coordination game in which players have a mixed strategy, and for which we consider the Nash equilibrium\footnote{A Nash equilibrium is a strategy point in a discrete non-collaborative game for which no player can improve his gain by changing his strategy~\cite{ordeshook1986game}.} for corresponding probabilities, which implies a competition between players;
	\item an heuristic according to which players take their decision following a discrete choice model. It implies only a maximization of the utility gain and an indirect competition only.
\end{itemize}
}{
Pour simplifier, nous supposerons les paramètres de coût redimensionnés à une accessibilité ce qui revient à avoir $\kappa = 1$. Nous verrons par ailleurs que seuls des différentiels d'utilité étant déterminants, le coût de construction $I$ ne joue finalement pas de rôle. Cette matrice de gain est utilisée dans deux jeux traduisant des processus complémentaires :
\begin{itemize}
	\item le jeu de coordination dans lequel les joueurs ont une stratégie mixte, et pour lequel nous considérons l'équilibre de Nash\footnote{Un équilibre de Nash est un point de stratégies dans un jeu discret non-collaboratif pour lequel aucun joueur ne peut améliorer son gain en changeant sa stratégie~\cite{ordeshook1986game}.} pour les probabilités correspondantes, qui implique une compétition entre les joueurs ;
	\item une heuristique selon laquelle les joueurs prennent leur décision suivant un modèle de choix discrets. Celle-ci implique uniquement une maximisation du gain d'utilité et une compétition indirecte seulement.
\end{itemize}
}

\bpar{
We write $p_i = \Pb{S_i = C}$ the probability of each player to collaborate.
}{
Notons $p_i = \Pb{S_i = C}$ la probabilité de chaque joueur de collaborer.
}

\paragraph{Nash equilibrium}{Equilibre de Nash}

\bpar{
We can solve the mixed strategy Nash Equilibrium for this coordination game in all generality. We detail the computation in Appendix~\ref{app:sec:lutecia}. By writing $U_i(S_i,S_{1-i})$ the full payoff matrix, we have the expression of probabilities
}{
L'équilibre de Nash à stratégie mixte pour ce jeu non-coopératif peut être obtenu en toute généralité. Nous détaillons le calcul en Annexe~\ref{app:sec:lutecia}. En écrivant $U_i(S_i,S_{1-i})$ la matrice de gain complète, on a l'expression des probabilités
}

\[
p_{1-i} = - \frac{U_i(C,NC) - U_i(NC,NC)}{\left(U_i(C,C) - U_i(NC,C)\right) - \left(U_i(C,NC) - U_i(NC,NC)\right)}
\]

\bpar{
What gives with the expression of utilities previously given,
}{
Ce qui donne avec les expressions des utilités données précédemment,
}

\begin{equation}
p_i = \frac{J}{\Delta X_{1 - i}{Z^{\star}_{C}} - \Delta X_{1 - i}{Z^{\star}_{1 - i}}}
\end{equation}

\bpar{
This expression can be interpreted the following way: in this competitive game, the likelihood of a player to cooperate will decrease as the other player gain increases, and somehow counterintuitively, will increase as collaboration cost increases. The realism of this assumption must thus be moderated, and we can assume that in practice the equilibrium is never reached.
}{
Cette expression peut être interprétée de la façon suivante : dans ce jeu compétitif, la chance qu'un joueur coopère décroit quand le gain de l'autre joueur augmente, et d'une certaine manière contre-intuitif, s'accroit quand le coût de collaboration augmente. Le réalisme de cette hypothèse est donc à modérer, et nous pouvons supposer que l'équilibre n'est en pratique jamais atteint.
}

\bpar{
It also forces feasibility conditions on $J$ and accessibility gains to keep a probability. These are
\begin{itemize}
	\item $ J \leq \Delta X_{1 - i}(Z^{\star}_{C}) - \Delta X_{1 - i}(Z^{\star}_{1 - i})$, what can be interpreted as a cost-benefits condition, i.e. that the gain induced by the common infrastructure must be larger than the collaboration cost;
	\item $\Delta X_{1 - i}(Z^{\star}_{C}) \leq \Delta X_{1 - i}(Z^{\star}_{1 - i})$, i.e. that the gain induced by the common infrastructure must be positive.
\end{itemize}
}{
Cela impose également des conditions de faisabilité pour $J$ et les gains d'accessibilité pour conserver une probabilité. Celles-ci sont :
\begin{itemize}
	\item $ J \leq \Delta X_{1 - i}(Z^{\star}_{C}) - \Delta X_{1 - i}(Z^{\star}_{1 - i})$, qui s'interprète comme une condition coût-bénéfices, c'est-à-dire que le gain induit par l'infrastructure commune doit être supérieur au coût de collaboration ;
	\item $\Delta X_{1 - i}(Z^{\star}_{C}) \leq \Delta X_{1 - i}(Z^{\star}_{1 - i})$, c'est-à-dire que le gain induit par l'infrastructure commune doit être positif.
\end{itemize}
}

\paragraph{Discrete choice decisions}{Décisions par choix discrets}

\bpar{
Using the same utility functions, a random utility model for a discrete choice allows also to obtain expressions for probabilities. We have for player $i$ the utility differential between the choice $C$ and the choice $NC$ given by
}{
Avec les mêmes fonctions d'utilité, un modèle d'utilité aléatoire pour un choix discret permet également d'obtenir des expressions des probabilités. On a pour le joueur $i$ le différentiel d'utilité entre le choix $C$ et le choix $NC$ donné par
}

\[
U_i(C) - U_i(NC) = p_{1 - i} \left( \Delta X_{i}{Z^{\star}_{C}} - \Delta X_{i}{Z^{\star}_{i}}\right) - J
\]

\bpar{
Under the classical assumption of a model with a random utility distributed following a Gumbel law~\cite{ben1985discrete}, we have $\Pb{S_i=C} = \frac{1}{1 + \exp{[-\beta_{DC}(U_i(C) - U_i(NC))]}}$, where $\beta_{DC}$ is the discrete choice parameter (that we will fix at a large value $\beta_{DC} = 400$, by supposing a certain determinism at this level, since there is then a second random level).
}{
Sous l'hypothèse classique d'un modèle d'utilité aléatoire distribuée en loi de Gumbel~\cite{ben1985discrete}, on a $\Pb{S_i=C} = \frac{1}{1 + \exp{[-\beta_{DC}(U_i(C) - U_i(NC))]}}$, où $\beta_{DC}$ est le paramètre de choix discrets (que nous fixerons grand $\beta_{DC} = 400$, en supposant un certain déterminisme à ce niveau, puisqu'on a ensuite un deuxième niveau aléatoire). 
}

\bpar{
We substitute the expression of $p_{i-1}$ in the expression of $p_i$, what leads $p_i$ to verify the following equation
}{
On substitue l'expression de $p_{1-i}$ dans l'expression de $p_i$, ce qui conduit $p_i$ à vérifier l'équation suivante
}

\begin{equation}
p_i = \frac{1}{1 + \exp{\left(-\beta_{DC}\cdot \left(\frac{\Delta X_{i}{Z^{\star}_{C}} - \Delta X_{i}{Z^{\star}_{i}}}{1 + \exp{\left(- \beta_{DC}(p_i \cdot (\Delta X_{1 - i}(Z^{\star}_{C}) - \Delta X_{\bar{i}}(Z^{\star}_{1 - i})) - J)\right)}} - J \right)\right)}}
\end{equation}

\bpar{
We demonstrate (see Appendix~\ref{app:sec:lutecia}) that there always exists a solution $p_i \in [0,1]$, and we solve it numerically in the model to determine the probability to cooperate.
}{
Nous démontrons (voir Annexe~\ref{app:sec:lutecia}) qu'il existe toujours une solution $p_i \in [0,1]$, et nous la résolvons numériquement dans le modèle pour déterminer la probabilité de collaboration.
}

\paragraph{Random decision}{Décision aléatoire}

\bpar{
We also consider a baseline mechanism, which does not assume negotiations, but which in the case of a local decision draws randomly a mayor, following an uniform law with probabilities proportional to the number of employments of each.
}{
Nous considérons également un mécanisme de référence, qui ne suppose pas de négociations, mais qui dans le cas d'une décision locale tire un maire au hasard, selon une loi uniforme avec probabilités proportionnelles au nombres d'emplois de chaque.
}

\subsubsection{Model implementation}{Implémentation du modèle}

\bpar{
All model parameters are recalled in Table~\ref{tab:lutecia:parameters}. We give here only the parameters which have not been explicitly fixed previously, and these will be the privileged parameters on which the exploration and the application of the model will be done. The bound $\sqrt{2}\cdot K$ corresponds to the diagonal of the world, and the one for $J$ has been empirically fixed according to the values of the bound given previously.
}{
L'ensemble des paramètres du modèle est rappelé en Table~\ref{tab:lutecia:parameters}. Nous ne donnons ici que les paramètres qui n'ont pas été fixés explicitement précédemment, et il s'agit des paramètres privilégiés sur lesquels l'exploration et l'application du modèle sera faite. La borne $\sqrt{2}\cdot K$ correspond à la diagonale du monde, et celle pour $J$ a été fixée empiriquement selon les valeurs observées de la borne donnée précédemment.
}

\begin{table}
\caption[Summary of LUTECIA model parameters]{\textbf{Summary of Lutecia model parameters.} We also give the corresponding processes, typical bounds of the variation range and their default values.\label{tab:lutecia:parameters}}
\bpar{
\begin{tabular}{|c|c|c|c|c|c|}
  \hline
 Sub-model & Parameter & Name & Process & Domain & Default\\
  \hline
\multirow{5}{*}{Land-use}& $\lambda$ & Accessibility range & Accessibility & $]0;1]$ & $0.001$ \\\cline{2-6}
 & $\gamma_A$ & Cobb-Douglas exponents actives & \multirow{2}{*}{Utility} & $[0;1]$ & $0.85$ \\\cline{2-3}\cline{5-6}
 & $\gamma_E$ & Cobb-Douglas exponents employments &  & $[0;1]$ & $0.85$ \\\cline{2-6}
 & $\beta$ & Discrete choices exponent & \multirow{2}{*}{Relocalization} & $[0;+\infty]$ & $1$ \\\cline{2-3}\cline{5-6}
 & $\alpha$ & Relocation rate &  & $[0;1]$ & $0.05$ \\\hline
Transport & $v_G$ & Network speed & Hierarchy & $[1;+\infty [$ & $5$ \\\hline
\multirow{2}{*}{Governance} & $J$ & Collaboration cost & \multirow{2}{*}{Planning} & $[0;0.005]$ & $0.001$ \\\cline{2-3}\cline{5-6}
 & $l_r$ & Infrastructure length &  & $]0;\sqrt{2}\cdot K [$ & $2$ \\\hline
\end{tabular}
}{
\begin{tabular}{|c|c|c|c|c|c|}
  \hline
 Sous-modèle & Paramètre & Nom & Processus & Domaine & Défaut\\
  \hline
\multirow{5}{*}{Usage du sol}& $\lambda$ & Portée de l'accessibilité & Accessibilité & $]0;1]$ & $0.001$ \\\cline{2-6}
 & $\gamma_A$ & Exposant de Cobb-Douglas actifs & \multirow{2}{*}{Utilité} & $[0;1]$ & $0.85$ \\\cline{2-3}\cline{5-6}
 & $\gamma_E$ & Exposant de Cobb-Douglas emplois &  & $[0;1]$ & $0.85$ \\\cline{2-6}
 & $\beta$ & Exposant choix discrets & \multirow{2}{*}{Relocalisation} & $[0;+\infty]$ & $1$ \\\cline{2-3}\cline{5-6}
 & $\alpha$ & Taux de relocalisation &  & $[0;1]$ & $0.05$ \\\hline
Transports & $v_G$ & Vitesse du réseau & Hiérarchie & $[1;+\infty [$ & $5$ \\\hline
\multirow{2}{*}{Gouvernance} & $J$ & Coût de collaboration & \multirow{2}{*}{Planification} & $[0;0.005]$ & $0.001$ \\\cline{2-3}\cline{5-6}
 & $l_r$ & Longueur de l'infrastructure &  & $]0;\sqrt{2}\cdot K [$ & $2$ \\\hline
\end{tabular}
}
\end{table}

\bpar{
The model is implemented in Netlogo, for ergonomics reasons given its level of complexity, and also the possibilities of interactive exploration. A particular care has been given to the following points.
\begin{itemize}
	\item Computation of distance matrices are necessary for each potential infrastructure segment, what makes the governance module very costly from the computational point of view. We use therefore a computation of shortest paths based on dynamic programming, inspired by~\cite{tretyakov2011fast}, updating directly the distance matrix instead of recomputing shortest paths each time.
	\item The network is for this reason represented in a dual way, under vector and raster forms. The correspondence between the two and their consistence is ensured.
	\item For the determination of the optimal infrastructure, the order of magnitude of the total number of infrastructures to explore is in $O(l_r\cdot N)$, if $N$ is the number of patches and assuming that all potential infrastructures have their extremities in the center of a patch\footnote{For each patch, we will have an infrastructure for each other patch in a radius $l_r$, what asymptotically corresponds to the perimeter of the circle $2\pi l_r$. Furthermore, as detailed in \ref{app:sec:lutecia}, we assume a snapping heuristic to existing infrastructures to keep a consistant network.}. This considerably increases the operational computational cost, and we use an heuristic exploring a fixed number $N_I$ of randomly chosen infrastructures.
\end{itemize}
}{
Le modèle est implémenté en Netlogo, pour des raisons d'ergonomie vu son niveau de complexité, ainsi que les possibilités d'exploration interactive. Une attention particulière a été portée aux points suivants.
\begin{itemize}
	\item Les calculs des matrices de distance sont nécessaires pour chaque segment d'infrastructure potentiel, ce qui rend le module de gouvernance très couteux sur le plan computationnel. Nous utilisons donc un calcul des plus courts chemins basé sur la programmation dynamique, inspiré de~\cite{tretyakov2011fast}, mettant à jour directement la matrice des distances plutôt que de recalculer les plus courts chemins à chaque fois.
	\item Le réseau est pour cette raison représenté de manière duale, sous forme vecteur et raster. Le passage de l'un à l'autre et leur cohérence est assuré.
	\item Pour la détermination de l'infrastructure optimale, l'ordre de grandeur du nombre total d'infrastructures à explorer est un $O(l_r\cdot N)$, si $N$ est le nombre de cellules et en supposant que l'ensemble des infrastructures potentielles ont leurs extrémités au centre d'une cellule\footnote{Pour chaque cellule, on aura une infrastructure pour chaque autre cellule à un rayon $l_r$, ce qui asymptotiquement revient au périmètre du cercle $2\pi l_r$. Par ailleurs, comme précisé en \ref{app:sec:lutecia}, nous supposons une heuristique d'accrochage aux infrastructures existantes pour garder un réseau cohérent.}. Cela augmente considérablement le coût computationnel opérationnel, et nous utilisons une heuristique explorant un nombre fixé $N_I$ d'infrastructures choisies aléatoirement.
\end{itemize}
}

\bpar{
More implementation details are given in Appendix~\ref{app:sec:lutecia}.
}{
Plus de détails d'implémentation sont donnés en Annexe~\ref{app:sec:lutecia}.
}

\subsubsection{Model validation}{Validation du modèle}

\bpar{
Different experiments allow us to validate the model to a certain extent. We follow a modular strategy, i.e. by relatively independent tests of sub-models to begin with. The idea is to proceed to elementary experiments by making either land-use, or network, or both, evolve, and studying the consequences on the different aspects.
}{
Différentes expériences nous permettent de valider le modèle dans une certaine mesure. Nous adoptons une stratégie modulaire, c'est-à-dire par tests relativement indépendants des sous-modèles pour commencer. L'idée est de monter des expériences élémentaires en faisant évoluer soit l'usage du sol, soit le réseau, soit les deux, en étudiant les conséquences sur les différents aspects.
}

\bpar{
We work on synthetic systems. Population and employment configurations follow exponential mixtures. We give in Appendix~\ref{app:sec:lutecia} details of initialization parameters.
}{
Nous travaillons sur des systèmes synthétiques. Les configurations de populations et d'emplois suivent des mélanges d'exponentielles. Nous donnons en Annexe~\ref{app:sec:lutecia} les détails des paramètres d'initialisation.
}

\paragraph{Land-use}{Usage du sol}

\bpar{
Land-use dynamics always converge towards an asymptotic state when network does not evolve. We demonstrate the existence of the equilibrium in~\ref{app:sec:lutecia}. Furthermore, numerical experiments show that the model converge relatively quickly. Experiments targeting land-use only and which are detailed in~\ref{app:sec:lutecia} give the following results.
\begin{itemize}
	\item A large diversity of morphological trajectories in time, i.e. the evolution of morphological indicators for the distribution of population and employments, is obtained by playing on parameters $\gamma_A, \gamma_E, \lambda, \beta$, and also on the structure of a static network.
	\item Similarly, these trajectories do not converge towards the same forms and we have thus a diversity of final forms obtained.
	\item It is possible to minimize, at fixed $\alpha = 1$, the total quantity of relocalization. We will however use this parameter to control the speed of urban sprawl, and will typically take values around $0.1$, what corresponds to 10\% of actives relocating at each time step, i.e. on a period of the order of the decade.
\end{itemize}
}{
Les dynamiques d'usage du sol convergent toujours vers un état asymptotique lorsque le réseau n'évolue pas. Nous démontrons l'existence de l'équilibre en~\ref{app:sec:lutecia}. Par ailleurs, les expériences numériques montrent que le modèle converge assez rapidement. Les expériences ciblant l'usage du sol uniquement et qui sont détaillées en~\ref{app:sec:lutecia} fournissent les résultats suivants.
\begin{itemize}
	\item Une grande diversité de trajectoires morphologiques dans le temps, c'est-à-dire l'évolution des indicateurs morphologiques pour la distribution de la population et des emplois, est obtenue en jouant sur les paramètres $\gamma_A, \gamma_E, \lambda, \beta$, ainsi que sur la structure d'un réseau statique.
	\item De même, ces trajectoires ne convergent pas vers les mêmes formes et on a donc une diversité des formes finales obtenues.
	\item Il est possible de minimiser à $\alpha = 1$ fixé la quantité totale de relocalisation. Nous jouerons toutefois sur ce paramètre pour contrôler la vitesse d'étalement urbain, et prendrons typiquement des valeurs autour de $0.1$, qui correspond à 10\% d'actifs se relocalisant à chaque pas de temps, c'est-à-dire sur une période de l'ordre de la dizaine d'années.
\end{itemize}
}

\paragraph{Governance}{Gouvernance}

\bpar{
In order to understand the influence of governance parameters on forms produced by the model, we proceed to a simple experiment in the case of a bicentric system, without an initial network. Parameters for the land-use model are fixed at standard values $\gamma_A = \gamma_E = 0.8, \beta = 2 ; \lambda = 0.001, \alpha = 0.16$ and the length of infrastructure segments is fixed to $l_r = 2$. We consider uniquely the discrete choice game. The reference situation is given by a fully regional decision level, corresponding to $\xi = 1$. We compare it to two situations in which the level of decision is fully local ($\xi = 0$) but for which we force the possibility of collaboration to extreme values by the intermediate of the cooperation cost, taken respectively as $J=0$ and $J=0.005$.
}{
Afin de comprendre l'influence des paramètres de gouvernance sur les formes produites par le modèle, nous menons une expérience simple dans le cas d'un système bicentrique, sans réseau initial. Les paramètres du modèle d'usage du sol sont fixés à des valeurs standard $\gamma_A = \gamma_E = 0.8, \beta = 2 ; \lambda = 0.001, \alpha = 0.16$ et la longueur des tronçons est fixée à $l_r = 2$. Nous considérons uniquement le jeu à choix discrets. La situation de référence est donnée par un niveau de décision uniquement régional, correspondant à $\xi = 1$. Nous la comparons à deux situations dans lesquelles le niveau de décision est uniquement local ($\xi = 0$) mais pour lequel nous forçons les probabilités de collaboration à des valeurs extrêmes par l'intermédiaire du coût de coopération, pris respectivement comme $J=0$ et $J=0.005$.
}

\begin{figure}
	\includegraphics[width=\linewidth]{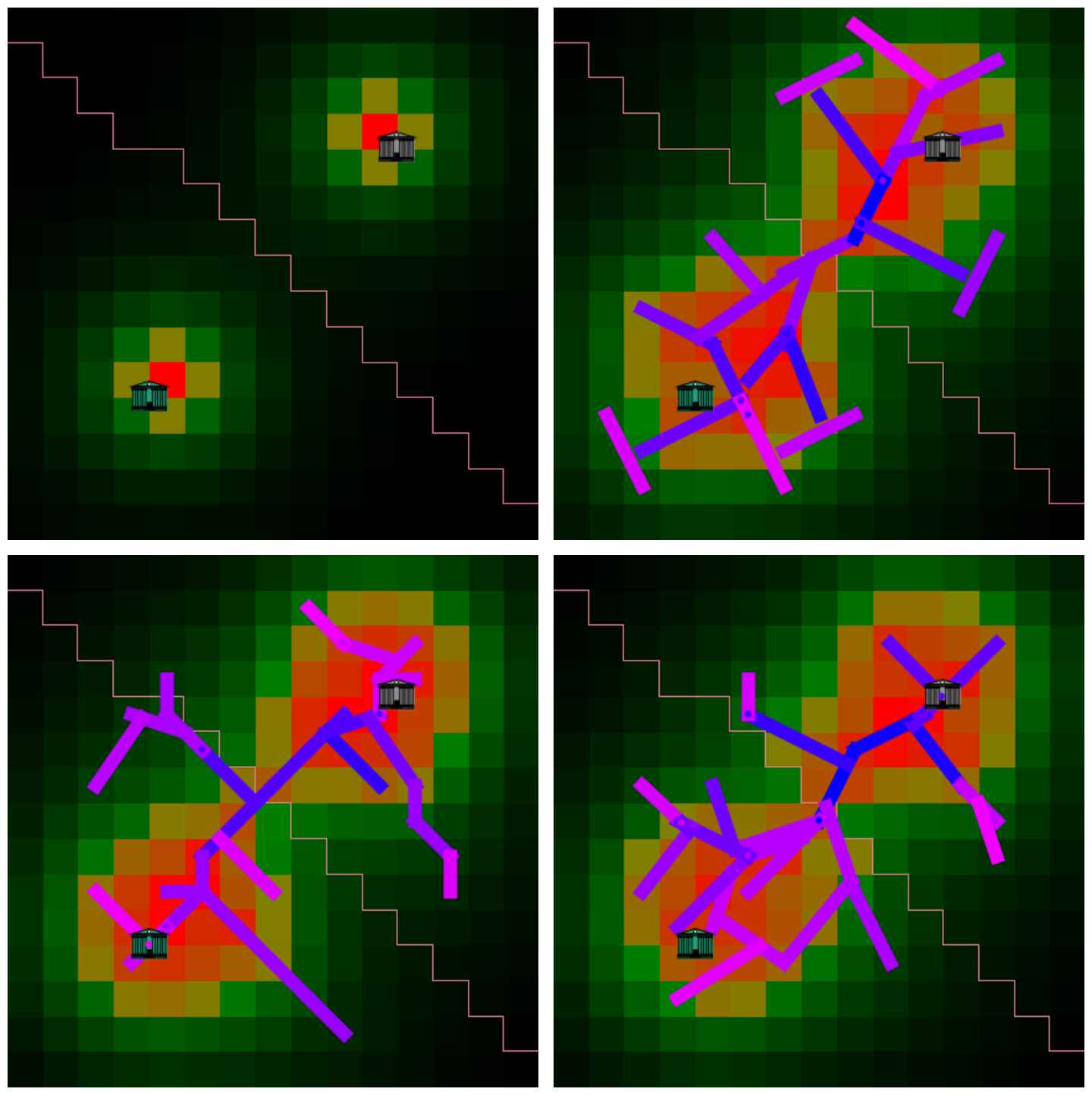}
	\caption[Network topologies obtained for different levels of governance]{\textbf{Network topologies obtained for different levels of governance.} The model is initialized on a symmetric synthetic configuration with two centers (\textit{Top Left}). Parameters for the evolution of land-use are $\gamma_A = \gamma_E = 0.8 ; \beta = 2 ; \lambda = 0.001 ; \alpha = 0.16$, and for network evolution $l_r = 2$ and a discrete choices game. The evolution is stopped at fixed stock $S = 50$ and the heuristic exploration done for $N_I = 200$. (\textit{Top Right}) Regional decision level ($\xi = 1$); (\textit{Bottom Left}) Local decision level ($\xi = 0$) and low level of collaboration, obtained with a high cost of cooperation $J=0.005$; (\textit{Bottom Right}) Local level and high level of collaboration, with $J=0$.\label{fig:lutecia:governance}}
\end{figure}

\bpar{
The initial configuration together with three examples of network shapes obtained for each configuration are shown in Fig.~\ref{fig:lutecia:governance}. Network shapes are visually\footnote{This preliminary experiment does not imply an intensive exploration, and it is thus impossible to translate these conclusions in a robust way in terms of indicators statistics.} different and witness particular structural characteristics. In the case of the regional decision, a structuring arc links the two centers, from which extensions branch, first perpendicularly and then in parallel. The structure obtained in the case of a collaborative local is also tree-like but has less branches, the extensions being in majority following the existing branches. Finally, as we could have expected, the non-collaborative network seems to be less optimal in terms of covering than the first two, and shows redundancies. Concerning the urban structure, we obtain that the local levels better conserve the bicentric structure compared to the regional level (see the position of final centers compared to their initial position): through the network, the decision at a regional level has more potential to create new centralities.
}{
La configuration initiale ainsi que trois exemples de formes de réseau obtenues pour chacune des configurations sont montrés en Fig.~\ref{fig:lutecia:governance}. Les formes de réseau sont visuellement\footnote{Cette expérience préliminaire n'implique pas d'exploration intensive, et il n'est donc pas possible de traduire ces conclusions de manière robuste en termes de statistiques des indicateurs.} différentes et témoignent de caractéristiques de structure particulière. Dans le cas de la décision régionale, un arc structurant relie les deux centres, à partir duquel se branchent des ramifications d'abord perpendiculaires puis parallèles. La structure obtenue dans le cas d'un local collaboratif est également arborescente mais comporte moins de branches, les prolongements se faisant majoritairement à la suite des branches existantes. Enfin, comme on pouvait s'y attendre, le réseau non-collaboratif parait moins optimal en termes de couverture que les deux premiers, et présente des redondances. Concernant la structure urbaine, on obtient que les niveaux locaux conservent plus la structure bicentrique en comparaison au niveau régional (voir la position des centres finaux par rapport à la position initiale) : via le réseau, la prise de décision au niveau régional a plus de potentiel pour créer de nouvelles centralités.
}

\paragraph{Co-evolution}{Co-évolution}

\bpar{
In a last stylized experiment, we propose to study more directly the effect of co-evolution, in particular on land-use variables. Therefore, we consider again the previous bi-centric configuration, with a disequilibrium of population and employments between the two centers (in practice with a rate of 2), and different proximities (close, at a distance of $0.4 \cdot K$, and far, at a distance of $K$). We fix a random local governance (choice of only one constructor with a probability proportional to employments) and the land-use and network parameters\footnote{We take here $\gamma_A = 0.9, \gamma_E = 0.65, \lambda = 0.005, \beta = 1.8, \alpha = 0.1, l_r = 1, v_0 = 6$.}, and we study the influence of the decision level $\xi$ on (i) the total accessibility gain between the initial and the final state, expressed as a rate $\frac{X(t_f)}{X(t_0)}$; and (ii) the evolution of relative accessibility between the two centers, given by $\frac{X_0(t_f)}{X_0(t_0)} / \frac{X_1(t_f)}{X_1(t_0)}$. The first indicators allows to understand the global benefit, whereas the second expresses the inequality between the centers (for example, is the weakest center drained by the more important, or does it benefit from it).
}{
Dans une dernière expérience stylisée, nous proposons d'étudier plus directement l'effet de la co-évolution, notamment sur les variables d'usage du sol. Pour cela, nous reprenons la configuration bicentrique précédente, avec un déséquilibre de population et d'emplois entre les deux centres (en pratique avec un rapport de 2), et une proximité variable (proches, à une distance de $0.4\cdot K$, et lointains, à une distance de $K$). Nous fixons une gouvernance locale aléatoire (choix d'un seul constructeur avec une probabilité proportionnelle aux emplois) et les paramètres d'usage du sol et de réseau\footnote{Nous prenons ici $\gamma_A = 0.9, \gamma_E = 0.65, \lambda = 0.005, \beta = 1.8, \alpha = 0.1, l_r = 1, v_0 = 6$.}, et nous étudions l'influence du niveau de décision $\xi$ sur (i) le gain d'accessibilité total entre l'instant initial et l'instant final, exprimé comme un rapport $\frac{X(t_f)}{X(t_0)}$ ; et (ii) l'évolution de l'accessibilité relative entre les deux centres, donnée par $\frac{X_0(t_f)}{X_0(t_0)} / \frac{X_1(t_f)}{X_1(t_0)}$. Le premier indicateur permet de comprendre le bénéfice global, tandis que le second exprime l'inégalité entre les centres (par exemple, le centre le plus faible est-il drainé par le plus important, ou bénéficie-t-il de celui-ci).
}

\begin{figure}
	\includegraphics[width=\linewidth]{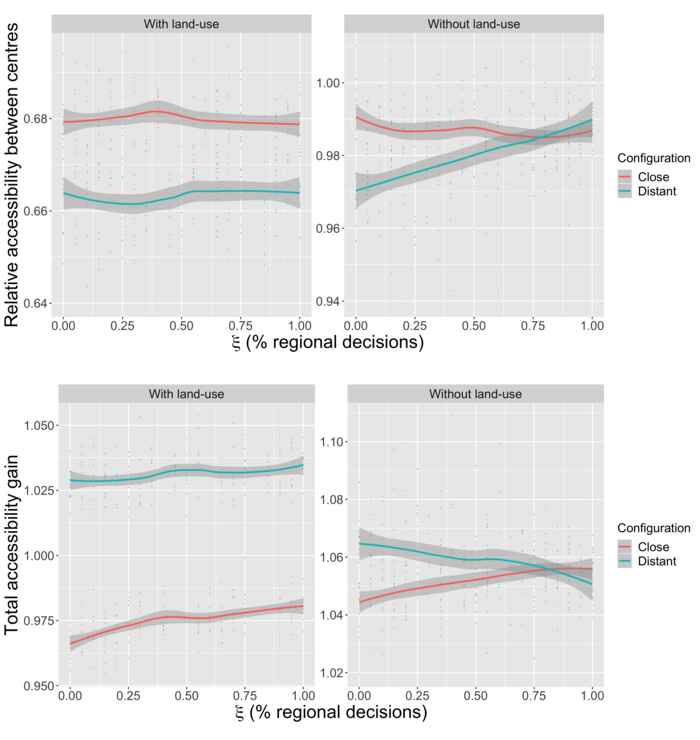}
	\caption[Impact of co-evolution on accessibility in the Lutecia model]{\textbf{Impact of co-evolution on accessibility in the Lutecia model.} We proceed to 10 repetitions with fixed parameters $\gamma_A = 0.9, \gamma_E = 0.65, \lambda = 0.005, \beta = 1.8, \alpha = 0.1, l_r = 1, v_0 = 6$, for a random local governance, and an evolution with constant stock $S=20$. We compare the evolution with network only (without land-use) and with co-evolution, for the close and distant configurations. (\textit{Top}) Evolution of the relative accessibility between centers, with and without land-use (columns) for the two configurations (colours); (\textit{Bottom}) Total accessibilty gain.\label{fig:lutecia:coevol}}
\end{figure}

\bpar{
Results of the experiment are given in Fig.~\ref{fig:lutecia:coevol}. The behavior of the accessibility gain unveil a direct effect of co-evolution processes: in the case of distant centers, the effect of $\xi$ on it is inverted when we add the evolution of land-use. In the case of a network evolving alone, a local decision is optimal for total accessibility, whereas in the case of a co-evolution of processes, the optimal is at a fully regional decision. We interpret this stylized fact as the existence of a need for coordination for the success of a coupled evolution of the transportation network and land-use, what can be put in correspondence with the concept of TOD seen in chapter~\ref{ch:thematic}. In the case of close centers, the regional decision is always optimal, corresponding then to a more integrated metropolitan area. The variation of the relative accessibility are to low to conclude on the evolution of inequalities between the centers in the case of a coupled evolution.
}{
Les résultats de l'expérience sont donnés en Fig.~\ref{fig:lutecia:coevol}. Le comportement du gain d'accessibilité révèle un effet direct des processus de co-évolution : dans le cas de centres distants, l'effet de $\xi$ sur celui-ci s'inverse lorsqu'on ajoute l'évolution de l'usage du sol. Dans le cas d'un réseau qui évolue seul, une décision locale est optimale pour l'accessibilité totale, tandis que dans le cas d'une co-évolution des processus, l'optimal est à une décision purement régionale. Nous interprétons ce fait stylisé comme l'existence d'un besoin de coordination pour la réussite d'une évolution couplée du réseau de transport et de l'usage du sol, ce qui peut être mis en relation avec le concept du TOD vu au chapitre~\ref{ch:thematic}. Dans le cas de centres proches, la décision régionale est toujours optimale, correspondant alors à une métropole plus intégrée. Les variations de l'accessibilité relative sont trop faibles pour conclure sur l'évolution des inégalités entre les centres dans le cas d'une évolution couplée.
}

\bpar{
Thus, this last experiment reveals indeed the existence of ``co-evolution effects'', in the emergence of a need for regional coordination in the case of a coupled evolution.
}{
Ainsi, cette dernière expérience révèle bien l'existence ``d'effets de co-évolution'', dans l'émergence d'un besoin de coordination régionale dans le cas d'une évolution couplée.
}

\subsection{Application to Pearl River Delta}{Application au Delta de la Rivière des Perles}

\bpar{
It was suggested by \cite{liao2017ouverture} that a sort of multi-level governance recently emerged in China, in the context of economic activities. We try with our model to test the relevance of this paradigm regarding the urban structure of the MCR.
}{
Il a été suggéré par \cite{liao2017ouverture} qu'une forme de gouvernance multi-niveau a récemment émergé en Chine, dans le contexte des activités économiques. Nous tentons par notre modèle de tester la pertinence de ce paradigme au regard de la structure urbaine de la MCR.
}

\subsubsection{Model setup}{Initialisation du modèle}

\begin{figure}
\includegraphics[width=\linewidth]{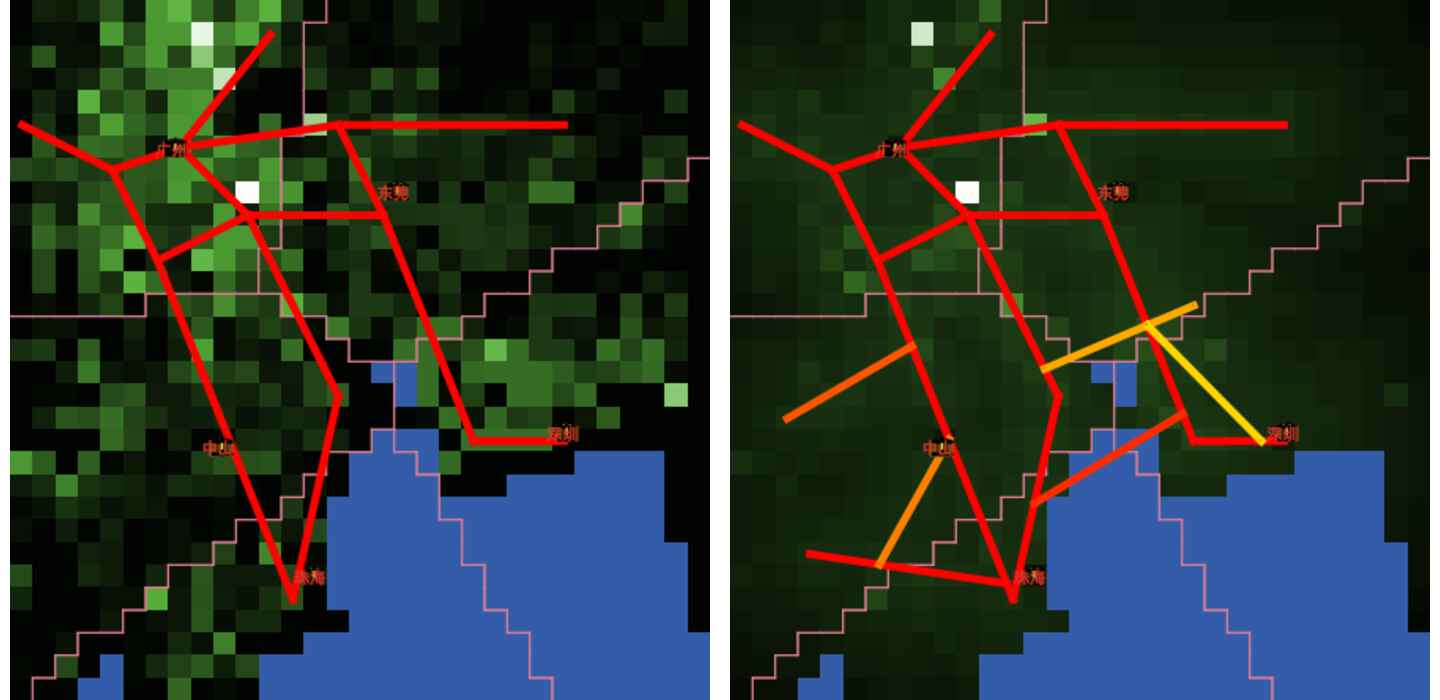}
\caption[Application of Lutecia to Pearl River Delta]{\textbf{Example of application to Pearl River Delta.} (\textit{Left}) Initialization with the 2010 population raster, aggregated at the 5km resolution, and the simplified freeway network; (\textit{Right}) State after 6 time steps ($\alpha = 1$).\label{fig:lutecia:ex-prd}}
\end{figure}

\bpar{
We work on a simplified raster configuration (5km cells) for population in Pearl River Delta, and on the stylized freeway network. We choose to consider only the road network since, following \cite{hou2011transport}, it has been the main driver of changes in accessibility patterns compared to railway network which accelerated development is recent. Networks are stylized from the plan given by~\cite{hou2011transport} which reproduces official documents of Guangdong province in 2010. We thus consider the freeway network in 2010 and the one planned at this time. Employment data are given for 2010 by~\cite{swerts2017database} at the level of cities. They are here uniformly distributed for each city in the simplified raster. The Fig.~\ref{fig:lutecia:ex-prd} illustrates the stylized configuration for Pearl River Delta.
}{
Nous travaillons sur une configuration raster simplifiée (cellules de 5km) pour la population du Delta de la Rivière des Perles, ainsi que sur le réseau d'autoroute stylisé. Nous considérons le réseau routier uniquement puisque, selon \cite{hou2011transport}, il s'agit du moteur principal des changements dans les motifs d'accessibilité en comparaison au réseau ferré dont le développement accéléré est récent. Les réseaux sont stylisés à partir du plan donné par~\cite{hou2011transport} qui reproduit les documents officiels de la province du Guangdong en 2010. Nous considérons ainsi le réseau autoroutier en 2010 et celui planifié à cette époque. Les données des emplois sont fournies en 2010 par~\cite{swerts2017database} au niveau des communes. Ils sont distribués ici uniformément pour chaque ville dans le raster simplifié. La Fig.~\ref{fig:lutecia:ex-prd} illustre la configuration stylisée pour le Delta de la Rivière des Perles.
}

%

\subsubsection{Calibration procedure}{Procédure de calibration}

\bpar{
To apply such a complex model to a semi-real situation, one must be particularly careful. It is important to choose the adequate processes and level of granularity to reproduce. In particular, our model is not aimed at producing particularly accurate land-use patterns, but uses their approximation as the basis of network growth, which qualitative evolution and the corresponding qualitative patterns in governance processes. We propose therefore to ``calibrate'' on the shape of a given infrastructure, in the sense of determining parameter configurations for which in probability the successive built pieces of infrastructure are the closest to pieces of the target infrastructure. 
}{
Lors de l'application d'un modèle si complexe à une situation semi-réelle, il faut rester vigilant. Il est important de choisir les processus adéquats ainsi que le niveau de granularité à reproduire. En particulier, notre modèle produit des motifs d'usage du sol relativement précis, mais utilise leur approximation comme base de la croissance du réseau, dont l'évolution qualitative permet d'informer sur les processus de gouvernance. Nous proposons pour cela de ``calibrer'' sur la forme d'une infrastructure donnée, au sens de déterminer des configurations de paramètres pour lesquelles en probabilité les morceaux successifs d'infrastructure sont les plus proches d'une infrastructure visée.
}

\bpar{
To calibrate on the network produced by the simulation, it must be compared to a reference network. This is however a difficult problem, as different proximity measures with different significations can be used. Geometrical measures focus on the spatial proximity of networks. For a network $(E,V)=((e_j),(v_i))$, a node-based distance is given by $\sum_{i \neq i'} d^2 \left(v_i,v_{i'}\right)$. A more accurate measure which is not biased by intermediate nodes is given by the cumulated area between each pair of edges $\sum_{j \neq j'} A \left(e_j,e_{j'}\right)$ (not a distance in the proper sense) where $A(e,e')$ is the area of the closed polygon formed by joining link extremities. We consider the latest for the calibration.
}{
Pour calibrer sur les réseaux produits par la simulation, il s'agit de comparer à un réseau de référence. C'est un problème difficile, puisque différentes mesures de proximité avec différentes significations peuvent être utilisées. Les mesures géométriques s'intéressent à la proximité spatiale des réseaux. Pour un réseau $(E,V)=((e_j),(v_i))$, une distance basée sur les noeuds est donnée par $\sum_{i \neq i'} d^2 \left(v_i,v_{i'}\right)$. Une mesure plus précise qui n'est pas biaisée par d'éventuels noeuds intermédiaires est donnée par l'aire cumulée entre chaque paire de liens $d_A = \sum_{j \neq j'} A \left(e_j,e_{j'}\right)$ (il ne s'agit pas d'une distance à proprement parler), où $A(e,e')$ est l'aire du polygone fermé constitué en reliant les sommets des liens. Nous considérerons cette dernière pour la calibration.
}

\subsubsection{Calibration}{Calibration}

\bpar{
The experiments we do are with a fixed land-use, since the required level of detail for more ancient or recent data, or even projections, for population and employments, is not allowed by the data we had access to.
}{
Les expériences que nous menons sont à usage du sol fixé, le niveau de détail requis pour des données plus anciennes et plus récentes, voir des projections, pour la population et les emplois n'étant pas permis par les données à notre disposition.
}

\bpar{
We make governance parameters vary, including the type of game, with a fixed $l_r = 2$, and explore a Latin Hypercube Sampling of 4000 points in this parameter space, with 10 repetitions of the model for each point. The two experiments we performed correspond to different target configurations:
\begin{itemize}
	\item no initial network and the 2010 network as a target, in the spirit of extrapolating the most probable governance configuration which led to the current configuration;
	\item initial network as the 2010 network, and planned network as target: extrapolation of the governance configuration for the planning.
\end{itemize}
}{
Nous faisons varier les paramètres de gouvernance, incluant le type de jeu, avec $l_r = 2$ fixé, et explorons un échantillonnage LHS de 4000 points dans l'espace de ces paramètres, avec 10 répétitions du modèle pour chaque point. Les deux expériences menées correspondent à des configurations cibles différentes :
\begin{itemize}
	\item aucun réseau initial et réseau de 2010 comme cible, dans l'esprit d'extrapoler la configuration de gouvernance la plus probable ayant mené à la configuration actuelle ;
	\item réseau 2010 initial, et réseau planifié comme cible : extrapolation de la configuration de gouvernance de la planification.
\end{itemize}
}

\begin{figure}
	\includegraphics[width=\linewidth]{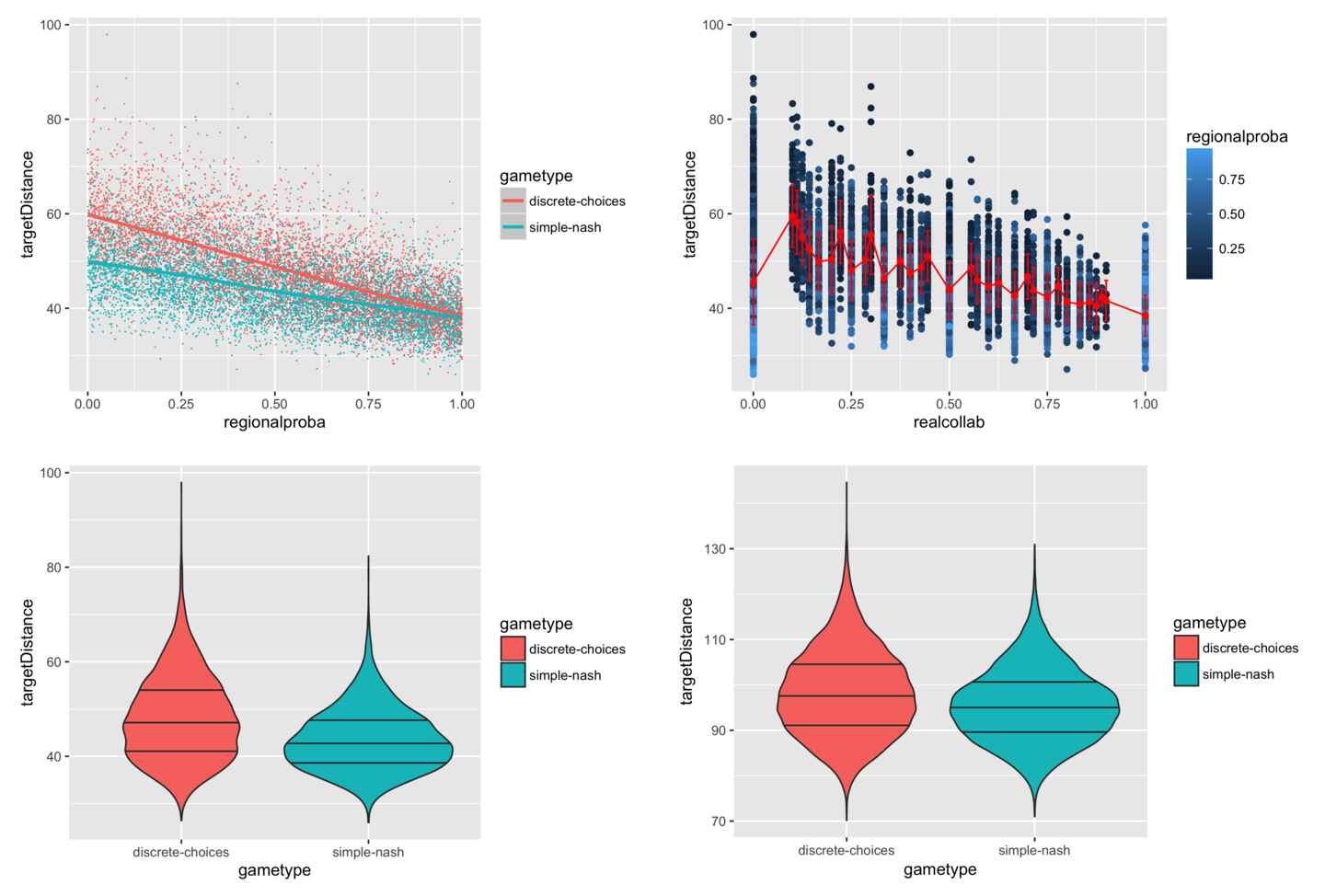}
	\caption[Calibration of the Lutecia model]{\textbf{Model calibration with fixed land use.} We take $\alpha = 0$ to make only network evolve, and sample the governance parameters space. (\textit{Top Left}) Distance $d_A$ to the target network (\texttt{targetDistance}), in the case of the real network, as a function of the regional decision probability $\xi$ (\texttt{regionalproba}), for the two types of games (colour). (\textit{Top Right}) Distance $d_A$ as a function of the observed collaboration probability (\texttt{realcollab}); the red curve gives the averages with standard errors. (\textit{Bottom Left}) Statistical distribution of distance as a function of the type of game, in the case of the real network; (\textit{Top Right}) in the case of the planned network. The difference between the types of games is larger in the case of the real network in comparison to the planned network.\label{fig:lutecia:calib}}
\end{figure}

\bpar{
We obtain qualitatively similar results for the two experiments, suggesting that there was no transition in the type of governance between the past network and the future network. Results are illustrated in Fig.~\ref{fig:lutecia:calib}. We obtain, by studying the graph of $d_A$ as a function of $\xi$, that the regional level is the most realistic to reproduce network shape. However, discrete choices and competition games have a different behavior, and the competitive game is the closest to reality when $\xi$ decreases: the relations between local actors would a priori be of a more competitive than an egoistic nature. When we study the variation of distance as a function of the observed collaboration level, we obtain an interesting inverted U-shape, i.e. that the most likely configurations are the ones where there is only collaboration, or the ones where there is no collaboration at all, but no intermediate situations. Finally, the comparison of statistical distributions of distances between target configurations and the types of games shows that the difference between the games is significant only for the real network but not for the planned network (what remains a conclusion difficult to interpret).
}{
Nous obtenons des résultats qualitativement similaires pour les deux expériences, suggérant qu'il n'y a pas eu de transition de type de gouvernance entre réseau passé et réseau futur. Les résultats sont illustrés en Fig.~\ref{fig:lutecia:calib}. On obtient, à l'étude du graphe de $d_A$ en fonction de $\xi$, que le niveau régional est le plus fidèle pour reproduire la forme du réseau. Par contre, les jeux de choix discrets et de compétition se comportent différemment, et le jeu compétitif est le plus proche de la réalité quand $\xi$ diminue : les relations entre acteurs locaux seraient a priori de nature plus compétitive qu'égoïste. Quand on étudie la variation de la distance en fonction du niveau de collaboration observé, on obtient une forme intéressante en cloche inversée, c'est-à-dire que les situations les plus probables sont soit celles où il n'y a que de la collaboration, soit celles où il n'y en a pas du tout, mais pas de situations intermédiaires. Enfin, la comparaison des distributions statistiques des distances entre les configurations cibles et les types de jeux montre que la différence entre les jeux n'est considérable que pour le réseau réel mais pas le réseau planifié (ce qui reste une conclusion difficile à interpréter).
}

\bpar{
We thus draw from this experiment the following conclusions, to be naturally taken with caution.
\begin{itemize}
	\item A competition between actors is less probables than an egoistic behavior in the case of local decisions, since the discrete choices game give better performances than the Nash for low values of $\xi$.
	\item Collaboration compromises correspond to less probable networks than situations with full collaboration or with no collaboration.
\end{itemize}
}{
Nous tirons donc de cette expérience les conclusions suivantes, à prendre bien sûr avec prudence.
\begin{itemize}
	\item Une compétition entre les acteurs est moins probable qu'un comportement égoïste dans le cas de décisions locales, puisque le jeu par choix discrets donne de meilleures performances que le Nash pour les faibles valeurs de $\xi$.
	\item Les compromis de collaboration forment des réseaux moins probables que les situations avec pleine collaboration ou avec aucune collaboration.
\end{itemize}
}

\bpar{
These conclusions can be put into perspective with the increased competition within the Delta revealed by~\cite{xu2005city}. Thus, this application of the model allows to indirectly infer governance processes.
}{
Ces conclusions peuvent être mises en perspective avec la compétition accrue au sein du Delta révélée par~\cite{xu2005city}. Ainsi, cette application du modèle permet d'inférer indirectement des processus de gouvernance.  
}

\subsubsection{Discussion}{Discussion}

\bpar{
Although the model must still be more deeply explored and for all its modules, some possible developments are worth of interest.
}{
Bien que le modèle doit encore être exploré plus en profondeur et pour l'ensemble de ses modules, certains développements possibles peuvent retenir notre attention.
}

\paragraph{Endogenous level of decision}{Niveau de décision endogène}

\bpar{
One relevant extension would be the study of the emergence of larger administrative zones by aggregation, i.e. the emergence of new levels of governance in polycentric metropolitan areas. The example of the \emph{M{\'e}tropole du Grand Paris} is a good illustration for it when considering it in a simplified way, since it is positioned between local collectivities and the Region but also the State~\cite{gilli2009paris}. An extension of the model with rules to merge entities is a potential direction to study this question.
}{
Une extension pertinente serait l'étude de l'émergence de zones administratives par agrégation, c'est-à-dire l'émergence de nouveaux niveaux de gouvernance dans une région métropolitaine polycentrique. L'exemple de la Métropole du Grand Paris en est une bonne illustration en la considérant de manière simplifiée, puisqu'elle se situe entre les collectivités locales et la région ainsi que l'État~\cite{gilli2009paris}. Une extension du modèle avec des règles de fusion des entités est une direction potentielle pour étudier cette question.
}

\paragraph{Competition for an external ressource}{Compétition pour une ressource externe}

\bpar{
The influence of external territories or of externalities on the evolution of a MCR is an open question. In the case of a common resource, localized within the spatial extent of the MCR, competition or collaboration dynamics can emerge between actors for its exploitation. This model is a solution to study this situation in a stylized way, and thus realize a controlled experiment on co-evolution dynamics, which would allow to answer more general questions concerning the role of territorial isolation in co-evolution processes.
}{
L'influence des territoires extérieurs ou des externalités sur l'évolution d'une MCR est une question ouverte. Dans le cas d'une ressource commune, localisée dans l'emprise de la MCR, des dynamiques de compétition ou de collaboration peuvent s'instituer entre acteurs pour son exploitation. Ce modèle est une solution pour étudier cette situation de manière stylisée, et réaliser ainsi une expérience contrôlée sur les dynamiques de co-évolution, qui permettrait de répondre à des questions plus générales quant au rôle de l'isolation territoriale dans les processus de co-évolution.
}

\stars

\bpar{
We have thus build the first bricks of models aiming at a more complex integration of co-evolution processes, by developing the Lutecia model which was then validated in a preliminary way and which potentialities have been demonstrated by the application to the case study of Pearl River Delta.
}{
Nous avons ainsi posé les premières briques de modèles visant à une intégration plus complexe des processus de co-évolution, en développant le modèle Lutecia qui a ensuite été validé de manière préliminaire et dont les potentialités ont été démontrées par l'application au cas d'étude du Delta de la Rivière des Perles.
}

\stars

%


\newpage

\section*{Chapter Conclusion}{Conclusion du Chapitre}

\bpar{
This second entry on co-evolution models, at the mesoscopic scale, has been the occasion to explore the coupling between urban form and functions through the coupling between territory and network. In comparison with macroscopic models, processes that are taken here into account are much more varied and complementary.
}{
Cette deuxième entrée sur les modèles de co-évolution, à l'échelle mesoscopique, a été l'occasion d'explorer le couplage entre forme urbaine et fonctions au travers du couplage entre territoire et réseau. En comparaison avec les modèles macroscopiques, les processus pris en compte ici sont beaucoup plus variés et complémentaires.
}

\bpar{
A first morphogenesis model includes different heuristics for network growth, which are necessary and complementary to capture all the possible range of generated network configurations. We show that the model is able to resemble observed situations, for the territorial form, network topology, and also for static correlations between these indicators, while requiring a compromise between these different objectives. In terms of causality regimes, and thus of capturing co-evolutive dynamics, the model is able to capture some in some precise situations, but we learn from that experiment a fundamental lesson for co-evolutive models: a fidelity to processes or static configurations is obtained at the price of less flexibility in produced dynamical regimes. This could be a structural effect of models, or more interesting, a restriction of existing regimes in real situations.
}{
Un premier modèle de morphogenèse inclut différentes heuristiques pour la croissance du réseau, qui sont nécessaires et complémentaires pour capturer toute l'étendue possible des configurations de réseau générées. Nous montrons que le modèle est capable de se rapprocher de situation observées, pour la forme territoriale, la topologie du réseau, ainsi que pour les corrélations statiques entre ces indicateurs, tout en nécessitant un compromis entre ces différents objectifs. En termes de régimes de causalité, et donc de capture de dynamiques co-évolutives, le modèle est capable d'en capturer dans certaines situations précises, mais on tire de cette expérience une leçon fondamentale pour les modèles de co-évolution : une fidélité des processus ou des configurations statiques doit se faire au prix de la flexibilité des régimes dynamiques produits. Cela peut être un effet structurel des modèles, ou plus intéressant, une restriction des régimes existants dans les situations réelles.
}


\bpar{
We have then made the bet to introduce a more complex model, including an ontology for governance processes for the evolution of the transportation network. We carry out first experiments for model validation on synthetic data, and propose an application to the case of Pearl River Delta, renewing the view we gave in~\ref{sec:casestudies}. We show for example that it is possible to extrapolate parameters linked to the level of collaboration between actors. This section allows thus to introduce a new approach to consider co-evolution, that takes into account the full conceptual frame developed in~\ref{ch:thematic}, and also opens numerous research directions.
}{
Nous avons ensuite fait le pari d'introduire un modèle plus complexe, incluant une ontologie pour les processus de gouvernance pour l'évolution du réseau de transport. Nous menons des premières expériences de validation du modèle sur données synthétiques, et proposons une application au cas du Delta de la Rivière des Perles, renouvelant le regard que nous en avons apporté en~\ref{sec:casestudies}. Nous montrons par exemple qu'il est possible d'extrapoler des paramètres liés au niveau de collaboration entre acteurs. Cette section permet ainsi d'introduire une nouvelle façon de considérer la co-évolution, prenant en compte l'intégralité du cadre conceptuel développé en~\ref{ch:thematic}, et ouvre également de nombreuses perspectives de recherche.
}

\stars




\bpar{
\chapter*{Conclusion of Part III: a complete view of co-evolution}
}{
\chapter*{Conclusion de la Partie III : une vue complète de la co-évolution}
}

\markboth{Conclusion}{Conclusion}


\bpar{
This part thus gave the first elements of the exploration of different entries on the modeling of co-evolution. We explored in chapter~\ref{ch:macrocoevolution} a co-evolution model at the macroscopic scale, which allows to isolate numerous causality regimes, which we can thus designate as co-evolution regimes for the ones exhibiting circular causalities, and which is calibrated on the French system of cities. We therein show that a simple representation and mechanisms already allow to synthetically and empirically capture co-evolution at this scale.
}{
Cette partie a ainsi donné des premiers éléments d'exploration de différentes entrées sur la modélisation de la co-évolution. Nous avons exploré dans le chapitre~\ref{ch:macrocoevolution} un modèle de co-évolution à l'échelle macroscopique, qui permet l'isolation de nombreux régimes de causalité, qu'on peut alors nommer régimes de co-évolution pour ceux présentant des causalités circulaires, et qui est calibré sur le système de villes français. Nous montrons ainsi que des mécanismes et une représentation simple permettent déjà de capturer synthétiquement et empiriquement la co-évolution à cette échelle.
}

\bpar{
We then explored models at a larger scale, implying an increasing complexity. A morphogenesis co-evolution model allows to couple urban form (distribution of population and network topology) with an abstraction of urban functions (centrality and accessibility measures within the network). The different heuristics for network evolution which have been tested appear to be complementary to approach real configurations. Finally, we introduced elements to take into account governance processes in the evolution of transportation networks.
}{
Nous avons ensuite exploré des modèles à une échelle plus grande, impliquant une complexité croissante. Un modèle de co-évolution par morphogenèse permet de coupler la forme urbaine (distribution de la population et topologie du réseau) à une abstraction des fonctions urbaines (mesures de centralité et d'accessibilité dans le réseau). Les différentes heuristiques d'évolution du réseau qui ont été testées se révèlent complémentaires pour s'approcher de configurations réelles. Enfin, nous avons introduit des pistes pour la prise en compte des processus de gouvernance dans l'évolution des réseaux de transport.
}

\subsection*{Processes in models}{Processus modélisés}


\bpar{
The models we developed have been so in a logic of parsimony, while seeking to effectively capture co-evolution processes at different scales and by being anchored into various disciplines: these constraints are payed by a price on the refinement of integrated mechanisms. We will come back on this compromise in~\ref{sec:contributions}.
}{
Les modèles que nous avons développés l'ont été dans une logique de parcimonie, tout en cherchant à effectivement capturer des processus de co-évolution à différentes échelles et en s'ancrant dans des disciplines variées : ces contraintes se paient par un prix en raffinement des mécanismes intégrés. Nous reviendrons sur ce compromis en~\ref{sec:contributions}.
}

\subsection*{A full view of co-evolution}{Une vue complète de la co-évolution}


\bpar{
At this stage we brought elements of answer to the two axis of our general problematic (how to define and characterize co-evolution, and how to model it). It is remarkable to note that these are articulated within the three knowledge domains of the conceptual (definition), of the empirical (characterization) and of modeling (models). These three aspect reciprocally auto-generate the others, and our viewpoint consists in a true trinity, i.e. a concept which is together unique and triple, in which none of the approaches can be ignored (the same way that \cite{morin2001methode} does for complex anthropology).
}{
Nous avons à ce stade apporté des éléments de réponse aux deux axes de notre problématique générale (comment définir et caractériser la co-évolution, et comment la modéliser). Il est remarquable de noter que ceux-ci s'articulent dans les trois domaines de connaissance du conceptuel (définition), de l'empirique (caractérisation) et de la modélisation (modèles). Ces trois aspects s'auto-génèrent l'un l'autre, et notre point de vue forme une véritable trinité, c'est-à-dire un concept à la fois unique et triple, dans lequel aucune des approches ne peut être ignorée (de la manière dont le fait \cite{morin2001methode} pour l'anthropologie complexe).
}

\bpar{
Thus, models contain the individual aspect of co-evolution (reciprocal interactions between entities), and in some cases the statistical aspect at the scale of a population. This conclusion is made possible through the operational characterization tool, which in its turn allows to reinforce the relevance of the definition.
}{
Ainsi, les modèles contiennent l'aspect individuel de la co-évolution (interactions réciproques entre entités), et dans certains cas l'aspect statistique au niveau d'une population. Cette conclusion est rendue possible par l'outil de caractérisation opérationnelle, celui-ci permettant par ailleurs de renforcer la pertinence de la définition.
}

\subsection*{Perspectives}{Perspectives}

\bpar{
Our point of view on co-evolution has naturally been reducing and limited, since the current state of our modes of knowledge production is still far from a paradigmatic integration of complexity~\cite{morin1991methode}, and that any tentative to apprehend a complex system combines with elegance analysis and synthesis, reductionism and holism, modularity and interdependency. In order to enrich our viewpoint, we finally propose an opening chapter.
}{
Notre point de vue sur la co-évolution a bien entendu été réducteur et limité, puisque l'état actuel de nos modes de production de connaissance est encore loin d'une intégration paradigmatique de la complexité~\cite{morin1991methode}, et que toute tentative d'appréhension d'un système complexe combine habilement analyse et synthèse, réductionnisme et holisme, modularité et interdépendance. Afin d'enrichir notre point de vue, nous proposons finalement un chapitre d'ouverture.
}


%
\ctparttext{A building is never used exactly the way it was designed: the integration of this reality sublimates the essence of the art of the architect. The effective functional use give sense to the form, while depending on it. It is the same for a knowledge construction. We know take a step back and open both theoretical and empirical perspectives.}
%
\part{Conclusion and Opening}
%


\markboth{Introduction}{Introduction}


\bpar{
\textit{An opening has the main feature of situating. Stating the present, future and past situations. While taking this step back, we imagine that this trajectory was not random, and that in the end this hell may be relieving, as did the shadow of this revisited Euridice fleeing to the pleasure of writing in the basement. There is this strange requirement of stating the desired career in the forms at the beginning of each school year: maybe their most useful function is retrospective as they help understanding the path-dependency of a trajectory. From train driver to cartographer, this ends up with a good trade-off. Computer science and the way and architectural detours also played their role. It is impossible to know if urban systems were there from the beginning, or if history is reinterpreted from the currently accepted dogma. But introspection sheds light on the current position and the future trajectory: it is indeed Hell here, but this is rather pleasant.}
}{
\textit{Une ouverture est principalement une mise en situation. Situation présente, situation future, situation passée. C'est en prenant ce recul qu'on s'imagine que cette trajectoire n'est pas fortuite, et qu'au fond, c'est peut être cet enfer qui nous délivrera, à l'image de l'ombre d'Euridice revisitée qui s'évade vers l'extase de la plume au dernier sous-sol. Il y a cette tradition incongrue de renseigner la profession souhaitée pour plus tard sur les fiches à chaque rentrée : finalement leur plus grand intérêt ne serait-il pas rétrospectivement, pour comprendre la dépendance au chemin de sa propre trajectoire. De conducteur de métro à cartographe, ce sera finalement un bon compromis. L'informatique qui passe par là est aussi crucial, les errances architecturales ont également joué leur rôle. Il sera sans doute impossible de dire si les systèmes urbains étaient là depuis le début, ou si l'histoire est réinterprétée à la lueur des idées triomphantes. Mais l'introspection illumine la position présente et la trajectoire future : finalement on est bien aux Enfers, mais on y est aussi pas si mal.}
}

\bigskip

\bpar{
An opening indeed feeds the construction of a meta viewpoint, and allows therein to considerably enrich the produced knowledge. The content of suggested elements conditions the underlying structure which one has to extract, allowing in turn some reflexivity. We can't reach personnal reflexivity levels as illustrated above, but aim at a certain level from the disciplinary and methdological viewpoints.
}{
Une ouverture amène en effet des éléments de construction d'un méta point de vue et permet ainsi d'enrichir considérablement la connaissance produite. La nature des éléments suggérés conditionne la structure sous-jacente qu'on cherchera alors à faire émerger, qui permet en retour une réflexivité. Nous n'atteindrons pas des niveaux de réflexivité personnels à l'image de l'illustration ci-dessus, mais chercherons un certain niveau du point de vue disciplinaire et méthodologique.
}

\bpar{
The last chapter (\ref{ch:theory}) thus brings some opening elements which act as a meta-synthesis when articulated within the global framework of this research. It includes both a thematic conclusion and a theoretical opening. It first puts into perspective our contributions and synthesises them. It then elaborates a theoretical link between the approaches we took, and finally by introducing a knowledge framework it allows putting into a global perspective all the work done until here.
}{
Le dernier chapitre (\ref{ch:theory}) apporte ainsi des éléments d'ouverture qui tiennent lieu de méta-synthèse lorsqu'on les articule dans le cadre global des recherches menées ici. Il propose ainsi à la fois une conclusion thématique et une ouverture théorique. Il met d'abord en perspective et synthétise nos contributions. Il élabore ensuite une articulation théorique des approches que nous avons prises, et enfin par la construction d'un cadre de connaissance permet une mise en perspective globale de l'ensemble du travail mené jusqu'à ce stade.
}


\stars










\bpar{
\chapter{Conclusion and theoretical opening}
}{
\chapter{Conclusion et Ouverture Théorique}
}

\label{ch:theory} 



\bigskip

\bpar{
One of the underlying implications of the work developed until here is the introduction of directions towards \emph{integrated theories}, i.e. being based on an horizontal and vertical integration in the sense of the complex system roadmap~\cite{2009arXiv0907.2221B}, but also allowing an integration of knowledge domains and a reflexivity. We develop in this chapter a theoretical opening at different levels. The corresponding framework emerges from the interaction of the different knowledge components developed until here.
}{
L'une des implications sous-jacentes du travail que nous avons mené jusqu'ici est l'introduction de pistes pour des \emph{théories intégrées}, c'est-à-dire s'appuyant sur une intégration horizontale et verticale au sens de la feuille de route des systèmes complexes~\cite{2009arXiv0907.2221B}, mais aussi permettant une intégration des domaines de connaissance et une réflexivité. Nous développons dans ce chapitre un ouverture théorique à plusieurs niveaux. Le cadre correspondant émerge de l'interaction des différentes composantes de la connaissance développées jusqu'ici.
}

\bpar{
We first propose to put into perspective our contributions on the subject of co-evolution of transportation network and territories, and thus to open in a thematic way potential developments.
}{
Nous proposons dans un premier temps de mettre en perspective nos contributions sur le sujet de la co-évolution des réseaux de transport et des territoires, et d'ouvrir ainsi de manière thématique des développements potentiels.
}

\bpar{
We then elaborate in a second section~\ref{sec:theory} a theoretical synthesis of the different approaches taken until here, allowing to make the link between the evolutive urban theory and morphogenesis, what yields a synthetic viewpoint on co-evolution.
}{
Nous élaborons ensuite dans une seconde section~\ref{sec:theory} une synthèse théorique des différentes approches prises jusqu'ici, permettant de faire le lien entre théorie évolutive des villes et morphogenèse, ce qui donne un point de vue synthétique sur la co-évolution.
}

\bpar{
Staying at a thematic level appears however to be not enough to obtain general guidelines on the type of methodologies and the approaches to use. More precisely, even if some theories imply a more natural use of some tools\footnote{To give a basic example, a theory emphasizing the complexity of relations between agents in a system will conduct generally to use agent-based modeling and simulation tools, whereas a theory based on a macroscopic equilibrium will favorise the use of exact mathematical derivations.}, at the subtler level of contextualization in the sense of the approach taken to implement the theory (as models or empirical analysis), the freedom of choice for objects and approaches in social sciences may mislead into unappropriated techniques or questionings (see the section~\ref{sec:computation} for the example of incautious use of big data and computation).
}{
Rester à un niveau thématique apparaît cependant ne pas être suffisant pour obtenir des lignes directrices générales sur le type de méthodologies et d'approches à utiliser. Plus précisément, même si certaines théories impliquent un usage plus naturel de certains outils\footnote{Pour donner un exemple basique, une théorie mettant l'emphase sur la complexité des relations entre agents dans un système conduira généralement à utiliser de la modélisation multi-agents et des outils de simulation, tandis qu'une théorie basée sur un équilibre macroscopique favorisera l'usage de dérivations mathématiques exactes.}, au niveau plus subtil de la mise en contexte au sens de l'approche prise pour implémenter la théorie (comme modèles ou analyses empiriques), la liberté de choix d'objets et d'approches en sciences sociales peut conduire à l'utilisation de techniques inappropriées ou des questionnements inadaptés (voir la section~\ref{sec:computation} pour l'exemple de l'usage inconsidéré des données massives et du calcul).
}

\bpar{
Therefore, we construct in a last section~\ref{sec:knowledgeframework} an applied knowledge framework aiming at making explicit knowledge production processes on complex systems. It is illustrated by a fine analysis of the genesis of the evolutive urban theory, and is then applied in a reflexive way on all our work. A possible formalization of this framework as an algebraic structure is suggested in~\ref{app:sec:csframework}.
}{
Ainsi, nous construisons dans une dernière section~\ref{sec:knowledgeframework} un cadre de connaissances appliqué visant à expliciter des processus de production de connaissance sur les systèmes complexes. Celui-ci est illustré par une analyse fine de la genèse de la Théorie Evolutive des Villes, puis est ensuite appliqué de manière réflexive à l'ensemble de notre travail. Une formalisation possible de ce cadre sous forme de structure algébrique est suggéré en~\ref{app:sec:csframework}.
}

\bpar{
This chapter must be read with caution since the theoretical constructions introduced are at a progressive abstraction level: in some sense, each theoretical level is a meta framework for the previous one. We therein link to the question of reflexivity, and to what extent theories can be applied to themselves\footnote{While keeping in mind that the separation between levels is not directly evident: for example the formal framework for socio-technical systems of~\ref{app:sec:csframework} could be applied as a formalization of the knowledge framework.}. Therefore, this synthesis allows us to simultaneously proceed to a synthesis and an opening.
}{
Ce chapitre doit être lu avec précaution car les constructions théoriques introduites sont à un niveau d'abstraction progressif : en quelque sorte, chaque niveau théorique est un cadre méta pour la précédente. On touche alors la question de la réflexivité, et dans quelle mesure les théories peuvent s'appliquer à elles-mêmes\footnote{En gardant à l'esprit que la séparation entre les niveaux n'est pas directement évidente : par exemple le cadre formel pour les systèmes socio-techniques de~\ref{app:sec:csframework} pourrait être appliqué comme une formalisation du cadre de connaissances.}. Ainsi, cette démarche nous permet de mener simultanément une synthèse et une ouverture.
}

\stars

\bpar{
\textit{The first section of this chapter is fully novel. The second uses elements from \cite{raimbault2018coevolution}. The third has been proposed by~\cite{raimbault:halshs-01505084} and then developed and applied in~\cite{raimbault2017applied}, and its reflexive application has been presented by~\cite{raimbault2017co}.}
}{
\textit{La première section de ce chapitre est entièrement inédite. La deuxième reprend des éléments de \cite{raimbault2018coevolution}. La troisième a été proposée par~\cite{raimbault:halshs-01505084} puis développée et appliquée dans~\cite{raimbault2017applied}, et son application réflexive a été présentée par~\cite{raimbault2017co}.}
}

%


\newpage

\section{Contributions and Perspectives}{Contributions et perspectives}

\label{sec:contributions}


\bpar{
We now propose to review our contributions in relation to the different existing contexts reviewed in the first part, and to suggest some perspectives they open. We do so in the logic of our general problematic, with in a first step our contributions on the definition and characterization of co-evolution, and in a second step the different modeling approaches of it.
}{
Nous proposons à présent de passer en revue nos contributions au regard des différents cadres existants revus en première partie, et de suggérer des perspectives qu'elles ouvrent. Nous le faisons dans la logique de notre problématique générale, avec dans un premier temps nos apports sur la définition et la caractérisation de la co-évolution, et dans un second temps les différentes approches de modélisation de celle-ci.
}

\subsection{Definition and characterisation of co-evolution}{Définition et caractérisation de la co-évolution}

\bpar{
The stage of defining and characterizing co-evolution relies on empirical, theoretical and methodological results.
}{
L'étape de définition et de caractérisation de la co-évolution se repose sur des résultats empiriques, théoriques et méthodologiques.
}

\subsubsection{Conceptual definition}{Définition conceptuelle}

\bpar{
One of our main contributions is the construction of a definition of co-evolution within territorial systems. As developed in~\ref{sec:epistemology}, geography uses this concept in a mostly fuzzy way, whereas some disciplines in which its usage may seem to be more mature, such as in the evolutionary current of economic geography (see~\ref{sec:epistemology}), do not agree on a precise use~\cite{schamp201020}.
}{
L'une de nos contributions principales est la construction d'une définition de la co-évolution au sein des systèmes territoriaux. Comme développé en~\ref{sec:epistemology}, la géographie utilise ce concept de manière très floue, tandis que des disciplines où son usage pourrait sembler plus mature, comme dans le courant évolutionnaire de l'économie géographique (voir~\ref{sec:epistemology}), ne s'accordent pas sur un usage précis~\cite{schamp201020}.
}

\bpar{
We therefore precise the definition which is taken in the evolutive urban theory (see for example \cite{paulus2004coevolution}), while conserving a compatibility. Our definition indeed relies on three axis:
\begin{enumerate}
	\item existence of transformation processes of components of the territorial system (\emph{evolution}\footnote{While knowing that a weak correspondence can be established with reproduction and mutation, in particular in the case of ``simple'' socio-economic components for which the principles of cultural evolution apply, but that the correspondance becomes conceptual when the entities considered are more complex, as indeed in our case of cities and transportation networks.});
	\item modalities of co-evolution at different levels: local, population, system\footnote{Which are hierarchically necessary: a relation at the level of the population implies one at the level of individuals, and the systemic view implies a co-evolution at the level of populations.};
	\item modularity in territorial subsystems: territorial entities are both the support and the object of co-evolution.
\end{enumerate}
}{
Nous précisons ainsi la définition qui en est prise dans la théorie évolutive des villes (voir par exemple \cite{paulus2004coevolution}), en gardant compatibilité. Notre définition repose en effet sur trois axes :
\begin{enumerate}
	\item existence de processus de transformation des composantes du système territorial (\emph{evolution}\footnote{Sachant qu'on peut établir une correspondance faible avec reproduction et mutation, notamment dans le cas de composantes socio-économiques ``simples'' pour lesquelles les principes de l'évolution culturelle s'appliquent, mais que la correspondance devient conceptuelle quand les entités considérées sont plus complexes, comme justement dans notre cas des villes et des réseaux de transport.}) ;
	\item modalités de co-évolution à différents niveaux : local, population, système\footnote{Qui sont hiérarchiquement nécessaires : une relation au niveau de la population en implique une au niveau des individus, et la vue systémique implique une co-évolution au niveau des populations.} ;
	\item modularité en sous-systèmes territoriaux : les entités territoriales sont à la fois le support et l'objet de la co-évolution.
\end{enumerate}
}

\bpar{
Our contribution regarding the literature in geography which uses the concept is a clarification, which furthermore allows in some cases to proceed to an empirical characterization. \cite{paulus2004coevolution} or \cite{bretagnolle1998space} start with the assumption that co-evolution necessarily exists within systems of cities, between the cities or between cities and transportation networks. Our approach leaves some place for some empirical verification and also extends the application to territories in a more general way.
}{
Notre apport par rapport à la littérature géographique mobilisant le concept est une clarification, qui permet par ailleurs la mise en place dans certains cas d'une caractérisation empirique. \cite{paulus2004coevolution} ou \cite{bretagnolle1998space} partent du postulat que la co-évolution existe nécessairement au sein des systèmes de villes, entre villes ou entre villes et réseaux de transport. Notre approche laisse une entrée à une vérification empirique et étend également l'application aux territoires de manière plus générale.
}

\bpar{
As a positioning regarding the literature in economic geography (see~\cite{schamp201020}), our approach provides a fundamentally multi-scale perspective, and thus more easily compatible with geographical positioning such as the one of the evolutive urban theory.
}{
En positionnement par rapport à la littérature en économie géographique (voir~\cite{schamp201020}), notre approche permet une vision fondamentalement multi-échelles, et donc plus facilement compatible avec les positionnements géographiques comme celui de la théorie évolutive des villes.
}

\bpar{
Finally, we have in particular studied the concept in the context of interactions between transportation networks and territories: we show that these are a type of territorial system for which it is particularly relevant and operational. We can even therein revisit the debate on structuring effects: the congruence of \cite{offner1993effets} can either be a spurious correlation, or a true co-evolution effect at the level of the population. A local manifestation (``expected'' local link between two entities) can but has no particular reason to manifest as a co-evolution at the level of the population of entities (and thus there is naturally no ``systematic effect''). But to qualify the approach to this question as a ``scientific mystification'' \cite{offner1993effets} corresponds to scientific reductionism, which our approach contributes to go beyond.
}{
Enfin, nous avons étudié particulièrement le concept dans le cadre des interactions entre réseaux de transports et territoires : nous montrons qu'il s'agit d'un type de système territorial pour lequel il est particulièrement pertinent et opérationnel. Nous pouvons par là même revisiter le débat des effets structurants : la congruence de \cite{offner1993effets} peut être soit une corrélation fortuite, soit un vrai effet de co-évolution au niveau de la population. Une manifestation locale (lien local ``attendu'' entre deux entités) peut mais n'a pas de raison particulière de se manifester en tant que co-evolution au niveau de la population des entités (et donc il n'y a bien sûr aucun ``effet systématique''). Mais qualifier les approches de cette question de ``mystification scientifique'' \cite{offner1993effets} relève du réductionnisme scientifique, que notre approche contribue à dépasser.
}

\subsubsection{Spatial scales and non-stationarity}{Echelles spatiales et non-stationnarité}

\bpar{
An empirical contribution, allowing to bring evidences for the characterization of co-evolution, is obtained from the work done in~\ref{sec:staticcorrelations}. The existence of different observable spatial scales in static correlations between properties of the network and of the territory, and also the spatial non-stationarity of these, suggests the verification of the last point of our definition, in particular the existence of territorial subsystems within which co-evolution could manifest itself.
}{
Une contribution empirique, permettant d'apporter des pistes pour la caractérisation de la co-évolution, est issue du travail mené en~\ref{sec:staticcorrelations}. L'existence de différentes échelles spatiales observables dans les corrélations statiques entre caractéristiques du territoire et celles du réseau, ainsi que la non-stationnarité spatiale de celles-ci, suggère la vérification du dernier point de notre définition, à savoir l'existence de sous-systèmes territoriaux au sein desquels la co-évolution pourrait se manifester.
}

\subsubsection{Co-evolution regimes}{Régimes de co-évolution}

\bpar{
Our fundamental contribution regarding the characterization of co-evolution is the method of causality regimes developed in~\ref{sec:causalityregimes}. We suggest that depending on the observable regimes, some are indeed co-evolution regimes, since they exhibit circular causal relationships statistically observed at the scale of a population. It corresponds thus to an empirical characterization of the intermediate level of co-evolution, which is furthermore particularly interesting since it coincides with the territorial subsystems\footnote{Giving then all its usefulness to the approach through morphogenesis, by making the link as we already suggested and will develop in the following, with the notion of ecological niche~\cite{holland2012signals}.}.
}{
Notre contribution fondamentale en termes de caractérisation de la co-évolution est la méthode des régimes de causalité développée en~\ref{sec:causalityregimes}. Nous suggérons que selon les régimes observables, certains sont en effet des régimes de co-évolution, puisque présentant des relations causales circulaires observées statistiquement au niveau d'une population. Il s'agit ainsi d'une caractérisation empirique du niveau intermédiaire de co-évolution, qui est par ailleurs particulièrement intéressant puisque coïncidant avec les sous-systèmes territoriaux\footnote{Qui donne alors toute sa puissance à l'approche par la morphogenèse, en faisant le lien comme nous l'avons déjà suggéré et le développerons par la suite, avec la notion de niche écologique~\cite{holland2012signals}.}.
}

\bpar{
We postulate that our measure is a relatively good proxy of a co-evolution, since its application is oriented towards the study of causal networks~\cite{seth2005causal}, i.e. a set of directed relations between variables. \cite{castellacci2013dynamics} for example applies a method similar to the one we use, but extended to a network of variables, to quantify the co-evolution between innovation and absorption capacity of territories.
}{
Nous pensons que notre mesure est un relativement bon proxy d'une co-évolution, puisque son application s'oriente vers l'étude des réseaux causaux~\cite{seth2005causal}, c'est-à-dire un ensemble de relations dirigées entre variables. \cite{castellacci2013dynamics} applique par exemple une méthode similaire à la notre, mais étendue à un réseau de variables, pour quantifier la co-évolution entre innovation et capacité d'absorption des territoires.
}

\bpar{
Our approach can be put into perspective with the view of \noun{Diderot} presented in chapter~\ref{ch:thematic}: if there exists a niche in which we isolate relations which are indeed circular, then on the long time the evolutionary drift in comparison with other niches will lead them to very different trajectories\footnote{We furthermore have considered this case in an indirect way in models, when they are calibrated on long time on successive periods: the evolution of parameters corresponds to evolutionary dynamics on long time.}. Hence the importance of our general multi-scale framework, which also allows to consider the system more globally, and within which the connection between subsystems will then complexify the co-evolution relations\footnote{There would be on that point a larger complexity of territorial systems in comparison to ``simple'' biological systems, i.e. the ones in which ecological niches are clearly identifiable and can be isolated, in the case where the connections between subsystems is limited.}.
}{
Notre approche est à remettre en perspective avec la vue de \noun{Diderot} présentée en chapitre~\ref{ch:thematic} : s'il existe une niche dans laquelle on isole des relations en effet circulaires, alors sur le temps long la dérive évolutionnaire (\emph{drift}) par rapport à d'autres niches les entrainera sur des trajectoires bien différentes\footnote{Nous avons par ailleurs considéré ce cas de manière indirecte dans les modèles, lorsqu'ils sont calibrés sur le temps long sur des périodes successives : l'évolution des paramètres correspond à des dynamiques évolutives sur le temps long.}. D'où l'importance de notre cadre général multi-scalaire, qui permet par ailleurs la considération du système plus globalement, et au sein duquel la mise en réseau des sous-systèmes complexifiera alors les relations de co-évolution\footnote{Il y aurait sur ce point une plus grande complexité des systèmes territoriaux par rapport aux systèmes biologiques ``simples'', c'est-à-dire ceux dans lesquels des niches écologiques sont clairement identifiables et isolables, dans le cas où la mise en réseau entre sous-systèmes est limitée.}.
}

\subsubsection{Empirical applicability}{Applicabilité empirique}

\bpar{
The different case studies we introduced however witness of the difficulty to put into practice the methods tested on synthetic data or only theoretical. The application of the method of causality regimes gives very diverse results. On the Ile-de-France data in~\ref{sec:casestudies}, at a short temporal scale and a restricted spatial range, its application suggests the existence of diverse regimes. On South Africa data in~\ref{sec:causalityregimes}, we are not able to classify the relations between different variables, in particular because of the autocorrelation of accessibility, but the method allows to study a sense of causality between population growth and average travel time decrease, what however gives concluding results. Finally, in the case of France in~\ref{sec:macrocoevol}, the signal obtained is very weak, with mostly no significant correlation for most of the dates from 1836 to 1946. We however obtain the interesting results of the intermediate scale of spatial stationarity, and also a temporal stationarity scale for the long range relations. Therefore in practice, the application of the method must be considered case by case, and results can come from auxiliary or preliminary analyses.
}{
Nos différents cas d'étude empiriques témoignent toutefois de la difficulté de mettre en place les méthodes testées sur des données synthétiques ou uniquement théoriques. L'application de la méthode des régimes de causalité donne des résultats très divers. Sur les données d'Ile-de-France en~\ref{sec:casestudies}, à une échelle temporelle courte et une portée spatiale restreinte, son application suggère l'existence de différents régimes. Sur les données sud-africaines en~\ref{sec:causalityregimes}, on n'est pas capable de classifier les relations entre différentes variables, notamment à cause de l'autocorrélation de l'accessibilité, mais la méthode permet l'étude d'un sens de causalité entre croissance de population et croissance de temps moyen de trajet, ce qui donne toutefois des résultats concluants. Enfin, dans le cas de la France en~\ref{sec:macrocoevol}, le signal obtenu est très faible, avec quasiment aucune corrélation significative pour la majorité des dates de 1836 à 1946. On dégage toutefois les résultats intéressant d'échelle intermédiaire de stationnarité spatiale, ainsi que d'une échelle de stationnarité temporelle pour les relations à longue distance. Ainsi en pratique, l'application de la méthode est à considérer au cas par cas, et les résultats peuvent provenir d'analyses annexes ou préliminaires.
}

\bpar{
In the case of analyses of static correlations, which could open the door to a finer analysis and to significant correlations, we already saw that the absence of temporal data forbids any perspective of analysis in that direction. The development of methods allowing a characterization of co-evolution (according to one level of our definition or to an other definition) from static data remains an open question.
}{
Dans le cas des analyses des corrélations statiques, qui pourraient ouvrir une porte à une analyse fine et des corrélations significatives, on a déjà vu que l'absence de données temporelles empêche toute perspective d'analyse dans ce sens. Le développement de méthodes permettant une caractérisation de la co-évolution (selon l'un des niveaux de notre définition ou selon une autre définition) à partir de données statiques reste une question ouverte.
}

\bpar{
To summarize, co-evolution remains difficult to identify empirically, because (i) either there is effectively no apparent dynamic, i.e. that observable variables can be assimilated to noise (this case rejoins a large part of literature which concludes to dynamics based on single cases); (ii) data are very poor and despite evidences suggesting the existence of co-evolution regimes, these are difficult to characterize.
}{
En résumé, la co-évolution reste difficile à caractériser empiriquement, car (i) soit il n'y a effectivement aucune dynamique apparente, c'est-à-dire que les variables observables sont assimilables à du bruit (ce cas rejoint une grande partie de la littérature qui conclut à des dynamiques au cas par cas) ; (ii) les données sont très pauvres et malgré des indices suggérant l'existence de régimes de co-évolution, ceux-ci sont difficiles à caractériser.
}

\subsubsection{Perspectives}{Perspectives}

\bpar{
The application of our approach must be lead carefully regarding the choice of scales, processes and objects of study. Typically, it will be not adapted to the quantification of spatio-temporal processes for which the temporal scale of diffusion if of the same order than the estimation window, as our stationarity assumption here stays basic. We could propose to proceed to estimations on moving windows but it would then require the elaboration of a spatial correspondence technique to follow the propagation of phenomena.
}{
L'application de notre méthode des régimes de causalité doit être menée précautionneusement concernant le choix des échelles, processus et objets d'étude. Typiquement, elle ne sera pas du tout adaptée à la quantification de processus spatio-temporels dont l'échelle temporelle de diffusion est de l'ordre de celle de la fenêtre d'estimation : l'hypothèse de stationnarité est basique. Nous pouvons proposer de procéder à des estimations par fenêtres glissantes, mais il faudrait ensuite élaborer une technique de correspondance spatiale pour traquer la propagation des phénomènes.
}

\bpar{
An example of concrete application with a strong thematic potential impact would be a characterization of a fundamental component of the evolutive urban theory which is the hierarchical diffusion of innovation between cities~\cite{pumain2010theorie}, by analyzing potential spatio-temporal dynamics of patents classifications such as the one introduced by~\cite{10.1371/journal.pone.0176310}, to revisit analyses such as \cite{co2002evolution} which studies the diffusion of innovations between States in the US, with a more refined viewpoint both on the geographical aspect and to characterize innovation. We also underline that these are rather open methodological questions, for which a concretisation is the potential link between the non-ergodic properties of urban systems~\cite{pumain2012urban} and a wave-based characterization of these processes.
}{
Un exemple d'application concrète à l'impact thématique fort serait une caractérisation d'une composante fondamentale de la théorie évolutive des villes, la diffusion hiérarchique de l'innovation entre les villes~\cite{pumain2010theorie}, en analysant les potentielles dynamiques spatio-temporelles des classifications de brevets comme celle introduite par~\cite{10.1371/journal.pone.0176310}, pour revisiter des analyses comme \cite{co2002evolution} qui étudie la diffusion des innovations entre les États aux États-Unis, avec un point du vue plus fin à la fois géographiquement et pour la caractérisation de l'innovation. Il faut noter toutefois qu'il s'agit de questions méthodologiques relativement ouvertes, dont une des manifestations est le lien potentiel entre le caractère non-ergodique des systèmes urbains~\cite{pumain2012urban} et une caractérisation ondulatoire de ces processus.
}

\bpar{
An other direction for developments and potential applications can be found when going to a more local scale, by exploring an hybridation with Geographically Weighted Regression techniques~\cite{brunsdon1998geographically}. The determination by cross-validation of Akaike criterion of an optimal spatial scale for the performance of these models, as done in~\ref{sec:staticcorrelations} and in~\ref{sec:energyprice}, could be adapted in our case to determine a local optimal scale on which lagged correlations would be the most significant, what would allow to tackle the question of non-stationarity by a mostly spatial approach.
}{
Une autre direction potentielle de développement se révèle en se tournant vers l'échelle plus locale, et d'explorer une hybridation avec les techniques de Regression Géographique Pondérée~\cite{brunsdon1998geographically}. La détermination par validation croisée du Critère d'Akaike d'une portée spatiale optimale pour la performance de ce type de modèles, comme fait en~\ref{sec:staticcorrelations} et en~\ref{sec:energyprice}, pourrait être adaptée dans notre cas pour déterminer une échelle locale optimale sur laquelle les corrélations retardées sont les plus significatives, ce qui permettrait de s'extraire du problème de la non-stationnarité prioritairement par l'aspect spatial.
}


\subsection{Modeling co-evolution}{Modélisation de la co-évolution}

\bpar{
Our second fundamental contribution consists in the construction of co-evolution models. We now detail our contributions obtained by the intermediate of modeling, following the two complementary axis followed.
}{
Notre deuxième contribution fondamentale se situe dans la construction de modèles de co-évolution. Nous détaillons à présent nos contributions obtenues par l'intermédiaire de la modélisation, selon les deux axes complémentaires suivis.
}

\bpar{
Processes included in models are, as we already highlighted, aimed at being relatively simple to allow for a certain generality and flexibility, and for example do not include elaborated economic processes such as in the model of~\cite{levinson2007co}. They however fulfil their objectives and cover a rather broad range of processes. These are synthesized in Table~\ref{tab:contributions:modeled}.
}{
Les processus pris en compte dans les modèles sont, comme nous l'avons déjà soulevé, voulus relativement simples pour permettre une certaine généralité et flexibilité, et n'incluent par exemple pas de processus économiques élaborés comme le modèle de~\cite{levinson2007co}. Ils remplissent toutefois leurs objectifs et couvrent un spectre assez large de processus. Ceux-ci sont synthétisés en Table~\ref{tab:contributions:modeled}. 
}

\begin{table}
\caption[Processes taken into account in the proposed models]{\textbf{List of different processes taken into account in co-evolution models.}\label{tab:contributions:modeled}}
\bpar{
\begin{tabular}[6pt]{|p{4cm}|c|p{4cm}|c|}
\hline
Process & Scales & Concept & Proposed models\\\hline
Preferential attachment/Gibrat  & Meso/Macro & Urban growth & Morphogenesis/Interactions \\\hline
Diffusion/Sprawl & Meso & Urban Form & Morphogenesis \\\hline
Closeness centrality/Accessibility & Meso/Macro & Accessibility & Morphogenesis/Interactions \\\hline
Direct flows & Macro & Interactions & Interactions\\\hline
Indirect flows/Tunnel effect/Betweenness centrality & Meso/Macro & Network effects & Morphogenesis/Interactions \\\hline
Network proximity & Meso & Accessibility & Morphogenesis \\\hline
Actives/employments relocations & Meso & Residential mobility & Lutecia\\\hline
Transportation governance & Meso & Governance & Lutecia\\\hline
\end{tabular}
}{
\begin{tabular}[6pt]{|p{4cm}|c|p{4cm}|c|}
\hline
Processus & Échelles & Concept & Modèles proposés\\\hline
Attachement préférentiel/Gibrat  & Meso/Macro & Croissance urbaine & Morphogenèse/Interactions \\\hline
Diffusion/Etalement & Meso & Forme Urbaine & Morphogenèse \\\hline
Centralité de proximité/Accessibilité & Meso/Macro & Accessibilité & Morphogenèse/Interactions \\\hline
Flux direct & Macro & Interactions & Interactions\\\hline
Flux indirect/Effet tunnel/Centralité de Chemin & Meso/Macro & Effet de réseau & Morphogenèse/Interactions \\\hline
Proximité au réseau & Meso & Accessibilité & Morphogenèse \\\hline
Relocalisations actifs/emplois & Meso & Mobilité résidentielle & Lutecia\\\hline
Gouvernance des Transports & Meso & Gouvernance & Lutecia\\\hline
\end{tabular}
}
\end{table}


\subsubsection{Systems of Cities and the macroscopic scale}{Systèmes de villes et échelle macroscopique}

\bpar{
We first consider in particular co-evolution of territories and transportation networks within systems of cities, at the macroscopic scale.
}{
Considérons en particulier la co-évolution des territoires et des réseaux de transport au sein des systèmes de villes, à l'échelle macroscopique.
}

\paragraph{Network effects}{Effets de réseau}

\bpar{
Our results of section~\ref{sec:interactiongibrat} support the hypothesis that physical transportation networks are necessary to explain the morphogenesis of territorial systems, in the sense that some dimensions of urban growth are contained within networks. We showed indeed on a relatively simple case that the integration of physical networks into some models effectively increase their explanative power even when controlling for overfitting. This can be understood as a direction to expand the evolutive urban theory, that consider networks as carriers of interactions in systems of cities but do not put particular emphasis on their physical aspect and the possible spatial patterns resulting from it such as bifurcations or network induced differentiations. The development of a sub-theory focusing on these aspect is an interesting direction suggested by these empirical and modeling results. We will explore this path in section~\ref{sec:theory}.
}{
Nos résultats de la section~\ref{sec:interactiongibrat} soutiennent l'hypothèse que les réseaux de transport sont nécessaires pour expliquer la morphogenèse des systèmes territoriaux, au sens où certaines dimensions de la croissance des villes sont contenues dans les réseaux. Nous avons montré en effet sur un cas relativement simple que l'intégration des réseaux physiques dans certains modèles améliore effectivement leur pouvoir explicatif pour les variables de population des villes, même lorsqu'on contrôle pour le sur-ajustement. Cela peut être compris comme une direction pour étendre la théorie évolutive des villes, qui considère les réseaux comme médiateurs des interactions dans les systèmes de villes mais ne met pas d'accent précis sur leur aspect physique et les possibles motifs spatiaux en résultant comme des bifurcations ou des différenciations induites par le réseau. Le développement d'une sous-théorie se concentrant sur ces aspects est une direction intéressante suggérée par ces résultats empiriques et de modélisation. Nous explorerons cette piste en section~\ref{sec:theory}.
}

\paragraph{Co-evolution at the macroscopic scale}{Co-évolution à l'échelle macroscopique}


\bpar{
Regarding co-evolution in itself, at the scale of the system of cities, our main contribution is a global understanding of possible trajectories and regimes in a simple co-evolution model, i.e. based on an abstract ontology for the network and taking into account with parsimony mechanisms of cities and network evolution based on flows between cities.
}{
Concernant la co-évolution en elle-même, à l'échelle du système de villes, notre contribution principale est la compréhension globale des trajectoires et régimes possibles dans un modèle de co-évolution simple, c'est-à-dire se basant sur une ontologie abstraite pour le réseau et prenant en compte avec parcimonie des mécanismes d'évolution des villes et du réseau se basant sur les flux entre villes.
}

\bpar{
We obtain the typical stylized facts such as the reinforcement of hierarchy for some parameters of self-reinforcement such as obtained by~\cite{baptistemodeling}. It is to the best of our knowledge the first time a co-evolution model between transportation and cities in a system of cities is systematically explored, and that its potential co-evolution regimes are established and interpreted. Our model is put in perspective with the one of~\cite{schmitt2014modelisation}: the latest is closer to reality in terms of microscopic processes and network representation, what however allows less flexibility in the production of co-evolution regimes.
}{
Nous retrouvons les faits stylisés typiques comme le renforcement de la hiérarchie pour certains paramètres d'auto-renforcement comme obtenu par~\cite{baptistemodeling}. Il s'agit à notre connaissance de la première fois qu'un modèle de co-évolution entre transport et villes dans un système de villes est exploré systématiquement, que ses régimes potentiels de co-évolution sont établis et interprétés. Notre modèle est mis en perspective avec celui de~\cite{schmitt2014modelisation} : ce dernier est plus fidèle à la réalité en termes de processus microscopiques et de représentation du réseau, ce qui permet toutefois moins de flexibilité dans la production de régimes de co-évolution.
}

\bpar{
For the application to the real case of the French system of cities, it is also to the best of our knowledge the first time that such a model is calibrated on observed data. It is difficult to know if co-evolution processes are indeed observable, since on the contrary to \cite{bretagnolle2003vitesse} we do not find a significant relation between city growth and accessibility. The calibration allows however to extrapolate the evolution of co-evolution parameters values in time.
}{
Pour l'application au cas réel du système de villes français, c'est également à notre connaissance la première fois qu'un tel modèle est calibré sur données observées. Il est difficile de dire si les processus de co-évolution sont effectivement observables, puisqu'au contraire de \cite{bretagnolle2003vitesse} nous ne trouvons pas de relation significative entre croissance des villes et accessibilité. La calibration permet toutefois d'extrapoler l'évolution de la valeur des paramètres de co-évolution dans le temps.
}

\paragraph{Perspectives}{Perspectives}


\bpar{
Our macroscopic models have not yet been tested on other urban systems and other temporalities, and further work should investigate which conclusions we obtained here are specific to the French Urban System on this periods, and which are more general and could be more generic in system of cities. Applying the model to other system of cities also recalls the difficulty of defining urban systems. In our case, a strong bias should arise from considering France only, as the insertion of its urban system into an European system is a reality that we had to neglect. The extent and scale of such models is always a delicate subject. We rely here on the administrative coherence and the consistence of the database, but sensitivity to system definition and extent should also be further tested.
}{
Nos modèles macroscopiques n'ont pas encore été testés sur d'autres systèmes urbains et d'autres étendues temporelles, et les développements futurs devront étudier quelles conclusions obtenues ici sont spécifiques au système de villes français sur ces périodes, et lesquelles sont plus générales et pourraient être plus génériques dans les systèmes de villes. L'application du modèle à d'autres systèmes de villes rappelle également la difficulté de définir les systèmes urbains. Dans notre cas, un fort biais doit être induit par le fait de considérer la France seule, puisque l'insertion de son système urbain dans un système européen est une réalité que nous avons dû négliger. L'étendue et l'échelle de tels modèles est toujours un sujet délicat. Nous reposons ici sur la cohérence administrative et celle de la base de données, mais la sensibilité à la définition du système et à son étendue doivent encore être testés.
}



\bpar{
Furthermore, the calibration used the rail network only for distances between cities. Considering one single transportation model is naturally reducing, and an immediate possibility of development is to test the model with real distance matrices for other types of networks, such as the freeway network which followed a considerable growth in France in the second half of the 20th century. This application requires to construct a dynamical database for the freeway network growth covering 1950 to 2015, since classical bases (IGN or OpenStreetMap) do not integrate the opening date. A natural extension of the modeling would then consist in integrating a multilayer network, which is a typical approach to represent multi-modal transportation networks~\cite{gallotti2014anatomy}. Each layer of the transportation network should have co-evolutive dynamics with populations, with possibly the existence of inter-layer dynamics.
}{
Par ailleurs, la calibration a utilisé le réseau ferré uniquement pour les distances entre villes. La considération d'un seul mode de transport est bien sûr réductrice, et une direction immediate de développement est le test du modèle avec des matrices de distance réelles pour d'autres types de réseaux, comme le réseau autoroutier qui a connu un essor considerable en France dans la seconde moitié du 20ème siècle. Cette application nécessite la mise en place d'une base dynamique pour la croissance du réseau autoroutier couvrant 1950 à 2015, les bases classiques (IGN ou OpenStreetMap) n'intégrant pas la date d'ouverture des tronçons. Une extension naturelle du modele consisterait alors en la mise en place d'un réseau multi-couches, approche typique pour représenter des systèmes de transport multi-modaux~\cite{gallotti2014anatomy}. Chaque couche du réseau de transport devrait avoir une dynamique co-évolutive avec les populations, avec possiblement l'existence d'une dynamique inter-couches.
}


\bpar{
Finally, one of our potential developments, taking into account the physical network in a finer way, is the object of~\cite{mimeur:tel-01451164}, which produces interesting results regarding the influence of the centralization of network investment decisions on final forms, but keeps static populations and does not produce co-evolution models. Similarly, the choice of indicators to quantify the distance of the simulated network to a real network is a delicate issue in this context: indicators such as the number of intersections taken by~\cite{mimeur:tel-01451164} corresponds to procedural modeling and not structural indicators. This is probably for the same reason that~\cite{schmitt2014modelisation} only focuses on population trajectories and not on network indicators: the conjunction and adjustment  of population and network dynamics at different scales seems to be a difficult problem.
}{
Enfin, l'un de nos développements potentiels, la prise en compte plus fine du réseau physique, est l'objet de~\cite{mimeur:tel-01451164}, qui produit des résultats intéressants quant à l'influence de la centralisation de la décision d'investissement dans le réseau sur les formes finales, mais garde des populations statiques et ne produit pas de modèle de co-évolution. De même, le choix des indicateurs pour quantifier la distance du réseau simulé à un réseau réel est un problème délicat dans ce contexte : des indicateurs comme le nombre d'intersections pris par~\cite{mimeur:tel-01451164} relève de la modélisation procédurale et non d'indicateurs de structure. C'est probablement pour la même raison que~\cite{schmitt2014modelisation} ne s'intéresse qu'aux trajectoires de population et pas aux indicateurs de réseau : la conjonction et l'ajustage des dynamiques de population et de réseau à des échelles différentes semble être un problème difficile.
}

\subsubsection{Territories and mesoscopic scale}{Territoires et échelle mesoscopique}

\bpar{
We propose now to develop our contributions for the modeling of co-evolution of territories and transportation networks at the mesoscopic scale.
}{
Nous proposons à présent de développer nos contributions pour la modélisation de la co-évolution des territoires et des réseaux de transport à l'échelle mesoscopique. 
}

\paragraph{Urban morphogenesis}{Morphogenèse urbaine}


\bpar{
First of all, the conceptual framework of morphogenesis developed in~\ref{sec:interdiscmorphogenesis} is a proper thematic contribution for urban modeling: we insist on the crucial role of urban form, and its strong link with urban functions. This framework allows furthermore to better situate some morphogenesis models such as the one by \cite{bonin2014modelisation} (which is to the best of our knowledge one of the only presented as morphogenetic having the required theoretical fundations) in an interdisciplinary context.
}{
Dans un premier temps, le cadre conceptuel de la morphogenèse développé en~\ref{sec:interdiscmorphogenesis} est un apport thématique propre pour la modélisation urbaine : nous appuyons le rôle crucial de la forme urbaine, et de son lien fort avec la fonction urbaine. Ce cadre permet par ailleurs de mieux situer des modèles de morphogenèse comme celui de \cite{bonin2014modelisation} (qui est à notre connaissance l'un des seuls modèles se présentant comme morphogénétiques ayant les fondements théoriques requis) dans un cadre interdisciplinaire.
}

\bpar{
It also allows to consider in a consistent way territorial subsystems, since the search for morphogenetic rules is common to the definition of more or less precise boundaries for the considered subsystem. This point remarkably rejoins the geographical isolation required for co-evolution, and we will do a theoretical link in the following in~\ref{sec:theory}.
}{
Il permet également de considérer de façon cohérente des sous-systèmes territoriaux, puisque la recherche de règles morphogénétiques est conjointe à la définition de limites plus ou moins précises au sous-système considéré. Ce point rejoint remarquablement l'isolation géographique requise pour la co-évolution, et nous ferons la jonction théorique par la suite en~\ref{sec:theory}.
}

\paragraph{Modeling co-evolution with morphogenesis}{Modélisation de la co-évolution par morphogenèse}


\bpar{
The contributions of our morphogenesis co-evolution model are multiple, and at least the following points can be mentioned:
\begin{itemize}
	\item comparison of multiple network generation heuristics within a co-evolution model;
	\item calibration on morphological indicators for population distribution and topological for the road network;
	\item calibration at the first and second order;
	\item study of co-evolution regimes produced by such a model.
\end{itemize}
}{
L'apport de notre modèle de co-évolution par morphogenèse est multiple et au moins les points suivants sont à noter :
\begin{itemize}
	\item comparaison de multiples heuristiques de génération au sein d'un modèle de co-évolution ;
	\item calibration sur indicateurs morphologiques pour la distribution de la population et topologiques pour le réseau routier ;
	\item calibration au premier et au second ordre ;
	\item étude des régimes de co-évolution produit par un tel modèle.
\end{itemize}
}

\bpar{
The coupled ontology between population distribution and network brings the strong coupling between form and function, and precisely to consider co-evolution processes. In comparison to \cite{barthelemy2009co} which only consider the network, our model allows for more flexibility in the processes taken into account, since it is then possible for example to add mechanisms proper to the evolution of population without artificially acting on network topology, and reciprocally.
}{
L'ontologie couplée distribution de population et réseau permet le couplage fort entre forme et fonction, et justement de considérer des processus de co-évolution. En comparaison à \cite{barthelemy2009co} qui ne considèrent que le réseau, notre modèle permet plus de flexibilité dans les processus pris en compte, puisqu'il est alors possible par exemple d'ajouter des mécanismes propres à l'évolution de la population sans agir artificiellement sur la topologie du réseau, et réciproquement.
}

\paragraph{Towards governance modeling}{Vers une modélisation de la gouvernance}


\bpar{
Finally, the Lutecia model is also a fundamental contribution towards the inclusion of more complex processes implied in co-evolution, such as transportation system governance. As we already mentioned, \cite{Xie2011} introduces an theoretical economic model focusing on similar issues, and \cite{xie2011governance} develops a simplified application on synthetic networks. We go further by considering an integration into a fully dunamical land-use transport interaction model, and implement a stylized application to the real case of Pearl River Delta. This models yield avenues to a new generation of models, that can potentially be operational in the case of regional systems with a very high evolution speed such as in the Chinese case.
}{
Enfin, le modèle Lutecia est également une contribution fondamentale vers la prise en compte de processus plus complexes impliqués dans la co-évolution, comme la gouvernance du système de transport. Comme nous l'avons déjà indiqué, \cite{Xie2011} introduit un modèle économique théorique s'intéressant à une problématique similaire, et \cite{xie2011governance} développe une application simplifiée sur réseau synthétique. Nous allons plus loin en considérant une intégration à un modèle entièrement dynamique d'interaction entre transport et usage du sol, et implémentons une application stylisée au cas réel du Delta de la Rivière des Perles. Ce modèle ouvre la porte à une nouvelle génération de modèles, pouvant être potentiellement opérationnels dans le cas de systèmes régionaux à très grande vitesse d'évolution comme dans le cas Chinois.
}

\paragraph{Perspectives}{Perspectives}

\bpar{
The question of the generic character of the morphogenesis model is also open, i.e. if it would work similarly when trying to reproduce Urban Forms on very different systems such as the United States or China. A first interesting development would be to test it on these systems and at slightly different scales (1km cell for example).
}{
La question du caractère générique du modèle de morphogenèse est également ouverte, c'est-à-dire s'il fonctionnerait de la même manière pour reproduire des formes urbaines sur des systèmes très différents comme les États-Unis ou la Chine. Un premier développement intéressant serait de le tester sur ces systèmes et à des échelles légèrement différentes (cellules de taille 1km par exemple).
}


\bpar{
Finally, we postulate that a significant insight into the non-stationarity of urban systems would be allowed by its integration into a multi-scale growth model. Urban growth patterns have been empirically shown to exhibit multi-scale behavior~\cite{zhang2013identifying}. Here at the meso-scale, total population and growth rates are fixed by exogenous conditions of processes occurring at the macro-scale. It is particularly the aim of spatial growth models such as the Favaro-pumain model~\cite{favaro2011gibrat} to determine such parameters through relations between cities as agents. One would condition the morphological development in each area to the values of the parameters determined at the level above. In that setting, one must be careful of the role of the bottom-up feedback: would the emerging urban form influence the macroscopic behavior in its turn? Such multi-scale complex model are promising but must be considered carefully for the required level of complexity and the way to couple scales.
}{
Enfin, nous pensons qu'un gain de connaissance important concernant la non-stationnarité des systèmes urbains serait rendu possible par son intégration dans un modèle de croissance multi-échelle. Les motifs de croissance urbaine ont été prouvés empiriquement exhibant un comportement multi-échelle~\cite{zhang2013identifying}. Ici à l'échelle mesoscopique, la population totale et le taux de croissance sont fixés par les conditions exogènes de processus se produisant à l'échelle macroscopique. C'est particulièrement le but des modèles spatiaux de croissance comme le modèle Favaro-Pumain~\cite{favaro2011gibrat} de déterminer de tels paramètres par les relations entre villes comme agents. Il serait alors possible de conditionner le développement morphologique de chaque zone aux valeurs des paramètres déterminés au niveau supérieur. Dans ce contexte, il faudrait être prudent sur le rôle de l'émergence : la forme urbaine émergente devrait-elle influencer le comportement macroscopique à son tour ? De tels modèles complexes multi-scalaires sont prometteurs mais doivent être considérés avec précaution pour le niveau de complexité requis et la manière de coupler les échelles.
}


\subsubsection{Position of models}{Position des modèles}

\bpar{
We do a synthesis of the position of different models regarding co-evolution in Table~\ref{tab:contributions:models}. We describe the models which have been a novelty in this work and also the external models which have been used, and the empirical studies. We thus see that it is immediate to introduce a co-evolution at the individual level within models, but that the co-evolution at the level of populations, i.e. the existence of circular causalities between network variables and territory variables, is more difficult to obtain in a direct manner. The structuring effects (existence of causal relationships in a sense or the other) are for themselves existing in most models. We recall that it is difficult to measure a co-evolution on empirical data.
}{
Nous faisons une synthèse de la position des différents modèles au regard de la co-évolution en Table~\ref{tab:contributions:models}. Nous donnons les modèles qui ont été nouvellement introduits dans ce travail ainsi que les modèles extérieurs ayant été utilisés, et les études empiriques. Nous voyons ainsi qu'il est direct d'introduire une co-évolution au niveau individuel dans les modèles, mais que la co-évolution au niveau de la population, c'est-à-dire l'existence de causalités circulaires entre variables de réseau et variables de territoire, est plus difficile à obtenir de façon marquée. Les effets structurants (existence de relations causales dans un sens ou dans l'autre) sont quant à eux présents dans presque la totalité des modèles. Nous rappelons qu'il est difficile de mesurer une co-évolution sur données empiriques.
}

\begin{table}
\caption[Behavior of models regarding co-evolution]{\textbf{Behavior of models regarding co-evolution.} For all the models we used or introduced, we give the positioning regarding different degrees of co-evolution: production of ``structuring effects'' (direct causality relationships between variables or exogenous aspect of a variable), existence of a co-evolution at the individual level (ontological specification of the model), existence of a co-evolution at the population level (existence of circular causalities) and existence of a coevolution at the systemic level (that we can not test). We also list the empirical studies. Modalities are the following: ``NA'' means that the effect has no reason to exist (for example systemic co-evolution for a mesoscopic model, or co-evolution for a static model); ``n.t.'' means that it was not tested (for practical reasons or as no test exist); ``x'' means that the model seems to be producing the effect but in a marginal way (or observed in a qualitative way for empirical studies); ``X'' means that the model produces the effect without doubt; ``o'' means that the effect is not produced or that the analysis is not conclusive.\label{tab:contributions:models}}
\bpar{
\begin{tabular}[6pt]{|p{4cm}|p{2.5cm}|p{2.5cm}|p{2.5cm}|p{2.5cm}|}
\hline
Model & Structuring effects & Individual co-evolution & Population co-evolution & Systemic co-evolution \\\hline
RBD \ref{sec:causalityregimes} & X & X & X & NA \\\hline
Interactions \ref{sec:interactiongibrat} & x & NA & NA & NA \\\hline
Weak coupling \ref{sec:correlatedsyntheticdata} & x & NA & NA & NA \\\hline
SimpopNet \ref{sec:macrocoevolexplo} & X & X & x & n.t. \\\hline
Macro co-evolution \ref{sec:macrocoevol} & X & X & X & n.t. \\\hline
Meso co-evolution \ref{sec:mesocoevolmodel} & X & X & x & NA\\\hline
Lutecia \ref{sec:lutecia} & n.t. & X & n.t. & NA\\\hline
Empirical: Grand Paris \ref{sec:casestudies} & X & x & o & NA\\\hline
Empirical: South Africa \ref{sec:causalityregimes} & X & x & o & n.t.\\\hline
Empirical: France \ref{sec:macrocoevol} & o & x & o & n.t.\\\hline
\end{tabular}
}{
\begin{tabular}[6pt]{|p{4cm}|p{2.5cm}|p{2.5cm}|p{2.5cm}|p{2.5cm}|}
\hline
Modèle & Effets structurants & Co-évolution individuelle &  Co-évolution population & Co-évolution systémique \\\hline
RBD \ref{sec:causalityregimes} & X & X & X & NA \\\hline
Interactions \ref{sec:interactiongibrat} & x & NA & NA & NA \\\hline
Couplage faible \ref{sec:correlatedsyntheticdata} & x & NA & NA & NA \\\hline
SimpopNet \ref{sec:macrocoevolexplo} & X & X & x & n.t. \\\hline
Macro co-évolution \ref{sec:macrocoevol} & X & X & X & n.t. \\\hline
Meso co-évolution \ref{sec:mesocoevolmodel} & X & X & x & NA\\\hline
Lutecia \ref{sec:lutecia} & n.t. & X & n.t. & NA\\\hline
Empirique : Grand Paris \ref{sec:casestudies} & X & x & o & NA\\\hline
Empirique : Afrique du Sud \ref{sec:causalityregimes} & X & x & o & n.t.\\\hline
Empirique : France \ref{sec:macrocoevol} & o & x & o & n.t.\\\hline
\end{tabular}
}
\end{table}

\subsection{Approaches of coevolution}{Approches de la co-évolution}

\bpar{
Finally, we propose to open broader perspectives on co-evolution approaches different from the one we took.
}{
Finalement, nous proposons d'ouvrir des perspectives plus larges d'approches de la co-évolution différentes de la notre.
}

\subsubsection{Alternative approaches}{Approches alternatives}

\bpar{
We made the choice of elementary characteristics of territories and networks to model their co-evolution: most of our models only consider population variables for territories, and several other possible dimensions (economic, political, institutional, social) are occulted. Dimensions where there potentially exists co-evolutive effects and where a modeling would be relevant can be summarized the following way: 
}{
Nous avons fait le choix de caractéristiques élémentaires des territoires et des réseaux pour la modélisation de leur co-évolution : la plupart de nos modèles ne considère que des variables de population pour les territoires, et de nombreuses autres dimensions possibles (économique, politique, institutionnelle, sociale) sont occultées. Des dimensions où il existe potentiellement des effets co-évolutifs et où une modélisation serait pertinente peuvent être regroupés de la manière suivante :
}

\bpar{
\begin{itemize}
	\item questions linked to the transportation system:
	\begin{itemize}
		\item role of transportation tolling and investments, already largely taken into account into economic models by \noun{Levinson} as~\cite{levinson2007co};
		\item more generally role of governance actors in the evolution of the transportation system, as we suggested with the Lutecia model in~\ref{sec:lutecia}, and as \cite{Xie2011} does in a more theoretical way;
		\item role of technological change in the relation between urban form and mobility \cite{brotchie1984technological};
	\end{itemize}
	\item questions linked to actors making the city:
	\begin{itemize}
		\item role of the different actors producing the city (real estate managers\footnote{We briefly evoked in~\ref{sec:casestudies} some variables linked to real estate transactions, and showed the potentialities to unveil strategies of anticipation of accessibility patterns by the new network, which confirms here the relevance of this viewpoint.} and local administrations for example \cite{le2010acteurs}) and of their strategies;
		\item in link with approaches of type Luti, be more precise on the role of location choices of actors (residential mobility or economic actors \cite{tannier2003trois}) in the production of the territory, in relation with networks (``scale of accessibility'') identified in chapter~\ref{ch:thematic};
	\end{itemize}
	\item finally, at the scale of daily mobility, mobility practices following socio-economic characteristics, is also a relevant territorial dimension to follow for the study of co-evolution (\cite{cerqueira2017inegalites} show for example the socio-economic differentiations in the link between accessibility and mobility), for which modeling directions has been for example proposed by~\cite{morency2005contributions} which construct through desegregation an integrated database coupling socio-economic characteristics of households and mobility data.
\end{itemize}
}{
\begin{itemize}
	\item problématiques liées au système de transport :
	\begin{itemize}
	\item rôle de la tarification des transports et des investissements, déjà largement pris en compte dans les modèles économiques de \noun{Levinson} comme~\cite{levinson2007co} ;
	\item plus généralement rôle des acteurs de gouvernance dans l'évolution du système de transport, comme nous avons esquissé avec le modèle Lutecia en~\ref{sec:lutecia}, et comme le fait de manière plus théorique~\cite{Xie2011} ;
	\item rôle du changement technologique dans la relation entre forme urbaine et mobilité \cite{brotchie1984technological} ;
	\end{itemize}
	\item problématiques liées aux acteurs faisant la ville :
	\begin{itemize}
	\item rôle des différents acteurs de production de la ville (promoteurs immobiliers\footnote{Nous avons abordé en~\ref{sec:casestudies} brièvement des variables liées aux transactions immobilières, et montré les potentialités pour la mise en valeur de stratégies d'anticipation de desserte par le nouveau réseau, ce qui confirme ici la pertinence de ce point de vue.} et collectivités locales par exemple \cite{le2010acteurs}) et de leurs stratégies ;
	\item en lien avec les approches de type Luti, approfondir le rôle des choix de localisation des acteurs (mobilités résidentielles ou acteurs économiques \cite{tannier2003trois}) dans la production territoriale, en relation aux réseaux (``échelle de l'accessibilité'') identifiée au chapitre~\ref{ch:thematic} ;
	\end{itemize}
	\item enfin, à l'échelle des mobilités quotidiennes, les pratiques de mobilité selon les caractéristiques socio-économiques, est également une dimension territoriale pertinente à creuser pour l'étude de la co-évolution (\cite{cerqueira2017inegalites} montre par exemple les différenciations socio-économiques dans le lien entre accessibilité et mobilité), pour laquelle des pistes de modélisation ont par exemple été proposées par~\cite{morency2005contributions} qui construit par désagrégation une base de donnée intégrée couplant caractéristiques socio-économiques des ménages et données de mobilité.
\end{itemize}
}

\bpar{
This list if naturally far from being exhaustive, but allows to grasp complementary dimensions which would also bring an entry on our general problematic.
}{
Cette liste est bien évidemment loin d'être exhaustive, mais permet de se rendre compte des dimensions complémentaires qui permettraient également une entrée sur notre problématique générale.
}

\bpar{
We are thus far from having exhausted the question of co-evolution, since it would require then to understand: (i) to what extent is our definition general and can be applied to other dimensions which were not initially conceived; (ii) to what extent our characterization method can be applied to the different dimensions and which alternative methods can be considered; (iii) if our model structures, which are relatively generic, can be extended to this connected issues.
}{
Nous sommes donc loin d'avoir épuisé la problématique de la co-évolution, puisqu'il s'agirait alors de savoir : (i) dans quelle mesure notre définition est générale et s'applique à des dimensions qui n'ont pas été initialement conçues ; (ii) dans quelle mesure notre méthode de caractérisation s'applique aux différentes dimensions et quelles méthodes alternatives sont envisageables ; (iii) si nos structures de modèles, relativement génériques, peuvent être étendues à ces problématiques connexes.
}

\subsubsection{Towards operational co-evolution models?}{Vers des modèles opérationnels de co-évolution ?}

\bpar{
Among the future challenges which open as a consequence of our work, we can mention the perspective of operational models. Is it possible to construct prospective or planning models similar to the ones we developed?
}{
Parmi les défis futurs qui s'ouvrent à la suite de notre travail, nous pouvons mentionner la perspective de modèles opérationnels. Est-il possible de construire des modèles de prospective ou de planification similaires à ceux que nous avons mis en place ?
}

\bpar{
First of all, we make the assumption that such models would effectively be efficient within particular contexts, in particular regarding the timescales considered: an application to the case of China where potentialities are relatively rapidly realized can be more relevant than an application of the Grand Paris metropolitan area which as we saw exhibits a strong complexity and thus longer time scales in the infrastructure evolution processes.
}{
Tout d'abord, nous subodorons que de tels modèles seraient effectivement efficaces dans des cadres particuliers, notamment au regard des échelles de temps concernées : une application au cas de la Chine où les potentialités sont réalisées assez rapidement peut être plus crédible qu'une application à la Métropole du Grand Paris qui comme nous l'avons vu présente une forte complexité et donc des échelles de temps plus longues dans les processus d'évolution des infrastructures.
}

\bpar{
Furthermore, we must keep in mind the difficulty of prospecting on long times: to the best of our knowledge it is impossible to integrate in an endogenous way in a model some structural changes of territorial systems\footnote{In the sense of a \emph{transition} of settlement systems as elaborated by~\cite{tannier:hal-01666491}.}. Therefore, the transition of the industrial revolution is exogenous in the Simpop2 model \cite{doi:10.1177/0042098010377366}. In our case of interactions between networks and territories, it is possible than currently non-existing practices (for example shared mobility within fleets of on-demand autonomous electric vehicles) totally change processes and drastically change the landscape of public transportation accessibility.
}{
Par ailleurs, il faut garder à l'esprit la difficulté d'une prospective sur le temps long : il est impossible à notre connaissance d'intégrer dans un modèle de manière endogène des changements structurels des systèmes territoriaux\footnote{Au sens d'une \emph{transition} des systèmes de peuplement comme élaboré par~\cite{tannier:hal-01666491}.}. Ainsi, la transition de la révolution industrielle est exogène dans le modèle Simpop2 \cite{doi:10.1177/0042098010377366}. Dans notre cas des interactions entre réseaux et territoires, il est possible que des pratiques de mobilité actuellement non existantes (par exemple mobilité partagée dans des flottes de véhicules électriques autonomes sur demande) changent totalement la donne et bouleversent le paysage d'accessibilité aux transports en commun.
}

\bpar{
Finally, as mentioned in chapter~\ref{ch:modelinginteractions}, it is possible that such a type of model would indeed be undesirable from the viewpoint of actors producing or managing territories, since their role of territorial scenarizing and prospecting would strongly be reduced by hypothetical operational models\footnote{Knowing that we also did not mention the aspect of the role of models in public decision making, and the way that models can take a place in the dialogue between science and society: the nature and relevance of operational models is also tightly linked to this question.}.
}{
Enfin, comme évoqué en chapitre~\ref{ch:modelinginteractions}, il est possible que ce type de modèle soit en fait indésirable du point de vue des acteurs produisant ou gérant les territoires, puisque leur rôle de scénarisation et de prospection territoriale serait grandement réduit par des hypothétiques modèles opérationnels\footnote{Sachant que nous n'avons également pas évoqué l'aspect de la place des modèles dans la décision publique, et la manière dont les modèles peuvent jouer une place dans le dialogue entre science et société : la nature et la pertinence de modèles opérationnels est également étroitement lié à cette question.}.
}

\stars

\bpar{
We could thus in this section take a step back on our contributions and put them into broader perspectives regarding the question of the co-evolution of transportation networks and territories.
}{
Nous avons ainsi pu dans cette section prendre du recul sur nos contributions et les mettre en perspective d'horizons plus vastes concernant la question de la co-évolution des réseaux de transport et des territoires.
}

\bpar{
The next section proposes an articulation of our different contributions from a theoretical viewpoint, as s synthesis making explicit some connexions which were implicit until here.
}{
La section suivante propose une articulation de nos différentes contributions d'un point de vue théorique, une synthèse permettant d'expliciter certaines connexions jusqu'alors relativement implicites.
}

\stars

%


\newpage


\section{A geographical theory}{Une théorie géographique}

\label{sec:theory}

\bpar{
\noun{Raffestin} highlights in his preface of~\cite{offner1996reseaux} that a geographical theory that articulates spaces, networks and territories has never been formulated in a consistent way, since each approach has a vision reduced to some components only and does not aim at constructing an integrative theory. A research direction we propose to introduce here is the conjunction of approaches of the evolutive urban theory and of morphogenesis, to produce a theory that is both multi-scalar and fully integrates networks and territories.
}{
\noun{Raffestin} souligne dans sa préface de~\cite{offner1996reseaux} qu'une théorie géographique articulant espaces, réseaux et territoires n'a jamais été formulée de manière cohérente, chaque approche ayant une vision réduite à certaines composantes seulement et ne visant pas à construire une théorie intégrée. Une piste que nous proposons d'introduire ici est la conjonction des approches de la théorie évolutive des villes et de la morphogenèse, pour produire une théorie à la fois multi-scalaire et intégrant pleinement réseaux et territoires.
}

\subsection{Foundations}{Fondations}

\bpar{
Our theoretical construction relies on four pillars that we will detail below\footnote{Or more precisely a funding horizontal pillar which gives fundamental objects, i.e. foundations introduced in Chapter~\ref{ch:thematic}, two vertical pillars for the structure, and an horizontal synthesis pillar allowing to link these two.}.
}{
Notre construction théorique repose sur quatre piliers que nous détaillons ci-dessous\footnote{Ou plutôt un pilier horizontal de fondement qui précise les objets fondamentaux, c'est-à-dire les fondations introduites en Chapitre~\ref{ch:thematic}, deux piliers verticaux de structure, et un pilier horizontal de synthèse permettant de faire le lien entre ceux-ci.}.
}

\subsubsection{Networked human territories}{Territoires humains en réseau}

\bpar{
Our first pillar corresponds to the theoretical construction elaborated in~\ref{sec:networkterritories}. We rely on the notion of \emph{Human Territory} elaborated by \noun{Raffestin} as the basis for a definition of a territorial system. It allows to capture complex human geographical systems in all the extent of their concrete and abstract characteristics, and also their representations. For example, a metropolitan territory can be apprehended simply by the functional extent of daily commuting flows, or by the perceived or lived space for different populations, the choice depending on the precise question that is considered.
}{
Notre premier pilier correspond à la construction théorique que nous avons élaborée en~\ref{sec:networkterritories}. Nous nous basons sur la notion de \emph{Territoire Humain} élaborée par \noun{Raffestin} comme la base de la définition d'un système territorial. Elle permet de capturer les systèmes complexes géographiques humains dans l'ensemble de leur caractéristiques concrètes et abstraites, ainsi que dans leur représentations. Par exemple, un territoire métropolitain peut être appréhendé simplement par l'étendue fonctionnelle des flux pendulaires journaliers, ou par l'espace perçu ou vécu des différentes populations, le choix dépendant de la question précise à laquelle on cherche à répondre.
}

\bpar{
The territory of \noun{Raffestin} indeed corresponds to a consistent system of \emph{synergetic inter-representation networks}, which are both a theory and a model for spatial cognition of individual and societies, constructed by \noun{Portugali} and \noun{Haken} (see~\cite{portugali2011sirn} for a synthetic presentation). It postulates that representations are the product of a strong coupling between individuals of cognitions and their individual and collective behaviors. This approach to the territory is of course a particular choice and other entries, possibly compatible, can be taken~\cite{murphy2012entente}.
}{
Le territoire de \noun{Raffestin} correspond en fait à un système cohérent de réseaux synergétiques d'inter-représentations (\emph{synergetic inter-representation networks}), qui sont à la fois une théorie et un modèle pour la cognition spatiale des individus et des sociétés, construite par \noun{Portugali} et \noun{Haken} (voir~\cite{portugali2011sirn} pour une présentation synthétique). Elle postule que les représentations sont le produit du couplage fort entre les individus des cognitions et de leurs comportements individuels et collectifs. Cette approche au territoire est bien sûr un choix particulier et d'autres entrées, possiblement compatibles, peuvent être prises~\cite{murphy2012entente}.
}

\bpar{
The concrete of this pillar in reinforced by the territorial theory of networks of \noun{Dupuy}, yielding the notion of networked human territory, as a human territory in which a set of potential transactional networks have been realized, which is in accordance with visions of the territory as networked places~\cite{champollion:halshs-00999026}. We will not use the implications of the development of the notion of \emph{place}, these being too sparse (see the definition of \cite{hypergeo}), and because of the redundancy with the territory in the vision of a complex link between representations and the physical reality. We will assume for this first pillar the fundamental assumption, already introduced in Chapter~\ref{ch:thematic}, that real networks are necessary ele;ents of territorial systems.
}{
Le ciment de ce pilier est renforcé par la théorie territoriale des réseaux de \noun{Dupuy}, fournissant la notion de territoire humain en réseau, comme un territoire humain dans lequel un ensemble de réseaux transactionnels potentiels ont été réalisés, ce qui s'accorde par ailleurs avec les visions du territoire comme un lieu des réseaux~\cite{champollion:halshs-00999026}. Nous n'utiliserons pas les implications du développement de la notion de \emph{lieu}, celles-ci étant trop éparses (voir définition de \cite{hypergeo}), et à cause de la redondance avec le territoire dans la vision de lien complexe entre représentation et réalité physique. Nous ferons pour ce premier pilier l'hypothèse fondamentale, déjà introduite en Chapitre~\ref{ch:thematic}, que les réseaux réels sont des éléments nécessaires des systèmes territoriaux.
}


\subsubsection{Evolutive urban theory}{Théorie évolutive des villes}

\bpar{
The second pillar of our theoretical construction is \noun{Pumain}'s evolutive urban theory, in close relation with the complex approach that we generally take. It has already been presented with details and its implications have been explored in Chapter~\ref{ch:evolutiveurban}. Here, this theory allows us to interpret territorial systems as complex adaptive systems and to introduce the co-evolution.
}{
Le second pilier de notre construction théorique est la théorie évolutive des villes de \noun{Pumain}, en relation étroite avec l'approche complexe que nous prenons de manière générale. Celle-ci a déjà été présenté en détails et ses implications ont été explorées en Chapitre~\ref{ch:evolutiveurban}. Ici, cette théorie nous permet d'interpréter les systèmes territoriaux comme systèmes complexes adaptatifs et d'introduire la co-évolution.
}

\subsubsection{Urban morphogenesis}{Morphogenèse Urbaine}





\bpar{
The notion of morphogenesis has been deeply explored and with an interdisciplinary point of view in~\ref{sec:interdiscmorphogenesis}. We recall here important axis and to what extent these contribute to the construction of our theory. Morphogenesis has been formalized especially by~\cite{turing1952chemical} which proposes to isolate elementary chemical rules that could lead to the emergence of the embryo and its form.
}{
Le concept de morphogenèse a été déjà exploré en profondeur et selon un point de vue interdisciplinaire en~\ref{sec:interdiscmorphogenesis}. Nous rappelons ici certains grands axes et dans quelle mesure ceux-ci contribuent à la construction de notre théorie. La morphogenèse a été formalisée notamment par~\cite{turing1952chemical} qui propose d'isoler des règles chimiques élémentaires qui pourraient mener à l'émergence de l'embryon et à sa forme.
}

\bpar{
The morphogenesis of a system consists in evolution rules that produce the emergence of its successives states, i.e. the precise definition of self-organization, with the additional property that an emergent architecture exists, in the sense of causal circular relations between the form and the function. Progresses towards the understanding of embryo morphogenesis (in particular the isolation of particular processes producing the differentiation of cells from an unique cell) have been made only recently with the use of complexity approaches in integrative biology~\cite{delile2016chapitre}.
}{
La morphogenèse d'un système consiste en des règles d'évolution qui produisent l'émergence de ses états successifs, i.e. la définition précise de l'auto-organisation, avec la propriété supplémentaire qu'une architecture émergente existe, au sens de relations causales circulaires entre la forme et la fonction. Les progrès vers la compréhension de la morphogenèse de l'embryon (en particulier l'isolation de processus particuliers induisant la différentiation de cellules à partir d'une unique) sont relativement récents grâce à l'application des approches complexes en biologie intégrative~\cite{delile2016chapitre}.
}

\bpar{
In the case of urban systems, the idea of urban morphogenesis, i.e. of self-consistent mechanisms that would produce the urban form, is more used in the field of architecture and urban design (as for example the generative grammar of ``Pattern Language'' of \cite{alexander1977pattern}), in relation with theories of urban form~\cite{moudon1997urban}. This idea can be pushed into very large scales such as the one of the building~\cite{whitehand1999urban} but we will use it more at a mesoscopic scale, in terms of land-use changes within an intermediate scale of territorial systems, with similar ontologies as the urban morphogenesis modeling literature (for example \cite{bonin2012modele} describes a model of urban morphogenesis with qualitative differentiation, whereas \cite{makse1998modeling} give a model of urban growth based on a mono-centric population distribution perturbed with correlated noises).
}{
Dans le cas des systèmes urbains, l'idée de morphogenèse urbaine, i.e. de mécanismes auto-cohérents qui produiraient la forme urbaine, est plutôt utilisé dans les champs de l'architecture et de l'urbanisme (comme par exemple la grammaire générative du ``Pattern Language'' de \cite{alexander1977pattern}), en relation avec des théories de la forme urbaine~\cite{moudon1997urban}. Cette idée peut être poussée jusqu'à de très grandes échelles comme celle du bâtiment~\cite{whitehand1999urban} mais nous la considérons à une échelle mesoscopique, en termes de changements d'usage du sol à une échelle intermédiaire des systèmes territoriaux, avec des ontologies similaires à la littérature de modélisation de la morphogenèse urbaine (par exemple \cite{bonin2012modele} décrit un modèle de morphogenèse urbaine avec différentiation qualitative, tandis que \cite{makse1998modeling} donne un modèle de croissance urbaine basé sur une distribution monocentrique de la population perturbée par des bruits corrélés).
}

\bpar{
The concept of morphogenesis is important in our theory in link with modularity and scale. Modularity of a complex system consists in its decomposition into relatively independent sub-modules, and the modular decomposition of a system can be seen as a way to disentangle non-intrinsic correlations~\cite{2015arXiv150904386K} (to have an idea, think of a block diagonalisation of a first order dynamical system). In the context of large-scale cyber-physical systems design and control, similar issues naturally raise and specific techniques are needed to scale up simple control methods~\cite{2017arXiv170105880W}. The isolation of a subsystem yields a corresponding characteristic scale. Isolating possible morphogenesis processes implies a controlled extraction (controlled boundary conditions e.g.) of the considered system, corresponding to a modularity level and thus a scale.
}{
Le concept de morphogenèse est important dans notre théorie en lien avec la modularité et l'échelle. La modularité d'un système complexe consiste en sa décomposition en sous-modules relativement indépendants, et la décomposition modulaire d'un système peut être vue comme un moyen de supprimer les corrélations non intrinsèques \cite{2015arXiv150904386K} (pour donner une image, penser à une diagonalisation par blocs d'un système dynamique du premier ordre). Dans le cadre de la conception et du contrôle de systèmes cyber-sociaux à grande échelle, des problèmes similaires surgissent naturellement et des techniques spécifiques sont nécessaires pour le passage à l'échelle des techniques simple de contrôle \cite{2017arXiv170105880W}. L'isolation d'un sous-système fournit une échelle caractéristique correspondante. Isoler des processus de morphogenèse possibles implique une extraction contrôlée (conditions au bord contrôlées par exemple) du système considéré, ce qui correspond à un niveau de modularité et donc à une échelle.
}

\bpar{
When local processes are not enough to explain the evolution of a system (with reasonable variations of initial conditions), a change of scale is necessary, caused by an underlying phase transition in modularity. The example of metropolitan growth is a good example: complexity of interactions within the metropolitan region will grow with size and the diversity of functions, leading to a change in the scale necessary to understand processes. The characteristic scales and the nature of processes for which these change occur can be precise questions investigated through modeling.
}{
Quand des processus locaux ne sont pas suffisants pour expliquer l'évolution d'un système (dans des variations raisonnables des conditions initiales), un changement d'échelle est nécessaire, causé par une transition de phase implicite dans la modularité. L'exemple de la croissance métropolitaine en est une très bonne illustration : la complexité des interactions au sein de la région métropolitaine sera croissante avec sa taille et la diversité des fonctions urbaines, ce qui conduit à un changement de l'échelle nécessaire pour comprendre les processus. Les échelles caractéristiques et la nature des processus pour lesquels ces changements ont lieu peuvent être des questions précisément approchées par l'angle de la modélisation.
}


\bpar{
Finally, it is important to remark as we did in~\ref{sec:interdiscmorphogenesis} that a territorial subsystem for which morphogenesis makes sense, which boundaries are well defined and which processes allow it to maintain itself as a network of processes, is close to an \emph{auto-poietic system} in the extended sense of \noun{Bourgine} in~\cite{bourgine2004autopoiesis}\footnote{Which are however not cognitive, making these morphogenetic systems not alive in the sense of auto-poietic and cognitive. Given the difficulty to define the delineation of cities for example, we will leave open the issue of the existence of auto-poietic territorial systems, and will consider in the following a less restrictive point of view on boundaries.}. These systems regulate then their boundary conditions, what underlines the importance of boundaries that we will finally develop.
}{
Enfin, il est important de noter comme nous l'avons fait en~\ref{sec:interdiscmorphogenesis} qu'un sous-système territorial pour lequel la morphogenèse prend sens, dont les frontières sont bien définies et dont les processus lui permettent de se maintenir en tant que réseau de processus, est proche d'un \emph{système auto-poiétique} au sens étendu de \noun{Bourgine} dans~\cite{bourgine2004autopoiesis}\footnote{Qui ne sont toutefois pas cognitifs, ne rendant pas ces systèmes morphogénétiques vivants au sens de auto-poiétique et cognitif. Vu la difficulté de définir la délimitation des villes par exemple, nous laisserons ouverte la question de l'existence de systèmes territoriaux auto-poiétiques, et considérerons par la suite un point de vue moins restrictif sur les frontières.}. Ces systèmes régulent alors leur conditions aux bords, ce qui souligne l'importance des frontières sur lesquelles nous allons finalement nous attarder.
}

\subsubsection{Co-evolution}{Co-évolution}


\bpar{
Our last pillar consists in an approach to the concept of \emph{co-evolution} complementary to the definition we already introduced. It is brought by \noun{Holland} which sheds a relevant light through an approach of complex adaptive systems (CAS) by a theory of CAS as agents which fundamental property is to process signals thanks to their boundaries~\cite{holland2012signals}.
}{
Notre dernier pilier consiste en une approche du concept de \emph{co-evolution} complémentaire à la définition que nous avons déjà introduite. Celle-ci est amenée par \noun{Holland} qui apporte un éclairage pertinent à travers son approche des systèmes complexes adaptatifs (CAS) par une théorie des CAS comme agents dont la propriété fondamentale est de traiter des signaux grâce à leur frontières~\cite{holland2012signals}.
}

\bpar{
In this theory, complex adaptive systems form aggregates at diverse hierarchical levels, which correspond to different level of self-organization, and boundaries are vertically and horizontally intricated in a complex way. That approach introduces the notion of \emph{niche} as a relatively independent subsystem in which ressources circulate (the same way as communities in a network as used in chapter~\ref{ch:modelinginteractions}): numerous illustrations such as economical niches or ecological niches can be given. Agents within a niche are then said to be \emph{co-evolving}.
}{
Dans cette théorie, les systèmes complexes adaptatifs forment des agrégats à différents niveaux hiérarchiques, qui correspondent à différents niveaux d'auto-organisation, et les frontières sont intriquées horizontalement et verticalement de manière complexe. Cette approche introduit la notion de \emph{niche} comme un sous-système relativement indépendant au sein duquel les ressources circulent (de la même façon que des communautés dans un réseau comme utilisé en chapitre~\ref{ch:modelinginteractions}) : de nombreuses illustrations telles les niches écologiques ou économiques peuvent être données. Les agents au sein d'une niche sont alors dits en \emph{co-évolution}.
}

\bpar{
Empirically, results obtained witness a co-evolution at the mesoscopic scale such as in~\ref{sec:causalityregimes}, confirming the existence of niches for some aspects of territorial systems. The co-evolution in that sense implies then strong interdependencies with circular causal processes (rejoining the definition we took) and a certain independence regarding the exterior of the niche.
}{
Empiriquement, les résultats obtenus témoignant d'une co-évolution à l'échelle mesoscopique comme en~\ref{sec:causalityregimes}, confirment l'existence de niches pour certains aspects des systèmes territoriaux. La co-évolution dans ce sens implique ainsi de fortes interdépendances avec des processus causaux circulaires (rejoignant la définition que nous en avons prise) et une certaine indépendance au regard de l'extérieur de la niche.
}

\bpar{
The notion is naturally flexible as it will depend on ontologies, on the resolution, on thresholds, etc. that we consider to define the system. We postulate given the clues of existence obtained in empirical results, but also models reproducing processes in a credible manner under a reasonable independence assumption, that this concept can easily be transmitted to the evolutive urban theory and corresponds to the notion of co-evolution we took (and in particular at the level of a population of entities): co-evolving agents in a system of cities consist in a niche with their own flows, signals and boundaries and thus co-evolving entities in the sense of \noun{Holland}.
}{
Le concept est flexible puisqu'il dépendra des ontologies, de la résolution, des seuils, etc. que l'on considère pour définir le système. Nous postulons vu les indices d'existence obtenus dans les résultats empiriques, mais aussi les modèles reproduisant les processus de manière crédible sous une hypothèse d'indépendance raisonnable, que ce concept peut se transmettre à la théorie évolutive urbaine et correspond à la notion de co-évolution que nous avons prise (et en particulier au niveau d'une population d'entités) : des agents co-évolutifs dans un système de villes consistent en une niche avec leur propres flux, signaux et limites et sont donc des entités co-évolutives au sens de \noun{Holland}.
}


\subsection{A theory of co-evolutive networked territorial systems}{Une théorie des systèmes territoriaux co-évolutifs en réseau}

\bpar{
We synthesize the different pillars as a geographical theory of territorial systems in which networks play a central role in the co-evolution of system components.
}{
Nous synthétisons les différents piliers en une théorie géographique des systèmes territoriaux pour lesquels les réseaux jouent un rôle central pour la co-évolution des composantes du système.
}


\medskip

\bpar{
\begin{definition}
\textbf{ - Territorial System.} A territorial system is a set of networked human territories, i.e. human territories in and between which real networks are materialized.
\end{definition}
}{
\begin{definition}
\textbf{ - Système Territorial.} Un système territorial est un ensemble de territoires humains en réseau, c'est-à-dire des territoires humains au sein desquels et entre lesquels des réseaux concrets sont matérialisés.
\end{definition}
}

\medskip

\bpar{
The territory is indeed an element of the territorial system, which more generally connects different territories with networks. At this stage complexity and the evolutive and dynamical character of territorial systems are implied by the positions taken but not an explicit part of the theory. We will assume to simplify a discrete definition of temporal, spatial and ontological dimensions, under modularity and local stationarity assumptions. This aspect, both for the discrete and the stationarity, corresponds to an ontological simplification of the assumption of a ``minimal scale'' at which subsystems give a simple modular decomposition of the global system.
}{
Le territoire est bien un élément du système territorial, qui de manière plus générale connecte différents territoires par les réseaux. À cette étape la complexité et le caractère évolutif et dynamique des systèmes territoriaux sont impliqués par les partis pris mais pas une partie explicite de la théorie. Nous supposerons pour simplifier une définition discrète des dimensions temporelles, spatiales et ontologiques, sous des hypothèses de modularité et de stationnarité locale. Cet aspect, à la fois pour le discret et la stationnarité, correspond à une simplification ontologique de la supposition d'une ``échelle minimale'' à laquelle les sous-systèmes fournissent une décomposition modulaire simple du système global.
}


\medskip


\bpar{
\begin{assumption}
\textbf{ - Discrete scales.} Assuming a discrete modular decomposition of a territorial system, the existence of a discrete set of temporal and functional scales for the territorial system is equivalent to the local temporal stationarity of a random dynamical system specification of the system.
\end{assumption}
}{
\begin{assumption}
\textbf{ - Echelle discrètes.} Supposant une décomposition modulaire discrète d'un système territorial, l'existence d'un ensemble discret d'échelles temporelles et fonctionnelles pour le système territorial est équivalent à la stationnarité temporelle locale d'une spécification par système dynamique stochastique du système.
\end{assumption}
}


\medskip

\bpar{
This proposition postulates a representation of system dynamics in time. Note that even in the absence of a modular representation, the system as a whole will verify the property. We will assume the case in which scales always exist, i.e. verifying one of the specifications of this assumption.
}{
Cette proposition postule une représentation des dynamiques du système dans le temps. Nous pouvons noter que même en l'absence de représentation modulaire, l'hypothèse s'appliquera au système dans son ensemble. Nous nous placerons dans le cadre où les échelles existent toujours, c'est-à-dire vérifiant l'une des spécifications de cette hypothèse.
}

\bpar{
This definition of scales allows to explicitly introduce feedback loops, since we can for example condition the evolution of a scale to the evolution of an other containing it, and thus emergence and complexity, making the theory compatible with the evolutive urban theory.
}{
Cette définition des échelles permet d'introduire explicitement des boucles de rétroaction, puisqu'on peut par exemple conditionner l'évolution d'une échelle à celle d'une autre qui la contient, et ainsi l'émergence et la complexité, rendant la théorie compatible avec la théorie évolutive urbaine.
}

\bpar{
\begin{assumption}
\textbf{ - Intrication of scales and subsystems. } Complex networks of feedbacks exist both between and within scales~\cite{bedau2002downward}. Furthermore, a horizontal and vertical imbrication of boundaries will not always be hierarchical.
\end{assumption}
}{
\begin{assumption}
\textbf{ - Imbrication des échelles et des sous-systèmes. } Des réseaux complexes d'interactions existent à la fois entre et à l'intérieur des échelles~\cite{bedau2002downward}. De plus, un emboîtement horizontal et vertical des limites ne sera généralement pas hiérarchique.
\end{assumption}
}


\bpar{
Within these complex subsystems intrications we can isolate co-evolving components using morphogenesis. The following proposition is a consequence of the equivalence between the independence of a niche and its morphogenesis. Morphogenesis provides the modular decomposition (under the assumption of local stationarity) necessary for the existence of scales, giving minimal vertically (scale) and horizontally (space) independent subsystems.
}{
Au sein de ces imbrications de sous-systèmes nous pouvons isoler des composantes en co-évolution en utilisant la morphogenèse. La proposition suivante est une conséquence de l'équivalence entre l'indépendance d'une niche et sa morphogenèse. La morphogenèse fournit la décomposition modulaire (sous hypothèse de stationnarité locale) nécessaire pour l'existence des échelles, donnant des sous-systèmes minimaux indépendants de manière verticale (échelle) et horizontale (espace).
}

\bpar{
\begin{assumption}
\textbf{ - Co-evolution of components. } Morphogenesis processes of a territorial system are an equivalent formulation of the existence of co-evolutive subsystems.
\end{assumption}
}{
\begin{assumption}
\textbf{ - Co-évolution des composantes. } Les processus morphogénétiques d'un système territorial sont une formulation équivalente de l'existence de sous-systèmes co-évolutifs.
\end{assumption}
}


\bpar{
Finally we make a key assumption putting real networks at the center of co-evolutive dynamics, introducing their necessity to explain dynamical processes of territorial systems.
}{
Nous formulons finalement la dernière hypothèse clé qui met les réseaux réels au centre des dynamiques co-évolutives, introduisant leur nécessité pour expliquer les processus dynamiques des systèmes territoriaux.
}

\bpar{
\begin{assumption}
\textbf{ - Necessity of networks. } The evolution of networks can not be explained only by the dynamics of other territorial components and reciprocally, i.e. co-evolving territorial subsystems include real networks. They can thus be at the origin of regime changes (transition between stationarity regimes) or more dramatic bifurcations in dynamics of the whole territorial system.
\end{assumption}
}{
\begin{assumption}
\textbf{ - Nécessité des réseaux. } L'évolution des réseaux ne peut pas être expliquée simplement par la dynamique des autres composantes territoriales et réciproquement, i.e. les sous-systèmes territoriaux co-évolutifs contiennent les réseaux réels. Ceux-ci peuvent ainsi être à l'origine de changements de régime (transitions entre régimes stationnaires) ou de bifurcations plus conséquentes dans les dynamiques de l'ensemble du système territorial.
\end{assumption}
}

\subsection{Contextualization}{Contextualisation}

\bpar{
Co-evolution is more or less easy to show empirically (see for example the debate on structuring effects) but we assume the existence of co-evolution processes at all scales of the system. Regional examples for the French system of cities may illustrate that aspect: Lyon has not the same interactions with Clermont-Ferrand than with Saint-Etienne, and network connectivity has probably a role in that (among intrinsic interaction dynamics, and distance for example). At a even larger scale, we speculate that effects are even less observable, but precisely because of the fact that co-evolution is stronger and local bifurcations will occur with stronger amplitude and greater frequency than in macroscopic systems where attractors are more stable and stationarity scales smaller. It is for this reason that we tried to identify bifurcations and phase transitions in toy models, hybrid models, and empirical analyses, at different scales, on different case studies and with different ontologies.
}{
La co-évolution est plus ou moins facile à mettre en évidence empiriquement (par exemple débat sur les effets structurants) mais nous supposons la présence de processus de co-évolution à toutes les échelles du système. Des exemples régionaux pour le système de villes français peuvent illustrer ce fait : Lyon n'a pas les mêmes interactions avec Clermont-Ferrand qu'avec Saint-Etienne, et la connectivité de réseau a probablement un rôle à y jouer (parmi les effets des dynamiques intrinsèques des interactions, et de la distance par example). À une plus grande échelle encore, nous partons du principe que les effets sont encore moins observables, mais précisément à cause du fait que la co-évolution est plus forte et les bifurcations locales se produisent avec une plus grande amplitude et une plus grande fréquence que dans les systèmes macroscopiques où les attracteurs sont plus stables et les échelles de stationnarité plus petites. C'est pour cela que nous avons tenté d'identifier des bifurcations ou des transitions de phase dans des modèles jouets, des modèles hybrides, et des analyses empiriques, à différentes échelles, sur différents cas d'études et avec différentes ontologies. 
}

\bpar{
One difficulty in our construction is the local stationarity assumption, which is essential to formulate models at the corresponding scale. Even if it seems a reasonable assumption on several scales and has already been observed in empirical data~\cite{sanders1992systeme}, we were able to verify it more or less in our empirical studies.
}{
Une difficulté dans notre construction est l'hypothèse de stationnarité locale, qui est essentielle pour formuler des modèles à l'échelle correspondante. Même si cela paraît une hypothèse raisonnable à plusieurs échelles et qui a déjà été observée avec des données empiriques~\cite{sanders1992systeme}, nous avons plus ou moins pu le vérifier dans nos études empiriques.
}

\bpar{
Indeed, this question is at the center of current research efforts to apply deep learning techniques to geographical systems: \noun{Paul Bourgine}\footnote{Personal communication, January 2016.} has recently proposed a framework to extract patterns from complex adaptive systems. Using a representation theorem~\cite{knight1975predictive}, any discrete stationary process is a \emph{Hidden Markov Model}. Given the definition of a causal state as as the set of states allowing an equivalent prediction of the future, the partition of system states induced by the corresponding equivalence relations allows to derive a \emph{Recurrent Network} that is sufficient to determine the next state of the system, as it is a \emph{deterministic} function of previous states and hidden states~\cite{shalizi2001computational}: $(x_{t+1},s_{t+1}) = F\left[(x_t,s_t)\right]$ if $x_t$ is the state of the system and $s_t$ the hidden states. The estimation of hidden states and of the recurrent function thus captures entirely through deep learning dynamical patterns of the system, i.e. full information on its dynamics and internal processes.
}{
En effet, cette question est au centre des efforts de recherche courants pour appliquer les techniques d'apprentissage profond aux systèmes géographiques : \noun{Paul Bourgine}\footnote{Communication personnelle, janvier 2016.} a récemment proposé un cadre pour extraire des motifs des systèmes complexes adaptatifs. En utilisant un théorème de représentation~\cite{knight1975predictive}, tout processus stationnaire discret est un Modèle de Markov Caché. Étant donné la définition d'un état causal comme l'ensemble des états permettant une prédiction équivalente du futur, la partition des états du système par la relation d'équivalence correspondante permet de produire un \emph{Réseau Récurrent} qui est suffisant pour déterminer l'état suivant du système, puisqu'il s'agit d'une fonction \emph{déterministe} des états précédents et des états cachés~\cite{shalizi2001computational} : $(x_{t+1},s_{t+1}) = F\left[(x_t,s_t)\right]$ si $x_t$ est l'état du système et $s_t$ les états cachés. L'estimation des états cachés et de la fonction récurrente capture ainsi entièrement par apprentissage profond le comportement dynamique du système, i.e. l'information complète sur ses dynamiques et les processus internes.
}

\bpar{
The issues that raise then are if the stationarity assumptions can be tackled through augmentation of system states, and if heterogeneous and asynchronous data can be used to bootstrap long enough time-series necessary for a correct estimation of the neural network or any other estimator. These issue are related to the stationarity assumption for the first and to non-ergodicity for the second.
}{
Les questions qui se posent ensuite sont si les hypothèses de stationnarité peuvent être réglés par augmentation des états du système, et si des données hétérogènes et asynchrones peuvent être utilisées pour initialiser des séries temporelles assez longues pour une estimation correcte du réseau de neurones ou de tout autre type d'estimateur. Ces questions sont reliées à l'hypothèse de stationnarité pour la première et à la non-ergodicité pour la seconde.
}

\stars

\bpar{
This section has thus given a theoretical opening, by proposing as hypothesis an articulation between the different complementary approaches that we developed. This articulation allows a global perspective and reinforces our definition of co-evolution.
}{
Cette section nous a ainsi permis une ouverture théorique, en proposant sous la forme d'hypothèses une articulation entre les différentes approches complémentaires que nous avons développées. Cette articulation permet une perspective globale et renforce notre définition de la co-évolution.
}

\bpar{
The next section concludes this opening from an epistemological point of view, by placing our work in the perspective of a knowledge framework, and opening thus reflexive approaches on it.
}{
La section suivante conclut cette ouverture d'un point de vue épistémologique, en plaçant notre travail dans une perspective d'un cadre de connaissance, et ouvrant ainsi des pistes de réflexivité sur celui-ci.
}

\stars

%


\newpage

\section{An applied knowledge framework}{Un cadre de connaissance appliqué}

\label{sec:knowledgeframework}

\bpar{
We propose to increase again the level of generality and to work at an epistemological level, by introducing a theoretical frame for the study of knowledge production processes.
}{
Nous proposons de monter encore en niveau de généralité et de nous placer à un niveau épistémologique, en introduisant un cadre théorique pour l'étude des processus de production de connaissance.
}

\bpar{
The complexity of knowledge production on complex systems is well-known, but there still lacks knowledge framework that would both account for a certain structure of knowledge production at an epistemological level and be directly applicable to the study and management of complex systems. We set a basis for such a framework, by first analyzing in detail a case study of the construction of a geographical theory of complex territorial systems, through mixed methods, namely qualitative interview analysis and quantitative citation network analysis. We can therethrough inductively build a framework that considers knowledge entreprises as perspectives, with co-evolving components within complementary knowledge domains. We finally discuss potential applications and developments.
}{
La complexité de la production de connaissance sur des systèmes complexes est bien connue, mais il n'existe toujours pas de cadre de connaissance qui rendrait à la fois compte d'une certaine structure de la production de connaissance à un niveau épistémologique et serait directement applicable à l'étude et à la gestion des systèmes complexes. Nous posons ici les bases d'un tel cadre, en commençant par analyser en détails l'étude de cas de la construction d'une théorie géographique des systèmes territoriaux complexes, au travers de méthodes mixtes, plus précisément des analyses qualitatives d'entretiens et une analyse quantitative de réseau de citations. Nous pouvons par cela construire de manière inductive un cadre qui considère les entreprises de production de connaissance comme des perspectives, dont les composantes sont en co-évolution au sein de domaines de connaissances complémentaires. Nous discutons finalement des applications et développements potentiels.
}

\bpar{
The understanding of processes and conditions of scientific knowledge production are still mainly open questions, to which monuments of epistemology such as Kant's Critique of Pure Reason, and more recently Kuhn's study of ``the structure of scientific revolutions''~\cite{kuhn1970structure} or Feyerabend's advocacy for a diversity of viewpoints \cite{feyerabend1993against}, have brought elements of answer from a philosophical approach. A more empirical point of view was brought also recently with quantitative studies of science, in a way a \emph{quantitative epistemology} that goes far beyond rough bibliometric indicators~\cite{cronin2014beyond}. Contributions harnessing complexity, i.e. studying complex systems in a very broad sense, can be shown to have produced very diverse frameworks that can be counted as building bricks contributing to answers to the above high-level question.
}{
La compréhension des processus et des conditions de production de la connaissance scientifique est une question toujours globalement ouverte, à laquelle des monuments de l'épistémologie comme la Critique de la Raison Pure de Kant, ou plus récemment l'étude par Kuhn de la ``structure des révolutions scientifiques''~\cite{kuhn1970structure} ou le positionnement de Feyerabend pour une diversité des approches~\cite{feyerabend1993against}, ont apporté des éléments de réponse d'un point de vue philosophique. Un matériau plus empirique a été apporté également récemment avec les analyses quantitatives de la science, dans un sens une \emph{épistémologie quantitative} qui va bien plus loin que des indicateurs bibliométriques purs~\cite{cronin2014beyond}. Les contributions s'intéressant à la complexité, c'est-à-dire étudiant des systèmes complexes en un sens très large, peuvent témoigner de la production de cadres de travail très divers qui peuvent être vus comme des éléments élémentaires de réponse à la question ci-dessus à un autre niveau.
}

\bpar{
We will in the following use the term Knowledge Framework, for any such framework having an epistemological component tackling the question of nature of knowledge or knowledge production. To illustrate this, we can mention such frameworks in different domains, at different levels and with different purposes. For example, \cite{durantin2017disruptive} explores the potentialities of coupling engineering with design paradigms to enhance disruptive innovation. Also in Knowledge Management, using the constraint of innovation as an advantage to understand to complex nature of knowledge, \cite{carlile2004transferring} introduces knowledge domains boundaries and production processes. Also introducing a meta-framework, but in the field of system engineering, \cite{gemino2004framework} recommends to use grammars to compare Conceptual Modeling Techniques. Meta-modeling frameworks can also be understood as Knowledge Frameworks. \cite{cottineau2015modular} describes a multi-modeling framework to test hypotheses in simulation of socio-technical complex systems. \cite{golden2012modeling} postulates a unified formulation of systems, including necessarily different types of knowledge on a system on its different description components.
}{
Nous utiliserons par la suite le terme \emph{cadre de connaissance}, pour tout cadre tel ayant une composante épistémologique s'intéressant à la nature de la connaissance et à sa production. Pour illustrer, nous pouvons mentionner de tels cadres dans différents domaines, à différents niveaux, et avec des objectifs différents. Par exemple, \cite{durantin2017disruptive} explorent les potentialités de coupler l'ingénierie avec des paradigmes du design pour encourager l'innovation disruptive. Toujours en gestion de connaissances, utilisant la contrainte de l'innovation comme un avantage pour appréhender la nature complexe de la connaissance, \cite{carlile2004transferring} introduit les notions de frontières des domaines de connaissance et de processus de production. Introduisant également un cadre au niveau méta, mais dans le champ de l'ingénierie des systèmes, \cite{gemino2004framework} recommandent l'utilisation de grammaires pour comparer les techniques de modélisation conceptuelle. Les cadres de méta-modélisation peuvent aussi être compris comme des cadres de connaissance. \cite{cottineau2015modular} décrivent un cadre de multi-modélisation pour le test d'hypothèses dans la simulation des systèmes complexes socio-techniques. \cite{golden2012modeling} postule une formulation unifiée de la notion de système, ce qui inclut nécessairement différents types de connaissance sur un système correspondant à la description de ses différents composants.
}

\bpar{
A possible explanation for this richness is the fundamental reflexive nature of the study of Complex Systems: because of the higher choice in methodology and what aspects of the system to put emphasis on, a significant part of a modeling or design entreprise is an investigation at a meta-level.  Furthermore, studies of knowledge production are mainly rooted in complexity, implying a reflexive nature of theories accounting knowledge on complexity, as Hofstadter had well highlighted in \cite{hofstadter1980godel} by noticing the importance of ``strange loops'', i.e. feedback loops allowing reflexivity such as a theory applying to itself, in what constitutes intelligence and the mind. Artificial intelligence is indeed a crucial field regarding our issues, as its progresses imply a finer understanding of the nature of knowledge. \cite{2017arXiv170401407M} introduces a meta-framework for a general typology of approaches in Artificial Intelligence, what is a Knowledge Framework not in the proper sense but in a specific applied case.
}{
Une explication possible pour une telle richesse est la nature fondamentalement réflexive de l'étude des systèmes complexes : à cause du choix plus grand pour la méthodologie et sur quels aspects du système mettre l'emphase, une partie significative d'une entreprise de modélisation ou de design est une exploration à un niveau méta. De plus, les études de la production de connaissance sont profondément ancrées dans la complexité, comme \noun{Hofstadter} a bien souligné dans \cite{hofstadter1980godel} en rappelant l'existence de ``boucles étranges'', c'est-à-dire de boucles de rétroaction permettant la réflexivité comme une théorie s'appliquant à elle-même, dans ce qui constitue l'intelligence et l'esprit. L'intelligence artificielle est de fait un champ crucial au regard de nos réflexions, comme ses progrès impliquent une compréhension plus fine de la nature de la connaissance. \cite{2017arXiv170401407M} introduit un meta-cadre pour une typologie générale des approches en intelligence artificielle, ce qui correspond à un cadre de connaissance non au sens propre mais dans un cas particulier d'application.
}

\bpar{
The level of frameworks described above may be very general but is conditioned to a certain field or discipline, and to a certain approach or methodology. There exists to our knowledge no framework realizing a difficult exercise, that is to capture a certain structure of knowledge production at an epistemological knowledge, but conjointly is thought in a very applied perspective, with direct consequence in the design and management of complex systems. The contribution of this paper attempts to set a basis for a Knowledge Framework realizing this in the case of Complex Systems.
}{
Le niveau des cadres présentés ci-dessus peut être très général mais reste conditionné à un certain champ ou discipline, et à une certaine approche ou méthodologie. Il n'existe à notre connaissance pas de cadre réalisant un exercice difficile, qui est de capturer une certaine structure de production de la connaissance à un niveau épistémologique, mais qui est conjointement pensé dans une perspective très appliquée, avec des conséquences directes pour la conception et la gestion de systèmes complexes. La contribution de cette section propose de poser les bases pour un cadre réalisant cela dans le cas des systèmes complexes.
}

\bpar{
To achieve that, we postulate that the tension between these two contradictory objective is an asset to avoid on one side an impossible overarching generality and on the other side a too restraining domain-specific specificity. Based on the idea of complementary Knowledge Domains introduced by~\cite{livet2010}, its central aspect is a cognitive approach to science inducing co-evolutive processes of knowledge domains and their carriers. A first sketch of this framework was presented by~\cite{raimbault:halshs-01505084}, in the specific case of complex territorial systems as studied by theoretical and quantitative geography. We choose to introduce it here with an inductive approach, i.e. starting from a concrete case study that has mainly inspired the construction of the framework to end with its generic description.
}{
Pour y parvenir, nous partons du postulat que la tension entre ces deux objectifs contradictoires est un atout pour éviter d'une part une généralité globale impossible et d'autre part une spécificité due à un domaine qui serait trop restrictive. En se basant sur l'idée des domaines de connaissance introduite par~\cite{livet2010}, son aspect central est une approche cognitive de la science qui implique des processus de co-évolution entre les domaines de connaissance et leur supports. Une première ébauche de ce cadre a été présentée par~\cite{raimbault:halshs-01505084}, dans le cas particulier des systèmes complexes territoriaux comme étudiés par la géographie théorique et quantitative. Nous proposons de l'introduire ici par une démarche inductive, c'est-à-dire en partant d'une étude de cas concrète qui a largement inspiré la construction du cadre, pour finir avec sa description générique.
}

\bpar{
The rest of the section is organized as follows: we first develop case studies, more precisely a detailed study of a geographical theory of complex urban systems: the evolutive urban theory, and a short example from engineering to illustrate the transferability of concepts. We then specify the definitions and formulate the epistemological framework. We finally discuss issues on applicability, and potential developments such as a mathematical version of the framework, and a reflexive application of the framework to our subject of study.
}{
La suite de cette section est organisée de la façon suivante : nous détaillons d'abord les études de cas, plus précisément une étude détaillée d'une théorie géographique des systèmes urbains complexes : la théorie évolutive des villes, puis un court exemple d'ingénierie qui permet d'illustrer les possibilités de transfert des concepts. Nous spécifions ensuite les définitions et formulons le cadre épistémologique. Nous discutons ensuite les questions d'applicabilité, des développements potentiels comme une version mathématique du cadre, puis une application réflexive du cadre à notre sujet d'étude. 
}

\subsection{Case studies}{Étude de cas}

\subsubsection{Genesis of the evolutive urban theory}{Genèse de la théorie évolutive des villes}

\bpar{
The first case study relates the construction of the \emph{evolutive urban theory}\footnote{The ambiguity of the adjective \emph{evolutive} brings subtlety to the theory, since it applied as much in the first sense i.e. to the urban entities studied, but also in a meta sense to the theory itself, what confirms a certain level of reflexivity of the theory which is essential as developed in~\ref{sec:epistemology}. To translate the term into English, \cite{pumain2006evolutionary} chose the term ``Evolutionary Urban Theory'', but ``Evolutive Urban Theory'' is also suitable, but in any case it seems difficult to transfer the ambiguity during the translation.}, a geographical theory considering territorial systems from a complexity perspective, that have been developed for around 20 years. We analyse its genesis using mixed methods, namely semi-directed interviews with main contributors, and quantitative bibliometric analysis of main publications.
}{
La première étude de cas rappelle la construction de la \emph{théorie évolutive des villes}\footnote{L'ambiguïté de l'adjectif \emph{évolutive} fait gagner la théorie en subtilité, puisqu'il s'applique aussi bien au sens premier c'est-à-dire aux entités urbaines étudiées, mais aussi à un sens méta à la théorie elle-même, ce qui confirme un certain niveau de réflexivité de la théorie qui est essentiel comme développé en~\ref{sec:epistemology}. Pour traduire le terme en anglais, il a été choisi ``Evolutionary Urban Theory'' par \cite{pumain2006evolutionary}, mais ``Evolutive Urban Theory'' convient aussi, mais il semble dans tous les cas difficile de transférer l'ambiguïté lors de la traduction.}, une théorie géographique qui considère les systèmes territoriaux par une perspective complexe, développée depuis une vingtaine d'années environ. Nous étudions sa genèse par l'utilisation de méthodes mixtes, c'est-à-dire à la fois des interviews semi-dirigées avec des contributeurs principaux, et une analyse bibliométrique quantitative des publications principales.
}

\bpar{
Interviews were done following methodological standards \cite{legavre1996neutralite} to ensure a limited interference of the interviewer's experiences but not make it fully disappear to ensure a precise context enhancing the fluency of the interviewed. We use here interviews\footnote{Both have a length of around 1h. Sound and transcript text are available under a CC Licence at \texttt{https://github.com/JusteRaimbault/Interviews}~\cite{raimbault2017entretiens}. Interviews are in French and translations here are done by the author.} with \noun{Pr. D. Pumain} who introduced and developed mainly the theory, and \noun{Dr. R. Reuillon}, whose research on intensive and distributed computation and model exploration has been a cornerstone of latest developments.
}{
Les interviews ont été menées en suivant les standards méthodologiques classiques~\cite{legavre1996neutralite} pour assurer une interférence limitée des expériences du sujet menant l'entretien, mais sans le faire disparaitre complètement afin de permettre un contexte précis favorable à la fluidité de la personne interrogée. Nous utilisons ici des entretiens\footnote{Toutes les deux d'une durée environ une heure. Le son et les transcripts sont disponibles sous une Licence CC à \url{https://github.com/JusteRaimbault/Interviews}~\cite{raimbault2017entretiens}. Les interviews sont en français et la traduction anglaise des passages cités dans l'article original est assurée par l'auteur.} avec \noun{Pr. D. Pumain} qui a introduit et développé majoritairement la théorie, et \noun{Dr. R. Reuillon}, dont la recherche sur le calcul intensif et distribué et l'exploration de modèles a été un élément essentiel des développements les plus récents.
}

\bpar{
To begin it is important to recall a brief overview of the content of the evolutive theory. Therefore, refer to the presentation done in introduction of chapter~\ref{ch:evolutiveurban}, which describes its main structure.
}{
Pour commencer il est important de rappeler un aperçu rapide du contenu de la théorie évolutive. Pour cela, se référer à la présentation qui en est faite en introduction du chapitre~\ref{ch:evolutiveurban}, qui en donne la substantifique moelle.
}

\bpar{
The striking feature in the construction of all this is the balance between the different \emph{types} of knowledge, of which a typology will be the starting point of our construction. The relation between theoretical considerations and empirical cases studies is fundamental. Indeed, the seminal article \cite{pumain1997pour} is already positioned as an ``advocacy for a less ambitious theory, but that does not neglects the back-and-forth with observation''\footnote{page 2, trad. author}.
}{
La caractéristique frappante dans la construction de la théorie évolutive des villes est l'équilibre entre les différents \emph{types} de connaissance, desquelles une typologie sera le point de départ de notre construction. La relation entre les considérations théoriques et les cas d'étude empiriques est fondamental. En effet l'article séminal \cite{pumain1997pour} est déjà positionné comme ``\textit{un plaidoyer pour une théorie [...] moins ambitieuse, mais qui ne néglige pas les aller-retours avec l'observation}''.
}

\bpar{
We shall now turn to interviews to better understand the implications of the intrication of types of knowledge. \noun{D. Pumain} traces back germinal ideas back to her graduate student work in 1968, when ``\textit{everything started with a question of data}''. The interest for cities, and \emph{change in cities}, was driven by the availability of a refined migration flow dataset at different dates. Also rapidly, ``\textit{[they] were frustrated that methods were missing}'', but the access to the computation center (\emph{technical tool}) allowed the test of newly introduced methods and models, linked to the \noun{Prigogine} approach to complexity. Methods were however still limited to grasp the heterogeneity of spatial interactions. A progressively specified need and a chance encounter, with ``\textit{a lady working on neural networks and agent-based modeling at the Sorbonne}'', led to a bifurcation and a new level of interaction between modeling, theory and empirical knowledge: in 1997, two seminal articles, one stating the theoretical basis and the other introducing the first Simpop model, were published.
}{
Nous pouvons maintenant nous tourner vers les entretiens pour mieux comprendre les implications de l'intrication des différents types de connaissance. \noun{D. Pumain} retrace les idées germinales à son travail de maîtrise en 1968, quand ``\textit{tout à commencé avec une question de données}''. L'intérêt pour les villes, et pour le \emph{changement dans les villes}, a été conduit par la disponibilité d'un jeu de données raffiné sur les flux migratoires à différentes dates. Également rapidement, est venue ``\textit{la frustration des méthodes qui manquaient}'', mais l'accès au centre de calcul (\emph{outil technique}) a permis le test de méthodes et modèles nouvellement introduits, liés à l'approche de la complexité par \noun{Prigogine}. Les méthodes restaient toutefois limitées pour capturer l'hétérogénéité des interactions spatiales. Un besoin progressivement spécifié et une rencontre fortuite, avec ``\textit{une dame qui travaillait sur les réseaux de neurones et les modèles agents à la Sorbonne}'', a conduit à une bifurcation et un nouveau niveau d'interaction entre modèles, théorie et connaissance empirique : en 1997, deux articles séminaux, l'un donnant la base théorique, l'autre introduisant le premier modèle Simpop, étaient publiés simultanément.
}

\bpar{
From this point, it was clear that all modeling entreprise was conditioned to empirical knowledge of geographical case studies and theoretical assumptions to test. Methods and technical tools took also a necessary role, when specific model exploration methods were developed together with the Software OpenMole. \noun{R. Reuillon} relates that a qualitative shift of knowledge was rapidly made possible when systematic model exploration methods were introduced to understand the behavior of the SimpopLocal model. Initially, geographers were not sure if the model worked at all, in the sense that it produced expected stylized facts such as the emergence of hierarchy in a system of cities. Satisfying trajectories were found for some parameter values through genetic algorithm calibration, with distributed computation on grid~\cite{schmitt2014half}. The existence of multiple candidate solutions for parameter values is a barrier for concrete questions of necessity or sufficiency of a given mechanism of the agent-based model. This need, coming from the domain of empirical and theoretical geographical knowledge, led to the design of a specific algorithm: the \emph{calibration profile}, which is a methodological advance in model exploration~\cite{reuillon2015}.
}{
À partir de ce point, il était clair que toute entreprise de modélisation était conditionnée à une connaissance empirique de cas d'étude géographiques et à des hypothèses théoriques à tester. Les méthodes et les outils techniques ont alors pris aussi un rôle nécessaire, par le biais de méthodes d'exploration de modèles spécifiques développées au sein du logiciel OpenMole. \noun{R. Reuillon} rappelle qu'un saut qualitatif de connaissances a été rendu rapidement possible quand les méthodes d'exploration systématiques ont été introduites pour comprendre le comportement du modèle SimpopLocal. À la base, les géographes n'étaient pas sûr si le modèle fonctionnait seulement, dans le sens où il produisait les faits stylisés attendus comme l'émergence de la hiérarchie d'un système de villes. Des trajectoires satisfaisantes ont été trouvées par l'utilisation d'algorithmes génétiques de calibration, en calcul distribué sur grille~\cite{schmitt2014half}. L'existence de multiples solutions équivalentes pour les valeurs des paramètres est une barrière pour des question concrètes de nécessité ou suffisance d'un mécanisme donné du modèle agent. Ce besoin, venant du domaine de la connaissance empirique et théorique géographique, a mené à la conception d'un algorithme spécifique : le \emph{Calibration Profile}, qui est une avancée méthodologique dans l'exploration de modèles~\cite{reuillon2015}.
}

\bpar{
This virtuous circle was continued with the Marius model family~\cite{cottineau2014evolution} and the Parameter Space Exploration algorithm~\cite{10.1371/journal.pone.0138212}. \noun{R. Reuillon} evaluates its impact from a Computer Scientist point of view: ``\textit{I'm not sure if [geographers] were immediately conscious of the amplitude of the result, that was really heavy, people working with us directly saw it.}'' This positive vision is confirmed by \noun{D. Pumain}, who highlights the benefits of these new methods for geographical knowledge, and that it was the first time that research led to publications at the edge of knowledge both in geography and computer science.
}{
Ce cercle vertueux a été continué avec la famille de modèles Marius~\cite{cottineau2014evolution} et l'algorithme \emph{Parameter Space Exploration}~\cite{10.1371/journal.pone.0138212}. \noun{R. Reuillon} évalue son impact du point de vue d'un informaticien : ``\textit{Je ne suis pas sûr si les géographes étaient immédiatement conscients de la portée du résultat, c'était du lourd, les gens qui bossaient avec nous l'ont directement vu.}'' Cette vision positive est confirmée par \noun{D. Pumain}, qui souligne les bénéfices de ces nouvelles méthodes pour la connaissance géographique, et que c'était la première fois qu'une recherche menait à des publications à la frontière de la connaissance à la fois en géographie et en informatique.
}

\bpar{
Taking a step back, emerges a typology of domains in which knowledge was created but also necessary for the other domains in the genesis of the Evolutive Urban Theory. The collection of data and construction of datasets is a first requirement for any further knowledge. From data are extracted empirical stylized facts, from which are induced theoretical hypotheses. Theory can then be tested for falsification, in the empirical domain but also through models, for example by doing targeted experiments in models of simulation. New methods are developed to better explore them. Tools are crucial at each step, to implement model, do data mining for example or collect and format data for example. The previous analysis reveals how these domains are interdependent, are in a sense \emph{co-evolutive}.
}{
En prenant du recul, émerge une typologie de domaines dans laquelle de la connaissance a été créée mais également nécessaire pour les autres domaines dans la genèse de la théorie évolutive des villes. La récolte des données et la construction de jeux de données est un premier pré-requis pour toute connaissance supplémentaire. À partir des données on extrait des faits stylisés empiriques, desquels sont déduites des hypothèses théoriques. La théorie peut être testée pour falsification, dans le domaine empirique mais aussi par les modèles, par exemple par des expériences ciblées dans les modèles de simulation. De nouvelles méthodes sont alors développées pour mieux les explorer. Les outils sont cruciaux à chaque étape, pour implémenter un modèle, faire de la fouille de données ou collecter et formater les données par exemple. L'analyse précédente montre comment ces domaines sont interdépendants, et sont dans un sens \emph{co-évolutifs}.
}

\begin{figure}[h!]
\includegraphics[width=\linewidth]{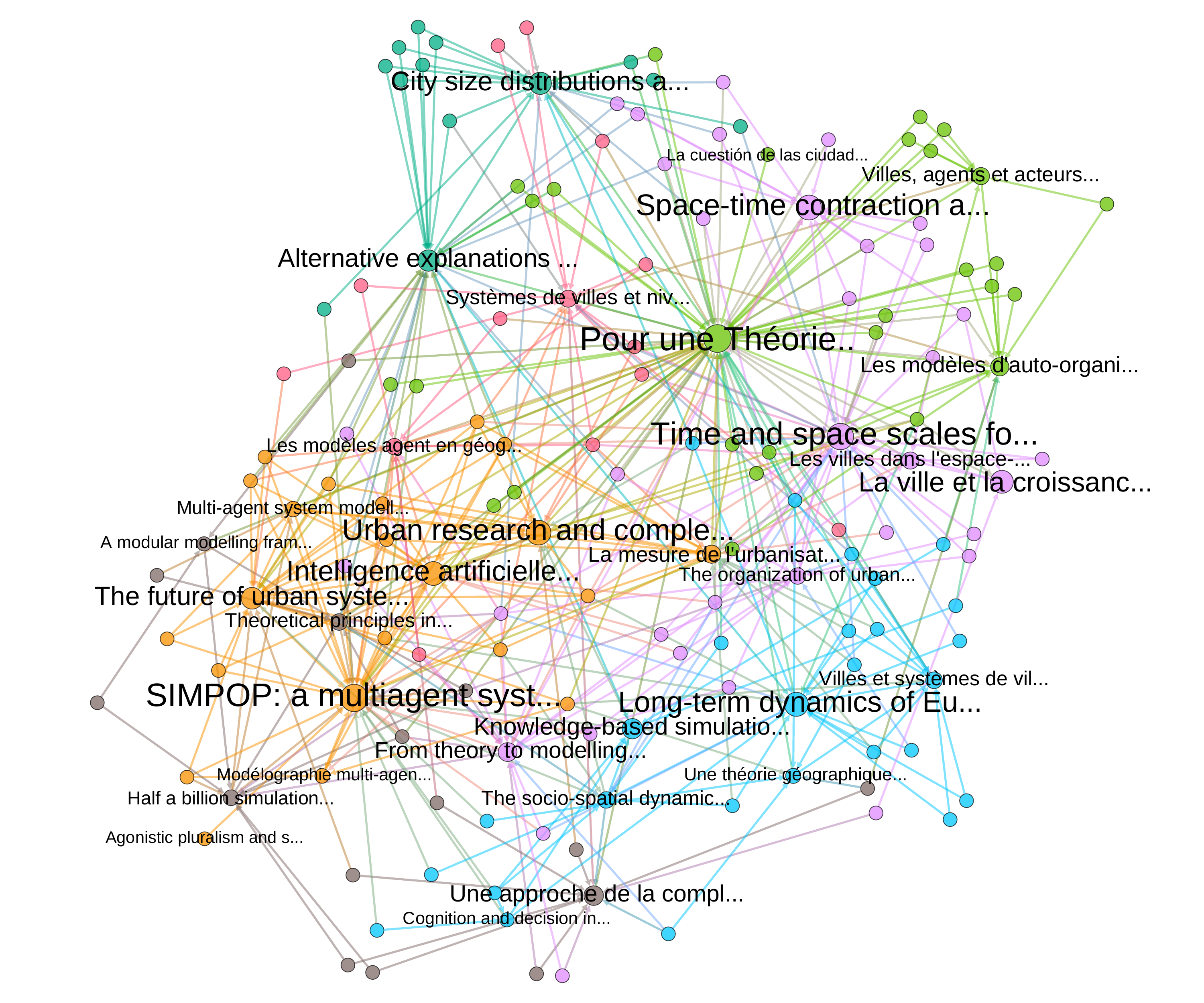}
\caption[Citation Network of main publications of Evolutive Urban Theory]{\textbf{Citation Network of main publications of Evolutive Urban Theory.} The network is constructed the following way: starting from the two seminal publications~\cite{pumain1997pour} and \cite{sanders1997simpop}, we get citing publications, filter conditionally to one of the main contributors, get again citing publications and filter. Nodes are publications ($\left|V\right|=155$), the size corresponding to eigenvector centrality, and edges are directed citation links ($\left|E\right|=449$). Colors are communities obtained with Louvain clustering algorithm (7 communities, modularity 0.39).\label{fig:knowledgeframework:citnw}}
\end{figure}

\bpar{
We back up now this qualitative analysis with a modest quantitative bibliometric analysis. The idea is to investigate the structure of the core citation network of main publications constructing the Evolutive Urban Theory. We construct the citation network as described in Fig.~\ref{fig:knowledgeframework:citnw}, by using the data collection tool provided by~\cite{raimbault2016indirect} and already used in~\ref{sec:quantepistemo}\footnote{all code and data are available at \texttt{https://github.com/JusteRaimbault/CityNetwork/tree/master/Models/QuantEpistemo}.}. Starting from the two seminal publications \cite{pumain1997pour} and \cite{sanders1997simpop}, the backward citation network is obtained at depth 2 (references citing these initial references, and the ones citing the citing), with filtering for the first step on authors to have at least one main contributor of the Theory (that we take as \emph{Pumain}, \emph{Sanders} and \emph{Bretagnolle}, according to the full Pumain's interview). We remove nodes of degree 1, to have the core structure only of the ego network. Note that we do not have missing links between nodes at the first level, because all citing links were retrieved.
}{
Nous supportons cette analyse qualitative par une analyse quantitative bibliométrique modeste. L'idée est d'étudier la structure du coeur du réseau de citations des publications principales construisant la théorie évolutive des villes. Nous construisons le réseau de citations comme décrit en Fig.~\ref{fig:knowledgeframework:citnw}, en utilisant l'outil de collection de données fourni par~\cite{raimbault2017exploration} et déjà utilisé en~\ref{sec:quantepistemo}\footnote{L'ensemble du code et des données pour cette analyse sont disponibles à \url{https://github.com/JusteRaimbault/CityNetwork/tree/master/Models/QuantEpistemo}.}. Partant des deux publications séminales \cite{pumain1997pour} et \cite{sanders1997simpop}, le réseau de citation inverse est obtenu à profondeur 2 (les références citant ces références initiales, et celles citant les citantes), en filtrant à la première étape sur les auteurs pour avoir au moins un des principaux contributeurs de la théorie (que nous prenons comme \emph{Pumain}, \emph{Sanders} et \emph{Bretagnolle}, en accord avec l'entretien avec \noun{D. Pumain}). Les noeuds de degré 1 sont supprimés, pour obtenir uniquement le coeur du réseau. Nous pouvons noter qu'il ne manque pas de lien entre les noeuds du premier niveau, puisque tous les liens citants ont été récupérés.
}

\bpar{
Network has a density of 0.019, what is rather high for a citation network, and the signature of a high level of dependency between publications. Starting from two separate nodes, we could have in theory distinct connected components, but as expected the network has only one because both aspects are strongly interconnected. To analyse the structure in a finer way, we detect communities using Louvain clustering algorithm, and evaluate the directed modularity of the partition as described by \cite{nicosia2009extending}.
}{
Le réseau a une densité de 0.019, ce qui est plutôt élevé pour un réseau de citation, et la signature d'un haut niveau de dépendance entre les publications. En partant de deux noeuds distincts, nous aurions pu avoir en théorie des composantes connexes distinctes, mais comme attendu le réseau n'en a qu'une de par la nature fortement interconnectée des deux aspects. Pour analyser la structure de manière plus fine, nous détectons les communautés en utilisant l'algorithme de clustering de Louvain, et évaluons la modularité dirigée de la partition comme donnée par \cite{nicosia2009extending}.
}

\bpar{
We show in Fig.~\ref{fig:knowledgeframework:citnw} a visualization of the network. We obtain 7 communities with a modularity value of 0.39. To ensure the significance of modularity, we proceed to Monte Carlo simulations and randomize citation links 100 times, computing each time the modularity of communities within the randomized network. We obtain an average directed modularity of $\bar{m} = 0.002 \pm 0.015$, making the modularity of the real network highly significant (more than 200 standard deviations).
}{
Nous montrons en Fig.~\ref{fig:knowledgeframework:citnw} une visualisation du réseau. Nous obtenons 7 communautés avec une valeur de modularité de 0.39. Pour assurer que cette valeur est significative, nous procédons à des simulations de Monte Carlo et distribuons de manière aléatoire les liens de citation 100 fois, en calculant à chaque fois la modularité des communautés dans le réseau aléatoire. Nous obtenons une modularité moyenne dirigée de $\bar{m} = 0.002 \pm 0.015$, rendant la modularité du réseau réel hautement significative (plus de 200 déviations standard).
}

\bpar{
We analyse the content of communities by looking at publications of the first level. We find that communities are roughly consistent with the typology of domains: one on methods, three on spatio-temporal modeling of urban systems that mixes empirical and modeling, one conceptual, one on Simpop models, and a last on scaling laws that is fully empirical. Data papers are not yet current practice in geography and specific papers tackling the Data domain cant be found in the network. An increased citation rate between papers of the same domain could be expected because of the scientific standard to always situate a contribution regarding similar works. The significant value of modularity confirms that domains are consistent regarding an certain endogenous structure of knowledge production.
}{
Nous analysons le contenu des communautés en examinant leur publications du premier niveau. Nous trouvons que les communautés sont globalement cohérentes avec la typologie des domaines : une pour les méthodes, trois sur la modélisation spatio-temporelle des systèmes urbains qui mélange empirique et modélisation, une conceptuelle, une sur les modèles Simpop, et une dernière sur les lois d'échelle qui est complètement empirique. Les \emph{Data Papers} ne sont pas encore une pratique courante en géographie et des articles spécifiques au domaine des données ne peuvent être trouvés dans le réseau. Un taux de citation accru entre papiers du meme domaine est dans tous les cas attendu à cause du standard scientifique de toujours situer une contribution au regard des travaux similaires. La valeur significative de la modularité confirme que les domaines sont cohérents au regard d'une certaine structure endogène de la production de connaissance.
}

\subsubsection{Engineering}{Ingénierie}

\bpar{
After the glance on domains of knowledge extracted in the previous case study, we propose to take the corresponding point of view on a rather different example more related to technology and engineering. We interpret thus issues of engineering related to Parisian metropolitan system through this prism of Knowledge Domains.
}{
Après l'aperçu sur les domaines de connaissance extraits dans l'étude de cas précédente, nous proposons de prendre un point de vue similaire sur un exemple assez différent plus en relation avec la technologie et l'ingénierie. Nous interprétons ainsi des questions d'ingénierie liées au système de transport métropolitain parisien au travers du prisme des domaines de connaissance.
}

\bpar{
Taking the example of the progressive automatization of line 1, considered widely as a technical achievement, several integrated empirical and modeling studies were preliminary conducted~\cite{belmonte2008automatisation}. The use and adaptation of particular methods such as agent-based modeling is crucial for the development of innovative autonomous transportation~\cite{balbo2016positionnement}. In this engineering problem, some technical solutions such as platform doors\footnote{Automatic doors separating the platform from the track, in particular necessary for the introduction of an automatic service.} may be seen as tools that also evolve, and are necessary for a new conceptual approach (\emph{automatic transportation}) to be implemented~\cite{foot2005faut}. But they may also have interactions with other aspects of conceptual knowledge, such as management and organisation within the operator~\cite{foot1994ratp}. The complex multi-dimensional aspect of innovation for such systems was already highlighted for a while as~\cite{hatchuel1988stations} shows. Other technical aspects, such as civil engineering issues~\cite{moreno2016etude}, are also put in line when developing such a new approach, and they necessitate at least empirical and modeling, if not more, Knowledge Domains. This rather short example is an illustration of how the interpretation of knowledge domains can be applied to the engineering and management of a complex industrial systems.
}{
En prenant l'exemple de l'automatisation progressive de la ligne 1, considérée largement comme une prouesse technique, de nombreuses études intégrant modélisation et études empiriques ont été conduites en préliminaire~\cite{belmonte2008automatisation}. L'utilisation et l'adaptation de méthodes particulières comme la modélisation multi-agents est cruciale pour le développement de transports autonomes innovants~\cite{balbo2016positionnement}. Dans ce problème d'ingénierie, des solutions techniques comme les portes palières de quai\footnote{Portes automatiques séparant le quai de la voie du métro, notamment nécessaires pour la mise en place d'un service automatique.} peuvent être vues comme des outils qui évoluent également, et sont nécessaires pour qu'une nouvelle approche conceptuelle (\emph{le transport automatique}) soit implémentée~\cite{foot2005faut}. Mais ils peuvent aussi interagir avec d'autres aspects de la connaissance conceptuelle, comme le management et l'organisation au sein de l'opérateur~\cite{foot1994ratp}. L'aspect multi-dimensionnel complexe de l'innovation pour de tels systèmes avait déjà été souligné depuis longtemps comme le montrent~\cite{hatchuel1988stations}. D'autres aspects techniques, comme des problèmes d'ingénierie civile~\cite{moreno2016etude}, sont aussi mis en jeu pour développer une telle approche, et ils nécessitent au moins les domaines empiriques et de modélisation, voire plus. Cet exemple relativement court illustre comment l'interprétation par domaines de connaissance peut être appliqué à l'ingénierie et au management de systèmes complexes industriels.
}

\bpar{
Specific details would be needed for a more in-depth application, but we claim to have a proof-of-concept here. We summarize in Table~\ref{tab:application} the engineering issues identified above, the corresponding knowledge domains, and the processes through which transferability may be achieved.
}{
Des détails spécifiques seraient nécessaires pour une application plus en profondeur, mais nous proposons ici une preuve de concept. Nous synthétisons en Table~\ref{tab:knowledgeframework:application} les questions d'ingénierie identifiées ci-dessus, les domaines de connaissance correspondants, et les processus par lesquels un transfert pourrait être réalisé.
}

\begin{table}[h!]
\caption[Illustration of Knowledge Framework Application]{Illustration of Knowledge Framework Application.\label{tab:knowledgeframework:application}}
\bpar{
\begin{tabular}[10pt]{|p{3.7cm}|p{3.7cm}|p{3.7cm}|p{4.5cm}|}
\hline
\textbf{Engineering problem} & \textbf{Knowledge domains} & \textbf{transferability} & \textbf{References}\\\hline
Autonomous transport & Empirical, Modeling & Integrated modeling & \cite{belmonte2008automatisation}\\\hline
Innovative modeling & Modeling, Methods & Method development & \cite{balbo2016positionnement}\\\hline
Functional specifications & Empirical, Tools & Ergonomic tools & \cite{foot2005faut}\\\hline
Social adaptation & Theoretical, Empirical & Stakeholders involvment & \cite{foot1994ratp}, \cite{hatchuel1988stations}\\\hline
Technical constraints & Empirical, Modeling & Integrated modeling & \cite{moreno2016etude}\\\hline
\end{tabular}
}{
\begin{tabular}[10pt]{|p{3.7cm}|p{3.7cm}|p{3.7cm}|p{4.5cm}|}
\hline
\textbf{Problème d'ingénierie} & \textbf{Domaines de connaissance} & \textbf{Transferabilité} & \textbf{Références}\\\hline
Transport autonome & Empirique, Modélisation & Modélisation intégrée & \cite{belmonte2008automatisation}\\\hline
Modélisation innovante & Modélisation, Méthodes & Développement de méthodes & \cite{balbo2016positionnement}\\\hline
Spécifications fonctionnelles & Empiriques, Outils & Outils ergonomiques & \cite{foot2005faut}\\\hline
Adaptation sociale & Théorique, Empirique & Implication des stakeholders & \cite{foot1994ratp}, \cite{hatchuel1988stations}\\\hline
Contraintes techniques & Empirique, Modélisation & Modélisation intégrée & \cite{moreno2016etude}\\\hline
\end{tabular}
}
\end{table}

\subsection{Knowledge Framework}{Cadre de Connaissances}

\bpar{
We can formulate now inductively the knowledge framework. As mentioned, it takes the idea of interacting domains of knowledge from the framework introduced by~\cite{livet2010}, but extends these domains and takes a novel epistemological position, focusing on co-evolutive dynamics of agents and knowledge.
}{
Nous pouvons à présent formuler le cadre de manière inductive. Comme déjà évoqué, il tire l'idée de domaines de connaissance en interactions du cadre introduit par~\cite{livet2010}, mais étend ces domaines et prend une nouvelle position épistémologique, se concentrant sur les dynamiques co-évolutives entre agents et connaissances.
}

\paragraph{Constraints}{Contraintes}

\bpar{
To be particularly fitted for the study and management of complexity, we postulate that the framework must meet certain requirements, especially to take into account and even favor the \emph{integrative nature of knowledge}, as illustrated by the importance of interdisciplinarity and diversity in the case studies. The framework must thus be favorable to the following:
\begin{itemize}
\item Integration of disciplines, as Complex Systems are by essence at the crossing of multiple fields
\item Integration of knowledge domains, i.e. that no particular type of knowledge must be privileged in the production process\footnote{this is not incompatible with very strict system specifications, as multiple paths are possible to obtain the same fixed final state}
\item Integration of methodology types, in particular breaking the artificial boundaries between ``quantitative'' and ''qualitative'' methods, which are particularly strong in classical social sciences and humanities.
\end{itemize}
}{
Pour être particulièrement adapté à l'étude et à la gestion de la complexité, nous postulons que le cadre doit répondre à certaines contraintes, en particulier pour prendre en compte et même favoriser la \emph{nature intégrative de la connaissance}, comme illustré par l'importance de l'interdisciplinarité et de la diversité dans les cas d'étude. Le cadre doit ainsi être favorable aux points suivants :
\begin{itemize}
\item intégration des disciplines, puisque les systèmes complexes sont par essence à la croisée de champs multiples ;
\item intégration des domaines de connaissance, c'est-à-dire qu'aucun type particulier de connaissance ne doit être privilégié dans le processus de production\footnote{Ce qui n'est pas incompatible avec des spécifications fonctionnelles très strictes, puisque des chemins divers sont possible pour atteindre le même état final fixé.} ;
\item intégration des types de méthodologie, en particulier dépasser les frontières artificielles entre méthodes ``quantitatives'' et ``qualitatives'', qui sont particulièrement fortes en sciences sociales et humanités classiques. 
\end{itemize}
}

\paragraph{Epistemological foundations}{Fondations épistémologiques}

\bpar{
The epistemological positioning of the framework is the one developed in the first section of~\ref{sec:epistemology}. We recall the importance of the \emph{perspective}~\cite{giere2010scientific}, composed by agents, the objects represented, the purpose and the medium (the model). The approach by agents is fundamental for the relevance of the framework.
}{
Le positionnement épistémologique du cadre est celui développé dans la première section de~\ref{sec:epistemology}. Nous rappelons l'importance de la \emph{perspective}~\cite{giere2010scientific}, composée des agents, des objets représentés, du but et du medium (le modèle). L'approche par agents est fondamentale pour la cohérence du cadre.
}

\paragraph{Knowledge domains}{Domaines de connaissance}

\bpar{
We postulate the following knowledge domains, with their definitions:
\begin{itemize}
\item \textbf{Empirical.} Empirical knowledge of real world objects.
\item \textbf{Theoretical.} More general conceptual knowledge, implying cognitive constructions.
\item \textbf{Modeling.} The model is the formalized \emph{medium} of the scientific perspective, as diverse as Varenne's classifications of models functions~~\cite{varenne2017theories}.
\item \textbf{Data.} Raw information that has been collected.
\item \textbf{Methods.} Generic structures of knowledge production.
\item \textbf{Tools.} Proto-methods (implementation of methods) and supports of others domains.
\end{itemize}
}{
Nous proposons les domaines de connaissance suivants, avec leurs définitions :
\begin{itemize}
\item \textbf{Empirique.} Connaissance empirique d'objets du monde réel.
\item \textbf{Théorique.} Connaissance conceptuelle plus générale, impliquant des constructions cognitives.
\item \textbf{Modélisation.} Le modèle est le \emph{medium} formalisé de la Perspective Scientifique, aussi divers que la classification de \noun{Varenne} des fonctions des modèles~\cite{varenne2017theories}.
\item \textbf{Données.} Information brute qui a été collectée.
\item \textbf{Méthodes.} Structures génériques de production de connaissance.
\item \textbf{Outils.} Proto-méthodes (implémentation des méthodes) et supports des autres domaines.
\end{itemize}
}

\bpar{
We choose to keep separate Methods and Tools, to insist on the support role of tools, and because development of both are related but not identical. The same way, Data domain and Empirical Domain are distinct, as new datasets do not systematically imply new knowledge of empirical facts, even if the construction of data collection tools often requires an empirical knowledge. The Modeling Domain has a central role as we postulate that \emph{any knowledge on a complex system requires a model}. 
}{
Nous prenons le parti de séparer Outils et Méthodes, pour insister sur le rôle de support des outils, et car le développement des deux est lié mais pas identique. De la même façon, le domaine des Données et le domaine Empirique sont distincts, car des nouveaux jeux de données n'impliquent pas systématiquement une nouvelle connaissance de faits empiriques, même si la construction des outils de captation de données souvent requière une connaissance empirique. Le domaine de la Modélisation a un rôle central puisque nous postulons que \emph{toute connaissance d'un système complexe nécessite un modèle}.
}

\paragraph{Co-evolution of knowledges}{Co-évolution des connaissances}

\bpar{
We can now formulate the central hypothesis of our framework, that is partially contained in the positioning within Perspectivism. We postulate that \emph{any scientific knowledge construction on a complex system}\footnote{We believe that this intricate aspect of knowledge production is necessary present for Complex Systems, in echo of the remark on reflexivity in introduction. Even \emph{simple models} of complex systems do imply a conceptual complexity that requires complexity of knowledge to be grasped. This last assumption may be related to the nature of complexity and to the relation between computational complexity and complexity in the sense of weak emergence, that is suggested for example by \cite{2014arXiv1403.7686B} that explains emergence and decoherence from the quantum level by the NP-completude of fundamental equations resolution. These considerations are far beyond the reach of this section (see~\ref{sec:epistemology} for a deepening of the question), and we take as an assumption that complex systems necessitate complex knowledge, whereas simple knowledge (in the sense of non co-evolving domains and agents) \emph{can} exist for simple systems.} is a perspective in the sense of Giere. It is composed of knowledge contents within each domain, that \emph{co-evolve} between themselves and with the other elements of the perspective, in particular the cognitive agents. The notion of co-evolution is taken in the sense of~\cite{holland2012signals}, i.e. of co-evolving entities being within strongly interdependent niches with circular causal relations and that have a certain independence with the exterior within their boundaries. We note the importance of weak emergence in the sense of Bedau~\cite{bedau2002downward} in the construction of the perspective from the co-evolution of its components, as it corresponds to an autonomous upper level that can be understood alone, as the scientific knowledge can be. Note that a perspective does not necessarily have components in all domains, but should generally have in most.
}{
Nous pouvons à présent formuler l'hypothèse centrale de notre cadre, qui est partiellement contenu dans le positionnement par rapport au perspectivisme. Nous postulons que \emph{toute construction de connaissance scientifique sur un système complexe}\footnote{Nous sommes convaincu que cet aspect intriqué de la production de connaissance est nécessairement présent pour les Systèmes Complexes, en écho à la remarque sur la réflexivité en introduction de la section. Même des \emph{modèles simples} de systèmes complexes impliquent une complexité conceptuelle qui nécessite que la complexité de la connaissance soit présente pour être traduite. Cette dernière hypothèse pourrait liée à la nature de la complexité et la relation entre la complexité computationnelle et la complexité au sens de l'émergence faible, qui est suggérée par exemple par~\cite{2014arXiv1403.7686B} qui explique l'émergence et la décohérence depuis le niveau quantique par la NP-complétude de la résolution des équations fondamentales. Ces considérations sont bien au delà de la portée de cette section (voir~\ref{sec:epistemology} pour une réflexion plus approfondie), et nous prenons comme une hypothèse que les systèmes complexes nécessitent de la connaissance complexe, tandis que de la connaissance simple (au sens de domaines et agents non co-évolutifs) \emph{peut} exister pour des systèmes simples.} est une perspective au sens de \noun{Giere}. Elle est composée de contenu de connaissance dans chacun des domaines, qui \emph{co-évoluent} entre eux et avec les autres éléments de la perspective, en particulier les agents cognitifs. La notion de co-évolution est prise au sens de~\cite{holland2012signals}, c'est-à-dire d'entités étant fortement interdépendantes au sein de niches avec des relations causales circulaires et qui ont une certaine indépendance avec l'extérieur dans leur frontières. Nous notons l'importance de l'émergence faible au sens de \noun{Bedau}~\cite{bedau2002downward} dans la construction de la perspective à partir de la co-évolution de ses composants, puisqu'il s'agit d'un niveau supérieur autonome qui peut être compris en lui-même, comme la connaissance scientifique peut l'être. Il faut aussi noter qu'une perspective n'a pas nécessairement des composants dans tous les domaines, mais devraient généralement en avoir dans la plupart.
}

\bpar{
The social aspect of knowledge production is not included in knowledge domains, but within agents and their relations. \cite{roth2010social} shows a co-evolution of social networks with semantic networks through the example of a scientific community in development biology and an environment of political blogs, what confirms in our case the co-evolution between agents and domains.
}{
L'aspect social de la production de connaissance n'est pas inclus dans les domaines de connaissance, mais dans les agents et leur relation. \cite{roth2010social} montre une co-évolution des réseaux sociaux et des réseaux sémantiques avec l'exemple d'une communauté scientifique en biologie du développement et un environnements de blogs politiques, ce qui confirme dans notre cas la co-évolution entre les agents et les domaines.
}

\paragraph{Application}{Application}

\begin{figure}[h!]
\includegraphics[width=\linewidth]{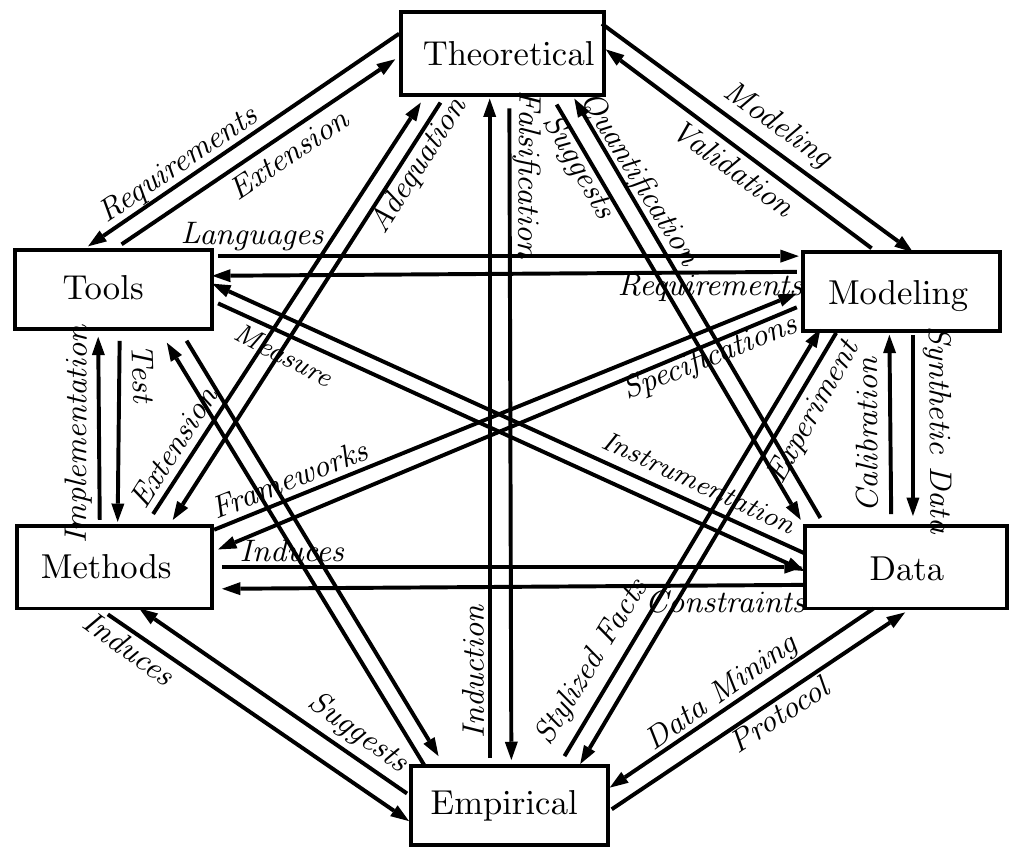}
\caption[Full network of knowledge domains]{\textbf{Projection of a perspective into a full network of knowledge domains.} To illustrate the domains and the interaction processes between them, we do the exercise of trying to qualify all possible binary relations between two given domains. This does not reflect the real structure of the framework, but is an aid to consider what interactions can be. Note that the nature of relations is not always the same here, some being constraints, other knowledge transfer, other processes within other domains such as synthetic data which is a methodology. This shows that some domains act as catalyzers for relations between others, in this network setting, what corresponds indeed to a situation of co-evolution.\label{fig:knowledgeframework:fwk}}
\end{figure}

\bpar{
The types of models to which our framework applies are supposed to be all possible models in a very loose sense, as Giere calls a model any medium of a perspective. A functional view of models as Varenne introduces~\cite{varenne2010simulations} (introducing a typology of models through functions, e.g. explicative models, simulation models, predictive models, comprehensive models, interactive models, etc.) is a way to grasp the variety. We can also see it in terms of more classical classifications, and apply it to mathematical, statistical, simulation, data or conceptuel models for example. Concerning the constraints given before, as all knowledge are co-evolving no domain is particularly privileged. No discipline either as these will have their different aspects be contained within the domains, and finally qualitative and quantitative methods are present and necessary in most. We show in Fig.~\ref{fig:knowledgeframework:fwk} a projection of knowledge domains as a complete network, to illustrate what relations between domains can be composed of.
}{
Les types de modèles auquel notre cadre s'applique sont supposés être tous les modèles possibles en un sens très large, puisque \noun{Gière} désigne par modèle tout \emph{medium} d'une perspective. Une vue fonctionnelle des modèles comme \noun{Varenne} introduit~\cite{varenne2017theories} (introduisant une typologie des modèles par leur fonctions, par exemple les modèles explicatifs, les modèles de simulation, les modèles prédictifs, les modèles de compréhension, les modèles interactifs, etc.) est un moyen d'appréhender leur variété. Il est aussi possible de le voir en termes de classifications plus classiques, et l'appliquer au modèles mathématiques, statistiques, de simulation, de données, ou conceptuels par exemple. Concernant les contraintes données précédemment, comme toutes les connaissances sont en co-évolution, aucun domaine n'est privilégié en particulier. Aucune discipline non plus, puisque celles-ci auront leur différents aspects contenus dans les domaines, et finalement les méthodes qualitatives et quantitatives seront présentes et nécessaire dans la majorité. Nous montrons en Fig.~\ref{fig:knowledgeframework:fwk} une projection des domaines de connaissance comme un réseau complet, pour illustrer de quoi peuvent être composées les relations entre domaines.
}

\subsection{Discussion}{Discussion}

\subsubsection{Application Range}{Portée d'application}

\bpar{
We insist that our framework does not pretend to introduce a general epistemology of scientific knowledge, but far from that is rather targeted towards reflexivity in the understanding of complex systems. The level of generality is at a very different level, but the aim to practical implication in the handling of complexity contributes to a certain generic character in applications. It is furthermore particularly suited to study Complex Systems, since more reductionist approaches can handle more compartmented production of knowledge, whereas integration of disciplines and scales and therefore domains of knowledge has been emphasized as crucial to study complexity.
}{
Nous insistons sur le fait que notre cadre ne prétend pas introduire une épistémologie générale de la connaissance scientifique, mais est plutôt ciblé vers une réflexivité dans la compréhension des systèmes complexes. Le niveau de généralité est à un niveau très différent, mais le but d'implications pratiques dans la compréhension de la complexité contribue à un certain caractère générique dans les applications. Il est de plus particulièrement adapté à l'étude des systèmes complexes, puisque des approches plus réductionnistes peuvent gérer des productions de connaissance plus compartimentées, tandis que l'intégration des disciplines et des échelles et donc des domaines de connaissance a été souligné comme crucial pour l'étude de la complexité.
}

\subsubsection{Towards a formalisation}{Vers une formalisation}

\bpar{
The knowledge framework stays at an epistemological level, and its application could be formalized in a more systematic way. Therefore, the framework developed in Appendix~\ref{app:sec:csframework} could be partly integrated. Let recall its main elements and how these can be articulated. The main aspect is the coupling of a formalization of the system model with the perspective. A perspective would be defined as a dataflow machine $M$ in the sense of~\cite{golden2012modeling} that gives a convenient way to represent it and to introduce timescales and data, to which is associated an ontology $O$ in the sense of~\cite{livet2010}, i.e. a set of elements each corresponds to an entity (which can be an object, an agent, a process, etc.) of the real world. Purpose and carrier of the perspective are contained in the ontology if they make sense for studying the system. Decomposing the ontology into atomic elements $O=(O_j)_j$ and introducing an order relation between ontology elements based on weak emergence ($O_j\succcurlyeq O_i$ if and only if $O_j$ weakly emerges of $0_i$) should yield a canonical decomposition of the perspective containing the structure of the system. The challenge would be then to link this decomposition with the canonical decomposition of the dataflow machine postulated by~\cite{golden2012modeling}, and then define knowledge domains within this coupling: data is in flows of the machine, modeling in the machine, empirical and theoretical in ontologies, methods in the structure of the tree. Such an enterprise with consistent operations is however totally beyond the scope of this paper, but would be a powerful development.
}{
Le cadre de connaissances reste à un niveau épistémologique, et son application pourrait être formalisée de manière plus systématique. Pour cela, il faudrait reprendre partiellement le cadre développé en Annexe~\ref{app:sec:csframework}. Rappelons-en les éléments clés et comment ceux-ci peuvent s'articuler. L'aspect principal est le couplage d'une formalisation du modèle du système avec celle de la perspective. Une perspective serait définie comme une \emph{Dataflow Machine} $M$ au sens de~\cite{golden2012modeling} qui donne un moyen pratique pour la représenter et pour introduire les échelles de temps et les données, à laquelle est associée une ontologie $O$ au sens de~\cite{livet2010}, i.e. un ensemble d'éléments dont chacun correspond à une entité (qui peut être un objet, un agent, un processus, etc.) du monde réel. Le motif et l'agent porteur de la perspective sont contenus dans l'ontologie s'ils font sens pour étudier le système. Décomposer l'ontologie en éléments atomiques $O=(O_j)_j$ et introduire une relation d'ordre entre les éléments des ontologies basée sur l'émergence faible ($O_j\succcurlyeq O_i$ si et seulement si $O_j$ émerge faiblement de $0_i$) devrait fournir une décomposition canonique de la perspective contenant la structure du système. Le défi serait alors de lier cette décomposition avec la décomposition canonique de la \emph{Dataflow Machine} postulée par~\cite{golden2012modeling}, et ensuite définir les domaines de connaissance au sein de ce couplage : les données sont dans les flux des machines, le modèle est la machine, l'empirique et le théorique dans les ontologies, les méthodes dans la structure de l'arbre. Une telle entreprise avec des opérations cohérentes entre les éléments est cependant hors de notre portée pour l'instant, mais serait un développement puissant.
}

\bpar{
We have studied with mixed methods the construction of a scientific theory in theoretical and quantitative geography, and from that inductively introduced a knowledge framework aiming at understanding the production of knowledge on complex system as a complex system itself, namely a perspective with co-evolving components within interdependent knowledge domains. Note that the approach is fully reflexive as several components were necessary. We postulate that our framework is a useful tool to study complexity and manage complex systems, since it explicits some choices and directions of developments that may otherwise be unconscious.
}{
Nous avons étudié par des méthodes mixtes la construction d'une théorie scientifique en géographie théorique et quantitative, et à partir de cela introduit de manière inductive un cadre de connaissance visant comprendre la production de connaissances sur un système complexe comme un système complexe elle-même, plus précisément une perspective avec des composantes co-évolutives au sein de domaines de connaissance interdépendants. On peut noter que cette approche est totalement réflexive puisque plusieurs de ces composantes ont été nécessaires dans la démarche de construction. Nous postulons que ce cadre peut être un outil utile pour étudier la complexité et gérer des systèmes complexes, puisqu'il explicite certains choix et directions de développements qui pourraient autrement être inconscients.
}


\subsubsection[Co-construction of theories and models][Co-construction des théories et modèles]{Co-construction of theories and models in quantitative geography: an synthesis of our contributions}{Co-construction des théories et modèles en géographie quantitative : une synthèse de nos contributions}

\bpar{
We conclude this opening chapter by putting into a consistent perspective the different contributions of the thesis, from the viewpoint of illustrating the co-evolution of knowledge within different domains, and to complete the loop by coming back on the construction of the geographical theory. As detailed in preamble, a linear reading mode would be too reducing, since most of the works mutually enrich themselves whatever their domain and their reach, and a linear synthesis, beside being intrinsically poor, is in a sense a lie by omission of all the complex interactions between the produced knowledge ensembles. Naturally the synthesis exercise and the ability to enter an imposed formatted template are important, or even desirable given the current state of scientific knowledge production.
}{
Nous concluons ce chapitre d'ouverture par une mise en perspective cohérente des diverses contributions de la thèse, du point de vue de l'illustration de la co-évolution des connaissances dans différents domaines, et de boucler la boucle par un retour sur la construction de la théorie géographique. Comme précisé en préambule, un mode de lecture linéaire serait trop réducteur, puisque la plupart des travaux s'enrichissent mutuellement quel que soit leur domaine et leur portée, et un compte-rendu linéaire, au delà d'être intrinsèquement appauvrissant, est en quelque sorte un mensonge par omission de l'ensemble des interactions complexes entre les pans de connaissance produite. Bien sûr l'exercice de synthèse et la capacité à faire rentrer dans un cadre formaté imposé, sont louables, voir souhaitables dans l'état actuel des conditions de production scientifiques.
}

\bpar{
But a fundamental position that we take and advocate all along this work is the one of an anarchist science proposed by \noun{Feyerabend}, which without being taken totally literally and put into context, is extremely useful to propose paradigms shifts and emancipate from \emph{mainstream} approaches which basis and legitimacy seem to increase despite the increasing critics. The writing of a strongly formatted monograph loose some interest through the strongly constraint aspect of the exercise, and seem to be relatively vain given the foreseeable destiny of underuse for most of works currently produced, without being saved by online availability given the imposed language\footnote{The initial manuscript has been written in French. This however relates to much more complex issues than the sole audience~\cite{tardy2004role} and the richness of scientific thoughts allowed by the use of different languages should not be discussed as the legitimacy of organisations such as ASRDLF. But it is indeed this audience which is an issue here and in that case it is as much old-fashioned for a graduate school to impose French as a writing language than the choice of a diplomat to impose a speech in French to a non-French-speaking assembly.}.
}{
Mais une posture fondamentale que nous prenons et défendons tout au long de ce travail est celle d'une science anarchiste proposée par \noun{Feyerabend}, qui sans être prise totalement littéralement et mise en contexte, est extrêmement fructifiante pour proposer des changements de paradigmes et s'émanciper de travaux \emph{mainstream} dont les bases et la légitimité semblent s'enrichir malgré les critiques croissantes. L'écriture d'une monographie extrêmement formatée perd en intérêt de par le caractère contraint de l'exercice, et parait relativement vaine vu la destinée de sous-utilisation pour une grande partie des travaux actuellement produits, sans être sauvée par la mise en ligne vu la langue imposée\footnote{Ce qui relève bien sûr par ailleurs d'une problématique bien plus complexe que la simple audience~\cite{tardy2004role} et la richesse des pensées scientifiques permises par l'utilisation de différentes langues n'est pas discutable ainsi que la légitimité d'organisations comme l'ASRDLF. Mais c'est bien cette audience qui nous pose problème ici et dans ce cas il est quasiment aussi vieux jeu pour une école doctorale d'imposer le français comme langue d'écriture qu'un choix de consul à imposer un discours en français à une audience non-francophone.}.
}


\bpar{
We can wonder on the possibilities of an entirely digital thesis, in which the path of the reader elaborated within the numerical environment would be at the origin of a multitude of possible visions, effectively rendering the complexity of the construction process, and numerous enriching perspectives through a feedback and an interaction with readers, i.e. going beyond the linear presentation mode, as already sustained in introduction. The invention of new scientific communication modes\footnote{Internal and external scientific communication is a challenge in itself, as recalls \cite{Martinez-Conde01082017} by proposing a guenuine \emph{storytelling} of scientific research results.} is an urgent challenge in itself, and our sketch of reflexivity developed in Appendix~\ref{app:reflexivity} aims at contributing to it.
}{
Nous nous prenons à rêver de la possibilité d'une thèse entièrement digitale et dont le cheminement du lecteur tracé dans le support numérique serait à l'origine d'une multitude de visions possibles, traduisant effectivement la complexité du processus de construction, et des perspectives d'enrichissement innombrables par une rétroaction et une interaction avec les lecteurs, c'est-à-dire sortir du mode de présentation linéaire, comme déjà soutenu en introduction. L'invention de nouveaux modes de communication scientifiques\footnote{La communication scientifique interne et externe est un défi à part entière, comme le rappelle \cite{Martinez-Conde01082017} qui propose un véritable \emph{storytelling} des résultats de la recherche scientifique.} est un défi urgent à part entière, et notre ébauche de réflexivité développée en Annexe~\ref{app:reflexivity} cherche à y contribuer.
}

\bpar{
The construction of geographical theories, in the context of a theoretical and quantitative geography, is achieved by iterations within a co-evolution dynamic with empirical and modeling efforts~\cite{livet2010}. Among the numerous examples, we can mention the evolutive urban theory (co-constructed through a spectrum of works ranging for example from the first propositions by \cite{pumain1997pour} until the mature results presented in~\cite{pumain2012multi}), the study of the fractal properties of urban structures (for example from \cite{frankhauser1998fractal} to \cite{frankhauser2008fractal}) or more recently the Transmondyn project \cite{sanders2017peupler} aiming at enriching the notion of settlement system transition. We propose here a synthesis of different empirical and modeling works jointly made in this work with the elaboration of theoretical constructions aiming at better understanding relations between territories and transportation networks.
}{
La construction de théories géographiques, dans le cadre d'une géographie théorique et quantitative, s'effectue par itérations dans une dynamique de co-évolution avec les efforts empiriques et de modélisation~\cite{livet2010}. Parmi les nombreux exemples, on peut citer la théorie évolutive des villes (co-construite par un spectre de travaux s'étendant par exemple des premières propositions de \cite{pumain1997pour} jusqu'aux résultats matures présentés dans~\cite{pumain2012multi}), l'étude du caractère fractal des structures urbaines (par exemple de \cite{frankhauser1998fractal} à \cite{frankhauser2008fractal}) ou plus récemment le projet Transmondyn \cite{sanders2017peupler} visant à enrichir la notion de transition des systèmes de peuplement. Nous proposons ici une synthèse de différents travaux empiriques et de modélisation menés conjointement dans ce travail avec l'élaboration d'appareils théoriques visant à mieux comprendre les relations entre territoires et réseaux de transports.
}


\paragraph{Why a theory and models of coevolution}{Pourquoi une théorie et des modèles de co-évolution}

\bpar{
Our first entry takes a quantitative epistemology viewpoint to attempt to explain the fact that, if the co-evolution between networks and territories has for example been shown by~\cite{bretagnolle:tel-00459720}, the literature is rather poor in simulation models endogeneizing this co-evolution. An algorithmic exploration of the literature has been done in \cite{raimbault2015models}, suggesting a partitioning of scientific domains interested in that subject. More elaborated methods and the corresponding tools (collection and analysis of data), coupling a semantic analysis to the citation network, have been developed to further investigate these preliminary conclusions~\cite{raimbault2016indirect}, and first results at the second order seem to confirm the hypothesis of an understudied domain since at the intersection of fields which do not necessarily easily communicate. These first results in quantitative epistemology confirm the relevance of a modeling coupling processes related to different scales and fields of study, and moreover the relevance of elaborating a proper theory.
}{
Notre première entrée prend un point de vue d'épistémologie quantitative pour tenter d'expliquer le fait que, si la co-évolution entre territoires et réseaux a par exemple été prouvée par~\cite{bretagnolle:tel-00459720}, la littérature est très pauvre en modèles de simulation endogenéisant cette co-évolution. Une exploration algorithmique de la littérature a été menée dans \cite{raimbault2015models}, suggérant un cloisonnement des domaines scientifiques s'intéressant à ce sujet. Des méthodes plus élaborées ainsi que les outils correspondants (collecte et analyse des données), couplant une analyse sémantique au réseau de citations, ont été développées pour renforcer ces conclusions préliminaires~\cite{raimbault2016indirect}, et les premiers résultats au second ordre semblent confirmer l'hypothèse d'un domaine peu défriché car à l'intersection de champs ne dialoguant pas nécessairement aisément. Ces premiers résultats d'épistémologie quantitative confirment l'intérêt d'une modélisation couplant des processus relevant de différentes échelles et domaines d'études, et surtout l'intérêt de l'élaboration d'une théorie propre.
}

\paragraph{Empirical studies}{Etudes empiriques}

\bpar{
The first axis for the developments in themselves consists in empirical analysis. A study of static spatial correlations between urban form measures (morphological indicators computed on the Eurostat population grid) and network form measures (topology of the road network obtained from OpenStreetMap), on the full Europe at different scales, allowed to suggest the non-stationarity and spatial multi-scalarity of their interactions~\cite{raimbault2016cautious}. This aspect has also been highlighted in space and time at a microscopic scale through the study of dynamics of a transportation system~\cite{raimbault2017investigating}, jointly with the heterogeneity of processes for an other type of system~\cite{raimbault2015hybrid}. These stylized facts validate the use of complex simulation models, for which first modeling efforts paved the road towards more elaborated models.
}{
Le premier axe pour les développements en eux-mêmes consiste en des analyses empiriques. Une étude des corrélations spatiales statiques entre mesures de forme urbaine (indicateurs morphologiques calculés sur la grille de population eurostat) et mesures de forme de réseau (topologie du réseau routier issu d'OpenStreetMap), sur l'ensemble de l'Europe à différentes échelles, a pu révéler la non-stationnarité et la multi-scalarité spatiale de leurs interactions~\cite{raimbault2016cautious}. Cet aspect a aussi été mis en évidence dans l'espace et le temps à une échelle microscopique lors de l'étude des dynamiques d'un système de transport~\cite{raimbault2017investigating}, conjointement avec l'hétérogénéité des processus pour un autre type de système~\cite{raimbault2015hybrid}. Ces faits stylisés valident l'utilisation de modèles de simulation complexes, pour lesquels des premiers efforts de modélisation ont ouvert la voie vers des modèles plus élaborés.
}

\paragraph{Modeling}{Modélisation}

\bpar{
At the mesoscopic scale, aggregation-diffusion processes have been shown sufficient to reproduce a large number of urban forms with a small number of parameters, calibrated on the whole spectrum of real values of urban form indicators for Europe. This simple model could, in the context of a methodological exercise exploring the possibility to control the structure of synthetic data at the second order~\cite{raimbault2016generation}, be weakly coupled to a network generation model, showing a broad latitude of potentially generated configurations. The exploration of different autonomous heuristics for network generation has furthermore been explored, to compare for example road network growth models based on local optimization to models inspired by biological networks: each exhibits a large variety of generated topologies. At the macroscopic scale, a simple model of urban growth dynamically calibrated on French cities from 1830 to 2000 (Pumain-Ined database) allowed to demonstrate the existence of a network effect through the increase of the model explicative power when adding an effect of flows going through a physical network, while correcting for the gain due to additional parameters by the construction of an empirical Akaike information criteria~\cite{raimbault2016models}. This ensemble of models is positioned within a logic of parsimony and a perspective to be applied within multi-modeling. In a more descriptive multi-agent approach and thus an more complex model, \cite{le2015modeling} describe a co-evolution model at the metropolitan scale (Lutecia model) which includes in particular governance processes for the development of transportation infrastructures. For this last model, the first studies of the dynamics suggest the importance of the multi-level aspect of the transportation network development to obtain complex patterns of network and collaboration between agents. All these modeling efforts support the theoretical foundations which were proposed as a consequence.
}{
A l'échelle mesoscopique, des processus d'agrégation-diffusion ont été montrés suffisant pour reproduire un grand nombre de formes urbaines avec un faible nombre de paramètres, calibrés sur l'ensemble du spectre des valeurs réelles des indicateurs de forme urbaine pour l'Europe. Ce modèle simple a pu, à l'occasion d'un exercice méthodologique explorant la possibilité de contrôle au second ordre de la structure de données synthétiques~\cite{raimbault2016generation}, être couplé faiblement à un modèle de génération de réseau, démontrant une grande latitude de configurations potentiellement générées. L'exploration de différentes heuristiques autonomes de génération de réseau a par ailleurs été menée, pour comparer par exemple des modèles de croissance de réseau routier basés sur l'optimisation locale à des modèles inspirés des réseaux biologiques : chacun présente une très grande variété de topologies générées. A l'échelle macroscopique, un modèle simple de croissance urbaine calibré dynamiquement sur les villes françaises de 1830 à 2000 (base Pumain-Ined) a permis de démontrer l'existence d'un effet réseau de par l'augmentation de pouvoir explicatif du modèle lors de l'ajout d'un effet des flux transitant par un réseau physique, tout en corrigeant le gain dû à l'ajout de paramètres par la construction d'un Critère d'Information d'Akaike empirique~\cite{raimbault2016models}. Cet ensemble de modèles se positionne avec un objectif de parcimonie et dans une perspective d'application en multi-modélisation. Dans une démarche multi-agents plus descriptive et donc par un modèle plus complexe, \cite{le2015modeling} décrivent un modèle de co-évolution à l'échelle métropolitaine (modèle Lutecia) qui inclut en particulier des processus de gouvernance pour le développement des infrastructures de transport. Pour ce dernier modèle, les premières études de la dynamique montrent l'importance du caractère multi-niveau du développement du réseau de transport pour obtenir des motifs complexes de réseaux et de collaboration entre agents. L'ensemble de ces efforts de modélisation supportent les fondements théoriques que nous avons proposé par la suite.
}

\paragraph{Construction of a geographical theory}{Construction d'une théorie géographique}

\bpar{
We revisit finally the theory constructed in~\ref{sec:theory} from the viewpoint of co-evolution of domains. We insist here on its integrative aspect which allows to link urban evolutive theory and morphogenesis. Based on the previous works, we propose to gather two entries for the construction of a geographical theory having a privileged focus on interactions between territories and networks. The first is through the notion of \emph{morphogenesis}, which has been explored from an interdisciplinary viewpoint in~\cite{antelope2016interdisciplinary}. In our context, morphogenesis consists in the emergence of the form and the function, through autonomous local processes within a system which then exhibits a self-organized architecture. The presence of a function and thus of an architecture distinguishes morphogenetic systems from simply self-organized systems (see~\cite{doursat2012morphogenetic}). Furthermore, the concepts of autonomy and locality can well be applied to territorial systems, for which we try to isolate the subsystems and relevant scales. The works on generation of calibrated urban form through autonomous processes, the first results on network generation through multiple processes also autonomous, and older works studying a simple model of urban morphogenesis which was sufficient to reproduce stylized patterns of forms~\cite{raimbault2014hybrid}, suggest the possible existence of such processes within territorial systems.
}{
Nous revoyons enfin sous l'oeil de la co-evolution des domaines la théorie construite en~\ref{sec:theory}. Nous insistons ici sur son caractère intégratif permettant de joindre théorie évolutive des villes et morphogenèse. En se basant sur les travaux précédents, nous proposons de joindre deux entrées pour la construction d'une théorie géographique ayant un focus privilégié sur les interactions entre territoires et réseaux. La première est par la notion de \emph{morphogénèse}, qui a été explorée d'un point de vue interdisciplinaire dans~\cite{antelope2016interdisciplinary}. Pour notre part, la morphogenèse consiste en l'émergence de la forme et de la fonction, via des processus locaux autonomes dans un système qui exhibe alors une architecture auto-organisée. La présence d'une fonction et donc d'une architecture distingue les systèmes morphogénétiques de systèmes simplement auto-organisés (voir~\cite{doursat2012morphogenetic}). De plus, les notions d'autonomie et de localité s'appliquent bien à des systèmes territoriaux, pour lesquels on essaye d'isoler les sous-systèmes et les échelles pertinentes. Les travaux sur la génération de forme urbaine calibrée par des processus autonomes, les premiers travaux sur la génération de réseaux par de multiples processus également autonomes, et des travaux plus anciens étudiant un modèle simple de morphogenèse urbaine qui suffisait à reproduire des motifs de forme stylisés~\cite{raimbault2014hybrid}, nous suggèrent la possible existence de tels processus au sein des systèmes territoriaux. 
}

\bpar{
Moreover, the framework of the evolutive urban theory is favoured by our empirical results, which show the non-stationary, heterogenous, multi-scalar aspects of urban systems. To remain as more general as possible, and as both our empirical and modeling results (generation of arbitrary forms by the aggregation-diffusion model for example) apply in general to territorial systems, we situate within the framework of human territories by \cite{raffestin1988reperes}, i.e. ``\textit{the conjunction of a territorial process with an informational process}'', which can be interpreted in our case as the complex socio-technical-enviromental system which is constituted by a territory and the agents and artefacts which interact within. The importance of networks is underlined by our results on the necessity of networks in the macroscopic growth model: we suggest then to consider \emph{networked complex territorial systems}, by adding to the insertion of the territory within the evolutive theory the particularity that there exists crucial components which are networks (transportation networks in particular), which origin can be explained by the territorial theory of networks by \cite{dupuy1987vers}.
}{
D'autre part, le cadre de théorie évolutive des villes est plébiscité par nos résultats empiriques, qui montrent le caractère non-stationnaire, hétérogène, multi-scalaire des systèmes urbains. Pour rester le plus général possible, et comme nos résultats à la fois empiriques et de modélisation (génération de formes quelconques par le modèle d'agrégation-diffusion par exemple) s'appliquent aux systèmes territoriaux en général, nous nous plaçons dans le cadre de territoires humains de \cite{raffestin1988reperes}, c'est-à-dire ``\textit{la conjonction d'un processus territorial avec un processus informationnel}'', qui peut être interprété dans notre cas comme le système complexe socio-techno-environmental que constitue un territoire et les agents et artefacts qui y interagissent. L'importance des réseaux est soulignée par nos résultats sur la nécessité du réseau dans le modèle de croissance macroscopique : nous proposons alors de parler de \emph{systèmes territoriaux complexes en réseau}, en ajoutant au plongement du territoire dans la théorie évolutive la particularité qu'il existe des composantes cruciales qui sont les réseaux (de transport en l'occurrence), dont l'origine peut être expliquée par la théorie territoriale des réseaux de \cite{dupuy1987vers}.
}

\bpar{
We then propose the following hypothesis to link our two approaches: \textit{the existence of morphogenetic processes within which networks play a crucial role is equivalent to the existence of subsystems within networked complex territorial systems, which are then defined as co-evolving}.
}{
Nous proposons alors l'hypothèse suivante afin de réconcilier nos deux approches : \textit{l'existence de processus morphogénétiques dans lesquels les réseaux ont un rôle crucial est équivalente à l'existence de sous-systèmes dans les systèmes territoriaux complexes en réseaux, qu'on définit alors comme co-évolutifs}.
}

\bpar{
This proposal has multiple implications, but typically guided in particular the modeling choices towards a modular methodology and towards multi-modeling in order to attempt to exhibit morphogenetic processes, and also the empirical analysis towards a more focused study of correlations, causalities (in the case of time series) and search for modular decompositions of systems.
}{
Cette proposition a de multiples implications, mais a typiquement guidé notamment les choix de modélisation vers une méthodologie modulaire et de multi-modélisation afin d'essayer d'exhiber des processus morphogénétiques, ainsi que les travaux empiriques vers une étude plus poussée des corrélations, causalités (dans le cas de séries temporelles) et recherche de décompositions modulaires des systèmes.
}

\stars

%


\newpage

\section*{Chapter Conclusion}{Conclusion du Chapitre}


\bpar{
This chapter thus allowed us to take a step back on our contributions and to put these into perspective. It indeed opens several doors, and recall the fact that the coverage of knowledge remain very low.
}{
Ce chapitre nous a permis ainsi de prendre du recul sur nos contributions et de les mettre en perspective. Il ouvre en fait de nombreuses portes, et fait prendre conscience que la portée des connaissances reste embryonnaire.
}


\bpar{
The questions raised by each of the levels are fundamental for the study of complex territorial systems but also of complex systems in general. The theory proposed in~\ref{sec:theory} again highlights the issue of spatio-temporal non-stationarity within a multi-scale context, that we postulate as crucial but under-explored in the case of territorial systems.We also distinguish the difficulty to integrate existing theories what implies an understanding of model coupling processes.
}{
Les questions soulevées par chacun des niveaux sont fondamentales pour l'étude des systèmes territoriaux complexes mais aussi des systèmes complexes en général. La théorie proposée en~\ref{sec:theory} pointe à nouveau la question de la non-stationnarité spatio-temporelle dans un contexte multi-échelle, que nous postulons cruciale mais peu explorée dans le cas des systèmes territoriaux. Nous distinguons également la difficulté d'intégration de théories existantes ce qui implique une compréhension des processus de couplage des modèles.
}

\bpar{
This issue is at the heart of the formal framework developed in the following~\ref{app:sec:csframework}, which also raises scale imbrication issues. The problem to obtain a consistent algebraic structure with a monoid action on data implies an integration of \noun{Krob}'s theory, what more generally questions the integration of system engineering approaches (``industrial'' complex systems) with the ones of natural complex systems.
}{
Ce problème est au coeur du cadre formel développé par la suite~\ref{app:sec:csframework}, qui soulève aussi des questions d'imbrication d'échelles. Le problème d'obtenir une structure algébrique cohérente avec une action de monoïde sur les données implique une intégration de la théorie de \noun{Krob}, ce qui questionne plus généralement l'intégration des approches d'ingénierie système (systèmes complexes ``industriels'') avec celles de systèmes complexes naturels.
}

\bpar{
The possibility of integrative theories is raised by the introduction of the knowledge framework~\ref{sec:knowledgeframework}, which also introduces more general questions of knowledge production and of the nature of complexity which was briefly evoked from an epistemological viewpoint in~\ref{sec:epistemology}.
}{
La possibilité de théories intégratives est soulevée par l'introduction du cadre de connaissance~\ref{sec:knowledgeframework}, qui pose également des problèmes plus généraux de production des connaissances et de nature de la complexité que nous avions brièvement abordé d'un point de vue épistémologique en~\ref{sec:epistemology}.
}

\bpar{
We propose to synthesize a part of these diverse open questions in a consistent research project on the long term, but which includes first immediate directions, and that we present in opening.
}{
Nous proposons de synthétiser une partie de ces diverses questions ouvertes dans un projet de recherche cohérent sur un long terme mais incluant des premières pistes concrètes immédiates, que nous présenterons en ouverture.
}

\stars



\bpar{
\chapter*{General opening}
}{
\chapter*{Ouvertures Générales}
}

\bpar{
\markboth{General opening}{General opening}
}{
\markboth{Ouvertures générales}{Ouvertures générales}
}

\label{ch:opening}



\bpar{
As we suggested before, the opening in fact allows to take a step back and in our case a clarification of the global frame. We therefore propose here the exercise to summarize already invoked opening works, questions opened by our work, and their synthesis within a long term research project.
}{
Comme nous l'avons suggéré précédemment, l'ouverture permet en fait une prise de recul et dans notre cas une clarification du cadre global. Nous proposons donc ici l'exercice de recension des travaux d'ouverture déjà menés, celle des problèmes ouverts par notre travail, et leur synthèse dans un projet de recherche à long terme.
}

\section*{Thematic and general perspectives}{Perspectives thématiques et générales}

\subsection*{Into a global perspective}{Mise en perspective globale}

\bpar{
A second reading of the thesis enlightened by the theoretical articulation proposed in~\ref{sec:theory} confirms us that (i) the morphogenetic approach was naturally induced by the constraint of ecological niche in the definition of co-evolution; (ii) the evolutive urban theory is thus refined for the precise cas of co-evolution; (iii) territorial systems must intrinsically induce such processes, since they are both support and objects of these. The question of network necessity to represent territorial systems remains open, and we have postulated it in our theoretical construction. Our results suggest the relevance to take them into account, and open the question of a demonstration of this postulate.
}{
Une relecture de la thèse à la lumière de l'articulation théorique proposée en~\ref{sec:theory} nous confirme que (i) l'approche morphogénétique était naturellement induite par la contrainte de niche écologique dans la définition de la co-évolution ; (ii) la théorie évolutive des villes est ainsi précisée pour le cas précis de la co-évolution ; (iii) les systèmes territoriaux doivent intrinsèquement induire de tels processus, puisqu'ils sont à la fois support et objets de ceux-ci. La question de la nécessité des réseaux pour représenter les systèmes territoriaux reste ouverte, et nous l'avons postulée dans notre construction théorique. Nos résultats suggèrent la pertinence de leur prise en compte, et ouvrent la question d'une démonstration de ce postulat.
}

\bpar{
Then, a third reading through knowledge domains allows to better understand the articulation between the different components: conceptual and empirical constructions of the first part yield a definition of co-evolution, and the elaboration of methods and models in a second part, which in turn feed back these other domains in the third part. We propose a brief quantitative analysis of these dynamics in~\ref{app:reflexivity}. Therefore, the interdependancy within the path taken, given by the diagram in introduction (Frame~\ref{frame:intro:organisation}), is indeed much more complex and not necessarily linear. Renewed readings of this monograph will thus be richer, through the emergence of implicit links. We propose in Frame~\ref{frame:opening:organisation} a possible new reading of the organisation of our work, in relation to the general problematic and knowledge domains. 
}{
Ensuite, une relecture par les domaines de connaissance permet de mieux comprendre l'articulation entre les différentes composantes : les constructions conceptuelles et empiriques de la première partie permettent une définition de la co-évolution, puis la mise en place de méthodes et de modèles en deuxième partie, qui en retour alimentent ces autres domaines en troisième partie. Nous proposons une analyse quantitative brève de ces dynamiques en~\ref{app:reflexivity}. Ainsi, l'interdépendance dans le cheminement, donnée par le diagramme en introduction (Encadré~\ref{frame:intro:organisation}), est en fait bien plus complexe et pas forcément linéaire. Une deuxième lecture de notre monographie sera ainsi plus riche, par émergence des liens implicites. Nous proposons en Encadré~\ref{frame:opening:organisation} une relecture possible de l'organisation de notre travail, au regard de la problématique générale et des domaines de connaissance.
}

\begin{figure}
	\begin{mdframed}
	\includegraphics[width=\linewidth]{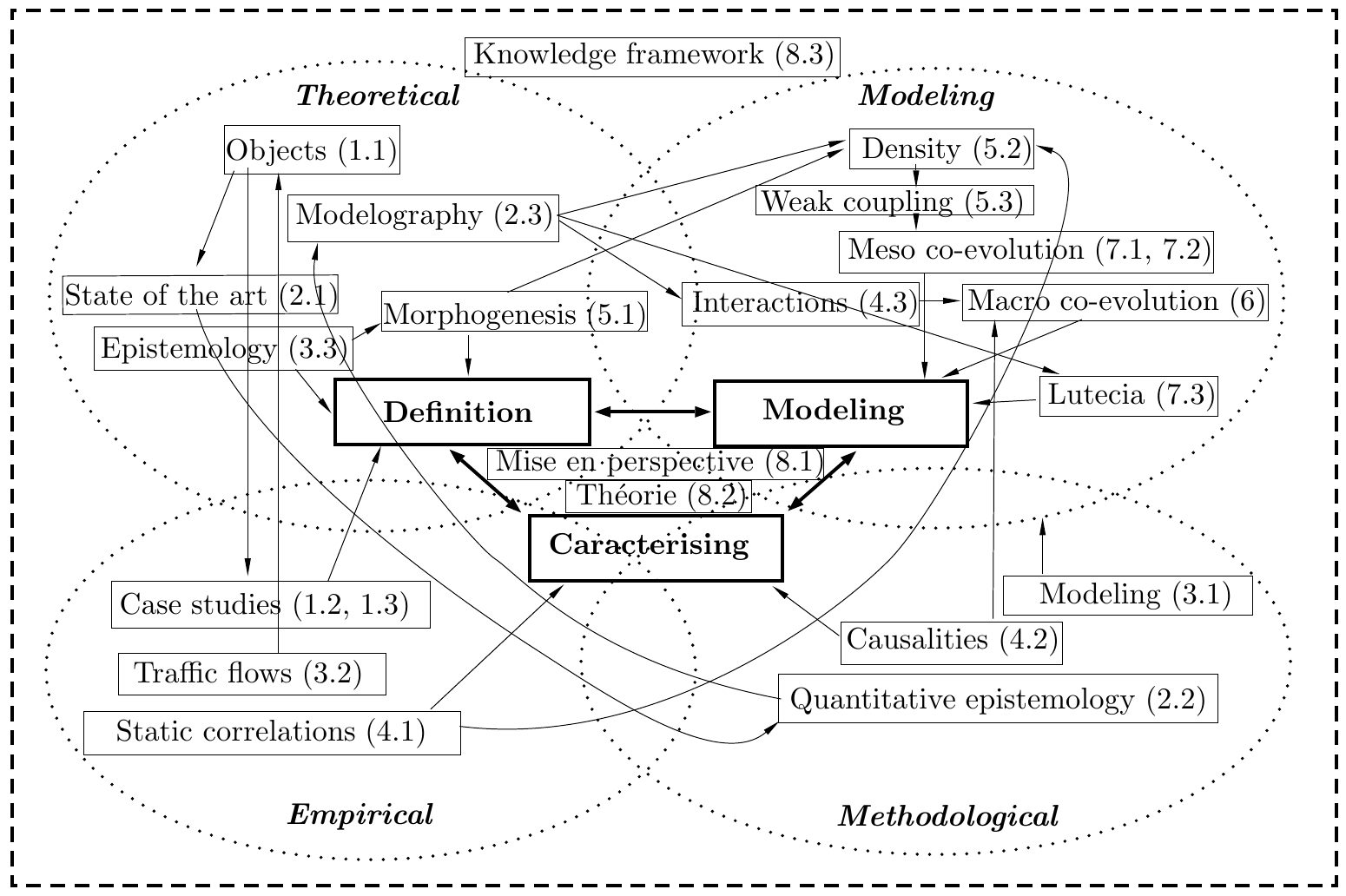}
	\medskip
	\framecaption{\textbf{Rereading of the organisation at the light of knowledge domains.} We put at the core the three axis coming from the general problematic, of defining, characterizing and modeling co-evolution. Different components irrigate its elements which can not be dissociated, and finally blended together by the conclusion and opening. Sections are located in an indicative way within the knowledge domain which corresponds the most (knowing that they are all across different domains). The domains of data and tools are here let aside to facilitate reading (and would necessitate a more precise description of the work). The relations between components are also given in an indicative way and are not exhaustive, but allow to grasp the complexity of the global articulation.\label{frame:opening:organisation}}{\textbf{Relecture de l'organisation à la lumière des domaines de connaissance.} Nous plaçons au coeur le triptyque issu de la problématique générale de définition, caractérisation et modélisation de la co-évolution. Différentes composantes irriguent ses éléments qui sont indissociables, et soudés finalement par la conclusion et ouverture. Les sections sont placées à titre indicatif dans le domaine de connaissance qui leur correspond le plus (sachant qu'elles sont toutes à cheval sur plusieurs domaines). Les domaines de données et outils sont laissés de côté ici pour faciliter la lecture (et nécessiteraient un découpage plus précis du travail). Les relations entre composantes sont également données à titre indicatif et ne sont pas exhaustives, mais permettent de se rendre compte de la complexité de l'articulation globale.\label{frame:opening:organisation}}
	\end{mdframed}
\end{figure}

\bpar{
Our work can also be inserted within a broader perspective. Let precise the ``meta-articulation'' of our work, i.e. the implicit structure of the diverse developments and openings and therefore the global frame in which the core is integrated (first three parts). The Frame~\ref{frame:opening:perspective} gives a schematic representation of this articulation. The core, which consists in the answer to the problematic, is constituted by three axis in strong interaction: the definition, the characterization and the modeling of co-evolution of transportation networks and territories. Each call in its way to developments in different fields\footnote{We do not use the term of domain here to avoid a confusion with knowledge domains, these being used in a different way as we will see in Appendix~\ref{app:reflexivity}.}: contributions in epistemology and quantitative epistemology, mainly linked to the aspect of definition; development of systemic frameworks, induced by issues linked to modeling; and thematic developments linked to the characterization.
}{
Notre travail peut également se placer dans une perspective plus large. Précisons la ``méta-articulation'' de notre travail, c'est-à-dire la structure implicite des divers développements et ouvertures et donc le cadre global dans lequel s'inscrit le coeur (trois premières parties). L'Encadré~\ref{frame:opening:perspective} schématise cette articulation. Le coeur, qui consiste en la réponse à la problématique, est constitué de trois axes en interaction forte : la définition, la caractérisation et la modélisation de la co-évolution des réseaux de transport et des territoires. Chacun appelle à sa manière des développements dans divers champs\footnote{Nous n'utilisons pas le terme domaine ici pour ne pas entrainer une confusion avec les domaines de connaissance, ceux-ci étant mobilisés différemment comme nous le verrons en Annexe~\ref{app:reflexivity}.} : des développements épistémologiques et en épistémologie quantitative, principalement liés à l'aspect de définition ; des développements de cadres systémiques, induits par les problématiques liées à la modélisation ; et des développements thématiques liés à la caractérisation.
}

\bpar{
We can detail the content of each of these developments, by linking them to the corresponding content mainly in Appendices:
\begin{enumerate}
	\item Quantitative epistemology: mostly in relation with methods and tools for systematic review and exploration of a scientific landscape in~\ref{ch:modelinginteractions}, we include the original case study which initiated the method, the corpus of the Cybergeo journal, in~\ref{app:sec:cybergeo}, and also an application to a massice patent corpus in~\ref{app:sec:patentsmining}.
	\item Epistemology: contextualizing the study of Cybergeo with other complementary approaches yields epistemological considerations in~\ref{app:sec:cybergeonetworks}; we also start a reflexion on the links between economics and geography in~\ref{app:sec:ecogeo}.
	\item Systemic frameworks: a knowledge framework, contributing to organize a complex knowledge, has already been proposed in~\ref{sec:knowledgeframework}; a framework formalizing the coupling of models of socio-technical systems, suggesting directions to formalize the knowledge framework, is developed in~\ref{app:sec:csframework}; a framework to study the robustness of multi-attribute evaluations is developed in~\ref{app:sec:robustness}.
	\item Thematical: the case studies of transportation systems achieved in \ref{sec:reproducibility} and in~\ref{sec:energyprice} provide indeed a confirmation of relevant scales; the study of the generation of synthetic data, in relation with the methodology developed in~\ref{sec:computation}, is done in~\ref{app:sec:syntheticdata} for the method and in~\ref{app:sec:syntheticdata-finance} for an example of application; the modeling of migration dynamics within Pearl River Delta sketched in~\ref{app:sec:migrationdynamics}, introduces elements for multi-scale models and refines interactions between cities at the level of individual flows.
\end{enumerate}
}{
Détaillons le contenu de chacun de ces développements, en les reliant au contenu correspondant principalement en Annexes :
\begin{enumerate}
	\item Epistémologie quantitative : principalement en lien avec les méthodes et outils de revue systématique et d'exploration d'un paysage scientifique en~\ref{ch:modelinginteractions}, nous incluons le cas d'étude original qui a initié la méthode, le corpus du journal Cybergeo, en~\ref{app:sec:cybergeo}, ainsi qu'une application à un corpus massif de brevets en~\ref{app:sec:patentsmining}.
	\item Epistémologie : la mise en contexte de l'étude de Cybergeo avec d'autres approches complémentaires permet une prise de recul épistémologique dans~\ref{app:sec:cybergeonetworks} ; nous amorçons également une réflexion sur les liens entre économie et géographie en~\ref{app:sec:ecogeo}.
	\item Cadres systémiques : un cadre de connaissance, contribuant à organiser une connaissance complexe, a déjà été proposé en~\ref{sec:knowledgeframework} ; un cadre formalisant le couplage des modèles des systèmes socio-techniques, suggérant des pistes de formalisation du cadre de connaissance, est développé dans~\ref{app:sec:csframework} ; un cadre pour l'étude de la robustesse des évaluations multi-attributs est développé dans~\ref{app:sec:robustness}.
	\item Thématique : les études de cas des systèmes de transport effectuées en \ref{sec:reproducibility} et en~\ref{sec:energyprice} permettent en l'occurence une confirmation des échelles pertinentes ; l'étude de la génération de données synthétiques, en lien avec la méthodologie développée en~\ref{sec:computation}, est faite en~\ref{app:sec:syntheticdata} pour la méthode et en~\ref{app:sec:syntheticdata-finance} pour un exemple d'application ; la modélisation des dynamiques migratoires au sein du Delta de la Rivière des Perles ébauchée en~\ref{app:sec:migrationdynamics}, introduit une piste de modèles multi-échelle et raffine les interactions entre villes au niveau des flux individuels.
\end{enumerate}
}

\begin{figure}
	\begin{mdframed}
	\includegraphics[width=\linewidth]{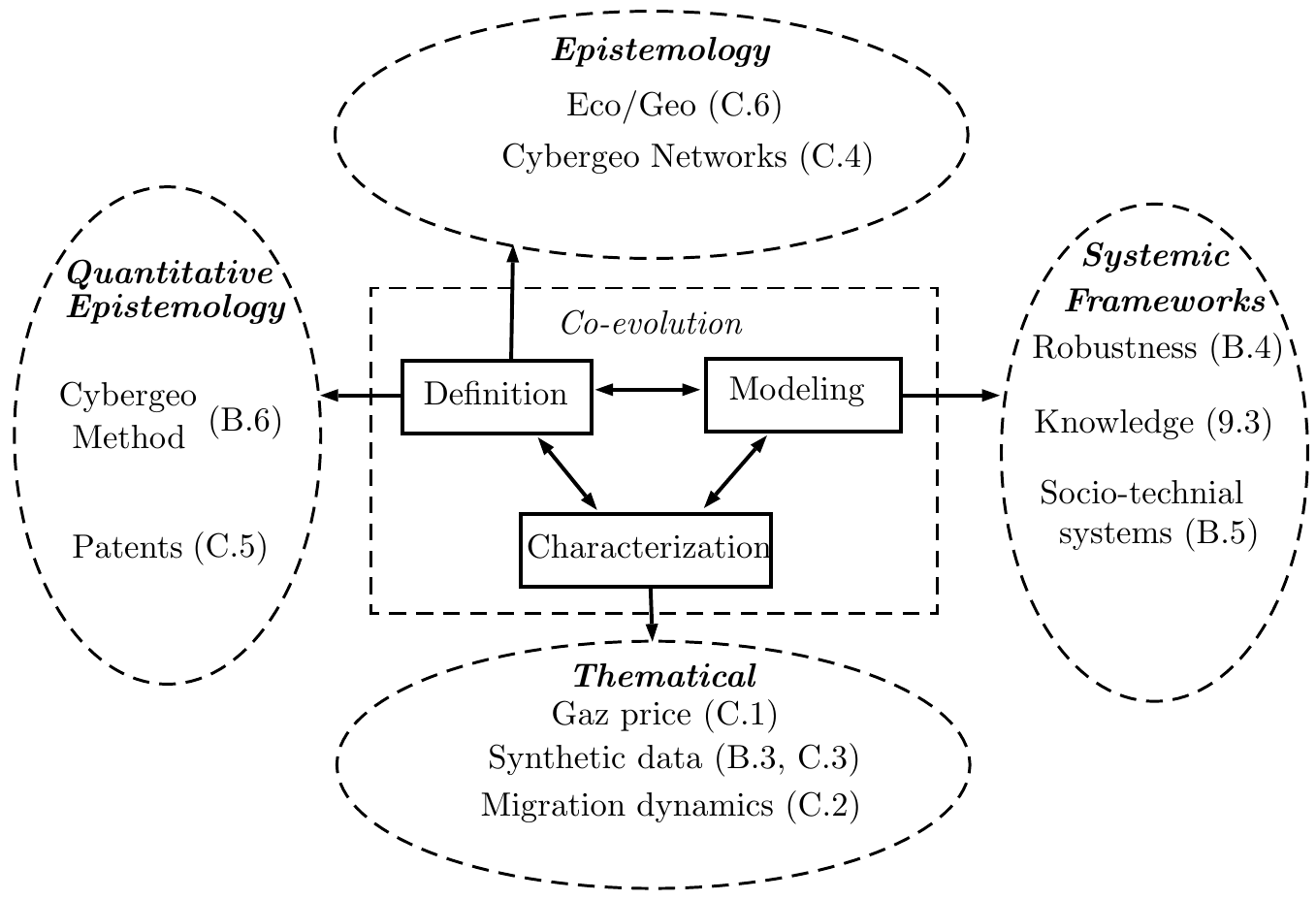}
	\medskip
	\framecaption{\textbf{Perspective on the global frame.} The core of the problematic, co-evolution, is composed by three components which call for extensions within diverse fields (their intersections being not represented to ease reading). We list here within each the different openings, mainly done in Appendices.\label{frame:opening:perspective}}{\textbf{Mise en perspective du cadre global.} Le coeur de la problématique, la co-évolution, est composé de trois composantes qui appellent des extensions dans des champs divers (leurs intersections n'étant pas représentées pour faciliter la lecture). Nous y listons dans chaque les différentes ouvertures, menées principalement en Annexes.\label{frame:opening:perspective}}
	\end{mdframed}
\end{figure}

\bpar{
These different fields have naturally non-empty intersections (the knowledge framework of~\ref{sec:knowledgeframework} corresponds for example both to a systemic framework and to epistemology, or the study of patents is an important thematic aspect in link with the evolutive urban theory) and are interacting: studies in quantitative epistemology inform epistemology, what guides thematic studies, which can be put into perspective in the systemic frameworks, which in their turn also depend on the epistemological positioning.
}{
Ces différents champs sont bien sûr à intersections non vides (le cadre de connaissance de~\ref{sec:knowledgeframework} relève par exemple à la fois du cadre systémique et de l'épistémologie, ou l'étude des brevets est un important aspect thématique en lien avec la théorie évolutive des villes) et en interactions : les études d'épistémologie quantitative informent l'épistémologie, qui guide les études thématiques, qui peuvent être mises en perspective dans les cadres systémiques, qui eux dépendent également du positionnement épistémologique.
}

\bpar{
Thus, we highlight a more global structure for our work, which partly sketches the structure of a research project that we will detail in the following.
}{
Ainsi, nous mettons en évidence une structure plus globale pour notre travail, qui dessine en partie la structure d'un projet de recherche que nous détaillerons par la suite.
}

\subsection*{Open questions}{Questions ouvertes}

\bpar{
We now develop fundamental questions which have been evoked or opened all along our work, which we classify into three axis: scientific practice (applied epistemology), modeling, and foundations of spatial complex systems.
}{
Nous développons à présent des questions fondamentales qui ont été abordées ou ouvertes tout au long de notre travail, que nous classons en trois axes : pratique scientifique (épistémologie appliquée), modélisation, et fondements des systèmes complexes spatiaux.
}

\subsubsection*{Applied epistemology}{Epistémologie appliquée}

\paragraph{For a completely open science}{Pour une science totalement ouverte}

\bpar{
A first crucial axis of development for all the ecosystem of knowledge production within which we are integrated (see chapter~\ref{ch:positioning}) is the contribution to a maximal opening of the scientific practice, i.e. the combination of all the approaches summarized by \cite{fecher2014open}, in particular the democratic and public aspects which foster the access of all to the production of knowledge and its results\footnote{Knowing that the opening of knowledge products is intrinsic in a complex perspective, since as \cite{morin1991methode} puts it, our ideas find a certain independence in the noosphere and do not belong to us.}, and the pragmatic and infrastructure aspects which focuses on the increased efficiency within an open framework.
}{
Un premier axe de développement crucial pour l'ensemble de l'écosystème de production de connaissance dans lequel nous nous inscrivons (voir chapitre~\ref{ch:positioning}) est la contribution à une ouverture maximale de la pratique scientifique, c'est-à-dire la combinaison de l'ensemble des approches résumées par \cite{fecher2014open}, en particulier les aspects démocratique et public qui encouragent l'accès de tous à la production de connaissance et à ses résultats\footnote{Sachant que l'ouverture des produits de la connaissance est évidente dans une perspective complexe, puisque comme le souligne \cite{morin1991methode}, nos idées prennent une certaine indépendance dans la noosphère et ne nous appartiennent pas.}, et les aspects pragmatique et d'infrastructure qui appuient l'efficience augmentée dans un cadre ouvert.
}

\bpar{
Transparency and availability of raw data or at least preprocessed data, and of the computer code producing simulation outputs or figures, seems to be more an exception than the rule in geography. As recalls \cite{banos2013pour} which consecrates one of its principles to it, ``\textit{the modeler is not the guardian of the proofed truth}'', and as recalled in~\ref{sec:reproducibility}, a perfect reproducibility of results is necessary for any value to be acknowledged by the scientific community, as a theory which does not provide falsification possibilities can not be considered as scientific in the sense of \noun{Popper}. Reviewing experiences for \emph{Cybergeo} have confirmed with unanimity this fundamental issue. We can recall that the journal \emph{PNAS} imposes to provide raw data and tables producing any figure, to prevent any visualization bias let it be voluntary (what is crippling and leads to a signalling) or not.
}{
La transparence et mise en disponibilité des données brutes ou au moins pré-traitées, et du code informatique produisant les sorties de simulation ou les figures, semble être plutôt l'exception que la règle en géographie. Comme le rappelle \cite{banos2013pour} qui y dédie l'un de ses principes, ``\textit{le modélisateur n'est pas le gardien de la vérité prouvée}'', et comme rappelé en~\ref{sec:reproducibility}, une reproductibilité parfaite des résultats est nécessaire pour une reconnaissance d'une quelconque valeur par la communauté scientifique, comme une théorie qui ne fournit pas de possibilité de falsification ne peut être considérée comme scientifique au sens de \noun{Popper}. Des experiences de revue pour \emph{Cybergeo} ont confirmé à l'unanimité ce problème fondamental. Rappelons que la revue \emph{PNAS} exige les données brutes et tableau produisant toute figure, pour prévenir tout biais de visualisation qu'il soit volontaire (ce qui est rédhibitoire et conduit à un signalement) ou non.
}

\bpar{
Furthermore, scientific communication is an important aspect of open science. The current mode of scientific publishing is far from being ideal. An article is not an understandable format nor really reproducible, and leads to bias. The writing of a paper while answering to the norms in order to be accepted can be assimilated to ``a game'' which rules are subtle and must be mastered to follow a carrier. According to our positioning, such a communication mode is contrary to the honesty and intellectual integrity which are necessary for an ethical and open science. Initiatives multiply to propose alternative models: post-publication review is one, the use of version control systems and public repositories is an other, or the flash publication of research directions\footnote{See for example the \emph{Journal of Brief Ideas} at \url{http://beta.briefideas.org/about}. Short descriptions of research directions are often delegated to the discussion of the conclusion of papers, which are written in a conventional way, often with a bias to justify a posteriori the interest of \emph{our new method} which unfortunately must be sold. We therefore build gigantic plans, propose developments with few connections, or application domains \emph{which will have an impact} (read which are fashionable or receive the most financing at the period of writing). This manuscript naturally falls under these critics, as do the associated papers.}. For example, \cite{bon2017novel} describes an experiment of dynamical articles evaluated in an open way by the community, with associated metrics allowing to make the works judged as interesting emerge.
}{
Par ailleurs, la communication scientifique est un aspect important de la science ouverte. Le mode actuel de publication scientifique est loin d'être idéal. Un article n'est pas un format compréhensible ni vraiment reproductible, et pousse au biais. L'écriture d'un article en répondant au normes de façon à être accepté peut être assimilé à ``un jeu'' dont les règles sont subtiles et qu'il faut maitriser pour faire carrière. Selon notre positionnement, un tel mode de communication est contraire à l'honnêteté et l'intégrité intellectuelle nécessaires à une science éthique et ouverte. Les initiatives se multiplient pour proposer des modèles alternatifs : la revue post-publication en est une, l'utilisation de systèmes de contrôle de version et de dépôts publics une autre, ou la publication éclair de pistes de recherche\footnote{Voir par exemple le \emph{Journal of Brief Ideas} à \url{http://beta.briefideas.org/about}. Les descriptions courtes de pistes de recherche sont souvent reléguées à la discussion ou la conclusion des articles, qui s'écrivent de manière conventionnelle, souvent avec un biais pour justifier a posteriori l'intérêt de \emph{sa nouvelle méthode} qu'il faut malheureusement vendre. On fait alors des plans sur la comète, propose des développements ayant peu de rapport, ou des domaines d'application \emph{qui auront un impact} (lire qui sont à la mode ou qui reçoivent le plus de financements à la période de l'écriture). Ce manuscrit tombe bien évidemment partiellement sous ces critiques, comme les articles qui lui sont associés.}. Par exemple, \cite{bon2017novel} décrit une expérience d'articles dynamiques évalués de manière ouverte par la communauté, avec des métriques associées permettant de faire émerger les travaux jugés intéressants.
}


\bpar{
Similarly, we claim that a linear presentation of a research project is too much reducing, and that the invention of alternative communication modes is a future issue for open science. We can for example imagine interactive networks, translating the structure of the underlying knowledge, and in which the reader can navigate between concepts and analyses, being directly redirected to data, models and analyses. The main interpretation schemes corresponding to the narration that would follow a linear explanation can then be superposed to the network to come back to a more classical reading mode. A communication through gaming is also a relevant alternative, particularly in the case of a communication for the public, and we give an illustration for an ecology problem in Appendix~\ref{app:sec:mediationecotox}.
}{
De la même façon, nous soutenons qu'une présentation linéaire d'un projet de recherche est trop fortement réducteur, et que l'invention de modes de communication alternatifs est un enjeu futur pour la science ouverte. On peut par exemple imaginer des réseaux interactifs, traduisant la structure de la connaissance sous-jacente, et dans lesquels le lecteur peut naviguer entre les concepts et les analyses, être renvoyé directement vers les données, modèles et analyses. Les grilles de lecture principales en accord avec l'argument que prendrait une explication linéaire peuvent alors être superposées au réseau pour revenir à un mode plus classique de lecture. Une communication par le jeu est également une alternative crédible, notamment dans le cas d'une communication pour le public, et nous en donnons une illustration pour un problème d'écologie en Annexe~\ref{app:sec:mediationecotox}.
}

\paragraph{For an evidence-based science}{Pour une science evidence-based}

\bpar{
We postulate that a fully \emph{evidence-based} science, whatever its subject, is possible and desirable in articulation with open science. The idea is to aim at disconnecting scientific knowledge from any dogmatism, any political a priori and any judgement about values\footnote{Knowing that these must moreover developed more than ever and thought to articulate science with society, but should interfere the least possible with the knowledge production process in itself. Following \cite{morin2004methode}, an ethic of knowledge and a complex thinking naturally induces a broader ethic, inducing the autonomy of scientific knowledge without making it inhuman.}. In the case of the study of subject linked to individuals or societies (i.e. social sciences and humanities), such a positioning is possible according to \cite{morin1991methode} only by going through the construction of a ``meta-viewpoint'', i.e. through a certain reflexivity which allows the subject building knowledge to understand its position and its own approach. We give directions for the construction of such viewpoints, under the form of what we call \emph{applied perspectivism}, in Appendix~\ref{app:sec:cybergeonetworks} and also in Appendix~\ref{app:sec:csframework} for a sketch of formalization.
}{
Nous postulons qu'une science entièrement \emph{evidence-based}, quel que soit son objet, est possible et souhaitable en articulation avec la science ouverte. L'idée est de chercher à déconnecter la connaissance scientifique de tout dogmatisme, de tout a priori politique et de tout jugement de valeur\footnote{Sachant que par ailleurs ceux-ci doivent être plus que jamais développés et réfléchis pour articuler la science avec la société, mais doivent le moins possible interférer avec le processus de production de connaissance en lui-même. Suivant \cite{morin2004methode}, une éthique de la connaissance et une pensée complexe induit naturellement une éthique plus large, permettant l'autonomie de la connaissance scientifique sans la rendre inhumaine.}. Dans le cas de l'étude de sujets en lien avec des individus ou des sociétés (c'est-à-dire les sciences humaines), un tel positionnement n'est possible selon \cite{morin1991methode} que par le passage par l'établissement d'un ``méta-point de vue'', c'est-à-dire par une certaine réflexivité qui permet au connaisseur de comprendre sa position et sa propre démarche. Nous donnons des pistes pour la construction de tels points de vue, sous la forme de ce que nous appelons \emph{perspectivisme appliqué}, en Annexe~\ref{app:sec:cybergeonetworks} ainsi qu'en Annexe~\ref{app:sec:csframework} pour une piste de formalisation.
}

\bpar{
This issue is directly linked to the recurrent question of the ``qualitative-quantitative'' dichotomy, which we consider of low relevance in the context of integrative sciences. Indeed, if the dichotomy is based on a difference between objective and subjective, we recall that any knowledge is subjective, and that the ones where the role of the subject is particularly determining can be ``made objective'' by taking the meta-viewpoint, for example through the coupling with other approaches, i.e. precisely by taking an integrative position. If it is based on an issue of the nature of data, it is only partly relevant since the boundary is fuzzy: an interview text can well be used for textual analysis while a regression must be interpreted qualitatively. Indeed, we postulate that there exists different methods more or less appropriated depending on the knowledge to produce (see for example \cite{gros2017quantifier} which criticizes the use of inferential statistics for an ethnographic corpus), but that there is no ``reserved area'' of some disciplines on a given method, and that couplings and transfers will be always more necessary in the future.
}{
Cette problématique est directement reliée à la question récurrente de la dichotomie ``qualitatif-quantitatif'', que nous jugeons peu pertinente dans le cadre de sciences intégratives. En effet, si la dichotomie se base sur une différence entre objectif et subjectif, nous rappelons que toute connaissance est subjective, et que celles où le rôle du sujet est particulièrement déterminant peuvent ``s'objectiver'' par la prise du méta-point de vue, par exemple par le couplage avec d'autres approches, c'est-à-dire précisément par la prise d'une position intégrative. Si elle se base sur une question de nature des données, elle n'est que partiellement pertinente puisque la limite est floue : un texte d'interview peut très bien faire l'objet d'analyse textuelle alors qu'une régression doit être interprétée qualitativement. En fait, nous pensons qu'il existe différentes méthodes plus ou moins appropriées selon la connaissance à produire (voir par exemple \cite{gros2017quantifier} qui fustige l'utilisation de statistiques inférentielles pour un corpus ethnographique), mais qu'il n'y a pas ``chasse gardée'' de telle discipline sur telle méthode et que les couplages et transferts seront toujours plus nécessaires à l'avenir.
}

\paragraph{Quantitative epistemology}{Epistémologie quantitative}

%

\bpar{
The previous points must be treated in coordination with the use of quantitative epistemology methods allowing for an increased reflexivity, such as for example the hypernetwork method used in~\ref{sec:quantepistemo}, applied to the Cybergeo corpus in~\ref{app:sec:cybergeo}, to a patent corpus in~\ref{app:sec:patentsmining} and to our own work in~\ref{app:reflexivity}. The CybergeoNetworks platform\footnote{Available at\url{http://shiny.parisgeo.cnrs.fr/cybergeonetworks/}.} is a collaboration in that direction, presented with details in~\ref{app:sec:cybergeonetworks}. It allows in particular an increased autonomy for authors but also for open journals which can then compete with the predator publishing companies which use corpus analyses for their own profit.
}{
Les points précédents doivent être traités conjointement avec l'utilisation de méthodes d'épistémologie quantitative permettant une réflexivité accrue, comme par exemple la méthode par hyperréseau utilisée en~\ref{sec:quantepistemo}, appliquée au corpus Cybergeo en~\ref{app:sec:cybergeo}, à un corpus de brevets en~\ref{app:sec:patentsmining} et à notre propre travail en~\ref{app:reflexivity}. La plateforme CybergeoNetworks\footnote{Accessible à \url{http://shiny.parisgeo.cnrs.fr/cybergeonetworks/}.} est une collaboration dans cette direction, présentée en détails en~\ref{app:sec:cybergeonetworks}. Elle permet notamment la prise d'autonomie par les auteurs mais également par les journaux libres qui peuvent alors rivaliser avec les entreprises prédatrices d'édition qui valorisent à leur profit les analyses de corpus.
}

\subsubsection*{Modeling}{Modélisation}

\bpar{
On the domain of modeling methodology, we give precise axis which are complementary to the ones implemented by~\cite{pumain2017urban} (multi-modeling, model exploration).
}{
Sur le plan de la méthodologie de la modélisation, nous donnons des axes précis complémentaires à ceux mis en place par~\cite{pumain2017urban} (multi-modélisation, exploration des modèles).
}

\paragraph{Model couplinng}{Couplage des modèles}


\bpar{
The definition of models or approaches coupling, and in particular of the degree of coupling (strong or weak coupling) depends on the frameworks used and does not necessarily have theoretical foundations. The construction of theories providing such a definition which would furthermore be operational remains an open question. A possible approach for example uses the difference between Kolmogorov complexities of the different concerned models. A formal approach is given in~\ref{app:sec:csframework} for the coupling of perspectives. This approach is deeply linked to epistemological questions, since it could be a way to formalize the logic of the knowledge framework.
}{
La définition du couplage de modèles ou d'approches, et notamment du degré de couplage (couplage fort ou couplage faible) dépend des cadres utilisés et n'a pas forcément de fondement théorique. La construction de théories permettant une telle définition qui serait par ailleurs opérationnelle est une question ouverte. Une approche possible utilise par exemple les rapports entre complexités de Kolmogorov des différents modèles concernés. Une approche formelle est donnée en~\ref{app:sec:csframework} pour le couplage de perspectives. Cette approche est profondément liée aux questions épistémologiques, puisqu'il pourrait s'agir d'une manière de formaliser la logique du cadre de connaissance.
}

\bpar{
The question of coupling heterogenous models is naturally linked: to what extent is it relevant to choose such or such type of model and how to couple them? \cite{banos2015coupling} illustrate it for an epidemiological model, coupling a classical differential equations model to a microsimulation model. The link between agent-based models and dynamical systems can be established in certain configurations, as we have done for the Simon model and the Gibrat model in~\ref{app:sec:stochurbgrowth}, but the question of classes of problems for which links would be systematic or not remains an open question.
}{
La question du couplage de modèles hétérogènes est bien sûr liée : dans quelle mesure est-il pertinent de choisir tel ou tel type de modèle et comment les coupler ? \cite{banos2015coupling} l'illustrent pour un modèle épidémiologique, couplant un modèle classique par équations différentielles à un modèle de microsimulation. Le lien entre modèles agents et systèmes dynamiques peut être établi dans certaines configurations, comme nous l'avons fait pour le modèle de Simon et le modèle de Gibrat en~\ref{app:sec:stochurbgrowth}, mais la question de classes de problèmes pour lesquels des liens seraient systématiques ou non reste une question ouverte.
}

\bpar{
Finally, the necessity to benchmark comparable models has been highlighted for quite a long time~\cite{axtell1996aligning}, but remains not much applied: the development of tools and methods facilitating such comparisons is also an important point.
}{
Enfin, la nécessité du benchmarking de modèles comparables a été soulevée depuis un certain temps \cite{axtell1996aligning}, mais reste très peu appliquée : le développement d'outils et de méthodes facilitant de telles comparaisons est également un point important.
}

\paragraph{Constructing validation tools for simulation models}{Construire des outils de validation pour les modèles de simulation}


\bpar{
Most of the enterprise around OpenMOLE is oriented towards this aim of constructing tools and methods for the validation of models. We contribute to this effort in our work, for example in~\ref{sec:interactiongibrat} through the construction of an overfitting criteria, or in~\ref{app:sec:robustness} through the elaboration of a measure of robustness to missing data. The study of model behavior regarding overfitting, in particular in the context of multi-modeling, is a fundamental issue for the future development of these approaches.
}{
L'essentiel de l'entreprise d'OpenMole est orientée dans ce but de construction d'outils et de méthodes pour la validation des modèles. Nous contribuons à cet effort dans notre travail, par exemple en~\ref{sec:interactiongibrat} par la construction d'un critère de sur-ajustement, ou en~\ref{app:sec:robustness} par l'élaboration d'une mesure de la robustesse d'un modèle aux données manquantes. L'étude du comportement des modèles par rapport au sur-ajustement, notamment dans le cadre de la multi-modélisation, est un enjeu fondamental pour le développement futur de ces approches.
}

\subsubsection*{Foundations of spatial complex systems}{Fondements des systèmes complexes spatiaux}

\bpar{
Some fundamental questions have been suggested regarding complex systems with a spatial structure.
}{
Certaines questions fondamentales ont été suggérées au sujet des systèmes complexes ayant une structure spatiale.
}

\paragraph{Non-stationarity, non-ergodicity and path-dependancy}{Non-stationnarité, non-ergodicité et dépendance au chemin}


\bpar{
The link between spatial and/or temporal non-stationarity and non-ergodicity, which can shed light on path-dependency properties, has to the best of our knowledge not been studied systematically, at least in the context of territorial systems. We suggest that increased relationships between geosimulation, spatial statistics and economic geography would contribute to the understanding of such questions.
}{
Le lien entre non-stationnarité spatiale et/ou temporelle et non-ergodicité, pouvant éclairer les propriétés de dépendance au chemin, n'a à notre connaissance pas été étudié systématiquement, au moins dans le cadre des systèmes territoriaux. Nous suggérons qu'un lien accru entre géosimulation, statistiques spatiales et économie géographique, contribuerait à la compréhension de ce type de question.
}

\paragraph{Multi-scale models}{Modèles multi-échelle}


\bpar{
As we already largely recalled, there are very few models of territorial systems which are effectively multi-scale, and their development at relevant scales and with a reasonable degree of complexity is also an important future challenge.
}{
Comme nous l'avons déjà amplement répété, il existe très peu de modèles des systèmes territoriaux effectivement multi-échelle, et leur développement à des échelles pertinentes et à un degré de complexité raisonnable, est également un défi futur important.
}


\paragraph{Methodological standards}{Standards méthodologiques}


\bpar{
Finally, a considerable effort should be made, particularly in geography, to respect minimal methodological standards: for example use of suited time-series clustering methods~\cite{liao2005clustering}, adjustment of power laws on empirical data following the standard method of \cite{clauset2009power} and not a simple least-square regression, use of non-linear models if needed.
}{
Enfin, un effort considérable doit être fait, particulièrement en géographie, pour respecter des standards méthodologiques a minima : par exemple utilisation de classifications de séries temporelles appropriées~\cite{liao2005clustering}, ajustement de loi puissances sur des données empiriques selon la méthode standard de \cite{clauset2009power} et non une simple régression des moindres carrés, utilisation de modèles non-linéaires si besoin.
}



%

\stars



\section*{Towards a research program}{Vers un programme de recherche}

\label{sec:researchprogram}

\subsection*{For an integrated geography}{Pour une géographie intégrée}


\bpar{
As already highlighted in~\ref{sec:epistemology}, the technical and methodological evolutions that a discipline can undergo are often followed by profound epistemological mutations, even of the nature of the discipline itself. It is impossible to judge if the current state of knowledge is transitory, and if it is what would be the stable regime which would end the transition is there exists one.
}{
Comme déjà souligné en~\ref{sec:epistemology}, les bouleversements techniques et méthodologiques qu'une discipline peut subir sont souvent accompagnés de profondes mutations épistémologiques, voire de la nature même de la discipline. Il est impossible de juger si l'état actuel des connaissances est transitoire, et s'il l'est quelle est le régime stable qui terminerait la transition s'il en existe un.
}

\bpar{
Speculating is the only way to have minimal insights, knowing that it furthermore will be necessarily self-fulfilling: introduce directions or research programs conditions means and interrogations. The theoretical gaps in physics, in the case for example of linking general relativity and quantum physics, i.e. the stochastic microscopic to the deterministic macroscopic, directs the propositions for the future of the discipline which in its turn condition the concrete actions which in this field are crucial (funding of the CERN or of the spatial gravitational waves interferometer LISA).
}{
La spéculation est le seul moyen de lever partiellement le voile, sachant que celle-ci sera nécessairement auto-réalisatrice : proposer des visions ou des programmes de recherche oriente les moyens et questions. L'incomplétude théorique en physique, lorsqu'il s'agit par exemple de lier relativité générale et physique quantique, c'est-à-dire le microscopique stochastique au macroscopique déterministe, oriente les visions du futur de la discipline qui elle-même conditionnent les actions concrètes qui dans ce domaine sont indispensables (financement du CERN ou de l'interféromètre d'ondes gravitationnelles spatial LISA).
}

\bpar{
In geography, even if the technical investments can not be compared, these exist (access to computation facilities, funding of integrated laboratories, etc.) and are also determined by the perspectives for the discipline. We propose here directions and a manifest for a new geography, which is already emerging and which basis are progressively constructed in a solid way. The adventure of the Geodivercity ERC project \cite{pumain2017urban} is an allegory of it, since it furthermore confirmed most of directions proposed by~\cite{banos2017knowledge}. The integration of theory, of the empirical, of modeling, but also of technics and of methodology, has never been so much investigated and reinforced than in the diverse developments of the project. Without the access to the computation grid and the new explorarion algorithms allowed by OpenMOLE, knowledge extracted from the SimpopLocal model would have been significantly less, but technical developments have also been conducted by thematic questions.
}{
En géographie, même si les investissements techniques sont incomparables, ceux-ci existent (accès aux moyens de calcul, financement de laboratoires intégrés, etc.) et sont déterminés également par les perspectives pour la discipline. Nous proposons ici une vision et un manifeste d'une nouvelle géographie, qui est déjà en train de se faire et dont les bases sont solidement construites petit à petit. L'aventure de l'ERC Geodivercity \cite{pumain2017urban} en est l'allégorie, d'autant plus qu'elle a confirmé la plupart des directions proposées par~\cite{banos2017knowledge}. L'intégration de la théorie, de l'empirique, de la modélisation, mais aussi de la technique et de la méthode, n'a jamais été aussi creusée et renforcée que dans les divers développements du projet. Sans l'accès à la grille de calcul et aux nouveaux algorithmes d'exploration permis par OpenMole, les connaissances tirées du modèle SimpopLocal auraient été moindres, mais les développements techniques ont aussi été conduits par la demande thématique.
}

\bpar{
We insist on the fact that the knowledge framework introduced in~\ref{sec:knowledgeframework} is particularly suitable to be applied as a contemporary continuity of theoretical and quantitative geography. This framework indeed fulfils the following constraints: (i) reach beyond the artificial boundaries between quantitative and qualitative; (ii) do not favor any particular component among the knowledge production means (as diverse as all the classical quantitative and qualitative methods, modeling methods, theoretical approaches, data, tools), but indeed the joint development of each component.
}{
Nous appuyons le fait que le cadre de connaissance proposé en~\ref{sec:knowledgeframework} est particulièrement favorable à une application à la continuité contemporaine de la géographie théorique et quantitative. Ce cadre permet en effet de répondre aux contraintes suivantes : (i) transcender les frontières artificielles entre quantitatif et qualitatif ; (ii) ne pas favoriser de composante particulière parmi les moyens de production de connaissance (aussi divers que l'ensemble des méthodes qualitatives et quantitatives classiques, les méthodes de modélisation, les approches théoriques, les données, les outils), mais bien le développement conjoint de chaque composante.
}

\bpar{
We recall that this framework extends the one proposed by~\cite{livet2010}, which describes the three domains as empirical, conceptual, and modeling, but we add the self-consistent domains which are methods, tools (which can be understood as proto-methods) and data. Any knowledge production approach, seen as a \emph{perspective} in the sense of~\cite{giere2010scientific}, is then a complex combination of the six domains, the knowledge fronts being within each co-evolving.
}{
Nous rappelons que le cadre étend celui de~\cite{livet2010}, qui consacre le triptyque des domaines empiriques, conceptuels et de la modélisation, en y ajoutant les domaines à part entière que sont les méthodes, les outils (qu'on peut voir comme des proto-méthodes) et les données. Toute démarche de production de connaissance, vue comme une \emph{perspective} au sens de~\cite{giere2010scientific}, est alors une combinaison complexe des six domaines, les fronts de connaissance dans chacun étant en co-évolution.
}

\bpar{
We postulate that the application of our knowledge framework coincides with the emergence of an \emph{integrated geography}, that we call so to highlight both the integration of the different domains but also of the quantitative and qualitative knowledge, since both are grounded within all of the domains.
}{
Nous postulons que l'application de notre cadre de connaissance est de mise avec l'émergence d'une \emph{géographie intégrée}, que nous nommons ainsi pour souligner à la fois l'intégration des différents domaines mais aussi des connaissances qualitatives et quantitatives, puisque les deux se fondent dans chacun des domaines.
}


\subsection*{Research project}{Projet de recherche}

\bpar{
We finally detail a long term research project which (i) comes as a continuity of this monograph; (ii) enters within the frame of an integrated geography, and more generally of a vertical and horizontal integration, but also of domains and types of knowledge; (iii) aims at tackling a broad range of open question previously described; and (iv) is intrinsically reflexive and complex.
}{
Nous détaillons finalement un projet de recherche à long terme qui (i) s'inscrit dans la continuité de cette monographie ; (ii) s'inscrit dans le cadre d'une géographie intégrée, et plus généralement d'une intégration verticale et horizontale, mais aussi des domaines et des types de connaissance ; (iii) s'attaque à un certain nombre de questions ouvertes mentionnées ci-dessus ; et (iv) est intrinsèquement réflexif et complexe.
}


\bpar{
The multi-scale coupling of urban models suggested in~\ref{sec:contributions} can indeed be projected within a more global problematic. The idea would be to tackle a multi-scale geographical problem, that is to understand how and when interdependencies between cities have built regional systems of cities and to identify the most probable scenario of their potential coalescence as a consequence of globalisation processes. These abstract questions have direct practical implications for measuring global and local inequalities and managing urban growth.
}{
Le couplage multi-échelle des modèles urbains suggéré en~\ref{sec:contributions} peut en fait se projeter dans une problématique plus globale. L'idée serait d'aborder un problème géographique multi-échelle, qui est de comprendre comment et quand les interdépendances entre les villes ont conduit à l'émergence de systèmes régionaux de villes, et d'identifier le scenario le plus probable de leur fusion potentielle comme conséquence des processus de globalisation. Ces questions abstraites ont des implications directes pour la mesure des inégalités globales et locales et la gestion de la croissance urbaine.
}

\bpar{
This question finds roots in the multi-scalar nature of territorial systems. Converging evidence suggest the relative independent historical development of regional urban systems across the world, and an increased interdependency between these in the processes of globalisation. Can we already quantify these at different scales ? How does the coupling and the opening of subsystems operate, and what are its most plausible consequences, from convergence of dynamics to an increase of inter- and intra-subsystems inequalities ?
}{
Cette question trouve sa source dans la nature multi-échelle des systèmes territoriaux. Des résultats convergents suggèrent une certaine indépendance du développement historique des systèmes urbains régionaux tout autour du monde, et une interdépendance accrue entre ceux-ci dans les processus de globalisation. Est-il possible de déjà les identifier à différentes échelles ? De quelle manière le couplage et l'ouverture des sous-systèmes s'opère, et quelles sont ses conséquences les plus plausibles, de la convergence des dynamiques à l'augmentation des inégalités inter et intra système ?
}

\bpar{
We postulate that a powerful entry to this research question is the construction of bridges between geographical theories of territorial systems in the spirit of the Evolutive Urban Theory~\cite{pumain1997pour} and Scaling Theories of Cities~\cite{west2017scale}. The first emphasize particularities of territorial entities whereas the second focuses on universal laws, and both provide credible explanations for scaling laws. A strategy to answer the question and combining both would consist in: (i) finding empirical modular decompositions of territorial systems and corresponding scales, and quantifying their universality through inter and intra scaling; (ii) modeling this multi-scalar system by coupling models of urban growth, that would be validated through scaling properties. The models developed here are good candidates as sub-models, since co-evolution inside and between scales is a characteristic feature of complex urban systems, as we showed.
}{
Nous postulons qu'une entrée puissante pour cette question de recherche est la construction de ponts entre les théories géographiques des systèmes territoriaux dans l'esprit de la théorie évolutive des villes~\cite{pumain1997pour} et de la théorie du \emph{Scaling}~\cite{west2017scale}. La première appuie les particularités des entités territoriales tandis que la seconde se concentre sur des lois universelles, et chacune fournit des explications crédibles aux lois d'échelles urbaines. Une stratégie pour répondre à la question tout en combinant les deux théories consisterait en : (i) la mise en évidence empirique de décompositions modulaires des systèmes territoriaux et des échelles correspondantes, et la quantification de leur universalité par le scaling intra- et inter-système ; (ii) la modélisation de ce système multi-échelle par le couplage de modèles de croissance urbaine, qui seraient validés par les propriétés de scaling. Les modèles que nous avons développé ici sont de bons candidats comme sous-modèles, puisque la co-évolution au sein et entre les échelles est une caractéristiques des systèmes urbains complexes, comme nous l'avons montré.
}

\bpar{
Two underlying research axis appear then as necessary for the global consistence of the project. The first consists in the exploration of potential relations between territorial systems and artificial intelligence. It comes as a corollary and is informative for the main question, for at least two very different reasons. The first is rather practical and linked to the emergence of ubiquitous information and computing in cities, that can be understood as taking part in the emergence of ``smart cities''  \cite{batty2018artificial}: the new large datasets available have been proven to be a powerful analysis tool as witness the numerous recent works by physicists on cities for example, and these new urban behavior may probably induce some regime changes partly because of of their self-fulfilling nature. The second is more difficult to grasp: the importance we gave to morphogenesis and the possible application of this concept at different levels such as knowledge production. Morphogenesis can be used to conceptualize both the evolution of territories and of ideas: to what extent the emergence of territories contains an endogenous intelligence? The use of slime mould network generation in~\ref{sec:networkgrowth}, which have been shown otherwise to be powerful computation tools, is an other clue of a possible connexion. We also refer to the contribution by \noun{White} synthesized in~\ref{app:sec:ecogeo} which suggests the construction of intelligent and autonomous models as the future of territorial systems modeling.
}{
Deux axes de recherches sous-jacents apparaissent alors comme nécessaires à la cohérence globale du projet. Le premier consiste en l'exploration des relations potentielles entre les systèmes territoriaux et l'intelligence artificielle. Celui-ci vient comme corollaire et enrichit la question principale pour au moins deux raisons très différentes. La première est plutôt pratique et liée à l'émergence de l'information et du calcul omniprésents dans les villes, qui peuvent être compris comme participant à l'émergence des \emph{smart cities} \cite{batty2018artificial} : les jeux de données massifs nouvellement disponibles se sont révélés être des outils efficaces comme en témoigne la quantité de travaux récents des physiciens sur la ville par exemple, et ces nouveaux comportements urbains peuvent probablement induire des changement de régime en partie à cause de leur nature auto-réalisatrice. La seconde est plus difficile à saisir : l'importance que nous avons donné à la morphogenèse et la possibilité d'appliquer ce concept à différents niveaux, notamment celui de la production de connaissance. La morphogenèse peut être utilisée pour conceptualiser à la fois l'évolution des territoires et des idées : dans quelle mesure l'émergence des territoires contient-elle une intelligence endogène ? L'utilisation du modèle de slime mould en~\ref{sec:networkgrowth}, qui est par ailleurs doté de capacités de calcul, et un autre indice d'une connexion possible. Nous notons également la contribution de \noun{White} rapportée en~\ref{app:sec:ecogeo} qui suggère la construction de modèles intelligents et autonomes comme le futur de la modélisation des systèmes territoriaux.
} 

\bpar{
A second auxiliary subject is the theoretical and applied study of knowledge production on complex systems. This axis is necessary, first to continue to enhance the reflexivity and interdisciplinarity through the further development of quantitative epistemology methods and tools such as more elaborated text-mining and meta-analysis tools, and secondly precisely because of reflexivity as concrete case studies such as cultural evolution (which can be studied through the intermediate of technological innovation in the continuity of~\ref{app:sec:patentsmining}, or of language evolution on which a collaboration is currently ongoing) apply to territorial systems which main components are cognitive agents.
}{
Un deuxième axe complémentaire est l'étude théorique et appliquée de la production de connaissance sur les systèmes complexes. Cet axe est nécessaire, d'une part pour continuer à favoriser la réflexivité et l'interdisciplinarité par le développement des méthodes et outils d'épistémologie quantitative comme des outils plus élaborés de fouille de textes complets et de méta-analyse, et dans un second temps, précisément grâce à la réflexivité, parce que des cas d'étude potentiels comme l'évolution culturelle (pouvant être étudiée par l'intermédiaire de l'innovation technologique dans la continuité de~\ref{app:sec:patentsmining}, ou de l'évolution du langage sur laquelle une collaboration est en cours) s'appliquent aux systèmes territoriaux dont les principales composantes sont des agents cognitifs.
}

\bpar{
The strongly coupled elaboration of these three different components, i.e. their co-evolution, in the exact spirit of what has been achieved until now, is necessary for the integrated nature of the project and achieve its objective to produce integrative theories of territorial systems.
}{
L'élaboration fortement couplée de ces trois différentes composantes, c'est-à-dire leur co-évolution, dans l'esprit de tout ce qui a été accompli jusque là, est nécessaire pour le caractère intégré du projet et l'atteinte de son objectif de production de théories intégratives des systèmes territoriaux.
}

\stars



\vfill

\stars

{\raggedleft
\textit{
Cet hiver a des airs de printemps\\
Des peuples ou de l'esprit, au diable l'âme.\\
Le vent se lève, ça faisait longtemps\\
Triste de s'enfermer pour quelques grammes.\\
}
}
\medskip

{\raggedright

\textit{
Cet avenir des airs de passé\\
S'il fallait juste trouver le régime,\\
Assassinée la complexité\\
Maintes perspectives se cachent en les crimes.\\
}
}

\medskip
{\raggedleft

\textit{
Pour une morphogenèse politique\\
Adieu le coron, ses tristes briques\\
Murs qui s'érigent tuent votre espérance.\\
}
}

\medskip
{\raggedright

\textit{
Perle de la mer, sirène hante la crique\\
Du haut des tours s'amuser du cirque\\
L'hiver d'idées qui peuple la France.\\
}
}

\stars

\vfill


\newpage

\bpar{
\chapter*{Conclusion}
}{
\chapter*{Conclusion}
}

\markboth{Conclusion}{Conclusion}

\bpar{
\headercit{Keeping on exploring geographical systems\ldots}{Arnaud Banos}{}
}{
\headercit{Explorer sans relâche les systèmes géographiques\ldots}{Arnaud Banos}{}
}


\bpar{
Our thesis is a complex system which exhibits an auxiliary deterministic finality: this conclusion by a citation of \noun{Banos}. Principles of its context, simple but efficient and deep, indeed flow through this work: the ``9 principles of Banos'' are implicitly present in most of the work done and of the perspectives opened. Even if an ideal application of these principles would be achieved by a ``Banos Deamon'', similar to the Laplace or Maxwell Deamon, which would be able to articulate interdisciplinary and disciplinary without being lost while respecting all the principles, their understanding as a scientific utopia, naturally reflexive and thus evolutive and adaptive, seems a powerful entry for new integrative approaches of territorial systems.
}{
Notre thèse est un système complexe qui exhibe une finalité auxiliaire déterministe : cette conclusion par un adage de \noun{Banos}. Les principes de son contexte, simples mais efficaces et profonds, traversent en effet ce travail : les ``9 principes de Banos'' sont implicitement présents dans la majorité des travaux menés et perspectives ouvertes. Même si une application idéale de ces principes relèverait d'un ``Démon de Banos'', à l'instar du Démon de Laplace ou de Maxwell, qui serait capable d'articuler interdisciplinaire et disciplinaire sans se perdre tout en respectant l'ensemble des principes, leur appréhension comme utopie scientifique, naturellement réflexive donc évolutive et adaptive, nous semble une entrée puissante pour de nouvelles approches intégratives des systèmes territoriaux. 
}

\bpar{
Our epistemological and methodological contribution in relation with these points is essential, even if it remains difficult to explicit and will necessitate a certain step back to be effectively understood. In a way, we brought an additional brick as a \emph{proof-of-concept} of the banosian system of principles, but also as an implementation and a deepening of it on some points. We have shown that their application is far from being simple, and that the risk of fall into reductionist is never far around the corner despite these fundamentally complex principles. The tenth commandment would it then be: \textit{aiming at applying these principles while always keeping in sight complexity and the role of reflexivity}?
}{
Notre contribution épistémologique, méthodologique en lien avec ces points est essentielle, même si celle ci est difficile à expliciter et nécessitera un certain recul pour être effectivement cernée. D'une certaine manière, nous avons apporté une brique supplémentaire comme \emph{proof-of-concept} du système de principes banosien, mais également comme implémentation et approfondissement de celui-ci sur certains points. Nous avons montré que leur application est loin d'être simple, et que toujours guette le risque de sombrer dans le réductionnisme malgré ces principes fondamentalement complexes. Le dixième commandement serait-il alors : \textit{S'efforcer à appliquer ces principes en ne perdant jamais de vue la complexité et le rôle de la réflexivité} ?
}

\bpar{
Our thematical contribution is not necessarily easy to situate and will necessitate a considerable step back to highlight its implications. Did we solve the gordian knot of co-evolution? Did we cut it? The most honest answer would be that we cut a part of it, the naive one including the definition from which we started in introduction or positioning of type ``chicken-and-egg'' typical of debates on structuring effects, but that we tied an other much more considerable, by unveiling the complexity of this concept and its manifestations.
}{
Notre contribution thématique n'est pas forcément facile à situer et nécessitera un recul considérable pour appréhender ses implications. Avons-nous résolu le noeud gordien de la co-évolution ? L'avons-nous tranché ? La réponse la plus fidèle serait que nous en avons tranché une partie, celle naïve comprenant la définition dont nous sommes parti en introduction ou les positionnement de type ``poule-et-oeuf'' typique des débats des effets structurants, mais que nous avons noué un autre bien plus considérable, en révélant la complexité de ce concept et de ses manifestations.
}

\bpar{
Coming back to our funding problematic, we recall that (i) we gave a definition of co-evolution proper to territorial systems and also an operational method of caracterization; (ii) we explored modeling directions at different scales, which articulate with a more global theoretical frame. Answer to this problematic furthermore allowed us to progressively build a broader frame and vast research perspectives.
}{
Revenant à notre problématique fondatrice, nous rappelons que (i) nous avons donné une définition de la co-évolution propre aux systèmes territoriaux ainsi qu'une méthode opérationnelle de caractérisation ; (ii) nous avons exploré des pistes de modélisation à différentes échelles, qui s'accordent avec un cadre théorique global. Répondre à cette problématique nous a permis par ailleurs de progressivement dégager un cadre plus large et de vastes perspectives de recherche.
}

\bpar{
Our modest mission is accomplished, and a fantastic journey only begins. The provisory accomplishment becomes the foundations of the ones to come. Cumulativity of knowledge is not improvised, and we suggest that the complex fabric of which we sew the first stitches will be robust enough to be inserted within. \textit{The road is long but the path is free.}
}{
Notre modeste mission est accomplie, et un fantastique voyage commence tout juste. L'accomplissement passager devient les fondations de ceux à venir. La cumulativité des connaissances ne s'improvise pas, et nous espérons que le tissu complexe dont nous avons cousu les premières mailles sera assez robuste pour s'y insérer. \textit{la route est longue mais la voie est libre.}
}

\stars






\begin{figure}
	\includegraphics[width=\textheight,angle=90]{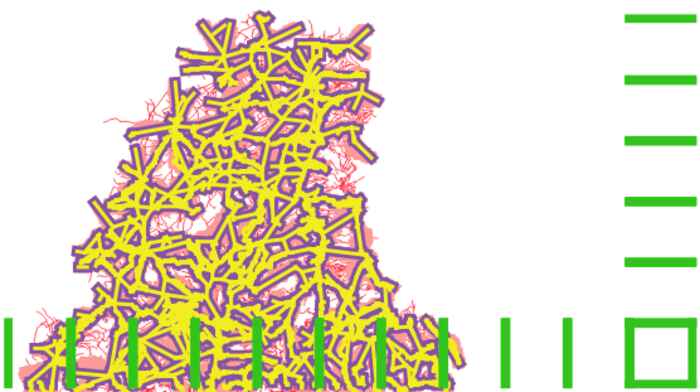}
\end{figure}



\cleardoublepage


\label{app:bibliography} 

\manualmark 
\markboth{\spacedlowsmallcaps{\bibname}}{\spacedlowsmallcaps{\bibname}} 
\phantomsection
\refstepcounter{dummy}

\addtocontents{toc}{\protect\vspace{\beforebibskip}} 
\addcontentsline{toc}{chapter}{\tocEntry{\bibname}}

\printbibliography

@article{guerois2009dynamique,
  title={La dynamique spatio-temporelle des prix immobiliers {\`a} diff{\'e}rentes {\'e}chelles: le cas des appartements anciens {\`a} Paris (1990-2003)},
  author={Gu{\'e}rois, Marianne and Le Goix, Renaud},
  journal={Cybergeo: European Journal of Geography},
  year={2009},
  publisher={CNRS-UMR G{\'e}ographie-cit{\'e}s 8504}
}

@article{pumain2010theorie,
  title={Une th{\'e}orie g{\'e}ographique des villes},
  author={Pumain, Denise},
  journal={Bulletin de la Soci{\'e}t{\'e} g{\'e}ographie de Li{\`e}ge},
  number={55},
  pages={5--15},
  year={2010}
}

@article{10.1371/journal.pone.0176310,
    author = {Bergeaud, Antonin AND Potiron, Yoann AND Raimbault, Juste},
    journal = {PLOS ONE},
    publisher = {Public Library of Science},
    title = {Classifying patents based on their semantic content},
    year = {2017},
    month = {04},
    volume = {12},
    url = {https://doi.org/10.1371/journal.pone.0176310},
    pages = {1-22},
    number = {4},
    doi = {10.1371/journal.pone.0176310}
}

@article{brunsdon1998geographically,
  title={Geographically weighted regression},
  author={Brunsdon, Chris and Fotheringham, Stewart and Charlton, Martin},
  journal={Journal of the Royal Statistical Society: Series D (The Statistician)},
  volume={47},
  number={3},
  pages={431--443},
  year={1998},
  publisher={Wiley Online Library}
}

@article{pumain2003approche,
  title={Une approche de la complexit{\'e} en g{\'e}ographie},
  author={Pumain, Denise},
  journal={Geocarrefour},
  volume={78},
  number={1},
  pages={25--31},
  year={2003},
  publisher={Association des amis de la Revue de G{\'e}ographie de Lyon}
}

@article{audretsch1996r,
  title={R\&D spillovers and the geography of innovation and production},
  author={Audretsch, David B and Feldman, Maryann P},
  journal={The American economic review},
  volume={86},
  number={3},
  pages={630--640},
  year={1996},
  publisher={JSTOR}
}

@article{cerqueira2017inegalites,
  title={Les in{\'e}galit{\'e}s d’acc{\`e}s aux ressources urbaines dans les franges p{\'e}riph{\'e}riques de Belo Horizonte (Br{\'e}sil): quelles {\'e}volutions?},
  author={Cerqueira, Eug{\^e}nia Viana},
  journal={EchoG{\'e}o},
  number={39},
  year={2017},
  publisher={P{\^o}le de recherche pour l'organisation et la diffusion de l'information g{\'e}ographique (CNRS UMR 8586)}
}

@incollection{durand2003geographes,
  TITLE = {{Les g{\'e}ographes et la notion de causalit{\'e}}},
  AUTHOR = {Durand-Dast{\`e}s, Fran{\c c}ois},
  URL = {https://halshs.archives-ouvertes.fr/halshs-00241841},
  BOOKTITLE = {{Enqu{\^e}te sur la notion de causalit{\'e}}},
  PUBLISHER = {{PUF}},
  PAGES = {145-160},
  YEAR = {2003},
  HAL_ID = {halshs-00241841},
  HAL_VERSION = {v1},
}

@article{claval1985causalite,
  title={Causalit{\'e} et g{\'e}ographie},
  author={Claval, Paul},
  journal={Espace g{\'e}ographique},
  volume={14},
  number={2},
  pages={109--115},
  year={1985},
  publisher={Pers{\'e}e-Portail des revues scientifiques en SHS}
}

@article{loi1985etude,
  title={Une {\'e}tude de la causalit{\'e} dans la g{\'e}ographie classique fran{\c{c}}aise.[L'exemple des premi{\`e}res th{\`e}ses r{\'e}gionales]},
  author={Loi, Daniel},
  journal={Espace g{\'e}ographique},
  volume={14},
  number={2},
  pages={121--125},
  year={1985},
  publisher={Pers{\'e}e-Portail des revues scientifiques en SHS}
}

@article{desjardins2010bataille,
  title={la bataille du Grand Paris},
  author={Desjardins, Xavier},
  journal={L'Information g{\'e}ographique},
  volume={74},
  number={4},
  pages={29--46},
  year={2010},
  publisher={Armand Colin}
}

@article{beaucire2013grand,
  title={{\guillemotleft}Grand Paris Express{\guillemotright}: un projet au service de la r{\'e}duction des in{\'e}galit{\'e}s d'accessibilit{\'e} entre l'Ouest et l'Est de la r{\'e}gion urbaine de Paris?},
  author={Beaucire, Francis and Drevelle, Matthieu},
  journal={Revue d’{\'E}conomie R{\'e}gionale \& Urbaine},
  number={3},
  pages={437--460},
  year={2013},
  publisher={Armand Colin}
}

@misc{sdrif2013,
  title = {Île-de-France 2030. Orientations réglementaires et carte de destination générale des différentes parties du territoire.},
  author = {SDRIF},
  year = {2013}
}

@misc{stif2007arc,
  title = {ArcExpress, débat public sur le métro de rocade. Dossier du Maitre d'Ouvrage.},
  author = {STIF},
  year={2010}
}

@inproceedings{hamerly2003learning,
  title={Learning the k in k-means},
  author={Hamerly, Greg and Elkan, Charles and others},
  booktitle={NIPS},
  volume={3},
  pages={281--288},
  year={2003}
}

@article{grimm2005pattern,
	Author = {Grimm, Volker and Revilla, Eloy and Berger, Uta and Jeltsch, Florian and Mooij, Wolf M and Railsback, Steven F and Thulke, Hans-Hermann and Weiner, Jacob and Wiegand, Thorsten and DeAngelis, Donald L},
	Journal = {science},
	Number = {5750},
	Pages = {987--991},
	Publisher = {American Association for the Advancement of Science},
	Title = {Pattern-oriented modeling of agent-based complex systems: lessons from ecology},
	Volume = {310},
	Year = {2005}}

@incollection{reynard2015application,
	Author = {Reynard, Emmanuel and Kaiser, Christian and Martin, Simon and Regolini, G{\'e}raldine},
	Booktitle = {Engineering Geology for Society and Territory-Volume 8},
	Pages = {265--268},
	Publisher = {Springer},
	Title = {An Application for Geosciences Communication by Smartphones and Tablets},
	Year = {2015}}

@article{morris2013gaming,
	Author = {Morris, Bradley J and Croker, Steve and Zimmerman, Corinne and Gill, Devin and Romig, Connie},
	Journal = {Frontiers in psychology},
	Publisher = {Frontiers Media SA},
	Title = {Gaming science: the ``Gamification'' of scientific thinking},
	Volume = {4},
	Year = {2013}}

@article{cain2015serious,
	Author = {Cain, Jeff and Piascik, Peggy},
	Journal = {American journal of pharmaceutical education},
	Number = {4},
	Publisher = {American Association of Colleges of Pharmacy},
	Title = {Are Serious Games a Good Strategy for Pharmacy Education?},
	Volume = {79},
	Year = {2015}}

@article{geman_stochastic_1984,
	title = {Stochastic {Relaxation}, {Gibbs} {Distributions}, and the {Bayesian} {Restoration} of {Images}},
	volume = {6},
	issn = {0162-8828},
	url = {http://dx.doi.org/10.1109/TPAMI.1984.4767596},
	doi = {10.1109/TPAMI.1984.4767596},
	number = {6},
	urldate = {2017-12-19},
	journal = {IEEE Trans. Pattern Anal. Mach. Intell.},
	author = {Geman, Stuart and Geman, Donald},
	month = nov,
	year = {1984},
	pages = {721--741}
}

@inproceedings{hofmann1999probabilistic,
  title={Probabilistic latent semantic indexing},
  author={Hofmann, Thomas},
  booktitle={Proceedings of the 22nd annual international ACM SIGIR conference on Research and development in information retrieval},
  pages={50--57},
  year={1999},
  organization={ACM}
}

@book{salton_introduction_1986,
	title = {Introduction to {Modern} {Information} {Retrieval}},
	isbn = {978-0-07-054484-0},
	publisher = {McGraw-Hill, Inc.},
	author = {Salton, Gerard and McGill, Michael J.},
	year = {1986}
}

@article{bouveyron2016stochastic,
  title={The stochastic topic block model for the clustering of vertices in networks with textual edges},
  author={Bouveyron, Charles and Latouche, Pierre and Zreik, Rawya},
  journal={Statistics and Computing},
  pages={1--21},
  year={2016},
  publisher={Springer}
}

@article{maisonobe2013diffusion,
  title={Diffusion et structuration spatiale d'une question de recherche en biologie mol{\'e}culaire},
  author={Maisonobe, Marion},
  journal={Mappe Monde},
  volume={110},
  number={2},
  pages={13202},
  year={2013}
}

@incollection{fecher2014open,
	Author = {Fecher, Benedikt and Friesike, Sascha},
	Booktitle = {Opening science},
	Pages = {17--47},
	Publisher = {Springer},
	Title = {Open science: one term, five schools of thought},
	Year = {2014}}

@article{nicosia2009extending,
	Author = {Nicosia, Vincenzo and Mangioni, Giuseppe and Carchiolo, Vincenza and Malgeri, Michele},
	Journal = {Journal of Statistical Mechanics: Theory and Experiment},
	Number = {03},
	Pages = {P03024},
	Title = {Extending the definition of modularity to directed graphs with overlapping communities},
	Volume = {2009},
	Year = {2009}}

@article{wilson2017good,
  title={Good enough practices in scientific computing},
  author={Wilson, G and Bryan, J and Cranston, K and Kitzes, J and
Nederbragt, L and Teal, TK},
  journal={PLoS Comput Biol},
  volume={13},
  number={6},
  pages={e1005510},
  year={2017},
  publisher={Public Library of Science}
}

@Article{10.12688/f1000research.11369.1,
AUTHOR = { Ross-Hellauer, T},
TITLE = {What is open peer review? A systematic review [version 1; referees: 1 approved, 2 approved with reservations]
},
JOURNAL = {F1000Research},
VOLUME = {6},
YEAR = {2017},
NUMBER = {588},
DOI = {10.12688/f1000research.11369.1}
}

@article{chavalarias2013phylomemetic,
  title={Phylomemetic patterns in science evolution—the rise and fall of scientific fields},
  author={Chavalarias, David and Cointet, Jean-Philippe},
  journal={PloS one},
  volume={8},
  number={2},
  pages={e54847},
  year={2013},
  publisher={Public Library of Science}
}

@book{cronin2014beyond,
  title={Beyond bibliometrics: Harnessing multidimensional indicators of scholarly impact},
  author={Cronin, Blaise and Sugimoto, Cassidy R},
  year={2014},
  publisher={MIT Press}
}

@book{jacobs2016death,
  title={The death and life of great American cities},
  author={Jacobs, Jane},
  year={2016},
  publisher={Vintage}
}

@article{shibata2008detecting,
	Author = {Shibata, Naoki and Kajikawa, Yuya and Takeda, Yoshiyuki and Matsushima, Katsumori},
	Journal = {Technovation},
	Number = {11},
	Pages = {758--775},
	Title = {Detecting emerging research fronts based on topological measures in citation networks of scientific publications},
	Volume = {28},
	Year = {2008}}

@article{querriau2004localisation,
  title={Localisation optimale d’unit{\'e}s de soins dans un pays en voie de d{\'e}veloppement: analyse de sensibilit{\'e}},
  author={Querriau, Xavier and Kissiyar, Mohamed and Peeters, Dominique and Thomas, Isabelle},
  journal={Cybergeo: European Journal of Geography},
  year={2004},
  publisher={CNRS-UMR G{\'e}ographie-cit{\'e}s 8504}
}

@article{vallee2009disparites,
  title={Les disparit{\'e}s spatiales de sant{\'e} en ville: l’exemple de Vientiane (Laos)},
  author={Vall{\'e}e, Julie},
  journal={Cybergeo: European Journal of Geography},
  year={2009},
  publisher={CNRS-UMR G{\'e}ographie-cit{\'e}s 8504}
}

@article{belizal2011quand,
  title={Quand l’al{\'e}a devient la ressource: l’activit{\'e} d’extraction des mat{\'e}riaux volcaniques autour du volcan Merapi (Indon{\'e}sie) dans la compr{\'e}hension des risques locaux},
  author={B{\'e}lizal, {\'E}douard de and Lavigne, Franck and Grancher, Delphine},
  journal={Cybergeo: European Journal of Geography},
  year={2011},
  publisher={CNRS-UMR G{\'e}ographie-cit{\'e}s 8504}
}

@article{putra2009einfluss,
  title={Der Einfluss ungesteuerter Urbanisierung auf die Grundwasserressourcen am Beispiel der indonesischen Millionenstadt Yogyakarta},
  author={Putra, Doni PE and Baier, Klaus},
  journal={Cybergeo: European Journal of Geography},
  year={2009},
  publisher={CNRS-UMR G{\'e}ographie-cit{\'e}s 8504}
}

@article{ackermann2003analysis,
  title={Analysis of built-up areas extension on the Petite C{\^o}te region (Senegal) by remote sensing},
  author={Ackermann, Gabriela and Mering, Catherine and Quensiere, Jacques},
  journal={Cybergeo: European Journal of Geography},
  volume={9},
  number={249},
  year={2003},
  publisher={CNRS-UMR G{\'e}ographie-cit{\'e}s 8504}
}

@article{devaux2007extraction,
  title={Extraction automatique d’habitations en milieu rural de PED {\`a} partir de donn{\'e}es THRS},
  author={Devaux, Nicolas and Fotsing, Jean-Marie and Ch{\'e}ry, Jean-Pierre},
  journal={Cybergeo: European Journal of Geography},
  year={2007},
  publisher={CNRS-UMR G{\'e}ographie-cit{\'e}s 8504}
}

@article{jelokhani2012web,
  title={A web 3.0-driven collaborative multicriteria spatial decision support system},
  author={Jelokhani-Niaraki, Mohammadreza and Malczewski, Jacek},
  journal={Cybergeo: European Journal of Geography},
  year={2012},
  publisher={CNRS-UMR G{\'e}ographie-cit{\'e}s 8504}
}

@article{escobar2000distribution,
  title={Distribution of Online Cartographic Products in Australia},
  author={Escobar, Francisco J and Polley, Iestyn S and Williamson, Ian P},
  journal={Cybergeo: European Journal of Geography},
  year={2000},
  publisher={CNRS-UMR G{\'e}ographie-cit{\'e}s 8504}
}

@article{santamaria2009schema,
  title={Le Sch{\'e}ma de d{\'e}veloppement de l’espace communautaire (SDEC): application d{\'e}faillante ou {\'e}laboration probl{\'e}matique?},
  author={Santamaria, Fr{\'e}d{\'e}ric},
  journal={Cybergeo: European journal of geography},
  year={2009},
  publisher={CNRS-UMR G{\'e}ographie-cit{\'e}s 8504}
}

@article{le2011consommation,
  title={Consommation d’{\'e}nergie et mobilit{\'e} quotidienne selon la configuration des densit{\'e}s dans 34 villes europ{\'e}ennes.},
  author={Le N{\'e}chet, Florent},
  journal={Cybergeo: European Journal of Geography},
  year={2011},
  publisher={CNRS-UMR G{\'e}ographie-cit{\'e}s 8504}
}

@article{lusso2009musees,
  title={Les mus{\'e}es, un outil efficace de r{\'e}g{\'e}n{\'e}ration urbaine? Les exemples de Mons (Belgique), Essen (Allemagne) et Manchester (Royaume-Uni)},
  author={Lusso, Bruno},
  journal={Cybergeo: European Journal of Geography},
  year={2009},
  publisher={CNRS-UMR G{\'e}ographie-cit{\'e}s 8504}
}

@article{ozer2005tsunami,
  title={Tsunami en Asie du Sud-Est: retour sur la gestion d’un cataclysme naturel apocalyptique},
  author={Ozer, Pierre and De Longueville, Florence},
  journal={Cybergeo: European Journal of Geography},
  year={2005},
  publisher={CNRS-UMR G{\'e}ographie-cit{\'e}s 8504}
}

@inproceedings{raimbault2016indirect,
  title={Indirect Bibliometrics by Complex Network Analysis},
  author={Raimbault, Juste},
  booktitle={20e Anniversaire de Cybergeo},
  year={2016}
}

@article{livingston_spaces_1995,
  title = {The spaces of knowledge: contributions toward a historical geography of science},
  journal = {{{Environment and planning D}}},
  volume = {13},
  issue = {1},
  pages = {13-42},
  year = {1995},
  author = {Livingstone, David N}
}

@book{livingston_science_2003,
	title = {Puttings science in its place: geographies of scientific knowledge},
	publisher = {The University of Chicago Press},
	author = {Livingstone, David N.},
	year = {2003}
}

@article{withers_place_2009,
  title = {Place and the spatial turn in geography and history},
  journal = {Journal of the History of Ideas},
  volume = {70},
  issue = {4},
  pages = {637-658},
  year = {2009},
  author = {Withers, Charles W. J.}
}

@book{livingston_geography_2005,
	title = {Geography and revolution},
	publisher = {The University of Chicago Press},
	editor = {Livingstone, David N. and Withers, Charles W. J.},
	year = {2005}
}

@Article{blondel_fast_2008,
  Title                    = {Fast unfolding of communities in large networks},
  Author                   = {Blondel, Vincent D. and Guillaume, Jean-Louis and Lambiotte, Renaud and Lefebvre. Emmanuel},
  Journal                  = {Journal of Statistical Mechanics: Theory and Experiment},
  Year                     = {2008},
  Number                   = {10},
  Pages                    = {10008}
}

@article{dong2010policing,
    author = {Han, Dong},
    title = {Policing and racialization of rural migrant workers in Chinese cities},
    journal = {Ethnic and Racial Studies},
    volume = {33},
    number={1},
    publisher = {Informa UK Limited},
    issn={1466-4356},
    pages={593-610},
    year={2010}
}

@article{IJUR:IJUR820,
author = {Xu, Xue-qiang and Li, Si-ming},
title = {China's open door policy and urbanization in the Pearl River Delta region},
journal = {International Journal of Urban and Regional Research},
volume = {14},
number = {1},
publisher = {Blackwell Publishing Ltd},
issn = {1468-2427},
url = {http://dx.doi.org/10.1111/j.1468-2427.1990.tb00820.x},
doi = {10.1111/j.1468-2427.1990.tb00820.x},
pages = {49--69},
year = {1990},
}

@article {IJUR:IJUR12437,
author = {Wu, Fulong},
title = {China's Emergent City-Region Governance: A New Form of State Spatial Selectivity through State-orchestrated Rescaling},
journal = {International Journal of Urban and Regional Research},
volume = {40},
number = {6},
issn = {1468-2427},
url = {http://dx.doi.org/10.1111/1468-2427.12437},
doi = {10.1111/1468-2427.12437},
pages = {1134--1151},
year = {2016},
}

@ARTICLE{2017arXiv170600682Y,
   author = {{Yang}, Y. and {Tan}, C. and {Liu}, Z. and {Wu}, F. and {Zhuang}, Y.
	},
    title = {Urban Dreams of Migrants: A Case Study of Migrant Integration in Shanghai},
  journal = {arXiv preprint arXiv:1706.00682},
archivePrefix = "arXiv",
   eprint = {1706.00682},
 primaryClass = "cs.CY",
     year = 2017,
    month = jun,
   adsurl = {http://adsabs.harvard.edu/abs/2017arXiv170600682Y}
}

@article{de2007netlogo,
	Author = {De Leon, FD and Felsen, M and Wilensky, U},
	Journal = {Center for Connected Learning and Computer-Based Modeling, Northwestern University, Evanston, IL},
	Title = {NetLogo Urban Suite-Tijuana Bordertowns model},
	Year = {2007}}

@article{fan2005modeling,
	Author = {Fan, C Cindy},
	Journal = {Eurasian Geography and Economics},
	Number = {3},
	Pages = {165--184},
	Title = {Modeling interprovincial migration in China, 1985-2000},
	Volume = {46},
	Year = {2005}}

@article{florida2008rise,
	Author = {Florida, Richard and Gulden, Tim and Mellander, Charlotta},
	Journal = {Cambridge Journal of Regions, Economy and Society},
	Number = {3},
	Pages = {459--476},
	Title = {The rise of the mega-region},
	Volume = {1},
	Year = {2008}}

@book{hall2006polycentric,
	Author = {Hall, Peter Geoffrey and Pain, Kathy},
	Publisher = {Routledge},
	Title = {The polycentric metropolis: learning from mega-city regions in Europe},
	Year = {2006}}

@article{reuillon2013openmole,
	Author = {Reuillon, Romain and Leclaire, Mathieu and Rey-Coyrehourcq, Sebastien},
	Journal = {Future Generation Computer Systems},
	Number = {8},
	Pages = {1981--1990},
	Title = {OpenMOLE, a workflow engine specifically tailored for the distributed exploration of simulation models},
	Volume = {29},
	Year = {2013}}

@article{wilensky1999netlogo,
	Author = {Wilensky, Uri},
	Title = {NetLogo},
	Year = {1999}}

@article{wu2006migrant,
	Author = {Wu, Weiping},
	Journal = {Housing Studies},
	Number = {5},
	Pages = {745--765},
	Title = {Migrant intra-urban residential mobility in urban China},
	Volume = {21},
	Year = {2006}}

@article{xie2007simulating,
	Author = {Xie, Yichun and Batty, Michael and Zhao, Kang},
	Journal = {Annals of the Association of American Geographers},
	Number = {3},
	Pages = {477--495},
	Title = {Simulating emergent urban form using agent-based modeling: Desakota in the Suzhou-Wuxian region in China},
	Volume = {97},
	Year = {2007}}

@article{zhang2003rural,
	Author = {Zhang, Kevin Honglin and Shunfeng, Song},
	Journal = {China Economic Review},
	Number = {4},
	Pages = {386--400},
	Title = {Rural--urban migration and urbanization in China: Evidence from time-series and cross-section analyses},
	Volume = {14},
	Year = {2003}}

@article{zhang2013measuring,
	Author = {Zhang, Junfu and Zhao, Zhong},
	Title = {Measuring the income-distance tradeoff for rural-urban migrants in China},
	Journal = {IZA Discussion Paper No. 7160},
	Year = {2013}}

@article{batty1994cells,
  title={From cells to cities},
  author={Batty, Michael and Xie, Yichun},
  journal={Environment and planning B: Planning and design},
  volume={21},
  number={7},
  pages={S31--S48},
  year={1994},
  publisher={SAGE Publications Sage UK: London, England}
}

@article{seto2012global,
  title={Global forecasts of urban expansion to 2030 and direct impacts on biodiversity and carbon pools},
  author={Seto, Karen C and G{\"u}neralp, Burak and Hutyra, Lucy R},
  journal={Proceedings of the National Academy of Sciences},
  volume={109},
  number={40},
  pages={16083--16088},
  year={2012},
  publisher={National Acad Sciences}
}

@article{tadmor2012review,
  title={A review of numerical methods for nonlinear partial differential equations},
  author={Tadmor, Eitan},
  journal={Bulletin of the American Mathematical Society},
  volume={49},
  number={4},
  pages={507--554},
  year={2012}
}

@article {GEAN:GEAN940,
author = {Xie, Yichun},
title = {A Generalized Model for Cellular Urban Dynamics},
journal = {Geographical Analysis},
volume = {28},
number = {4},
publisher = {Blackwell Publishing Ltd},
issn = {1538-4632},
url = {http://dx.doi.org/10.1111/j.1538-4632.1996.tb00940.x},
doi = {10.1111/j.1538-4632.1996.tb00940.x},
pages = {350--373},
year = {1996},
}

@article{zhang2013identifying,
  title={Identifying determinants of urban growth from a multi-scale perspective: A case study of the urban agglomeration around Hangzhou Bay, China},
  author={Zhang, Zhonghao and Su, Shiliang and Xiao, Rui and Jiang, Diwei and Wu, Jiaping},
  journal={Applied Geography},
  volume={45},
  pages={193--202},
  year={2013},
  publisher={Elsevier}
}

@article{ward2000stochastically,
  title={A stochastically constrained cellular model of urban growth},
  author={Ward, Douglas P and Murray, Alan T and Phinn, Stuart R},
  journal={Computers, Environment and Urban Systems},
  volume={24},
  number={6},
  pages={539--558},
  year={2000},
  publisher={Elsevier}
}

@article{frankhauser1998fractal,
  title={Fractal geometry of urban patterns and their morphogenesis},
  author={Frankhauser, Pierre},
  journal={Discrete Dynamics in Nature and Society},
  volume={2},
  number={2},
  pages={127--145},
  year={1998},
  publisher={Hindawi Publishing Corporation}
}

@book{pichon_riviere_processus_2004,
	title = {Le processus groupal},
	isbn = {978-2-7492-0317-1},
	publisher = {Érès},
	author = {Pichon Rivière, Enrique},
	year = {2004}
}

@article{seidl_luhmanns_2004,
	title = {Luhmann’s theory of autopoietic social systems},
	urldate = {2017-01-01},
	journal = {Ludwig-Maximilians-Universität München-Munich School of Management},
	author = {Seidl, David},
	year = {2004}
}

@article{hart_held_2013,
	title = {Held in mind, out of awareness. {Perspectives} on the continuum of dissociated experience, culminating in dissociative identity disorder in children},
	volume = {39},
	issn = {0075-417X, 1469-9370},
	doi = {10.1080/0075417X.2013.846577},
	number = {3},
	urldate = {2017-01-02},
	journal = {Journal of Child Psychotherapy},
	author = {Hart, Carolyn},
	month = dec,
	year = {2013},
	pages = {303--318}
}

@book{piers_self-organizing_2007,
	title = {Self-{Organizing} {Complexity} in {Psychological} {Systems}},
	isbn = {978-1-4616-3065-4},
	publisher = {Jason Aronson, Incorporated},
	author = {Piers, Craig and Muller, John P. and Brent, Joseph},
	month = apr,
	year = {2007}
}

@book{_issues_2013,
	title = {Issues in {Neuroscience} {Research} and {Application}: 2013 {Edition}},
	isbn = {978-1-4901-0778-3},
	shorttitle = {INRA},
	author = {INRA},
	publisher = {ScholarlyEditions},
	month = may,
	year = {2013}
}

@article{archer_margaret_1999,
  title={Morphogenesis versus structuration: on combining structure and action},
  author={Archer, Margaret S},
  journal={The British Journal of Sociology},
  volume={61},
  number={s1},
  pages={225--252},
  year={2010},
  publisher={Wiley Online Library}
}

@article{de_luca_picione_processes_2016,
	title = {The processes of meaning making, starting from the morphogenetic theories of {Rene} {Thom}},
	volume = {22},
	issn = {1354-067X, 1461-7056},
	doi = {10.1177/1354067X15576171},
	number = {1},
	urldate = {2017-01-02},
	journal = {Culture \& Psychology},
	author = {Picione, R.L. and Freda, M.F.},
	month = mar,
	year = {2016},
	pages = {139--157}
}

@book{freud_totem_1989,
	address = {New York},
	series = {The {Standard} edition of the complete psychological works of {Sigmund} {Freud}},
	title = {Totem and taboo: some points of agreement between the mental lives of savages and neurotics},
	isbn = {978-0-393-00143-3},
	shorttitle = {Totem and taboo},
	language = {eng},
	publisher = {W.W. Norton},
	author = {Freud, Sigmund and Strachey, James and Freud, Sigmund},
	year = {1989}
}

@article{Wolpert1969,
author = {Wolpert, Lewis},
journal = {J Theor Biol},
pages = {1--47},
title = {{Positional Information and the Spatial Pattern of Cellular Differentiation}},
year = {1969}
}

@article{Lecuit2007,
author = {Lecuit, Thomas and Lenne, Pierre-fran{\c{c}}ois},
doi = {10.1038/nrm2222},
journal = {Nat Rev Mol Cell Biol},
number = {August},
pages = {633--644},
title = {{Cell surface mechanics and the control of cell shape, tissue patterns and morphogenesis}},
volume = {8},
year = {2007}
}

@article{Wolpert2011,
author = {Wolpert, L},
doi = {10.1016/j.jtbi.2010.10.034},
isbn = {1095-8541 (Electronic)
0022-5193 (Linking)},
journal = {J Theor Biol},
number = {1},
pages = {359--365},
pmid = {21044633},
title = {{Positional information and patterning revisited}},
url = {http://www.ncbi.nlm.nih.gov/pubmed/21044633 http://ac.els-cdn.com/S0022519310005795/1-s2.0-S0022519310005795-main.pdf?{\_}tid=5a5ab764-a7fa-11e5-a834-00000aacb35d{\&}acdnat=1450713120{\_}44ebcc31cb7683da9cfaa05d97abdb37 http://ac.els-cdn.com/S0022519310005795/1-s2},
volume = {269},
year = {2011}
}

@article{Rogers2011,
author = {Rogers, Katherine W and Schier, Alexander F},
doi = {10.1146/annurev-cellbio-092910-154148},
journal = {Annu Rev Cell Dev Biol},
title = {{Morphogen Gradients : From Generation to Interpretation}},
year = {2011}
}

@article{Heisenberg2013,
author = {Heisenberg, Carl Philipp and Bella{\"{i}}che, Yohanns},
doi = {10.1016/j.cell.2013.05.008},
isbn = {0092-8674},
issn = {00928674},
journal = {Cell},
number = {5},
pmid = {23706734},
title = {{Forces in tissue morphogenesis and patterning}},
volume = {153},
year = {2013}
}

@article{Guillot2013,
author = {Guillot, C. and Lencuit, T.},
doi = {10.1126/science.1235249},
isbn = {1095-9203 (Electronic)$\backslash$n0036-8075 (Linking)},
issn = {1095-9203},
journal = {Science},
number = {June},
pages = {1185--1189},
pmid = {23744939},
title = {{Mechanics of Epithelial Tissue}},
volume = {340},
year = {2013}
}

@article{levy2005formes,
	Author = {L{\'e}vy, Albert},
	Journal = {Espaces et soci{\'e}t{\'e}s},
	Number = {3},
	Pages = {25--48},
	Title = {Formes urbaines et significations: revisiter la morphologie urbaine},
	Year = {2005}}

@article{desmarais1992premisses,
  title={Des pr{\'e}misses de la th{\'e}orie de la forme urbaine au parcours morphog{\'e}n{\'e}tique de l’{\'e}tablissement humain},
  author={Desmarais, Ga{\"e}tan},
  journal={Cahiers de g{\'e}ographie du Qu{\'e}bec},
  volume={36},
  number={98},
  pages={251--273},
  year={1992},
  publisher={D{\'e}partement de g{\'e}ographie de l'Universit{\'e} Laval}
}

@book{holland2012signals,
	Author = {Holland, John H},
	Publisher = {Mit Press},
	Title = {Signals and boundaries: Building blocks for complex adaptive systems},
	Year = {2012}}

@article{whitehand1999urban,
	Author = {Whitehand, JWR and Morton, NJ and Carr, CMH},
	Journal = {Environment and Planning B: Planning and Design},
	Number = {4},
	Pages = {503--515},
	Title = {Urban morphogenesis at the microscale: how houses change},
	Volume = {26},
	Year = {1999}}

@article{dollens2014alan,
	Author = {Dollens, Dennis},
	Journal = {Leonardo},
	Number = {3},
	Pages = {249--254},
	Title = {Alan Turing's Drawings, Autopoiesis and Can Buildings Think?},
	Volume = {47},
	Year = {2014}}

@article{bedau2002downward,
	Author = {Bedau, Mark},
	Journal = {Principia: an international journal of epistemology},
	Number = {1},
	Pages = {5--50},
	Title = {Downward causation and the autonomy of weak emergence},
	Volume = {6},
	Year = {2002}}

@phdthesis{ceccarini2001essai,
  title={Essai de formalisation dynamique de la cath{\'e}drale gothique: morphog{\'e}n{\`e}se et mod{\'e}lisation de la Basilique Saint-Denis: les relations entre th{\'e}ologie, sciences et architecture au XIII{\`e}me si{\`e}cle {\`a} Saint-Denis},
  author={Ceccarini, Patrice},
  year={2001},
  school={Paris, EHESS}
}

@article{jun2005formation,
  title={Formation and structure of ramified charge transportation networks in an electromechanical system},
  author={Jun, Joseph K and H{\"u}bler, Alfred H},
  journal={Proceedings of the National Academy of Sciences of the United States of America},
  volume={102},
  number={3},
  pages={536--540},
  year={2005},
  publisher={National Acad Sciences}
}

@article{niizato2010model,
	Author = {Niizato, Takayuki and Shirakawa, Tomohiro and Gunji, Yukio-Pegio},
	Journal = {Biosystems},
	Number = {2},
	Pages = {108--112},
	Title = {A model of network formation by Physarum plasmodium: interplay between cell mobility and morphogenesis},
	Volume = {100},
	Year = {2010}}

@incollection{cussat2012synthesis,
	Author = {Cussat-Blanc, Sylvain and Pascalie, Jonathan and Mazac, S{\'e}bastien and Luga, Herv{\'e} and Duthen, Yves},
	Booktitle = {Morphogenetic Engineering},
	Pages = {353--381},
	Publisher = {Springer},
	Title = {A synthesis of the Cell2Organ developmental model},
	Year = {2012}}

@inproceedings{doursat2008programmable,
	Author = {Doursat, Ren{\'e}},
	Booktitle = {ALIFE},
	Pages = {181--188},
	Title = {Programmable Architectures That Are Complex and Self-Organized-From Morphogenesis to Engineering.},
	Year = {2008}}

@misc{crosato2014self,
	Author = {E. Crosato},
	HowPublished = {Working Paper. Vrije Universiteit Amsterdam},
	Title = {Artificial Self-Assembly : Literature Review},
	Year = {2014}}

@book{bourgine2010morphogenesis,
	Author = {Bourgine, Paul and Lesne, Annick},
	Publisher = {Springer Science \& Business Media},
	Title = {Morphogenesis: origins of patterns and shapes},
	Year = {2010}}

@article{chapman1998morphogenesis,
  title={Morphogenesis by symbiogenesis},
  author={Chapman, Michael J and Margulis, Lynn},
  journal={International Microbiology},
  volume={1},
  number={4},
  year={1998}
}

@article{doursat2013review,
  title={A review of morphogenetic engineering},
  author={Doursat, Ren{\'e} and Sayama, Hiroki and Michel, Olivier},
  journal={Natural Computing},
  volume={12},
  number={4},
  pages={517--535},
  year={2013},
  publisher={Springer}
}

@article{gierer1972theory,
  title={A theory of biological pattern formation},
  author={Gierer, Alfred and Meinhardt, Hans},
  journal={Kybernetik},
  volume={12},
  number={1},
  pages={30--39},
  year={1972},
  publisher={Springer}
}

@article{kondo2010reaction,
  title={Reaction-diffusion model as a framework for understanding biological pattern formation},
  author={Kondo, Shigeru and Miura, Takashi},
  journal={science},
  volume={329},
  number={5999},
  pages={1616--1620},
  year={2010},
  publisher={American Association for the Advancement of Science}
}

@article{bourgine2004autopoiesis,
	Author = {Bourgine, Paul and Stewart, John},
	Journal = {Artificial life},
	Number = {3},
	Pages = {327--345},
	Title = {Autopoiesis and cognition},
	Volume = {10},
	Year = {2004}}

@article{mehaffy2007notes,
	Author = {Mehaffy, Michael W},
	Journal = {Urban Design International},
	Number = {1},
	Pages = {41--49},
	Title = {Notes on the genesis of wholes: Christopher Alexander and his continuing influence},
	Volume = {12},
	Year = {2007}}

@article{renfrew1978trajectory,
	Author = {Renfrew, Colin},
	Journal = {American Antiquity},
	Pages = {203--222},
	Title = {Trajectory discontinuity and morphogenesis: the implications of catastrophe theory for archaeology},
	Year = {1978}}

@article{achibet2014model,
	Author = {Achibet, Merwan and Balev, Stefan and Dutot, Antoine and Olivier, Damien},
	Journal = {Procedia Computer Science},
	Pages = {828--833},
	Title = {A Model of Road Network and Buildings Extension Co-evolution},
	Volume = {32},
	Year = {2014}}

@article{thom1974stabilite,
	Author = {Thom, Ren{\'e}},
	Journal = {Poetics},
	Number = {2},
	Pages = {7--19},
	Title = {Stabilit{\'e} structurelle et morphog{\'e}n{\`e}se},
	Volume = {3},
	Year = {1974}}

@article{makse1998modeling,
	Author = {Makse, Hern{\'a}n A and Andrade, Jos{\'e} S and Batty, Michael and Havlin, Shlomo and Stanley, H Eugene and others},
	Journal = {Physical Review E},
	Number = {6},
	Pages = {7054},
	Title = {Modeling urban growth patterns with correlated percolation},
	Volume = {58},
	Year = {1998}}

@inproceedings{bonin2012modele,
	Author = {Bonin, Olivier and Hubert, Jean-Paul and others},
	Booktitle = {49{\`e} colloque de l'ASRDLF},
	Title = {Mod{\`e}le de morphog{\'e}n{\`e}se urbaine: simulation d'espaces qualitativement diff{\'e}renci{\'e}s dans le cadre du mod{\`e}le de l'{\'e}conomie urbaine},
	Year = {2012}}

@article{moudon1997urban,
	Author = {Moudon, Anne Vernez},
	Journal = {Urban morphology},
	Number = {1},
	Pages = {3--10},
	Title = {Urban morphology as an emerging interdisciplinary field},
	Volume = {1},
	Year = {1997}}

@phdthesis{burke1972dublin,
	Author = {Burke, Nuala T},
	Title = {Dublin 1600-1800: a study in urban morphogenesis},
	Year = {1972}}

@article{olsen1982urban,
	Author = {Olsen, Sherry},
	Journal = {Urban Geography},
	Number = {2},
	Pages = {87--109},
	Title = {Urban metabolism and morphogenesis},
	Volume = {3},
	Year = {1982}}

@article{abercrombie1977concepts,
  title={Concepts in morphogenesis},
  author={Abercrombie, Michael},
  journal={Proceedings of the Royal Society of London B: Biological Sciences},
  volume={199},
  number={1136},
  pages={337--344},
  year={1977},
  publisher={The Royal Society}
}

@article{gilbert2003morphogenesis,
  title={The morphogenesis of evolutionary developmental biology},
  author={Gilbert, Scott F},
  journal={International Journal of Developmental Biology},
  volume={47},
  number={7-8},
  pages={467},
  year={2003}
}

@book{thompson1942growth,
  title={On growth and form},
  author={Thompson, Darcy Wentworth},
  year={1942},
  publisher={Cambridge University Press}
}

@article{nakamasu2009interactions,
  title={Interactions between zebrafish pigment cells responsible for the generation of Turing patterns},
  author={Nakamasu, Akiko and Takahashi, Go and Kanbe, Akio and Kondo, Shigeru},
  journal={Proceedings of the National Academy of Sciences},
  volume={106},
  number={21},
  pages={8429--8434},
  year={2009},
  publisher={National Acad Sciences}
}

@article{goryachev2008dynamics,
  title={Dynamics of Cdc42 network embodies a Turing-type mechanism of yeast cell polarity},
  author={Goryachev, Andrew B and Pokhilko, Alexandra V},
  journal={FEBS letters},
  volume={582},
  number={10},
  pages={1437--1443},
  year={2008},
  publisher={Wiley Online Library}
}

@article{brand2013accelerating,
  title={Accelerating the transformation to a low carbon passenger transport system: The role of car purchase taxes, feebates, road taxes and scrappage incentives in the UK},
  author={Brand, Christian and Anable, Jillian and Tran, Martino},
  journal={Transportation Research Part A: Policy and Practice},
  volume={49},
  pages={132--148},
  year={2013},
  publisher={Elsevier}
}

@article{fullerton2002can,
  title={Can taxes on cars and on gasoline mimic an unavailable tax on emissions?},
  author={Fullerton, Don and West, Sarah E},
  journal={Journal of Environmental Economics and Management},
  volume={43},
  number={1},
  pages={135--157},
  year={2002},
  publisher={Elsevier}
}

@inproceedings{lee2009determinants,
  title={Determinants of crime incidence in Korea: a mixed GWR approach},
  author={Lee, SeongWoo and Kang, DongWoo and Kim, MiYoung},
  booktitle={World conference of the spatial econometrics association},
  pages={8--10},
  year={2009}
}

@article{chen2012using,
  title={Using multilevel modeling and geographically weighted regression to identify spatial variations in the relationship between place-level disadvantages and obesity in Taiwan},
  author={Chen, Duan-Rung and Truong, Khoa},
  journal={Applied Geography},
  volume={32},
  number={2},
  pages={737--745},
  year={2012},
  publisher={Elsevier}
}

@article{tsai2005quantifying,
  title={Quantifying urban form: compactness versus' sprawl'},
  author={Tsai, Yu-Hsin},
  journal={Urban studies},
  volume={42},
  number={1},
  pages={141--161},
  year={2005},
  publisher={Sage Publications Sage UK: London, England}
}

@article{gautier2015dynamics,
  title={The dynamics of gasoline prices: Evidence from daily French micro data},
  author={Gautier, Erwan and Saout, Ronan Le},
  journal={Journal of Money, Credit and Banking},
  volume={47},
  number={6},
  pages={1063--1089},
  year={2015},
  publisher={Wiley Online Library}
}

@article{tan2013social,
  title={Social-network-sourced big data analytics},
  author={Tan, Wei and Blake, M Brian and Saleh, Iman and Dustdar, Schahram},
  journal={IEEE Internet Computing},
  volume={17},
  number={5},
  pages={62--69},
  year={2013},
  publisher={IEEE}
}

@article{brunsdon1996geographically,
  title={Geographically weighted regression: a method for exploring spatial nonstationarity},
  author={Brunsdon, Chris and Fotheringham, A Stewart and Charlton, Martin E},
  journal={Geographical analysis},
  volume={28},
  number={4},
  pages={281--298},
  year={1996},
  publisher={Wiley Online Library}
}

@article{dupuy2015sciences,
  title={Sciences urbaines: interdisciplinarit{\'e}s passive, na{\"\i}ve, transitive, offensive},
  author={Dupuy, Gabriel and Benguigui, Lucien Gilles},
  journal={M{\'e}tropoles},
  number={16},
  year={2015},
  publisher={ENTPE}
}

@article{batty2013big,
	Author = {Batty, Michael},
	Journal = {Dialogues in Human Geography},
	Number = {3},
	Pages = {274--279},
	Publisher = {SAGE Publications Sage UK: London, England},
	Title = {Big data, smart cities and city planning},
	Volume = {3},
	Year = {2013}}

@article{combes2005transport,
	Author = {Combes, Pierre-Philippe and Lafourcade, Miren},
	Journal = {Journal of Economic Geography},
	Number = {3},
	Pages = {319--349},
	Publisher = {Oxford Univ Press},
	Title = {Transport costs: measures, determinants, and regional policy implications for France},
	Volume = {5},
	Year = {2005}}

@article{gregg2009temporal,
	Author = {Gregg, Jay S and Losey, London M and Andres, Robert J and Blasing, TJ and Marland, Gregg},
	Journal = {Journal of Applied Meteorology and Climatology},
	Number = {12},
	Pages = {2528--2542},
	Title = {The temporal and spatial distribution of carbon dioxide emissions from fossil-fuel use in North America},
	Volume = {48},
	Year = {2009}}

@article{macharis2010decision,
	Author = {Macharis, Cathy and Van Hoeck, Ellen and Pekin, Ethem and Van Lier, Tom},
	Journal = {Transportation Research Part A: Policy and Practice},
	Number = {7},
	Pages = {550--561},
	Publisher = {Elsevier},
	Title = {A decision analysis framework for intermodal transport: Comparing fuel price increases and the internalisation of external costs},
	Volume = {44},
	Year = {2010}}

@article{rietveld2005fuel,
	Author = {Rietveld, Piet and van Woudenberg, Stefan},
	Journal = {Energy Economics},
	Number = {1},
	Pages = {79--92},
	Publisher = {Elsevier},
	Title = {Why fuel prices differ},
	Volume = {27},
	Year = {2005}}

@article{rietveld2001spatial,
	Author = {Rietveld, Piet and Bruinsma, FR and Van Vuuren, DJ},
	Journal = {Transportation Research Part A: Policy and Practice},
	Number = {5},
	Pages = {433--457},
	Publisher = {Elsevier},
	Title = {Spatial graduation of fuel taxes; consequences for cross-border and domestic fuelling},
	Volume = {35},
	Year = {2001}}

@inproceedings{zhu2013scalable,
  title={Scalable text and link analysis with mixed-topic link models},
  author={Zhu, Yaojia and Yan, Xiaoran and Getoor, Lise and Moore, Cristopher},
  booktitle={Proceedings of the 19th ACM SIGKDD international conference on Knowledge discovery and data mining},
  pages={473--481},
  year={2013},
  organization={ACM}
}

@article{li2014disambiguation,
  title={Disambiguation and co-authorship networks of the US patent inventor database (1975--2010)},
  author={Li, Guan-Cheng and Lai, Ronald and D’Amour, Alexander and Doolin, David M and Sun, Ye and Torvik, Vetle I and Amy, Z Yu and Fleming, Lee},
  journal={Research Policy},
  volume={43},
  number={6},
  pages={941--955},
  year={2014},
  publisher={Elsevier}
}

@article{gao2017complex,
  title={Complex network analysis of time series},
  author={Gao, Zhong-Ke and Small, Michael and Kurths, J{\"u}rgen},
  journal={EPL (Europhysics Letters)},
  volume={116},
  number={5},
  pages={50001},
  year={2017},
  publisher={IOP Publishing}
}

@article{gao2015multiscale,
  title={Multiscale complex network for analyzing experimental multivariate time series},
  author={Gao, Zhong-Ke and Yang, Yu-Xuan and Fang, Peng-Cheng and Zou, Yong and Xia, Cheng-Yi and Du, Meng},
  journal={EPL (Europhysics Letters)},
  volume={109},
  number={3},
  pages={30005},
  year={2015},
  publisher={IOP Publishing}
}

@book{ziman2003technological,
	Author = {Ziman, John},
	Publisher = {Cambridge University Press},
	Title = {Technological innovation as an evolutionary process},
	Year = {2003}}

@article{ARCHIBUGI199279,
	Author = {Daniele Archibugi and Mario Pianta},
	Doi = {http://dx.doi.org/10.1016/0048-7333(92)90028-3},
	Issn = {0048-7333},
	Journal = {Research Policy},
	Number = {1},
	Pages = {79 - 93},
	Title = {Specialization and size of technological activities in industrial countries: The analysis of patent data},
	Url = {http://www.sciencedirect.com/science/article/pii/0048733392900283},
	Volume = {21},
	Year = {1992}}

@article{clauset2004finding,
	Author = {Clauset, Aaron and Newman, Mark EJ and Moore, Cristopher},
	Journal = {Physical review E},
	Number = {6},
	Pages = {066111},
	Title = {Finding community structure in very large networks},
	Volume = {70},
	Year = {2004}}

@article{de2015ranking,
	Author = {De Domenico, Manlio and Sol{\'e}-Ribalta, Albert and Omodei, Elisa and G{\'o}mez, Sergio and Arenas, Alex},
	Journal = {Nature communications},
	Publisher = {Nature Publishing Group},
	Title = {Ranking in interconnected multilayer networks reveals versatile nodes},
	Volume = {6},
	Year = {2015}}

@article{gerken2012new,
	Author = {Gerken, Jan M and Moehrle, Martin G},
	Journal = {Scientometrics},
	Number = {3},
	Pages = {645--670},
	Title = {A new instrument for technology monitoring: novelty in patents measured by semantic patent analysis},
	Volume = {91},
	Year = {2012}}

@article{kaplan2015double,
	Author = {Kaplan, Sarah and Vakili, Keyvan},
	Journal = {Strategic Management Journal},
	Number = {10},
	Pages = {1435--1457},
	Publisher = {Wiley Online Library},
	Title = {The double-edged sword of recombination in breakthrough innovation},
	Volume = {36},
	Year = {2015}}

@article{sorenson2006complexity,
	Author = {Sorenson, Olav and Rivkin, Jan W and Fleming, Lee},
	Journal = {Research policy},
	Number = {7},
	Pages = {994--1017},
	Publisher = {North-Holland},
	Title = {Complexity, networks and knowledge flow},
	Volume = {35},
	Year = {2006}}

@misc{oecdpatentmanual,
	Author = {OECD},
	Doi = {http://dx.doi.org/10.1787/9789264056442-en},
	Publisher = {OECD Publishing},
	Title = {OECD Patent Statistics Manual},
	Url = {/content/book/9789264056442-en},
	Year = {2009}}

@article{bruck2016recognition,
	Author = {Bruck, P{\'e}ter and R{\'e}thy, Istv{\'a}n and Szente, Judit and Tobochnik, Jan and {\'E}rdi, P{\'e}ter},
	Journal = {Scientometrics},
	Number = {3},
	Pages = {1465--1475},
	Publisher = {Springer},
	Title = {Recognition of emerging technology trends: class-selective study of citations in the US Patent Citation Network},
	Volume = {107},
	Year = {2016}}

@article{kay2014patent,
	Author = {Kay, Luciano and Newman, Nils and Youtie, Jan and Porter, Alan L and Rafols, Ismael},
	Journal = {Journal of the Association for Information Science and Technology},
	Number = {12},
	Pages = {2432--2443},
	Publisher = {Wiley Online Library},
	Title = {Patent overlay mapping: Visualizing technological distance},
	Volume = {65},
	Year = {2014}}

@techreport{RePEc:nbr:nberwo:3301,
	Author = {Zvi Griliches},
	Institution = {National Bureau of Economic Research},
	Month = Mar,
	Number = {3301},
	Title = {{Patent Statistics as Economic Indicators: A Survey}},
	Type = {NBER Working Papers},
	Url = {https://ideas.repec.org/p/nbr/nberwo/3301.html},
	Year = 1990}

@techreport{Hall2001,
	Author = {Hall, Bronwyn H and Jaffe, Adam B and Trajtenberg, Manuel},
	Month = Dec,
	Number = {3094},
	Title = {{The NBER Patent Citations Data File: Lessons, Insights and Methodological Tools}},
	Type = {CEPR Discussion Papers},
	Url = {https://ideas.repec.org/p/cpr/ceprdp/3094.html},
	Year = 2001}

@article{romer1990,
	Author = {Romer, Paul M},
	Journal = {Journal of Political Economy},
	Month = {October},
	Number = {5},
	Pages = {S71-102},
	Title = {{Endogenous Technological Change}},
	Url = {https://ideas.repec.org/a/ucp/jpolec/v98y1990i5ps71-102.html},
	Volume = {98},
	Year = 1990}

@article{lerner2015use,
	Author = {Lerner, Josh and Seru, Amit},
	Journal = {Booth/Harvard Business School Working Paper},
	Title = {The use and misuse of patent data: Issues for corporate finance and beyond},
	Year = {2015}}

@article{aghionhowitt1992,
	Author = {Aghion, Philippe and Howitt, Peter},
	Journal = {Econometrica},
	Month = {March},
	Number = {2},
	Pages = {323-51},
	Title = {{A Model of Growth through Creative Destruction}},
	Url = {https://ideas.repec.org/a/ecm/emetrp/v60y1992i2p323-51.html},
	Volume = {60},
	Year = 1992}

@article{AAKnetwork2016,
  title={Innovation network},
  author={Acemoglu, Daron and Akcigit, Ufuk and Kerr, William R},
  journal={Proceedings of the National Academy of Sciences},
  volume={113},
  number={41},
  pages={11483--11488},
  year={2016},
  publisher={National Acad Sciences}
}

@inproceedings{bird2006nltk,
	Author = {Bird, Steven},
	Booktitle = {Proceedings of the COLING/ACL on Interactive presentation sessions},
	Organization = {Association for Computational Linguistics},
	Pages = {69--72},
	Title = {NLTK: the natural language toolkit},
	Year = {2006}}

@inproceedings{yang2000improving,
	Author = {Yang, Yiming and Ault, Tom and Pierce, Thomas and Lattimer, Charles W},
	Booktitle = {Proceedings of the 23rd annual international ACM SIGIR conference on Research and development in information retrieval},
	Organization = {ACM},
	Pages = {65--72},
	Title = {Improving text categorization methods for event tracking},
	Year = {2000}}

@misc{nltk,
	Author = {NLTK},
	Title = {Natural Language Toolkit, Stanford Univeristy},
	Year = {2015}}

@article{iacovacci2015mesoscopic,
  title={Mesoscopic structures reveal the network between the layers of multiplex data sets},
  author={Iacovacci, Jacopo and Wu, Zhihao and Bianconi, Ginestra},
  journal={Physical Review E},
  volume={92},
  number={4},
  pages={042806},
  year={2015},
  publisher={APS}
}

@article{abbas2014literature,
	Author = {Abbas, Assad and Zhang, Limin and Khan, Samee U},
	Journal = {World Patent Information},
	Pages = {3--13},
	Publisher = {Elsevier},
	Title = {A literature review on the state-of-the-art in patent analysis},
	Volume = {37},
	Year = {2014}}

@article{Adams2010text,
	Author = {Adams, Stephen},
	Journal = {World Patent Information},
	Month = {March},
	Number = {1},
	Pages = {22-29},
	Title = {{The text, the full text and nothing but the text: Part 1 - Standards for creating textual information in patent documents and general search implications}},
	Url = {https://ideas.repec.org/a/eee/worpat/v32y2010i1p22-29.html},
	Volume = {32},
	Year = 2010}

@misc{akcigit2013mechanics,
	Author = {Akcigit, Ufuk and Kerr, William R and Nicholas, Tom},
	Title = {The Mechanics of Endogenous Innovation and Growth: Evidence from Historical US Patents},
	HowPublished = {Working Paper},
	Year = {2013}}

@article{Bloom2005distance,
	Author = {Nicholas Bloom and Mark Schankerman and John Van Reenen},
	Journal = {Econometrica},
	Month = {07},
	Number = {4},
	Pages = {1347-1393},
	Title = {{Identifying Technology Spillovers and Product Market Rivalry}},
	Url = {https://ideas.repec.org/a/ecm/emetrp/v81y2013i4p1347-1393.html},
	Volume = {81},
	Year = 2013}

@article{choi2014patent,
	Author = {Choi, Jinho and Hwang, Yong-Sik},
	Journal = {Technological Forecasting and Social Change},
	Pages = {170--182},
	Publisher = {Elsevier},
	Title = {Patent keyword network analysis for improving technology development efficiency},
	Volume = {83},
	Year = {2014}}

@article{fattori2003text,
	Author = {Fattori, Michele and Pedrazzi, Giorgio and Turra, Roberta},
	Journal = {World Patent Information},
	Number = {4},
	Pages = {335--342},
	Publisher = {Elsevier},
	Title = {Text mining applied to patent mapping: a practical business case},
	Volume = {25},
	Year = {2003}}

@article{tseng2007text,
	Author = {Tseng, Yuen-Hsien and Lin, Chi-Jen and Lin, Yu-I},
	Journal = {Information Processing \& Management},
	Number = {5},
	Pages = {1216--1247},
	Publisher = {Elsevier},
	Title = {Text mining techniques for patent analysis},
	Volume = {43},
	Year = {2007}}

@article{Furman2011shoulders,
	Author = {Furman, Jeffrey L. and Stern, Scott},
	Doi = {10.1257/aer.101.5.1933},
	Journal = {American Economic Review},
	Month = {August},
	Number = {5},
	Pages = {1933-63},
	Title = {Climbing atop the Shoulders of Giants: The Impact of Institutions on Cumulative Research},
	Url = {http://www.aeaweb.org/articles?id=10.1257/aer.101.5.1933},
	Volume = {101},
	Year = {2011}}

@article{curran2011patent,
	Author = {Curran, Clive-Steven and Leker, Jens},
	Journal = {Technological Forecasting and Social Change},
	Number = {2},
	Pages = {256--273},
	Publisher = {Elsevier},
	Title = {Patent indicators for monitoring convergence--examples from NFF and ICT},
	Volume = {78},
	Year = {2011}}

@article{katz1996remarks,
	Author = {Katz, Michael L},
	Journal = {Industrial and Corporate Change},
	Number = {4},
	Pages = {1079--1095},
	Publisher = {Oxford Univ Press},
	Title = {Remarks on the economic implications of convergence},
	Volume = {5},
	Year = {1996}}

@article{preschitschek2013,
	Author = {Nina Preschitschek and Helen Niemann and Jens Leker and Martin G. Moehrle},
	Day = {11},
	Doi = {10.1108/fs-10-2012-0075},
	Issn = {1463-6689},
	Journal = {Foresight},
	Month = {11},
	Number = {6},
	Pages = {446-464},
	Publisher = {Emerald},
	Title = {Anticipating industry convergence: Semantic analyses vs IPC co-classification analyses of patents},
	Volume = {15},
	Year = {2013}}

@article{yoon2004text,
	Author = {Yoon, Byungun and Park, Yongtae},
	Journal = {The Journal of High Technology Management Research},
	Number = {1},
	Pages = {37--50},
	Publisher = {Elsevier},
	Title = {A text-mining-based patent network: Analytical tool for high-technology trend},
	Volume = {15},
	Year = {2004}}

@article{Youn:2015fk,
	Author = {Youn, Hyejin and Strumsky, Deborah and Bettencourt, Luis M. A. and Lobo, Jos{\'e}},
	Doi = {10.1098/rsif.2015.0272},
	Isbn = {1742-5662},
	Issn = {1742-5689},
	Journal = {Journal of The Royal Society Interface},
	Number = {106},
	Publisher = {The Royal Society},
	Title = {Invention as a combinatorial process: evidence from US patents},
	Volume = {12},
	Year = {2015}}

@article{yoon2011detecting,
	Author = {Yoon, Janghyeok and Kim, Kwangsoo},
	Journal = {Scientometrics},
	Number = {2},
	Pages = {445--461},
	Publisher = {Akad{\'e}miai Kiad{\'o}, co-published with Springer Science+ Business Media BV, Formerly Kluwer Academic Publishers BV},
	Title = {Detecting signals of new technological opportunities using semantic patent analysis and outlier detection},
	Volume = {90},
	Year = {2011}}

@article{park2014semantic,
	Author = {Park, Inchae and Yoon, Byungun},
	Journal = {Technology Analysis \& Strategic Management},
	Number = {8},
	Pages = {855--874},
	Publisher = {Taylor \& Francis},
	Title = {A semantic analysis approach for identifying patent infringement based on a product--patent map},
	Volume = {26},
	Year = {2014}}

@techreport{martin2015,
	Author = {Antoine Dechezlepratre and Ralf Martin and Myra Mohnen},
	Institution = {Centre for Economic Performance, LSE},
	Month = Sep,
	Number = {dp1300},
	Title = {{Knowledge Spillovers from Clean and Dirty Technologies}},
	Type = {CEP Discussion Papers},
	Url = {https://ideas.repec.org/p/cep/cepdps/dp1300.html},
	Year = 2014}

@article{mimeur:hal-01616746,
	Author = {Mimeur, Christophe and Queyroi, Fran{\c c}ois and Banos, Arnaud and Th{\'e}venin, Thomas},
	Journal = {{Historical Methods: A Journal of Quantitative and Interdisciplinary History}},
	Title = {{Revisiting the structuring effect of transportation infrastructure: an empirical approach with the French Railway Network from 1860 to 1910}},
	Year = {2017}}

@article{jacobs2016transport,
  title={Transport link scanner: simulating geographic transport network expansion through individual investments},
  author={Jacobs-Crisioni, Chris and Koopmans, Carl C},
  journal={Journal of Geographical Systems},
  volume={18},
  number={3},
  pages={265--301},
  year={2016},
  publisher={Springer}
}

@article{li2016integrated,
  title={Integrated co-evolution model of land use and traffic network design},
  author={Li, Tongfei and Wu, Jianjun and Sun, Huijun and Gao, Ziyou},
  journal={Networks and Spatial Economics},
  volume={16},
  number={2},
  pages={579--603},
  year={2016},
  publisher={Springer}
}

@article{xie2009jurisdictional,
  title={Jurisdictional control and network growth},
  author={Xie, Feng and Levinson, David},
  journal={Networks and Spatial Economics},
  volume={9},
  number={3},
  pages={459--483},
  year={2009},
  publisher={Springer}
}

@article{levinson2012forecasting,
	Author = {Levinson, David and Xie, Feng and Oca, Norah M},
	Journal = {Networks and Spatial Economics},
	Number = {2},
	Pages = {239--262},
	Title = {Forecasting and evaluating network growth},
	Volume = {12},
	Year = {2012}}

@article{liao2017ouverture,
  title={L’ouverture au march{\'e} en Chine (ann{\'e}es 1980-2000) et le d{\'e}veloppement {\'e}conomique local: une forme de gouvernance multi-niveaux?},
  author={Liao, Liao and Gaudin, Jean Pierre},
  journal={Cybergeo: European Journal of Geography},
  year={2017},
  publisher={CNRS-UMR G{\'e}ographie-cit{\'e}s 8504}
}

@article{xie2009modeling,
  title={Modeling the growth of transportation networks: a comprehensive review},
  author={Xie, Feng and Levinson, David},
  journal={Networks and Spatial Economics},
  volume={9},
  number={3},
  pages={291--307},
  year={2009},
  publisher={Springer}
}

@article{courtat2011mathematics,
	Author = {Courtat, Thomas and Gloaguen, Catherine and Douady, Stephane},
	Journal = {Physical Review E},
	Number = {3},
	Pages = {036106},
	Title = {Mathematics and morphogenesis of cities: A geometrical approach},
	Volume = {83},
	Year = {2011}}

@phdthesis{schmitt2014modelisation,
  title={Mod{\'e}lisation de la dynamique des syst{\`e}mes de peuplement: de SimpopLocal {\`a} SimpopNet.},
  author={Schmitt, Clara},
  year={2014},
  school={Paris 1}
}

@article{raimbault2016models,
  title={Models Coupling Urban Growth and Transportation Network Growth: An Algorithmic Systematic Review Approach},
  author={Raimbault, Juste},
  journal={arXiv preprint arXiv:1605.08888},
  year={2016}
}

@article{hou2011transport,
  title={Transport infrastructure development and changing spatial accessibility in the Greater Pearl River Delta, China, 1990--2020},
  author={Hou, Quan and Li, Si-Ming},
  journal={Journal of Transport Geography},
  volume={19},
  number={6},
  pages={1350--1360},
  year={2011},
  publisher={Elsevier}
}

@article{innes2010strategies,
  title={Strategies for megaregion governance: Collaborative dialogue, networks, and self-organization},
  author={Innes, Judith E and Booher, David E and Di Vittorio, Sarah},
  journal={Journal of the American Planning Association},
  volume={77},
  number={1},
  pages={55--67},
  year={2010},
  publisher={Taylor \& Francis}
}

@phdthesis{bretagnolle:tel-00459720,
	Author = {Bretagnolle, Anne},
	Month = Jun,
	School = {Universit{\'e} Panth{\'e}on-Sorbonne - Paris I},
	Title = {{Villes et r{\'e}seaux de transport : des interactions dans la longue dur{\'e}e, France, Europe, {\'E}tats-Unis}},
	Type = {HDR},
	Year = {2009}}

@article{xu2005city,
	Author = {Xu, Jiang and Yeh, Anthony GO},
	Journal = {International Journal of Urban and Regional Research},
	Number = {2},
	Pages = {283--308},
	Publisher = {Wiley Online Library},
	Title = {City repositioning and competitiveness building in regional development: New development strategies in Guangzhou, China},
	Volume = {29},
	Year = {2005}}

@book{lowry1964model,
  title={A model of metropolis},
  author={Lowry, Ira S},
  year={1964},
  publisher={Rand Corporation Santa Monica, CA}
}

@book{ordeshook1986game,
	Author = {Ordeshook, Peter C},
	Publisher = {Cambridge University Press},
	Title = {Game theory and political theory: An introduction},
	Year = {1986}}

@article{wegener2004land,
  title={Land-use transport interaction: state of the art},
  author={Wegener, Michael and F{\"u}rst, Franz},
  year={2004}
}

@article{Roumboutsos2008209,
	Author = {Athena Roumboutsos and Seraphim Kapros},
	Journal = {Transport Policy},
	Number = {4},
	Pages = {209 - 215},
	Title = {A game theory approach to urban public transport integration policy},
	Volume = {15},
	Year = {2008}}

@inproceedings{tretyakov2011fast,
	Author = {Tretyakov, Konstantin and Armas-Cervantes, Abel and Garc{\'\i}a-Ba{\~n}uelos, Luciano and Vilo, Jaak and Dumas, Marlon},
	Booktitle = {Proceedings of the 20th ACM international conference on Information and knowledge management},
	Organization = {ACM},
	Pages = {1785--1794},
	Title = {Fast fully dynamic landmark-based estimation of shortest path distances in very large graphs},
	Year = {2011}}

@article{le2015modeling,
	Author = {Le N{\'e}chet, Florent and Raimbault, Juste},
	Journal = {Plurimondi. An International Forum for Research and Debate on Human Settlements},
	Number = {15},
	Title = {Modeling the emergence of metropolitan transport autorithy in a polycentric urban region},
	Volume = {7},
	Year = {2015}}

@phdthesis{le2010approche,
	Author = {Le Nechet, Florent},
	School = {Universit{\'e} Paris-Est},
	Title = {Approche multiscalaire des liens entre mobilit{\'e} quotidienne, morphologie et soutenabilit{\'e} des m{\'e}tropoles europ{\'e}ennes: cas de Paris et de la r{\'e}gion Rhin-Ruhr},
	Year = {2010}}

@article{kwan1998space,
	Author = {Kwan, Mei-Po},
	Journal = {Geographical analysis},
	Number = {3},
	Pages = {191--216},
	Publisher = {Wiley Online Library},
	Title = {Space-time and integral measures of individual accessibility: a comparative analysis using a point-based framework},
	Volume = {30},
	Year = {1998}}

@article{dragomir1999ostrowski,
	Author = {Dragomir, SS},
	Journal = {Computers \& Mathematics with Applications},
	Number = {11},
	Pages = {33--37},
	Publisher = {Elsevier},
	Title = {The Ostrowski's integral inequality for Lipschitzian mappings and applications},
	Volume = {38},
	Year = {1999}}

@article{ram2013git,
	Author = {Ram, Karthik},
	Journal = {Source code for biology and medicine},
	Number = {1},
	Pages = {7},
	Title = {Git can facilitate greater reproducibility and increased transparency in science.},
	Volume = {8},
	Year = {2013}}

@article{qgis2011quantum,
	Author = {QGis, DT},
	Journal = {Open Source Geospatial Foundation Project},
	Title = {Quantum GIS geographic information system},
	Year = {2011}}

@book{bennett2010openstreetmap,
	Author = {Bennett, Jonathan},
	Publisher = {Packt Publishing Ltd},
	Title = {OpenStreetMap},
	Year = {2010}}

@article{tivadar2014oasis,
	Author = {Tivadar, Mihai and Schaeffer, Yves and Torre, Andr{\'e} and Bray, Fr{\'e}d{\'e}ric},
	Journal = {Cybergeo: European Journal of Geography},
	Publisher = {CNRS-UMR G{\'e}ographie-cit{\'e}s 8504},
	Title = {OASIS--un Outil d'Analyse de la S{\'e}gr{\'e}gation et des In{\'e}galit{\'e}s Spatiales},
	Year = {2014}}

@book{dick2010digital,
	Author = {Dick, Josef and Pillichshammer, Friedrich},
	Publisher = {Cambridge University Press},
	Title = {Digital nets and sequences: Discrepancy Theory and Quasi--Monte Carlo Integration},
	Year = {2010}}

@book{picon2013smart,
	Author = {Picon, Antoine},
	Publisher = {B2},
	Title = {Smart cities: th{\'e}orie et critique d'un id{\'e}al auto-r{\'e}alisateur},
	Year = {2013}}

@article{haken2003face,
	Author = {Haken, Herman and Portugali, Juval},
	Journal = {Journal of Environmental Psychology},
	Number = {4},
	Pages = {385--408},
	Publisher = {Elsevier},
	Title = {The face of the city is its information},
	Volume = {23},
	Year = {2003}}

@article{newman2011complex,
	Author = {Newman, MEJ},
	Journal = {arXiv preprint arXiv:1112.1440},
	Reading = {A},
	Title = {Complex systems: A survey},
	Usage = {GSB},
	Year = {2011}}

@article{marler2004survey,
	Author = {Marler, R Timothy and Arora, Jasbir S},
	Journal = {Structural and multidisciplinary optimization},
	Number = {6},
	Pages = {369--395},
	Publisher = {Springer},
	Title = {Survey of multi-objective optimization methods for engineering},
	Volume = {26},
	Year = {2004}}

@article{carver1991integrating,
	Author = {Carver, Stephen J},
	Journal = {International Journal of Geographical Information System},
	Number = {3},
	Pages = {321--339},
	Publisher = {Taylor \& Francis},
	Title = {Integrating multi-criteria evaluation with geographical information systems},
	Volume = {5},
	Year = {1991}}

@article{dobbie2013robustness,
	Author = {Dobbie, Melissa J and Dail, David},
	Journal = {Ecological Indicators},
	Pages = {270--277},
	Publisher = {Elsevier},
	Title = {Robustness and sensitivity of weighting and aggregation in constructing composite indices},
	Volume = {29},
	Year = {2013}}

@article{deb2006introducing,
	Author = {Deb, Kalyanmoy and Gupta, Himanshu},
	Journal = {Evolutionary Computation},
	Number = {4},
	Pages = {463--494},
	Publisher = {MIT Press},
	Title = {Introducing robustness in multi-objective optimization},
	Volume = {14},
	Year = {2006}}

@inproceedings{1688537,
	Author = {Barrico, C. and Antunes, C.H.},
	Booktitle = {Evolutionary Computation, 2006. CEC 2006. IEEE Congress on},
	Doi = {10.1109/CEC.2006.1688537},
	Pages = {1887-1892},
	Title = {Robustness Analysis in Multi-Objective Optimization Using a Degree of Robustness Concept},
	Year = {2006}}

@article{jegou2012evaluation,
	Author = {J{\'e}gou, Anne and Augiseau, Vincent and Guyot, C{\'e}cile and Jud{\'e}aux, C{\'e}cile and Monaco, Fran{\c{c}}ois-Xavier and Pech, Pierre and others},
	Journal = {Cybergeo: European Journal of Geography},
	Publisher = {CNRS-UMR G{\'e}ographie-cit{\'e}s 8504},
	Title = {L'{\'e}valuation par indicateurs: un outil n{\'e}cessaire d'am{\'e}nagement urbain durable?. R{\'e}flexions {\`a} partir de la d{\'e}marche parisienne pour le g{\'e}ographe et l'am{\'e}nageur},
	Year = {2012}}

@book{launer2014robustness,
	Author = {Launer, Robert L and Wilkinson, Graham N},
	Publisher = {Academic Press},
	Title = {Robustness in statistics},
	Year = {2014}}

@article{niederreiter1972discrepancy,
	Author = {Niederreiter, H},
	Journal = {Annali di matematica pura ed applicata},
	Number = {1},
	Pages = {89--97},
	Publisher = {Springer},
	Title = {Discrepancy and convex programming},
	Volume = {93},
	Year = {1972}}

@phdthesis{varet2010developpement,
	Author = {Varet, Suzanne},
	School = {Universit{\'e} Paul Sabatier-Toulouse III},
	Title = {D{\'e}veloppement de m{\'e}thodes statistiques pour la pr{\'e}diction d'un gabarit de signature infrarouge},
	Year = {2010}}

@book{mangin1999projet,
	Author = {Mangin, David and Panerai, Philippe},
	Publisher = {Parenth{\`e}ses},
	Reading = {A},
	Title = {Projet urbain},
	Year = {1999}}

@article{wang2009review,
	Author = {Wang, Jiang-Jiang and Jing, You-Yin and Zhang, Chun-Fa and Zhao, Jun-Hong},
	Journal = {Renewable and Sustainable Energy Reviews},
	Number = {9},
	Pages = {2263--2278},
	Publisher = {Elsevier},
	Title = {Review on multi-criteria decision analysis aid in sustainable energy decision-making},
	Volume = {13},
	Year = {2009}}

@book{souami2012ecoquartiers,
	Author = {Souami, Taoufik},
	Publisher = {Scrineo},
	Title = {Ecoquartiers: secrets de fabrication},
	Year = {2012}}

@article{franco20092,
	Author = {Franco, Jessica and Dupuy, Delphine and Roustant, Olivier and Jourdan, Astrid},
	Journal = {Designs of Computer Experiments},
	Pages = {2},
	Title = {DiceDesign-package},
	Year = {2009}}

@article{newman2003structure,
  title={The structure and function of complex networks},
  author={Newman, Mark EJ},
  journal={SIAM review},
  volume={45},
  number={2},
  pages={167--256},
  year={2003},
  publisher={SIAM}
}

@book{west2017scale,
  title={Scale: The Universal Laws of Growth, Innovation, Sustainability, and the Pace of Life in Organisms, Cities, Economies, and Companies},
  author={West, Geoffrey},
  year={2017},
  publisher={Penguin}
}

@article{jacomy2014forceatlas2,
  title={ForceAtlas2, a continuous graph layout algorithm for handy network visualization designed for the Gephi software},
  author={Jacomy, Mathieu and Venturini, Tommaso and Heymann, Sebastien and Bastian, Mathieu},
  journal={PloS one},
  volume={9},
  number={6},
  pages={e98679},
  year={2014},
  publisher={Public Library of Science}
}

@article{blondel2008fast,
	Author = {Blondel, Vincent D and Guillaume, Jean-Loup and Lambiotte, Renaud and Lefebvre, Etienne},
	Journal = {Journal of statistical mechanics: theory and experiment},
	Number = {10},
	Pages = {P10008},
	Title = {Fast unfolding of communities in large networks},
	Volume = {2008},
	Year = {2008}}

@article{redner1998popular,
  title={How popular is your paper? An empirical study of the citation distribution},
  author={Redner, Sidney},
  journal={The European Physical Journal B-Condensed Matter and Complex Systems},
  volume={4},
  number={2},
  pages={131--134},
  year={1998},
  publisher={Springer}
}

@article{pumain2015adapting,
  title={Adapting the model of scientific publishing},
  author={Pumain, Denise},
  journal={Cybergeo: European Journal of Geography},
  year={2015},
  publisher={CNRS-UMR G{\'e}ographie-cit{\'e}s 8504}
}

@book{bracken2016interdisciplinarity,
  title={Interdisciplinarity and Geography},
  author={Bracken, Louise J},
  year={2016},
  publisher={Wiley Online Library}
}

@article{rinia2002impact,
  title={Impact measures of interdisciplinary research in physics},
  author={Rinia, Ed and van Leeuwen, Thed and van Raan, Anthony},
  journal={Scientometrics},
  volume={53},
  number={2},
  pages={241--248},
  year={2002},
  publisher={Akad{\'e}miai Kiad{\'o}, co-published with Springer Science+ Business Media BV, Formerly Kluwer Academic Publishers BV}
}

@article{porter2009science,
  title={Is science becoming more interdisciplinary? Measuring and mapping six research fields over time},
  author={Porter, Alan and Rafols, Ismael},
  journal={Scientometrics},
  volume={81},
  number={3},
  pages={719--745},
  year={2009},
  publisher={Akad{\'e}miai Kiad{\'o}, co-published with Springer Science+ Business Media BV, Formerly Kluwer Academic Publishers BV}
}

@article{porter2007measuring,
  title={Measuring researcher interdisciplinarity},
  author={Porter, Alan and Cohen, Alex S and Roessner, J David and Perreault, Marty},
  journal={Scientometrics},
  volume={72},
  number={1},
  pages={117--147},
  year={2007},
  publisher={Springer}
}

@article{leydesdorff2007betweenness,
  title={Betweenness centrality as an indicator of the interdisciplinarity of scientific journals},
  author={Leydesdorff, Loet},
  journal={Journal of the Association for Information Science and Technology},
  volume={58},
  number={9},
  pages={1303--1319},
  year={2007},
  publisher={Wiley Online Library}
}

@article{lariviere2010relationship,
  title={On the relationship between interdisciplinarity and scientific impact},
  author={Larivi{\`e}re, Vincent and Gingras, Yves},
  journal={Journal of the Association for Information Science and Technology},
  volume={61},
  number={1},
  pages={126--131},
  year={2010},
  publisher={Wiley Online Library}
}

@article{austin1996defining,
  title={Defining interdisciplinarity},
  author={Austin, Timothy R and Rauch, Alan and Blau, Herbert and Yudice, George and van Den Berg, Sara and Robinson, Lillian S and Henkel, Jacqueline and Murray, Timothy and Schoenfield, Mark and Traub, Valerie and others},
  journal={Publications of the Modern Language Association of America},
  pages={271--282},
  year={1996},
  publisher={JSTOR}
}

@article{huutoniemi2010analyzing,
  title={Analyzing interdisciplinarity: Typology and indicators},
  author={Huutoniemi, Katri and Klein, Julie Thompson and Bruun, Henrik and Hukkinen, Janne},
  journal={Research Policy},
  volume={39},
  number={1},
  pages={79--88},
  year={2010},
  publisher={Elsevier}
}

@article{moreno2016uncertainty,
  title={On the uncertainty of interdisciplinarity measurements due to incomplete bibliographic data},
  author={Moreno, Mar{\'\i}a del Carmen Calatrava and Auzinger, Thomas and Werthner, Hannes},
  journal={Scientometrics},
  volume={107},
  number={1},
  pages={213--232},
  year={2016},
  publisher={Springer}
}

@article{morin1986methode,
  title={La M{\'e}thode, tome 3. La connaissance de la connaissance},
  author={Morin, Edgar},
  journal={Le Seuil},
  year={1986}
}

@inproceedings{baldwin2010language,
  title={Language identification: The long and the short of the matter},
  author={Baldwin, Timothy and Lui, Marco},
  booktitle={Human Language Technologies: The 2010 Annual Conference of the North American Chapter of the Association for Computational Linguistics},
  pages={229--237},
  year={2010},
  organization={Association for Computational Linguistics}
}

@book{bais2010praise,
  title = {In Praise of Science: Curiosity, Understanding, and Progress},
  author = {Bais, Sanders},
  year= {2010},
  publisher={MIT Press}
}

@article{lariviere201410,
  title={Measuring Interdisciplinarity},
  author={Larivi{\`e}re, Vincent and Gingras, Yves},
  journal={Beyond bibliometrics: Harnessing multidimensional indicators of scholarly impact},
  pages={187},
  year={2014},
  publisher={MIT Press}
}

@article{nichols2014topic,
  title={A topic model approach to measuring interdisciplinarity at the National Science Foundation},
  author={Nichols, Leah G},
  journal={Scientometrics},
  volume={100},
  number={3},
  pages={741--754},
  year={2014},
  publisher={Springer}
}

@article{natureInterdisc,
    title = {Interdisciplinarity, Nature Special Issue},
    author = {Nature},
    journal={Nature},
    month={September},
    volume={525},
    number = {7569},
    pages = {289--418},
    year = {2015},
    publisher={Nature} 
}

@misc{torpool,
  title = {TorPool v1.0, DOI : 10.5281/zenodo.53739},
  author = {Raimbault, Juste},
  year = {2016}
}

@misc{mendeley,
	Author = {Mendeley},
	Howpublished = {\url{http://www.mendeley.com}},
	Title = {Mendeley Reference Manager},
	Year = {2015}}

@article{pumain2005cumulativite,
	Author = {Pumain, Denise},
	Journal = {Revue europ{\'e}enne des sciences sociales. European Journal of Social Sciences},
	Number = {XLIII-131},
	Pages = {5--12},
	Publisher = {Librairie Droz},
	Title = {Cumulativit{\'e} des connaissances},
	Year = {2005}}

@article{2009arXiv0907.2221B,
	Adsurl = {http://adsabs.harvard.edu/abs/2009arXiv0907.2221B},
	Archiveprefix = {arXiv},
	Author = {{Bourgine}, P. and {Chavalarias}, D. and al.},
	Eprint = {0907.2221},
	Isipubdate = {and {Perrier}, E. and {Amblard}, F. and {Arlabosse}, F. and {Auger}, P. and {Baillon}, J.-B. and {Barreteau}, O. and {Baudot}, P. and {Bouchaud}, E. and {Ben Amor}, S. and {Berry}, H. and {Bertelle}, C. and {Berthod}, M. and {Beslon}, G. and {Biroli}, G. and {Bonamy}, D. and {Bourcier}, D. and {Brodu}, N. and {Bui}, M. and {Burnod}, Y. and {Chapron}, B. and {Christophe}, C. and {Clement}, B. and {Coatrieux}, J.-L. and {Cointet}, J.-P. and {Dagrain}, V. and {Dauchot}, K. and {Dauchot}, O. and {Daviaud}, F. and {De Monte}, S. and {Deffuant}, G. and {Degond}, P. and {Delahaye}, J.-P. and {Doursat}, R. and {D'Ovidio}, F. and {Dubois}, M. and {Dubruelle}, B. and {Dutreix}, M. and {Faivre}, R. and {Farge}, E. and {Flandrin}, P. and {Franceschelli}, S. and {Gaucherel}, C. and {Gaudin}, J.-P. and {Ghil}, M. and {Giavitto}, J.-L. and {Ginelli}, F. and {Ginot}, V. and {Houllier}, F. and {Hubert}, B. and {Jensen}, P. and {Jullien}, L. and {Kapoula}, Z. and {Krob}, D. and {Ladieu}, F. and {Lang}, G. and {Lavelle}, C. and {Le Bivic}, A. and {Leca}, J.-P. and {Lecerf}, C. and {Legrain}, P. and {L'hote}, D. and {Loireau}, M. and {Mangin}, J.-F. and {Monga}, O. and {Morvan}, M. and {Muller}, J.-P. and {Negrutiu}, I. and {Peyreiras}, N. and {Pumain}, D. and {Radulescu}, O. and {Sallantin}, J. and {Sanchis}, E. and {Schertzer}, D. and {Schoenauer}, M. and {Sebag}, M. and {Simonet}, E. and {Six}, A. and {Tarissan}, F. and {Vincent}, P.},
	Journal = {ArXiv e-prints},
	Month = jul,
	Primaryclass = {nlin.AO},
	Title = {{French Roadmap for complex Systems 2008-2009}},
	Year = 2009}

@inproceedings{schmid1994probabilistic,
	Author = {Schmid, Helmut},
	Booktitle = {Proceedings of the international conference on new methods in language processing},
	Pages = {44--49},
	Title = {Probabilistic part-of-speech tagging using decision trees},
	Volume = {12},
	Year = {1994}}

@article{noruzi2005google,
	Author = {Noruzi, Alireza},
	Journal = {Libri},
	Number = {4},
	Pages = {170--180},
	Title = {Google Scholar: The new generation of citation indexes},
	Volume = {55},
	Year = {2005}}

@article{10.1371/journal.pone.0147913,
	Author = {Wicherts, Jelte M.},
	Doi = {10.1371/journal.pone.0147913},
	Journal = {PLoS ONE},
	Month = {01},
	Number = {1},
	Pages = {e0147913},
	Publisher = {Public Library of Science},
	Title = {Peer Review Quality and Transparency of the Peer-Review Process in Open Access and Subscription Journals},
	Volume = {11},
	Year = {2016}}

@incollection{pumain2012urban,
	Author = {Pumain, Denise},
	Booktitle = {Complexity Theories of Cities Have Come of Age},
	Pages = {91--103},
	Publisher = {Springer},
	Title = {Urban systems dynamics, urban growth and scaling laws: The question of ergodicity},
	Year = {2012}}

@misc{li2009netlogo,
  title={NetLogo Sugarscape 3 Wealth Distribution model},
  author={Li, J and Wilensky, U},
  year={2009},
  publisher={Evanston, IL: Center for Connected Learning and Computer-Based Modeling, Northwestern Institute on Complex Systems, Northwestern University. http://ccl. northwestern. edu/netlogo/models/Sugarscape3WealthDistribution}
}

@book{gell1995quark,
  title={The Quark and the Jaguar: Adventures in the Simple and the Complex},
  author={Gell-Mann, Murray},
  year={1995},
  publisher={Macmillan}
}

@article{rubner2000earth,
  title={The earth mover's distance as a metric for image retrieval},
  author={Rubner, Yossi and Tomasi, Carlo and Guibas, Leonidas J},
  journal={International journal of computer vision},
  volume={40},
  number={2},
  pages={99--121},
  year={2000},
  publisher={Springer}
}

@article{visser2006map,
  title={The map comparison kit},
  author={Visser, Hans and De Nijs, T},
  journal={Environmental Modelling \& Software},
  volume={21},
  number={3},
  pages={346--358},
  year={2006},
  publisher={Elsevier}
}

@article{10.1371/journal.pone.0178165,
    author = {Koch, Julian AND Stisen, Simon},
    journal = {PLOS ONE},
    publisher = {Public Library of Science},
    title = {Citizen science: A new perspective to advance spatial pattern evaluation in hydrology},
    year = {2017},
    month = {05},
    volume = {12},
    url = {https://doi.org/10.1371/journal.pone.0178165},
    pages = {1-20},
    number = {5},
    doi = {10.1371/journal.pone.0178165}
}

@inproceedings{cottineau2016back,
	Author = {Cottineau, Cl{\'e}mentine and Rey, S{\'e}bastien and Reuillon, Romain},
	Booktitle = {Royal Geographical Society-Annual Conference 2016-Session: Geocomputation, the Next 20 Years (1)},
	Title = {Back to the Future of Multimodeling},
	Year = {2016}}

@article{AllenSanglier1979,
	title = { A dynamic model of growth in a central place system},
	volume = {11},
	journal = {Geographical Analysis},
	author = {Allen, P. and Sanglier, M.},
	year = {1979},
	pages = {256-272}
}

@book{EpsteinAxtell1996,
  title={Growing artificial societies: Social science from the bottom up (complex adaptive systems)},
  author={Epstein, Joshua M and Axtell, Robert L},
  year={1996},
  publisher={Brookings Institution Press MIT Press}
}

@article{FotheringhamWong1991,
	title = {The modifiable areal unit problem in multivariate statistical analysis},
	volume = {23},
    number = {7},
	journal = {Environment and Planning A},
	author = {Fotheringham, A. S. and Wong, D. W. S.},
	year = {1991},
	pages = {1025-1044}
}

@article{Kwan2012,
	title = {The uncertain geographic context problem},
	volume = {102},
    number = {5},
	journal = {Annals of the Association of American Geographers},
	author = {Kwan, M.P.},
	year = {2012},
	pages = {958-968}
}

@article{LeTexierCaruso2017,
	title = {Assessing geographical effects in spatial diffusion processes: The case of euro coins},
	journal = {Computer, Environment and Urban Systems},
	author = {Le Texier, Marion and Caruso, Geoffrey},
	year = {2017},
	volume = {61},
    number = {A},
    pages = {81-93}
}

@book{Openshaw1984,
  title={The Modifiable Areal Unit Problem},
  author={Openshaw, Stan},
  year={1984},
  publisher={Geo Books},
  address={Norwich, UK}
}

@article{Thomasetal2017,
  title={City delineation in European applications of LUTI models: review and tests},
  author={Thomas, Isabelle and Jones, Jonathan and Caruso, Geoffrey and Gerber, Philippe},
  journal={Transport Reviews},
  volume={38},
  number={1},
  pages={6--32},
  year={2018},
  publisher={Taylor \& Francis}
}

@book{Wilson1981,
  title={Catastrophe theory and bifurcation: Application to Urban and Regional System},
  author={Wilson, A.},
  year={1981},
  publisher={Croom Helm}
}

@book{mantegna1999introduction,
  title={Introduction to econophysics: correlations and complexity in finance},
  author={Mantegna, Rosario N and Stanley, H Eugene},
  year={1999},
  publisher={Cambridge university press}
}

@article{nitsch2005zipf,
  title={Zipf zipped},
  author={Nitsch, Volker},
  journal={Journal of Urban Economics},
  volume={57},
  number={1},
  pages={86--100},
  year={2005},
  publisher={Elsevier}
}

@article{storper2009rethinking,
  title={Rethinking human capital, creativity and urban growth},
  author={Storper, Michael and Scott, Allen J},
  journal={Journal of economic geography},
  volume={9},
  number={2},
  pages={147--167},
  year={2009},
  publisher={Oxford University Press}
}

@article{masucci2013gravity,
  title={Gravity versus radiation models: On the importance of scale and heterogeneity in commuting flows},
  author={Masucci, A Paolo and Serras, Joan and Johansson, Anders and Batty, Michael},
  journal={Physical Review E},
  volume={88},
  number={2},
  pages={022812},
  year={2013},
  publisher={APS}
}

@article{batty1972calibration,
  title={The calibration of gravity, entropy, and related models of spatial interaction},
  author={Batty, Michael and Mackie, S},
  journal={Environment and Planning A},
  volume={4},
  number={2},
  pages={205--233},
  year={1972},
  publisher={SAGE Publications Sage UK: London, England}
}

@article{dietterich1995overfitting,
  title={Overfitting and undercomputing in machine learning},
  author={Dietterich, Tom},
  journal={ACM computing surveys (CSUR)},
  volume={27},
  number={3},
  pages={326--327},
  year={1995},
  publisher={ACM}
}

@article{guerin1990150,
  title={150 ans de croissance urbaine},
  author={Gu{\'e}rin-Pace, France and Pumain, Denise},
  journal={Economie et statistique},
  volume={230},
  number={1},
  pages={5--16},
  year={1990},
  publisher={Pers{\'e}e-Portail des revues scientifiques en SHS}
}

@article{pumain1986fichier,
	Author = {Pumain, Denise and Riandey, Beno{\^\i}t},
	Journal = {Espace, populations, soci{\'e}t{\'e}s},
	Number = {2},
	Pages = {269--277},
	Publisher = {Pers{\'e}e-Portail des revues scientifiques en SHS},
	Title = {Le Fichier de l'Ined},
	Volume = {4},
	Year = {1986}}

@book{sanders1992systeme,
	Author = {Sanders, Lena},
	Publisher = {Economica},
	Title = {Syst{\`e}me de villes et synerg{\'e}tique},
	Year = {1992}}

@incollection{taylor2016polymath,
	Author = {Taylor, Peter J},
	Booktitle = {Sir Peter Hall: Pioneer in Regional Planning, Transport and Urban Geography},
	Pages = {11--20},
	Publisher = {Springer},
	Title = {A Polymath in City Studies},
	Year = {2016}}

@article{krugman1998space,
	Author = {Krugman, Paul},
	Journal = {The Journal of Economic Perspectives},
	Number = {2},
	Pages = {161--174},
	Publisher = {JSTOR},
	Title = {Space: the final frontier},
	Volume = {12},
	Year = {1998}}

@article{favaro2011gibrat,
	Author = {Favaro, Jean-Marc and Pumain, Denise},
	Journal = {Geographical Analysis},
	Number = {3},
	Pages = {261--286},
	Publisher = {Wiley Online Library},
	Title = {Gibrat Revisited: An Urban Growth Model Incorporating Spatial Interaction and Innovation Cycles.},
	Volume = {43},
	Year = {2011}}

@article{bretagnolle2000long,
	Author = {Bretagnolle, Anne and Mathian, H{\'e}l{\`e}ne and Pumain, Denise and Rozenblat, C{\'e}line},
	Journal = {Cybergeo: European Journal of Geography},
	Publisher = {CNRS-UMR G{\'e}ographie-cit{\'e}s 8504},
	Title = {Long-term dynamics of European towns and cities: towards a spatial model of urban growth},
	Year = {2000}}

@incollection{pumain2012multi,
	Author = {Pumain, Denise},
	Booktitle = {Agent-based models of geographical systems},
	Pages = {721--738},
	Publisher = {Springer},
	Reading = {A},
	Title = {Multi-agent system modelling for urban systems: The series of SIMPOP models},
	Usage = {GSB},
	Year = {2012}}

@article{chicheportiche2013nested,
	Author = {Chicheportiche, R{\'e}my and Bouchaud, Jean-Philippe},
	Journal = {arXiv preprint arXiv:1309.3102},
	Title = {A nested factor model for non-linear dependences in stock returns},
	Year = {2013}}

@article{pigozzi1980interurban,
	Author = {Pigozzi, Bruce Wm},
	Journal = {Geographical Analysis},
	Number = {4},
	Pages = {340--352},
	Publisher = {Wiley Online Library},
	Title = {Interurban linkages through polynomially constrained distributed lags},
	Volume = {12},
	Year = {1980}}

@article{girres2010quality,
	Author = {Girres, Jean-Fran{\c{c}}ois and Touya, Guillaume},
	Journal = {Transactions in GIS},
	Number = {4},
	Pages = {435--459},
	Publisher = {Wiley Online Library},
	Title = {Quality assessment of the French OpenStreetMap dataset},
	Volume = {14},
	Year = {2010}}

@article{offner1996reseaux,
	Author = {Offner, Jean-Marc and Pumain, Denise},
	Publisher = {Editions de l'Aube},
	Title = {R{\'e}seaux et territoires-significations crois{\'e}es},
	Year = {1996}}

@article{sanders1997simpop,
	Author = {Sanders, Lena and Pumain, Denise and Mathian, H{\'e}lene and Gu{\'e}rin-Pace, France and Bura, Stephane},
	Journal = {Environment and Planning B},
	Pages = {287--306},
	Publisher = {Pion Ltd},
	Title = {SIMPOP: a multiagent system for the study of urbanism},
	Volume = {24},
	Year = {1997}}

@phdthesis{potiron2016estimating,
  title={Estimating the integrated parameter of the locally parametric model in high-frequency data},
  author={Potiron, Yoann},
  year={2016},
  school={The University of Chicago}
}

@incollection{banos2012towards,
	Author = {Banos, Arnaud and Genre-Grandpierre, Cyrille},
	Booktitle = {Agent-based models of geographical systems},
	Pages = {627--641},
	Publisher = {Springer},
	Title = {Towards new metrics for urban road networks: Some preliminary evidence from agent-based simulations},
	Year = {2012}}

@article{barndorff2011multivariate,
	Author = {Barndorff-Nielsen, Ole E and Hansen, Peter Reinhard and Lunde, Asger and Shephard, Neil},
	Journal = {Journal of Econometrics},
	Pages = {149--169},
	Publisher = {Elsevier},
	Title = {Multivariate realised kernels: consistent positive semi-definite estimators of the covariation of equity prices with noise and non-synchronous trading},
	Volume = {162},
	Year = {2011}}

@article{2001PhyA..299...16B,
	Adsurl = {http://adsabs.harvard.edu/abs/2001PhyA..299...16B},
	Author = {{Bonanno}, G. and {Lillo}, F. and {Mantegna}, R.~N.},
	Eprint = {cond-mat/0104369},
	Journal = {Physica A Statistical Mechanics and its Applications},
	Month = oct,
	Pages = {16-27},
	Title = {{Levels of complexity in financial markets}},
	Volume = 299,
	Year = 2001}

@article{potiron2015estimation,
	Author = {Potiron, Yoann and Mykland, Per},
	Journal = {arXiv preprint arXiv:1507.01033},
	Title = {Estimation of integrated quadratic covariation between two assets with endogenous sampling times},
	Year = {2015}}

@article{ramsey2002wavelets,
	Author = {Ramsey, James B},
	Journal = {Studies in Nonlinear Dynamics \& Econometrics},
	Title = {Wavelets in economics and finance: Past and future},
	Volume = {6},
	Year = {2002}}

@article{bouchaud2000apparent,
	Author = {Bouchaud, J-P and Potters, Marc and Meyer, Martin},
	Journal = {The European Physical Journal B-Condensed Matter and Complex Systems},
	Number = {3},
	Pages = {595--599},
	Publisher = {Springer},
	Title = {Apparent multifractality in financial time series},
	Volume = {13},
	Year = {2000}}

@article{jarrow1999honor,
	Author = {Jarrow, Robert A},
	Journal = {The Journal of Economic Perspectives},
	Pages = {229--248},
	Publisher = {JSTOR},
	Title = {In Honor of the Nobel Laureates Robert C. Merton and Myron S. Scholes: A Partial Differential Equation that Changed the World},
	Year = {1999}}

@manual{Tsay:2015xy,
	Author = {Ruey S. Tsay},
	Note = {R package version 0.33},
	Title = {MTS: All-Purpose Toolkit for Analyzing Multivariate Time Series (MTS) and Estimating Multivariate Volatility Models},
	Url = {http://CRAN.R-project.org/package=MTS},
	Year = {2015}}

@article{ye2011investigation,
	Author = {Ye, Xin},
	Journal = {Transportation Research Record: Journal of the Transportation Research Board},
	Number = {2254},
	Pages = {36--43},
	Publisher = {Transportation Research Board of the National Academies},
	Title = {Investigation of Underlying Distributional Assumption in Nested Logit Model Using Copula-Based Simulation and Numerical Approximation},
	Year = {2011}}

@article{van2006syntren,
	Author = {Van den Bulcke, Tim and Van Leemput, Koenraad and Naudts, Bart and van Remortel, Piet and Ma, Hongwu and Verschoren, Alain and De Moor, Bart and Marchal, Kathleen},
	Journal = {BMC bioinformatics},
	Number = {1},
	Pages = {43},
	Publisher = {BioMed Central Ltd},
	Title = {SynTReN: a generator of synthetic gene expression data for design and analysis of structure learning algorithms},
	Volume = {7},
	Year = {2006}}

@article{abadie2010synthetic,
	Author = {Abadie, Alberto and Diamond, Alexis and Hainmueller, Jens},
	Journal = {Journal of the American Statistical Association},
	Number = {490},
	Title = {Synthetic control methods for comparative case studies: Estimating the effect of California's tobacco control program},
	Volume = {105},
	Year = {2010}}

@article{tumminello2005tool,
	Author = {Tumminello, Michele and Aste, Tomaso and Di Matteo, Tiziana and Mantegna, Rosario N},
	Journal = {Proceedings of the National Academy of Sciences of the United States of America},
	Pages = {10421--10426},
	Publisher = {National Acad Sciences},
	Title = {A tool for filtering information in complex systems},
	Volume = {102},
	Year = {2005}}

@article{2009arXiv0910.1205B,
	Adsurl = {http://adsabs.harvard.edu/abs/2009arXiv0910.1205B},
	Archiveprefix = {arXiv},
	Author = {{Bouchaud}, J.~P. and {Potters}, M.},
	Eprint = {0910.1205},
	Journal = {arXiv preprint arXiv:0910.1205},
	Month = oct,
	Primaryclass = {q-fin.ST},
	Title = {{Financial Applications of Random Matrix Theory: a short review}},
	Year = 2009}

@inproceedings{pritchard2009advances,
	Author = {Pritchard, David R and Miller, Eric J},
	Booktitle = {Transportation Research Board 88th Annual Meeting},
	Number = {09-1686},
	Title = {Advances in agent population synthesis and application in an integrated land use and transportation model},
	Year = {2009}}

@inproceedings{moeckel2003creating,
	Author = {Moeckel, Rolf and Spiekermann, Klaus and Wegener, Michael},
	Booktitle = {Proceedings of the 8th International Conference on Computers in Urban Planning and Urban Management (CUPUM)},
	Title = {Creating a synthetic population},
	Year = {2003}}

@article{bolon2013review,
	Author = {Bol{\'o}n-Canedo, Ver{\'o}nica and S{\'a}nchez-Maro{\~n}o, Noelia and Alonso-Betanzos, Amparo},
	Journal = {Knowledge and information systems},
	Number = {3},
	Pages = {483--519},
	Publisher = {Springer},
	Title = {A review of feature selection methods on synthetic data},
	Volume = {34},
	Year = {2013}}

@article{brown2009models,
  title={Models and perspectives on stage: remarks on Giere’s scientific perspectivism},
  author={Brown, Matthew J},
  journal={Studies in History and Philosophy of Science Part A},
  volume={40},
  number={2},
  pages={213--220},
  year={2009},
  publisher={Elsevier}
}

@article{chavalarias2016s,
  title={What's wrong with Science?},
  author={Chavalarias, David},
  journal={Scientometrics},
  pages={1--23},
  year={2016},
  publisher={Springer}
}

@book{hofstadter1980godel,
  title={G{\"o}del, Escher, Bach: An Eternal Golden Braid;[a Metaphoric Fugue on Minds and Machines in the Spirit of Lewis Carroll].},
  author={Hofstadter, Douglas H},
  year={1980},
  publisher={Penguin Books}
}

@ARTICLE{2017arXiv170401407M,
   author = {{Moulin-Frier}, C. and {Puigb{\`o}}, J.-Y. and {Arsiwalla}, X.~D. and 
	{Sanchez-Fibla}, M. and {Verschure}, P.~F.~M.~J.},
    title = {Embodied Artificial Intelligence through Distributed Adaptive Control: An Integrated Framework},
  journal = {arXiv preprint arXiv:1704.01407},
archivePrefix = {arXiv},
   eprint = {1704.01407},
 primaryClass = {cs.AI},
     year = {2017},
    month = {apr},
   adsurl = {http://adsabs.harvard.edu/abs/2017arXiv170401407M}
}

@book{kuhn1970structure,
  title={The structure of scientific revolutions},
  author={Kuhn, Thomas S},
  year={1970},
  publisher={The University of Chicago Press}
}

@preamble{"\newcommand{\noopsort}[1]{} " #
   "\newcommand{\printfirst}[2]{#1} " #
   "\newcommand{\singleletter}[1]{#1} " #
   "\newcommand{\switchargs}[2]{#2#1} "}

@string{acm = {The OX Association for Computing Machinery}}


\appendix


\captionsetup{list=no}

\ctparttext{\bpar{Appendices are organized in the logic of knowledge domains: after a linear presentation of diverse supplementary materials for each section of main text, we introduce methodological developments (domain of methods), thematical developments (empirical domain), a synthesis of developed softwares (domain of tools), a synthesis of constructed datasets (domain of data). We conclude with a short reflexive analysis of the content of this memoire.}{Les annexes sont organisées dans la logique des domaines de connaissance : après une présentation linéaire des diverses informations supplémentaires pour chaque section du texte principal, nous introduisons des développements méthodologiques (domaine des méthodes), des développements thématiques (domaine empirique), une synthèse des logiciels développés (domaine des outils), une synthèse des jeux de données construits (domaine des données). Nous concluons par une courte analyse reflexive du contenu de ce mémoire.}}

\part{Appendices}


%

\chapter{Informations supplémentaires}

\markboth{\thechapter\space Informations supplémentaires}{\thechapter\space Informations supplémentaires}

\label{app:supplementary} 


\bpar{
This appendix gathers various supplementary materials, necessary for the robustness but not necessary to the main argument. It includes for example in the case of simulation models more precise explorations and sensitivity analyses.
}{
Cette annexe regroupe divers matériaux supplémentaires, nécessaire à la robustesse des études mais pas à l'argumentaire général. Elle inclut par exemple dans le cas des modèles de simulation des explorations plus précises et des analyses de sensibilité.
}


\bpar{
It includes in particular the following points:
\begin{itemize}
	\item Fieldwork observations in China in~\ref{app:sec:qualitative}, for the qualitative results presented in Chapter~\ref{ch:thematic}.
	\item Precisions for the quantitative epistemology of~\ref{sec:quantepistemo} in~\ref{app:sec:quantepistemo}.
	\item Complete results for the modelography of~\ref{sec:modelography} in~\ref{app:sec:modelography}.
	\item For the static correlations of~\ref{sec:staticcorrelations}: results for China, sensitivity analyses, network simplification algorithm, analytical derivations for the multi-scale aspect in~\ref{app:sec:staticcorrelations}.
	\item Derivations for the expression of lagged correlations on synthetic data of~\ref{sec:causalityregimes} in~\ref{app:sec:causalityregimes}.
	\item Behavior of the model and semi-analytical study of the aggregation-diffusion model of~\ref{sec:densitygeneration} in~\ref{app:sec:density}.
	\item Feasible correlations for the weak coupling of~\ref{sec:correlatedsyntheticdata} in~\ref{app:sec:correlatedsyntheticdata}.
	\item Extended figures for the exploration of the SimpopNet model of~\ref{sec:macrocoevolexplo} in~\ref{app:sec:macrocoevolexplo}.
	\item Extended figures for the exploration of the macroscopic co-evolution model of~\ref{sec:macrocoevol} in~\ref{app:sec:macrocoevolexplo}.
	\item Details of the \emph{slime mould} model used in~\ref{sec:networkgrowth}, and extended figures in~\ref{app:sec:networkgrowth}.
	\item Second order calibration process for the mesoscopic co-evolution model of~\ref{sec:mesocoevolmodel} in~\ref{app:sec:mesocoevolmodel}.
	\item For the Lutecia model of~\ref{sec:lutecia}, study of the land-use model, derivation of cooperation probabilities, implementation and initialization details in~\ref{app:sec:lutecia}.
\end{itemize}
}{
Elle inclut notamment les points suivants :
\begin{itemize}
	\item Relevés de terrain en Chine en~\ref{app:sec:qualitative}, pour les résultats qualitatifs présentés en Chapitre~\ref{ch:thematic}.
	\item Précisions pour l'épistémologie quantitative de~\ref{sec:quantepistemo} en~\ref{app:sec:quantepistemo}.
	\item Résultats complets pour la modélographie de~\ref{sec:modelography} en~\ref{app:sec:modelography}.
	\item Pour les corrélations statiques de~\ref{sec:staticcorrelations} : résultats pour la Chine, analyses de sensibilité, algorithme de simplification de réseau, dérivation analytiques pour le caractère multi-échelle en~\ref{app:sec:staticcorrelations}.
	\item Dérivations pour l'expression des corrélations retardées sur données synthétiques de~\ref{sec:causalityregimes} en~\ref{app:sec:causalityregimes}.
	\item Comportement du modèle et étude semi-analytique du modèle d'agrégation-diffusion de~\ref{sec:densitygeneration} en~\ref{app:sec:density}.
	\item Corrélations faisable pour le couplage faible de~\ref{sec:correlatedsyntheticdata} en~\ref{app:sec:correlatedsyntheticdata}.
	\item Figures étendues pour l'exploration du modèle SimpopNet de~\ref{sec:macrocoevolexplo} en~\ref{app:sec:macrocoevolexplo}.
	\item Figures étendues pour l'exploration du modèle macroscopique de co-évolution de~\ref{sec:macrocoevol} en~\ref{app:sec:macrocoevolexplo}.
	\item Détails du modèle \emph{slime mould} utilisé en~\ref{sec:networkgrowth}, et figures étendues en~\ref{app:sec:networkgrowth}
	\item Processus de calibration au second ordre du modèle mesoscopique de co-évolution de~\ref{sec:mesocoevolmodel} en~\ref{app:sec:mesocoevolmodel}.
	\item Pour le modèle Lutecia de~\ref{sec:lutecia}, étude du modèle d'usage du sol, dérivations de probabilités de coopération, détails d'implémentation et d'initialisation en~\ref{app:sec:lutecia}.
\end{itemize}
}

\stars

%


\newpage

\section{Fieldwork Elements}{Élements de terrain}

\label{app:sec:qualitative}

\subsection{Localization of fieldworks in China}{Localisation des terrains en Chine}

\bpar{
We precise the geographical localization of territories and places evoked in~\ref{sec:casestudies} and in~\ref{sec:qualitative} in the following maps. We give:
\begin{itemize}
	\item A map in Fig.~\ref{fig:app:casestudies:nanfang} at the scale of South China, which allows to locate Pearl River Delta (which includes Guangzhou and Zhuhai), Chengdu and Leshan, and also Yangshuo.
	\item A map in Fig.~\ref{fig:app:casestudies:prd} at the scale of Pearl River Delta, which allows to locate the main cities: Guangzhou/Foshan, Dongguan, Zhongshan, Zhuhai and Shenzhen (SEZ), and also Hong-Kong and Macao (SAR).
	\item A map in Fig.~\ref{fig:app:casestudies:zhuhai} at the scale of Zhuhai, which allows to locate the different districts of Zhuhai: Gongbei, Xiangzhou, Tangjia, and also Zhuhai Bei railway station, the HZMB bridge and the \emph{New Territories} in Hong-Kong (we designate by district here not administrative districts, since for example Tangjia is included in Xinwan district, but perceived districts).
\end{itemize}
}{
Nous précisons la localisation géographique des territoires et lieux évoqués en~\ref{sec:casestudies} et en~\ref{sec:qualitative} dans les cartes suivantes. Nous donnons :
\begin{itemize}
	\item Une carte en Fig.~\ref{fig:app:casestudies:nanfang} à l'échelle du sud de la Chine, qui permet de localiser le Delta de la Rivière des Perles (qui inclut Guangzhou et Zhuhai), Chengdu et Leshan, ainsi que Yangshuo.
	\item Une carte en Fig.~\ref{fig:app:casestudies:prd} à l'échelle du Delta de la Rivière des Perles, qui permet de localiser les principales villes : Guangzhou/Foshan, Dongguan, Zhongshan, Zhuhai et Shenzhen (ZES), ainsi que Hong-Kong et Macao (ZAS).
	\item Une carte en Fig.~\ref{fig:app:casestudies:zhuhai} à l'échelle de Zhuhai, qui permet de localiser les différents quartiers de Zhuhai : Gongbei, Xiangzhou, Tangjia, ainsi que la gare de Zhuhai Bei, le pont HZMB et les \emph{New Territories} à Hong-Kong (nous désignons par quartier ici non pas des districts administratifs, puisque par exemple Tangjia fait partie du district de Xinwan, mais des quartiers vécus).
\end{itemize}
}

\begin{figure}
	\includegraphics[width=\linewidth]{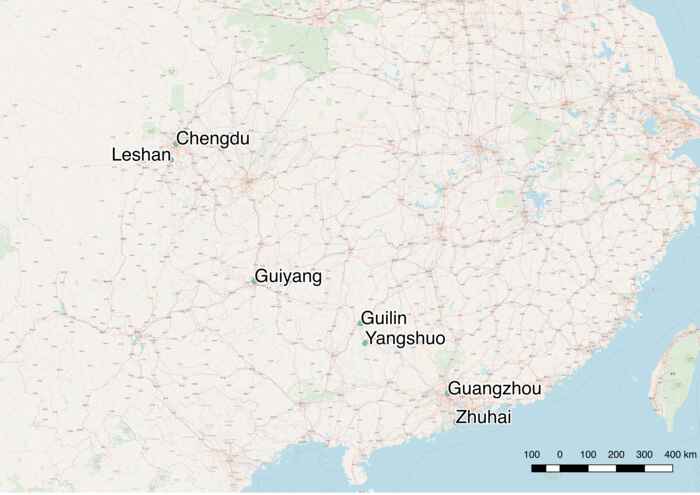}	
	\appcaption{\textbf{Localization of fieldwork places, } at the scale of South China. Source: OpenStreetMap.\label{fig:app:casestudies:nanfang}}{\textbf{Localisation des lieux de terrain, } à l'échelle du sud de la Chine. Source : OpenStreetMap.\label{fig:app:casestudies:nanfang}}
\end{figure}

\begin{figure}
	\includegraphics[width=\linewidth]{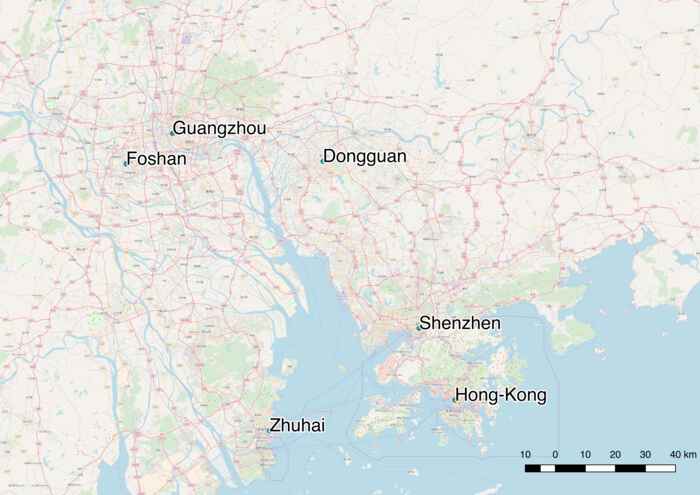}	
	\appcaption{\textbf{Localization of fieldwork places, } at the scale of Pearl River Delta. Source: OpenStreetMap.\label{fig:app:casestudies:prd}}{\textbf{Localisation des lieux de terrain, } à l'échelle du Delta de la Rivière des Perles. Source : OpenStreetMap.\label{fig:app:casestudies:prd}}
\end{figure}

\begin{figure}
	\includegraphics[width=\linewidth]{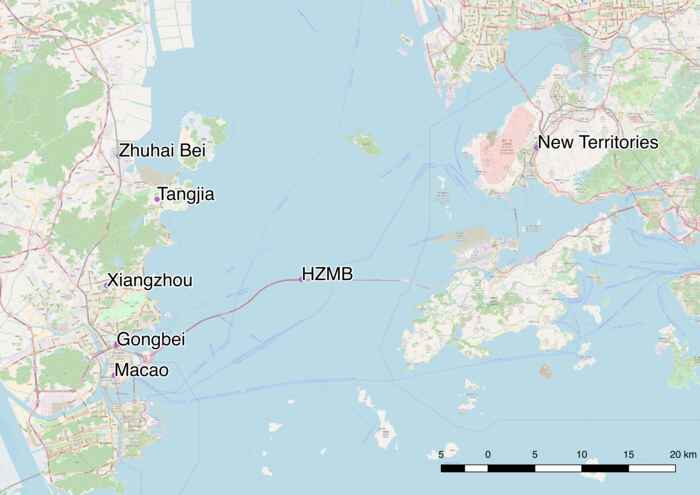}	
	\appcaption{\textbf{Localization of fieldwork places, } at the scale of Zhuhai. Source: OpenStreetMap.\label{fig:app:casestudies:zhuhai}}{\textbf{Localisation des lieux étudiés, } à l'échelle de Zhuhai. Source : OpenStreetMap.\label{fig:app:casestudies:zhuhai}}
\end{figure}


\subsection{Fieldwork Notebook}{Carnet de Terrain}

\bpar{
We summarize here in a synthetic way the different fieldwork journeys that feed section~\ref{sec:qualitative}. If it is not a priori a standard practice to give in a raw and open form the content of fieldwork notebooks, \cite{goffman1989fieldwork} highlights that this content can be a research material in itself. Raw reports and photographs are openly available at \url{https://github.com/JusteRaimbault/CityNetwork/tree/master/Data/Fieldwork}.
}{
Nous rendons compte ici de manière synthétique les différentes sorties de terrain alimentant la section~\ref{sec:qualitative}. S'il n'est a priori pas standard de fournir de manière brute et ouverte le contenu des carnets de terrain, \cite{goffman1989fieldwork} souligne que celui-ci peut être un matériau de recherche en lui-même. Les compte-rendus bruts et les photos sont disponibles de manière ouverte à \url{https://github.com/JusteRaimbault/CityNetwork/tree/master/Data/Fieldwork}.
}

\bpar{
Below are summarized the contexts and main observations of fieldworks. Places are localized in maps from Fig.~\ref{fig:app:casestudies:nanfang} to Fig~\ref{fig:app:casestudies:zhuhai}. Fieldworks were achieved alone, except when it is detailed otherwise for some.
}{
Ci-dessous sont résumés les contextes et observations principaux des sorties. Les lieux sont localisés dans les cartes de Fig.~\ref{fig:app:casestudies:nanfang} à Fig.~\ref{fig:app:casestudies:zhuhai}. Les sorties sont effectuées seul sauf si précisé pour certaines d'entre elles.
}


\paragraph{29/10/2016}{29/10/2016}

\bpar{
Trip in Zhuhai (Xiangzhou et Gongbei), with C. Losavio as a guide and interpreter. Nature in the city and use of the parks by inhabitants.
}{
Sortie à Zhuhai (Xiangzhou et Gongbei), avec C. Losavio pour guide et interprétation. Nature en ville et utilisation des parcs par les habitants.
}

\paragraph{06/11/2016}{06/11/2016}

\bpar{
Trip in Macao through Gongbei, with C. Losavio. Daily flows through the boundary of the SAR.
}{
Sortie à Macao par Gongbei, avec C. Losavio. Flux journaliers par la frontière de la ZAS.
}

\paragraph{07/11/2016}{07/11/2016}

\bpar{
Return trip Zhuhai-Hong-Kong. Observable relation of Zhuhai inhabitants to the SAR.
}{
Aller-retour Zhuhai-Hong-Kong. Relation apparente des habitants de Zhuhai à la ZAS.
}

\paragraph{16/01/2017}{16/01/2017}

\bpar{
Tentative to reach Guangzhou from Tangjia using city buses, full day. Final itinerary was Tangjia-Zhongshan-Xiaolan-Zhuahaibei. Local transport and urban fringes.
}{
Tentative de relier Tangjia à Guangzhou par bus de ville, journée. Itinéraire final Tangjia-Zhongshan-Xiaolan-Zhuahaibei. Transports locaux et franges urbaines.
}

\paragraph{11/12/2016}{11/12/2016}

\bpar{
From Beijing to Shenzhen through Guangzhou and Dongguan. Transportation, accessibility difficulties.
}{
De Pekin à Shenzhen par Guangzhou et Dongguan. Transports, difficultés d'accessibilité.
}

\paragraph{8/06/2017}{8/06/2017}

\bpar{
From Hong-Kong to Tangjia trough Zhuhai. Transportation.
}{
De Hong-Kong à Tangjia par Zhuhai. Transports.
}

\paragraph{19/06/2017}{19/06/2017}

\bpar{
Official fieldwork visit in the frame of the Medium Conference, Guangzhou, supervized by guides and interprets engaged by the SYSU. Urban renewal, urban projects, heritage. 
}{
Visite de terrain officielle dans le cadre de la Conférence Medium, Guangzhou, encadrée par guides et interprètes engagés par l'université SYSU. Rénovation Urbaine, projets urbains, patrimoine.
}

\paragraph{09/07/2017}{09/07/2017}

\bpar{
Visit of the New Territories in Hong-Kong: heavy rail transport from Kwoloon, then different tramway lines locally. Back with Shenzhen metro then by ferry until Zhuhai.
}{
Visite des New Territories à Hong-Kong : transport lourd depuis Kwoloon puis différentes lignes de tramway sur place. Retour par le métro de Shenzhen puis par ferry jusqu'à Zhuhai. 
}

\paragraph{11/07/2017}{11/07/2017}

\bpar{
Return trip Tangjia-Guangzhou. Transportation congestion (road and sharing bikes).
}{
Aller-retour Tangjia-Guangzhou. Congestion des transports (routier et vélos libre-service).
}

\paragraph{24/07/2017}{24/07/2017}

\bpar{
Outing in Tangjia. Local socio-economic discontinuities.
}{
Sortie à Tangjia. Discontinuités socio-économiques locales.
}

\paragraph{31/07/2017}{31/07/2017}

\bpar{
Outing in Xiangzhou. Test of the tramway, Line 2.
}{
Sortie à Xiangzhou. Test du Tramway, Ligne 2.
}

\paragraph{09/08/2017}{09/08/2017}

\bpar{
Outing in Xiangzhou then Tangjia. TOD operation: west end of the tram line; bus to Tangjia station following the high speed line.
}{
Sortie à Xiangzhou puis Tangjia. Opération de TOD : terminus ouest Tram ; bus pour la gare de Tangjia le long de la ligne à grande vitesse.
}

\paragraph{13/08/2017}{13/08/2017}

\bpar{
From Yangshuo (Guangxi) to GuangzhouNan with the high speed rail.
}{
De Yangshuo (Guangxi) à GuangzhouNan par le Train à Grande Vitesse.
}

\paragraph{17/08/2017}{17/08/2017}

\bpar{
Bureau of the planning committee of Zhuhai High Tech zone. Administration and bureaucracy. 
}{
Bureau du Comité de Planification de la zone High Tech de Zhuhai. Administration et bureaucratie.
}

\paragraph{20/08/2017}{20/08/2017}

\bpar{
Round-trip through Leshan (Sichuan) in bus. Transportation and tourism.
}{
Traversée de Leshan (Sichuan) en bus, aller-retour. Transports et Tourisme.
}

\paragraph{21/08/2017}{21/08/2017}

\bpar{
From Guangzhou Baiyun to Zhongshan Daxue (South campus of SYSU) then Tangjia. Transportation, urban village.
}{
De Guangzhou Baiyun à Zhongshan Daxue (campus sud de l'université SYSU) puis Tangjia. Transports, village urbain.
}





\subsection{Interviews}{Entretiens}

\bpar{
The ``interviews'' done correspond to unstructured active interviews~\cite{holstein2004active} in a jointly lived situation. Linguistic difficulties on both sides may have complicated the dialogues and we give here a synthesis of the information acquired. Names have been modified when the explicit agreement of the interviewed has not been obtained. In this narrative and subjective synthesis, the first person designate the author.
}{
Les ``entretiens'' menés relèvent de l'entretien actif non-structuré~\cite{holstein2004active} lors d'une mise en situation vécue conjointement. Les difficultés linguistiques de part et d'autre ont pu rendre compliqué les dialogues et nous donnons ici une synthèse des informations acquises. Les noms ont été modifiés lorsque l'accord explicite de l'interviewé n'a pas été obtenu. Dans cette synthèse narrative et subjective, la première personne désigne l'auteur.
}



\paragraph{12/08/2018}{12/08/2018}

\bpar{
\textit{Lin is an inhabitant of Guangzhou, coming from Guangxi. We met in the back of the last bus coming back to Yangshuo after a visit in Pingxi. The inebriety state facilitates the contact and the reciprocal understanding of my very bad mandarin and of her bad english. They came for a team-building week-end with her working team in a new technologies startup. A colleague helps for the interpretation whereas two others are absolutely absorbed in a Dota2 game on their cellphone. This city is the new trending destination since it is less than two hours in time from Guangzhou with the high speed line, it is apparently less frequented than Guilin.}
}{
\textit{Lin est une habitante de Guangzhou, originaire du Guangxi. Nous nous rencontrons au fond du dernier bus retournant à Yangshuo après une visite à Pingxi. Un état d'ébriété facilite la prise de contact et la compréhension réciproque de mon très mauvais mandarin et de son mauvais anglais. Ils sont venus en week-end de team-building avec son équipe d'une start-up numérique. Une collègue aide à l'interprétation tandis que deux autres sont absolument absorbés dans une partie de Dota2 sur leur portable. Cette ville est la nouvelle destination tendance depuis qu'elle est à moins de deux heures de Guangzhou par la ligne à grande vitesse, elle est parait-il moins fréquentée que Guilin.}
}

\bpar{
\textit{We meet again later in the center, after they got rid of their colleagues who were desperately seeking for a fixed internet access for a new game. We discuss on the touristic aspect of the city center. A crowd of consumers is filling pseudo-authentic alleys. Even the illuminated karstic peaks seem to be fake at this stage. Communist scouts sell some lentil ice-creams, they tell me to be cautious about these and that ice-creams would surely give me a stomach pain. We later discuss sceptically on occidental bars which flourish in this kind of town, they tell me that they are frequented by ``a certain kind of person'' (sociological prejudice that I did not manage to interpret).}
}{
\textit{Nous nous retrouvons plus tard dans le centre, après qu'elles se soient débarrassées de leur collègues qui cherchaient désespérément un poste internet fixe pour une nouvelle partie. Nous parlons de l'aspect touristique de ce centre-ville. Une foule de consommateurs se presse dans des ruelles pseudo-authentiques. Même les pics karstiques illuminés semblent faux à ce point. Des scouts communistes vendent des glaces aux lentilles, elles me disent qu'elles s'en méfient et que les glaces me donneront surement mal à l'estomac. Nous critiquons plus tard les bars à l'occidentale qui fleurissent dans ce genre de villes, elles me disent qu'ils sont fréquentés par ``un certain type de personnes'' (préjugé sociologique que je n'ai pas réussi à interpréter).}
}

\paragraph{16/08/2016}{16/08/2016}


\bpar{
\textit{I meet Zexian at the restaurant near the Rencai Gongyu residence in Tangjia, where live in particular the teachers of Zhongshan University. Shops associated to this complex are not only used by local inhabitants, and people (often new rich given the price) specially come for the brand new KTV (karaoke). She suggest me we should go there later. She tells me that she studies in a linguistic institute near the South gate of the campus. She studies in particular English, and would like that we stay in contact so that she can practice, we then exchange the Weixin contacts (Wechat).}
}{
\textit{Je rencontre Zexian au restaurant en bas de la résidence Rencai Gongyu à Tangjia, où sont logés notamment les professeurs de l'université Zhongshan. Les commerces associés à ce complexe ne sont pas uniquement utilisés par les habitants locaux, et les gens (souvent des nouveaux riches vu le prix) viennent spécialement pour le tout nouveau KTV (karaoke). Elle me propose d'y aller à la suite. Elle me raconte qu'elle est étudiante dans un institut de langues à proximité de la porte sud du campus. Elle étudie en particulier l'anglais, et voudrait qu'on reste en contact pour qu'elle puisse s'entraîner, nous échangeons alors les contact Weixin (Wechat).}
}

\bpar{
\textit{She tells me that her family lives in the South of Zhongshan, nearby then, but that it is very complicated to come back. The bus indeed makes the connection but several changes are necessary. The train connects Zhongshan to Zhuhai Bei or Tangjia but stations have a low accessibility, stops are not that frequent at these intermediate stops, and the reservation of a ticket is complicated. She most often takes a taxi on demand via the Didi application.}
}{
\textit{Elle m'explique que sa famille habite au sud de Zhongshan, à proximité donc, mais qu'il est très compliqué de rentrer. Le bus fait bien la connexion mais plusieurs changements sont nécessaires. Le train connecte Zhongshan à Zhuhai Bei ou Tangjia mais les gares sont peu accessibles, les horaires peu fréquentes à ces arrêts intermédiaires, et la réservation d'un billet compliquée. Elle prend le plus souvent un taxi sur demande via l'application Didi.}
}

\paragraph{19-20/08/2018}{19-20/08/2018}

\bpar{
\textit{Xing is a young inhabitant of Beijing having nearly 30 years, that I met at the entrance of Emeishan National Park. Leaving behind the unbelievable crowd of the area accessible by car, not so many people want to fully accomplish the mythic climb, and we naturally speak on the way. She explains to me the significance of this mountain and symbolic aspect of the climb. After visiting one or two temples, we are lost.}
}{
\textit{Xing est une jeune pékinoise d'une trentaine d'année rencontrée à l'entrée du Parc National d'Emeishan. Passé le délire de foule de la zone accessible aux voitures, peu de personnes souhaitent accomplir l'ascension initiatique intégralement, et nous nous parlons naturellement sur le chemin. Elle m'explique la signification de cette montagne et la portée symbolique de son ascension. Après la visite d'un ou deux temples, nous nous perdons.}
}

\bpar{
\textit{She works in Beijing in an Industrial Design company, it is her first job that she began a few months ago. The company sent her spend a month in Chengdu for a formation. She studied at Beijing Ligong Daxue (Beijing University of Technology) and would have liked to go to Europe to study, but the field was too selective. She speaks German and did a summer school there a few years ago. She is a marathonian but confirms the difficulties to train in Beijing, because of the pollution. Having her hometown in Hebei, she does not like to live in Beijing but her work obliges her to. Living conditions are not particularly nice and traffic issues are tiring.}
}{
\textit{Elle travaille à Pékin dans une entreprise de Design Industriel, c'est son premier emploi qu'elle a commencé il y a quelques mois. Son entreprise l'a envoyée passer un mois à Chengdu pour une formation. Elle a étudié à la Beijing Ligong Daxue (Université technologique de Beijing) et aurait souhaité partir étudier en Europe, mais les filières du domaine étaient trop sélectives. Elle parle allemand et y a fait une école d'été il a quelques années. Elle est marathonienne mais confirme les difficultés à s'entrainer à Beijing, à cause de la pollution. Originaire du Hebei, elle n'aime pas vivre à Beijing mais son travail l'y oblige. Le cadre de vie n'est pas particulièrement agréable et les problèmes de traffic sont pesants.}
}

\bpar{
\textit{She confirms to me the cultural aspect of \emph{Jingye}, one of the Core Values of Socialism promoted by the Party propaganda, which can be translated as dedication to work, but she is not satisfied about a lack of open-mindedness and creativity.}
}{
\textit{Elle me confirme l'aspect culturel du \emph{Jingye}, l'une des Valeurs Centrales du Socialisme promues par la propagande du Parti qui se traduit par la dévouement au travail, mais se désole d'un manque d'ouverture d'esprit et d'inventivité.}
}


\stars

%


\section{Quantitative Epistemology}{Épistémologie quantitative}

\label{app:sec:quantepistemo}


\subsection{Algorithmic systematic review}{Revue systématique algorithmique}

\paragraph{Algorithm description}{Description de l'algorithme}

\bpar{
Let $A$ be an alphabet (an arbitrary set of symbols), $A^{\ast}$ corresponding words (strings of finite length on it). Texts of finite length on it are then $T = \cup_{k\in \mathbb{N}} {A^{\ast}}^k$. What we call a reference is for the algorithm a record with text fields representing title, abstract and keywords. Set of references at iteration $n$ will be denoted $\mathcal{C}_n \subset T^3$: it is a subset of text triplets. We assume the existence of a set of keywords $\mathcal{K}_n$, initial keywords being $\mathcal{K}_0$, specified by the user\footnote{We could also start from a corpus $\mathcal{C}_0$, but it is more the spirit of the methodology presented in the next sub-section. We remain here for this preliminary exploration by assuming the necessarily arbitrary biased aspect of this specification. The choice of the initial corpus must thus be done with a good knowledge of existing domains, and necessarily done after the literature review of~\ref{sec:modelingsa}.}. An iteration proceeds the following way:
}{
Soit $A$ un alphabet (un ensemble arbitraire de symboles), $A^{\ast}$ les mots correspondants (chaînes de longueur finie sur l'alphabet). Les textes de longueur finie sur celui-ci sont donc $T = \cup_{k\in \mathbb{N}} {A^{\ast}}^k$. Ce qu'on nomme une référence est pour l'algorithme un enregistrement avec des champs textuels représentant le titre, le résumé et les mots-clés. L'ensemble de références à l'itération $n$ est ainsi noté $\mathcal{C}_n \subset T^3$ : il s'agit d'un sous-ensemble de triplets de textes. Nous supposons l'existence d'un ensemble de mots-clés $\mathcal{K}_n$, les mots-clés initiaux étant $\mathcal{K}_0$, spécifiés par l'utilisateur\footnote{On pourrait également partir d'un corpus $\mathcal{C}_0$, mais il s'agit plutôt de l'esprit de la méthodologie présentée dans la sous-section suivante. Nous nous en tiendrons ici pour cette exploration préliminaire en assumant le caractère arbitraire forcément biaisé de cette spécification. Le choix du corpus initial doit donc être fait en bonne connaissance des domaines existants, et fait nécessairement suite à la revue de littérature de~\ref{sec:modelingsa}.}. Une itération procède de la manière suivante :
}

\bpar{
\begin{enumerate}
\item A raw intermediate corpus $\mathcal{R}_n$ is obtained through a catalog request\footnote{The catalog is a function providing references as an answer to a request composed by regular expressions of keywords. In practice, we use the online bibliographic catalog Mendeley. Catalog dependency should surely introduce a bias which can not be controlled, since a sensitivity analysis or a cross-search through diverse catalogs being out of the scope of this exploratory analysis.} to which we provide the previous keywords $\mathcal{K}_{n-1}$.
\item Overall corpus is actualized by $\mathcal{C}_n = \mathcal{C}_{n-1} \cup \mathcal{R}_n$.
\item The new keywords $\mathcal{K}_n$ are extracted from corpus through Natural Language Processing (NLP) treatment, given a parameter $N_k$ fixing the number of keywords extracted at this stage.
\end{enumerate}
}{
\begin{enumerate}
\item Un corpus intermédiaire brut $\mathcal{R}_n$ est obtenu par une requête à un catalogue\footnote{Le catalogue est une fonction fournissant des références en réponse à une requête composée d'expression régulières de mots-clés. En pratique, nous utilisons le catalogue bibliographique en ligne Mendeley. La dépendance au catalogue devant sûrement introduire un biais que nous ne pouvons contrôler, une analyse de sensibilité ou le croisement de divers catalogues étant hors de propos pour cette analyse exploratoire.}
 auquel on fourni les mots-clés précédents $\mathcal{K}_{n-1}$.
\item Le corpus total est actualisé par $\mathcal{C}_n = \mathcal{C}_{n-1} \cup \mathcal{R}_n$.
\item Les nouveaux mot-clés $\mathcal{K}_n$ sont extraits du corpus par Traitement du Language Naturel (NLP), étant donné un paramètre fixé $N_k$ donnant le nombre de mot-clés extraits à cette étape.
\end{enumerate}
}

\bpar{
The algorithm stops when corpus size becomes stable (experiments on tested requests shows that for these, the corpus does not contain new references after a certain number of iterations) or when a maximal number of iterations defined by the user is reached. Fig.~\ref{fig:quantepistemo:algo} synthesizes the global workflow.
}{
L'algorithme s'arrête quand la taille du corpus ne varie plus (l'expérience sur les requêtes testées montre pour celles-ci que le corpus ne contient plus de nouvelles références après un certain nombre d'itérations) ou quand un nombre maximal d'itérations défini par l'utilisateur est atteint. La figure~\ref{fig:quantepistemo:algo} synthétise le processus général.
}

\paragraph{Implementation}{Implémentation}

\bpar{
Because of the heterogeneity of operations required by the algorithm (references organisation, catalog requests, text processing), it was found a reasonable choice to implement it in Java. Source code is available on the open repository of the project\footnote{at the adress \texttt{https://github.com/JusteRaimbault/CityNetwork/tree/master/Models/QuantEpsitemo/AlgoSR}}. Catalog requests, consisting in retrieving a set of references from a set of keywords, are done using the Mendeley software API \cite{mendeley} as it allows an open access to a large database. Keyword extraction is done by Natural Language Processing (NLP) techniques, following the workflow given in~\cite{chavalarias2013phylomemetic}, through a Python script that uses \cite{bird2006nltk}.
}{
De par l'hétérogénéité des opérations requises par l'algorithme (organisation des références, requêtes au catalogue, analyse textuelle), le language Java s'est présenté comme une alternative raisonnable. Le code source est disponible sur le dépôt ouvert du projet\footnote{à l'adresse \url{https://github.com/JusteRaimbault/CityNetwork/tree/master/Models/QuantEpistemo/AlgoSR}}. Les requêtes au catalogue, qui consistent à récupérer un ensemble de références à partir d'un ensemble de mots-clés, sont faites via l'API du logiciel Mendeley~\cite{mendeley} qui permet un accès ouvert à une base de données conséquente. L'extraction des mots-clés est effectuée par techniques d'Analyse Textuelle (NLP) selon le processus donné dans~\cite{chavalarias2013phylomemetic}, via un script Python qui utilise~\cite{bird2006nltk}.
}

\paragraph{Convergence and sensitivity analysis}{Convergence et analyse de sensibilité}

\bpar{
A formal proof of algorithm convergence is not possible as it will depend on the empirical unknown structure of request results and keywords extraction. We need thus to study empirically its behavior. Good convergence properties but various sensitivities to $N_k$ were found as presented in Fig.~\ref{fig:app:quantepistemo:sensitivity-algosr}. We also study the internal lexical consistence of final corpuses as a function of keywords number. As expected, small number yields more consistent corpuses, but the variability when increasing stays reasonable.
}{
Une preuve formelle de convergence de l'algorithme n'est guère envisageable puisque qu'elle dépendra de la structure empirique inconnue des résultats de requête et d'extraction de mots-clés. Il est donc nécessaire d'étudier le comportement de l'algorithme de manière empirique. Comme présenté en Fig.~\ref{fig:app:quantepistemo:sensitivity-algosr}, l'algorithme a de bonnes propriétés de convergence mais diverses sensibilités à $N_k$. Nous étudions également la cohérence lexicale interne des corpus finaux et fonction du nombre de mots-clés. Comme attendu, des valeurs faibles produisent des corpus plus cohérents, mais la variabilité lorsque qu'elles augmentent reste raisonnable.
}

\bpar{
We take the weakest assumption for the parameter $N_k=100$. Indeed, the larger $N_k$ is, the less constrained the explored domain will be, what increases the chances of overlap between two corpuses originating from different initial requests. In this case, a small final distance between corpuses will be more significant for larger $N_k$ values.
}{
Nous prenons l'hypothèse la plus faible pour le paramètre $N_k=100$. En effet, plus $N_k$ est grand, moins le domaine exploré sera restreint, ce qui augmente les chances de recouvrement de deux corpus provenant de requêtes initiales différentes. Dans ce cas, une faible distance finale entre corpus sera plus significative pour des valeurs de $N_k$ grandes.
}

\begin{figure}
\includegraphics[width=\linewidth,height=0.85\textheight]{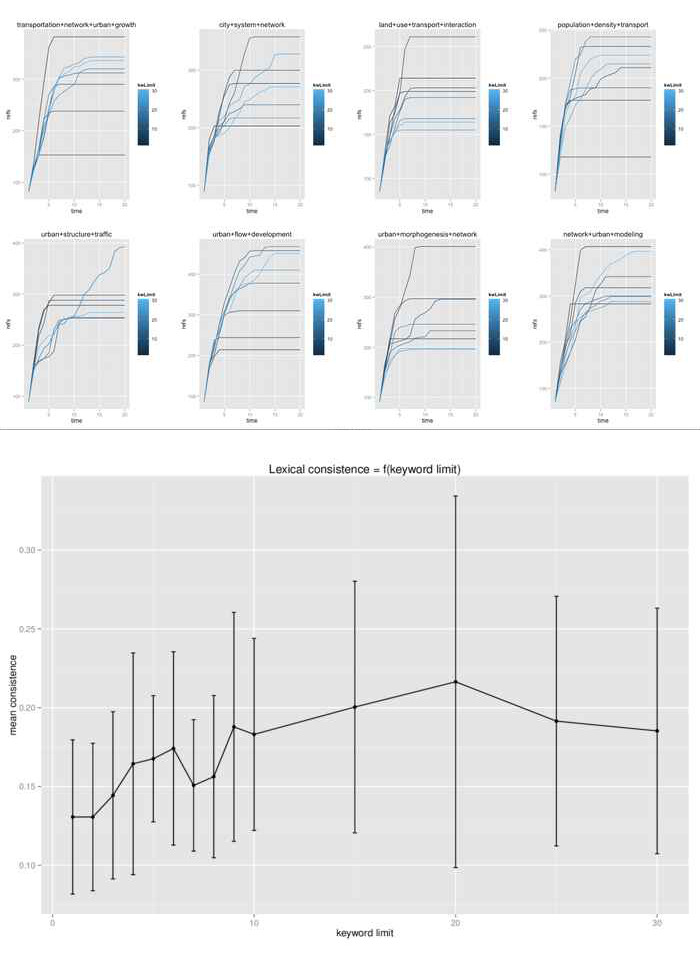}
\appcaption{\textbf{Convergence and sensitivity analysis of the algorithmic systematic review.} (\textit{Top}) Plots of number of references as a function of iteration, for various queries linked to our theme (see further), for various values of $N_k$ (from 2 to 30). We obtain a rapid convergence for most cases, around 10 iterations needed. Final number of references appears to be very sensitive to keyword number depending on queries, what confirms a strong variability of the encountered landscape depending on terms. (\textit{Bottom}) Mean lexical consistence and standard error bars for various queries, as a function of $n_k$. Lexical consistence is defined though co-occurrences of keywords by, with $N$ final number of keywords, $f$ final step, and $c(i)$ co-occurrences in references, $k = \frac{2}{N(N-1)}\cdot \sum_{i,j \in \mathcal{K}_f}{\left| c(i) - c(j) \right|}$. The stability confirms the consistence of final corpuses.\label{fig:app:quantepistemo:sensitivity-algosr}}{\textbf{Convergence et analyse de sensibilité de l'algorithme de revue systématique.} (\textit{Haut}) Graphes des nombres de références en fonction de l'itération, pour différentes requêtes liées à notre thème, et pour différentes valeurs de $N_k$ (de 2 à 30, couleur). On obtient une convergence rapide dans la majorité des cas, autour de 10 itérations étant nécessaires. Le nombre final de références semble très sensible au nombre de mots-clés selon les requêtes, ce qui confirme une forte variabilité du paysage rencontré selon les termes. (\textit{Bas}) Consistence lexicale moyenne et déviation standard sur différentes requêtes, en fonction de $N_k$. La consistence lexicale est définie par les co-occurrences des mots-clés, comme $k = \frac{2}{N_k(N_k-1)}\cdot \sum_{i,j \in \mathcal{K}_f}{\left| c(i) - c(j) \right|}$, avec $f$ temps final, $c(i)$ co-occurrence des mots dans les références. La stabilisation confirme la consistence des corpus finaux.\label{fig:app:quantepistemo:sensitivity-algosr}}
\end{figure}


\subsection{Indirect bibliometrics}{Bibliométrie indirecte}

\paragraph{Initial corpus}{Corpus initial}

\bpar{
The Table~\ref{tab:app:quantepistemo:corpus} gives the composition of the initial corpus for the construction of the citation network.
}{
Le tableau~\ref{tab:app:quantepistemo:corpus} donne la composition du corpus initial pour la construction du réseau de citation.
}

\begin{table}
\apptabcaption{\textbf{Composition of the initial corpus for the construction of the citation network.}\label{tab:app:quantepistemo:corpus}}{\textbf{Composition du corpus initial pour la construction du réseau de citation.}\label{tab:app:quantepistemo:corpus}}
\bpar{
\begin{tabular}{|l|p{6cm}|l|}
	\hline
	Discipline & Title & Reference \\\hline
	Political science & \textit{Les effets structurants du transport: mythe politique, mystification scientifique} & \cite{offner1993effets} \\\hline 
	Interdisciplinary & \textit{Réseaux et territoires-significations croisées} & \cite{offner1996reseaux} \\\hline
	Geography & \textit{Villes et réseaux de transport: des interactions dans la longue durée (France, Europe, Etats-Unis)} & \cite{bretagnolle:tel-00459720} \\\hline
	Transportation & Land-use transport interaction: state of the art & \cite{wegener2004land} \\\hline
	Economics & The co-evolution of land use and road networks & \cite{levinson2007co} \\\hline
	Economics & Modeling the growth of transportation networks: a comprehensive review & \cite{xie2009modeling} \\\hline
	Physics & Co-evolution of density and topology in a simple model of city formation & \cite{barthelemy2009co} \\\hline
	\end{tabular}
}{
	\begin{tabular}{|l|p{6cm}|l|}
	\hline
	Domaine & Titre & Référence \\\hline
	Sciences politiques & Les effets structurants du transport: mythe politique, mystification scientifique & \cite{offner1993effets} \\\hline 
	Interdisciplinaire & Réseaux et territoires-significations croisées & \cite{offner1996reseaux} \\\hline
	Géographie & Villes et réseaux de transport: des interactions dans la longue durée (France, Europe, Etats-Unis) & \cite{bretagnolle:tel-00459720} \\\hline
	Transports & Land-use transport interaction: state of the art & \cite{wegener2004land} \\\hline
	Économie & The co-evolution of land use and road networks & \cite{levinson2007co} \\\hline
	Économie & Modeling the growth of transportation networks: a comprehensive review & \cite{xie2009modeling} \\\hline
	Physique & Co-evolution of density and topology in a simple model of city formation & \cite{barthelemy2009co} \\\hline
	\end{tabular}
}
\end{table}

\paragraph{Sensitivity analysis}{Analyse de sensibilité}

\bpar{
The sensitivity analysis allowing to fix the optimal parameters for the semantic network is shown in Fig.~\ref{fig:app:quantepistemo:sensitivity}.
}{
L'analyse de sensibilité permettant de fixer les paramètres optimaux pour le réseau sémantique est montrée en Fig.~\ref{fig:app:quantepistemo:sensitivity}.
}


\begin{figure}
\includegraphics[width=\linewidth]{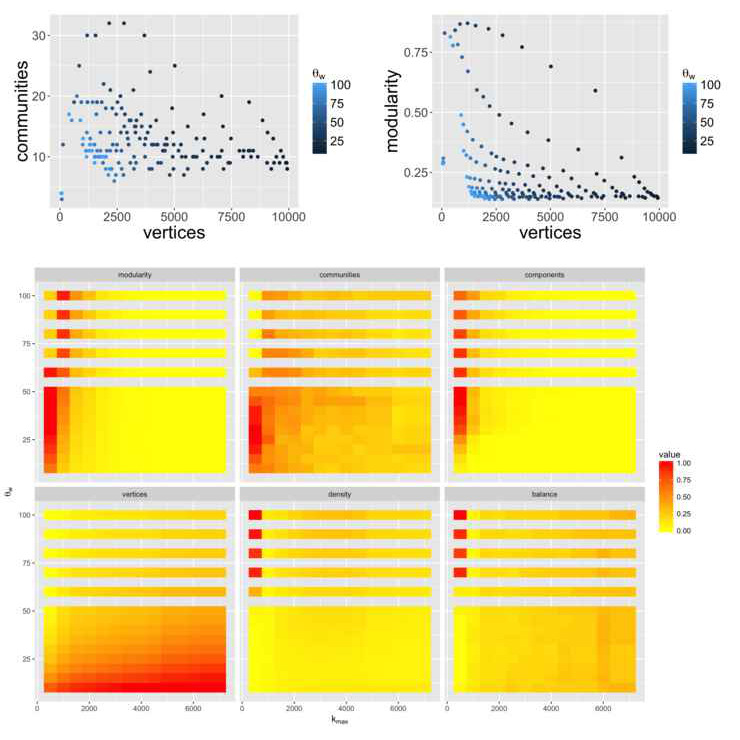}
\appcaption{\textbf{Sensitivity analysis of modular properties of the semantic network as a function of filtering parameters.} \textit{(Top Left)} Pareto front of the number of communities and the number of vertices (two objectives to be maximized), the color giving the value of $\theta_w$; \textit{(Top Right)} Pareto front of the modularity as a function of number of vertices, for varying $\theta_w$; (\textit{Bottom}) Values of possible objectives (modularity, number of communities, number of connnected components, number of vertices, density, size balance between communities), each objective being normalized in $\left[0;1\right]$, as a function of parameters $\theta_w$ and $k_{max}$.\label{fig:app:quantepistemo:sensitivity}}{\textbf{Analyse de sensibilité des propriétés modulaires du réseau sémantique en fonction des paramètres de filtrage.} \textit{(Haut Gauche)} Front de Pareto du nombre de communauté et du nombre de sommets (deux objectifs à maximiser), la couleur donnant la valeur de $\theta_w$ ; \textit{(Haut Droite)} Front de Pareto de la modularité en fonction du nombre de sommets, pour $\theta_w$ variant ; \textit{(Bas)} Valeurs des objectifs possibles (modularité, nombre de communautés, nombre de composantes connexes, nombre de sommets, densité, équilibre de taille entre communautés), chaque objectif étant normalisé dans $\left[0;1\right]$, en fonction des paramètres $\theta_w$ et $k_{max}$.\label{fig:app:quantepistemo:sensitivity}}
\end{figure}

\paragraph{Semantic network}{Réseau sémantique}


\bpar{
A visualization of the semantic network is given in Fig~\ref{fig:app:quantepistemo:semanticnw}.
}{
Une visualisation du réseau sémantique est donnée en Fig.~\ref{fig:app:quantepistemo:semanticnw}.
}

\begin{figure}
\includegraphics[width=\linewidth]{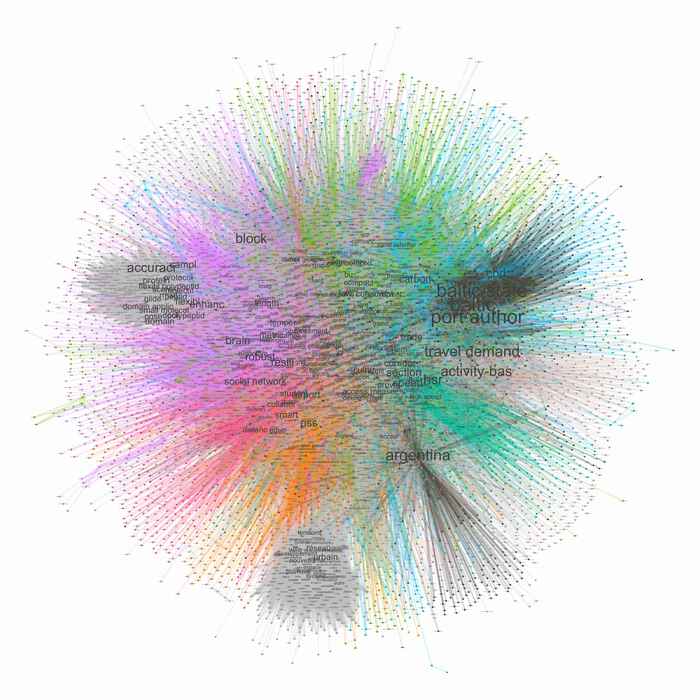}
\appcaption{\textbf{Semantic network of domains.} The color of links gives the community and the size of keywords is fixed by their degree.\label{fig:app:quantepistemo:semanticnw}}{\textbf{Réseau sémantique des domaines.} La couleur des liens donne la communauté et la taille des mots-clés est fixée par leur degré.\label{fig:app:quantepistemo:semanticnw}}
\end{figure}

\stars


\newpage

\section{Modelography}{Modélographie}

\label{app:sec:modelography}


\subsection{Systematic review methodology}{Méthodologie de la revue systématique}

\bpar{
For the choice of initial keywords for the indirect construction (through semantic request), a possible alternative is to extract the relevant keywords for each sub-community of the citation network, and then select the most relevant for each domain. We make the choice of extracting them on the complete corpus, and then to collect them by sub-community thereafter. For a small corpus, the second option is more suitable, since the notion of relevance is less importance than for very large corpuses, in which some relevant words may be drowned and less relevant to emerge in a spurious way. In other words, the keyword selection method appears to be more robust on smaller corpuses, as suggest the comparison of this application with the one done on the Cybergeo journal and the one done on the patent corpus (see~\ref{app:sec:patentsmining}).
}{
Pour le choix des mots-clés initiaux pour la construction indirecte (via requête sémantique), une alternative possible est d'extraire les mots-clés pertinents par sous-communautés du réseau de citations, puis sélectionner les plus pertinents ensuite pour chaque domaine. Nous faisons le choix de les extraire sur le corpus complet, puis de les récupérer par sous-communautés ensuite. Pour un petit corpus, la deuxième option est plus souhaitable, puisque la notion de pertinence moins importante que pour des très grands corpus, ou certains mots pertinents pourront être noyés et des moins pertinents ressortir de manière fortuite. En d'autre termes, la méthode de selection des mots-clés parait plus robuste sur des petits corpus, comme le suggère la comparaison de cette application avec celle faite sur le journal Cybergeo et celle faite sur le corpus de brevets (voir~\ref{app:sec:patentsmining}).
}

\subsubsection{First corpus review}{Première revue du corpus}

\bpar{
The methods used do not allow to be cleared from a ``noise'', i.e. of articles that are not relevant to the subject, even with a very low tolerance threshold. We obtained for example articles as absurd as gender and car use, colorectal cancer in Texas, vibration mechanics at the passage of a high speed train, protein transport within the cell, public space in Beyrouth, spatial patterns of \emph{street gangs} in Los Angeles, urban geology in Brussels. This confirms that the manual filtering stage is essential.
}{
Les méthodes utilisées ne permettent pas de s'affranchir d'un ``bruit'', c'est-à-dire d'article ne relevant a priori pas même de loin à la thématique. Nous avons obtenu par exemple des articles aussi divers qu'incongrus sur le genre et l'usage de la voiture, le cancer colorectal au Texas, la mécanique des vibrations au passage d'un train à grande vitesse, le transport des protéines dans la cellule, l'espace public à Beyrouth, les motifs spatiaux des \emph{street gangs} à Los Angeles, la géologie urbaine à Bruxelles. Cela confirme que l'étape de filtrage manuel est essentielle.
}

\bpar{
This noise can for example be due to:
\begin{itemize}
	\item Effective citations for diverse reasons, but having only a low relevance in the citing article.
	\item Noise intrinsic to the keyword search.
	\item Catalog classification errors.
\end{itemize}
}{
Ce bruit peut être du par exemple à :
\begin{itemize}
	\item Des citations effectives pour diverses raisons, mais n'ayant que peu de pertinence dans l'article citant.
	\item Du bruit intrinsèque à la recherche par mots-clés.
	\item Des erreurs de classification du catalogue.
\end{itemize}
}

\subsubsection{Remarks on manual screening}{Remarques sur la classification manuelle}

\bpar{
During the manual classification achieved when screening abstracts, the following points appear: 
}{
Lors de la classification manuelle opérée lors de l'inspection des résumés, les points suivants ressortent :
}

\bpar{
\begin{itemize}
	\item The ``a priori'' disciplines are judged based on the journal in which the article was published. In particular, we operate the following particular choices (for other journals such as physics journals there is no ambiguity): Journal of Transport Geography, Environment and Planning B: Geography; Journal of Transport and Land-use, Transportation Research: Transportation.
	\item Geography in our sense includes urbanism and urban studies is these are not too close from planning (urban sustainability for example).
\end{itemize}
}{
\begin{itemize}
	\item Les disciplines ``a priori'' sont jugées par le journal dans lequel l'article a été publié. En l'occurence, nous opérons les choix particuliers suivants (pour d'autres journaux comme des journaux de physique il n'y a pas d'ambiguïté) : Journal of Transport Geography, Environment and Planning B : geography ; Journal of Transport and Land-Use, Transportation Research : Transportation.
	\item La géographie en notre sens inclut l'urbanisme et les études urbaines si celles-ci ne sont pas trop proches de la planification (urbain durable par exemple).
\end{itemize}
}

\subsection{Meta-analysis}{Meta-analyse}

\bpar{
We give here the full numerical results of statistical analysis linking model characteristics and explicative variables.
}{
Nous donnons ici les résultats numériques complets des analyses statistiques reliant caractéristiques de modèles et variables explicatives.
}

\subsubsection{Modalities of variables}{Modalités des variables}

\bpar{
We recall here the variables used in the meta-analysis and their modalities. These are:
}{
Rappelons ici les variables utilisées dans la méta-analyse et leur modalités. Celles-ci sont :
}

\bpar{
\begin{itemize}
	\item Type of model (\texttt{TYPE}): strong, territory, network.
	\item Publication year (\texttt{YEAR}), integer number.
	\item Citation community (\texttt{CITCOM}), defined within the citation network: Accessibility, Geography, Infra Planning, LUTI, Networks, TOD.
	\item A priori discipline (\texttt{DISCIPLINE}): biology, computer science, economics, engineering, environment, geography, physics, planning, transportation.
	\item Semantic community (\texttt{SEMCOM}): brt, complex networks, hedonic, hsr, infra planning, networks, tod.
	\item Methodology used: ca (Cellular Automaton), eq (analytical equations), map (cartography), mas (Multi-agent simulation), ro (operations research), sem (Structural Equation Modeling), sim (simulation), stat (statistics).
	\item Interdisciplinarity index (\texttt{INTERDISC}): real number in $[0,1]$.
	\item Temporal scale (\texttt{TEMPSCALE}): given in years, is set to 0 for static analyses.
	\item Spatial scale (\texttt{SPATSCALE}): continent (10000), country (1000), region (100), metro (10). These modalities are numerically transformed in km by the values given in parenthesis (stylized scales).
\end{itemize}
}{
\begin{itemize}
	\item Type de modèle (\texttt{TYPE}) : strong, territory, network.
	\item Année de publication (\texttt{YEAR}), nombre entier.
	\item Communauté de citation (\texttt{CITCOM}), définies par le réseau de citations : Accessibility, Geography, Infra Planning, LUTI, Networks, TOD.
	\item Discipline a priori (\texttt{DISCIPLINE}) : biology, computer science, economics, engineering, environment, geography, physics, planning, transportation.
	\item Communauté sémantique (\texttt{SEMCOM}) : brt, complex networks, hedonic, hsr, infra planning, networks, tod.
	\item Méthodologie utilisée : ca (\emph{Cellular Automaton}), eq (équations analytiques), map (cartographie), mas (\emph{Multi-agent simulation}), ro (recherche opérationnelle), sem (\emph{Structural Equation Modeling}), sim (simulation), stat (statistiques).
	\item Indice d'interdisciplinarité (\texttt{INTERDISC}) : réel dans $[0,1]$.
	\item Echelle temporelle (\texttt{TEMPSCALE}) : donnée en année, vaut 0 pour les analyses statiques.
	\item Echelle spatiale (\texttt{SPATSCALE}) : continent (10000), country (1000), region (100), metro (10). Ces modalités sont transformées numériquement en km par les valeurs données entre parenthèses (échelles stylisées).
\end{itemize}
}


\subsubsection{Model selection}{Sélection des modèles}

\bpar{
Regarding model selection, it is not achieved following a unique criteria, because of the low number of observations for some models, but by the optimization in the Pareto sense of contradictory objectives of adjustment (adjusted $R^2$, to be maximized) and of the overfitting (corrected Akaike criterion AICc, to be minimized), while controlling the number of observation points. The Fig.~\ref{fig:app:quantepistemo:regressions} gives for each variable to be explained the localization of the set of potential models within the objective space, and also the corresponding number of observations. For interdisciplinarity, two point clouds correspond to different compromises, and we select the two optimal models (one for each cloud). For the spatial scale, we postulate a positive $R^2$, and a single optimal model then emerges. For the temporal scale, we have as for interdisciplinarity two compromise models. Finally for the year, the AICc gain between the two potential optima is negligible in comparison to the $R^2$ loss, and we thus select the optimal model such that $R^2>0.25$ and AICc$<600$. The results of models are given in the following.
}{
Concernant la sélection des modèles, celle-ci n'est pas opérée en critère unique, de par le faible nombre d'observations pour certains modèles, mais par l'optimisation au sens de Pareto des objectifs contradictoires de l'ajustement ($R^2$ ajusté, à maximiser) et du sur-ajustement (critère d'Akaike corrigé AICc, à minimiser), tout en contrôlant le nombre de points d'observation. La Fig.~\ref{fig:app:quantepistemo:regressions} donne pour chaque variable à expliquer la localisation de l'ensemble des modèles potentiels dans l'espace des objectifs, ainsi que le nombre d'observations correspondantes. Pour l'interdisciplinarité, deux nuages de points correspondent à des compromis différents, et nous sélectionnons les deux modèles optimaux (un pour chaque nuage). Pour l'échelle d'espace, nous postulons un $R^2$ positif, et un seul modèle optimal émerge alors. Pour l'échelle de temps, on a comme pour l'interdisciplinarité deux modèles compromis. Enfin, pour l'année, le gain en AICc entre les deux optimaux potentiels est négligeable en comparaison à la perte en $R^2$, et nous sélectionnons donc le modèle optimal tel que $R^2>0.25$ et AICc$<600$. Les résultats des modèles sont donnés par la suite.
}

\begin{figure}
\includegraphics[width=\linewidth]{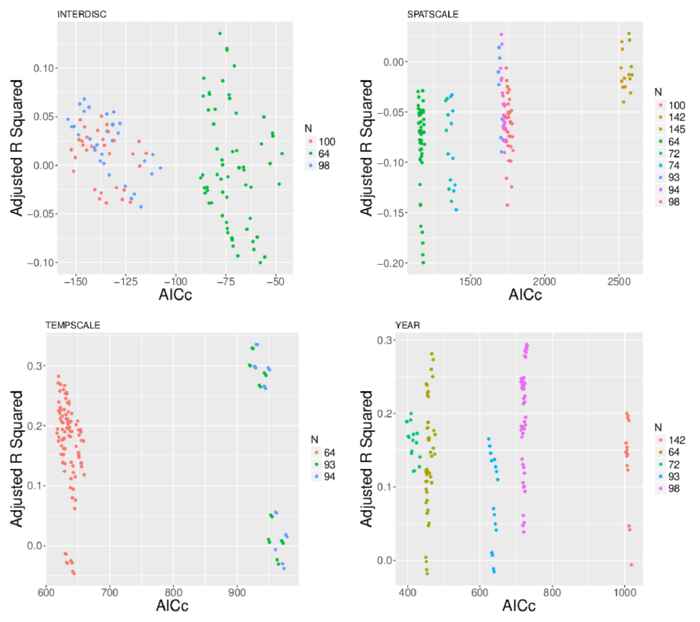}
\appcaption{\textbf{Multi-objective selection of linear models.} For each variable to be explained, we represent the position of all linear models in the objective space (corrected Akaike criterion AICc and adjusted $R^2$). The color of points gives the number of observations.\label{fig:app:quantepistemo:regressions}}{\textbf{Sélection multi-objectif des modèles linéaires.} Pour chaque variable à expliquer, nous représentons la position de l'ensemble des modèles linéaires dans l'espace des objectifs (critère d'Akaike corrigé AICc et $R^2$ ajusté). La couleur des points donne le nombre d'observations.\label{fig:app:quantepistemo:regressions}}
\end{figure}

\subsubsection{Model fitting}{Ajustement des modèles}

\paragraph{Interdisciplinarity}{Interdisciplinarité}

\bpar{
Interdisciplinarity is adjusted according to the linear models presented in Table~\ref{tab:app:modelography:interdisc}.
}{
L'interdisciplinarité est ajustée selon les modèles linéaires présentés en Table~\ref{tab:app:modelography:interdisc}.
}

\begin{table}
  \apptabcaption{\textbf{Linear models for interdisciplinarity}\label{tab:app:modelography:interdisc}}{\textbf{Modèles linéaires pour l'interdisciplinarité.}\label{tab:app:modelography:interdisc}}
\bpar{
\begin{tabular}{@{\extracolsep{5pt}}lcc} 
\footnotesize
\\[-1.8ex]\hline 
\hline \\[-1.8ex] 
\\[-1.8ex] & \multicolumn{2}{c}{INTERDISC} \\ 
\\[-1.8ex] & (1) & (2)\\ 
\hline \\[-1.8ex] 
 YEAR & $-$0.004 ($-$0.008, $-$0.00002), p = 0.055$^{*}$ & $-$0.002 ($-$0.005, 0.0001), p = 0.061$^{*}$ \\ 
  TEMPSCALE & $-$0.0003 ($-$0.001, 0.001), p = 0.615 &  \\ 
  DISCIPLINEengineering & 0.144 ($-$0.082, 0.371), p = 0.218 &  \\ 
  DISCIPLINEenvironment & 0.092 ($-$0.132, 0.316), p = 0.425 &  \\ 
  DISCIPLINEgeography & 0.036 ($-$0.043, 0.114), p = 0.378 &  \\ 
  DISCIPLINEphysics & $-$0.103 ($-$0.287, 0.080), p = 0.275 &  \\ 
  DISCIPLINEplanning & $-$0.047 ($-$0.135, 0.041), p = 0.30 &  \\ 
  DISCIPLINEtransportation & 0.062 ($-$0.025, 0.149), p = 0.169 &  \\ 
  TYPEstrong &  & $-$0.026 ($-$0.134, 0.081), p = 0.633 \\ 
  TYPEterritory &  & 0.044 ($-$0.026, 0.114), p = 0.222 \\ 
  SEMCOMcomplex networks &  & $-$0.217 ($-$0.522, 0.087), p = 0.166 \\ 
  SEMCOMhedonic & $-$0.179 ($-$0.407, 0.049), p = 0.130 & $-$0.184 ($-$0.400, 0.032), p = 0.100$^{*}$ \\ 
  SEMCOMhsr & $-$0.100 ($-$0.361, 0.162), p = 0.459 & $-$0.122 ($-$0.357, 0.112), p = 0.309 \\
  SEMCOMinfra planning & $-$0.032 ($-$0.273, 0.209), p = 0.797 & $-$0.096 ($-$0.321, 0.128), p = 0.404  \\ 
  SEMCOMnetworks & $-$0.038 ($-$0.272, 0.195), p = 0.750 & $-$0.107 ($-$0.324, 0.109), p = 0.335 \\ 
  SEMCOMtod & $-$0.105 ($-$0.332, 0.121), p = 0.366 & $-$0.152 ($-$0.364, 0.060), p = 0.165 \\ 
  Constant & 8.962 (0.776, 17.147), p = 0.037$^{**}$ & 5.531 (0.575, 10.487), p = 0.032$^{**}$ \\
 \hline \\[-1.8ex] 
Observations & 64 & 98 \\ 
R$^{2}$ & 0.314 & 0.155 \\ 
Adjusted R$^{2}$ & 0.136 & 0.068 \\ 
Residual Std. Error & 0.109 (df = 50) & 0.107 (df = 88) \\ 
F Statistic & 1.761$^{*}$ (df = 13; 50) & 1.789$^{*}$ (df = 9; 88) \\ 
\hline 
\hline \\[-1.8ex] 
\textit{Note:}  & \multicolumn{2}{r}{$^{*}$p$<$0.1; $^{**}$p$<$0.05; $^{***}$p$<$0.01} \\ 
\end{tabular}
}{
\begin{tabular}{@{\extracolsep{5pt}}lcc} 
\footnotesize
\\[-1.8ex]\hline 
\hline \\[-1.8ex] 
\\[-1.8ex] & \multicolumn{2}{c}{INTERDISC} \\ 
\\[-1.8ex] & (1) & (2)\\ 
\hline \\[-1.8ex] 
 YEAR & $-$0.004 ($-$0.008, $-$0.00002), p = 0.055$^{*}$ & $-$0.002 ($-$0.005, 0.0001), p = 0.061$^{*}$ \\ 
  TEMPSCALE & $-$0.0003 ($-$0.001, 0.001), p = 0.615 &  \\ 
  DISCIPLINEengineering & 0.144 ($-$0.082, 0.371), p = 0.218 &  \\ 
  DISCIPLINEenvironment & 0.092 ($-$0.132, 0.316), p = 0.425 &  \\ 
  DISCIPLINEgeography & 0.036 ($-$0.043, 0.114), p = 0.378 &  \\ 
  DISCIPLINEphysics & $-$0.103 ($-$0.287, 0.080), p = 0.275 &  \\ 
  DISCIPLINEplanning & $-$0.047 ($-$0.135, 0.041), p = 0.30 &  \\ 
  DISCIPLINEtransportation & 0.062 ($-$0.025, 0.149), p = 0.169 &  \\ 
  TYPEstrong &  & $-$0.026 ($-$0.134, 0.081), p = 0.633 \\ 
  TYPEterritory &  & 0.044 ($-$0.026, 0.114), p = 0.222 \\ 
  SEMCOMcomplex networks &  & $-$0.217 ($-$0.522, 0.087), p = 0.166 \\ 
  SEMCOMhedonic & $-$0.179 ($-$0.407, 0.049), p = 0.130 & $-$0.184 ($-$0.400, 0.032), p = 0.100$^{*}$ \\ 
  SEMCOMhsr & $-$0.100 ($-$0.361, 0.162), p = 0.459 & $-$0.122 ($-$0.357, 0.112), p = 0.309 \\
  SEMCOMinfra planning & $-$0.032 ($-$0.273, 0.209), p = 0.797 & $-$0.096 ($-$0.321, 0.128), p = 0.404  \\ 
  SEMCOMnetworks & $-$0.038 ($-$0.272, 0.195), p = 0.750 & $-$0.107 ($-$0.324, 0.109), p = 0.335 \\ 
  SEMCOMtod & $-$0.105 ($-$0.332, 0.121), p = 0.366 & $-$0.152 ($-$0.364, 0.060), p = 0.165 \\ 
  Constant & 8.962 (0.776, 17.147), p = 0.037$^{**}$ & 5.531 (0.575, 10.487), p = 0.032$^{**}$ \\
 \hline \\[-1.8ex] 
Observations & 64 & 98 \\ 
R$^{2}$ & 0.314 & 0.155 \\ 
R$^{2}$ ajusté & 0.136 & 0.068 \\ 
Erreur Std. Résiduelle & 0.109 (df = 50) & 0.107 (df = 88) \\ 
Statistique F & 1.761$^{*}$ (df = 13; 50) & 1.789$^{*}$ (df = 9; 88) \\ 
\hline 
\hline \\[-1.8ex] 
\textit{Note:}  & \multicolumn{2}{r}{$^{*}$p$<$0.1; $^{**}$p$<$0.05; $^{***}$p$<$0.01} \\ 
\end{tabular}
}
\end{table} 

\paragraph{Spatial scale}{Echelle d'espace}

\bpar{
The spatial scale is adjusted following the linear model which adjustment is given in Table~\ref{tab:app:modelography:spatscale}.
}{
L'échelle spatiale est ajustée selon le modèle linéaire dont l'ajustement est donné en Table~\ref{tab:app:modelography:spatscale}.
}


\begin{table}
  \apptabcaption{\textbf{Linear model for the spatial scale.}\label{tab:app:modelography:spatscale}}{\textbf{Modèle linéaire pour l'échelle spatiale.}\label{tab:app:modelography:spatscale}}
\bpar{
\begin{tabular}{@{\extracolsep{5pt}}lc} 
\\[-1.8ex]\hline 
\hline \\[-1.8ex] 
\\[-1.8ex] & SPATSCALE \\ 
\hline \\[-1.8ex] 
 TEMPSCALE & $-$5.179 ($-$16.259, 5.901) \\ 
  & p = 0.363 \\ 
  DISCIPLINEengineering & $-$154.461 ($-$3,003.326, 2,694.405) \\ 
  & p = 0.916 \\ 
  DISCIPLINEenvironment & $-$5.878 ($-$3,977.974, 3,966.219) \\ 
  & p = 0.998 \\ 
  DISCIPLINEgeography & 1,445.457 (389.349, 2,501.565) \\ 
  & p = 0.009$^{***}$ \\ 
  DISCIPLINEphysics & 292.559 ($-$2,717.659, 3,302.777) \\ 
  & p = 0.850 \\ 
  DISCIPLINEplanning & $-$143.554 ($-$1,361.357, 1,074.249) \\ 
  & p = 0.818 \\ 
  DISCIPLINEtransportation & 568.329 ($-$606.167, 1,742.826) \\ 
  & p = 0.346 \\ 
  Constant & 235.357 ($-$458.201, 928.914) \\ 
  & p = 0.508 \\ 
 \hline \\[-1.8ex]
Observations & 94 \\ 
R$^{2}$ & 0.100 \\ 
Adjusted R$^{2}$ & 0.027 \\ 
Residual Std. Error & 1,995.272 (df = 86) \\ 
F Statistic &  1.369 (df = 7; 86) \\ 
\hline 
\hline \\[-1.8ex] 
\textit{Note:}  & \multicolumn{1}{r}{$^{*}$p$<$0.1; $^{**}$p$<$0.05; $^{***}$p$<$0.01} \\ 
\end{tabular}
}{
\begin{tabular}{@{\extracolsep{5pt}}lc} 
\\[-1.8ex]\hline 
\hline \\[-1.8ex] 
\\[-1.8ex] & SPATSCALE \\ 
\hline \\[-1.8ex] 
 TEMPSCALE & $-$5.179 ($-$16.259, 5.901) \\ 
  & p = 0.363 \\ 
  DISCIPLINEengineering & $-$154.461 ($-$3,003.326, 2,694.405) \\ 
  & p = 0.916 \\ 
  DISCIPLINEenvironment & $-$5.878 ($-$3,977.974, 3,966.219) \\ 
  & p = 0.998 \\ 
  DISCIPLINEgeography & 1,445.457 (389.349, 2,501.565) \\ 
  & p = 0.009$^{***}$ \\ 
  DISCIPLINEphysics & 292.559 ($-$2,717.659, 3,302.777) \\ 
  & p = 0.850 \\ 
  DISCIPLINEplanning & $-$143.554 ($-$1,361.357, 1,074.249) \\ 
  & p = 0.818 \\ 
  DISCIPLINEtransportation & 568.329 ($-$606.167, 1,742.826) \\ 
  & p = 0.346 \\ 
  Constant & 235.357 ($-$458.201, 928.914) \\ 
  & p = 0.508 \\ 
 \hline \\[-1.8ex]
Observations & 94 \\ 
R$^{2}$ & 0.100 \\ 
R$^{2}$ ajusté & 0.027 \\ 
Erreur Std. Résiduelle & 1,995.272 (df = 86) \\ 
Statistique F & 1.369 (df = 7; 86) \\ 
\hline 
\hline \\[-1.8ex]
\textit{Note:}  & \multicolumn{1}{r}{$^{*}$p$<$0.1; $^{**}$p$<$0.05; $^{***}$p$<$0.01} \\ 
\end{tabular}
}
\end{table} 

\paragraph{Time scale}{Echelle de temps}

\bpar{
The temporal scale is adjusted according to the linear models presented in Table~\ref{tab:app:modelography:tempscale}.
}{
L'échelle de temps est ajustée selon les modèles linéaires présentés en Table~\ref{tab:app:modelography:tempscale}.
}


\begin{table}
    \apptabcaption{\textbf{Linear models for the temporal scale.}\label{tab:app:modelography:tempscale}}{\textbf{Modèles linéaires pour l'échelle temporelle.}\label{tab:app:modelography:tempscale}}
\bpar{
\begin{tabular}{@{\extracolsep{5pt}}lcc} 
\\[-1.8ex]\hline 
\hline \\[-1.8ex] 
\\[-1.8ex] & \multicolumn{2}{c}{TEMPSCALE} \\ 
\\[-1.8ex] & (1) & (2)\\ 
\hline \\[-1.8ex] 
 YEAR & 0.674 ($-$0.294, 1.643) &  \\ 
  & p = 0.179 &  \\ 
  TYPEstrong &  & 100.271 (58.312, 142.230) \\ 
  &  & p = 0.00002$^{***}$ \\ 
  TYPEterritory & $-$38.933 ($-$64.249, $-$13.617) & $-$14.988 ($-$37.411, 7.435) \\ 
  & p = 0.004$^{***}$ & p = 0.194 \\ 
  DISCIPLINEengineering & $-$52.107 ($-$110.950, 6.735) & $-$9.609 ($-$55.841, 36.624) \\ 
  & p = 0.089$^{*}$ & p = 0.685 \\ 
  DISCIPLINEenvironment & 17.110 ($-$37.350, 71.569) & 17.886 ($-$45.319, 81.090) \\ 
  & p = 0.541 & p = 0.581 \\ 
  DISCIPLINEgeography & 3.640 ($-$15.364, 22.644) & 9.126 ($-$7.590, 25.843) \\ 
  & p = 0.709 & p = 0.288 \\ 
  DISCIPLINEphysics & 46.879 (0.638, 93.120) & 77.897 (28.225, 127.570) \\ 
  & p = 0.053$^{*}$ & p = 0.003$^{***}$ \\ 
  DISCIPLINEplanning & 1.304 ($-$19.336, 21.945) & 4.553 ($-$14.865, 23.971) \\ 
  & p = 0.902 & p = 0.648 \\ 
  DISCIPLINEtransportation & $-$14.718 ($-$34.978, 5.543) & 8.753 ($-$9.864, 27.371) \\ 
  & p = 0.161 & p = 0.360 \\ 
  INTERDISC & 2.357 ($-$59.200, 63.915) &  \\ 
  & p = 0.941 &  \\ 
  Constant & $-$1,305.126 ($-$3,252.499, 642.247) & 22.103 ($-$0.951, 45.156) \\ 
  & p = 0.195 & p = 0.064$^{*}$ \\ 
 \hline \\[-1.8ex] 
Observations & 64 & 94 \\ 
R$^{2}$ & 0.385 & 0.393 \\ 
Adjusted R$^{2}$ & 0.282 & 0.336 \\ 
Residual Std. Error & 26.984 (df = 54) & 31.747 (df = 85) \\ 
F Statistic & 3.755$^{***}$ (df = 9; 54) & 6.871$^{***}$ (df = 8; 85) \\ 
\hline 
\hline \\[-1.8ex] 
\textit{Note:}  & \multicolumn{2}{r}{$^{*}$p$<$0.1; $^{**}$p$<$0.05; $^{***}$p$<$0.01} \\ 
\end{tabular} 
}{
\begin{tabular}{@{\extracolsep{5pt}}lcc} 
\\[-1.8ex]\hline 
\hline \\[-1.8ex] 
\\[-1.8ex] & \multicolumn{2}{c}{TEMPSCALE} \\ 
\\[-1.8ex] & (1) & (2)\\ 
\hline \\[-1.8ex] 
 YEAR & 0.674 ($-$0.294, 1.643) &  \\ 
  & p = 0.179 &  \\ 
  TYPEstrong &  & 100.271 (58.312, 142.230) \\ 
  &  & p = 0.00002$^{***}$ \\ 
  TYPEterritory & $-$38.933 ($-$64.249, $-$13.617) & $-$14.988 ($-$37.411, 7.435) \\ 
  & p = 0.004$^{***}$ & p = 0.194 \\ 
  DISCIPLINEengineering & $-$52.107 ($-$110.950, 6.735) & $-$9.609 ($-$55.841, 36.624) \\ 
  & p = 0.089$^{*}$ & p = 0.685 \\ 
  DISCIPLINEenvironment & 17.110 ($-$37.350, 71.569) & 17.886 ($-$45.319, 81.090) \\ 
  & p = 0.541 & p = 0.581 \\ 
  DISCIPLINEgeography & 3.640 ($-$15.364, 22.644) & 9.126 ($-$7.590, 25.843) \\ 
  & p = 0.709 & p = 0.288 \\ 
  DISCIPLINEphysics & 46.879 (0.638, 93.120) & 77.897 (28.225, 127.570) \\ 
  & p = 0.053$^{*}$ & p = 0.003$^{***}$ \\ 
  DISCIPLINEplanning & 1.304 ($-$19.336, 21.945) & 4.553 ($-$14.865, 23.971) \\ 
  & p = 0.902 & p = 0.648 \\ 
  DISCIPLINEtransportation & $-$14.718 ($-$34.978, 5.543) & 8.753 ($-$9.864, 27.371) \\ 
  & p = 0.161 & p = 0.360 \\ 
  INTERDISC & 2.357 ($-$59.200, 63.915) &  \\ 
  & p = 0.941 &  \\ 
  Constant & $-$1,305.126 ($-$3,252.499, 642.247) & 22.103 ($-$0.951, 45.156) \\ 
  & p = 0.195 & p = 0.064$^{*}$ \\ 
 \hline \\[-1.8ex] 
Observations & 64 & 94 \\ 
R$^{2}$ & 0.385 & 0.393 \\ 
R$^{2}$ ajusté & 0.282 & 0.336 \\ 
Erreur Std. Résiduelle & 26.984 (df = 54) & 31.747 (df = 85) \\ 
Statistique F & 3.755$^{***}$ (df = 9; 54) & 6.871$^{***}$ (df = 8; 85) \\
\hline 
\hline \\[-1.8ex] 
\textit{Note:}  & \multicolumn{2}{r}{$^{*}$p$<$0.1; $^{**}$p$<$0.05; $^{***}$p$<$0.01} \\ 
\end{tabular} 
}
\end{table} 

\paragraph{Year}{Année}

\bpar{
The publications year is adjusted following the linear model which adjustement is given in Table~\ref{tab:app:modelography:year}.
}{
L'année de publication est ajustée selon le modèle linéaire dont l'ajustement est donné en Table~\ref{tab:app:modelography:year}.
}

\begin{table}
  \apptabcaption{\textbf{Linear model for the publication year.}\label{tab:app:modelography:year}}{\textbf{Modèle linéaire pour l'année de publication.}\label{tab:app:modelography:year}}
\bpar{
\begin{tabular}{@{\extracolsep{5pt}}lc} 
\footnotesize
\\[-1.8ex]\hline 
\hline \\[-1.8ex] 
\\[-1.8ex] & YEAR \\ 
\hline \\[-1.8ex] 
 TYPEterritory & 10.898 (3.045, 18.750), p = 0.010$^{***}$ \\ 
  TEMPSCALE & 0.035 ($-$0.033, 0.103), p = 0.320 \\ 
  FMETHODeq & $-$6.224 ($-$20.162, 7.714), p = 0.387 \\ 
  FMETHODmap & 4.747 ($-$7.595, 17.089), p = 0.456 \\ 
  FMETHODro & 6.128 ($-$11.694, 23.950), p = 0.504 \\ 
  FMETHODsem & 1.009 ($-$16.659, 18.676), p = 0.912 \\ 
  FMETHODsim & 5.153 ($-$6.809, 17.114), p = 0.404 \\ 
  FMETHODstat & $-$0.357 ($-$10.925, 10.211), p = 0.948 \\ 
  DISCIPLINEengineering & 13.486 ($-$7.238, 34.210), p = 0.210 \\ 
  DISCIPLINEenvironment & $-$3.668 ($-$21.605, 14.269), p = 0.691 \\ 
  DISCIPLINEgeography & 1.121 ($-$4.528, 6.769), p = 0.700 \\ 
  DISCIPLINEphysics & 3.392 ($-$8.461, 15.245), p = 0.578 \\ 
  DISCIPLINEplanning & $-$2.850 ($-$8.873, 3.173), p = 0.359 \\ 
  DISCIPLINEtransportation & 5.503 (0.006, 11.000), p = 0.057$^{*}$ \\ 
  INTERDISC & $-$12.876 ($-$29.567, 3.815), p = 0.138 \\ 
  SEMCOMhedonic & $-$5.769 ($-$19.931, 8.393), p = 0.430 \\ 
  SEMCOMhsr & 6.135 ($-$9.889, 22.159), p = 0.458 \\ 
  SEMCOMinfra planning & $-$4.123 ($-$18.910, 10.663), p = 0.588 \\ 
  SEMCOMnetworks & 4.711 ($-$9.736, 19.158), p = 0.527 \\ 
  SEMCOMtod & $-$1.653 ($-$15.837, 12.532), p = 0.821 \\ 
  Constant & 2,004.945 (1,981.531, 2,028.359), p = 0.000$^{***}$ \\ 
 \hline \\[-1.8ex]
Observations & 64 \\ 
R$^{2}$ & 0.510 \\ 
Adjusted R$^{2}$ & 0.281 \\ 
Residual Std. Error & 6.617 (df = 43) \\ 
F Statistic & 2.234$^{**}$ (df = 20; 43) \\ 
\hline 
\hline \\[-1.8ex] 
\textit{Note:}  & \multicolumn{1}{r}{$^{*}$p$<$0.1; $^{**}$p$<$0.05; $^{***}$p$<$0.01} \\ 
\end{tabular}
}{
\begin{tabular}{@{\extracolsep{5pt}}lc} 
\footnotesize
\\[-1.8ex]\hline 
\hline \\[-1.8ex] 
\\[-1.8ex] & YEAR \\ 
\hline \\[-1.8ex] 
 TYPEterritory & 10.898 (3.045, 18.750), p = 0.010$^{***}$ \\ 
  TEMPSCALE & 0.035 ($-$0.033, 0.103), p = 0.320 \\ 
  FMETHODeq & $-$6.224 ($-$20.162, 7.714), p = 0.387 \\ 
  FMETHODmap & 4.747 ($-$7.595, 17.089), p = 0.456 \\ 
  FMETHODro & 6.128 ($-$11.694, 23.950), p = 0.504 \\ 
  FMETHODsem & 1.009 ($-$16.659, 18.676), p = 0.912 \\ 
  FMETHODsim & 5.153 ($-$6.809, 17.114), p = 0.404 \\ 
  FMETHODstat & $-$0.357 ($-$10.925, 10.211), p = 0.948 \\ 
  DISCIPLINEengineering & 13.486 ($-$7.238, 34.210), p = 0.210 \\ 
  DISCIPLINEenvironment & $-$3.668 ($-$21.605, 14.269), p = 0.691 \\ 
  DISCIPLINEgeography & 1.121 ($-$4.528, 6.769), p = 0.700 \\ 
  DISCIPLINEphysics & 3.392 ($-$8.461, 15.245), p = 0.578 \\ 
  DISCIPLINEplanning & $-$2.850 ($-$8.873, 3.173), p = 0.359 \\ 
  DISCIPLINEtransportation & 5.503 (0.006, 11.000), p = 0.057$^{*}$ \\ 
  INTERDISC & $-$12.876 ($-$29.567, 3.815), p = 0.138 \\ 
  SEMCOMhedonic & $-$5.769 ($-$19.931, 8.393), p = 0.430 \\ 
  SEMCOMhsr & 6.135 ($-$9.889, 22.159), p = 0.458 \\ 
  SEMCOMinfra planning & $-$4.123 ($-$18.910, 10.663), p = 0.588 \\ 
  SEMCOMnetworks & 4.711 ($-$9.736, 19.158), p = 0.527 \\ 
  SEMCOMtod & $-$1.653 ($-$15.837, 12.532), p = 0.821 \\ 
  Constant & 2,004.945 (1,981.531, 2,028.359), p = 0.000$^{***}$ \\ 
 \hline \\[-1.8ex]
Observations & 64 \\ 
R$^{2}$ & 0.510 \\ 
R$^{2}$ ajusté & 0.281 \\ 
Erreur Std. Résiduelle & 6.617 (df = 43) \\ 
Statistique F & 2.234$^{**}$ (df = 20; 43) \\ 
\hline 
\hline \\[-1.8ex] 
\textit{Note:}  & \multicolumn{1}{r}{$^{*}$p$<$0.1; $^{**}$p$<$0.05; $^{***}$p$<$0.01} \\ 
\end{tabular}
}
\end{table} 

\stars

%


\newpage

\section{Static Correlations}{Corrélations statiques}

\label{app:sec:staticcorrelations}

\subsection{Morphological Measures}{Mesures morphologiques}

\bpar{
We compute for China, from the population grid with a 1km resolution~\cite{fu1km}, the morphological indicators. We consider areas of width 100km, in order to have a reasonable number of points for the estimation, with an offset of 50km. Corresponding maps are given in Fig.~\ref{fig:app:staticcorrelations:morphocn}. The distribution of some indicators such as the entropy $\mathcal{E}$ seems to be conditioned to province boundaries as in Sichuan, the uniformity of the dataset must possibly be questioned.
}{
Nous calculons pour la Chine, à partir de la grille de population à 1km~\cite{fu1km}, les indicateurs morphologiques. Nous prenons des zones de 100km de côté, afin d'avoir un nombre raisonnable de points pour l'estimation, avec un décalage de 50km. Les cartes correspondantes sont données en Fig.~\ref{fig:app:staticcorrelations:morphocn}. La distribution de certains indicateurs comme l'entropie $\mathcal{E}$ semblant conditionnée aux limites de province comme pour le Sichuan, l'uniformité du jeu de données est éventuellement à questionner.
}

\begin{figure}
\includegraphics[width=\linewidth]{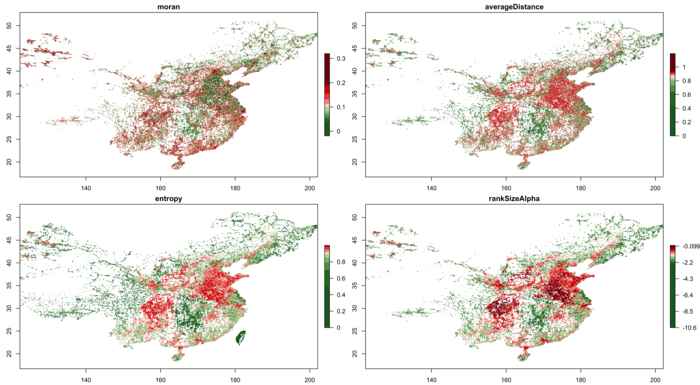}
\appcaption{\textbf{Morphological indicators for China.} We give for areas where a population and the network are simultaneously defined, the Moran index $I$ (\texttt{moran}), the average distance $\bar{d}$ (\texttt{averageDistance}), the entropy $\mathcal{E}$ (\texttt{entropy}) and the hierarchy $\gamma$ (\texttt{rankSizeAlpha}).\label{fig:app:staticcorrelations:morphocn}}{\textbf{Indicateurs morphologiques pour la Chine.} Nous donnons pour les zones où une population et le réseau sont simultanément définis, l'indice de Moran $I$ (\texttt{moran}), la distance moyenne $\bar{d}$ (\texttt{averageDistance}), l'entropie $\mathcal{E}$ (\texttt{entropy}) et la hiérarchie $\gamma$ (\texttt{rankSizeAlpha}).\label{fig:app:staticcorrelations:morphocn}}
\end{figure}

\subsection{Road network}{Réseau routier}

\subsubsection{Network Simplification Algorithm}{Algorithme de simplification du réseau}


\bpar{
We detail here the road network simplification algorithm from OpenStreetMap data. The general workflow is the following: (i) data import by selection and spatial aggregation at the raster resolution; (ii) simplification to keep only the topological network, processed in parallel through \emph{split/merge}.
}{
Nous détaillons ici l'algorithme de simplification du réseau routier à partir des données OpenStreetMap. La logique générale est la suivante : (i) import des données par sélection et agrégation spatiale à la résolution du raster ; (ii) simplification pour conserver le réseau topologique uniquement, opérée en parallèle par \emph{split/merge}.
}

\bpar{
OSM data are imported into a \texttt{pgsql} database (\texttt{Postgis} extension for the management of geometries and to have spatial indexes). The import is done using the software \texttt{osmosis}~\cite{osmosis}, from an image in compressed \texttt{pbf} format of the OpenStreetMap database\footnote{Dumps were retrieved from \url{http://download.geofabrik.de}, in July 2016 for Europe, and July 2017 for China.}. We filter at this stage the links (\texttt{ways}) which posses the tag \texttt{highway}, and keep the corresponding nodes.
}{
Les données OSM sont importées dans une base de données \texttt{pgsql} (extension \texttt{Postgis} pour la gestion des géométries et avoir des index spatiaux). L'import est fait en utilisant le logiciel \texttt{osmosis}~\cite{osmosis}, à partir d'une image en format compressé \texttt{pbf} de la base OpenStreetMap\footnote{Les dumps ont été récupérés à partir de \url{http://download.geofabrik.de}, en juillet 2016 pour l'Europe, et juillet 2017 pour la Chine.}. Nous filtrons à cette étape les liens (\texttt{ways}) qui possèdent le tag \texttt{highway}, et conservons les noeuds correspondants.
}

\bpar{
The network is first aggregated at a 100m granularity in order to be consistently used with population grids. It furthermore allows to be robust to local coding imperfections or to very local missing data. For this step, roads are filtered on a relevant subset of tags\footnote{That we take within \texttt{motorway,trunk,primary,secondary,tertiary,unclassified,residential}.}. For the set of segments of corresponding lines, a link is created between the origin and the destination cell, with a real length computed between the center of cells and a speed taken as the speed of the line if it is available.
}{
Le réseau est d'abord agrégé à une granularité de 100m pour pouvoir être utilisé de manière cohérente avec les grilles de population. Cela permet par ailleurs d'être robuste aux imperfections locales de codage ou données très locales manquantes. Pour cette étape, les routes sont filtrées sur un sous-ensemble de tags pertinents\footnote{Que nous prenons dans \texttt{motorway,trunk,primary,secondary,tertiary,unclassified,residential}.}. Pour l'ensemble des segments des lignes correspondantes, un lien est créé entre la cellule d'origine et de destination, avec longueur réelle calculée entre les centres des cellules et vitesse prise comme la vitesse de la ligne si elle est disponible.
}

\bpar{
The simplification is then operated the following way:
\begin{enumerate}
\item The whole geographical coverage is cut into areas on which computations will be partly done through parallel computation (\emph{split} paradigm). Areas have a fixed size in number of cells of the base raster (200 cells).
\item On each sub-area, a simplification algorithm is applied the following way: as long as there still are vertices of degree 2, successive sequences of such vertices are determined, and corresponding links are replaced by a unique link with real length and speed computed by cumulation on the deleted links.
\item As the simplification algorithm keeps the links having an intersection with the border of areas, a fusion followed by a simplification of resulting graphs is necessary. To keep a reasonable computational cost, the size of merged areas has to stay low: we take merge areas composed by two contiguous areas. A paving by four sequences of independent merging allows then to cover the full set of joints between areas\footnote{In the very rare cases of a link between two non-contiguous areas, the remaining link is not simplified. This case was not observed in practice in our data.}, these sequences being executed sequentially. The Frame \ref{frame:app:staticcorrelations:merging} shows the covering of joints by merging areas.
\end{enumerate}
}{
La simplification est alors opérée de la façon suivante :
\begin{enumerate}
\item L'ensemble de la couverture géographique est découpée en zones sur lesquelles les traitement seront effectués en partie par du calcul parallèle (paradigme \emph{split}). Les zones sont de côté fixe en nombre de cellules du raster de base (200 cellules).
\item Sur chaque sous-zone, un algorithme de simplification est effectué de la façon suivante : tant qu'il reste des sommets de degré 2, les séquences successives de tels sommets sont déterminées, et les liens correspondants sont remplacés par un lien unique avec longueur et vitesse réels calculés par cumul sur les liens supprimés.
\item Comme l'algorithme de simplification conserve les liens ayant une intersection avec la bordure des zones, une fusion puis simplification des graphes fusionnés est nécessaire. Pour garder un coût computationnel raisonnable, la taille des zones fusionnées doit rester modeste : nous prenons des zones de fusion composées de deux zones contiguës. Un pavage par quatre séquences de fusion indépendantes permet alors de couvrir l'ensemble des jointures entre zones\footnote{Dans les cas très rares d'un lien entre deux zones non contiguës, le lien restant n'est pas simplifié. Ce cas n'a pas été observé en pratique dans nos données.}, ces séquences pouvant être exécutées à la suite. L'encadré \ref{frame:app:staticcorrelations:merging} montre le recouvrement des jointures par les zones de fusion.
\end{enumerate}
}

\bpar{
We have then at our disposition a topological graph given by the links between cells of the base raster, having distance and speed attributes corresponding to the underlying real links.
}{
Nous disposons alors d'un graphe topologique donné par les liens entre cellules du raster de base, ayant des attributs de distance et de vitesse correspondant aux liens réels sous-jacents.
}

\bpar{
Graphs for Europe and China are available as open databases (see Appendix~\ref{app:data}).
}{
Les graphes pour l'Europe et la Chine sont disponibles en base de données ouverte (voir Annexe~\ref{app:data}).
}

\begin{figure}[h!]
\begin{mdframed}

	\includegraphics[width=\linewidth]{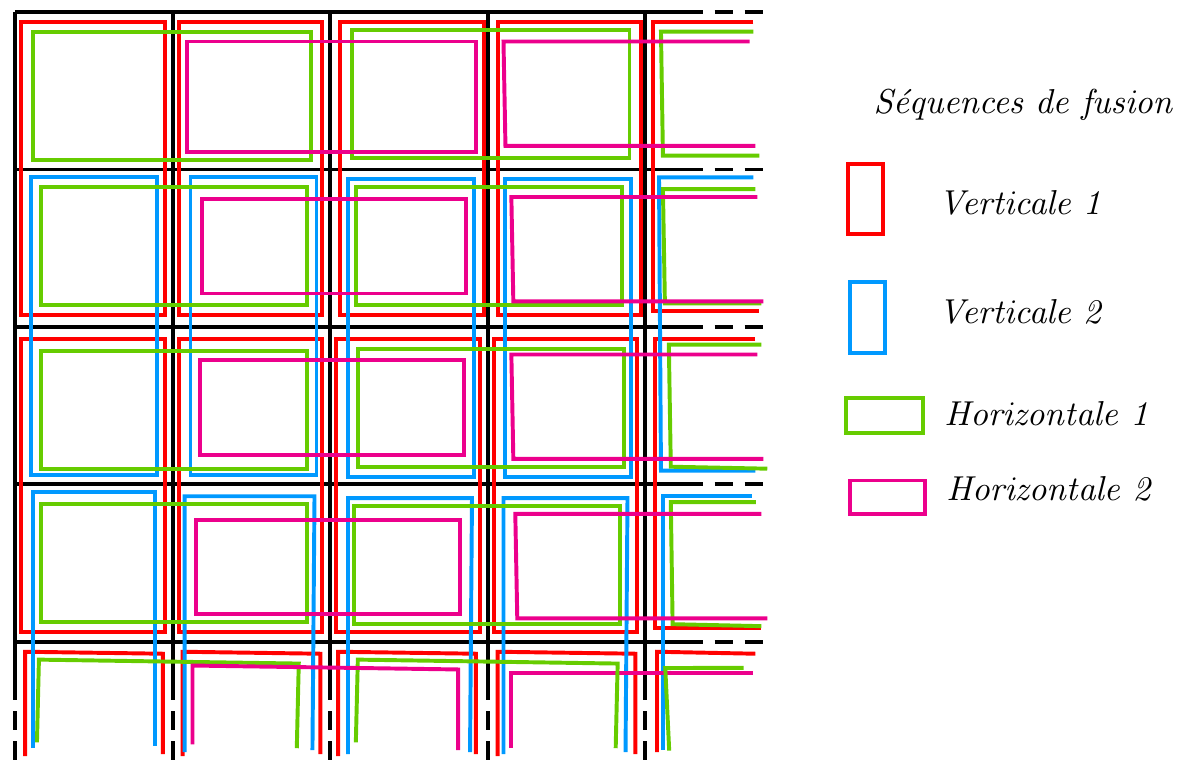}

	\medskip

	\framecaption{\textbf{Illustration of merging sequences.} The four independent sequences (horizontally and vertically) allow the coverage of all joints between areas.\label{frame:app:staticcorrelations:merging}}{\textbf{Illustration des séquences de fusion.} Les quatre séquences indépendantes (horizontales et verticales) permettent le recouvrement de l'ensemble des jointures entre zones.\label{frame:app:staticcorrelations:merging}}
\end{mdframed}
\end{figure}





\subsubsection{Network Indicators}{Indicateurs de réseau}

\bpar{
We give in Fig.~\ref{fig:app:staticcorrelations:networkcn} a sample of network indicators for China.
}{
Nous donnons en Fig.~\ref{fig:app:staticcorrelations:networkcn} un échantillon d'indicateurs de réseau pour la Chine.
}

\begin{figure}
\includegraphics[width=\linewidth]{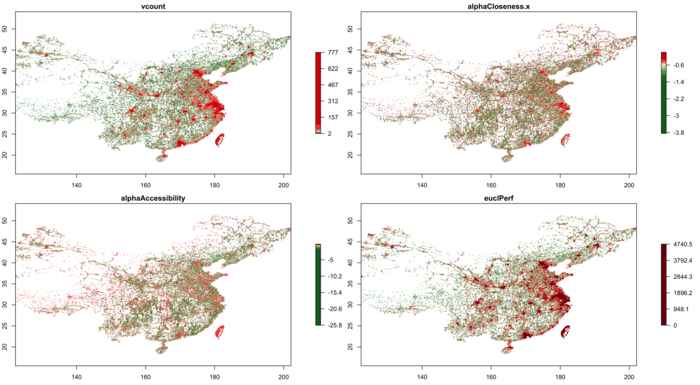}
\appcaption{\textbf{Network indicators for China.} We show a selection of network indicators: number of nodes $\left|V\right|$ (\texttt{vcount}), closeness hierarchy $\alpha_{cl}$ (\texttt{alphaCloseness.x}), accessibility hierarchy $\alpha_Z$ (\texttt{alphaAccessibility}), euclidian performance $v_0$ (\texttt{euclPerf}).\label{fig:app:staticcorrelations:networkcn}}{\textbf{Indicateurs de réseau pour la Chine.} Nous montrons une sélection d'indicateurs de réseau : nombre de noeuds $\left|V\right|$ (\texttt{vcount}), hiérarchie de \emph{closeness} $\alpha_{cl}$ (\texttt{alphaCloseness.x}), hiérarchie de l'accessibilité $\alpha_Z$ (\texttt{alphaAccessibility}), performance euclidienne $v_0$ (\texttt{euclPerf}).\label{fig:app:staticcorrelations:networkcn}}
\end{figure}

\subsection{Sensitivity to resolution}{Sensibilité à la résolution}

\bpar{
We evaluate here the sensitivity of indicators to grid size. We show in Fig.~\ref{fig:app:staticcorrelations:sensitivity-maps-morpho} morphological indicators and in Fig.~\ref{fig:app:staticcorrelations:sensitivity-maps-network} some network indicators, mapped for France, for different grid sizes. The sizes taken here, in correspondance to the 50km scale used in main results, are at similar magnitudes: we test windows of size 30km and 100km. The offsets are in each case half of the window (15km and 50km respectively). It is possible to see with eyeball validation that some indicators have a low sensitivity, the change in scale resembling a smoothing of the finer field: for example for morphology in the case of Moran, entropy and hierarchy. Average distance, which is indeed rather noisy at the smaller scale, is necessarily sensitive to aggregation, what is consistent with a sensitivity expected at smoothing. Network indicators are relatively robust to window size.
}{
Nous évaluons ici la sensibilité des divers indicateurs à la taille de la grille. Nous montrons en Fig.~\ref{fig:app:staticcorrelations:sensitivity-maps-morpho} les indicateurs morphologiques et en Fig.~\ref{fig:app:staticcorrelations:sensitivity-maps-network} certains indicateurs de réseau, cartographiés pour la France, pour des tailles différentes de grille. Les tailles données ici, en écho à celle de 50km utilisée dans les résultats principaux, sont dans des ordres de grandeur équivalents : nous testons des fenêtres de taille 30km et 100km. Les décalages sont à chaque fois de la moitié de la fenêtre (15km et 50km respectivement). Il est possible de voir ``à l'oeil'' que certains indicateurs sont peu sensibles, le changement d'échelle ressemblant à un lissage du champ le plus fin : par exemple dans le cas morphologique pour l'indice de Moran, l'entropie et la hiérarchie. La distance moyenne, en fait très bruitée à l'échelle la plus faible, est nécessairement sensible à l'agrégation, ce qui est cohérent avec une sensibilité attendue au lissage. Les indicateurs de réseau sont relativement robustes à la taille de la fenêtre.
}


\begin{figure}
	\includegraphics[width=\linewidth,height=0.95\textheight]{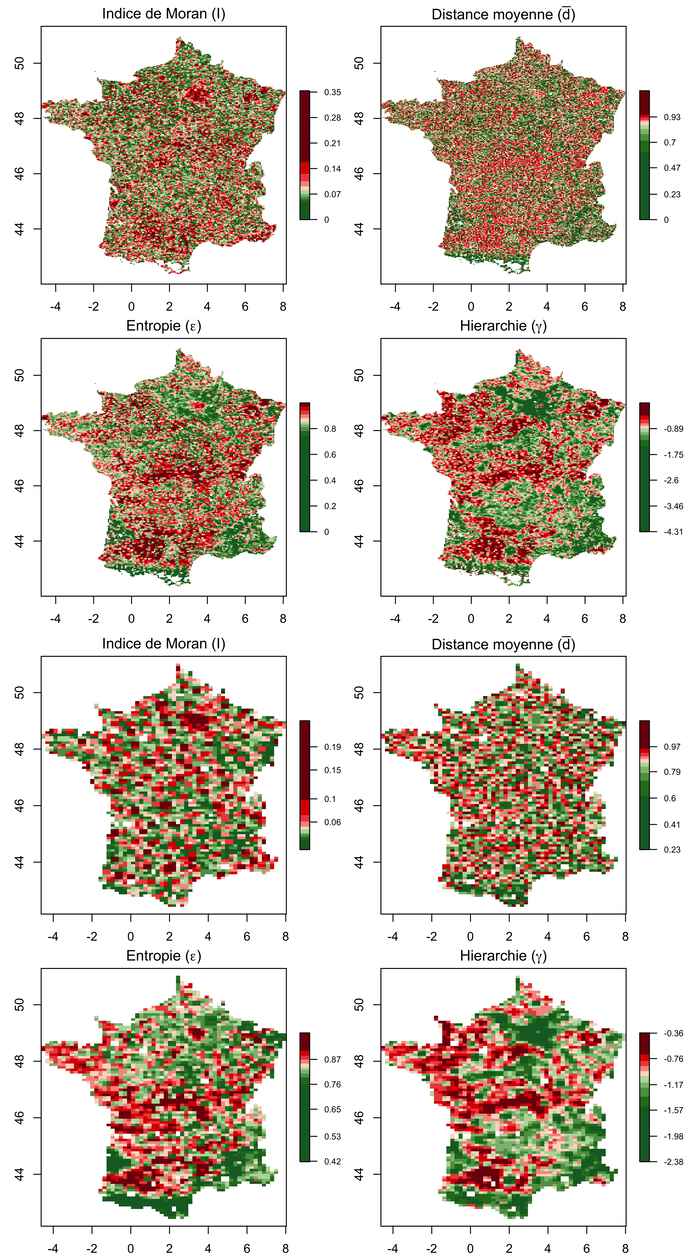}
	\appcaption{\textbf{Morphological indicators for different grid sizes.} The first four maps show the indicators computed on a window of size 30km, the last four maps with a window of size 100km.\label{fig:app:staticcorrelations:sensitivity-maps-morpho}}{\textbf{Indicateurs morphologiques pour différentes tailles de grille.} Les 4 premières cartes montrent les indicateurs calculés avec une fenêtre de taille 30km, les 4 dernières avec une fenêtre de taille 100km.\label{fig:app:staticcorrelations:sensitivity-maps-morpho}}
\end{figure}

\begin{figure}
	\includegraphics[width=\linewidth,height=0.95\textheight]{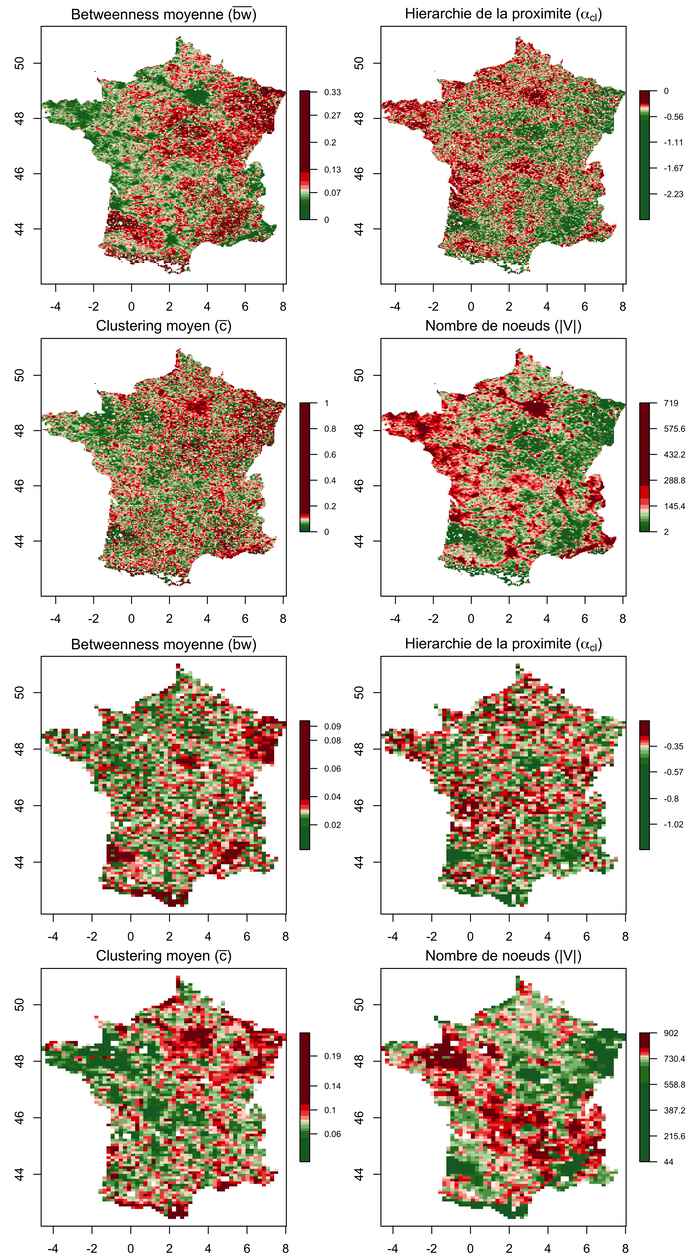}
	\appcaption{\textbf{Sample of network indicators for different grid sizes.} The first four maps give the indicators computed with a window of size 30km, the last four maps with a window of size 100km.\label{fig:app:staticcorrelations:sensitivity-maps-network}}{\textbf{Echantillon des indicateurs de réseau pour différentes tailles de grille.} Les 4 premières cartes montrent les indicateurs calculés avec une fenêtre de taille 30km, les 4 dernières avec une fenêtre de taille 100km.\label{fig:app:staticcorrelations:sensitivity-maps-network}}
\end{figure}

\bpar{
This comparison, on the one hand is to be taken cautiously because of the difficulty to directly compare scales for indicators, and on the other hand stays limited. We propose then a method to quantify the variability of indicators to window size. Let $X_D$ and $X_d$ two spatial fields corresponding to two spatial scales $D > d$ (that we take as characteristic distances). The fields are assumed discrete at points respectively denoted by $\left(\vec{x}_i^{(D)}\right)_{1 \leq i \leq N_D}$ and $\left(\vec{x}_j^{(d)}\right)_{1 \leq i \leq N_d}$. The idea is to compare a smoothing of the finer field to the field with the larger scale: if the correlation between these two values is high, it is possible to deduce one field from the other by aggregation and the scale of computation does not influence final results in an other way than the final resolution. Let $W_{ij} = \left( \exp{ - d_{ij} / d_0} \right)_{ij}$ a matrix of spatial weights computed with euclidian distances $d_{ij}$ between the points $\vec{x}_i^{(D)}$ and $\vec{x}_j^{(d)}$. Then with $W'_{ij} = W_{ij} / \sum_j W_{ij}$, we can compute the spatial smoothing of $X_d$ at the points $\vec{x}_i^{(D)}$, with the matrix product
\[
\tilde{X}_d (\vec{x}_i^{(D)}) = W' \ast \vec{x}_j^{(d)}
\]
The correlation is then given by $\rho\left[\tilde{X}_d,X_D\right]$ estimated on all $\vec{x}_i^{(D)}$ points.
}{
Cette comparaison, d'une part est à prendre avec précaution de par la non-comparabilité directe des échelles pour les indicateurs, et d'autre part reste limitée. Nous proposons alors une méthode pour quantifier la variabilité des indicateurs à la taille de la fenêtre. Soit $X_D$ et $X_d$ deux champs spatiaux correspondant à deux échelles spatiales $D > d$ (qu'on prend comme des distances caractéristiques), qu'on suppose discrets en des points respectifs $\left(\vec{x}_i^{(D)}\right)_{1 \leq i \leq N_D}$ et $\left(\vec{x}_j^{(d)}\right)_{1 \leq i \leq N_d}$. L'idée est de comparer un lissage du champ le plus fin au champ le moins fin : si la corrélation entre ces deux valeurs est forte, il est possible de passer d'un champ à l'autre par agrégation et l'échelle de calcul n'influe ainsi pas autrement que sur la résolution finale. Soit $W_{ij} = \left( \exp{ - d_{ij} / d_0} \right)_{ij}$ une matrice de poids spatiaux calculée par les distances euclidiennes $d_{ij}$ entre les points $\vec{x}_i^{(D)}$ et $\vec{x}_j^{(d)}$. Alors avec $W'_{ij} = W_{ij} / \sum_j W_{ij}$, on peut calculer le lissage spatial de $X_d$ aux points $\vec{x}_i^{(D)}$, par le produit matriciel
\[
\tilde{X}_d (\vec{x}_i^{(D)}) = W' \ast \vec{x}_j^{(d)}
\]

La corrélation est alors donnée par $\rhob{\tilde{X}_d}{X_D}$ évaluée sur l'ensemble des points $\vec{x}_i^{(D)}$.
}

\bpar{
The Fig.~\ref{fig:app:staticcorrelations:sensitivity-corrs} gives the variation of this correlation for all $(D,d)$ couples, with a variable $d_0$ for smoothing. We generally observe the existence of a maximum, which corresponds to the optimal smoothing level to deduce the larger scale from the finer. The largest correlations on all indicators are obtained for $D=50$km and $d=30$km, what means that indicators are not very sensitive to small variations in small window sizes. As expected, the lowest correlations are obtained for the largest scale difference (100/30km). Morphological indicators have the same qualitative behavior across combinations, and we find the behavior suggested by the previous maps (entropy and hierarchy being the less sensitive, Moran index and average distance a bit more sensitive). For the network, some indicators such as $\alpha_{bw}$ show a significant transition depending on $D-d$: there exists for this indicator a large sensitivity in small sizes. For all indicators, the sensitivity remains however reasonable. Finally, a smoothing of both fields yields asymptotic maximal correlations with very high values: the computation window size does not matter if we consider smoothed fields.
}{
La Fig.~\ref{fig:app:staticcorrelations:sensitivity-corrs} donne les variations de cette corrélation pour l'ensemble des couples $(D,d)$, avec le lissage effectué avec $d_0$ variable à chaque fois. Nous observons généralement l'existence d'un maximum, qui correspond au niveau de lissage optimal pour déduire la plus grande échelle à partir de la petite. Les meilleures corrélations sur l'ensemble des indicateurs sont obtenues pour $D=50$km et $d=30$km, ce qui signifie que les indicateurs sont peu sensibles aux faibles variations dans les petites tailles. Logiquement, les plus basses corrélations sont obtenues pour l'écart d'échelle le plus grand (100/30km). Les indicateurs morphologiques ont le même comportement qualitatif selon les combinaisons, et on retrouve le comportement observé à l'oeil sur les cartes (entropie et hiérarchie les moins sensibles, Moran et distance moyenne plus sensibles). Pour le réseau, certains indicateurs comme $\alpha_{bw}$ présentent une forte transition selon $D-d$ : il existe pour cet indicateur une forte sensibilité dans les petites tailles. Sur l'ensemble des indicateurs, la sensibilité reste tout de même raisonnable. Enfin, un lissage des deux champs donne des corrélations maximales asymptotiquement et très forte : la taille de calcul importe peu si on considère des champs lissés.
}


\begin{figure}
	\includegraphics[width=\linewidth,height=0.9\textheight]{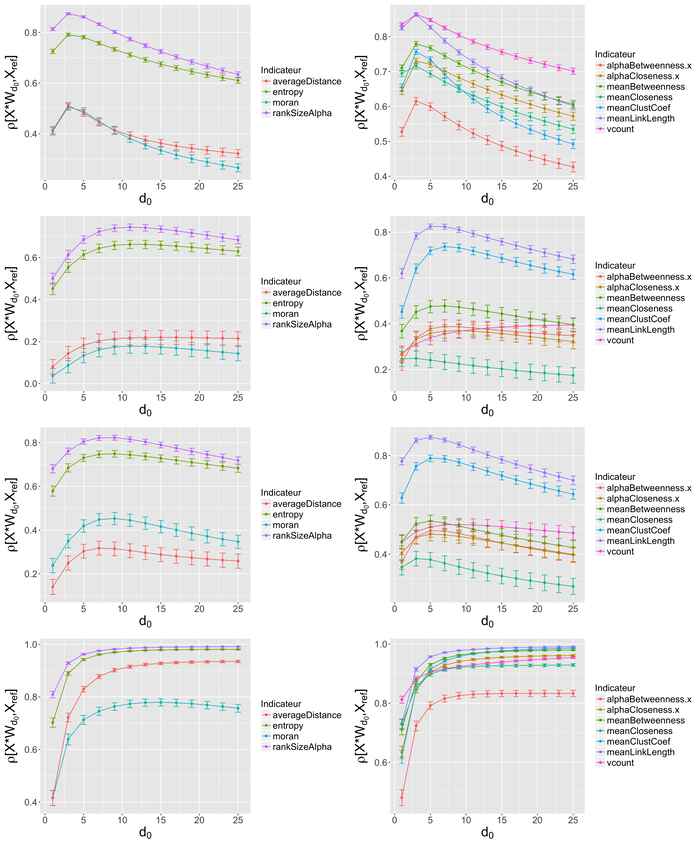}
	\appcaption{\textbf{Correlations between indicators computed at different scales.} From top to bottom (left column giving morphological indicators and right column network indicators), $(d=30,D=50)$, $(d=30,D=100)$, $(d=50,D=100)$, and the last row gives the correlation between the two fields $d_1=30$ and $d_2=50$ both smoothed at the characteristic distance of $d_0$.\label{fig:app:staticcorrelations:sensitivity-corrs}}{\textbf{Corrélations entre indicateurs à différentes échelles.} Dans l'ordre de haut en bas (colonne de gauche indicateurs morphologiques, colonne de droite indicateurs de réseau), $(d=30,D=50)$, $(d=30,D=100)$, $(d=50,D=100)$, et la dernière ligne donne la corrélation entre les deux champs $d_1=30$ et $d_2=50$ tous les deux lissés à la taille $d_0$.\label{fig:app:staticcorrelations:sensitivity-corrs}}
\end{figure}

\subsection{Spatial Correlations}{Corrélations Spatiales}

\bpar{
The Fig.~\ref{fig:app:staticcorrelations:overallcorrs} gives the correlation matrix estimated for $\delta = \infty$. To have an idea of the robustness of the estimation, we investigate the relative size of confidence intervals at the 95\% level (Fisher method) given by $\frac{\left|\rho_+ - \rho_-\right|}{\left|\rho\right|}$, for correlations such that $\left|\rho\right|>0.05$. The median of this rate is at $0.04$, the ninth decile at $0.12$ and the maximum at $0.19$, what means that the estimation is always relatively good compared to the value of correlations.
}{
La Fig.~\ref{fig:app:staticcorrelations:overallcorrs} donne la matrice de corrélation estimée pour $\delta = \infty$. Pour se donner une idée de la robustesse de l'estimation, nous regardons les tailles relatives des intervalles de confiance à 95\% (méthode de Fisher) données par $\frac{\left|\rho_+ - \rho_-\right|}{\left|\rho\right|}$, pour les corrélations telles que $\left|\rho\right|>0.05$. La médiane de ce rapport est à $0.04$, le neuvième décile à $0.12$ et le maximum à $0.19$, ce qui veut dire que l'estimation est toujours relativement bonne par rapport à la valeur des corrélations.
}



\begin{figure}
	\includegraphics[width=\linewidth]{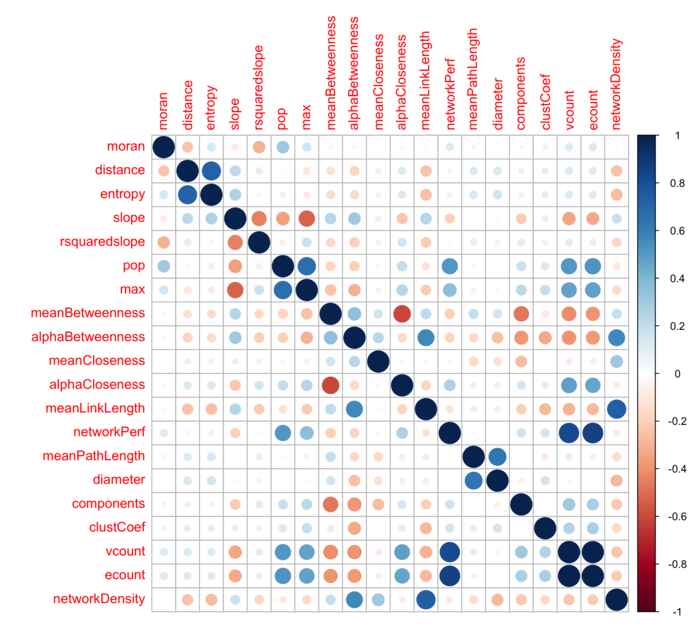}
	\appcaption{\textbf{Correlation matrix.} The matrix is estimated here on all indicator values for Europe, what is equivalent to take $\delta=\infty$.\label{fig:app:staticcorrelations:overallcorrs}}{\textbf{Matrice de corrélation.} La matrice est ici estimée sur l'ensemble des valeurs des indicateurs pour l'Europe, ce qui est équivalent à prendre $\delta=\infty$.\label{fig:app:staticcorrelations:overallcorrs}}
\end{figure}

\bpar{
The Fig.~\ref{fig:app:staticcorrelations:europe-correlations} gives the spatial distribution for all Europe, of a sample of correlations between indicators: $\rhob{\alpha_{cl}}{I}$, $\rhob{\gamma}{\alpha}$, $\rhob{\bar{bw}}{\gamma}$, $\rhob{\alpha_{bw}}{\alpha_{cl}}$, $\rhob{\left|V\right|}{\bar{l}}$, $\rhob{\gamma}{r_{\gamma}}$ (with $r_{\gamma}$ adjustment coefficient for $\gamma$). We see interesting structures emerging, such as the hierarchy and its adjustment which present an area of strong correlation in the center of Europe and negative correlation areas, or the number of nodes and the path length which correlate in mountains and along the coasts (what is expected since roads then do several detours) and have a negative correlation otherwise.
}{
La Fig.~\ref{fig:app:staticcorrelations:europe-correlations} donne la distribution spatiale pour l'ensemble de l'Europe, d'un échantillon de corrélations entre indicateurs : $\rhob{\alpha_{cl}}{I}$, $\rhob{\gamma}{\alpha}$, $\rhob{\bar{bw}}{\gamma}$, $\rhob{\alpha_{bw}}{\alpha_{cl}}$, $\rhob{\left|V\right|}{\bar{l}}$, $\rhob{\gamma}{r_{\gamma}}$ (avec $r_{\gamma}$ coefficient d'ajustement pour $\gamma$). On voit apparaître des structures intéressantes, comme la hiérarchie et son ajustement qui présentent une zone de forte corrélation au centre de l'Europe et des zones de corrélation négatives, ou le nombre de noeuds et la longueur de chemin qui se corrèlent en montagne et le long de côtes (ce qui est logique car les routes font alors de nombreux détours) et ont une corrélation négative ailleurs.
}

\begin{figure}
	\includegraphics[width=\linewidth]{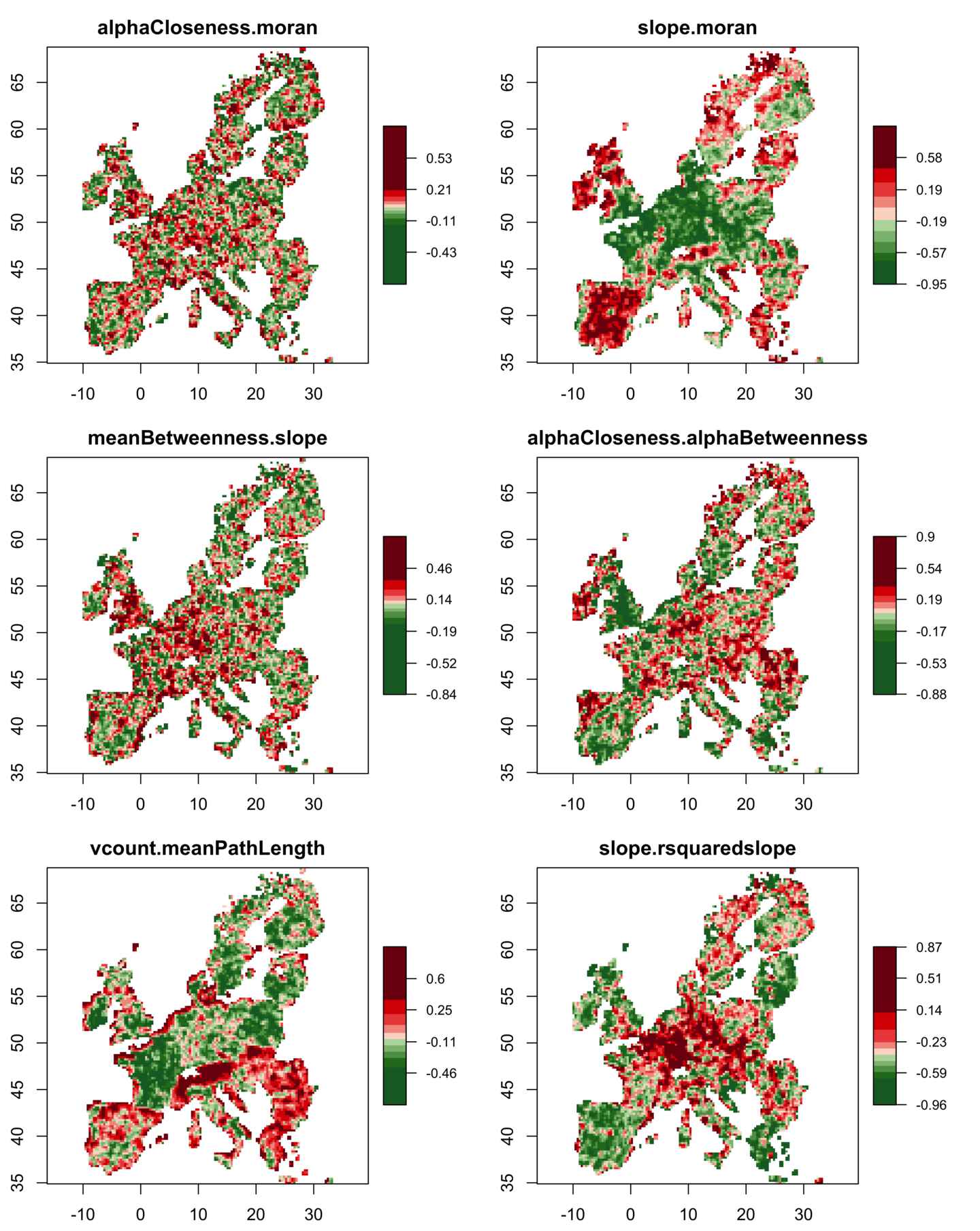}
\appcaption{\textbf{Spatial correlations for Europe.} The estimation is done here with $\delta = 12$.\label{fig:app:staticcorrelations:europe-correlations}}{\textbf{Correlations spatiales pour l'Europe.} L'estimation est faite ici avec $\delta = 12$.\label{fig:app:staticcorrelations:europe-correlations}}
\end{figure}

\bpar{
The Fig.~\ref{fig:app:staticcorrelations:corr-distribs} gives the statistical distributions of estimated correlations on all the areas, for different values of $\delta$. We distinguish there the different blocks in the correlation matrix, i.e. correlations between morphological indicators, the ones between network indicators, and also cross-correlations. The latest have rather symmetrical distributions, whereas network and morphological correlations are dissymmetrical. We also give point clouds allowing to make a link between these different components.
}{
La Fig.~\ref{fig:app:staticcorrelations:corr-distribs} donne les distributions statistiques des correlations estimées sur l'ensemble des zones, pour différentes valeurs de $\delta$. Nous y distinguons les différents blocs dans la matrice de corrélation, c'est-à-dire les corrélations entre les indicateurs morphologiques, celles entre les indicateurs de réseau, ainsi que les corrélation croisées. Ces dernières ont des distributions plutôt symétriques, tandis que les corrélations de réseau et morphologiques sont dissymétriques. Nous donnons également des nuages de points permettant de faire le lien entre ces différentes composantes.
}

\begin{figure}
\includegraphics[width=\linewidth,height=0.9\textheight]{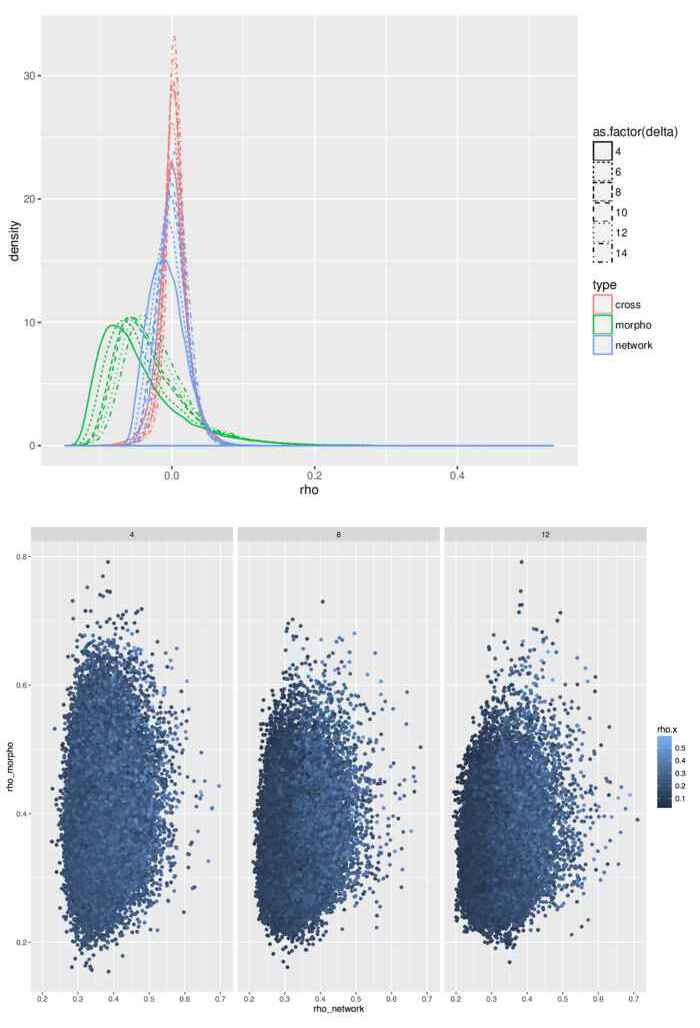}
\appcaption{\textbf{Distribution of correlations.} \textit{(Top)} Statistical distribution of correlations, for the different morphological, network and cross-correlations blocks (color), for different values of $\delta$ (line type); \textit{(Bas)} Average absolute correlations for the network as a function of correlations for morphology, the color level giving the cross-correlation, for different values of $\delta$ (columns).\label{fig:app:staticcorrelations:corr-distribs}}{\textbf{Distribution des corrélations.} \textit{(Haut)} Distribution statistique des corrélations, pour les différents blocs morphologiques, réseau et corrélations croisées (couleur), pour différentes valeurs de $\delta$ (type de ligne) ; \textit{(Bas)} Correlation absolues moyennes pour le réseau en fonction des correlations pour la morphologie, le niveau de couleur donnant la corrélation croisée, pour différentes valeur de $\delta$ (colonnes).\label{fig:app:staticcorrelations:corr-distribs}}
\end{figure}


%

\subsection{Multiscalarity}{Multi-scalarité}

\subsubsection{Estimation of correlations for a multi-scalar process}{Estimation des corrélations pour un processus multi-scalaire}

\bpar{
We propose here to link the multi-scalar character of a spatio-temporal stochastic process with the estimation of its correlation matrix. To simplify and in the framework in which this result is used in main text, we consider static correlations estimated in space. To also simplify, we consider processes with two characteristic scales which linearly superpose, i.e. which can be written as
}{
Nous proposons ici de relier le caractère multi-scalaire d'un processus stochastique spatio-temporel avec l'estimation de sa matrice de correlation. Pour simplifier et dans le cadre où ce résultat est utilisé en texte principal, nous considérons des corrélations statiques estimées dans l'espace. Pour simplifier également, considérons des processus ayant deux échelles caractéristiques se superposant linéairement, c'est-à-dire s'écrivant sous la forme
}
\[
X_i = X_i^{(0)} + \tilde{X}_i
\]
\bpar{
with $X_i^{(0)}$ trend at the small scales with a characteristic evolution distance $d_0$, and $\tilde{X}_i$ signal evolving at a characteristic distance $d \ll d_0$.
}{
avec $X_i^{(0)}$ tendance aux petites échelles ayant une distance caractéristique d'évolution $d_0$, et $\tilde{X}_i$ signal évoluant à une distance caractéristique $d \ll d_0$.
}

\bpar{
We can then compute the decomposition of the correlation between two processes, in a manner similar to what is done in~\ref{app:sec:syntheticdata-finance}. Assuming that $\Covb{X_i^{(0)}}{\tilde{X}_j} = 0$ for all $i,j$, and denoting $\varepsilon_i = \frac{\sigma\left[X_i^{(0)}\right]}{\sigma\left[\tilde{X}_i\right]}$ the rate between standard deviations of trend and signal, we have
}{
Nous pouvons alors calculer la décomposition de la corrélation entre deux processus, de manière similaire à ce qui est fait en~\ref{app:sec:syntheticdata-finance}. En supposant que $\Covb{X_i^{(0)}}{\tilde{X}_j} = 0$ pour tous $i,j$, et en notant $\varepsilon_i = \frac{\sigma\left[X_i^{(0)}\right]}{\sigma\left[\tilde{X}_i\right]}$ le rapport des écarts type entre tendance et signal, il y a
}
\[
\begin{split}
	\rho\left[X_1,X_2\right] & = \rho\left[X_1^{(0)} + \tilde{X}_1,X_2^{(0)} + \tilde{X}_2\right]\\
	& = \frac{\Covb{\tilde{X}_1}{\tilde{X}_2} + \Covb{X_1^{(0)}}{X_2^{(0)}}}{\sqrt{\left(\Varb{X_1^{(0)}} + \Varb{\tilde{X}_1}\right)\left(\Varb{X_2^{(0)}} + \Varb{\tilde{X}_2}\right)}}\\
	& = \frac{\varepsilon_1 \varepsilon_2\rho\left[X_1^{(0)},X_2^{(0)}\right] + \rho\left[\tilde{X}_1,\tilde{X}_2\right]}{\sqrt{\left(1 + \varepsilon_1^2\right)\left(1 + \varepsilon_2^2\right)}}
\end{split}
\]

\bpar{
By supposing $\varepsilon_i \ll 1$, we can develop this expression at the first order and obtain
}{
En supposant $\varepsilon_i \ll 1$, on peut développer cette expression au premier ordre et obtenir
}

\begin{equation}
	\rho\left[X_1,X_2\right] = \left( \varepsilon_1 \varepsilon_2\rho\left[X_1^{(0)},X_2^{(0)}\right] + \rho\left[\tilde{X}_1,\tilde{X}_2\right]\right)\cdot\left(1 - \frac{1}{2}(\varepsilon_1^2 + \varepsilon_2^2)\right)
\end{equation}

\bpar{
The addition of the trend to the signal thus introduces a correction on the correlation, on the one hand by the direct accounting of the attenuated correlation between trends, and on the other hand by the interference term as a multiplicative factor. 
}{
L'ajout de la tendance au signal introduit ainsi une correction sur la corrélation, d'une part par la prise en compte directe de la corrélation atténuée entre tendances, et d'autre part par le terme d'interférence en facteur.
}

\bpar{
To apply this result to our problematic, we assume that $d \simeq l_0$, $l_0$ being the minimal distance to estimate correlations. We furthermore have the stationarity scale $d_s$ which corresponds to the scale of variation of correlations, and according to the empirical results verifies $d_s > l_0$, significantly at least for some indicators (for example hierarchy and Moran, for which it is of the order of magnitude of the country). Finally, we denote by $\delta_0 = d_0/d$ the scale of the trend in terms of $\delta$. We therefore assume
}{
Pour appliquer ce résultat à notre problématique, supposons que $d \simeq l_0$, $l_0$ étant la distance minimale d'estimation des corrélations. On a par ailleurs l'échelle de stationnarité $d_s$ qui correspond à l'échelle de variation des corrélations, et selon les résultats empiriques vérifie $d_s > l_0$, significativement au moins pour certains indicateurs (par exemple hiérarchie et Moran, pour laquelle elle est de l'ordre du pays). Enfin, notons $\delta_0 = d_0/d$ l'échelle de la tendance en termes de $\delta$. On suppose donc
}

\[
d < d_s < d_0
\]

\bpar{
For $\delta$ values such that $\delta \cdot d < d_s$, we should have $\hat{\Cov}_{\delta}\left[\tilde{X}_1,\tilde{X}_2\right] \simeq \hat{\Cov}_{\delta = 1}\left[\tilde{X}_1,\tilde{X}_2\right]$ if $\hat{\Cov}_{\delta}$ is the estimator on the area of size $\delta$.
}{
Pour les valeurs de $\delta$ telles que $\delta \cdot d < d_s$, on devrait avoir $\hat{\Cov}_{\delta}\left[\tilde{X}_1,\tilde{X}_2\right] \simeq \hat{\Cov}_{\delta = 1}\left[\tilde{X}_1,\tilde{X}_2\right]$ si $\hat{\Cov}_{\delta}$ est l'estimateur sur la zone de taille $\delta$.
}

\bpar{
Furthermore, we can reasonably assume that $\hat{\Var}_{\delta = 1}\left[X_i^{(0)}\right] \ll \hat{\Var}_{\delta = d_s / d}\left[X_i^{(0)}\right]$, i.e. that the trend is constant at the largest scale in comparison to variation at the intermediate scales.
}{
Par ailleurs, on peut supposer raisonnablement que $\hat{\Var}_{\delta = 1}\left[X_i^{(0)}\right] \ll \hat{\Var}_{\delta = d_s / d}\left[X_i^{(0)}\right]$, c'est-à-dire que la tendance est constante à la plus grande échelle en comparaison des variations aux échelles intermédiaires.
}

\bpar{
Under these assumptions, the estimator of $\rho$ should vary as a function of $\delta$ according to the variations of $\varepsilon_i$ as a function of $\delta$. Finally, under the assumption that trends have a very low correlation (independent structural effects), we keep the correction by interferences in the expression of $\rho$, and thus that $\rho(\delta)$ decreases for low values of $\delta$.
}{
Sous ces hypothèses, l'estimateur de $\rho$ devrait varier en fonction de $\delta$ selon les variations de $\varepsilon_i$ en fonction de $\delta$. En supposant enfin les tendances très peu corrélées (effets structurels indépendants), on conserve la correction d'interférence dans l'expression de $\rho$, et donc que $\rho(\delta)$ décroit pour des faibles valeurs de $\delta$.
}

\bpar{
We have thus demonstrated that a simple multi-scalar structure of the process implies a variation of the estimated correlation as a function of $\delta$, under a certain number of assumptions. The reciprocal has no reason a priori to be true. The link we establish here is thus an illustration to reinforce an hypothesis, which is furthermore also sustained by results on the variation of the confidence interval described in the following.
}{
Nous avons ainsi démontré qu'une structure simple multi-scalaire du processus implique une variation de la corrélation estimée en fonction de $\delta$, sous un certain nombre d'hypothèses. La réciproque n'a a priori pas de raison d'être vraie. Le lien que nous opérons ici est ainsi une illustration pour renforcer une hypothèse, qui est par ailleurs également soutenue par les résultats sur la variation de l'intervalle de confiance décrits par la suite.
}

\subsubsection{Confidence interval for the correlation}{Intervalle de confiance pour la corrélation}

\bpar{
We derive here the behavior of the correlation estimator as a function of the size of the sample. Under the assumption of a normal distribution of two random variables $X,Y$, then the Fisher transform of the Pearson estimator $\hat{\rho}$ such that $\hat{\rho} = \tanh (\hat{z})$ has a normal distribution. If $z$ is the transform of the real correlation $\rho$, then a confidence interval for $\rho$ is of size
}{
Nous dérivons ici le comportement de l'estimateur de corrélation en fonction de la taille de l'échantillon. Sous l'hypothèse de distribution normale de deux variables aléatoires $X,Y$, alors la transformée de Fisher de l'estimateur de Pearson $\hat{\rho}$ telle que $\hat{\rho} = \tanh (\hat{z})$ a une distribution normale. Si $z$ est la transformée de la corrélation réelle $\rho$, alors un intervalle de confiance pour $\rho$ est de taille
}

\[
\rho_{+} - \rho_{-} = \tanh (z + k / \sqrt{N}) - \tanh (z - k / \sqrt{N})
\]

\bpar{
where $k$ is a constant. As $\tanh{z} = \frac{\exp (2z) - 1}{\exp (2z) + 1}$, we can develop this expression and reduce it, to obtain
}{
où $k$ est une constante. Comme $\tanh{z} = \frac{\exp (2z) - 1}{\exp (2z) + 1}$, on peut développer puis réduire cette expression, pour obtenir
}

\[
\begin{split}
	\rho_{+} - \rho_{-} & = 2\cdot \frac{\exp (2k/\sqrt{N})-\exp (-2k/\sqrt{N})}{\exp (2z)-\exp (-2z) + \exp (2k/\sqrt{N}) + \exp (-2k/\sqrt{N})}\\
	& = 2\cdot \frac{\sinh{(2k/\sqrt{N})}}{\cosh{(2z)} + \cosh{(2k/\sqrt{N})}}
\end{split}
\]

\bpar{
Using the fact that $\cosh u \sim_0 1 + u^2/2$ and that $\sinh u \sim_0 u$, we indeed obtain that $\rho_{+} - \rho_{-} \sim_{N\gg 0} k' / \sqrt{N}$.\qed
}{
En utilisant le fait que $\cosh u \sim_0 1 + u^2/2$ et que $\sinh u \sim_0 u$, on obtient bien que $\rho_{+} - \rho_{-} \sim_{N\gg 0} k' / \sqrt{N}$.\qed
}

\stars


\newpage

\section{Causality regimes}{Régimes de causalité}

\label{app:sec:causalityregimes}

\subsection{Synthetic data}{Données Synthétiques}

\subsubsection{Time series}{Séries temporelles}


\bpar{
We calculate here the theoretical values of lagged correlations for a simple auto-regressive process. We recall the framework, namely $\vec{X}(t)$ which is a stochastic process following the auto-regression equation
}{
Calculons ici les valeurs théoriques des corrélations retardées pour un processus auto-régressif simple. Nous rappelons le cadre, à savoir $\vec{X}(t)$ qui est un processus stochastique suivant l'équation d'auto-régression
}

\[
\vec{X}(t) = \sum_{\tau > 0} \mathbf{A}(\tau) \cdot \vec{X}(t - \tau ) + \vec{\varepsilon}(t)
\]
\bpar{
and we situate in the case where $\mathbf{A}(\tau) = 0$ for $\tau \neq \tau_0$ and
}{
et nous nous plaçons dans le cas où $\mathbf{A}(\tau) = 0$ pour $\tau \neq \tau_0$ et 
}
\[
\mathbf{A}(\tau_0) = \left( {\begin{array}{cc} 0 & a \\ a & 0 \\ \end{array}} \right)\]
\bpar{
with $-1<a<1$. We furthermore assume $\vec{\varepsilon}$ white noise and write $\vec{\varepsilon} = (\varepsilon_X,\varepsilon_Y)$ and assume $\Varb{\varepsilon_X} = \Varb{\varepsilon_Y} = \sigma^2$.
}{
avec $-1<a<1$. Nous supposons de plus $\vec{\varepsilon}$ bruit blanc et notons $\vec{\varepsilon} = (\varepsilon_X,\varepsilon_Y)$ et supposons $\Varb{\varepsilon_X} = \Varb{\varepsilon_Y} = \sigma^2$.
}

\bpar{
By writing $\vec{X} = (X,Y)$, the process is specified by
}{
En notant $\vec{X} = (X,Y)$, le processus est spécifié par
}
\[
\begin{cases}
	X(t) = a\cdot Y(t-\tau_0) + \varepsilon_X \\
	y(t) = a\cdot X(t-\tau_0) + \varepsilon_Y
\end{cases}
\]

\bpar{
By considering the variance in the two equations and taking the difference, we obtain that necessarily $\Varb{X} = \Varb{Y}$ since $\alpha^2 \neq 1$. The sum then gives $\Varb{X} = \Varb{Y} = \frac{\sigma^2}{1 - a^2}$.
}{
En prenant la variance dans les deux équations et en faisant la différence, on obtient que nécessairement $\Varb{X} = \Varb{Y}$ car $\alpha^2 \neq 1$. La somme donne alors $\Varb{X} = \Varb{Y} = \frac{\sigma^2}{1 - a^2}$.
}

\bpar{
We then compute
}{	
Nous calculons alors 
}

\[
\begin{split}
	\rhob{X(t)}{Y(t-\tau_0)} & = \rhob{aY(t-\tau_0)+\varepsilon_X}{Y(t-\tau_0)}\\
	& = \frac{\Covb{aY(t-\tau_0)+\varepsilon_X}{Y(t-\tau_0)}}{\sqrt{(a^2\Varb{Y} + \sigma^2)\Varb{Y}}} \\
	& = \frac{a\Varb{Y}}{\left|a\right|\Varb{Y}\sqrt{1 + \frac{\sigma^2}{a^2\Varb{Y}}}} = \frac{a}{\left|a\right|\sqrt{1 + \frac{1 - a^2}{a^2}}}\\
	& = a
\end{split}
\]

\bpar{
It is in fact possible to compute the lagged correlation for an arbitrary $\tau$. By stationarity of the process, we have for $\tau > 0$, $\rhob{X(t)}{Y(t-\tau)} = \rhob{X(\tau)}{Y(0)}$.
}{
Il est en fait possible de calculer la corrélation retardée pour $\tau$ quelconque. Par stationnarité du processus, on a pour $\tau > 0$, $\rhob{X(t)}{Y(t-\tau)} = \rhob{X(\tau)}{Y(0)}$.
}

\bpar{
In a similar way than previously, we develop for $\tau > 0$
}{
De la même manière que précédemment, nous développons pour $\tau > 0$
}

\[
\begin{split}
	\rhob{X(\tau)}{Y(0)} & = \rhob{aY(\tau - \tau_0) + \varepsilon_X}{Y(0)}\\
	 & = \rhob{a^2 X(\tau - 2\tau_0) + a \varepsilon_Y + \varepsilon_X}{Y(0)}\\
	 & = \frac{a^2\Covb{X(\tau - 2\tau_0)}{Y(0)}}{\sqrt{(a^4\Varb{X} + (1+a^2)\sigma^2)\Varb{Y}}}\\
	 & = \frac{\rhob{X(\tau-2\tau_0)}{Y(0)}}{\sqrt{1 + (1+a^2)(1-a^2)/a^4}}
	 & = a^2\cdot \rhob{X(\tau-2\tau_0)}{Y(0)}
\end{split}
\]

\bpar{
and thus by recurrence, for $k\in \mathbb{N}$,
}{
et donc par récurrence, pour $k\in \mathbb{N}$, 
}

\[
\rhob{X(\tau)}{Y(0)} = a^{2k}\cdot \rhob{X(\tau-2k\tau_0)}{Y(0)}
\]

\bpar{
If $\tau \notin (2 \mathbb{N} + 1) \tau_0$, we go down to $\rhob{X(\tau')}{Y(0)}$ such that $\tau' < \tau_0$ and the correlation therefore vanishes.
}{
Si $\tau \notin (2 \mathbb{N} + 1) \tau_0$, on descend à $\rhob{X(\tau')}{Y(0)}$ tel que $\tau' < \tau_0$ et la corrélation est donc nulle.
}

\bpar{
If $\tau \in (2 \mathbb{N} + 1) \tau_0$, we have then
}{
Si $\tau \in (2 \mathbb{N} + 1) \tau_0$, on a alors
}

\[
\rhob{X((2k+1)\tau_0)}{Y(0)} = a^{2k+1}
\]

\bpar{
For $\tau < 0 $, the computation is similar with an exchange of variables.
}{
Pour $\tau < 0 $, le calcul est similaire en échangeant les variables.
}

\bpar{
This simple auto-regressive model allows thus simply controlling the lagged correlations at given orders.
}{
Ce modèle simple auto-régressif permet ainsi de contrôler simplement les corrélations retardées à des ordres donnés.
}

\subsubsection{Urban morphogenesis}{Morphogenèse urbaine}

\bpar{
The Fig.~\ref{fig:app:causalityregimes:clustering} gives, for the unsupervised analysis done of features issued from lagged correlations, the behavior of clustering results as a function of number of cluster $k$, which allows reading a transition as a function of $k$. We also give the repartition of clusters in a principal plan for $k=6$.
}{
La Fig.~\ref{fig:app:causalityregimes:clustering} donne, pour l'analyse non-supervisée menée sur les caractéristiques issues des corrélations retardées, le comportement des résultats du clustering en fonction du nombre de cluster $k$, qui permet de lire une transition en fonction de $k$. Nous donnons aussi la répartition des clusters dans un plan principal pour $k=6$.
}

\begin{figure}
\includegraphics[width=\linewidth]{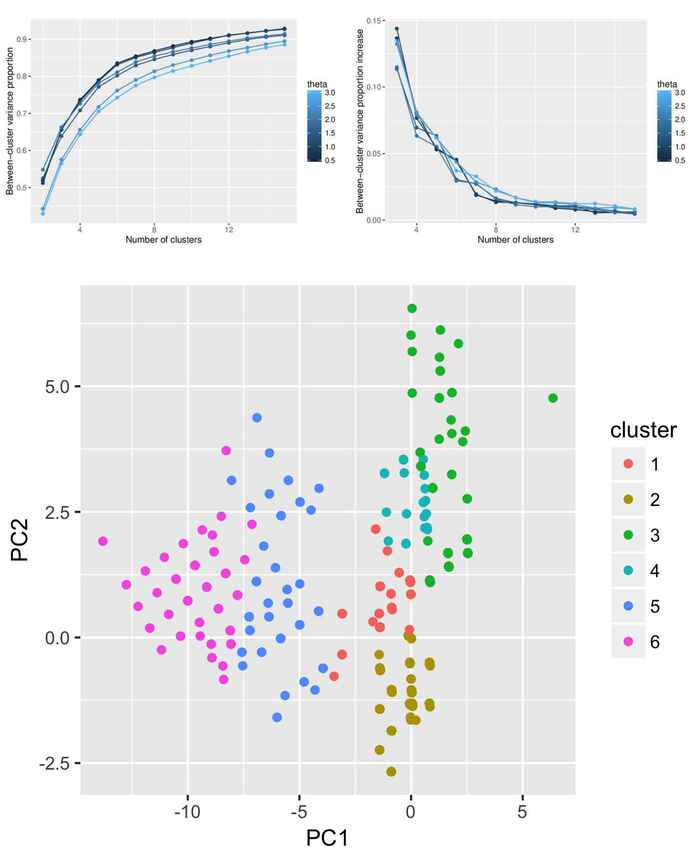}
\appcaption{\textbf{Identification of endogenous interaction regimes through unsupervised classification.} (\textit{Top Left}) Inter-cluster variance as a function of number of clusters. (\textit{Top Right}) Derivative of the inter-cluster variance. (\textit{Bottom}) Features in a principal plan (81\% of explained variance by the two first components).\label{fig:app:causalityregimes:clustering}}{\textbf{Identification de régimes d'interactions endogènes par classification non-supervisée.} (\textit{Haut Gauche}) Variance inter-cluster comme fonction du nombre de clusters. (\textit{Haut Droite}) Dérivée de la variance inter-cluster. (\textit{Bas}) \emph{Features} dans un plan principal (81\% de variance expliquée par les deux premières composantes).\label{fig:app:causalityregimes:clustering}}
\end{figure}

\subsection{South Africa}{Afrique du Sud}

\bpar{
The Fig.~\ref{fig:app:causalityregimes:sudafcorrs} gives the behavior of estimated correlations, in terms of average absolute correlation, and of proportion of significant correlations, as a function of $d_0$ and of $T_W$. It also gives the lagged correlation profiles for the weighted accessibilities, at the origin and at the destination.
}{
La Fig.~\ref{fig:app:causalityregimes:sudafcorrs} donne le comportement des corrélations estimées, en termes de corrélation absolue moyenne, et de proportion de corrélations significatives, en fonction de $d_0$ et de $T_W$. Elle donne également les profils de corrélations retardées pour les accessibilité pondérées, à l'origine et à la destination.
}

\begin{figure}
\includegraphics[width=\linewidth]{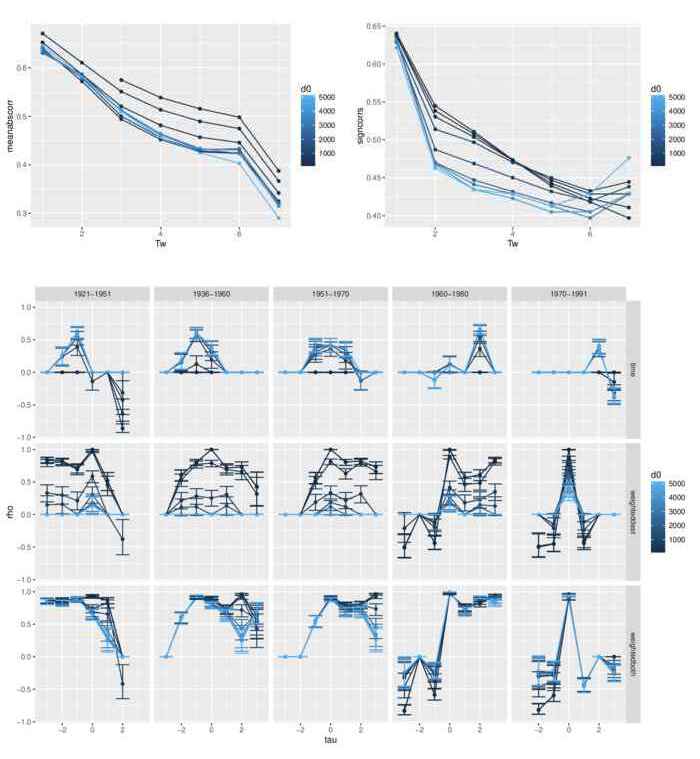}
\appcaption{\textbf{Behavior of empirical correlations in the case of South Africa.} \textit{(Top left)} Average absolute correlations for all delays, as a function of time window duration $T_W$ (in number of temporal observations), for different values of the decay parameter $d_0$; \textit{(Top Right)} Proportion of significant correlations, as a function of $T_W$ for varying $d_0$; \textit{(Bottom)} Lagged correlations as a function of delay $\tau$, for the optimal size $T_W=3$, on the different successive periods (columns), for different levels of weighting (first row $w_i=1$, second row $w_i = 1,w_j=P_j/\sum_k P_k$, third row $w_i = P_i/\sum_k P_k,w_j=P_j/\sum_k P_k$), and for varying $d_0$ (color). \label{fig:app:causalityregimes:sudafcorrs}}{\textbf{Comportement des corrélations empiriques dans le cas de l'Afrique du Sud.} \textit{(Haut Gauche)} Corrélations absolues moyennes sur l'ensemble des retards, en fonction de la taille de la fenêtre temporelle $T_W$ (en nombre d'observations temporelles), pour différentes valeurs du paramètre de décroissance $d_0$ ; \textit{(Haut Droite)} Proportion de corrélations significatives, en fonction de $T_W$ pour $d_0$ variable ; \textit{(Bas)} Corrélations retardées en fonction du délai $\tau$, pour la taille optimale $T_W=3$, sur les différentes périodes successives (colonnes), pour les différents degrés de pondérations (première ligne $w_i=1$, deuxième ligne $w_i = 1,w_j=P_j/\sum_k P_k$, troisième ligne $w_i = P_i/\sum_k P_k,w_j=P_j/\sum_k P_k$), et pour $d_0$ variable (couleur).\label{fig:app:causalityregimes:sudafcorrs}}
\end{figure}

\stars

\newpage

\section{Aggregation-diffusion morphogenesis}{Morphogenèse par agrégation-diffusion}

\label{app:sec:density}

\subsection{Extended figures for model exploration}{Figures supplémentaires pour l'exploration du modèle}

\subsubsection{Convergence}{Convergence}

\bpar{
Histograms for the 81 parameters points for which we did 100 repetitions are given in Fig.~\ref{fig:app:density:histograms}, for Moran index and slope indicators. Other indicators showed similar convergence patterns. The visual exploration of histograms confirms the numerical analysis done in main text for statistical convergence.
}{
Les histogrammes pour les 81 points de paramètres pour lesquelles 100 répétitions ont été menées sont donnés en Fig.~\ref{fig:app:density:histograms}, pour l'index de Moran et la hiérarchie. Les autres indicateurs témoignent de propriétés de convergence similaires. L'exploration visuelle des histogrammes confirme l'analyse numérique menée dans le texte principal pour la convergence statistique.
}




\begin{figure}
\includegraphics[width=\linewidth]{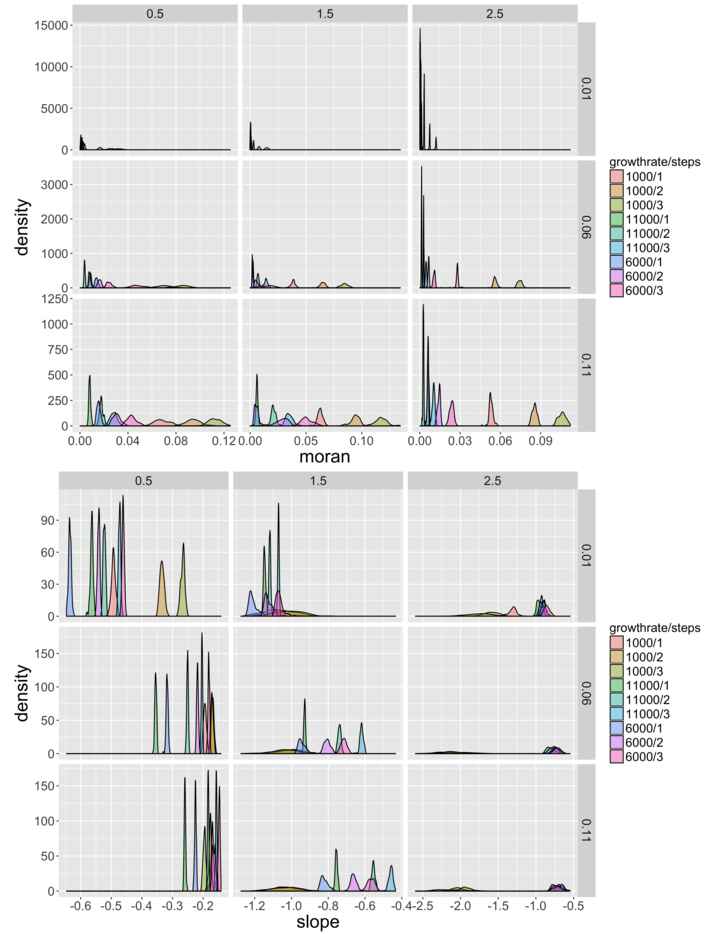}
\appcaption{Histograms for Moran index (top) and slope (bottom), for varying $\alpha$ (columns), $\beta$ (rows), $N_G$ and $n_d$ (colors).\label{fig:app:density:histograms}}{\textit{(Haut)} Distributions de l'Index de Moran, pour des valeurs variables de $\alpha$ (colonnes), $\beta$ (lignes), $N_G$ et $n_d$ (couleurs).\label{fig:app:density:histograms}}
\end{figure}

\subsubsection{Indicators}{Indicateurs}


\bpar{
We show in Fig.~\ref{fig:app:density:moran} to Fig.~\ref{fig:app:density:entropy} the full behavior of all indicators, with all parameters varying, obtained through the extensive exploration, from which the plots in main text have been extracted. Because of the complex nature of emergent urban form, one can not predict output values without referring to this ``exhaustive'' parameter sweep.
}{
Nous donnons en Fig.~\ref{fig:app:density:moran} à Fig.~\ref{fig:app:density:entropy} le comportement exhaustif des indicateurs, pour l'ensemble des paramètres variant. Ceux-ci ont été obtenus par l'exploration intensive, et les graphiques en texte principal en sont des cas particuliers. A cause de la nature complexe de la forme urbaine émergente, il n'est pas possible de prédire les valeurs de sorties sans référer à cette exploration ``exhaustive'' de l'espace des paramètres.
}

\begin{figure}
\includegraphics[width=\linewidth]{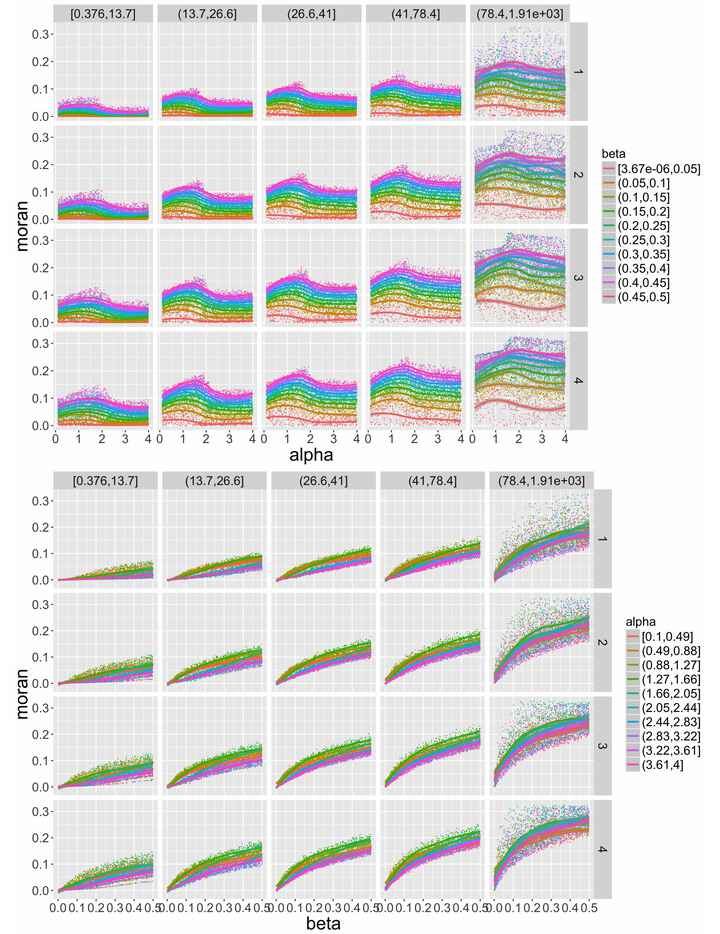}
\appcaption{Moran index as a function of $\alpha$ (Top) and $\beta$ (Bottom) for varying $\beta$ (resp. $\alpha$) given by color, and varying $n_d$ (rows) and $N_G$ (columns).\label{fig:app:density:moran}}{Indice de Moran en fonction de $\alpha$ (Haut) et $\beta$ (Bas) pour $\beta$ variable (resp. $\alpha$) donné par la couleur, et $n_d$ (lignes) et $N_G$ (colonnes) variables.\label{fig:app:density:moran}}
\end{figure}

\begin{figure}
\includegraphics[width=\linewidth]{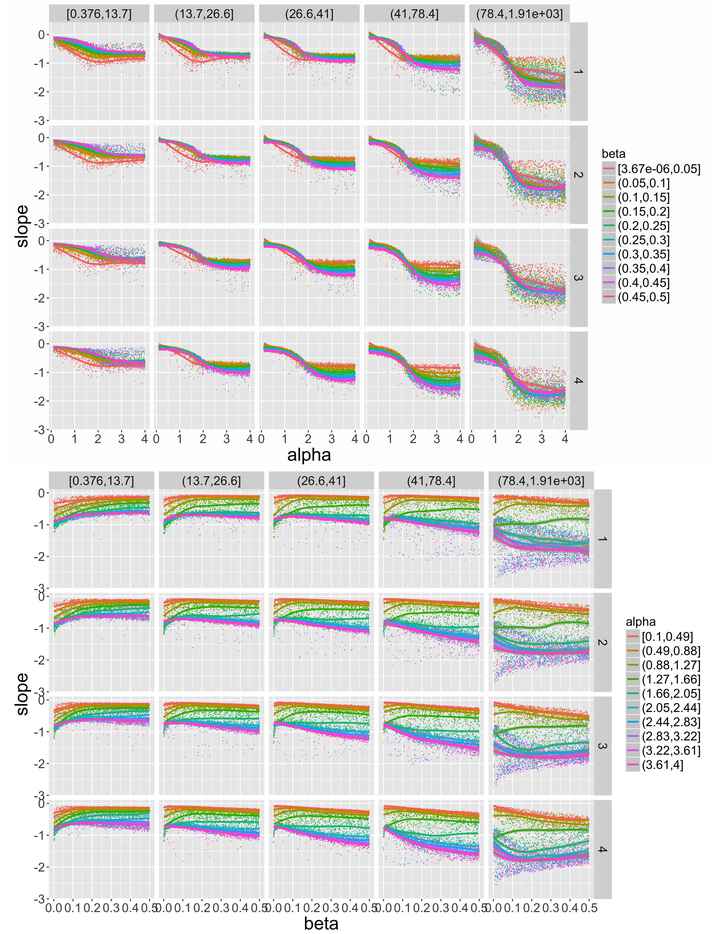}
\appcaption{Slope as a function of $\alpha$ (Top) and $\beta$ (Bottom) for varying $\beta$ (resp. $\alpha$) given by color, and varying $n_d$ (rows) and $N_G$ (columns).\label{fig:app:density:slope}}{Hiérarchie en fonction de $\alpha$ (Haut) et $\beta$ (Bas) pour $\beta$ variable (resp. $\alpha$) donné par la couleur, et $n_d$ (lignes) et $N_G$ (colonnes) variables.\label{fig:app:density:slope}}
\end{figure}

\begin{figure}
\includegraphics[width=\linewidth]{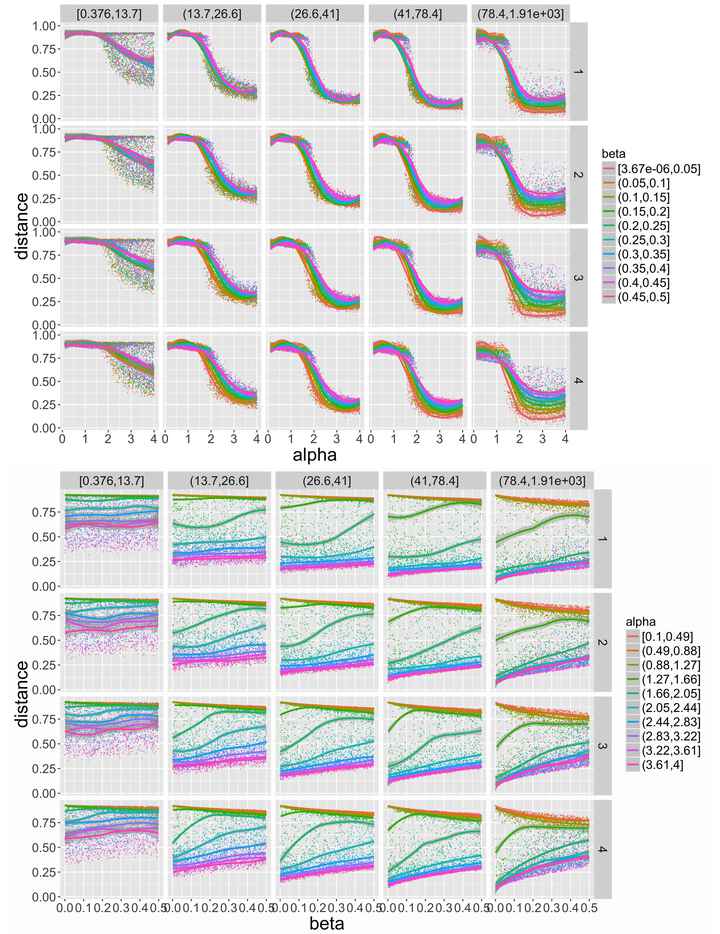}
\appcaption{Average distance index as a function of $\alpha$ (Top) and $\beta$ (Bottom) for varying $\beta$ (resp. $\alpha$) given by color, and varying $n_d$ (rows) and $N_G$ (columns).\label{fig:app:density:distance}}{Distance moyenne en fonction de $\alpha$ (Haut) et $\beta$ (Bas) pour $\beta$ variable (resp. $\alpha$) donné par la couleur, et $n_d$ (lignes) et $N_G$ (colonnes) variables.\label{fig:app:density:distance}}
\end{figure}

\begin{figure}
\includegraphics[width=\linewidth]{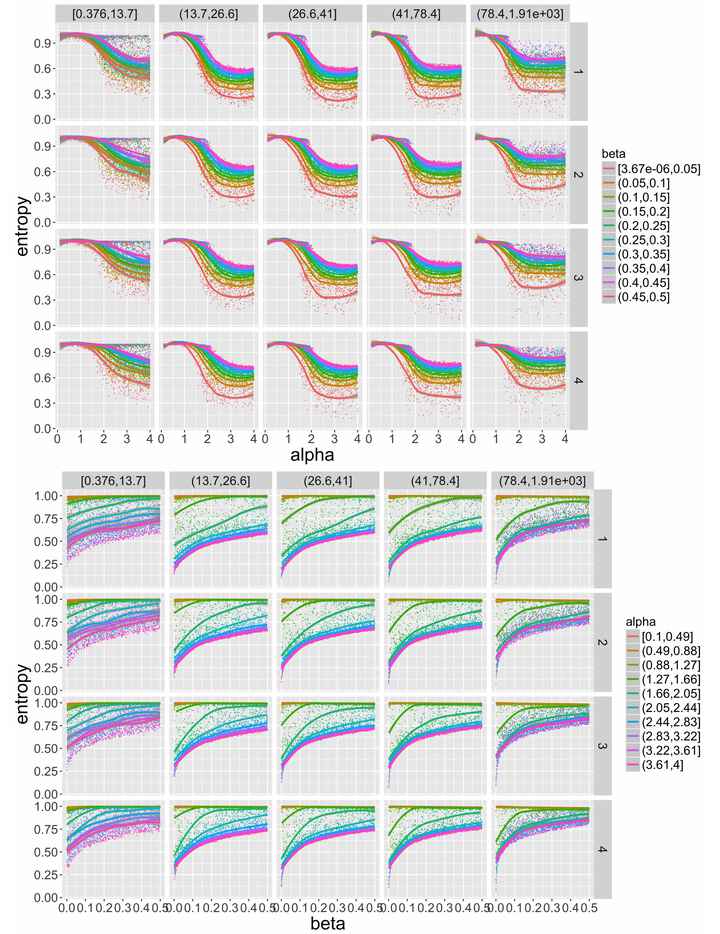}
\appcaption{Entropy as a function of $\alpha$ (Top) and $\beta$ (Bottom) for varying $\beta$ (resp. $\alpha$) given by color, and varying $n_d$ (rows) and $N_G$ (columns).\label{fig:app:density:entropy}}{Entropie en fonction de $\alpha$ (Haut) et $\beta$ (Bas) pour $\beta$ variable (resp. $\alpha$) donné par la couleur, et $n_d$ (lignes) et $N_G$ (colonnes) variables.\label{fig:app:density:entropy}}
\end{figure}

\subsubsection{Indicators scatterplots}{Scatterplot des indicateurs}


\bpar{
We show finally the full scatterplots of indicators, with real data points, in Fig.~\ref{fig:app:density:densityscatter}. These are preliminary step of the calibration on principal components, and we can see on these on which dimensions the model fails relatively to fit real data (in particular average distance).
}{
Nous montrons finalement les nuages de points complets des indicateurs, avec les points observés, en Fig.~\ref{fig:app:density:densityscatter}. Il s'agit de l'étape préliminaire à la calibration sur les composantes principales, et nous pouvons voir ici sur quelles dimensions le modèle échoue particulièrement à s'approcher des données observées (en particulier la distance moyenne).
}

\begin{figure}
\includegraphics[width=\linewidth]{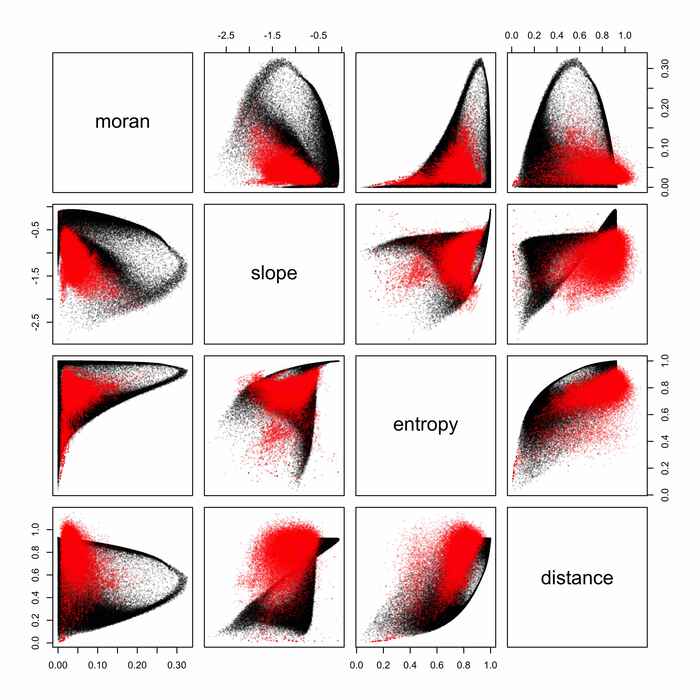}
\appcaption{Scatterplots of indicators distribution in the sampled hypercube of the parameter space. Red points correspond to real data.\label{fig:app:density:densityscatter}}{Nuages de points des indicateurs dans l'hypercube échantillonné de l'espace des paramètres. Les points rouges correspondent aux données réelles.\label{fig:app:density:densityscatter}}
\end{figure}

\subsection{Semi-analytical analysis of the simplified model}{Analyse semi-analytique du modèle simplifié}

\subsubsection{Partial differential equation}{Equation aux dérivées partielles}

\bpar{
We propose to derive the PDE in a simplified setting. To recall the configuration given in main text, the system has one dimension, such that $x\in \mathbb{R}$ with $1/\delta x$ cells of size $\delta x$, and we use the expected values of cell population $p(x,t) = \Eb{P(x,t)}$. We furthermore take $n_d=1$. Larger values would imply derivatives at an order higher than 2 but the following results on the existence of a stationary solution should still hold. 
}{
Nous proposons de dériver l'EDP dans un cadre simplifié. Pour rappeler la configuration donnée en texte principal, le système a une dimension, tel que $x\in \mathbb{R}$ avec $1/\delta x$ cellules de taille $\delta x$, et nous utilisons les valeurs attendues des populations des cellules $p(x,t) = \Eb{P(x,t)}$. Nous prenons de plus $n_d=1$. Des valeurs plus grandes devraient impliquer des dérivées à un ordre supérieur à deux, mais les résultats qui suivent sur l'existence d'une solution stationnaire devraient être conservés.
}

\bpar{
Denoting $\tilde{p}(x,t)$ the intermediate populations obtained after the aggregation stage, we have
}{
En écrivant $\tilde{p}(x,t)$ les populations intermédiaires obtenues après l'étape d'agrégation, nous avons
}

\[
\tilde{p}(x,t) = p(x,t) + N_g\cdot \frac{p(x,t)^{\alpha}}{\sum_x p(x,t)^{\alpha}}
\]

\bpar{
since all populations units are added independently. If $\delta x \ll 1$ then $\sum_x p^{\alpha} \simeq \int_x p(x,t)^{\alpha}dx$ and we write this quantity $P_{\alpha}(t)$. We furthermore write $p=p(x,t)$ and $\tilde{p} = \tilde{p}(x,t)$ in the following for readability.
}{
puisque toutes les unités de population sont ajoutés indépendamment. Si $\delta x \ll 1$ alors $\sum_x p^{\alpha} \simeq \int_x p(x,t)^{\alpha}dx$ et nous écrivons cette quantité $P_{\alpha}(t)$. Nous notons de plus $p=p(x,t)$ et $\tilde{p} = \tilde{p}(x,t)$ par la suite pour faciliter la lecture.
}

\bpar{
The diffusion step is then deterministic, and for any cell not on the border ($0<x<1$), if $\delta t$ is the interval between two time steps, we have
}{
L'étape de diffusion est ensuite déterministe, et pour toute cellule qui n'est pas au bord ($0<x<1$), si $\delta t$ est l'intervalle entre deux pas de temps, nous avons
}

\[
\begin{split}
p(x,t+\delta t) & = (1 - \beta) \cdot \tilde{p} + \frac{\beta}{2} \left[\tilde{p}(x-\delta x,t) + \tilde{p}(x+\delta x,t)\right]\\
& = \tilde{p} + \frac{\beta}{2} \left[\left(\tilde{p}(x+\delta x,t) - \tilde{p}\right) - \left(\tilde{p} - \tilde{p}(x-\delta x,t)\right)\right]
\end{split}
\]

\bpar{
Assuming the partial derivatives exist, and as $\delta x \ll 1$, we make the approximation $\tilde{p}(x+\delta x,t) - \tilde{p} \simeq \delta x\cdot \frac{\partial \tilde{p}}{\partial{x}}(x,t)$, what gives 
}{
Sous l'hypothèse que les dérivées partielles existent, et comme $\delta x \ll 1$, nous faisons l'approximation $\tilde{p}(x+\delta x,t) - \tilde{p} \simeq \delta x\cdot \frac{\partial \tilde{p}}{\partial{x}}(x,t)$, ce qui donne
}

\[
\left(\tilde{p}(x+\delta x,t) - \tilde{p}\right) - \left(\tilde{p} - \tilde{p}(x-\delta x,t)\right) = \delta x \cdot \left(\frac{\partial \tilde{p}}{\partial{x}}(x,t) - \frac{\partial \tilde{p}}{\partial{x}}(x - \delta x,t)\right)
\]

\bpar{
and therefore at the second order
}{
et donc au second ordre
}

\[
p(x,t+\delta t) = \tilde{p} + \frac{\beta \delta x^2}{2} \cdot \frac{\partial^2 \tilde{p}}{\partial x^2}
\]

\bpar{
Substituting $\tilde{p}$ gives
}{
Le remplacement de $\tilde{p}$ donne
}

\[
\begin{split}
\frac{\partial^2 \tilde{p}}{\partial x^2} & = \frac{\partial^2 p}{\partial x^2} + \frac{N_G}{P_\alpha}\cdot \frac{\partial}{\partial x}\left[\alpha \frac{\partial p}{\partial x} p^{\alpha - 1}\right]\\
& = \frac{\partial^2 p}{\partial x^2} + \alpha \frac{N_G}{P_\alpha} \left[\frac{\partial^2 p}{\partial x^2} p^{\alpha - 1} + (\alpha - 1) \left( \frac{\partial p}{\partial x}\right)^2 p^{\alpha - 2}\right]
\end{split}
\]

\bpar{
By supposing that $\frac{\partial p}{\partial t}$ exists and that $\delta t$ is small, we have $p(x,t+\delta t) - p(x,t) \simeq \delta t\frac{\partial p}{\partial t}$, what finally yields, by combining the results above, the partial differential equation
}{
En supposant que $\frac{\partial p}{\partial t}$ existe et que $\delta t$ est petit, nous avons $p(x,t+\delta t) - p(x,t) \simeq \delta t\frac{\partial p}{\partial t}$, ce qui donne finalement, par combinaison des résultats ci-dessus, l'équation aux dérivées partielles
}

\begin{equation}\label{eq:pde}
\delta t \cdot \frac{\partial p}{\partial t} = \frac{N_G \cdot p^{\alpha}}{P_{\alpha}(t)} + \frac{\alpha \beta (\alpha - 1) \delta x^2}{2}\cdot \frac{N_G \cdot p^{\alpha-2}}{P_{\alpha}(t)} \cdot \left(\frac{\partial p}{\partial x}\right)^2 + \frac{\beta \delta x^2}{2} \cdot \frac{\partial^2 p}{\partial x^2} \cdot\left[ 1 + \alpha \frac{N_G p^{\alpha - 1}}{P_{\alpha(t)}} \right]
\end{equation}

\bpar{
Initial conditions should be specified as $p_0(x) = p(x,t_0)$. To have a well-posed problem similar to more classical PDE problems, we need to assume a domain and boundary conditions. A finite support is expressed by $p(x,t)=0$ for all $t$ and $x$ such that $\left|x\right|>x_m$.
}{
Les conditions initiales sont spécifiées par $p_0(x) = p(x,t_0)$. Pour obtenir un problème bien posé comme dans des formulations PDE plus classiques, nous devons supposer un domaine et des conditions au bord. Un support fini est traduit par $p(x,t)=0$ pour tout $t$ et $x$ tel que $\left|x\right|>x_m$.
}


\subsubsection{Stationary solution for density}{Solution stationnaire pour la densité}

\bpar{
The non-linearity and the integral terms making the equation above out of the scope for analytical resolution, we study its behavior numerically in some cases. Taking a simple initial condition $p_0(0)=1$ and $p_0(x)=0$ for $x\neq 0$, we show that on a finite domain, density $d(x,t)$ always converge to a stationary solution for large $t$, for a large set of values of $(\alpha,\beta)$ with fixed $N_G=10$ ($\alpha\in \left[0.4,1.5\right]$ varying with step $0.025$ and $\log\beta \in \left[-1,-0.5\right]$ with step $0.1$). We show in Fig.~\ref{fig:app:density:stationary} the corresponding trajectories on a typical subset. The variation of the asymptotic distribution as a function of $\alpha$ and $\beta$ are not directly visible, as they depend on very low values of the outward flows at boundaries. We give in Fig.~\ref{fig:app:density:pmax} their behavior, by showing the value of the maximum of the distribution. Low values of $\beta$ give an inversion in the effect of $\alpha$, whereas high values of $\beta$ give comparable values for all $\alpha$.
}{
La non-linéarité et les termes intégraux rendant l'équation ci-dessus hors d'atteinte d'une résolution analytique, nous étudions son comportement de manière numérique pour certaines configurations. Prenant une condition initiale simple $p_0(0)=1$ et $p_0(x)=0$ pour $x\neq 0$, nous montrons que sur un domaine fini, la densité $d(x,t)$ converge toujours vers une solution stationnaire pour les grandes valeurs de $t$, pour un grand nombre de valeurs pour $(\alpha,\beta)$ avec $N_G=10$ fixé ($\alpha\in \left[0.4,1.5\right]$ variant avec un pas de $0.025$ et $\log\beta \in \left[-1,-0.5\right]$ avec un pas de $0.1$). Nous montrons en Fig.~\ref{fig:app:density:stationary} les trajectoires correspondantes sur un sous-ensemble typique. La variation des distributions asymptotiques comme fonction de $\alpha$ et $\beta$ ne sont pas directement observables, puisqu'elle dépendent des valeurs très faibles des flux sortants aux bords. Nous donnons en Fig.~\ref{fig:app:density:pmax} leur comportement, en donnant la valeur du maximum de la distribution. Les valeurs faibles de $\beta$ mènent à une inversion de l'effet de $\alpha$, tandis que les fortes valeurs de $\beta$ donnent des valeurs comparables pour tous les $\alpha$.
}

\begin{figure}[h!]
\includegraphics[width=\linewidth]{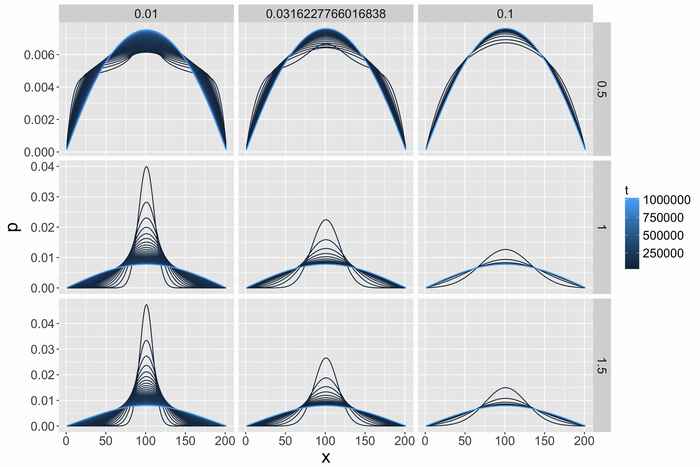}
\appcaption{Trajectories of densities as a function of the spatial dimension, for varying $\beta$ (columns) and $\alpha$ (rows). Color level gives time.\label{fig:app:density:stationary}}{Trajectoires des densités en fonction de la coordonnée spatiale, pour $\beta$ variable (colonnes) et $\alpha$ variable (lignes). Le niveau de couleur donne le temps.\label{fig:app:density:stationary}}
\end{figure}

\begin{figure}[h!]
\includegraphics[width=\linewidth]{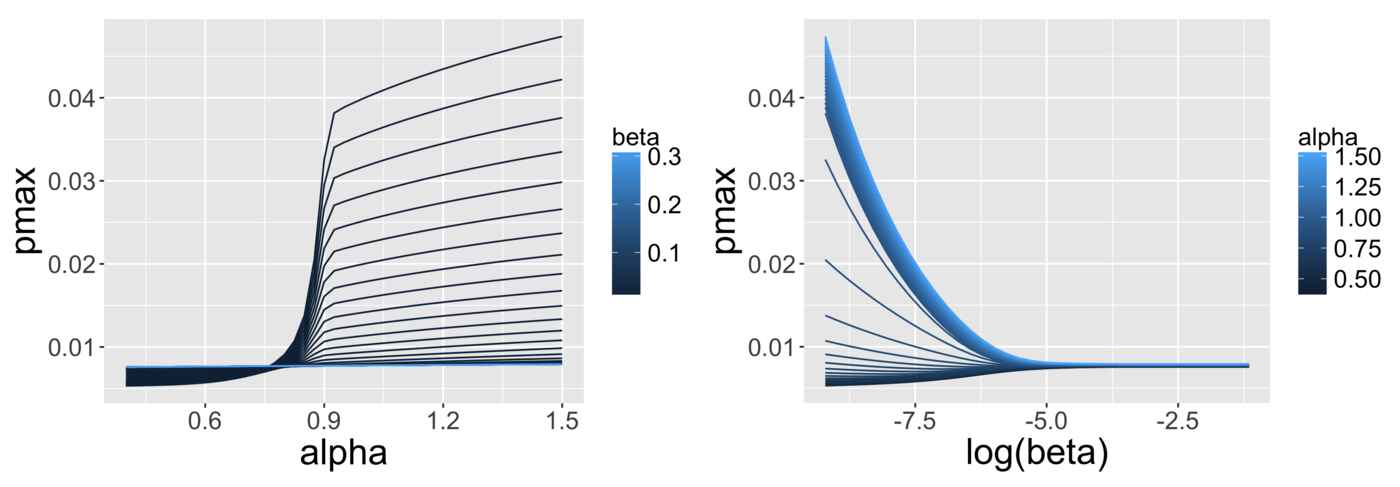}
\appcaption{Dependency of $\max d(t\rightarrow \infty )$ to $\alpha$ and $\beta$.\label{fig:app:density:pmax}}{Dépendance de $\max d(t\rightarrow \infty )$ à $\alpha$ et $\beta$.\label{fig:app:density:pmax}}
\end{figure}


\newpage

\section{Correlated synthetic data}{Données synthétiques corrélées}

\label{app:sec:correlatedsyntheticdata}

\bpar{
For the simulation of the weak coupling between a density configuration and a network generation model, the Fig.~\ref{fig:app:correlatedsyntheticdata:correlations} gives the errors on feasible correlations shown in Fig.~\ref{fig:correlatedsyntheticdata:densnwcor}, and also the amplitude of correlation for the full matrix, i.e. both the maximal absolute correlation $c_{ij}=\max_k\left| \rho_{ij}^{k} \right|$ and the total amplitude $a_{ij}=\max_k{\rho_{ij}^{(k)}}-\min_k{\rho_{ij}^{(k)}}$.
}{
Pour la simulation du couplage faible entre génération d'une configuration de densité et modèle de génération de réseau, la Fig.~\ref{fig:app:correlatedsyntheticdata:correlations} donne les erreurs sur les corrélations faisables montrées en Fig.~\ref{fig:correlatedsyntheticdata:densnwcor}, ainsi que l'amplitude des corrélations pour l'ensemble de la matrice, c'est-à-dire à la fois la corrélation absolue maximale $c_{ij}=\max_k\left| \rho_{ij}^{k} \right|$ et l'amplitude totale $a_{ij}=\max_k{\rho_{ij}^{(k)}}-\min_k{\rho_{ij}^{(k)}}$.
}

\begin{figure}
\includegraphics[width=\linewidth]{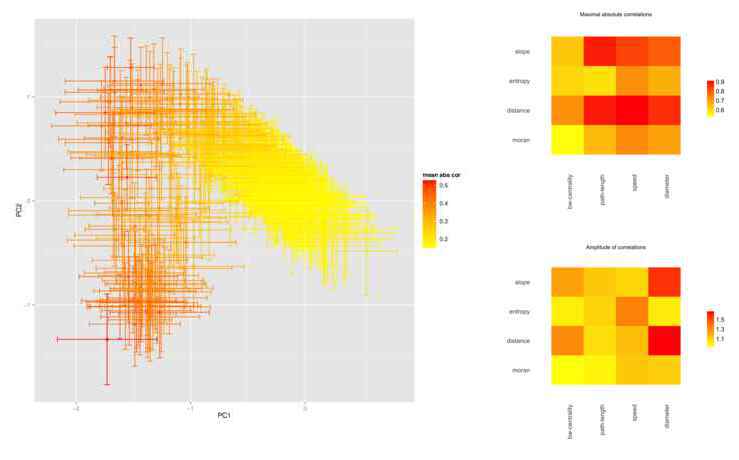}
\appcaption{\textbf{Space of feasible correlations.} \textit{(Left)} Projection of correlation matrices in a principal plan obtained by Principal Component Analysis on matrix population (cumulated variances: PC1=38\%, PC2=68\%). Error bars are initially computed as 95\% confidence intervals on each matrix element (by standard Fisher asymptotic method), and upper bounds after transformation are taken in principal plan; \textit{(Right)} Heatmaps for amplitude of correlations, defined as $a_{ij}=\max_k{\rho_{ij}^{(k)}}-\min_k{\rho_{ij}^{(k)}}$ and maximal absolute correlation, defined as $c_{ij}=\max_k\left| \rho_{ij}^{k} \right|$. Scale color gives mean absolute correlation on full matrices.\label{fig:app:correlatedsyntheticdata:correlations}}{\textbf{Espace des corrélations faisables.} \textit{(Gauche)} Projection des matrices de correlations dans un plan principal obtenu par analyse en composantes principales sur la population des matrices (variances cumulées PC1=38\%, PC2=68\%, s'agissant de corrélations les données sont elles-mêmes corrélées d'où la structure du nuage de points); les barres d'erreur sont calculées initialement comme les intervalles de confiance à 95\% sur chaque matrice (par méthode asymptotique de Fisher standard), et les bornes supérieures après transformation sont prises dans le plan principal; \textit{(Droite)} Amplitude des correlations, définie comme $a_{ij}=\max_k{\rho_{ij}^{(k)}}-\min_k{\rho_{ij}^{(k)}}$ et corrélation maximale absolue, définie comme $c_{ij}=\max_k\left| \rho_{ij}^{k} \right|$ ; l'échelle de couleur donne la corrélation moyenne absolue sur les matrices entières.\label{fig:app:correlatedsyntheticdata:correlations}}
\end{figure}

%


\newpage

\section{Exploration of the SimpopNet model}{Exploration du modèle SimpopNet}

\label{app:sec:macrocoevolexplo}

\bpar{
We give here supplementary figures that allow rendering the sensitivity of results to parameters not presented in main text.
}{
Nous donnons ici des figures supplémentaires permettant de se rendre compte de la sensibilité des résultats aux paramètres non présentés en texte principal.
}

\bpar{
The Fig~\ref{fig:app:macrocoevolexplo:closeness} allows visualizing the sensitivity of the entropy of centralities $\varepsilon \left[\mu_i\right]$ as a function of $d_G$, $\theta_N$ and $\gamma_G$. The shape of temporal curves is mainly sensitive to $\gamma_G$.
}{
La Fig.~\ref{fig:app:macrocoevolexplo:closeness} permet de visualiser la sensibilité de l'entropie des centralités $\varepsilon \left[\mu_i\right]$ en fonction de $d_G$, $\theta_N$ et $\gamma_G$. La forme des courbes temporelles est principalement sensible à $\gamma_G$.
}

\begin{figure}
    \includegraphics[width=\linewidth]{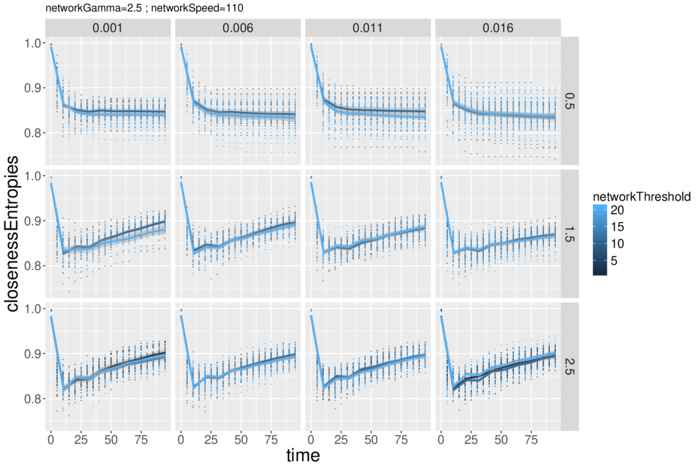}
\appcaption{\textbf{Entropy of closeness centralities.} We give $\varepsilon \left[\mu_i\right]$ as a function of time $t$, for $\theta_N$ variable (color), $d_G$ variable (columns) and $\gamma_G$ variable.\label{fig:app:macrocoevolexplo:closeness}}{\textbf{Entropie des centralités de proximité.} Nous donnons $\varepsilon \left[\mu_i\right]$ en fonction du temps $t$, pour $\theta_N$ variable (couleur), $d_G$ variable (colonnes) et $\gamma_G$ variable.\label{fig:app:macrocoevolexplo:closeness}}
\end{figure}

\bpar{
The Fig.~\ref{fig:app:macrocoevolexplo:rankcorrpop} gives the variations of $\rho_r$ as a function of $d_G$ and $\gamma_G$, for variable values of $\theta_N$ and of $\gamma_N$. We see that the regularity observed as a function of $d_G$ and of $\gamma_G$ appears not to be sensitive to the variations of $\theta_N$ and of $\gamma_N$.
}{
La Fig.~\ref{fig:app:macrocoevolexplo:rankcorrpop} donne les variations de $\rho_r$ en fonction de $d_G$ et $\gamma_G$, pour des valeurs variables de $\theta_N$ et de $\gamma_N$. Nous constatons que la régularité observée en fonction de $d_G$ et de $\gamma_G$ n'est pas visiblement sensible aux variations de $\theta_N$ et de $\gamma_N$.
}

\begin{figure}
    \includegraphics[width=\linewidth]{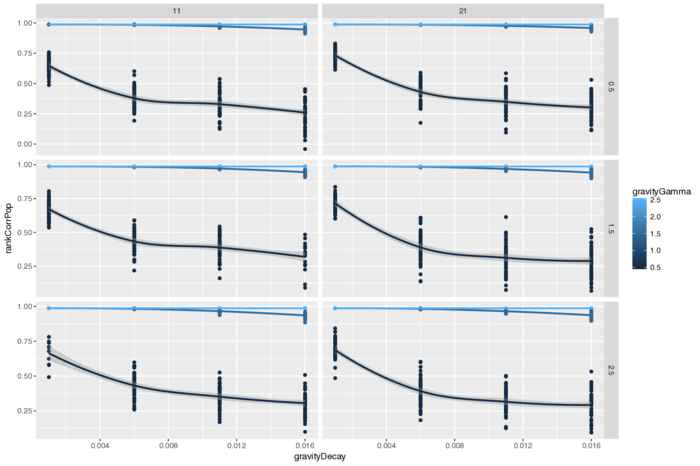}
\appcaption{\textbf{Population rank correlations.} We give $\rho_r \left[\mu_i\right]$ as a function of $d_G$, for $\gamma_G$ variable (color), $\theta_N$ variable (columns) and $\gamma_N$ variable (rows).\label{fig:app:macrocoevolexplo:rankcorrpop}}{\textbf{Corrélations de rang pour la population.} Nous donnons $\rho_r \left[\mu_i\right]$ en fonction de $d_G$, pour $\gamma_G$ variable (couleur), $\theta_N$ variable (colonnes) et $\gamma_N$ variable (lignes).\label{fig:app:macrocoevolexplo:rankcorrpop}}
\end{figure}

\bpar{
The Fig.~\ref{fig:app:macrocoevolexplo:distcorrs} gives correlations $\rho_d$ as a function of distance for all couples of variables, for varying $d_G$ and $\gamma_G$. We obtain qualitatively the same behaviors than with $d_G = 0.016$, at the exception of a very low growth for the largest distances, for the correlation between population and accessibility, at $d_G=0.001$ and $\gamma_G = 0.5$, which remains difficult to interpret.
}{
La Fig.~\ref{fig:app:macrocoevolexplo:distcorrs} donne les corrélations $\rho_d$ en fonction de la distance pour l'ensemble des couples de variables, pour $d_G$ et $\gamma_G$ variables. Nous retrouvons qualitativement les mêmes comportements que avec $d_G = 0.016$, à l'exception d'une très légère croissance pour les plus grande distances, pour la corrélation entre la population et l'accessibilité, à $d_G=0.001$ et $\gamma_G = 0.5$, qui reste difficile à interpréter.
}

\begin{figure}
	\includegraphics[width=\linewidth]{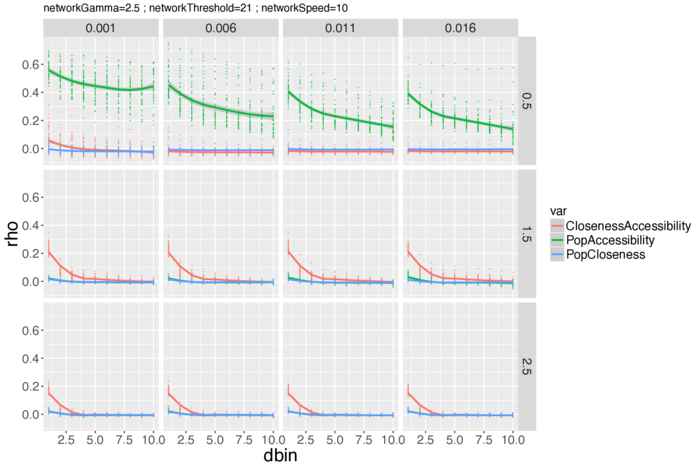}
	\appcaption{\textbf{Correlations as a function of distance.} We give correlations $\rho_d$ as a function of distance decile, for all couples of variables (color), for $d_G$ variable (columns) and $\gamma_G$ variable (rows), at fixed $\gamma_N = 2.5$, $\theta_N=21$ and $v_0 = 10$.\label{fig:app:macrocoevolexplo:distcorrs}}{\textbf{Corrélations en fonction de la distance.} Nous donnons les corrélations $\rho_d$ en fonction du décile de la distance, pour l'ensemble des couples de variables (couleur), pour $d_G$ variable (colonnes) et $\gamma_G$ variable (lignes), à $\gamma_N = 2.5$, $\theta_N=21$ et $v_0 = 10$ fixés.\label{fig:app:macrocoevolexplo:distcorrs}}
\end{figure}

\bpar{
Finally, we give in Fig.~\ref{fig:app:macrocoevolexplo:laggedcorrs} the lagged correlations $\rho_{\tau}$ between all couples of variables, for varying $d_G$ and $\gamma_G$. Similarly, qualitative behaviors are globally stable for other parameters than $\gamma_G$.
}{
Enfin, nous donnons en Fig.~\ref{fig:app:macrocoevolexplo:laggedcorrs} les corrélations retardées $\rho_{\tau}$ entre l'ensemble des couples de variables, pour $d_G$ et $\gamma_G$ variables. De même, les comportements qualitatifs sont globalement stables pour les paramètres autres que $\gamma_G$.
}

\begin{figure}
	\includegraphics[width=\linewidth]{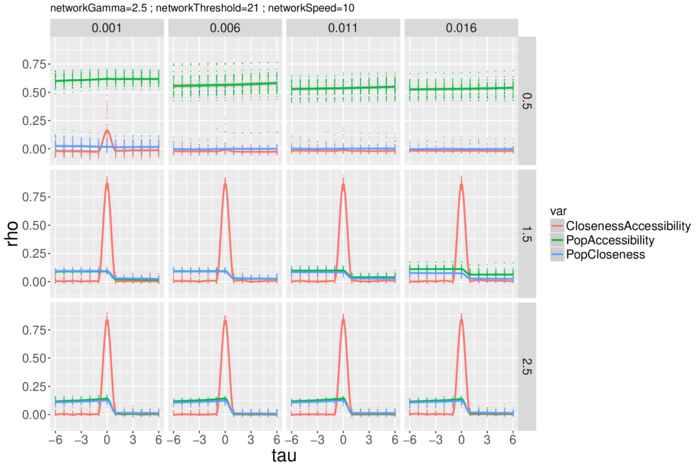}
	\appcaption{\textbf{Lagged correlations.} We give the lagged correlations $\rho_{\tau}$ as a function of the delay $\tau$, for all couples of variables (color), for $d_G$ variable (columns) and $\gamma_G$ variable (rows), at fixed $\gamma_N = 2.5$, $\theta_N=21$ and $v_0 = 10$.\label{fig:app:macrocoevolexplo:laggedcorrs}}{\textbf{Corrélations retardées.} Nous donnons les corrélations retardées $\rho_{\tau}$ en fonction du délai $\tau$, pour l'ensemble des couples de variables (couleur), pour $d_G$ variable (colonnes) et $\gamma_G$ variable (lignes), à $\gamma_N = 2.5$, $\theta_N=21$ et $v_0 = 10$ fixés.\label{fig:app:macrocoevolexplo:laggedcorrs}}
\end{figure}


\newpage

\section{Macroscopic co-evolution model}{Modèle de co-évolution macroscopique}

\label{app:sec:macrocoevol}

\subsection{Synthetic data}{Données synthétiques}

\subsubsection{Exploration}{Exploration}

\bpar{
we give in Fig.~\ref{fig:app:macrocoevol:behavior-time} the sensitivity of temporal indicators for the co-evolution model on synthetic data, in particular $\bar{c_i}(t)$ and $\varepsilon\left[\mu_i\right](t)$, for variations of $d_G$, $\gamma_G$ and $\phi_0$. The behavior of $\bar{c_i}$  is sensitive to $\gamma_G$ et $\phi_0$ but not much to $d_G$. The one of $\varepsilon\left[\mu_i\right]$ does depend only on $\gamma_G$ for its average behavior, and on $d_G$ for its dispersion in low $d_G$ values.
}{
Nous donnons en Fig.~\ref{fig:app:macrocoevol:behavior-time} la sensibilité des indicateurs temporels pour le modèle de co-évolution sur données synthétiques, en particulier $\bar{c_i}(t)$ et $\varepsilon\left[\mu_i\right](t)$, pour des variations de $d_G$, $\gamma_G$ et $\phi_0$. Le comportement de $\bar{c_i}$ est sensible à $\gamma_G$ et $\phi_0$ mais très peu à $d_G$. Celui de $\varepsilon\left[\mu_i\right]$ ne dépend que de $\gamma_G$ pour son comportement moyen, et de $d_G$ pour sa dispersion dans les faibles valeurs de $d_G$.
}

\begin{figure}
\includegraphics[width=\linewidth,height=0.9\textheight]{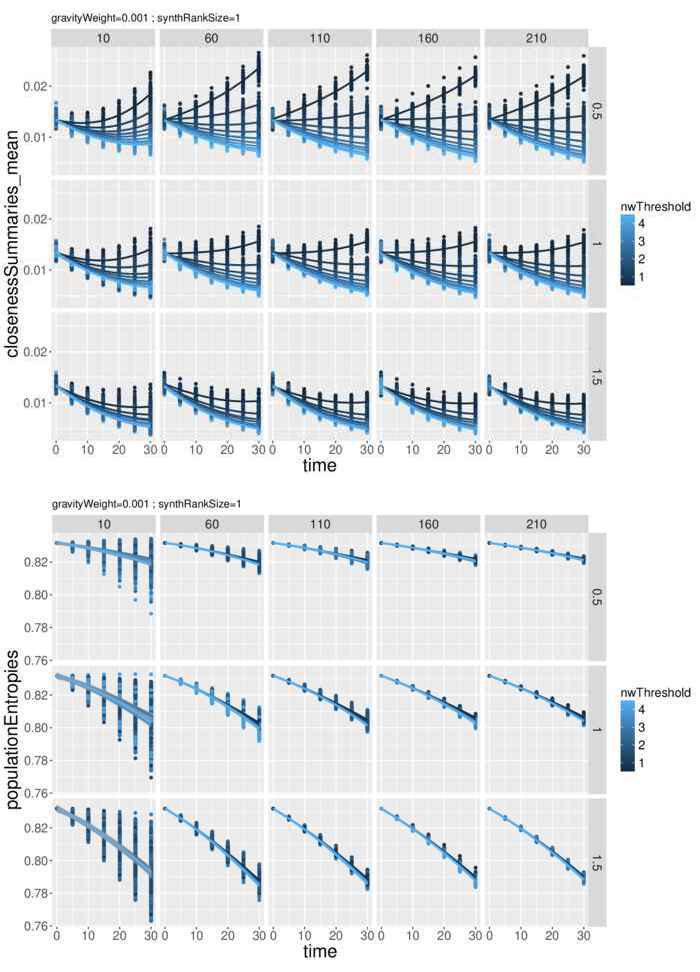}
\appcaption{\textbf{Behavior of temporal indicators for the co-evolution model at the macroscopic scale.} \textit{(Top)} Average of closeness centralities, as a function of time, for varying $d_G$ (columns), $\gamma_G$ (rows) and $\phi_0$ (color), at fixed $w_G = 0.001$; \textit{(Bottom)} Entropy of populations, as a function of time, for varying $d_G$ (columns), $\gamma_G$ (rows) and $\phi_0$ (color), at fixed $w_G = 0.001$.\label{fig:app:macrocoevol:behavior-time}}{\textbf{Comportement d'indicateurs temporels pour le modèle de co-évolution à l'échelle macroscopique.} \textit{(Haut)} Moyenne des centralités de proximité, en fonction du temps, pour $d_G$ (colonnes), $\gamma_G$ (lignes) et $\phi_0$ (couleur) variables, à $w_G = 0.001$ fixé ; \textit{(Bas)} Entropie des populations, en fonction du temps, pour $d_G$ (colonnes), $\gamma_G$ (lignes) et $\phi_0$(couleur) variables, à $w_G = 0.001$ fixé.\label{fig:app:macrocoevol:behavior-time}}
\end{figure}

\bpar{
We give in Fig.~\ref{fig:app:macrocoevol:behavior-aggreg} the behavior of aggregated indicators, namely $C\left[Z_i\right]$ and $\rho_r \left[Z_i\right]$. The complexity of accessibility trajectories mostly varies according to $d_G$, $\gamma_G$ and $\phi_0$ for the low values. The rank correlation of accessibilities is in its turn only sensitive to $d_G$ and $\gamma_G$, what means that differences in network evolution do not perturb the dynamics of the hierarchy of accessibilities.
}{
Nous donnons en Fig.~\ref{fig:app:macrocoevol:behavior-aggreg} le comportement d'indicateurs agrégés, à savoir $C\left[Z_i\right]$ et $\rho_r \left[Z_i\right]$. La complexité des trajectoires d'accessibilité varie principalement selon $d_G$, $\gamma_G$ et $\phi_0$ pour les faibles valeurs. La corrélation de rang des accessibilités est quant à elle uniquement sensible à $d_G$ et $\gamma_G$, ce qui veut dire que des différences d'évolution du réseau ne perturbent pas la dynamique de la hiérarchie des accessibilités.
}

\begin{figure}
\includegraphics[width=\linewidth,height=0.9\textheight]{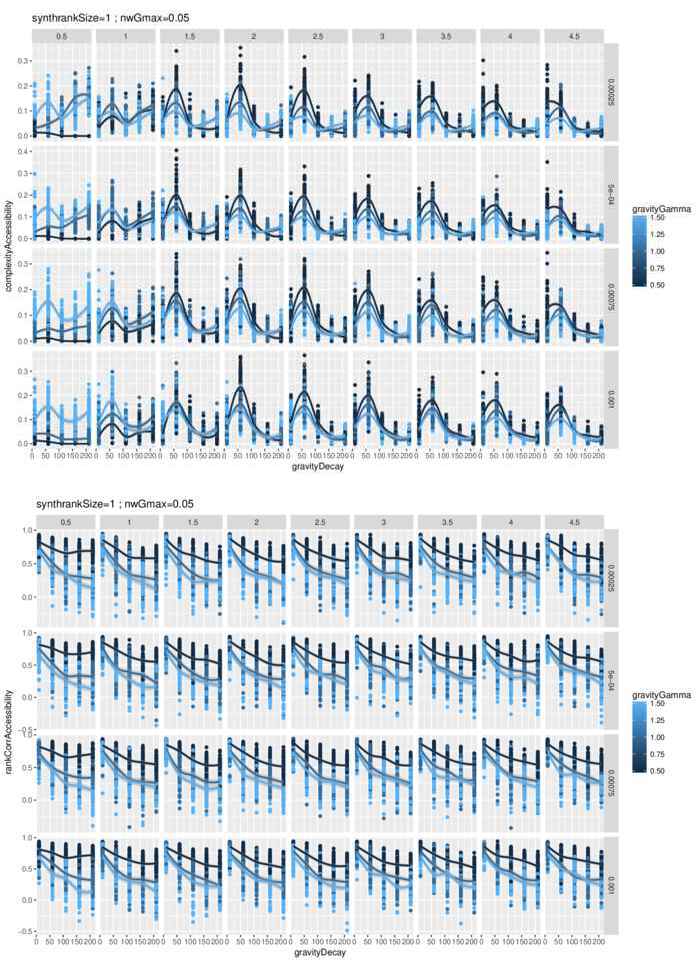}
\appcaption{\textbf{Behavior of aggregated indicators behavior for the model of coevolution at the macroscopic scale.} \textit{(Top)} Complexxity of accessibilities, as a function of $d_G$, for varying $\phi_0$ (columns), $w_G$ (rows) and $\gamma_G$ (color); \textit{(Bottom)} Rank correlations of accessibilities, for the same parameters.\label{fig:app:macrocoevol:behavior-aggreg}}{\textbf{Comportement d'indicateurs agrégés pour le modèle de co-évolution à l'échelle macroscopique.} \textit{(Haut)} Complexité des accessibilités, en fonction de $d_G$, pour $\phi_0$ (colonnes), $w_G$ (lignes) et $\gamma_G$ (couleur) variables ; \textit{(Bas)} Corrélations de rang des accessibilités, pour les mêmes paramètres.\label{fig:app:macrocoevol:behavior-aggreg}}
\end{figure}

\bpar{
The Fig.~\ref{fig:app:macrocoevol:distcorrs} gives the correlations $\rho_d$ as a function of distance deciles for all couples of variables. The strong values of $d_G$ give zero correlations for all values of distance, while $d_G=10$ exhibits local regimes. A constant correlation between centrality and accessibility emerges for an intermediate value $d_G = 60$, what could possibly be put in correspondence with the maximum of complexity for accessibilities which was obtained before.
}{
La Fig.~\ref{fig:app:macrocoevol:distcorrs} donne les corrélations $\rho_d$ en fonction des déciles de distance pour l'ensemble des couples de variables. Les fortes valeurs de $d_G$ donnent des corrélations nulles pour l'ensemble des valeurs de la distance, tandis que $d_G = 10$ témoigne de régimes locaux. Une corrélation constante entre centralité et accessibilité émerge pour une valeur intermédiaire $d_G = 60$, qui est éventuellement à mettre en correspondance avec le maximum de complexité pour les accessibilités obtenu précédemment.
}

\begin{figure}
\includegraphics[width=\linewidth]{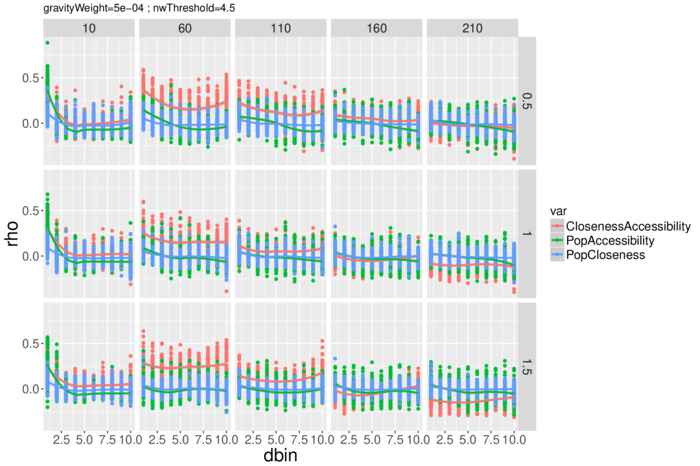}
\appcaption{\textbf{Correlations as a function of distance.} Correlation $\rho_d$ between couples of variables (given by color), as a function of distance $d$ (discretized into deciles), for varying $d_G$ (columns) and varying $\gamma_G$ (rows), at $w_G = 5e-4$ et $\phi_0 = 4.5$.\label{fig:app:macrocoevol:distcorrs}}{\textbf{Corrélations en fonction de la distance.} Correlation $\rho_d$ entre couples de variables (donné par la couleur), en fonction de la distance $d$ (discretisée en déciles), pour $d_G$ variable (colonnes) et $\gamma_G$ variable (lignes), à $w_G = 5e-4$ et $\phi_0 = 4.5$.
\label{fig:app:macrocoevol:distcorrs}}
\end{figure}

\bpar{
Finally, the Fig.~\ref{fig:app:macrocoevol:laggedcorrs} gives the lagged correlations $\rho_{\tau}$ for all variable couples. The variations of $\gamma_G$ do not influence much the regimes obtained, on the contrary to $d_G$, for which we observe a continuous variation of the qualitative shape of profiles.
}{
Enfin, la Fig.~\ref{fig:app:macrocoevol:laggedcorrs} donne les corrélations retardées $\rho_{\tau}$ pour l'ensemble des couples de variables. Les variations de $\gamma_G$ influencent peu les régimes obtenus, contrairement à $d_G$, pour lequel on observe une variation continue de la forme qualitative des profils.
}

\bpar{
More precisely, we observe that the correlation between population and accessibility is globally constant, possibly because of the auto-correlation, and does not play a role in the definition of regimes. For large values of $d_G$, we observe a positive deviation of correlations for positive and negative delays for accessibility and centrality. There is in that case circular causality and the model captures a co-evolution in that sense. Accessibility strongly causes centrality for $d_G = 10$, and the trend is inverted for large $d_G$. For $d_G = 10$, we observe a single direction relation of population towards the network. For intermediate regimes, there is directly circularity between population and centrality. Finally, for $d_G . 110$ there is ``indirect circularity'' between population and accessibility, since accessibility causes centrality which causes population.
}{
Plus précisément, nous observons que la corrélation entre population et accessibilité est globalement constante, probablement du fait de l'auto-corrélation, et n'entre pas en jeu dans la définition des régimes. Pour des grandes valeurs de $d_G$, on observe une déviation positive des corrélations pour les délais positifs et négatifs pour accessibilité et centralité. Il y a dans ce cas causalité circulaire et le modèle capture une co-évolution dans ce sens. L'accessibilité cause fortement la centralité pour $d_G = 10$, puis la tendance s'inverse pour les grands $d_G$. Pour $d_G = 10$, nous observons une relation à sens unique de la population vers le réseau. Pour les régimes intermédiaires, il y a circularité directement entre population et centralité. Enfin, pour $d_G > 110$ il y a ``circularité indirecte'' entre population et accessibilité, puisque accessibilité cause centralité qui cause population.
}

\bpar{
This visual exploration is preliminary and is continued by the statistical validation of the different regimes in main text.
}{
Cette exploration visuelle est préliminaire et est continuée par la validation statistique des différents régimes en texte principal.
}

\begin{figure}
\includegraphics[width=\linewidth]{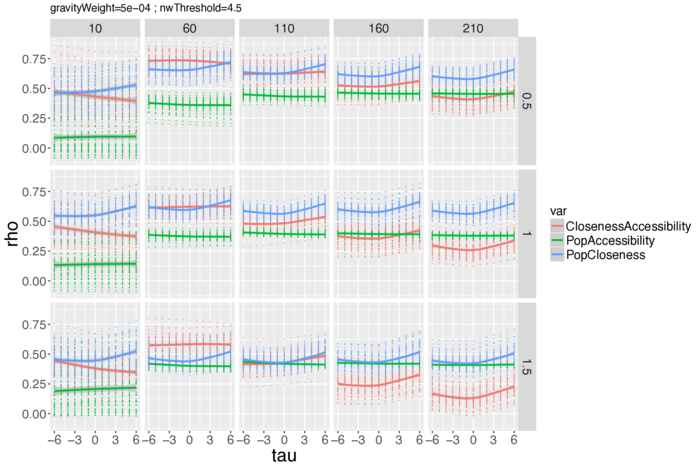}
\appcaption{\textbf{Lagged correlations.} Lagged correlations $\rho_{\tau}$ as a function of the delay $\tau$, in a similar way for varying $d_G$ (columns) and $\gamma_G$ (rows), at $w_G = 5e-4$ et $\phi_0 = 4.5$.\label{fig:app:macrocoevol:laggedcorrs}}{\textbf{Corrélations retardées.} Correlations retardées $\rho_{\tau}$ en fonction du retard $\tau$, de manière similaire pour $d_G$ variable (colonnes) et $\gamma_G$ variable (lignes), à $w_G = 5e-4$ et $\phi_0 = 4.5$.
\label{fig:app:macrocoevol:laggedcorrs}}
\end{figure}

\subsubsection{Application of the PSE algorithm}{Application de l'algorithme PSE}

\bpar{
The algorithm has been precisely applied with the objectives $\rho_{\tau_{\pm}}\left[x_i,x_j\right] - \rho_0$ with $x_i$ the 3 variables we considered and $i<j$, the estimated correlation being zero if it is non-significant or lower than $\rho_0$ in absolute value. Objectives vary in $\left[-0.2,0.2\right]$ with a step of $0.01$ (in practice, the quasi-totality of obtained values are lower in absolute value to 0.1, since the first and last centiles are lower, except two cases at 0.12 and 0.16).
}{
L'algorithme a été précisément appliqué avec les objectifs $\rho_{\tau_{\pm}}\left[x_i,x_j\right] - \rho_0$ avec $x_i$ les 3 variables considérées et $i<j$, la corrélation estimée étant nulle si non significative ou moins forte que $\rho_0$. Les objectifs varient dans $\left[-0.2,0.2\right]$ avec un pas de $0.01$ (en pratique, la quasi-totalité des valeurs obtenues sont inférieures en valeur absolue à 0.1, puisque les premiers et derniers centiles y sont inférieurs, sauf deux exceptions à 0.12 et 0.16).
}

\bpar{
The algorithm is launched on the grid with 300 parallel islands, each island having a lifetime of 2 hours, for a total of 616 generations.
}{
L'algorithme est lancé sur grille avec 300 îles en parallèle, chaque île ayant une durée de vie de 2 heures, pour un total de 616 générations.
}

\bpar{
Results of the population obtained are shown as a point cloud in Fig.~\ref{fig:app:macrocoevol:pse}. We observe that the correlation for which the distribution is the most dispersed is $\rho_{\tau_+}\left[\mu_i,c_i\right]$. Furthermore, each couple of correlations has non-reachable quadrants, suggesting impossible behaviors in the model: for example, there is close to no point with a negative causality between population and centrality and a negative causality between centrality and accessibility, these two links being thus not compatible. The couple with which it seems the hardest to extend direct circularities is accessibility and centrality, what suggests a domination of centrality in comparison to population in the expression of accessibility since the link with population has a higher range of freedom.
}{
Les résultats de la population obtenue sont montrés sous forme de nuage de points en Fig.~\ref{fig:app:macrocoevol:pse}. Nous constatons que la corrélation dont la distribution est la plus dispersée est $\rho_{\tau_+}\left[\mu_i,c_i\right]$. Par ailleurs, chaque couple de corrélations possède des cadrants impossibles à atteindre, suggérant des comportements impossibles du modèle : par exemple, il n'y a quasiment aucun point avec une causalité négative entre population et centralité et une causalité négative entre centralité et accessibilité, ces deux liens étant alors incompatibles. Le couple avec lequel il semble le plus dur d'étendre les circularités directes est accessibilité et centralité, ce qui suggère une domination de la centralité par rapport à la population dans l'expression de l'accessibilité puisque le lien avec population possède une plus grande étendue de liberté.
}

\bpar{
Basically, the algorithm unveils a richness of behaviors, extending again the one obtained with the simple exploration.
}{
Principalement, l'algorithme révèle une richesse de comportements étendant encore celle obtenue par l'exploration simple.
}


\begin{figure}
	\includegraphics[width=\linewidth]{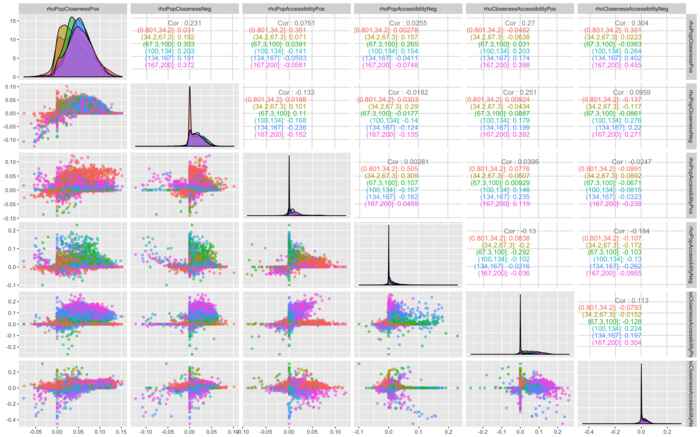}
	\appcaption{\textbf{Application of the PSE algorithm to the macroscopic model}. We give point clouds of optimal lagged correlations, for each couple of variables and the sign of the delay. The color of points gives the value of the $d_G$ parameter, numerical inserts the stratified values of correlations between correlations, and histograms the statistical distribution of each correlation.\label{fig:app:macrocoevol:pse}}{\textbf{Application de l'algorithme PSE au modèle macroscopique}. Nous donnons les nuages de points des corrélations retardées optimales, pour chaque couple de variable et le signe du délai. La couleur des points donne la valeur du paramètre $d_G$, les encarts numériques les valeurs stratifiées des corrélations entre corrélations, et les histogrammes la distribution statistique de chaque corrélation.\label{fig:app:macrocoevol:pse}}
\end{figure}

\subsection{Real data}{Données réelles}

\bpar{
We give in Fig.~\ref{fig:app:macrocoevol:pareto} the Pareto fronts for the model calibration on real data with the objectives $(\varepsilon_G,\varepsilon_L)$, which are similar to the ones given in~\ref{fig:macrocoevol:pareto}, but here with color giving the value of the $d_G$ parameter. We observe a dichotomy between large values of $d_G$ and low values, for example within the 1946 period, the decrease corresponding to a significant gain for the fit on population. In this case, long-range interactions better correspond to an adjustment of the distance, while population rather follows local processes.
}{
Nous donnons en Fig.~\ref{fig:app:macrocoevol:pareto} les fronts de Pareto pour la calibration du modèle sur données réelles selon $(\varepsilon_G,\varepsilon_L)$, similaires à ceux donnés en~\ref{fig:macrocoevol:pareto}, mais ici avec la couleur donnant la valeur du paramètre $d_G$. Nous constatons une dichotomie entre des grandes valeurs de $d_G$ et des faibles, par exemple au sein de la période 1946, la diminution correspondant à un gain considérable pour la population. Dans ce cas, les interactions lointaines correspondent mieux à un ajustement de la distance, tandis que la population suit plutôt une logique locale.
}

\begin{figure}
\includegraphics[width=\linewidth]{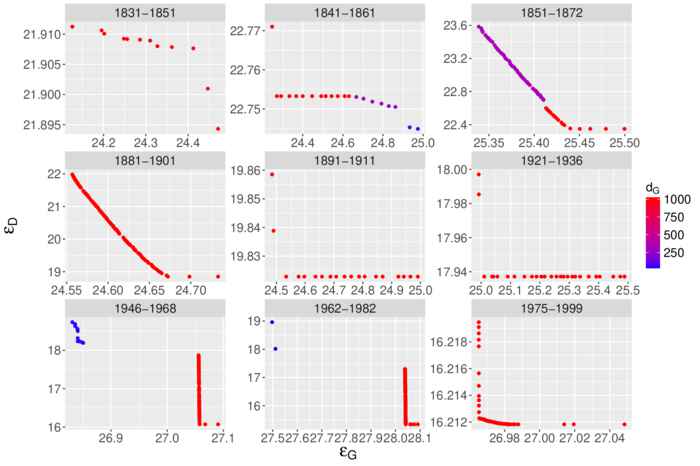}
\appcaption{\textbf{Pareto fronts for the bi-objective calibration with population and distance.} The fronts are given for each calibration period, and colored as a function of $d_G$.\label{fig:app:macrocoevol:pareto}}{\textbf{Fronts de Pareto pour la calibration bi-objectif population et distance.} Les fronts sont donnés pour chaque période de calibration, et colorés en fonction de $d_G$.\label{fig:app:macrocoevol:pareto}}
\end{figure}

%


\newpage

\section{Network generation heuristics}{Heuristiques de génération de réseau}

\label{app:sec:networkgrowth}

\subsection{Slime mould model}{Modèle de slime mould}

\bpar{
We recall here the procedure of type \emph{slime mould} to evolve the biological network, based on~\cite{tero2007mathematical}. The network is composed by nodes characterized by their pressure $p_i$ and by links characterized by their length $L_{ij}$, their diameter $D_{ij}$, an impedance $Z_{ij}$ and the flow traversing them $\phi_{ij}$. The relation analogous to Ohm's law for links writes
}{
Nous rappelons ici la procédure d'évolution du réseau biologique de type \emph{slime mould}, à partir de~\cite{tero2007mathematical}. Le réseau est composé de noeuds caractérisés par leur pression $p_i$ et de liens caractérisés par leur longueur $L_{ij}$, leur diamètre $D_{ij}$, une impédance $Z_{ij}$ et le flux les traversant $\phi_{ij}$. La relation analogue à la loi d'Ohm pour les liens s'écrit
}

\[
\phi_{ij} = \frac{D_{ij}}{Z_{ij}\cdot L_{ij}} \left(p_i - p_j\right)
\]

\bpar{
Furthermore, the conservation of flows at each node (Kirchoff's law) imposes
}{
Par ailleurs, la conservation des flux à chaque noeud (loi de Kirchoff) impose
}

\[
\sum_i \phi_{ij} = 0
\]

\bpar{
for all $j$ except the source and the sink, that we assume at indices $j_+$ and $j_-$, such that $\sum_i \phi_{ij_+} = I_0$ and $\sum_i \phi_{ij_-} = -I_0$ with $I_0$ initial flow parameter.
}{
pour tout $j$ sauf pour la source et le puit, que nous supposons aux indices $j_+$ et $j_-$, tel que $\sum_i \phi_{ij_+} = I_0$ et $\sum_i \phi_{ij_-} = -I_0$ avec $I_0$ paramètre de flux initial.
}

\bpar{
The combination of above constraints gives for all $j$
}{
La combinaison des contraintes ci-dessus donne pour tout $j$
}

\[
\sum_i \frac{D_{ij}}{Z_{ij}\cdot L_{ij}} (p_i - p_j) = \mathbbm{1}_{j=j_+} I_0 - \mathbbm{1}_{j=j_-} I_0 
\]

\bpar{
what simplifies into a matrix equation, by denoting $\mathbf{Z} = \left(\frac{\frac{D_{ij}}{Z_{ij}\cdot L_{ij}}}{\sum_i \frac{D_{ij}}{Z_{ij}\cdot L_{ij}}}\right)_{ij}$, and also $\vec{k} = \frac{\mathbbm{1}_{j=j_+} I_0 - \mathbbm{1}_{j=j_-} I_0 }{\sum_i \frac{D_{ij}}{Z_{ij}\cdot L_{ij}}}$ and $\vec{p} = p_i$, what simplifies into
}{
ce qui se simplifie en une équation matricielle, en notant $\mathbf{Z} = \left(\frac{\frac{D_{ij}}{Z_{ij}\cdot L_{ij}}}{\sum_i \frac{D_{ij}}{Z_{ij}\cdot L_{ij}}}\right)_{ij}$, ainsi que $\vec{k} = \frac{\mathbbm{1}_{j=j_+} I_0 - \mathbbm{1}_{j=j_-} I_0 }{\sum_i \frac{D_{ij}}{Z_{ij}\cdot L_{ij}}}$ et $\vec{p} = p_i$, qui se simplifie en
}

\[
\left(Id - \mathbf{Z}\right) \vec{p} = \vec{k}
\]

\bpar{
The system admits a solution when $\left(Id - \mathbf{Z}\right)$ is invertible. The space of invertible matrices being dense in $\mathcal{M}_n(\mathbb{R})$, by multilinearity of the determinant, an infinitesimal perturbation of the position of nodes allows to invert the matrix if it is indeed singular. We obtain thus the pressures $p_i$ and as a consequence the flows $\phi_{ij}$.
}{
Le système admet une solution lorsque $\left(Id - \mathbf{Z}\right)$ est inversible. L'espace des matrices inversible étant dense dans $\mathcal{M}_n(\mathbb{R})$, par multilinéarité du déterminant, une perturbation infinitésimale de la position des noeuds permet d'inverser la matrice si celle-ci est effectivement singulière. On obtient donc les pressions $p_i$ et par conséquent les flux $\phi_{ij}$.
}


\bpar{
The evolution of the diameter $D_{ij}$ between two equilibrium stages is a function of the flow at equilibrium, through the equation
}{
L'évolution du diamètre $D_{ij}$ entre deux étapes d'équilibre est fonction du flux à l'équilibre, par l'équation 
}

\[
D_{ij} (t+1) - D_{ij} = \delta t \left[ \frac{\phi_{ij}(t)^\gamma}{1 + \phi_{ij}(t)^\gamma} - D_{ij}(t)\right]
\]

\bpar{
We take to simplify $\gamma = 1.8$, following the configuration used by~\cite{tero2010rules} for the generation of a network in a real configuration. We furthermore take $\delta t = 0.05$ and $I_0 = 10$.
}{
Nous prenons pour simplifier $\gamma = 1.8$, suivant la configuration utilisée par~\cite{tero2010rules} pour la génération d'un réseau dans une configuration réelle. Nous prenons par ailleurs $\delta t = 0.05$ et $I_0 = 10$.
}

\bpar{
The generation of a network can be achieved from an initial network, until a convergence criteria is reached, for example $\sum_{ij} \Delta D_{ij} (t) < \varepsilon$ with $\varepsilon$ fixed threshold parameter. We will use this model with a criteria of a number of iterations, and proceed to an iteration to obtain final networks with a reasonable number of links.
}{
La génération d'un réseau peut s'effectuer à partir d'un réseau initial, jusqu'à atteindre un critère de convergence, par exemple $\sum_{ij} \Delta D_{ij} (t) < \varepsilon$ avec $\varepsilon$ paramètre de seuil fixé. Nous utilisons ce modèle avec un critère de nombre d'itérations, et procédons à une itération pour obtenir des réseaux finaux avec un nombre raisonnable de liens.
}

\subsection{Results}{Résultats}

\bpar{
In the experiment exploring the distance to real networks, the initialization of the density is done according to 50 density grids classified into 5 morphological classes (10 grids per class). The Table~\ref{app:tab:networkgrowth:morpho} gives the composition of centers of classes in terms of morphological indicators. Classes can be interpreted the following way:
\begin{itemize}
	\item Class 5: lowest Moran, high distance, hierarchy and entropy; numerous population centers that are localized and dispersed.
	\item Class 4: highest entropy and hierarchy; a small number of localized centers.
	\item Class 3: lowest distance and entropy; diffuse population.
	\item Class 2: highest Moran; one or a few centers with consequent size.
	\item Class 1: intermediate values for all indicators; a certain number of centers of intermediate size.
\end{itemize}
}{
Dans l'expérience explorant la distance aux réseaux réels, l'initialisation de la densité est faite selon 50 grilles classées dans 5 classes morphologiques (10 grilles par classe). La Table~\ref{app:tab:networkgrowth:morpho} donne la composition des centres des classes en termes d'indicateurs morphologiques. Les classes peuvent être interprétées de la façon suivante :
\begin{itemize}
	\item Classe 5 : plus bas Moran, distance, hiérarchie et entropie élevées ; nombreux foyers de peuplement localisés et dispersés.
	\item Classe 4 : plus fortes entropie et hiérarchie ; un petit nombre de foyers localisés.
	\item Classe 3 : plus basse distance et entropie ; population diffuse.
	\item Classe 2 : plus haut Moran ; un ou quelques centres de taille conséquente.
	\item Classe 1 : valeurs intermédiaires pour tous les indicateurs ; un certain nombre de centres de taille intermédiaire.
\end{itemize}
}

\begin{table}
\apptabcaption{\textbf{Morphological indicators for centers of classes for initial density grids.}\label{app:tab:networkgrowth:morpho}}{\textbf{Indicateurs morphologiques pour les centres des classes des grilles de densité initiales.}\label{app:tab:networkgrowth:morpho}}
\begin{tabular}{|c|c|c|c|c|}
\hline
\bpar{
Class & Moran $I$ & Distance $\bar{d}$ & Entropy $\mathcal{E}$ & Hierarchy $\gamma$ \\\hline 
}{
Classe & Moran $I$ & Distance $\bar{d}$ & Entropie $\mathcal{E}$ & Hiérarchie $\gamma$ \\\hline 
}
1 & 0.23 & 0.66 & 0.76 & 0.62 \\\hline 
2 & 0.47 & 0.50 & 0.75 & 0.53 \\\hline 
3 & 0.21 & 0.42 & 0.57 & 0.65 \\\hline 
4 & 0.24 & 0.75 & 0.90 & 0.87 \\\hline 
5 & 0.15 & 0.76 & 0.84 & 0.72 \\\hline 
\end{tabular}
\end{table}

\bpar{
Topological spaces of networks generated in~\ref{sec:networkgrowth} can be conditioned to morphological classes for initial density distribution. This conditioning is shown in Fig.~\ref{fig:app:networkgrowth:feasiblespace_bymorph}. We also give feasible spaces with real points. Classes 1 and 5 seem to be the ones for which being close to real points is the easiest, in terms of extreme points.
}{
Les espaces topologiques des réseaux générés en~\ref{sec:networkgrowth} peuvent être conditionnés aux classes morphologiques pour la distribution de densité initiale. Ce conditionnement est montré en Fig.~\ref{fig:app:networkgrowth:feasiblespace_bymorph}. Nous donnons également les espaces faisables avec les points réels. Les classes 1 et 5 semblent être celles pour laquelle le rapprochement aux points réels est le plus facile, en termes de points extrêmes.
}

\begin{figure}
\includegraphics[width=0.9\linewidth]{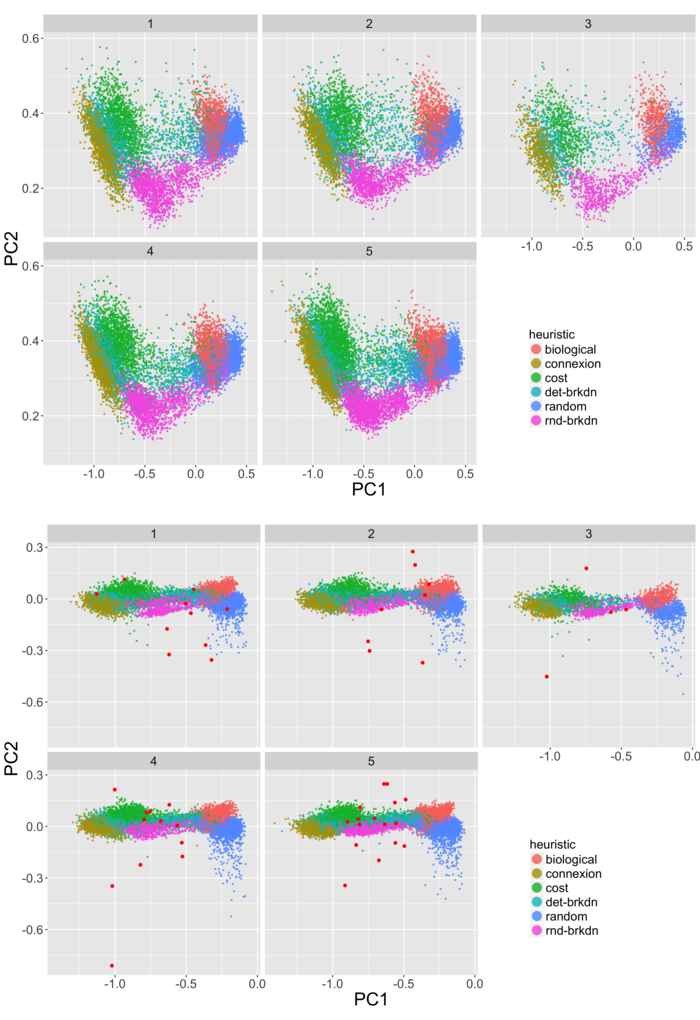}
\appcaption{\textbf{Conditioning of results to morphological classes for density.} (\textit{Top}) Topological feasible space for the different generation heuristics, conditioned to the morphological density class. (\textit{Bottom}) Same plots with real points in red.\label{fig:app:networkgrowth:feasiblespace_bymorph}}{\textbf{Conditionnement des résultats aux classes morphologiques pour la densité.} (\textit{Haut}) Espace topologique faisable pour les différentes heuristiques de génération, conditionné à la classe morphologique de densité. (\textit{Bas}) Mêmes graphiques avec les points réels en rouge.\label{fig:app:networkgrowth:feasiblespace_bymorph}}
\end{figure}


\newpage

\section{Co-evolution at the mesoscopic scale}{Co-évolution à l'échelle mesoscopique}

\label{app:sec:mesocoevolmodel}

\subsection{Calibration}{Calibration}

\bpar{
In order to justofy the aggregation of distances for indicators and for correlations, we have visually controlled the shape of Pareto fronts for these two objectives for around twenty simulated points. An example for two points is given in Fig.~\ref{fig:app:mesocoevolmodel:paretodists}. It appears that these fronts are close to be not existing, i.e. that there almost exist a global optimum.
}{
Afin de justifier l'agrégation des distances pour les indicateurs et pour les corrélations, nous avons contrôlé visuellement la forme des fronts de Pareto pour ces deux objectifs pour une vingtaine de points simulés. Un example pour deux points est donné en Fig.~\ref{fig:app:mesocoevolmodel:paretodists}. Il apparait que ces fronts sont quasi-inexistants, c'est-à-dire qu'il existe presque un optimum global.
}

\bpar{
Let illustrate to what extent a linear aggregation with equal coefficients can be relevant in the case of a Pareto front which is close to being vertical/horizontal. The function
}{
Illustrons dans quelle mesure une agrégation linéaire à coefficient égaux peut être pertinente dans le cas d'un front de Pareto quasiment vertical/horizontal. La fonction
}
\[
f_{\alpha} : x \mapsto \frac{1}{(x+1)^\alpha}
\]
\bpar{
takes this form in a neighborhood of 0 when $\alpha$ becomes large. We then consider the two objectives $o_1(x) = x$ and $o_2(x) = f_{\alpha}(x)$, which can either be considered for a bi-objective minimization, or in the frame of a linear aggregation through the minimization of $o(x) = \beta x + (1-\beta) \frac{1}{(x+1)^{\alpha}}$. That latest is minimal in $x = \left(\frac{\beta}{\alpha (1-\beta)}\right)^{\frac{1}{\alpha + 1}} - 1$, term which can be developed into
}{
prend cette forme dans un voisinage de 0 lorsque $\alpha$ devient grand. Considérons alors les deux objectifs $o_1(x) = x$ et $o_2(x) = f_{\alpha}(x)$, qui peuvent soit être considérés pour une minimisation bi-objectifs, soit dans le cadre d'une agrégation linéaire par minimisation de $o(x) = \beta x + (1-\beta) \frac{1}{(x+1)^{\alpha}}$. Cette dernière est minimale en $x = \left(\frac{\beta}{\alpha (1-\beta)}\right)^{\frac{1}{\alpha + 1}} - 1$, terme qui se développe en 
}
\[
x = \frac{\ln\left(\beta (1-\beta)\right)}{\alpha + 1} + \frac{\ln\alpha}{\alpha + 1} + o(\frac{1}{\alpha})
\]

\bpar{
Furthermore, let consider that in the frame of a bi-objective optimization, we take the compromise at which the variations of $o_1$ equalize the ones of $o_2$, what is equivalent to take $x$ such that $\frac{\partial f}{\partial x} = \frac{\partial f^{-1}}{\partial x}$. This equation leads to $\frac{x^{\frac{1}{\alpha}}}{x + 1} = \frac{1}{\alpha^{\frac{2}{\alpha + 1}}}$. We can then develop at the second order on each side to obtain
}{
Par ailleurs, considérons que dans le cadre d'une optimisation bi-objectifs, nous prenions le compromis auquel les variations de $o_1$ égalent celles de $o_2$, ce qui revient à prendre $x$ tel que $\frac{\partial f}{\partial x} = \frac{\partial f^{-1}}{\partial x}$. Cette équation conduit à $\frac{x^{\frac{1}{\alpha}}}{x + 1} = \frac{1}{\alpha^{\frac{2}{\alpha + 1}}}$. On peut alors développer au second ordre de chaque côté pour obtenir
}
\[
\frac{\ln x}{\alpha} = x \left[1 - 2 \frac{\ln \alpha}{\alpha + 1} + o(\frac{1}{\alpha})\right] - 2 \frac{\ln \alpha}{\alpha + 1} + o(\frac{1}{\alpha}) 
\]
\bpar{
We indeed necessarily have $x\rightarrow_{\alpha \rightarrow \infty} 0$, since if $x \rightarrow K \neq 0$, we have a contradiction in the previous equation since $1/(1+K) \neq 0$. It implies that $\frac{\ln x}{\alpha} = o(\frac{1}{\alpha})$, and thus that
}{
Or on a nécessairement $x\rightarrow_{\alpha \rightarrow \infty} 0$, puisque si $x \rightarrow K \neq 0$, on a une contradiction dans l'équation précédente car $1/(1+K) \neq 0$. Cela implique que $\frac{\ln x}{\alpha} = o(\frac{1}{\alpha})$, et donc que
}
\[
x = 2 \frac{\ln \alpha}{\alpha + 1} + o(\frac{1}{\alpha})
\]
\bpar{
In order thus to have the same order of magnitude for the solutions to the two approaches, we need to eliminate the term in $1/(\alpha + 1)$ in the first, what is equivalent to take $\ln \left(\beta (1- \beta)\right) = 0$ and therefore $\beta = 1/2$.
}{
Pour avoir donc les mêmes ordres de grandeur pour les solutions aux deux approches, il faut éliminer le terme en $1/(\alpha + 1)$ dans la première, ce qui revient à prendre $\ln \left(\beta (1- \beta)\right) = 0$ et donc $\beta = 1/2$.
}

\bpar{
Thus, there is equivalence of orders of magnitude in $\alpha$ for the two approaches if an only if $\beta = 1/2$. Given the shape of our Pareto fronts, we consider that the solution is analogous and consider thus the sum of the two distances.
}{
Ainsi, il y a équivalence des ordres de grandeurs en $\alpha$ pour les deux approches si et seulement si $\beta = 1/2$. Vu la forme de nos fronts de Pareto, nous considérons la solution analogue et considérons ainsi la somme des deux distances.
}

\begin{figure}
	\includegraphics[width=\linewidth]{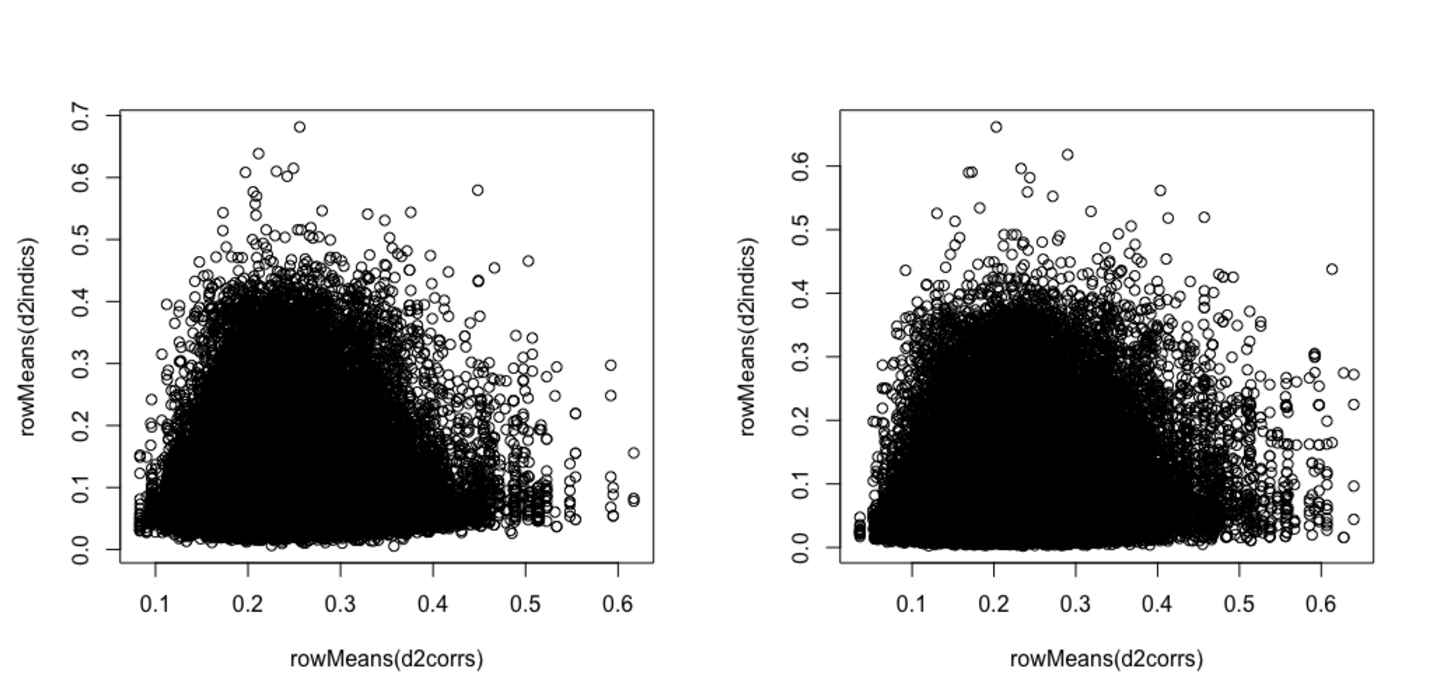}
	\appcaption{\textbf{Example of Pareto fronts for the calibration at the first and second order.} We give for two particular simulation points, the distances to indicators $d_I^2$ and the distances to correlations $d_C^2$ for all the real points.\label{fig:app:mesocoevolmodel:paretodists}}{\textbf{Exemples de fronts de Pareto pour la calibration au premier et au second ordre.} Nous donnons pour deux points particuliers de simulation, les distances aux indicateurs $d_I^2$ et les distances aux corrélations $d_C^2$ pour l'ensemble des points réels.\label{fig:app:mesocoevolmodel:paretodists}}
\end{figure}


\newpage

\section{Transportation system governance modeling}{Modélisation de la gouvernance du système de transport}

\label{app:sec:lutecia}

\subsection{Land-use model}{Modèle d'usage du sol}

\subsubsection{Convergence}{Convergence}

\bpar{
We study here the issue of the convergence in time of the distribution of activities, with a fixed infrastructure.
}{
Nous étudions ici la question de la convergence dans le temps de la distribution des activités, à infrastructure fixe.
}

\bpar{
Let consider a very simple case: by taking $\lambda = 0$ the problem is made not spatial and by taking $\gamma_A = 1$ we achieve the decoupling between population and employments. By denoting $\beta' = \sum_j E_j \cdot \beta$ and $P_0 = \alpha \cdot \sum_i P_i$, the existence of a fixed point for populations is equivalent to the resolution of
}{
Considérons un cas très simple : en prenant $\lambda = 0$ on déspatialise le problème et en prenant $\gamma_A = 1$ on finit de découpler population et emplois. En posant $\beta' = \sum_j E_j \cdot \beta$ et $P_0 = \alpha \cdot \sum_i P_i$, l'existence d'un point fixe pour les populations se ramène à la résolution de
}
\[
P_i = P_0 \cdot \frac{\exp\left(\beta' \cdot P_i\right)}{\sum \exp\left(\beta' \cdot P_i\right)}
\]

\bpar{
The function is indeed continuous in $P_i$ and variation ranges for population are $[0,\sum_i P_i]$, it therefore admits a fixed point through the Brouwer fixed point theorem.
}{
La fonction est bien continue en les $P_i$ et les plages de variations de la population sont $[0,\sum_i P_i]$, elle admet donc un point fixe par le Théorème du Point Fixe de Brouwer. 
}

\bpar{
Indeed, in all generality, if we write
}{
En fait, en toute généralité, si on écrit
}

\[
(\vec{P}(t+1),\vec{E}(t+1)) = f(\vec{P}(t),\vec{E}(t))
\]

\bpar{
for arbitrary parameter values, the function $f$ is also continuous in each component, and takes its values with a bounded closed interval (employments being also limited) therefore a compact. The same way that \cite{leurent2014user} establishes it for a model of traffic flows, we also have a fixed point in our case, what corresponds to an equilibrium point. The unicity is however not trivial and there is no reason for it to be a priori verified. We empirically verify the systematic convergence at fixed infrastructure (see below the exploration of the parameter space).
}{
pour des valeurs des paramètres arbitraires, la fonction $f$ est également continue en chaque composante, et prend ses valeurs dans un fermé borné (les emplois étant également limités) donc compact. De la même manière que \cite{leurent2014user} l'établit pour un modèle de flux de traffic, on a aussi un point fixe dans notre cas, ce qui correspond à un point d'équilibre. L'unicité n'est cependant pas triviale et il n'y a pas de raison qu'elle soit vérifiée a priori. On vérifie empiriquement la convergence systématique à infrastructure fixe (voir ci-dessous l'exploration de l'espace des paramètres).
}

\subsubsection{Exploration}{Exploration}

\bpar{
We proceed to an exploration of the behavior of the land-use model alone, i.e. at fixed infrastructure, in order to understand the influence of parameters on the urban form. We fix $\alpha = 1$ here to study the model in an extreme case.
}{
Nous procédons à une exploration du comportement du modèle d'usage du sol seul, i.e. à infrastructure fixe, afin de comprendre l'influence des paramètres sur la forme urbaine. Nous fixons $\alpha = 1$ ici pour étudier le modèle dans un cas extrême.
}

\bpar{
We follow the urban form indicators defined in~\ref{sec:staticcorrelations}, for the distribution of population and employments, in time and until the model has converged. We reduce the morphological space of the spatial distribution of actives in a principal plan, such that $PC_1 = -0.98 \cdot I - 0.13 \cdot \mathcal{E} + 0.05 \bar{d} - 0.13 \cdot \gamma $ and $PC_2 = -0.19 \cdot I + 0.57 \cdot \mathcal{E} - 0.16 \bar{d} + 0.77 \cdot \gamma $. The first component expresses a level of dispersion and the second a hierarchical aggregation.
}{
Nous suivons les indicateurs de forme urbaine définis en~\ref{sec:staticcorrelations}, pour la distribution de la population et pour les emplois, dans le temps et jusqu'après convergence. Nous réduisons l'espace morphologique de la distribution spatiale des actifs dans un plan principal, tel que $PC_1 = -0.98 \cdot I - 0.13 \cdot \mathcal{E} + 0.05 \bar{d} - 0.13 \cdot \gamma $ et $PC_2 = -0.19 \cdot I + 0.57 \cdot \mathcal{E} - 0.16 \bar{d} + 0.77 \cdot \gamma $. La première composante exprime un niveau de dispersion et la seconde une agrégation hiérarchique.
}


\bpar{
The Fig.~\ref{fig:app:lutecia:morphotrajs} gives temporal trajectories in the plan $(PC_1,PC_2)$ for $\gamma_A = 0.9$, $\gamma_E = 0.6$, $v_0 = 6$, for different values of $\lambda$ and $\beta$ and also for different initial networks. We observe that increasing $\beta$ has the tendency to make trajectories uniform. For $\beta = 1$, the shape of the network strongly conditions trajectories conjointly to $\lambda$: we switch for example from a decreasing dispersion and a u-shaped hierarchy to a stable dispersion and an increasing hierarchy for low values of $\lambda$, between no network and a spider network. 
}{
La Fig.~\ref{fig:app:lutecia:morphotrajs} donne des trajectoires temporelles dans le plan $(PC_1,PC_2)$ pour $\gamma_A = 0.9$, $\gamma_E = 0.6$, $v_0 = 6$, pour différentes valeurs de $\lambda$ et de $\beta$ ainsi que pour différents réseaux initiaux. On constate qu'augmenter $\beta$ a tendance à uniformiser les trajectoires. Pour $\beta = 1$, la forme du réseau conditionne fortement les trajectoire conjointement à $\lambda$ : on passe par exemple d'une dispersion décroissante et d'une hiérarchie en cloche à une dispersion stable et une hiérarchie croissante pour les valeurs faibles de $\lambda$, entre aucun réseau et un réseau araignée.
}


\begin{figure}
	\includegraphics[width=\linewidth,height=0.9\textheight]{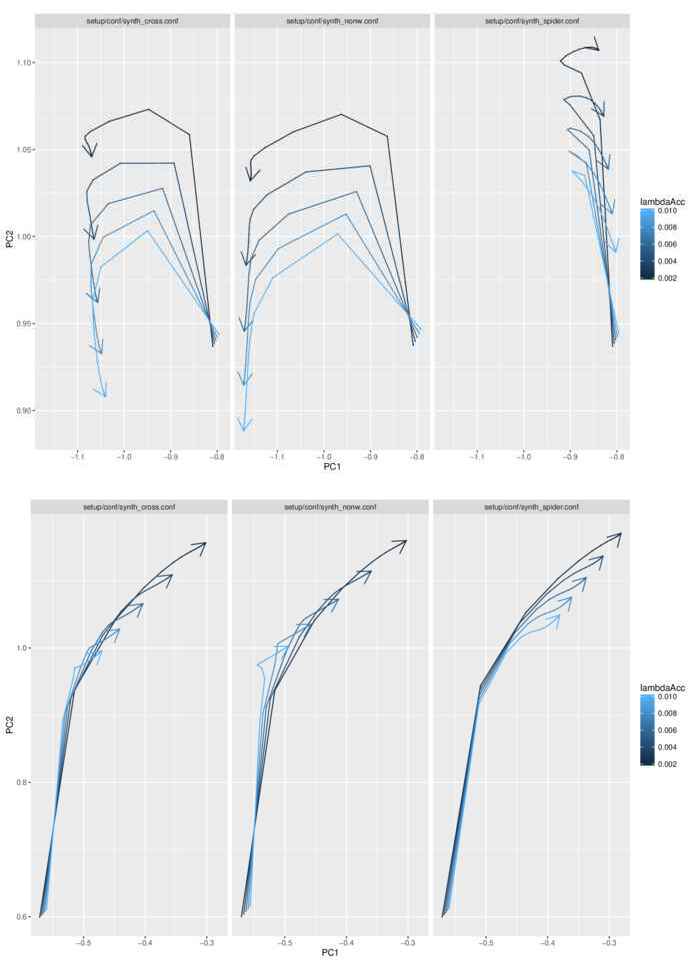}
	\appcaption{\textbf{Morphological trajectories for the distribution of population.} We fix here $\gamma_A = 0.9$ and $\gamma_E = 0.6$. (\textit{Top}) Trajectories in the space $(PC_1,PC_2)$ for $\beta = 1$, with variable $\lambda$ (color), and for three different network configurations (columns): cross network, no network, cross network with branches (spider). (\textit{Bottom}) Same plots, for $\beta = 2$.\label{fig:app:lutecia:morphotrajs}}{\textbf{Trajectoires morphologiques pour la distribution de population.} On fixe ici $\gamma_A = 0.9$ et $\gamma_E = 0.6$. (\textit{Haut}) Trajectoires dans l'espace $(PC_1,PC_2)$ pour $\beta = 1$, avec $\lambda$ variable (couleur), et pour trois configurations de réseau différentes (colonnes) : réseau en croix, pas de réseau, réseau en croix avec ramifications (spider). (\textit{Bas}) Mêmes graphes, pour $\beta = 2$.\label{fig:app:lutecia:morphotrajs}}
\end{figure}

\bpar{
The Fig.~\ref{fig:app:lutecia:morphosens} gives the value of $PC_1$ for the final configuration on all the space of explored parameters. We thus observe the variability of forms (here in terms of dispersion) as a function of all parameters: for example, for large $\beta$ values, complex diagrams emerge. For low $\beta$ values, we have a diagonal privileged for dispersion within concentrated configurations.
}{
La Fig.~\ref{fig:app:lutecia:morphosens} donne la valeur de $PC_1$ pour la configuration finale sur l'ensemble de l'espace des paramètres exploré. Nous constatons ainsi la variabilité des formes (ici en termes de dispersion) en fonction de l'ensemble des paramètres : par exemple, pour des grandes valeurs de $\beta$, des diagrammes complexes émergent. Pour les faibles valeurs de $\beta$, on a une diagonale privilégiée pour la dispersion au sein de configurations concentrées.
}

\begin{figure}
	\includegraphics[width=\linewidth]{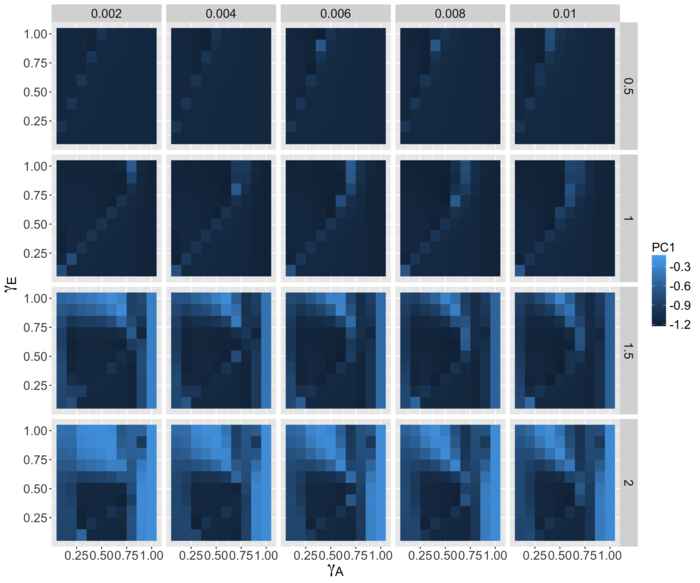}
	\appcaption{\textbf{Sensitivity of the urban form.} For the distribution of populations, without initial network, value of $PC_1$ as a function of $(\gamma_A,\gamma_E)$, with variable $\lambda$ (columns) and variable $\beta$ (rows).\label{fig:app:lutecia:morphosens}}{\textbf{Sensibilité de la forme urbaine.} Pour la distribution des populations, sans réseau initial, valeur de $PC_1$ en fonction de $(\gamma_A,\gamma_E)$, avec $\lambda$ variable (colonnes) et $\beta$ variable (lignes).\label{fig:app:lutecia:morphosens}}
\end{figure}

\bpar{
Finally, in order to understand the influence of parameters on total mobility within a complete trajectory, we study in Fig.~\ref{fig:app:lutecia:ludiff} the cumulated variation of actives given by $\tilde{\Delta} = \sum_t \sum_k \left|\Delta A_k (t)\right|$. We see that high values of $\gamma_A$, for a high $\beta$, allow to minimize the total quantity of relocalization, which have a very low dependence in $\gamma_E$. It is therefore possible to optimize, even at fixed $\alpha$, the total quantity of urban sprawl.
}{
Enfin, afin de comprendre l'influence des paramètres sur la mobilité totale au cours d'une trajectoire complète, nous étudions en Fig.~\ref{fig:app:lutecia:ludiff} la variation cumulée des actifs donnée par $\tilde{\Delta} = \sum_t \sum_k \left|\Delta A_k (t)\right|$. On voit que des valeurs fortes de $\gamma_A$, pour $\beta$ élevé, permettent de minimiser la quantité totale de relocalisation, qui ne dépend que très faiblement de $\gamma_E$. Il est ainsi possible d'optimiser, même à $\alpha$ fixé, la quantité totale d'étalement urbain.
}

\begin{figure}
	\includegraphics[width=\linewidth]{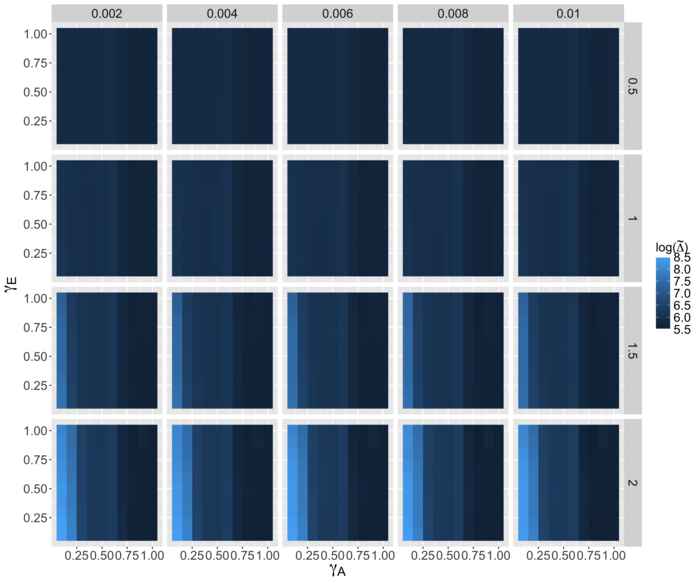}
	\appcaption{\textbf{Cumulated variability of urban configurations.} Value of $\ln\tilde{\Delta}$, without initial network, as a function of $(\gamma_A,\gamma_E)$, with variable $\lambda$ (columns) and variable $\beta$ (rows).\label{fig:app:lutecia:ludiff}}{\textbf{Variabilité cumulée des configurations urbaines.} Valeur de $\ln\tilde{\Delta}$, sans réseau initial, en fonction de $(\gamma_A,\gamma_E)$, avec $\lambda$ variable (colonnes) et $\beta$ variable (lignes).\label{fig:app:lutecia:ludiff}}
\end{figure}

\subsection{Transportation model}{Modèle de transport}

\bpar{
We did not take into account transportation flows in our implementation of the model, assuming that the constructed infrastructures have a sufficient capacity to be significantly not sensitive to congestion.
}{
Nous n'avons pas pris en compte les flux de transport dans notre implémentation du modèle, supposant que les infrastructures construites sont de capacités suffisantes pour ne pas être significativement sensibles à la congestion.
}

\bpar{
For the computation of flows between cells, the operation is the following: flows $\phi_{ij}$ are computed by a solving on $p_i,q_j$ through a fixed point method (Furness algorithm), of the system of gravity flows:
}{
Pour le calcul des flux entre cellules, l'opération est la suivante : les flux $\phi_{ij}$ sont calculés par résolution sur $p_i,q_j$ par une méthode de point fixe (algorithme de Furness), du système des flux gravitaires :
}

\[
\begin{cases}
\phi_{ij} = p_i q_j A_i E_j \exp{\left(-\lambda_{tr} d_{ij}\right)}\\
\sum_k \phi_{kj} = E_j\\
\sum_k \phi_{ik} = A_i\\
p_i = \frac{1}{\sum_k{q_k E_k \exp{(-\lambda_{tr}d_{ik})}}}\\
q_j = \frac{1}{\sum_k{p_k A_k \exp{(-\lambda_{tr}d_{kj})}}} 
\end{cases}
\]

\bpar{
where $\lambda_{tr}$ is a parameter giving the spatial reach of daily flows. The iteration of the last two equations rapidly converges starting from equal weights, by maintaining at each stage normalized weights.
}{
où $\lambda_{tr}$ est un paramètre donnant la portée spatiale des flux journaliers. L'itération des deux dernières équations converge rapidement à partir de poids égaux, en maintenant à chaque étape des poids normalisés.
}

\bpar{
In order to implement the stage of flows distribution within the network, when flows between cells are known, we should for example determine flows of the Static User Equilibrium with an appropriated algorithm. An assignment by shortest paths is implemented with the computation of flows in the model, but we deactivate this process in order to simplify the study of the model.
}{
Pour implémenter l'étape de distribution des flux dans le réseau, une fois les flux entre cellules connus, il faudrait par exemple déterminer les flux de l'Equilibre Utilisateur Statique avec un algorithme approprié. Une affectation par plus courts chemins est implémentée avec le calcul des flux dans le modèle, mais nous désactivons ce processus pour simplifier l'étude du modèle.
}

\bpar{
Congestion can be computed as a ratio to capacity, as $c/c_{max}$ if $c$ is the flow and $c_{max}$ the capacity. The speed is obtained with a BPR function of the form $v(c) = v_0 \left(1 - \frac{c}{c_max}\right)^{\gamma_c}$. Our configuration os equivalent to assuming an infinite capacity $c_{max} = \infty$.
}{
La congestion peut être calculée comme un rapport à la capacité, comme $c/c_{max}$ si $c$ est le flux et $c_{max}$ la capacité. La vitesse est obtenue par une fonction BPR sous la forme $v(c) = v_0 \left(1 - \frac{c}{c_{max}}\right)^{\gamma_c}$. Notre configuration revient à supposer une capacité infinie $c_{max} = \infty$.
}

\subsection{Probabilities to cooperate}{Probabilités de coopération}

\subsubsection{Nash equilibrium}{Equilibre de Nash}

\bpar{
The equilibrium assumption implies that conditional expectancies of each player are equal given their two choices, i.e. that
}{
L'hypothèse d'équilibre implique que les espérances conditionnelles de chaque joueur sont égales étant donné leur deux choix, i.e. que 
}
\[
\Eb{U_i|S_i=C} = \Eb{U_i|S_i=NC}
\] 

\bpar{
It is indeed equivalent in that case to maximize $\Eb{U_i}$ as a function of $p_i$, since by conditioning we have$\Eb{U_i} = p_i \Eb{U_i|S_i = C} + (1 - p_i) \Eb{U_i|S_i = NC}$, and thus $\frac{\partial \Eb{U_i}}{\partial p_i} = \Eb{U_i|S_i = C} - \Eb{U_i| S_i = NC}$.
}{
Cela revient en effet dans ce cas à maximiser $\Eb{U_i}$ par rapport à $p_i$, puisque en conditionnant on a $\Eb{U_i} = p_i \Eb{U_i|S_i = C} + (1 - p_i) \Eb{U_i|S_i = NC}$, et donc $\frac{\partial \Eb{U_i}}{\partial p_i} = \Eb{U_i|S_i = C} - \Eb{U_i| S_i = NC}$.
}

\bpar{
We have then
}{
On a alors
}

\[
\Eb{U_i|S_i=C} = p_{1-i} U_i(S_i=C,S_{1-i}=C) + (1- p_{1-i}) U_i (S_i=C,S_{1-i}=NC)
\]

\bpar{
and thus
}{
et donc 
}

\begin{equation*}
\hspace{-1cm}
\begin{split}
	p_{1-i} & U_i(S_i=C,S_{1-i}=C) + (1- p_{1-i}) U_i (S_i=C,S_{1-i}=NC) \\
	& = p_{1-i} U_i(S_i=NC,S_{1-i}=C) + (1- p_{1-i}) U_i (S_i=NC,S_{1-i}=NC)
\end{split}
\end{equation*}

\bpar{
what gives
}{
ce qui donne
}

\[
p_{1-i} = - \frac{U_i(C,NC) - U_i(NC,NC)}{\left(U_i(C,C) - U_i(NC,C)\right) - \left(U_i(C,NC) - U_i(NC,NC)\right)}
\]

\bpar{
By substituting the expressions of utilities from the payoff matrix, we obtain the expression of $p_i$ as a function of collaboration cost $J$ and of the difference of accessibility differentials.
}{
En substituant les expressions des utilités à partir de la matrice de gain, on obtient l'expression de $p_i$ en fonction du coût de collaboration $J$ et de la différence des différentiels d'accessibilité.
}

\subsubsection{Discrete choice coordination}{Coordination par choix discrets}

\bpar{
To determine the probability to cooperate in the discrete choice case, we have to solve $f(p_i) = 0$ with
}{
Pour déterminer la probabilité de coopération dans le cas des choix discrets, il s'agit de résoudre $f(p_i) = 0$ avec
}

\[
f(x) = \frac{1}{1+\exp\left[-\beta_{DC}\frac{\Delta_i}{1 + \exp(-\beta_{DC}(x \Delta_{1-i} - J))} - J\right]} - x
\]

\bpar{
where we wrote $\Delta_i = \Delta X_{i}(Z^{\star}_{C}) - \Delta X_{\bar{i}}(Z^{\star}_{i})$.
}{
où nous avons noté $\Delta_i = \Delta X_{i}(Z^{\star}_{C}) - \Delta X_{\bar{i}}(Z^{\star}_{i})$.
}

\bpar{
We immediately have $f(0) > 0 $ and $f(1) < 0$ and $f$ is continuous, there therefore always exists a solution $x\in [0,1]$ by the theorem of intermediate values.
}{
On a immédiatement $f(0) > 0 $ et $f(1) < 0$ et $f$ est continue, il existe donc toujours une solution $x\in [0,1]$ par le théorème des valeurs intermédiaires.
}

\bpar{
Regarding uniqueness, it can be shown under some assumptions. A computation of $\frac{\partial f}{\partial x}$ gives
}{
Concernant l'unicité, il est possible de la montrer sous certaines conditions. Un calcul de $\frac{\partial f}{\partial x}$ donne 
}

\[
\frac{\partial f}{\partial x} = 2 (\cosh u(x) - 1) + \beta^2 \Delta_i \Delta_{1-i} \frac{\exp(-\beta_{DC}(x \Delta_{1-i} - J))}{(1 + \exp(-\beta_{DC}(x \Delta_{1-i} - J)))^2}
\]

\bpar{
where $u(x) = -\beta_{DC} (\frac{\Delta_i}{1 + \exp(-\beta_{DC}(x \Delta_{1-i} - J))} - J)$.
}{
où $u(x) = -\beta_{DC} (\frac{\Delta_i}{1 + \exp(-\beta_{DC}(x \Delta_{1-i} - J))} - J)$.
}

\bpar{
As $\cosh u \geq 1$, we have $\frac{\partial f}{\partial x} > 0$ if $\Delta_i \Delta_{1-i} > 0$. The function is in this case strictly increasing and we have a unique solution.
}{
Comme $\cosh u \geq 1$, on a $\frac{\partial f}{\partial x} > 0$ si $\Delta_i \Delta_{1-i} > 0$. La fonction est dans ce cas strictement croissante et on a une unique solution.
}

\bpar{
In practice, the solution is determined with the Brent algorithm, with boundaries $[0,1]$ and a tolerance of $0.01$.
}{
En pratique, la solution est déterminée par algorithme de Brent, avec les bornes $[0,1]$ et une tolérance de $0.01$.
}

\subsection{Implementation details}{Détails d'implémentation}

\subsubsection{Distance matrix}{Matrice des distances}

\bpar{
The distance matrix is updated in a dynamical way because of execution time issues (given the number of network updates), and the following way:
\begin{enumerate}
	\item The euclidian distance matrix $d(i,j)$ is computed analytically
	\item The shortest paths between intersections of links (between cells of the corresponding raster network) are updated in a dynamical way (step of complexity $O(N_{inters}^3$):
	\begin{itemize}
		\item For each new intersection, shortest paths towards all other intersections are computed using the old matrix and the new link.
		\item For all former shortest paths, they are updated if needed after checking potential shortcuts using the new link.
		\item The correspondence between basic network cells and intersections is updated.
	\end{itemize}
	\item Connected components and the distances between them are updated (complexity in $O(N_{nw}^2)$)
	\item Distances within the network between network cells are updated, with the heuristic of minimal connections only (a unique shortest link between each cluster) (complexity in $O(N_{nw}^2)$)
	\item Effective distances between all cells (taking speed into account and congestion if it is implemented) are computed as the minimum between euclidian distance and \[\min_{C,C'}{d(i,C)+d_{nw}(p_C(i),p_C'(j))+d(C',j)}\] that we approximate by taking $\min_C$ only in the implementation, what is consistent with relatively interaction ranges that we consider. The complexity is in $O(N_{clusters}^2\cdot N^2)$.
\end{enumerate}
}{
La matrice des distances est mise à jour de manière dynamique pour des questions de rapidité d'execution (vu le nombre de mises à jour du réseau), de la façon suivante :
\begin{enumerate}
	\item La matrice de distance euclidienne $d(i,j)$ est calculée analytiquement
	\item Les plus courts chemins entre les intersections des liens (entre les cellules du réseau raster correspondant) sont mis à jour de manière dynamique (étape de complexité $O(N_{inters}^3$) :
	\begin{itemize}
		\item Pour chaque nouvelle intersection, les plus courts chemins vers l'ensemble des autres intersections sont calculés par l'ancienne matrice et le nouveau lien.
		\item Pour l'ensemble des anciens plus courts chemins, ils sont mis à jour si besoin après vérification des éventuels raccourcis par le nouveau lien.
		\item La correspondance entre les cellules quelconques du réseau et les intersections est mise à jour.
	\end{itemize}
	\item Les composantes connexes et les distances entre celles-ci sont mises à jour (complexité en $O(N_{nw}^2)$)
	\item Les distances par le réseau entre les cellules du réseau sont mises à jour, avec l'heuristique des connexions minimales uniquement (un lien unique le plus court entre chaque cluster) (complexité en $O(N_{nw}^2)$)
	\item Les distances effectives entre l'ensemble des cellules (prenant la vitesse et la congestion en compte si celle-ci est implémentée) sont calculées comme le minimum entre la distance euclidienne et 
	\[\min_{C,C'}{d(i,C)+d_{nw}(p_C(i),p_C'(j))+d(C',j)}\]
	dont nous prenons une approximation avec $\min_C$ uniquement dans l'implémentation, ce qui est consistant avec les portées d'interaction relativement faibles considérées. La complexité est en $O(N_{clusters}^2\cdot N^2)$.
\end{enumerate}
}

\subsubsection{Network growth}{Croissance du réseau}

\bpar{
The potential infrastructures, with a count of $N_I$ during the heuristic search of an optimal infrastructure, are randomly drawn among all possible infrastructures having an extremity at the center of a cell. If the extremity is at a distance lower than a threshold of an already existing link of the network, it is replaced by its projection on the corresponding link. This is a snapping step which allows to obtain a network with a reasonable form locally. In correspondance with the raster representation of the network, we take $\theta_I = 1$, what corresponds to the size of a cell.
}{
Les infrastructures potentielles, au nombre de $N_I$ lors de la recherche heuristique d'une infrastructure optimale, sont tirées aléatoirement parmi l'ensemble des infrastructures possibles ayant une extrémité au centre d'une cellule. Si l'extrémité est à une distance inférieure à un seuil $\theta_I$ d'un lien déjà existant du réseau, celle-ci est remplacée par sa projection sur le lien correspondant. Il s'agit de l'étape d'accrochage permettant d'obtenir un réseau de forme raisonnable localement. En cohérence avec la représentation raster du réseau, nous prenons $\theta_I = 1$, ce qui correspond à la taille d'une cellule. 
}

\subsection{Setup}{Initialisation}

\subsubsection{Synthetic setup}{Initialisation synthétique}

\bpar{
We describe here the details of the synthetic setup.
}{
Nous décrivons ici les détails de l'initialisation synthétique.
}

\bpar{
Initial distributions of actives and employments in the synthetic setup are taken around governance centers (mayors) at positions $\vec{x}_i$ using exponential kernels by
}{
Les distributions initiales des actifs et des emplois dans la configuration synthétique sont pris autour des centres de gouvernance (maires) aux positions $\vec{x}_i$ avec des noyaux exponentiels par
}

\[
A(\vec{x}) = A_{max} \cdot \exp{\left(\frac{\norm{\vec{x}-\vec{x}_i}}{r_A}\right)} ; 
E(\vec{x}) = E_{max} \cdot \exp{\left(\frac{\norm{\vec{x}-\vec{x}_i}}{r_E}\right)}
\]

\subsubsection{Setup on a real configuration}{Initialisation sur configuration réelle}

\bpar{
We show in Fig.~\ref{fig:app:lutecia:realsetup} the population and networks on which experiments on real data are done: with a fixed land-use,
\begin{itemize}
	\item an experiment without initial network, and with target calibration network the network in 2010;
	\item an experiment with the initial network of 2010, and with target network the planned network.
\end{itemize}
}{
Nous montrons en Fig.~\ref{fig:app:lutecia:realsetup} la population et les réseaux sur lesquels les expériences sur données réelles sont menées : à usage du sol fixe,
\begin{itemize}
	\item une expérience sans réseau initial, et avec pour réseau cible de calibration le réseau de 2010 ;
	\item une expérience avec le réseau initial de 2010, et pour réseau cible le réseau planifié.
\end{itemize} 
}

\begin{figure}
	\includegraphics[width=\linewidth]{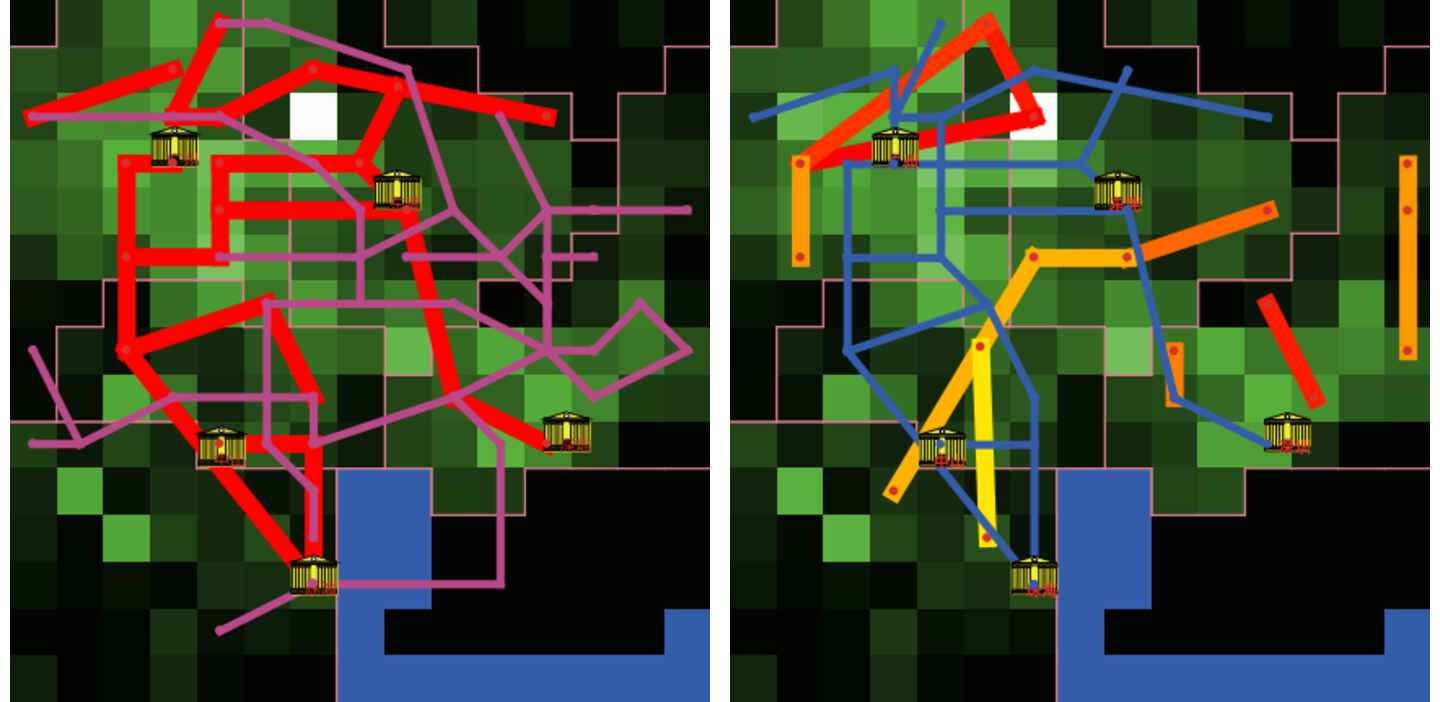}
	\appcaption{\textbf{Setup on real data used for model application.} Because of computational performance issues, the number of cells is here lower in comparison to the illustration in main text. (\textit{Left}) Initial networks, in red the initial network corresponding to the network in 2010, in purple and tight the target network for calibration, corresponding to the planned network. (\textit{Right}) Result obtained with $\alpha = 0$ at $t_f = 11$ after a setup without network; in blue the target network, which corresponds to the network in 2010.\label{fig:app:lutecia:realsetup}}{\textbf{Initialisation sur données réelles utilisée lors de l'application.} Pour des raisons de performances computationnelle, le nombre de cellules est ici diminué par rapport à l'illustration en texte principal. (\textit{Gauche}) Réseaux à l'initialisation, en rouge le réseau initial correspondant au réseau en 2010, en violet fin le réseau cible pour la calibration, correspondant au réseau planifié. (\textit{Droite}) Résultat obtenu avec $\alpha = 0$ à $t_f = 11$ après une initialisation sans réseau ; en bleu le réseau cible, qui correspond au réseau de 2010.\label{fig:app:lutecia:realsetup}}
\end{figure}


%


\bpar{
\chapter{Methodological Developments}
\markboth{\thechapter\space Methodological Developments}{\thechapter\space Methodological Developments}
}{
\chapter{Développements méthodologiques}
\markboth{\thechapter\space Développements Méthodologiques}{\thechapter\space Développements Méthodologiques}
}

\label{app:methodology} 





\bpar{
This appendix gathers different methodological developments which are indirectly used, or allows deepening connected issues but which are not crucial for our main argumentation.
}{
Cette annexe rassemble différents développements méthodologiques qui sont utilisés indirectement, ou permet de creuser des questions liées mais non centrales à notre fil principal.
}

\bpar{
The first three sections relate to questions raising in the study of urban or territorial systems.
\begin{enumerate}
	\item A formal link between different stochastic models of urban growth allows introducing a general frame for this kind of approach, and illustrate the implicit link between our mesoscopic approach and our macroscopic approach.
	\item The sensitivity of scaling laws to the definition of the city is analytically studied for a simple model of urban system. This perspective reinforces the methodology of sensitivity analysis of models to the spatial configuration introduced in~\ref{sec:computation}.
	\item The literature and formal context of the notion of synthetic data also allows to situate it.
\end{enumerate}
}{
Les trois premières sections traitent des questions se posant particulièrement lors de l'étude des systèmes urbains ou territoriaux.
\begin{enumerate}
	\item Un lien formel entre différents modèles stochastiques de croissance urbaine permet de poser un cadre général pour ce genre d'approche, et d'illustrer le lien implicite entre notre approche mesoscopique et notre approche macroscopique.
	\item La sensibilité des lois d'échelles à la définition de la ville est étudiée analytiquement pour un modèle simple de système urbain. Cette perspective renforce la méthodologie d'analyse de sensibilité des modèles à la configuration spatiale introduite en~\ref{sec:computation}.
	\item Le contexte bibliographique et formel de la notion de données synthétiques permet également de situer celle-ci.
\end{enumerate}
}

\bpar{
We then develop general methodological frameworks linked to the study of systems.
}{
Nous développons ensuite des cadres méthodologiques généraux liés à l'étude des systèmes.
}

\bpar{
\begin{enumerate}\setcounter{enumi}{4}
	\item In the context of systems including multi-attribute optimizations, a method for the sensitivity analysis to the structure of data is introduced. It is not directly applied in our work but suggests directions for the application of mesoscopic morphogenesis models, since these are based on a similar optimization by agents.
	\item a general framework for the modeling of socio-technical complex systems, sets the first bricks on the one hand of a formalization of the \emph{applied perspectivism} but also of the formalization of the knowledge framework suggested in~\ref{sec:knowledgeframework}.
\end{enumerate}
}{
\begin{enumerate}\setcounter{enumi}{4}
	\item Dans le cadre de systèmes incluant des optimisation multi-attributs, une méthode d'analyse de sensibilité à la structure des données, est introduite. Elle n'est pas directement appliquée dans notre travail mais suggère des pistes pour l'application des modèles mesoscopiques de morphogenèse, puisque ceux-ci se basent sur une telle optimisation par les agents.
	\item Un cadre général pour la modélisation des systèmes complexes socio-techniques, pose les premières bases d'une part d'une formalisation du \emph{perspectivisme appliqué} mais également de la formalisation du cadre de connaissances suggérée en~\ref{sec:knowledgeframework}.
\end{enumerate}
}

\bpar{
Finally, the last development is focused on quantitative epistemology methods.
}{
Enfin, le dernier développement concerne les méthodes d'épistémologie quantitative.
}

\bpar{
\begin{enumerate}\setcounter{enumi}{4}
	\item The technical details of the method used in~\ref{sec:quantepistemo} are developed in the context of an application to the corpus of the journal \emph{Cybergeo}. The considerations are fundamentally methodological, and must also be put in perspective with the companion thematical paper that we adapt in~\ref{app:sec:cybergeonetworks}.
\end{enumerate}
}{
\begin{enumerate}\setcounter{enumi}{4}
	\item Les détails techniques de la méthode utilisée en~\ref{sec:quantepistemo} sont développés dans le cadre d'application au corpus de la revue \emph{Cybergeo}. Les considérations sont fondamentalement méthodologiques, et doivent être également mises en perspective avec l'article thématique companion que nous adaptons en~\ref{app:sec:cybergeonetworks}.
\end{enumerate}
}

\newpage

\section{Stochastic models of urban growth}{Modèles stochastiques de croissance urbaine}

\label{app:sec:stochurbgrowth}

\bpar{
The different stochastic models of urban growth that we have developed follow the same logic of autonomous rules to reproduce the dynamics of urban systems. We propose here from a methodological viewpoint to highlight the links between the different frameworks, in order to formulate a unified framework.
}{
Les différents modèles stochastiques de croissance urbaine que nous avons développé suivent la même logique de règles autonomes pour reproduire les dynamiques des systèmes urbains. Nous proposons ici d'un point de vue méthodologique de mettre en valeur les liens entre les différents cadres, afin d'en formuler un cadre unifié.
}

\subsection{Introduction}{Introduction}

\bpar{
Diverse stochastic models of urban growth aiming at reproducing population trajectories, or stylized facts on these, often on long time scales or large spatial spans (systems of cities) have been proposed by the literature in various fields, from economics or physics to geography (see for example~\ref{sec:interactiongibrat} and \ref{sec:densitygeneration} for reviews at different scales). We propose here a general approach which allows making a link between different existing models, more particularly the Gibrat model, the Simon model, and the preferential attachment model.
}{
Divers modèles stochastiques de croissance urbaine visant à reproduire des trajectoires de population, ou des faits stylisés sur celles-ci, souvent sur de longues échelles de temps et de grandes étendues spatiales (systèmes de villes) ont été proposé par la littérature dans des champs variés, de l'économie ou la physique à la géographie (voir par exemple~\ref{sec:interactiongibrat} et \ref{sec:densitygeneration} pour des revues à différentes échelles). Nous proposons ici une approche générale permettant de faire le lien entre plusieurs modèles existant, plus particulièrement les modèles de Gibrat, de Simon et d'attachement préférentiel.
}

\bpar{
Seminal models of urban growth are the Gibrat model (see~\ref{sec:interactiongibrat}) and the Simon model~\cite{simon1955class} (which has more recently been generalized, see e.g. \cite{haran1973modified}). Many extensions have been given across disciplines. \cite{benguigui2007dynamic} give an equation-based dynamical model, whereas \cite{gabaix1999zipf} shows that the Gibrat model produces Zipf's law in a stationary state. \cite{Gabaix20042341} reviews urban growth approaches in economics. A model adapted from evolutive urban theory is described in~\cite{favaro2011gibrat} and extends the Gibrat model by adding propagation of innovation between cities. The question of empirical scales at which it is consistent to study urban growth was also tackled in the particular case of France~\cite{bretagnolle2002time}, which shows that long time scales (more than a few decade) are appropriate to study dynamics of urban systems at a small spatial scale.
}{
Des modèles fondamentaux de croissance urbaine sont les modèles de Gibrat (voir~\ref{sec:interactiongibrat}) et le modèle de Simon~\cite{simon1955class} (qui a plus récemment été généralisé, voir par exemple~\cite{haran1973modified}). Diverses extensions ont été données selon les disciplines. \cite{benguigui2007dynamic} donne un modèle de système dynamique, tandis que \cite{gabaix1999zipf} montre que le modèle de Gibrat produit la loi de Zipf pour la distribution de la taille des villes dans l'état stationnaire. Les approches en économie sont revues par \cite{Gabaix20042341}. Un modèle inspiré par la Théorie Evolutive des Villes est décrit dans~\cite{favaro2011gibrat} et étend le modèle de Gibrat par l'addition de la propagation de l'innovation entre les villes. La question des échelles empiriques auxquelles ce type d'approche est pertinent a été traité dans le cas particulier de la France par~\cite{bretagnolle2002time}, qui montre que de longues échelles de temps (supérieures à quelques décades) sont appropriées pour étudier la dynamique des systèmes urbains à une petite échelle spatiale.
}

\subsection{Framework}{Cadre de Travail}


\bpar{
The framework we introduce can be understood as a meta-model, in the sense that eaach model can be understood as an extension or a limit case of an other model.
}{
Le cadre que nous introduisons peut se comprendre comme un meta-modèle, au sens où chaque modèle peut être compris comme extension ou cas limite d'un autre modèle. 
}








\subsection{Derivations}{Dérivations}

\subsubsection{Generalization of Preferential Attachment}{Généralisation de l'Attachement Préférentiel}

\bpar{
\cite{yamasaki2006preferential} give a generalization of the classical preferential attachment model for network growth, as a birth and death model with evolving entities. More precisely, network nodes gain and lose population units at fixed probabilities, and new nodes can be created at an also fixed rate.
}{
\cite{yamasaki2006preferential} donne une généralisation du modèle classique d'attachement préférentiel pour la croissance des réseaux, comme un modèle de vie et mort avec des entités évolutives. Plus précisément, les noeuds du réseau gagnent et perdent des unités de population à des probabilités fixes, et de nouveaux noeuds peuvent être ajoutés à un taux également fixe.
}

\subsubsection{Link between Gibrat and Preferential Attachment Models}{Lien entre Gibrat et Attachement Préférentiel}

\bpar{
Let consider a strictly positive growth Gibrat model given by $P_i(t)=R_i(t)\cdot P_{i}(t-1)$ with $R_i(t)>1$, $\mu_i(t)=\Eb{R_i(t)}$ and $\sigma_i(t)=\Eb{R_i(t)^2}$. The $P_i$ are the populations of cities while $R_i$ are random growth rates. On the other hand, we take a simple preferential attachment, with fixed attachment probability $\lambda \in [0,1]$ and new arrivants number $m>0$, what gives in expectancy $\mu_i(t+1) - \mu_i(t) = m\cdot \lambda$. We derive that Gibrat model can be statistically equivalent to a limit of the preferential attachment model, assuming that the moment-generating functions of $R_i(t)$ exist. Classical distributions that could be used in that case, e.g. log-normal distribution, are entirely defined by two first moments, making this assumption reasonable.
}{
Considérons un modèle de croissance strictement positive de Gibrat donnée par $P_i(t)=R_i(t)\cdot P_{i}(t-1)$ avec $R_i(t)>1$, $\mu_i(t)=\Eb{P_i(t)}$, $\lambda_i(t)=\Eb{R_i(t)}$ et $\sigma_i(t)=\Eb{R_i(t)^2}$. Les $P_i$ sont les populations des villes tandis que $R_i$ sont des taux de croissance aléatoires. D'autre part, soit un modèle simple d'attachement préférentiel, avec une probabilité d'attachement $\lambda \in [0,1]$ et un nombre de nouveau arrivants $m>0$, ce qui revient en espérance à $\mu_i(t+1) - \mu_i(t) = m\cdot \lambda$. Il est possible de dériver que le Gibrat est statistiquement équivalent à une limite de l'attachement préférentiel, sous l'hypothèse que toutes les fonctions génératrices des moments de $R_i(t)$ existent. Les distributions classiques qui peuvent être utilisées dans ce cas, e.g. une distribution normale ou log-normale, sont entièrement déterminées par leur deux premiers moments, ce qui rend cette hypothèse raisonnable.
}


\bpar{
\begin{lemma}
The limit of a Preferential Attachment model when $\lambda \ll 1$ is a linear-growth Gibrat model, with limit parameters $\mu_i(t)=1+\frac{\lambda}{m\cdot (t-1)}$.
\end{lemma}
}{
\begin{lemma}
	La limite d'un modèle d'attachement préférentiel quand $\lambda \ll 1$ est un modèle de croissance de Gibrat linéaire, avec le paramètres limites $\lambda_i(t)=1+\frac{\lambda}{m\cdot (t-1)}$.
\end{lemma}
}

\begin{proof}

\bpar{
Starting with first moment, we denote $\bar{P}_i(t)=\Eb{P_i(t)}$. Independence of Gibrat growth rates yields directly $\bar{P}_i(t)=\Eb{R_i(t)}\cdot \bar{P}_i(t-1)$. Starting for the preferential attachment model, we have $\bar{P}_i(t) = \Eb{P_i(t)} = \sum_{k=0}^{+\infty}{k\Pb{P_i(t)=k}}$. But on the other hand,
}{
S'intéressant au premier moment, nous notons $\bar{P}_i(t)=\mu_i(t)=\Eb{P_i(t)}$. L'indépendance entre les taux de croissance de Gibrat donne directement $\bar{P}_i(t)=\Eb{R_i(t)}\cdot \bar{P}_i(t-1)$. En partant du modèle d'attachement préférentiel, nous avons $\bar{P}_i(t) = \Eb{P_i(t)} = \sum_{k=0}^{+\infty}{k\Pb{P_i(t)=k}}$. Mais par ailleurs,
}
\[
\{P_i(t)=k\}=\bigcup_{\delta=0}^{\infty}{\left(\{P_i(t-1)=k-\delta\}\cap \{P_i\leftarrow P_i + 1\}^{\delta}\right)}
\]
\bpar{
where the second event corresponds to city $i$ being increased $\delta$ times between $t-1$ and $t$ (note that events are empty for $\delta \geq k$). Thus, being careful on the conditional nature of preferential attachment formulation, stating that $\Pb{\{P_i\leftarrow P_i + 1\} | P_i(t-1)=p} = \lambda\cdot\frac{p}{P(t-1)}$ (total population $P(t)$ assumed deterministic), we obtain
}{
où le second évènement correspond à la ville $i$ étant augmentée $\delta$ fois entre $t-1$ et $t$ (avec la convention que les évènement sont vides pour $\delta \geq k$). Ainsi, en prenant en compte la formulation conditionnelle de l'attachement préférentiel, qui postule que $\Pb{\{P_i\leftarrow P_i + 1\} | P_i(t-1)=p} = \lambda\cdot\frac{p}{P(t-1)}$ (la population totale $P(t)$ étant déterministe), nous obtenons
}
\begin{equation*}
\begin{split}
\Pb{\{P_i\leftarrow P_i + 1\}} & = \sum_{p}{\Pb{\{P_i\leftarrow P_i + 1\} | P_i(t-1)=p}\cdot \Pb{P_i(t-1)=p}}\\
&=\sum_{p}{\lambda\cdot\frac{p}{P(t-1)}\Pb{P_i(t-1)=p}}=\lambda\cdot\frac{\bar{P}_i(t-1)}{P(t-1)}\\
\end{split}
\end{equation*}

\bpar{
\noindent It gives therefore, knowing that $P(t-1)=P_0 + m\cdot (t-1)$ and denoting $q=\lambda\cdot\frac{\bar{P}_i(t-1)}{P_0 + m\cdot (t-1)}$
}{
\noindent Ce qui donne, sachant que $P(t-1)=P_0 + m\cdot (t-1)$ et en notant $q=\lambda\cdot\frac{\bar{P}_i(t-1)}{P_0 + m\cdot (t-1)}$
}

\[
\begin{split}
\bar{P}_i(t) & =\sum_{k=0}^{\infty}{\sum_{\delta=0}^{\infty}{k\cdot \left(\lambda\cdot\frac{\bar{P}_i(t-1)}{P_0 + m\cdot (t-1)}\right)^{\delta}\cdot \Pb{P_i(t-1)=k-\delta}}}\\
& = \sum_{\delta^{\prime}=0}^{\infty}{\sum_{k^{\prime}=0}^{\infty}{\left(k^\prime + \delta^{\prime}\right)\cdot q^{\delta^{\prime}} \cdot \Pb{P_i(t-1)=k^\prime}}}\\
& = \sum_{\delta^{\prime}=0}^{\infty}{q^{\delta^{\prime}}\cdot \left(\delta^{\prime} + \bar{P}_i(t-1)\right)} = \frac{q}{(1-q)^2} + \frac{\bar{P}_i(t-1)}{(1-q)}\\
& = \frac{\bar{P}_i(t-1)}{1-q}\left[1 + \frac{1}{\bar{P}_i(t-1)}\frac{q}{(1-q)}\right]
\end{split}
\]


\bpar{
As it is not expected to have $\bar{P}_i(t)\ll P(t)$ (fat tail distributions), a limit can be taken only through $\lambda$. Taking $\lambda \ll 1$ yields, since $0 < \bar{P}_i(t)/P(t) < 1$, we obtain $q=\lambda\cdot\frac{\bar{P}_i(t-1)}{P_0 + m\cdot (t-1)} \ll 1$ and thus we can expand in first order of $q$. It finally yields
}{
On s'attend à ce que pour la majorité des villes, $\bar{P}_i(t)\ll P(t)$ (distributions fortement dissymétriques), la limite peut être prise pour $\lambda$ uniquement. En prenant $\lambda \ll 1$, comme $0 < \bar{P}_i(t)/P(t) < 1$, nous obtenons $q=\lambda\cdot\frac{\bar{P}_i(t-1)}{P_0 + m\cdot (t-1)} \ll 1$, qui peut être développée au premier ordre en $q$. Cela donne finalement
}

\[
\bar{P}_i(t)=\bar{P}_i(t-1)\cdot \left[1 + \left(1+\frac{1}{\bar{P}_i(t-1)}\right)q + o(q))\right]
\]

\bpar{
and therefore
}{
et donc
}

\[
\bar{P}_i(t) \simeq \left[1 + \frac{\lambda}{P_0 + m\cdot (t-1)}\right]\cdot \bar{P}_i(t-1)
\]

\bpar{
This means that this limit is equivalent in expectancy to a Gibrat model with $\mu_i(t) = \mu(t)=1 + \frac{\lambda}{P_0 + m\cdot (t-1)}$.
}{
Cela signifie que cette limite est équivalente en espérance à un modèle de Gibrat avec $\mu_i(t) = \mu(t)=1 + \frac{\lambda}{P_0 + m\cdot (t-1)}$.
}

\bpar{
For the second moment, we can do a similar computation. We have still \[\Eb{P_i(t)^2} = \Eb{R_i(t)^2}\cdot \Eb{P_i(t-1)^2}\]
and
\[\Eb{P_i(t)^2}=\sum_{k=0}^{+\infty}{k^2 \Pb{P_i(t)=k}}\] 
}{
Pour le second moment, on peut faire un calcul similaire. On a toujours  \[\Eb{P_i(t)^2} = \Eb{R_i(t)^2}\cdot \Eb{P_i(t-1)^2}\]
et
\[\Eb{P_i(t)^2}=\sum_{k=0}^{+\infty}{k^2 \Pb{P_i(t)=k}}\] 
}

\bpar{
We obtain the same way 
}{
On obtient ainsi de la même façon
}

\[
\begin{split}
\Eb{P_i(t)^2} & = \sum_{\delta^{\prime}=0}^{\infty}{\sum_{k^{\prime}=0}^{\infty}{\left(k^\prime + \delta^{\prime}\right)^2\cdot q^{\delta^{\prime}} \cdot \Pb{P_i(t-1)=k^\prime}}}\\ 
& = \sum_{\delta^{\prime}=0}^{\infty}{q^{\delta^{\prime}}\cdot \left(\Eb{P_i(t-1)^2}+2\delta^{\prime}\bar{P}_i(t-1) + {\delta^{\prime}}^2\right)}\\
& = \frac{\Eb{P_i(t-1)^2}}{1-q} + \frac{2 q \bar{P}_i(t-1)}{(1-q)^2} + \frac{q(q+1)}{(1-q)^3}\\
& = \frac{\Eb{P_i(t-1)^2}}{1-q}\left[1 + \frac{q}{\Eb{P_i(t-1)^2}}\left(\frac{2\bar{P}_i(t-1)}{1-q} + \frac{(1+q)}{(1-q)^2}\right)\right]
\end{split}
\]

\bpar{
We have therefore an equivalence between the Gibrat model as a continuous formulation of a Preferential Attachment (or Simon model) in the limit given before. \qed
}{
On a ainsi équivalence entre le modèle de Gibrat et une formulation continue de l'Attachement préférentiel (ou du modèle de Simon) dans la limite donnée ci-dessus. \qed
}

\end{proof}

\subsubsection{Link between Simon and Preferential Attachment}{Lien entre Simon et Attachement Préférentiel}

\bpar{
A rewriting of Simon model yields a particular case of the generalized preferential attachment, in particular by vanishing death probability.
}{
Une reformulation du modèle de Simon le présente comme un cas particulier de l'attachement préférentiel généralisé, en particulier avec la probabilité de mort nulle.
}

\subsubsection{Link between Favaro-Pumain and Gibrat}{Lien entre Favaro-Pumain et Gibrat}

\bpar{
\cite{favaro2011gibrat} generalizes Gibrat models with innovation propagation dynamics. Theoretically, a microsocpic equivalent should be formulated if we consider all models in a typology by ontology and by paradigm. The MMaruis models~\cite{cottineau2014evolution} correspond to a Gibrat paradigm, and should also have their counterpart in terms of microscopic formulation. 
}{
\cite{favaro2011gibrat} généralise le modèle de Gibrat avec les dynamiques de propagation de l'innovation. En théorie, un équivalent microscopique devrait pouvoir être formulé si on considère l'ensemble des modèles dans une typologie par ontologie et par paradigme. 
Les modèles Marius~\cite{cottineau2014evolution} correspondent à un paradigme de Gibrat, et devraient aussi avoir leur contrepartie en termes de formulation microscopique.
}








\stars


\newpage

\section{Sensitivity of urban scaling laws}{Sensibilité des lois d'échelle urbaines}

\label{app:sec:scaling}

\bpar{
At the center of evolutive urban theory are hierarchy and associated scaling laws. We develop here a brief methodological development on the sensitivity of scaling laws to the definition of cities.
}{
Au centre de la théorie évolutive des villes se trouvent la hiérarchie et les lois d'échelle associées. Nous proposons ici un bref développement méthodologique sur la sensibilité des lois d'échelle à la définition de la ville. 
}


\bpar{
Scaling laws have been shown to be universal of urban systems at many scales and for many socio-economic or technical indicators (GDP, education, employment, crime, infrastructure stock, housing stock). Recent studies question however the consistence of scaling exponents determination, as their value can vary significantly depending on thresholds used to define urban entities on which quantities are integrated, even crossing the qualitative border of linear scaling, from infra-linear to supra-linear scaling. We use a simple theoretical model of spatial distribution of densities and urban functions to show analytically that such behavior can be derived as a consequence of the type of spatial distribution and the method used.
}{
Les lois d'échelle ont été montrées universelles des systèmes urbains à de nombreuses échelles et pour différents indicateurs socio-économiques ou techniques (PIB, éducation, emploi, crime, stock d'infrastructures, stock de logement). Des études récentes questionnent toutefois la cohérence de la détermination des exposants d'échelle, puisque leur valeur peut varier significativement selon les seuils utilisés pour définir les entités urbaines sur lesquelles les quantités urbaines sont intégrées, franchissant même dans certains cas la barrière qualitative de l'échelle linéaire, d'une loi infra-linéaire à une loi super-linéaire. Nous utilisons un modèle théorique simple de distribution spatiale des densités et des fonctions urbaines pour montrer analytiquement qu'un tel comportement peut être dérivé comme conséquence du type de distribution spatiale et de la méthode utilisée. 
}

\bpar{
Scaling laws for urban systems, starting from the well-known rank-size Zipf's law for city size distribution~\cite{gabaix1999zipf}, have been shown to be a recurrent feature of urban systems, at many scales and for many types of indicators. They reside in the empirical constatation that indicators computed on elements of an urban system, that can be cities for system of cities, but also smaller entities at a smaller scale, do fit relatively well a power-law distribution as a function of entity size, i.e. that for entity $i$ with population $P_i$, we have for an integrated quantity $A_i$, the relation $A_i \simeq A_0\cdot \left(\frac{P_i}{P_0}\right)^{\alpha}$. Scaling exponent $\alpha$ can be smaller or greater than 1, leading to infra or supralinear effects. Various thematic interpretation of this phenomena have been proposed, typically under the form of processes analysis. The economic literature has produced abundant work on the subject (see~\cite{Gabaix20042341} for a review), but that are generally weakly spatial, thus of poor interest to our approach that deals precisely with spatial organization. Simple economic rules such as energetic equilibria can lead to simple power-laws~\cite{bettencourt2008large} but are difficult to fit empirically. A interesting proposition by \noun{Pumain} is that they are intrinsically due to the evolutionary character of city systems, and that these laws correspond to different maturity levels in innovation cycles with hierarchically propagate within systems of cities~\cite{pumain2006evolutionary}. Although a tempting parallel can be done with self-organizing biological or physical systems, \cite{pumain2012urban} insists on the fact that the ergodicity assumption (see~\ref{sec:staticcorrelations}) for such systems is not reasonable in the case of geographical systems and that the analogy can difficultly be exploited in the case of physical systems. Other explanations have been proposed at other scales, such as the urban growth model at the mesoscopic scale (city scale) given in~\cite{louf2014congestion} which shows that the congestion within transportation networks may be one reason for city shapes and corresponding scaling laws. We can note that ``classical'' urban growth models such as Gibrat model~\cite{favaro2011gibrat} do provide first order approximation of scaling systems, but that interactions between agents have to be incorporated into the model to obtain better fit on real data, such as the Favaro-Pumain model for innovation cycles propagation proposed in~\cite{favaro2011gibrat}, which generalize a Gibrat model for French cities with an ontology similar to Simpop models.
}{
Les lois d'échelle pour les systèmes urbains, en commençant par la bien connue loi rang-taille de Zipf pour la distribution des tailles des villes~\cite{gabaix1999zipf}, sont une caractéristique récurrente des systèmes urbains, à différentes échelles et pour différents types d'indicateurs. Elles reposent sur la constatation empirique que des indicateurs calculés sur des éléments du système urbain, qui peuvent être les villes dans le cas d'un système de villes, mais aussi des entités plus petites à une plus petite échelle, suivent relativement bien une distribution en loi de puissance en fonction de la taille de l'entité, i.e. pour l'entité $i$ avec population $P_i$, on a pour une quantité intégrée $A_i$, la relation $A_i \simeq A_0\cdot \left(\frac{P_i}{P_0}\right)^{\alpha}$. Les exposants d'échelle $\alpha$ peuvent être plus petits ou plus grands que 1, menant à des effets infra ou supra-linéaires. Diverses interprétations thématiques de ce phénomène ont été proposées, typiquement sous la forme d'analyse des processus. La littérature économique contient une production abondante sur le sujet (voir~\cite{Gabaix20042341} pour une revue), mais est généralement faiblement spatiale, donc de faible intérêt pour notre approche qui s'intéresse particulièrement à l'organisation spatiale. Des règles économiques simples comme un équilibre énergétique peut conduire à de simples lois d'échelles~\cite{bettencourt2008large} mais sont difficiles à ajuster empiriquement. Une proposition intéressante par \noun{Pumain} est qu'elles sont intrinsèquement dues au caractère évolutionnaire des systèmes de villes, et que ces lois correspondent à différents niveaux de maturité dans les cycles d'innovation qui se diffusent hiérarchiquement dans les systèmes de villes~\cite{pumain2006evolutionary}. Même si un parallèle tentant peut être fait avec les systèmes biologiques ou physiques auto-organisés, \cite{pumain2012urban} insiste sur le fait que l'hypothèse d'ergodicité (voir~\ref{sec:staticcorrelations}) pour de tels systèmes n'est pas raisonnable dans le cas de systèmes géographiques et que l'analogie est difficilement exploitable dans le cas des systèmes physiques. D'autres explications ont été proposées à d'autres échelles, comme le modèle de croissance urbaine à échelle mesoscopique (échelle de la ville) donné dans~\cite{louf2014congestion} qui montre que la congestion dans les réseaux de transport pourrait être une raison de la forme des villes et des lois d'échelle correspondantes. On peut noter que les modèles ``classiques'' de croissance urbaine comme le modèle de Gibrat~\cite{favaro2011gibrat} fournissent une approximation au premier ordre des systèmes 
 exhibant des lois d'échelles, mais que les interactions entre agents doivent être incorporées dans le modèle pour obtenir un résultat plus fidèle aux données réelles, comme le modèle de Favaro-Pumain pour la propagation des cycles d'innovation proposé dans~\cite{favaro2011gibrat}, qui généralise un modèle de Gibrat pour la croissance des villes françaises avec une ontologie similaire à celle des modèles Simpop.
}



\bpar{
However, the incautious application of scaling exponents computations was recently pointed as misleading in most cases, as~\cite{arcaute2015constructing} shows the variability of computed exponents to the parameters defining urban areas, such as density thresholds. \cite{cottineau2015scaling} studies empirically for France the influence of 3 parameters playing a role in city definition, that are a density threshold $\theta$ to delimitate boundaries of an urban area, a number of commuters threshold $\theta_c$ that is the proportion of commuters going to core area over which the unity is considered belonging to the area, and a cut-off parameter $P_c$ below which entities are not taken into account for the linear regression providing the scaling exponent. Significant results are that exponents can move from infra-linear to supra-linear when threshold varies. A systematic exploration of parameter space produces phase diagrams of exponents for various quantities. One question raising immediately is how these variations can be explained by the features of spatial distribution of variables. Do they result from intrinsic mechanisms present in the system or can they be explained more simply by the fact that the system is spatialized in a particular way? We prove on a toy analytical model that even simple distributions can lead to such significant variations in the exponents, along one dimension of parameters (density threshold), directing the response towards the second explanation.
}{
Cependant, l'application sans vergogne de l'estimation des exposants de lois d'échelle a été récemment rappelé comme pouvant mener à des interprétations divergentes, comme~\cite{arcaute2015constructing} qui montre la variabilité des exposants calculés aux paramètres définissant les aires urbaines françaises, comme le seuil de densité. \cite{cottineau2015scaling} étudie empiriquement pour la France l'influence des 3 paramètres jouant un rôle dans la définition de la ville, qui sont un seuil de densité $\theta$ pour délimiter les limites d'une aire urbaine, un seuil du nombre de navetteurs $\theta_c$ qui correspond à la proportion de ceux-ci devant travailler dans la zone centrale pour que la zone considérée y soit associée, et un paramètre de cut-off $P_c$ en dessous duquel les entités ne sont pas prises en compte pour la régression linéaire fournissant l'exposant d'échelle. Un résultat significatif est que les exposants peuvent varier d'un comportement sous-linéaire à un comportement super-linéaire que les seuils varient. Une exploration systématique de l'espace des paramètres produit les diagrammes de phase des exposants pour diverses quantités. Une question qui est directement soulevée est la manière dont ces variations peuvent être expliquées par les caractéristiques de la distribution spatiale des variables. Résultent-elles de mécanismes intrinsèques au système ou peuvent-elles être expliquées simplement par le fait que le système est spatialisé d'une façon particulière ? Nous montrons avec un exemple analytique simplifié que même des distributions spatiales élémentaires induisent une variation significative des exposants le long d'une dimension des paramètres (seuil de densité), suggérant une réponse positive à la deuxième hypothèse.
}



\bpar{
We derive in the following the expression of the variation of scaling exponents in the simple case of an exponential mixture distribution.
}{
Nous dérivons par la suite l'expression de la variation des exposants d'échelle dans le cas simple d'une distribution en mixture d'exponentielle.
}

\bpar{
We formalize the simple theoretical context in which we will derive the sensitivity of scaling to city definition. Let consider a polycentric city system, which spatial density distributions can be reasonably constructed as the superposition of monocentric fast-decreasing spatial kernels, such as an exponential mixture model~\cite{anas1998urban}. Taking a geographical space as $\mathbb{R}^2$, we take for any $\vec{x}\in\mathbb{R}^2$ the density of population as
}{
Formalisons le contexte théorique simple dans lequel nous dérivons la sensibilité des lois d'échelle à la définition de la ville. Considérons ainsi un système de villes polycentrique, dont la distribution spatiale des densité de population peut raisonnablement être estimé par la superposition de noyaux spatiaux rapidement décroissants, comme par exemple un modèle à mixture d'exponentielles~\cite{anas1998urban}. Prenant l'espace géographique comme $\mathbb{R}^2$, nous prenons pour tout $\vec{x}\in\mathbb{R}^2$ la densité de population comme
}

\[
d(\vec{x}) = \sum_{i=1}^{N}{d_i(\vec{x})} = \sum_{i=1}^{N}{d_i^0\cdot \exp{\left(\frac{-\norm{\vec{x}-\vec{x}_i}}{r_i}\right)}}
\]

\bpar{
where $r_i$ are span parameters of kernels, $d_i^0$ densities at origin points, $\vec{x}_i$ positions of centers. We furthermore assume the following constraints:
}{
où $r_i$ sont les paramètres d'étendue des noyaux, $d_i^0$ les densités aux points d'origine et $\vec{x}_i$ les positions des centres. Nous supposons de plus les contraintes suivantes :
}

\bpar{
\begin{enumerate}
\item To simplify, cities are monocentric, in the sense that for all $i\neq j$, we have $\norm{\vec{x}_i - \vec{x}_j}\gg r_i$.
\item It allows to impose ``structural'' scaling in the urban system by the simple constraint on city populations $P_i$. One can compute by integration that $P_i=2\pi d_i^0 r_i^2$, what gives by injection into the scaling hypothesis $\ln{P_i}=\ln{P_{max}}-\alpha \ln{i}$, the following relation between parameters: $\ln{\left[d_i^0 r_i^2\right]}=K' - \alpha \ln{i}$.
\end{enumerate}
}{
\begin{enumerate}
	\item Pour simplifier, chaque ville est supposée monocentrique, au sens que pour tout $i\neq j$, nous avons $\norm{\vec{x}_i - \vec{x}_j}\gg r_i$.
	\item Cela permet d'imposer une loi d'échelle ``structurelle'' au système urbain en contraignant les populations $P_i$. On obtient immédiatement par intégration que $P_i=2\pi d_i^0 r_i^2$, ce qui donne par insertion dans l'hypothèse de loi d'échelle donnée par $\ln{P_i}=\ln{P_{max}}-\alpha \ln{i}$, la relation suivante entre paramètres : $\ln{\left[d_i^0 r_i^2\right]}=K' - \alpha \ln{i}$.
\end{enumerate}
}

\bpar{
To study scaling relations, we consider a random scalar spatial variable $a(\vec{x})$ representing one aspect of the city, that can be everything but has the dimension of a spatial density, such that the indicator $A(D)=\Eb{\iint_D{a(\vec{x})d\vec{x}}}$ represents the expected quantity of $a$ in area $D$. We make the assumption that $a\in \{0;1\}$ (``counting'' indicator) and that its law is given by $\Pb{a(\vec{x})=1}=f(d(\vec{x}))$. Following the empirical work done in~\cite{cottineau2015scaling}, the integrated indicator on city $i$ as a function of $\theta$ is given by
}{
Pour étudier les relations de lois d'échelles, nous considérons une variable aléatoire scalaire dans l'espace $a(\vec{x})$ représentant un aspect de la ville, qui peut être n'importe lequel mais a la dimension physique d'une densité spatiale, de telle façon que l'indicateur $A(D)=\Eb{\iint_D{a(\vec{x})d\vec{x}}}$ représente la quantité espérée de $a$ dans la zone $D$. Nous faisons l'hypothèse que $a\in \{0;1\}$ (indicateur de comptage) et que sa loi est donnée par $\Pb{a(\vec{x})=1}=f(d(\vec{x}))$. Suivant le travail empirique fait par~\cite{cottineau2015scaling}, l'indicateur intégré sur la ville $i$ en fonction de $\theta$ est donné par
}
\[
A_i(\theta) = A(D(\vec{x}_i, \theta))
\]
\bpar{
where $D(\vec{x}_i, \theta)$ is the area centered in $\vec{x}_i$ where $d(\vec{x})>\theta$. Assumption 1 above ensures that the area are roughly disjoint circles. We take furthermore a simple amenity such that it follows a local scaling law in the sense that $f(d)=\lambda\cdot d^\beta$. It seems a reasonable assumption since it was shown that many urban variables follow a fractal behavior at the intra-urban scale~\cite{keersmaecker2003using} what implies a power-law distribution~\cite{chen2010characterizing}. We make the additional assumption that $r_i=r_0$ does not depend on $i$, what is reasonable if the urban system is considered at a small scale. The estimated scaling exponent $\alpha(\theta)$ is then the result of the log-regression\footnote{We do not situate within the context of a refined estimation of scaling laws, which assumes a cut-off~\cite{newman2005power}.} of $(A_i(\theta))_i$ against $(P_i(\theta))_i$ where $P_i(\theta)=\iint_{D(\vec{x}_i,\theta)}{d}$.
}{
où $D(\vec{x}_i, \theta)$ est la zone centrée en $\vec{x}_i$ telle que $d(\vec{x})>\theta$. L'hypothèse 1 ci-dessus assure que les zones sont des cercles relativement disjoint. Nous considérons de plus une aménité qui suit implement une loi d'échelle locale au sens que $f(d)=\lambda\cdot d^\beta$. Cette hypothèse semble raisonnable puisqu'il a été montré que de nombreuses variables urbaines suivent un comportement fractal à l'échelle intra-urbaine~\cite{keersmaecker2003using} ce qui implique une distribution en loi puissance~\cite{chen2010characterizing}. Nous faisons l'hypothèse supplémentaire que $r_i=r_0$ ne dépend pas de $i$, ce qui est raisonnable si le système urbain est considéré à une petite échelle. L'exposant d'échelle estimé $\alpha(\theta)$ est alors pris comme le résultat de la régression logarithmique\footnote{Nous ne nous plaçons pas dans le cas d'une estimation raffinée des lois d'échelle, qui suppose un cut-off~\cite{newman2005power}.} de $(A_i(\theta))_i$ contre $(P_i(\theta))_i$ où $P_i(\theta)=\iint_{D(\vec{x}_i,\theta)}{d}$.
}

\subsection{Analytical derivation of sensitivity}{Dérivation analytique de la sensibilité}

\bpar{
With above notations, let derive the expression of estimated exponent for quantity $a$ as a function of density threshold parameter $\theta$. The quantity computed for a given city $i$ is, thanks to the monocentric assumption and in a spatial range and a range for $\theta$ such that $\theta \gg \sum_{j\neq i}{d_j(\vec{x})}$, allowing to approximate $d(\vec{x})\simeq d_i(\vec{x})$ on $D(\vec{x}_i,\theta)$, is computed by
}{
Avec les notations précédentes, dérivons l'expression de l'exposant estimé pour la quantité $a$ en fonction du paramètre de seuil de densité $\theta$. La quantité calculée pour une ville donnée $i$ est, grâce à l'hypothèse monocentrique et dans une portée spatiale et des bornes pour $\theta$ telles que $\theta \gg \sum_{j\neq i}{d_j(\vec{x})}$, permettant d'approximer $d(\vec{x})\simeq d_i(\vec{x})$ sur $D(\vec{x}_i,\theta)$, est donnée par
}

\[
\begin{split}
A_i(\theta) & = \lambda\cdot \iint_{D(\vec{x}_i,\theta)}{d^\beta} = 2\pi\lambda {d_i^0}^{\beta} \int_{r=0}^{r_0 \ln{\frac{d_i^0}{\theta}}}{r\exp{\left(-\frac{r\beta}{r_0}\right)}dr}\\
& = \frac{2\pi {d_i^0}^\beta r_0^2}{\beta^2} \left[1 + \beta \ln{\frac{\theta}{d_i^0}\left(\frac{\theta}{d_i^0}\right)^\beta} - \left(\frac{\theta}{d_i^0}\right)^\beta\right]
\end{split}
\]

\bpar{
We obtain in a similar way the expression of $P_i(\theta)$
}{
Nous obtenons de la même manière l'expression de $P_i(\theta)$
}

\[
P_i(\theta) = 2\pi d_i^0 r_0^2 \left[1 + \ln{\left[\frac{\theta}{d_i^0}\right]}\frac{\theta}{d_i^0} - \frac{\theta}{d_i^0}\right]
\]

\bpar{
The Ordinary-Least-Square estimation, solving the problem $\inf_{\alpha,C}\norm{(\ln{A_i(\theta)} - C - \alpha \ln{P_i(\theta)})_i}^2$, gives the value $\alpha(\theta) = \frac{\Covb{(\ln{A_i(\theta)})_i}{(\ln{P_i(\theta)})_i}}{\Varb{(\ln{P_i(\theta)})_i}}$. As we work on city boundaries, the threshold is expected to be significantly smaller than center density, i.e. $\theta / d_i^0 \ll 1$. We can develop the expression in the first order of $\theta / d_i^0$ and use the global scaling law for city sizes, what gives
}{
L'estimation par Moindres-carrés, pour résoudre le problème $\inf_{\alpha,C}\norm{(\ln{A_i(\theta)} - C - \alpha \ln{P_i(\theta)})_i}^2$, donne la valeur $\alpha(\theta) = \frac{\Covb{(\ln{A_i(\theta)})_i}{(\ln{P_i(\theta)})_i}}{\Varb{(\ln{P_i(\theta)})_i}}$. Comme nous travaillons aux limites de la ville, le seuil est supposé être significativement plus petit que la densité au centre, i.e. $\theta / d_i^0 \ll 1$. Nous pouvons développer l'expression au premier ordre en $\theta / d_i^0$ et utiliser la loi d'échelle globale pour les tailles des villes, ce qui donne
}
\[
\ln{A_i(\theta)} \simeq K_A - \alpha \ln{i} + (\beta - 1)\ln{d_i^0} + \beta \ln{\frac{\theta}{d_i^0}\left(\frac{\theta}{d_i^0}\right)^\beta}
\]
\bpar{and}{et}
\[
\ln{P_i(\theta)} = K_P - \alpha \ln{i} + \ln{\left[\frac{\theta}{d_i^0}\right]}\frac{\theta}{d_i^0}
\]
\bpar{
Developing the covariance and variance gives finally an expression of the scaling exponent as a function of $\theta$, where $k_j,{k_j}'$ are constants obtained in the development :
}{
Le développement des variances et covariances donne finalement une expression de l'exposant d'échelle en fonction de $\theta$, où $k_j,{k_j}'$ sont des constantes obtenues dans le développement :
}

\begin{equation*}
\label{eq:th}
\hspace{-2.5cm}
\alpha(\theta) = \frac{k_0 + k_1 \theta + k_2 \theta^\beta + k_3 \theta^{\beta + 1} +  k_4 \theta \ln{\theta} + k_5 \theta^\beta \ln{\theta} + k_6 \theta^\beta (\ln{\theta})^2 + k_7 \theta^{\beta + 1}(\ln{\theta})^2 + k_8 \theta^{\beta + 1}\ln{\theta}}{k_0'+k_1' \ln{\theta} + k_2' \theta \ln{\theta} + k_3' \theta^2 + k_4' \theta^2\ln{\theta} + k_5' \theta^2 (\ln{\theta})^2}
\end{equation*}



\bpar{
This rational fraction in $\theta$ and $\ln\theta$ gives the theoretical expression of the scaling exponent when the threshold varies.
}{
Cette fraction rationnelle en $\theta$ et $\ln\theta$ donne l'expression théorique de l'évolution des exposants d'échelle quand le seuil varie.
}

\newpage

\section{Generation of correlated synthetic data}{Génération de données synthétiques corrélées}

\label{app:sec:syntheticdata}


\bpar{
\textit{This section corresponds to the introduction and the formalization of~\cite{raimbault2016generation}.}
}{
\textit{Cette section correspond à l'introduction et la formalisation de~\cite{raimbault2016generation}.}
}

\stars

\bpar{
Generation of hybrid synthetic data resembling real data to some criteria is an important methodological and thematic issue in most disciplines which study complex systems. Interdependencies between constituting elements, materialized within respective relations, lead to the emergence of macroscopic patterns. Being able to control the dependance structure and level within a synthetic dataset is thus a source of knowledge on system mechanisms. We propose a methodology consisting in the generation of synthetic datasets on which correlation structure is controlled. The method is illustrated on financial time-series in~\ref{app:sec:syntheticdata-finance} and allows to understand the role of interferences between components at different scales on performances of a predictive model. The section~\ref{sec:correlatedsyntheticdata} furthermore proposes an application to a geographical system, in which the weak coupling between a population density model and a network morphogenesis model allows simulating territorial configurations. The intensive exploration of the model unveils a large spectrum of feasible correlations between morphological and network measures. We demonstrate therein the various application possibilities and the potentialities of the method.
}{
La génération de données synthétiques hybrides similaires à des données réelles présente des enjeux méthodologiques et thématiques pour la plupart des disciplines dont l'objet est l'étude de systèmes complexes. Comme l'interdépendance entre les éléments constitutifs d'un système, matérialisée par leur relations, conduit à l'émergence de ses propriétés macroscopiques, une possibilité de contrôle de l'intensité des dépendances dans un jeu de données synthétiques est un instrument de connaissance du comportement du système. Nous proposons une méthodologie de génération de données synthétiques hybrides sur lequel la structure de correlation est contrôlée. La méthode est illustrée sur des séries temporelles financières en~\ref{app:sec:syntheticdata-finance} et permet l'étude de l'interférence entre composantes à différentes fréquences sur la performance d'un modèle prédictif, en fonction des correlations entre composantes à différentes échelles. La section~\ref{sec:correlatedsyntheticdata} propose par ailleurs une application à un système géographique, dans laquelle le couplage faible d'un modèle de distribution de densité de population avec un modèle de génération de réseau permet la simulation de configurations territoriales. L'exploration intensive du modèle permet l'obtention d'un large spectre de valeurs pour la matrice de correlation entre mesures morphologiques et mesures du réseau. On démontre ainsi les possibilités d'applications variées et les potentialités de la méthode.
}

\subsection{Context}{Contexte}

\bpar{
The use of synthetic data, in the sense of statistical populations generated randomly under constraints of patterns proximity to the studied system, is a widely used methodology, and more particularly in disciplines related to complex systems such as therapeutic evaluation~\cite{abadie2010synthetic}, territorial science~\cite{moeckel2003creating,pritchard2009advances}, machine learning~\cite{bolon2013review} or bio-informatics~\cite{van2006syntren}. It can consist in data desegregation by creation of a microscopic population with fixed macroscopic properties, or in the creation of new populations at the same scale than a given sample, with criteria of proximity to the real sample. The level of this criteria will depend on expected applications and can for example vary from a restrictive statistical fit on given indicators, to weaker assumptions of similarity in aggregated patterns, i.e. the existence of similar macroscopic patterns. In the case of chaotic systems, or systems where emergence plays a strong role, a microscopic property does not directly imply given macroscopic patterns, which reproduction is indeed one aim of modeling and simulation practices in complexity science. With the rise of new computational paradigms~\cite{arthur2015complexity}, data (simulated, measured or hybrid) shape our understanding of complex systems. Methodological tools for data-mining and modeling and simulation (including the generation of synthetic data) are therefore crucial to be developed.
}{
L'utilisation de données synthétiques, au sens de populations statistiques d'individus générées aléatoirement sous la contrainte de reproduire certaines caractéristiques du système étudié, est une pratique méthodologique largement répandue dans de nombreuses disciplines, et particulièrement pour des problématiques liées aux systèmes complexes, telles que par exemple l'évaluation thérapeutique~\cite{abadie2010synthetic}, l'étude des systèmes territoriaux~\cite{moeckel2003creating,pritchard2009advances}, l'apprentissage statistique~\cite{bolon2013review} ou la bio-informatique~\cite{van2006syntren}. Il peut s'agir d'une désagrégation par création d'une population au niveau microscopique présentant des caractéristiques macroscopiques données, ou bien de la création de nouvelles populations au même niveau d'agrégation qu'un échantillon donné avec un critère de ressemblance aux données réelles. Le niveau de ce critère dépendra des applications attendues et peut par exemple aller de la fidélité des distributions statistiques pour un certain nombre d'indicateurs à des contraintes plus faibles de valeurs pour des indicateurs agrégés, c'est-à-dire l'existence de motifs macroscopiques similaires. Dans le cas de systèmes chaotiques ou présentant de fortes caractéristiques d'émergence, une contrainte microscopique n'implique pas nécessairement le respect des motifs macroscopiques, et arriver à les reproduire est justement un des enjeux des pratiques de modélisation et simulation en sciences de la complexité. La donnée, qu'elle soit simulée, mesurée ou hybride est au coeur de l'étude des systèmes complexes de par la maturation de nouvelles approches computationelles~\cite{arthur2015complexity}, il est donc essentiel d'étudier des procédures d'extraction d'information des données (fouille de données) et de simulation d'une information similaire (génération de données synthétiques).
}

\bpar{
Whereas first order (in the sense of distribution moments) is generally well used, it is not systematic nor simple to control generated data structure at the second order, i.e. the covariance structure between generated variables. Some specific examples can be found, such as in~\cite{ye2011investigation} where the sensitivity of discrete choices models to the distributions of inputs and to their dependance structure is examined. It is also possible to interpret complex networks generative models~\cite{newman2003structure} as the production of an interdependence structure for a system, contained within link topology. We introduce here a generic method taking into account interdependence structure for the generation of synthetic datasets, under the form of correlations.
}{
Si le premier ordre est de manière générale bien maîtrisé, il n'est pas systématique ni aisé de contrôler le second ordre, c'est-à-dire les structures de covariance entre les variables générées, même si des exemples spécifiques existent, comme dans~\cite{ye2011investigation} où la sensibilité des sorties de modèles de choix discrets à la forme des distributions des variables aléatoires ainsi qu'à leur structures de dépendance. Il est également possible d'interpréter les modèles de génération de réseaux complexes~\cite{newman2003structure} comme la création d'une structure d'interdépendance au sein d'un système, représentée par la topologie des liens. Nous proposons ici une méthode générique prenant en compte l'interdépendance lors de la génération de données synthétiques, sous la forme de correlations.
}

\bpar{
Domain-specific methods aforementioned are too broad to be summarized into a same formalism. We propose here a generic formulation which does not depend on the application domain, centered on the control of correlations structure in synthetic data.
}{
L'ensemble des méthodologies mentionnées en introduction sont trop variées pour être résumées par un même formalisme. Nous proposons ici une formulation générique ne dépendant pas du domaine d'application, ciblée sur le contrôle de la structure de correlation des données synthétiques.
}

\subsection{Formalization}{Formalisation}

\bpar{
Let $\vec{X}_I$ a multidimensional stochastic process (that can be indexed e.g. with time in the case of time-series, but also space, or discrete set abstract indexation). We assume given a real dataset $\mathbf{X}=(X_{i,j})$, interpreted as a set of realizations of the stochastic process. We propose to generate a statistical population $\mathbf{\tilde{X}}=\tilde{X}_{i,j}$ such that
\begin{enumerate}
\item a given criteria of proximity to data is verified, i.e. given a precision $\varepsilon$ and an indicator $f$, we have $\norm{f(\mathbf{X})-f(\mathbf{\tilde{X}})} < \varepsilon$
\item level of correlation is controlled, i.e. given a matrix $R$ fixing the covariance structure, $\Varb{(\tilde{X}_i)} = R$, where the variance/covariance matrix is estimated on the synthetic population.
\end{enumerate}
}{
Soit un processus stochastique multidimensionnel $\vec{X}_I$ (l'ensemble d'indexation pouvant être par exemple le temps dans le cas de séries temporelles, l'espace, ou une indexation quelconque). On se propose, à partir d'un jeu de réalisations $\mathbf{X}=(X_{i,j})$, de générer une population statistique $\mathbf{\tilde{X}}=\tilde{X}_{i,j}$ telle que
\begin{enumerate}
\item d'une part un certain critère de proximité aux données est vérifié, i.e. étant donné une précision $\varepsilon$ et un indicateur $f$, $\norm{f(\mathbf{X})-f(\mathbf{\tilde{X}})} < \varepsilon$
\item d'autre part le niveau de correlation est controlé, i.e. étant donné une matrice fixant une structure de covariance $R$, $\Varb{(\tilde{X}_i)} = R$, où la matrice de variance/covariance est estimée sur la population synthétique.
\end{enumerate}
}

\bpar{
The second requirement will generally be conditional to parameter values determining generation procedure, either generation models being simple or complex. Formally, synthetic processes are parametric families $\tilde{X}_i[\vec{\alpha}]$. We propose to apply the methodology on very different examples, both typical of complex systems: financial high-frequency time-series and territorial systems. We illustrate the flexibility of the method, and claim to help building interdisciplinary bridges by methodology transposition and reasoning analogy. In the first case, proximity to data is the equality of signals at a fundamental frequency, to which higher frequency synthetic components with controlled correlations are superposed. It follows a logic of hybrid data for which hypothesis or model testing is done on a more realistic context than on purely synthetic data. This example is presented in Appendix~\ref{app:sec:syntheticdata-finance}. In the second case, morphological calibration of a population density distribution model allows to respect real data proximity. Correlations of urban form with transportation network measures are empirically obtained by exploration of coupling with a network morphogenesis model. The control is in this case indirect as feasible space is empirically determined.
}{
La satisfaction du deuxième point sera généralement conditionnée par la valeur de paramètres, dont dépendra la procédure de génération, qu'il s'agisse de modèles simples ou complexes. Formellement, les processus synthétiques sont des familles paramétriques $\tilde{X}_i[\vec{\alpha}]$. Nous proposons de décliner cette méthode sur deux exemples très différents mais tous deux typiques des systèmes complexes : des séries temporelles financières à haute fréquence, et les systèmes territoriaux. On illustre ainsi la flexibilité de la logique, ouvrant des portes interdisciplinaires par l'exportation de méthodes ou raisonnements par exemple. Dans le premier cas, la proximité aux données est l'égalité des signaux à une fréquence fondamentale, auxquels on superpose des composantes synthétiques dont il est facile de contrôler le niveau de correlation. On se place dans une logique de données hybrides, pour tester des hypothèses ou modèles dans un contexte plus proche de la réalité que sur des données purement synthétiques. Cet exemple est présenté en Appendice~\ref{app:sec:syntheticdata-finance}. Dans le deuxième cas, la calibration morphologique d'un modèle de distribution de densité de peuplement permet de respecter le critère de proximité aux données. Les correlations de la forme urbaine avec celle d'un réseau de transport sont ensuite obtenues empiriquement par exploration du couplage avec un modèle de génération de réseau. Leur contrôle est dans ce cas indirect puisque constaté empiriquement.
}

\stars

\newpage

\section{Robustness of a multi-attribute evaluation}{Robustesse d'une évaluation multi-attributs}

\label{app:sec:robustness}


\bpar{
Multidimensionality is a fundamental aspect of the behavior of complex systems, in particular in their optimization processes. Most of explorations and calibrations we achieved are multi-objective, but model ontologies often imply agents with multiple objectives. Furthermore, the issue of the sensitivity of models to data has already been evoked in~\ref{sec:computation}. We do here the junction between these two problems by studying the robustness of multi-objective evaluations to data structure, in the particular case of multi-attribute evaluations. This work opens application perspectives to the models we developed, such as for example for the mesoscopic morphogenesis models for which the agents use a multi-attribute utility function for the attribution of new locations.
}{
La multidimensionalité est un aspect fondamental du comportement des systèmes complexes, notamment dans leur processus d'optimisation. La plupart des explorations et calibrations que nous avons mené sont multi-objectif, mais les ontologies des modèles impliquent souvent des agents dont les objectifs sont multiples. Par ailleurs, la question de la sensibilité des modèles aux données a déjà été soulevée en~\ref{sec:computation}. Nous faisons ici la jonction entre ces deux problèmes en étudiant la robustesse d'évaluations multi-objectifs à la structure des données, dans le cas particulier des évaluations multi-attributs. Ce travail ouvre des perspectives d'application aux modèles que nous avons développé, comme par exemple pour les modèles de morphogenèse mesoscopique pour lesquels les agents utilisent une fonction d'utilité multi-attribut pour l'attribution des nouvelles localisations.
}

\stars

\bpar{
\textit{This section has been published in English as~\cite{raimbault2017discrepancy}. It is here adapted.}
}{
\textit{Cette section a été publiée en anglais comme~\cite{raimbault2017discrepancy}. Elle est ici traduite et adaptée.}
}

\stars

\bpar{
Multi-objective evaluation is a necessary aspect when managing complex systems, as the intrinsic complexity of a system is generally closely linked to the potential number of optimization objectives. However, an evaluation makes no sense without its robustness being given (in the sense of its reliability). Statistical robustness computation methods are highly dependent of underlying statistical models. We propose a formulation of a model-independent framework in the case of integrated aggregated indicators (multi-attribute evaluation), that allows to define a relative measure of robustness taking into account data structure and indicator values. We implement and apply it to a synthetic case of urban systems based on Paris districts geography, and to real data for evaluation of income segregation for Greater Paris metropolitan area. First numerical results show the potentialities of this new method. Furthermore, its relative independence to system type and model may position it as an alternative to classical statistical robustness methods.
}{
Les évaluations multi-objectifs sont un aspect essentiel de la gestion de systèmes complexes, puisque la complexité intrinsèque d'un système est généralement étroitement liée au nombre d'objectifs d'optimisation potentiels. Cependant, une évaluation ne fait pas sens si sa robustesse, au sens de sa fiabilité, n'est pas donnée. Les méthodes statistiques usuelles fournissant une mesure de robustesse sont très dépendantes des modèles sous-jacents. Nous proposons une formulation d'un cadre indépendant du modèle, dans le cas d'indicateurs intégrés et agrégés (évaluation multi-attributs), qui permet de définir une mesure de robustesse relative prenant en compte la structure des données et les valeurs des indicateurs. La méthode est testée sur données urbaines synthétiques associées aux arrondissements de Paris, et à des données réelles de revenus pour l'évaluation de la ségrégation urbaine dans la région métropolitaine du Grand Paris. Les premiers résultats numériques montrent les potentialités de cette nouvelle méthode. De plus, sa relative indépendance au type de système et au modèle pourrait la positionner comme une alternative aux méthodes statistiques classiques d'évaluation de la robustesse.
}

\subsection{Introduction}{Introduction}

\subsubsection{General context}{Contexte général}

\bpar{
Multi-objective problems are organically linked to the complexity of underlying systems. Indeed, either in the field of \emph{Complex Industrial Systems}, in the sense of engineered systems, where construction of Systems of Systems (SoS) by coupling and integration often leads to contradictory objectives~\cite{marler2004survey}, or in the field of \emph{Natural Complex Systems}, in the sense of non engineered physical, biological or social systems that exhibit emergence and self-organization properties, where objectives can e.g. be the result of heterogeneous interacting agents (see~\cite{newman2011complex} for a large survey of systems concerned by this approach), multi-objective optimization can be explicitly introduced to study or design the system but is often already implicitly ruling the internal mechanisms of the system. The case of socio-technical Complex Systems is particularly interesting as, following~\cite{haken2003face}, they can be seen as hybrid systems embedding social agents into ``technical artifacts'' (sometimes to an unexpected degree creating what \noun{Picon} describes as \emph{cyborgs}~\cite{picon2013smart}), and thus cumulate propensity to be at the origin of multi-objective issues\footnote{We design by \emph{Multi-Objective Evaluation} all practices including the computation of multiple indicators of a system (it can be multi-objective optimization for system design, multi-objective evaluation of an existing system, multi-attribute evaluation ; our particular framework corresponds to the last case).}. The new notion of \emph{eco-districts}~\cite{souami2012ecoquartiers} is a typical example where sustainability implies contradictory objectives. The example of transportation systems, which conception shifted during the second half of the 20th century from cost-benefit analysis to multi-criteria decision-making, is also typical of such systems~\cite{bavoux2005geographie}. Geographical system are now well studied from such a point of view in particular thanks to the integration of multi-objective frameworks within Geographical Information Systems~\cite{carver1991integrating}. As for the micro-case of eco-districts, meso and macro urban planning and design may be made sustainable through indicators evaluation~\cite{jegou2012evaluation}.
}{
Les problèmes multi-objectifs sont organiquement liés à la complexité des systèmes sous-jacents. En effet, que ce soit dans le champ des \emph{Systèmes Complexes Industriels}, dans le sens de systèmes conçus par ingénierie, où la construction de Systèmes de Systèmes (SoS) par couplage et intégration induit souvent des objectifs contradictoires~\cite{marler2004survey}, ou dans le champ des \emph{Systèmes Complexes Naturels}, au sens de systèmes non désignés, physiques, biologiques ou sociaux, qui présentent des propriétés d'émergence et d'auto-organisation, pour lesquels les objectifs peuvent e.g. être le résultat de l'interaction d'agents hétérogènes (voir~\cite{newman2011complex} pour une revue étendue des types de systèmes concernés par cette approche), l'optimisation multi-objectifs peut être explicitement introduite pour étudier ou désigner le système, mais régit généralement déjà implicitement les mécanismes internes du système. Le cas des Systèmes Complexes Sociaux-techniques est particulièrement intéressant puisque selon Haken~\cite{haken2003face}, ils peuvent être vus comme des systèmes hybrides embarquant des agents sociaux dans des ``artefacts techniques'' (parfois jusqu'à un niveau inattendu, créant ce que \noun{Picon} décrit comme \emph{cyborgs}~\cite{picon2013smart}), et cumulent ainsi la potentialité d'être à l'origine de problèmes multi-objectifs\footnote{Nous désignons ici par \emph{Evaluation Multi-objectifs} toutes les pratiques incluant le calcul de multiples indicateurs d'un système (il peut s'agir d'optimisation multi-objectif pour un design de système, une évaluation multi-objectif d'un système existant, une évaluation multi-attributs ; notre cadre particulier correspondra au dernier cas).}. La notion récente d'\emph{éco-quartier}~\cite{souami2012ecoquartiers} est un exemple typique pour lequel la durabilité implique des objectifs contradictoires. L'exemple des systèmes de transport, dont la conception a glissé durant la seconde moitié du 20ème siècle d'analyses coût-bénéfices à la price de décision multi-critères, est également typique de tels systèmes~\cite{bavoux2005geographie}. Les systèmes géographiques sont à présent bien étudiés d'un tel point de vue, en particulier grâce à l'intégration des cadres multi-objectifs au sein des Systèmes d'Information Géographiques~\cite{carver1991integrating}. Comme dans le cas microscopique des éco-quartiers, la planification et le design urbains mésoscopiques et macroscopiques peuvent être rendus durables grâce aux évaluations par indicateurs~\cite{jegou2012evaluation}.
}

\bpar{
A crucial aspect of an evaluation is a certain notion of its reliability, that we call here \emph{robustness}. 
Statistics naturally include this notion since the construction and estimation of statistical models give diverse indicators of the consistence of results~\cite{launer2014robustness}. The first example that comes to mind is the application of the law of large numbers to obtain the \emph{p-value} of a model fit, that can be interpreted as a confidence measure of estimates. Besides, confidence intervals and \emph{beta-power} are other important indicators of statistical robustness. Bayesian inference provide also measures of robustness when distribution of parameters are sequentially estimated. Concerning multi-objective optimization, in particular through heuristic algorithms (for example genetic algorithms, or operational research solvers), the notion of robustness of a solution concerns more the stability of the solution on the phase space of the corresponding dynamical system. Recent progresses have been done towards unified formulation of robustness for a multi-objective optimization problem, such as~\cite{deb2006introducing} where robust Pareto-front as defined as solutions that are insensitive to small perturbations. In~\cite{1688537}, the notion of degree of robustness is introduced, formalized as a sort of continuity of other solutions in successive neighborhood of a solution.
}{
Un aspect crucial de l'évaluation est une certaine notion de sa fiabilité, que nous nommerons ici \emph{robustesse}. Les méthodes statistiques incluent naturellement cette notion puisque la construction et l'estimation de modèles statistiques donne divers indicateurs de la consistence des résultats~\cite{launer2014robustness}. Le premier exemple venant à l'esprit est l'application de la loi des grands nombres pour obtenir la \emph{p-valeur} d'une estimation de modèle, qui peut être interprété comme une mesure de confiance en les valeurs estimées. D'autre part, les intervalles de confiance et le  \emph{beta-power} sont d'autres indicateurs importants de robustesse statistique. L'inférence bayésienne fournit également des mesures de robustesse quand la distribution des paramètres est estimée de manière séquentielle. Concernant les optimisations multi-objectifs, en particulier par des algorithmes heuristiques (comme par exemple les algorithmes génétiques, ou les solveurs de recherche opérationelle), la notion de robustesse d'une solution consiste plus en la stabilité de la solution dans l'espace des phases du système dynamique correspondant. Des progrès récents ont été faits vers une formulation unifiée de la robustesse pour les problèmes d'optimisation multi-objectifs, comme dans~\cite{deb2006introducing} où les fronts de Pareto robustes sont définis comme des solutions insensibles aux petites perturbations. Dans~\cite{1688537}, la notion de degré de robustesse est introduite, formalisée comme une sorte de continuité des autres solutions dans des voisinages successifs d'une solution.
}

\bpar{
However, there still lack generic methods to estimate robustness of an evaluation that would be model-independent, i.e. that would be extracted from data structure and indicators but that would not depend on the method used. Some advantages could be for example an \emph{a priori} estimation of potential robustness of an evaluation and thus to decide if the evaluation is worth doing. We propose here a framework answering this issue in the particular case of Multi-attribute evaluations, i.e. when the problem is made unidimensional by objectives aggregation. It is data-driven and not model-driven in the sense that robustness estimation does not depend on how indicators are computed, as soon as they respect some assumptions that will be detailed in the following.
}{
Cependant, il n'existe pas de méthode générique qui permettrait une évaluation de la robustesse de façon indépendante au modèle, i.e. qui serait extraite de la structure des données et des indicateurs mais ne dépendrait pas de la méthode utilisée. Un avantage serait par exemple une estimation \emph{a priori} de la robustesse potentielle d'une évaluation et de décider ainsi si elle vaut la peine d'être faite. Nous proposons un cadre répondant à cette contrainte dans le cas particulier des évaluations multi-attributs, i.e. quand le problème est rendu unidimensionnel par agrégation des objectifs. Il est basé sur les données et non sur les modèles, au sens ou l'estimation de la robustesse ne dépendra pas de la manière dont les indicateurs sont calculés, tant qu'ils respectent certaines hypothèses détaillées par la suite.
}

\subsubsection{Proposed approach}{Approche proposée}

\paragraph{Objectives as spatial integrals}{Objectifs comme intégrales spatiales}

\bpar{
We assume that objectives can be expressed as spatial integrals, so it should apply to any territorial system and our application cases are urban systems. It is not that restrictive in terms of possible indicators if one uses suitable variables and integrated kernels : in a way analog to the method of geographically weighted regression~\cite{brunsdon1998geographically}, any spatial variable can be integrated against regular kernels of variable size and the result will be a spatial aggregation which sense depends on kernel size. The example we use in the following such as conditional means or sums suit well the assumption. Even an already spatially aggregated indicator can be interpreted as a spatial indicator by using a Dirac distribution on the centroid of the corresponding area.
}{
Nous supposons que les objectifs peuvent être exprimés comme intégrales spatiales, ce qui devrait s'appliquer à tout système territorial, et nos cas d'application sont des systèmes urbains. Ce n'est pas si restrictif en terme d'indicateurs possibles si l'on utilise les bonnes variables et noyaux intégrés : de façon analogue à la méthode de Regression Géographique Pondérée~\cite{brunsdon1998geographically}, toute variable spatiale peut être intégrée contre des noyaux réguliers de taille variable et le résultats sera une agrégation spatiale dont la signification dépendra de l'étendue du noyau. Les exemples utilisés par la suite comme des moyennes conditionnelles ou des sommes vérifient parfaitement cette hypothèse. Même un indicateur déjà agrégé dans l'espace peut être interprété comme une intégrale spatiale en utilisant une distribution de Dirac au centroïde de la zone correspondante. 
}

\paragraph{Linearly aggregated objectives}{Objectifs agrégés linéairement}

\bpar{
A second assumption we make is that the multi-objective evaluation is done through linear aggregation of objectives, i.e. that we are tackling a multi-attribute optimization problem. If $(q_i(\vec{x}))_i$ are values of objectives functions, then weights $(w_i)_i$ are defined in order to build the aggregated decision-making function $q(\vec{x})=\sum_i{w_i q_i(\vec{x})}$, which value determines then the performance of the solution. It is analog to aggregated utility techniques in economics and is used in many fields. The subtlety lies in the choice of weights, i.e. the shape of the projection function, and various approaches have been developed to find weights depending on the nature of the problem. Recent work~\cite{dobbie2013robustness} proposed to compare robustness of different aggregation techniques through sensitivity analysis, performed by Monte-Carlo simulations on synthetic data. Distribution of biases where obtained for various techniques and some showed to perform significantly better than others. Robustness assessment still depended on models used in that work.
}{
Une seconde hypothèse que nous faisons est que l'évaluation multi-objectifs est effectuée par agrégation linéaire des objectif, c'est-à-dire qu'on se place dans le cadre d'un problème d'optimisation multi-attributs. Si $(q_i(\vec{x}))_i$ sont les valeurs des fonctions objectifs, on définit alors des poids $(w_i)_i$ afin de construire la fonction de prise de décision $q(\vec{x})=\sum_i{w_i q_i(\vec{x})}$, dont la valeur détermine ensuite la performance d'une solution. Cette approche est analogue aux utilités agrégées en économie et est utilisée dans de nombreux domaines. La subtilité réside dans le choix des poids, i.e. de la forme de la fonction de projection, et différentes solutions ont été dévelopées pour obtenir des poids selon la nature du problème. Récemment, \cite{dobbie2013robustness} a proposé de comparer la robustesse des différentes techniques d'agrégation par une analyse de sensibilité, effectuée par simulations de Monte-Carlo pour produire des données synthétiques, ce qui permet d'obtenir la distribution des biais pour les différentes techniques, certaines étant significativement plus performantes que d'autres. Toutefois, la quantification de la robustesse dépend toujours des modèles utilisés dans ce travail.
}


\bpar{
The rest of this section is organized as follows: subsection 2 describes intuitively and mathematically the proposed framework; subsection 3 then details implementation, data collection for case studies and numerical results for an artificial intra-urban case and a metropolitan real case ; subsection 4 finally discuss limitations and potentialities of the method.
}{
Le reste de cette section est organisé de la façon suivante : la sous-section 2 décrit intuitivement puis mathématiquement le cadre proposé ; la sous-section 3 détaille ensuite l'implémentation, la collecte des données pour les cas d'étude et les résultats numériques pour une évaluation intra-urbaine synthétique et un cas réel métropolitain ; la sous-section 4 discute finalement les limitations et les potentialités de la méthode.
}

\subsection{Framework description}{Description du cadre}

\subsubsection{Intuitive description}{Description Intuitive}

\bpar{
We describe now the abstract framework allowing theoretically to compare robustnesses of evaluations of two different urban systems.
 Intuitively, it relies on empirical base resulting from the following axioms:
}{
Nous décrivons à présent le cadre proposé pour permettre théoriquement de comparer la robustesse d'évaluation de deux systèmes urbains différents. 
 Intuitivement, la base empirique se base sur les principes suivants :
}

\bpar{
\begin{itemize}
\item Urban systems can be seen from the information available, i.e. raw data describing the system. As a data-driven approach, this raw data is the basis of our framework and robustness will be determined by its structure.
\item From data are computed indicators (objective functions). We assume that a choice of indicators is an intention to translate particular aspects of the system, i.e. to capture a realization of an ``urban fact'' (\emph{fait urbain}) in the sense of \noun{Mangin}~\cite{mangin1999projet} - a sort of stylized fact in terms of processes and mechanisms, having various realizations on spatially distinct systems, depending on each precise context.
\item Given many systems and associated indicators, a common space can be built to compare them. In that space, data represents more or less well real systems, depending e.g. on initial scale, precision of data, missing data. We precisely propose to capture that through the notion of point cloud discrepancy, which is a mathematical tool coming from sampling theory expressing how a dataset is distributed in the space it is embedded in~\cite{dick2010digital}. 
\end{itemize}
}{
\begin{itemize}
\item Les systèmes urbains peuvent être vus selon l'information disponible, i.e. les données brutes décrivant le système. Dans une approche basée sur les données, celles-ci sont la base de notre cadre et la robustesse sera déterminée par leur structure.
\item A partir des données sont capturés des indicateurs (fonctions objectifs). Nous supposons qu'un choix d'indicateurs est une intention particulière de traduire des aspects particuliers du système, i.e. de capturer une réalisation d'un ``fait urbain'' au sens de \noun{Mangin}~\cite{mangin1999projet} - une sorte de fait stylisé en terme de processus et de mécanismes, ayant différentes réalisations sur des systèmes distincts dans l'espace, dépendant de chaque contexte géographique précis.
\item Etant donné plusieurs systèmes et indicateurs associés, un espace commun peut être construit pour les comparer. Dans cet espace, les données représentent plus ou moins bien le système réel, c'est-à-dire qu'elles sont imprécises en fonction de l'échelle initiale, de la précision effective des données. Nous proposons de capturer exactement ces différents aspects au travers de la notion de discrépance d'un nuage de points, qui est un outil mathématique provenant des théories d'échantillonnage, permettant d'exprimer la façon dont un jeu de données rempli l'espace dans lequel il s'insère~\cite{dick2010digital}.
\end{itemize}
}

\bpar{
Synthesizing these requirements, we propose a notion of \emph{Robustness} of an evaluation that captures both, by combining data reliability with relative importance,
}{
Synthétisant ces contraintes, nous proposons une notion de \emph{Robustesse} d'une évaluation qui capture à la fois, en combinant la fiabilité des données à l'importance relative des indicateurs,
}

\bpar{
\begin{enumerate}
\item \emph{Missing Data} : an evaluation based on more refined datasets will naturally be more robust.
\item \emph{Indicator importance} : indicators with more relative influence will weight more on the total robustness.
\end{enumerate}
}{
\begin{enumerate}
\item \emph{Données manquantes} : une évaluation se basant sur des jeux de données plus raffinés sera naturellement plus robuste.
\item \emph{Importance des indicateurs} : les indicateurs avec plus d'importance relative pèseront plus dans la robustesse totale.
\end{enumerate}
}

\subsubsection{Formal description}{Description formelle}

\paragraph{Indicators}{Indicateurs}

\bpar{
Let $(S_{i})_{1\leq i\leq N}$ be a finite number of geographically disjoints territorial systems, that we assume described through raw data and intermediate indicators, yielding $S_{i}=(\mathbf{X}_{i},\mathbf{Y}_{i})\in\mathcal{X}_{i}\times\mathcal{Y}_{i}$ with $\mathcal{X}_{i}=\prod_{k}\mathcal{X}_{i,k}$ such that each subspace contain real matrices : $\mathcal{X}_{i,k}=\mathbb{R}^{n_{i,k}^{X}p_{i,k}^{X}}$ (the same holding for $\mathcal{Y}_{i}$). We also define an ontological index function $I_{X}(i,k)$ (resp. $I_{Y}(i,k)$) taking integer values which coincide if and only if the two variables have the same ontology in the sense of~\cite{livet2010}, i.e. they are supposed to represent the same real object. We distinguish ``raw data'' $\mathbf{X}_{i}$ from which indicators are computed via explicit deterministic functions, from ``intermediate indicators'' $\mathbf{Y}_{i}$ that are already integrated and can be e.g. outputs of elaborated models simulating some aspects of the urban system. We define the partial characteristic space of the ``urban fact'' by 
}{
Soit $(S_{i})_{1\leq i\leq N}$ un nombre fini de systèmes territoriaux, que nous supposons décrits par les données brutes et des indicateurs intermédiaires, donnés par $S_{i}=(\mathbf{X}_{i},\mathbf{Y}_{i})\in\mathcal{X}_{i}\times\mathcal{Y}_{i}$ avec $\mathcal{X}_{i}=\prod_{k}\mathcal{X}_{i,k}$ tel que chaque sous-espace contient des matrices réelles : $\mathcal{X}_{i,k}=\mathbb{R}^{n_{i,k}^{X}p_{i,k}^{X}}$ (de la même façon pour $\mathcal{Y}_{i}$). Nous définissons également une fonction d'indice ontologique $I_{X}(i,k)$ (resp. $I_{Y}(i,k)$) prenant des valeurs entières qui coincident si et seulement si les deux variables ont même ontologie au sens de~\cite{livet2010}, c'est-à-dire qu'elles sont supposées représenter le même objet réel. On distingue les ``données brutes'' $\mathbf{X}_{i}$ à partir desquelles les indicateurs sont calculés généralement par des fonctions déterministes explicites, 
 des ``indicateurs intermédiaires'' $\mathbf{Y}_{i}$ qui sont déjà intégrés et peuvent être par exemple les sorties de modèles élaborés simulant certains aspects du système urbain. Nous définissons l'espace caractéristique du ``fait urbain'' par
}

\begin{equation}
(\mathcal{X},\mathcal{Y}) \underset{def}{=} \left(\prod\tilde{\mathcal{X}}_{c}\right)\times\left(\prod\tilde{\mathcal{Y}}_{c}\right) = \left(\prod_{\mathcal{X}_{i,k}\in\mathcal{D}_{\mathcal{X}}}\mathbb{R}^{p_{i,k}^{X}}\right)\times\left(\prod_{\mathcal{Y}_{i,k}\in\mathcal{D}_{\mathcal{Y}}}\mathbb{R}^{p_{i,k}^{Y}}\right)
\end{equation}

\bpar{
with $\mathcal{D}_{\mathcal{X}}=\{\mathcal{X}_{i,k}|I(i,k)\textrm{ distincts},n_{i,k}^{X}\mbox{ maximal}\}$
(the same holding for $\mathcal{Y}_{i}$). It is indeed the abstract space on which indicators are integrated. The indices $c$ introduced as a definition here correspond to different indicators across all systems. This space is the minimal space common to all systems allowing a common definition for indicators on each.
}{
avec $\mathcal{D}_{\mathcal{X}}=\{\mathcal{X}_{i,k}|I(i,k)\textrm{ distincts},n_{i,k}^{X}\mbox{ maximal}\}$
(de même pour $\mathcal{Y}_{i}$). Il s'agit en fait de l'espace abstrait sur lequel les indicateurs sont intégrés. Les indices $c$ introduit par définition correspondent aux différents indicateurs au sein des différents systèmes. Cette espace est l'espace minimal commun à tous les systèmes permettant une définition commune des indicateurs pour tous.
}

\bpar{
Let $\mathbf{X}_{i,c}$ be the data canonically projected in the corresponding subspace, well defined for all $i$ and all $c$. We make the key assumption that all indicators are computed by integration against a certain kernel, i.e. that for all $c$, there exists $H_{c}$ space of real-valued functions on $(\tilde{\mathcal{X}}_{c},\tilde{\mathcal{Y}}_{c})$, such that for all $h\in H_{c}$:
}{
Soit $\mathbf{X}_{i,c}$ les données projetées canoniquement sur le sous-espace correspondant, bien définies pour tout $i$ et tout $c$. Nous faisons donc l'hypothèse clé que tous les indicateurs sont calculés par intégration contre un noyau donné, i.e. pour tout $c$ il existe $H_{c}$ espace de fonctions à valeurs réelles sur $(\tilde{\mathcal{X}}_{c},\tilde{\mathcal{Y}}_{c})$, tel que pour tout $h\in H_{c}$ :
}

\bpar{
\begin{enumerate}
\item $h$ is ``enough'' regular (tempered distributions e.g.)
\item $q_c=\int_{(\tilde{\mathcal{X}}_{c},\tilde{\mathcal{Y}}_{c})}h$ is a function describing the ``urban fact'' (the indicator in itself)
\end{enumerate}
}{
\begin{enumerate}
\item $h$ est ``suffisamment'' régulière (distribution tempérée par exemple)
\item $q_c=\int_{(\tilde{\mathcal{X}}_{c},\tilde{\mathcal{Y}}_{c})}h$ est une fonction décrivant le ``fait urbain'' (l'indicateur en lui-même)
\end{enumerate}
}

\bpar{
Typical concrete example of kernels can be:
}{
Des exemples typiques de noyaux peuvent être :
}

\bpar{
\begin{itemize}
\item A mean of rows of $\mathbf{X}_{i,c}$ is computed with $h(x)=x\cdot f_{i,c}(x)$ where $f_{i,c}$ is the density of the distribution of the assumed underlying variable.
\item A rate of elements respecting a given condition $C$, $h(x)=f_{i,c}(x)\chi_{C(x)}$ 
\item For already aggregated variables $\mathbf{Y}$, a Dirac distribution allows to express them also as a kernel integral. 
\end{itemize}
}{
\begin{itemize}
\item Une moyenne des lignes de $\mathbf{X}_{i,c}$ est calculée par $h(x)=x\cdot f_{i,c}(x)$ où $f_{i,c}$ est la densité de la distribution de la variable sous-jacente.
\item Un taux d'éléments du jeu de données respectant une condition donnée $C$, $h(x)=f_{i,c}(x)\chi_{C(x)}$.
\item Pour des variables déjà agrégées $\mathbf{Y}$, une distribution de Dirac permet des les exprimer également comme des intégrales de noyaux.
\end{itemize}
}

\paragraph{Aggregation}{Agrégation}

\bpar{
Weighting objectives in multi-attribute decision-making is indeed the crucial point of the processes, and numerous methods are available (see~\cite{wang2009review} for a review for the particular case of sustainable energy management). Let define weights for the linear aggregation. We assume the indicators normalized, i.e. $h_c \in [0,1]$, for a more simple construction of relative weights. For $i,c$ and $h_{c}\in H_{c}$ given, the weight $w_{i,c}$ is simply constituted by the relative importance of the indicator $w_{i,c}^{L}=\frac{\hat{q}_{i,c}}{\sum_{c}\hat{q}_{i,c}}$ where $\hat{q}_{i,c}$ is an estimator of $q_{c}$ for data $\mathbf{X}_{i,c}$ (i.e. the effectively calculated value). Note that this step can be extended to any sets of weight attributions, by taking for example $\tilde{w}_{i,c} = w_{i,c} \cdot w'_{i,c}$ if $\mathbf{w}'$ are the weights attributed by the decision-maker. We focus here on the relative influence of attributes and thus choose this simple form for weights.
}{
La détermination des poids est en fait le point crucial des processus de prise de décision multi-attributs, et de nombreuses méthodes sont disponibles (voir~\cite{wang2009review} pour une revue dans le cas particulier de la gestion de l'énergie durable). Définissons les poids pour l'agrégation linéaire. Nous supposons les indicateurs normalisés, i.e. $h_c \in [0,1]$, pour une construction plus simple des poids relatifs. 
Pour $i,c$ et $h_{c}\in H_{c}$ donnés, le poids $w_{i,c}$ est simplement constitué par l'importance relative de l'indicateur $w_{i,c}^{L}=\frac{\hat{q}_{i,c}}{\sum_{c}\hat{q}_{i,c}}$ où $\hat{q}_{i,c}$ est un estimateur de $q_{c}$ pour les données $\mathbf{X}_{i,c}$ (i.e. la valeur calculée effectivement). On peut noter que cette étape n'est pas contraignante et que cela peut être étendu à tout ensemble d'attribution de poids, en prenant par exemple $\tilde{w}_{i,c} = w_{i,c} \cdot w'_{i,c}$ si $\mathbf{w}'$ sont les poids fixés par le preneur de décisions. Nous nous concentrons sur l'influence relative des attributs et pour cela choisissons cette forme simple pour les poids. 
}

\paragraph{Robustness estimation}{Estimation de la robustesse}

\bpar{
The scene is now set up to be able to estimate the robustness of the evaluation done through the aggregated function. Therefore, we apply an integral approximation method similar to methods introduced in~\cite{varet2010developpement}, since the integrated form of indicators indeed brings the benefits of such powerful theoretical results. Let $\mathbf{X}_{i,c}=(\vec{X}_{i,c,l})_{1\leq l\leq n_{i,c}}$ and $D_{i,c}=Disc_{\tilde{\mathcal{X}}_{c},L^2}(\mathbf{X}_{i,c})$ the discrepancy of data points cloud\footnote{The discrepancy is defined as the $L2$-norm of local discrepancy which is for normalized data points $\mathbf{X}=(x_{ij})\in \left[0,1\right]^d$, a function of $\mathbf{t}\in \left[0,1\right]^d$ comparing the number of points falling in the corresponding hypercube with its volume, by $disc(\mathbf{t}) = \frac{1}{n}\sum_i \mathbbm{1}_{\prod_j x_{ij}<t_j} - \prod_j t_j$. It is a measure of how the point cloud covers the space.}~\cite{niederreiter1972discrepancy}. With $h\in H_{c}$, we have the upper bound on the integral approximation error
}{
La scène est à présent apprêtée pour construire une estimation de la robustesse d'une évaluation faite par la fonction d'agrégation. Pour cela, nous appliquons un théorème d'approximation d'intégrale similaire au méthodes introduites dans~\cite{varet2010developpement}, puisque la forme intégrée des indicateurs permet justement de bénéficier de tels résultats théoriquement puissant. Soit $\mathbf{X}_{i,c}=(\vec{X}_{i,c,l})_{1\leq l\leq n_{i,c}}$ et $D_{i,c}=Disc_{\tilde{\mathcal{X}}_{c},L^2}(\mathbf{X}_{i,c})$ le discrépance du jeu de données\footnote{La discrépance est définie comme la norme-$L2$ de la discrépance locale qui est pour des points de données normalisés $\mathbf{X}=(x_{ij})\in \left[0,1\right]^d$, une fonction de $\mathbf{t}\in \left[0,1\right]^d$ comparant le nombre de points compris dans le volume de l'hypercube correspondant, donné par $disc(\mathbf{t}) = \frac{1}{n}\sum_i \mathbbm{1}_{\prod_j x_{ij}<t_j} - \prod_j t_j$. C'est une mesure de la manière dont le nuage de points couvre l'espace.}~\cite{niederreiter1972discrepancy}. Avec $h\in H_{c}$, on a la borne supérieure sur l'erreur d'approximation de l'intégrale
}

\[
\left\Vert \int h_{c}-\frac{1}{n_{i,c}}\sum_{l}h_{c}(\vec{X}_{i,c,l})\right\Vert \leq K\cdot\left|\left|\left|h_{c}\right|\right|\right|\cdot D_{i,c}
\]

\bpar{
where $K$ is a constant independent of data points and objective function. It directly yields
}{
où $K$ est une constante indépendante des points de données et des fonctions objectifs. Cela donne directement
}

\[
\left\Vert \int\sum w_{i,c}h_{c}-\frac{1}{n_{i,c}}\sum_{l}w_{i,c}h_{c}(\vec{X}_{i,c,l})\right\Vert \leq K\sum_{c}\left|w_{i,c}\right|\left|\left|\left|h_{c}\right|\right|\right|\cdot D_{i,c}
\]

\bpar{
Assuming the error reasonably realized (``worst case'' scenario for knowledge of the theoretical value of aggregated function), we take this upper bound as an approximation of its magnitude. Furthermore, taking normalized indicators implies $\left|\left|\left|h_c\right|\right|\right| = 1$. We propose then to compare error bounds between two evaluations. They depend only on data distribution (equivalent to \emph{statistical robustness}) and on indicators chosen (sort of \emph{ontological robustness}, i.e. do the indicators have a real sense in the chosen context and do their values make sense), and are a way to combine these two type of robustnesses into a single value.
}{
En supposant l'erreur réalisée de manière raisonnable (scénario du ``pire de cas'' pour la connaissance de la valeur théorique de la fonction agrégée), nous prenons cette borne supérieure comme une approximation de sa magnitude. De plus, la normalisation des indicateurs implique que $\left|\left|\left|h_c\right|\right|\right| = 1$. Nous proposons alors de comparer les bornes d'erreurs entre deux évaluations. Elle dépendent seulement de la distribution des données (équivalence à la \emph{robustesse statistique}) et des indicateurs choisis (sorte de \emph{robustesse ontologique}, i.e. est-ce que les indicateurs ont un sens réel dans le contexte choisi et est-ce que leur valeur fait sens), et sont un moyen de combiner ces deux types de robustesse dans une seule valeur.
}

\bpar{
We thus define a \emph{robustness ratio} to compare the robustness of two evaluations by
}{
Nous définissons ainsi un \emph{ratio de robustesse} pour comparer la robustesse de deux évaluations par 
}

\begin{equation}
R_{i,i'}=\frac{\sum_{c}w_{i,c}\cdot D_{i,c}}{\sum_{c}w_{i',c}\cdot D_{i',c}}
\end{equation}

\bpar{
The intuitive sense of this definition is that one compares robustness of evaluations by comparing the highest error done in each based on data structure and relative importance.
}{
L'interprétation intuitive de cette définition est que l'on compare la robustesse des évaluations en comparant la plus grande erreur faite dans chaque cas selon la structure des données et l'importance relative.
}

\bpar{
By taking then an order relation on evaluations by comparing the position of the ratio to one, it is obvious that we obtain a complete order on all possible evaluations. This ratio should theoretically allow to compare any evaluation of an urban system. To keep an ontological sense to it, it should be used to compare disjoints sub-systems with a reasonable proportion of indicators in common, or the same sub-system with varying indicators. Note that it provides a way to test the influence of indicators on an evaluation by analyzing the sensitivity if the ratio to their removal. On the contrary, finding a ``minimal'' number of indicators each making the ratio strongly vary should be a way to isolate essential parameters ruling the sub-system.
}{
En construisant une relation d'ordre sur les évaluations en comparant la position du ratio par rapport à un, il est clair qu'on obtient un ordre complet sur l'ensemble des évaluations possibles. Ce ratio devrait en théorie permettre de comparer n'importe quelle évaluation d'un système urbain. Afin de garder un sens ontologique à cela, il devrait être utilisé pour comparer des sous-systèmes disjoints avec une proportion raisonnable d'indicateurs en commun, ou le même sous-système avec des indicateurs différents. On peut noter que cela fournit un moyen de tester l'influence des indicateurs sur une évaluation, en analysant la sensibilité du ratio à leur suppression. Au contraire, la détermination d'un nombre ``minimal'' d'indicateurs faisant chacun varier le ratio fortement pourrait être un moyen d'isoler des paramètres essentiels régissant le sous-système.
}

\subsection{Results}{Résultats}


\paragraph{Implementation}{Implémentation}

\bpar{
Preprocessing of geographical data is made through QGIS~\cite{qgis2011quantum} for ergonomy reasons. Core implementation of the framework is done in R~\cite{R-Core-Team:2015fk} for the flexibility of data management and statistical computations. Furthermore, the package \texttt{DiceDesign}~\cite{franco20092} written for numerical experiments and sampling purposes, allows an efficient and direct computation of discrepancies. Last but not least, all source code is openly available on the \texttt{git} repository of the project\footnote{at \url{https://github.com/JusteRaimbault/RobustnessDiscrepancy}} for reproducibility and reuse purposes~\cite{ram2013git}.
}{
Le pré-traitement des données géographiques est fait via QGIS~\cite{qgis2011quantum} pour des raisons d'ergonomie. L'implémentation du coeur est faite en R~\cite{R-Core-Team:2015fk} pour la flexibilité de la gestion des données et du traitement statistique. De plus, le package \texttt{DiceDesign}~\cite{franco20092} conçu pour les expériences numériques et l'échantillonnage, permet un calcul efficient et direct des discrépances. Enfin, tout aussi important, l'ensemble du code source est disponible de manière ouverte sur le dépôt \texttt{git}du projet\footnote{à \url{https://github.com/JusteRaimbault/RobustnessDiscrepancy}} pour permettre la reproductibilité et la réutilisation~\cite{ram2013git}.
}

\subsubsection{Implementation on synthetic data}{Implémentation sur données synthétiques}

\bpar{
We propose in a first time to illustrate the implementation with an application to synthetic data and indicators, for intra-urban quality indicators in the city of Paris.
}{
Nous proposons dans un premier temps d'illustrer l'implémentation par une application à des données et indicateurs synthétiques, pour des indicateurs de qualité de vie intra-urbaine pour la ville de Paris.
}

\paragraph{Data collection}{Collecte des données}

\bpar{
We base our virtual case on real geographical data, in particular for \emph{arrondissements} of Paris. We use open data available through the OpenStreetMap project~\cite{bennett2010openstreetmap} that provides accurate high definition data for many urban features. We use the street network and position of buildings within the city of Paris. 
Limits of \emph{arrondissements}, used to overlay and extract features when working on single districts, are also extracted from the same source. We use centroids of buildings polygons, and segments of street network. Dataset overall consists of around $200k$ building features and $100k$ road segments.
}{
Le cas virtuel se base sur des données géographiques réelles, en particulier pour les arrondissements parisiens. Nous utilisons les données disponibles par le projet OpenStreetMap~\cite{bennett2010openstreetmap} qui fournit déjà des données précises en haute définition pour de nombreux aspects urbains. Nous utilisons le réseau de rues et la position des bâtiments dans la ville de Paris. Les limites des arrondissements, utilisées pour agréger et extraire les features lorsqu'on travaille sur un seul district, sont aussi pris de la même source. Nous utilisons les centroïdes des polygones des bâtiments et les segments du réseau de rues. Le jeu de données brutes consiste d'environ $200k$ bâtiments et $100k$ segments de rues.
}

\paragraph{Virtual case}{Cas Virtuel}

\bpar{
We work on each district of Paris (from the 1st to the 20th) as an evaluated urban system. We construct random synthetic data associated to spatial features, so each district has to be evaluated many time to obtain mean statistical behavior of toy indicators and robustness ratios. The indicators chosen need to be computed on residential and street network spatial data. We implement two mean kernels and a conditional mean to show different examples, linked to environmental sustainability and quality of life, that are required to be maximized. Note that these indicators have a real meaning but no particular reason to be aggregated, they are chosen here for the convenience of the toy model and the generation of synthetic data. With $a\in \{1\ldots 20\}$ the number of the district, $A(a)$ corresponding spatial extent, $b\in B$ building coordinates and $s\in S$ street segments, we take
}{
Nous travaillons sur chaque arrondissement de Paris (du 1er au 20ème) comme un système urbain évalué. Des données synthétiques aléatoires sont associées aux features spatiales, chaque arrondissement pouvant alors être évalué de manière stochastique, et des répétitions permettent d'obtenir le comportement statistique moyen des indicateurs jouets et des ratios de robustesse. Les indicateurs choisis doivent être calculés comme des indicateurs résidentiels et du réseau de rues. Pour montrer différents exemples, nous implémentons deux kernels moyens et une moyenne conditionnelle, tous liés à la durabilité environnementale et la qualité de vie, chacun devant être maximisés. On peut noter que ces indicateurs ont un sens réel mais pas de raison particulière d'être agrégés, ils sont ici choisis pour l'aspect pratique du modèle jouet et de la génération de données synthétiques. Avec $a\in \{1\ldots 20\}$ le nombre d'arrondissements, $A(a)$ l'aire spatiale correspondante à chacun, $b\in B$ les coordonnées des bâtiments et $s\in S$ les segments de rues, nous prenons
}

\bpar{
\begin{itemize}
\item Complementary of the average daily distance to work with car per individual, approximated by, with $n_{cars}(b)$ number of cars in the building (randomly generated by associated of cars to a number of building proportional to motorization rate $\alpha_m ~ 0.4$ in Paris), $d_w$ distance to work of individuals (generated from the building to a uniformly generated random point in spatial extent of the dataset), and $d_{max}$ the diameter of Paris area, $\bar{d}_w = 1 - \frac{1}{|b\in A(a)|} \cdot \sum_{b\in A(a)}{n_{cars}(b)\cdot \frac{d_w}{d_{max}}}$
\item Complementary of average car flows within the streets in the district, approximated by, with $\varphi(s)$ relative flow in street segment $s$, generated through the minimum of 1 and a log-normal distribution adjusted to have $95\%$ of mass smaller than 1 what mimics the hierarchical distribution of street use (corresponding to betweenness centrality), and $l(s)$ segment length, $\bar{\varphi} = 1 - \frac{1}{|s\in A(a)|} \cdot \sum_{s \in A(a)}{\varphi(s)\cdot \frac{l(s)}{\max{(l(s))}}}$
\item Relative length of pedestrian streets $\bar{p}$, computed through a randomly uniformly generated dummy variable adjusted to have a fixed global proportion of segments that are pedestrian.
\end{itemize}
}{
\begin{itemize}
\item Le complémentaire de la distance journalière moyenne au travail en voiture par individu, approché par, avec $n_{cars}(b)$ nombre de voiture dans le bâtiment (généré aléatoirement en associant des voiture à bâtiments proportionnel au taux de motorisation attendu $\alpha_m ~ 0.4$ à Paris), $d_w$ distance des individus à leur travail (généré à partir du bâtiment vers un point aléatoire distribué uniformément dans l'étendue spatiale du jeu de données), et $d_{max}$ le diamètre de l'aire de Paris, $\bar{d}_w = 1 - \frac{1}{|b\in A(a)|} \cdot \sum_{b\in A(a)}{n_{cars}(b)\cdot \frac{d_w}{d_{max}}}$
\item Le complémentaire des flots moyens de voitures des rues dans la zone, approché par, avec $\varphi(s)$ flot relatif dans le segment de rue $s$, généré par le minimum entre 1 et une distribution log-normale ajustée pour avoir $95\%$ de masse plus petite que 1, ce qui mimique la distribution hiérarchique de l'utilisation des rues (qui correspond à la centralité de chemin), et $l(s)$ longueur du segment, $\bar{\varphi} = 1 - \frac{1}{|s\in A(a)|} \cdot \sum_{s \in A(a)}{\varphi(s)\cdot \frac{l(s)}{\max{(l(s))}}}$
\item Longueur relative de rues piétonnes $\bar{p}$, calculé vie une dummy variable aléatoire uniforme ajustée pour obtenir une proportion fixée de segments pédestre.
\end{itemize}
}


\begin{table}[h!]
\hspace{-1cm}
\apptabcaption{Numerical results of simulations for each district with $N=50$ repetitions. Each toy indicator value is given by mean on repetitions and associated standard deviation. Robustness ratio is computed relative to first district (arbitrary choice). A ratio smaller than 1 means that integral bound is smaller for upper district, i.e. that evaluation is more robust for this district.
\label{tab:robustness:paris}}{Résultats numériques des simulations pour chaque arrondissement avec $N=50$ répétitions. Chaque valeur des indicateurs factice est donnée par sa moyenne sur les répétitions et la déviation standard associée. Le ratio de robustesse est calculé par rapport au premier arrondissement (choix arbitraire). Un ratio inférieur à 1 signifie que la borne de l'intégrale est plus petite pour le premier système, i.e. que l'évaluation est plus robuste pour celui-ci.\label{tab:robustness:paris}}
\begin{tabular}[6pt]{|c|c|c|c|c|}\hline
Arrdt & $<\bar{d}_w> \pm \sigma (\bar{d}_w)$ & $<\bar{\varphi}> \pm \sigma (\bar{\varphi})$ & $<\bar{p}> \pm \sigma (\bar{p})$ & $R_{i,1}$ \\[3pt]
\hline
1 th & 0.731655 $\pm$ 0.041099 & 0.917462 $\pm$ 0.026637 & 0.191615 $\pm$ 0.052142 & 1.000000 $\pm$ 0.000000\\[3pt]
\hline
2 th & 0.723225 $\pm$ 0.032539 & 0.844350 $\pm$ 0.036085 & 0.209467 $\pm$ 0.058675 & 1.002098 $\pm$ 0.039972\\[3pt]
\hline
3 th & 0.713716 $\pm$ 0.044789 & 0.797313 $\pm$ 0.057480 & 0.185541 $\pm$ 0.065089 & 0.999341 $\pm$ 0.048825\\[3pt]
\hline
4 th & 0.712394 $\pm$ 0.042897 & 0.861635 $\pm$ 0.030859 & 0.201236 $\pm$ 0.044395 & 0.973045 $\pm$ 0.036993\\[3pt]
\hline
5 th & 0.715557 $\pm$ 0.026328 & 0.894675 $\pm$ 0.020730 & 0.209965 $\pm$ 0.050093 & 0.963466 $\pm$ 0.040722\\[3pt]
\hline
6 th & 0.733249 $\pm$ 0.026890 & 0.875613 $\pm$ 0.029169 & 0.206690 $\pm$ 0.054850 & 0.990676 $\pm$ 0.031666\\[3pt]
\hline
7 th & 0.719775 $\pm$ 0.029072 & 0.891861 $\pm$ 0.026695 & 0.209265 $\pm$ 0.041337 & 0.966103 $\pm$ 0.037132\\[3pt]
\hline
8 th & 0.713602 $\pm$ 0.034423 & 0.931776 $\pm$ 0.015356 & 0.208923 $\pm$ 0.036814 & 0.973975 $\pm$ 0.033809\\[3pt]
\hline
9 th & 0.712441 $\pm$ 0.027587 & 0.910817 $\pm$ 0.015915 & 0.202283 $\pm$ 0.049044 & 0.971889 $\pm$ 0.035381\\[3pt]
\hline
10 th & 0.713072 $\pm$ 0.028918 & 0.881710 $\pm$ 0.021668 & 0.210118 $\pm$ 0.040435 & 0.991036 $\pm$ 0.038942\\[3pt]
\hline
11 th & 0.682905 $\pm$ 0.034225 & 0.875217 $\pm$ 0.019678 & 0.203195 $\pm$ 0.047049 & 0.949828 $\pm$ 0.035122\\[3pt]
\hline
12 th & 0.646328 $\pm$ 0.039668 & 0.920086 $\pm$ 0.019238 & 0.198986 $\pm$ 0.023012 & 0.960192 $\pm$ 0.034854\\[3pt]
\hline
13 th & 0.697512 $\pm$ 0.025461 & 0.890253 $\pm$ 0.022778 & 0.201406 $\pm$ 0.030348 & 0.960534 $\pm$ 0.033730\\[3pt]
\hline
14 th & 0.703224 $\pm$ 0.019900 & 0.902898 $\pm$ 0.019830 & 0.205575 $\pm$ 0.038635 & 0.932755 $\pm$ 0.033616\\[3pt]
\hline
15 th & 0.692050 $\pm$ 0.027536 & 0.891654 $\pm$ 0.018239 & 0.200860 $\pm$ 0.024085 & 0.929006 $\pm$ 0.031675\\[3pt]
\hline
16 th & 0.654609 $\pm$ 0.028141 & 0.928181 $\pm$ 0.013477 & 0.202355 $\pm$ 0.017180 & 0.963143 $\pm$ 0.033232\\[3pt]
\hline
17 th & 0.683020 $\pm$ 0.025644 & 0.890392 $\pm$ 0.023586 & 0.198464 $\pm$ 0.033714 & 0.941025 $\pm$ 0.034951\\[3pt]
\hline
18 th & 0.699170 $\pm$ 0.025487 & 0.911382 $\pm$ 0.027290 & 0.188802 $\pm$ 0.036537 & 0.950874 $\pm$ 0.028669\\[3pt]
\hline
19 th & 0.655108 $\pm$ 0.031857 & 0.884214 $\pm$ 0.027816 & 0.209234 $\pm$ 0.032466 & 0.962966 $\pm$ 0.034187\\[3pt]
\hline
20 th & 0.637446 $\pm$ 0.032562 & 0.873755 $\pm$ 0.036792 & 0.196807 $\pm$ 0.026001 & 0.952410 $\pm$ 0.038702\\[3pt]
\hline
\end{tabular}
\end{table}


\bpar{
As synthetic data are stochastic, we run the computation for each district $N=50$ times, what was a reasonable compromise between statistical convergence and time required for computation. Table~\ref{tab:robustness:paris} shows results (mean and standard deviations) of indicator values and robustness ratio computation. Obtained standard deviation confirm that this number of repetitions give consistent results. Indicators obtained through a fixed ratio show small variability what may a limit of this toy approach. However, we obtain the interesting result that a majority of districts give more robust evaluations than 1st district, what was expected because of the size and content of this district : it is indeed a small one with large administrative buildings, what means less spatial elements and thus a less robust evaluation following our definition of the robustness.
}{
Comme les données synthétiques sont stochastiques, les simulations sont lancées pour chaque quartier $N=50$ fois, ce qui était un compromis raisonnable entre convergence statistique et temps nécessaire au calcul. La table~\ref{tab:robustness:paris} montre les résultats (moyennes et déviations standard) des valeurs des indicateurs et le calcul du ratio de robustesse. Les déviations standard obtenues confirment que ce nombre de simulations donnent des résultats consistants. Les indicateurs obtenus en fixant un ratio fixe montre peu de variabilité, ce qui peut être une limite de cette approche jouet. On obtient toutefois le résultat intéressant que la majorité des arrondissements donne des évaluations plus robustes que le 1er arrondissement, ce qui était attendu par la taille et la fonction de ce quartier: il s'agit en effet d'un petit quartier avec de grand bâtiment administratifs, ce qui implique moins d'éléments spatiaux et pour cela une évaluation moins robuste selon la définition qu'on en a donnée.
}

\subsubsection{Application to a real case: metropolitan segregation}{Application à un cas réel : ségrégation métropolitaine}

\bpar{
The first example was aimed to show potentialities of the method but was purely synthetic, hence yielding no concrete conclusion nor implications for policy. We propose now to apply it to real data for the example of metropolitan segregation.
}{
Le premier exemple avait pour but de montrer les potentialités de la méthode mais était purement synthétique, ne pouvant pour cela fournir pas de conclusion concrete ni d'implications pour la gouvernance. Nous proposons maintenant de l'appliquer à des données réelles dans le cas de la ségrégation métropolitaine.
}

\paragraph{Data}{Données}

\bpar{
We work on income data available for France at an intra-urban level (basic statistical units IRIS) for the year 2011 under the form of summary statistics (deciles if the area is populated enough to ensure anonymity), provided by INSEE\footnote{\texttt{http://www.insee.fr}}. Data are associated with geographical extent of statistical units, allowing computation of spatial analysis indicators. 
}{
Nous travaillons sur les données de revenus, disponible pour la France à un niveau intra-urbain (unités statistiques élémentaires IRIS) pour l'année 2011 sous la forme de résumé statistiques (déciles uniquement si la zone est peuplée suffisamment pour assurer l'anonymat), fournies par l'INSEE\footnote{\texttt{http://www.insee.fr}}. Les données sont associées à l'étendue géographique des unités statistiques, permettant le calcul d'indicateurs d'analyse spatiale.
}

\paragraph{Indicators}{Indicateurs}

\bpar{
We use here three indicators of segregation integrated on a geographical area. Let assume the area divided into covering units $\mathcal{S}_i$ for $1\leq i \leq N$ with centroids $(x_i,y_i)$. Each unit has characteristics of population $P_i$ and median income $X_i$. We define spatial weights used to quantify strength of geographical interactions between units $i,j$, with $d_{ij}$ euclidian distance between centroids : $w_{ij} = \frac{P_i P_j}{\left(\sum_k P_k\right)^2}\cdot \frac{1}{d_{ij}}$ if $i\neq j$ and $w_{ii} = 0$. The normalized indicators are the following
}{
Nous utilisons ici trois indicateurs de ségrégation intégrés sur une zone géographique. Supposons la zone divisée en unités couvrantes $\mathcal{S}_i$ pour $1\leq i \leq N$ avec pour centroïdes $(x_i,y_i)$. Chaque unité a des caractéristiques de population $P_i$ et de revenu médian $X_i$. On définit des poids spatiaux utilisés pour quantifier l'intensité des interactions géographiques entre unités $i,j$, avec $d_{ij}$ distance euclidienne entre centroïdes: $w_{ij} = \frac{P_i P_j}{\left(\sum_k P_k\right)^2}\cdot \frac{1}{d_{ij}}$ si $i\neq j$ 
 et $w_{ii} = 0$. Les indicateurs normalisés sont les suivants
}

\bpar{
\begin{itemize}
\item Spatial autocorrelation Moran index, defined as weighted normalized covariance of median income by $\rho = \frac{N}{\sum_{ij}w_{ij}}\cdot \frac{\sum_{ij}w_{ij}\left(X_i - \bar{X}\right)\left(X_j - \bar{X}\right)}{\sum_i \left(X_i - \bar{X}\right)^2}$
\item Dissimilarity index (close to Moran but integrating local dissimilarities rather than correlations), given by $d =  \frac{1}{\sum_{ij}w_{ij}} \sum_{ij} w_{ij} \left|\tilde{X}_i - \tilde{X}_j\right|$\\ with $\tilde{X}_i = \frac{X_i - \min(X_k)}{\max(X_k) - \min(X_k)}$
\item Complementary of the entropy of income distribution that is a way to capture global inequalities $\varepsilon = 1 + \frac{1}{\log(N)} \sum_i \frac{X_i}{\sum_k X_k} \cdot \log\left(\frac{X_i}{\sum_k X_k}\right)$
\end{itemize}
}{
\begin{itemize}
\item Indice d'autocorrelation spatiale de Moran, défini comme la covariance pondérée normalisée du revenu médian par $\rho = \frac{N}{\sum_{ij}w_{ij}}\cdot \frac{\sum_{ij}w_{ij}\left(X_i - \bar{X}\right)\left(X_j - \bar{X}\right)}{\sum_i \left(X_i - \bar{X}\right)^2}$
\item Indice de dissimilarité (proche du Moran mais intégrant les dissimilarités locales plutôt que les corrélations), donné par $d =  \frac{1}{\sum_{ij}w_{ij}} \sum_{ij} w_{ij} \left|\tilde{X}_i - \tilde{X}_j\right|$\\ avec $\tilde{X}_i = \frac{X_i - \min(X_k)}{\max(X_k) - \min(X_k)}$
\item Le complémentaire de l'entropie de la distribution des revenus, qui est une façon de capturer des inégalités globales $\varepsilon = 1 + \frac{1}{\log(N)} \sum_i \frac{X_i}{\sum_k X_k} \cdot \log\left(\frac{X_i}{\sum_k X_k}\right)$
\end{itemize}
}

\bpar{
Numerous measures of segregation with various meanings and at different scales are available, as for example at the level of the unit by comparison of empirical wage distribution with a theoretical null model~\cite{louf2016patterns}. The choice here is arbitrary in order to illustrate our method with a reasonable number of dimensions.
}{
De nombreuses mesures de ségrégation avec différentes signification à différentes échelles existent, comme par exemple à l'échelle d'une unité spatiale élémentaire par comparaison de la distribution de revenus empirique avec un modèle nul~\cite{louf2016patterns}. Le choix est ici arbitraire, afin d'illustrer la méthode avec un nombre raisonnable de dimensions.
}

\begin{figure}
\includegraphics[width=\linewidth]{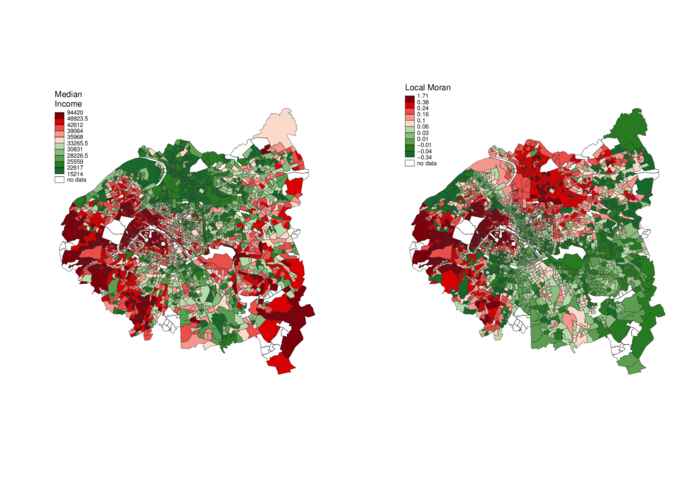}
\appcaption{\textbf{Maps of Metropolitan Segregation.} Maps show yearly median income on basic statistical units (IRIS) for the three departments constituting mainly the Great Paris metropolitan area, and the corresponding local Moran spatial autocorrelation index, defined for unit $i$ as $\rho_i = N/\sum_{j}w_{ij} \cdot \frac{\sum_{j} w_{ij} (X_j - \bar{X})(X_i - \bar{X})}{\sum_i (X_i - \bar{X})^2}$. The most segregated areas coincide with the richest and the poorest, suggesting an increase of segregation in extreme situations.\label{fig:robustness:segreg}}{\textbf{Cartes de ségrégation métropolitaine.} Les cartes montrent le revenu annuel médian pour les unités statistiques élémentaires (IRIS) pour les trois départements correspondant globalement à la métropole du Grand Paris, et l'index local d'autocorrelation spatiale de Moran correspondant, défini pour l'unité $i$ par $\rho_i = N/\sum_{j}w_{ij} \cdot \frac{\sum_{j} w_{ij} (X_j - \bar{X})(X_i - \bar{X})}{\sum_i (X_i - \bar{X})^2}$. Les zones les plus ségréguées coincident avec les plus riches et les plus pauvres, suggérant une augmentation de la ségrégation dans les cas extremes.\label{fig:robustness:segreg}}
\end{figure}

\paragraph{Results}{Résultats}

\bpar{
We apply our method with these indicators on the Greater Paris area, constituted of four \emph{d{\'e}partements} that are intermediate administrative units. The recent creation of a new metropolitan governance system~\cite{gilli2009paris} underlines interrogations on its consistence, and in particular on its relation to intermediate spatial inequalities. We show in Fig.~\ref{fig:robustness:segreg} maps of spatial distribution of median income and corresponding local index of autocorrelation. We observe the well-known West-East opposition and district disparities inside Paris as they were formulated in various studies, such as~\cite{guerois2009dynamique} through the analysis of real estate transactions dynamics. We then apply our framework to answer a concrete question that has implications for urban policy: \textit{how are the evaluation of segregation within different territories sensitive to missing data ?} To do so, we proceed to Monte Carlo simulations (75 repetitions) during which a fixed proportion of data is randomly removed, and the corresponding robustness index is evaluated with renormalized indicators. Simulations are done on each \emph{department} separately, each time relatively to the robustness of the evaluation of full Greater Paris. Results are shown in Fig.~\ref{fig:robustness:sensitivity}. All areas present a slightly better robustness than the reference, what could be explained by local homogeneity and thus more fiable segregation values. Implications for policy that can be drawn are for example direct comparisons between areas : a loss of 30\% of information on 93 area corresponds to a loss of only 25\% in 92 area. The first being a deprived area, the inequality is increased by this relative lower quality of statistical information. The study of standard deviations suggest further investigations as different response regimes to data removal seem to exist.
}{
La méthode est appliquée avec ces indicateurs à la zone du Grand Paris, constitué de 4 département qui sont des niveaux administratifs intermédiaires. La création récente d'un nouveau système de gouvernance métropolitaine~\cite{gilli2009paris} met en évidence des interrogations sur sa pertinence, notamment sur ses capacités d'atténuer les inégalités spatiales. On peut voir en Fig.~\ref{fig:robustness:segreg} les cartes de la distribution spatiale du revenu médian et de l'index local d'autocorrelation spatiale correspondant. La dichotomie bien connue entre est et ouest est retrouvée ainsi que la disparité des quartiers intra-muros, comme cela été présenté par diverses études, comme~\cite{guerois2009dynamique} à travers l'analyse des dynamiques des transactions immobilières. Notre cadre d'étude est ensuite appliqué à une question concrète ayant des implications pour la prise de décision : \textit{dans quelle mesure une évaluation de la ségrégation au sein de différents territoires est sensible aux données manquantes ?} Pour cela, on procède à des simulations de Monte-Carlo (75 répétitions) pour lesquelles une proportion fixe de données est supprimée aléatoirement, et l'indice de robustesse correspondant est évalué avec les indicateurs normalisés. Les simulations sont faites sur chaque département de façon indépendante, à chaque fois pour une robustesse relative à l'évaluation du Grand Paris complet. Les résultats sont présentés en Fig.~\ref{fig:robustness:sensitivity}. Toutes les zones ont une robustesse légèrement meilleure que la référence, ce qui pourrait être expliqué par une homogénéité locale et donc des indices de ségrégation plus fiables. Les implications pour la prise de décision qui peuvent être par exemple tirées sont des comparaisons directes entre les zones : une perte de 30\% de l'information sur le 93 correspond à une perte de seulement 25\% pour le 92. La première zone étant déjà défavorisée socio-économiquement, l'inégalité est augmentée par cette qualité moindre de l'information statistique. L'étude des déviations standard suggère des études plus approfondies comme différents régimes de réponse à la suppression de données semblent exister.
}

\begin{figure}
\includegraphics[width=\linewidth,height=0.85\textheight]{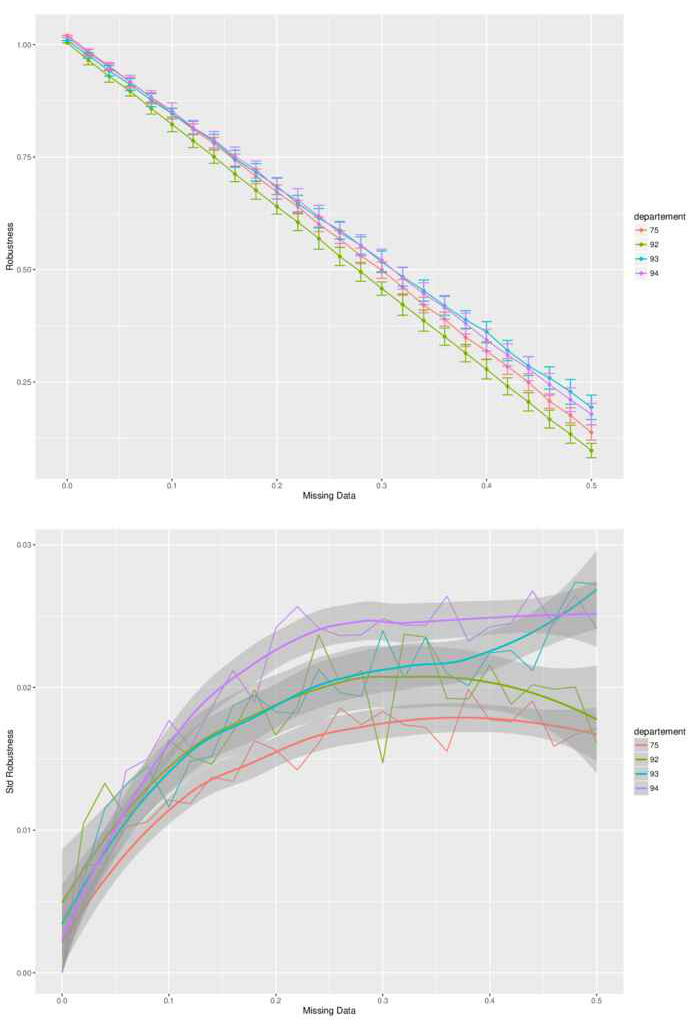}
\appcaption{\textbf{Sensitivity of robustness to missing data.} (\textit{Left}) For each department, Monte Carlo simulations (N=75 repetitions) are used to determine the impact of missing data on robustness of segregation evaluation. Robustness ratios are all computed relatively to full metropolitan area with all available data. Quasi-linear behavior translates an approximative linear decrease of discrepancy as a function of data size. The similar trajectory of poorest departments (93,94) suggest the correction to linear behavior being driven be segregation patterns. (\textit{Right}) Corresponding standard deviations of robustness ratios. Different regimes (in particular 93 against others) unveil phase transitions at different levels of missing data, meaning that the evaluation in 94 is from this point of view more sensitive to missing data.\label{fig:robustness:sensitivity}}{\textbf{Sensibilité de la robustesse aux données manquantes.} (\textit{Gauche}) Pour chaque département, des simulations de Monte-Carlo (N=75 répétitions) sont utilisées pour déterminer l'impact des données manquantes sur la robustesse de l'évaluation de la ségrégation. Les ratios de robustesse sont tous calculés relativement à la région métropolitaine complète avec toutes les données disponibles. Le comportement quasi-linéaire traduit une décroissance approximativement linéaire de la discrépance en fonction de la taille des données. Les trajectoires similaires des départements les plus pauvres (93,94) suggère que la correction au comportement linéaire est fonction des motifs de ségrégation. (\textit{Droite}) Déviations standard des ratios de robustesse. Les différents régimes (en particulier le 93 contre les autres) révèlent des transitions de phase à différents niveaux de données manquantes, signifiant que l'évaluation dans le 94 est de ce point de vue plus sensible aux données manquantes.\label{fig:robustness:sensitivity}}
\end{figure}


\subsection{Discussion}{Discussion}

\subsubsection{Applicability to real situations}{Applicabilité à des situations réelles}

\paragraph{Implications for decision-making}{Implications pour la prise de décision}

\bpar{
The application of our method to concrete decision-making can be thought in different ways. First in the case of a comparative multi-attribute decision process, such as the determination of a transportation corridor, the identification of territories on which the evaluation may be flawed (i.e. has a poor relative robustness) could allow a more refined focus on these and a corresponding revision of datasets or an adapted revision of weights. In any case the overall decision-making process should be made more reliable. A second direction lays in the spirit of the real application we have proposed, i.e. the sensitivity of evaluation to various parameters such as missing data. If a decision appears as reliable because data have few missing points, but the evaluation is very sensitive to it, one will be more careful in the interpretation of results and taking the final decision. Further work and testing will however be needed to understand framework behavior in different contexts and be able to pilot its application in various real situations.
}{
L'application de notre méthode à des situations concrètes de prise de décision peu être pensée de différentes manières. Tout d'abord dans le cas d'un processus multi-attributs à but comparatif, comme la détermination d'un corridor pour une nouvelle infrastructure de transport, l'identification des territoires sur lesquels l'évaluation pourrait être biaisée (i.e. avec une mauvaise robustesse relative) devrait permettre une attention particulière pour ceux-ci, et l'adaptation des jeux de données ou la révision des points en conséquence. Dans tous les cas le processus total devrait être plus fiable. Une autre possibilité ressemble à l'application réelle que nous avons développé, i.e. la sensibilité de l'évaluation à divers paramètres comme les données manquantes. Si une décision parait fiable car la taille de données est grande, mais que l'évaluation est très sensible à la suppression de données, il faudra être prudent pour l'interprétation des résultats et pour la prise de décision finale. Un travail approfondi et de test sera cependant nécessaire pour comprendre le comportement du cadre dans différents contextes et pouvoir piloter son application dans des situations réelles diverses.
}

\paragraph{Integration within existing frameworks}{Intégration au sein de cadres existants}

\bpar{
The applicability of the method on real cases will directly depend on its potential integration within existing framework. Beyond technical difficulties that will surely appear when trying to couple or integrate implementations, more theoretical obstacles could occur, such as fuzzy formulations of functions or data types, consistency issues in databases, etc. Such multi-criteria framework are numerous. Further interesting work would be to attempt integration into an open one, such as e.g. the one described in~\cite{tivadar2014oasis} which calculates various indices of urban segregation, as we have already illustrated the application on metropolitan segregation indexes.
}{
L'applicabilité de la méthode à des cas réels dépendra directement de son intégration potentielle dans des environnements existants. Au delà des difficultés techniques qui apparaissent nécessairement en essayant de coupler ou d'intégrer des implémentations existantes, des obstacles plus théoriques pourraient émerger, comme des formulations floues des fonctions ou des types de données, la cohérence des bases de données, etc. De tels cadres multi-critères sont nombreux. Un développement possible serait l'intégration dans un cadre open-source, comme par exemple celui décrit dans~\cite{tivadar2014oasis} qui calcule divers indices de ségrégation urbaine, comme on l'a déjà illustré pour l'application à la ségrégation métropolitaine.
}

\paragraph{Availability of raw data}{Disponibilité des données brutes}

\bpar{
In general, sensitive data such as transportation questionnaires, or very fine granularity census data are not openly available but provided already aggregated at a certain level (for instance French Insee Data are publicly available at basic statistical unit level or larger areas depending on variables and minimal population constraints, more precise data is under restricted access). It means that applying the framework may imply complicated data research procedure, its advantage to be flexible being thus reduced through additional constraints.
}{
De manière générale, des données sensibles comme des questionnaires de transport, ou des données de sondage à granularité très fine, ne sont pas disponibles de manière ouverte, mais fournis de manière déjà agrégée à un certain niveau (comme par exemple les données françaises de l'Insee sont disponibles publiquement au niveau des unités statistiques élémentaires ou pour des zones plus grandes selon les variables et des contraintes de population minimale, les données plus précises étant à accès restreint). Cela signifie que l'application de notre cadre peut impliquer une procédure de recherche de données laborieuse, l'avantage d'être flexible étant alors compensé par ces contraintes additionnelles.
}

\subsubsection{Validity of theoretical assumptions}{Validité des hypothèses théoriques}

\bpar{
A possible limitation of our approach is the validity of the assumption formulating indicators as spatial integrals. Indeed, many socio-economic indicators are not necessarily depending explicitly on space, and trying to associate them with spatial coordinates may become a slippery slope (e.g. associate individual economic variables with individual residential coordinates will have a sense only if the use of the variable has a relation with space, otherwise it is a non-legitimate artifact). Even indicators which have a spatial value may derive from non-spatial variables, as~\cite{kwan1998space} points out concerning accessibility, when opposing integrated accessibility measures with individual-based non necessarily spatial-based (e.g. individual decisions) measures. Constraining a theoretical representation of a system to fit a framework by changing some of its ontological properties (always in the sense of real meaning of objects) can be understood as a violation of a fundamental rule of modeling and simulation in social science given in~\cite{banos2013pour}, that is that there can be an universal ``language'' for modeling and some can not express some systems, having for consequence misleading conclusion due to ontology breaking in the case of an over-constrained formulation.
}{
Une limitation possible de notre approche est la validité de l'hypothèse qui formule les indicateurs comme des intégrales spatiales. En fait, de nombreux indicateurs socio-économiques ne dépendent pas nécessairement directement de l'espace, et essayer de les associer à des coordonnées peut entraîner sur une pente glissante (par exemple, associer des variables économiques individuelles à des coordonnées résidentielles aura un sens seulement si la variable à une relation à l'espace, autrement un devient un artefact superflu). Même des indicateurs qui ont une valeur spatiale peuvent dériver de variables non-spatiales, comme~\cite{kwan1998space} le souligne au sujet de l'accessibilité, en opposant les mesures d'accessibilité intégrée aux mesures individu-centrées mais pas forcément basée sur l'espace (comme par exemple des décisions individuelles). Contraindre une représentation théorique d'un système pour le faire rentrer dans un cadre en changeant certaines de ses propriétés ontologiques (toujours dans le sens de la signification réelle des objets) peut être compris comme une violation d'une des règles pour la modélisation et la simulation en sciences sociales données par~\cite{banos2013pour}, car cela impliquerait qu'il pourrait exister un langage universel pour la modélisation, malgré qu'il ne puisse retranscrire certains systèmes, ayant pour conséquences des conclusions errantes à cause d'une rupture d'ontologie dans le cas d'une formulation sur-contrainte.
}

\subsubsection{Framework generality}{Généralité du cadre}

\bpar{
We argue that the fundamental advantage of the proposed framework is its generality and flexibility, since robustness of the evaluations are obtained only through data structure if ones relaxes constraints on the value of weight. Further work should go towards a more general formulation, suppressing for example the linear aggregation assumption. Non-linear aggregation functions would require however to present particular properties regarding integral inequalities. For example, similar results could search in the direction of integral inequalities for Lipschitzian functions such as the one-dimensional results of~\cite{dragomir1999ostrowski}.
}{
Nous soutenons qu'un des avantages fondamentaux de notre cadre est sa généralité et sa flexibilité, puisque la robustesse des évaluations est obtenue seulement par la structure des données si l'on relaxe les hypothèses sur les valeurs des poids. Des approfondissement pourraient inclure une formulation plus générale, en supprimant par exemple l'hypothèse d'agrégation linéaire. Des fonctions d'agrégation non-linéaires demanderaient toutefois de vérifier certaines propriétés regardant les inégalités intégrales. Par exemple, des résultats similaires pourraient être obtenus en s'orientant vers des inégalités intégrales pour fonctions Lipschitziennes, comme les résultats en une dimension de~\cite{dragomir1999ostrowski}.
}

\subsection{Conclusion}{Conclusion}

\bpar{
We have proposed a model-independent framework to compare the robustness of multi-attribute evaluations between different urban systems. Based on data discrepancy, it provide a general definition of relative robustness without any assumption on model for the system, but with limiting assumptions that are the need of linear aggregation and of indicators being expressed through spatial kernel integrals. We propose a toy implementation based on real data for the city of Paris, numerical results confirming general expected behavior, and an implementation on real data for income segregation on Greater Paris metropolitan areas, giving possible insights into concrete policy questions. Further work should be oriented towards sensitivity analysis of the method, application to other real cases and theoretical assumptions relaxation, i.e. the relaxation of linear aggregation and spatial integration.
}{
Nous avons proposé un cadre indépendant du modèle pour comparer la robustesse d'évaluations multi-attributs entre différents systèmes urbains. A partir de la discrépance des données, on fournit une définition générale de la robustesse relative sans aucune hypothèse de modèle pour le système, mais en supposant une agrégation linéaire des objectifs et des indicateurs exprimés comme des intégrales à noyaux. Nous proposons une première implémentation preuve de concept pour la ville de Paris pour laquelle les résultats numériques confirment la tendance générale attendue, et une implémentation sur des données réelles pour la ségrégation de revenus pour la région métropolitaine du Grand Paris, fournissant des réponses possibles à des questions de prise de décision plus concrètes. Des développements possibles peuvent inclure une analyse de sensibilité de la méthode, des applications à d'autres cas réels et une relaxation des hypothèses théoriques, c'est-à-dire de l'agrégation linéaire et de la formulation comme intégrale spatiale.
}


\stars



%


\newpage

\section{A framework for socio-technical systems}{Un cadre pour les systèmes socio-techniques}

\label{app:sec:csframework}


%



\bpar{
We develop here a formal framework for the modeling of socio-technical systems. More precisely, it implements the idea of applied perspectivism to unveil a possible structure of perspective coupling operations. It can furthermore also be understood as a preliminary work for the formalization of the knowledge framework suggested in~\ref{sec:knowledgeframework} (still without including the algebraic structure for the operation on data).
}{
Nous développons ici un cadre formel pour la modélisation des systèmes socio-techniques. Plus précisément, celui-ci implémente l'idée de perspectivisme appliqué pour éclairer une structure possible des opérations de couplages de perspectives. Il peut par ailleurs également être compris comme un travail préliminaire pour la formalisation du cadre de connaissance suggérée en~\ref{sec:knowledgeframework} (sans encore inclure la structure algébrique pour l'opération sur les données).
}

\subsection{Context}{Contexte}

\subsubsection{Scientific Context}{Contexte Scientifique}

\bpar{
The structural misunderstandings between Social Sciences and Humanities on one side, and so-called Exact Sciences on the other side, far from being a generality, seems to have however a significant impact on the structure of scientific knowledge: \cite{2015arXiv151103981H} show how sociology and physics have developed very similar network analysis methods with a very low inter-fertilization. These can be due to the epistemological divergences which themselves are consequences of fundamental differences in the objects studied: humans are naturally not particles. In particular, as we develop here different theoretical frameworks, it is important to focus on the role of theory. Theory, and in fact the signification of the term itself, has a totally different role in the construction of knowledge, partly because of different \emph{perceived complexities}\footnote{We used the term \emph{perceived} as most of systems studied by physics might be described as simple whereas they are intrinsically complex and indeed not well understood~\cite{laughlin2006different}.} of studied objects. For example, mathematical constructions and by extent theoretical physics are \emph{simple} in the sense that they are generally analytically solvable (or at least semi-analytically)\footnote{We take the position here that analytically solvable implies simplicity, since the system then does not exhibit weak emergence (see~\ref{sec:epistemology}).}, whereas Social Science subjects such as humans or society (to give a \emph{clich{\'e}} exemple) are \emph{complex} in the sense of complex systems. This implies a stronger need of a constructed theoretical (generally empirically based) framework to identify and define the objects of research that are necessarily more arbitrary in the framing of their boundaries, relations and processes, because of the multitude of possible viewpoints: \noun{Pumain} suggests indeed in~\cite{pumain2005cumulativite} a new approach to complexity deeply rooted in social sciences that ``would be measured by the diversity of disciplines needed to elaborate a notion''. These differences in backgrounds are naturally desirable in the spectrum of science, but things can get difficult when playing on overlapping terrains, typically complex systems problematics as already detailed, as the exemple of geographical urban systems has recently shown~\cite{dupuy2015sciences}. Complex System Science\footnote{That we deliberately call that way although there is a running debate on wether it can be seen as a science in itself or more as a different way to do science.} is presented by some as a ``new kind of Science''~\cite{wolfram2002new}, and would at least be a symptom of a shift in scientific practices, from analytical and ``exact'' approaches to computational and evidence-based approaches~\cite{arthur2015complexity}, but what is sure is that it brings, together with new methodologies, new scientific fields in the sense of converging interests of various disciplines on transversal questions or of integrated approaches on a particular field~\cite{2009arXiv0907.2221B}. Our work particularly enters that context and would have no sense if it would be disconnected from these aspects, precisely the computational ones (see~\ref{sec:computation}).
}{
Les malentendus structurels entre les Sciences Sociales et Humanités d'une part, et les dénommées Sciences Exactes d'autre part, comme celui maintes fois évoqué déjà entre physiciens et géographes, loin d'être une règle nécessaire, semble toutefois avoir un impact conséquent sur la structure de la connaissance scientifique : \cite{2015arXiv151103981H} montre comment la sociologie et la physique ont développé des méthodes d'analyse de réseau très similaire avec une inter-fertilisation faible. Ceux-ci peuvent être dus aux divergences épistémologiques qui elles-mêmes découlent de différences fondamentales dans les objets étudiés : les humains ne sont bien sûr pas des particules. Plus particulièrement, comme nous développons ici différents cadres théoriques, il est important de s'intéresser au rôle de la théorie. La théorie, et en fait la signification elle-même du terme, a une place complètement différente dans l'élaboration de la connaissance, en partie à cause de différentes \emph{complexités perçues}\footnote{Nous utilisons le terme \emph{perçu} car la plupart des systèmes étudiés en physique peuvent être décrits comme simple alors qu'ils sont intrinsèquement complexe et finalement mal compris~\cite{laughlin2006different}.} des objets étudiés. Par exemple, de nombreuses constructions mathématiques et par extension certaines en physique théorique sont \emph{simples} au sens où elles sont résolubles de manière analytique (ou au moins semi-analytique)\footnote{Nous prenons ici le parti que soluble analytiquement implique la simplicité, puisque le système n'exhibe alors pas d'émergence faible (voir~\ref{sec:epistemology}).}, tandis que les sujets des Sciences Sociales tels les humains ou la société (pour prendre un exemple préconçu) sont \emph{complexes} au sens de systèmes complexes. Cela implique un besoin accru d'une construction théorique (qui se base généralement sur l'empirique) pour identifier et définir qui sont nécessairement plus arbitraires dans la définition de leur limites, relations et processus, de par la multitude des points de vue possibles : \noun{Pumain} suggère en effet dans~\cite{pumain2005cumulativite} une nouvelle approche de la complexité qui serait profondément ancrée dans les sciences sociales et qui serait ``mesurée par la diversité des disciplines nécessaires pour élaborer une notion''. Ces différences de fond sont naturellement bénéfiques pour la diversité scientifique, mais les choses peuvent se corser quand les terrains d'étude se chevauchent, typiquement dans le cas de problématiques liées aux systèmes complexes comme déjà détaillé, comme l'exemple géographique des systèmes urbains a récemment montré~\cite{dupuy2015sciences}. La Science des Systèmes Complexes\footnote{Que nous appelons délibérément ainsi même si des débats existent sur le fait de considérer comme une science en elle-même ou comme une façon différente de faire de la science.} est présentée par certains comme ``un nouveau type de science''~\cite{wolfram2002new}, et serait au moins symptomatique d'un changement de paradigme des pratiques, des approches analytiques ``exactes'' vers des approches computationnelles et \emph{evidence-based}~\cite{arthur2015complexity}, mais il est certain que cela permet de faire émerger, conjointement avec de nouvelles méthodologies, des nouveaux champs scientifiques au sens d'intérêts convergents de disciplines variées sur des questions transversales ou d'approches intégrées d'un champ particulier~\cite{2009arXiv0907.2221B}. Notre travail s'ancre particulièrement dans ce cadre et n'aurait pas de sens s'il était déconnecté de ces aspects notamment computationnels (voir~\ref{sec:computation}).
}

\subsubsection{Objectives}{Objectifs}

\bpar{
Within that scientific context, the study of what we will call \emph{Socio-technical Systems}, which we define in a rather broad way as hybrid complex systems including social agents or objects that interact with technical artifacts and/or a natural environment\footnote{geographical systems in the sense of \cite{dollfus1975some} are the archetype of such systems, but that definition may cover other type of systems such as an extended transportation system, social systems taken with an environmental context, complicated industrial systems taken with users, etc.}, lies precisely between social sciences and hard sciences. The example of urban systems is the best example, as already before the arrival of approaches claiming to be ``more exact'' than soft approaches (typically by physicists, see e.g. the positioning of~\cite{louf2014scaling}, but also by scientists coming from social sciences such as \noun{Batty}~\cite{batty2013new}), many diverse aspects of urban systems were already in the field of exact sciences, among which we can cite without any particular hierarchy, urban hydrology, urban climatology or technical aspects of transportation systems, whereas the core of their study relied in social sciences such as geography, urbanism, sociology, economy. Therefore a necessary place of theory in their study, given its role as knowledge domain for the knowledge of complex systems (see the framework introduced in~\ref{sec:knowledgeframework}).
}{
Dans ce contexte scientifique, l'étude de ce que nous désignons par \emph{Systèmes socio-techniques}, que nous définissons de manière assez large comme des systèmes complexes hybrides qui incluent des agents ou objets sociaux qui interagissent avec des artefacts techniques et/ou un environnement naturel\footnote{les systèmes géographiques au sens de \cite{dollfus1975some} sont l'archetype de tels systèmes, mais cette définition peut couvrir d'autres types de systèmes comme un système de transport étendu, des systèmes sociaux pris dans un contexte environnemental, des systèmes industriels compliqués considérés avec leur utilisateurs, etc.}, se situent précisément entre sciences sociales et sciences dures. L'exemple des systèmes urbains est relativement représentatif, puisque même avant l'arrivée de nouvelles approches prétendant être ``plus exactes'' que les approches des sciences sociales (typiquement par des physiciens, voir e.g. le positionnement de~\cite{louf2014scaling}, mais aussi par des chercheurs venant des sciences sociales comme \noun{Batty}~\cite{batty2013new}), une multitude d'aspects de l'étude des systèmes urbains étaient déjà traités dans des sciences dures très diverse, parmi lesquelles on peut citer sans hiérarchie particulière, l'hydrologie urbaine, la climatologie urbaine ou les aspects techniques des systèmes de transport, tandis que le centre de leur attention se reposait sur des sciences sociales comme la géographie, l'urbanisme, la sociologie, l'économie. D'où une place nécessaire de la théorie dans leur étude, vu son rôle comme domaine de connaissance pour la connaissance des systèmes complexes (voir le cadre introduit en~\ref{sec:knowledgeframework}).
}

\bpar{
We propose in this section to construct a theory, or rather a theoretical framework, that would ease some aspects of the study of such systems. Many theories already exist in all fields related to this kind of problems, and also at higher levels of abstraction concerning methods such as agent-based modeling e.g., but there is to our knowledge no theoretical framework including all of the following aspects that we consider as being crucial (and that can be understood as an informal basis of our theory):
\begin{enumerate}
\item a precise definition and emphasis on the notion of coupling between subsystems, in particular allowing to qualify or quantify a certain degree of coupling: dependence, interdependence, etc. between components.
\item a precise definition of scale, including timescale and scales for other dimensions.
\item as a consequence of the previous points, a precise definition of what is a system.
\item the inclusion of the notion of emergence in order to capture multi-scale aspects of systems. 
\item a central place of ontology in the definition of systems, i.e. of the sense in the real world given to its objects\footnote{as already explained before, this positioning along with the importance of structure may be related to Ontic Structural Realism~\cite{frigg2011everything} in further developments.}.
\item taking into account heterogeneous aspects of the same system, that could be heterogeneous components but also complementary intersecting views.
\end{enumerate}
}{
Nous proposons dans cette section de construire une théorie, ou plutôt un cadre théorique, pour faciliter certains aspects de l'étude de tels systèmes. De nombreuses théories existent déjà dans l'ensemble des champs liés à ce type de questionnement, et aussi à de plus haut niveaux d'abstraction concernant des méthodes comme e.g. la modélisation basée agent, mais il n'existe à notre connaissance pas de cadre théorique qui incluraient l'ensemble des points suivants que nous jugeons cruciaux (et qui peuvent être compris comme une base informelle de notre théorie) :
\begin{enumerate}
\item une définition précise et une emphase particulière sur la notion de couplage entre sous-systèmes, en particulier permettant de qualifier ou quantifier un certain niveau de couplage : dépendance, interdépendance, etc. entre composantes.
\item une précise définition de l'échelle, incluant l'échelle temporelle et l'échelle pour d'autres dimensions.
\item en conséquence des points précédents, une définition précise de ce qu'est un système.
\item la prise en compte de la notion d'émergence pour capturer les aspects multi-scalaires des systèmes.
\item une place centrale de l'ontologie dans la définition des systèmes, i.e. du sens dans le monde réel donné à ses objets\footnote{comme déjà expliqué précédemment, ce positionnement combiné à l'importance de la structure pourrait être relié au \emph{Réalisme Structurel Ontologique}~\cite{frigg2011everything} dans des approfondissements.}.
\item la prise en compte d'aspects hétérogènes du même système, qui peuvent être des composantes hétérogènes mais aussi différents points de vue sur le système qui se complètent.
\end{enumerate}
}

\bpar{
The rest of this section is organized as follows: we construct the theory in the following subsection, staying at an abstract level, and propose a first application to the question of co-evolving subsystems. We then discuss positioning regarding existing theories, and possible developments and concrete applications.
}{
La suite de cette section est organisée de la façon suivante : nous construisons la théorie dans la sous-section suivante en restant à un niveau abstrait, et proposons une première application à la question des sous-systèmes co-évolutifs. Nous discutons ensuite le positionnement au regard de théories existantes, ainsi que les développements possibles et des applications concrètes.
}

\subsection{Construction of the theory}{Construction de la Théorie}

\subsubsection{Perspectives and ontologies}{Perspectives et ontologies}

\bpar{
The starting point of the theory construction is a perspectivist epistemological approach on systems introduced by \noun{Giere}~\cite{giere2010scientific}. To summarize, it interprets any scientific approach as a perspective, in which someone pursues some objective and uses what is called \emph{a model} to reach it. The model is nothing more than a scientific medium. \noun{Varenne} developed~\cite{varenne2010framework} a functional model typology that can be interpreted as a refinement of this theory. Let for now relax this possible precision and use perspectives as proxies of the undefined objects and concepts. Indeed, different views on the same object (being complementary or diverging) have the property to share at least the object in itself, thus the proposition to define objects (and more generally systems) from a set of perspectives on them, that verify some properties that we formalize in the following.
}{
Le point de départ pour construire la théorie est une approche épistémologique perspectiviste des systèmes introduite par \noun{Giere}~\cite{giere2010scientific}. Pour résumer, cette position interprète toute démarche scientifique comme une perspective, au sein de laquelle chacun poursuit certains objectifs et utilise ce qui est appelé \emph{un modèle} pour les atteindre. Le modèle n'est alors rien de plus qu'un medium scientifique. \noun{Varenne} a développé~\cite{varenne2010framework} une typologie fonctionnelle des modèles qui peut être interprété comme un raffinement de cette théorie. Relâchons dans un premier temps cette précision potentielle et utilisons les perspectives comme des approximations des objets et concepts indéfinis. En effet, diverses visions du même objet (pouvant être complémentaires ou divergentes) ont la propriété de partager au moins l'objet lui-même, d'où notre proposition de définir les objets (et plus généralement les systèmes) à partir d'un ensemble de perspectives sur ceux-ci, qui vérifient certaines propriétés que nous formalisons par la suite.
}


\bpar{
A perspective is defined in our case as a dataflow machine $M$ (that corresponds to the model as medium) in the sense of~\cite{golden2012modeling} that gives a convenient way to represent it and to introduce timescales and data, to which is associated an ontology $O$ in the sense of~\cite{livet2010}, i.e. a set of elements each corresponds to an entity (which can be an object, an agent, a process, etc.) of the real world. We include only two aspect (the model and the objects represented) of Giere's theory, making the assumption that purpose and producer of the perspective are indeed contained in the ontology if they make sense for studying the system.
}{
Une perspective est définie dans notre cas comme une \emph{Dataflow Machine} $M$ au sens de~\cite{golden2012modeling}, que nous considérons comme une boîte noire transformant un flux de données d'entrée en flux de sortie à une échelle de temps associée, et qui correspond au model comme medium. Celle-ci fournit un moyen adapté de représenter un modèle et d'y associer échelle de temps et données. On y associe un ontologie $O$ au sens de~\cite{livet2010}, i.e. un ensemble d'éléments qui correspondent à une entité (qui peut être un objet, un agent, un processus, un état, un concept, c'est-à-dire tout élément modulaire formalisable) du monde réel. Nous incluons seulement ces deux aspects (le modèle et les objets représentés) de la théorie de Giere, en faisant l'hypothèse que le but et le producteur de la perspective sont en fait contenus dans l'ontologie s'ils font sens pour l'étude du système: par exemple, dans le cas des sondages subjectifs en anthropologie ou sociologie, le sondeur est un élément clé est sera nécessairement inclut dans l'ontologie. De même pour l'objectif poursuivi, tout particulièrement en sciences humaines où la recherche n'est jamais neutre comme nous l'avons vu en~\ref{ch:positioning}. Formalisons cette définition :
}

\bpar{
\begin{definition}
A \emph{perspective on a system} is given by a dataflow machine $M = (i,o,\mathbb{T})$ and an associated ontology $O$. We assume that the ontology can be decomposed into atomic elements $O=(O_j)_j$.
\end{definition}
}{
\begin{definition}
Une \emph{perspective sur un système} est donnée par une \emph{Dataflow Machine} $M = (i,o,\mathbb{T})$ et une Ontologie associée $O$. Nous supposons que l'ontologie peut être décomposée de manière discrete en éléments atomiques $O=(O_j)_j$.
\end{definition}
}

\bpar{
The atomic elements of the ontology can be particular elements such as agents or components of the system, but also processes, interactions, states, or concepts for example. The ontology can be seen as the exhaustive and rigorous description of the content of the perspective. The assumption of a dataflow machine implies that possible inputs and outputs can be quantified, what is not necessarily restrictive to quantitative perspectives, as most of qualitative approaches can be translated into discrete variables as soon as the set of possibles is known or assumed. 
}{
Les éléments atomiques de l'ontologie peuvent être des constituants particuliers du systèmes, comme des agents ou des composantes, mais aussi des processus, interactions, états ou concepts par exemple. L'ontologie peut être vue comme la description exhaustive et rigoureuse du contenu de la perspective. L'hypothèse d'une \emph{Dataflow Machine} implique que les entrées et sorties potentielles peuvent être quantifiées, ce qui n'est pas nécessairement restrictif aux perspectives quantitatives, puisque la plupart des approches qualitatives peuvent être traduites en variables discrètes à partir du moment où l'ensemble des possibles est connu ou supposé.
}

\bpar{
The system is then defined ``reversely'', i.e. from a set of perspectives on what would constitute then the system:

\begin{definition}
A \emph{system} is a set of \emph{perspectives on a system}: $S = (M_i,O_i)_{I\in I}$, where $I$ may be finite or not.
\end{definition}

We denote by $\mathcal{O} = (O_{j,i})_{j,i\in I}$ the set of all elements within ontologies.
}{
Nous définissons alors le système de manière ``réciproque'', i.e. à partir d'un ensemble de perspectives sur ce qui constitue alors le système :

\begin{definition}
Un \emph{système} est un ensemble de \emph{perspectives sur un système} :
 $S = (M_i,O_i)_{I\in I}$, où $I$ n'est pas nécessairement fini.
\end{definition}

Nous désignons par $\mathcal{O} = (O_{j,i})_{j,i\in I}$ l'ensemble des elements dans les ontologies.
}

\bpar{
Note that at this level of construction, there is not necessarily any structural consistence in what we call a system, as given our broad definition could allow for example to consider as a system a perspective on a car together with a perspective on a system of cities what makes reasonably no sense at all. Further definitions and developments will allow to be closer from classical definition of a system (interacting entities, designed artifacts, etc.). The same way, the definition of a subsystem will be given further. The introduced elements of our approach help to tackle so far points three, five and six of the requirements.
}{
Comme on part des perspectives sur un système pour définir le système dans son ensemble, il n'y a pas de contradiction. On peut noter qu'à ce stade de la construction, il n'existe pas nécessairement de cohérence structurelle, au sens d'une correspondance avec une structure réelle, sur ce qu'on appelle un système, puisque étant donné notre définition très large nous pourrions par exemple considérer un système comme une perspective sur un véhicule conjointement à une perspective sur un système de villes, ce qui ne fait pas raisonnablement sens. Des définitions approfondies et développements doivent permettre de se rapprocher des définitions classiques d'un système (entités en interaction, artefacts précisément définis, etc. ). De la même manière, la définition d'un sous-système sera donnée plus loin. Les éléments de l'approche déjà introduits permettent jusqu'ici de répondre aux points trois, cinq et six des recommandations.
}

\paragraph{Precision on the recursive aspect of the theory}{Précision sur l'aspect récursif de la théorie}

\bpar{
One direct consequence of these definitions must be detailed: the fact that they can be applied recursively. Indeed, one could imagine taking as perspective a system in our sense, therefore a set of perspectives on a system, and do that at any order. If ones takes a system in any classical sense, then the first order can be understood as an epistemology of the system, i.e. the study of diverse perspectives on a system. A set of perspectives on related systems may in some conditions be a domain or a field, thus a set of perspectives on various related systems the epistemology of a field. These are more analogies to give the idea behind the recursive character of the theory. It is indeed crucial for the meaning and consistence of the theory because of the following arguments:
\begin{itemize}
\item The choice of perspectives in which a system consists is necessarily subjective and therefore understood as a perspective, and a perspective on a system if we are able to build a general ontology.
\item We will use relations between ontologies in the following, which construction based on emergence is also subjective and seen as perspectives.
\end{itemize}
}{
Une conséquence directe de ces définitions doit être détaillée : le fait qu'elles peuvent être appliquées de manière récursive. En effet, on peut imaginer prendre comme perspective un système dans notre sens, c'est-à-dire un ensemble de perspectives sur un système, et le faire à tout ordre. Si on considère un système à n'importe quel sens classique, alors le premier ordre peut être interprété comme une épistémologie du système, i.e. l'étude de perspectives sur un système. Une ensemble de perspectives sur des systèmes en relation peut sous certaines conditions être un domaine ou un champ d'étude, et donc un ensemble de perspectives sur diverses perspectives l'épistémologie d'un champ. On peut proposer des analogies supplémentaires pour traduire l'idée derrière le caractère récursif de la théorie. C'est en effet crucial pour la signification et la cohérence de la théorie, notamment pour les raisons suivantes : (i) le choix des perspectives qui constituent un système est nécessairement subjectif et peut donc être compris comme une perspective en lui-même, et ainsi une perspective sur un système si l'on est en mesure de construire une ontologie générale ; (ii) nous utiliserons des relations entre ontologies par la suite, dont la construction est basée sur l'émergence est également subjective et vue comme perspectives. Ces aspects de réflexivité sont fondamentaux, en écho à la discussion de~\ref{sec:epistemology} sur la production de connaissance et la nature de la complexité.
}

\subsubsection{Ontological Graph}{Graphe Ontologique}


\bpar{
We propose then to capture the structure of the system by linking ontologies. This approach could eventually be linked to structural realism epistemological positioning~\cite{frigg2011everything} as knowledge of the world is partly contained here in structure of models. 
 Therefore, we choose to emphasize the role of emergence as we believe that it may be one practical minimalist way to capture quite well complex systems structure\footnote{what of course can not been presented as a provable claim as it depends on system definition, etc.}. We follow on that point the approach of \noun{Bedau} on different type of emergences, in particular his definition of weak emergence given in~\cite{bedau2002downward}. Let recall briefly definitions we will use in the following. \noun{Bedau} starts from defining emerging properties and then extends it to phenomena, entities, etc. The same way, our framework is not restricted to objects or properties and wraps thus the generalized definitions into emergence between ontologies. We will apply the notion of emergence under the two following forms\footnote{the third form \noun{Bedau} recalls, \emph{Strong emergence} will not be used, as we need only to capture dependance and autonomy, and weak emergence is more satisfying in terms of complex systems, as it does not assume ``irreducible causal powers'' to objects of upper scales at a given level. Nominal emergence is used to capture inclusion between ontologies.}:
\begin{itemize}
\item \emph{Nominal emergence}: one ontology $O'$ is included in an other $O$ but the aspect of $O$ that is said to be nominally emergent regarding $O'$ does not depend on $O'$.
\item \emph{Weak emergence}: one part of an ontology $O$ can be \emph{computationnaly} derived by aggregation of elements and interactions between elements of an ontology $O'$.
\end{itemize}
}{
Nous proposons ensuite la structure du système en reliant les ontologies. Cette approche pourrait éventuellement être mise en perspective par rapport à un positionnement épistémologique de réalisme structurel~\cite{frigg2011everything}, c'est-à-dire que les théories tendent à capturer une certaine structure existante du monde réel, puisqu'une connaissance du monde est ici partiellement contenue dans la structure des modèles, tout en gardant à l'esprit que notre position s'en éloigne en partie de par la conjugaison des perspectives qui induit un certain ``degré de constructivisme'' comme expliqué en~\ref{sec:epistemology}. Pour cette raison, nous faisons le choix d'appuyer le rôle de l'émergence, suivant l'intuition qu'il pourrait s'agir d'un outil pratique minimaliste pour capturer de façon raisonnable la structure d'un système complexe\footnote{ce qui bien sûr ne peut être formulé comme une affirmation prouvable car cela dépendra de la définition d'un système, etc.}. Nous prenons pour cet aspect le positionnement de \noun{Bedau} sur les différents types d'émergence déjà présenté plusieurs fois, en particulier sa définition de l'émergence faible donnée dans~\cite{bedau2002downward}. Rappelons brièvement les définitions que nous utiliserons par la suite. \noun{Bedau} commence par définir les propriétés émergentes puis étend le concept aux phénomènes, entités, etc. De la même manière, notre cadre n'est pas restreints aux objets ou propriétés et inclut ainsi les définitions généralisées comme lien entre ontologies. Nous appliquons la notion d'émergence sous les deux formes suivantes\footnote{la troisième forme rappelée par \noun{Bedau}, \emph{l'émergence forte}, ne sera pas utilisée, car nous avons besoin de capturer rien de plus des relations de dépendance et d'autonomie, et l'émergence faible est plus adéquate en termes de systèmes complexes, puisqu'elle n'assume pas ``des pouvoirs causaux irréductibles'' aux objets des échelles supérieures à un niveau donné. L'émergence nominale est utilisée pour capturer des relations d'inclusion entre les ontologies.} :
\begin{itemize}
\item \emph{Emergence nominale} : une ontologie $O'$ est inclue dans une autre ontologie $O$ mais l'aspect de $O$ qui est dit nominalement émergent en rapport à $O'$ ne dépend pas de $O'$.
\item \emph{Emergence faible} : une partie d'une ontologie $O$ peut être dérivée \emph{de manière computationnelle} par agrégation et interactions entre les éléments d'une ontologie $O'$.
\end{itemize}
}

\bpar{
As developed before, the presence of emergence, and especially weak emergence, will consist in itself in a perspective. It can be conceptual and postulated as an axiom within a thematic theory, but also experimental if clues of weak emergence are effectively measured between objects. In any case, the relation between ontologies must be encoded within an ontology, which was not necessarily introduced in the initial definition of the system.
}{
Comme développé précédemment, la présence d'émergence, et spécifiquement d'émergence faible, constitue une perspective en soi. Elle peut être conceptuelle et postulée comme un axiome dans une théorie thématique, mais aussi expérimentale si des traces d'émergence faible sont effectivement mesurées entre objets. Dans tous les cas, la relation entre ontologies doit être encodée dans une ontologie, ce qui n'était pas nécessairement introduit dans la définition initiale d'un système. Ainsi pour simplifier, les perspectives permettent de décomposer le système en briques ontologiques spécifiant une description ``complète''.
}


\bpar{
We make therefore the following assumption for next developments:
\begin{assumption}
A system can be partially structured by extending it with an ontology that contains (not necessarily only) relations between elements of ontologies of its perspectives. We name it the \emph{coupling ontology} and assume its existence in the following. We assume furthermore its atomicity, i.e. if $O$ is in relation with $O'$, then any subsets of $O,O'$ can not be in relation, what is not restrictive as a decomposition into several independent subsets ensures it if it is not the case.
\end{assumption}
}{
Nous faisons pour cette raison l'hypothèse suivante importante par la suite :
\begin{assumption}
Un système peut être partiellement structuré par son extension avec une ontologie qui contient (pas nécessairement uniquement) des relations entre les éléments des ontologies de ses perspectives. Nous la désignons \emph{ontologie de couplage} et supposons son existence par la suite. Nous postulons de plus son atomicité, i.e. si $O$ est en relation avec $O'$, alors tout sous-ensemble de $O,O'$ ne peuvent être en relation, ce qui n'est pas contraignant puisqu'une décomposition en des sous-ensembles indépendants assurera cette propriété si elle n'est pas vérifiée initialement.
\end{assumption}
Cette hypothèse revient concrètement qu'il est possible de coupler des perspectives, c'est-à-dire souvent des modèles en pratique, et que ce couplage peut être représenté de façon similaire. Notre expérience pratique du couplage tout au long de nos travaux nous pousse à faire cette hypothèse : tant que les systèmes considérés sont ``raisonnables'' (choisi raisonnablement l'un par rapport à l'autre, et donc choisi pour être couplés en quelque sorte), il est toujours possible de les coupler.
}

\bpar{
It allows to exhibit emergence relations not only within a perspective itself but also between elements of different perspectives. We define then pre-order relations between subsets of ontologies:
}{
Cela nous permet d'exhiber des relations d'émergence pas seulement au sein d'une perspective elle-même, mais également entre les éléments de différentes perspectives. Nous définissons ensuite des relations de pré-ordre entre les sous-ensemble des ontologies :
}

\bpar{
\begin{proposition}
The following binary relationships are pre-orders on $\mathcal{P(O)}$:
\begin{itemize}
\item Emergence (based on Weak Emergence): $O' \preccurlyeq O$ if and only if $O$ weakly emerges from $O'$.
\item Inclusion (based on Nominal Emergence): $O' \Subset O$ if and only if $O$ nominally emerges from $O'$.
\end{itemize}
\end{proposition}
}{
\begin{proposition}
Les relations binaires suivantes sont des pré-ordres sur $\mathcal{P(O)}$ :
\begin{itemize}
\item Emergence (basée sur l'émergence faible) : $O' \preccurlyeq O$ si et seulement si $O$ émerge faiblement de $O'$.
\item Inclusion (basée sur l'émergence nominale) : $O' \Subset O$ si et seulement si $O$ émerge nominalement de $O'$.
\end{itemize}
\end{proposition}
}

\bpar{
\begin{proof}
With the convention that it can be said that an object emerges from itself, we have reflexivity (if such a convention seems absurd, we can define the relationships as \emph{$O$ emerges from $O'$ or $O=O'$ }). Transitivity is clearly contained in definitions of emergence.
\end{proof}
}{
Avec la convention qu'il peut être admis qu'un objet émerge de lui-même, on a réflexivité (si une telle convention parait absurde, on peut définir les relations comme \emph{$O$ émerge de $O'$ ou $O=O'$}). La transitivité est clairement contenue dans la définition de l'émergence.
}

\medskip

\bpar{
Note that the inclusion relation is more general than an inclusion between sets, as it translates an inclusion ``inside'' the elements of the ontology.
}{
Notons que la relation d'inclusion est plus général qu'une inclusion entre ensembles, puisqu'elle traduit une inclusion ``au sein'' des éléments de l'ontologie. Par exemple, une ontologie peut supposer un couplage fort non-décomposable (qui serait une hypothèse de la perspective en elle-même), et une autre perspective contenir l'un des éléments de ce couplage. Nous allons voir que ces relations d'ordre vont nous permettre de définir un graphe par l'algorithme de réduction qui suit.
}

\bpar{
These relations are the basis for the construction of a graph called the \emph{ontological graph} :

\begin{definition}
The \emph{ontological graph} is constructed by induction the following way:
\begin{enumerate}
\item A graph is constricted, with vertices elements of $\mathcal{P(O)}$ and edges of two types: $E_W = \{(O,O') | O' \preccurlyeq O \}$ and $E_N = \{(O,O') | O' \Subset O \}$
\item Nodes are reduced\footnote{the reduction procedure aims to delete redundancy, keeping an entity at the higher level at which it exists.} by: if $o \in O,O'$ and ($O' \preccurlyeq O$ or $O' \Subset O$) but not ($O \preccurlyeq O'$ or $O \Subset O'$), then $O' \leftarrow O' \setminus o$
\item Nodes with intersecting sets are merged, keeping edges linking merged nodes. This step ensures non-overlapping nodes.
\end{enumerate}
\end{definition}
}{
\begin{definition}
Le \emph{graphe ontologique} est construit par induction de la manière suivante :
\begin{enumerate}
\item Un graphe est construit, avec pour noeuds des éléments de $\mathcal{P(O)}$ et des liens de deux types : $E_W = \{(O,O') | O' \preccurlyeq O \}$ et $E_N = \{(O,O') | O' \Subset O \}$
\item Les noeuds sont réduits\footnote{la procédure de réduction vise à supprimer la redondance, gardant une entité au plus haut niveau où elle existe.} par : si $o \in O,O'$ et ($O' \preccurlyeq O$ ou $O' \Subset O$) mais pas ($O \preccurlyeq O'$ or $O \Subset O'$), alors $O' \leftarrow O' \setminus o$
\item Les noeuds avec des ensemble se recoupant sont fusionnés, en gardant les liens liant des noeuds fusionnés. Cette étape assure des noeuds ne se recoupant pas.
\end{enumerate}
\end{definition}
}



\subsubsection{Minimal Ontological Tree}{Arbre Ontologique Minimal}


\bpar{
The topological structure of the graph, that contains in a way the \emph{structure of the system}
, can be reduced into a minimal tree that captures hierarchical structure essential to the theory.
}{
La structure topologique du graphe, qui contient en un sens la \emph{structure du système}, peut être réduite en un arbre minimal qui capture la structure hiérarchique essentielle pour la théorie.
}

\bpar{
We need first to give consistence to the system:

\begin{definition}
A consistent part of the ontological graph is a weakly connected component of the graph. We assume for now to work on a consistent part.
\end{definition}
}{
Nous devons d'abord donner cohérence au système :
\begin{definition}
Une partie cohérente du graphe ontologique est une composante du graphe faiblement connectée au sens d'un graphe dirigé. Nous assumons pour la suite travailler sur une partie cohérente.
\end{definition}
}


\bpar{
The notion of consistent system, together with subsystem or nodes timescales that will be defined later, requires to reconstruct perspectives from ontological elements, i.e. the inverse operation of what was done in our deconstruction procedure.
}{
La notion de système cohérent, ainsi que de sous-système ou d'échelle de temps des noeuds qui seront définies par la suite, nécessite de reconstruire des perspectives à partir des éléments ontologiques, i.e. l'opération inverse de ce qui a été fait dans notre procédure qui peut être vue comme une deconstruction.
}

\bpar{
\begin{assumption}
There exists $\mathcal{O}' \subset \mathcal{P(O)}$ such that for any $O \subset \mathcal{O}'$, there exists a corresponding dataflow machine $M$ such that the corresponding perspective is consistent with initial elements of the system (i.e. machines are equivalent on ontology overlaps). If $\Phi : M \mapsto O$ is the initial mapping, we denote this extended reciprocal construction by $M' = \Phi^{<-1>}(O)$.
\end{assumption}
}{
\begin{assumption}
Il existe $\mathcal{O}' \subset \mathcal{P(O)}$ tel que pour tout $O \subset \mathcal{O}'$, il existe une \emph{Dataflow Machine} $M$ correspondante telle que la perspective correspondante est cohérente avec les éléments initiaux du système (i.e. les machine sont équivalentes sur les parties communes des ontologies). Si $\Phi : M \mapsto O$ est la correspondance initiale, nous notons cette construction réciproque étendue par $M' = \Phi^{<-1>}(O)$.
\end{assumption}
}

\paragraph{Remark}{Remarque}

\bpar{
This assumption could eventually be changed into a provable proposition, assuming that the coupling ontology indeed corresponds to a coupling perspective, which dataflow machine part is consistent with coupled entities. Therein, the decomposition postulate of~\cite{golden2012modeling} should allow to identify basic components corresponding to each element of the ontology, and then construct the new perspective by induction. We find however these assumptions too restrictive, as for example various ontological elements may be modeled by an irreducible machine, as a differential equations with aggregated variables. We prefer to be less restrictive and postulate the existence of the reverse mapping on some sub-ontologies, that should be in practice the ones where couplings can be effectively modeled.
}{
Cette hypothèse pourrait éventuellement être changée en une proposition prouvable, en supposant que l'ontologie de couplage correspond effectivement à une perspective de couplage, dont la composante \emph{Dataflow Machine} est cohérente avec les entités couplées. Ainsi, le postulat de décomposition de~\cite{golden2012modeling} devrait permettre d'identifier des composantes de base correspondantes à chaque élément de l'ontologie, et construire ainsi la nouvelle perspective par induction. Nous trouvons toutefois ces hypothèses trop restrictives, puisque par exemple divers éléments de l'arbre ontologique peuvent être modélisés par la même machine irréductible, à l'image d'une équation différentielle aux variables agrégées. Nous préférons être moins restrictifs et postuler l'existence de la correspondance inverse sur certaines sous-ontologies, qui devraient être en pratique celles sur lesquelles le couplage peut effectivement être modélisé.
}

\bpar{
Given this assumption, we can define the consistent system as the reciprocal image of the consistent part of the ontological graph. It ensures system connectivity what is a requirement for tree construction.
}{
Grace à l'hypothèse ci-dessus, on peut définir le système cohérent comme l'image réciproque de la partie cohérente du graphe ontologique. Cela permet la connectivité du système qui est un pré-requis pour la construction de l'arbre. 
}

\bpar{
\begin{proposition}
The tree decomposition of the ontological graph in which nodes contains strongly connected components is unique. The reduced tree, that corresponds to the ontological graph in which strongly connected components have been merged with edges kept, is called the \emph{Minimal Ontological Tree}.
\end{proposition}
}{
\begin{proposition}
La décomposition arborescente du graphe ontologique 
 dans laquelle les noeuds contiennent les composantes fortement connexes est unique. L'arbre réduit, qui correspond au graphe ontologique les composantes fortement connexes ont été fusionnées et les liens gardés, est nommé \emph{Arbre Ontologique Minimal}.
\end{proposition}
}

\bpar{
\begin{proof}
(sketch of) The unicity is obtained as nodes are fixed as strongly connected components. It is trivially a tree decomposition as in a directed graph, strongly connected components do not intersect, thus the consistence of the decomposition.
\end{proof}
}{
\begin{proof}
(esquisse) L'unicité découle de la définition univoque puisque les noeuds sont fixés comme les composantes fortement connexes. Il s'agit trivialement d'une décomposition en arbre puisque dans un graphe dirigé, les composantes fortement connexes ne se recoupent pas, d'où la cohérence de la décomposition. 
\end{proof}
}

\bpar{
Any loop $O \rightarrow O' \rightarrow \ldots \rightarrow O$ in the ontological graph assumes that all its elements are equivalent in the sense of $\preccurlyeq$. This equivalence loops should help to define the notion of strong coupling as an application of the theory (see applications).
}{
Toute boucle $O \rightarrow O' \rightarrow \ldots \rightarrow O$ dans le graphe ontologique suppose que tous ses éléments sont équivalent au sens de $\preccurlyeq$. 
Ces boucles d'équivalence devrait aider à définir la notion de couplage fort comme une application de la théorie, avec cependant un caractère qualitatif dans la nature du couplage, ne permettant pas une définition fine de la force de couplage par exemple.
}

\bpar{
The Minimal Ontological Tree (MOT) is a tree in the undirected sense but a forest in the directed sense. Its topology contains a sort of system hierarchy. Consistent subsystems are defined from the set $\mathcal{B}$ of branches of the forest, as $(\Phi^{<-1>}(\mathcal{B}),\mathcal{B})$. The timescale of a node, and by extension of a subsystem, is the union of timescales of corresponding machines. Levels of the tree are defined from root nodes, and the emergence relations between nodes implies a vertical inclusion between timescales.
}{
L'Arbre Minimal Ontologique (MOT) est un arbre au sens non-dirigé, mais une forêt au sens dirigé. Sa topologie contient une représentation des hiérarchies du système. Les sous-systèmes cohérents sont définis à partir de l'ensemble $\mathcal{B}$ des branches de la forêt, comme $(\Phi^{<-1>}(\mathcal{B}),\mathcal{B})$. L'échelle de temps d'un noeud, et par extension d'un sous-système, est l'union est échelles de temps des machines correspondantes. Les niveaux de l'arbre sont définis à partir des noeuds racine, et les relations d'émergence entre les noeuds implique une inclusion verticale entre échelles de temps.
}


\subsubsection{Action on Data}{Action sur des Données}



De la même manière que les actions de groupes permettent de donner structure à l'utilisation d'un groupe sur un ensemble (généralement de données), une piste de développement puissante serait l'ajout à la théorie de l'aspect essentiel de relation à la réalité par une action des noeuds de l'arbre ontologique sur des ensembles de données. Cette opération est hors de propos pour l'instant car nous n'avons pas encore exploité la structure interne des \emph{dataflow machines}. Une piste, que nous confirmons comme ouverture dans la section suivante~\ref{sec:knowledgeframework}, impliquerait le couplage de ce cadre avec le cadre de connaissances qui y est introduit.

\subsubsection{Scales}{Echelles}

\bpar{
Finally, we propose to define scales associated to a system. Following~\cite{manson2008does}, an epistemological continuum of visions on scale is a consequence of differences between disciplines in the way we developed in the introduction. This proposition is indeed compatible with our framework, as the construction of scales for each level of the ontological tree results in a broad variety of scales.
}{
Enfin, nous proposons de définir les échelles associées à un système. Suivant~\cite{manson2008does}, un continuum épistémologique de visions sur l'échelle est une conséquence des différences propres à chaque discipline, comme nous avons développé en introduction. Cette proposition est en fait compatible avec notre cadre, puisque la construction d'échelles pour chaque niveau de l'arbre ontologique résulte en une grande variété d'échelles.
}

\bpar{
Let $(M,O)$ a subsystem and $\mathbb{T}$ the corresponding timescale. We propose to define the ``thematic scale'' (for example spatial scale) assuming a representation theorem, i.e. that an aspect (thematic aspect) of the machine can be represented as a dynamic state variable $\vec{X}(t)$. Assuming a scale operator\footnote{that can be of various nature: extent, probabilistic extent, spectral scales, stationarity scales, etc.} $\norm{\cdot}_{S}$ and that the state variable has a certain level of differentiability, the \emph{thematic scale} if defined as $\norm{(d^k \vec{X}(t))_k}_S$.
}{
Soit $(M,O)$ un sous-système et $\mathbb{T}$ l'échelle de temps correspondante. Nous proposons de définir ``l'échelle thématique'' (par exemple l'échelle spatiale) en supposant un théorème de représentation, i.e. qu'un aspect (aspect thématique) de la machine peut être représenté par une variable d'état dynamique $\vec{X}(t)$. Etant donné un opérateur d'échelle\footnote{qui peut être de nature variée : étendue, étendue probabiliste, échelles spectrales, échelles de stationnarité, etc.} $\norm{\cdot}_{S}$ et que la variable d'état est différentiable à un certain niveau, \emph{l'échelle thématique} pour cet aspect, c'est-à-dire l'échelle typique à laquelle les agents ou processus correspondants opèrent (pouvant être multiple si l'opérateur est multidimensionnel), est définie par $\norm{(d^k \vec{X}(t))_k}_S$.
}

\subsection{Application and discussion}{Applications et discussion}

\subsubsection{The particular case of geographical systems}{Le cas particulier des systèmes géographiques}




\bpar{
In~\cite{dollfus1975some} 
 \noun{Durand-Dast{\`e}s} proposes a definition of geographical structure and system, structure would be the spatial container for systems viewed as complex open interacting systems (elements with attributes, relations between elements and inputs/outputs with external world). For a given system, its definition is a perspective, completed by structure to have a system in our sense. Depending on the way to define relations, it may be more or less easy to extract ontological structure.
}{
Dans~\cite{dollfus1975some}, \noun{Durand-Dast{\`e}s} introduit une définition des systèmes et structures géographiques, la structure étant le contenant spatial des systèmes vus comme des systèmes complexes ouverts en interaction (donné par ses éléments et leur attributs, les relations entre éléments et les entrée/sorties avec le monde extérieur). Pour un système donné, sa définition est une perspective, complété par la structure pour avoir un système selon notre sens. Selon la manière dont les relations sont définies, cela peut être plus ou moins aisé d'extraire la structure ontologique.
}


\subsubsection*{Modularity and co-evolving subsystems}{Modularité et sous-systèmes en co-évolution}

\bpar{
For the example of Urban Systems, urban evolutionary theory enters this framework using our previous thematic theory. The decomposition into uncorrelated subsystems yields precisely strongly coupled components as co-evolving components. The correlation between subsystems should be in a certain way positively correlated with topological distance in the tree. If we define elements of a node before merging as \emph{strongly coupled elements}, in the case of dynamic ontologies, it provides a definition of \emph{co-evolution} and co-evolving subsystems equivalent to the thematic definition.
}{
Pour l'exemple des systèmes urbains, la théorie évolutive des villes entre dans ce cadre en utilisant notre théorie thématique développée dans la section précédente. La décomposition en sous-systèmes décorrélés fournit précisément des composantes fortement couplées comme des composantes en co-évolution. La correlation entre sous-systèmes devrait d'une certaine façon être corrélée à la distance topologique dans l'arbre. Si on définit les éléments d'un noeud avant réduction comme \emph{éléments fortement couplés}, dans le cas d'ontologies dynamiques, cela fournit une définition de la \emph{co-évolution} et de sous-systèmes en co-évolution, équivalente à la définition thématique.
}

\subsubsection{Discussion}{Discussion}

\paragraph{Link with existing frameworks}{Lien avec des cadres existants}

\bpar{
A link with the Cottineau-Chapron framework for multi-modeling~\cite{10.1371/journal.pone.0138212} may be done in the case they add the bibliographical layer, which would correspond to the reconstruction of perspectives. \cite{reymond2013logique} proposes the notion of ``interdisciplinary coupling'' what is close to our notion of coupling perspectives. A correspondance with System of Systems approaches (see e.g. \cite{luzeaux2015formal} for a recent general framework englobing system modeling and system description) may be also possible as our perspectives are constructed as dataflow machines, but with the significant difference that the notion of emergence is central.
}{
Un lien avec le cadre de Cottineau-Chapron pour la multi-modélisation~\cite{10.1371/journal.pone.0138212} pourrait être fait dans le cas où ils ajouteraient la couche bibliographique, qui correspondrait à la reconstruction des perspectives. \cite{reymond2013logique} propose la notion de ``couplage interdisciplinaire'' qui est proche de notre notion de coupler des perspectives. Une correspondance avec les approches de Système de Systèmes (voir e.g. \cite{luzeaux2015formal} pour un cadre récent englobant la modélisation et la description des systèmes) pourrait être également possible puisque nos perspectives sont construites comme des \emph{Dataflow Machines}, mais avec la différence cruciale que la notion d'émergence est centrale dans notre cas.
}

\paragraph{Contributions to the study of complex systems}{Contribution à l'étude des systèmes complexes}

\bpar{
We do not claim to provide a theory of systems (beware of cybernetics, systemics etc. that could not model everything), but more a framework to guide research questions (e.g. in our case the direct outcomes will be quantitative epistemology that comes from system construction as perspectives ; empirical to construct robust ontologies for perspectives ; targeted thematic to unveil causal relationship/emergence for construction of ontological network ; study of coupling as possible processes containing co-evolution ; study of scales ; etc.). It may be understood as meta-theory which application gives a theory, the thematic theory developed before being a specific implementation to territorial networked systems. We emphasize the notion of socio-technical system, crossing a social complex system approach (ontologies) with a description of technical artifacts (dataflow machines), taking the ``best of both worlds''.
}{
Nous ne prétendons pas exhiber une théorie des systèmes (il faut généralement se méfier de la cybernétique, la systémique etc. qui ne peuvent pas tout modéliser), mais plutôt un cadre majoritairement axiomatique et la structure associée pour guider les questions de recherche (e.g. dans notre cas les conséquences directes sont les études d'épistémologie quantitative qui vient de la construction des systèmes comme perspectives ; les études empiriques pour construire des ontologies robustes pour les perspectives ; des études thématiques ciblées pour révéler des relations causales ou l'émergence pour la construction des réseaux ontologiques ; l'étude des couplages comme processus contenant possiblement de la co-évolution ; l'étude des échelles ; etc.). Cela peut être compris comme une meta-théorie dont l'application donne une théorie, la théorie thématique qui précède étant une implémentation aux systèmes territoriaux en réseau. Nous appuyons la notion de système socio-technique, croisant une approche des systèmes sociaux complexes (ontologies) avec une description des artefacts techniques (\emph{Dataflow Machines}), prenant ``le meilleur des deux mondes''.
}

\subsubsection{Reflexivity}{Réflexivité}

\bpar{
We can learn from the application of this framework to our work, i.e. from some reflexivity, a clarification of research directions followed until here, and thus of the co-construction of answers to these questions with the different theoretical frameworks.
}{
Nous pouvons tirer de l'application de ce cadre à notre travail, c'est-à-dire d'une réflexivité, une clarification des directions de recherche menées jusqu'ici, et donc de la co-construction des réponses à ces questions avec les différents cadres théoriques.
}

\bpar{
\begin{enumerate}
\item The perspectivist approach implies a broad undertsanding of existing perspectives on a system, and of coupling possibilities between these; thus an emphasis on quantitative epistemology including algorithmic systematic review (exploration of knowledge space), knowledge mapping (description of its structure) and datamining bibliographic content (refinment at the micro level of scientific knowledge) which are included in \ref{sec:quantepistemo}.
\item At a more refined level, the knowledge of perspectives corresponds to some knowledge of empirical stylised facts, as for example the ones found in~\ref{sec:computation} for trafic flows, on gas price in \ref{sec:energyprice}, on urban form and road networks \ref{sec:staticcorrelations}.
\end{enumerate}
}{
\begin{enumerate}
\item L'approche perspectiviste implique une compréhension large des perspectives existantes sur un système, et des possibilités de couplage entre celle-ci ; d'où une emphase sur l'épistémologie quantitative qui inclue la revue systématique algorithmique (exploration de l'espace des connaissances), la cartographie des connaissances (extraction de sa structure) et de possibilités de fouille de contenu (raffinement au niveau atomique de la connaissance scientifique) qui correspondent au travail de~\ref{sec:quantepistemo}.
\item A un niveau plus fin de particularité, la connaissance des perspectives signifie une connaissance des faits stylisés empiriques, comme par exemple ceux pour le traffic routier~\ref{sec:computation}, les prix des carburants~\ref{sec:energyprice}, les formes urbaines et de réseau~\ref{sec:staticcorrelations}.
\end{enumerate}
}

\stars

%

\newpage

\section{Exploration of an Interdisciplinary Scientific Landscape}{Exploration d'un paysage scientifique interdisciplinaire}

\label{app:sec:cybergeo}


\bpar{
The methodological and technical constructions enabling the epistemological analysis of~\ref{sec:quantepistemo} have been done within a broader context, in particular starting with the analysis of corpuses linked to the \textit{Cybergeo} journal. We detail here the methodological aspect of these analyses.
}{
Les constructions méthodologiques et techniques rendant possible l'analyse épistémologique de~\ref{sec:quantepistemo} ont été menées dans un cadre plus large, notamment débutant avec l'analyse de corpus construits à partir de la revue \textit{Cybergeo}. Nous détaillons ici l'aspect méthodologique de ces analyses.
}

\stars

\bpar{
\textit{The content of this appendix has been elaborated in the context of a common project to quantitatively analyse publications in Cybergeo (see \ref{app:sec:cybergeonetworks} for the common production), initiated for the 20th anniversary of the journal in May 2016. Preliminary results have been presented as~\cite{raimbault2016indirect} at the anniversary conference, and the text of this appendix is extracted from~\cite{raimbault2017exploration}.}
}{
\textit{Le contenu de cette annexe a été élaboré dans le cadre du projet commun d'analyse quantitative des publications de Cybergeo (voir \ref{app:sec:cybergeonetworks} pour la production commune), initié pour les 20 ans de la revue en mai 2016. Les résultats préliminaires ont été présentés comme~\cite{raimbault2016indirect} à la conférence anniversaire, et le texte de cette annexe est extrait et traduit de~\cite{raimbault2017exploration}.}
}

\stars


\bpar{
Patterns of interdisciplinarity in science can be quantified through diverse complementary dimensions. This paper studies as a case study the scientific environment of a generalist journal in Geography, \emph{Cybergeo}, in order to introduce a novel methodology combining citation network analysis and semantic analysis. We collect a large corpus of around 200,000 articles with their abstracts and the corresponding citation network that provides a first citation classification. Relevant keywords are extracted for each article through text-mining, allowing us to construct a semantic classification. We study the qualitative patterns of relations between endogenous disciplines within each classification, and finally show the complementarity of classifications and of their associated interdisciplinarity measures. The tools we develop accordingly are open and reusable for similar large scale studies of scientific environments.
}{
Les motifs d'interdisciplinarité en science peuvent être quantifiés au travers de diverses dimensions complémentaires. Cette monographie étudie comme cas d'étude l'environnement scientifique d'un journal généraliste en géographie, \emph{Cybergeo}, afin d'introduire une nouvelle méthodologie qui combine analyse du réseau de citation et analyse sémantique. Nous construisons un corpus massif d'environ 200,000 articles avec leur résumés et le réseau de citation correspondant qui fournit une première classification par citations. Les mots-clés pertinents sont extraits pour chaque article par analyse textuelle, permettant la construction d'une classification sémantique. Nous étudions les motifs qualitatifs de relations entre disciplines endogènes au sein de chaque classification, et montrons finalement la complémentarité des classifications et des mesures d'interdisciplinarité associées. Les outils développés en conséquence sont ouverts et réutilisables pour des études similaires à grande échelle d'environnements scientifiques.
}

\subsection{Introduction}{Introduction}

\bpar{
We develop in this paper a case study coupling citation network exploration and analysis with text-mining, aiming at mapping the scientific landscape in the neighborhood of a particular journal. We choose to study an electronic journal in Geography, named \textit{Cybergeo}\footnote{\texttt{http://cybergeo.revues.org/}}, that publishes articles within all subfields of Geography and is in that way multidisciplinary. The choice is initially due to data availability, but ensures several constraints making it highly relevant to the context given above. First of all, the ``discipline'' of Geography is very broad and by essence interdisciplinary~\cite{bracken2016interdisciplinarity} : the spectrum ranges from Human and Critical geography to physical geography and geomorphology, and interactions between these subfields are numerous. Secondly, bibliographical data is difficult to obtain, raising the concern of how the perception of a scientific landscape may be shaped by actors of the dissemination and thus far from objective, and making technical solutions as the ones we will consequently develop here crucial tools for an open and neutral science. Finally it makes a particularly interesting case study as the editorial policy is generalist and concerned with open science issues such as peer-review ethics transparency~\citep{10.1371/journal.pone.0147913}, open data and model practices, as recalled by~\cite{pumain2015adapting}, and this work contributes to these by fostering the opening of reflexivity.
}{
Cette section développe un cas d'étude qui couple exploration et analyse du réseau de citation avec analyse textuelle, dans le but de cartographier le paysage scientifique du voisinage d'un journal particulier. Nous choisissons d'étudier un journal électronique en Géographie, appelé \textit{Cybergeo}\footnote{\texttt{http://cybergeo.revues.org/}}, qui publie des articles dans l'ensemble des champs de la Géographie et est de cette façon multi-disciplinaire. Le choix est initialement dû à la disponibilité des données, mais répond à différentes contraintes le rendant particulièrement pertinent dans le contexte décrit précédemment. Tout d'abord, la ``discipline'' de la Géographie est très large et par essence interdisciplinaire~\cite{bracken2016interdisciplinarity}: le spectre s'étend de la Géographie Humaine et Critique à la Géographie Physique et la géomorphologie, et les interactions entre ces sous-champs sont nombreuses. Dans un second temps, les données bibliographiques sont difficiles à obtenir, soulevant la difficulté de la façon dont un paysage scientifique est perçu peut être influencée par les acteurs de la dissémination et donc loin d'être objectif, ce qui rend ainsi des solutions techniques comme celle que nous développerons en conséquence ici des outils cruciaux pour une science ouverte et neutre. Enfin, il s'agit d'un cas d'étude particulièrement intéressant puisque la politique éditoriale est généraliste et préoccupée par des questions de science ouverte comme la transparence de l'éthique de revue par les pairs~\cite{10.1371/journal.pone.0147913}, les pratiques d'ouverture des données et des modèles, comme rappelé par~\cite{pumain2015adapting}, et ce travail contribue à celles-ci en encourageant l'ouverture de la reflexivité.
}

\bpar{
Our contribution is original and significant on at least two aspects :
\begin{enumerate}
	\item we combine endogenous classifications in a network multilayer fashion, using semantic information ;
	\item a large dataset is constructed from scratch to study a journal not referenced in main databases, tackling both data retrieval and large scale data processing issues.
\end{enumerate}
}{
Le contexte méthodologique dans lequel ce travail se situe est développé en~\ref{sec:quantepistemo}. Cette contribution est originale et significative sur au moins deux aspects :
\begin{enumerate}
	\item nous combinons des classifications endogènes dans l'idée d'un réseau multi-couches, en utilisant l'information sémantique ;
	\item un jeux de données massif est construit à partir de rien pour étudier un journal non référencé dans les bases principales, répondant ainsi à des problèmes à la fois liés à la collecte de données et au traitement de données massives.
\end{enumerate}
}

\bpar{
The rest of the paper is organized as follows : we describe in the next section the dataset used and the data collection procedure. We then study properties of the citation network and describe the procedure to construct the semantic classification through text-mining. We finally study complementary measures of interdisciplinarity obtained with the different classifications.
}{
Le reste de cette section est organisée de la façon suivante : nous décrivons d'abord le jeu de données utilisé et la procédure de collecte des données. Nous étudions ensuite les propriétés du réseau de citation et décrivons la procédure pour construire la classification sémantique par analyse textuelle. Nous étudions finalement des mesures complémentaires d'interdisciplinarité obtenues par les différentes classifications.
}

\subsection{Database Construction}{Construction de la base de données}

\bpar{
Our approach imposes some requirements on the dataset used, namely: (i) cover a certain neighborhood of the studied journal in the citation network in order to have a consistent view on the scientific landscape; (ii) have at least a textual description for each node. For these to be met, we need to gather and compile data from heterogeneous sources. We use therefore an application specifically designed, which general architecture is given in Fig.~\ref{fig:cybergeo:fig1}. Source code of the application and all scripts used in this paper are available on the open \texttt{git} repository of the project\footnote{at \texttt{https://github.com/JusteRaimbault/HyperNetwork}}. Raw and processed data are also openly available on Dataverse\footnote{at \texttt{http://dx.doi.org/10.7910/DVN/VU2XKT}}. We recall that an important contribution of this paper is the construction of such an hybrid dataset from heterogeneous sources, and the development of associated tools that can be reused and further developed for similar purposes.
}{
Notre approche impose des contraintes sur le jeu de données utilisé, à savoir : (i) couvrir un certain voisinage du journal étudié dans le réseau de citation pour avoir une vue cohérente sur le paysage scientifique ; (ii) avoir au moins une description textuelle pour chaque noeud. Pour satisfaire celles-ci, nous devons rassembler et compiler des données de sources hétérogènes. Nous utilisons pour cela une application spécifiquement conçue, dont l'architecture générale est donnée en Fig.~\ref{fig:cybergeo:fig1}. Le code source de l'application et l'ensemble des scripts utilisés ici sont disponibles sur le dépôt \texttt{git} ouvert du projet\footnote{à \texttt{https://github.com/JusteRaimbault/HyperNetwork}}. Les données brutes et traitées sont également disponibles de manière ouverte sur Dataverse\footnote{à \texttt{http://dx.doi.org/10.7910/DVN/VU2XKT}}. Nous rappelons qu'une importante contribution de ce travail est la construction d'un tel jeu de données hybride à partir de sources hétérogènes, et le développement d'outils associés qui peuvent être utilisés et étendus dans des cadres similaires.
}

\begin{figure}
\includegraphics[width=\linewidth]{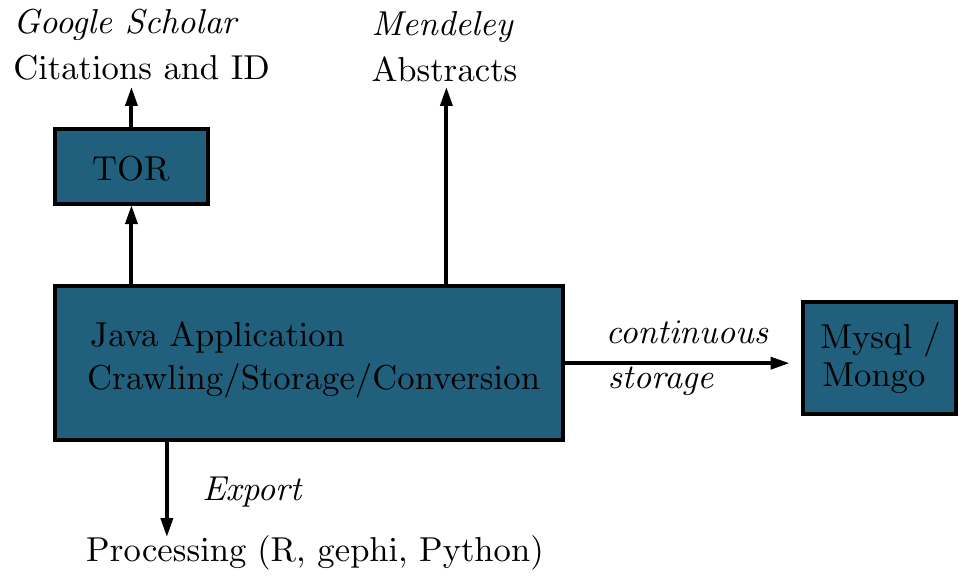}
\appcaption{\textbf{Heterogeneous Bibliographical Data Collection and processing.} Architecture of the application for content (semantic data), metadata and citation data collection. The heterogeneity of tasks requires the use of multiple languages : data collection and management is done in Java, and data stored in databases (Mysql and MongoDB) ; data processing is done in python for Natural Language Processing and in R for statistical and network analyses; graph visualizations are done with Gephi software.\label{fig:cybergeo:fig1}}{\textbf{Collecte et traitement de données bibliographiques hétérogènes.} Architecture de l'application pour la récolte des données de contenu (données sémantiques), des métadonnées et des données de citation. L'hétérogénéité des tâches requière l'utilisation de multiples langages : la collecte et la gestion des données sont faites en Java, et les données sont stockées dans des bases de données (Mysql et MongoDB) ; le traitement des données est effectué en python pour le traitement du langage naturel et en R pour les analyses statistiques et de réseaux ; les visualisations de graphes sont effectuées avec le logiciel Gephi.\label{fig:cybergeo:fig1}}
\end{figure}

\subsubsection{Initial Corpus}{Corpus initial}

\bpar{
The production database of \textit{Cybergeo} (snapshot taken in February 2016, provided by the editorial board), provides after pre-processing the initial database of articles, with basic information (title, abstract, publication year, authors). The processed version used is available together with the full database constructed, as a \texttt{mysql} dump, at the address given above. This base provide also bibliographical records of articles that give all references cited by the initial base (\emph{forward citations} for the initial corpus).
}{
La base de production de \textit{Cybergeo} (snapshot pris en février 2016, fournit par l'équipe éditoriale), permet d'obtenir après pré-traitement la base initiale des articles, avec information élémentaire (titre, résumé, année de publication, auteurs). La version traitée que l'on utilise est disponible en même temps que l'ensemble des données construites, comme dump \texttt{mysql}, à l'adresse donnée précédemment. Cette base fournit également les enregistrements de la bibliographie des articles, qui fournissent l'ensemble des références citées par la base initiale (citations \emph{données} par le corpus initial).
}

\subsubsection{Citation Data}{Données de citation}

\bpar{
Citation data is collected from \texttt{Google Scholar}, that is the only source for incoming citations~\citep{noruzi2005google} in our case as the journal is poorly referenced in other databases\footnote{or was just added as in the case of \textit{Web of Science}, indexing \textit{Cybergeo} since May 2016 only}. We are aware of the possible biaises using this single source (see e.g.~\cite{bohannon2014scientific})\footnote{or \texttt{http://iscpif.fr/blog/2016/02/the-strange-arithmetic-of-google-scholars}}, but these critics are more directed towards search results or possible targeted manipulations than the global structure of the citation network. The automatic collection requires the use of a crawling software to pipe requests, namely \texttt{TorPool}~\citep{torpool} that provides a Java API allowing an easy integration into our application of data collection. A crawler can therethrough retrieve html pages and get backward citation data, i.e. all citing articles for a given initial article. We retrieve that way two sub-corpuses: references citing papers in \textit{Cybergeo} and references \emph{citing the ones cited} by \textit{Cybergeo}. At this stage, the full corpus contains around $4\cdot10^5$ references.
}{
Les données de citation sont collectées à partir de \texttt{Google Scholar}, qui est la seule source pour les citations entrantes~\cite{noruzi2005google} dans notre cas puisque le journal n'est pas référencé dans les autres bases\footnote{ou vient d'y être ajouté comme dans le cas du \emph{Web of Science} qui n'indexe \textit{Cybergeo} que seulement depuis mai 2016}. Nous sommes conscient des possibles biais de l'utilisation de cette source unique (voir par exemple~\cite{bohannon2014scientific})\footnote{ou \texttt{http://iscpif.fr/blog/2016/02/the-strange-arithmetic-of-google-scholars}}, mais ces critiques sont plutôt dirigées vers les résultats de recherche ou de possibles manipulations ciblées que envers la structure globale du réseau de citation. La collecte automatique demande l'utilisation d'un logiciel de collecte pour transférer les requêtes, à savoir \texttt{TorPool}~\cite{torpool} qui fournit une API Java permettant une intégration aisée au sein de notre application de collecte. Un crawler peut ainsi récupérer les pages html et obtenir les citations inverses, c'est-à-dire l'ensemble des articles citant un article initial donné. Nous récupérons de cette manière deux sous-corpus : les références citant des articles dans \textit{Cybergeo} et les références citant celles citées par \textit{Cybergeo}. A ce stade, le corpus complet contient autour de $4\cdot10^5$ références.
}

\bpar{
For the sake of simplicity, we will denote by \emph{reference} any standard scientific production that can be cited by another (journal paper, book, book chapter, conference paper, communication, etc.) and contains basic records (title, abstract, authors, publication year). We work in the following on networks of references, linked by citations.
}{
Pour simplifier, nous appellerons \emph{référence} toute production scientifique standard qui peut être citée par une autre (article de journal, livre, chapitre de livre, article de conférence, communication, etc.) et contient des champs basiques (titre, résumé, auteurs, année de publication). Nous travaillons par la suite sur le réseau de références, reliées par des citations.
}

\subsubsection{Text Data}{Données textuelles}

\bpar{
A textual description for all references is necessary for a complete semantic analysis. We use for this an other source of data, that is the online catalog of \textit{Mendeley} reference manager software~\cite{mendeley}. It provides a free API allowing to get various records under a structured format. Although not complete, the catalog provides a reasonable coverage in our case, around 55\% of the full citation network. This yields a final corpus with full abstracts of size $2.1\cdot 10^5$. The structure and descriptive statistics of the corresponding citation network is recalled in Fig.~\ref{fig:cybergeo:fig2}.
}{
Une description textuelle pour l'ensemble des références est nécessaire pour une analyse sémantique complète. Nous utilisons pour cela une autre source de données, qui est le catalogue en ligne du logiciel de gestion bibliographique \textit{Mendeley}~\cite{mendeley}. Celui fournit une API gratuite permettant de récupérer divers champs sous un format structuré. Même s'il n'est pas complet, le catalogue fournit une couverture raisonnable dans notre cas, autour de 55\% du réseau de citation complet. Cela correspond à un corpus final avec résumés complets de taille $2.1\cdot 10^5$. La structure et les statistiques descriptives du réseau de citation correspondant sont rappelés en Fig.~\ref{fig:cybergeo:fig2}.
}


\begin{figure}
\includegraphics[width=\linewidth]{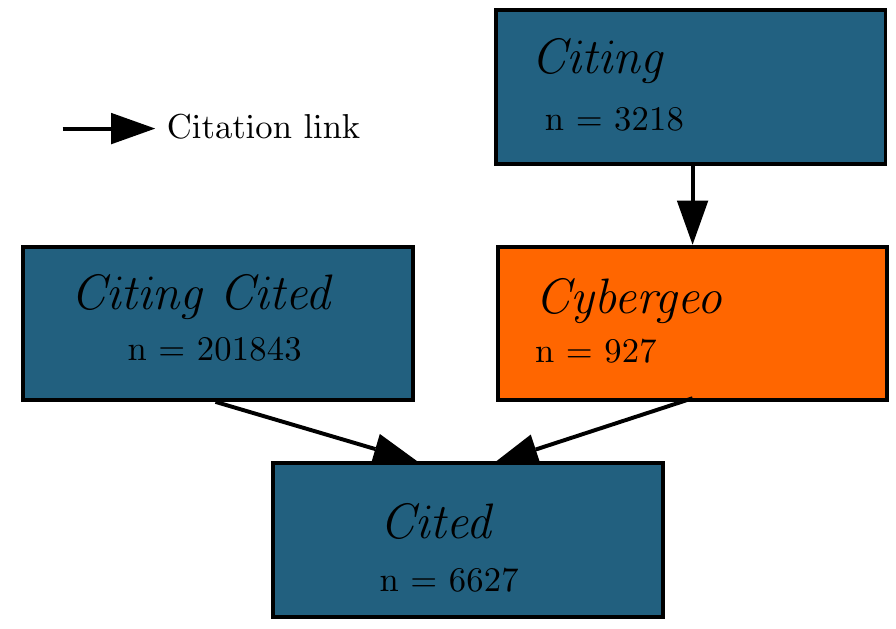}
\appcaption{\textbf{Structure and content of the citation network.} The original corpus of \emph{Cybergeo} is composed by 927 articles, themselves cited by a slightly larger corpus (yielding a stationary impact factor of around 3.18), cite $\simeq 6600$ references, themselves co-cited by more than $2\cdot 10^5$ works for which we have a textual description.\label{fig:cybergeo:fig2}}{\textbf{Structure et contenu du réseau de citation.} Le corpus initial de \emph{Cybergeo} est composé de 927 articles, eux-mêmes cités par un corpus légèrement plus grand (donnant un facteur d'impact stationnaire autour de 3.18), cite $\simeq 6600$ références, elles-mêmes co-citées par plus de $2\cdot 10^5$ travaux pour lesquels nous avons une description textuelle.\label{fig:cybergeo:fig2}}
\end{figure}

\subsection{Methods and Results}{Méthodes et Résultats}

\subsubsection{Citation Network Properties}{Propriétés du réseau de citation}

\paragraph{Properties}{Propriétés}


\bpar{
As detailed above, we are able by the reconstruction of the citation network at depth $\pm 1$ from the original $927$ references of the journal to retrieve around $4\cdot 10^5$ references, on which $2.1\cdot 10^5$ have an abstract text allowing semantic analysis. A first glance on citation network properties provides useful insights. Mean in-degree (that can be interpreted as a stationary integrated impact factor) on references for which it can be defined has a value of $\bar{d}=121.6$, whereas for articles in \textit{Cybergeo} we have $\bar{d}=3.18$. This difference suggests a variety for status of references, from old classical works (the most cited has 1051 incoming citations) to recent less influential works.
}{
Comme détaillé précédemment , nous sommes en mesure de récupérer autour de $4\cdot 10^5$ références par reconstruction du réseau de citation à profondeur $\pm 1$ à partir des $927$ références initiales du journal, parmi lesquelles $2.1\cdot 10^5$ ont un texte de résumé permettant une analyse sémantique. Un premier regard sur les propriétés du réseau de citation fournit des informations utiles. Le degré moyen entrant (qui peut être interprété comme un facteur d'impact stationnaire intégré) pour les références pour lesquelles il peut être défini a une valeur de $\bar{d}=121.6$, tandis que pour les articles de \textit{Cybergeo} nous avons $\bar{d}=3.18$. Cette différence suggère une variété de status de références, de travaux anciens classiques (le plus cité a 1051 citations entrantes) à des travaux plus récents moins influents.
}

\begin{figure}
\includegraphics[width=\linewidth]{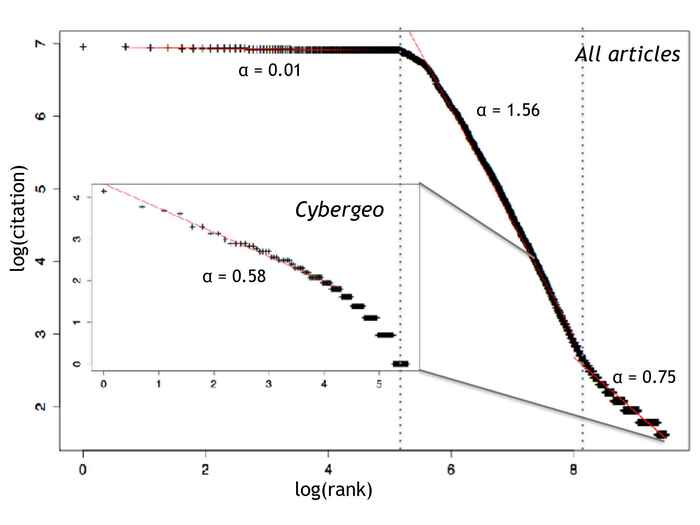}
\appcaption{\textbf{Rank-size plot of citations received.} The plot unveils three superposed citations regimes, corresponding to power laws with different levels of hierarchy. The references in \textit{Cybergeo} (inset plot) are themselves in the tail and less hierarchical.\label{fig:cybergeo:fig3}}{\textbf{Graphe rang-taille des citations reçues.} La courbe dévoile trois régimes de citation superposés, correspondant à des loin puissance avec différents niveaux de hiérarchie. Les références dans \textit{Cybergeo} (graphe en insert) sont elles-mêmes dans la queue et moins hiérarchiques.\label{fig:cybergeo:fig3}}
\end{figure}

\bpar{
This diversity is confirmed by the hierarchical organisation examined in Fig.~\ref{fig:cybergeo:fig3} that unveils three superposed regimes. More precisely, we look at the rank-size plot, given by the logarithm of the number of citations received as a function of the rank of the paper. We find, as expected~\citep{redner1998popular}, localized power-law behaviors. A first set of around 150 references shows a very low hierarchy (rank-size exponent $\alpha = 0.01$) and corresponds to classical references in different disciplines. A second regime ($\alpha = 1.56$) is much more hierarchized, followed by a last regime less hierarchical ($\alpha = 0.75$) containing more recent papers (average publication year mid-2005, against mid-1998 for the second and 1983 for the first).
}{
Cette diversité est confirmée par l'organisation hiérarchique examinée en Fig.~\ref{fig:cybergeo:fig3} qui révèle trois régimes superposés. Plus précisément, nous nous intéressons à la courbe rang-taille, donnée par le logarithme du nombre de citations reçues comme fonction du logarithme du rang de l'article. Nous obtenons, comme attendu \cite{redner1998popular}, des comportements de loi puissance localisés. Un premier ensemble d'environ 150 références présente une hiérarchie très faible (exposant rang-taille $\alpha = 0.01$) et correspond aux références classiques dans différentes disciplines. Un second régime ($\alpha = 1.56$) est bien plus hiérarchisé, suivi par un dernier régime moins hiérarchique ($\alpha = 0.75$) contenant des articles plus récents (année moyenne de publication mi-2005, contre mi-1998 pour le second et 1983 pour le premier).
}

\bpar{
Other topological properties reveal typical patterns of citation practices: for example, the existence of high-order cliques (complete sub-networks) implies citation practices which compatibility with the cumulative nature of knowledge may be questionable~\cite{pumain2005cumulativite}, since these need always to source back the production of knowledge in the most recent works. An exemple of such a clique in shown in Fig.~\ref{fig:cybergeo:fig4}.
}{
D'autres propriétés topologiques révèlent des motifs typiques de pratiques de citation : par exemple, l'existence de cliques (sous-graphes complets) de fort ordre implique des pratiques de citation dont la compatibilité avec la nature cumulative de la connaissance peut être remise en question~\cite{pumain2005cumulativite}, puisque celles-ci doivent toujours trouver la source de la production de connaissance dans les travaux les plus récents. Un exemple de telle clique est montré en Fig.~\ref{fig:cybergeo:fig4}.
}

\begin{figure}
\includegraphics[width=\linewidth]{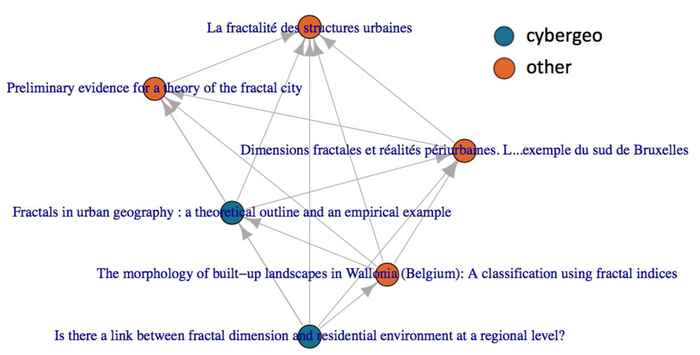}
\appcaption{Example of a maximal clique in the citation network, paper of \texttt{Cybergeo} being in blue. Such topological structure reveal citation practices such as here a systematic citation of previous works in the research niche.\label{fig:cybergeo:fig4}}{\textbf{Exemple d'une clique maximale dans le réseau de citation.} Les articles de \textit{Cybergeo} sont en bleu. Une telle structure topologique révèle certaines pratiques de citation comme ici une citation systématique des travaux précédents dans la niche de recherche.\label{fig:cybergeo:fig4}}
\end{figure}

\paragraph{Citation communities}{Communautés de citation}

\bpar{
The citation network is a first opportunity to construct endogenous disciplines, by extracting citation communities. More precisely, this step aims at finding recurrent patterns in citations that would define a field by its citation practices. In order to be consistent with the particular data structure we have (missing incoming citations for sub-corpuses at maximal depth), we filter the network by removing all nodes with degree smaller than one. This ensures that kept nodes are either at least cited by an other node (and thus there are no missing edges for these nodes) or cite at least two other nodes, what can make ``bridges'' between sub-communities. The resulting network has a size of $\left|V\right| = 107164$ nodes and $\left|E\right| = 309778$ edges. It is visualized in Fig.~\ref{fig:cybergeo:fig5}.
}{
Le réseau de citation est une première opportunité pour construire des disciplines endogènes, par extraction des communautés de citation. Plus précisément, cette étape vise à identifier des motifs récurrents dans les citations, qui définiraient une discipline par ses pratiques de citation. Afin de rester cohérent avec la structure particulière de données que nous avons (citations entrantes manquantes pour les sous-corpus à profondeur maximale), nous filtrons le réseau en supprimant tous les noeuds de degré inférieur à 1. Cela assure que les noeuds restants sont soit au moins cités par un autre noeud (et donc il n'y a pas de liens manquants pour ces noeuds) ou citent au moins deux autre noeuds, ce qui peut faire des ``ponts'' entre sous-communautés. Le réseau obtenu a une taille de $\left|V\right| = 107164$ noeuds et $\left|E\right| = 309778$ liens. Celui-ci est visualisé en Fig.~\ref{fig:cybergeo:fig5}.
}

\begin{figure}
\includegraphics[width=\linewidth]{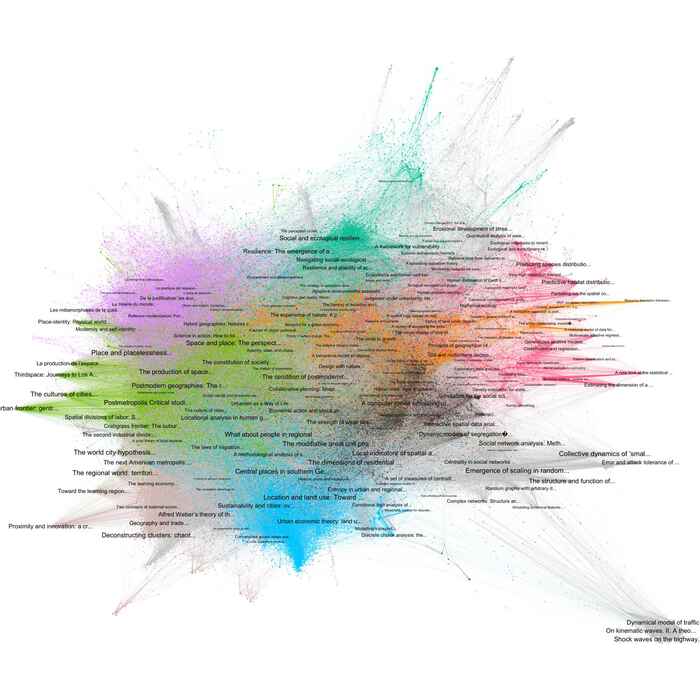}
\appcaption{\textbf{Citation Network.} We show only the ``core'' of the citation network, composed by references with a degree larger than one ($\left|V\right| = 107164$ and $\left|E\right| = 309778$). The community detection algorithm provides 29 communities with a modularity of 0.71. Nodes and edges color gives the main communities (for example ecology in magenta, GIS in orange, Socio-ecology in turquoise, Social geography in green, Spatial analysis in blue). Node labels give shortened titles of most cited papers, size is scaled according to their in-degree. The graph is spatialized using a Force-Atlas algorithm.\label{fig:cybergeo:fig5}}{\textbf{Réseau de citations.} Nous visualisons uniquement le ``coeur'' du réseau de citation, composé des références avec un degré plus grand que 1 ($\left|V\right| = 107164$ et $\left|E\right| = 309778$). L'algorithme de détection de communautés fournit 29 communautés avec une modularité de 0.71. Les couleurs des noeuds et des liens donnent les communautés principales (par exemple l'écologie en magenta, GIS en orange, la Socio-écologie en turquoise, la géographie sociale en vert, l'analyse spatiale en bleu). Les labels des noeuds donnent les titres raccourcis des références les plus citées, la taille étant à l'échelle de leur degré entrant. Le graphe est spatialisé par l'utilisation d'un algorithme Force-Atlas.\label{fig:cybergeo:fig5}}
\end{figure}

\bpar{
We use a standard modularity optimization algorithm to identify communities~\citep{blondel2008fast} in this citation network. It provides 29 communities with a modularity of 0.71. In comparison, a bootstrap of 100 randomisations of links in the network gives an average modularity of $-1.0\cdot 10^{-4} \pm 4.4\cdot 10^{-4}$ which means that communities are highly significant.
}{
Nous utilisons un algorithme standard d'optimisation de modularité pour identifier des communautés~\citep{blondel2008fast} dans ce réseau de citation. Il fournit 29 communautés avec une modularité de 0.71. En comparaison, un bootstrap de 100 tirages aléatoires des liens dans le réseau donne une modularité moyenne de $-1.0\cdot 10^{-4} \pm 4.4\cdot 10^{-4}$ ce qui signifie une forte significativité des communautés.
}

\bpar{
We name the communities by inspection of the titles of most cited references in each. The 14 communities that have a size larger than 2.5\% of the network are : Complex Networks, Ecology, Social Geography, Sociology, GIS, Spatial Analysis, Agent-based Modeling and Simulation (ABMS), Socio-ecology, Urban Networks, Urban Simulation, Urban Studies, Economic Geography, Accessibility/Land-use, Time Geography. These categories do not directly correspond to well-defined disciplines, as some correspond more to methods (ABMS), objects of study (Urban Studies), or paradigms (Complex Networks). Some are ``specializations'' of others : most papers in Urban Studies can also be classified as Critical and Social geography. This way, we construct endogenous disciplines that correspond to \emph{scientific practices} (what is cited) more than their representation (the ``official'' disciplines). The relative positioning of communities in Fig.~\ref{fig:cybergeo:fig5}, obtained with a Force-Atlas algorithm, tells a lot about their respective relations : for example, social geography makes a bridge between Urban Studies and Economic Geography, whereas the connection between Socio-ecology and Urban simulations is done by GIS (what can be expected as geomatics is an interdisciplinary field). GIS also separates and connects two subfield of Ecology, on one side more thematic studies on ecological habitats, and on the other sides statistical methods. These relations already inform qualitatively patterns of interdisciplinarity, in the sense of integration measures. We will also in the following use these communities to situate the semantic classification.
}{
Nous nommons les communautés par inspection des titres des références les plus citées dans chaque. Les 14 communautés qui ont une taille plus grande que 2.5\% du réseau sont : \textit{Complex Networks, Ecology, Social Geography, Sociology, GIS, Spatial Analysis, Agent-based Modeling and Simulation (ABMS), Socio-ecology, Urban Networks, Urban Simulation, Urban Studies, Economic Geography, Accessibility/Land-use, Time Geography}. Ces catégories ne correspondent par directement à des disciplines bien définies, puisque certaines correspondent plus à des méthodes (\textit{ABMS}), des objets d'étude (\textit{Urban Studies}), ou des paradigmes (\textit{Complex Networks}). Certains sont des ``spécialisations'' d'autres : la plupart des travaux en \textit{Urban Studies} peuvent aussi être classifiés comme géographie critique ou sociale. De cette façon, nous construisons des disciplines endogènes qui correspondent à des \emph{pratiques scientifiques} (ce qui est cité) plus qu'à leur représentations (les disciplines ``officielles''). Le positionnement relatif des communautés en Fig.~\ref{fig:cybergeo:fig5}, obtenu par un algorithme de Force-Atlas, en dit long sur leurs relations respectives : par exemple la géographie sociale fait une pont entre les \textit{Urban Studies} et l'économie géographique, tandis que la connection entre socio-écologie et les simulations urbaines est fait par le GIS (ce qui pouvait être attendu car la géomatique est un champ interdisciplinaire). Le GIS sépare également et connecte deux sous-champs de l'écologie, d'une part des études plus thématiques sur les habitats écologiques, et d'autre part des méthodes statistiques. Ces relations informent déjà qualitativement des motifs d'interdisciplinarité, au sens de mesures d'intégration. Nous allons par la suite utiliser ces communautés pour situer la classification sémantique.
}

\subsubsection{Semantic Communities Construction}{Construction des communautés sémantiques}

\bpar{
We now turn to the methodological details for the construction of the semantic classification. This step adapts the methodology described by~\cite{bergeaud2017classifying}, who construct a semantic classification on patent data.
}{
Nous présentons à présent les détails méthodologiques de la construction de la classification sémantique. Cette étape adopte la méthodologie décrite par~\cite{bergeaud2017classifying}, qui construit une classification sémantique sur données de brevets.
}

\paragraph{Relevant Keywords Extraction}{Extraction des Mots-clés pertinents}

\bpar{
We recall that our corpus with available text consists of around $2\cdot 10^5$ abstracts of publications at a topological distance shorter than 2 from the journal \textit{Cybergeo} in the citation network. The first important step is to extract relevant keywords from abstracts. Text processing is done with the python library \texttt{nltk}~\citep{bird2006nltk}. We add a particular treatment to the method of~\cite{bergeaud2017classifying}, as our corpus is multilingual: language detection is done with the technique of \emph{stop-words}~\citep{baldwin2010language}. We also use a specific tagger (the function allowing the attribution of grammatical function to words), \texttt{TreeTagger}~\citep{schmid1994probabilistic}, for languages other than English.
}{
Nous rappelons que notre corpus avec des textes disponibles consiste en environ $2\cdot 10^5$ résumés des publications à une distance topologique plus petite que 2 du journal \textit{Cybergeo} dans le réseau de citation. La première étape importante est l'extraction de mots-clés pertinents à partir des résumés. L'analyse textuelle est effectué avec la bibliothèque python \texttt{nltk}~\cite{bird2006nltk}. Nous ajoutons un traitement particulier à la méthode de~\cite{bergeaud2017classifying}, puisque notre corpus est multilingue : la détection de langue est faite par technique des \emph{stop-words}~\citep{baldwin2010language}. Nous utilisons également un tagger spécifique (la fonction permettant l'attribution de fonctions grammaticales aux mots), \texttt{TreeTagger}~\citep{schmid1994probabilistic}, pour les langues autres que l'anglais.
}

\bpar{
To summarize, the keyword extraction workflow goes through the following steps :
}{
En résumé, le flux d'extraction des mots-clés suit les étapes suivantes :
}

\bpar{
\begin{enumerate}
\item Language detection is done using \textit{stop-words}
\item Pos-tagging (detection of word functions) and stemming (extraction of the \emph{stem}) are done differently depending on language :
\begin{itemize}
\item English : \texttt{nltk} built-in pos-tagger, combined to a \emph{PorterStemmer}
\item French or other : use of \texttt{TreeTagger}~\citep{schmid1994probabilistic}
\end{itemize}
\item Selection of potential \textit{n-grams} (keywords of length $n$ with $1 \leq n \leq 4$) following the given grammatical rules: for English $\bigcap \{NN \cup VBG \cup JJ \}$, and for French $\bigcap \{NOM \cup ADJ\}$. Other languages are a negligible proportion of the corpus and are discarded.
\item Estimation of the relevance \textit{n-grams}, by attributing a score following the deviation of the statistical distribution of co-occurrences to a random distribution.
\end{enumerate}
}{
\begin{enumerate}
	\item La détection de la langue est faite par utilisation des \textit{stop-words}
	\item Le \emph{pos-tagging} (détection des fonctions des mots) et le \emph{stemming} (extraction du \emph{stem}) sont effectués différemment selon la langue :
	\begin{itemize}
		\item Anglais : \emph{pos-tagger} intégré à \texttt{nltk}, combiné à un \emph{PorterStemmer}
		\item Français ou autre : utilisation de \texttt{TreeTagger}~\citep{schmid1994probabilistic}
	\end{itemize}
	\item Sélection de \textit{n-grams} potentiels (mots-clés de longueur $n$ avec $1 \leq n \leq 4$) suivant les règles grammaticales données : pour l'anglais $\bigcap \{NN \cup VBG \cup JJ \}$, et pour le français $\bigcap \{NOM \cup ADJ\}$. Les autres langues sont une proportion négligeable du corpus et ne sont pas pris en compte.
	\item Estimation de la pertinence des \textit{n-grams}, par attribution d'un score suivant la déviation des distribution statistiques des co-occurrences à une distribution aléatoire.
\end{enumerate}
}

\paragraph{Semantic Network}{Réseau sémantique}

\bpar{
We keep at this stage a fixed number $K_W$ of \textit{n-grams}, based on their relevance score, that will be designated as the relevant keywords. We find that for large values of $K_W$, results are not sensitive to the total number of keywords, and take a reasonably large value for computational performance, $K_W = 50,000$. We construct the co-occurrence matrix of the relevant keywords. This co-occurrence matrix provides the semantic network as its adjacency matrix : nodes are keywords, and they are linked according to their co-occurrences.
}{
Nous conservons à cette étape un nombre fixé $K_W$ de \textit{n-grams}, en se basant sur leur score de pertinence, qui seront désignés comme mots-clés pertinents. Nous observons que pour des grandes valeurs de $K_W$, les résultats ne sont pas sensibles au nombre total de mots-clés, et prenons ainsi une grande valeur raisonnable pour la performance computationnelle, $K_W = 50,000$. Nous construisons la matrice de co-occurrence des mots-clés pertinents. Cette matrice de co-occurrence fournit le réseau sémantique comme sa matrice d'adjacence : les noeuds sont les mots-clés, et ils sont reliés en fonction de leurs co-occurrences.
}

\paragraph{Sensitivity Analysis}{Analyse de sensibilité}

\bpar{
We observe the same phenomenon than in~\cite{bergeaud2017classifying}, that is the existence of nodes with large degree and not specific to a particular field : for example \texttt{model} and \texttt{space} are used in most of subfields of Geography. We also adapt the original filtering procedure, as we do not have here an exogenous information to calibrate parameters. We assume the highest degree terms do not carry specific information on particular classes and can be thus filtered given a maximal degree threshold $k_{max}$. We keep the second filter on a minimal edge weight threshold $\theta_w$. We add the supplementary constraint that keywords are also filtered on a document frequency window $\left[ f_{min},f_{max} \right]$ (number of references in which they appear), what is slightly different from network filtering.
}{
Nous observons un phénomène similaire à celui observé par~\cite{bergeaud2017classifying}, qui est la présence de noeuds avec un grand degré mais non spécifiques à un champ particulier : par exemple \texttt{model} et \texttt{space} sont utilisés dans la majorité des sous-champs de la Géographie. Nous adaptons également la procédure de filtration originale, puisque nous ne disposons pas ici d'information exogène pour calibrer les paramètres. Nous supposons que les termes de plus haut degré ne portent pas d'information sur des classes en particulier et peuvent ainsi être filtrés étant donné un degré maximal $k_{max}$. Nous gardons le second filtre sur un seuil de poids minimal des liens $\theta_w$. Nous ajoutons la contraintes supplémentaires que les mots-clés sont aussi filtrés par leur fréquence d'apparition dans les documents sur une fenêtre $\left[ f_{min},f_{max} \right]$ (nombre de références dans lesquelles ils apparaissent), ce qui est légèrement différent de la filtration du réseau.
}

\bpar{
A sensitivity analysis of resulting network topology to these four parameters is presented in Fig.~\ref{fig:cybergeo:fig6}. Given a filtered network, we detect communities using modularity optimization as before for the citation network. Various properties of the network can be optimized, and we look in particular at its size (number of keywords after filtering), the optimal modularity, the number of communities, and the balance between their sizes (defined as a concentration index $\sum_k s_k^2 / (\sum_k s_k)^2$). This multi-objective optimization problem does not have a unique solution as objectives are contradictory in a complex way, and a compromise point must be chosen. We take a compromise point between modularity and network size, with a high balance and a reasonable number of communities, given by $k_{max} = 1200, \theta_w = 100, f_{min} = 50, f_{max} = 10000$. These values give a network of size 2868, with 18 communities and a modularity of 0.57.
}{
Une analyse de sensibilité de la topologie du réseau résultant à ces quatre paramètres est présentée en Fig.~\ref{fig:cybergeo:fig6}. Etant donné un réseau filtré, nous détectons les communautés en utilisant une optimisation de la modularité comme ci-dessus pour le réseau de citation. Diverses propriétés du réseau peuvent être optimisées, et nous nous intéressons en particulier à sa taille (nombre de mots-clés après filtrage), la modularité optimale, le nombre de communautés, et l'équilibre entre leurs tailles (défini comme un indice de concentration $\sum_k s_k^2 / (\sum_k s_k)^2$). Ce problème d'optimisation multi-objectif ne possède pas de solution unique car les objectifs sont contradictoires de manière complexes, et un point de compromis doit être choisit. Nous prenons un point compromis entre modularité et taille du réseau, avec un fort équilibre et un nombre raisonnable de communautés, donné par $k_{max} = 1200, \theta_w = 100, f_{min} = 50, f_{max} = 10000$. Ces valeurs donnent un réseau de taille 2868, avec 18 communautés et une modularité de 0.57.
}

\bpar{
Note that the small proportion of keywords in French is always separated from the rest of the network as they cannot co-occur with English keywords, and that with these parameter settings no French keywords are kept. All communities described in the following therefore contain only keywords in English.
}{
Notons que la petite proportion de mots-clés en français est toujours séparée du reste du réseau puisqu'ils ne peuvent pas coïncider avec des mots-clés anglais, et qu'avec ces valeurs des paramètres aucun mot-clé français n'est conservé. L'ensemble des communautés décrites par la suite ne contient pour cette raison que des mots-clés en anglais.
}


\begin{figure}
\includegraphics[width=\linewidth]{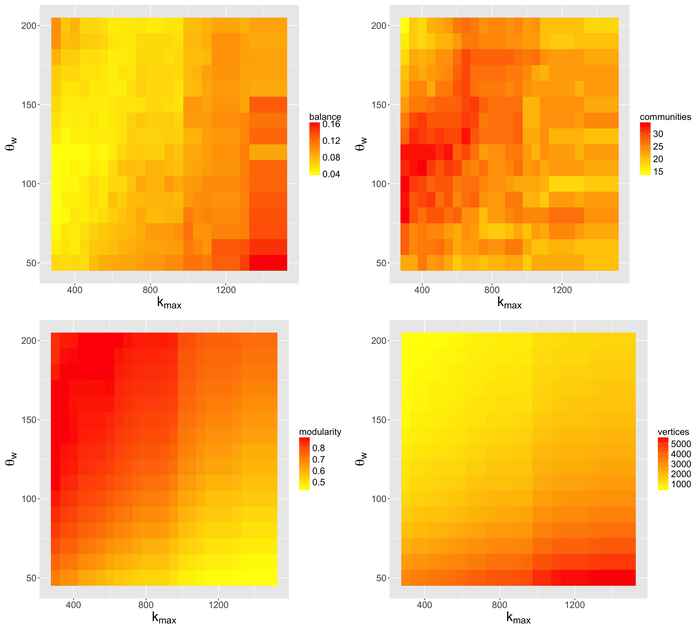}
\appcaption{\textbf{Sensitivity analysis of network indicators to filtering parameters.} We show here 4 indicators (balance between community sizes, modularity of the decomposition, number of communities, number of vertices), as a function of parameters $k_{max}$ and $\theta_w$, at fixed $f_{min} = 50, f_{max} = 10000$. Close values for these two last parameters (in a reasonable range) give similar behavior.\label{fig:cybergeo:fig6}}{\textbf{Analyse de sensibilité des indicateurs de réseau aux paramètres de filtrage.} Nous donnons ici 4 indicateurs (équilibre entre les tailles des communautés, modularité de la décomposition, nombre de communautés, nombre de noeuds), comme fonction des paramètres $k_{max}$ et $\theta_w$, à des valeurs fixées $f_{min} = 50, f_{max} = 10000$. Des valeurs proches pour ces deux derniers paramètres (dans des bornes raisonnables) donne un comportement similaire.\label{fig:cybergeo:fig6}}
\end{figure}

\paragraph{Semantic Communities}{Communautés sémantiques}

\bpar{
We obtain therein communities in the semantic network with the optimized filtering parameters. At the exception of a small proportion apparently resulting from noise (representing less than 10 keywords in 3 communities that we remove, i.e. 0.33\% of keywords), communities correspond to well-defined scientific fields, domains, or approaches. Naming is also done by inspection of the most relevant keywords in each community, in order to stick here to a certain level of supervision.
}{
Nous obtenons ainsi des communautés dans le réseau sémantique avec les paramètres de filtration optimisés. A l'exception d'une petite proportion s'apparentant apparemment à du bruit (représentant moins de 10 mots-clés dans 3 communautés que nous supprimons, i.e. 0.33\% des mots-clés), les communautés correspondent à des champ scientifiques, domaines, ou approches bien définis. La dénomination est également faite par inspection des mots-clés les plus pertinents dans chaque communauté, afin de se tenir ici à un certain niveau de supervision.
}

\begin{table}
\appcaption{\textbf{Semantic communities reconstructed from community detection in the semantic network.}\label{tab:cybergeo:domains}}{\textbf{Composition des communautés sémantiques.} Elles sont construites par détection de communautés dans le réseau sémantique.\label{tab:cybergeo:domains}}
\hspace{-1cm}
\begin{tabular}{lll}
\hline\noalign{\smallskip}
Name & Size & Keywords  \\
\noalign{\smallskip}\hline\noalign{\smallskip}
Political sciences/critical geography & 535 & \texttt{decision-mak, polit ideolog, democraci, stakehold}\\
Biogeography & 394 & \texttt{plant densiti, wood, wetland, riparian veget} \\
Economic geography & 343 &  \texttt{popul growth, transact cost, socio-econom, household}\\
Environnment/climate & 309 & \texttt{ice sheet, stratospher, air pollut, climat model} \\
Complex systems & 283 & \texttt{scale-fre, multifract, agent-bas model, self-organ} \\
Physical geography & 203 & \texttt{sedimentari, digit elev model, geolog, river delta} \\
Spatial analysis & 175 & \texttt{spatial analysi, princip compon analysi, heteroscedast}\\
Microbiology & 118 & \texttt{chromosom, phylogenet, borrelia} \\
Statistical methods & 88 & \texttt{logist regress, classifi, kalman filter, sampl size} \\
Cognitive sciences & 81 & \texttt{semant memori, retrospect, neuroimag} \\
GIS & 75 & \texttt{geograph inform scienc, softwar design, volunt gi}\\
Traffic modeling & 63 & \texttt{simul model, lane chang, traffic flow, crowd behavior} \\
Health & 52 & \texttt{epidem, vaccin strategi, acut respiratori syndrom}\\
Remote sensing & 48 & \texttt{land-cov, landsat imag, lulc} \\
Crime & 17 & \texttt{crimin justic system, social disorgan, crime} \\
\noalign{\smallskip}\hline
\end{tabular}
\end{table}

\begin{figure}
\includegraphics[width=\linewidth]{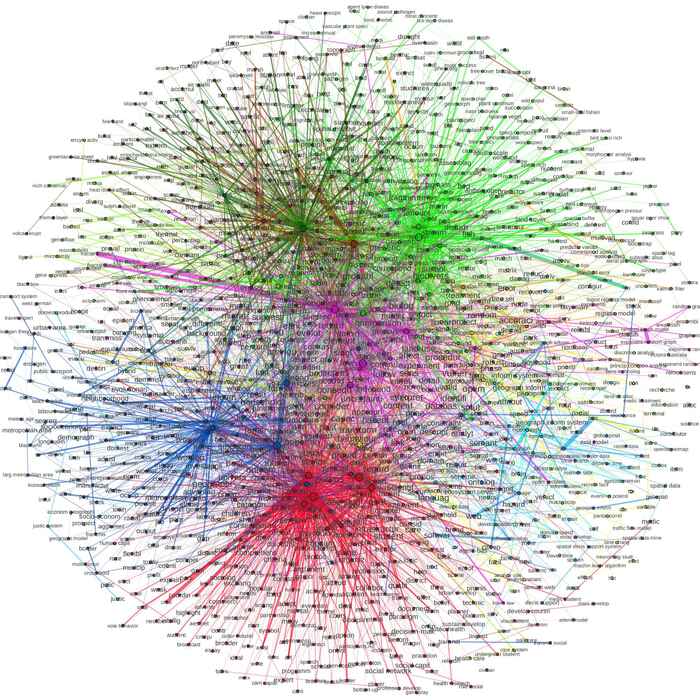}
\appcaption{\textbf{Visualization of the semantic network.} Network is constructed by co-occurrences of most relevant keywords. Filtering parameters are here taken according to the multi-objective optimization done in Fig.~\ref{fig:cybergeo:fig6}, i.e. $(k_{max}=1200,\theta_w=100,f_{min}=50,f_{max}=10000)$. The graph spatialization algorithm (Fruchterman-Reingold), despite its stochastic and path-dependent character, unveils information on the relative positioning of communities.\label{fig:cybergeo:fig7}}{\textbf{Visualisation du réseau sémantique.} Le réseau est construit par co-occurrences des mots-clés les plus pertinents. Les paramètres de filtrage sont pris ici selon l'optimisation multi-objectifs faite en Fig.~\ref{fig:cybergeo:fig6}, i.e. $(k_{max}=1200,\theta_w=100,f_{min}=50,f_{max}=10000)$. L'algorithme de spatialisation du graphe (Fruchterman-Reingold), malgré son caractère stochastique et dépendant au chemin, révèle de l'information sur le positionnement relatif des communautés.\label{fig:cybergeo:fig7}}
\end{figure}

\begin{figure}
\includegraphics[width=\linewidth]{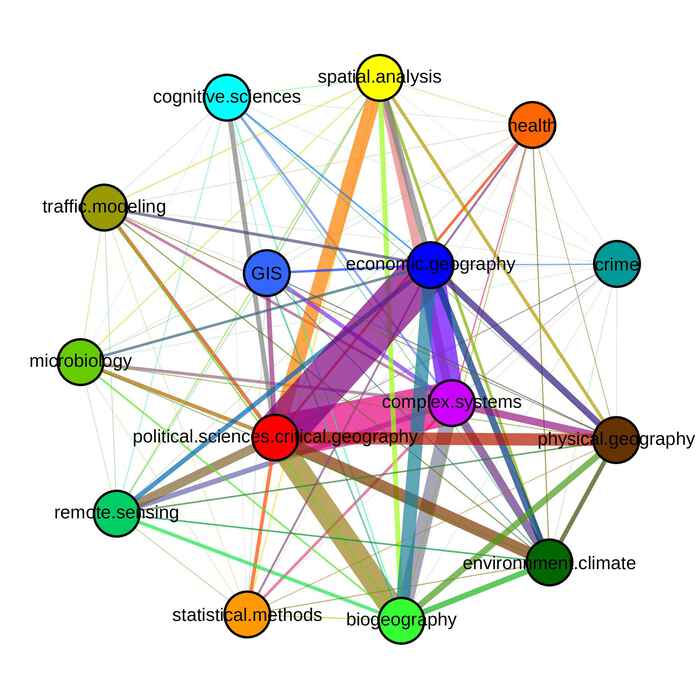}
\appcaption{\textbf{Synthesis of semantic communities and their links.} Weights of links are computed as probabilities of co-occurrences of corresponding keywords within references.\label{fig:cybergeo:fig8}}{\textbf{Synthèse des communautés sémantiques et de leurs liens.} Les poids des liens sont calculés comme les probabilités de co-occurrence des mots-clés correspondants au sein des références.\label{fig:cybergeo:fig8}}
\end{figure}

\bpar{
Table~\ref{tab:cybergeo:domains} summarizes the communities, giving their names, sizes, and corresponding keywords. The most important community is related to issues in political science and critical geography, what could have been expected as several previously obtained citations communities (Social geography, Urban studies) deal with these issues. We then obtain a large cluster of terms related to biogeography, that must correspond to publications in Ecology and Socio-ecology identified before, together with a community in Environment and Climate.
}{
La Table~\ref{tab:cybergeo:domains} résume les communautés, donnant leurs noms, tailles, et mots-clés correspondants. La communauté la plus importante est en rapport avec des questions de Sciences politiques et de Géographie critique, ce qui pouvait être attendu puisque plusieurs communautés de citation obtenues précédemment (\textit{Social Geography}, \textit{Urban Studies}) s'occupent de ces questions. Nous obtenons ensuite un groupement conséquent en termes de biogéographie, qui doit correspondre aux publications en écologie et socio-écologie identifiées précédemment, avec une communauté en environnement et climat.
}

\bpar{
In a way similar to the citation communities, but more pronounced here, we obtain endogenous ``disciplines'' that can correspond to real disciplines, to methodologies, to object of studies. This classification thus also unveil \emph{effective scientific practices}, here in terms of semantic content. A class here related to complex systems can be associated to a paradigm and various approaches that were separated in the citation communities : agent-based models and complex networks for example. On the contrary, some studies that were gathered in a large domain before can be precisely differentiated in the semantic network, such as microbiology and health here that are used by studies related to socio-ecology or ecology in the citation network. Some very specific domains appear here as they have very few connections in their actual semantic content : for example, Geography of crime is very precise and disconnected from other communities.
}{
De manière similaire au communautés de citation, mais plus prononcée ici, nous obtenons des ``disciplines'' endogènes qui peuvent correspondre à des vraies disciplines, à des méthodologies, à des objets d'étude. Cette classification révèle pour cela également des \emph{pratiques scientifiques effectives}, ici en termes de contenu sémantique. Une classe ici en rapport avec les Systèmes Complexes peut être associées à un paradigme et à différentes approches qui étaient séparées dans les communautés de citation : les modèles basés-agent et les réseaux complexes par exemple. Au contraire, certaines études qui étaient rassemblées dans de larges domaines précédemment peuvent être différenciées précisément dans le réseau sémantique, comme la microbiologie et la santé ici qui sont utilisées par des études en relation à l'écologie et la socio-écologie dans le réseau de citation. Certains domaines très spécifiques apparaissent ici puisqu'ils ont très peu de connections avec les autres dans leur contenu sémantique : par exemple, la géographie du crime est très précise et déconnectée des autres communautés.
}

\bpar{
We show in Fig.~\ref{fig:cybergeo:fig7} a visualisation of the semantic network, in which the positioning of communities, induced by a Fruchterman-Reingold algorithm (that we use here to have a more precise layout in the relative positioning compared to Force Atlas~\citep{jacomy2014forceatlas2}). The bridging between distant disciplines is done quite differently compared to the citation network, and reveals thus qualitatively an other dimension of interdisciplinarity, i.e. the semantics shared by disciplines. Here, the communities corresponding to Economic Geography (blue) and to Critical Geography (red) are close as in the citation network, but are linked to ecology and geomorphology (green and brown) by Complex Systems (magenta), although these were not present as a community in the citation network. Complexity methodologies such as Fractals, Scaling~\citep{west2017scale} or Networks~\citep{newman2003structure} are indeed widely used both in social sciences and in physics or biology. The semantic analysis reveals thus that very distant disciplines, that are distant in their citation patterns, are finally close in terms of actual content.
}{
Nous montrons en Fig.~\ref{fig:cybergeo:fig7} une visualisation du réseau sémantique, dans lequel le positionnement des communautés, induit par un algorithme de Fruchterman-Reingold (que nous utilisons ici pour avoir un positionnement plus précis dans le positionnement relatif en comparaison du Force-Atlas~\citep{jacomy2014forceatlas2}). Les ponts entre disciplines distantes est effectué différemment en comparaison du réseau de citation, et révèle ainsi qualitativement une autre dimension de l'interdisciplinarité, i.e. la sémantique partagée par les disciplines. Ici, les communautés correspondant à l'Economie Géographique (bleu) et à la Géographie critique (rouge) sont proches dans le réseau de citation, mais sont liés à l'écologie et la géomorphologie (vert et marron) par les Systèmes Complexes (magenta), bien que ceux-ci n'étaient pas présents comme communauté dans le réseau de citation. Les méthodologies de la complexité comme les fractales, les loi d'échelles~\cite{west2017scale}, ou les réseaux~\cite{newman2003structure} sont en effet largement utilisés à la fois en sciences sociales et en physique ou biologie. L'analyse sémantique montre ainsi que des disciplines très distantes, qui sont distantes dans leur motifs de citation, sont finalement proches en termes de contenu observé.
}

\bpar{
In terms of overlaps between communities, in the sense of co-occurrences of corresponding keywords within texts of references, we show a synthesis of links between semantic communities in Fig.~\ref{fig:cybergeo:fig8}. We see that communities such as Critical Geography and Biogeography are not totally disconnected and share still a certain number of co-occurrences. More isolated communities can be spotted such as Health and Crime Geographies. Surprisingly, Statistical Methods does not share strong links with other communities, what could mean that articles dealing with methodological issues in this field are rather disconnected from the field of application, or at least do not describe it extensively. On the contrary, methods in Complex Systems are organically integrated with the thematic issues they tackle.
}{
En termes de chevauchements entre les communautés, au sens des co-occurrences des mots-clés correspondants dans les textes des références, nous montrons une synthèse des liens entre communautés sémantiques en Fig.~\ref{fig:cybergeo:fig8}. Nous voyons que les communautés comme la Géographie critique et la biogéographie ne sont pas totalement déconnectées et partagent finalement un certain nombre de co-occurrences. Des communautés plus isolées peuvent être identifiées comme les géographies de la Santé et du Crime. De manière surprenante, les méthodes statistiques ne partagent pas de liens forts avec d'autres communautés, ce qui pourrait signifier que des articles traitant de questions méthodologiques dans ce champ sont plutôt déconnectées du champ d'application, ou au moins ne le décrivent pas en détail. Au contraire, les méthodes en Systèmes Complexes sont organiquement intégrées avec les questions thématiques qu'ils traitent.
}

\subsubsection{Semantic composition of citation communities}{Composition sémantique des communautés de citation}

\begin{figure}
\includegraphics[width=\linewidth]{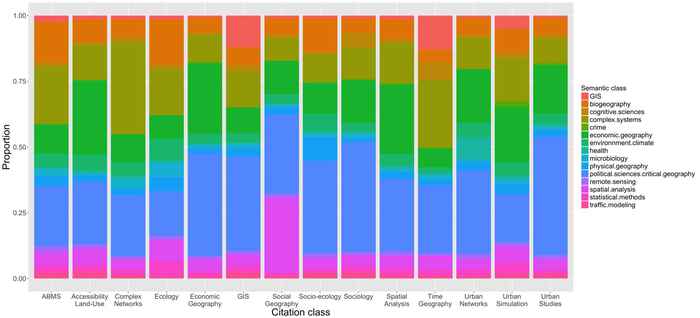}
\appcaption{\textbf{Composition of citation communities in terms of semantic content.} For each citation class (horizontally), the bar is decomposed as the proportions of each semantic class (given by color).\label{fig:cybergeo:fig9}}{\textbf{Composition des communautés de citation en termes de contenu sémantique.} Pour chaque classe de citation (horizontalement), la barre est décomposée entre les proportions de chaque classe sémantique (données par la couleur).\label{fig:cybergeo:fig9}}
\end{figure}

\bpar{
We can now turn to the study of the relation between classifications. First, a simple way to link them is to look at the semantic content of citation communities. Each reference has a given proportion of keywords within each semantic class, and an average composition in terms of semantic classes for each citation class can thus be computed. We show these composition in Fig.~\ref{fig:cybergeo:fig9}. Some expected results are obtained, such as Complex Networks (citation) having the largest part in Complex Systems (semantic), or GIS (citation) the largest in GIS (semantic), and similarly for Economic Geography.
}{
Nous pouvons à présent nous tourner vers l'étude des relations entre classifications. Tout d'abord, une façon simple de les relier est d'étudier le contenu sémantique des communautés de citations. Chaque référence a une proportion donnée de mots-clés dans chaque classe sémantique, et une composition moyenne en termes de classes sémantiques pour chaque classe de citation peut ainsi être calculée. Nous montrons ces compositions en Fig.~\ref{fig:cybergeo:fig9}. Des résultats attendus sont obtenus, comme \textit{Complex Networks} (citation) ayant la proportion la plus forte en \textit{Complex Systems} (sémantique), ou le GIS (citation) le plus fort en GIS (sémantique), et de même pour l'économie géographique.
}

\bpar{
But the study of patterns that could have not been expected is very informative, and unveils practices of interdisciplinarity. For example, Time Geography (citation) uses as much GIS (semantic) as GIS (citation), what means that they should be using the corresponding methods and tools to study the thematic question of spatio-temporal trajectories of geographical agents. The most important in terms of political science (semantic) are Urban Studies, what suggest a convergence of the City as an object of study and of the disciplines of Political Science and Critical Geography. Also interestingly, the citation communities using most biogeography are Ecology (what could have been expected) and ABMS, confirming again the role of the thematic application in complex systems methodologies.
}{
Mais l'étude de motifs qui auraient pu ne pas être attendus est très informatif, et dévoile des pratiques d'interdisciplinarité. Par exemple, la \textit{Time Geography} (citation) utilise quasiment autant de GIS (sémantique) que le GIS (citation), ce qui signifie qu'ils doivent utiliser les méthodes et outils correspondants pour étudier la question thématique des trajectoires spatio-temporelles des agents géographiques. Le plus important en termes de sciences politiques (sémantique) sont les \textit{Urban Studies}, ce qui suggère une convergence de la Ville comme objet d'étude et des disciplines de Sciences politiques et de Géographie critique. Egalement de manière intéressante, les communautés de citation utilisant le plus la biogéographie sont l'écologie (ce qui pouvait être attendu) et les ABMS, confirmant ainsi le rôle de l'application thématique dans les méthodologies des Systèmes Complexes.
}

\subsubsection{Measuring interdisciplinarity}{Mesure de l'interdisciplinarité}

\begin{figure}
\includegraphics[width=\linewidth,height=0.9\textheight]{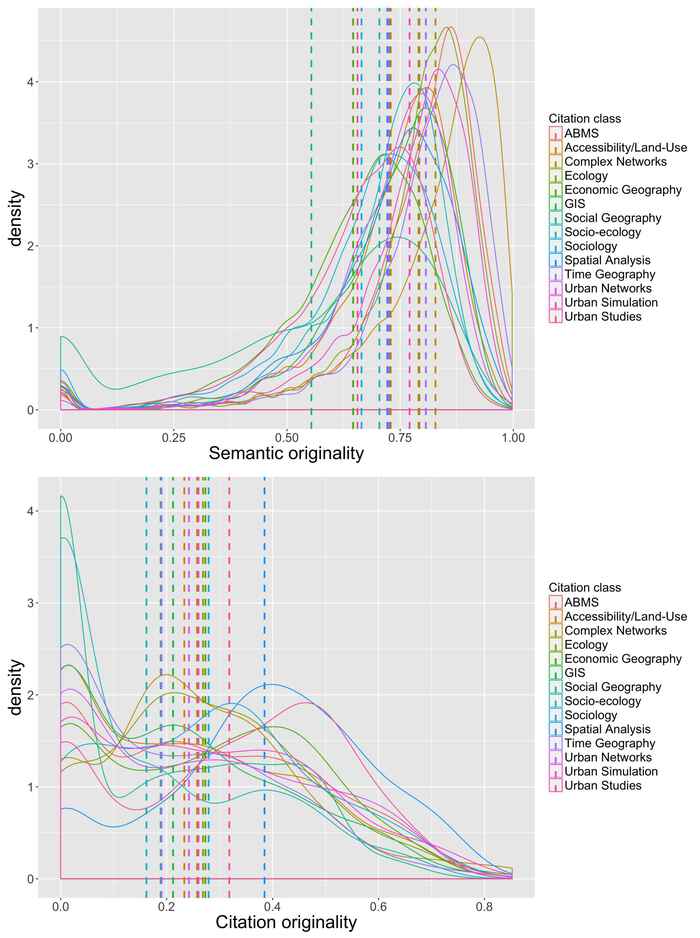}
\appcaption{\textbf{Statistical distribution of originalities.} We show the smoothed probability densities of originality indexes, by citation class (given by color), for the Semantic originality $o^{(Semantic)}$ (top plot) and for the Citation originality $o^{(Citation)}$ (bottom plot). Dashed lines give the mean for each distribution, with the corresponding color.\label{fig:cybergeo:fig10}}{\textbf{Distribution statistique des originalités.} Nous donnons les densités de probabilité lissées des indices d'originalité, par classe de citation (donnée par la couleur), pour l'originalité sémantique $o^{(Semantic)}$ (\textit{Haut}) et pour l'originalité de citation $o^{(Citation)}$ (\textit{Bas}). Les lignes pointillées donnent la moyenne pour chaque distribution, avec la couleur correspondante.\label{fig:cybergeo:fig10}}
\end{figure}

\bpar{
We had up to now a qualitative view on interdisciplinarity patterns, by looking at the relative localisation of communities within the citation and semantic classifications, and the relation between the classifications. We propose now to look at quantitative measures of interdisciplinarity, for each classification.
}{
Nous avons eu jusqu'à présent une vue qualitative sur les motifs d'interdisciplinarité, en s'intéressant à la localisation relative des communautés au sein des classifications de citation et sémantique, et la relation entre les classifications. Nous proposons à présent de regarder des mesures quantitatives de l'interdisciplinarité, pour chaque classification.
}

\bpar{
More precisely, for a given classification $C \in \{ Citation,Semantic\}$ a reference $i$ can be viewed as a probability vector $(p_{ij}^{(C)})_j$ on classes $j$ that give for each class the probability to belong to it. Given this setting, we measure interdisciplinarity of one reference using Herfindhal concentration index~\citep{porter2009science}, that can also be called an originality index. We define originality as
}{
Plus précisément, pour une classification donnée $C \in \{ Citation,Semantic\}$ une référence $i$ peut être représentée par un vecteur de probabilités $(p_{ij}^{(C)})_j$ sur les classes $j$ qui donne pour chaque classe la probabilité d'y appartenir. Etant donné cette configuration, nous mesurons l'interdisciplinarité d'une référence en utilisant une concentration d'Herfindhal~\cite{porter2009science}, qui peut aussi être appelée indice d'originalité. Nous définissons l'originalité par
}
\[
o_i^{(C)} = 1 - \sum_j {p_{ij}^{(C)}}^2
\]
\bpar{
For the semantic classification, probabilities are defined as the proportion of keywords of the abstract within each semantic class. With the deterministic citation classification, each reference has only one class and the originality index is always 0. Therefore in order to be able to compare the two classification, we associate a probability to each citation class for each article as the proportion of citations received from this class. The induced index is original, and measures interdisciplinarity as \emph{how a reference is used} by different disciplines in its lifetime.
}{
Pour la classification sémantique, les probabilités sont définies comme la proportion de mots-clés du résumé dans chaque classe sémantique. Avec la classification de citation déterministe, chaque référence a une classe unique et l'index d'originalité est toujours nul. Pour cette raison, afin de pouvoir comparer les deux classifications, nous associons une probabilité à chaque classe de citation pour chaque article comme la proportion de citations reçues de cette classe. L'indice induit est original, et mesure l'interdisciplinarité comme la façon dont une référence \emph{est utilisée} par différentes disciplines pendant sa vie.
}

\bpar{
We show in Fig.~\ref{fig:cybergeo:fig10} the statistical distribution for both indexes $o^{(Semantic)}$ and $o^{(Citation)}$, stratified by citation class. This allow a direct comparison between the two and also an indirect comparison by the variation of semantic distribution between citation classes. For the distribution of semantic originalities, all citation classes exhibit a similar pattern, that is a peak around large values and a smaller peak at zero. It means that either references are highly specialized and have keywords in one class only, or they use keywords from different classes in a quite even manner (for comparison, an abstract with half keywords in a class and half in an other gives an originality of 0.5). The most original, i.e. the most mixed, citation class, is Complex Networks, with a distribution clearly detached from others, what would confirm their use as a method with a lot of different problems. Social Geography is from far the less original, with a large number of single class references, and an average far lower than other classes, what would mean an increased presence of compartmentalization within the associated disciplines.
}{
Nous montrons en Fig.~\ref{fig:cybergeo:fig10} la distribution statistique des deux indices $o^{(Semantic)}$ et $o^{(Citation)}$, stratifiées par classe de citation. Cela permet une comparaison directe entre les deux et également une comparaison indirecte par la variation des distributions sémantiques selon les classes de citations. Pour la distribution des originalités sémantiques, l'ensemble des classes de citation présentent un motif similaire, qui correspond à un pic autour d'une grande valeur et un pic plus faible à zéro. Cela signifie que soit les références sont fortement spécialisées et n'ont des mots-clés que dans une seule classe, ou bien elles utilisent des mots-clés de différentes classes de façon relativement équilibrée (pour comparaison, un résumé avec moitié de mots dans une classe et moitié dans une autre donne une originalité de 0.5). La classe de citation la plus originale, i.e. la plus mélangée, est \textit{Complex Networks}, avec une distribution clairement détachée des autres, ce qui confirmerait leur utilisation comme une méthode pour un certain nombre de problèmes différents. La Géographie sociale est de très loin la moins originale, avec un grand nombre de références à classe unique, et une moyenne bien plus basse que celle des autres classes, ce qui signifierait une présence accrue d'isolation dans les disciplines associées.
}

\bpar{
In terms of citation originality index, the global picture is fundamentally different, as average originality indexes are all lower than 0.4 and most of distributions show their mode in 0, meaning that most references are only cited by their own citation class. Again, Social Geography is the less original, confirming a similar behavior in terms of citation practice than in terms of research content. The most original classes in average, with a peak in large values, are Spatial Analysis and Urban Simulation: this corresponds to the fact that these class feature quite generic methods that can be applied in several fields and are cited accordingly. Complex Networks do not reach the same level, but however exhibit a peak around 0.2 and no peak in 0, together with Ecology, suggesting disciplines having still significant impact in other disciplines.
}{
En termes d'indice d'originalité de citation, le comportement général est fondamentalement différent, les indices d'originalité moyens étant tous inférieurs à 0.4 et la plupart des distributions ont leur mode en 0, ce qui signifie que la majorité des références sont citées uniquement par leur propre classe de citation. A nouveau, la Géographie sociale est la moins originale, confirmant un comportement similaire en termes de pratiques de citation qu'en termes de contenu de la recherche. Les classes les plus originales en moyenne, avec un pic dans les grandes valeurs, sont \textit{Spatial Analysis} et \textit{Urban Simulation} : cela correspond au fait que ces classes contiennent des méthodes relativement génériques qui peuvent être appliquées dans différents champs et sont citées de manière appropriée. Les \textit{Complex Networks} n'atteignent pas le même niveau, mais présentent cependant un pic autour de 0.2 et pas de pic en 0, de la même manière que l'écologie, suggérant des disciplines ayant toujours un impact significatif sur les autres disciplines.
}

\bpar{
To summarize, we show (i) different patterns of interdisciplinarity, depending on disciplines, in terms of scientific content (semantic) and of scientific impact (citation); and (ii) a strong qualitative difference in behavior of originalities between the two classifications, what suggests their complementarity.
}{
En résumé, nous montrons (i) différents motifs d'interdisciplinarité, selon les disciplines, en termes de contenu scientifique (sémantique) et d'impact scientifique (citation) ; et (ii) une forte différence qualitative de comportement des originalités entre les deux classifications, ce qui suggère leur complémentarité.
}

\subsubsection{Correlation between classifications}{Corrélation entre classifications}

\bpar{
In order to strengthen the idea of a complementarity of classifications, that would each capture different dimensions of processes of knowledge production, we finally look at the correlation matrix between classifications. We use this time effective class probabilities for the citation classification, i.e. a vector of zeros except with a one at the index of the class of the reference. We compute a Pearson correlation coefficient between classes $k$ (in semantic) and $k'$ (in citation) as
}{
Afin de renforcer l'idée d'une complémentarité des classifications, qui captureraient chacune différentes dimensions des processus de production de connaissance, nous examinons finalement la matrice des corrélations entre les classifications. Nous utilisons cette fois les probabilités de classes effectives pour la classification de citation, i.e. un vecteur de zeros à l'exception d'un un à l'index de la classe de la référence. Nous calculons un coefficient de corrélation de Pearson entre les classes $k$ (sémantique) et la classe $k'$ (citation) comme
}
\[
\rho_{k,k'} = \frac{Cov \left[ (p^{(Sem)}_{ik})_i , (p^{(Cit)}_{ik'})_i \right]}{\sqrt{\Var\left[(p^{(Sem)}_{ik})_i\right]\Var\left[(p^{(Sem)}_{ik})_i\right]}}
\]

\bpar{
\noindent where the covariance is estimated with the unbiased estimator.
}{
\noindent où la covariance est estimée avec l'estimateur non biaisé.
}

\bpar{
The structure of the correlation matrix recalls the conclusions obtained when studying the semantic composition of citation communities, such as GIS being strongly correlated with GIS ($\rho=0.26$), or Sociology with Political Science ($\rho=0.16$). More importantly for our question are summary statistics of the overall matrix. It has a minimum of $-0.16$ (Ecology (citation) against Political Sciences (semantic)), an average of $-0.002$ and a maximum of $0.33$ (Social geography (citation) and Spatial Analysis (semantic)). The ``high'' values are highly skewed, as the first decile is at $-0.06$ and the last at $0.09$, what means that 80\% of coefficient lie within that interval, corresponding to low correlations. In a nutshell, classifications are consistent as highest correlations are observed where one can expect them, but most of classes are uncorrelated, meaning that the classifications are quite orthogonal and therefore complementary.
}{
La structure de la matrice de correlation rejoint les conclusions obtenues lors de l'étude de la composition sémantique des communautés de citation, comme le GIS étant fortement corrélé au GIS ($\rho=0.26$), ou la Sociologie à la Science politique ($\rho=0.16$). Plus importantes pour notre questions sont les statistiques de synthèse de la matrice entière. Elle a un minimum de $-0.16$ (\textit{Ecology} (citation) contre les sciences politiques (sémantique)), une moyenne de $-0.002$ et un maximum de $0.33$ (\textit{Social geography} (citation) et \textit{Spatial Analysis} (sémantique)). Les valeurs ``fortes'' sont fortement isolées, comme le premier décile est à $-0.06$ et le dernier à $0.09$, ce qui signifie que 80\% des coefficients tombent dans cet intervalle, correspondant à des corrélations faibles. En résumé, les classifications sont cohérentes puisque les corrélations les plus fortes sont observées où on peut les attendre, mais la majorité des classes ne sont pas corrélées, ce qui signifie que les classifications sont relativement orthogonales et ainsi complémentaires.
}

\subsection{Discussion}{Discussion}

\bpar{
We have this way shown the complementarity of classifications in the qualitative patterns they unveil, but also quantitatively in terms of interdisciplinarity measures and quantitatively in terms of correlations. Our work can be extended regarding several aspects, of which we give some suggestions below.
}{
Nous avons ainsi montré la complémentarité des classifications dans les motifs qualitatifs qu'elles dévoilent, mais aussi quantitativement en termes de mesures d'interdisciplinarité et quantitativement en termes de corrélations. Notre travail peut être étendu selon différents aspects, desquels nous donnons certaines suggestions par la suite.
}

\subsubsection{Further Developments}{Développements}

\bpar{
A first development consists in the comparison of journals. The starting point for construction of the scientific environment, the journal \textit{Cybergeo}, was the entry point but not the subject of our study. A development more focused on journals, trying for example to answer comparative issues, or to classify journals according to their effective level of interdisciplinarity regarding different dimensions, would be potentially interesting. The collection of precise data on the origin of references is however a first step that need to be solved first.
}{
Un premier développement consiste en la comparaison de journaux. Le point de départ pour la construction de l'environnement scientifique, le journal \textit{Cybergeo}, était le point d'entrée mais pas le sujet principal de notre étude. Un développement se concentrant plus sur les journaux, essayant par exemple de répondre à des questions comparatives, ou de classifier les journaux selon leur niveau effectif d'interdisciplinarité selon différentes dimensions, serait potentiellement intéressant. La collection de données précises sur l'origine des références est cependant un premier point qui doit d'abord être résolu.
}

\bpar{
The performance of the semantic classification was also not quantified here. A further validation of the relevance of using complementary information contained in both classifications could be done by the analysis of modularities within the citation network, as done in~\cite{bergeaud2017classifying}. This would however require a baseline classification to compare with, which is not available in the type of data we use. Open repository such as arXiv (for physics mainly) or Repec (for Economics) provide API to access metadata including abstracts, and could be starting points for such targeted case studies.
}{
La performance de la classification sémantique n'a également pas été quantifiée ici. Une validation approfondie de la pertinence en utilisant l'information complémentaire contenue dans les deux classifications pourrait être menée par l'analyse des modularités dans le réseau de citation, comme fait dans~\cite{bergeaud2017classifying}. Cela nécessiterait cependant une classification de référence pour comparaison, qui n'est pas disponible avec le type de données que nous utilisons. Des archives ouvertes comme arXiv (principalement pour la physique) ou Repec (pour l'économie) fournissent des API pour accéder aux métadonnées incluant les résumés, et pourraient être des points de départ pour de telles études ciblées.
}

\subsubsection{Applications}{Applications}

\bpar{
A first potential application of our methodology relies on the facts that both classifications unveils thematic domains (objects of study), classical disciplines, methodological communities. These different types of communities can indeed be understood as different \emph{Knowledge Domains}. \cite{raimbault2017applied} postulates co-evolving Knowledge Domains in every process of scientific knowledge production, that are Theoretical, Empirical, Modeling, Methodology, Tools and Data domains. Most of them are necessary for any process, and investigations within one conditions the advances in others. A refinement of classifications, associated with supervised classification to associate knowledge domains to some communities (potentially using full texts to have more precise information on the proportion of each knowledge domains involved in each), would allow to quantify relations between domains. Furthermore, using temporal data with the date of publications, would yield an effective quantification of the \emph{co-evolution} of domains in the sense of patterns of temporal correlations (e.g. Granger causality).
}{
Une première application potentielle de notre méthodologie se base sur le fait que les deux classifications dévoilent à la fois des domaines thématiques (objets d'étude), des disciplines classiques, des communautés méthodologiques. Ces différents types de communautés peuvent en effet être comprises comme différents \emph{Domaines de Connaissance}. \cite{raimbault2017applied} postule des Domaines de Connaissance en co-évolution dans tout processus de production de connaissance scientifique, qui sont les domaines Théorique, Empirique, des Modèles, Méthodologique, des Outils et des Données. La majorité sont nécessaire pour tout processus, et les investigations dans l'un conditionne les avances dans les autres. Un raffinement des classifications, associé à une classification supervisée pour associer des domaines de connaissance à des communautés (en utilisant potentiellement les textes complets pour avoir une information plus précise sur la proportion de chaque domaine de connaissance impliqué dans chaque), permettrait de quantifier les relations entre domaines. De plus, l'utilisation de données temporelles avec les dates des publications, fournirait une quantification effective de la \emph{co-évolution} des domaines au sens des motifs de corrélations temporelles (e.g. causalité de Granger).
}

\bpar{
An other interesting direction is the application of our classifications to the quantification of spatial diffusion of knowledge, as \cite{maisonobe2013diffusion} does for the diffusion of a specific question in genetics. It is not clear if different dimensions of knowledge diffuse the same way: for example citation practices can be correlated to social networks and thus exhibit different patterns than effective research contents. Therefore, our work would allow to study such questions from complementary point of views.
}{
Une autre direction intéressante est l'application de nos classifications à la quantification de la diffusion spatiale de la connaissance, comme \cite{maisonobe2013diffusion} fait pour la diffusion d'un question spécifique en génétique. Il n'est pas évident si des dimensions différentes de la connaissance se diffusent de la même façon : par exemple les pratiques de citation peuvent être corrélées aux réseaux sociaux et peuvent ainsi montrer différents motifs que les contenus effectifs de la recherche. Ainsi, notre travail permettrait d'étudier de telles questions de points de vue complémentaires.
}

\bpar{
Finally, we believe the tool we developed can contribute to an increased empowerment of authors and to the development of open science practices. Among the various visions of Open Science~\citep{fecher2014open}, the opening of data is always an important aspect, together with a development of reflexivity in all disciplines, beyond the sole Social Sciences to which it is classically associated. The first point is dealt with by our open tools for dataset construction, whereas the second is implied by the new knowledge of the different dimensions of the scientific environment we studied.
}{
Enfin, nous affirmons que les outils que nous avons développé peuvent contribuer à une autonomisation accrue des auteurs et au développement de pratiques de science ouverte. Au sein des différentes visions de la Science Ouverte~\cite{fecher2014open}, l'ouverture des données est toujours un aspect important, en même temps d'un développement de la réflexivité dans toutes les disciplines, au delà des seules sciences sociales auxquelles elle est classiquement associée. Le premier point est traité par nos outils ouverts pour la construction de jeux de données, tandis que le second est impliqué par la connaissance nouvelle des différentes dimensions de l'environnement scientifique que nous avons étudié.
}

\subsection{Conclusion}{Conclusion}

\bpar{
We have introduced a multi-dimensional approach to the understanding of interdisciplinarity, based on citation network and semantic network analysis. Starting from a generalist journal in Geography, we construct a large corpus of the citation neighborhood, from which we extract relevant keywords to elaborate a semantic classification. We then show qualitatively and quantitatively the complementarity of classifications. The methodology and associated tools are open and can be reused in similar studies for which data is difficult to access or poorly referenced in classical databases.
}{
Nous avons introduit une approche multi-dimensionnelle pour la compréhension de l'interdisciplinarité, basée sur les analyses du réseau de citation et du réseau sémantique. A partir d'un journal généraliste en Géographie, nous construisons un vaste corpus du voisinage de citation, duquel nous extrayons les mots-clés pertinents pour élaborer une classification sémantique. Nous montrons ensuite qualitativement et quantitativement la complémentarité des classifications. La méthodologie et les outils associés sont ouverts et peuvent être réutilisés dans des études similaires pour lesquelles les données sont difficiles à obtenir ou faiblement référencées dans les bases classiques. 
}

\stars




%


\newpage

\chapter{Thematic Developments}

\markboth{\thechapter\space Thematic Developments}{\thechapter\space Thematic Developments}

\label{app:thematic} 


\bpar{
This appendix includes thematic developments, i.e. corresponding to the empirical, conceptual and modeling domains. They can seem relatively far from our main preoccupations, but are necessary for the demonstration of precise points.
}{
Cette annexe regroupe des développements thématiques, c'est-à-dire qui tombent dans les domaines empiriques, conceptuels et de modélisation. Elles peuvent être relativement éloignées à première vue de nos préoccupations principales, mais sont nécessaires pour la démonstration de points précis.
}

\bpar{
The first three developments are important regarding empirical, modelling and methodlogical questions, approached from a particular thematic viewpoint.
\begin{enumerate}
	\item An empirical study of the geography of fuel prices in the United States which provides, under the assumption that it captures processes at the interface of road networks and territories, to highlight two typical scales for these processes and also the superposition of governance processes with neighbourhood effects.
	\item A multi-scale model for migration dynamics at the metropolitan scale is presented with the first results from its exploration.
	\item Methods for correlated synthetic data, in link with~\ref{sec:computation} and \ref{sec:correlatedsyntheticdata}, and presented as an abstract methodological perspective in~\ref{app:sec:syntheticdata}, is here applied to a quantitative finance question.
\end{enumerate}
}{
Les trois premiers développements sont importants quant à des questions empiriques, de modélisation et de méthodologie, abordées d'un point de vue thématique précis.
\begin{enumerate}
	\item Une étude empirique de la géographie des prix du carburant aux Etats-unis, permet, sous l'hypothèse que celle-ci capture des processus à l'interface du réseau routier et des territoires, de mettre en valeur deux échelles typiques pour ces processus ainsi que la superposition de processus de gouvernance avec des effets de voisinage.
	\item Un modèle multi-échelles de dynamiques de migrations résidentielles à l'échelle métropolitaine est présenté avec les premiers résultats issus de son exploration.
	\item Les méthodes de données synthétiques corrélées, en lien avec~\ref{sec:computation} et \ref{sec:correlatedsyntheticdata}, et présentée de dans la perspective méthodologique abstraite en~\ref{app:sec:syntheticdata}, est ici appliquée à une question de finance quantitative.
\end{enumerate}
}

\bpar{
The following developments correspond to epistemological questions, mostly in link with interdisciplinarity.
\begin{enumerate}\setcounter{enumi}{3}
	\item The concept of applied perspectivism is introduced in the presentation of the \textit{CybergeoNetworks} application, which allows analysing scientific corpuses through the combination of diverse approaches. It also also crucial regarding Open Science questions.
	\item The semantic analysis method used in~\ref{sec:quantepistemo} and already described in~\ref{app:sec:cybergeo} is applied to a patent corpus, what allows us to deploy it on massive data, and also to develop the question of innovation, which is a crucial thematic aspect for the evolutionary theory.
	\item A report of the special session Economy and Geography at ECTQG 2017 allows on the one hand exploring the role of models in interdisciplinary approaches, and on the other hand illustrating the applied perspectivism approach.
	\item The question of tools for scientific mediation is directly studied by presenting a project aimed at exploring tools based on these games, in the case of environmental research questions linked to freshwater ecology.
\end{enumerate}
}{
Les développements suivants se rapportent à des questions épistémologiques, principalement en lien avec l'interdisciplinarité.
\begin{enumerate}\setcounter{enumi}{3}
	\item Le concept de perspectivisme appliqué est introduit dans la présentation de l'application \textit{CybergeoNetworks}, qui permet l'analyse de corpus scientifiques par la combinaison de différentes approches. Celle-ci est également cruciale quant aux questions de Science Ouverte.
	\item La méthode d'analyse sémantique utilisée en~\ref{sec:quantepistemo} et déjà présentée en~\ref{app:sec:cybergeo} est appliquée à un corpus de brevets, ce qui nous permet de la déployer sur données massives, et également de développer la question de l'innovation, aspect thématique crucial pour la théorie évolutive.
	\item Un compte rendu de la session spéciale Economie et Géographie à l'ECTQG 2017 permet d'une part d'explorer le rôle des modèles dans les démarches interdisciplinaires, et d'autre part d'illustrer la démarche du perspectivisme appliqué.
	\item La question des outils de la médiation scientifique est abordée directement par la présentation d'un projet d'exploration d'outils basés sur les jeux dans le cas des questions environnementales liées aux écosystèmes d'eau douce.
\end{enumerate}
}

\stars

\bpar{
\textit{Publications or communications corresponding to the content of these appendices are detailed for each, with the explicit contribution of each collaborator.}
}{
\textit{Les publications ou communications correspondant au contenu de ces annexes sont détaillées pour chacune, avec le détail des contributions des différents collaborateurs.}
}






%

\newpage


\section{Road network and price drivers}{Transport routier et déterminants des coûts}

\label{sec:energyprice}

\bpar{
Interactions between networks and territories can indirectly manifest themselves within local economic properties of territories: therefore, the price of energy strongly conditions the impedance of a road network, and thus its impact on territories, and reciprocally this price is locally produced by sub-markets with are fully part of territories. The geography of gaz prices is thus an indirect signal of interactions. For example, \cite{orfeuil2012grand} (p.~307) suggests a link between fuel price and real estate crisis in the Parisian region.
}{
Les interactions entre réseaux et territoires peuvent se manifester indirectement au sein de propriétés économiques locales de territoires : ainsi, le prix de l'énergie conditionne fortement l'impédance d'un réseau routier, et donc son impact sur les territoires, et réciproquement ce prix est localement produit par des sous-marchés qui sont partie intégrante des territoires. La géographie des prix du carburant est donc un marqueur indirect des interactions. Par exemple, \cite{orfeuil2012grand} (p.~307) suggère un lien entre prix de l'essence et crise immobilière en région parisienne.
}

\stars

\bpar{
\textit{This appendix has been realized in collaboration with the economist \noun{Dr. A. Bergeaud} (Banque de France), in the context of a convergence of problematics between energy markets and indirect observation of interactions between networks and territories. It has been presented at the EWGT 2017 conference as \cite{raimbault2017cost}.}
}{
\textit{Cette annexe a été réalisée en collaboration avec l'économiste \noun{Dr. A. Bergeaud} (Banque de France), dans le cadre d'une convergence des problématiques entre marchés de l'énergie et observation indirecte des interactions entre réseaux et territoires. Elle a été présenté à la conférence EWGT 2017 comme \cite{raimbault2017cost}.}
}

\stars

\bpar{
The geography of fuel prices has many various implications, from its significant impact on accessibility to being an indicator of territorial equity and transportation policy. In this section, we study the spatio-temporal patterns of fuel price in the US at a very high resolution using a newly constructed dataset collecting daily oil prices for two months, on a significant proportion of US gas facilities (around 70\% of existing stations). These data have been collected using a specifically-designed large scale data crawling technology that we describe.
}{
La géographie des prix du carburant a de nombreuses applications variées, de son impact significatif sur l'accessibilité à son rôle comme indicateur d'équité territoriale et de politique de transports. Dans cette section, nous étudions les variations spatio-temporelles des prix du carburant aux États-Unis à une résolution très fine, par l'utilisation d'un nouveau jeu de données, donnant les prix journaliers sur deux mois pour une proportion significative des stations essence (autour de 70\% des stations existantes). Les données ont été collectées par l'intermédiaire d'une technologie de crawling à grande échelle élaborée spécifiquement, que l'on décrira.
}

\bpar{
We study the influence of socio-economic variables, by using complementary methods: Geographically Weighted Regression to take into account spatial non-stationarity, and linear econometric modeling to condition at the state and test county level characteristics. The former yields an optimal spatial range roughly corresponding to stationarity scale, and significant influence of variables such as median income or wage per job, with a non-simple spatial behavior that confirms the importance of geographical particularities. On the other hand, multi-level modeling reveals a strong state fixed effect, while county specific characteristics still have significant impact. Through the combination of such methods, we unveil the superposition of a governance process with a local socio-economical spatial process. We discuss one important application that is the elaboration of locally parametrized car-regulation policies.
}{
Nous étudions l'influence de variables socio-économiques, en utilisant des méthodes complémentaires : la Régression Géographique Pondérée pour tenir compte de la non-stationnarité spatiale, et une modélisation économétrique linéaire pour conditionner à l'État et tester des caractéristiques au niveau du comté. La première fournit une portée spatiale endogène qui correspond globalement à l'échelle de stationnarité, et une influence significative des variables comme le revenu moyen ou le salaire par travail, avec un comportement spatial dont la non simplicité confirme l'importance des particularités géographiques. D'autre part, la modélisation multi-niveaux révèle un très fort effet État, alors que les caractéristiques spécifiques au comté gardent un impact significatif. A travers la combinaison de ces méthodes, nous démontrons la superposition d'un processus de gouvernance avec un processus spatial socio-économique local. Nous discutons une application potentielle importante qui est l'élaboration de politiques de régulation automobiles localement paramétrisées.
}

\subsection{Context}{Contexte}

\bpar{
What drives the price of fuel? Using a new database on oil price at a gas station level collected during two months, we explore its variability across time and space. Variation in the cost of fuel can have many causes, from the crude oil price to local tax policy and geographical features, all having heterogeneous effect in space and time. If the evolution of the average fuel price in time is an indicator that is carefully followed and analyzed by many financial institution, its variability across space remain a rather unexplored topic in the literature. Yet, such differences can reflect variation in more indirect socio-economic indicators such as territorial inequalities and geographical singularities or consumer preferences.
}{
Quels sont les déterminants des prix du carburant ? Par l'utilisation d'une nouvelle base de données des prix du carburant au niveau de la station essence, collectée pendant deux mois, nous explorons leur variabilité dans le temps et l'espace. Une variation du coût du carburant peut avoir de nombreuses causes, du prix brut du pétrole au politiques fiscales locales et aux caractéristiques géographiques, chacun ayant des effets hétérogènes dans l'espace et le temps. Bien que l'évolution du prix moyen du carburant dans le temps soit un indicateur suivi avec attention et analysé par de nombreuses institutions financières, sa variabilité dans l'espace reste relativement non-explorée dans la littérature. Cependant, de telles différences peuvent refléter des variations dans des indicateurs socio-économiques plus indirects comme des inégalités territoriales, des singularités géographiques ou des préférences des consommateurs.
}

\bpar{
There exists to our knowledge no systematic mapping in space and time of retail fuel prices for a country with a high resolution\footnote{In France, data at the station level are available as open data at \url{https://www.prix-carburants.gouv.fr}. The website however does not allow interactive mapping in time and space.}. The main reason is probably that the availability of data have been a significant obstacle. It is also likely that the nature of the problem may also have influence, as it lies at the crossroad of several disciplines. While economists study price elasticity and measurement in different markets, transportation geography with method such as transportation prices in spatial models, puts more emphasis on spatial distribution than on precise market mechanisms.
}{
Il n'existe à notre connaissance pas de cartographie systématique dans le temps et l'espace des prix de vente à l'échelle d'un pays à une résolution fine\footnote{En France, les données au niveau de la station sont mises à disposition en données ouvertes, à \url{https://www.prix-carburants.gouv.fr}. Le site ne permet toutefois pas de cartographie interactive dans le temps et l'espace.}. La raison principale est probablement que la disponibilité des données a pu être un obstacle important. Il est aussi probable que la nature de la question joue un rôle, puisque celle-ci se trouve à l'interface de plusieurs disciplines. Alors que les économistes étudient l'élasticité des prix et leur mesure dans différents marchés, la géographie et la socio-économie des transports, par des méthodes comme les prix des transports intégrés aux modèles spatiaux, met une emphase plus grande sur la distribution spatiale que sur des mécanismes précis de marché. 
}

\bpar{
Nevertheless, examples of somehow related works can be found. For example,~\cite{rietveld2001spatial} studies the impact of cross-border differences in fuel price and the implications for gradual spatial taxation in  Netherlands. At the country-level, \cite{rietveld2005fuel} provides statistical models to explain fuel price variability across European countries. \cite{macharis2010decision} models the impact of spatial fuel price variation on patterns of inter-modality, implying that the spatial heterogeneity of fuel prices has a strong impact on user behavior. With a similar view on the geography of transportation, \cite{gregg2009temporal} studies spatial distribution of gas emission at the US-state level.
}{
Toutefois, des exemples de travaux relativement liés peuvent être trouvés. Par exemple, \cite{rietveld2001spatial} étudient l'impact de différences de prix transfrontalières et leur implications pour une taxation spatiale graduelle aux Pays-Bas. À l'échelle du pays, \cite{rietveld2005fuel} fournit des modèles statistiques pour expliquer les variabilité des prix entre les pays Européens. \cite{macharis2010decision} modélise l'impact d'une variation spatiale des prix sur les motifs d'intermodalité, en faisant l'hypothèse que l'hétérogénéité spatiale des prix du carburant a un impact sur le comportement des utilisateurs. Avec une approche similaire par la géographie des transports, \cite{gregg2009temporal} étudie la distribution spatiale des émissions à l'échelle des États américains.
}

\bpar{
The geography of fuel prices also have important implications on effective costs, as shows \cite{combes2005transport} by determining accurate transportation costs across urban areas for France. More closely related to our work, and using very similar daily open data for France, \cite{gautier2015dynamics} investigate dynamics of transmission from crude oil prices to fuel retail prices. However, they do not introduce an explicit spatial model of prices diffusion and do not study spatio-temporal dynamics.
}{
La géographie des prix du carburant a également d'importantes répercussions sur les coûts effectifs de transport, comme le montrent \cite{combes2005transport} en déterminant les coûts réels de transport pour les différentes aires urbaines françaises. De façon plus proche de notre travail, et en utilisant des données similaires en Accès Ouvert pour la France, \cite{gautier2015dynamics} étudient les dynamiques de transmission des prix bruts du pétrole aux prix de vente. Toutefois, ils n'introduisent pas de modèle spatial explicite de diffusion des prix et n'étudient pas de dynamiques spatio-temporelles.
}

\bpar{
In this section we adopt a different approach by proceeding to exploratory spatial analysis on US fuel prices. We show that most of the variation occurs between counties and not across time, although crude oil price was not constant during the period considered. We therefore turn to a spatial analysis of the distribution of fuel prices. Our main findings are twofold: first we show that there are significant spatial pattern in some large US regions, second we show that even if most of the observed variation can be explained by state level policies, and especially the level of tax, some county level characteristics are still significant.
}{
Dans cette section nous adoptons une approche qui se distingue de cette littérature en procédant à une analyse spatiale exploratoire de la variation des prix du carburant aux États-Unis. Nous montrons que la majorité des variations s'observent entre les comtés et non dans le temps, malgré les évolutions du baril brut pendant la période considérée. Pour cela, notre analyse de la distribution des prix se concentre sur la dimension spatiale. Les résultats majeurs obtenus sont les suivants : d'une part nous montrons l'existence de motifs spatiaux significatifs dans des grandes régions américaines, d'autre part nous montrons que même si la majorité des variations observées par les politiques des États, et en particulier le niveau de taxation, certaines caractéristiques à l'échelle du Comté restent significatives.
}

\subsubsection{Dataset}{Données}

\bpar{
Our dataset contain daily information on fuel price at the gas station level for the whole US mainland territory. These information have been constructed from self-reported fuel price and span around 70\% of gas station in the US. We start by describing data collection and then give some statistics about this new dataset.
}{
Notre jeu de données contient l'information journalière des prix des carburants à l'échelle de la station essence pour l'ensemble du territoire des États contigus (\textit{US mainland}). Ces informations sont construites à partir des prix reportés par les utilisateurs et couvrent environ 70\% des station essence aux États-Unis. Nous commençons par décrire la collection des données et donnons des statistiques de ce jeu de données nouveau.
}

\subsubsection{Collecting large scale heterogeneous data}{Collection de données hétérogènes à grande échelle}

\bpar{
The availability of new type of data has induced consequent changes in various disciplines from social science (e.g. online social network analysis~(\cite{tan2013social})) to geography (e.g. new insights into urban mobility or perspectives on ``smarter'' cities~(\cite{batty2013big})) including economics where the availability of exhaustive individual or firm level data is seen as a revolution of the field. Most studies involving these new data are at the interface of implied disciplines, what is both an advantage but also a source of difficulties. For example misunderstandings between physics and urban sciences described in \cite{dupuy2015sciences} are in particular caused by different attitudes towards unconventional data or divergent interpretations and ontologies of it. Collection and use of new data has therefore become a crucial stack in social-science.
}{
La disponibilité de nouveaux types de données a conduit à des évolutions significatives dans de nombreuses disciplines (e.g. l'analyse des réseaux sociaux en ligne (\cite{tan2013social})) à la géographie (e.g. les nouvelles approches de la mobilité urbaine ou les perspectives de ville plus ``intelligentes'' (\cite{batty2013big})) en incluant l'économie pour laquelle la disponibilité de données exhaustives à l'échelle individuelle ou de l'entreprise est vu comme une révolution dans le champ. La plupart des études impliquant ces nouvelles données sont à l'interface des disciplines concernées, ce qui est à la fois un avantage mais aussi une source de complications. Par exemple les malentendus entre physique et sciences urbaines décrites par~\cite{dupuy2015sciences} sont en particulier causées par des attitudes différentes au regard des données non conventionnelles ou des interprétations et ontologies différentes pour celles-ci. La collection et l'utilisation des nouvelles données est donc devenu un enjeu essentiel en sciences sociales.
}

\bpar{
The construction of such datasets is however far from straightforward because of the incomplete and noisy nature of data. Specific technical tools have to be implemented but have often been designed to overcome one specific problem and are difficult to generalize. We develop such a tool that fills the following constraints that are typical of large scale data collection: (i) reasonable level of flexibility and generality; (ii) optimized performance through parallel collection jobs; (iii) anonymity of collection jobs to avoid any possible bias in the behavior of the data source\footnote{Since multiple requests from the same address may induce blocking for example.}. The architecture, at a high level, has the following structure:
}{
La construction des tels jeux de données est cependant loin d'être évidente de par la nature incomplète et bruitée de la donnée. Des outils techniques spécifiques doivent être implémentés mais sont souvent conçus pour surmonter un problème donné et sont difficiles à généraliser. Nous développons un tel outil qui remplit les contraintes suivantes typiques de la collection de données à grande échelle : (i) flexibilité et généralité ; (ii) une performance optimisée par une collection en parallèle ; (iii) l'anonymat des tâches de collection pour éviter le plus possible tout biais dans le comportement de la source de données\footnote{Puisque des requêtes multiples depuis une même adresse peuvent amener à un blocage par exemple.}. L'architecture, à un assez haut niveau, a la structure suivante : 
}

\bpar{
\begin{itemize}
\item An independent pool of tasks runs continuously socket proxies to pipe requests through \texttt{tor}.
\item A manager monitors current collection tasks, split collection between subtasks and launches new ones when necessary.
\item Subtasks can be any callable application taken as argument destination urls, they proceed to the crawling, parsing and storage of collected data.
\end{itemize}
The application is open and its modules are reusable: source code is available on the repository of the project.\footnote{at \texttt{https://github.com/JusteRaimbault/EnergyPrice}} We constructed our dataset by using the tool continuously in time during two months to collect crowdsourced data available from various online sources.
}{
\begin{itemize}
\item Un ensemble indépendant des tâches fait tourner en continu des proxies socks pour envoyer les requêtes via \texttt{tor}.
\item Un manager suit les tâches de collection en cours, réparti la collection entre les sous-tâches et en lance des nouvelles lorsque cela s'avère nécessaire.
\item Les sous-tâches peuvent être toute application prenant comme argument les adresses de destination, elles procèdent à la collecte, au parsage et au stockage des données collectées.
\end{itemize}
L'application est ouverte et ses modules sont réutilisables : le code source est disponible sur le dépôt du projet.\footnote{à \texttt{https://github.com/JusteRaimbault/EnergyPrice}} Nous avons construit notre jeu de données en utilisant l'outil en continu pendant deux mois pour collecter des données crowdsourcées disponibles de diverses sources en ligne.
}

\subsubsection{Dataset}{Jeu de données}

\bpar{
Our dataset comprises around $41\cdot 10^6$ unique observations of retail fuel prices at the station level, spanning the period starting the $10^{th}$ of January 2017 and ending the $19^{th}$ of March 2017 and corresponds to 118,573 unique retail stations. For each of these stations, we associate a precise geographical location (city resolution). On average we have 377 price information by station. Prices correspond to a unique purchase mode (credit card, other modes such as cash being less than 10\% in test datasets, they were discarded in the final dataset) and four possible fuel types: Diesel (18\% of observations), Regular (34\%), Midgrade (24\%) and Premium (24\%). The best coverage of stations is for Regular fuel type with on average 4,629 price information by county. We therefore choose to focus the study to this type of fuel, keeping in mind that further developments with the dataset may include comparative analysis on fuel types.
}{
Le jeu de données contient autour de $41\cdot 10^6$ observations uniques des prix de vente au niveau de la station, s'étendant sur une période du 10 janvier 2017 au 19 mars 2017, correspondant à 118,573 station service uniques. Pour chacune, nous disposons d'une localisation géographique précise (résolution à la ville). En moyenne nous avons 377 informations de prix par station. Les prix correspondent à un mode d'achat unique (par carte de crédit, les autres modes comme l'argent liquide représentant moins de 10\% sur des jeux tests, ils ont été abandonnés dans le jeu de données final car ils correspondent à des prix différents) et quatre types de carburant possibles : Diesel (18\% des observations), Regular (34\%), Midgrade (24\%) et Premium (24\%). La meilleure couverture des stations est pour le carburant Regular avec en moyenne 4,629 données de prix par comté. Nous choisissons pour cette raison de concentrer l'étude sur ce type de carburant, en gardant à l'esprit que des développement futurs avec le jeu de données pourraient inclure des analyses comparatives des types de carburant.
}

\bpar{
Our final dataset thus contains 14,192,352 observations from 117,155 gas station, followed during 68 days. We further aggregate these data by day, taking the average of the observed price per gallon, to obtain a panel of 5,204,398 gas station - day observations.\footnote{The panel is not balanced as prices are not reported every day in every station. The average gas station has information on price for 44 days (over 68).}  Table \ref{tab:energyprice:stat_desc} gives some basic descriptive statistics of on price data showing that the distribution of oil price is highly concentrated with a small skewness (the ratio of the $99^{th}$ to the $1^{st}$ percentile is 1.6). Finally, in the spatial analysis, we will also use socio-economic data at the county level, available from the US Census Bureau. We shall use the latest available, which most of the time implies relying to the 2010 Census.
}{
Notre jeu de données final contient ainsi 14,192,352 observations provenant de 117,155 stations service, suivies pendant 68 jours. Nous agrégeons de plus les données par jour, en prenant la moyenne du prix observé par gallon, pour obtenir un panel de 5,204,398 observations station-jour.\footnote{Le panel n'est pas équilibré puisque les prix ne sont pas reportés chaque jour pour chaque station. Une station moyenne possède l'information de prix pour 44 jours (sur 68).} La table \ref{tab:energyprice:stat_desc} donne des statistiques descriptives basiques sur les données de prix, montrant que la distribution des prix est fortement concentrée avec une faible skewness (le ratio du $99^{th}$ au $1^{st}$ quantiles est 1.6). Enfin, dans l'analyse spatiale, nous utiliserons également des données socio-économiques au niveau du comté, disponible par le US Census Bureau. Nous utiliserons les plus récentes disponibles (ce qui dans la plupart des cas implique d'utiliser le Census de 2010).
}

\begin{table}
\appcaption{\textbf{Descriptive statistics on Fuel Price.} The price is given in \$ per gallon ($1 \textrm{gallon} = 3,78541 \textrm{L}$).\label{tab:energyprice:stat_desc}}{\textbf{Statistiques descriptives des prix du carburant.} Le prix est donné en \$ par gallon ($1 \textrm{gallon} = 3,78541 \textrm{L}$).\label{tab:energyprice:stat_desc}}
\label{tab:stat_desc}
\bpar{
\begin{tabular}{ccccccc}
\textbf{Average} & \textbf{Std. Dev.} & \textbf{p10} & \textbf{p25} & \textbf{p50} & \textbf{p75} & \textbf{p90} \\
\hline
\cr
2.28 & 0.27 &  2.02  &  2.09  &  2.21  &  2.39  &  2.65  \\
\hline
\end{tabular}
}{
\begin{tabular}{ccccccc}
\textbf{Moyenne} & \textbf{Dev. Std.} & \textbf{p10} & \textbf{p25} & \textbf{p50} & \textbf{p75} & \textbf{p90} \\
\hline
\cr
2.28 & 0.27 &  2.02  &  2.09  &  2.21  &  2.39  &  2.65  \\
\hline
\end{tabular}
}
\end{table}

\subsection{Results}{Résultats}

\subsubsection{Spatio-temporal Patterns of Prices}{Motifs spatio-temporels des prix}

\bpar{
Before moving to a more systematic study of the variation of fuel price, we propose a first exploratory introduction to give insight about its spatio-temporal structure. This exercise is a crucial stage to guide further analyses, but also to understand their implications in a geographical context. To explore the data, we built a simple web application which allow to map the data in space and time. This application is available at \url{http://shiny.parisgeo.cnrs.fr/fuelprice/}.
}{
Avant de se consacrer à une étude plus systématique de la variation des prix des carburants, nous proposons une exploration pour donner une idée de sa structure spatio-temporelle. Cette exercice est une étape cruciale pour guider les analyses suivantes, mais aussi pour comprendre leurs implications dans le contexte géographique. Afin d'explorer les données, nous construisons une application web permettant de cartographier les données dans l'espace et le temps. Elle est disponible à \url{http://shiny.parisgeo.cnrs.fr/fuelprice/}.
}

\bpar{
We also show one example of mapping the data at the county level in Figure \ref{fig:energyprice:map_price} where we used average price over the whole period. We clearly see regional patterns with the Southcentral and Southeast regions having the lowest prices and the Pacific cost and Northeast the highest prices. Of course, plotting aggregated data over the whole period does not bring much information about the time variation of the data. As we will show more in detail below most of the variation of fuel price occurs across space. A variance decomposition of fuel price yields only 11\% of the total variance is explained by within gas station variations. Similarly, the Spearman's rank correlation coefficient between the gas station price of regular fuel in the first day of dataset and in the last day is 0.867, and the null hypothesis that these two information are independent is strongly rejected.
}{
Nous montrons également une carte au niveau du comté en Fig.~\ref{fig:energyprice:map_price} pour le prix moyen sur l'ensemble de la période. On voit clairement apparaître des motifs régionaux, avec les régions du centre sud et du sud est ayant les prix les plus bas et la côte Pacifique et le nord est les prix les plus hauts. Bien évidement , une carte agrégée sur l'ensemble de la période n'apporte guère d'information sur les variations temporelles des données. Comme nous allons le montrer plus en détails par la suite, la majorité des variations des prix des carburants a lieu dans l'espace. une décomposition de la variance des prix donne seulement 11\% de la variance totale expliquée par les variations intra-station. De la même manière, le coefficient de corrélation de rang de Spearman pour le prix des stations entre le premier jour du jeu de données et le dernier jour est de 0.867, et l'hypothèse nulle que ces deux informations sont indépendantes est fortement rejetée.
}

\begin{figure}
\includegraphics[width=\linewidth]{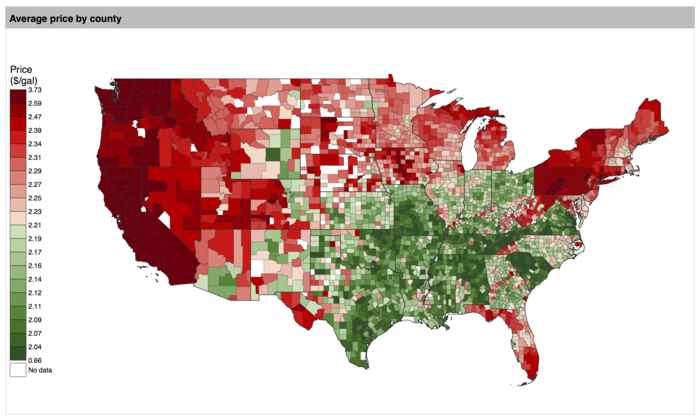}
\appcaption{\textbf{Map of mean price for counties.} The price is given for regular fuel, averaged over the whole period.\label{fig:energyprice:map_price}}{\textbf{Carte du prix moyen par comté.} Le prix est donné pour du carburant régulier, et la moyenne temporelle est prise sur l'ensemble de la période.\label{fig:energyprice:map_price}}
\end{figure}

\bpar{
Since most of the variation in oil price is between gas station, we now focus mainly on spatial correlations. We will conduct the analysis at the county level for various reasons. First it appears that a variance decomposition of fuel price between and within county shows that more than 85\% of the variance is between-county, second because the localization of gas station is not reliable enough to allow for a smaller granularity and third because we have many socio-economic information at this level. We therefore study the spatial autocorrelation of prices at the county level, as already specified several times with a decay parameter $d_0$. We show in Fig.~\ref{fig:energyprice:moran} its variations for each day and also as a function of the decay parameter. 
}{
Puisque la majorité de la variation des prix est entre les stations, nous nous intéressons maintenant principalement aux corrélations spatiales. Nous conduisons l'analyse à l'échelle du comté pour diverses raisons. D'une part une décomposition des prix des carburants inter et intra-comté montre que plus de 85\% de la variance est inter-comté, d'autre part car la localisation des stations n'est pas assez fiable pour permettre une granularité plus fine, et enfin car la majorité des variables socio-économiques est à ce niveau. Nous étudions donc l'autocorrelation spatiale des prix à l'échelle du comté, comme déjà spécifiée plusieurs fois avec un paramètre de décroissance $d_0$. Nous montrons en Fig.~\ref{fig:energyprice:moran} ses variations pour chaque jour ainsi que comme fonction du paramètre de décroissance.
}


\bpar{
The fluctuations in time of the daily Moran index for low and medium spatial range, confirms geographical specificities in the sense of locally changing correlation regimes. These are logically smoothed for long ranges, as price correlations drop down with distance. The behavior of spatial autocorrelation with decay distance is particularly interesting: we observe a first regime change around 10km (from constant to piecewise linear regime), and a second important one around 1000km, both consistent across weekly time windows. We postulate that these correspond to typical spatial scales of the involved processes: the low regime would be local specificities and the middle one the state level processes.
}{
Les fluctuations dans le temps de l'index de Moran journalier pour les valeurs basses et moyennes du paramètre de décroissance, confirme les spécificités géographiques au sens de régimes de corrélation changeant localement. Celles-ci sont logiquement atténuées pour les longues portées, puisque les correlations des prix diminuent avec la distance. Le comportement de l'autocorrelation spatiale en fonction du paramètre de decay est particulièrement intéressant : nous observons une premier changement de régime atour de 10km (d'un régime constant à un régime linéaire par morceau), et une seconde transition importante autour de 1000km, les deux transitions se retrouvant sur l'ensemble des fenêtres temporelles à la semaine. Nous postulons que celles-ci correspondent aux échelles spatiales typiques des phénomènes observés: le régime bas serait les spécificités locales et l'intermédiaire le processus au niveau de l'État.
}

\bpar{
This behavior confirms that prices are non-stationary in space, and that therefore appropriate statistical techniques must be used to study potential drivers at different level. The two next subsections follow this idea and investigate potential explicative variables of local fuel prices, using two different techniques corresponding to two complementary paradigms: geographically weighted regression that puts the emphasis on neighborhood effects, and multi-level regression taking into account administrative boundaries.
}{
Ce comportement confirme que les prix sont non-stationnaires dans l'espace, et que pour cette raison des techniques statistiques appropriées doivent être utilisées pour étudier les variables jouant un rôle à différents niveaux. Les deux parties suivantes suivent cette idée et étudient des variables explicatives potentielles des prix locaux du carburant, utilisant deux techniques différentes qui correspondent à deux paradigmes complémentaires : la régression géographique pondérée qui se concentre sur les effets de voisinage, et des régressions multi-niveaux prenant en compte les limites administratives, capturant notamment l'effet des politiques de taxation par État.
}

\begin{figure}
\includegraphics[width=\linewidth]{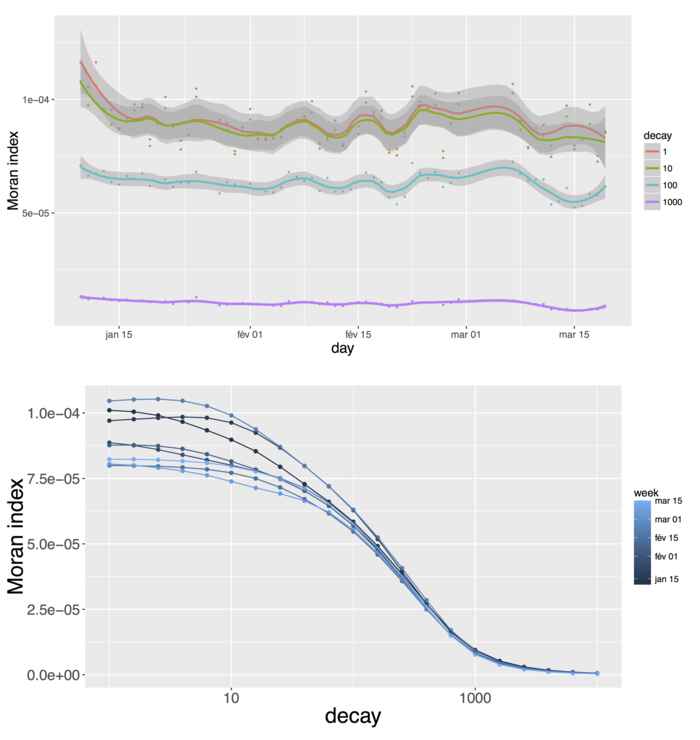}
\appcaption{\textbf{Behavior of Moran spatial-autocorrelation index.} (\textit{Left}) Evolution in time of Moran index computed on daily time windows, for different decay parameter values. (\textit{Right}) Moran index as a function of decay parameter, computed on weekly time windows.\label{fig:energyprice:moran}}{\textbf{Comportement de l'index d'autocorrelation spatiale de Moran.} (\textit{Gauche}) Evolution dans le temps de l'indice de Moran, calculé sur des fenêtres journalières, pour différentes valeurs du paramètre de décroissance. (\textit{Droite}) Indice de Moran en fonction du paramètre de décroissance, calculé sur des fenêtres hebdomadaires.\label{fig:energyprice:moran}}
\end{figure}

\subsubsection{Geographically Weighted Regression}{Régression Géographique Pondérée}

\bpar{
The issue of spatial non-stationarity of geographical processes has always been a source of biased aggregated analyses or misinterpretations when applying general conclusions to local cases. To take it into account into statistical models, numerous techniques have been developed, among which the simple but very elegant Geographically Weighted Regression (GWR), that estimates non-stationary regressions by weighting observations in space similarly to kernel estimation methods. This was introduced in a seminal paper by~\cite{brunsdon1996geographically} and has been subsequently used and matured since then. The significant advantage of this technique is that an optimal spatial range in the sense of model performance can be inferred to derive a model that yields the effect of variables varying in space, thus revealing local effects that can occur at different spatial scales or across boundaries.
}{
La question de la non-stationnarité des processus géographiques a toujours été une source d'analyses agrégées biaisées ou de mauvaises interprétations lorsque des conclusions générales sont appliquées à des cas locaux (erreur écologique). Pour le prendre en compte dans les modèles statistiques, de nombreuses techniques ont été proposées, parmi lesquelles la méthode de la Régression Géographique Pondérée (GWR), qui estime des régressions non-stationnaires en pondérant les observations dans l'espace de manière similaire aux techniques d'estimation de densité par noyaux. Elle a été introduite dans un article séminal par \cite{brunsdon1996geographically} et a été utilisée et développée en conséquence depuis. L'avantage considérable de cette technique est qu'une portée spatiale optimale au sens de la performance du modèle peut être déduite pour dériver un modèle qui traduit des effets des variables variant dans l'espace, révélant ainsi des effets locaux qui peuvent se produire à différentes échelles spatiales ou à travers les frontières.
}

\bpar{
We proceed to multi-modeling to find the best model and associated kernel and spatial range. More specifically, we do the following: (i) we generate all possible linear models from the five potential variables (income $inc_i$, population $pop_i$, wage per job $wage_i$, jobs per capita $jc_i$, jobs $job_i$); (ii) for each model and each candidate kernel shape (exponential, gaussian, bisquare, step), we determine the optimal bandwidth in the sense of both cross-validation and corrected Akaike Information Criterion (AICc) which quantifies information included in the model; (iii) we fit the models with this bandwidth. We choose the model with the best overall AICc, namely 
\begin{equation}
price_i = \beta\cdot\left( inc_i, wage_i, jc_i\right)
\end{equation}
 for a bandwidth of 22 neighbors and a gaussian kernel,\footnote{note that the kernel shape does not have much influence as soon as gradually decaying functions are used} with an AICc of $2,900$. The median AICc difference with all other models tested is 122. The global R-squared is 0.27, what is relatively good also compared to the best R-squared of 0.29 (obtained for the model with all variables, which clearly overfits with an AICc of 3010; furthermore, effective dimension is less than 5 as 90\% of variance is explained by the three first principal components for the normalized variables).
}{
Nous procédons à une multi-modélisation pour trouver le meilleur modèle ainsi que le noyau et la portée spatiale optimaux associés, c'est-à-dire que nous procédons à une comparaison systématique d'un ensemble de modèles possibles. Plus précisément, nous suivons les étapes suivantes : (i) tous les modèles linéaires potentiels à partir des cinq variables candidates sont générés (revenu $inc_i$, population $pop_i$, salaire par emploi $wage_i$, emploi par tête $jc_i$, emplois $job_i$); (ii) pour chaque modèle et chaque forme de noyau candidate (exponentiel, gaussien, bisquare, escalier), nous déterminons la portée optimale au sens à la fois de la cross-validation et du critère d'Information d'Akaike corrigé (AICc) qui quantifie l'information contenue dans le modèle; (iii) nous ajustons les modèles avec cette portée. Nous choisissons le modèle avec le meilleur AICc, en l'occurence 
\begin{equation}
price_i = \beta\cdot\left( inc_i, wage_i, jc_i\right)
\end{equation}
 pour une portée de 22 voisins et un noyau Gaussien,\footnote{Nous notons que la forme du noyau n'a pas plus d'influence tant que des fonctions décroissant graduellement sont utilisées.} avec un AICc de $2900$. La différence médiane d'AICc avec l'ensemble des autres modèles est 122, ce qui permet de sélectionner le modèle sans équivoque par rapport au modèle médian (coefficient d'Akaike proche de 1). Le coefficient de détermination global est 0.27, ce qui est relativement bon en comparaison du meilleur R$^2$ de 0.29 (obtenu pour le modèle avec l'ensemble des variables, qui sur-ajuste clairement avec un AICc de 3010; de plus la dimension effective est inférieure à 5 puisque 90\% de la variance est expliquée par les trois premières composantes principales pour les variables normalisées).
}

\bpar{
The coefficients and local R-squared for the best model are shown in Fig.~\ref{fig:gwr}. The spatial distribution of residuals (not shown here) seems globally randomly distributed, which confirms in a way the consistency of the approach. Indeed, if a distinguishable geographical structure had been found in the residuals, it would have meant that the geographical model or the variable considered had failed to translate spatial structure.
}{
Les coefficients et le R$^2$ local pour le meilleur modèle sont montrés en Fig.~\ref{fig:energyprice:gwr}. La distribution spatiales des résidus (qui n'est pas montrée ici), semble globalement distribuée aléatoirement, ce qui confirme d'une certaine façon la cohérence de l'approche. En effet, si une structure géographique distinguable était trouvée dans les résidus, cela signifierait que le modèle géographique ou les variables considérées ont échoués à traduire la structure spatiale.
}

\bpar{
Let now turn to an interpretation of the spatial structures we obtain. First of all, the spatial distribution of the model performance reveals that regions where these simple socio-economic factors explain do a good job in explaining prices are mostly located on the west coast, the south border, the north-east region from lakes to the east coast, and a stripe from Chicago to the south of Texas. The corresponding coefficients have different behaviors across the areas, suggesting different regimes.\footnote{We comment their behavior in areas where the model has a minimal performance, that we fix arbitrarily as a local R-squared of 0.5.}.
}{
Nous pouvons à présent proposer une interprétation des structures spatiales obtenues. Tout d'abord, la distribution spatiale de la performance du modèle révèle des régions où ces indicateurs socio-économiques simples expliquent relativement bien les prix, et celles-ci sont localisées sur la côte ouest, la frontière sud, la région nord-est des lacs à la côte est, et une bande de Chicago au sud du Texas. Les coefficients correspondants ont des comportements différents selon les zones, suggérant différents régimes\footnote{Nous commentons leur comportement dans les zones où le modèle a une performance minimale, que nous fixons arbitrairement à un R2 local de 0.5.}.
}

\bpar{
For example, the influence of income in each region seems to be inverted when the distance to the coast increases (from north to south-east in the west, south to north in Texas, east to west in the east), what may be a fingerprint of different economic specializations. On the contrary, the regime shifts for wage show a clear cut between west (except around Seattle) and middle/east, that does not correspond to state-policies only as Texas splits in two. The same way, jobs per capita show an opposition between east and west, what could be due for example to cultural differences.
}{
Par exemple, l'influence du revenu dans chaque région semble s'inverser quand la distance à la côte augmente (du nord au sud-est dans l'Ouest, du sud au nord au Texas, de l'Est à l'ouest dans l'Est), ce qui pourrait témoigner de différentes spécialisations économiques. Au contraire, le changement de régime pour les salaires montre une rupture notable entre l'Ouest (sauf autour de Seattle) et le Centre et l'Est, qui ne correspond pas directement à des politiques d'État locales puisque le Texas est coupé en deux par exemple. De la même façon, l'influence des emplois par tête montrent une opposition entre Est et Ouest, qui pourrait être due par exemple à des différences culturelles.
}

\bpar{
These results are difficult to interpret directly, and must be understood as a confirmation that geographical particularities matters, as regions differ in regimes of role for each of the simple socio-economic-variables. Further precise knowledge could be obtained through targeted geographical studies including qualitative field studies and quantitative analyses, that are beyond the scope of this exploratory study and left for potential further research.
}{
Ces résultats sont toutefois difficiles à interpréter directement, et doivent être compris comme la confirmation que les particularités géographiques importent, puisque les régions diffèrent dans le régime du rôle de chacune des variables socio-économiques simples. Une connaissance plus précise pourrait être obtenue par des études géographiques ciblées incluant des études de terrain qualitatives et des analyses quantitatives, qui sont au delà de la portée de cette étude exploratoire et laissée à une éventuelle recherche future.
}

\bpar{
Finally, we extract the spatial scale of the studied processes, that is, by computing the distribution of distance to nearest neighbors with the optimal bandwidth. It yields roughly a log-normal distribution, of median 77km and interquartile 30km. We interpret this scale as the spatial stationarity scale of price processes in relation with economic agents, which can also be understood as a range of coherent market competition between gas stations.
}{
Enfin, nous extrayons l'échelle spatiale des processus étudies, c'est-à-dire en calculant la distribution de la distance aux plus proches voisins avec la portée optimale. On obtient approximativement une distribution log-normale, de médiane 77km et d'interquartile 30km. Nous interprétons cette échelle comme l'échelle de stationnarité spatiale du processus de prix en relation avec les agents économiques, qui peut également être comprise comme la portée des marchés cohérents de compétition entre les stations service.
}

\begin{figure}
\includegraphics[width=\linewidth]{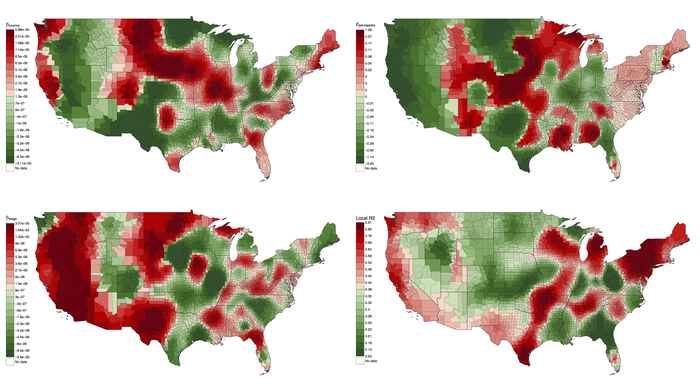}
\appcaption{\textbf{Results of GWR analyses.} For the best model in the sense of AICc, we map the spatial distribution of fitted coefficient, in order from left to right and top to bottom, $\beta_{income}$, $\beta_{percapjobs}$, $\beta_{wage}$, and finally the local R-squared values.\label{fig:energyprice:gwr}}{\textbf{Résultats des analyses GWR.} Pour le meilleur modèle au sens de l'AICc, les cartes donnent la distribution spatiale des coefficients estimés, dans l'ordre de gauche à droite et de haut en bas, $\beta_{income}$, $\beta_{percapjobs}$, $\beta_{wage}$, et finalement les valeurs du R2 local.\label{fig:energyprice:gwr}}
\end{figure}

\subsubsection{Multi-level Regression}{Régressions multi-niveaux}

\bpar{
Since our initial database enables to look at the level of variable $x_{i,s,c,t}$, the fuel price in day $t$, in gas station $i$, in state $s$ and in county $c$, we start by running high dimensional fixed effect regressions following the model:
}{
Comme notre base initiale permet de regarder au niveau des variables $x_{i,s,c,t}$, le prix du carburant au jour $t$, dans la station $i$, dans l'État $s$ et dans le comté $c$, nous commençons par estimer des régressions à effets fixes en grande dimension, suivant les modèles :
}

\begin{eqnarray}
x_{i,s,c,t} &=& \beta_s + \varepsilon_{i,s,c,t} \\
x_{i,s,c,t} &=& \beta_c + \varepsilon_{i,s,c,t} \\
x_{i,s,c,t} &=& \beta_i + \varepsilon_{i,s,c,t}
\end{eqnarray}

\bpar{
Where $\varepsilon_{i,s,c,t}$ contains an idiosyncratic error\footnote{i.e. which is proper to each individual.} and a day fixed effect. This first analysis confirm that most of the variance can be explained by a state fixed effect and that integrating more accurate levels has only small effect on the fit of our model as measured by the R-squared.
}{
Où $\varepsilon_{i,s,c,t}$ contient une erreur idiosyncrasique\footnote{C'est-à-dire étant propre à chaque individu.} et un effet fixe jour. Cette première analyse confirme que la majorité de la variance peut être expliquée par un effet fixe au niveau de l'État et que d'intégrer des niveaux plus fins a un effet négligeable sur la performance du modèle mesurée par le R2.
}

\bpar{
We now turn to a different analysis, aiming at capturing the explanatory variables that account for spatial price variation of fuel. We consider the following linear model:
}{
Nous nous tournons à présent vers une analyse différente, visant à capturer les variables explicatives qui rendent compte des variations spatiales du carburant. Nous considérons le modèle linéaire suivant :
}

\begin{equation}
\label{eq:energyprice:reg}
log(x_{i}) = \beta_0 + X_{i}\beta_1 + \beta_{s(i)} + \varepsilon_{i},
\end{equation}

\bpar{
where $x_{i}$ denotes average measured fuel price in county $i$ aggregated across all days, $X_{i}$ is a set of county specific variables and $s(i)$ is the state to which the county belongs so that $\beta_{s(i)}$ capture all state specific variation. Finally $\varepsilon_{i}$ is an error term satisfying $Cov(\varepsilon_{i}, \varepsilon_{j}) = 0$ if $s(i) \neq s(j)$. This clustering of standard error at the state level is motivated by finding of the previous section, showing that spatial autocorrelation of fuel price at the state level is still potentially strong. This specification aims at capturing the effect of various socio-economic variable at the county level after a state fixed effect has been removed.
}{
où $x_{i}$ dénote le prix moyen mesuré du carburant dans le comté $i$ agrégé sur l'ensemble des jours, $X_{i}$ est un ensemble de variables spécifiques au comté et $s(i)$ est l'état dans lequel se trouve le comté de telle façon que $\beta_{s(i)}$ capture toute la variation spécifique aux États. Enfin $\varepsilon_{i}$ est un terme d'erreur satisfaisant $Cov(\varepsilon_{i}, \varepsilon_{j}) = 0$ si $s(i) \neq s(j)$. ce regroupement de l'erreur standard au niveau de l'état est motivé par les résultats de la partie précédente, montrant que l'autocorrélation spatiale des prix du carburant au niveau de l'État est forte. Cette spécification vise à capturer les effets de variables socio-économiques variées au niveau du comté après que l'effet fixe État ait été retiré.
}

\bpar{
We present our results in Table~\ref{tab:energyprice:reg}. The first column shows that regressing the log of price on a state fixed-effect is already enough to explain 74\% of the variance. This is mostly due to tax on fuel which are set at the state level in the US. In fact, when we regress the log of oil price on the level of state tax, we find a R-squared of 0.33\%. The remaining explanatory variables show that dense urban counties have higher fuel price, but this price decreases with population. This result seems sensible, desert areas have on average higher oil price. Fuel price increases with total income, decreases with poverty and decrease with the extent to which a county has voted for a Republican candidate. This last finding suggests a circular link: counties that use car the most tend to vote to politician that promote pro car policies. Adding these explanatory variables slightly increase the R-squared, suggesting that even after having removed a state fixed-effect, the price of fuel can be explained by local socio-economic features.	
}{
Les résultats sont présentés en Table~\ref{tab:energyprice:reg}. La première colonne montre que la regression du logarithme des prix sur un effet fixe État est déjà suffisant pour expliquer 74\% de la variance. Cela est majoritairement du aux taxes sur les carburants qui sont fixées au niveau de l'État aux États-Unis. En fait, une régression du log-prix sur le niveau de taxe donne un R-squared de 0.33\%. Les variables explicatives restantes montrent que les comtés urbains denses ont des prix plus élevés, mais que le prix décroit avec la population. Ce résultat paraît raisonnable, les zones désertiques ayant en moyenne des prix plus hauts. Les prix augmentent avec le revenu total, décroissent avec le niveau de pauvreté et décroisse avec le niveau de vote pour un candidat républicain. Ce dernier point suggère un lien circulaire : nous pouvons faire l'hypothèse que les comtés qui utilisent beaucoup la voiture auront tendance à voter pour un politicien qui promouvra des politiques favorable à son usage. L'ajout de ces variables explicatives augmente légèrement le R-squared, ce qui suggère que même après avoir enlevé l'effet fixe État, la prix du carburant peut être expliqué par des caractéristiques socio-économiques locales.
}

\begin{table}[htbp]
\vspace{-0.1cm}
\begin{center}
{
\begin{threeparttable}
\apptabcaption{Regressions at the county level\label{tab:energyprice:reg}}{\textbf{Régressions multi-niveau au niveau du comté.}\label{tab:energyprice:reg}}
\begin{tabular}{lccccc}
 \toprule
 \hline
 \cr

  & (1) & (2) & (3) & (4) & (5) \\ 
\cmidrule(r){2-6}
Density      &               &       0.016***&       0.016***&       0.016***&       0.015***\\
                    &               &     (0.002)   &     (0.001)   &     (0.001)   &     (0.001)   \\ \cr
Population (log)             &               &      -0.007***&      -0.040***&      -0.041***&      -0.039***\\                    &               &     (0.001)   &     (0.011)   &     (0.011)   &     (0.010)   \\ \cr
Total Income (log)            &               &               &       0.031***&       0.031***&       0.027***\\
                    &               &               &     (0.010)   &     (0.010)   &     (0.009)   \\ \cr
Unemployment        &               &               &       0.001   &       0.000   &       0.000   \\
                    &               &               &     (0.001)   &     (0.001)   &     (0.001)   \\ \cr
Poverty   &               &               &      -0.028** &      -0.030***&      -0.029** \\
                    &               &               &     (0.011)   &     (0.011)   &     (0.011)   \\ \cr
Percentage Black    &               &               &               &       0.000***&      -0.000   \\
                    &               &               &               &     (0.000)   &     (0.000)   \\ \cr
Vote GOP      &               &               &               &               &      -0.072***\\
                    &               &               &               &               &     (0.015)   \\ \cr 
\cmidrule{2-6}
R-squared           &       0.743   &       0.767   &       0.774   &       0.776   &       0.781   \\
N                   &        3,066   &        3,011   &        3,011   &        3,011   &        3,011   \\
\cr
\hline
\bottomrule
\end{tabular}
 \begin{tablenotes}
       \item \bpar{\protect\scriptsize{\textbf{Notes}: This table plots results from an Ordinary Least Square regression of model presented in equation (\ref{eq:energyprice:reg}). Density is measured as the number of inhabitant by square miles and total income is given in dollars. Poverty is measured as the number of people below the poverty threshold per inhabitant. Percentage black is the percentage of black people living in the county and vote GOP is the share of people having voted for Donald Trump in the 2016 elections. Regression includes a state fixed effect. Robust standard errors clustered at the state level are reported in parenthesis. ***, ** and * respectively indicate 0.01, 0.05 and 0.1 levels of significance.}}{\protect\scriptsize{\textbf{Notes} : Cette table donne les résultats d'une régression des Moindres Carrés Ordinaire pour le modèle présenté en équation  (\ref{eq:energyprice:reg}). La densité est mesurée comme le nombre d'habitants au mile-carré et le revenu total est donné en dollars. La pauvreté est mesurée comme le nombre de personnes sous le seuil de pauvreté par habitants. On étudie aussi l'influence du pourcentage de personnes noires et de la part de personnes ayant voté pour Donald Trump aux élections de 2016. La régression inclut un effet fixe État. Les erreurs standard robustes, agrégées au niveau de l'état, sont données entre parenthèses. ***, ** and * indiquent respectivement les niveau de significativité 0.01, 0.05 and 0.1.}}
    \end{tablenotes}
  \end{threeparttable}
}
\end{center}
\end{table}

\subsection{Discussion}{Discussion}

\subsubsection{On the complementarity of Econometric and Spatial Analysis methods}{Sur la complémentarité des méthodes économétriques et des méthodes d'analyse spatiale}

\bpar{
One important aspect of our contribution is methodological. We show that to explore a new panel dataset, geographers and economists have different approach, leading to similar generic conclusion but with different path. Some studies have already combined GWR and multi-level regressions (\cite{chen2012using}), or compared them in terms of model fit or robustness (\cite{lee2009determinants}). We take here a multi-disciplinary point of view and combines approaches answering to different questions, GWR aiming at finding precise explicative variables and to measure the extent of spatial correlation, whereas econometric models explain with more accuracy the effect of factors at different levels (state, county) but take these geographical characteristics as exogenous. We claim that both are necessary to understand all dimensions of the studied phenomenon.
}{
Un aspect important de cette contribution est méthodologique. Nous montrons que pour explorer un nouveau panel de données, les géographes et les économistes prennent des approches différentes, menant à des conclusions génériques similaires par des chemins différents. Des études ont déjà combiné les GWR et les régressions multi-niveau (\cite{chen2012using}), ou les ont comparées en terme de performance de modèle ou de robustesse (\cite{lee2009determinants}). Nous prenons ici un point de vue multi-disciplinaire et combinons des approches répondant à des questions différentes, GWR ayant pour but de trouver des variables explicatives précises et de mesurer le role de l'auto-corrélation spatiale, tandis que les modèles économétriques expliquent plus précisément les effets des différents facteurs à plusieurs niveaux (État, comté) mais prennent ces caractéristiques géographiques comme exogènes. Nous postulons que les deux sont nécessaires pour comprendre toutes les dimensions du phénomène étudié.
}

\subsubsection{Designing localized car-regulation policies}{Proposition de politiques de régulation localisées}

\bpar{
Another application of such analysis is to help better designing car-regulation policies. Environmental and health issues nowadays require a reasoned use of cars, in cities with the problem air pollution but also overall to reduce carbon emissions. \cite{fullerton2002can} showed that a taxation of fuel and cars can be equivalent to a taxation on emissions. \cite{brand2013accelerating} highlight the role of incentives for the transition towards a low carbon transportation. However, such measures can't be uniform across states or even counties for obvious reasons of territorial equity: areas with different socio-economic characteristics or with different amenities shall contribute regarding their capabilities and preferences. Knowing local prices dynamics and their drivers, in which our study is a preliminary step, may be a path to localized policies taking into account the socio-economic configuration and include an equity criterion.
}{
Une autre application de ce type d'analyse est d'aider à une meilleure conception de politiques de régulation de la voiture. Les problèmes environnementaux et de santé requièrent de nos jours un usage raisonné de celle-ci, dans les villes avec le problème de la pollution atmosphérique, mais aussi globalement pour réduire les emissions de CO\textsubscript{2}. \cite{fullerton2002can} montre qu'une taxation des carburants et des voitures peut être équivalente à une taxation des emissions. \cite{brand2013accelerating} souligne le rôle des incitations pour une transition vers des transports décarbonés. Cependant, de telles mesures ne peuvent pas être uniformes d'un État à l'autre ou même entre les comtés, pour des raisons évidentes d'équité territoriale : des zones avec des caractéristiques socio-économiques différentes ou avec différentes aménités doivent contribuer selon leur possibilité et préférences. La connaissance des dynamiques locales des prix et leur déterminants, ce en quoi notre étude est une étape préliminaire, peut être une voie vers des régulations localisées prenant en compte la configuration socio-économique et inclure un critère d'équité.
}

\subsubsection{Conclusion}{Conclusion}

\bpar{
We have described a first exploratory study of US fuel prices in space and time, using a new database at the gas station level spanning two months. We our first result is to show the high spatial heterogeneity of price processes, using interactive data exploration and auto-correlation analyses. We proceed with two complementary studies of potential drivers: GWR unveils spatial structures and geographical particularities, and yields a characteristic scale of processes around 75km; multi-level regressions show that even though most of the variation can be explained by state level characteristics, and mostly by the level of the tax on fuel that is set by the state, there are still socio-economic specificities at the county level that can explain spatial variation of fuel price.
}{
Nous avons décrit une première étude exploratoire des prix des carburants aux US dans le temps et l'espace, utilisant une nouvelle base de données au niveau de la station s'étendant sur deux mois. Notre premier résultat est de montrer la grande hétérogénéité spatiale des processus de prix, par une exploration interactive des données et des analyses d'auto-corrélation. Nous procédons à deux études complémentaires des déterminants potentiels : GWR révèle des structures spatiales et des particularités géographiques, and fournit une échelle caractéristique des processus autour de 75km ; les régressions multi-niveaux montrent que même si la majorité des variations sont expliquées par les caractéristiques des États, et majoritairement par le niveau de taxation fixé par l'État, il existe toujours des spécificités socio-économiques au niveau du comté qui peuvent expliquer la variation spatiale des prix du carburant.
}

\subsubsection{Perspective}{Mise en perspective}

\bpar{
In the perspective of our global problematic, this second empirical opening is thus relevant for several reasons: (i) the spatial distribution of prices is an object of study at the intersection of territories and networks, since local markets are closely linked to territorial properties, but are also determined by the properties of the road network (for example accessibility) and by its use patterns; (ii) the two endogenous scales identified correspond to our mesoscopic and macroscopic scales, confirming through an other approach the necessity to indeed take the two into account; (iii) the unveiling of the superposition of a governance process to local geographical processes rejoins our last development on taking into account governance in models, since this complexity is effectively observed in the empirical analysis here.
}{
Dans la perspective de notre problématique générale, cette deuxième ouverture empirique est ainsi pertinente pour différentes raisons : (i) la distribution spatiale des prix est un objet d'étude à la croisée des territoires et des réseaux, puisque les marchés locaux sont intimement liés aux caractéristiques territoriales, mais également déterminés par les propriété du réseau routier (par exemple accessibilité) et par ses motifs d'utilisation ; (ii) les deux échelles endogènes identifiées correspondent à nos échelles mesoscopique et macroscopique, confirmant par une autre approche la nécessité de prendre bien les deux en compte ; (iii) la mise en évidence de la superposition d'un processus de gouvernance à des processus géographiques locaux rejoint notre dernier développement sur la prise en compte de la gouvernance dans les modèles, puisque cette complexité est effectivement observée empiriquement ici.
}

\stars

%

\newpage

\section{Multi-scalar modeling of residential dynamics}{Modélisation multi-scalaire des dynamiques résidentielles} 

\label{app:sec:migrationdynamics} 



\bpar{
We have evoked in the introducing chapter issues on mobility (daily and residential) as similar processes to the ones that we studied all along this work, at an other scale and with other ontologies. We have furthermore suggested the opening towards multi-scale models as a privileged development and a relatively immediate application of the preliminary bricks we forged here. This appendix briefly describes a work precisely developing these two points, in the case of residential dynamics of rural migrants in Pearl River Delta in China.
}{
Nous avons effleuré dans le chapitre introductif les questions de mobilité (quotidienne et résidentielle) comme processus voisins de ceux qui nous ont occupé tout au long de ce travail, à une autre échelle et avec d'autres ontologies. Nous avons d'autre part suggéré l'ouverture vers des modèles multi-échelles comme un développement privilégié et une application relativement immédiate des briques préliminaires que nous avons forgé ici. Cet annexe présente brièvement un travail développant précisément ces deux points, dans le cas des dynamiques résidentielles des migrants ruraux dans le Delta de la rivière des Perles en Chine.
}

\stars


\bpar{
\textit{This work is the product of an interdisciplinary collaboration with the sociologist and sinologist \noun{Cinzia Losavio} (UMR CNRS 8504 Géographie-cités), in the context of the MEDIUM project. The text written in collaboration is here adapted. These results have been presented at the Urban China 2017 international conference as \cite{losavio2017modeling}}.
}{
\textit{Ce travail est le fruit d'une collaboration interdisciplinaire avec la sociologue et sinologue \noun{Cinzia Losavio} (UMR CNRS 8504 Géographie-cités), dans le cadre du projet MEDIUM. Le texte produit en collaboration est ici adapté et traduit. Ces résultats ont été présenté à la conférence internationale Urban China 2017 comme \cite{losavio2017modeling}}.
}

\stars


\bpar{
This section introduces an agent-based model of regional migration dynamics, applied to the Mega-city Region of Pearl River Delta. It focuses on residential dynamics of migrants workers and aims at taking into account the variety of migrant's profiles, based on qualitative fieldwork observations. The extensive exploration of the model, both on synthetic and real-world configurations, yield several stylized facts of migration processes and specific effects of the regional geography, which can be used to inform migration policies. We postulate that such integrated modeling approaches will be more and more appropriate to study cities in China.
}{
Cette section introduit un modèle basé agent des dynamique de migration intra-régionales, appliqué à la Méga-city Region du Delta de la Rivière des Perles. Il se concentre sur les dynamiques résidentielles de travailleurs migrants et vise à prendre en compte la variété des profils des migrants, en se basant sur des observations qualitatives de terrain. L'exploration intensive du modèle, à la fois sur des configurations synthétiques et réelles, fournit des faits stylisés sur les processus de migration, et des effets spécifiques de la géographie régionale, qui pourraient être appliqués aux régulations des migrations. Nous supposons que de telles approches de modélisation intégrées seront par l'avenir de l'étude des villes chinoises.
}

\subsection{Introduction}{Introduction}

\paragraph{Context}{Contexte}

\bpar{
Over the last three decades, rural-to-urban migrant-workers have been a driving force for China's economy, raising attention on associated socio-economical issues. However, the importance of their economic diversity and social mobility has been poorly considered in the analysis of urban development strategy.
}{
Ces dernières décades, les travailleurs migrants du rural vers l'urbain ont été une force motrice pour l'économie chinoise, attirant l'attention sur les questions socio-économiques associées. Cependant, l'importance de leur diversité économique et la mobilité sociale ont été peu considérés dans l'analyse des stratégies de développement urbaines.
}

\bpar{
We use an agent-based model to simulate residential dynamics of migrants in Pearl River Delta (PRD) mega city region, taking into account the full range of migrants’ socio-economical status and their evolution. Mega-city regions have become a new scale of Chinese State regulation, and PRD represent the most prosperous and dynamic one in term of migration waves, standing as an ideal unit of analysis.
}{
Nous utilisons un modèle basé agent pour simuler les dynamiques résidentielles des migrants dans la méga-région urbaine du Delta de la Rivière des Perles (PRD), prenant en compte l'ensemble des statuts socio-économique des migrants et leur évolution. Les méga-régions urbaines sont devenues une nouvelles échelle de régulation pour l'Etat Chinois, et le PRD représente le plus prospère et dynamique en termes de vagues de migrations. Cela en fait ainsi une parfaite unité d'analyse.
}

\paragraph{Mega-city Regions}{Méga-régions urbaines}

\bpar{
Mega-city regions (MCRs) as defined by \cite{florida2008rise} are ``integrated sets of cities and their surrounding suburban hinterlands across which labour and capital can be reallocated at very low cost''. This urban configuration recalls what \cite{gottmann1961megalopolis} defined as \emph{megalopolis} in reference to the north-east coast of the United States. Despite this affinity in their spatial and functional configuration, MCRs perform on a different scale than megalopolis: they operate at a regional as well as at a global scale. Indeed, one of the main characteristics of MCRs is their ``connectivity'': spatially, they branch out into nearby rural and metropolitan areas, and economically they grow beyond their physical border, becoming international. These densely populated regions do not have a single barycenter but merging into one another they turn into highly networked spaces connected through multiple nodes. The high density of connections and the polycentrism characterizing these new economic units facilitate migrations flows and encourage regional integration.
}{
Les \emph{Mega-city Regions} (MCR) sont définies par \cite{florida2008rise} comme ``\textit{des ensembles de villes intégrés et leur territoires suburbains environnants, au sein desquels le travail et le capital peuvent être réalloués à moindre coût}''. Cette configuration urbaine correspond à ce que \cite{gottmann1961megalopolis} définit comme \emph{megalopolis} en référence à la cote nord-est des Etats-unis. Malgré cette similarité dans leur configuration spatiale et fonctionnelle, les MCRs fonctionnent à une échelle différentes de celle de la megapolis : elles opèrent à la fois à une échelle régionale et à une échelle globale. En effet, l'une des caractéristiques principales des MCRs est leur connectivité : spatialement elles s'étendent aux régions urbaines et métropolitaines proches, et économiquement elles ont un impact international, bien au delà de leur frontière physique. Ces régions densément peuplées n'ont pas un unique barycentre, mais consistent en de multiples centres fortement connectés. La forte densité de connections et le polycentrisme caractérisant ces nouvelles unités économiques facilitent les flux de migration et encouragent l'intégration régionale.
}

\bpar{
In China, the development of mega-city regions has started right after the implementation of the Open Door Policy in 1978. But it is the gradual decentralization of the State power - which occurred in the beginning of 1990 – that promote cities and more recently mega-city regions as a new scale of Chinese State regulation~\cite{IJUR:IJUR12437}. The process of rapid economic growth and urban development molds new densely populated and industrially dynamic mega-city regions, of which the Pearl River Delta (PRD)\footnote{The PRD Mega City Region consists of nine cities: the core cities are Guangzhou and Shenzhen, surrounded by Dongguan, Foshan, Zhongshan, Zhuhai, Huizhou, Jiangmen, and Zhaoqing. The model does not include Hong Kong and Macau, which are part of the PRD Mega Urban Region but are not in mainland China.} is the most obvious example. The area was designed in 1988 as a ``comprehensive economic reform area'', and was granted many ``one step ahead'' policies to attract foreign capital. Evolving into the most important exporter center since the economic reform, the Pearl River Delta represents the most dynamic MCR in terms of migration waves~\cite{IJUR:IJUR820}.
}{
En Chine, le développement des méga-régions urbaines a commencé juste après l'implémentation de la politiques des portes ouvertes en 1978. Mais c'est la décentralisation progressive du pouvoir de l'Etat, qui a eu lieu au début des années 1990, qui promeut les villes et plus récemment les méga-régions urbaines comme une nouvelle échelle de régulation de l'Etat Chinois~\cite{IJUR:IJUR12437}. Le processus de croissance économique rapide et le développement urbain conduisent à de nouvelles méga-régions urbaines densément peuplées et industriellement dynamiques, parmi lesquelles le Delta de la Rivière des Perles (PRD)\footnote{La \emph{Mega City Region} du PRD est composée de neuf villes : les villes coeur sont Guangzhou et Shenzhen, entourées de Dongguan, Foshan, Zhongshan, Zhuhai, Huizhou, Jiangmen, et Zhaoqing. Le modèle n'inclut pas Hong-Kong et Macao, qui font partie de la méga-région urbaine du PRD mais ne sont pas en Chine continentale.} est l'exemple le plus représentatif. La zone a été choisie en 1988 comme une ``zone de réforme économique complète'', et il lui a été accordé de nombreuse politiques de ``pas en avant'' pour attirer le capital étranger. Evoluant vers le centre d'exportation le plus important depuis les réformes économiques, le Delta de la Rivière des Perles représente la MCR la plus dynamique en termes de vagues de migration~\cite{IJUR:IJUR820}.
}

\paragraph{Migrant Workers}{Travailleurs migrants}

\bpar{
Taking the PRD as the spatial unit of the model, we aim to reproduce migrant workers’ residential patterns taking into account the full range of their socio-economical status. Migration patterns and key related issues have extensively been studied from very diverse perspectives, ranging for example from racialization issues~\cite{dong2010policing} to big data analysis of their spatio-temporal behavior~\cite{2017arXiv170600682Y}. However, migrant workers are generally considered and treated as a uniform category, which stand at the bottom of the urban society, carrying the stigma of the rural household registration system. The rural-urban dual structure has been for years the only approach to define and understand migrant-workers, but the process of rapid economic growth China have been experiencing accelerated social transformation. We postulate that studying migrant workers, by merely considering their \emph{hukou} status and place of registration is not sufficient anymore to apprehend such a complex and diversified social category. Others aspects such as migrant workers economical, cultural and human capital should be taken into account.
}{
Considérant le PRD comme l'unité spatiale d'étude, nous visons à reproduire les motifs résidentiels des travailleurs migrants, en prenant en compte l'ensemble des possibilités de leur statut socio-économique. Les motifs de migration et les questions essentielles qui y sont rattachées ont été largement étudiés selon diverses perspectives, s'étendant par exemple de questions ethniques~\cite{dong2010policing} aux analyses par données massives de leur comportement spatio-temporel~\cite{2017arXiv170600682Y}. Cependant, les travailleurs migrants sont généralement considérés et traités comme une catégorie uniforme, qui est placée au bas de la société urbaine, portant les stigmates du système d'enregistrement rural. La structure duale rurale-urbain a depuis des années été l'unique approche pour définir et comprendre les travailleurs migrants, mais le processus de croissance économique rapide que la Chine a expérimenté a accéléré la transformation sociale. Nous postulons que l'étude des travailleurs migrants en considérant seulement leur statut de \emph{Hukou} et le lieu d'enregistrement n'est plus suffisant pour appréhender une telle catégorie sociale complexe et diversifiée. D'autres aspects comme le capital économique, culturel et humain des travailleurs migrants doit être pris en compte.
}

\bpar{
Especially three dimensions can help differentiate number of migrant-workers sub-categories: (i) the professional dimension, which not only determines migrants’ economical situation but also influences their trajectory and the duration of their staying in the city as well as their residential choice; (ii) the residential dimension which impacts all aspects of migrants’ urban lives – patterns of urban settlement, housing choices, residential conditions, relation with the city, neighborhood activities etc; (iii) the generational dimension.\footnote{The generational dimension is not taken into account in the model, as simulated dynamics correspond to rather short time scales, between 10 and 20 years.}
}{
En particulier, trois dimensions peuvent être utiles pour différencier un certain nombre de sous-catégories de travailleurs migrants : (i) la dimension professionnelle, qui détermine non-seulement la situation économique des migrants mais influence également leur trajectoires et la durée de leur séjour dans la ville ainsi que leurs choix résidentiels ; (ii) la dimension résidentielle qui influe sur l'ensemble des aspects des vies urbaines des migrants : établissements urbains, choix de logement, conditions résidentielles, relations à la ville, activités de voisinage, etc. ; (iii) la dimension de la génération\footnote{La dimension génération n'est pas prise en compte dans le modèle, puisque les dynamiques simulées correspondent à des échelles temporelles plutôt courtes, entre 10 et 20 ans.}.
}

\bpar{
All these sub-categories have different mobility patterns, that we simulate in the model. Considering this diversity and translating it in qualitative stylized facts that correspond to precise patterns of synthetic data, this model aims at establishing a new perspective for understanding China’s urban and regional mobility employing a more qualitative approach, specifying the mechanisms through which Party-State shape the parameters of migrants’ choices. 
}{
Toutes ce sous-catégories ont différents motifs de mobilité, que nous simulons dans le modèle. Considérant cette diversité et la traduisant en faits stylisés qualitatifs qui correspondent à des motifs précis de données synthétiques, ce modèle vise à établir une nouvelle perspective pour comprendre la mobilité résidentielle urbaine et régionale en Chine, en utilisant une approche plus qualitative par la spécification des mécanismes par lesquels l'Etat-Parti contrôle les paramètres des choix des migrants.
}

\subsection{Model}{Modèle}

\paragraph{Modeling Rural-urban migrations in China}{Modélisation des migrations rural-urbain en Chine}





\bpar{
Existing works in rural-urban migration modeling in China are mainly econometric studies, relying on census or on survey data. \cite{zhang2013measuring} estimate discrete choice models to study the trade-off between migration distance and earning difference. \cite{fan2005modeling} shows that gravity-based models can explain well inter-provincial migratory patterns, implying an underlying strong dominant aggregation processes. The positive association between wage gap and migration rates was obtained from time-series analysis in~\cite{zhang2003rural}. An empirical study of intra-urban migrants residential dynamics is done by~\cite{wu2006migrant}. \cite{xie2007simulating} uses an agent-based model to simulate the emergence of Urban Villages. To the best of our knowledge, there was no previous attempt in the literature to model regional migrations in China from an agent-based perspective.
}{
La plupart des travaux existant sur la modélisation de la migration rural-urbain en Chine sont principalement des études économétriques, qui se basent sur des données des sondages ou d'études ciblées. \cite{zhang2013measuring} estime des modèles de choix discrets pour étudier le trade-off entre distance de migration et différence de salaire. \cite{fan2005modeling} montre que des modèles gravitaires peuvent bien expliquer les motifs de migration inter-provinciaux, impliquant de forts processus dominants d'agrégation sous-jacents. L'association positive entre les écarts salariaux et les taux de migration a été obtenue à partir d'analyse de séries temporelles dans~\cite{zhang2003rural}. Une étude empirique des dynamiques résidentielles intra-urbaines des migrants est faite par~\cite{wu2006migrant}. \cite{xie2007simulating} utilise un modèle basé agent pour simuler l'émergence des villages urbains. Au meilleur de notre connaissance, il n'existe pas de tentative précédente dans la littérature pour modéliser les migrations régionales en Chine à partir d'une perspective basée agent.
}





\paragraph{Model}{Modèle}

\bpar{
The model is designed to include targeted stylized facts and experiments, in particular the role of the socio-economic structure of migrant population. More precisely, a recent shift in socio-economic structure of migrating population was observed, including a rise of middle-income migrants and a relativisation of the role of \emph{Hukou} in migration dynamics. The core of the model is thus centered on the exploration of the impact of a varying population economic structure for migrants on system dynamics, and the influence of government migration policies.
}{
Le modèle est conçu pour inclure des faits stylisés précis et des expériences associées, en particulier le rôle de la structure socio-économique de la population de migrants. Plus précisément, un changement récent dans la structure socio-économique de la population migrante a été observée, incluant une augmentation du nombre de migrants aux salaires médians et une relativisation du rôle du \emph{Hukou} dans les dynamiques migratoires. Le coeur du modèle est pour cette raison centré sur l'exploration de l'impact d'une variation de la structure économique de la population des migrants sur les dynamiques du système, et l'influence des politiques migratoires gouvernementales.
}

\bpar{
 The region is represented in the model by $N$ patches, characterized by their population $P_i(t)$ and an economic structure $E_i^{(c)}(t)$ giving a potential number of jobs for socio-economic classes $c$. The associated effective number of workers is denoted by $W_i^{(c)}(t)$. For the sake of simplicity, we assume a discrete number of classes. At initial time, the variables are initialized either following a synthetic data generation process (see below), or from real geographical data (abstracted and simplified to fit our context). 
}{
La région est représentée dans le modèle par $N$ patches, caractérisé par leur population $P_i(t)$ et une structure économique $E_i^{(c)}(t)$ qui représente un nombre potentiel d'emplois pour une classe socio-économique $c$. Le nombre effectif de travailleurs associés est noté $W_i^{(c)}(t)$. Dans un souci de simplicité, nous supposons un ensemble discret de classes. A l'instant initial, les variables sont initialisées soit selon un processus de génération de données synthétiques (voir ci-dessous), ou à partir de données géographiques réelles (abstraites et simplifiées pour répondre à notre contexte).
}

\bpar{ 
Urban Centers are characterized by aggregated population $\tilde{P}_k(t)$ and corresponding economic variables $\tilde{E}_k^{c}(t)$. An agent is a household of migrants, with location for residence and job. Socio-economic structure of the population is captured by the distribution of wealth $g(w)$, which are then stratified into categories. At a given time, the utility difference between not moving and moving to cell $j$ from cell $i$, for a category $c$ is given by
}{
Les centres urbains sont caractérisés par une population agrégée $\tilde{P}_k(t)$ et les variables économiques correspondantes $\tilde{E}_k^{c}(t)$. Un agent est un foyer de migrants, avec une localisation résidentielle et pour le travail. La structure socio-économique de la population est capturée par la distribution de richesse $g(w)$, qui est ensuite stratifiée en catégories. A un instant donné, la différence d'utilité entre l'action de relocalisation de la cellule $j$ vers la cellule $i$ et rester sur place, pour une catégorie $c$, est donnée par
}

\[
\Delta U_{i,j}^{(c)}(t) = \frac{Z_j^{(c)}- Z_i^{(c)}}{Z_0} + \gamma \cdot \frac{C_i^{(c)}- C_j^{(c)}}{C_0} - u_i^{(c)} - h_j^{(c)}
\]

\bpar{
where $Z_i^{(c)}$ is a measure of generalized accessibility given by 
}{
avec $Z_i^{(c)}$ une mesure d'accessibilité généralisée définie comme
}

\[
Z_i^{(c)} = P_i \cdot \sum_k \left[E_k^{(c)}-W_k^{(c)}\right]\cdot \exp{\left(\frac{-d_{ij}}{d_0}\right)}
\]

\bpar{
with $d_{ij}$ effective travel distance\footnote{As the model does not focus on the role of transportation, we take euclidian distance, and $d_0$ captures typical commuting distance in both public transportation or car. A more complicated model could include an explicit transportation network and modal choice depending on socio-economic category.} and $d_0$ commuting characteristic distance ; the parameter $\gamma$ is the ratio giving the relative importance of life cost compared to accessibility in the migration decisions ; $C_i^{(c)}$ is the cost of life which is a function of cell and city variables, that will be taken as $C_i^{(c)} \propto P_i^{\alpha_0}\cdot  \tilde{P}_i^{\alpha_1}$ ; $u_i^{(c)}$ a baseline aversion to move and $h_j^{(c)}$ an exogenous variable corresponding to regulation policies; $Z_0$ and $C_0$ dimensioning parameters.
}{
avec $d_{ij}$ une distance effective de trajet\footnote{Comme le modèle ne se concentre pas sur le rôle des transports, nous prenons la distance euclidienne, et $d_0$ capture une distance typique de migration pendulaire à la fois par transport public et voiture. Un modèle plus compliqué pourrait inclure un réseau de transport explicite ainsi que des choix modaux dépendants des catégories socio-économiques.} et $d_0$ distance de migration pendulaire typique ; le paramètre $\gamma$ est un rapport donnant l'importance relative du coût de la vie en comparaison à l'accessibilité dans les décisions migratoires ; $C_i^{(c)}$ est le coût de la vie qui est une fonction à la fois de la cellule et des variables de la ville, que nous prenons comme $C_i^{(c)} \propto P_i^{\alpha_0}\cdot  \tilde{P}_i^{\alpha_1}$ ; $u_i^{(c)}$ est une aversion au mouvement de référence et $h_j^{(c)}$ une variable exogène correspondant aux politiques de régulation ; $Z_0$ et $C_0$ sont des paramètres de dimensionnement.
}
 

\bpar{
At each time step, the system evolves sequentially according to the following rules: 
\begin{enumerate}
	\item cities-level variables are updated and distributed across patches variables (in our first experiments, we will assume short time scale and skip this step)
	\item new migrants enter the region and lean on social network to settle
	\item migration occur within the region, randomly drawn from discrete choice probabilities obtained with the above utility difference between two patches
	\item Migrants update their wealth and eventually economic category, according to an abstract ``quality of place'' that we associate to per-capita GDP which follows a scaling law of population.
\end{enumerate}
}{
A chaque pas de temps, le système évolue séquentiellement selon les règles suivantes :
\begin{enumerate}
	\item les variables au niveau de la ville sont mises à jour et distribuées aux variables de patch (dans les premières expériences, nous supposerons une courte échelle temporelle et sauterons cette étape) ;
	\item de nouveaux migrants entrent dans la région et se fient à leur réseau social pour s'établir ;
	\item les migrations ont lieu dans la région, tirées aléatoirement à partir des probabilités de choix discrets obtenues avec les différences d'utilités entre patches données ci-dessus ;
	\item les migrants mettent à jour leur richesse et possiblement leur catégorie économique, selon une ``qualité du lieu'' abstraite que nous associons au GPD par tête qui suit une loi d'échelle de la population.
\end{enumerate}
}

\subsection{Results}{Résultats}

\bpar{
The model is implemented in NetLogo, the open source implementation being available with results at \url{https://www.github.com/JusteRaimbault/MigrationDynamics}. We explore the model on synthetic city systems first, to isolate results due to processes from results due to geographical configuration. With such a random model where many parameters cannot be given directly a real-world value, it is necessary to explore intensively the parameter space to obtain robust conclusions. Using the software OpenMole~\cite{reuillon2013openmole}, we proceed to 1,599,495 simulations of the model on computation grid, achieving 15 years of equivalent CPU in around 2 days. We validate the model internally by checking the statistical convergence of indicators.
}{
Le modèle est implémenté en NetLogo, l'implémentation ouverte étant disponible avec les résultats à \url{https://www.github.com/JusteRaimbault/MigrationDynamics}. Le modèle est d'abord exploré sur des systèmes de villes synthétiques, afin d'isoler les résultats dus aux processus en eux-même des résultats dus à la configuration géographique. Avec un tel modèle stochastique pour lequel de nombreux paramètres ne peuvent être fixé par des valeurs observées, il est nécessaire d'explorer de manière extensive l'espace des paramètres pour obtenir des conclusions robustes. Grâce au logiciel OpenMole~\cite{reuillon2013openmole}, nous procédons à 1,599,495 simulations du modèle sur grille de calcul, réalisant autour de 15 ans de calcul en équivalent CPU en environ 2 jours. Nous validons le modèle de manière interne en vérifiant la convergence statistique des indicateurs.
}

\bpar{
From the baseline experiments (reference parameter values) we learn the following stylized facts on intrinsic dynamics at the core of the model:
\begin{enumerate}
	\item When migrants have a high propensity to move, the spatial repartition of jobs becomes suboptimal in intermediate regimes of stochasticity, corresponding to a regime where congestion dominates.
	\item This congestion regime corresponds to a linear decrease of job distance with randomness, meaning that social determinism creates spatial inequalities.
	\item Changing the relative importance of accessibility does not affect much the aggregated dynamics: an increased gain in mobility produced by policies such as individual transportation subsidies will have no effect on migrations patterns.
	\item Configurations with an intermediate value of move aversion (in which real configurations fall) yield a negative feedback effect of time, witnessing a progressive saturation. In a ``U-shape'' manner, very mobile or very fixed configurations yield positive feedback of time (increase in the number of migrations).
\end{enumerate}
We then turn to targeted experiments.
}{
Les expériences de contrôle (valeur de référence des paramètres) fournissent les faits stylisés suivants sur les dynamiques intrinsèques au coeur du modèle :
\begin{enumerate}
	\item Lorsque les migrants ont une forte potentialité de mobilité, la répartition spatiale des emplois devient sous-optimale dans des régimes intermédiaires de stochasticité (au sens des valeur prises par le paramètre de choix discrets), ce qui correspond à un régime où la congestion domine.
	\item Ce régime de congestion implique une décroissance linéaire de la distance à l'emploi avec la diminution du caractère aléatoire, ce qui signifie que le déterminisme social crée des inégalités spatiales.
	\item L'importance relative de l'accessibilité influe très peu les dynamiques agrégées. Ainsi, un gain croissant en mobilité (c'est-à-dire une importance accrue pour l'accessibilité), qui peut être encouragé par des politiques locales telles des subventions, n'aura que très peu d'effet sur les motifs de migration.
	\item Les configurations avec des valeurs intermédiaires de l'aversion au mouvement (dans lesquelles la situation réelle se trouve) induisent un effet de feedback négatif du temps au cours des trajectoires, témoignant d'une saturation progressive. Dans un comportement ``en-U'', les configurations très mobiles et celles très stables donnent un effet positif du temps (augmentation du nombre de migrations).
\end{enumerate}
Nous étudions ensuite des expériences ciblées.
}

\bpar{
Adding categorization does not change the qualitative behavior of the model. The lower category appears more vulnerable to spatial inequalities created by social determinism. Concerning the influence of economic parameters, namely income inequality and income growth, we find that : (i) larger income inequalities yield stronger spatial inequities in job accessibility; (ii) larger enrichments when migrating induces a suboptimal regime for the upper category.
}{
L'ajout des catégories socio-économiques ne change pas fondamentalement le comportement qualitatif du modèle. La catégorie la plus basse semble cependant plus vulnérable aux inégalités spatiales induites par le déterminisme spatial. Concernant l'influence des paramètres économiques, en particulier les inégalités de revenu et la croissance des revenus, nous trouvons que : (i) de plus fortes inégalités de revenus induisent de plus fortes inégalités spatiales dans l'accessibilité à l'emploi ; (ii) une croissance des revenus plus forte (un enrichissement plus grand) lors d'une migration conduit à un régime sous-optimal pour la catégorie supérieure.
}

\bpar{
The application of the model on the real population and economic configuration of Pearl River Delta slightly changes conclusions: we witness for example the emergence of optimal behavior ranges for the commuting distance indicators. It means that incentives for migrations have to be specifically tuned depending on the region configuration. Other conclusions mainly hold and are therefore process-specific.
}{
L'application du modèle sur la configuration observée pour la population et les emplois dans le delta de la rivière des Perles change légèrement les conclusions : celle-ci témoigne par exemple de plages optimales pour les paramètres comportementaux au regard des indicateurs de distance de mobilité pendulaire. Cela signifie que les incitations aux migrations doivent être spécifiquement conçues selon la configuration de la région. La plupart des conclusions tirées précédemment tiennent toujours, et sont ainsi spécifiques au processus considérés.
}

\paragraph{Discussion}{Discussion}

\bpar{
A last application we are currently developing is testing the impact of localized regulation policies, i.e. having the term $h_j^{(c)}$ varying across cities and across categories, what corresponds to policies effectively observed in practice. This various stylized facts listed above may furthermore inform more general policies, such as the impact of mobility or the existence of optimal regimes for intermediate values of randomness. Further work may consist in a calibration of the model on migration trajectories with appropriate datasets, but also in a feedback of simulation results on qualitative fieldwork, trying to compare to concrete real situations.
}{
Une dernière application en cours de développement est l'exploration de l'impact de politiques de régulation localisées, i.e. en ayant le terme $h_j^{(c)}$ qui varie selon les villes et selon les catégories socio-économiques, ce qui correspondrait à des politiques effectivement observées en pratique. Les divers faits stylisés décrits précédemment peuvent de plus être porteurs de sens pour l'élaboration de politiques plus générales, comme l'impact d'une mobilité accrue, ou bien l'existence de régimes optimaux pour des valeurs intermédiaires du caractère aléatoire. Un développement futur pourra consister en une calibration du modèle sur des trajectoires de migration avec les jeux de données appropriés, mais également en un retour des résultats de simulation sur le travail de terrain qualitatif, en les comparant aux situation concrètement observées.
}


\bpar{
This modeling entreprise is aimed at being integrated, as the model is initially built with taking in consideration qualitative observations from fieldwork\footnote{To recall the context in the more general frame of the thesis, this is not our fieldwork described in~\ref{sec:qualitative}, but the one by \noun{Cinzia Losavio} realized in the context of her current thesis (see contributions above).}, and as its outputs shall in return inform qualitative research. We believe that such integrated modeling approaches will be important tools in the future of Urban China research, in particular because of the emergence of new urban regimes in Chinese cities that were never observed somewhere else before, making difficult the use of some of previous empirical knowledge on cities.
}{
Cette entreprise de modélisation vise à être intégrée, puisque le modèle est initialement construit en prenant en considération des observation qualitative du travail de terrain\footnote{Pour rappeler le contexte dans le cadre plus général de la thèse, il ne s'agit pas du travail de terrain décrit en~\ref{sec:qualitative}, mais de celui de \noun{Cinzia Losavio} réalisé dans le cadre de sa thèse en cours (voir contributions ci-dessus).}, et ses sorties devraient en retour être utiles pour la recherche qualitative. Nous sommes convaincus que de telles approches de modélisation intégrée seront des outils importants pour le futur de la recherche Urbaine en Chine, en particulier en lien avec l'émergence de nouveaux régimes urbain dans les villes chinoises qui n'ont jamais été observés ailleurs auparavant, rendant difficile l'utilisation de certaines des connaissances empiriques précédentes sur les villes.
}

\stars

\begin{figure}
\includegraphics[width=\linewidth]{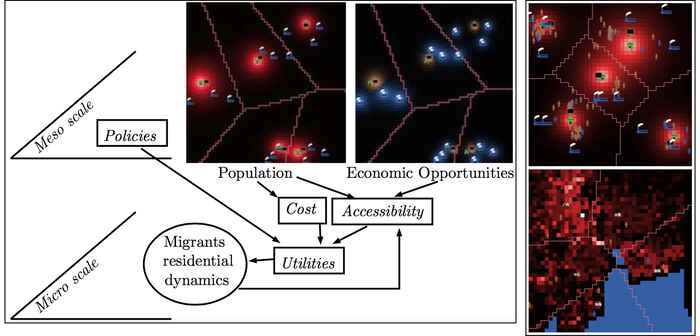}
\appcaption{\textbf{Structure of the intra-regional migration model.}(\textit{Left}) Multi-scale schema of processes included in the model. (\textit{Right}) Examples of regional population configuration, for a synthetic city system (top) and Pearl River Delta (bottom).\label{fig:app:migrationdynamics:model}}{\textbf{Structure du modèle de migrations intra-régionales.} (\textit{Gauche}) Schéma des processus pris en compte et des agents, dans une perspective multi-scalaire. (\textit{Droite}) Exemples de configurations régionales de populations, pour un système de villes synthétique (haut) et le Delta de la Rivière des Perles (bas).\label{fig:app:migrationdynamics:model}}
\end{figure}

\begin{figure}
\includegraphics[width=\linewidth]{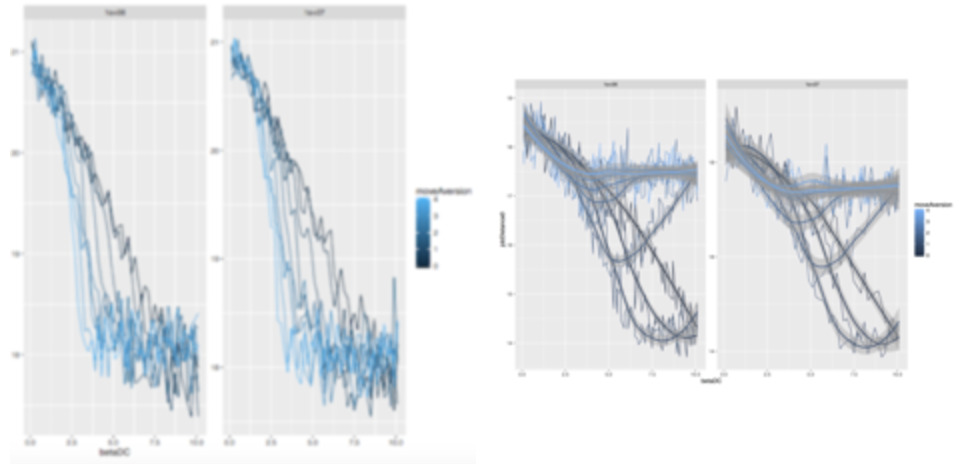}
\appcaption{\textbf{Example of stylized facts obtained with the model.} Comparison of average distance to jobs for the lowest economic category, as a function of the randomness parameter $\beta$, between synthetic city systems (\textit{two left plots}) and real configuration (\textit{two right plots}). The color gives the constant move aversion $u_i^{(c)}$ and plots are given for two values of cost-accessibility ratio $\gamma$. We witness the apparition of optimal values of $\beta$ in the real situation, probably caused by the geography.\label{fig:app:migrationdynamics:results}}{\textbf{Exemple de faits stylisés obtenus par le modèle.} Nous comparons la distance moyenne aux emplois pour la catégorie économique la plus basse, en fonction du paramètre d'aléatoire $\beta$, entre des systèmes de villes synthétiques (\textit{Deux graphes de gauche}) et la configuration réelle (\textit{Deux graphes de droite}). La couleur donne l'aversion au mouvement constante $u_i^{(c)}$ et les graphes sont donnés pour deux valeurs du ratio entre coût et accessibilité $\gamma$. Nous voyons émerger des valeurs optimales pour $\beta$ dans la situation réelle, probablement causées par la géographie.\label{fig:app:migrationdynamics:results}}
\end{figure}

\newpage

\section{Correlated Synthetic Data: financial time series}{Données synthétiques corrélées : séries temporelles financières} 

\label{app:sec:syntheticdata-finance} 


\subsection*{Context}{Contexte}

\bpar{
A field of application for the synthetic data method presented in~\ref{app:sec:syntheticdata} is that of financial complex systems, of which captured signals, financial time-series, are heterogeneous, multi-scalar and highly non-stationary~\cite{mantegna1999introduction}. Correlations have already been the object of a broad bunch of related literature. For example, Random Matrix Theory allows to undress signal of noise, or at least to estimate the proportion of information undistinguishable from noise, for a correlation matrix computed for a large number of asset with low-frequency signals (daily returns mostly)~\cite{2009arXiv0910.1205B}. Similarly, Complex Network Analysis on networks constructed from correlations, by methods such as Minimal Spanning Tree~\cite{2001PhyA..299...16B} or more refined extensions developed for this purpose~\cite{tumminello2005tool}, yielded promising results such as the reconstruction of economic sectors structure. At high frequency, the precise estimation of of interdependence parameters in the framed of fixed assumptions on asset dynamics, has been extensively studied from a theoretical point of view aimed at refinement of models and estimators~\cite{barndorff2011multivariate}. Theoretical results must be tested on synthetic datasets as they ensure a control of most parameters in order to check that a predicted effect is indeed observable \emph{all things equal otherwise}. For example, \cite{potiron2015estimation} obtains a bias correction for the \emph{Hayashi-Yoshida} estimator (used to estimate integrated covariation between two brownian at high frequency in the case of asynchronous observation times) by deriving a central limit theorem for a general model that endogeneize observation times. Empirical confirmation of estimator improvement is obtained on a synthetic dataset at a fixed correlation level.
}{
Un domaine d'application proposé pour la méthode de données synthétiques présentée en~\ref{app:sec:syntheticdata} est celui des séries temporelles financières, signaux typiques de systèmes complexes hétérogènes et multiscalaires~\cite{mantegna1999introduction} et pour lesquels les corrélations ont fait l'objet d'abondants travaux. Ainsi, l'application de la théorie des matrices aléatoires peut permettre de débruiter, ou du moins d'estimer la part de signal noyée dans le bruit, une matrice de correlations pour un grand nombre d'actifs échantillonnés à faible fréquence (retours journaliers par exemple)~\cite{2009arXiv0910.1205B}. De même, l'analyse de réseaux complexes construits à partir des corrélations, selon des méthodes type arbre couvrant minimal~\cite{2001PhyA..299...16B} ou des extensions raffinées pour cette application précise~\cite{tumminello2005tool}, ont permis d'obtenir des résultats prometteurs, tels la reconstruction de la structure économique des secteurs d'activités. A haute fréquence, l'estimation précise de paramètres d'interdépendance dans le cadre d'hypothèses fixées sur la dynamique, fait l'objet d'importants travaux théoriques dans un but de raffinement des modèles et des estimateurs~\cite{barndorff2011multivariate}. Les résultats théoriques doivent alors être testés sur des jeux de données synthétiques, qui permettent de contrôler un certain nombre de paramètres et de s'assurer qu'un effet prédit par la théorie est bien observable \emph{toutes choses égales par ailleurs}. Par exemple, \cite{potiron2015estimation} dérive une correction du biais de l'estimateur de \emph{Hayashi-Yoshida} qui est un estimateur de la covariance de deux browniens corrélés à haute fréquence dans le cas de temps d'observation asynchrones, par démonstration d'un théorème de la limite centrale pour un modèle généralisé endogénéisant les temps d'observations. La confirmation empirique de l'amélioration de l'estimateur est alors obtenue sur un jeu de données synthétiques à un niveau de corrélation fixé.
}

\subsection*{Formalization}{Formalisation}

\subsubsection*{Framework}{Cadre}

\bpar{
We consider a network of assets $(X_i(t))_{1\leq i \leq N}$ sampled at high-frequency (typically 1s). We use a multi-scalar framework (used e.g. in wavelet analysis approaches~\cite{ramsey2002wavelets} or in multi-fractal signal processing~\cite{bouchaud2000apparent}) to interpret observed signals as the superposition of components at different time scales : $X_i=\sum_{\omega}{X_i^{\omega}}$. We denote by $T_i^{\omega} = \sum_{\omega' \leq \omega} X_i^{\omega}$ the filtered signal at a given frequency $\omega$. A recurrent problem in the study of complex systems is the prediction of a trend at a given scale. It can be viewed as the identification of regularities and their distinction from components considered as random\footnote{see~\cite{gell1995quark} for an extended discussion on the construction of \emph{schema} to study complex adaptive systems (by complex adaptive systems).}. For the sake of simplicity, we represent such a process as a trend prediction model at a given temporal scale $\omega_1$, formally an estimator $M_{\omega_1} : (T_i^{\omega_1}(t'))_{t'<t} \mapsto \hat{T_i}^{\omega_1}(t)$ which aims to minimize error on the real trend $\norm{T_i^{\omega_1} - \hat{T}_i^{\omega_1}}$. In the case of autoregressive multivariate estimators, the performance will depend among other parameters on respective correlations between assets. It is thus interesting to apply the method to the evaluation of performance as a function of correlation at different scales. We assume a Black-Scholes dynamic for assets~\cite{jarrow1999honor}, i.e. $dX = \sigma\cdot dW$, with $W$ Wiener process. Such a dynamic model allows an easy modulation of correlation levels.
}{
Considérons un réseau d'actifs $(X_i(t))_{1\leq i \leq N}$ échantillonnés à haute fréquence (typiquement 1s). On se place dans un cadre multi-scalaire (utilisé par exemple dans les approches par ondelettes~\cite{ramsey2002wavelets} ou analyses multifractales du signal~\cite{bouchaud2000apparent}) pour interpréter les signaux observés comme la superposition de composantes à des multiples échelles temporelles : $X_i=\sum_{\omega}{X_i^{\omega}}$. On notera $T_i^{\omega} = \sum_{\omega' \leq \omega} X_i^{\omega}$ le signal filtré à une fréquence $\omega$ donnée. Prédire l'évolution d'une composante à une échelle donnée est alors un problème caractéristique de l'étude des systèmes complexes, pour lequel l'enjeu est l'identification de régularités et leur distinction des composantes considérées comme stochastiques en comparaison\footnote{voir~\cite{gell1995quark} pour une discussion étendue sur la construction de \emph{schema} pour l'étude de systèmes complexes adaptatifs (par des systèmes complexes adaptatifs).}. Dans un souci de simplicité, on représente un tel processus par un modèle de prédiction de tendance à une échelle temporelle $\omega_1$ donnée, formellement un estimateur $M_{\omega_1} : (T_i^{\omega_1}(t'))_{t'<t} \mapsto \hat{T_i}^{\omega_1}(t)$ dont l'objectif est la minimisation de l'erreur sur la tendance réelle $\norm{T_i^{\omega_1} - \hat{T}_i^{\omega_1}}$. Dans le cas d'estimateurs auto-regressifs multivariés, la performance dépendra entre autre des correlations respectives entre actifs et il est alors intéressant d'utiliser la méthode pour évaluer celle-ci en fonction de niveaux de correlation à plusieurs échelles. On assume une dynamique de Black-Scholes~\cite{jarrow1999honor} pour les actifs, i.e. $dX = \sigma\cdot dW$ avec $W$ processus de Wiener, ce qui permettra d'obtenir facilement des niveaux de correlation voulus.
}

\subsubsection*{Data generation}{Génération des données}

\bpar{
We can straightforward generate $\tilde{X}_i$ such that $\Varb{\tilde{X}_i^{\omega_1}}=\Sigma R \Sigma$ (with $\Sigma$ estimated standard deviations and $R$ fixed correlation matrix) and verifying $X_i^{\omega \leq \omega_0} = \tilde{X}_i^{\omega \leq \omega_0}$ (data proximity indicator : components at a lower frequency than a fundamental frequency $\omega_0 < \omega_1$ are identical). We use therefore the simulation of Wiener processes with fixed correlation. Indeed, if $dW_1 \indep dW_1^{\indep}$ (and $\sigma_1 < \sigma_2$ indicatively, assets being interchangeable), then $W_2 = \rho_{12}W_1 + \sqrt{1-\frac{\sigma_1^2}{\sigma_2^2}\cdot\rho_{12}^2}\cdot W_1^{\indep}$ is such that $\rho(dW_1,dW_2)=\rho_{12}$. Next signals are constructed the same way by Gram orthonormalization. We isolate the component at the desired frequency $\omega_1$ by filtering the signal, i.e. $\tilde{X}_i^{\omega_1} = W_i - \mathcal{F}_{\omega_0}[W_i]$ (with $\mathcal{F}_{\omega_0}$ low-pass filter with cut-off frequency $\omega_0$). We reconstruct then the hybrid synthetic signals by $\tilde{X}_i = T_i^{\omega_0} + \tilde{X}_i^{\omega_1}$.
}{
Il est alors aisé de générer $\tilde{X}_i$ tel que $\Varb{\tilde{X}_i^{\omega_1}}=\Sigma R$ ($\Sigma$ variance estimée et $R$ matrice de corrélation fixée), par la simulation de processus de Wiener au niveau de corrélation fixé et tel que $X_i^{\omega \leq \omega_0} = \tilde{X}_i^{\omega \leq \omega_0}$ (critère de proximité au données : les composantes à plus basse fréquence qu'une fréquence fondamentale $\omega_0 < \omega_1$ sont identiques). En effet, si $dW_1 \indep dW_1^{\indep}$ (et $\sigma_1 < \sigma_2$ pour fixer les idées, quitte à échanger les actifs), alors $W_2 = \rho_{12}W_1 + \sqrt{1-\frac{\sigma_1^2}{\sigma_2^2}\cdot\rho_{12}^2}W_1^{\indep}$ est tel que $\rho(dW_1,dW_2)=\rho_{12}$. Les signaux suivants sont construits de la même manière par orthonormalisation de Gram. On isole alors la composante à la fréquence $\omega_1$ voulue par filtrage, c'est-à-dire $\tilde{X}_i^{\omega_1} = W_i - \mathcal{F}_{\omega_0}[W_i]$ (avec $\mathcal{F}_{\omega_0}$ filtre passe-bas à fréquence de coupage $\omega_0$), puis on reconstruit les signaux synthétiques par $\tilde{X}_i = T_i^{\omega_0} + \tilde{X}_i^{\omega_1}$.
}

\subsection*{Results}{Résultats}

\subsubsection*{Methodology}{Méthodologie}

\bpar{
The method is tested on an example with two assets from foreign exchange market (EUR/USD and EUR/GBP), in a six month period from June 2015 to November 2015. Data\footnote{Obtained from \texttt{http://www.histdata.com/}, without specified licence. For the respect of copyright, only cleaned and filtered at $\omega_m$ data are made openly available.} cleaning, starting from original series sampled at a frequency around 1s, consists in a first step to the determination of the minimal common temporal range (missing sequences being ignored, by vertical translation of series, i.e. $S(t):=S(t)\cdot \frac{S(t_{n})}{S(t_{n-1})}$ when $t_{n-1},t_n$ are extremities of the ``hole'' and $S(t)$ value of the asset, what is equivalent to keep the constraint to have returns at similar temporal steps between assets). We study then \emph{log-prices} and \emph{log-returns}, defined by $X(t):=\log{\frac{S(t)}{S_0}}$ and $\Delta X (t) = X(t) - X(t-1)$. Raw data are filtered at a maximal frequency $\omega_m = 10\textrm{min}$ (which will be the maximal frequency for following treatments) for concerns of computational efficiency. We use a non-causal gaussian filter of total width $\omega$. We fix the fundamental frequency $\omega_0=24\textrm{h}$ and we propose to construct synthetic data at frequencies $\omega_1 = 30\textrm{min},1\textrm{h},2\textrm{h}$. See Fig.~\ref{fig:syntheticdata:example_signal} for an example of signal structure at these different scales.
}{
La méthode est testée sur un exemple de deux actifs du marché des devises (EUR/USD et EUR/GBP), sur une période de 6 mois de juin 2015 à novembre 2015. Le nettoyage des données\footnote{Obtenues depuis \url{http://www.histdata.com/}, sans licence spécifiée, les données nettoyées et filtrées à $\omega_m$ uniquement sont mises en accessibilité pour respect du copyright.}, originellement échantillonnées à l'ordre de la seconde, consiste dans un premier temps à la détermination du support temporel commun maximal (les séquences manquantes étant alors ignorées, par translation verticale des séries, i.e. $S(t):=S(t)\cdot \frac{S(t_{n})}{S(t_{n-1})}$ lorsque $t_{n-1},t_n$ sont les extrémités du ``trou'' et $S(t)$ la valeur de l'actif, ce qui revient à garder la contrainte d'avoir des retours à pas de temps similaires entre actifs). On étudie alors les \emph{log-price} et \emph{log-returns}, définis par $X(t):=\log{\frac{S(t)}{S_0}}$ et $\Delta X (t) = X(t) - X(t-1)$. Les données brutes sont filtrées à une fréquence $\omega_m = 10\textrm{min}$ (qui sera la fréquence maximale d'étude) pour un souci de performance computationnelle. On utilise un filtre gaussien non causal de largeur totale $\omega$. On fixe $\omega_0=24\textrm{h}$ et on se propose de construire des données synthétiques aux fréquences $\omega_1 = 30\textrm{min},1\textrm{h},2\textrm{h}$. Voir la figure~\ref{fig:syntheticdata:example_signal} pour un exemple de la structure du signal à ce différentes échelles.
}

\begin{figure}
\includegraphics[width=\linewidth]{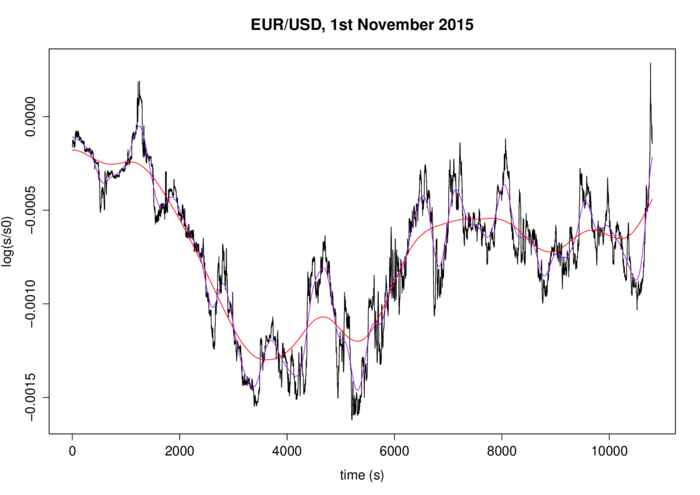}
\appcaption{\textbf{Example of the multi-scalar structure of the signal.} It is  the basis of the construction of synthetic signals. \emph{Log-prices} are represented on a time window of around 3h for November 1st 2015 for asset EUR/USD, with 10min (purple) and 30min trends.\label{fig:syntheticdata:example_signal}}{\textbf{Exemple de la nature multi-scalaire des signaux.} Celle-ci est à la base de la construction des signaux synthétiques. Les \emph{log-prices} sont représentés sur une fenêtre d'environ 3h le 1er novembre 2015 pour l'actif EUR/USD, avec en violet la tendance à 10min et en rouge à 30min.\label{fig:syntheticdata:example_signal}}
\end{figure}

\bpar{
It is crucial to consider the interference between $\omega_0$ and $\omega_1$ frequencies in the reconstructed signal : correlation indeed estimated is 
\[
\rho_{e} = \rho \left[ \Delta \tilde{X}_1 , \Delta \tilde{X}_2 \right] = \rho \left[ \Delta T_1^{\omega_0} + \Delta \tilde{X}_1^{\omega} , \Delta T_2^{\omega_0} + \Delta \tilde{X}_2^{\omega}\right]
\]
what yields in the reasonable limit $\sigma_1 \gg \sigma_0$ (fundamental frequency small enough), when $\Covb{\Delta \tilde{X}_i^{\omega_1}}{\Delta X_j^{\omega}}=0$ for all $i,j,\omega_1 > \omega$ and returns centered at any scale, by writing $\rho_0 = \rho \left[ \Delta T_1^{\omega_0} , \Delta T_2^{\omega_0} \right]$, $\rho = \rho \left[  \Delta \tilde{X}_1^{\omega_1} , \Delta \tilde{X}_2^{\omega_1} \right]$, et $\varepsilon_i = \frac{\sigma (\Delta T_i^{\omega_0})}{\sigma \left( \Delta \tilde{X}_i^{\omega_1}\right)}$, the correction on effective correlation due to interferences: we have at first order the expression of effective correlation

\begin{equation}
\label{eq:eff_corr}
\rho_e = \left[ \varepsilon_1 \varepsilon_2 \rho_0 + \rho \right] \cdot \left[ 1 - \frac{1}{2}\left(\varepsilon_1^2 + \varepsilon_2^2 \right) \right]
\end{equation}

{\noindent}what gives the correlation that we can effectively simulate in synthetic data.
}{
Il est crucial de noter l'interférence entre les fréquences $\omega_0$ et $\omega_1$ dans le signal construit : la correlation effectivement estimée est
\[
\rho_{e} = \rho \left[ \Delta \tilde{X}_1 , \Delta \tilde{X}_2 \right] = \rho \left[ \Delta T_1^{\omega_0} + \Delta \tilde{X}_1^{\omega} , \Delta T_2^{\omega_0} + \Delta \tilde{X}_2^{\omega}\right]
\]
ce qui conduit à dériver dans la limite raisonnable $\sigma_1 \gg \sigma_0$ (fréquence fondamentale suffisamment basse), lorsque $\Covb{\Delta \tilde{X}_i^{\omega_1}}{\Delta X_j^{\omega}}=0$ pour tous $i,j,\omega_1 > \omega$, et les retours d'espérance nulle à toutes échelles, en notant $\rho_0 = \rho \left[ \Delta T_1^{\omega_0} , \Delta T_2^{\omega_0} \right]$, $\rho = \rho \left[  \Delta \tilde{X}_1^{\omega_1} , \Delta \tilde{X}_2^{\omega_1} \right]$, et $\varepsilon_i = \frac{\sigma (\Delta T_i^{\omega_0})}{\sigma \left( \Delta \tilde{X}_i^{\omega_1}\right)}$, la correction sur la correlation effective due aux interférences : la correlation effective est alors au premier ordre

\begin{equation}
\label{eq:syntheticdata:eff_corr}
\rho_e = \left[ \varepsilon_1 \varepsilon_2 \rho_0 + \rho \right] \cdot \left[ 1 - \frac{1}{2}\left(\varepsilon_1^2 + \varepsilon_2^2 \right) \right]
\end{equation}

{\noindent}ce qui donne l'expression de la correlation que l'on pourra effectivement simuler dans les données synthétiques.
}

\bpar{
Correlation is estimated by Pearson method, with estimator for covariance corrected for bias, i.e.
\[
\hat{\rho}[X1,X2] = \frac{\hat{C}[X1,X2]}{\sqrt{\hat{\Var{}}[X1]\hat{\Var{}}[X2]}}\]
, where $\hat{C}[X1,X2] = \frac{1}{(T-1)}\sum_{t} X_1(t)X_2(t) - \frac{1}{T\cdot (T-1)} \sum_t X_1(t) \sum_t X_2(t)$ and $\hat{\Var{}}[X] = \frac{1}{T}\sum_t{X^2(t)}-\left(\frac{1}{T}\sum_tX(t)\right)^2$.
}{
La correlation est estimée par méthode de Pearson, avec l'estimateur de la covariance au biais corrigé, c'est-à-dire 
\[
\hat{\rho}[X1,X2] = \frac{\hat{C}[X1,X2]}{\sqrt{\hat{\Var{}}[X1]\hat{\Var{}}[X2]}}
\]
, où $\hat{C}[X1,X2] = \frac{1}{(T-1)}\sum_{t} X_1(t)X_2(t) - \frac{1}{T\cdot (T-1)} \sum_t X_1(t) \sum_t X_2(t)$ et $\hat{\Var{}}[X] = \frac{1}{T}\sum_t{X^2(t)}-\left(\frac{1}{T}\sum_tX(t)\right)^2$.
}

\bpar{
The tested predictive model $M_{\omega_1}$ is a simple \emph{ARMA} for which parameters $p=2,q=0$ are fixed (as we do not create lagged correlation, we do not expect large orders of auto-regression as these kind of processes have short memory for real data; furthermore smoothing is not necessary as data are already filtered). It is however applied in an adaptive way\footnote{adaptation level staying low, as parameters $T_W,p,q$ and model type do not vary. We are positioned within the framework of~\cite{potiron2016estimating} which assumes a locally parametric dynamic but for which meta-parameters are fixed. We could imagine a variable $T_W$ which would adapt for the best local fit, the same way parameters are estimated in bayesian signal processing by augmentation of the state with parameters.}. More precisely, given a time window $T_W$, we estimate for any $t$ the model on $[t-T_W+1,t]$ in order to predict signals at $t+1$.
}{
Le modèle de prédiction $M_{\omega_1}$ testé est simplement un modèle \emph{ARMA} pour lequel on fixe les paramètres $p=2,q=0$ (on ne créée pas de correlation retardée, on ne s'attend donc pas à de grand ordre d'auto-regression, les signaux originaux étant à mémoire relativement courte ; de plus le lissage n'est pas nécessaire puisqu'on travaille sur des données filtrées), appliqué de manière adaptative\footnote{il s'agit d'un niveau d'adaptation relativement faible, les paramètres $T_W,p,q$ et même le type de modèle restant fixés. On se place ainsi dans le cadre de~\cite{potiron2016estimating} qui suppose une dynamique localement paramétrique, mais pour lequel on fixe les méta-paramètres de la dynamique. On pourrait imaginer estimer un $T_W$ variable qui s'adapterait pour une meilleure estimation locale, à l'image de l'estimation de paramètres en traitement du signal Bayesien effectuée via augmentation de l'état par les paramètres.}. Plus précisément, étant donné une fenêtre temporelle $T_W$, on estime pour tout $t$ le modèle sur $[t-T_W+1,t]$ afin de prédire les signaux à $t+1$.
}

\paragraph{Implementation}{Implémentation}

\bpar{
Experiments are implemented in \texttt{R} language, using in particular the \texttt{MTS}~\cite{Tsay:2015xy} library for time-series models. Cleaned data and source code are openly available on the \texttt{git} repository of the project\footnote{at \texttt{https://github.com/JusteRaimbault/SynthAsset}}. 
}{
L'implémentation est faite en language R, utilisant en particulier la bibliothèque \texttt{MTS}~\cite{Tsay:2015xy} pour les modèles de séries temporelles. Les données nettoyées et le code source sont disponibles de manière ouverte sur le dépôt \texttt{git} du projet\footnote{À \url{https://github.com/JusteRaimbault/SynthAsset}.}.
}

\paragraph{Results}{Résultats}

\bpar{
Figure~\ref{fig:syntheticdata:effective_corrs} gives effective correlations computed on synthetic data. For standard parameter values (for example $\omega_0=24\textrm{h}$, $\omega_1=2\textrm{h}$ and $\rho=-0.5$), we find $\rho_0\simeq 0.71$ et $\varepsilon_i \simeq 0.3$ what yields $\left| \rho_e - \rho \right|\simeq 0.05$. We observe a good agreement between observed $\rho_e$ and values predicted by~\ref{eq:eff_corr} in the interval $\rho \in [-0.5,0.5]$. On the contrary, for larger absolute values, a deviation increasing with $\left|\rho\right|$ and as $\omega_1$ decreases : it confirms the intuition that when frequency decreases and becomes closer to $\omega_0$, interferences between the two components are not negligible anymore and invalidate independence assumptions for example. 
}{
La figure~\ref{fig:syntheticdata:effective_corrs} donne les correlations effectives calculées sur les données synthétiques. Pour des valeurs standard des paramètres (par exemple pour $\omega_0=24\textrm{h}$, $\omega_1=2\textrm{h}$ et $\rho=-0.5$), on a $\rho_0\simeq 0.71$ et $\varepsilon_i \simeq 0.3$ et ainsi $\left| \rho_e - \rho \right|\simeq 0.05$. On constate dans l'intervalle $\rho \in [-0.5,0.5]$ un bon accord entre la valeur $\rho_e$ prédite ci-dessus et les valeurs observées, et une déviation pour de plus grandes valeurs absolues, d'autant plus grande que $\omega_1$ est petit : cela confirme l'intuition que lorsque la fréquence descend et se rapproche de $\omega_0$, les interférences entre les deux composantes vont devenir non négligeables et invalider les hypothèses d'indépendance par exemple.
}

\bpar{
We apply then the predictive model described above to synthetic data, in order to study its mean performance as a function of correlation between signals. Results for $\omega_1 = 1\textrm{h},1\textrm{h}30,2\textrm{h}$ are shown in Fig.~\ref{fig:syntheticdata:model_perf}. The a priori counter-intuitive result of a maximal performance at vanishing correlation for one of the assets confirms the role of synthetic data to better understand system mechanisms : the study of lagged correlations shows an asymmetry in the real data that we can understand at a daily scale as an increased influence of EUR/GBP on EUR/USD with a rough two hours lag. The existence of this \emph{lag} allows a ``good'' prediction of EUR/USD thanks to fundamental component. This predictive power is perturbed by added noises in a way that increases with their correlation. The more noises correlated are, the more he model will take them into account and will make false predictions because of the markovian character of simulated brownian\footnote{the model used has theoretically no predictive power at all on pure brownian}.
}{
On applique ensuite le modèle prédictif décrit ci-dessus aux données synthétiques, afin d'étudier sa performance moyenne en fonction du niveau de correlation des données. Les résultats pour $\omega_1 = 1\textrm{h},1\textrm{h}30,2\textrm{h}$ sont présentés en figure~\ref{fig:syntheticdata:model_perf}. Le résultat a priori contre-intuitif d'une performance maximale à correlation nulle pour l'un des actifs confirme l'intérêt d'une génération de données hybrides : l'étude des correlations décalées (\emph{lagged correlations}) montre une dissymétrie présente dans les données réelles, interprété à l'échelle journalière comme une influence augmentée de EURGBP sur EURUSD à 2h de décalage environ. L'existence de ce \emph{lag} permet une ``bonne'' prédiction de EURUSD due à la fréquence fondamentale, perturbée par le bruit ajouté, de façon proportionnelle à sa correlation : plus les bruits sont corrélés, plus le modèle les prendra en compte et se trompera plus à cause du caractère markovien des browniens simulés\footnote{En théorie le modèle utilisé n'a aucun pouvoir prédictif sur des browniens purs}.
}

\bpar{
This case study stays a \emph{toy-model} and has no direct practical application, but demonstrates however the relevance of using simulated synthetic data. Further developments can be directed towards the simulation of more realistic data (presence of consistent \emph{lagged correlation} patterns, more realistic models than Black-Scholes) and apply it on more operational models.
}{
L'exemple présenté ici est un \emph{modèle jouet} et n'a pas d'application pratique, mais démontre l'intérêt de l'utilisation des données synthétiques simulées. On peut imaginer simuler des données plus proches de la réalité (existence de motifs réalistes de \emph{lagged correlation} par exemple, modèles plus réalistes que le Black-Scholes) et appliquer la méthode sur des modèles plus opérationnels.
}

\stars

\begin{figure}
\includegraphics[width=\linewidth]{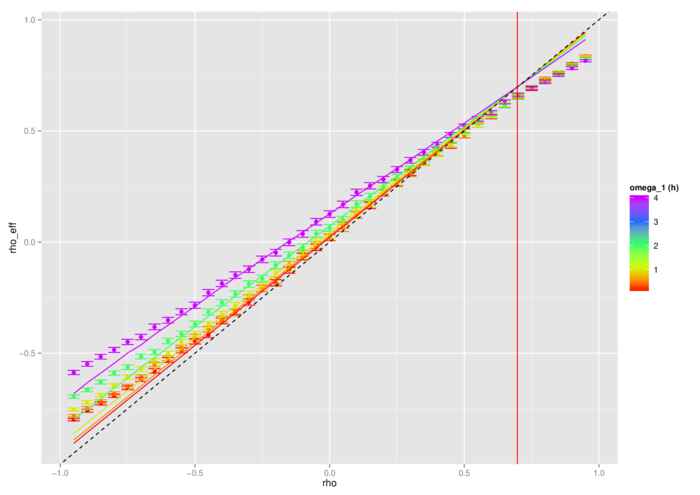}
\appcaption{\textbf{Effective correlations obtained on synthetic data.} Dots represent estimated correlations on a synthetic dataset corresponding to 6 months between June and November 2015 (error-bars give 95\% confidence intervals obtained with standard Fisher method); scale color gives filtering frequency $\omega_1=10\textrm{min},30\textrm{min},1\textrm{h},2\textrm{h},4\textrm{h}$ ; solid lines give theoretical values for $\rho_e$ obtained by~\ref{eq:syntheticdata:eff_corr} with estimated volatilities (dotted-line diagonal for reference); vertical red line position is the theoretical value such that $\rho = \rho_e$ with mean values for $\varepsilon_i$ on all points. We observe for high absolute correlations values a deviation from corrected values, what should be caused by non-verified independence and centered returns assumptions. Asymmetry is caused by the high value of $\rho_0 \simeq 0.71$.\label{fig:syntheticdata:effective_corrs}}{\textbf{Corrélations Effectives obtenues sur données synthétiques.} Les points donnents les corrélations estimées sur un jeu de données synthétiques basé sur 6 mois entre juin et novembre 2015 (les barres d'erreur donnent les intervalles de confiance à 95\% obtenus par méthode de Fisher standard) ; l'échelle de couleur donne la fréquence de filtrage $\omega_1=10\textrm{min},30\textrm{min},1\textrm{h},2\textrm{h},4\textrm{h}$ ; les lignes pleines donnent les valeurs théoriques obtenues par l'équation~\ref{eq:syntheticdata:eff_corr} avec les volatilités estimées (la diagonale en pointillé donne la référence) ; la ligne verticale rouge est à la position de la valeur théorique telle que $\rho = \rho_e$ avec les valeurs moyennes de $\varepsilon_i$ sur l'ensemble des points. Nous observons pour les fortes valeurs de corrélations absolues une déviation des valeurs corrigées, qui devrait être dues à la non-vérification des hypothèses d'indépendance et de centrage des retours. L'asymétrie est due à la forte valeur de $\rho_0 \simeq 0.71$.\label{fig:syntheticdata:effective_corrs}}
\end{figure}


\begin{figure}
\includegraphics[width=\linewidth]{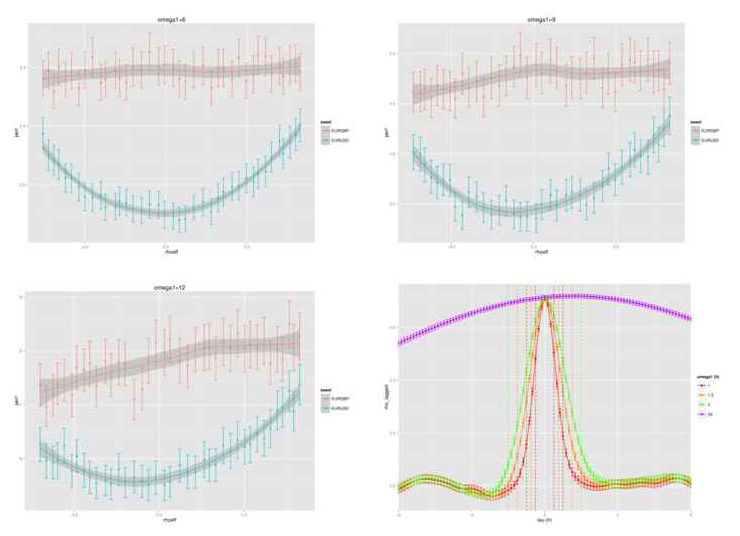}
\appcaption{\textbf{Performance of a predictive model as a function of simulated correlations.} From left to right and top to bottom, three first graphs show for each asset the normalized performance of an ARMA model ($p=2,q=0$), defined as $\pi = \left(\frac{1}{T}\sum_t\left(\tilde{X}_i(t) - M_{\omega_1}\left[\tilde{X}_i\right](t)\right)^2 \right) / \sigma \left[ \tilde{X}_i \right]^2$ (95\% confidence intervals computed by $\pi = \bar{\pi} \pm (1.96\cdot \sigma [\pi])/\sqrt{T}$, local polynomial smoothing to ease reading). It is interesting to note the U-shape for EUR/USD, due to interference between components at different scales. Correlation between simulated noises deteriorates predictive power. The study of \emph{lagged correlations} (here $\rho [\Delta X_{\textrm{EURUSD}}(t),\Delta X_{\textrm{EURGBP}}(t-\tau)]$) on real data clarifies this phenomenon: the fourth graph show an asymmetry in curves at any scale compared to zero lag $(\tau = 0)$ what leads fundamental components to increase predictive power for the dollar, amelioration then perturbed by correlations between simulated components. Dashed lines show time steps (in equivalent $\tau$ units) used by the ARMA at each scale, what allows to read the corresponding lagged correlation on fundamental components.\label{fig:syntheticdata:model_perf}}{\textbf{Performance d'un modèle prédictif en fonction des corrélations simulées.} De gauche à droite et de haut en bas, les trois premier graphes donnent pour chaque actif la performance normalisée d'un modèle ARMA ($p=2,q=0$), définie par $\pi = \left(\frac{1}{T}\sum_t\left(\tilde{X}_i(t) - M_{\omega_1}\left[\tilde{X}_i\right](t)\right)^2 \right) / \sigma \left[ \tilde{X}_i \right]^2$ (intervalles de confiance à 95\% calculés par $\pi = \bar{\pi} \pm (1.96\cdot \sigma [\pi])/\sqrt{T}$, le lissage polynomial local est pour l'aide à la lecture). Il est intéressant de noter la forme de U pour EUR/USD, due aux interférences entre composantes à différentes échelles. La corrélation entre les bruits simulés détériore le pouvoir de prédiction. L'étude des corrélation retardées (ici données par $\rho [\Delta X_{\textrm{EURUSD}}(t),\Delta X_{\textrm{EURGBP}}(t-\tau)]$) sur les données réelles clarifie ce phénomène : le quatrième graphe présente une asymétrie des courbes à toutes les échelles par rapport au décalage nul $(\tau = 0)$ ce qui conduit les composantes fondamentales à accroitre le pouvoir prédictif pour le dollar, amélioration qui est ensuite perturbée par les corrélations entre les composantes simulées. Les lignes pointillées donnent les pas de temps (en équivalent en unités de $\tau$) utilisés par les ARMA à chaque échelle, ce qui permet de lire la corrélation retardée correspondante sur les composantes fondamentales.\label{fig:syntheticdata:model_perf}}
\end{figure}

%


\newpage

\section{CybergeoNetworks: a multi-dimensional and spatialized bibliometric analysis}{CybergeoNetworks : une analyse bibliométrique multi-dimensionnelle spatialisée}

\label{app:sec:cybergeonetworks}

\bpar{
The analysis of the \emph{Cybergeo} corpus has also been an opportunity for reflexivity and to test more thoroughly the idea of applied perspectivism through the combination of methodological approaches. This appendix shows their complementarity and the new knowledge which can be produced by their coupling, in particular here through their spatialisation.
}{
L'analyse du corpus de \emph{Cybergeo} a également été occasion de réflexivité et de creuser l'idée de perspectivisme appliqué par la combinaison d'approches méthodologiques. Cette annexe montre leur complémentarité et les connaissances nouvelles qui peuvent être produites par leur couplage, par en particulier ici leur spatialisation.
}

\stars

\bpar{
\textit{This appendix is the result of a collaboration in the context of the 20 years anniversary of the Cybergeo journal: initiated by \noun{D. Pumain} (Université Paris 1) and \noun{A. Banos} (Université Paris 1), an interdisciplinary team composed by \noun{C. Cottineau} (University College London), \noun{P.-O. Chasset} (LISER), \noun{H. Commenges} (Université Paris 1), has lead an analysis by multiple and complementary methods of the Cybergeo journal corpus. The corresponding paper (submitted to Journal of Informetrics) is here adapted.}
}{
\textit{Cette annexe est le fruit d'une collaboration dans le cadre des 20 ans de la revue Cybergeo : initiée par \noun{D. Pumain} (Université Paris 1) et \noun{A. Banos} (Université Paris 1), une équipe interdisciplinaire composée de \noun{C. Cottineau} (University College London), \noun{P.-O. Chasset} (LISER), \noun{H. Commenges} (Université Paris 1), a mené une analyse par méthodes multiples et complémentaires du corpus de la revue Cybergeo. L'article correspondant (soumis à Journal of Informetrics) est ici traduit et adapté.}
}

\stars


\bpar{
Bibliometrics have become commonplace and widely used by authors and journal to monitor, evaluate and identify their readership in an ever-increasing publishing scientific world. With this contribution, we aim to move from the near-real time counts to investigate the semantic proximities and evolution of the papers published in the online journal Cybergeo since its creation in 1996. We compare three strategies for building semantic networks, using keywords (self-declared themes), citations (areas of research using the papers published in Cybergeo) and full-texts (themes derived from the words used in writing). We interpret these networks and semantic proximities with respect to their temporal evolution as well as spatial expressions, by considering the countries studied in the papers under inquiry. Finally, we compare the three methods and conclude that their complementarity can help go beyond simple statistics to better understand the epistemological evolution of a scientific community and the readership target of the journal.
}{
La bibliométrie est devenue monnaie courante et largement utilisée par les auteurs et les journaux pour suivre, évaluer et identifier le lectorat dans un contexte de publications toujours accrues. Cette contribution vise à se détacher des comptages en temps réel pour s'intéresser les proximités sémantiques et l'évolution des articles publiés dans le journal en ligne Cybergeo depuis sa création en 1996. Nous comparons trois stratégies pour construire des réseaux sémantiques, en utilisant les mots-clés (thématiques auto-déclarées), les citations (aires de recherche utilisant les articles publiés dans Cybergeo) et les textes complets (thèmes dérivés des mots utilisés dans l'écriture). Nous interprétons ces réseaux et les proximités sémantiques selon leur évolution temporelle ainsi que leur inscription spatiale, em considérant les pays étudiés dans les articles considérés. Enfin, nous comparons les trois méthodes et concluons que leur complémentarité peut contribuer à dépasser les simples statistiques pour mieux comprendre l'évolution épistémologique d'une communauté scientifique et de l'audience visée du journal.
}

\subsection{Introduction}{Introduction}

\bpar{
Since the seminal work of \noun{Kuhn} in the early 1960s the development of science studies has been based on three disciplinary pillars: history of science, philosophy of science, sociology of science. In the 1980s political science grew in importance studying the links between knowledge production and knowledge utilisation. This ``political turn'' began with the creation of the journal \textit{Knowledge} in 1979. Since late 1990s, science studies has been affected by a ``spatial turn'' and eventually emerged a geography of science \citep{livingston_spaces_1995,livingston_science_2003,livingston_geography_2005,withers_place_2009}. Our work follows this trend: we propose in this paper a spatialised bibliometrics approach.
}{
Depuis le travail fondateur de \noun{Kuhn} au début des années 1960, le développement des études de la science s'est basée sur trois pilier disciplinaires : l'histoire des sciences, la philosophie de la science et la sociologie de la science. Dans les années 1980, les sciences politiques ont pris une importance grandissante en étudiant les liens entre la production de la connaissance et l'utilisation de la connaissance. Ce ``tournant politique'' a commencé avec la création du journal \textit{Knowledge} en 1979. Depuis la fin des années 1990, les études de la science ont pris un ``tournant spatial'' et ont vu l'émergence d'une géographie de la science~\cite{livingston_spaces_1995,livingston_science_2003,livingston_geography_2005,withers_place_2009}. Ce travail se positionne dans cette tendance : cet article propose une approche de bibliométrie spatialisée.
}

\bpar{
Faced with the increasing number of articles, journals and channels of publication used by researchers in an open access and digital world, journals need tools to identify their readership and authors need this information to better reach their target audience, using the right keywords, vocabulary and citations. This paper provides a set of complementary digital tools which meet three requirements: 1) to go beyond the usual citation metrics and give semantic and network analytics directly from the scientific contents of the papers; 2) to situate the position of sets of papers according to the semantic fields of their topics; 3) to identify the significant variations in research topics that may be linked with the geographical origin of authors or to the country they choose to analyse. This last point is especially interesting for our first case of study which is a journal of geography.
}{
Faisant face à un nombre croissant d'articles, de journaux et de canaux de publication utilisés par les chercheurs dans un monde digital et d'accès ouvert, les journaux ont besoin d'identifier leur lectorat et les auteurs ont besoin de cette information pour mieux atteindre leur audience cible, en adaptant les mots-clés, le vocabulaire et les citations adaptés. Ce travail fournit un ensemble d'outils digitaux complémentaires qui satisfont trois pré-requis : 1) d'aller au delà les métriques de citation classiques et de proposer des analyses sémantiques et de réseaux extraits directement des contenus scientifiques des articles ; 2) de situer la position des ensemble d'articles selon les champs sémantiques de leurs thèmes ; 3) d'identifier les variations significatives dans les thématiques de recherche qui peuvent être reliées à l'origine géographique des auteurs ou aux pays étudiés. Ce dernier point est particulièrement intéressant dans notre cas comme notre cas d'étude est un journal en géographie.
}

\bpar{
The 20-year anniversary of the first journal exclusively digital in social science – Cybergeo –, was the occasion to analyse a consistent corpus of over 700 articles published in 7 languages, with respect to the geography of its authorship and readership. We performed a quantitative epistemology analysis of the scientific papers published since 1996 to measure their similarities according three types of textual indicators: their keywords (the way authors advertise their research), their citation network (the way the paper is used by other fields and disciplines), or their full-text (the vocabulary used to write the paper and present the research).
}{
L'anniversaire des 20 ans du premier journal exclusivement digital en sciences sociales, Cybergeo, a été l'occasion d'analyser un corpus conséquent de plus de 700 articles publiés dans 7 langages, selon la géographie des auteurs et du lectorat. Nous effectuons une analyse en épistémologie quantitative des articles scientifiques publiés depuis 1996 pour mesurer leur similarité selon trois types d'indicateurs textuels : leur mots-clés (la façon dont les auteurs situent leur recherche), leur réseau de citation (la façon dont l'article est utilisé par d'autres champs et disciplines), ou leur textes complets (le vocabulaire utilisé pour écrire l'article et présenter la recherche).
}

\bpar{
These analyses are complementary and show the evolution of a journal towards emergent themes of research. It also highlights the need for Cybergeo to keep extending its authorship base beyond the French-speaking community, in order to match its ambition of a European Journal of Geography. Our contribution consists in these specific epistemological conclusions, but also in a broader methodological and technical input on handling interactively large-scale heterogeneous scientific corpus. We show how the coupling of complementary views can create a second order knowledge: the spatial embedding of the three classification methods unveils unexpected patterns. Furthermore, the dedicated tool that we designed is available as an open source software, that can be used by journals for a collective scientific reflexivity, but also by institutions and individual scientists for a bottom-up empowerment of Open Science.
}{
Ces analyses sont complémentaires et montrent l'évolution d'un journal vers des thèmes de recherche émergents. Elles montrent aussi le besoin pour Cybergeo d'étendre sa base d'auteurs au delà de la communauté francophone, afin de remplir son ambition de journal Européen en géographie. Cette contribution consiste en ces conclusions épistémologiques spécifiques, mais aussi en une entrée méthodologique et technique plus large pour gérer de manière interactive des corpus scientifiques hétérogènes à grande échelle. Nous montrons dans quelle mesure le couplage de vues complémentaires peut créer une connaissance au second ordre : la contextualisation spatiale des trois méthodes de classification révèle des motifs inattendus. De plus, les outils dédiés que nous avons construit sont disponibles comme un logiciel open source, qui peut être utilisé par les journaux pour une réflexivité scientifique accrue, mais aussi par les institutions et les scientifiques eux-mêmes pour une autonomisation bottom-up de la Science Ouverte.
}

\bpar{
The rest of this section is organized as follows: we first review similar initiatives tackling heterogeneous or multidimensional approaches to bibliometrics, and describe the case study we work on. We then develop technical details of the different methods used, and how these are coupled through interactive spatial data exploration ; describe results at the first order (each method) and at the second order (achieved through coupling) ; and finally discuss broader implication for quantitative epistemology and reflexivity for Open Science.
}{
Le reste de cet section est organisé de la façon suivante : nous revoyons d'abord les approches similaires s'intéressant à la bibliométrie de manière hétérogène ou multidimensionnelle, et décrivons l'étude de cas sur laquelle nous travaillerons. Nous développons ensuite les détails techniques des différentes méthodes utilisées, et la manière dont celles-ci sont couplées par exploration interactive de données spatiales ; puis nous décrivons les résultats au premier ordre (chaque méthode) et au second ordre (obtenus par couplage) ; et nous discutons finalement des implications plus larges pour l'épistémologie et la réflexivité en Science Ouverte.
}

\subsubsection{Bibliographic context}{Contexte bibliographique}

\bpar{
Studies in bibliometrics having as a main focus the complementarity of different approaches are rather sparse. \cite{omodei2017evaluating} shows that taking into account citation and discipline data into a multilayer network is useful to understand patterns of interdisciplinarity. \cite{cronin2014beyond} is an attempt of an overview of the complex nature of measuring scientific publications and the intrinsic multidimensional nature of knowledge production. It provides both recent technical contributions with critical approaches. It insists on the ``Janus-faced nature of metrics'', confirming that reducing knowledge production to a few dimensions is not only wrong but also dangerous for science. The geographical dimension of science has been studied by numerous targeted studies, such as \cite{maisonobe2013diffusion} that investigates the diffusion of specific questions and practices in molecular biology across the world.
}{
Les études en bibliométrie ayant pour objet principal la complémentarité de différentes approches sont plutôt rares. \cite{omodei2017evaluating} montre que la prise en compte des données de citation et de disciplines dans un réseau multicouches permet de comprendre les motifs d'interdisciplinarité. \cite{cronin2014beyond} est une tentative d'un aperçu de la nature complexe de la mesure des publications scientifiques et de la nature multidimensionnelle de la production de connaissance. Il fournit à la fois des contributions techniques récentes et des approches critiques, et insiste sur la nature ``à-tête-de-Janus'' des métriques, confirmant que la reduction de la production de connaissance à peu de dimensions n'est pas seulement trompeur mais aussi dangereux pour la science. La dimension géographique de la science a été étudiée par de nombreuses études de cas, comme \cite{maisonobe2013diffusion} qui étudie la diffusion de questions et pratiques spécifiques en biologie moléculaire autour du monde. 
}

\subsubsection{Cybergeo as a case study}{Cybergeo comme cas d'étude}

\bpar{
Cybergeo was founded in 1996 as a digital-only European journal of geography. Since then, 737 scientific articles have been published (until May 2016) by 1351 authors from 51 countries. These articles have generated 2710 citations altogether over the last twenty years, which corresponds to half the number of other articles cited in Cybergeo (5545). 
}{
Cybergeo a été fondé en 1996 comme une journal européen de Géographie entièrement digital. Depuis, 737 articles scientifiques ont été publiés (jusqu'en mai 2016) par 1351 auteurs de 51 pays. Ces articles sont à l'origine de 2710 citations au total sur les 20 dernières années, ce qui correspond à la moitié des autres articles cités dans Cybergeo (5545).
}

\bpar{
Most contributions come from a French institution (561), although French-speaking countries (35 papers from an author affiliated in Canada, 21 in Switzerland) and neighbouring countries (UK: 23 contributions, Italy: 18) are well represented too (Fig.~\ref{fig:app:cybergeonetworks:authoring}). The geographical subjects of the articles themselves show a larger diversity, as the world is almost fully covered (Fig.~\ref{fig:app:cybergeonetworks:authoring}). However, France and neighbouring countries such as Spain and Germany are the main focus of the majority of articles, although the United States are the 5th most studied single country. By linking authors to their geographical subject (Fig.~\ref{fig:app:cybergeonetworks:who}), we find different patterns:
\begin{itemize}
\item European and North American countries studying each other through Cybergeo articles;
\item American countries being studied by authors affiliated in Europe and North-America;
\item African and Asian countries being studied mainly by Europeans and marginally by Americans and themselves;
\item Russia and Australia being studied by Western authors and studying their own hinterland.
\end{itemize}
}{
La majorité des contributions proviennent d'une institution française (561), bien que des pays francophones (35 articles comprennent un auteur affilié au Canada, 21 en Suisse) et des pays voisins (23 contributions pour le Royaume-Uni, 18 pour l'Italie) soient également bien représentés (Fig.~\ref{fig:app:cybergeonetworks:authoring}). Les sujets géographiques des articles aux-mêmes présentent une plus grande diversité, comme le monde est quasiment entièrement couvert (Fig. \ref{fig:app:cybergeonetworks:authoring}). Toutefois, la France et des pays voisins comme l'Espagne ou l'Allemagne sont le sujet principal de la majorité des articles, bien que les Etats-unis soient le 5ème pays le plus étudié. En reliant les auteurs à leur sujet géographique (Fig.~\ref{fig:app:cybergeonetworks:who}), différentes tendances peuvent être mises en valeur :
\begin{itemize}
	\item Des pays européens et nord-américains s'étudiant mutuellement au travers des articles de Cybergeo;
	\item Les pays d'Amérique sont étudiés par des auteurs affiliés en Europe et Amérique du Nord;
	\item Les pays asiatiques et africains sont principalement étudiés par les européens, et de façon marginale par les américains et eux-mêmes;
	\item La Russie et l'Australie sont étudiés par des auteurs occidentaux et étudient leur propre territoire.
\end{itemize}
}

\bpar{
Finally, the temporal evolution shows an accelerated growth of the number of authors – although the number of articles by 5-year period remains stable –, a spread of geographical coverage – with more articles published about emerging countries and extra-European territories –, along with a growing connexion in citation networks. There is a reinforcing bias towards a French-speaking authorship, revealed by the origin of authors as well as by the share of papers published in French.
}{
Enfin, l'évolution temporelle montre une croissance accélérée du nombre d'auteurs - même si le nombre d'articles sur des périodes de 5 ans reste stable, une extension de la couverture géographique, avec plus d'articles publiés sur les pays émergents et les territoires extra-européens, ainsi que des connexions croissantes dans les réseaux de citation. Il existe un biais de renforcement en faveur d'auteurs francophone, révélé par l'origine des auteurs ainsi que la part des articles publiés en français.
}

\begin{figure}
\includegraphics[width=\linewidth]{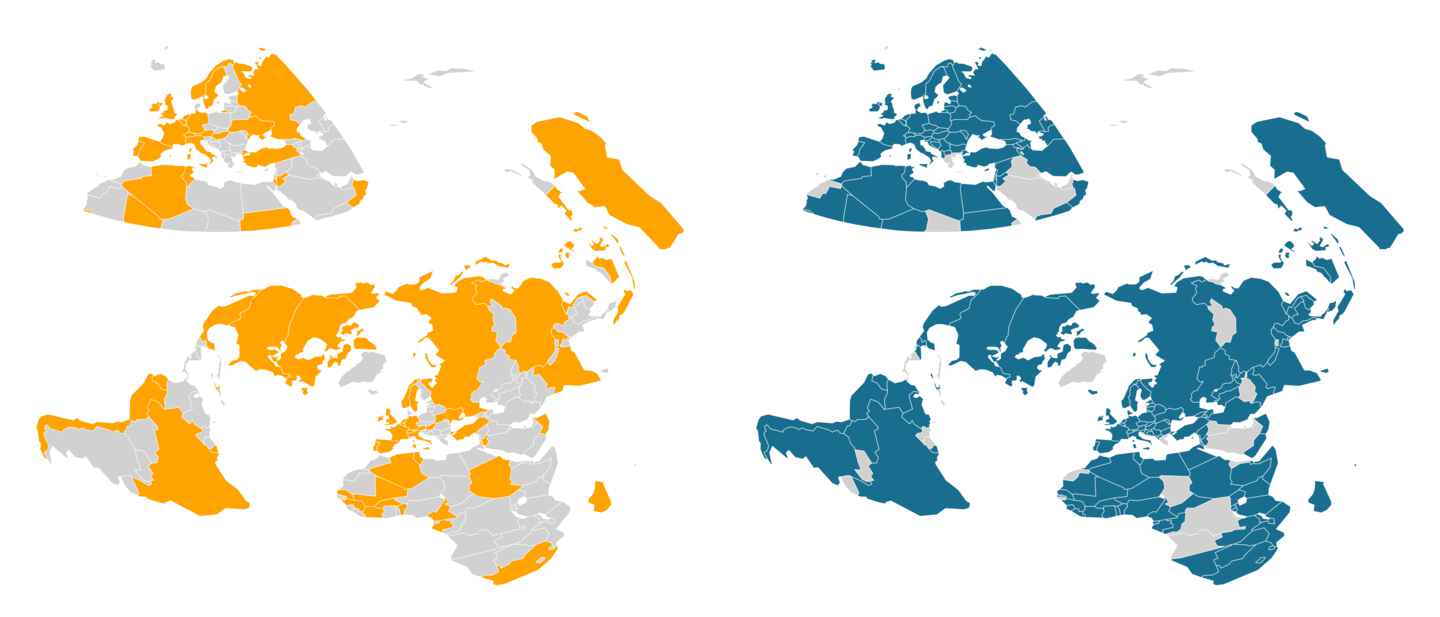}
\appcaption{\textbf{Countries with at least one author, 1996-2015} (\textit{Left}), \textbf{Countries studied at least once, 1996-2015} (\textit{Right})\label{fig:app:cybergeonetworks:authoring}}{\textbf{Pays avec au moins un auteur, 1996-2015} (\textit{Gauche}) ; \textbf{Pays étudiés au moins une fois, 1996-2015} (\textit{Droite})\label{fig:app:cybergeonetworks:authoring}} 
\end{figure} 

\begin{figure}
\includegraphics[width=\linewidth]{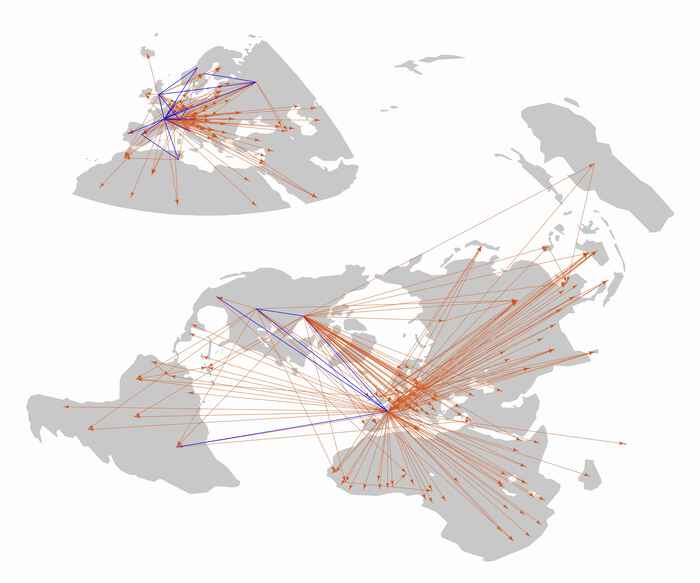}
\appcaption{\textbf{Geographical origins and destinations of papers, 1996-2015}\label{fig:app:cybergeonetworks:who}}{\textbf{Origine géographique et destination des articles, 1996-2015}\label{fig:app:cybergeonetworks:who}}
\end{figure} 

\subsection{Methods}{Méthodes}

\bpar{
One main aspect of our contribution is the complementary combination of different methodologies, each having its potentialities and pitfalls, but also specific questions and objects of study. We detail in this section the different methods and how they are coupled to produce new knowledge.
}{
Un aspect principal de cette contribution est la combinaison complémentaire de différentes méthodologies, chacune ayant ses potentialités et limitations, mais aussi des questions et des objets d'étude spécifiques. Nous détaillons dans cette section les différentes méthodes et comme celles-ci sont couplées pour produire une nouvelle connaissance.
}

\subsubsection{Internal semantic network}{Réseau sémantique interne}

\bpar{
The first exploration method is based on the set of keywords declared by the articles' authors in the \textit{Cybergeo} journal. We consider articles and keywords as a bipartite network. This network can be decomposed in two simple networks: a network of articles (vertices) linked by common keywords (edges); a network of keywords (vertices) linked by common articles (edges) \citep{roth2010social}. We consider the second one as a semantic network. 
}{
La première méthode d'exploration est basée sur l'ensemble des mots-clés déclarés par les auteurs des articles du journal \textit{Cybergeo}. Deux réseaux peuvent être construits à partir des articles et des mots-clés : un réseau d'articles reliés par les mots-clés communs, et un réseau de mots-clés reliés par les articles communs~\cite{roth2010social}. Ce dernier sera le réseau sémantique.
}

\bpar{
The vertices have a frequency variable corresponding to the number of articles in which they appear, and also their degree defined by their number of links. Starting from degrees, one can compute the probability of two nodes to be connected, what gives an expected weight to the corresponding link. The observed weight being the number of articles citing the two keywords, we define the \textit{modal weight} as a ratio between the observed weight and the squared-root expected weight of the edge. This modal weight can be used as a preferential attachment measure. 
}{
Les noeuds possèdent une variable de fréquence correspondant au nombre d'articles dans lequel ils apparaissent, ainsi que leur degré défini par leur nombre de liens. A partir des degrés, on peut calculer la probabilité de deux noeuds d'être connectés, ce qui donne un poids attendu au lien correspondant. Le poids observé étant le nombre d'articles citant les deux mots-clés, on défini le \emph{poids modal} comme le ratio entre le poids observé et la racine du poids attendu. Ce poids modal peut être utilisé comme une mesure d'attachement préférentiel.
}

\bpar{
Based on this preferential attachment measure two kinds of visualisations are proposed: semantic fields and communities. The semantic field shows a any given keyword at the centre of the plot and all its neighbours at a distance inversely proportional to the modal weight. The communities are computed with the Louvain algorithm \citep{blondel_fast_2008}. This community detection method is chosen among others because it is based on the modularity such as the modal weight above defined.
}{
A partir de cette mesure d'attachement préférentiel deux types de visualisations sont utilisées : les champs sémantiques et les communautés. Le champ sémantique représente un mot-clé donné au centre d'un graphe avec l'ensemble de ses voisins à une distance inversement proportionnelle au poids modal. Les communautés sont obtenues avec l'algorithme de Louvain~\cite{blondel2008fast}. Cette méthode de détection de communauté est choisie en particulier car elle est basée sur une mesure de modularité similaire au poids modal défini ci-dessus.
}

\subsubsection{External semantic network}{Réseau sémantique externe}

\bpar{
The second methodological development focuses on the combination of citation network exploration with network semantic analysis. The method applied for this development is described in details by~\cite{raimbault2017exploration}. Citation Networks have been widely used in studies of science, for example as a predictive tool for the success of a paper~\citep{newman2014prediction}, or to unveil emerging research fronts~\citep{shibata2008detecting}. Indeed, the bibliography of a paper contains a sort of scientific positioning, as a heritage to which it aims to contribute and which fields it is based on. The other way, reverse citations, i.e. contributions citing a given paper, up to a given level, shows how the knowledge produced was understood, interpreted and used, and in particular by which field (on this point the interesting example of \cite{jacobs2016death}, heavily cited today by most of quantitative studies of the city by physicists, shows how unexpected the type of the audience can be).
}{
Le deuxième développement méthodologique se concentre sur la combinaison de l'exploration du réseau de citations avec l'analyse du réseau sémantique. La méthode appliquée pour ce développement est décrite en détails par~\cite{raimbault2017exploration}. Les réseaux de citation ont été largement utilisés dans les études de la science, par exemple comme outil prédictif pour le succès d'un article~\cite{newman2014prediction}, ou pour révéler des fronts de recherche émergents~\cite{shibata2008detecting}. En effet, la bibliographie d'un article contient d'une certaine façon un positionnement scientifique, comme un héritage auquel il cherche à contribuer et les champs sur lesquels il se base. Dans l'autre sens, les citations inverses, i.e. les contributions citant un article donné, jusqu'à un certain niveau, montre dans quelle mesure la connaissance produite a été comprise, interprétée et utilisée, et en particulier par quel champ (sur ce point l'exemple intéressant de \cite{jacobs2016death}, cité en masse aujourd'hui par les études quantitatives de la ville par les physiciens, montre comment le type d'audience peut être inattendu).
}



\bpar{
We define the citation neighborhood of our corpus as all the articles citing articles published in \textit{Cybergeo}, all the articles citing the ones cited by \textit{Cybergeo}, and all the articles citing these ones. (having thus a network at depth 2). The citation data is collected using automatic data collection similarly to~\cite{raimbault2017exploration}.
}{
Nous définissons le voisinage de citation de notre corpus comme tous les articles citant les articles publiés dans \textit{Cybergeo}, tous les articles citant ceux cités par \textit{Cybergeo}, et tous les articles citant ceux-ci (obtenant ainsi un réseau à profondeur 2). Les données de citation sont collectées en utilisant une collecte automatiques de données de façon similaire à~\cite{raimbault2017exploration}.
}

\bpar{
Having constructed this citation neighborhood, we introduce a method to analyze its content through text mining. More precisely, we focus on the \emph{relevant} keywords of abstracts, in a precise sense, which was introduced by~\cite{chavalarias2013phylomemetic} to study the evolution of scientific fields, and later refined and scaled to big data on a Patent database by~\cite{bergeaud2017classifying}. Using co-occurrences of $n$-grams (keywords with multiple components, obtained after a first text cleaning and filtering), the deviation from an uniform distribution across texts using a chi-squared test gives a measure of keyword relevance, on which a fixed number $N_k = 50,000$ is filtered. The weighted co-occurrence network between relevant keywords captures their second order relationship and we assume that its topology contains information on the structure of disciplines that are present in the citation network. A sensitivity analysis of community structure to network filtering parameters (minimal edge weight, minimal and maximal document occurrence for nodes) yield a robust network with optimal community structure, what allows to associate to each paper a list of keywords and corresponding disciplines. These are complementary to the declared keywords and the full text themes presented in the next subsection, as they reveal how authors position in the semantic landscape associated to the citation neighborhood, or what are their ``cultural backgrounds''.
}{
Après avoir construit ce voisinage de citation, nous introduisons une méthode pour analyser son contenu par analyse textuelle. Plus précisément, nous nous intéressons aux mots-clés \emph{pertinents} des résumés, en un sens précis tel qu'introduit par~\cite{chavalarias2013phylomemetic} pour étudier l'évolution des champs scientifiques, et plus tard raffiné et étendu à des données massives pour une base de données de brevets par~\cite{bergeaud2017classifying}. En utilisant les co-occurrences des $n$-grams (mots-clés avec de multiples composantes, obtenus après un premier filtrage et nettoyage de texte), la déviation à une distribution uniforme entre les textes par un test du chi-deux donne une mesure de la pertinence des mots-clés, sur laquelle un nombre fixe $N_k = 50,000$ est filtré. Le réseau de co-occurrence pondéré correspondant entre les les mots-clés pertinents capture leur relation au second ordre et nous supposons que sa topologie contient une information sur la structure des disciplines présentes dans le réseau de citation. Une analyse de sensibilité de la structure des communautés aux paramètres de filtrage du réseau (poids minimal sur les liens, degré maximal pour les noeuds, fréquence minimal et maximale de l'occurence dans les documents) fournit un réseau robuste avec une structure de communautés optimale, qui permet d'associer à chaque article une liste de mots-clés et de disciplines correspondantes. Ceux-ci sont complémentaires aux mots-clés déclarés et aux thèmes des textes complets présentés ci-dessous, comme ils révèlent la manière dont les auteurs se positionnent dans le paysage sémantique associé au voisinage de citation, ou ce qu'est leur ``bagage culturel''.
}

\subsubsection{Topics allocation using full text documents}{Attribution de thématiques avec les textes complets}

\bpar{
The third and last exploration method details the allocation of topics in full text documents, and is thus complementary to the previous ones that used declared keywords and relevant keywords within abstracts of the citation neighborhood. Topic classification of texts documents is an intense field of research, that have developed several algorithms. In this field, a topic is considered as a set of words frequently used together in the same document, and a text document as a mixture of topics. Following a long standing development in natural language processing from the weighting scheme of words called Term frequency-inverse document frequency (tfidf) introduced by \cite{salton_introduction_1986} to first generative probabilistic model of \cite{hofmann1999probabilistic}, \cite{blei2003latent} have lastly proposed an evolution with the Latent Dirichlet Allocation model (LDA).
}{
La troisième et dernière méthode d'exploration détaille la construction de thèmes à partir des textes complets, et se révèle ainsi complémentaires aux analyses précédentes. L'extraction de thèmes de documents textuels est un champ de recherche intense. Un thème est défini comme un ensemble de mots fréquemment utilisés conjointement dans les documents, et les documents sont des mélanges de thèmes. Les premières méthodes se basaient sur une pondération des mots, en particulier la méthode \textit{term frequency inverse document frequency} (\textit{tf-idf}) introduite par~\cite{salton_introduction_1986}. \cite{hofmann1999probabilistic} a plus récemment proposé les premiers modèles probabilistes génératif, qui ont conduit à la méthode de la \textit{Latent Dirichlet Allocation} (LDA) \cite{blei2003latent}.
}

\bpar{
The LDA method ignores the structure of texts and considers articles as bags of words. In order to keep a certain level of structure, we use here the \emph{part-of-speech tagging} developed by \cite{schmid1994probabilistic} which provides the function of words in sentences and extracts their stem. Nouns, pronouns and verbs are filtered and weighted by their \textit{tf-idf} statistic. We then use the LDA method to produce the composition of documents in terms of themes and the composition of themes in terms of keywords. Given the a-priori matrix $\beta$ of composition of themes in terms of keywords, documents are generated following different probability laws. In an iterative way, distribution parameters, including $\beta$ are estimated, through the use of a Gibbs sampling algorithm \cite{geman_stochastic_1984}. We can then analyze $\beta$ to explain the themes contained in the corpus. The number of themes is a fixed parameter. An optimal number can be obtained by minimizing perplexity and maximizing entropy of themes in the corpus, as proposed by \cite{blei2003latent}.
}{
La méthode LDA déstructure les textes et considère les articles comme des ensembles de mots. Afin de garder un certain niveau de structure, nous utilisons ici le \textit{part-of-speech tagging} développé par \cite{schmid1994probabilistic} qui fournit la fonction des mots dans les phrases et extrait leur racine. Les noms, pronoms et verbes sont filtrés et pondérés par leur statistique \textit{tf-idf}. On utilise la méthode LDA pour alors produire la composition des documents en termes de thèmes et la composition des thèmes en termes de mots-clés. Etant donné la matrice a priori $\beta$ de composition des thèmes en termes de mots-clés, les documents sont générés suivant diverses lois de probabilités. De manière itérative, les paramètres des distributions, incluant $\beta$ sont estimés, par l'utilisation d'un échantillonnage de Gibbs \cite{geman_stochastic_1984}. On peut alors analyser $\beta$ pour expliquer les thèmes présents dans le corpus. Le nombre de thèmes est un paramètre fixé. Un nombre optimal peut être obtenu par minimalisation de la perplexité et maximisation de l'entropie des thèmes dans le corpus, comme proposé par \cite{blei2003latent}.
}

\subsubsection{Geographical aggregation of semantic profiles}{Aggregation géographique des profils sémantiques}

\bpar{
In order to produce the maps in Fig.~\ref{fig:app:cybergeonetworks:authoring} and Fig.~\ref{fig:app:cybergeonetworks:who} and analyses at the country level, articles have been geo-tagged in two ways. Firstly, the country of affiliation of the author(s) has been coded following the 2-letter identifiers of the International Organization for Standardization. Secondly, the articles were read one by one to extract the major geographical subjects. Articles were tagged with a country if this country or a sub-region of it constituted the focus of the study. In the case of European countries, different sets of countries were associated with the publication, depending on the perimeter of the subject (for instance: EU15, EU25, Schengen area, EuroMed, etc.). Given a semantic characterisation of articles (using keywords, citations or full-texts), it is then possible to determine two semantic profiles of countries: one using countries as authoring origins and one using countries as subject 'destination'. This semantic profile of a given country is made of the mean share of themes present in articles authoring from or studying this country. All in all, given the three semantic characterisations of articles (using keywords, citations and full-texts) and the two geographical allocation of articles (authoring or studied), each country has a maximum of six distinct semantic profiles. We use these semantic profiles to cluster countries. The clustering method applied is an ascending hierarchical clustering algorithm using the Ward criterion of distance maximisation. When analysing authoring clusters, we consider groups of countries from which a certain geography is made and written. This option is interesting in a reflexive aspect but practically more hazardous because of the high concentration of emission and the consequently low number of emitting countries. Therefore, in the results section, we base our clustering on studied countries. When analysing clusters of studied countries, we consider how certain groups of territories are studied, what words authors use to talk about them and in which research areas the papers about them are used. 
}{
Pour pouvoir produire les cartes des Fig.~\ref{fig:app:cybergeonetworks:authoring} et Fig.~\ref{fig:app:cybergeonetworks:who} ainsi que les analyses au niveau du pays, les articles ont été géocodés de deux façons. Dans un premier temps, la pays d'affiliation de ou des auteur(s) a été codé par les identifiants ISO à deux lettres. Dans un second temps, les articles ont été lus un par un pour extraire les sujets géographiques principaux. Les articles ont alors été codés par un pays si le pays ou une sous-région de celui-ci constituaient l'objet principal de l'étude. Dans le cas des pays européens, différents ensembles de pays ont été associés à la publication, selon le périmètre du sujet (par exemple EU15, EU25, espace Schengen, EuroMed, etc.). Etant donné une caractérisation sémantiques des articles (en utilisant les mots-clés, les citations ou les textes complets), il est ensuite possible de déterminer deux profils sémantiques pour les pays : l'un utilisant les pays comme origine des auteurs, l'autre comme la destination des sujets. Le profil sémantique d'un pays est constitué de la part moyenne des thèmes présent dans les articles dont l'auteur en provient ou étudiant celui-ci. Au total, étant donné les trois caractérisations sémantiques des articles et les deux attributions géographiques, chaque pays a au maximum six profils sémantiques distincts. Ces profils peuvent être utilisés pour regrouper les pays. La méthode de clustering utilisée est un algorithme de classification ascendante hiérarchique avec le critère de maximisation de distance de Ward. En analysant les clusters des profils d'auteurs, on considère les groupes de pays dans lesquels une certaine géographie est développée et écrite. Cet aspect est intéressant du point de vue de la réflexivité mais en pratique relativement aléatoire à cause de la forte concentration des émissions et par conséquent du faible nombre de pays écrivant. Pour cette raison, les résultats présentés seront basés sur les pays étudiés. Pour cela, nous considérons la manière dont un certain groupe de territoires est étudiés, quel mots-clés les auteurs utilisent pour les désigner et dans quel domaine de recherche les articles les concernant sont utilisés.
}

\subsubsection{Open data + interactivity = reproducibility \& transparency}{Open Data + interactivité = reproductibilité \& transparence}

\bpar{
Last but not least, our methodological contribution is also closely linked to issues of reflexivity, transparency and reproducibility in the process of knowledge production. It is now a well sustained idea that all these aspects are closely linked and that their strong coupling participates to a virtuous circle enhancing and accelerating knowledge production, as seen in the various approaches of Open Science~\citep{fecher2014open}. For example, open peer review is progressively emerging as an alternative way to the rigid and slow classical canons of scientific communication: \cite{10.12688/f1000research.11369.1} proceeds to a systematic review on the notion to give an unified definition and understand its potential benefits and pitfalls. In the domain of computational science, tools are numerous to ensure reproducibility and transparency but require a strict discipline of use and are not easily accessible~\citep{wilson2017good}. Open Science suggest transparency of the knowledge production process itself, but also of the knowledge communication patterns: on this point we claim that interactive exploration of quantitative epistemological patterns are necessary. We build therefore an interactive application to allow the exploration of heterogeneous scientific corpuses.
}{
Enfin, notre contribution méthodologique est également étroitement liée aux questions de réflexivité, transparence et reproductibilité dans le processus de production de connaissance. Il s'agit d'une idée acceptée que ces aspects sont en interaction et que leur couplage participe à un cercle vertueux encourageant et accélérant la production de connaissances, comme on peut le voir dans les différentes approches de la Science Ouverte~\citep{fecher2014open}. Par exemple, la revue par les pairs ouverte émerge progressivement comme une pratique alternative aux canons rigides et lents de la communication scientifique : \cite{10.12688/f1000research.11369.1} procède à une revue systématique des approches de la notion pour en donner une définition unifiée et comprendre ses bénéfices ou défauts potentiels. Dans le domaine des sciences computationnelles, les outils pour assurer une reproductibilité et une transparence sont nombreux mais requièrent une discipline stricte d'utilisation et ne sont pas toujours facilement accessibles~\citep{wilson2017good}. La Science Ouverte suggère une transparence du processus de production de connaissance lui-même, mais aussi des motifs de communication des connaissances : sur ce point nous appuyons l'idée que l'exploration interactive des motifs épistémologiques quantitatifs sont nécessaires. Nous construisons pour cela une application interactive pour permettre l'exploration de corpus scientifiques hétérogènes.
}

\bpar{
The web application is available online at \url{http://shiny.parisgeo.cnrs.fr/CybergeoNetworks/}. Source code and data, both for analyses and the web application, are available on the open \texttt{git} repository of the project at \url{https://github.com/AnonymousAuthor3/cybergeo20}.
}{
L'application web est disponible en ligne à \url{http://shiny.parisgeo.cnrs.fr/CybergeoNetworks/}. Le code source et les données, à la fois pour les analyses et pour l'application web, sont disponibles sur le dépôt \texttt{git} ouvert du projet à \url{https://github.com/AnonymousAuthor3/cybergeo20}.
}


\subsection{Results}{Résultats}

\subsubsection{Internal semantic network}{Réseau sémantique interne}
\paragraph{Communities and semantic fields}{Communautés et champs sémantiques}

\bpar{
The community detection algorithm finds a modularity optimum with 10 clusters: mobility and transportation; imagery and GIS; climate and environment; history and epistemology; sustainability, risk, planning; Economic geography; Territory and population; urban dynamics; statistics and modelling; emotional geography. Some clusters concentrate a large number of keywords and articles, such as ``imagery and GIS'' or ``statistics and modelling''. This result was expected because of the original aim and scope of the journal. Beside the main clusters and a set of medium-sized clusters, two small and totally unexpected clusters emerged: "emotional geography" and "climate and environment". The CybergeoNetworks application proposes a set of visualisation parameters to draw the communities, as presented in Fig.~\ref{fig:app:cybergeonetworks:commintern}, such as setting the size of vertices and edges according to different variables (degree, number of articles, modal weight).
}{
L'algorithme de detection de communauté trouve une modularité optimale qui donne 10 clusters : mobilité et transports ; télédétection et SIG ; climat et environnement ; histoire et épistémologie ; soutenabilité, risques, planification ; géographie économique ; territoires et populations ; dynamiques urbaines ; statistiques et modélisation ; géographie des émotions. Certains clusters concentrent un grand nombre de mots-clés et d'article, comme ``télédétection et SIG'' ou ``statistiques et modélisation''. Ce résultat était prévisible vu le but et la portée originaux du journal. A côté des clusters principaux et d'un ensemble de clusters de taille moyenne, deux clusters de petite taille et totalement inattendus émergent : ``géographie des émotions'' et ``environnement et climat''. L'application \textit{CybergeoNetworks} propose un ensemble de paramètres de visualisation pour tracer les communautés, comme présenté en Fig.~\ref{fig:app:cybergeonetworks:commintern}, comme la taille des noeuds et des liens pouvant être variée selon différentes variables (degré, fréquence, poids modal).
}

\bpar{
The above explained modal weight metrics can be used to draw semantic fields. The CybergeoNetworks application proposes the full list of keywords. The user chooses one keyword from that list, the word is placed is the centre of the plot and all its neighbours are arranged at a distance inversely proportional to the preferential attachment (modal weight). The application proposes visualisation parameters such as setting the character size according to the weight of the keywords (number of articles or degree in the network), as shown in Fig.~\ref{fig:app:cybergeonetworks:commintern}). Some proximities are expected (``urban'' is closely linked to ``city''), some are expected knowing the original scope of the journal in the field of theoretical and quantitative geography (``model'' or ``spatial statistics'' are linked to ``city''). Some proximities are totally unexpected: for ``city'' the preferential attachment of keywords like ``movie'', ``web'', ``virtual''.
}{
Les métriques de poids modal expliquée précédemment peuvent être utilisée pour tracer les champs sémantiques. L'application \textit{CybergeoNetworks} propose la liste complète des mots-clés. L'utilisateur choisit un mot-clé dans cette liste, celui-ci est placé au centre du graphe, et l'ensemble de ses voisins sont placés à une distance inversement proportionnelle à l'attachement préférentiel (poids modal). L'application propose des paramètres de visualisation comme la taille des caractères en fonction du poids des mots-clés (fréquence dans les articles ou degré dans le réseau), comme montré en Fig.~\ref{fig:app:cybergeonetworks:commintern}. Certaines proximités sont attendues (``urban'' est étroitement relié à ``city''), d'autres sont attendus sachant l'étendue originale du journal dans le champ de la géographie théorique et quantitative (``model'' ou ``spatial statistics'' sont reliés à ``city''). Certaines sont inattendues, comme pour ``city'' l'attachement préférentiel de mots-clés comme ``movie'', ``web'', ``virtual''.
}

\begin{figure}
\includegraphics[width=\linewidth]{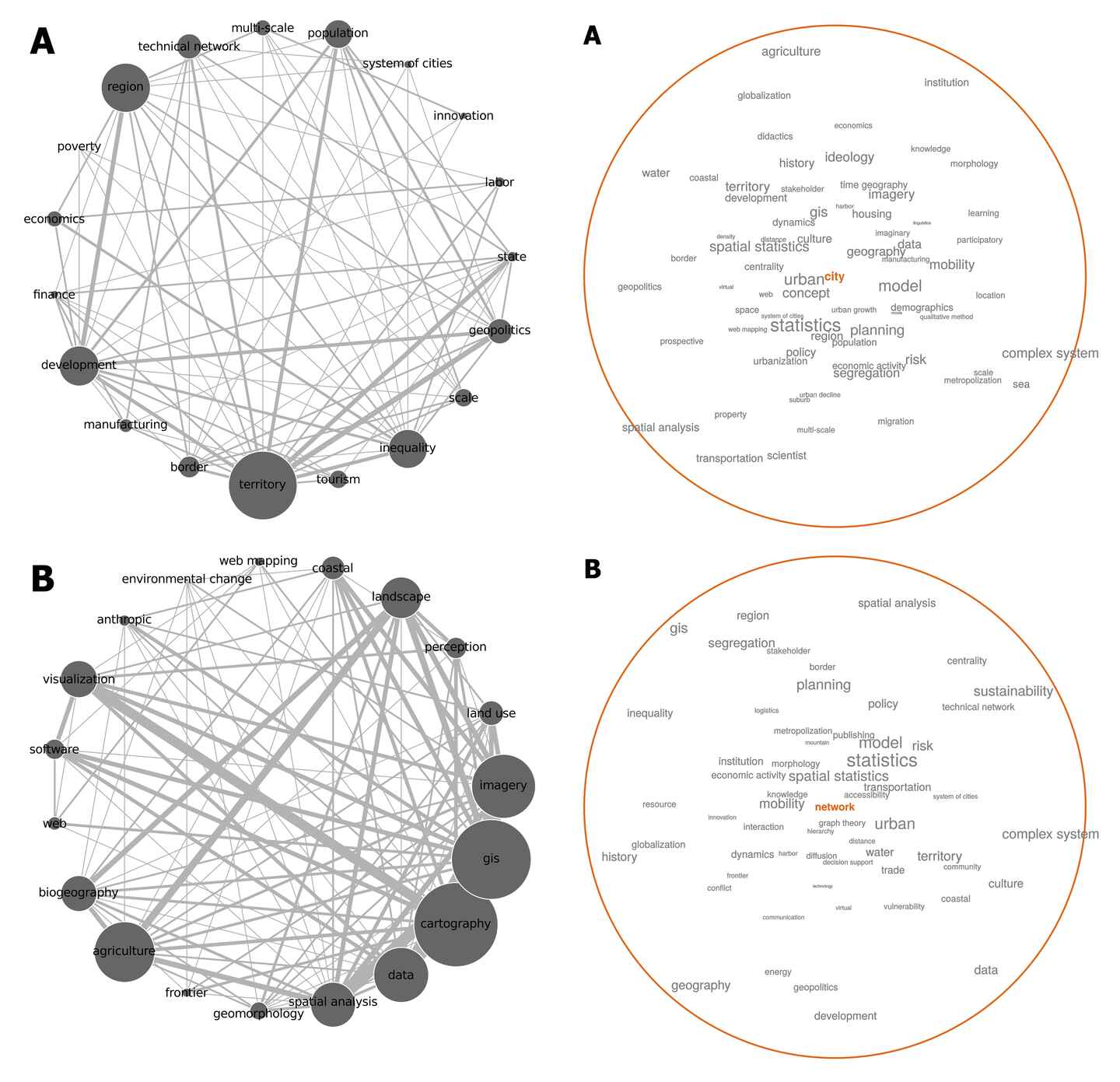}
\appcaption{\textbf{Community structure of the internal semantic network : } A-Territory; B-Imagery \& GIS (\textit{Left}); \textbf{Semantic fields : } A-City; B-Network (\textit{Right}).\label{fig:app:cybergeonetworks:commintern}}{\textbf{Structure de communauté du réseau sémantique interne : } A-Territory, B-Imagery \& GIS (\textit{Gauche}) ; \textbf{Champs sémantiques : } A-City, B-Network (\textit{Droite}).\label{fig:app:cybergeonetworks:commintern}}
\end{figure}

\paragraph{Spatialised communities}{Communautés spatialisées}

\bpar{
Using the keywords distributions to draw the semantic profile of the 128 countries studied in a Cybergeo article, we obtain a clustering in 4 groups representing 16.5\% of the initial inertia. Its geographical distribution is shown in Fig.~\ref{fig:app:cybergeonetworks:cluster_hadri} with the average profile of each group. 
}{
En utilisant les distributions des mots-clés pour déterminer les profils sémantiques des 128 pays étudiés dans les articles de \textit{Cybergeo}, nous obtenons un clustering en 4 groupes représentant 16.5\% de l'inertie initiale. Sa distribution géographique est montrée en Fig.~\ref{fig:app:cybergeonetworks:cluster_hadri} avec le profil moyen de chaque groupe.
}

\begin{figure}
\includegraphics[width=\linewidth]{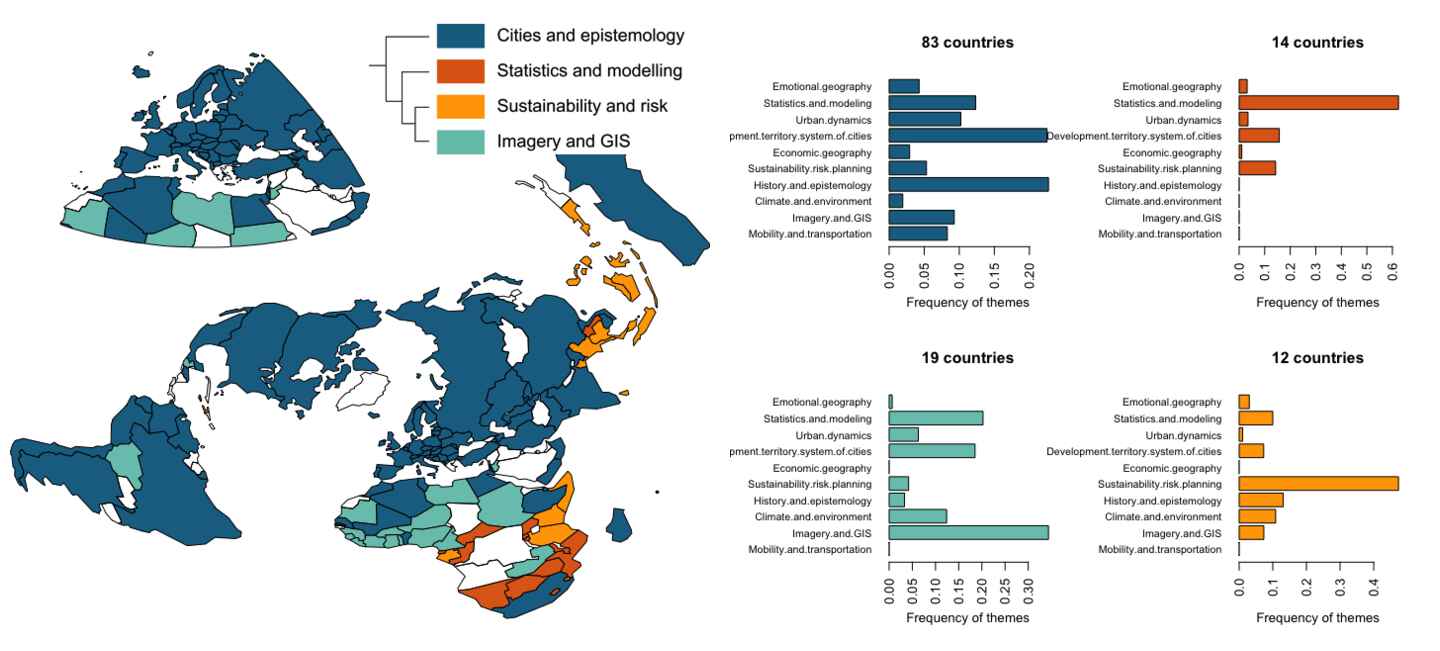}
\appcaption{\textbf{Geographical communities of declared interest} (\textit{Left}) ;  \textbf{Corresponding semantic profile of groups} (\textit{Right}) \label{fig:app:cybergeonetworks:cluster_hadri}}{\textbf{Communautés géographiques d'intérêt déclaré} (\textit{Gauche}) ; \textbf{Profil sémantique des groupes correspondants} (\textit{Droite}). \label{fig:app:cybergeonetworks:cluster_hadri}}
\end{figure} 

\bpar{
Countries are differentiated firstly by whether or not the articles studying them also declare keywords related to transport and mobility, history and epistemology, urban systems and/or emotional geography. Indeed, the first group of 83 countries (in blue, Fig.~\ref{fig:app:cybergeonetworks:cluster_hadri}) is defined by these themes. The corresponding countries are the most developed and rich territories of the world, including emergent countries such as the BRICS. The keywords used to advertise the articles about them follow the fashions of geography, with mentions of emotions and mobility for instance. 
}{
Les pays sont premièrement différentiés par le fait ou non que les pays les étudiant déclarent également des mots-clés en lien avec la mobilité et le transport, l'histoire et l'épistémologie, les systèmes urbains, la géographie émotionnelle. En effet, le premier groupe de 83 pays (en bleu, Fig.~\ref{fig:app:cybergeonetworks:cluster_hadri}) est défini par ces thèmes. Les pays correspondants sont les territoires les plus développés et les plus riches du monde, incluant les pays émergents comme les BRICS. Les mots-clés utilisés pour mettre en valeur les articles sur ceux-ci suivent les modes de la géographie, avec des mentions des émotions et de la mobilité par exemple.
}

\bpar{
The countries of the other groups over-represent the keywords related to:
\begin{itemize}
\item methods (in orange) such as statistics and modelling. The countries associated with these keywords are all located in central and southern Africa, with the exception of Lao. These countries are studied by a small number of articles which focus on methodological approaches. For example, the only article studying Rwanda \citep{querriau2004localisation} relates to an optimal location problem whereas \citet{vallee2009disparites} uses 'multilevel modelling' as a keyword for the only article about Lao.
\item sustainability and risks (in yellow). This is the case of articles about Indonesia for example, which all relate to aleas and vulnerability: to tsunamis \citep{ozer2005tsunami}, to volcanoes \citep{belizal2011quand} and to water scarcity \citep{putra2009einfluss}.
\item Finally, 19 countries are associated with keywords related to imagery and GIS (teal colour). They are located primarily in Saharan Africa. In many cases, this happens because the articles present a methodology which uses aerial and satellite images to substitute missing socioeconomic data  \citep{ackermann2003analysis, devaux2007extraction}.
\end{itemize}
}{
Les pays des autres groupes sur-représentent les mots-clés en relation avec :
\begin{itemize}
	\item les méthodes (en orange) comme les statistiques et la modélisation. Les pays associés à ces mots-clés sont tous situés en Afrique centrale et du sud, à l'exception du Laos. Ces pays sont étudiés par un petit nombre d'articles qui se concentrent sur les approches méthodologiques. Par exemple, le seul article étudiant le Rwanda \cite{querriau2004localisation} est en relation avec un problème de localisation optimale tandis que \cite{vallee2009disparites} utilise le mot-clé ``multilevel modelling'' pour le seul article traitant du Laos.
	\item La soutenabilité et les risques (en jaune). C'est le cas des articles sur l'Indonésie par exemple, qui sont tous en relation avec les aléas et la vulnérabilité : des tsunamis \cite{ozer2005tsunami}, aux volcans \cite{belizal2011quand} et à la rareté des ressources en eau \cite{putra2009einfluss}.
	\item Enfin, 19 pays sont associés à des mots-clés en relation avec la télédétection et les SIG (en turquoise). Ils sont situés principalement en Afrique saharienne. Dans de nombreux cas, il s'agit d'articles qui présente une méthodologie qui exploite les images aériennes ou satellites pour extrapoler des données socio-économiques absentes \cite{ackermann2003analysis,devaux2007extraction}.
\end{itemize}
}

\bpar{
Thus, drawing communities of declared interest, we find an interesting dichotomy between rich countries on the one hand, which are studied extensively in the literature and for which authors use trendy keywords to singularise themselves from past and concurrent work; and developing countries on the other hand, which are associated with more technical keywords reflecting a narrower spectrum of domains and specific data challenges.
}{
Ainsi, en construisant les communautés d'intérêt déclaré, nous trouvons une dichotomie intéressante entre les pays riche d'une part, qui sont étudiés de manière intensive dans la littérature et pour lesquels les auteurs utilisent des mots-clés à la mode pour se détacher des travaux existants ou concurrents ; et les pays en développement d'autre part, qui sont associés à des mots-clés plus techniques témoignant un spectre des domaines plus étroit et des problématiques spécifiques aux données.
}

\subsubsection{External semantic network}{Réseau sémantique externe}

\bpar{
The application allows to explore the citation neighborhood of chosen articles, in terms of semantic contents (the visualisation of full networks are technically not feasible as the full corpus contains around 200,000 articles). Wordclouds give the content of the article and the content of the articles in the neighborhood, with each word being associated to the semantic communities. The user can therefore situate a work within a semantic context, and we expect that unanticipated connexions can be made with this tools, as authors may not be aware of similar works in alien disciplines.
}{
L'application permet l'exploration du voisinage de citation d'un article choisi, en termes de contenu sémantique (la visualisation des réseaux complets n'est pas faisable techniquement comme le corpus complet contient autour de 200,000 articles). Les nuages de mots donnent le contenu de l'article et le contenu des articles dans le voisinage, chaque mot étant associé aux communautés sémantiques. L'utilisateur peut ainsi situer un travail dans son contexte sémantique, et nous nous attendons à ce que des liens non anticipés puissent être faits avec ces outils, comme les auteurs ne sont pas forcément au courant de travaux similaires dans des disciplines étrangères.
}

\paragraph{Communities structures}{Structure des communautés}

\bpar{
As explained before, the raw semantic network is optimized for modularity and size, taking a compromise between these two opposite objectives, when edge and node filtering parameter vary. This provides 12 communities, that can correspond to existing disciplines, to methodological issues, or to very precise thematic subjects. The communities are, in order of importance in terms of proportion of total keywords : Political Science/Communication; Biogeography; Social and Economic Geography; Climate; Physical Geography; Commerce; Spatial Analysis; Microbiology; Neuroscience; GIS; Agriculture; Health. This method has the characteristic of grouping keywords by co-occurrences, revealing thus the actual structure of abstracts contents: it is both an advantage when revealing links as for the large field of Social and Economic Geography, but can also blur information by grouping more detailed communities. Very precise small communities such as Health Geography appear as they are strongly isolated from the rest of the communities. This structure is particular, and shows a dimension of knowledge that for example classical citation analysis do not reveal.
}{
Comme expliqué précédemment, la réseau sémantique brut est optimisé pour la modularité et la taille, en prenant un compromis entre ces deux objectifs opposés, tandis que les paramètres de filtrage des liens et des noeuds varient. Cela permet d'obtenir 12 communautés, qui peuvent correspondre à des disciplines existantes, à des questions méthodologiques, ou à des sujets thématiques très précis. Les communautés sont, par ordre d'importance en termes de proportion de l'ensemble des mots-clés : Science politiques/communication ; Biogéographie ; Géographie Economique et Sociale ; Climat ; Géographie physique ; commerce ; analyse spatiale ; microbiologie ; neurosciences ; SIG ; agriculture ; santé. Cette méthode a la caractéristique de grouper les mots-clés par co-occurrences, révélant ainsi la structure effective du contenu des résumés : il s'agit simultanément d'un avantage en révélant des liens comme pour le champ très large de la Géographie Economique et Sociale, mais peut aussi bruiter l'information en groupant des communautés plus détaillées. Des communautés de taille modeste et très précises apparaissent, comme elles sont très isolées du reste des communautés. Cette structure est particulière, et témoigne d'une dimension de la connaissance qui par exemple n'est pas révélée par les analyses de citation classiques.
}

\paragraph{Spatialised communities}{Communautés spatialisées}

\bpar{
Using the previous networks to draw the semantic profile of the 128 countries studied in a Cybergeo article, we obtain a clustering in 5 groups representing 19.3\% of the initial inertia. Its geographical distribution is shown in Fig.~\ref{fig:app:cybergeonetworks:cluster_juste} with the average profile of each group. 
}{
A l'aide des précédents réseaux pour construire le profil sémantique des 128 pays étudiés dans un article de Cybergeo, nous obtenons une classification en 5 groupes qui représentent 19.3\% de l'inertie initiale. Sa distribution géographique est montrée en Fig.~\ref{fig:app:cybergeonetworks:cluster_juste} ainsi que le profil moyen de chaque groupe.
}

\begin{figure}
\includegraphics[width=\linewidth]{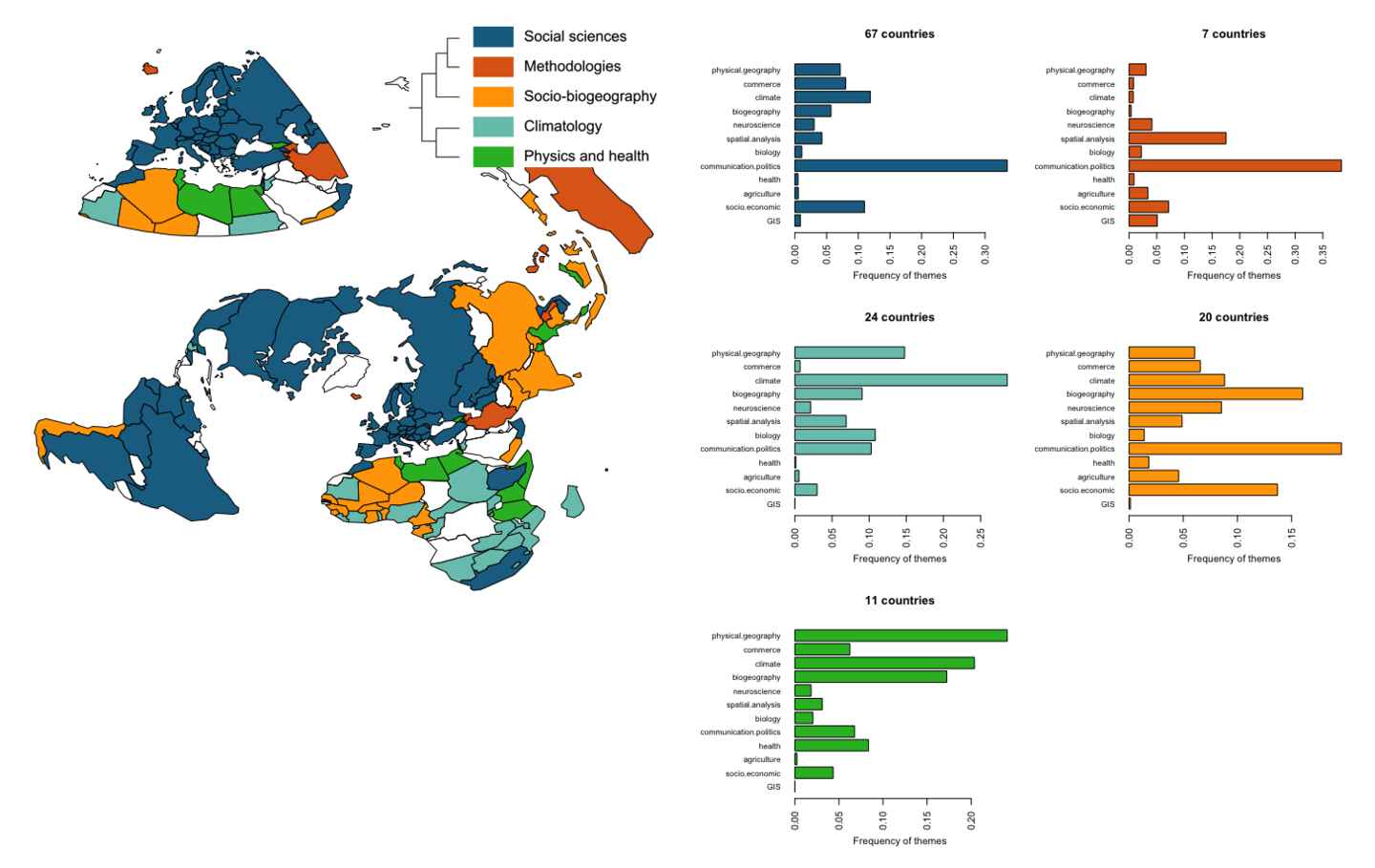}
\appcaption{\textbf{Geographical communities of bibliographical use} (\textit{Left}); \textbf{Corresponding semantic profile of groups} (\textit{Right}).\label{fig:app:cybergeonetworks:cluster_juste}}{\textbf{Communautés géographiques d'usage bibliographique} (\textit{Gauche}) ; \textbf{Profil sémantique correspondant.} (\textit{Droite}).\label{fig:app:cybergeonetworks:cluster_juste}}
\end{figure} 

\bpar{
The largest group of countries largely overlaps with the largest cluster of keywords communities presented in the previous section. Indeed, rich and emergent countries are studied in articles used in similar ways in citation networks. There are further divides among this group. A first subgroup (in blue) of countries is studied by Cybergeo articles cited preferentially in the fields of commerce, socio-economic and politics analysis. These correspond to articles mostly in Economics and Social Sciences.
}{
Le plus grand groupe de pays recoupe largement le cluster le plus large des communautés de mots-clés présentées dans la section précédente. En effet, des pays riches et émergents sont étudiés dans des articles utilisés de manière similaire dans les réseaux de citation. Il existe des sous-divisions au sein de ce groupe. Un premier sous-groupe de pays (en bleu) est étudié par les articles de Cybergeo cités préférentiellement dans les champs du commerce et des analyses socio-économiques et politiques. Ceux-ci correspondent principalement à des articles en Economie et Sciences Sociales.
}

\bpar{
The nearest subgroup of countries (in orange) comprises Australia, Azerbaijan, Iran, Lao, the Philippines and Iceland. It corresponds to countries treated by articles cited preferentially in methodological fields (spatial analysis and GIS). Indeed, the only article about Iran presents a collaborative decision support system \citep{jelokhani2012web} while the only article about Australia reviews online cartographic products \citep{escobar2000distribution}. This kind of articles then tends to stay in the citation clique of geomatics.
}{
Le second sous-groupe de pays (en orange) comprend l'Australie, l'Azerbaijan, l'Iran, le Laos, les Philippines et l'Islande. Il correspond à des pays traités par des articles cités préférentiellement dans des champs méthodologiques (analyse spatiale et GIS). En effet, le seul article à propos de l'Iran présente un système collaboratif d'aide à la décision \citep{jelokhani2012web} tandis que le seul article sur l'Australie est une revue de produits cartographiques en ligne \citep{escobar2000distribution}. Ce type d'article tend ensuite à rester dans la clique de citation de la géomatique.
}

\bpar{
The third subgroup refers to countries of South-East Asia, Western Africa, Yemen and Chile. The articles studying them are cited preferentially in the fields of biogeography and socioeconomic studies, although they match the average profile.
}{
Le troisième sous-groupe correspond à des pays d'Asie du Sud-est, d'Afrique occidentale, le Yemen et le Chili. Les articles les étudiants sont cités en préférence dans les champs de la biogéographie et des études socio-économiques, bien qu'il correspondent au profil moyen.
}

\bpar{
In the second group of countries, we find a first subgroup of sub-Saharan countries (in teal) associated with papers cited in the climatology citation community. The second subgroup is composed of East African, North African and South-East Asian countries (in green) associated with papers cited in the fields of physical geography and health.
}{
Dans le second groupe de pays, nous trouvons un premier sous-groupe de pays sub-Saharien (en turquoise) associés à des articles cités dans la communauté de citation de la climatologie. Le second sous-groupe est composé de pays d'Afrique de l'Est, du Nord et d'Asie du Sud-est (en vert) associés à des articles cités dans les champs de la géographie physique et de la santé.
}

\bpar{
Thus, drawing communities of bibliographical use, we find an interesting  dichotomy between rich countries on the one hand, which are associated with papers cited in broad communities, including topical and methodological fields; and poor and developing countries on the other hand, which are associated with papers cited mainly in relation to natural hazards, health and risks in the literature.
}{
Ainsi, en construisant les communauté d'usage bibliographique, nous trouvons une dichotomie intéressante entre les pays riches d'une part, qui sont associés à des articles cités dans des communautés très larges, incluant des champs thématiques et méthodologiques ; et des pays pauvres ou en développement d'autre part, qui sont associés à des articles cités principalement en relation dans la littérature aux catastrophes naturelles, la santé et les risques.
}

\subsubsection{Themes with full texts}{Thèmes avec textes complets}

\paragraph{Evolution of topics in the corpus}{Evolution des thèmes du corpus}

\bpar{
The destructuration of documents and filtering yield a dictionary with around $1.4\cdot 10^5$ words. The LDA parameters are estimated for a number of themes varying between 2 and 200, with different resolutions in particular between 20 and 40. Stochasticity is taken into account by repeating 10 times each estimation for a given number of themes. As shown in Fig.~\ref{fig:app:cybergeonetworks:perplexity}, the number of 20 themes is optimal regarding the perplexity and entropy indicators.
}{
La déstructuration des documents et le filtrage donne un dictionnaire d'environ $1.4\cdot 10^5$ mots. Les paramètres LDA sont estimés pour un nombre de thèmes variant de 2 à 200, avec différentes résolutions notamment entre 20 et 40. La stochasticité est prise en compte en répétant 10 fois chaque estimation pour un nombre donné de thèmes. Comme montré en Fig.~\ref{fig:app:cybergeonetworks:perplexity}, le nombre de 20 thèmes est optimal au regard des indicateurs de perplexité et d'entropie.
}

\begin{figure}
	\includegraphics[width=\linewidth]{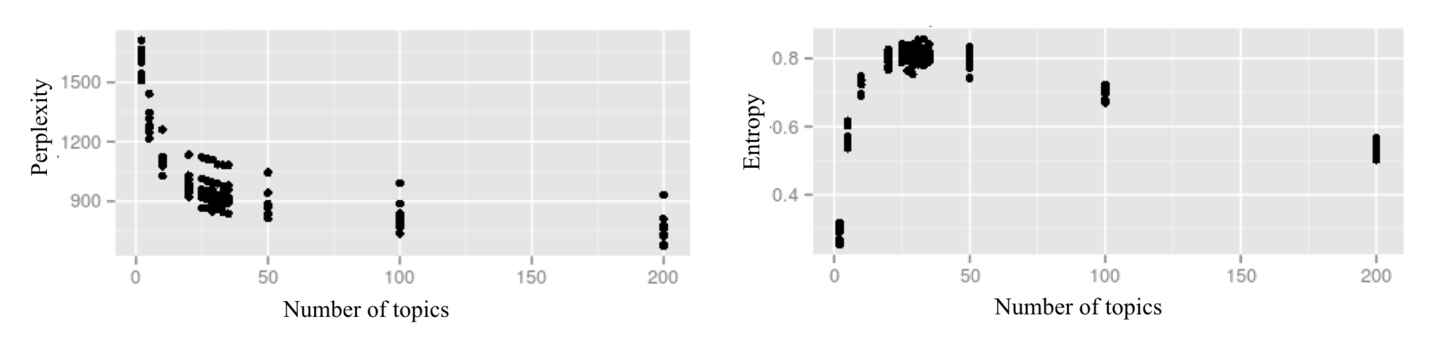}
	\appcaption{\textbf{Perplexity and Entropy of the LDA model as a function of number of topics.}\label{fig:app:cybergeonetworks:perplexity}}{\textbf{Perplexité et entropie du modèle LDA par nombre de thèmes.}\label{fig:app:cybergeonetworks:perplexity}}
\end{figure} 

\bpar{
The 20 themes obtained, classified by importance order (in terms of frequency of occurence in all documents) can be synthesized as corresponding to: housing and neighborhoods; mobility and accessibility; remote sensing; planning and governance; risks and vulnerability; health; garbage words; complex systems modeling; cities; water resources; cartography; history of geography; education; urban agglomerations; geopolitics; electoral geography; administration; geomorphology; landscape; maritime geography. We can observe the evolution of themes by year, as shown in Fig.~\ref{fig:app:cybergeonetworks:topics-evolution}. Different evolution profiles can be distinguished: decreasing, localized, constant and increasing. The article on cartography (11) are decreasing in number. Articles on remote sensing (3) have mostly been produced in 2000, the same way as articles on water resources (10) in 2004 and 2011. Articles on urban agglomerations are produced regularly. Themes such as neighborhood (1) and mobility (2) have a tendency to increase.
}{
Les 20 thèmes obtenus, classés par ordre d'importance (en terme de fréquence d'occurence dans l'ensemble des documents) peuvent être synthétisés comme correspondant à : logement et voisinages ; mobilité et accessibilité ; télédétection ; planification et gouvernance ; risques et vulnérabilité ; santé ; mots de liaison ; modélisation des systèmes complexes ; ville ; ressources en eau ; cartographie ; histoire de la géographie ; éducation ; agglomérations urbaines ; géopolitique ; géographie électorale ; administration ; géomorphologie ; paysage ; géographie maritime. Nous pouvons observer l'utilisation des thèmes par année, comme présenté en Fig.~\ref{fig:app:cybergeonetworks:topics-evolution}. Différents profils d'évolution se distinguent : décroissant, ponctuel, constant et croissant. Les articles sur la cartographie (11) sont en nombre décroissant. Les articles sur la télédétection (3) ont été majoritairement produits en 2000, de la même manière que les articles sur les ressources en eau (10) en 2004 et 2011. Les articles sur les agglomérations urbaines sont produits de façon régulière. Des thèmes comme le voisinage (1) et la mobilité (2) ont tendance à augmenter.
}

\begin{figure}
\includegraphics[width=\linewidth]{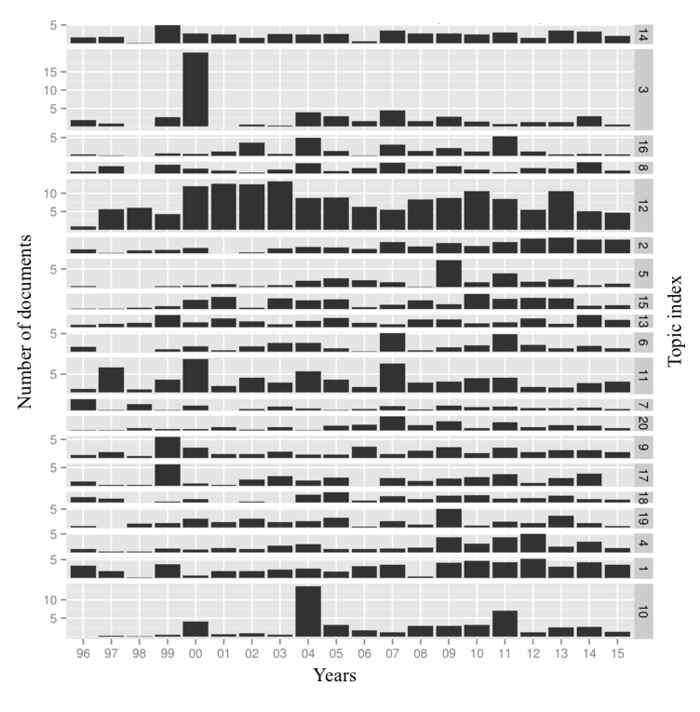}
\appcaption{\textbf{Number of documents addressing a topic per year, between 1996 and 2015.}\label{fig:app:cybergeonetworks:topics-evolution}}{\textbf{Nombre de documents traitant d'un thème par années, entre 1996 et 2015.}\label{fig:app:cybergeonetworks:topics-evolution}} 
\end{figure} 

\paragraph{Spatialised full-text communities}{Communautés de texte complet spatialisées}

\bpar{
Using the full texts to draw the semantic profile of the 128 countries studied in a \textit{Cybergeo} article, we obtain a clustering in 4 groups representing 13.4\% of the initial inertia. Its geographical distribution is shown in figure \ref{fig:app:cybergeonetworks:cluster_poc} with the average profile of each group. 
}{
En utilisant les textes complets pour construire les profils sémantiques des 128 pays étudiés dans les articles de \textit{Cybergeo}, nous obtenons une classification en 4 groupes qui représentent 13.4\% de l'inertie totale. Sa distribution géographique est montrée en Fig.~\ref{fig:app:cybergeonetworks:cluster_poc} avec le profil moyen de chaque groupe.
}

\begin{figure}
	\includegraphics[width=\linewidth]{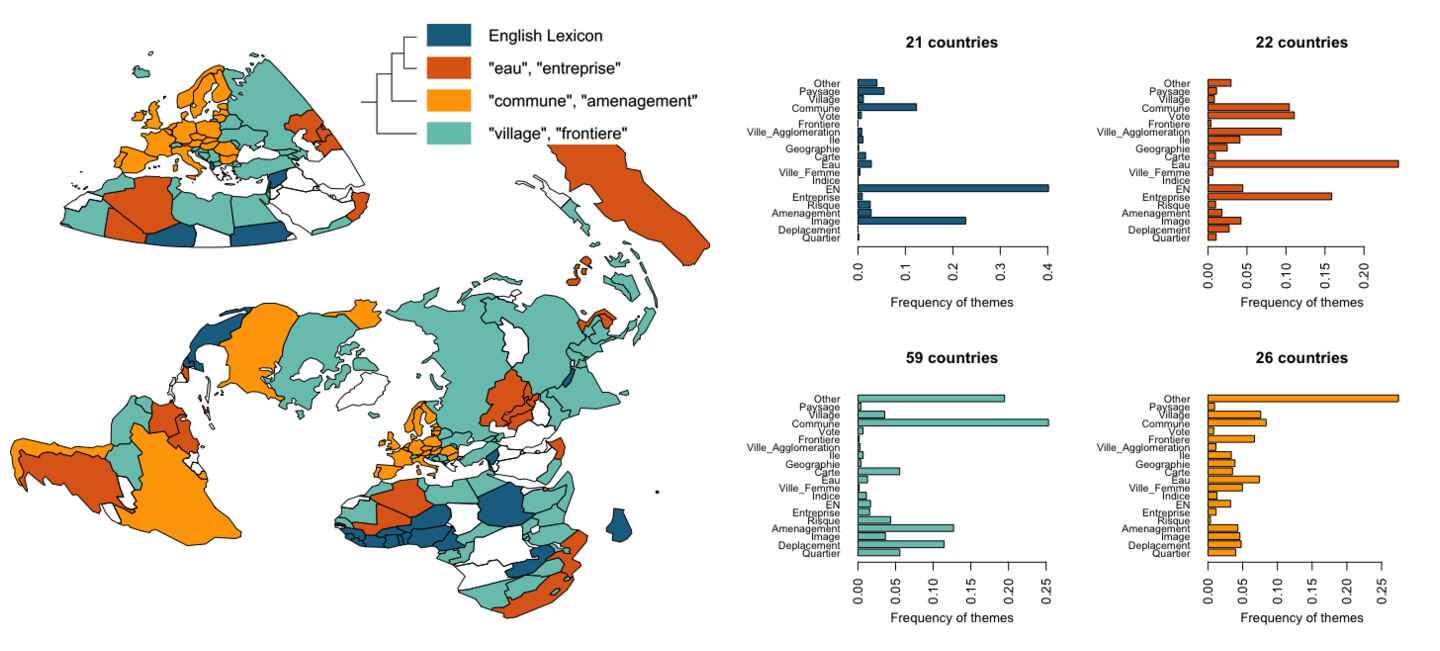}
	\appcaption{\textbf{Geographical communities of writing practice} (\textit{Left}); \textbf{Corresponding semantic profile of groups} (\textit{Right})\label{fig:app:cybergeonetworks:cluster_poc}}{\textbf{Communautés géographiques des pratiques d'écriture} (\textit{Gauche}) ; \textbf{Profile sémantique des groupes correspondants} (\textit{Droite})\label{fig:app:cybergeonetworks:cluster_poc}}
\end{figure}

\bpar{
In this clustering analysis, we do not find the dichotomy of countries based on their wealth and economic development levels. The link between semantic and geographical proximity is also less obvious at the world level, although one region is strikingly revealed: the institutional boundaries of Europe. The group of countries included in the EU27 plus the USA, Brazil and Chile (in yellow) appear strongly similar in terms of vocabulary used to talk about them. In particular, themes related to issues of administrative boundaries (``\textit{communes}'') and regional planning (``\textit{amenagement}'') describe these countries well (for example: \citep{santamaria2009schema, lusso2009musees, le2011consommation}). Two subgroups neighbour this cluster in the clustering tree. The first one includes countries studied by papers written in English. The second subgroup includes countries from all continents and corresponds to papers written preferentially with words such as ``\textit{eau}'' (water) and ``\textit{entreprise}'' (enterprise). Finally, 59 countries are distant from these groups in that words used to write about them refer to villages and borders (``\textit{frontiere}''), in contexts as diverse as Canada, Ecuador, Malaysia or Zimbabwe. The communities of vocabulary and writing practice thus appear less straightforward and less linked to geographical proximity. The main result lays in the fact that there is a specific set of words used to write about the European Union, a sort of EU27 Novlang made of words like ``Eurovision'', ``subsidiarity'' and ``Spatial Development Perspectives''.
}{
Dans cette analyse de classification, nous ne retrouvons pas les dichotomies des pays selon leur richesse et leur niveau de développement économique. Le lien entre la proximité sémantique et géographique est également moins évidente au niveau mondial, bien qu'une région soit particulièrement révélée : les limites institutionnelles de l'Europe. Le groupe de pays qui contient l'EU27 et les USA, le Brésil et le Chili (en jaune) apparaissent fortement similaires en termes de vocabulaire utilisé pour les décrire. En particulier, les thèmes en relation avec les questions de frontières administratives (``\textit{communes}'') et la planification régionale (``\textit{amenagement}'') décrivent bien ces pays (par exemple : \cite{santamaria2009schema, lusso2009musees, le2011consommation}). Deux sous-groupes sont voisins de ce cluster dans l'arbre de classification. Le premier inclut les pays étudiés par les articles écrits en Anglais. Le second sous-groupe inclut des pays de tous les continents et correspond aux articles qui utilisent préférentiellement des mots comme ``\textit{eau}'' et ``\textit{entreprise}''. Enfin, 59 pays sont distants de ces groupes en ce que les mots utilisés pour les décrire réfèrent aux villages et aux frontières ("\textit{frontière}"), dans des contextes aussi divers que le Canada, l'Equateur, la Malaisie ou le Zimbabwe. Les communautés de vocabulaire et de pratiques d'écriture apparaissent ainsi moins évidentes et moins liées à la proximité géographique. Le résultat principal consiste en le fait qu'il existe un ensemble spécifique de mots pour écrire à propos de l'Union Européenne, une sorte de novlangue de l'EU27 composée de mots comme ``Eurovision'', ``subventions'' ou ``Perspectives de développement spatial''.
}

\subsection{Discussion}{Discussion}

\subsubsection{Evaluating the complementarity of approaches}{Evaluation de la complémentarité des approches}

\bpar{
This section backs up the previous qualitative comparison of approaches through their spatialization by quantitative measures of their complementarity. Although we have seen that the communities obtained from the three different methods are semantically and geographically distinct, we do not know precisely how they complement each other. The overlapping analysis is complicated by the fact that articles belong simultaneously to several clusters for each classification. Therefore, we compare the methods 2 by 2 by computing the share of articles classified simultaneously in each possible pair of clusters from the two methods. In other words, if a method $M_1$ (for ex. based on citation communities) is composed of $n$ categories and a method $M_2$ (for ex. based on keywords communities) is composed of $m$ categories, we compute for each article $n\cdot m$ products of co-occurrences and then sum these products into $flows$ for the whole \textit{Cybergeo} corpus. If the methods were equivalent ways of describing and clustering articles, we would expect all the flows between communities to be $1:1$, $n:1$ or $1:n$, given that the methods do not give the same number of clusters. If the methods were completely orthogonal, we should find that each flow is proportional to the size of the origin cluster and of the size of the destination clusters. The fact that we find $n:n$ flows and that they are not determined entirely by the size of the clusters at origin and destination means that our three methods of semantic clustering are not equivalent nor orthogonal (Fig.~\ref{fig:app:cybergeonetworks:complementarity}). On the contrary, they shed different lights on the journal corpus.
}{
Cette section supporte les comparaisons qualitatives précédentes des approches par leur spatialisation par des mesures quantitatives de leur complémentarité. Bien que nous ayons vu que les communautés obtenues par les trois méthodes différentes sont distinctes sémantiquement et géographiquement, nous ne savons pas précisément comment elles se complètent. Une analyse de recouvrement est rendu difficile par le fait que les articles appartiennent simultanément à différents clusters pour chaque classification. Pour cela, nous comparons les méthodes 2 à 2 en calculant la part des articles classifiés simultanément dans chaque paire possible de clusters des deux méthodes. En d'autres termes, si une méthode $M_1$ (par exemple basée sur les communautés de citation) est composée de $n$ catégories et une méthode $M_2$ (par exemple basée sur les communautés de mots-clés) est composée de $m$ catégories, nous calculons pour chaque article $n\cdot m$ produits de co-occurrences et sommons ensuite ces produits en flux sur l'ensemble du corpus de \textit{Cybergeo}. Si les méthodes étaient des moyens équivalents pour décrire et classifier les articles, nous devrons obtenir uniquement des flux de la forme $1:1$, $n:1$ ou $1:n$, comme les méthodes ne donnent pas les mêmes nombres de clusters. Si les méthodes étaient complètement orthogonales, chaque flux devrait être proportionnel à la taille du cluster d'origine et à celle du cluster de destination. Le fait que nous trouvions $n:n$ flux et qu'ils ne sont pas entièrement déterminés par les tailles des clusters d'origine et de destination signifie que nos trois méthodes de classification sont ni équivalentes ni orthogonales (Fig.~\ref{fig:app:cybergeonetworks:complementarity}). Au contraire, ils donnent différents éclairages sur le corpus du journal.
}

\begin{figure}
\includegraphics[width=\linewidth]{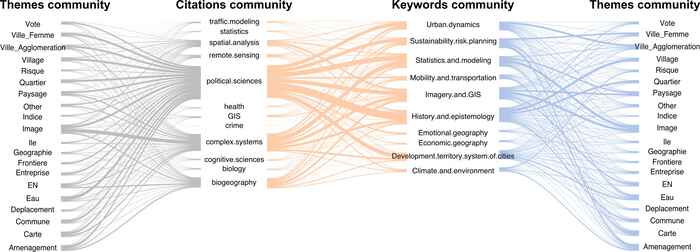}
\appcaption{\textbf{Overlap between semantic communities.}\label{fig:app:cybergeonetworks:complementarity}}{\textbf{Recouvrement entre les communautés sémantiques.}\label{fig:app:cybergeonetworks:complementarity}}
\end{figure}

\bpar{
For instance, there are clear preferential positive and negative relations between some citations communities and keywords communities (Fig.~\ref{fig:app:cybergeonetworks:complementarity}). On the one hand, 35\% of the Cybergeo articles cited by papers in the GIS cluster are characterized by keywords identified as ``Imagery and GIS''. On the other hand, there is no article cited by papers in the ``crime'' cluster which have keywords of the ``Climate and environment'' community. These relationships make sense, because the way a paper is advertised by its keywords is one of the first elements indicating the potential reader that the paper is relevant or not. Interestingly, the ``complex systems'' citation community is characterized by a variety of keywords communities (27\% of the articles cited by this community are tagged in the ``statistics and modelling'' cluster, 17\% in ``Imagery and GIS'' cluster, 13\% in ``history and epistemology'', 11\% in ``urban dynamics''). This suggests that the field of complex systems, being unified by methods rather than objects of inquiry, are more open to diverse topics than other citation communities. It could also mean that within \textit{Cybergeo}, authors of articles relevant to the complex systems community advertise their paper with keywords from the discipline of geography rather than methods only in order to attract topical reader as well.
}{
Par exemple, il existe clairement des relations préférentielles positives et négatives entre certaines communautés de citation et communautés de mots-clés (Fig.~\ref{fig:app:cybergeonetworks:complementarity}). D'une part, 35\% des articles de Cybergo dans le cluster GIS des citations sont caractérisés par des mots-clés identifiés comme ``télédétection et SIG''. D'autre part, il n'existe pas d'article dans la communauté de citation ``crime'' qui ont des mots-clés dans la communauté de mots-clés ``Climate and Environment''. Ces relations font sens, puisque la manière dont un papier est promu dans ses mots-clés est l'un des premiers éléments indiquant au lecteur potentiel si le papier est pertinent ou non. De façon intéressante, la communauté de citation ``complex systems'' est caractérisée par une variété de communautés de mots-clés (27\% des articles sont marqués dans ``statistics and modelling'', 17\% dans ``Imagery and GIS'', 13\% dans ``History and Epistemology'', 11\% dans ``Urban Dynamics''). Cela suggère que les champs des systèmes complexes, unifiés par des méthodes plutôt que par des objets de recherche, sont plus ouverts à des thèmes divers en comparaison aux autres communautés de citation. Cela peut également signifier qu'au sein de \textit{Cybergeo}, les auteurs des articles pertinents pour le champ des systèmes complexes présentent leurs articles avec des mots-clés de la discipline géographique plutôt que les méthodes afin d'attirer des lecteurs thématiques également.
}

\bpar{
Looking at the relations between keywords communities and themes communities, we find that some topics require specific words to write about them. For example, ``Imagery and GIS''-tagged articles use more words from the ``EN'' theme category, which corresponds to English words (rather than French). Urban studies are distinguished between its quantitative side (advertised by keywords around ``urban dynamics'' and using words such as ``agglomeration'') and its qualitative side (advertised by keywords around ``sustainability, risk and planning'' and using words such as ``\textit{femme}'':  woman). Interestingly, the words like ``risk'' (\textit{risque}) are used themselves more in articles tagged around ``Climate and environment'' than around ``sustainability, risk and planning''. Finally, the flows between themes communities and citations communities appear roughly proportional to the size of clusters at origin and destination, suggesting that citations are rather independent of the vocabulary used in the articles. This is reflected in the quantitative analysis below (Fig.~\ref{fig:app:cybergeonetworks:modularities}), this pair having the smallest mean absolute correlation. In short, the words that count in a citation strategy are much more the keywords than the actual content of the paper.
}{
S'intéressant aux relations entre les communautés de mots-clés et celles de thèmes, nous trouvons que certains thèmes nécessitent des mots spécifiques pour les désigner. Par exemple, les articles étiquetés comme ``Imagery and GIS'' utilisent plus de mots de la catégorie thématique ``EN'', qui correspond aux mots en anglais (plutôt qu'en français). Les études urbaines se distinguent entre leur côté quantitatif (présenté par des mots-clés autour de ``urban dynamics'' et utilisant des mots comme ``agglomeration'') et son côté qualitatif (présenté par des mots-clés autour de ``sustainability, risk and planning'' et utilisant des mots comme ``\textit{femme}''). De manière intéressante, les mots comme ``\textit{risque}'' sont eux-mêmes utilisés plus fréquemment dans des articles étiquetés comme ``Climate and Environment'' plutôt que ``sustainability, risk and planning''. Enfin, les flux entre les communautés thématiques et les communautés de citation apparaissent globalement proportionnels aux tailles des clusters à l'origine et à la destination, suggérant que les citations sont relativement indépendantes du vocabulaire utilisé dans les articles. Cela se reflète dans l'analyse quantitative ci-dessous (Fig.~\ref{fig:app:cybergeonetworks:modularities}), cette paire ayant la corrélation absolue moyenne la plus faible. En résumé, les mots qui importent dans la stratégie de citation sont plus les mots-clés que le contenu effectif de l'article.
}

\bpar{
We synthesize the flow relations between classifications by looking at their covariance structure in an aggregated way. More precisely, given the probability matrices $(p_{ki}) = (P_i)$ and $(p_{kj}) = (P_j)$ summarizing two classifications, where articles are indexed by rows, we estimate the correlation matrix between their columns $\rho_{ij} = \hat{\rho}\left[P_i,P_j\right]$ using a standard Pearson correlation estimator. We look then at aggregated measures, namely minimal correlation, maximal correlation and mean absolute correlation. In order to have a reference to interpret the values of these correlations, we compare them to two null models obtained by bootstrapping random corpuses. The estimate for the lower null model ($\rho_0$) is expected to minimize correlation and is obtained by shuffling all rows of one of the two matrices, which is done successively on both to ensure symmetry. The upper null model ($\rho_+$) is constructed by computing correlations between one matrix and the same where a fixed proportion of rows have been shuffled. We set this proportion to 50\%, which is a rather high level of similarity, and compute the model for both matrices each time. Average and standard deviations are computed for null models on $b=10000$ bootstrap repetitions. Table~\ref{tab:app:cybergeonetworks:cors} summarizes the results. We find that the maximal correlation for the Cybergeo corpus, which can be interpreted as a maximum overlap between approaches of semantic clustering, is always significantly smaller (around $5\cdot \sigma$) than for the upper null model. This confirms that our three classifications are highly independent of one another in their main components. It is interesting to note that for Keywords/Themes, the mean absolute correlation is within the standard error range of the mean absolute correlation of the upper null model, suggesting that these two must be rather close on small overlaps. They are actually closer than with Citations for all indicators. We also confirm that Themes/Citations has the lowest mean absolute overlap.
}{
Nous synthétisons les relations de flux entre classifications en examinant leur structure de covariance de manière agrégée. Plus précisément, étant donné les matrices de probabilité $(p_{ki}) = (P_i)$ et $(p_{kj}) = (P_j)$ résumant deux classifications, dans lesquelles les articles sont indexés par la ligne, nous estimons la matrice de corrélation entre leurs colonnes $\rho_{ij} = \hat{\rho}\left[P_i,P_j\right]$ en utilisant un estimateur de corrélation de Pearson standard. Nous regardons ensuite des mesures agrégées, qui sont la corrélation minimale, la corrélation maximale et la corrélation absolue moyenne. Afin de disposer d'une référence pour interpréter les valeurs de ces corrélations, nous les comparons à deux modèles nul obtenus en par un bootstrap de corpus aléatoires. L'estimation pour le modèle nul inférieur ($\rho_0$) doit minimiser la corrélation et est obtenu par permutation de l'ensemble des lignes d'une des deux matrices, ce qui est fait successivement sur chaque pour assurer la symétrie. Le modèle nul supérieur ($\rho_+$) est construit est construit par calcul des corrélations entre une matrice et l'identique dont une proportion fixée de lignes ont été permutées. Nous fixons cette proportion à 50\%, ce qui est un assez haut niveau de similarité, et calculons le modèle pour chaque matrice à chaque fois. Les moyennes et deviation standard sont calculées sur $b=10000$ répétitions de bootstrap. La Table~\ref{tab:app:cybergeonetworks:cors} résume les résultats. Nous trouvons que la corrélation maximale pour le corpus Cybergeo, qui peut être interprétée comme un recouvrement maximal entre les approches de classification sémantique, est toujours significativement plus petite (autour de $5\cdot \sigma$) que pour le modèle nul supérieur. Cela confirme que nos trois classifications sont fortement indépendantes l'une de l'autre dans leurs composantes principales. Il est intéressant de relever que pour la relation Mots-Clés/Thèmes, la corrélation absolue moyenne est dans la déviation standard de la corrélation absolue moyenne du modèle nul supérieur, suggérant que celles-ci doivent être plutôt proches sur les faibles recouvrements. Elles sont effectivement plus proches qu'avec Citation pour l'ensemble des indicateurs. Nous confirmons aussi que Thèmes/Citation ont le plus bas recouvrement absolu moyen.
}

\begin{table}
\begin{threeparttable}
\appcaption{\textbf{Correlations between classifications.}\label{tab:app:cybergeonetworks:cors}}{\textbf{Corrélations entre les classifications.}\label{tab:app:cybergeonetworks:cors}}
\begin{tabular}{lccccccccc}
 \toprule
 \hline\cr
 & $\min \rho$ & $\min \rho_0$ & $\min \rho_+$ & $\max \rho$ & $\max \rho_0$ &$\max \rho_+$ & $< \left| \rho \right|>$ & $< \left| \rho_0 \right|>$ & $< \left| \rho_+ \right|>$ \\\cr\hline\cr
 Themes/Citations & -0.30 &$-0.12$&$-0.17$&0.36&$0.21$&$0.69$&0.059&$0.043$&$0.073$\\
  &  &$\pm 0.019$&$\pm 0.071$&&$\pm 0.042$&$\pm 0.070$&&$\pm 0.0021$&$\pm 0.012$\\\cr
 Citations/Keywords & -0.26 & $-0.096$ & $-0.20$ & 0.30 & $ 0.13$ & $0.64$ & 0.070 & $ 0.034$ & $ 0.092$\\
  &  & $\pm 0.015$ & $\pm 0.047$ && $\pm 0.027 $ & $\pm0.068$ & & $\pm 0.0026 $ & $\pm 0.0081$\\\cr
 Keywords/Themes &-0.20&$-0.11$&$-0.13$&0.51&$0.17$&$0.66$&0.091&$0.040$&$0.080$\\
  &&$\pm0.013$&$\pm0.030$&&$\pm0.032$&$\pm0.075$&&$\pm0.0022$&$\pm0.020$\\\cr
 \hline
\bottomrule
\end{tabular}
 \begin{tablenotes}
       \item \protect\scriptsize{\bpar{\textbf{Notes}: For each pair of classification and measure, we also give average and standard deviation for lower ($\rho_0$) and upper ($\rho_+$) null models, obtained by bootstrapping $b=10000$ random corpuses.}{\textbf{Notes} : Pour chaque paire de classification et chaque mesure, nous donnons également la moyenne et la déviation standard pour les modèles nuls inférieur ($\rho_0$) et supérieur ($\rho_+$), obtenus par bootstrap de $b=10000$ corpus aléatoires.}
       }
    \end{tablenotes}
  \end{threeparttable}
\end{table}

\bpar{
To make these conclusions more robust, we complement the analysis with a network modularity analysis, which is a widely applied method to evaluate the relevance of a classification within a network. To be able to compare two classifications, since the citation network is too sparse for any analysis as mentioned, we evaluate the modularity of a classification within the network induced by the other. More precisely, given a distance threshold $\theta$ and two documents given by their probabilities within a classification $\vec{p}_i^{(c)},\vec{p}_j^{(c)}$, we consider the network with documents as nodes linked if and only if $d(\vec{p}_i^{(c)},\vec{p}_j^{(c)})<\theta$ with $d$ euclidian distance. We can then compute the multi-class modularity of the other classification in the sense of~\cite{nicosia2009extending}. We show in Fig.~\ref{fig:app:cybergeonetworks:modularities}, for different thresholds, the modularities normalized by the modularity of the network classification within its own network. The closest the measure is to 1, the closer are the classifications. Most of couple have low values for large ranges of $\theta$, confirming the previous conclusions of orthogonality. Furthermore, the different behavior as a function of $\theta$ (increasing or decreasing) suggests different \emph{internal structures} of classification, what is consistent with the fact that they rely on different processes to classify data.
}{
Pour rendre ces conclusions plus robustes, nous complémentons l'analyse par une analyse de modularité de réseau, qui est une méthode largement appliquée pour évaluer la pertinence d'une classification dans un réseau. Pour être en mesure de comparer deux classifications, comme le réseau de citations est trop creux pour toute analyse comme déjà mentionné, nous évaluons la modularité d'une classification au sein du réseau induit par l'autre. Plus précisément, étant donné un seuil de distance $\theta$ et deux documents donnés par leur probabilités dans une classification $\vec{p}_i^{(c)},\vec{p}_j^{(c)}$, nous considérons le réseau avec les documents comme noeuds, liés si et seulement si $d(\vec{p}_i^{(c)},\vec{p}_j^{(c)})<\theta$ avec $d$ distance euclidienne. On peut ensuite calculer la modularité multi-classes de l'autre classification au sens de~\cite{nicosia2009extending}. Nous montrons en Fig.~\ref{fig:app:cybergeonetworks:modularities}, pour différents seuils, les modularités normalisées par la modularité de la classification du réseau dans son propre réseau. Le plus proche de 1 est la mesure, le plus proches sont les classifications. La plupart des couples ont de faibles valeurs pour de larges plages de $\theta$, confirmant les conclusions précédentes d'orthogonalité. De plus, les différents comportements en fonction de $\theta$ (croissant ou décroissant) suggère différentes \emph{structures internes} des classifications, ce qui est cohérent avec le fait qu'elles reposent sur des processus différents pour classifier les données.
}

\begin{figure}
\includegraphics[width=\linewidth]{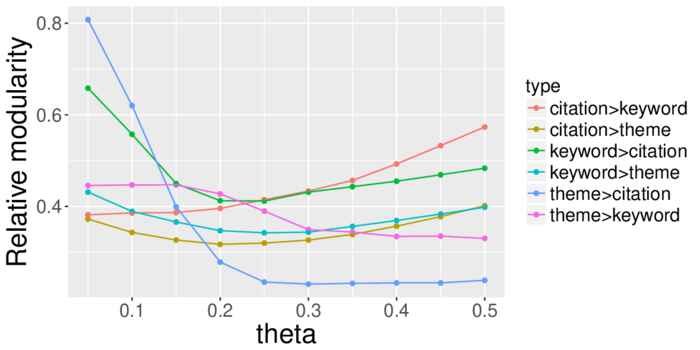}
\appcaption{\textbf{Evaluating the complementarity of classification through network modularities.} The plot gives the relative modularity of the first classification in the network induced by the second with the threshold $\theta$ (see text), for each couple of classifications (color).\label{fig:app:cybergeonetworks:modularities}}{\textbf{Evaluation de la complémentarité des classifications par la modularités des réseaux.} Le graphe donne la modularité relative de la première classification dans le réseau induit par la seconde avec le seuil $\theta$ (voir texte), pour chaque couple de classifications (couleurs).\label{fig:app:cybergeonetworks:modularities}}
\end{figure}

\bpar{
Together with the visual diagrams, these analyses show the complementarity of classifications in the exploration of semantic diversity of publication in a 20 year old journal.
}{
En combinaison aux diagrammes visuels, ces analyses montrent la complémentarité des classifications dans l'exploration de la diversité sémantique des publications dans un journal qui fête ses 20 ans.
}

\subsubsection{Applied perspectivism}{Perspectivisme appliqué}

\bpar{
Our approach can be understood as an ``applied perspectivism'', which we believe is a way to enhance second order knowledge creation and to ensure reflexivity. Perspectivism is an epistemological position defended by~\cite{giere2010scientific}, that aims at going beyond the constructivism-reductivism debates. Focusing on scientific agents as carriers of knowledge creation, any scientific entreprise is a certain \emph{perspective} on the world, taken by the agent for a given purpose and through a given medium that is considered as the \emph{model}. Perspectives are necessary complementary as they result from different approaches to the same objects, even if the definition of objects and research questions will not necessarily be the same. Coupling perspectives should thus be a typical feature of interdisciplinarity. We position our work as a deliberate attempt to couple complementary perspectives on the same corpus. \cite{varenne2017theories} recalls that one of the various function of models is to foster coupling between theories through coupling of models themselves, allowing the creation of novel knowledge within the virtuous spiral between disciplinarity and interdisciplinarity coined by~\cite{banos2013pour}. Our work aims precisely at accelerating and improving such processes.
}{
Notre approche peut être comprise comme un ``perspectivisme appliqué'', que nous postulons comme une manière de contribuer à la création d'une connaissance au second ordre et d'assurer une réflexivité. Le perspectivisme est une position épistémologique défendue par~\cite{giere2010scientific}, qui vise à sortir par le haut des débats entre constructivisme et réalisme. En se concentrant sur les agents scientifiques comme supports de la création de connaissances, toute entreprise scientifique est une certaine \emph{perspective} sur le monde, prise par l'agent dans un certain but et grâce à un certain medium qui est considéré comme étant son \emph{modèle}. Les perspectives sont généralement complémentaires puisqu'elles résultent d'approches différentes sur les mêmes objets, bien que les définitions des objets et les questions de recherche ne seront pas nécessairement les mêmes. Le couplage de perspectives devrait ainsi être une caractéristique essentielle de l'interdisciplinarité. Nous positionnons notre travail comme une tentative délibérée de coupler des perspectives complémentaires sur le même corpus. \cite{varenne2017theories} rappelle que l'une des nombreuses fonctions des modèles est de favoriser le couplage entre théories par le couplage des modèles eux-mêmes, permettant la création d'une connaissance nouvelle au sein de la spirale vertueuse entre disciplinarité et interdisciplinarité mise en valeur par~\cite{banos2013pour}. Notre travail vise précisément à accélérer et améliorer de tels processus.
}

\subsubsection{Fostering Open Science and Reflexivity}{Encourager la Science ouverte et la Réflexivité}

\bpar{
The open tools and software we provide participate to a larger effort of reflexivity tools in the context of Open Science. It is aimed at being complementary to existing platforms, like the Community Explorer for the community of Complex Systems developed by ISCPIF\footnote{available at \url{https://communityexplorer.org}} that provides an interactive visualisation of social research networks combined to semantic networks based on self-declared keywords provided by researchers. An other example closer to what we developed is Gargantext\footnote{\url{https://gargantext.org/}} that provides corpus exploration functionalities. Linkage\footnote{\url{https://linkage.fr/}} is a similar tool with different methods, using latent topic allocation for networks with textual annotations~\citep{bouveyron2016stochastic}. We differentiate from these by exploring simultaneously multiple dimensions of semantic classification and more importantly by adding the geographical aspect. Furthermore, in comparison to various tools that private publishers are beginning to introduce, the open and collaborative nature of our work is crucial. For example, \cite{bohannon2014scientific} suggests that one must stay careful when using search results from a popular academic search engine, as the mechanisms of the ranking algorithm and thus the multiple biases are unknown. The comparison is similar with text-mining paying services provided by private companies, as we suggest that a subtle synergy between knowledge content and knowledge production processes (that is allowed by open tools only) can be more beneficial to both.
}{
Les outils ouverts et les logiciels que nous produisons participent à un effort plus large de conception d'outils de réflexivité dans le contexte de la Science Ouverte. Ils visent à être complémentaires aux plateformes existantes, comme le \textit{Community Explorer} développé pour la communauté des Systèmes Complexes développé par l'ISCPIF\footnote{disponible à \url{https://communityexplorer.org}} qui fournit une visualisation interactive des réseaux sociaux de la recherche en combinaison aux réseaux sémantiques basés sur les mots-clés auto-déclarés donnés par les chercheurs. Un autre exemple plus proche de ce que nous avons développé est l'outil Gargantext\footnote{\url{https://gargantext.org/}} qui fournit des fonctions d'exploration de corpus. Linkage\footnote{\url{https://linkage.fr/}} est un outil similaire avec des méthodes différentes, qui utilise un allocation latente de thèmes pour les réseaux avec annotations textuelles~\citep{bouveyron2016stochastic}. Nous nous différentions de ceux-ci en explorant simultanément de multiples dimensions de classification sémantique et de façon plus importante en y ajoutant la composante géographique. De plus, en comparaison aux outils variés que les éditeurs privés commencent à introduire, la nature ouverte et collaborative de notre travail est essentielle. Par exemple, \cite{bohannon2014scientific} suggère qu'il faut rester précautionneux quant à l'utilisation des résultats de recherche issus d'un engin académique de requêtes populaire, comme les mécanismes de l'algorithme de classement et ainsi les biais multiples restent inconnus. La comparaison est similaire avec les services payant de fouille de texte fournis par des entreprises privées, puisque nous suggérons qu'une synergie subtile entre le contenu de la connaissance et les processus de production de connaissance (qui est permise par des outils ouverts de manière privilégiée) peut être plus bénéfique aux deux.
}


\bpar{
We have studied a scientific corpus of a journal in Geography, combining multiple points of view through their embedding in the geographical space. This work is therefore in itself reflexive, illustrating the kind of new approach to science it aims at promoting. We believe that the open tools we develop in this context will contribute to the empowerment of authors within Open Science.
}{
Nous avons étudié un corpus scientifique d'un journal en Géographie, combinant des points de vue multiples par leur plongement dans l'espace géographique. Ce travail est pour cela en lui-même réflexif, illustrant le type d'approche nouvelle à la science qu'il cherche à promouvoir. Nous sommes convaincus que les outils ouverts que nous développons dans ce contexte contribueront à la prise d'autonomie des auteurs dans la Science Ouverte.
}

\stars

%


\newpage

\section{Semantic classification of patents}{Classification sémantique des brevets}

\label{app:sec:patentsmining}

\bpar{
The classifications through hyper-network obtained in Chapter~\ref{ch:modelinginteractions} have been made possible thanks to different complementary technical contributions. The construction of the semantic network, including the extraction of keywords and the quantification of their relevance, and then its analysis, have initially been launched in the context of the analysis of the \emph{Cybergeo} journal (see~\ref{app:sec:cybergeo}). The application to massive corpuses and the method to extract optimal sub-networks through Pareto optimization have been developed in the context of the application which is presented here, to the corpus of patents filled in the United States between 1976 and 2013.
}{
Les classifications par hyper-réseau obtenues au Chapitre~\ref{ch:modelinginteractions} ont été permises par diverses contributions techniques complémentaires. La construction du réseau sémantique, incluant l'extraction des mots-clés et la quantification de leur pertinence, puis son analyse, ont été initialement lancés dans le cadre de l'analyse de la revue \emph{Cybergeo} (voir~\ref{app:sec:cybergeo}). L'application à des corpus massifs et la méthode d'extraction de sous-réseau optimal par optimisation de Pareto ont été développé dans le cadre de l'application qui est présentée ici, au corpus de brevets déposés aux Etats-unis de 1976 à 2013.
}

\bpar{
This appendix has thus a crucial role regarding methods and tools (even if we classify it into technical developments due to its autonomous thematic nature), but also for its proper content regarding the possible future developments, as we suggested for a quantitative characterization of the diffusion of innovation in the framework of an empirical investigation of hypothesis of the evolutive urban theory.
}{
Cette annexe a ainsi un caractère essentiel du point de vue des méthodes et des outils (même si nous la classons dans les développements thématiques de par son caractère thématique autonome), mais également pour son contenu propre au vu des futurs développements possibles, comme nous avons suggéré pour une caractérisation quantitative de la diffusion de l'innovation dans le cadre d'une investigation empirique des hypothèses de la Théorie Evolutive des Villes.
}

\stars

\bpar{
\textit{This appendix is the result of a collaboration with \noun{Dr A. Bergeaud} (Paris School of Economics) and \noun{Dr Y. Potiron} (Keyo University), in the context of a convergence of respective research questions of semantic analysis of massive corpuses and of an endogenous characterization of innovation. It has been published as~\cite{10.1371/journal.pone.0176310}. It is here adapted.}
}{
\textit{Cette annexe résulte d'une collaboration avec \noun{Dr A. Bergeaud} (Paris School of Economics) et \noun{Dr Y. Potiron} (Keyo University), dans le cadre d'une convergence des problématiques respectives d'analyse sémantique de corpus massifs et de caractérisation endogène de l'innovation. Elle a été publiée comme~\cite{10.1371/journal.pone.0176310}. Elle est ici traduite et adaptée.}
}

\stars

\bpar{
Therefore, we extend some usual techniques of classification resulting from a large-scale data-mining and network approach. This new technology, which in particular is designed to be suitable to big data, is used to construct an open consolidated database from raw data on 4 million patents taken from the US patent office from 1976 onward. To build the pattern network, not only do we look at each patent title, but we also examine their full abstract and extract the relevant keywords accordingly. We refer to this classification as \emph{semantic approach} in contrast with the more common \emph{technological approach} which consists in taking the topology when considering US Patent office technological classes. Moreover, we document that both approaches have highly different topological measures and strong statistical evidence that they feature a different model. This suggests that our method is a useful tool to extract endogenous information.    
}{
Ainsi, nous étendons ici les techniques de classification usuelles par une approche d'analyse de réseau et de fouille de données à grande échelle. Ces nouvelles techniques, qui sont conçues particulièrement pour être adaptées aux données massives, sont utilisées pour construire une base consolidée ouverte à partir de données brutes de 4 millions de brevets issus de l'US Patent Office (USPTO) depuis 1976. Pour construire le réseau, nous utilisons les titres, mais aussi les résumés complets desquels nous extrayons les mots-clés pertinents. Nous désignons cette classification comme \emph{approche sémantique}, en comparaison avec la plus classique \emph{approche technologique} qui consiste à considérer la topologie induite par les classes technologiques de l'USPTO. De plus, nous obtenons des mesures topologiques fortement différentes pour les deux approches. Cela suggère que cette méthode est un outil puissant pour extraire une information endogène.
}

\subsection*{Introduction}{Introduction}

\bpar{
Innovation and technological change have been described by many scholars as the main drivers of economic growth as in \cite{aghionhowitt1992} and \cite{romer1990}. \cite{RePEc:nbr:nberwo:3301} advertised the use of patents as an economic indicator and as a good proxy for innovation. Subsequently, the easier availability of comprehensive databases on patent details and the increasing number of studies allowing a more efficient use of these data (e.g.~\cite{Hall2001}) have opened the way to a very wide range of analysis. Most of the statistics derived from the patent databases relied on a few key features: the identity of the inventor, the type and identity of the rights owner, the citations made by the patent to prior art and the technological classes assigned by the patent office post patent's content review. Combining this information is particularly relevant when trying to capture the diffusion of knowledge and the interaction between technological fields as studied in \cite{Youn:2015fk}. With methods such as citation dynamics modeling discussed in~\cite{newman2014prediction} or co-authorship networks analysis in~\cite{sarigol2014predicting}, a large body of the literature such as~\cite{sorenson2006complexity} or~\cite{kay2014patent} has studied patents citation network to understand processes driving technological innovation, diffusion and the birth of technological clusters. Finally, \cite{bruck2016recognition} look at the dynamics of citations from different classes to show that the laser/ink-jet printer technology resulted from the recombination of two different existing technologies. 
}{
L'innovation et le changement technologique ont été présentés par de nombreux chercheurs comme des moteurs principaux de la croissance économique comme dans \cite{aghionhowitt1992} et \cite{romer1990}. \cite{RePEc:nbr:nberwo:3301} a proposé l'utilisation des brevets comme un indicateur économique et comme un bon proxy pour l'innovation. Par conséquent, la disponibilité accrue de bases de données plus complètes et détaillées des brevets et l'augmentation du nombre d'études qui ont permis une utilisation plus efficace de ces données (par exemple~\cite{Hall2001}) ont ouvert la voie à des analyses très variées. La majorité des statistiques issues des bases de données sur les brevets se basent sur un petit nombre de caractéristiques essentielles : l'identité de l'inventeur, le type et l'identité du possesseur des droits, les citations faites par le brevet à des travaux antérieurs et les classes technologiques attribuées par le bureau des brevets après une revue du contenu du brevet. La combinaison des ces informations est particulièrement pertinent pour tenter de capturer la diffusion des connaissances et les interactions entre champs technologiques comme étudié dans \cite{Youn:2015fk}. Par des méthodes comme la modélisation dynamique des citations, étudiée par~\cite{newman2014prediction}, ou les analyses de réseaux de co-auteurs dans~\cite{sarigol2014predicting}, une partie conséquente de la littérature comme~\cite{sorenson2006complexity} ou~\cite{kay2014patent} s'est intéressée aux réseaux de citations de brevets pour comprendre les processus à l'origine de l'innovation technologique, de la diffusion et de la naissance de clusters technologiques. Enfin, \cite{bruck2016recognition} étudie par exemple la dynamique des citations depuis différentes classes pour montrer que la technologique de l'imprimante laser jet d'encre est le résultat de la recombination de deux technologies existantes.
}

\bpar{
Consequently, technological classification combined with other features of patents can be a valuable tool for researchers interested in studying technologies throughout history and to predict future innovations by looking at past knowledge and interaction across sectors and technologies. But it is also crucial for firms that face an ever changing demand structure and need to anticipate future technological trends and convergence (see, e.g., \cite{curran2011patent}) to adapt to the resulting increase in competition discussed in~\cite{Katz1996remarks} and to maintain market share. Curiously, and in spite of the large number of studies that analyze interactions across technologies~\cite{Furman2011shoulders}, little is known about the underlying ``innovation network'' (e.g. ~\cite{AAKnetwork2016}). 
}{
Par conséquent, la classification technologique combinée à d'autres caractéristiques des brevets peut être un outil efficace pour l'étude des technologies à travers l'histoire et pour prédire des innovations futures en se basant sur la connaissance passée et les interactions entre secteurs et technologies. Mais il est également crucial pour les entreprises qui font face à une structure toujours changeante de la demande et qui ont besoin d'anticiper les tendances technologiques futures et les convergences (voir par exemple \cite{curran2011patent}) pour s'adapter à la compétition accrue qui en ressort, discutée dans \cite{Katz1996remarks} et pour maintenir une part de marché. De manière surprenante, et malgré le grand nombre d'études qui analysent les interactions entre technologies~\cite{Furman2011shoulders}, la connaissance du ``réseau d'innovation''~\cite{AAKnetwork2016} sous-jacent est relativement faible.
}

\bpar{
We propose here an alternative classification based on semantic network analysis from patent abstracts and explore the new information emerging from it. In contrast with the regular technological classification which results from the choice of the patent reviewer, semantic classification is carried automatically based on the content of the patent abstract. Although patent officers are experts in their fields, the relevance of the existing classification is limited by the fact that it is based on the state of technology at the time the patent was granted and cannot anticipate the birth of new fields. To correct for this, the USPTO regularly make changes in its classification in order to adapt to technological change (for example, the ``nanotechnology'' class (977) was established in 2004 and retroactively to all relevant previously granted patents). In contrast we don't face this issue with the semantic approach. The semantic links can be clues of one technology taking inspiration from another and  good predictors of future technology convergence (e.g.~\cite{preschitschek2013} study semantic similarities from the whole text of 326 US-patents on \textit{phytosterols} and show that semantic analysis have a good predicting power of future technology convergence). One can for instance consider the case of the word \textit{optic}. Until more recently, this word was often associated with technologies such as photography or eye surgery, while it is now almost exclusively used in a context of semi-transistor design and electro-optic. This semantic shift did not happen by chance but contains information on the fact that modern electronic extensively uses technologies that were initially developed in optic. 
}{
Nous proposons ici une classification alternative basée sur l'analyse sémantique des résumés des brevets et explorons la nouvelle information qui em émerge. En opposition avec la classification technologique classique qui est produite par les choix du relecteur du brevet, la classification sémantique est opérée automatiquement en se basant sur le contenu du résumé du brevet. Bien que les \emph{patent officers} (experts chargés de la classification) soit experts dans leur champs, la pertinence de la classification existante est limitée par le fait qu'elle est basée sur l'état de la technologie au moment où le brevet est accordé, et ne peut pas anticiper la naissance de nouveau champs. Pour corriger cela, l'USPTO change régulièrement ses classifications afin de les adapter au changement technologique (par exemple, la classe ``nanotechnologie'' (977) a été établie en 2004 et de manière retroactive à tous les brevets antérieurs pour lesquels elle était pertinente). Au contraire, ce problème n'est pas présent dans l'approche sémantique. Les liens sémantiques peuvent être des indices d'une technologie s'inspirant d'une autre et de bons prédicteurs de futures convergences technologiques (par exemple~\cite{preschitschek2013} étudie des similarités sémantiques dans l'ensemble du texte de 326 brevets sur les \textit{phytosterols} et montre que l'analyse sémantique a un bon pouvoir prédictif d'une convergence technologique future). On peut par exemple considérer le cas du mot \textit{optic}. Jusqu'à récemment, ce mot était souvent associé aux technologies comme la photographie et la chirurgie de l'oeil, alors qu'il est maintenant presque exclusivement utilisé dans le contexte de la conception des semi-transistors et de l'optique électronique. Cette dérive sémantique ne s'est pas opérée par hasard mais contient de l'information sur le fait que l'électronique moderne utilise de manière intensive des technologies qui ont été initialement développées en optique.
}

\bpar{
Previous research has already proposed to use semantic networks to study technological domains and detect novelty. \cite{yoon2004text} was one of the first to enhance this approach with the idea of visualizing keywords network illustrated on a small technological domain. The same approach can be used to help companies identifying the state of the art in their field and avoid patent infringement as in \cite{park2014semantic} and \cite{yoon2011detecting}. More closely related to our methodology,~\cite{gerken2012new} develop a method based on patent semantic analysis of patent to vindicate the view that this approach outperform others in the monitoring of technology and in the identification of novelty innovation. Semantic analysis has already proven its efficiency in various fields, such as in technology studies (e.g.~\cite{choi2014patent} and~\cite{fattori2003text}) and in political science (e.g.~\cite{2015arXiv151003797G}).
}{
Des travaux existants ont déjà proposé d'utiliser les réseaux sémantiques pour étudier les domaines technologiques et détecter la nouveauté. \cite{yoon2004text} était parmi les premiers à utiliser cette approche dans l'idée de visualiser les réseaux de mots-clés, illustrée sur un champ technologique restreint. La même approche peut être utilisée pour aider les entreprises à identifier l'état de l'art dans leur domaine et éviter les violations de brevets, comme dans \cite{park2014semantic} et \cite{yoon2011detecting}. Plus proche de notre méthodologie, \cite{gerken2012new} développe une méthode basée sur l'analyse sémantique des brevets pour appuyer l'idée que cette approche est plus performante pour suivre la technologie et pour l'identification de la nouveauté dans les innovations. L'analyse sémantique a déjà montré ses capacités dans des champs variés, comme l'étude des technologies (par exemple~\cite{choi2014patent} et~\cite{fattori2003text}) et en sciences politiques~\cite{2015arXiv151003797G}.
}

\bpar{
Building on such previous research, we make several contributions by fulfilling some shortcomings of existing studies, such as for example the use of frequency-selected single keywords. First of all, we develop and implement a novel fully-automatized methodology to classify patents according to their semantic abstract content, which is to the best of our knowledge the first of its type. This includes the following refinements for which details can be found in the following section: (i) use of multi-stems as potential keywords; (ii) filtering of keywords based on a second-order (co-occurrences) relevance measure and on an external independent measure (technological dispersion); (iii) multi-objective optimization of semantic network modularity and size. The use of all this techniques in the context of semantic classification is new and essential from a practical perspective. 
}{
En se basant sur ces travaux, nous faisons diverses contributions en dépassant certaines limitations des études existantes, comme par exemple l'utilisation de mots-clés simples sélectionnés par leur fréquence. Tout d'abord, nous développons et implémentons une nouvelle méthodologie totalement automatisée pour classifier les brevets selon le contenu sémantique de leur résumés, qui est à notre connaissance la première de ce type. Elle inclut les raffinements suivants pour lesquels des détails peuvent être trouvés dans la section suivante : (i) utilisation de racines multiples comme mots-clés potentiels ; (ii) filtration des mots-clés selon une mesure de pertinence au second ordre (cooccurrences) et sur une mesure externe indépendante (dispersion technologique) ; (iii) optimisation multi-objectifs de la modularité et de la taille du réseau sémantique. L'utilisation de l'ensemble de ces techniques dans le contexte de la classification sémantique est nouveau et essentiel dans une perspective appliquée.
}

\bpar{
Furthermore, most of the existing studies rely on a subsample of patent data, whereas we implement it on the full US Patent database from 1976 to 2013. This way, a general structure of technological innovation can be studied. We draw from this application promising qualitative stylized facts, such as a qualitative regime shift around the end of the 1990s, and a significant improvement of citation modularity for the semantic classification when comparing to the technological classification. These thematic conclusions validate our method as a useful tool to extract endogenous information, in a complementary way to the technological classification.
}{
De plus, la majorité des études existantes se basent sur un échantillon des données de brevet, alors que nous implémentons ici notre méthode sur l'ensemble de la base USPTO de 1976 à 2013. De cette manière, une structure générale de l'innovation technologique peut être étudiée. Nous tirons de cette application divers faits stylisés qualitatifs, comme un changement qualitatif de régime autour de la fin des années 1990, et une amélioration significative de la modularité de citation pour la classification sémantique en comparaison à la classification technologique. Ces conclusions thématiques valident notre méthode comme un outil utile pour extraire de l'information endogène, de manière complémentaire à la classification technologique.
}

\bpar{
Thanks to this complementarity, we believe that patent officers could benefit very much from looking at the semantic network when considering potential citation candidates of a patent in review.
}{
Grace à cette complémentarité, nous postulons que les \emph{patent officers} pourraient bénéficier significativement de la considération du réseau sémantique lors de la considération des citations potentielles pour un brevet en revue.
}

\bpar{
The work is organized as follows. We first describe the patent data, the existing classification and provide details about the data collection process. We then explain the construction of the semantic classes. Their relevance is then tested by providing exploratory results. Finally, we discuss potential further developments in conclusion. 
}{
Ce travail est organisé de la façon suivante : nous présentons d'abord les données, la classification existante, et donnons des détails sur la procédure de collection des données. Nous expliquons ensuite la procédure de construction des classes sémantiques. La pertinence de celles-ci est alors testée par des analyses exploratoires. Enfin, nous discutons des développements potentiels en conclusion.
}

\subsection*{Background}{Contexte}
 \label{app:subsec:data}

\bpar{
In our analysis, we will consider all utility patents granted in the United States Patent and Trademark Office (USPTO) from 1976 to 2013. A clearer definition of utility patent is given in Appendix of \cite{10.1371/journal.pone.0176310}. Also, additional information on how to correctly exploit patent data can be found in \cite{Hall2001} and \cite{lerner2015use}.
}{
Dans l'analyse, nous considérons l'ensemble des brevets accordés par l'USPTO de 1976 à 2013. Une définition plus précise d'un brevet d'utilité (\emph{utility patent}) est donnée en Annexe de \cite{10.1371/journal.pone.0176310}. De plus, des informations supplémentaires sur la manière d'exploiter correctement les données de brevet est donnée dans \cite{Hall2001} et \cite{lerner2015use}.
}

\subsubsection*{An existing classification: the USPC system}{Une classification existante : le système USPC}

\bpar{
Each USPTO patent is associated with a non-empty set of technological classes and subclasses. There are currently around 440 classes and over 150,000 subclasses constituting the United State Patent Classification (USPC) system. While a technological class corresponds to the technological field covered by the patent, a subclass stands for a specific technology or method used in this invention. A patent can have multiple technological classes, on average in our data a patent has 1.8 different classes and 3.9 pairs of class/subclass. At this stage, two features of this system are worth mentioning: (i) classes and subclasses are not chosen by the inventors of the patent but by the examiner during the granting process based on the content of the patent; (ii) the classification has evolved in time and continues to change in order to adapt to new technologies by creating or editing classes. When a change occurs, the USPTO reviews all the previous patents so as to create a consistent classification.
}{
Chaque brevet USPTO est associé avec un ensemble non vide de classes et de sous-classes technologiques. Il existe actuellement autour de 440 classes et plus de 150000 sous-classes, qui constituent le système \emph{United States Patent Classification} (USPC). Alors qu'une classe technologique correspond au champ technologique couvert par le brevet, une sous-classe correspond à une technologie spécifique ou une méthode utilisée dans l'invention. Un brevet peut avoir de multiples classes technologiques : en moyenne dans nos données un brevet a 1.8 classes différentes et 3.9 paires de classe/sous-classe. A cette étape, deux caractéristiques de ce système valent la peine d'être mentionnées : (i) la classe et la sous-classe ne sont pas choisies par l'inventeur du brevet mais par l'examinateur pendant le processus de revue, en fonction du contenu du brevet ; (ii) la classification a évolué au cours du temps et continue à évoluer afin de s'adapter aux nouvelles technologies par la création ou l'édition de classes. Lorsqu'un changement a lieu, l'USPTO revoit l'ensemble des brevets précédents afin de créer une classification cohérente.
}

\subsubsection*{A bibliographical network between patents: citations}{Un réseau bibliographique entre brevets : les citations}
\label{app:subsubsec:citation}

\bpar{
As with scientific publications, patents must give reference to all the previous patents which correspond to related prior art. They therefore indicate the past knowledge which relates to the patented invention. Yet, contrary to scientific citations, they also have an important legal role as they are used to delimit the scope of the property rights awarded by the patent. One can consult~\cite{oecdpatentmanual} for more details about this. Failing to refer to prior art can lead to the invalidation of the patent (e.g.~\cite{martin2015}). Another crucial difference is that the majority of the citations are actually chosen by the  examiners and not by the inventors themselves. From the USPTO, we gather information of all citations made by each patent (backward citations) and all citations received by each patent as of the end of 2013 (forward citations). We can thus build a complete network of citations that we will use later on in the analysis.
}{
De la même manière que les publications scientifiques, les brevets doivent faire référence à l'ensemble des brevet précédent qui correspondent à une connaissance antérieure nécessaire. Les citations indiquent donc la connaissance passée en relation avec l'invention brevetée. Toutefois, au contraire des publications scientifiques, elles ont aussi un important rôle légal puisqu'elles sont utilisées pour délimiter la portée des droit propriétaires donnés par le brevet. On peut consulter~\cite{oecdpatentmanual} pour plus de détail sur cette procédure. Un manque de références aux brevets passés peut mener à une invalidation du brevet~\cite{martin2015}. Une autre différence cruciale est que la majorité des citations est en fait choisie par les examinateurs et non les inventeurs eux-mêmes. A partir de l'USPTO, nous rassemblons l'information de l'ensemble des citations faites par chaque brevet (citations données) et toutes les citations reçues par chaque brevet à la fin de 2013 (citation reçues). Nous pouvons ainsi construire un réseau complet de citations qui sera utilisé par la suite dans l'analyse.
}

\bpar{
Turning to the structure of the lag between the citing and the cited patent in terms of application date, we see that the mean of this lag is 8.5 years and the median is 7 years. This distribution is highly skewed, the $95^{th}$ percentile is 21 years. We also report 164,000 citations with a negative time lag. This is due to the fact that some citations can be added during the examination process and some patents require more time to be granted than others.
}{
En se tournant vers la structure du délai entre les brevets citants et les brevets cités en termes de date de soumission, nous constatons que la moyenne de ce délai est de 8.5 ans et la médiane 7 ans. Cette distribution est fortement dissymétrique, le $95^{ème}$ centile étant 21 ans. Nous relevons également 164000 citations avec un délai négatif. Cela est du au fait que des citations peuvent être ajoutées pendant le processus de revue et que des brevets requièrent plus de temps que d'autres pour être accordés.
}

\bpar{
In what follows, we choose to restrict attention to pairs of citations with a lag no larger than 5 years. We impose this restriction for two reasons. First, the number of citations received peaks 4-5 years after application. Second, the structure of the citation lag is necessarily biased by the truncation of our sample: the more recent patents mechanically receive less citations than the older ones. As we are restricting to citations received no later than 5 years after the application date, this effect will only affect patents with an application date after 2007.
}{
Dans la suite, nous choisissons de restreindre notre attention aux paires de citations avec un délai inférieur ou égal à 5 ans. Cette restriction est imposée pour deux raisons. D'abord, le nombre de citations reçues présente un pic 4-5 ans après l'application. Ensuite, la structure du délai de citation est nécessairement biaisée par la restriction de l'échantillon : les brevets les plus récents reçoivent mécaniquement moins de citations que les plus anciens. Comme nous nous restreignons aux citations reçues en moins de 5 ans après la date de soumission, cet effet n'affectera que les brevets avec une date de soumission postérieure à 2007.
}

\subsubsection*{Data collection and basic description}{Collecte des données et description élémentaire}

\bpar{
Each patent contains an abstract and a core text which describe the invention. To see what a patent looks like in practice, one can refer to the USPTO patent full-text database \url{http://patft.uspto.gov/netahtml/PTO/index.html} or to Google patent which publishes USPTO patents in $pdf$ format at \url{https://patents.google.com}. Although including the full core texts would be natural and probably very useful in a systematic text-mining approach as done in~\cite{tseng2007text}, they are too long to be included and thus we consider only the abstracts for the analysis. Indeed, the semantic analysis counts more than 4 million patents, with corresponding abstracts with an average length of 120.8 words (and a standard deviation of $62.4$), a size that is already challenging in terms of computational burden and data size. In addition, abstracts are aimed at synthesizing purpose and content of patents and must therefore be a relevant object of study (see~\cite{Adams2010text}). The USPTO defines a guidance stating that an abstract should be ``a summary of the disclosure as contained in the description, the claims, and any drawings; the summary shall indicate the technical field to which the invention pertains and shall be drafted in a way which allows the clear understanding of the technical problem, the gist of the solution of that problem through the invention, and the principal use or uses of the invention'' (PCT Rule 8). 
}{
Chaque brevet contient un résumé et un texte qui décrit l'invention. Pour voir à quoi ressemble un brevet en pratique, on peut se référer à la base USPTO des textes complets \url{http://patft.uspto.gov/netahtml/PTO/index.html} ou à Google Patent qui publie les brevets USPTO au format $pdf$ à \url{https://patents.google.com}. Même si l'inclusion des textes complets serait naturelle et probablement très utile dans une approche d'analyse textuelle systématique comme faite dans~\cite{tseng2007text}, ceux-ci sont trop longs pour être inclus ici et nous ne considérons donc que les résumés pour l'analyse. En effet, l'analyse sémantique porte sur plus de 4 million de brevets, avec des résumés correspondants avec une longueur moyenne de $120.8$ mots (et une déviation standard de $62.4$), une taille qui est déjà un défi en termes de charge computationnelle et de taille des données. De plus, les résumés visent à synthétiser le but et le contenu des brevets et doivent pour cela être un objet d'étude pertinent (voir~\cite{Adams2010text}). L'USPTO défini une ligne directrice qui précise que le résumé doit être ``un résumé de l'information comme contenue dans la description, les positions et les figures ; le résumé doit également indiquer le champ technique dans lequel l'invention se situe et doit être rédigé d'une manière permettant une compréhension claire du problème technique, le coeur de la solution au problème apporté par l'invention et le ou les usages principaux de l'invention'' (PCT Règle 8).
}

\bpar{
We construct from raw data a unified database. Data is collected from USPTO patent redbook bulk downloads, that provides as raw data (specific \texttt{dat} or \texttt{xml} formats) full patent information, starting from 1976. Detailed procedure of data collection, parsing and consolidation are available as supplementary material of~\cite{10.1371/journal.pone.0176310}. The latest dump of the database in \texttt{Mongodb} format is available at \url{http://dx.doi.org/10.7910/DVN/BW3ACK}. Collection and homogenization of the database into a directly usable database with basic information and abstracts was an important task as USPTO raw data formats are involved and change frequently.
}{
Nous construisons à partir des données brutes une base unifiée. Les données sont récupérées à partir de la page des téléchargement du \emph{redbook} des brevets de l'USPTO, qui fournit sous format brut (formats \texttt{dat} ou \texttt{xml} spécifiques) l'ensemble des informations des brevets, à partir de 1976. La procédure détaillée de collection des données, de parsing et de consolidation sont disponible en Annexe de~\cite{10.1371/journal.pone.0176310}. L'image la plus récente de la base est disponible au format \texttt{Mongodb} à \url{http://dx.doi.org/10.7910/DVN/BW3ACK}. La collection et l'homogénéisation des données en une base directement utilisable avec les informations de base et les résumés est une étape importante, puisque des formats USPTO bruts sont impliqués et changent fréquemment.
}

\bpar{
We count 4,666,365 utility patents with an abstract granted from 1976 to 2013. A very small number of patents have a missing abstract, these are patents that have been withdrawn and we do not consider them in the analysis. The number of patents granted each year increases from around 70,000 in 1976 to about 278,000 in 2013. When distributed by the year of application, the picture is slightly different. The number of patents steadily increase from 1976 to 2000 and remains constant around 200,000 per year from 2000 to 2007. Restricting our sample to patent with application date ranging from 1976 to 2007, we are left with 3,949,615 patents. These patents cite 38,756,292 other patents with the empirical lag distribution that has been extensively analyzed in~\cite{Hall2001}. Conditioned on being cited at least once, a patent receives on average 13.5 citations within a five-year window. 270,877 patents receive no citation during the next five years following application, 10\% of patents receive only one citation and 1\% of them receive more than 100 citations. A within class citation is defined as a citation between two patents sharing at least one common technological class. Following this definition, 84\% of the citations are within class citations. 14\% of the citations are between two patents that share the exact same set of technological classes.
}{
Nous dénombrons 4,666,365 brevets d'utilité avec un résumé obtenus entre 1976 et 2013. Un très faible nombre de brevet ont un résumé manquant, ils correspondent aux brevets qui ont été annulés et nous ne les considérons pas dans l'analyse. Le nombre de brevets accordés chaque année de environ 70000 en 1976 à environ 278000 en 2013. Lorsqu'ils sont distribués par année d'application, le schéma est légèrement différent. Le nombre de brevets s'accroit à taux constant de 1976 à 2000 et reste constant à environ 200,000 par an de 2000 à 2007. En restreignant l'échantillon aux brevets dont la date de soumission est entre 1976 et 2007, il reste 3,949,615 brevets. Ces brevets en citent 38,756,292 autres avec le délai empirique qui a été étudié de manière détaillée par~\cite{Hall2001}. Conditionnellement à être cité au moins une fois, un brevet reçoit en moyenne 13.5 citations sur une fenêtre de 5 ans. 270,877 brevets ne reçoivent aucune citation pendant les 5 ans qui suivent la soumission, 10\% des brevets ne reçoivent qu'une seule citation et 1\% reçoit plus de 100 citations. Une citation interne à une classe est définie comme une citation entre deux brevets partageant au moins une classe technologique. Suivant cette définition, 84\% des citations sont internes aux classes. 14\% des citations sont entre des brevets qui partagent exactement le même ensemble de classes technologiques.
}

\subsubsection*{Towards a complementary classification}{Vers une classification complémentaire}

\bpar{
Potentialities of text-mining techniques as an alternative way to analyze and classify patents are documented in~\cite{tseng2007text}. The author's main argument, in support of an automatic classification tool for patent, is to reduce the considerable amount of human effort needed to classify all the applications. The work conducted in the field of natural language processing and/or text analysis has been developed in order to improve search performance in patent databases, build technology map or investigate the potential infringement risks prior to developing a new technology (see~\cite{abbas2014literature} for a review). Text-mining of patent documents is also widely used as a tool to build networks which carry additional information to the simplistic bibliographic connections model as argued in~\cite{yoon2004text}. As far as the authors know, the use of text-mining as a way to build a global classification of patents remains however largely unexplored. One notable exception can be found in~\cite{preschitschek2013} where semantic-based classification is shown to outperform the standard classification in predicting the convergence of technologies even in small samples. Semantic analysis reveals itself to be more flexible and more quickly adaptable to the apparition of new clusters of technologies. Indeed, as argued in~\cite{preschitschek2013}, before two distinct technologies start to clearly converge, one should expect similar words to be used in patents from both technologies.
}{
La potentialité des techniques de fouille textuelle comme un moyen alternatif pour analyser et classifier les brevets est documenté dans~\cite{tseng2007text}. L'argument principal de l'auteur, en support d'un outil de classification automatique pour les brevets, est la réduction de la quantité de travail humain considérable nécessaire pour classer toutes les soumissions. Le travail conduit dans le champ du traitement naturel du langage et/ou de l'analyse textuelle a été développé afin d'améliorer la performance des recherches dans les bases de brevets, construire des cartes des technologies ou investiguer de potentiels risque de violation de brevets avant le développement d'une nouvelle technologie (voir~\cite{abbas2014literature} pour une revue). La fouille de texte des documents de brevets est également largement utilisée comme outil pour construire des réseaux qui contiennent une information complémentaire au modèles simplistes de connection bibliographique comme argumenté dans~\cite{yoon2004text}. D'après les auteurs, l'utilisation de fouille de texte comme moyen de construction d'une classification globale des brevets reste cependant largement sous explorée. Une exception notable peut être trouvée dans~\cite{preschitschek2013} où une classification sémantique est montrée plus performante que la classification standard dans la prédiction de la convergence technologique même sur de petits échantillons. L'analyse sémantique se révèle plus flexible et plus rapidement adaptable à l'apparition de nouveaux clusters technologiques. En effet, comme décrit dans~\cite{preschitschek2013}, avant que deux technologies distinctes commencent à converger clairement, on peut s'attendre à ce que des termes similaires soient utilisés dans des brevets des deux technologies.
}

\bpar{
Finally, a semantic classification where patents are gathered based on the fact that they share similar significant keywords has the advantage of including a network feature that cannot be found in the USPC case, namely that each patent is associated with a vector of probability to belong to each of the semantic classes (more details on this feature can be found below). Using co-occurrence of keywords, it is then possible to construct a network of patents and to study the influence of some key topological features. As reviewed previously, the use of co-occurrences is the usual way to construct a semantic network. Other hybrid technique such as bipartite semantic/authors networks, do not have the nice feature of relying solely on endogenous semantic information contained in data.
}{
Enfin, une classification sémantique dans laquelle les brevets sont rassemblés en se basant sur le fait qu'ils partagent des mots-clés significatifs similaires a l'avantage d'inclure une caractéristique du réseau qui ne se retrouve pas dans le cas de l'USPC, c'est-à-dire que chaque brevet est associé à un vecteur de probabilités d'appartenir à chacune des classes sémantiques (plus de détails sur cette caractéristique sont donnés plus loin). En utilisant les co-occurrences des mots-clés, il est ensuite possible de construire un réseau de brevets et d'étudier l'influence de caractéristiques topologiques clés. Comme rappelé précédemment, l'utilisation des co-occurrences est la manière usuelle de construire un réseau sémantique. D'autres techniques hybrides comme les réseaux bipartites auteur/sémantique n'ont pas la caractéristique agréable de se baser uniquement sur l'information sémantique endogène contenue dans les données.
}

\subsection*{Semantic classification construction}{Construction de la classification sémantique}
\label{app:subsec:keywords}

\bpar{
In this section, we describe methods and empirical analysis leading to the construction of semantic network and the corresponding classification. 
}{
Dans cette section, nous décrivons la méthode et les analyses empiriques menant à la construction du réseau sémantique et la classification correspondante.
}

\subsubsection*{Keywords extraction}{Extraction des mots-clés}
\label{app:subsec:keywordsextraction}

\bpar{
Let $\mathcal{P}$ be the set of patents, we first assign to a patent $p\in \mathcal{P}$ a set of potentially significant keywords $K(p)$ from its text ${\mathcal{A}}(p)$ (that corresponds to the concatenation of its own title and abstract).
$K(p)$ are extracted through a similar procedure as the one detailed in~\cite{chavalarias2013phylomemetic}: 
}{
Soit $\mathcal{P}$ l'ensemble des brevets. Nous assignons pour commencer à un brevet $p\in \mathcal{P}$ un ensemble de mots-clés potentiellement significatifs $K(p)$ à partir de son texte ${\mathcal{A}}(p)$ (qui correspond à la concaténation de son propre titre et de son résumé). Les éléments de $K(p)$ sont extraits selon une procédure similaire à celle détaillée dans~\cite{chavalarias2013phylomemetic} :
}

\bpar{
\begin{enumerate}
\item Text parsing and Tokenization: we transform raw texts into a set of words and sentences, reading it (parsing) and splitting it into elementary entities (words organized in sentences).
\item Part-of-speech tagging: attribution of a grammatical function to each of the tokens defined previously.
\item Stem extraction: families of words are generally derived from a unique root called stem (for example \texttt{compute}, \texttt{computer}, \texttt{computation} all yield the same stem \texttt{comput}) that we extract from tokens. At this point the abstract text is reduced to a set of stems and their grammatical functions.
\item Multi-stems construction: these are the basic semantic units used in further analysis. They are constructed as groups of successive stems in a sentence which satisfies a simple grammatical function rule. The length of the group is between 1 and 3 and its elements are either nouns, attributive verbs or adjectives. We choose to extract the semantics from such nominal groups in view of the technical nature of texts, which is not likely to contain subtle nuances in combinations of verbs and nominal groups.
\end{enumerate}
}{
\begin{enumerate}
	\item Parsing du texte et mise en tokens : nous transformons les textes bruts en un ensemble de mots et phrases, qui sont lues (parsing) et séparés en entités élémentaires (mots organisés en phrases).
	\item Attribution des tags \emph{part-of-speech} : attribution d'une fonction grammaticale à chacun des tokens définis précédemment.
	\item Extraction des racines (stems) : les familles de mots sont généralement dérivées d'une unique racine appelée \emph{stem} (par exemple \texttt{compute}, \texttt{computer}, \texttt{computation} dérivent tous de la même racine \texttt{comput}) que nous extrayons des tokens. A ce point, le texte du résumé est réduit à un ensemble de \emph{stems} et leur fonctions grammaticales.
	\item Construction des multi-stems : Ce seront les unités sémantiques de base utilisées dans les analyses par la suite. Ils sont construits comme groupes de stems successifs dans une phrase, qui satisfont une règle de fonction grammaticale simple. La longueur du groupe doit être entre 1 et 3 et ses éléments être soit des noms, soit des verbes attributifs, soit des adjectifs. Nous choisissons d'extraire la sémantique à partir de tels groupes nominaux à la vue de la nature technique des textes, qui ont peu de probabilité de contenir des nuances subtiles dans les combinaisons entre verbes et groupes nominaux. 
\end{enumerate}
}

\bpar{
Text processing operations are implemented in \texttt{python} in order to use built-in functions \texttt{nltk} library~\cite{nltk} for most of above operations. This library supports most of state-of-the-art natural language processing operations. Source code is openly available on the repository of the project at \url{https://github.com/JusteRaimbault/PatentsMining}.
}{
Les opérations d'analyse textuelle sont implémentées en \texttt{python} afin d'utiliser les fonctions intégrées à la bibliothèque \texttt{nltk}~\cite{nltk} pour la majorité des opérations ci-dessus. Cette bibliothèque supporte la plupart des opérations de traitement du langage naturel au niveau de l'état de l'art. Le code source est disponible de manière ouverte sur le dépôt du projet à \url{https://github.com/JusteRaimbault/PatentsMining}.
}

\subsubsection*{Keywords relevance estimation}{Estimation de la pertinence des mots-clés}
\label{app:subsubsec:keywords_est}
\paragraph{Relevance definition}{Estimation de la pertinence}

\bpar{
Following the heuristic in~\cite{chavalarias2013phylomemetic}, we estimate relevance score in order to filter multi-stem. The choice of the total number of keywords to be extracted, which we shall denote $K_w$, is important, too small a value would yield similar network structures but including less information whereas very large values tend to include too many irrelevant keywords. We choose to set this parameter to $K_w = 100,000$. We first consider the filtration of $k\cdot K_w$ (with $k=4$) to keep a large set of potential keywords but still have a reasonable number of co-occurrences to be computed. This step has only very marginal effects on the nature of the final keywords but is necessary for computational purposes. The filtration is done on the \emph{unithood} $u_i$, defined for keyword $i$ as $u_i = f_i\cdot \log{(1 + l_i)}$ where $f_i$ is the multi-stem's number of apparitions over the whole corpus and $l_i$ its length in words. A second filtration of $K_w$ keywords is done on the \emph{termhood} $t_i$, where the formal definition can be found in Eq.~\ref{app:eq:termhood}. It is computed as a chi-squared score on the distribution of the stem's co-occurrences and then compared to a uniform distribution within the whole corpus. Intuitively, uniformly distributed terms will be identified as plain language and they are thus not relevant for the classification. More precisely, we compute the co-occurrence matrix $(M_{ij})$, where $M_{ij}$ is defined as the number of patents where stems $i$ and $j$ appear together. The \emph{termhood} score $t_i$ is defined as
}{
Suivant l'heuristique de~\cite{chavalarias2013phylomemetic}, nous estimons un score de pertinence pour filtrer les multi-stems. Le choix du nombre total de mots-clés extraits, que nous notons $K_w$ est important, puisque des valeurs trop faibles donnent des structures de réseau similaires mais qui contiennent moins d'information, tandis que des très grandes valeurs tendent à inclure trop de mots-clés non pertinents. Nous choisissons de fixer ce paramètre à $K_w = 100,000$. Nous considérons pour commencer une filtration de $k\cdot K_w$ (avec $k=4$) mots, pour garder un grand nombre de mots-clés potentiels mais calculer un nombre raisonnable de co-occurrences. Cette étape n'a que des effets marginaux sur la nature des mots-clés finaux mais est nécessaire pour des raisons computationnelles. La filtration est faite sur la mesure \emph{unithood} $u_i$, définie pour le mot-clé $i$ par $u_i = f_i\cdot \log{(1 + l_i)}$ où $f_i$ est le nombre d'apparitions du multi-stem dans l'ensemble du corpus et $l_i$ sa longueur en nombre de mots. Une seconde filtration de $K_w$ mots-clés est faite sur la \emph{termhood} $t_i$, dont la définition formelle est donnée en Eq.~\ref{app:eq:termhood}. Elle est calculée comme un score du chi-deux sur la distribution des co-occurrences du stem et comparée à une distribution uniforme sur l'ensemble du corpus. Intuitivement, les termes distribués uniformément seront identifiés comme du langage courant, et ne sont donc par pertinents pour la classification. Plus précisément, nous calculons la matrice de co-occurrence $(M_{ij})$, où $M_{ij}$ est défini comme le nombre de brevets où les stems $i$ et $j$ apparaissent simultanément. Le score de \emph{termhood} $t_i$ est alors défini par
}

\begin{eqnarray}
\label{app:eq:termhood}
t_i = \sum_{j\neq i}\frac{\left( M_{ij} - \sum_{k}M_{ik} \sum_{k} M_{jk}\right)^2}{\sum_{k}M_{ik} \sum_{k} M_{jk}}.
\end{eqnarray}

\paragraph{Moving window estimation}{Estimation sur fenêtre glissante}

\bpar{
The previous scores are estimated on a moving window with fixed time length following the idea that the present relevance is given by the most recent context and thus that the influence vanishes when going further into the past. Consequently, the co-occurrence matrix is chosen to be constructed at year $t$ restricting to patent which applied during the time window $\big[ t - T_0 ; t \big]$. Note that the causal property of the window is crucial as the future cannot play any role in the current state of keywords and patents. This way, we will obtain semantic classes which are exploitable on a $T_0$ time span. For example, this enables us to compute the modularity of classes in the citation network as below. In the following, we take $T_0 = 4$ (which corresponds to a five year window) consistently with the choice of maximum time lag for citations made in Section \ref{sub:citation}. Accordingly, the sensitivity analysis for $T_0=2$ can be found in Appendix of~\cite{10.1371/journal.pone.0176310}.
}{
Les scores précédents sont estimés sur une fenêtre glissante de longueur temporelle fixe, suivant l'idée que la pertinence présente est donnée par le contexte le plus récent et que l'influence diminue lorsqu'on remonte dans le passé. Par conséquent, la matrice de co-occurrence est construite pour l'année $t$ en se restreignant aux brevets qui ont été soumis pendant la fenêtre temporelle $\big[ t - T_0 ; t \big]$. Notons que la propriété causale de la fenêtre est essentielle puisque le futur ne peut jouer aucun rôle dans l'état courant des mots-clés et des brevets. De cette manière, nous obtiendrons des classes sémantiques exploitables sur une durée $T_0$. Par exemple, cela nous permettra plus loin de calculer la modularité des classes dans le réseau de citations. Dans la suite, nous prenons $T_0 = 4$ (ce qui correspond à une fenêtre temporelle de 5 ans), en cohérence avec le choix du délai maximal pour les citations fait précédemment. Nous présentons l'analyse de sensibilité effectuée pour $T_0 = 2$ en Annexe de~\cite{10.1371/journal.pone.0176310}.
}

\subsubsection*{Construction of the semantic network}{Construction du réseau sémantique}
\label{app:subsubsec:construction}

\bpar{
We keep the set of most relevant keywords $\mathcal{K}_W$ and obtain their co-occurrence matrix as defined previously. This matrix can be directly interpreted as the weighted adjacency matrix of the semantic network. At this stage, the topology of raw networks does not allow the extraction of clear communities. This is partly due to the presence of hubs that correspond to frequent terms common to many fields (e.g. \texttt{method}, \texttt{apparat}) which are wrongly filtered as relevant. We therefore introduce an additional measure to correct the network topology: the concentration of keywords across technological classes, defined as: 
}{
Nous conservons l'ensemble des $\mathcal{K}_W$ mots-clés les plus pertinents et obtenons leur matrice de co-occurrence comme défini précédemment. Cette matrice peut directement être interprétée comme une matrice d'adjacence pondérée du réseau sémantique. A cette étape, la topologie du réseau brut ne permet pas l'extraction de communautés claires. Cela est partiellement dû à la présence de hubs qui correspondent à des termes fréquents communs à de nombreux champs (par exemple \texttt{method}, \texttt{apparat}) qui sont considérés pertinents mais sont des faux positifs. Nous introduisons donc une mesure supplémentaire pour corriger la topologie du réseau : la concentration des mots-clés au sein des classes technologiques, définie par
}

$$c_{tech}(s) = \displaystyle \sum_{j=1}^{N^{(tec)}} \frac{k_j(s)^2}{ \left(\sum_i k_i(s)\right)^2},$$  

\bpar{
where $k_j(s)$ is the number of occurrences of the $s$th keyword in each of the $j$th technological class taken from one of the $N^{(tec)}$ USPC classes. The higher $c_{tech}$, the more specific to a technological class the node is. For example, the terms \texttt{semiconductor} is widely used in electronics and does not contain any significant information in this field. We use a threshold parameter, defined as $\theta_c$, and keep nodes with $c_{tech}(s) > \theta_c$. Likewise, edges with low weights correspond to rare co-occurrences and are considered to be noise. To account for this we define the threshold parameter for edges $\theta_w$, and we filter edges with a weight below $\theta_w$, following the rationale that two keywords are not linked ``by chance'' if they appear simultaneously a minimal number of time. To control for size effect, we normalize by taking $\theta_w = \theta_w^{(0)}\cdot N_P$ where $N_P$ is the number of patents in the corpus ($N_P = \left|\mathcal{P} \right|$). $\theta_w^{(0)}$ is thus a varying parameter interpreted as a noise threshold \emph{per patent}. Communities are then extracted using a standard modularity maximization procedure as described in~\cite{clauset2004finding} to which we add the two constraints captured by $\theta_w$ and $\theta_c$, namely that edges must have a weight greater than $\theta_w$ and nodes a concentration greater than $\theta_c$. At this stage, both parameters $\theta_c$ and $\theta_w^{(0)}$ are unconstrained and their choice is not straightforward. Indeed, many optimization objectives are possible, such as the modularity, network size or number of communities. We find that modularity is maximized at a roughly stable value of $\theta_w$ across different $\theta_c$ for each year, corresponding to a stable $\theta_w^{(0)}$ across years, which leads us to choose $\theta_w^{(0)} = 4.1\cdot 10^{-5}$. Then for the choice of $\theta_c$, different candidates points lie on a Pareto front for the bi-objective optimization on number of communities and network size. There is a priori no reason to choose any specific point among the different optimums. Consequently, we have tried the analysis with all the candidate values for $\theta_c$ and found that the results are the most reasonable when taking $\theta_c = 0.06$ (see Fig.~{\ref{fig:app:patentsmining:networksensitivity}}). We show in Fig.~\ref{fig:app:patentsmining:rawnetwork} an example of semantic network visualization.
}{
où $k_j(s)$ est le nombre d'occurrences du mot-clé $s$ dans la classe technologique $j$ prise parmi les $N^{(tec)}$ classes USPC. Plus $c_{tech}$ est grand, plus le noeud est spécifique à une classe technologique. Par exemple, le terme \texttt{semiconductor} est largement utilisé en électronique et ne contient pas d'information significative au regard de ce champ. Nous utilisons un paramètre de seuil, défini comme $\theta_c$, et conservons les noeuds tels que $c_{tech}(s) > \theta_c$. De la même manière, les liens avec des poids faibles correspondent à des co-occurrences rares et sont considérées comme du bruit. Pour prendre cela en compte, nous définissons le paramètre $\theta_w$ pour les liens, et nous filtrons les liens qui ont un poids en dessous de $\theta_w$, suivant la logique que deux mots-clés ne sont pas connectés ``par chance'' s'ils apparaissent simultanément un nombre minimal de fois. Pour contrôler l'effet de taille, nous normalisons en prenant $\theta_w = \theta_w^{(0)}\cdot N_P$ où $N_P$ est le nombre de brevets dans le corpus ($N_P = \left|\mathcal{P} \right|$). $\theta_w^{(0)}$ est ainsi un paramètre variable qui peut s'interpréter comme un seuil de bruit \emph{par brevet}. Les communautés sont ensuite extraites par l'utilisation d'une procédure standard de maximisation de modularité comme décrite par~\cite{clauset2004finding}, à laquelle nous ajoutons les deux contraintes capturées par $\theta_w$ et $\theta_c$, c'est-à-dire que les liens doivent avoir un poids plus grand que $\theta_w$ et les noeuds une concentration plus grande que $\theta_c$. A cette étape, les deux paramètres $\theta_c$ et $\theta_w^{(0)}$ ne sont pas contraints et leur choix n'est pas direct. En effet, différents objectifs d'optimisation sont possibles, comme la modularité, la taille du réseau ou le nombre de communautés. Nous trouvons que la modularité est maximisée à une valeur relativement stable de $\theta_w$ pour les différentes valeurs de $\theta_c$ pour chaque année, ce qui correspond à une valeur stable de $\theta_w^{(0)}$ au cours des années, et ce qui amène le choix $\theta_w^{(0)} = 4.1\cdot 10^{-5}$. Ensuite, pour le choix de $\theta_c$, différents points candidats se trouvent sur un front de Pareto pour l'optimisation bi-objectif sur le nombre de communautés et la taille du réseau. Il n'y a a priori pas de raisons de choisir un point spécifique entre les différents optima. Par conséquent, nous avons testé l'analyse pour toutes les valeurs candidates pour $\theta_c$ et trouvé que les résultats sont les plus raisonnables avec $\theta_c = 0.06$ (voir Fig.~\ref{fig:app:patentsmining:networksensitivity}). Nous montrons en Fig.~\ref{fig:app:patentsmining:rawnetwork} un exemple de visualisation du réseau sémantique.
}

\begin{figure}
\includegraphics[width=\linewidth]{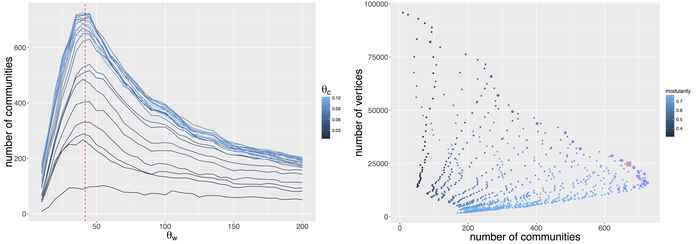}
\appcaption{\textbf{Sensitivity analysis of network community structure to filtering parameters.} We consider a specific window 2000-2004 and the obtained plots are typical. \textit{(Left panel)} We plot the number of communities as a function of the edge threshold parameter $\theta_w$ for different values of the node threshold parameter $\theta_c$. The maximum is roughly stable across $\theta_c$ (dashed red line). \textit{(Right panel)} To choose $\theta_c$, we do a Pareto optimization on communities and network size: the compromise point (red overline) on the Pareto front (purple overline: possible choices after having fixed $\theta_w^{(0)}$; blue level gives modularity) corresponds to $\theta_c = 0.06$.\label{fig:app:patentsmining:networksensitivity}}{\textbf{Analyse de sensibilité de la structure des communautés du réseau aux paramètres de filtrage.} Nous considérons une fenêtre spécifique 2000-2004. Les graphes obtenus sont typiques. \textit{(Gauche)} Le graphe donne le nombre de communautés en fonction du paramètre de seuil pour les liens $\theta_w$ pour différentes valeurs du paramètre de seuil des noeuds $\theta_c$. Le maximum est globalement stable pour les différents $\theta_c$ (ligne rouge pointillée). \textit{(Droite)} Pour choisir $\theta_c$, nous faisons une optimisation de Pareto sur les communautés et la taille du réseau : le point de compromis (surligné en rouge) sur le front de Pareto (surligné en violet : les choix possibles après avoir fixé $\theta_w^{(0)}$ ; le niveau de bleu donne la modularité) correspond à $\theta_c = 0.06$. \label{fig:app:patentsmining:networksensitivity}}
\end{figure}

\begin{figure}
\includegraphics[width=\linewidth]{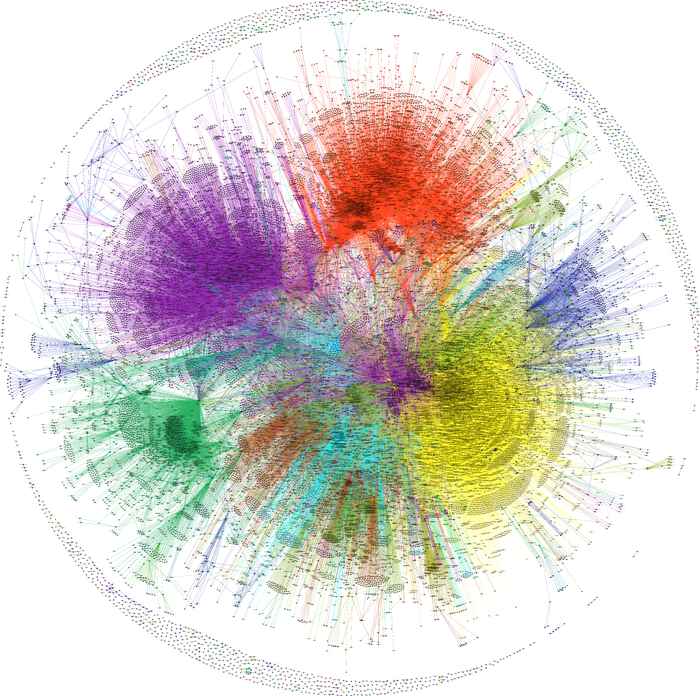}
\appcaption{\textbf{An example of semantic network visualization.} We show the network obtained for the window 2000-2004, with parameters $\theta_c = 0.06$ and $\theta_w = \theta_w^{(0)}\cdot N_P = 4.5e^{-5} \cdot 9.1e^{5}$. The corresponding file in a vector format (\texttt{.svg}), that can be zoomed and explored, is available as supplementary material of~\cite{10.1371/journal.pone.0176310}.\label{fig:app:patentsmining:rawnetwork}}{\textbf{Un exemple de visualisation du réseau sémantique.} Nous montrons le réseau obtenu pour la fenêtre 2000-2004, avec les paramètres $\theta_c = 0.06$ et $\theta_w = \theta_w^{(0)}\cdot N_P = 4.5e^{-5} \cdot 9.1e^{5}$. Le fichier correspondant sous un format vectoriel (\texttt{.svg}), qui peut être zoomé et exploré, est disponible en Annexe de~\cite{10.1371/journal.pone.0176310}.\label{fig:app:patentsmining:rawnetwork}}
\end{figure}

\subsubsection*{Characteristics of semantic classes}{Caractéristiques des classes sémantiques}
\label{app:subsubsec:characteristics}

\bpar{
For each year $t$, we define as $N^{(sem)}_t$ the number of semantic classes which have been computed by clustering keywords from patents appeared during the period $\big[ t-T_0, t \big]$ (we recall that we have chosen $T_0=4$). Each semantic class $k =  1, \cdots, N^{(sem)}_t$ is characterized by a set of keywords $K(k,t)$ which is a subset of $\mathcal{K}_W$ selected as described in previous sections. The cardinal of $K(k, t)$ distribution across each semantic class $k$ is highly skewed with a few semantic classes containing over $1,000$ keywords, most of them with roughly the same number of keywords. In contrast, there are also many semantic classes with only two keywords. There are around 30 keywords by semantic class on average and the median is 2 for any $t$. Fig.~\ref{fig:app:patentsmining:mean_K} shows that the average number of keywords is relatively stable from 1976 to 1992 and then picks around 1996 prior to going down.
}{
Pour chaque année $t$, nous définissons comme $N^{(sem)}_t$ le nombre de classes sémantiques qui ont été calculées par la detection de communautés pour les brevets soumis sur la période $\big[ t-T_0, t \big]$ (nous rappelons que nous avons pris $T_0=4$). Chaque classe sémantique $k =  1, \cdots, N^{(sem)}_t$ est caractérisée par un ensemble de mots-clés $K(k,t)$ qui est un sous-ensemble de $\mathcal{K}_W$ sélectionné comme décrit dans les précédentes sections. La distribution des cardinaux de $K(k, t)$ pour l'ensemble des classes sémantiques $k$ a une très longue queue avec quelques classes sémantiques contenant plus de 1000 mots-clés, la plupart avec approximativement le même nombre de mots-clés. A l'opposé, il existe de nombreuses classes qui ne contiennent que deux mots. Il y a en moyenne 30 mots-clés par classe sémantique et la médiane est 2 pour tout $t$. La Fig.~\ref{fig:app:patentsmining:mean_K} montre que le nombre moyen de mots-clés est relativement stable de 1976 à 1992 et ensuite présente un pic atour de 1996 avant de redescendre à nouveau.
}

\begin{figure}
\includegraphics[width=\linewidth]{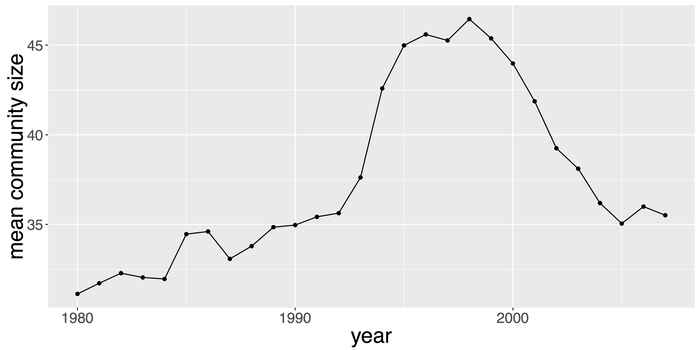}
\appcaption{\textbf{This figure plots the average number of keywords by semantic class for each time window $\left[t-4; t\right]$} from $t=1980$ to $t=2007$.\label{fig:app:patentsmining:mean_K}}{\textbf{Cette figure montre le nombre moyen de mots-clés par classe sémantique pour chaque fenêtre $\left[t-4; t\right]$} de $t=1980$ à $t=2007$.\label{fig:app:patentsmining:mean_K}}
\end{figure}

\paragraph*{Title of semantic classes}{Titre des classes sémantiques}

\bpar{
USPC technological classes are defined by a title and a highly accurate definition which help retrieve patents easily. The title can be a single word (e.g.: class 101: ``Printing'') or more complex (e.g.: class 218: ``High-voltage switches with arc preventing or extinguishing devices''). As our goal is to release a comprehensive database in which each patent is associated with a set of semantic classes, it is necessary to give an insight on what these classes represent by associating a short description or a title as in~\cite{tseng2007text}. In our case, such description is taken as a subset of keywords taken from $K(k,t)$. For the vast majority of semantic classes that have less than 5 keywords, we decide to keep all of theses keywords as a description. For the remaining classes which feature around 50 keywords on average, we rely on the topological properties of the semantic network. \cite{yang2000improving} suggest to retain only the most frequently used terms in $K(k,t)$. Another possibility is to select 5 keywords based on their network centrality with the idea that very central keywords are the best candidates to describe the overall idea captured by a community. For example, the largest semantic class in 2003-2007 is characterized by the keywords: \texttt{Support Packet; Tree Network; Network Wide; Voic Stream; Code Symbol Reader}.
}{
Les classes technologiques USPC sont définies par un titre et une définition très précise qui aide à retrouver les brevets facilement. Le titre peut être un simple mot (par exemple, la classe 101 : ``Printing'') ou plus compliqué (par exemple, classe 218 : ``High-voltage switches with arc preventing or extinguishing devices''). Notre but étant de produire une base exhaustive dans laquelle chaque brevet est associé à un ensemble de classes sémantiques, il est nécessaire de donner un aperçu de ce que ces classes représentent en leur associant une courte description ou un titre comme dans~\cite{tseng2007text}. Dans notre cas, cette description est prise comme un sous-ensemble de mots-clés pris dans $K(k,t)$. Pour la grande majorité des classes sémantiques qui ont moins de 5 mots-clés, nous décidons de garder l'ensemble de ces mots-clés comme description. Pour les classes restantes qui ont en moyenne une cinquantaine de mots-clés, nous nous reposons sur les propriétés topologiques du réseau sémantique. \cite{yang2000improving} suggère de ne garder que les termes les plus fréquents dans $K(k,t)$. Une autre possibilité est de sélectionner 5 mots-clés en se basant sur leur centralité dans le réseau, en suivant l'idée que les mots-clés très centraux sont les meilleurs candidats pour décrire le thème général capturé par une communauté. Par exemple, la plus grande classe sémantique en 2003-2007 est caractérisée par les mots-clés : \texttt{Support Packet; Tree Network; Network Wide; Voic Stream; Code Symbol Reader}.
}

\paragraph*{Size of technological and semantic classes}{Tailles des classes sémantiques et technologiques}

\bpar{
We consider a specific window of observations (for example 2000-2004), and we define $Z$ the number of patents which appeared during that time window. For each patent $i=1, \cdots, Z$ we associate a vector of probability where each component $p_{ij}^{(sem)} \in \big[ 0,1 \big]$, with  $j = 1, \cdots, N{(sem)}$ and where
$$\displaystyle \sum_{j=1}^{N^{(sem)}} p_{ij}^{(sem)} = 1$$ (when there is no room for confusion, we drop the subscript $t$ in $N_t^{(sem)}$). On average across all time windows, a patent is associated to 1.8 semantic classes with a positive probability. Next we define the size of a semantic class as $$S_j^{(sem)} = \displaystyle \sum_{i=1}^Z p_{ij}^{(sem)}.$$ 
Correspondingly, we aim to provide a consistent definition for technological classes. For that purpose, we follow the so-called ``fractional count'' method, which was introduced by the USPTO and consists in dividing equally the patents between all the classes they belong to. Formally, we define the number of technological classes as
$N^{(tec)}$  (which is not time dependent contrary to the semantic case) and for $j = 1, \cdots, N^{(tec)}$ the corresponding matrix of probability is defined as
\[
 p_{ij}^{(tec)} = \frac{B_{ij}}{\displaystyle \sum_{k=1}^{N^{(tec)}}{B_{ik}}},
\]
where $B_{ij}$ equals $1$ if the $i$th patent belongs to the $j$th technological class and $0$ if not. When there is no room for confusion, we will drop the exponent part and write only $p_{ij}$ when referring to either the technological or semantic matrix. Empirically, we find that both classes exhibit a similar hierarchical structure in the sense of a power-law type of distribution of class sizes as shown in Fig.~\ref{fig:patentsmining:class-sizes}. This feature is important, it suggests that a classification based on the text content of patents has some separating power in the sense that it does not divide up all the patents in one or two communities. 
}{
Nous considérons une fenêtre spécifique d'observations (par exemple 2000-2004), et nous définisson $Z$ le nombre de brevets qui apparaissent dans cette fenêtre temporelle. Pour chaque brevet $i=1, \cdots, Z$, nous lui associons un vecteur de probabilités où chaque composante $p_{ij}^{(sem)} \in \big[ 0,1 \big]$, avec $j = 1, \cdots, N{(sem)}$ et où $$\displaystyle \sum_{j=1}^{N^{(sem)}} p_{ij}^{(sem)} = 1$$ (lorsqu'il n'y a pas risque de confusion, nous oublions l'indice $t$ dans $N_t^{(sem)}$). En moyenne sur l'ensemble des fenêtres temporelles, un brevet est associé à 1.8 classes sémantiques avec une probabilité strictement positive. Nous définissons alors la taille d'une classe sémantique comme $$S_j^{(sem)} = \displaystyle \sum_{i=1}^Z p_{ij}^{(sem)}.$$ De manière correspondante, nous proposons d'introduire une définition cohérente pour les classes technologiques. Pour cela, nous suivons la méthode dite du ``compte fractionnel'', qui a été introduite par l'USPTO et consiste en la division égale des brevets entre l'ensemble des classes auxquelles ils appartiennent. Formellement, nous définissons le nombre de classes technologiques comme $N^{(tec)}$ (qui ne dépend pas du temps contrairement au cas sémantique) et pour $j = 1, \cdots, N^{(tec)}$ la matrice de probabilité correspondante définie par
\[
 p_{ij}^{(tec)} = \frac{B_{ij}}{\displaystyle \sum_{k=1}^{N^{(tec)}}{B_{ik}}},
\]
où $B_{ij}$ est égal à 1 si le $i$ème brevet appartient à la $j$ème classe technologique et 0 sinon. Quand il n'y a pas risque de confusion, nous oublions l'exposant et écrivons uniquement $p_{ij}$ pour référer soit à la matrice sémantique soit à la matrice technologique. Empiriquement, nous obtenons que les deux classifications présentent une structure hiérarchique similaire, ayant toutes les deux une distribution de type loi puissance pour la taille des classes comme montré en Fig.~\ref{fig:patentsmining:class-sizes}. Cette caractéristique est importante, puisqu'elle suggère qu'une classification basée sur le contenu sémantique des brevets a un certain pouvoir de séparation au sens qu'elle ne divise pas les brevets en uniquement une ou deux communautés.
}

\begin{figure}
\includegraphics[width=\linewidth]{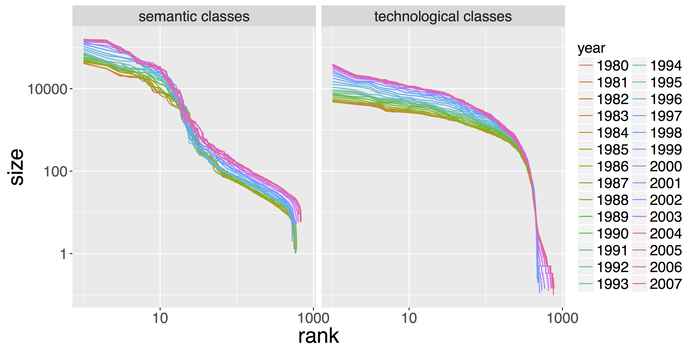}
\appcaption{\textbf{Sizes of classes.} Yearly from $t = 1980$ to $t =2007$, we plot the size of semantic classes (left-side) and technological classes (right-side) for the corresponding time window $[t-4, t]$, from the biggest to the smallest. The formal definition of size can be found in section~\ref{app:subsubsec:characteristics}. Each color corresponds to one specific year. Yearly semantic classes and technological classes present a similar hierarchical structure which confirms the comparability of the two classifications. Over time, curves are translated and levels of hierarchy stays roughly constant.\label{fig:patentsmining:class-sizes}}{\textbf{Tailles des classes.} Pour chaque année de $t = 1980$ à $t =2007$, nous montrons la taille des classes sémantiques \textit{(Gauche)} et des classes technologiques \textit{(Droite)} sur la fenêtre temporelle correspondante $[t-4, t]$, de la plus grande à la plus petite. La définition formelle des tailles est donnée en section~\ref{app:subsubsec:characteristics}. Chaque couleur correspond à une année donnée. Les classes sémantique annuelles et les classes technologiques présente une structure hiérarchique similaire, ce qui confirme la comparabilité des deux classifications. Dans le temps, les courbes sont translatées et le niveau de hiérarchie reste globalement constant.\label{fig:patentsmining:class-sizes}}
\end{figure}

\subsubsection*{Potential refinements of the method}{Extensions possibles de la méthode}

\bpar{
Our semantic classification method could be refined by combining it with other techniques such as Latent Dirichlet Allocation which is a widely used topic detection method (e.g.~\cite{blei2003latent}), already used on patent data as in~\cite{kaplan2015double} where it provides a measure of idea novelty and the counter-intuitive stylized facts that breakthrough invention are likely to come out of local search in a field rather than distant technological recombination. Using this approach should first help further evaluate the robustness of our qualitative conclusions (external validation). Also, depending on the level of orthogonality with our classification, it can potentially bring an additional feature to characterize patents, in the spirit of multi-modeling techniques where neighbor models are combined to take advantage of each point of view on a system.
}{
Notre classification sémantique pourrait être améliorée en la combinant à d'autres techniques comme l'Allocation de Dirichlet Latente qui est une méthode répandue en détection de thèmes (voir par exemple \cite{blei2003latent}), déjà utilisée sur les données de brevets comme dans \cite{kaplan2015double} où elle fournit une mesure de la nouveauté d'une idée et le fait stylisé contre-intuitif que les innovations de rupture ont plus de chance d'émerger de recherches locales dans un champ plutôt que de recombinaisons de technologies distantes. L'utilisation de cette approche devrait dans un premier temps permettre d'évaluer plus précisément la robustesse de nos conclusions qualitatives (validation externe). Selon le niveau d'orthogonalité avec notre classification, cela pourrait également potentiellement constituer une caractéristique supplémentaire pour la caractérisation des brevets, dans l'esprit de la multi-modélisation où les modèles voisins sont combinés pour tirer parti de l'ensemble des points de vue sur le système.
}

\bpar{
Our use of network analysis can also be extended using newly developed techniques of hyper-network analysis. Indeed, patents and keywords can for example be nodes of a bipartite network, or patents be links of an hyper-network, in the sense of multiple layers with different classification links and citation links. The combination of citation network modeling by Stochastic Block Modeling with topic modeling was studied for scientific papers by \cite{zhu2013scalable}, outperforming previous link prediction algorithms. \cite{iacovacci2015mesoscopic} provide a method to compare macroscopic structures of the different layers in a multilayer network that could be applied as a refinement of the overlap, modularity and statistical modeling studied in this paper. Furthermore, is has recently been shown that measures of multilayer network projections induce a significant loss of information compared to the generalized corresponding measure~\cite{de2015ranking}, which confirms the relevance of such development that we left for further research.
}{
Notre utilisation d'analyse de réseau pourrait aussi être étendue en utilisant des techniques d'analyse par hyper-réseau récemment développées. En effet, les brevets et les mots clés peuvent par exemple être les noeuds d'un réseau bipartite, ou les brevets être les liens d'un hyper-réseau, au sens de multiples couches avec différents liens de classification et de citation. La combinaison de la modélisation du réseau de citations par de la Modélisation Stochastique par Blocs, à la modélisation des thèmes, a été étudiée dans le cas des articles scientifiques par~\cite{zhu2013scalable}, ayant une performance supérieure au algorithmes de prédiction de liens précédents. \cite{iacovacci2015mesoscopic} propose une méthode pour comparer les structures macroscopiques des différentes couches d'un réseau multi-couches, qui pourrait être appliqué comme un raffinement des analyses de l'intersection et de la modularité que nous faisons ici. De plus, il a récemment été montré que les mesures calculées sur des projections des réseaux multi-couches induisent une perte d'information non-négligeable en comparaison au mesures correspondantes généralisées~\cite{de2015ranking}, ce qui confirme la pertinence de tels développements que nous laissons pour des recherches futures.
}

\bpar{
An other potential research development would be to further exploit the temporal structure of our dataset. Indeed, large progress have recently been made in complex network analysis of time-series data (see~\cite{gao2017complex} for a review). For example, \cite{gao2015multiscale} develops a method to construct multiscale network from time series, which could in our case be a solution to identify structures in patents trajectories at different levels, and be an alternative to the single scale modularity analysis we use.
}{
Un autre développement de recherche potentiel serait une exploitation approfondie de la nature temporelle du jeu de données. En effet, des progrès conséquents on récemment été faits en analyse de réseau de données de séries temporelles (voir~\cite{gao2017complex} pour une revue). Par exemple, \cite{gao2015multiscale} développe une méthode pour construire des réseaux multi-échelles à partir de séries temporelles, ce qui pourrait dans notre cas être une solution pour identifier des structures dans les trajectoires des brevets à différents niveaux, et être une alternative à l'analyse de modularité à une seule échelle que nous utilisons.
}

\subsection*{Results}{Résultats}
\label{app:subsec:result}

\bpar{
In this section, we present some key features of our resulting semantic classification showing both complementary and differences with the technological classification. We first present several measures derived from this semantic classification at the patent level: diversity, originality, generality and overlapping for the classes. We then show that the two classifications show highly different topological measures.
}{
Dans cette partie, nous présentons les caractéristiques principales de notre classification sémantique obtenue, qui présente à la fois une complémentarité et des différences avec la classification technologique. Nous présentons d'abord diverses mesures dérivées de cette classification sémantique au niveau du brevet : diversité, originalité, généralité, et recouvrement pour les classes. Nous montrons ensuite des les deux classifications présentent des mesures topologiques fondamentalement différentes.
}

\subsubsection*{Patent level measures}{Mesures pour les brevets}
\label{app:subsubsec:orig-gene}

\bpar{
Given a classification system (technological or semantic classes), and the associated probabilities $p_{ij}$ for each patent $i$ to belong to class $j$ (that were defined previously), one can define a patent-level diversity measure as one minus the Herfindhal concentration index on $p_{ij}$ by
}{
Etant donné un système de classification (classes sémantiques ou technologiques), et les probabilités associées $p_{ij}$ pour chaque brevet $i$ d'appartenir à la classe $j$ (comme définies précédemment), on peut définir une mesure de diversité au niveau du brevet comme le complémentaire dans un de l'indice de concentration de Herfindhal sur les $p_{ij}$ par
}

\[
D_i^{(z)} = 1 - \sum_{j =1}^{N^{(z)}} {p_{ij}^2}, \text{ with } z \in \{tec, sem\}.
\]

\bpar{
We show in Fig.~\ref{fig:patentsmining:patent-level-orig} the distribution over time of semantic and technological diversity with the corresponding mean time-series. This is carried with two different settings, namely including/not including patents with zero diversity (i.e. single class patents). We call other patents ``complicated patents'' in the following. First of all, the presence of mass in small probabilities for semantic but not technological diversity confirms that the semantic classification contains patent spread over a larger number of classes. More interestingly, a general decrease of diversity for complicated patents, both for semantic and technological classification systems, can be interpreted as an increase in invention specialization. This is a well-known stylized fact as documented in~\cite{ARCHIBUGI199279}. Furthermore, a qualitative regime shift on semantic classification occurs around 1996. This can be seen whether or not we include patents with zero diversity. The diversity of complicated patents stabilizes after a constant decrease, and the overall diversity begins to strongly decrease. This means that on the one hand the number of single class patents begins to increase and on the other hand complicated patents do not change in diversity. It can be interpreted as a change in the regime of specialization, the new regime being caused by more single-class patents.
}{
Nous montrons en Fig.~\ref{fig:patentsmining:patent-level-orig} la distribution dans le temps de la diversité sémantique et technologique avec les séries temporelles correspondantes pour les moyennes. L'analyse est faite pour deux configurations différentes, c'est-à-dire en incluant ou non les brevets avec diversité nulle (les brevets n'appartenant qu'à une seule classe). Nous désignons les autres brevets comme ``brevets compliqués'' par la suite. Tout d'abord, la présence de masse dans les faibles probabilités pour la diversité sémantique mais non celle technologique confirme que la classification sémantique contient des brevet qui recouvrent un grand nombre de classes. D'autre part, une décroissance générale de la diversité pour les brevets compliqués, à la fois pour les systèmes de classification sémantique et technologique, peut être interprété comme une augmentation de la spécialisation des inventions. Il s'agit d'un fait stylisé bien connu comme documenté dans~\cite{ARCHIBUGI199279}. De plus, un changement de régime qualitatif sur la classification s'opère autour de 1996. Celui-ci s'observe que l'on inclue ou non les brevets à diversité nulle. La diversité des brevets compliqués se stabilise après une décroissance constante, et la diversité globale commence à décroitre fortement. Cela signifie d'une part que le nombre de brevet n'ayant qu'une seule classe commence à augmenter et d'autre part les brevets compliqués ne changent pas en diversité. Ce phénomène peut s'interpréter comme un changement dans le régime de spécialisation, le nouveau régime étant causé par une augmentation du nombre de brevets à classe unique.
}

\bpar{
More commonly used in the literature are the measures of originality and generality. These measures follow the same idea than the above-defined diversity in quantifying the diversity of classes (whether technological or semantic) associated with a patent. But instead of looking at the patent's classes, they consider the classes of the patents that are cited or citing. Formally, the originality $O_i$ and the generality $G_i$ of a patent $i$ are defined as
}{
Plus classiques dans la littérature sont les mesures d'originalité et de généralité. Ces mesures correspondent à la même idée que la diversité définie précédemment pour quantifier la diversité des classes (qu'elles soient technologiques ou sémantiques) associées à un brevet. Mais plutôt que de considérer les classes du brevet, celles-ci considèrent les classes des brevets qui sont cités ou citants. Formellement, l'originalité $O_i$ et la généralité $G_i$ d'un brevet $i$ sont définis par
}

\[
O_i^{(z)} = \displaystyle 1 - \sum_{j =1}^{N^{(z)}}{\left(\frac{\displaystyle \sum_{i' \in I_i}{p_{i'j}}}{\displaystyle \sum_{k =1}^{N^{(z)}}{\displaystyle \sum_{i' \in I_i}{p_{i'k}}}}\right)^2} \text{ and } G_i^{(z)} = \displaystyle 1 - \sum_{j =1}^{N^{(z)}}{\left(\frac{\displaystyle \sum_{i' \in \tilde{I}_i}{p_{i'j}}}{\displaystyle \sum_{k =1}^{N^{(z)}}{\displaystyle \sum_{i' \in \tilde{I}_i}{p_{i'k}}}}\right)^2}, 
\]

\bpar{
where $z \in \{tec, sem\}$, $I_i$ denotes the set of patents that are cited by the $i$th patent within a five year window (i.e. if the $i$th patent appears at year $t$, then we consider patents on $[t-T_0, t]$) when considering the originality and $\tilde{I}_i$ the set of patents that cite patent $i$ after less than five years (i.e. we consider patents on $[t ,t + T_0]$) in the case of generality. Note that the measure of generality is forward looking in the sense that $G_i^{(z)}$ used information that will only be available 5 years after patent applications. Both measures are lower on average based on semantic classification than on technological classification. Fig.~\ref{fig:patentsmining:orig-gene} plots the mean value of $O_i^{(sem)}$, $O_i^{(tec)}$, $G_i^{(sem)}$ and $G_i^{(tec)}$.
}{
où $z \in \{tec, sem\}$, $I_i$ dénote l'ensemble des brevet qui sont cités par le $i$ème brevet sur une fenêtre temporelle de cinq ans (i.e. si le $i$ème brevet apparaît en année $t$, alors nous considérons les brevets sur $[t-T_0, t]$) pour calculer l'originalité et $\tilde{I}_i$ l'ensemble des brevets qui citent le brevet $i$ après moins de cinq ans (i.e. nous considérons les brevets sur $[t ,t + T_0]$) dans le cas de la généralité. Il est important de noter que la mesure de généralité est anticipative au sens où $G_i^{(z)}$ utilise l'information qui ne sera disponible que 5 ans après la soumission du brevet. Les deux mesures sont inférieures à la moyenne lorsqu'on se base sur la classification sémantique par rapport à la classification technologique. Fig.~\ref{fig:patentsmining:orig-gene} donne les valeurs moyennes de $O_i^{(sem)}$, $O_i^{(tec)}$, $G_i^{(sem)}$ and $G_i^{(tec)}$.
}

\begin{figure}
\includegraphics[width=\linewidth]{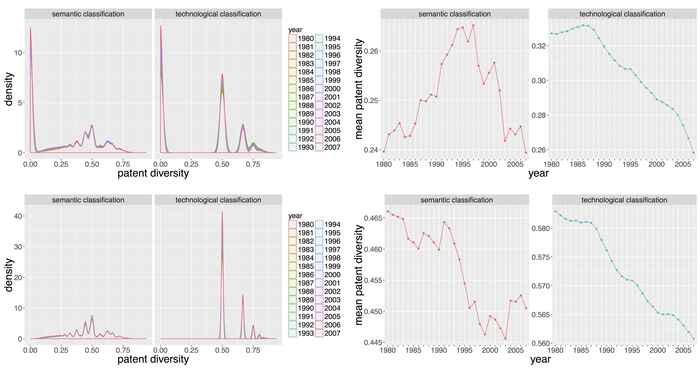}
\appcaption{\textbf{Patent level diversities.} Distributions of diversities \textit{(Left column)} and corresponding mean time-series \textit{(Right column)} for $t=1980$ to $t=2007$ (with the corresponding time window $[t-4,t]$). The first row includes all classified patents, whereas the second row includes only patents with more than one class (i.e. patents with diversity greater than 0).\label{fig:patentsmining:patent-level-orig}}{\textbf{Diversité au niveau du brevet.} Distribution des diversités \textit{(Colonne de gauche)} et séries temporelles correspondantes pour leur moyenne \textit{(Colonne de droite)} pour $t=1980$ à $t=2007$ (avec la fenêtre temporelle correspondante $[t-4,t]$). La première ligne inclut l'ensemble des brevets classifiés, tandis que la deuxième ligne inclut uniquement les brevets avec plus d'une classe (c'est-à-dire les brevets avec une diversité supérieure à 0). \label{fig:patentsmining:patent-level-orig}}
\end{figure}

\begin{figure}
\includegraphics[width=\linewidth]{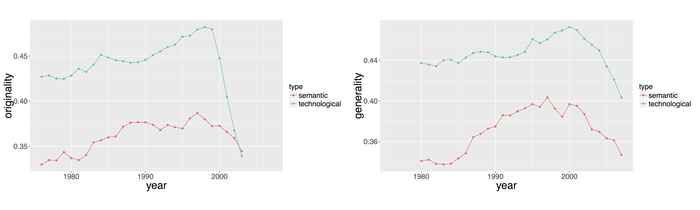}
\appcaption{\textbf{Patent level originality} (left hand side) and \textbf{generality} (right hand side)  for $t=1980$ to $t=2007$ (with the corresponding time window $[t-4,t]$) as defined in subsection \nameref{subsec:orig-gene}.\label{fig:patentsmining:orig-gene}}{\textbf{Originalité (Gauche) et généralité (Droite) au niveau du brevet de $t=1980$ à $t=2007$ (avec la fenêtre temporelle correspondante $[t-4,t]$) comme défini ci-dessus.}\label{fig:patentsmining:orig-gene}}
\end{figure}

\subsubsection*{Classes overlaps}{Intersection des classes} 
\label{app:subsubsec:overlaps}

\bpar{
A proximity measure between two classes can be defined by their overlap in terms of patents. Such measures could for example be used to construct a metrics between semantic classes. Intuitively, highly overlapping classes are very close in terms of technological content and one can use them to measure distance between two firms in terms of technology as done in~\cite{Bloom2005distance}. Formally, recalling the definition of $\left(p_{ij}\right)$ as the probability for the $i$th patent to belong to the $j$th class and $N_P$ as the number of patents it writes 
}{
Une mesure de proximité entre deux classes peut être définie comme leur recouvrement en nombre de brevets. De telles mesures peuvent par exemple être utilisées pour construire une métrique entre les classes sémantique. Intuitivement, des classes qui ont un très fort recouvrement seront très proche en terme de contenu technologique et on peut alors les utiliser pour mesurer la distance entre entreprises en terme de technologie développée comme fait dans~\cite{Bloom2005distance}. Formellement, en rappelant la définition de $\left(p_{ij}\right)$ comme la probabilité du $i$ème brevet d'appartenir à la $j$ème classe et $N_P$ le nombre de brevets, le recouvrement est donné par
}

\begin{eqnarray}
\label{overlap}
Overlap_{jk} = \frac{1}{N_P}\cdot \sum_{i=1}^{N_P} p_{ij} p_{ik}. 
\end{eqnarray}

\bpar{
The overlap is normalized by patent count to account for the effect of corpus size: by convention, we assume the overlap to be maximal when there is only one class in the corpus. A corresponding relative overlap is computed as a set similarity measure in the number of patents common to two classes A and B, given by $o(A,B)=2\cdot \frac{\left|A\cap B\right|}{\left|A\right| + \left|B\right|}$.
}{
Le recouvrement est normalisé par le nombre de brevets pour prendre en compte l'effet de la taille du corpus : par convention, nous supposons que le recouvrement est maximal quand il y a une unique classe dans le corpus. Un recouvrement relatif correspondant est calculé comme une mesure de similarité d'ensemble sur le nombre de brevets communs à deux classes $A$ et $B$, donné par $o(A,B)=2\cdot \frac{\left|A\cap B\right|}{\left|A\right| + \left|B\right|}$
}

\paragraph{Intra-classification overlaps}{Recouvrement intra-classifications}

\bpar{
The study of distributions of overlaps inside each classification, i.e. between technological classes and between semantic classes separately, reveals the structural difference between the two classification methods, suggesting their complementary nature. Their evolution in time can furthermore give insights into trends of specialization. We show in Fig.~\ref{fig:patentsmining:intra-classif-overlap} distributions and mean time-series of overlaps for the two classifications. The technological classification globally always follow a decreasing trend, corresponding to more and more isolated classes, i.e. specialized inventions, confirming the stylized fact obtained in previous subsection. For semantic classes, the dynamic is somehow more intriguing and supports the story of a qualitative regime shift suggested before. Although globally decreasing as technological overlap, normalized (resp. relative) mean overlap exhibits a peak (clearer for normalized overlap) culminating in 1996 (resp. 1999). Looking at normalized overlaps, classification structure was somewhat stable until 1990, then strongly increased to peak in 1996 and then decrease at a similar pace up to now. Technologies began to share more and more until a breakpoint when increasing isolation became the rule again. An evolutionary perspective on technological innovation~\cite{ziman2003technological} could shed light on possible interpretations of this regime shift: as species evolve, the fitness landscape first would have been locally favorable to cross-insemination, until each fitness reaches a threshold above which auto-specialization becomes the optimal path. It is very comparable to the establishment of an ecological niche~\cite{holland2012signals}, the strong interdependency originating here during the mutual insemination resulting in a highly path-dependent final situation. 
}{
L'étude des distributions des recouvrements à l'intérieur de chaque classification, c'est-à-dire entre les classes technologiques et les classes sémantiques séparément, révèle des différences structurelles entre les deux méthodes de classification, suggérant leur nature complémentaire. Leur évolution dans le temps donne de plus des indications sur les tendances de spécialisation. Nous montrons en Fig.~\ref{fig:patentsmining:intra-classif-overlap} les distributions et les séries temporelles moyennes des recouvrements pour les deux classifications. La classification technologique suit globalement toujours une tendance décroissante, correspondant à des classes de plus en plus isolées, i.e. des inventions spécialisées, confirmant le fait stylisé obtenu précédemment. Pour les classes sémantiques, la dynamique est en quelque sorte plus intrigante et soutient l'hypothèse d'un changement de régime qualitatif suggérée précédemment. Même si celle-ci décroit globalement comme pour le recouvrement technologique, le recouvrement normalisé (respectivement relatif) moyen présente un maximum (plus clair pour le recouvrement normalisé) correspondant à l'année 1996 (resp. 1999). En étudiant les recouvrements normalisés, on constate que la structure de classification a été relativement stable jusqu'en 1990, puis a fortement augmenté pour culminer en 1996 puis décroitre à une allure similaire jusqu'à aujourd'hui. Les technologies ont commencé par partager de plus en plus jusqu'à un point de rupture quand une isolation croissante est devenue à nouveau la règle. Une perspective évolutionnaire sur l'évolution technologique~\cite{ziman2003technological} pourrait éclaircir sur de possibles interprétations de ce changement de régime : quand une espèce évolue, l'environnement de fitness aurait d'abord été localement favorable à des inséminations réciproques, jusqu'à ce que chaque fitness dépasse un seuil au dessus duquel l'auto-spécialisation devient le chemin optimal. Ce phénomène est très comparable à l'établissement d'une niche écologique~\cite{holland2012signals}, la forte interdépendance qui a son origine ici dans l'insémination mutuelle, résultant dans une situation finale très fortement dépendante au chemin.
}

\begin{figure}[!ht]
\includegraphics[width=\linewidth]{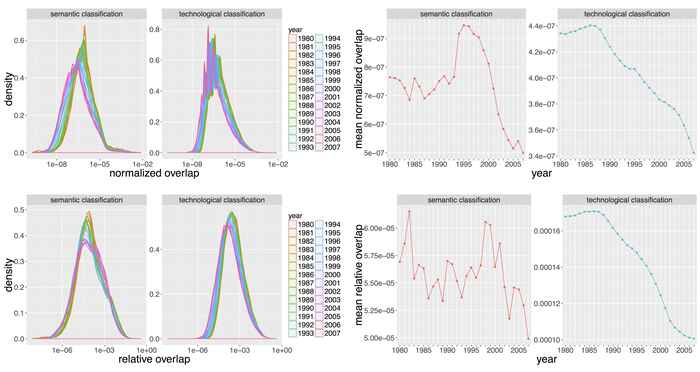}
\appcaption{\textbf{Intra-Classification overlaps.} \textit{(Left column)} Distribution of overlaps $O_{ij}$ for all $i\neq j$ (zero values are removed because of the log-scale). \textit{(Right column)} Corresponding mean time-series. \textit{(First row)} Normalized overlaps. \textit{(Second row)} Relative overlaps.\label{fig:patentsmining:intra-classif-overlap}}{\textbf{Recouvrements intra-classification.} \textit{(Colonne de gauche)} Distribution des recouvrements $O_{ij}$ pour tous $i\neq j$ (les valeurs nulles sont supprimé par l'échelle logarithmique). \textit{(Colonne de droite)} Séries temporelles moyennes correspondantes. \textit{(Première ligne)} Recouvrements normalisés. \textit{(Deuxième ligne)} Recouvrements relatifs.\label{fig:patentsmining:intra-classif-overlap}}
\end{figure}

\paragraph{Inter-classification overlaps}{Correspondance entre les classifications}

\bpar{
Overlaps \emph{between} classifications are defined as previsouly, but with $j$ standing for the $j$th technological class and $k$ for the $k$th semantic class: $p_{ij}$ are technological probabilities and $p_{ik}$ semantic probabilities. They describe the relative correspondence between the two classifications and are a good indicator to spot relative changes, as shown in Fig.~\ref{fig:patentsmining:inter-classif-overlap}. Mean inter-classification overlap clearly exhibits two linear trends, the first one being constant from 1980 to 1996, followed by a constant decrease. Although difficult to interpret directly, this stylized fact clearly unveils a change in the \emph{nature} of inventions, or at least in the relation between content of inventions and technological classification. As the tipping point is at the same time as the ones observed in the previous section and since the two statistics are different, it is unlikely that this is a mere coincidence. Thus, these observations could be markers of a hidden underlying structural changes in processes. 
}{
Les recouvrements \emph{entre} les classifications sont définis comme précedemment, mais avec $j$ désignant la $j$ème classe technologique et $k$ la $k$ème classe sémantique : $p_{ij}$ sont les probabilités technologiques et $p_{ik}$ les probabilités sémantiques. Ils décrivent la correspondance relative entre les deux classifications et sont un bon indicateur pour détecter des changements relatifs, comme montré en Fig.~\ref{fig:patentsmining:inter-classif-overlap}. Le recouvrement inter-classifications moyen présente clairement deux tendances linéaires, la première constante de 1980 à 1996, suivie par une décroissance constante. Même s'il est difficile à interpréter directement, ce fait stylisé dévoile clairement un changement dans la \emph{nature} des inventions, ou au moins dans la relations entre le contenu des inventions et la classification technologique. Comme le point de rupture est à la même date que ceux observés précédemment et comme les indicateurs sont différents, il est peu probable qu'il s'agisse d'une pure coincidence. Ainsi, ces observations pourraient être des marqueurs d'un changement structurel sous-jacent caché des processus.
}

\begin{figure}[!ht]
\includegraphics[width=\linewidth]{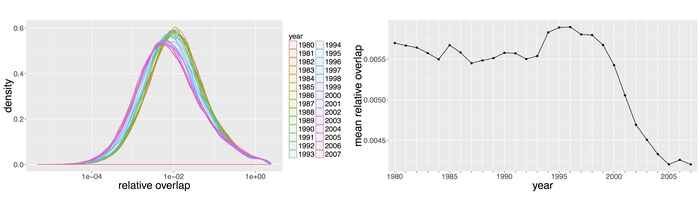}
\appcaption{\textbf{Distribution of relative overlaps between classifications.} \textit{(Left)} Distribution of overlaps at all time steps; \textit{(Right)} Corresponding mean time-series. The decreasing trend starting around 1996 confirms a qualitative regime shift in that period.\label{fig:patentsmining:inter-classif-overlap}}{\textbf{Distribution des recouvrements relatifs entre les classifications.} \textit{(Gauche)} Distributions des recouvrements à chaque date ; \textit{(Droite)} Série temporelle moyenne correspondante. La tendance décroissante commençant autour de 1996 confirme un changement qualitatif de régime à cette période.\label{fig:patentsmining:inter-classif-overlap}}
\end{figure}

\subsubsection*{Citation modularity}{Modularité de citation}
\label{app:subsubsec:citationmodularity}

\bpar{
An exogenous source of information on relevance of classifications is the citation network described previously. The correspondence between citation links and classes should provide a measure of accuracy of classifications, in the sense of an external validation since it is well-known that citation homophily is expected to be quite high (see, e.g.~\cite{AAKnetwork2016}). This section studies empirically modularities of the citation network regarding the different classifications. Modularity is a simple measure of how communities in a network are well clustered (see \cite{clauset2004finding} for the accurate definition). Although initially designed for single-class classifications, this measure can be extended to the case where nodes can belong to several classes at the same time, in our case with different probabilities as introduced in \cite{nicosia2009extending}. The simple directed modularity is given in our case by
}{
Une source exogène d'information concernant la pertinence des classifications est le réseau de citations décrit précédemment. La correspondance entre les liens de citation et les classes devrait fournir une mesure de la précision des classifications, au sens d'une validation externe puisqu'il est bien connu que l'homophilie de citation doit avoir de fortes valeurs (voir par exemple~\cite{AAKnetwork2016}). Cette section étudie empiriquement les modularités du réseau de citation pour les différentes classifications. La modularité est une mesure simple de la manière dont la partition des noeuds d'un réseau correspond plus ou moins bien à de plus fréquentes connexions internes aux communautés (voir \cite{clauset2004finding} pour la définition formelle). Bien qu'initialement conçue pour des classifications univoques, cette mesure peut être étendue au cas où les noeuds peuvent appartenir simultanément à plusieurs classes, dans notre cas avec différentes probabilités comme introduit dans \cite{nicosia2009extending}. La modularité dirigée simple est donnée dans notre cas par
}
\[
Q_d^{(z)} = \displaystyle \frac{1}{N_P}\sum_{1\leq i,j\leq N_P}\left[A_{ij} - \frac{k_{i}^{in}k_{j}^{out}}{N_P}\right]\delta(c_i,c_j),
\]
\bpar{
with $A_{ij}$ the citation adjacency matrix (i.e. $A_{ij} = 1$ if there is a citation from the $i$th patent to the $j$th patent, and $A_{ij}=0$ if not), $k_i^{in}=\left| I_i\right|$ (resp. $k_i^{out}= \left|\tilde{I}_i \right|$) in-degree (resp. out-degree) of patents (i.e. the number of citations made by the $i$th patent to others and the number of citations received by the $i$th patent). $Q_d$ can be defined for each of the two classification systems: $z \in \{tec, sem\}$. If $z=tec$, $c_i$ is defined as the main patent class, which is taken as the first class whereas if $z=sem$, $c_i$ is the class with the largest probability.
}{
avec $A_{ij}$ la matrice d'adjacence du réseau de citations (i.e. $A_{ij} = 1$ s'il existe une citations du $i$ème brevet vers le $j$ème brevet, et $A_{ij}=0$ sinon), $k_i^{in}=\left| I_i\right|$ (resp. $k_i^{out}= \left|\tilde{I}_i \right|$) degré entrant (resp. degré sortant) des brevets (i.e. le nombre de citations faites par le $i$ème brevets à des autres et le nombre de citations reçues par le $i$ème brevet). $Q_d$ peut être définie pour chacun des systèmes de classification : $z \in \{tec, sem\}$. Si $z=tec$, $c_i$ est défini comme la classe principale du brevet, qui est prise comme la première classe tandis que si $z=sem$, $c_i$ est la classe avec la probabilité la plus forte.
}

\bpar{
Multi-class modularity in turns is given by
}{
La modularité multi-classes est quant à elle donnée par
}

\[
\displaystyle Q_{ov}^{(z)} = \frac{1}{N_P} \sum_{c = 1}^{N^{(z)}} \sum_{1\leq i,j \leq N_P}\left[F(p_{ic},p_{jc})A_{ij} - \frac{\beta_{i,c}^{out}k_i^{out}\beta_{j,c}^{in}k_j^{in}}{N_P}\right],
\]
\bpar{
where
}{
où
}
\[
 \beta_{i,c}^{out} =   \frac{1}{N_P} \displaystyle \sum_j F(p_{ic},p_{jc}) \text{ and } \beta_{j,c}^{in} =  \frac{1}{N_P} \displaystyle \sum_i F(p_{ic},p_{jc}).
\]
\bpar{
We take $F(p_{ic},p_{jc}) = p_{ic}\cdot p_{jc}$ as suggested in \cite{nicosia2009extending}. Modularity is an aggregated measure of how the network deviates from a null model where links would be randomly made according to node degree. In other words it captures the propensity for links to be inside the classes. Overlapping modularity naturally extends simple modularity by taking into account the fact that nodes can belong simultaneously to many classes.
}{
Nous prenons $F(p_{ic},p_{jc}) = p_{ic}\cdot p_{jc}$ comme suggéré par \cite{nicosia2009extending}. La modularité est une mesure agrégée de la façon dont le réseau dévie d'un modèle nul où les liens sont attribués de manière aléatoire en respectant les degrés. En d'autres terms elle capture la tendance qu'on les liens d'être à l'intérieur des classes. La modularité recouvrante étend naturellement la modularité simple en prenant en compte le fait que les noeuds peuvent appartenir simultanément à plusieurs classes.
}

\bpar{
We document in Fig.~\ref{fig:patentsmining:modularities} both simple and multi-class modularities over time. For simple modularity, $Q_d^{(tec)}$ is low and stable across the years whereas $Q_d^{(sem)}$ is slightly greater and increasing. These values are however low and suggest that single classes are not sufficient to capture citation homophily. Multi-class modularities tell a different story. First of all, both classification modularities have a clear increasing trend, meaning that they become more and more adequate with citation network. The specializations revealed by both patent level diversities and classes overlap is a candidate explanation for this growing modularities. Secondly, semantic modularity dominates technological modularity by an order of magnitude (e.g. 0.0094 for technological against 0.0853 for semantic in 2007) at each time. This discrepancy has a strong qualitative significance. Our semantic classification fits better the citation network when using multiple classes. As technologies can be seen as a combination of different components as shown by~\cite{Youn:2015fk}, this heterogeneous nature is most likely better taken into account by our multi-class semantic classification.
}{
Nous donnons en Fig.~\ref{fig:patentsmining:modularities} à la fois les modularités simple et multi-classes dans le temps. Pour la modularité simple, $Q_d^{(tec)}$ est bas et stable dans le temps tandis que $Q_d^{(sem)}$ est légèrement supérieure et s'accroit. Ces valeurs sont cependant faibles est suggèrent que les classes uniques ne sont pas suffisantes pour capturer l'homophilie de citation. Les modularités multi-classes donnent des résultats différents. Tout d'abord, les modularités pour les deux classifications ont une claire tendance croissante, signifiant qu'elles deviennent de plus en plus adéquates au réseau de citation. Les spécialisations dévoilées à la fois par les diversités au niveau du brevet et les recouvrement des classes sont une explication potentielle pour l'accroissement de ces modularités. Ensuite, la modularité sémantique est plus grande que la modularité sémantique par un ordre de grandeur (par exemple 0.0094 pour la technologique contre 0.0853 pour la sémantique en 2007) à chaque date. Cette différence a une forte signification qualitative. Notre classification sémantique correspond mieux au réseau de citations avec des classes multiples. Comme les technologies peuvent être comprises comme une combinaison de différentes composantes comme montré par~\cite{Youn:2015fk}, cette nature hétérogène est probablement mieux prise en compte par notre classification sémantique multi-classes.
}

\begin{figure}[!ht]
\centering
\includegraphics[width=\linewidth]{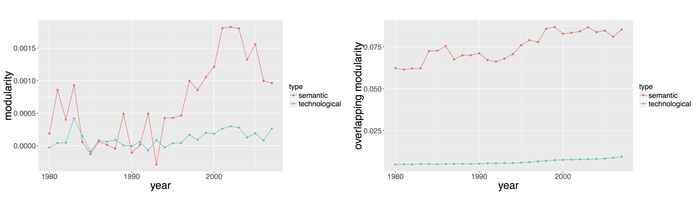}
\appcaption{\textbf{Temporal evolution of semantic and technological modularities of the citation network.} \textit{(Left)} Simple directed modularity, computed with patent main classes (main technological class and semantic class with larger probability). \textit{(Right)} Multi-class modularity, computed following~\cite{nicosia2009extending}.\label{fig:patentsmining:modularities}}{\textbf{Evolution temporelle des modularités sémantique et technologique du réseau de citation.} \textit{(Gauche)} Modularité dirigée simple, calculée avec les classes principales des brevets (classe technologique principale et classe sémantique avec la plus grande probabilité). \textit{(Droite)} Modularité multi-classes, calculée selon~\cite{nicosia2009extending}.\label{fig:patentsmining:modularities}}
\end{figure}

\subsection*{Perspectives}{Perspectives}
\label{app:subsec:discussion}

\bpar{
The main contribution of this study was twofold. First we have defined how we built a network of patents based on a classification that uses semantic information from abstracts. We have shown that this classification share some similarities with the traditional technological classification, but also have distinct features. Second, we provide researchers with materials resulting from our analysis, which include: (i) a database linking each patent with its set of semantic classes and the associated probabilities; (ii) a list of these semantic classes with a description based on the most relevant keywords; (iii) a list of patent with their topological properties in the semantic network (centrality, frequency, degree, etc.). The availability of this data suggests new avenues for further research. Linking our dataset with existing open ones can lead to various powerful developments. For example, using it together with the disambiguated inventor database provided by~\cite{li2014disambiguation} could be a way to study semantic profiles of inventors, or of cities as inventor addresses are provided. The investigation of spatial diffusion of innovation between cities, which is a key component of Pumain's Evolutive Urban Theory~\cite{pumain2010theorie}, would be made possible.
}{
La contribution principale de cette étude repose sur deux points. Tout d'abord nous avons précisé comment construire un réseau de brevets produisant une classification qui utilise l'information sémantique des résumés. Nous avons montré que cette classification partage des similarités avec la classification technologique traditionnelle, mais qu'elles ont aussi des caractéristiques distinctes. Deuxièmement, nous fournissons de manière ouverte les données résultantes de notre analyse, qui incluent : (i) une base de données reliant chaque brevet avec son ensemble de classes sémantiques et les probabilités associées ; (ii) une liste de ces classes sémantiques avec une description basée sur les mots clés les plus pertinents ; (iii) une liste des brevets avec leurs propriétés topologiques dans le réseau sémantique (centralité, fréquence, degré, etc.). La disponibilité de ces données suggère de nouvelles directions pour la recherche future. La jointure de notre base avec des bases ouvertes existantes peut mener à des développements potentiellement prometteurs. Par exemple, son utilisation simultanée avec la base des inventeurs désambiguisée fournie par~\cite{li2014disambiguation} pourrait être une manière d'étudier les profils sémantiques des inventeurs, ou de villes puisque les adresses des inventeurs sont fournies. L'étude de la diffusion spatiale de l'innovation entre les villes, qui est une composante cruciale de la Théorie Evolutive des Villes de \noun{Pumain}~\cite{pumain2010theorie}, serait rendue possible.
}

\bpar{
A first potential application is to use the patents' topological measures inherited from their relevant keywords. The fact that these measures are backward-looking and immediately available after the publication of the patent information is an important asset. It would for example be very interesting to test their predicting power to assess the quality of an innovation, using the number of forward citations received by a patent, and subsequently the future effect on the firm's market value. 
}{
Une première application potentielle est l'utilisation des mesures topologiques des brevets qu'ils héritent de leurs mots-clés pertinents. La fait que ces mesures remontent dans le passé et sont immédiatement disponibles après la publication de l'information du brevet est un point important. Il serait par exemple très intéressant de tester leur pouvoir prédictif pour juger la qualité d'une innovation, en utilisant le nombre de citations reçues par un brevet, et par conséquent l'effet futur sur la valeur de marché de l'entreprise.
}

\bpar{
Regarding firm innovative strategy, a second extension could be to study trajectories of firms in the two networks: technological and semantic. Merging these information with data on the market value of firms can give a lot of insight about the more efficient innovative strategies, about the importance of technology convergence or about acquisition of small innovative firms. It will also allow to observe innovation pattern over a firm life cycle and how this differ across technology field.
}{
Au sujet de la stratégie d'innovation des entreprises, une deuxième extension serait d'étudier les trajectoires des entreprises dans les deux réseaux : technologique et sémantique. La combinaison de ces informations avec des données sur la valeur de marché des entreprises peut apporter beaucoup d'information sur les stratégies d'innovation les plus efficaces, sur l'importance de la convergence des technologies ou sur l'acquisition de petites entreprises innovantes. Cela permettra aussi d'observer les motifs d'innovation sur l'ensemble du cycle de vie d'une entreprise et comment ceux-ci diffèrent selon les champs technologiques.
}

\bpar{
A third extension would be to use dig further into the history of innovation. USPTO patent data have been digitized from the first patent in July 1790. However, not all of them contain a text that is directly exploitable. We consider that the quality of patent's images is good enough to rely on Optical Character Recognition techniques to retrieve plain text from at least 1920. With such data, we would be able to extend our analysis further back in time and to study how technological progress occurs and combines in time. \cite{akcigit2013mechanics} conduct a similar work by looking at recombination and apparition of technological subclasses. Using the fact that communities are constructed yearly, one can construct a measure of proximity between two successive classes. This could give clear view on how technologies converged over the year and when others became obsolete and replaced by new methods.
}{
Une dernière extension serait d'étudier plus en profondeur l'histoire de l'innovation. Les données des brevets USPTO ont été digitalisées depuis le premier brevet en juillet 1790. Cependant, ils ne contiennent pas tous un texte qui est directement exploitable. Nous considérons que la qualité des images des brevets est assez fiable pour permettre l'utilisation de techniques d'\emph{Optical Character Recognition} pour récupérer les textes jusqu'au moins 1920. Avec de telles données, il serait possible d'étendre notre analyse en remontant plus loin dans le temps et d'étudier comment le progrès technologique se produit et se combine dans le temps. \cite{akcigit2013mechanics} procède à un travail similaire en étudiant les recombinaisons et l'apparition des sous-classes technologiques. En se basant sur le fait que les communautés sont construites chaque année, on peut construire une mesure de proximité entre deux classes successives. Cela pourrait fournir une vision plus claire sur la manière dont des technologies ont convergé dans le temps et quand d'autres sont devenues obsolètes et remplacées par des nouvelles méthodes.
}

\stars

%


\newpage

\section{Bridges between Economics and Geography}{Ponts entre géographie et économie}

\label{app:sec:ecogeo}

\bpar{
This section accounts of a first experiment in ``applied perspectivism'', i.e. the attempt to couple perspectives on common objects to create bridges between disciplines. In that spirit, a special session has been organized, together with \noun{B. Carantino} (Paris School of Economics) at the \emph{European Colloquium in Theoretical and Quantitative Geography} (York, September 2017) to question the links between Geography and Economy. The question of bridges within models, i.e. the way that models allow using concepts from economics in geography or reciprocally, has been particularly studied. The frame~\ref{frame:app:ecogeo:abstract} below gives the call for papers. The session gathered 11 contributions\footnote{The program is available at \url{http://www.geog.leeds.ac.uk/ectqg17/programme.html}.}, one of which the initiative was by economists and two others in collaboration with economists: the effort to interest economists in a geography congress has difficultly been fruitful.
}{
Cette section rend compte d'une première expérience en ``perspectivisme appliqué'', c'est-à-dire la tentative de couplage de perspectives sur des objets communs pour créer des ponts entre disciplines. Dans cet esprit, une session spéciale a été organisée, conjointement avec \noun{B. Carantino} (Paris School of Economics) à l'\emph{European Colloquium in Theoretical and Quantitative Geography} (York, septembre 2017) pour questionner les liens entre Géographie et Economie. La question de ponts au sein des modèles, c'est-à-dire de la façon dont les modèles permettent d'utiliser des concepts économiques en géographie ou réciproquement, a été particulièrement étudiée. L'encadré~\ref{frame:app:ecogeo:abstract} ci-dessous présente l'appel à communication. La session a rassemblé 11 contributions\footnote{Le programme est disponible à \url{http://www.geog.leeds.ac.uk/ectqg17/programme.html}.}, dont une à l'initiative d'économistes et deux autres en collaboration avec des économistes : l'effort pour intéresser des économistes à un congrès de géographie a difficilement porté ses fruits.
}

\begin{figure}[h!]
\begin{mdframed}
	
	As Krugman points out, space is for Economic Geography the final frontier, whereas Geographical analyses are somehow far from an advanced integration of economical concepts. What are the existing and potential links? Is there unsurmountable epistemic divergences making bridging approaches irrelevant? For example, the assumptions regarding equilibrium, but also the concepts of equilibrium itself in each discipline may be irreconciliable. This session aims at giving element of answers from a modeling perspective. It is open to case studies of models at the interface and from both disciplines, integrating both elements of spatial analysis and geosimulation together with concepts and methods from economics. It is also open to theoretical or conceptual contributions, in order to bring a broader point of view. An alternative way to study the question is through quantitative epistemology studies, in order to extract empirical endogenous information on the modeling practices themselves. The diversity of views will shed light on potential enrichments on both sides, but also on recurrent difficulties and epistemological divergences, as should illustrate the study of the same objects from totally different perspectives.
	
	\medskip
	
	\framecaption{\textbf{ECTQG 2017 Special Session : bridges between economics and geography.} Abstract of the call for papers for the special session.\label{frame:app:ecogeo:abstract}}{\textbf{ECTQG 2017 Special Session : bridges between economics and geography.} Résumé de l'appel à communication pour la session spéciale.\label{frame:app:ecogeo:abstract}}
\end{mdframed}
\end{figure}

\subsection*{Synthesis of contributions}{Synthèse des contributions}


\bpar{
The contributions to the session allowed shedding lights on the question at different levels and within different domains of knowledge. Modeling studies allowed showing the compromise that has always to be done between spatialization of the model and relevance of economic mechanisms, let it be in the case of a stylized model (contribution by \noun{M. Bida} et al.) or in the case of operational models of land-use evolution (contribution by \noun{E. Koomen} and \noun{D. Vasco}). This compromise can be found again at the theoretical level, but is also complicated by epistemological divergences, for example on the role to give to evolutionary dynamics (contribution by \noun{D. Pumain}) or to desequilibrium (contribution by \noun{R. White} et al.), which can be found in the effective relations between the disciplines, as observed by a bibliometric analysis (contribution by \noun{J. Raimbault}).
}{
Les contributions à la session ont permis d'apporter des éclairages sur la question à différents niveaux et selon différents domaines de connaissance. Des études de modélisation ont permis de montrer le compromis qu'il faut toujours faire entre spatialisation du modèle et pertinence des mécanismes économiques, que ce soit dans le cas d'un modèle stylisé (contribution de \noun{M. Bida} et al.) ou dans le cas de modèles opérationnels d'évolution de l'usage du sol (contribution de \noun{E. Koomen} et \noun{D. Vasco}). Ce compromis se retrouve au niveau théorique, mais est compliqué également par des divergences épistémologiques, par exemple sur le rôle à donner aux dynamiques évolutionnaires (contribution de \noun{D. Pumain}) ou au déséquilibre (contribution de \noun{R. White} et al.), qui se retrouvent dans les relations effectives entre disciplines, comme observé par une analyse bibliométrique (contribution de \noun{J. Raimbault}).
}

\bpar{
A concrete example of object studied according to diverse viewpoints illustrates these considerations: the trajectories of firms. From a purely economic viewpoint, internal factors and the characteristics of real estate induce the location changes of firms (contribution by \noun{A. Bergeaud} and \noun{S. Ray}), whereas the spatial dynamics of these can be understood through their spatial relationships and aggregation effects (contribution by \noun{C. Cottineau} et al.). At a smaller scale, the spatialization of the economic activity of transnational firms allows drawing conclusions of the structure of the geographical system (contribution by \noun{O. Finance}).
}{
Un exemple concret d'objet étudié selon divers point de vue illustre ces considérations : les trajectoires de firmes. Du point de vue purement économique, des facteurs internes et les caractéristiques des locaux induisent les déménagements des entreprises (contribution de \noun{A. Bergeaud} et \noun{S. Ray}), tandis que les dynamiques spatiales de celles-ci peuvent être appréhendées par leur relations spatiales et des effets d'agrégation (contribution de \noun{C. Cottineau} et al.). A une plus petite échelle, la spatialisation de l'activité économique des firmes transnationales permet de tirer des conclusions sur la structure du système géographique (contribution de \noun{O. Finance}).
}

\bpar{
Finally, the empirical studies presented show how combining economic data, such as land-use (contribution by \noun{J. Delloye} et al.), online transactions (contribution by \noun{J. Beckers} et al.) or housing locations (contribution by \noun{Z. Shabrina} et al.), and spatialized models such as an accessibility model or a density distribution model.
}{
Enfin, les études empiriques présentées montrent comment croiser données économiques, comme usage du sol (contribution de \noun{J. Delloye} et al.), transactions en ligne (contribution de \noun{J. Beckers} et al.) ou locations de logements (contribution de \noun{Z. Shabrina} et al.), et modèles spatialisés comme modèle d'accessibilité ou de distribution de densité.
}

\bpar{
The final discussions highlighted the following points: (i) epistemological divergences are not necessarily fundamental if they are contextualized; (ii) differences in behavior regarding the models of different disciplines are also linked to the demand formulated to these disciplines, such as public policy recommendations for economics, and relax the disciplinary standards could help to communicate; (iii) the bibliographic isolation, combined to difficulties to be intelligible, is a crucial point on which considerable progresses are possible, in particular by using new data and methods in textual analysis and datamining.
}{
Les discussions finales ont fait ressortir les points suivants : (i) les divergences épistémologiques ne sont pas nécessairement fondamentales si elles sont contextualisées ; (ii) les différences de comportement face au modèles des différentes disciplines sont aussi liées à la demande qui est faite à ces disciplines, comme des recommandations d'action publique pour l'économie, et relaxer les standards disciplinaires pourrait aider à la communication ; (iii) le cloisonnement bibliographique, combiné à des difficultés d'intelligibilité, est un point crucial sur lequel des progrès considérables sont possibles, notamment par l'utilisation des nouvelles données et méthodes en analyse textuelle et datamining.
}

\bpar{
Therefore, potential bridges are indeed present, and tools and methods that allow facilitating their realization are only waiting to be developed. An example of application fostering reflexivity and thus the interdisciplinary dialogue is given in~\ref{app:sec:cybergeonetworks}.
}{
Ainsi, les ponts potentiels sont bien présents, et les outils et méthodes permettant de faciliter leur concrétisation ne demandent qu'à être développés. Un exemple d'application favorisant la réflexivité et donc le dialogue interdisciplinaire est donné en~\ref{app:sec:cybergeonetworks}.
}


\stars

%


\newpage

\section{Scientific communication through gamification}{Communication scientifique par la gamification}

\label{app:sec:mediationecotox}


\bpar{
The issue of scientific communication, in particular between agents producing knowledge, has been a recurrent theme in our work. It also plays a role in the interface with the public for scientific mediation, and the development of a mediation can in return inform interdisciplinary enterprises. We develop here two models as games, with a similar objective to transmit freshwater ecology concepts. This reinforces the idea of the model as a crucial instrument of scientific mediation.
}{
La question de la communication scientifique, notamment entre les agents producteurs de connaissance, a été un thème récurrent de notre travail. Celle-ci intervient également dans le contact avec le public comme médiation scientifique, et le développement d'une médiation peut en retour informer les entreprises d'interdisciplinarité. Nous développons ici deux modèles sous forme de jeux, ayant un objectif similaire de transmettre des concepts d'écologie aquatique. Cela renforce l'idée du modèle comme instrument crucial de la médiation scientifique.
}

\stars

\bpar{
\textit{This section is the output of an interdisciplinary collaboration with the ecotoxicologist \noun{Dr. Hélène Serra} (Université de Bordeaux and Ineris) and was presented at the SETAC 2016 conference as~\cite{serra:halshs-01322860}.}
}{
\textit{Cette section est le fruit d'une collaboration interdisciplinaire avec l'écotoxicologue \noun{Dr. Hélène Serra} (Université de Bordeaux et Ineris) et a été présentée à la conférence SETAC 2016 comme~\cite{serra:halshs-01322860}.}
}

\stars


\subsection{Introduction}{Introduction}

\bpar{
There is an increasing expectation on people to be aware and to get involved in the environmental issues that our world is facing. However, expert knowledge is often required to understand most of these issues. One of the challenges in science today lies in explaining complex issues in a simple and understandable way to an unspecialized audience. Games can turn out to be a good medium for scientific vulgarization. Indeed, the first form of learning we all experienced was by playing. Games are very popular, and from an educational point of view, they present many advantages. They are dynamic and interactive. Therefore, the player engagement increases, as well as its knowledge retention. In addition, the player is immerged into a new world and discovers a virtual environment where he needs to develop strategies and to identify crucial processes. Those characteristics can be wisely used to spread scientific topics, and gamification has already been proposed as a tool for an easier propagation of scientific thinking~\cite{morris2013gaming} such as in pharmacology~\cite{cain2015serious} or geosciences~\cite{reynard2015application}. In this context, our project aims at developing game-based tools to transmit the basic concepts of freshwater ecology. We choose to focus on a classical board game and on a computer based game because they are complementary in the targeted audience (groups versus online gamers) and the possibilities offered, in particular regarding the interactions between players and the system dynamics. 
}{
L'attente de prise de conscience et d'implication du public concernant les questions environnementales est croissante. Toutefois, une connaissance experte est souvent nécessaire pour comprendre le enjeux sous-jacents à la plupart de ces problèmes. L'un des défis de la science aujourd'hui réside dans le fait d'expliquer des questions complexes de façon simple et compréhensible à une audience non-spécialisée. Les jeux apparaissent comme un moyen pertinent pour la vulgarisation scientifique. En effet, la première forme d'apprentissage est en général par le jeu. Les jeux sont très populaires et présentent divers avantages d'un point de vue éducatif. Ceux-ci sont dynamiques et interactifs. Ainsi, l'engagement du joueur est augmenté, ainsi que sa rétention de connaissances. De plus, le joueur est immergé dans un monde nouveau et découvre un environnement virtuel où il doit développer des stratégies et identifier les processus fondamentaux. Ces caractéristiques peuvent être aisément utilisées pour transmettre des concepts scientifiques, et la gamification a déjà été proposée comme un outil pour une meilleure propagation de la pensée scientifique~\cite{morris2013gaming} comme en pharmacologie~\cite{cain2015serious} ou les géosciences~\cite{reynard2015application}. Dans ce contexte, ce projet vise à développer des outils basés sur les jeux pour transmettre des concepts basiques en écologie aquatique. Nous nous intéressons à un jeu de plateau classique et à un jeu informatique car ceux-ci sont complémentaires dans l'audience visée (joueurs en groupe et joueurs en ligne) et dans les possibilités offertes, en particulier concernant les interactions entre joueurs et les dynamiques du système. 
}



\subsection{Methods}{Méthodologie}

\bpar{
The general methodology is divided in five steps: (1) selection of species; (2) definition of the instructions (object, game board, rules); (3) incorporation of environmental stressors (biotic and abiotic), (4) design and construction of interfaces (board and computer model); (5) test with players. All steps are necessarily interdependent and are tackled in parallel during the development of the games.
}{
La méthodologie pour la conception des deux types de jeux est divisée de manière similaire en 5 étapes : (1) sélection des espèces ; (2) définition des instructions (objets, environnement du jeu, règles) ; (3) inclusion des stress environnementaux (biotiques et abiotiques) ; (4) conception et construction des interfaces (plateau et implémentation informatique) ; (5) test avec des joueurs. L'ensemble des étapes sont interdépendantes et sont menées en parallèle pendant le développement des jeux.
}


\bpar{
While the board game is inspired by past experiences of players, the computer game is based on a model of simulation of the ecosystem. In order to introduce notions of equilibrium and its perturbations that occur at a larger time scale than on the board game, we propose to implement an agent-based model (ABM) and to couple its dynamics with gaming actions. ABM have already been widely used in ecology~\cite{grimm2005pattern}. Therefore, we selected a trophic chain dynamic model (extended prey-predator model) that can capture fish behavioral rules and spatially heterogeneous environment. It is particularly suitable for the game implementation: fish behaviors are influenced by players whereas the ecosystem is disturbed by external events. 
}{
Tandis que le jeu de plateau est inspiré d'experiences de joueurs, le jeu informatique se base sur un modèle de simulation de l'écosystème. De manière à introduire les notions d'équilibre et ses perturbations qui surviennent à une échelle de temps plus longue que celle du jeu de plateau, nous proposons d'implémenter un modèle basé-agent (ABM) et de coupler sa dynamique avec des actions de jeu. Les ABM sont déjà largement utilisés en écologie~\cite{grimm2005pattern}. Ainsi, nous choisissons un modèle dynamique de chaine trophique (modèle proie-prédateur étendu) qui est capable d'inclure des règles comportementales pour les poissons et un environnement spatial hétérogène. Un tel modèle est particulièrement adapté pour l'implémentation du jeu : les comportements des poissons sont influencés par les joueurs tandis que l'écosystème est perturbé par des évènements extérieurs.
}

\subsection{Results}{Résultats}


\bpar{
Both games are based on the same general rules, even if slight modifications have to be expected according to the type of game. The objective of the game is to ensure the stability of an ecological community in the lake. Therefore, each player must adapt the behavior of its fish population accordingly. External perturbations are illustrated by ``events'' that are supposed to reflect abiotic (e.g. water temperature, light, water scarcity) and biotic (e.g. chemicals, parasites, fisherman) stressors. The rationale behind lies in maximizing interactions between players (predation and competition, see Fig.~\ref{fig:app:mediationecotox:boardgame}) and to illustrate feeding and reproduction strategies from different perspectives (from a big solitary fish to a shoal fish, including a invasive fish species).
}{
Les deux jeux sont basés sur les mêmes règles générales, même si des adaptations sont nécessaires selon le type. L'objectif du jeu est de garantir la stabilité d'une communauté écologique dans un lac. Pour cela, chaque joueur doit adapter le comportement de sa population de poissons en circonstance. Des perturbations externes sont illustrées par des évènements qui reflètent des facteurs de stress abiotiques (par exemple température de l'eau, lumière, pénurie de ressources) et biotiques (par exemple produits chimiques, parasites, prédateurs humains). Le principe du jeu se base les interactions entre individus (prédation et compétition, voir Fig.~\ref{fig:app:mediationecotox:boardgame}) et illustre également les stratégies de reproduction selon différentes perspectives (dans le cas du jeu de plateau, d'un poisson solitaire à un poisson en banc, incluant une espèce invasive).
}

\subsubsection{The board game}{Jeu de plateau}



\bpar{
To maintain the populations in the board game, each player has to find resources accordingly to his fish species. The resources are converted into ``units'' that can be used thereafter by the player for different purposes, such as reproduction, juvenile growth, to escape a predator or to attack a pray.
}{
Pour maintenir les populations dans le jeu de plateau, chaque joueur doit trouver les ressources selon son espèce de poisson. Les ressources sont converties en ``unités'' qui peuvent être utilisées par la suite par le joueur pour différents motifs, qui sont la reproduction, la croissance juvénile, échapper à un prédateur ou attaquer une proie.
}

\bpar{
The conceptual version of the game includes four players, each of them being a different species, namely the roach (\textit{Rutilus rutilus}), the pumpkinseed (\textit{Lepomis gibbosus}), the zander (\textit{Sander lucioperca}), and the bleak (\textit{Alburnus alburnus}). The current implementation of the game has been reduced to two species for simplicity reasons, as described in Fig.~\ref{fig:app:mediationecotox:boardgame}.
}{
La version conceptuelle du jeu inclut quatre joueurs, chacun étant une espèce différente, à savoir le gardon (\textit{Rutilus rutilus}), la perche soleil (\textit{Lepomis gibbosus}), le sandre (\textit{Sander lucioperca}) et le \emph{bleak} (\textit{Alburnus alburnus}). L'implémentation actuelle du jeu a été réduite à deux espèces pour des raisons de simplicité, comme décrit en Fig.~\ref{fig:app:mediationecotox:boardgame}.
}

\bpar{
The board is basically composed of boxes. Each of them represents a type of resource (e.g. crustacean, plants, insects), and some boxes are combined with an ``event'' to include the external perturbations in the game. The player has 2 token on the board (one male and one female) and is moving them by throwing dice. The ecological characteristics of each species are kept on a record paper by each player. It describes the species-specific rules (feeding preferences, time and resources needed to reproduce, how to escape/attack etc). The board represents to shore of a lake. A first prototype is currently being tested to determine and adjust the board game design, the ecological characteristics of each species and the characterization of events, in particular their impacts on players.
}{
Le plateau se compose principalement de cases. Chacune représente un type de ressource (e.g. crustacés, plantes, insectes), et certaines cases sont combinées à un ``évènement'' pour inclure les perturbations externes dans le jeu. Le joueur a deux figurines sur le plateau (un mâle et un femelle) et les déplace à l'aide d'un dé. Les caractéristiques écologiques de chaque espèce sont gardées sur une carte par chaque joueur. Elles décrivent les règles spécifiques à chaque espèce (préférences alimentaires, temps et ressources nécessaires à la reproduction, comment s'échapper ou attaquer, etc). Le plateau représente le bord d'un lac. Un premier prototype est en cours de test pour ajuster la conception du plateau, les caractéristiques écologiques de chaque espèce et la caractérisation des évènements, en particulier leur impact sur les joueurs.
}

\begin{figure}
	\includegraphics[width=\linewidth]{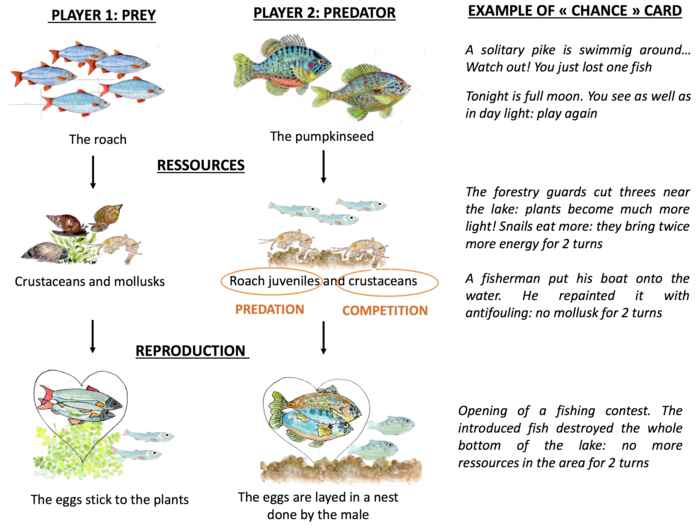}
	\appcaption{\textbf{Principles of the board game.} Species illustrated here are two common European fishes of small size, the roach (\textit{Rutilus rutilus}) as prey and the pumpkinseed (\textit{Lepomis gibbosus}) as predator. We also give examples of external perturbations (``chance'' cards).\label{fig:app:mediationecotox:boardgame}}{\textbf{Principes du jeu de plateau.} Les espèces illustrées ici sont deux poissons européens commun de petite taille, le gardon (\textit{Rutilus rutilus}) comme proie et la perche soleil (\textit{Lepomis gibbosus}) comme prédateur. Nous donnons également des exemples de perturbations extérieures (cartes ``chance'').\label{fig:app:mediationecotox:boardgame}}
\end{figure}


\subsubsection{Computer-based game}{Jeu pour ordinateur}

\bpar{
In the case of the computer game, the players\footnote{The number of player is not specified, since the aim is to maintain the stability of the total ecosystem. Two players can then distribute the roles of prey and predator, each playing on the parameters controlled to stabilize the ecosystem.} control an ecosytem with preys (the roach) and predators (the pumpkinseed). The objective of the game is to maintain the stability of the ecosystem and the concepts illustrated are population dynamic and ecosystem resilience.
}{
Dans le cas du jeu informatique, les joueurs\footnote{Le nombre de joueurs n'est pas spécifié, puisque le but est le maintien de la stabilité de l'ensemble de l'écosystème. Deux joueurs peuvent alors se répartirent les rôles de proie et prédateur, chacun jouant sur les paramètres qu'il contrôle pour stabiliser l'écosystème.} contrôlent un écosystème avec des proies (le gardon) et des prédateurs (la perche soleil). L'objectif est de maintenir la stabilité de l'écosystème et les concepts illustrés sont les dynamiques de population et la résilience d'un écosystème.
}

\bpar{
An agent-based model for a simple prey-predator system is proposed as a basis for the computer game. The ABM simulates the behavior and interactions between agents (fish) to reconstruct the population dynamic (bottom-up approach). Stochasticity is included with spatialized interactions (random encounters between smoothed brownian motions), illustrating the randomisation of prey-predator interactions. Discrete dynamics consist in the following steps: (a) wandering of species; (b) trophic interactions; (c) renewing of population (reproduction). The model parameters include reproduction rate and predation rate, and survival rate for the predator, and also movement parameters.
}{
Un modèle-basé agent pour un système proie-prédateur est proposé comme la base du jeu. L'ABM simule le comportement et les interactions entre agents (poissons) pour reconstruire la dynamique de population (approche \emph{bottom-up}). La stochasticité est prise en compte via les interactions spatialisées (rencontre aléatoires entre des mouvements browniens lissés), illustrant le caractère aléatoire des interactions proie-prédateur. La dynamique discrète est composée par les points suivants : (a) mouvement des individus ; (b) interactions trophiques ; (c) renouvellement de la population (reproduction). Les paramètres du modèle incluent taux de reproduction et de prédation, et de survie pour le prédateur, ainsi que les paramètres de mouvement.
}

\bpar{
The model is implemented in NetLogo, which allows its online use by integrating it into NetLogoweb\footnote{The open implementation is available on the repository of the project at \url{https://github.com/JusteRaimbault/MediationEcotox}.}. The model is explore by using OpenMole~\cite{reuillon2013openmole}, in order to verify the theoretical position of attractors and the average trajectories in the phase space. We obtain on a grid of the parameter space (prey reproduction rate, predation rate, predator survival rate) a good correspondance between the theoretical attractors and the simulated attractors. The Fig.~\ref{fig:app:mediationecotox:phasediags} illustrates the phase diagrams obtained through simulation. The knowledge of attractors allows utilizing the model for the game.
}{
Le modèle est implémenté en NetLogo, ce qui permet son utilisation en ligne par l'intermédiaire de NetLogoweb\footnote{L'implémentation ouverte est disponible sur le dépôt du projet à \url{https://github.com/JusteRaimbault/MediationEcotox}.}. Le modèle est exploré par l'intermédiaire d'OpenMole~\cite{reuillon2013openmole}, afin de vérifier la position théorique des attracteurs et des trajectoires moyennes dans l'espace des phases. Nous obtenons sur une grille de l'espace des paramètres (taux de reproduction de la proie, taux de prédation, taux de survie du prédateur) une bonne correspondance entre les attracteurs théoriques et les attracteurs simulés. La Fig.~\ref{fig:app:mediationecotox:phasediags} illustre des diagrammes de phase obtenus par les simulations. La connaissance des attracteurs permet d'utiliser le modèle pour le jeu.
}

\begin{figure}
	\includegraphics[width=\linewidth]{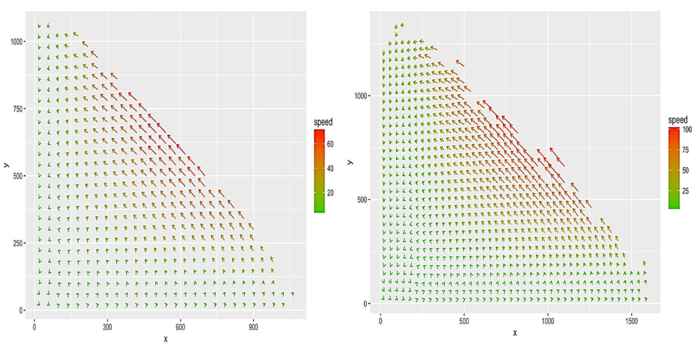}
	\appcaption{\textbf{Examples of phase diagrams of the predator-prey model.} The systematic exploration allows verifying the theoretical expression of average trajectories in the phase space. The plots give the phase portraits of the two populations (x/y), for two points of the parameter space.\label{fig:app:mediationecotox:phasediags}}{\textbf{Exemples de diagrammes de phase du modèle proie-prédateur.} L'exploration systématique permet de vérifier l'expression théorique des trajectoires moyennes dans l'espace des phases. Les graphes donnent les portraits de phase des deux populations (x/y), pour deux points de l'espace des paramètres.\label{fig:app:mediationecotox:phasediags}}
\end{figure}


\bpar{
Indeed, the game starts with an ecosystem at the equilibrium, i.e. that population values are fixed at the non-zero attractor. A button to play a turn makes the ecosystem evolve for 50 time steps. The player then observes the trajectory of populations. The trajectory can then be corrected by the player by acting on model parameters (predator survival, prey reproduction, predation) and thus the position of the attractor. External random events perturbate the populations, and jointly with the noise contribute to destabilize the ecosystem, which can switch to orbits closer to collapse (disparition of one species). The game includes 5 levels of difficulty, based on the strength of perturbations.
}{
En effet, le jeu commence à un écosystème à l'équilibre, c'est-à-dire que les valeurs des populations sont fixées à l'attracteur non nul. Un bouton pour jouer un tour fait évoluer l'écosystème sur 50 pas de temps. Le joueur observe alors la trajectoire des populations. La trajectoire peut alors être corrigée par le joueur par action sur les paramètres du modèle (survie du prédateur, reproduction de la proie, prédation) et donc la position de l'attracteur. Des évènements extérieur aléatoires perturbent les populations, et conjointement au bruit contribuent à déstabiliser l'écosystème, qui peut passer sur des orbites plus proches de l'effondrement (disparition d'une espèce). Le jeu inclut 5 niveaux de difficultés, basés sur la force des perturbations.
}


\begin{figure}
	\includegraphics[width=\linewidth]{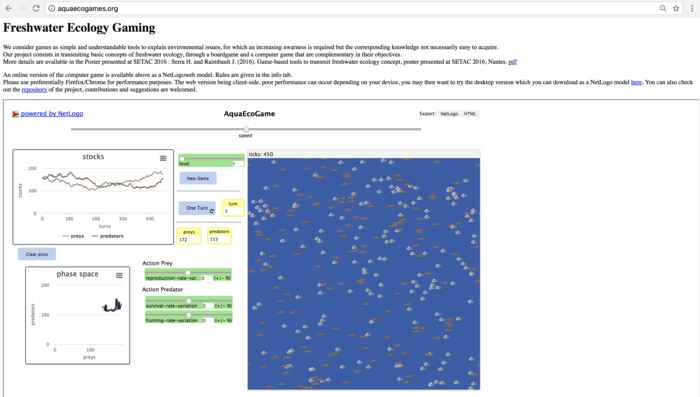}
	\appcaption{\textbf{Screenshot of the web application implementing the computer game.} The context, the documentation and links to resources are briefly recalled, and NetLogoweb is included in the page for the interface of the game.\label{fig:app:mediationecotox:webpage}}{\textbf{Capture de l'application web qui implémente le jeu informatique.} Le contexte, la documentation et les liens vers les ressources sont brièvement rappelés, et NetLogoweb est inclus dans la page pour l'interface du jeu.\label{fig:app:mediationecotox:webpage}}
\end{figure}

\bpar{
The NetLogoweb version of the game (which contains only minimal plots dur to restrictions in comparison to the native version of NetLogo) is available online at \url{http://aquaecogames.org/}. A screenshot of the web application is shown in Fig.~\ref{fig:app:mediationecotox:webpage}.
}{
La version NetLogoweb du jeu (qui ne contient que les graphiques minimaux à cause des restrictions par rapport à NetLogo natif) est disponible en ligne à \url{http://aquaecogames.org/}. Une capture d'écran de l'application web est montrée en Fig.~\ref{fig:app:mediationecotox:webpage}.
}

\subsection{Discussion}{Discussion}



\bpar{
A prototype of each game is currently available for testing and refinements are expected while experiencing the games. In a short term, next versions of the games will be developed after player feedback and will include the aesthetic design of the games and refined processes parameters. Mid-term and long-term objectives are oriented towards a native version of the web application and the use of crowdfunding platforms to diffuse the board game.
}{
Un prototype pour chaque jeu est disponible actuellement pour des tests et des ajustements sont prévus en fonction des retours d'expérience. A court terme, les prochaines versions des jeux seront développées selon le retour des joueurs et incluront une conception esthétique ainsi que des processus plus fins. Les objectifs à moyen et long terme s'orientent vers un développement natif de l'application web et l'utilisation de plateformes de financement participatif pour diffuser le jeu de plateau.
}

\bpar{
One must keep in mind that the ludic rather than pedagogical aspects are central in the success of such game-based media. If players forget that the game is about ecology, our precise objective is reached, since it would mean that the underlying scientific concepts are clearly understood.
}{
Il faut garder à l'esprit que le caractère ludique plutôt que pédagogique est central dans le succès de tels media basés sur les jeux. Si les joueurs oublient que le jeu est à propos d'un problème écologique, l'objectif est précisément atteint, puisque cela signifierait que les concepts scientifiques sous-jacents sont clairement compris.
}

\stars


%

\bpar{
\chapter{Datasets}
\markboth{\thechapter\space Datasets}{\thechapter\space Datasets}

}{
\chapter{Données}
\markboth{\thechapter\space Données}{\thechapter\space Données}
}

\label{app:data} 




\bpar{
This appendix lists and describes the different open datasets that we were brought to create and use in the thesis. Data are indeed a proper knowledge domain, and collection and consolidation operations are a scientific stage in itself.
}{
Cette annexe liste et décrit les différents jeux de données ouvertes que nous avons été amenés à créer et à utiliser dans la thèse. Les données sont en effet bien un domaine de connaissance propre, et les opérations de collecte et de consolidation sont une étape scientifique à part entière.
}

\section{Grand Paris traffic data}{Données de trafic du Grand Paris}


\subsection{Description}{Description}

\bpar{
This dataset, used on two months for the analysis of~\ref{sec:reproducibility}, finally extends on two years from February 2016 to February 2018. It is constituted by travel times on main freeway segments of the Parisian metropolitan area, at a time granularity of 2 minutes.
}{
Ce jeu de données, utilisé sur deux mois pour l'analyse de~\ref{sec:reproducibility}, s'étend finalement sur deux ans de février 2016 à février 2018. Il est constitué des temps de parcours sur les axes autoroutiers principaux de la métropole parisienne, à une granularité temporelle de 2 minutes.
}

\subsection{Specification}{Spécification}

\paragraph{Citation}{Citation}

Raimbault J., 2018, Replication Data for: Investigating the empirical existence of static user equilibrium, doi:10.7910/DVN/X22ODA, Harvard Dataverse, V1

\paragraph{Type and format}{Type et format}

\bpar{
List of road links, with effective time and theoretical travel time, and the time of observation (timestamps); as a sqlite3 format.
}{
Liste des liens routiers, avec temps effectif et temps théorique de parcours, et le moment d'observation (timestamps) ; au format sqlite3.
}

\paragraph{License}{Licence}

\bpar{
Public domain CC0.
}{
Domaine public CC0.
}

\paragraph{Availability}{Disponibilité}

\bpar{
The database is available on the Harvard Dataverse at \url{http://dx.doi.org/10.7910/DVN/X22ODA}.
}{
La base est disponible sur le Harvard Dataverse à \url{http://dx.doi.org/10.7910/DVN/X22ODA}.
}









\section{Topological graphs of road networks}{Graphes topologiques des réseaux routiers}

\subsection{Description}{Description}

\bpar{
The simplification of road networks, achieved at a large scale for Europe and China on OpenStreetMao data, yields the corresponding topological graphs as described in~\ref{sec:staticcorrelations} and in~\ref{app:sec:staticcorrelations}.
}{
La simplification des réseaux routiers, opérée à grande échelle pour l'Europe et la Chine sur les données d'OpenStreetMap, produit les graphes topologiques correspondants comme décrit en~\ref{sec:staticcorrelations} et en~\ref{app:sec:staticcorrelations}.
}

\bpar{
The relevance of this dataset is the possibility to directly use it to study graph measures of road networks, on any spatial extent. Indeed, the creation of the topological network at the scale considered required a considerable computational effort, which is not necessarily accessible to anyone.
}{
L'intérêt de ce jeu de données est la possibilité d'utilisation directe pour l'étude de mesures de graphes des réseaux routiers, sur une étendue spatiale quelconque. En effet, la création du réseau topologique à l'échelle considérée a requis un effort computationnel considérable, pas forcément accessible au plus grand nombre.
}

\subsection{Specification}{Spécification}

\paragraph{Citation}{Citation} 

Raimbault, Juste, 2018, "Simplified road networks, Europe and China", doi:10.7910/DVN/RKDZMV, Harvard Dataverse, V1

\paragraph{Type and format}{Type et format}

\bpar{
Data are as a list of links, as an compressed extraction from postgresql (dump).
}{
Les données sont sous forme de liste des liens, au format extraction compressée de postgresql (dump).
}

\paragraph{License}{Licence}

\bpar{
Public domain CC0.
}{
Domaine public CC0.
}

\paragraph{Availability}{Disponibilité}

\bpar{
The database is available on the Harvard Dataverse at \url{http://dx.doi.org/10.7910/DVN/RKDZMV}.
}{
La base est disponible sur le Harvard Dataverse à \url{http://dx.doi.org/10.7910/DVN/RKDZMV}.
}




\section{Interviews}{Entretiens}

\label{app:sec:interviews}


\bpar{
A research material which would be more ``qualitative'' in the classical sense, has no reason to be less open than ``quantitative'' databases. In the case of interviews, the opening of transcripts is essential for reproducibility since it is the last (and the first) stage before the non-reproductible translation into interpretations. We also think that it is crucial to exploit their full potential, the opening allowing their reuse and thus possibly reactions or debates. Initiatives in this direction begin to emerge, such as the \emph{Qualitative Data Repository}\footnote{https://qdr.syr.edu/} which allows archiving and presenting in a consistent way a qualitative corpus, often described only partly and jointly to the analyses in the papers~\cite{elman_kapiszewski_2018}.
}{
Un matériau de recherche qui serait plus ``qualitatif'' au sens classique, n'a pas de raison d'être moins ouvert que des bases de données ``quantitatives''. Dans le cas d'entretiens, l'ouverture des retranscriptions est essentielle pour la reproductibilité puisqu'il s'agit du dernier (et du premier) stade avant la traduction non reproductible en interprétations. Nous pensons également qu'elle est cruciale pour exploiter l'ensemble de leur potentiel, l'ouverture permettant leur réutilisation et donc possiblement réactions ou débats. Des initiatives dans cette direction commencent à émerger, comme le \emph{Qualitative Data Repository}\footnote{https://qdr.syr.edu/} qui permet d'archiver et de présenter de manière cohérente un corpus qualitatif, souvent décrit de façon parcellaire et conjointement aux analyses dans les articles~\cite{elman_kapiszewski_2018}.
}

\subsection{Description}{Description}

\subsubsection{Interview with Denise Pumain, 2017/03/31}{Entretien avec Denise Pumain, 2017/03/31}

\bpar{
This interview was conducted in the context of collecting empirical materials for the redaction of~\cite{raimbault2017applied}, which furthermore allowed the construction of the knowledge framework developed in~\ref{sec:knowledgeframework}. The interview is mostly centered on the genesis of the evolutive urban theory.
}{
Cet entretien est intervenu dans le contexte d'une collecte de matériau empirique pour la rédaction de~\cite{raimbault2017applied}, qui a permis entre autre la construction du cadre de connaissances développé en~\ref{sec:knowledgeframework}. L'entretien est principalement centré sur la genèse de la théorie évolutive des villes.
}

\subsubsection{Interview with Romain Reuillon, 2017/04/11}{Entretien avec Romain Reuillon, 2017/04/11}

\bpar{
This interview was conducted in the same context than the previous one, aiming at bringing a new vision from the viewpoint of methods and tools. In particular, it describes the genesis of OpenMole.
}{
Cet entretien intervient dans le même contexte que le précédent, en cherchant à apporter un éclairage du point de vue des méthodes et outils. Il retrace en particulier la genèse d'OpenMole.
}

\subsubsection{Interview with Clémentine Cottineau, 2017/05/05}{Entretien avec Clémentine Cottineau, 2017/05/05}

\bpar{
This interview aims at understanding the viewpoint of a geographer at the interdisciplinary interface (participation of the Geodivercity ERC project) on the evolutive urban theory and its elaboration in terms of knowledge domains.
}{
Cet entretien vise à comprendre le point de vue d'une géographe à l'interface interdisciplinaire (participation à l'ERC Geodivercity) sur la théorie évolutive des villes et son élaboration en termes de domaines de connaissance.
}

\subsubsection{Interview with Denise Pumain, 2017/12/15}{Entretien avec Denise Pumain, 2017/12/15}

\bpar{
This second interview with \noun{D. Pumain} concentrates more particularly on the structuring effects of transportation infrastructures and co-evolution, from the viewpoint of geography.
}{
Ce deuxième entretien avec \noun{D. Pumain} se concentre plus particulièrement sur les effets structurants des infrastructures de transport et co-évolution, du point de vue de la géographie.
}

\subsubsection{Interview with Alain Bonnafous, 2018/01/09}{Entretien avec Alain Bonnafous, 2018/01/09}

\bpar{
This interview focuses on the structuring effects of transportation infrastructures, from the viewpoint of transportation economics, and also to the interdisciplinary positioning of transportation economics.
}{
Cet entretien s'intéresse aux effets structurants des infrastructures de transport, du point de vue de l'économie des transports, ainsi qu'au positionnement interdisciplinaire de l'économie des transport.
}

\subsection{Specification}{Spécification}

\paragraph{Citation}{Citation}

Raimbault J., 2017. JusteRaimbault/Entretiens v0.2 (Version v0.2). Zenodo. http://doi.org/10.5281/zenodo.556331

\paragraph{Type and format}{Type et format}

\bpar{
Transcripts of interviews in text format.
}{
Transcription des entretiens au format texte.
}

\paragraph{License}{Licence}

Creative commons CC-BY-NC.

\paragraph{Availability}{Disponibilité}

\bpar{
Interviews are available on the dedicated git repository at \url{https://github.com/JusteRaimbault/Entretiens}, and the successive versions are accessible at \url{https://doi.org/10.5281/zenodo.596954}.
}{
Les entretiens sont disponibles sur le dépôt git dédié à \url{https://github.com/JusteRaimbault/Entretiens}, et les versions successives sont accessibles à \url{https://doi.org/10.5281/zenodo.596954}.
}

\section{Synthetic data and simulation results}{Données synthétiques et résultats de simulations}

\bpar{
Computation results or simulation results used for all the results presented are available in an open way, either on the git repository or on a dedicated dataverse repository in the case of autonomous papers or massive files. The links are the following for the dedicated repositories:
}{
Les résultats de calculs ou de simulations utilisés pour l'ensemble des résultats présentés sont disponibles de manière ouverte, soit sur le dépôt git soit sur un dépôt dataverse dédié dans le cas d'articles autonomes ou de fichiers massifs. Les liens sont les suivants pour les dépôts particuliers :
}

\bpar{
\begin{itemize}
	\item Results of the exploration of the Cybergeo corpus \url{http://dx.doi.org/10.7910/DVN/VU2XKT}; quantitative epistemology and modelography \url{https://github.com/JusteRaimbault/CityNetwork/tree/master/Models/QuantEpistemo/HyperNetwork/data}
	\item Morphological and topological indicators for Europe and China \url{http://dx.doi.org/10.7910/DVN/RHLM5Q}
	\item Simulation of synthetic data with the RBD model to identify spatio-temporal causality regimes \url{http://dx.doi.org/10.7910/DVN/KGHZZB}
	\item Calibration of the macroscopic interaction model \url{https://github.com/JusteRaimbault/CityNetwork/tree/master/Results/NetworkNecessity/InteractionGibrat/calibration}
	\item Simulation and calibration of the morphogenesis model for density \url{http://dx.doi.org/10.7910/DVN/WSUSBA}
	\item Simulation of the weak coupling of density and network growth models \url{http://dx.doi.org/10.7910/DVN/UIHBC7}
	\item Simulation of the SimpopNet model \url{http://dx.doi.org/10.7910/DVN/RW8S36}
	\item Simulations of the macroscopic co-evolution model \url{http://dx.doi.org/10.7910/DVN/TYBNFQ} and \url{https://github.com/JusteRaimbault/CityNetwork/tree/master/Models/MacroCoevol/MacroCoevol/calibres} for the calibration
	\item Simulations of the mesoscopic co-evolution model \url{http://dx.doi.org/10.7910/DVN/OBQ4CS}
	\item Simulations of the Lutecia model \url{http://dx.doi.org/10.7910/DVN/V3KI2N}
\end{itemize}
}{
\begin{itemize}
	\item Résultats de l'exploration du corpus Cybergeo \url{http://dx.doi.org/10.7910/DVN/VU2XKT} ; Epistémologie quantitative et modélographie \url{https://github.com/JusteRaimbault/CityNetwork/tree/master/Models/QuantEpistemo/HyperNetwork/data}
	\item Indicateurs morphologiques et topologiques pour l'Europe et la Chine \url{http://dx.doi.org/10.7910/DVN/RHLM5Q}
	\item Simulation de données synthétiques par le modèle RBD pour l'identification de régimes de causalité spatio-temporelle \url{http://dx.doi.org/10.7910/DVN/KGHZZB}
	\item Calibration du modèle macroscopique d'interactions \url{https://github.com/JusteRaimbault/CityNetwork/tree/master/Results/NetworkNecessity/InteractionGibrat/calibration}
	\item Simulation et calibration du modèle de morphogenèse pour la densité \url{http://dx.doi.org/10.7910/DVN/WSUSBA}
	\item Simulation du couplage faible des modèles de densité et de croissance de réseau \url{http://dx.doi.org/10.7910/DVN/UIHBC7}
	\item Simulation du modèle SimpopNet \url{http://dx.doi.org/10.7910/DVN/RW8S36}
	\item Simulations du modèle de co-évolution macroscopique \url{http://dx.doi.org/10.7910/DVN/TYBNFQ} et \url{https://github.com/JusteRaimbault/CityNetwork/tree/master/Models/MacroCoevol/MacroCoevol/calibres} pour la calibration
	\item Simulations du modèle de co-évolution mesoscopique \url{http://dx.doi.org/10.7910/DVN/OBQ4CS}
	\item Simulations du modèle Lutecia \url{http://dx.doi.org/10.7910/DVN/V3KI2N}
\end{itemize}
}

\stars


%

\bpar{
\chapter{Tools}
\markboth{\thechapter\space Tools}{\thechapter\space Tools}
}{
\chapter{Outils}
\markboth{\thechapter\space Outils}{\thechapter\space Outils}
}

\label{app:tools} 


\bpar{
This appendix accounts for the tools developed and used for all the analyses. As we described in section~\ref{sec:knowledgeframework}, tools generally correspond to the implementation of methods (being then some \emph{proto-methods}, at least in our case where we do not use any physical measuring apparatus), but indeed correspond to a knowledge domain in itself and with a certain independence.
}{
Cet annexe rend compte des outils développés et utilisés pour l'ensemble des analyses. Comme nous l'avons précisé en section~\ref{sec:knowledgeframework}, les outils correspondent généralement à l'implémentation de méthodes (étant alors des \emph{proto-méthodes}, du moins dans notre cas où nous n'utilisons pas d'appareillage physique de mesure), mais correspondent bien à un domaine de connaissance à part entière et avec une certaine indépendance.
}

\bpar{
We distinguish and describe here:
\begin{itemize}
	\item the \emph{packages} or \emph{softwares} developed in the context of this work, but which can fulfil much larger functions and can be distributed in an autonomous way;
	\item the implementation of simulation models and of data mining algorithms;
	\item tools or practices which particularly facilitate a fluid and open science.
\end{itemize}
}{
Nous distinguons et décrivons ici :
\begin{itemize}
	\item les \emph{packages} ou \emph{logiciel} développés dans le cadre de ce travail, mais qui peuvent remplir des fonctions bien plus larges et peuvent être distribués de manière autonome ;
	\item l'implémentation des modèles de simulation et des algorithmes de traitement des données ;
	\item des outils ou pratiques facilitant particulièrement une science ouverte et fluide.
\end{itemize}
}

\stars
%


\section{Softwares and packages}{Packages et logiciels} 

\label{app:sec:packages} 




\bpar{
This section describes the significant software contributions, which were the object of a packaging in the spirit of an open science. It is difficult to decide at which time an implementation and possibly a library developed in a particular context can be made generic and distributed in an autonomous way. We made the choice (i) of relatively general functions; (ii) of a strong potential impact; and (iii) of a certain level of maturity in the packaging.
}{
Cette section recense les contributions logicielles significatives, qui ont fait l'objet d'un \emph{packaging} dans l'esprit d'une science ouverte. Il est difficile de décider à quel moment une implémentation et éventuellement une bibliothèque développées dans un cadre particulier peuvent être rendus génériques et distribués de manière autonome. Nous avons fait le choix (i) de fonctions relativement générales ; (ii) d'un fort impact potentiel ; et (iii) d'un certain niveau de maturité au niveau du packaging.
}

\subsection{largeNetwoRk: network import and simplification for R}{largeNetwoRk : import de réseau et simplification pour R}

\paragraph{Description}{Description}

\bpar{
The \texttt{largeNetwoRk} package for the \texttt{R} language is aimed at the import and the simplification of massive transportation networks. It is constructed in particular for the import of OpenStreetMap data, but can tackle other formats such as shp. The objective is to allow analyses of networks on large surfaces while having access to modest computational capabilities, and to make transparent the import of spatial data into a topological graph.
}{
Le package \texttt{largeNetwoRk} pour le langage \texttt{R} est destiné à l'import et la simplification des réseaux de transport massifs. Il est particulièrement construit pour l'import de données OpenStreetMap, mais peut traiter d'autres formats comme shp. L'objectif est de permettre des analyses des réseaux sur de grandes surfaces tout en disposant de capacités de calcul modestes, et de rendre transparent l'import des données spatiales en un graphe topologique.
}

\paragraph{Characteristics}{Caractéristiques}


\bpar{
The package is fully written in \texttt{R}, and requires a connection with a \texttt{PostgreSQL} database (the the PostGis extension installed). Source code is available at \url{https://github.com/JusteRaimbault/CityNetwork/tree/master/Models/TransportationNetwork/NetworkSimplification} with the documentation.
}{
Le package est intégralement écrit en \texttt{R}, et requiert une connexion avec une base \texttt{PostgreSQL} (avec extension PostGis installée). Le code source est disponible à \url{https://github.com/JusteRaimbault/CityNetwork/tree/master/Models/TransportationNetwork/NetworkSimplification} avec la documentation.
}

\paragraph{Functions}{Fonctions}

\bpar{
The main functions implemented are the following:
\begin{itemize}
	\item \texttt{constructLocalGraph}: construct a topological graph from spatial lines queried from the postgis database (in a given spatial extent)
	\item \texttt{graphFromSpdf}: constructs a topological graph from a spatial data structure (allows for example to import from a \texttt{shp} file)
	\item \texttt{mergeGraphs}: merge two graphs neighbors in space
	\item \texttt{simplifyGraph}: simplification of a graph (see algorithm in~\ref{app:sec:staticcorrelations})
	\item \texttt{connexify}: gives a connected graph from an arbitrary graph, through the addition of connectors
	\item \texttt{exportGraph}: exports a topological graph in the database
\end{itemize}
}{
Les principales fonctions suivantes sont implémentées :
\begin{itemize}
	\item \texttt{constructLocalGraph} : construit un graphe topologique à partir des lignes spatiales issues de la base postgis (dans une étendue spatiale précisée)
	\item \texttt{graphFromSpdf} : construit un graphe topologique à partir d'une structure de données spatiale (permet d'importer depuis un fichier \texttt{shp} par exemple)
	\item \texttt{mergeGraphs} : fusion de deux graphes voisins dans l'espace
	\item \texttt{simplifyGraph} : simplification d'un graphe (voir algorithme en~\ref{app:sec:staticcorrelations})
	\item \texttt{connexify} : donne un graphe connecté à partir d'un graphe quelconque, par l'ajout de connecteurs
	\item \texttt{exportGraph} : exporte un graphe topologique dans la base de données
\end{itemize}
}

\bpar{
A complete script allows moreover to execute the \emph{split and merge} algorithm described in~\ref{app:sec:staticcorrelations} for the simplification of large spatial extents.
}{
Un script complet permet par ailleurs l'exécution de l'algorithme \emph{split and merge} décrit en~\ref{app:sec:staticcorrelations} pour la simplification de grandes étendues spatiales.
}

\paragraph{Particularities}{Particularités}

\bpar{
The use on massive data requires a parallel processing. Furthermore, the external program \texttt{osmosis} is used for the initial conversion of OpenStreetMap data (\texttt{osm pbf} for example) and their import into the postgis database.
}{
L'utilisation sur données massives requière un traitement en parallèle. De plus, le programme externe \texttt{osmosis} est utilisé pour la conversion initiale des données OpenStreetMap (\texttt{osm pbf} par exemple) et leur import dans la base postgis.
}

\subsection{Transportation networks and accessibility in R}{Réseaux de transports et accessibilité en R}

\paragraph{Description}{Description}

\bpar{
The package \texttt{tRansport} for the \texttt{R} language provides transparent primitives for computing indicators for public transportation networks and the associated accessibility computations. Starting from datasets including lines and stations for different transportation modes, it allows constructing a multimodal topological network and to compute different measures given geographical variables.
}{
Le package \texttt{tRansport} pour le langage \texttt{R} rend transparent les calculs d'indicateurs pour les réseaux de transport en commun et les calculs d'accessibilité associés. A partir de jeux de données comprenant lignes et stations pour différents modes de transports en commun, il permet de construire un réseau topologique multimodal et de calculer différentes mesures étant donné des variables géographiques.
}

\paragraph{Characteristics}{Caractéristiques}

\bpar{
The package is written in language \texttt{R} and produces graphs following the \texttt{igraph} package structure. Source code and documentation are available at \url{https://github.com/JusteRaimbault/CityNetwork/tree/master/Models/TransportationNetwork/NetworkAnalysis}.
}{
Le package est écrit en langage \texttt{R} et produit des graphes selon la structure du package \texttt{igraph}. Le code source et la documentation sont disponibles à \url{https://github.com/JusteRaimbault/CityNetwork/tree/master/Models/TransportationNetwork/NetworkAnalysis}.
}

\paragraph{Functions}{Fonctions}

\bpar{
The main following functions are available:
\begin{itemize}
	\item \texttt{addTransportationLayer}: constructs a graph from a layer of the network, or adds a layer to an existing network, from a shapefile description of links and nodes (stations) of the network.
	\item \texttt{addPointsLayer}: adds a layer of points, which can then be origin or destination of itineraries. They are linked to the closest station by connector links which speed is specified.
	\item \texttt{addAdministrativeLayer}: similar function, which connects the centroids of a polygon layer typically representing administrative areas, keeping their attributes as node attributes.
	\item \texttt{computeAccess}: computes the accessibility between points of the transportation network, following different specifications: travel time, weighting at the origin and/or destination by specified data.
\end{itemize}
}{
Les principales fonctions suivantes sont disponibles :
\begin{itemize}
	\item \texttt{addTransportationLayer} : construit un graphe à partir d'une couche du réseau, ou ajoute une couche à un réseau existant, à partir d'une description shapefile des liens et des noeuds (gares) du réseau.
	\item \texttt{addPointsLayer} : ajoute une couche de points, qui peuvent être alors origine ou destination d'itinéraires. Ils sont reliés à la gare la plus proche par des connecteurs dont la vitesse est spécifiée.
	\item \texttt{addAdministrativeLayer} : fonction similaire, qui connecte les centroïdes d'une couche de polygones représentant typiquement des zones administratives, conservant leur attributs comme attributs de noeuds.
	\item \texttt{computeAccess} : calcule l'accessibilité entre des point du réseau de transport, selon plusieurs spécifications : temps de trajet, pondération à l'origine et/ou à la destination par des données spécifiées.
\end{itemize}
}

\subsection{morphology: a NetLogo extension to measure urban form}{morphology : extension NetLogo pour mesurer la forme urbaine}

\label{app:subsec:morphologyextension}

\paragraph{Description}{Description}

\bpar{
The \texttt{morphology} extension for NetLogo5 allows computing in an efficient and transprent way the morphological indicators introduced in~\ref{sec:staticcorrelations} (Moran index, entropy, average distance, hierarchy), for the spatial distribution of an arbitrary patch variable.
}{
L'extension \texttt{morphology} pour NetLogo5 permet de calculer de manière efficiente et transparente les indicateurs morphologiques introduits en~\ref{sec:staticcorrelations} (indice de Moran, entropie, distance moyenne, hiérarchie), pour la distribution spatiale d'une variable de patch quelconque.
}

\paragraph{Characteristics}{Caractéristiques}

\bpar{
The extension is written in \texttt{scala} and is compatible the version 5 of NetLogo. It is available at \url{https://github.com/JusteRaimbault/nl-spatialmorphology}.
}{
L'extension est écrite en \texttt{scala} et est compatible avec les versions 5 de NetLogo. Elle est disponible à \url{https://github.com/JusteRaimbault/nl-spatialmorphology}.
}

\paragraph{Particularities}{Particularités}

\bpar{
The indicators implying a convolution (Moran index, average distance) are implemented with a fast Fourier transform, allowing decreasing the complexity from a $O(N^4)$ to a $O(N^2\cdot \log^2 N)$ if $N$ is the width of the grid.
}{
Les indicateurs impliquant une convolution (indice de Moran, distance moyenne) sont implémentés par transformée de Fourier rapide, permettant de faire passer la complexité d'un $O(N^4)$ à un $O(N^2\cdot \log^2 N)$ si $N$ est la taille d'un côté de la grille. 
}

\subsection{TorPool}{TorPool}

\paragraph{Description}{Description}

\bpar{
\texttt{TorPool} is a java wrapper for the tor software, which allows maintaining a pool of instance in parallel, and to renew these instances on demand. An interface with TorPool is available with java with a dedicated library. This tool allows in particular facilitating the automatic collection of data. It is available as source and executable at \url{https://github.com/JusteRaimbault/TorPool}.
}{
\texttt{TorPool} est un wrapper java du logiciel tor, qui permet de maintenir une équipe d'instances en parallèle, et de renouveler ces instances sur demande. Une interface avec TorPool est disponible avec java par une bibliothèque dédiée. Cet utilitaire permet entre autres de faciliter la collection automatique de données. Il est disponible comme source et sous forme executable à \url{https://github.com/JusteRaimbault/TorPool}.
}

\paragraph{Functions}{Fonctions}

\bpar{
The software is launched as a \texttt{jar} executable, and opens a specified range of ports as local \texttt{socks5} proxies to the Tor network.
}{
Le logiciel se lance sous forme d'executable \texttt{jar}, et ouvre une plage spécifiée de ports en proxy \texttt{socks5} local vers le réseau Tor.
}

\bpar{
The associated java library allows to (i) establish a connexion with the proxies, (ii) to ask for a renewal of instances, allowing a change of circuit in the network.
}{
La bibliothèque java associée permet (i) d'établir une connexion avec les proxys (ii) de demander un renouvellement des instances, permettant un changement de circuit dans le réseau.
}

\subsection{Scientific corpus mining}{Fouille de corpus scientifique}

\paragraph{Description}{Description}

\bpar{
The tools developed in the context of Chapter~\ref{ch:modelinginteractions}, and of Appendices~\ref{app:sec:cybergeo} and~\ref{app:sec:patentsmining} allow in a general way the mining of scientific corpuses, from the viewpoint of the citation network and the semantic network.
}{
Les outils développés dans le cadre du Chapitre~\ref{ch:modelinginteractions}, et des Annexes~\ref{app:sec:cybergeo} et~\ref{app:sec:patentsmining} permettent de manière générale la fouille de corpus scientifiques, du point de vue du réseau de citation et du réseau sémantique.
}

\paragraph{Characteristics}{Caractéristiques}

\bpar{
As recalled in~\ref{app:sec:cybergeo}, the tasks required are relatively heterogenous, and different languages are therefore used: Java for data collection, python for textual analysis, R for network analysis. The version of the different scripts used for Chapter~\ref{ch:modelinginteractions} is available at \url{https://github.com/JusteRaimbault/CityNetwork/tree/master/Models/QuantEpistemo/HyperNetwork}.
}{
Comme rappelé en~\ref{app:sec:cybergeo}, les tâches requises sont assez hétérogènes, et différents langages sont alors mobilisés : Java pour la collection des données, python pour l'analyse textuelle, R pour les analyses de réseau. La version des différents scripts utilisés pour le Chapitre~\ref{ch:modelinginteractions} est disponible à \url{https://github.com/JusteRaimbault/CityNetwork/tree/master/Models/QuantEpistemo/HyperNetwork}.
}

\paragraph{Functions}{Fonctions}

\bpar{
The following functions are ensured: (i) collection of the citation network from an initial corpus, collection of abstracts for a corpus; (ii) extraction of keywords as n-grams, estimation of the relevance of keywords; (iii) construction of semantic and citation networks.
}{
Les fonctions suivantes sont assurées : (i) collecte du réseau de citation à partir d'un corpus initial, collecte des résumés d'un corpus ; (ii) extraction des mots-clés sous forme de n-grams, estimation de la pertinence des mots clés ; (iii) construction des réseaux sémantiques et de citation.
}

%

\section{Description of algorithms and simulation models implementations}{Description des implémentations des algorithmes et des modèles de simulation}

\label{app:code} 







\bpar{
It is in our sense not particularly relevant to make the main text less readable with code listing as soon as there are no algorithmic details requiring a particular focus. As soon as the implementation biases are avoided, the architecture and the source code of the implementation of a simulation model should be independent of its formal description (but naturally provided with it, as we developed in~\ref{sec:reproducibility}).
}{
Il n'est à notre sens pas particulièrement pertinent d'alourdir un corps de texte avec du listing de code s'il n'y a pas de détails algorithmique nécessitant une attention particulière. Tant que les biais d'implémentation sont évités, l'architecture et le code source de l'implémentation d'un modèle de simulation devraient être indépendants de sa description formelle (mais naturellement fournis avec celle-ci, comme nous l'avons développé en~\ref{sec:reproducibility}).
}

\bpar{
We give thus in this section the list and a minimal description of simulation models and algorithms implementations we used. The language and the size (in terms of lines of code) are given, and also particular details when they are worth noticing. All models and analyses are gathered at \url{https://github.com/JusteRaimbault/CityNetwork/tree/master/Models}.
}{
Nous donnons ainsi dans cette section la liste et une description minimale des implémentations des modèles de simulation et des algorithmes que nous avons utilisé. Le langage et la taille (en termes de nombre de lignes de code) sont fournis, ainsi que les détails à noter le cas échéant. L'ensemble des modèles et des analyses sont regroupés à \url{https://github.com/JusteRaimbault/CityNetwork/tree/master/Models}.
}



\subsection{Algorithmic systematic review}{Revue systématique algorithmique}

\paragraph{Objectives}{Objectifs}

\bpar{
Implementation of the systematic literature review algorithm.
}{
Implémentation de l'algorithme de revue systématique.
}

\paragraph{Location}{Localisation}

\url{https://github.com/JusteRaimbault/CityNetwork/tree/master/Models/QuantEpistemo/AlgoSR/AlgoSRJavaApp}

\paragraph{Characteristics}{Caractéristiques}

\bpar{
\begin{itemize}
\item Language: \texttt{Java}
\item Size: 7116
\end{itemize}
}{
\begin{itemize}
\item Langage : \texttt{Java}
\item Taille : 7116
\end{itemize}
}

\paragraph{Particularities}{Particularités}

\bpar{
The HashConsing technique is used to keep unique bibliographic objects.
}{
La technique de HashConsing est utilisée pour conserver des objets bibliographiques uniques.
}

\paragraph{Architecture}{Architecture}

\bpar{
See the diagram in~\ref{fig:quantepistemo:algo}.
}{
Voir le diagramme en~\ref{fig:quantepistemo:algo}.
}

\paragraph{Additional scripts}{Scripts additionnels}

\bpar{
Exploration of results (R).
}{
Exploration des résultats (R).
}


\subsection{Indirect bibliometrics}{Bibliométrie indirecte}

\paragraph{Objectives}{Objectifs}

\bpar{
Analysis through citation network and semantic network of scientific corpuses: corpus of~\ref{sec:quantepistemo}, Cybergeo journal (\ref{app:sec:cybergeo}); modelography (section~\ref{sec:modelography}).
}{
Analyse par réseau de citation et réseau sémantique de corpus scientifiques : corpus de~\ref{sec:quantepistemo}, journal Cybergeo (\ref{app:sec:cybergeo}) ; modélographie (section~\ref{sec:modelography}).
}

\paragraph{Location}{Localisation}


\url{https://github.com/JusteRaimbault/CityNetwork/tree/master/Models/QuantEpistemo/HyperNetwork}

\paragraph{Characteristics}{Caractéristiques}

\bpar{
\begin{itemize}
\item Language: \texttt{Python}, \texttt{R} and \texttt{Java}.
\item Size: 2210
\end{itemize}
}{
\begin{itemize}
\item Langage : \texttt{Python}, \texttt{R} and \texttt{Java}.
\item Taille : 2210
\end{itemize}
}

\paragraph{Particularities}{Particularités}


\bpar{
Uses different databases, sqlite, sql or Mongodb depending on operations.
}{
Utilise des bases de données sqlite, sql ou Mongodb selon les opérations.
}

\paragraph{Architecture}{Architecture}

\bpar{
See Fig.~\ref{fig:cybergeo:fig1} in Appendix~\ref{app:sec:cybergeo}.
}{
Voir Fig.~\ref{fig:cybergeo:fig1} en Annexe~\ref{app:sec:cybergeo}.
}

















\subsection{Static correlations}{Corrélations statiques}

\paragraph{Objective}{Objectif}

\bpar{
Computation of morphological indicators, of network indicators, and of their correlations.
}{
Calcul des indicateurs morphologiques, indicateurs de réseau et de leur corrélations.
}

\paragraph{Location}{Localisation}

\url{https://github.com/JusteRaimbault/CityNetwork/tree/master/Models/StaticCorrelations}

\paragraph{Characteristics}{Caractéristiques}

\bpar{
\begin{itemize}
\item Language: \texttt{R}
\item Size: 1862
\end{itemize}
}{
\begin{itemize}
\item Langage : \texttt{R}
\item Taille : 1862
\end{itemize}
}







\subsection{Spatio-temporal causalities}{Causalités spatio-temporelles}

\paragraph{Objective}{Objectif}

\bpar{
Causality regimes, synthetic data (arma and rbd model) and empirical analyses (Grand Paris, South Africa, France).
}{
Régimes de causalité, données synthétiques (arma et modèle rbd) et analyses empiriques (Grand Paris, Afrique du Sud, France).
}

\paragraph{Location}{Localisation}

\url{https://github.com/JusteRaimbault/CityNetwork/tree/master/Models/SpatioTempCausalities}

\paragraph{Characteristics}{Caractéristiques}

\bpar{
\begin{itemize}
\item Language: \texttt{R}
\item Size: 8627
\end{itemize}
}{
\begin{itemize}
\item Langage : \texttt{R}
\item Taille : 8627
\end{itemize}
}







\subsection{Macroscopic interaction model}{Modèle d'interaction macroscopique}

\paragraph{Objective}{Objectif}

\bpar{
Macroscopic interaction model, section~\ref{sec:interactiongibrat}
}{
Modèle d'interaction macroscopique, section~\ref{sec:interactiongibrat}
}

\paragraph{Location}{Localisation}

\url{https://github.com/JusteRaimbault/CityNetwork/tree/master/Models/InteractionGibrat}

\paragraph{Characteristics}{Caractéristiques}

\bpar{
\begin{itemize}
\item Language: \texttt{NetLogo}, \texttt{scala}, \texttt{R}
\item Size: 5918
\end{itemize}
}{
\begin{itemize}
\item Langage : \texttt{NetLogo}, \texttt{scala}, \texttt{R}
\item Taille : 5918
\end{itemize}
}

\paragraph{Particularities}{Particularités}

\bpar{
The model is implemented in different languages for reasons of complementarity: \texttt{NetLogo} for the interactive exploration, \texttt{R} for the integration with statistical tests, \texttt{scala} for the calibration with OpenMole.
}{
Le modèle est implémenté dans différents langages pour des raisons complémentaires : \texttt{NetLogo} pour l'exploration interactive, \texttt{R} pour intégration avec les tests statistiques, \texttt{scala} pour la calibration par OpenMole.
}





\subsection{Density morphogenesis}{Morphogenèse de la densité}

\paragraph{Objective}{Objectif}

\bpar{
Morphogenesis model for density (section~\ref{sec:densitygeneration}).
}{
Modèle de morphogenèse pour la densité (section~\ref{sec:densitygeneration}).
}

\paragraph{Location}{Localisation}

\url{https://github.com/JusteRaimbault/CityNetwork/tree/master/Models/Synthetic/Density}

\paragraph{Characteristics}{Caractéristiques}

\bpar{
\begin{itemize}
\item Language: \texttt{NetLogo}, \texttt{scala}, \texttt{R}
\item Size: 5065
\end{itemize}
}{
\begin{itemize}
\item Langage : \texttt{NetLogo}, \texttt{scala}, \texttt{R}
\item Taille : 5065
\end{itemize}
}


\subsection{Correlated synthetic data generation}{Génération des données synthétiques corrélées}

\paragraph{Objectives}{Objectifs}

\bpar{
Weak coupling of density generation and network generation.
}{
Couplage faible de la génération de densité et de la génération de réseau.
}

\paragraph{Location}{Localisation}

\url{https://github.com/JusteRaimbault/CityNetwork/tree/master/Models/Synthetic/Network}

\paragraph{Characteristics}{Caractéristiques}

\bpar{
\begin{itemize}
\item Language: \texttt{NetLogo} (network) and \texttt{scala} (density)
\item Size: 3188
\end{itemize}
}{
\begin{itemize}
\item Langage : \texttt{NetLogo} (réseau) and \texttt{scala} (densité)
\item Taille : 3188
\end{itemize}
}

\paragraph{Particularities}{Particularités}

\bpar{
Network heuristic are more naturally implemented and explored in NetLogo.
}{
Les heuristiques de réseau sont plus naturelles à implémenter et explorer en NetLogo.
}

\paragraph{Architecture}{Architecture}

\bpar{
The weak coupling between modules is realized through the intermediate of an OpenMole script.
}{
Le couplage faible entre les modules est réalisé par l'intermédiaire d'un script OpenMole.
}




\subsection{Co-evolution at the macroscopic scale}{Co-évolution à l'échelle macroscopique}

\paragraph{Objective}{Objectif}

\bpar{
Implementation of the co-evolution model at the macroscopic scale (section~\ref{sec:macrocoevol}).
}{
Implémentation du modèle de co-évolution à l'échelle macroscopique (section~\ref{sec:macrocoevol}).
}

\paragraph{Location}{Localisation}

\url{https://github.com/JusteRaimbault/CityNetwork/tree/master/Models/MacroCoevol}

\paragraph{Characteristics}{Caractéristiques}

\bpar{
\begin{itemize}
\item Language: \texttt{NetLogo}
\item Size: 4950
\end{itemize}
}{
\begin{itemize}
\item Langage : \texttt{NetLogo}
\item Taille : 4950
\end{itemize}
}

\paragraph{Particularities}{Particularités}

\bpar{
Dual representation of the network with links and distance matrix.
}{
Représentation duale liens/matrice de distance du réseau.
}


\paragraph{Data used}{Données utilisées}

\bpar{
Population of French urban areas 1830-1999
}{
Population des aires urbaines Françaises 1830-1999
}

\paragraph{Additional scripts}{Scripts additionnels}

\bpar{
Exploration and calibration (oms), exploration of results (R)
}{
Exploration et calibration (oms), exploration des résultats (R)
}


\subsection{Co-evolution by morphogenesis}{Co-évolution par morphogenèse}

\paragraph{Objective}{Objectif}

\bpar{
Implementation of the co-evolution model at the mesoscopic scale (sections~\ref{sec:networkgrowth} and~\ref{sec:mesocoevolmodel}).
}{
Implémentation du modèle de co-évolution à l'échelle mesoscopique (sections~\ref{sec:networkgrowth} et~\ref{sec:mesocoevolmodel}).
}

\paragraph{Location}{Localisation}

\url{https://github.com/JusteRaimbault/CityNetwork/tree/master/Models/MesoCoevol}

\paragraph{Characteristics}{Caractéristiques}

\bpar{
\begin{itemize}
\item Language: \texttt{NetLogo}
\item Size: 5386
\end{itemize}
}{
\begin{itemize}
\item Langage : \texttt{NetLogo}
\item Taille : 5386
\end{itemize}
}



\paragraph{Additional scripts}{Scripts additionnels}

\bpar{
Exploration and calibration (oms), exploration of results (R)
}{
Exploration et calibration (oms), exploration des résultats (R)
}


\subsection{Lutecia model}{Modèle Lutecia}

\paragraph{Objective}{Objectif}

\bpar{
Implementation of the Lutecia model, (section~\ref{sec:lutecia}).
}{
Implémentation du modèle Lutecia (section~\ref{sec:lutecia}).
}

\paragraph{Location}{Localisation}

\url{https://github.com/JusteRaimbault/CityNetwork/tree/master/Models/Governance/Lutecia/Lutecia}

\paragraph{Characteristics}{Caractéristiques}

\bpar{
\begin{itemize}
\item Language: \texttt{NetLogo}
\item Size: 8866
\end{itemize}
}{
\begin{itemize}
\item Langage : \texttt{NetLogo}
\item Taille : 8866
\end{itemize}
}

\paragraph{Particularities}{Particularités}

\bpar{
The matrix of effective distances is updated through dynamical programming.
}{
La matrice des distances effectives est mise à jour par programmation dynamique.
}


\paragraph{Additional scripts}{Scripts additionnels}

\bpar{
Exploration/calibration of the model (oms), exploration of results (\texttt{R}).
}{
Exploration/calibration du modèle (oms), exploration des résultats (\texttt{R}).
}


\subsection{Static User Equilibrium}{Equilibre Utilisateur Statique}

\paragraph{Objective}{Objectif}

\bpar{
Collection and analysis of traffic data for the greater Paris metropolitan area (section~\ref{sec:reproducibility}).
}{
Collecte et analyse des données de trafic pour la métropole du Grand Paris (section~\ref{sec:reproducibility}).
}

\paragraph{Location}{Localisation}

\url{https://github.com/JusteRaimbault/TransportationEquilibrium/tree/master/Models}

\paragraph{Characteristics}{Caractéristiques}

\bpar{
\begin{itemize}
\item Language: \texttt{python}, \texttt{R}
\item Size: $\simeq$ 300
\end{itemize}
}{
\begin{itemize}
\item Langage : \texttt{python}, \texttt{R}
\item Taille : $\simeq$ 300
\end{itemize}
}






\subsection{Geography of fuel prices}{Géographie des prix du carburant}

\paragraph{Objective}{Objectif}

\bpar{
Collection and analysis of fuel price data in the United States (section~\ref{sec:energyprice}).
}{
Collecte et analyse des données des prix du carburant aux Etats-unis (section~\ref{sec:energyprice}).
}

\paragraph{Location}{Localisation}

\url{https://github.com/JusteRaimbault/EnergyPrice/tree/master/Models}

\paragraph{Characteristics}{Caractéristiques}

\bpar{
\begin{itemize}
\item Language: \texttt{python}, \texttt{R}
\item Size: 1469
\end{itemize}
}{
\begin{itemize}
\item Langage : \texttt{python}, \texttt{R}
\item Taille : 1469
\end{itemize}
}

\paragraph{Particularities}{Particularités}

\bpar{
Use of the TorPool software (see~\ref{app:sec:packages}) for data collection.
}{
Utilisation du logiciel TorPool (voir~\ref{app:sec:packages}) pour la collecte des données.
}



\stars

%

\section{Tools and workflow for an open reproducible research}{Outils et pratiques pour une recherche ouverte et reproductible}

\label{app:workflow} 



%
%
%

\bpar{
We briefly evoke here tools, practices, and development directions for a more transparent, free, open and fluid research.
}{
Nous décrivons ici brièvement des outils, des pratiques et des pistes de développement pour une recherche plus transparente, libre, ouverte et fluide.
}

\subsection{NetLogo documentation generator}{Générateur de Documentation Netlogo}

\bpar{
Documentation generation is central for reproducibility, as it can automatize the description of a model implementation. NetLogo does not provide a documentation generator. We implemented a \texttt{Doxygen} software (generation of documentation for different languages including Java) for its application to the NetLogo language. It basically consists in generating intermediate Java code, mirror of the NetLogo code in its object structures and containing the comment blocks of the NetLogo code. An experimental version is available at \url{https://github.com/JusteRaimbault/CityNetwork/tree/master/Models/Doc}.
}{
La génération de documentation est centrale pour la reproductibilité, permettant d'automatiser la description de l'implémentation d'un modèle. NetLogo ne fournit pas de générateur de documentation. Nous avons implémenté un wrapper du logiciel \texttt{Doxygen} (génération de documentation pour divers langages dont Java) pour son application au langage NetLogo. Il repose sur le principe basique de génération d'un code Java intermédiaire, miroir du code NetLogo dans ses structures objet et reprenant les blocs de commentaires dans le code NetLogo. Une version expérimentale est disponible à \url{https://github.com/JusteRaimbault/CityNetwork/tree/master/Models/Doc}.
}

\subsection{git as a reproducibility tool}{git comme outil de reproductibilité}

\bpar{
The use if \texttt{git} as a reproducibility and transparency tool has been emphasized by~\cite{ram2013git}, which list numerous advantages such as the exact tracking of the history of the knowledge production process, an immediate cloning (in combination with public repositories, for which collaborative sites exist such as github or gitlab), a possibility to branch from past commits.
}{
L'utilisation de \texttt{git} comme outil de reproductibilité et de transparence a été mis en valeur par~\cite{ram2013git}, qui soulignent de nombreux avantages tels le suivi exact de l'historique du processus de production de connaissance, un clonage immédiat (en combinaison avec des dépôts publics, pour lesquels existent des sites collaboratifs comme github ou gitlab), une possibilité de branchage à partir de commits passés.
}

\bpar{
This tool furthermore allows facilitating the individual workflow, providing for example an automatic backup, an organisational support, the following of experiments.
}{
Cet outil permet d'autre part de faciliter le flux de travail individuel, permettant par exemple le backup automatique, l'organisation, le suivi des expériences.
}

\subsection{Open review}{Revue ouverte}

\bpar{
The review process of this manuscript has experimentally tested an open review, through the use of the git repository and specific \LaTeX commands. The basic \texttt{{\textbackslash}comment} command allows the reviewers to insert their comments in the appropriate place (and is then placed as a margin annotation of the manuscript) and allows a discussion up to 5 consecutive answers through optional arguments. A \emph{pull request} from the reviewer branch allows integrating the feedbacks. Other commands for example allow marking changes or inserting lists of tasks.
}{
Le processus de revue de ce manuscrit a expérimentalement testé la revue ouverte, par l'utilisation du dépôt git et de commandes \LaTeX spécifiques. La commande basique \texttt{{\textbackslash}comment} permet aux relecteurs d'insérer leur commentaires à l'endroit approprié (et se place alors en annotation de marge dans le manuscrit) et permet une discussion jusqu'à 5 réponses successives par des arguments optionnels. Une \emph{pull request} depuis la branche du relecteur permet d'intégrer les retours. D'autres commandes permettent par exemple de marquer les changements ou d'insérer des listes de tâches.
}

\bpar{
One of the advantages of this approach is that it is a posteriori possible to reconstruct the review process, and that it is totally open (for a potential review of the review). The automation by traversing the network of the git repository history is even easily considerable. 
}{
L'un des intérêts de cette démarche est qu'il est possible a posteriori de reconstruire le processus de revue, et que celui-ci est entièrement ouvert (pour une éventuelle revue de la revue). L'automatisation par parcours du réseau de l'historique du dépôt git est même facilement envisageable.
}





\subsection{Towards a git-compatible metadata handler}{Vers un gestionnaire de métadonnées compatible avec git}

\bpar{
The issue of conserving metadata for figures is crucial for reproducibility, since it is often difficult to keep trace of the full configuration having generated a figure, and also of the corresponding code, since it can be modified by older versions. The use of script environments such as R or python can also build some traps since variables can be modified without modifying the code, and the full history of executed commands must then be kept.
}{
La question de la conservation des métadonnées pour les figures est cruciale pour la reproductibilité, puisqu'il est souvent difficile de garder une trace de l'ensemble de la configuration ayant généré une figure, ainsi que le code correspondant, celui-ci pouvant être modifié par des versions antérieures. L'utilisation d'environnements de scripts comme R ou python peuvent également être piégeurs puisque les variables peuvent être modifiées sans modification du code, et il faut garder alors l'ensemble de l'historique des commandes exécutées.
}

\bpar{
The exhaustive storage of data, the environment, code and the history which led to the generation of a precise figure are a necessary condition for an exact reproducibility. A direction to answer this issue is the construction of a tool compatible with git which would automatically generate these metadata, for example by creating a proper branch and conserving the commit hash associated to the figure. The final idea would be to have for each figure a unique identifier linking it to the exact environment having produced it, also implying an automation of the index system within the documents using them.
}{
Le stockage exhaustif des données, de l'environnement, du code et de l'historique qui a conduit à la génération d'une figure précise sont une condition nécessaire pour une reproductibilité exacte. Une piste pour répondre à ce problème est l'élaboration d'un outil compatible avec git qui générerait automatiquement ces métadonnées, par exemple en créant une branche propre et en conservant le hash du commit associé à la figure. L'idée finale serait d'avoir pour chaque figure un identifiant unique la reliant à l'environnement exact l'ayant produite, impliquant également une automatisation du système d'indexation au sein des documents les utilisant.
}


\stars


%


\bpar{
\chapter{Reflexive analysis}
}{
\chapter{Analyse réflexive}
}

\bpar{
\markboth{\thechapter\space Reflexivity}{\thechapter\space Reflexivity}
}{
\markboth{\thechapter\space Réflexivité}{\thechapter\space Réflexivité}
}

\label{app:reflexivity} 


\bpar{
We have throughout all this work highlighted the crucial role of a reflexivity in the research process. It has played a role for the definition of objects or questions asked, in a concrete way in works directly based on fieldwork, or in the elaboration of theories with a recursive aspect. Without pretending having exhaustively constructed a ``meta-viewpoint'' as recommended by~\cite{morin1991methode} for the construction of a complex thinking, we suggest to have brought preliminary elements of answer. 
}{
Nous avons vu tout au long de ce travail le rôle crucial d'une réflexivité dans la démarche de recherche. Celle-ci a pu jouer pour la définitions des objets ou les questions posées, concrètement dans les travaux directement basés sur un travail de terrain, ou dans l'élaboration de théories ayant un aspect récursif. Sans prétendre avoir exhaustivement construit un ``méta-point de vue'' comme préconisé par~\cite{morin1991methode} pour la construction d'une pensée complexe, nous suggérons avoir apporté des éléments de réponse de manière préliminaire.
}

\bpar{
We propose here, as a ``meta-conclusion'', to proceed to a quantitative analysis for reflexivity, by applying the methods we developed to our work itself. We do in a first part the analysis of the scientific landscape from the corpus of our bibliography, and then analyze in a second part the evolution of the knowledge produced in terms of projects and of knowledge domains.
}{
Nous proposons ici, en guise de ``méta-conclusion'', de mener une analyse quantitative pour la réflexivité, en appliquant les méthodes développées à notre travail lui-même. Nous menons dans un premier temps l'analyse du paysage scientifique à partir du corpus de notre bibliographie, puis analysons dans un second temps l'évolution de la connaissance produite en termes de projets et de domaines de connaissance.
}

\bpar{
This approach is particularly original since to the best of our knowledge there exist no monograph explicitly including its own analysis using quantitative tools. We defend a more systematic use of such approaches, to foster the development of a knowledge at the second order.
}{
Cette démarche est particulièrement originale puisqu'il n'existe à notre connaissance pas de monographie incluant explicitement sa propre analyse par des outils quantitatifs. Nous prenons le parti d'une utilisation plus systématique de ce genre de démarche, pour encourager le développement d'une connaissance de second ordre.
}

\section{Hypernetwork analysis}{Analyse par hyperréseau}

\bpar{
We apply here the methodology using citation and semantic networks developed in Chapter~\ref{ch:modelinginteractions}. The initial corpus is constituted by all our bibliography\footnote{Fixed at the 27/11/2017, and available at \url{https://github.com/JusteRaimbault/CityNetwork/raw/master/Models/Reflexivity/data/CityNetwork_20171127.bib}.} which includes 834 references.
}{
Nous appliquons ici la méthodologie par réseau de citation et réseau sémantique développée en Chapitre~\ref{ch:modelinginteractions}. Le corpus initial est constitué de l'ensemble de notre bibliographie\footnote{Figée au 27/11/2017, et disponible à \url{https://github.com/JusteRaimbault/CityNetwork/raw/master/Models/Reflexivity/data/CityNetwork_20171127.bib}.} qui comporte 834 références.
}

\subsection{Citation Network}{Réseau de citation}

\bpar{
We reconstruct the citation network at depth two from this initial corpus, and obtain a consequent network ($\left|V\right| = 177428$, $\left|E\right| = 203317$), of average degree $2.29$ (average in-degree $1.15$). The core of the network, constituted by vertices with a degree larger than or equal to 2 for the largest connected component (which covers 98\% of the network), has a size of $\left|V\right| = 19714$ and $\left|E\right| = 47348$.
}{
Nous reconstruisons le réseau de citation à profondeur deux à partir de ce corpus initial, et obtenons un réseau conséquent ($\left|V\right| = 177428$, $\left|E\right| = 203317$), de degré moyen $2.29$ (degré moyen entrant $1.15$). Le coeur du réseau, constitué des sommets de degré supérieur ou égal à 2 pour la plus grande composante connexe (qui couvre 98\% du réseau), est de taille $\left|V\right| = 19714$ et $\left|E\right| = 47348$.
}



\bpar{
A detection of communities using the Louvain algorithm gives a directed modularity of 0.74 for 19 communities with a size larger than 10. We interpret the communities by the tags given in Table~\ref{tab:app:reflexivity:citationcoms}. We recover domains that are directly covered and used in our work (Urban Systems, Spatial Models of Urban Growth), and other neighbors mentioned but not directly used (Fractals, Economic Geography, Space Syntax).
}{
Une détection de communautés par algorithme de Louvain donne une modularité dirigée de 0.74 pour 19 communautés de taille supérieure à 10. Nous interprétons les communautés par les étiquettes données en Table~\ref{tab:app:reflexivity:citationcoms}. Nous retrouvons des domaines directement couverts et utilisés dans notre travail (Systèmes Urbains, Modèles Spatiaux de Croissance Urbaine), et d'autres voisins évoqués mais pas directement utilisés (Fractales, Economie Géographique, Space Syntax).
}


\begin{table}
\apptabcaption{\textbf{Citation communities.} The size of communities is given as a proportion of the size of the core of the network.\label{tab:app:reflexivity:citationcoms}}{\textbf{Communautés de citation.} La taille des communautés est donnée en proportion de la taille du coeur du réseau.\label{tab:app:reflexivity:citationcoms}}
\begin{tabular}{|l|l|}
\hline
\bpar{
Community & Size \\ \hline
}{
Communauté & Taille \\ \hline
}
Economic Geography & 12.4 \% \\
Power Laws & 9.1 \% \\
Networks & 7.92 \% \\
Spatial Urban Growth Models & 7.67 \% \\
Physics of Cities & 7.43 \% \\
ABM & 7.37 \% \\
Complexity & 7.19 \% \\
LUTI & 7.16 \% \\
Urban Systems & 5.15 \% \\
Spatial Statistics & 5.13 \% \\
Evolutionary Economic Geography & 5.03 \% \\
Spatio-temporal data & 3.18 \% \\
Datamining & 2.81 \% \\
Quantitative Epistemology & 2.43 \% \\
Space Syntax/Procedural modeling & 2.43 \% \\
Fractals & 2.02 \% \\
VGI & 1.8 \% \\
Biological Networks & 1.33 \% \\
Chaos & 0.624 \% \\\hline
\end{tabular}
\end{table}

\begin{figure}
	\includegraphics[width=\linewidth]{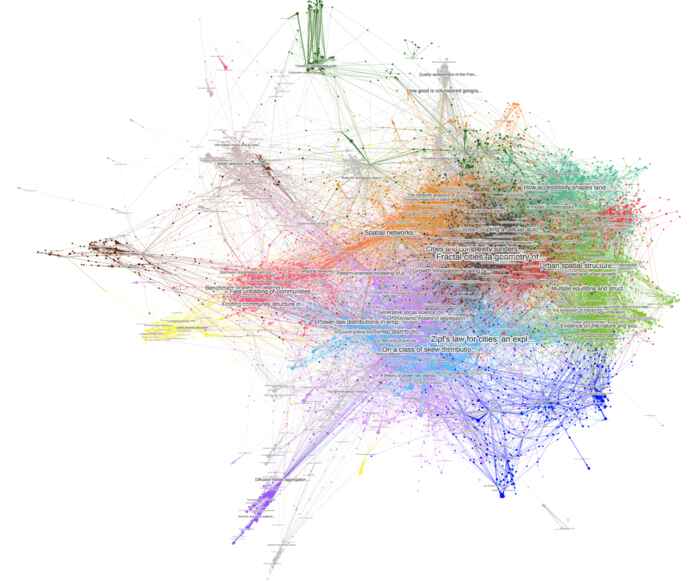}
	\appcaption{\textbf{Citation network.} We visualize only the core of the network, constituted here by nodes with a degree larger than or equal to 2. The network is spatialized with the algorithm Force Atlas 2. The size of labels is proportional to the degree of nodes, and the color gives the community.\label{fig:app:reflexivity:citnw}}{\textbf{Réseau de citation.} Nous visualisons uniquement le coeur du réseau, constitué ici des noeuds de degré supérieur ou égal à 2. Le réseau est spatialisé par algorithme Force Atlas 2. La taille des labels est proportionnelle au degré des noeuds, et la couleur donne la communauté.\label{fig:app:reflexivity:citnw}}
\end{figure}

\bpar{
The citation network is visualized in Fig.~\ref{fig:app:reflexivity:citnw}. The position of communities is very instructive to situate our work, which forms bridges between different domains depending on the viewpoint chosen. If we take the point of view of urban systems, the corresponding community (in red) makes a bridge between LUTI models (turquoise) and urban growth models (black) on one side, and economic geography on the other side (in green). If we take the viewpoint of simulation models (ABM community, purple), the link is established between Power Laws (light blue) and networks and spatial networks (magenta and orange). Auxiliary communities are attached at the periphery: Quantitative Epistemology (yellow) is close to network analysis, whereas the analysis of spatio-temporal processes (dark green) is relatively independent. Communities within which our models can be thematically classified (growth models and urban systems) are located at the core of the compact part of the network: this confirms that the direction explored are not auxiliary, and that all principal auxiliary domains evoked where ``necessary'' in the sense of a strong connection between communities here.
}{
Le réseau de citation est visualisé en Fig.~\ref{fig:app:reflexivity:citnw}. Le positionnement des communautés est très instructif pour situer notre travail, qui forme des ponts entre différents domaines selon le point de vue choisi. Si nous prenons le parti des systèmes urbains, la communauté correspondante (en rouge) fait le pont entre modèles LUTI (turquoise) et modèles de croissance urbaine (noir) d'une part, et économie géographique (en vert). Si on se place du point de vue des modèles de simulation (communauté ABM, violet), le lien est fait entre Lois Puissances (bleu clair) et les réseaux et réseaux spatiaux (magenta et orange). Des communautés annexes se rattachent à la périphérie : Epistémologie Quantitative (jaune) est proche de l'analyse de réseau, tandis que l'analyse des processus spatio-temporels (vert foncé) est relativement indépendante. Les communautés dans lesquelles nos modèles peuvent être classifiés de manière thématique (modèles de croissance et systèmes urbains) sont situées au coeur de la partie compacte du réseau : cela confirme que les directions explorées ne sont pas périphériques, et que l'ensemble des domaines principaux périphériques évoqués étaient ``nécessaires'' au sens d'une forte connexion entre communautés ici.
}

\subsection{Semantic network}{Réseau sémantique}


\bpar{
After collecting the abstracts, we obtain 91412 references on which it is possible to proceed to the semantic analysis. The construction of the raw co-occurrences network, after filtering links with a weight smaller than 5, and for a number of keywords $K_W = 50000$, yields a semantic network with $\left|E\right|\simeq 16\cdot 10^6$. The sensitivity analysis to filtering parameters suggests to choose $k_{min} = 0$, $k_{max}=500$, $f_{max} = 10000$, $\theta_w = 5$, what produces a semantic network of size $\left|V\right| = 37482$ and $\left|E\right| = 218926$, with 26 communities and a modularity of $0.78$. Main communities can be labeled as: toxicology, chemistry, political sciences, theoretical ecology, urban systems, sustainability, innovation economics, spatial analysis, physiology, physics, networks, bio-anthropology, health, statistics, microbiology, transportation, biological networks, health geography, botany, evolution, ecology, genetics.
}{
Après collecte des résumés, nous obtenons 91412 références sur lesquelles il est possible de procéder à l'analyse sémantique. La construction du réseau de co-occurrences brut, en filtrant les liens de poids inférieur à 5, et pour un nombre de mots-clés $K_W = 50000$, donne un réseau sémantique avec $\left|E\right|\simeq 16\cdot 10^6$. L'analyse de sensibilité aux paramètres de filtrage nous suggère de choisir $k_{min} = 0$, $k_{max}=500$, $f_{max} = 10000$, $\theta_w = 5$, ce qui fournit un réseau sémantique de taille $\left|V\right| = 37482$ et $\left|E\right| = 218926$, avec 26 communautés et une modularité de $0.78$. Les communautés principales peuvent être labellisées comme : toxicologie, chimie, sciences politiques, écologie théorique, systèmes urbains, soutenabilité, économie de l'innovation, analyse spatiale, physiologie, physique, réseaux, bio-anthropologie, santé, statistiques, microbiologie, transports, réseaux biologiques, géographie de la santé, botanique, évolution, écologie, génétique.
}


\bpar{
It is less evident to use this typology to understand our work, in comparison with the citation network, since remote domains (toxicology, chemistry, physiology, botany) can be found in relatively small amount in our citation corpus (coming from common citations on morphogenesis or ecology for example) but form then communities that are particularly isolated in the semantic network. We give in Fig.~\ref{fig:app:reflexivity:interdisc} the distribution of semantic interdisciplinarities for each citation community. At the exception of evolutionary economic geography which is relatively flat (and thus rather closed) and voluntary geographical information (VGI) which exhibits a peak at 0 (which is expected for such a specific domain), citation communities have fundamentally the same interdisciplinarity profile.
}{
Il est moins évident d'utiliser ce découpage pour comprendre notre travail, en comparaison au réseau de citation, puisque des domaines lointains (toxicologie, chimie, physiologie, botanique) se retrouvent en nombre relativement faible dans notre corpus de citation (provenant de citations communes sur la morphogenèse ou l'écologie par exemple) mais forment ensuite des communautés particulièrement isolées dans le réseau sémantique. Nous donnons en Fig.~\ref{fig:app:reflexivity:interdisc} la distribution des interdisciplinarités sémantiques par communautés de citation. À l'exception de l'économie géographique évolutionnaire qui est relativement plate (et donc assez cloisonnée) et de l'information géographique volontaire (VGI) qui présente un pic à 0 (attendu d'un domaine si spécifique), les communautés de citation ont fondamentalement le même profil d'interdisciplinarité.
}

\begin{figure}
	\includegraphics[width=\linewidth]{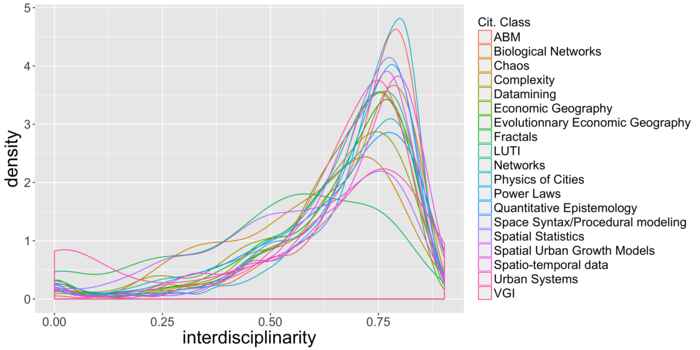}
	\appcaption{\textbf{Distribution of interdisciplinarities for each citation community.}\label{fig:app:reflexivity:interdisc}}{\textbf{Distribution des interdisciplinarités par communauté de citation.}\label{fig:app:reflexivity:interdisc}}
\end{figure}

\section{Interaction between projects}{Interaction entre projets}

\bpar{
We propose here to quantify the evolution of the different projects and of their interactions, and also of associated knowledge domains. A table of the time spent on each project, with a precision of half an hour, has been held between the 16/02/2015 and the 16/02/2018\footnote{It is available at \url{https://github.com/JusteRaimbault/CityNetwork/raw/master/Docs/Organisation/Projects.ods}. For the analyses here, we use the version frozen at the 02/12/2017.}. A project is defined as a minimal consistent entity, either by its thematic (for example: morphogenesis model of~\ref{sec:densitygeneration}) either by its content (case studies, geographical theory). These have been defined progressively in time, and some overlap or are the precursors of others: we have thus built a classification a posteriori under the form of ``macro-projects'' which globally correspond to the final articulation. We also associate to them a main knowledge domain\footnote{Knowing that there is a non-negligible bias in the fact of attributing a unique domain to a project, since domains are generally intimately linked at the core of knowledge production itself. The constraint of data collection however leads to this segmentation which is relatively reductionist.} and the main section of this memoir to which they are attached.
}{
Nous proposons ici de quantifier l'évolution des différents projets et de leur interactions, ainsi que des domaines de connaissance associés. Un tableau du temps passé sur chaque projet, à la demi-heure près, a été tenu entre le 16/02/2015 et le 16/02/2018\footnote{Celui-ci est disponible à \url{https://github.com/JusteRaimbault/CityNetwork/raw/master/Docs/Organisation/Projects.ods}. Pour les analyses ici, nous utilisons la version figée au 02/12/2017.}. Un projet est défini comme une entité minimale cohérente, soit par sa thématique (par exemple : modèle de morphogenèse de~\ref{sec:densitygeneration}) soit par son contenu (cas d'études, théorie géographique). Ceux-ci ayant été défini au cours du temps, certains se recoupent ou sont précurseurs d'autres : nous leur avons donc associé une classification a posteriori sous forme de ``macro-projets'' qui correspondent globalement à l'articulation finale. Nous leur attribuons également un domaine de connaissance majoritaire\footnote{Tout en sachant qu'il existe un biais non-négligeable dans le fait d'attribuer un domaine unique à un projet, puisque les domaines sont généralement intimement liés au coeur même de la production de connaissance. La contrainte de captation des données pousse toutefois à cette segmentation relativement réductionniste.} et la section principale de ce mémoire auxquels ils se rattachent.
}

\bpar{
The list of projects is given in Table~\ref{tab:app:reflexivity:projects}, with the macro-project, the knowledge domain and the cumulated time. The Fig.~\ref{fig:app:reflexivity:time} gives the temporal distribution according to these different modalities, in time. We confirm a non-linear organization, most of projects and chapters being treated in parallel. For example, the chapter~\ref{ch:mesocoevolution} has been the object of a first preliminary exploration in the first months, and a resurgence when converging as the intellectual maturity had been acquired. Methods regularly punctuate the distribution, but culminate just before the first half. Modeling projects, similarly to empirical studies, are also regularly distributed, whereas the conceptual takes more time in the end, what confirms that it necessitates the other domains and a thorough reflection. 
}{
La liste des projets est donnée en Table~\ref{tab:app:reflexivity:projects}, avec le macro-projet, le domaine et connaissance et le temps cumulé. La Fig.~\ref{fig:app:reflexivity:time} donne la répartition temporelle selon ces différentes modalités, dans le temps. Nous confirmons une organisation non-linéaire, la plupart des projets et chapitres étant menés en parallèle. Par exemple, le chapitre~\ref{ch:mesocoevolution} a fait l'objet d'une première exploration préliminaire dans les premiers mois, puis une résurgence à la convergence lorsque la maturité intellectuelle a été acquise. Les méthodes ponctuent régulièrement la répartition, mais culminent juste avant la première moitié. Les projets de modélisation, comme l'empirique, sont également régulièrement répartis, tandis que le conceptuel prend plus de place à la fin, ce qui confirme que celui-ci nécessite les autres domaines et une réflexion approfondie.
}

\begin{table}
	\apptabcaption{\textbf{Description of projects.} The section links to the part of the memoir where the project is mainly used. Global generic tasks are not taken into account in the chapter cumulated count (Memoire: writing of this memoir; Academic: academic life; Bibliography: general readings).\label{tab:app:reflexivity:projects}}{\textbf{Descriptif des projets.} La section renvoie à la partie du mémoire où le projet est principalement utilisé. Des tâches génériques globales ne sont pas prises en compte dans le cumul par chapitre (Memoire : écriture de ce document ; Academic : vie académique ; Bibliography : lectures générales).\label{tab:app:reflexivity:projects}}
\begin{tabular}{|l|l|l|l|l|}
\hline
\bpar{
Project & Macro-project & Section & Domain & Time (h) \\\hline
}{
Projet & Macro-projet & Section & Domaine & Time (h) \\\hline
}
CaseStudies & Thematic & \ref{sec:casestudies} & Empirical & 5.5 \\
Modelography & QuantEpistemo & \ref{sec:quantepistemo} & Empirical & 20 \\
QuantEpistemology & QuantEpistemo & \ref{sec:quantepistemo} & Empirical & 32\\
MacroCoEvol & MacroCoEvol & \ref{sec:macrocoevol} & Modeling & 72 \\
SpatioTempCausality & CausalityRegimes & \ref{sec:causalityregimes} & Methods & 37.5 \\
Entretiens & Thematic & \ref{app:sec:interviews} & Data & 13 \\
MesoCoEvol & MesoCoEvol & \ref{sec:mesocoevolmodel} & Modeling & 60.5 \\
Fieldwork & Thematic & \ref{sec:qualitative} & Empirical & 27.5 \\
EnergyPrice & Empirical & \ref{sec:energyprice} & Empirical & 72.5 \\
Morphogenesis & Morphogenesis & \ref{sec:interdiscmorphogenesis} & Conceptual & 24.5 \\
NetworkNecessity & InteractionGibrat & \ref{sec:interactiongibrat} & Modeling & 158 \\
Memoire & Memoire & - & Conceptual & 489.5 \\
SpatialStatistics & CausalityRegimes & \ref{sec:causalityregimes} & Methods & 44 \\
BPCaseStudy & CausalityRegimes & \ref{sec:casestudies} & Empirical & 12 \\
Perspectivism & Epistemology & \ref{sec:knowledgeframework} & Conceptual & 8.5 \\
RealEstate & CausalityRegimes & \ref{sec:casestudies} & Empirical & 18 \\
Theory & Thematic & \ref{sec:networkterritories}, \ref{sec:theory} & Conceptual & 136\\
CorrelatedSyntheticData & Methods & \ref{sec:correlatedsyntheticdata} & Methods & 128 \\
MediationEcotox & Methods & \ref{app:sec:mediationecotox} & Methods & 59 \\
DensityGeneration & DensityGeneration & \ref{sec:densitygeneration} & Modeling & 84.5 \\
PatentsMining & Methods & \ref{app:sec:patentsmining} & Methods & 349.5 \\
CyberGeo & Methods & \ref{app:sec:cybergeo}, \ref{app:sec:cybergeonetworks} & Methods & 332 \\
SpaceMatters & Methods & \ref{sec:computation} & Methods & 100.5 \\
NetworkDensityStatistics & Empirical & \ref{sec:staticcorrelations} & Empirical & 176.5\\
NetLogoUtils & Tools & - & Tools & 10 \\
StochasticUrbanGrowth & Methods & \ref{app:sec:stochurbgrowth} & Methods & 13 \\
TransportationEquilibrium & Empirical & \ref{sec:computation} & Empirical & 56.5 \\
BiologicalNetwork & MesoCoEvol & \ref{sec:networkgrowth} & Modeling & 5 \\
Discrepancy & Methods & \ref{app:sec:robustness} & Methods & 54 \\
Governance & Governance & \ref{sec:lutecia} & Modeling & 228 \\
SyntheticData & Methods & \ref{sec:densitygeneration} & Methods & 99 \\
Reproduction & MacroCoEvol & \ref{sec:macrocoevolexplo} & Modeling & 46 \\
AlgorithmicReview & QuantEpistemo & \ref{sec:quantepistemo} &  Empirical & 75.5 \\
Tools & Tools & - & Tools & 137 \\
Academic & Acad & - & NA & 1388 \\
Bibliography & Biblio & - & Conceptual & 312 \\
\hline
\end{tabular}
\end{table}

\begin{figure}
	\includegraphics[width=\linewidth,height=\textheight]{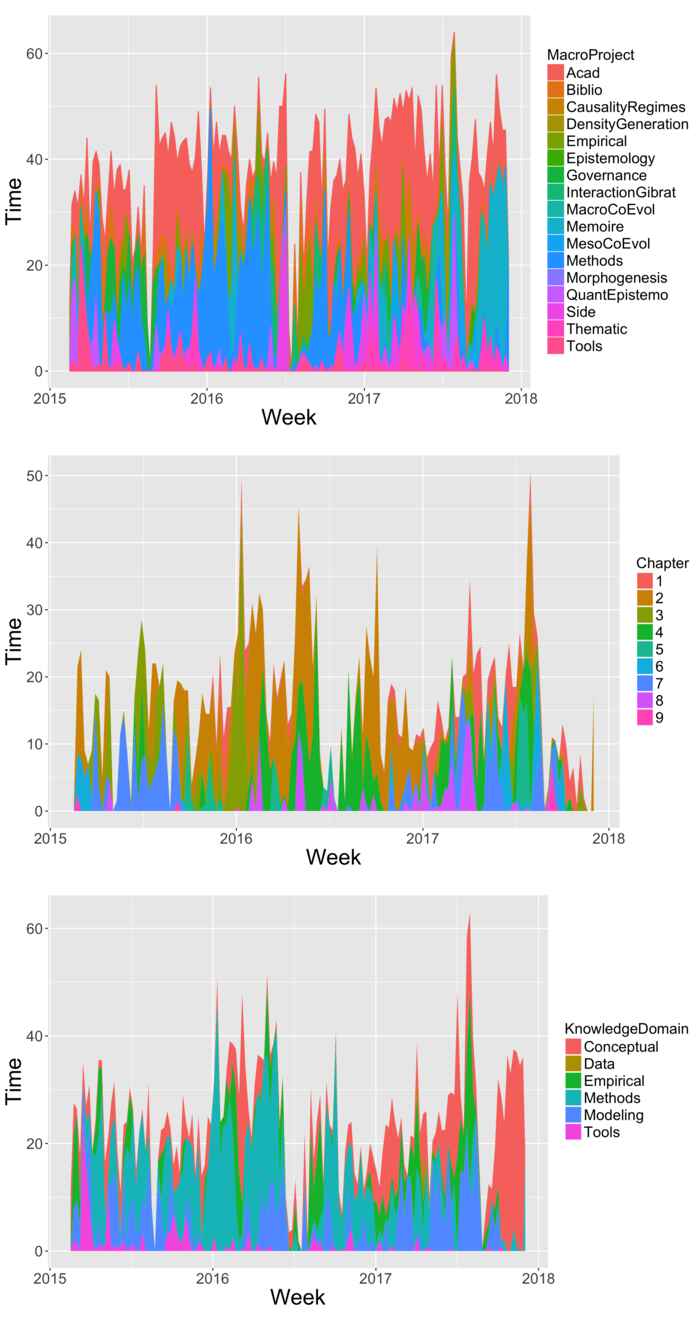}
	\appcaption{\textbf{Temporal distribution.} Times are aggregated at the level of the week and areas in color give the temporal distribution for macro-projects (\textit{first row}), chapters (\textit{second row}) and knowledge domains (\textit{third row}).\label{fig:app:reflexivity:time}}{\textbf{Répartition temporelle.} Les temps sont agrégés à la semaine et les aires en couleur donnent la répartition temporelle pour les macro-projets (\textit{première ligne}), les chapitres (\textit{deuxième ligne}) et les domaines de connaissance (\textit{troisième ligne}).\label{fig:app:reflexivity:time}}
\end{figure}

\bpar{
It is then possible to construct interaction graphs between macro-projects or knowledge domains, assuming simplifying hypotheses.
}{
Il est possible ensuite de construire des graphes d'interaction entre macro-projets ou domaines de connaissance, moyennant des hypothèses simplificatrices.
}

\bpar{
A first index of simultaneous interaction is based on an apparition at the same time. We denote $T_{i,t}$ the time for the entity $i$ (macro-project or knowledge domain) on the temporal unit $t$ (that we take as the week). The probability of simultaneous occurrence between $i$ and $j$ is at time $t$ given by $\frac{T_{i,t}T_{j,t}}{\left(\sum_i T_{i,t}\right)^2}$, and we can sum them in time to obtain an index of interaction between entities:
}{
Un premier indice d'interaction simultanée se base sur l'apparition au même instant. Notons $T_{i,t}$ le temps pour l'entité $i$ (macro-projet ou domaine de connaissance) sur l'unité temporelle $t$ (que nous prendrons à la semaine). La probabilité d'occurrence simultanée entre $i$ et $j$ est à l'instant $t$ donnée par $\frac{T_{i,t}T_{j,t}}{\left(\sum_i T_{i,t}\right)^2}$, et nous pouvons les cumuler dans le temps pour avoir un indice d'interaction entre entités :
}
\[
I_{i,j} = \sum_t \frac{T_{i,t}T_{j,t}}{\left(\sum_i T_{i,t}\right)^2}
\]

\bpar{
The matrix $(I_{i,j})$ allows then to construct a network. A similar index based uniquely on co-occurrence is given by
}{
La matrice $(I_{i,j})$ permet alors de construire un réseau. Un indice similaire se basant uniquement sur la co-occurrence est donné par
}

\[
C_{i,j} = \sum_t \mathbbm{1}_{T_{i,t} > 0} \mathbbm{1}_{T_{j,t} > 0}
\]

\bpar{
We also look at lagged interactions, under the assumption that an entity at time $t$ can trigger the one at time $t+1$, the non-symmetrical index being then
}{
Nous regardons également des interactions retardées, sous l'hypothèse qu'une entité à l'instant $t$ puisse déclencher celle à l'instant $t+1$, l'indice non symétrique étant alors
}
\[
\tilde{I}_{i \rightarrow j} = \sum_t \frac{T_{i,t}T_{j,t+1}}{\sum_i T_{i,t}\sum_j T_{j,t+1}}
\]

\bpar{
and the same index of lagged co-occurrence
}{
et le même indice de co-occurence retardé
}

\[
\tilde{C}_{i\rightarrow j} = \sum_t \mathbbm{1}_{T_{i,t} > 0} \mathbbm{1}_{T_{j,t+1} > 0}
\]

\bpar{
We show the corresponding graphs for the macro-projects in Fig.~\ref{fig:app:reflexivity:projects}. Regarding macro-projects, it appears that the core of the simultaneous co-occurence network is constituted by bibliography, academic life, and methods: these elements are present at closely to any moment and structure the rest of research. Then, the different thematic projects can be led relatively independently, and gravitate at the periphery of the network. With the directed graph of lagged interactions, it is difficult to extract supplementary information, the flows being close to symmetric: either the weekly aggregation is not relevant, either the lag may be different, either there is indeed reciprocity, the latest hypothesis being reasonable given the intrication of projects.
}{
Nous montrons les graphes correspondants pour les macro-projets en Fig.~\ref{fig:app:reflexivity:projects}. Concernant les macro-projets, il apparait que le coeur du réseau de co-occurrence simultanée est constitué par la bibliographie, la vie académique et les méthodes : ces éléments sont quasiment présents à chaque instant et structurent le reste de la recherche. Ensuite, les différents projets thématiques peuvent être menés relativement indépendamment, et gravitent à la périphérie du réseau. Avec le graphe dirigé des interactions retardées, il n'est guère possible de tirer une information supplémentaire, les flux étant quasi-symétriques : soit l'agrégation à la semaine n'est pas pertinente soit le délai peut être différent, soit il y a effectivement réciprocité, cette dernière hypothèse étant crédible vu l'intrication des projets.
}

\begin{figure}
	\includegraphics[width=\linewidth]{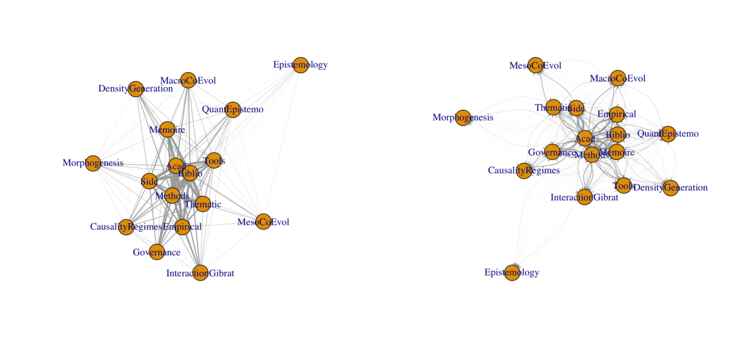}
	\appcaption{\textbf{Interaction graphs between macro-projects.} (\textit{Left}) Graph of simultaneous interaction by co-occurence, given by the adjacency matrix $C_{i,j}$; (\textit{Right}) Graph of lagged interaction, given here by flows $\tilde{I}_{i \rightarrow j}$.\label{fig:app:reflexivity:projects}}{\textbf{Graphes d'interaction entre macro-projets.} (\textit{Gauche}) Graphe d'interaction simultanée par co-occurrence, donné par la matrice d'adjacence $C_{i,j}$ ; (\textit{Droite}) Graphe d'interaction retardée, donné ici par les flux $\tilde{I}_{i \rightarrow j}$. \label{fig:app:reflexivity:projects}}
\end{figure}

\bpar{
The graphs for knowledge domains is given in Fig.~\ref{fig:app:reflexivity:kd}. Beside the fact that data are relatively on the periphery, what was expected given their low importance and their integration within other projects, we do not observe any particular pattern in these graphs: all domains are used at most times. These is also all the reciprocal relations in the directed graph, what suggests possibly a co-evolution between knowledge domains, what we will verify in the following.
}{
Les graphes pour les domaines de connaissance est donné en Fig.~\ref{fig:app:reflexivity:kd}. Outre que les données sont relativement périphériques, ce qui est attendu vu leur faible importance et leur intégration au sein d'autres projets, nous n'observons pas de motifs particuliers dans ces graphes : l'ensemble des domaines est mobilisé à la plupart des instants. Il y a également l'ensemble des relations réciproques dans le graphe dirigé, ce qui suggère éventuellement une co-évolution entre les domaines de connaissance, ce que nous allons vérifier par la suite.
}

\begin{figure}
	\includegraphics[width=\linewidth]{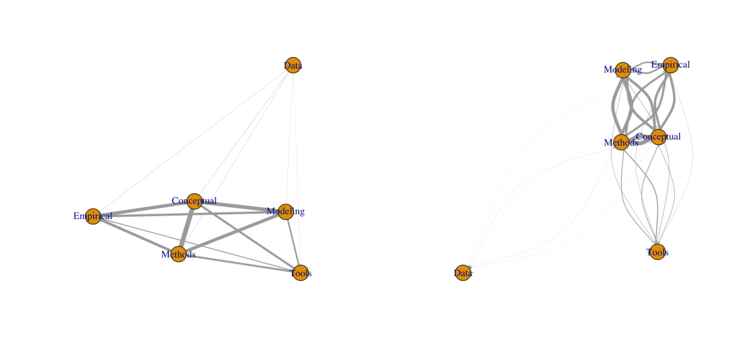}
	\appcaption{\textbf{Interaction graphs between knowledge domains.} (\textit{Left}) Graph of simultaneous interaction by co-occurence; (\textit{Right}) Lagged interaction graph.\label{fig:app:reflexivity:kd}}{\textbf{Graphes d'interaction entre domaines de connaissance.} (\textit{Gauche}) Graphe d'interaction simultanée par co-occurrence ; (\textit{Droite}) Graphe d'interaction retardée.\label{fig:app:reflexivity:kd}}
\end{figure}


\bpar{
We estimated for each couple of knowledge domains $i,j$ (excluding the data domain which only cumulates in total 13h and thus too few variations to estimate a correlation) the lagged correlations between differences $\rhob{\Delta T_{i,t-\tau}}{T_{j,t}}$ for $-4 \leq \tau \leq 4$ (maximal lag of one month). We keep the correlations if $p<0.05$ and select the maximal absolute correlation for each couple of variables if it exists.
}{
Nous estimons pour chaque couple de domaine de connaissance $i,j$ (hors du domaine données qui ne cumule que 13h au total donc trop peu de variations pour estimer une corrélation) les corrélations retardées entre les différences $\rhob{\Delta T_{i,t-\tau}}{T_{j,t}}$ pour $-4 \leq \tau \leq 4$ (délai maximal d'un mois). Nous conservons les corrélations si $p<0.05$ et sélectionnons la corrélation absolue maximale pour chaque couple de variable si elle existe.
}

\begin{figure}
	\includegraphics[width=\linewidth]{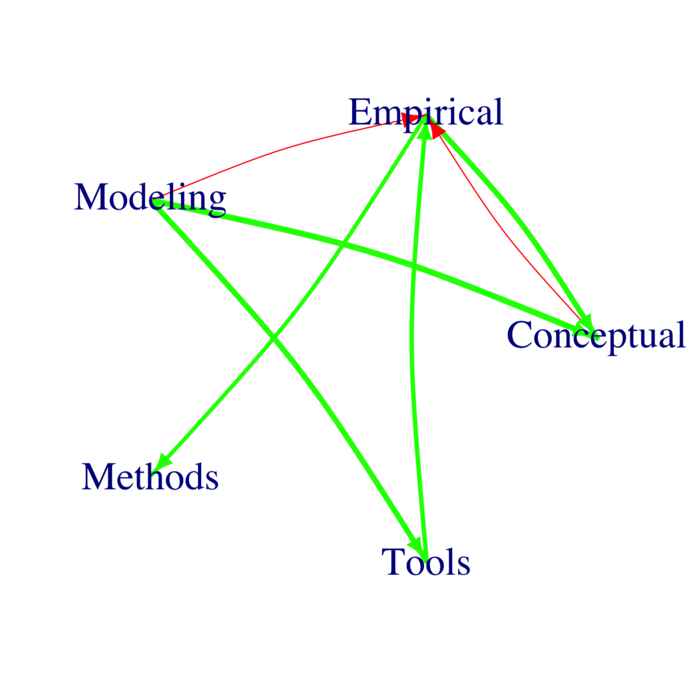}
	\appcaption{\textbf{Graph of lagged correlations between knowledge domains.} The color gives the sign of the correlation (green: positive, red: negative) and the link width its value. Links exist if and only if  $p<0.05$ for a Fisher test.\label{fig:app:reflexivity:laggedcorrs}}{\textbf{Graphe de corrélations retardées entre domaines de connaissances.} La couleur donne le signe de la corrélation (vert : positif, rouge : négatif) et l'épaisseur du lien sa valeur. Les liens existent si et seulement si $p<0.05$ pour un test de Fisher.\label{fig:app:reflexivity:laggedcorrs}}
\end{figure}

\bpar{
The graph of lagged correlations is given in Fig.~\ref{fig:app:reflexivity:laggedcorrs}. There exists a certain number of significant links, and even a circular relation between empirical and conceptual, i.e. a co-evolution in the proper sense between these domains. Modeling and empirical induce works in the conceptual domain, which can be interpreted as an induction of theories. However, the conceptual has a negative influence on the empirical, which could be symptomatic of a too large lack of connection with concrete issues sometimes.
}{
Le graphe des corrélations retardées est donné en Fig.~\ref{fig:app:reflexivity:laggedcorrs}. Il existe un certain nombre de liens significatifs, et même une relation circulaire entre empirique et conceptuel, c'est-à-dire une co-évolution à proprement parler entre ces domaines. La modélisation et l'empirique induisent des travaux dans le domaine conceptuel, ce qui peut s'interpréter comme une induction des théories. Par contre, le conceptuel diminue l'empirique, ce qui pourrait être symptôme d'une trop grande déconnexion avec le concret parfois.
}

\bpar{
Therefore, even if these results are naturally to be taken with caution given the intrinsic biases in the data (difficulty to give a label, reduction within projects, etc.), we suggest an intrication of knowledge domains and a co-evolution for some. This can be put in correspondance with the fundamental hypothesis of the knowledge framework developed in~\ref{sec:knowledgeframework}, which implies that a complex knowledge necessitates a co-evolution of domains. Finally, the application of the own tools of our work to itself suggests an hologramatic dimension~\cite{morin1986methode}, recalling the link between complexity and knowledge production suggested in~\ref{sec:epistemology}.
}{
Ainsi, même si ces résultats sont bien sûr à prendre avec précaution vu les biais intrinsèques aux données (difficulté de donner un label, réduction au sein de projets, etc.), nous suggérons une intrication des domaines de connaissance et une co-évolution pour certains. Cela peut être mis en écho avec l'hypothèse fondamentale du cadre de connaissance développé en~\ref{sec:knowledgeframework}, qui implique qu'une connaissance complexe nécessite co-évolution des domaines. Enfin, l'application des propres outils de notre travail à lui-même suggère une dimension hologrammatique~\cite{morin1986methode}, rappelant le lien entre complexité et production de connaissance suggéré en~\ref{sec:epistemology}. 
}

\stars













\newcommand*\cleartoleftpage{%
  \clearpage
  \ifodd\value{page}\hbox{}\newpage\fi
}

\cleartoleftpage

\backgroundsetup{%
    scale=1.7,
    angle=0,
    contents={%
        \begin{tikzpicture}
            \node[opacity=0.35] {\includegraphics[width=\linewidth]{Figures/cover2}};
        \end{tikzpicture}
    }
}

\markboth{}{}

\pagenumbering{gobble}

\begin{adjustwidth*}{-1.8cm}{-4.0cm}

\small

\bpar{
\noindent\textbf{Characterizing and modeling the co-evolution of transportation networks and territories}
}{
\noindent\textbf{Caractérisation et modélisation de la co-évolution des réseaux de transport et des territoires}
}

\bigskip

\bpar{
\noindent\textbf{Keywords: } Territories; Transportation Networks; Co-evolution; Morphogenesis; Evolutive Urban Theory; Quantitative Epistemology; Systems of Cities; Urban Morphology; Greater Paris; Pearl River Delta
}{
\noindent\textbf{Mots-clés : } Territoires ; Réseaux de Transport ; Co-évolution ; Morphogenèse ; Théorie Évolutive des Villes ; Épistémologie Quantitative ; Systèmes de Villes ; Morphologie Urbaine ; Grand Paris ; Delta de la Rivière des Perles
}

\bigskip

\bpar{
\noindent
The identification of structuring effects of transportation infrastructure on territorial dynamics remains an open research problem. This issue is one of the aspects of approaches on complexity of territorial dynamics, within which territories and networks would be co-evolving. The aim of this thesis is to challenge this view on interactions between networks and territories, both at the conceptual and empirical level, by integrating them in simulation models of territorial systems. The intrinsically multidisciplinary nature of the question requires first to proceed to a quantitative epistemology analysis, that allow us to draw a map of the scientific landscape and to give a description of common features and specificities of models studying the co-evolution between network and territories within each discipline. We propose consequently a definition of co-evolution and an empirical method for its characterization, based on spatio-temporal correlation analysis. Two complementary modeling approaches, that correspond to different scales and ontologies, are then explored. At the macroscopic scale, we build a family of models inheriting from interaction models within system of cities, developed by the Evolutive Urban Theory (Pumain, 1997). Their exploration shows that they effectively capture co-evolutionary dynamics, and their calibration on demographic data for the French system of cities (1830-1999) quantifies the evolution of interaction processes such as the tunnel effect or the role of centrality. At the mesoscopic scale, a morphogenesis model captures the co-evolution of the urban form and of network topology. It is calibrated on corresponding indicators for local form and topology, computed for all Europe. Multiple network evolution processes are shown complementary to reproduce the large variety of observed configurations, at the level of indicators but also interactions between indicators. These results suggest new research directions for urban models integrating co-evolutive dynamics in a multi-scale perspective.
}{
\noindent
L'identification d'effets structurants des infrastructures de transports sur la dynamique des territoires reste un défi scientifique ouvert. Cette question est une des facettes de recherches sur la complexité des dynamiques territoriales, au sein desquelles territoires et réseaux de transport seraient en co-évolution. L'objectif de cette thèse est de mettre à l'épreuve cette vision des interactions entre réseaux et territoires, autant sur le plan conceptuel que sur le plan empirique, en les intégrant au sein de modèles de simulation des systèmes territoriaux. La nature intrinsèquement pluri-disciplinaire de la question nous conduit à mener un travail d'épistémologie quantitative, qui permet de dresser une carte du paysage scientifique et une description des éléments communs et des spécificités des modèles traitant la co-évolution entre réseaux et territoires dans chaque discipline. Nous proposons ensuite une définition de la co-évolution, ainsi qu'une méthode de caractérisation empirique, basée sur une analyse de corrélations spatio-temporelles. Deux pistes complémentaires de modélisation, correspondant à des ontologies et des échelles différentes sont alors explorées. A l'échelle macroscopique, nous construisons une famille de modèles dans la lignée des modèles d'interaction au sein des systèmes de villes développés par la Théorie Evolutive des Villes (Pumain, 1997). Leur exploration montre qu'ils capturent effectivement des dynamiques de co-évolution, et leur calibration sur des données démographiques pour le système de villes français (1830-1999) quantifie l'évolution des processus d'interaction comme l'effet tunnel ou le rôle de la centralité. A l'échelle mésoscopique, un modèle de morphogenèse capture la co-évolution de la forme urbaine et de la topologie du réseau. Il est calibré sur les indicateurs correspondants pour la forme et la topologie locales calculés pour l'ensemble de l'Europe. De multiples processus d'évolution du réseau s'avèrent être complémentaires pour reproduire la grande variété des configurations observées, au niveau des indicateurs ainsi que des interactions entre indicateurs. Ces résultats suggèrent de nouvelles pistes d'exploration des modèles urbains intégrant les dynamiques co-évolutives dans une perspective multi-échelles.
}

\vspace{1.3cm}

\bpar{
\noindent\textbf{Caractérisation et modélisation de la co-évolution des réseaux de transport et des territoires}
}{
\noindent\textbf{Characterizing and modeling the co-evolution of transportation networks and territories}
}

\bigskip

\bpar{
\noindent\textbf{Mots-clés : } Territoires ; Réseaux de Transport ; Co-évolution ; Morphogenèse ; Théorie Évolutive des Villes ; Épistémologie Quantitative ; Systèmes de Villes ; Morphologie Urbaine ; Grand Paris ; Delta de la Rivière des Perles
}{
\noindent\textbf{Keywords: } Territories; Transportation Networks; Co-evolution; Morphogenesis; Evolutive Urban Theory; Quantitative Epistemology; Systems of Cities; Urban Morphology; Greater Paris; Pearl River Delta
}

\bigskip

\bpar{
\noindent
L'identification d'effets structurants des infrastructures de transports sur la dynamique des territoires reste un défi scientifique ouvert. Cette question est une des facettes de recherches sur la complexité des dynamiques territoriales, au sein desquelles territoires et réseaux de transport seraient en co-évolution. L'objectif de cette thèse est de mettre à l'épreuve cette vision des interactions entre réseaux et territoires, autant sur le plan conceptuel que sur le plan empirique, en les intégrant au sein de modèles de simulation des systèmes territoriaux. La nature intrinsèquement pluri-disciplinaire de la question nous conduit à mener un travail d'épistémologie quantitative, qui permet de dresser une carte du paysage scientifique et une description des éléments communs et des spécificités des modèles traitant la co-évolution entre réseaux et territoires dans chaque discipline. Nous proposons ensuite une définition de la co-évolution, ainsi qu'une méthode de caractérisation empirique, basée sur une analyse de corrélations spatio-temporelles. Deux pistes complémentaires de modélisation, correspondant à des ontologies et des échelles différentes sont alors explorées. A l'échelle macroscopique, nous construisons une famille de modèles dans la lignée des modèles d'interaction au sein des systèmes de villes développés par la Théorie Evolutive des Villes (Pumain, 1997). Leur exploration montre qu'ils capturent effectivement des dynamiques de co-évolution, et leur calibration sur des données démographiques pour le système de villes français (1830-1999) quantifie l'évolution des processus d'interaction comme l'effet tunnel ou le rôle de la centralité. A l'échelle mésoscopique, un modèle de morphogenèse capture la co-évolution de la forme urbaine et de la topologie du réseau. Il est calibré sur les indicateurs correspondants pour la forme et la topologie locales calculés pour l'ensemble de l'Europe. De multiples processus d'évolution du réseau s'avèrent être complémentaires pour reproduire la grande variété des configurations observées, au niveau des indicateurs ainsi que des interactions entre indicateurs. Ces résultats suggèrent de nouvelles pistes d'exploration des modèles urbains intégrant les dynamiques co-évolutives dans une perspective multi-échelles.
}{
\noindent
The identification of structuring effects of transportation infrastructure on territorial dynamics remains an open research problem. This issue is one of the aspects of approaches on complexity of territorial dynamics, within which territories and networks would be co-evolving. The aim of this thesis is to challenge this view on interactions between networks and territories, both at the conceptual and empirical level, by integrating them in simulation models of territorial systems. The intrinsically multidisciplinary nature of the question requires first to proceed to a quantitative epistemology analysis, that allow us to draw a map of the scientific landscape and to give a description of common features and specificities of models studying the co-evolution between network and territories within each discipline. We propose consequently a definition of co-evolution and an empirical method for its characterization, based on spatio-temporal correlation analysis. Two complementary modeling approaches, that correspond to different scales and ontologies, are then explored. At the macroscopic scale, we build a family of models inheriting from interaction models within system of cities, developed by the Evolutive Urban Theory (Pumain, 1997). Their exploration shows that they effectively capture co-evolutionary dynamics, and their calibration on demographic data for the French system of cities (1830-1999) quantifies the evolution of interaction processes such as the tunnel effect or the role of centrality. At the mesoscopic scale, a morphogenesis model captures the co-evolution of the urban form and of network topology. It is calibrated on corresponding indicators for local form and topology, computed for all Europe. Multiple network evolution processes are shown complementary to reproduce the large variety of observed configurations, at the level of indicators but also interactions between indicators. These results suggest new research directions for urban models integrating co-evolutive dynamics in a multi-scale perspective.
}

\end{adjustwidth*}




\end{document}